%% file: report.tex
\documentclass[11pt,twoside,a4paper]{cernrep}
\usepackage{lineno}
\usepackage{rep_common}
\usepackage{rotating}
\usepackage{braket}
\newboolean{uprightparticles}
\setboolean{uprightparticles}{false} 
\usepackage{upgreek}
\usepackage{xspace}
\usepackage{wrapfig}
\usepackage{xcolor}
\usepackage{yfonts}
\usepackage{placeins}
\usepackage[colorlinks, urlcolor=blue, linkcolor=blue, citecolor=blue, plainpages=false]{hyperref}
\def\bibfiles{\main/bib/chapter,\main/bib/misc,\main/bib/LHCb-PAPER,\main/bib/LHCb-CONF,\main/section1/bib/section,\main/section2/bib/section,\main/section3/bib/section,\main/section4/bib/section,\main/section5/bib/section,\main/section6/bib/section,\main/section7/bib/section,\main/section8/bib/section,\main/section9/bib/section,\main/section10/bib/section,\main/section11/bib/section,\main/section12/bib/section,\main/section13/bib/section}
\providecommand{\biblio}{\nocite{article-minimal}\bibliographystyle{report}\clearpage\bibliography{\bibfiles}}  
\input{lhcb-symbols-def}

\DeclareMathOperator{\br}{Br}
\let\Re\undefined
\let\Im\undefined
\DeclareMathOperator{\Re}{Re}
\DeclareMathOperator{\Im}{Im}

\newcommand{\mcO}{\mathcal{O}}

\newcommand{\cL}{\mathcal{L}}
\newcommand{\cR}{\mathcal{R}}

\newcommand{\cc}{\ensuremath{\cquark\bar{\cquark}}\xspace}

\newcommand{\msd}{\ensuremath{m_\text{SD}}\xspace}
\newcommand{\hgg}{\ensuremath{h\to\PGg\PGg}\xspace}
\newcommand{\zz}{\ensuremath{\PZ\PZ}\xspace}
\newcommand{\hzz}{\ensuremath{h\to\zz}\xspace}
\newcommand{\bb}{\ensuremath{b\bar b}\xspace}
\newcommand{\kappat}{\ensuremath{\kappa_{t}}\xspace}

\newcommand{\hbb}{\ensuremath{h\to\bb}\xspace}
\newcommand{\kappab}{\ensuremath{\kappa_{b}}\xspace}
\newcommand{\kappac}{\ensuremath{\kappa_{c}}\xspace}
\newcommand{\fbinv}{\mbox{\ensuremath{\,\text{fb}^\text{$-$1}}}\xspace}

\newcommand{\ptmiss}{\ensuremath{\textbf{p}_{\mathrm{T}}^{\mathrm{miss}}}\xspace}
\newcommand{\delphes}[1]{\textsc{Delphes#1}}
\newcommand{\met}{E_{\mathrm{T}}^{\mathrm{miss}}}
\newcommand{\abinv}{\mbox{\ensuremath{\,\text{ab}^\text{$-$1}}}\xspace}
\newcommand{\intLumi}{3\abinv}
\newcommand{\GeV}{\ensuremath{\rm~GeV}\xspace}
\newcommand{\TeV}{\ensuremath{\rm~TeV}\xspace}
\newcommand*{\tauh}{\ensuremath{\tau_h}}
\newcommand{\kt}{\ensuremath{k_{\mathrm{T}}}\xspace}

\definecolor{Blu}{rgb}{0.,0.,1.}
\newcommand{\Blu}[1]{{{#1}}}
\newcommand{\vtwo}[1]{#1}

\begin{document}
\newcommand{\main}{.}
\def\biblio{}
\input{titlepage}

\setcounter{tocdepth}{2}
\tableofcontents
\newpage

\subfile{\main/section1/section}
\clearpage
\newpage

\subfile{\main/section2/section}

\clearpage
\newpage

\subfile{\main/section3/section}

\clearpage
\newpage

\subfile{\main/section4/section}

\clearpage
\newpage

\subfile{\main/section5/section}

\clearpage
\newpage

\subfile{\main/section6/section}

\clearpage
\newpage

\subfile{\main/section7/section}

\clearpage
\newpage

\subfile{\main/section8/section}
\clearpage
\newpage

\subfile{\main/section9/section}

\clearpage
\newpage

\subfile{\main/section10/section}
\clearpage
\newpage

\subfile{\main/section11/section}

\clearpage
\newpage

\subfile{\main/section12/section}

\newpage
\subfile{\main/acknowledgments}

\clearpage
\appendix

\subfile{\main/section13/section}

\clearpage


\bibliographystyle{report}
\bibliography{\bibfiles}

\end{document}

%% file: lhcb-symbols-def.tex
\usepackage{xspace} 
\usepackage{upgreek}

\def\lhcb {\mbox{LHCb}\xspace}

\def\babar  {\mbox{BaBar}\xspace}
\def\belle  {\mbox{Belle}\xspace}

\def\lhc    {\mbox{LHC}\xspace}

\def\upgradeone {\mbox{Upgrade~I}\xspace}
\def\upgradetwo {\mbox{Upgrade~II}\xspace}
\def\belletwo   {\mbox{Belle~II}\xspace}
\def\bfactories {\mbox{\B-Factories}\xspace}

\def\MagUp {\mbox{\em Mag\kern -0.05em Up}\xspace}

\ifthenelse{\boolean{uprightparticles}}%
{

 \def\Peta        {\ensuremath{\upeta}\xspace}

 \def\Pmu         {\ensuremath{\upmu}\xspace}                 
 \def\Pnu         {\ensuremath{\upnu}\xspace}                 
                  
 \def\Ppi         {\ensuremath{\uppi}\xspace}                 
                  
 \def\Prho        {\ensuremath{\uprho}\xspace}                 
                  
 \def\Ptau        {\ensuremath{\uptau}\xspace}                 
                  
 \def\Pphi        {\ensuremath{\upphi}\xspace}                 
                  
 \def\Pchi        {\ensuremath{\upchi}\xspace}                 
 \def\Ppsi        {\ensuremath{\uppsi}\xspace}                 
 \def\Pomega      {\ensuremath{\upomega}\xspace}                 

 \def\PDelta      {\ensuremath{\Delta}\xspace}                 
 \def\PXi      {\ensuremath{\Xi}\xspace}                 
 \def\PLambda      {\ensuremath{\Lambda}\xspace}                 
 \def\PSigma      {\ensuremath{\Sigma}\xspace}                 
 \def\POmega      {\ensuremath{\Omega}\xspace}                 
 \def\PUpsilon      {\ensuremath{\Upsilon}\xspace}                 
 
 \def\PB      {\ensuremath{\mathrm{B}}\xspace}                 
                  
 \def\PD      {\ensuremath{\mathrm{D}}\xspace}

 \def\PH      {\ensuremath{\mathrm{H}}\xspace}                 
                  
 \def\PJ      {\ensuremath{\mathrm{J}}\xspace}                 
 \def\PK      {\ensuremath{\mathrm{K}}\xspace}

 \def\PN      {\ensuremath{\mathrm{N}}\xspace}                 
                  
 \def\PP      {\ensuremath{\mathrm{P}}\xspace}                 
 \def\PQ      {\ensuremath{\mathrm{Q}}\xspace}

 \def\PW      {\ensuremath{\mathrm{W}}\xspace}                 
 \def\PX      {\ensuremath{\mathrm{X}}\xspace}                 
                  
 \def\PZ      {\ensuremath{\mathrm{Z}}\xspace}                 
 \def\Pa      {\ensuremath{\mathrm{a}}\xspace}                 
 \def\Pb      {\ensuremath{\mathrm{b}}\xspace}                 
 \def\Pc      {\ensuremath{\mathrm{c}}\xspace}                 
 \def\Pd      {\ensuremath{\mathrm{d}}\xspace}                 
 \def\Pe      {\ensuremath{\mathrm{e}}\xspace}                 
 \def\Pf      {\ensuremath{\mathrm{f}}\xspace}                 
 \def\Pg      {\ensuremath{\mathrm{g}}\xspace}                 
 \def\Ph      {\ensuremath{\mathrm{h}}\xspace}                 
 \def\Pi      {\ensuremath{\mathrm{i}}\xspace}

 \def\Pl      {\ensuremath{\mathrm{l}}\xspace}                 
                  
 \def\Pn      {\ensuremath{\mathrm{n}}\xspace}                 
                  
 \def\Pp      {\ensuremath{\mathrm{p}}\xspace}                 
 \def\Pq      {\ensuremath{\mathrm{q}}\xspace}                 
                  
 \def\Ps      {\ensuremath{\mathrm{s}}\xspace}                 
                  
 \def\Pu      {\ensuremath{\mathrm{u}}\xspace}

}
{

 \def\Peta        {\ensuremath{\eta}\xspace}

 \def\Pmu         {\ensuremath{\mu}\xspace}                 
 \def\Pnu         {\ensuremath{\nu}\xspace}                 
                  
 \def\Ppi         {\ensuremath{\pi}\xspace}                 
                  
 \def\Prho        {\ensuremath{\rho}\xspace}                 
                  
 \def\Ptau        {\ensuremath{\tau}\xspace}                 
                  
 \def\Pphi        {\ensuremath{\phi}\xspace}                 
                  
 \def\Pchi        {\ensuremath{\chi}\xspace}                 
 \def\Ppsi        {\ensuremath{\psi}\xspace}                 
 \def\Pomega      {\ensuremath{\omega}\xspace}                 
 \mathchardef\PDelta="7101
 \mathchardef\PXi="7104
 \mathchardef\PLambda="7103
 \mathchardef\PSigma="7106
 \mathchardef\POmega="710A
 \mathchardef\PUpsilon="7107
                  
 \def\PB      {\ensuremath{B}\xspace}                 
                  
 \def\PD      {\ensuremath{D}\xspace}

 \def\PH      {\ensuremath{H}\xspace}                 
                  
 \def\PJ      {\ensuremath{J}\xspace}                 
 \def\PK      {\ensuremath{K}\xspace}

 \def\PN      {\ensuremath{N}\xspace}                 
                  
 \def\PP      {\ensuremath{P}\xspace}                 
 \def\PQ      {\ensuremath{Q}\xspace}

 \def\PW      {\ensuremath{W}\xspace}                 
 \def\PX      {\ensuremath{X}\xspace}                 
                  
 \def\PZ      {\ensuremath{Z}\xspace}                 
 \def\Pa      {\ensuremath{a}\xspace}                 
 \def\Pb      {\ensuremath{b}\xspace}                 
 \def\Pc      {\ensuremath{c}\xspace}                 
 \def\Pd      {\ensuremath{d}\xspace}                 
 \def\Pe      {\ensuremath{e}\xspace}                 
 \def\Pf      {\ensuremath{f}\xspace}                 
 \def\Pg      {\ensuremath{g}\xspace}                 
 \def\Ph      {\ensuremath{h}\xspace}                 
 \def\Pi      {\ensuremath{i}\xspace}

 \def\Pl      {\ensuremath{l}\xspace}                 
                  
 \def\Pn      {\ensuremath{n}\xspace}                 
                  
 \def\Pp      {\ensuremath{p}\xspace}                 
 \def\Pq      {\ensuremath{q}\xspace}                 
                  
 \def\Ps      {\ensuremath{s}\xspace}                 
                  
 \def\Pu      {\ensuremath{u}\xspace}

}

\makeatletter
\ifcase \@ptsize \relax
  \newcommand{\miniscule}{\@setfontsize\miniscule{4}{5}}
\or
  \newcommand{\miniscule}{\@setfontsize\miniscule{5}{6}}
\or
  \newcommand{\miniscule}{\@setfontsize\miniscule{5}{6}}
\fi
\makeatother

\DeclareRobustCommand{\optbar}[1]{\shortstack{{\miniscule (\rule[.5ex]{1.25em}{.18mm})}
  \\ [-.7ex] $#1$}}

\def\en         {{\ensuremath{\Pe^-}}\xspace}   
\def\ep         {{\ensuremath{\Pe^+}}\xspace}

\def\epem       {{\ensuremath{\Pe^+\Pe^-}}\xspace}

\def\muon       {{\ensuremath{\Pmu}}\xspace}
\def\mup        {{\ensuremath{\Pmu^+}}\xspace}
\def\mun        {{\ensuremath{\Pmu^-}}\xspace} 
\def\mupm       {{\ensuremath{\Pmu^\pm}}\xspace} 
\def\mump       {{\ensuremath{\Pmu^\mp}}\xspace} 
\def\mumu       {{\ensuremath{\Pmu^+\Pmu^-}}\xspace}

\def\taum       {{\ensuremath{\Ptau^-}}\xspace}
\def\taupm      {{\ensuremath{\Ptau^\pm}}\xspace}
\def\taump      {{\ensuremath{\Ptau^\mp}}\xspace}

\def\ellell     {\ensuremath{\ell^+ \ell^-}\xspace}

\def\neu        {{\ensuremath{\Pnu}}\xspace}
\def\neub       {{\ensuremath{\overline{\Pnu}}}\xspace}
\def\neum       {{\ensuremath{\neu_\mu}}\xspace}
\def\neumb      {{\ensuremath{\neub_\mu}}\xspace}

\def\g      {{\ensuremath{\Pgamma}}\xspace}

\def\quark     {{\ensuremath{\Pq}}\xspace}
\def\quarkbar  {{\ensuremath{\overline \quark}}\xspace}

\def\uquark    {{\ensuremath{\Pu}}\xspace}
\def\uquarkbar {{\ensuremath{\overline \uquark}}\xspace}

\def\dquark    {{\ensuremath{\Pd}}\xspace}
\def\dquarkbar {{\ensuremath{\overline \dquark}}\xspace}

\def\squark    {{\ensuremath{\Ps}}\xspace}
\def\squarkbar {{\ensuremath{\overline \squark}}\xspace}
\def\ssbar     {{\ensuremath{\squark\squarkbar}}\xspace}
\def\cquark    {{\ensuremath{\Pc}}\xspace}
\def\cquarkbar {{\ensuremath{\overline \cquark}}\xspace}
\def\ccbar     {{\ensuremath{\cquark\cquarkbar}}\xspace}
\def\bquark    {{\ensuremath{\Pb}}\xspace}
\def\bquarkbar {{\ensuremath{\overline \bquark}}\xspace}
\def\bbbar     {{\ensuremath{\bquark\bquarkbar}}\xspace}

\def\hadron {{\ensuremath{\Ph}}\xspace}
\def\pion   {{\ensuremath{\Ppi}}\xspace}
\def\piz    {{\ensuremath{\pion^0}}\xspace}

\def\pip    {{\ensuremath{\pion^+}}\xspace}
\def\pim    {{\ensuremath{\pion^-}}\xspace}
\def\pipm   {{\ensuremath{\pion^\pm}}\xspace}
\def\pimp   {{\ensuremath{\pion^\mp}}\xspace}

\def\rhomeson {{\ensuremath{\Prho}}\xspace}
\def\rhoz     {{\ensuremath{\rhomeson^0}}\xspace}

\def\rhopm    {{\ensuremath{\rhomeson^\pm}}\xspace}

\def\kaon    {{\ensuremath{\PK}}\xspace}
  \def\Kbar    {{\kern 0.2em\overline{\kern -0.2em \PK}{}}\xspace}

\def\KorKbar    {\kern 0.18em\optbar{\kern -0.18em K}{}\xspace}
\def\Kz      {{\ensuremath{\kaon^0}}\xspace}
\def\Kzb     {{\ensuremath{\Kbar{}^0}}\xspace}
\def\Kp      {{\ensuremath{\kaon^+}}\xspace}
\def\Km      {{\ensuremath{\kaon^-}}\xspace}
\def\Kpm     {{\ensuremath{\kaon^\pm}}\xspace}

\def\KS      {{\ensuremath{\kaon^0_{\mathrm{ \scriptscriptstyle S}}}}\xspace}

\def\Kstarz  {{\ensuremath{\kaon^{*0}}}\xspace}
\def\Kstarzb {{\ensuremath{\Kbar{}^{*0}}}\xspace}
\def\Kstar   {{\ensuremath{\kaon^*}}\xspace}
\def\Kstarb  {{\ensuremath{\Kbar{}^*}}\xspace}
\def\Kstarp  {{\ensuremath{\kaon^{*+}}}\xspace}

\def\Kstarpm {{\ensuremath{\kaon^{*\pm}}}\xspace}

\newcommand{\etaz}{\ensuremath{\Peta}\xspace}
\newcommand{\etapr}{\ensuremath{\Peta^{\prime}}\xspace}

\newcommand{\omegaz}{\ensuremath{\Pomega}\xspace}

  \def\Dbar    {{\kern 0.2em\overline{\kern -0.2em \PD}{}}\xspace}
\def\D       {{\ensuremath{\PD}}\xspace}

\def\DorDbar    {\kern 0.18em\optbar{\kern -0.18em D}{}\xspace}
\def\Dz      {{\ensuremath{\D^0}}\xspace}
\def\Dzb     {{\ensuremath{\Dbar{}^0}}\xspace}
\def\Dp      {{\ensuremath{\D^+}}\xspace}
\def\Dm      {{\ensuremath{\D^-}}\xspace}

\def\Dstar   {{\ensuremath{\D^*}}\xspace}

\def\Dstarzb {{\ensuremath{\Dbar{}^{*0}}}\xspace}

\def\Dstarp  {{\ensuremath{\D^{*+}}}\xspace}
\def\Dstarm  {{\ensuremath{\D^{*-}}}\xspace}
\def\Dstarpm {{\ensuremath{\D^{*\pm}}}\xspace}

\def\Ds      {{\ensuremath{\D^+_\squark}}\xspace}
\def\Dsp     {{\ensuremath{\D^+_\squark}}\xspace}
\def\Dsm     {{\ensuremath{\D^-_\squark}}\xspace}

\def\Dssp    {{\ensuremath{\D^{*+}_\squark}}\xspace}

\def\B       {{\ensuremath{\PB}}\xspace}
\def\Bbar    {{\ensuremath{\kern 0.18em\overline{\kern -0.18em \PB}{}}}\xspace}
\def\Bb      {{\ensuremath{\Bbar}}\xspace}
\def\BorBbar    {\kern 0.18em\optbar{\kern -0.18em B}{}\xspace}
\def\Bz      {{\ensuremath{\B^0}}\xspace}
\def\Bzb     {{\ensuremath{\Bbar{}^0}}\xspace}
\def\Bu      {{\ensuremath{\B^+}}\xspace}
\def\Bub     {{\ensuremath{\B^-}}\xspace}
\def\Bp      {{\ensuremath{\Bu}}\xspace}
\def\Bm      {{\ensuremath{\Bub}}\xspace}
\def\Bpm     {{\ensuremath{\B^\pm}}\xspace}

\def\Bd      {{\ensuremath{\B^0}}\xspace}
\def\Bs      {{\ensuremath{\B^0_\squark}}\xspace}
\def\Bsb     {{\ensuremath{\Bbar{}^0_\squark}}\xspace}

\def\Bdb     {{\ensuremath{\Bbar{}^0}}\xspace}
\def\Bc      {{\ensuremath{\B_\cquark^+}}\xspace}
\def\Bcp     {{\ensuremath{\B_\cquark^+}}\xspace}

\def\Bcpm    {{\ensuremath{\B_\cquark^\pm}}\xspace}
\def\Bds     {{\ensuremath{\B_{(\squark)}^0}}\xspace}
\def\Bdsb    {{\ensuremath{\Bbar{}_{(\squark)}^0}}\xspace}

\def\jpsi     {{\ensuremath{{\PJ\mskip -3mu/\mskip -2mu\Ppsi\mskip 2mu}}}\xspace}
\def\psitwos  {{\ensuremath{\Ppsi{(2S)}}}\xspace}

\def\chiczero {{\ensuremath{\Pchi_{\cquark 0}}}\xspace}
\def\chicone  {{\ensuremath{\Pchi_{\cquark 1}}}\xspace}

  \def\Y#1S{\ensuremath{\PUpsilon{(#1S)}}\xspace}
\def\OneS  {{\Y1S}}

\def\proton      {{\ensuremath{\Pp}}\xspace}
\def\antiproton  {{\ensuremath{\overline \proton}}\xspace}
\def\neutron     {{\ensuremath{\Pn}}\xspace}

\def\Xires       {{\ensuremath{\PXi}}\xspace}
\def\Xiresbar    {{\ensuremath{\overline \Xires}}\xspace}
\def\Lz          {{\ensuremath{\PLambda}}\xspace}
\def\Lbar        {{\ensuremath{\kern 0.1em\overline{\kern -0.1em\PLambda}}}\xspace}
\def\LorLbar    {\kern 0.18em\optbar{\kern -0.18em \PLambda}{}\xspace}
\def\Lambdares   {{\ensuremath{\PLambda}}\xspace}

\def\Sigmares    {{\ensuremath{\PSigma}}\xspace}

\def\Omegares    {{\ensuremath{\POmega}}\xspace}

\def\Lb      {{\ensuremath{\Lz^0_\bquark}}\xspace}

\def\Lc      {{\ensuremath{\Lz^+_\cquark}}\xspace}
\def\Lcbar   {{\ensuremath{\Lbar{}^-_\cquark}}\xspace}
\def\Xib     {{\ensuremath{\Xires_\bquark}}\xspace}
\def\Xibz    {{\ensuremath{\Xires^0_\bquark}}\xspace}
\def\Xibm    {{\ensuremath{\Xires^-_\bquark}}\xspace}

\def\Xicz    {{\ensuremath{\Xires^0_\cquark}}\xspace}
\def\Xicp    {{\ensuremath{\Xires^+_\cquark}}\xspace}

\def\Omegac    {{\ensuremath{\Omegares^0_\cquark}}\xspace}

\def\Omegab    {{\ensuremath{\Omegares^-_\bquark}}\xspace}

\def\Xicc      {{\ensuremath{\Xires_{\cquark\cquark}}}\xspace}

\def\Xiccp     {{\ensuremath{\Xires^+_{\cquark\cquark}}}\xspace}
\def\Xiccpp    {{\ensuremath{\Xires^{++}_{\cquark\cquark}}}\xspace}

\def\Omegacc     {{\ensuremath{\Omegares^+_{\cquark\cquark}}}\xspace}

\def\BF         {{\ensuremath{\mathcal{B}}}\xspace}

\def\BR         {\BF}
\newcommand{\decay}[2]{\mbox{\ensuremath{#1\!\to #2}}\xspace}         

\def\to                 {\ensuremath{\rightarrow}\xspace}

\def\grpsuthree {{\ensuremath{\mathrm{SU}(3)}}\xspace}

\def\order   {{\ensuremath{\mathcal{O}}}\xspace}

\def\qsq       {{\ensuremath{q^2}}\xspace}

\def\CP                {{\ensuremath{C\!P}}\xspace}

\def\rhobar {{\ensuremath{\overline \rho}}\xspace}
\def\etabar {{\ensuremath{\overline \eta}}\xspace}

\def\Vus  {{\ensuremath{V_{\uquark\squark}}}\xspace}

\def\Vub  {{\ensuremath{V_{\uquark\bquark}}}\xspace}
\def\Vcb  {{\ensuremath{V_{\cquark\bquark}}}\xspace}

\def\KorKbar    {\kern 0.18em\optbar{\kern -0.18em K}{}\xspace}

\newcommand{\dm}{{\ensuremath{\Delta m}}\xspace}

\newcommand{\DG}{{\ensuremath{\Delta\Gamma}}\xspace}
\newcommand{\DGs}{{\ensuremath{\Delta\Gamma_{\squark}}}\xspace}

\newcommand{\Gs}{{\ensuremath{\Gamma_{\squark}}}\xspace}

\newcommand{\phid}{{\ensuremath{\phi_{\dquark}}}\xspace}

\newcommand{\phis}{{\ensuremath{\phi_{\squark}}}\xspace}

\def\BdTopipi     {\decay{\Bd}{\pip\pim}}

\def\BsToKK       {\decay{\Bs}{\Kp\Km}}

\def\Cpipi        {\ensuremath{C_{\pip\pim}}\xspace}
\def\Spipi        {\ensuremath{S_{\pip\pim}}\xspace}
\def\CKK          {\ensuremath{C_{\Kp\Km}}\xspace}
\def\SKK          {\ensuremath{S_{\Kp\Km}}\xspace}
\def\ADGKK        {\ensuremath{A^{\DG}_{\Kp\Km}}\xspace}

\def\FL       {\ensuremath{F_{\mathrm{L}}}\xspace}
\def\AT#1     {\ensuremath{A_{\mathrm{T}}^{#1}}\xspace}           

\def\Bsmm     {\decay{\Bs}{\mup\mun}}
\def\Bdmm     {\decay{\Bd}{\mup\mun}}

\def\C#1      {\ensuremath{\mathcal{C}_{#1}}\xspace}                       
\def\Cp#1     {\ensuremath{\mathcal{C}_{#1}^{'}}\xspace}                    
\def\Ceff#1   {\ensuremath{\mathcal{C}_{#1}^{\mathrm{(eff)}}}\xspace}        
\def\Cpeff#1  {\ensuremath{\mathcal{C}_{#1}^{'\mathrm{(eff)}}}\xspace}       
\def\Ope#1    {\ensuremath{\mathcal{O}_{#1}}\xspace}                       
\def\Opep#1   {\ensuremath{\mathcal{O}_{#1}^{'}}\xspace}                    

\def\agamma     {\ensuremath{A_{\Gamma}}\xspace}

\newcommand{\tev}{\ensuremath{\mathrm{\,Te\kern -0.1em V}}\xspace}
\newcommand{\gev}{\ensuremath{\mathrm{\,Ge\kern -0.1em V}}\xspace}
\newcommand{\mev}{\ensuremath{\mathrm{\,Me\kern -0.1em V}}\xspace}
\newcommand{\kev}{\ensuremath{\mathrm{\,ke\kern -0.1em V}}\xspace}
\newcommand{\ev}{\ensuremath{\mathrm{\,e\kern -0.1em V}}\xspace}
\newcommand{\gevc}{\ensuremath{{\mathrm{\,Ge\kern -0.1em V\!/}c}}\xspace}
\newcommand{\mevc}{\ensuremath{{\mathrm{\,Me\kern -0.1em V\!/}c}}\xspace}
\newcommand{\gevcc}{\ensuremath{{\mathrm{\,Ge\kern -0.1em V\!/}c^2}}\xspace}
\newcommand{\gevgevcccc}{\ensuremath{{\mathrm{\,Ge\kern -0.1em V^2\!/}c^4}}\xspace}
\newcommand{\mevcc}{\ensuremath{{\mathrm{\,Me\kern -0.1em V\!/}c^2}}\xspace}

\def\nb {\ensuremath{\mathrm{ \,nb}}\xspace}

\def\pb {\ensuremath{\mathrm{ \,pb}}\xspace}
\def\invpb {\ensuremath{\mbox{\,pb}^{-1}}\xspace}
\def\fb   {\ensuremath{\mbox{\,fb}}\xspace}
\def\invfb   {\ensuremath{\mbox{\,fb}^{-1}}\xspace}
\def\ab   {\ensuremath{\mbox{\,ab}}\xspace}
\def\invab   {\ensuremath{\mbox{\,ab}^{-1}}\xspace}

\def\ps   {\ensuremath{{\mathrm{ \,ps}}}\xspace}
\def\fs   {\ensuremath{\mathrm{ \,fs}}\xspace}

\newcommand{\stat}{\ensuremath{\mathrm{\,(stat)}}\xspace}
\newcommand{\syst}{\ensuremath{\mathrm{\,(syst)}}\xspace}

\def\order{{\ensuremath{\mathcal{O}}}\xspace}
\newcommand{\chisq}{\ensuremath{\chi^2}\xspace}

\def\deriv {\ensuremath{\mathrm{d}}}

\def\gsim{{~\raise.15em\hbox{$>$}\kern-.85em
          \lower.35em\hbox{$\sim$}~}\xspace}
\def\lsim{{~\raise.15em\hbox{$<$}\kern-.85em
          \lower.35em\hbox{$\sim$}~}\xspace}

\newcommand{\Real}{\ensuremath{\mathcal{R}e}\xspace}
\newcommand{\Imag}{\ensuremath{\mathcal{I}m}\xspace}

\def\sqs   {\ensuremath{\protect\sqrt{s}}\xspace}

\def\pt         {\ensuremath{p_{\mathrm{ T}}}\xspace}

\def\degrees{\ensuremath{^{\circ}}\xspace}

\def\mrad{\ensuremath{\mathrm{ \,mrad}}\xspace}
\def\rad{\ensuremath{\mathrm{ \,rad}}\xspace}

\def\tell1  {TELL1\xspace}
\def\ukl1   {UKL1\xspace}

\def\Xibz    {\ensuremath{\Xires^0_\bquark}\xspace}
\def\Xibzbar {\ensuremath{\Xiresbar^0_\bquark}\xspace}
\def\XibPrimeMinus  {{\ensuremath{\Xires^{\prime -}_\bquark}}\xspace}
\def\XibStarMinus  {{\ensuremath{\Xires^{*-}_\bquark}}\xspace}

\def\Xiccpp  {\ensuremath{\PXi_{\cquark\cquark}^{++}}\xspace}

\newcommand{\phisPP}{\ensuremath{{\phi_{s}^{\squark\squarkbar\squark}}}\xspace}

\newcommand{\figref}[1]{Fig.~\ref{#1}}
\newcommand{\tabref}[1]{Table~\ref{#1}}

%% file: titlepage.tex
\title{{\normalfont\bfseries\boldmath\huge
\begin{center}
Opportunities in Flavour Physics\\
at the HL-LHC and HE-LHC\\
\begin{normalsize} 
\href{http://lpcc.web.cern.ch/hlhe-lhc-physics-workshop}{Report from Working Group 4 on the Physics of the HL-LHC, and Perspectives at the HE-LHC} 
\end{normalsize}
\end{center}\vspace*{0.2cm}
}}

\input{\main/authors}

\begin{titlepage}

\vspace*{-1.8cm}

\noindent
\begin{tabular*}{\linewidth}{lc@{\extracolsep{\fill}}r@{\extracolsep{0pt}}}
\vspace*{-1.2cm}\mbox{\!\!\!\includegraphics[width=.14\textwidth]{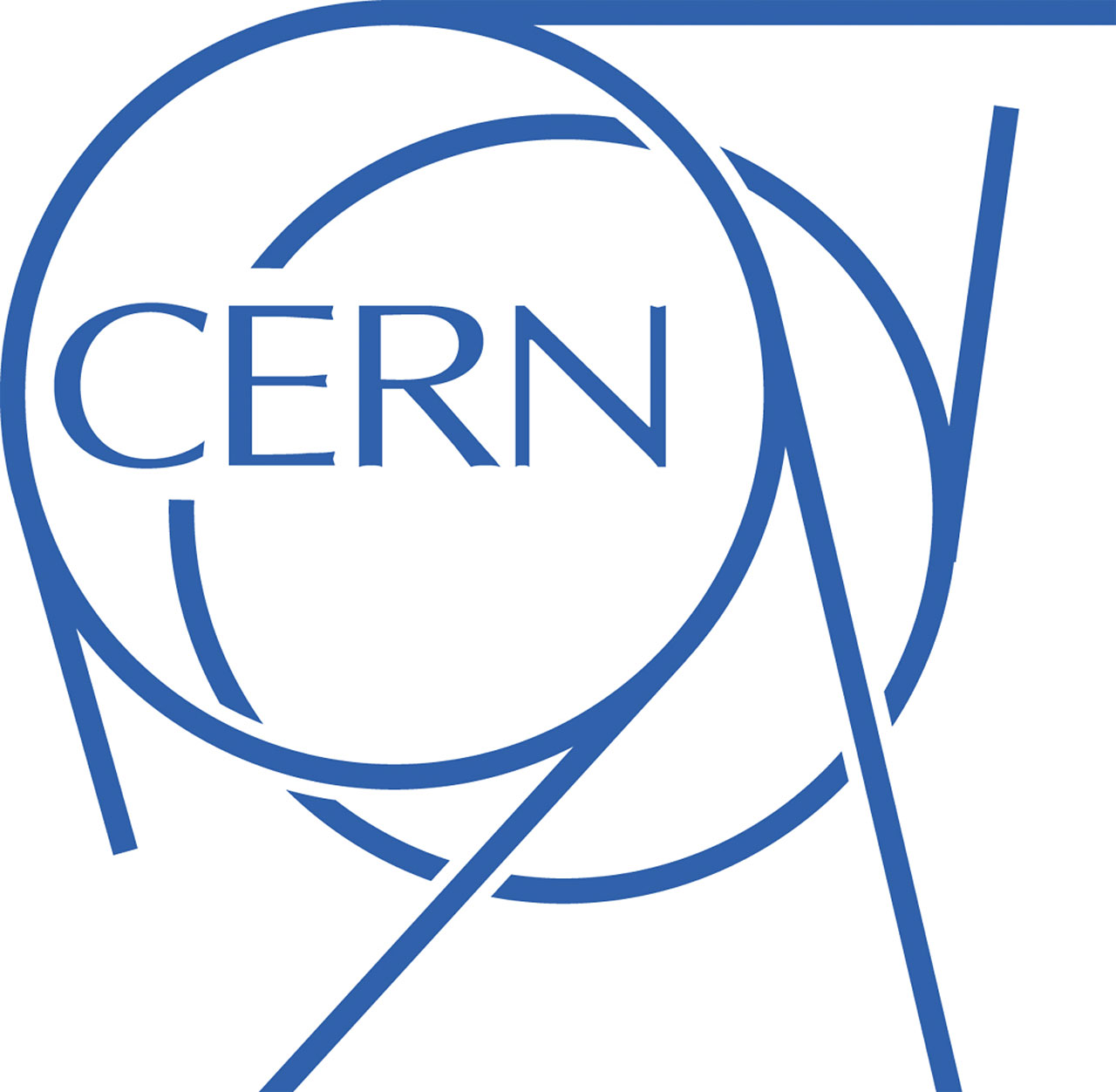}} & & \\
 & & CERN-LPCC-2018-06 \\  
 & & \today \\ 
 & & \\
\hline
\end{tabular*}

\vspace*{0.3cm}
%
%
\maketitle
\vspace{\fill}

\newpage
\begin{center}
    \begin{abstract}
  \noindent
  
\end{abstract}
\end{center}
  Motivated by the success of the flavour physics programme carried out over the last decade at the Large Hadron Collider (LHC), we characterize in detail the physics potential of its High-Luminosity and High-Energy upgrades in this domain of physics. We document the extraordinary breadth of the HL/HE-LHC programme enabled by a putative Upgrade II of the dedicated flavour physics experiment LHCb and the evolution of the established flavour physics role of the ATLAS and CMS general purpose experiments. We connect the dedicated flavour physics programme to studies of the top quark, Higgs boson, and direct high-$p_T$ searches for new particles and force carriers. We discuss the complementarity of their discovery potential for physics beyond the Standard Model, affirming the necessity to fully exploit the LHC's flavour physics potential throughout its upgrade eras.
\vspace*{2.0cm}
\vspace{\fill}

\end{titlepage}

%% file: authors.tex
\author{Editors: \\\href{http://inspirehep.net/record/1021757}{A.~Cerri}$^{1}$
, \href{http://inspirehep.net/record/1057204}{V.V.~Gligorov}$^{2}$
, \href{http://inspirehep.net/record/999046}{S.~Malvezzi}$^{3}$
, \href{http://inspirehep.net/record/1056267}{J.~Martin~Camalich}$^{4,5}$
, \href{http://inspirehep.net/record/1020959}{J.~Zupan}$^{6}$
\\ \vspace*{4mm} 
Contributors: 
\\ \href{http://inspirehep.net/record/1073335}{S.~Akar}$^{6}$
, \href{http://inspirehep.net/record/1068305}{J.~Alimena}$^{7}$
, \href{http://inspirehep.net/record/1018633}{B.C.~Allanach}$^{8}$
, \href{http://inspirehep.net/record/1048366}{W.~Altmannshofer}$^{9}$
, \href{http://inspirehep.net/record/1189488}{L.~Anderlini}$^{10}$
% ...
\vspace*{1cm} 
}
\small\institute{$^{1}$Department of Physics and Astronomy, University of Sussex, Brighton, United Kingdom, $^{2}$LPNHE, Sorbonne Universit{\'e}, Paris Diderot Sorbonne Paris Cit{\'e}, CNRS/IN2P3, Paris, France, $^{3}$INFN Sezione di Milano-Bicocca, Milano, Italy, $^{4}$Universidad de La Laguna, Facultad de F\'isica, Avda. Astrof\'isico Fco. Sanchez s/n, 38206, La Laguna, Tenerife, Spain, $^{5}$Instituto de Astrof\'isica de Canarias (IAC) C/ V\'ia Lactea, s/n E-38205, La Laguna, Tenerife, Spain, $^{6}$University of Cincinnati, Cincinnati, OH, United States, $^{7}$The Ohio State University, Columbus, USA, $^{8}$Department of Applied Mathematics and Theoretical Physics, University of Cambridge, Wilberforce Road, United Kingdom, CB1 3BZ, $^{9}$Santa Cruz Institute for Particle Physics (SCIPP), 1156 High Street, Santa Cruz, CA 95064, USA, $^{10}$INFN Sezione di Firenze, Firenze, Italy, $^{11}$Nikhef National Institute for Subatomic Physics, Amsterdam, Netherlands, $^{12}$INFN Sezione di Padova, Padova, Italy, $^{13}$Durham University, Institute for Particle Physics Phenomenology, Ogden Centre for Fundamental Physics, South Road, Durham DH1 3LE, United Kingdom, $^{14}$Imperial College, London, United Kingdom, $^{15}$Physics Department, Lancaster University, Lancaster, United Kingdom, $^{16}$Institute of Theoretical and Experimental Physics (ITEP), Moscow, Russia, $^{17}$Cavendish Laboratory, University of Cambridge, Cambridge, United Kingdom, $^{18}$Saha Institute of Nuclear Physics, HBNI, Kolkata, India, $^{19}$University of Birmingham, Birmingham, United Kingdom, $^{20}$Fakult{\"a}t Physik, Technische Universit{\"a}t Dortmund, Dortmund, Germany, $^{21}$Deutsches Elektronen-Synchrotron (DESY), Notkestrasse 85, 22607 Hamburg, Germany, $^{22}$Department of Physics, University of Warwick, Coventry, United Kingdom, $^{23}$Syracuse University, Syracuse, NY, United States, $^{24}$Skobeltsyn Institute of Nuclear Physics, Lomonosov Moscow State University, Moscow, Russia, $^{25}$Physikalisches Institut, Ruprecht-Karls-Universit{\"a}t Heidelberg, Heidelberg, Germany, $^{26}$INFN Sezione di Ferrara, Ferrara, Italy, $^{27}$European Organization for Nuclear Research (CERN), Geneva, Switzerland, $^{28}$Universit\`{a} di Padova, Padova, Italy, $^{29}$Henryk Niewodniczanski Institute of Nuclear Physics  Polish Academy of Sciences, Krak{\'o}w, Poland, $^{30}$TUM-IAS, Lichtenbergstr. 2a, D-85748 Garching, Germany, $^{31}$University of Florida, Gainesville, USA, $^{32}$INFN Sezione di Bologna, Bologna, Italy, $^{33}$Universit{\`a} di Bologna, Dipartimento di Fisica, Bologna, Italy, $^{34}$Theoretical Physics Department, Fermilab, Batavia, IL 60510, USA, $^{35}$Enrico Fermi Institute and Kavli Institute for Cosmological Physics, University of Chicago, Chicago, IL 60637, USA, $^{36}$PRISMA Cluster of Excellence and Institute for Physics (THEP), Johannes Gutenberg University Mainz, D-55099 Mainz, Germany, $^{37}$Ludwig-Maximilians-Universit{\"a}t (LMU) M{\"u}chen, $^{38}$CERN, European Organization for Nuclear Research, Geneva, Switzerland, $^{39}$Centro de Investigaciones Energ\'{e}ticas Medioambientales y Tecnol\'{o}gicas (CIEMAT), Madrid, Spain, $^{40}$Imperial College, London, UK, $^{41}$Institute for Particle Physics Phenomenology, University of Durham, Durham, UK, $^{42}$Aix-Marseille Univ, Universit{\`e} de Toulon, CNRS, CPT, Marseille, France, $^{43}$National Taiwan University (NTU), Taipei, Taiwan, $^{44}$Instituto Galego de F{\'\i}sica de Altas Enerx{\'\i}as (IGFAE), Spain, $^{45}$Theoretical Division, Los Alamos National Laboratory, MS B283, Los Alamos, NM 87545, USA, $^{46}$INFN Sezione di Roma Tre, Via della Vasca Navale 84, I-00146 Roma, Italy, $^{47}$Aix Marseille Univ, CNRS/IN2P3, CPPM, Marseille, France, $^{48}$Institute for Theoretical Physics, Albert Einstein Center for Fundamental Physics, University of Bern, Sidlerstrasse 5, 3012 Bern, Switzerland, $^{49}$INFN Sezione di Cagliari, Monserrato, Italy, $^{50}$Universit{\`a} di Torino, Torino, Italy, $^{51}$INFN Sezione di Torino, Torino, Italy, $^{52}$School of Physics and Astronomy, University of Edinburgh, Edinburgh, United Kingdom, $^{53}$Paul Scherrer Institut, CH-5232 Villigen PSI, Switzerland, $^{54}$Istituto Nazionale di Fisica Nucleare (INFN), Sezione di Napoli,via Cintia, I-80126 Napoli, $^{55}$Oliver Lodge Laboratory, University of Liverpool, Liverpool, United Kingdom, $^{56}$University of Louisville; United States of America, $^{57}$School of Physics and Astronomy, University of Manchester, Manchester, United Kingdom, $^{58}$INFN Roma Tor Vergata, Roma, Italy, $^{59}$Universit\`a di Roma Tor Vergata, Dipartimento di Fisica, Roma, Italy, $^{60}$Michigan State University, Department of Physics and Astronomy; United States of America, $^{61}$University of California (UC), Department of physics, 9500 Gilman Dr. La Jolla, CA 92093-0319, USA, $^{62}$IRFU, CEA, Universit\'e Paris-Saclay, Gif-sur-Yvette, France, $^{63}$CP3-Origins, Syddansk Unversitet, Campusvej 55, DK-5230 Odense M, Denmark, $^{64}$Department of Physics, Yale University, New Haven CT, United States of America, $^{65}$National Research University Higher School of Economics, Moscow, Russia, $^{66}$Universit'{e} Clermont Auvergne, CNRS/IN2P3, LPC, F-63000 Clermont-Ferrand, France , $^{67}$Laboratoire de Physique Th\'{e}orique (UMR8627), CNRS, Univ. Paris-Sud, Universit\'{e} Paris-Saclay, 91405 Orsay, France, $^{68}$Oliver Lodge Laboratory, University of Liverpool, Liverpool, United Kingdom, $^{69}$Universit\`{a} di Milano-Bicocca, Milano, Italy, $^{70}$INFN Sezione di Trieste, Trieste, Italy, $^{71}$Department of Physics, Technion, Haifa 32000, Israel, $^{72}$Theoretical Particle Physics Laboratory (LPTP), Institute of Physics, EPFL, Lausanne, Switzerland, $^{73}$International Center for Advanced Studies (ICAS), 25 de Mayo y Francia, San Martin, Pcia. de Buenos Aires, Argentina, $^{74}$Jo\v{z}ef Stefan Institute, Jamova 39, P.O.B. 3000, SI-1001 Ljubljana, Slovenia, $^{75}$University of Ljubljana, Faculty of Mathematics and Physics, Jadranska ulica 19, SI-1000 Ljubljana, Slovenia, $^{76}$INFN Sezione di Pisa, Pisa, Italy, $^{77}$Scuola Normale Superiore di Pisa, Pisa, Italy, $^{78}$Princeton, Institute for Advanced Study (IAS), Einstein Drive Princeton, NJ, 08540, USA , $^{79}$CAFPE and Departamento de F\'{\i}sica Te\'orica y del Cosmos, Universidad de Granada, Granada, Spain , $^{80}$Dipartimento di Fisica, Universita di Torino and INFN, Sezione di Torino, I-10125 Torino, Italy, $^{81}$Physik-Institut, Universit{\"a}t Z{\"u}rich, Z{\"u}rich, Switzerland, $^{82}$School of Physics, Beihang University, Beijing 100191, China, $^{83}$CERN, TH Department, Geneva, Switzerland, $^{84}$Deutsches Elektronen-Synchrotron, Hamburg, Germany, $^{85}$Department of Physics, LEPP, Cornell University, Ithaca, NY 14853, USA, $^{86}$LAPTh, 9 Chemin de Bellevue, F-74941 Annecy Cedex, France, $^{87}$CAS Key Laboratory of Theoretical Physics, Institute of Theoretical Physics, Chinese Academy of Sciences, Zhong Guan Cun East Street 55, Beijing 100190, China, $^{88}$School of Physical Sciences, University of Chinese Academy of Sciences, Beijing 100049, China , $^{89}$University of Hamburg, Hamburg, Germany, $^{90}$University of Maryland, College Park, MD, United States, $^{91}$University of Pittsburgh, $^{92}$Department of Physics, University of Oxford, Oxford, United Kingdom, $^{93}$Fakult\"at Physik, TU Dortmund, Otto-Hahn-Str.4, D-44221 Dortmund, Germany, $^{94}$RWTH Aachen University, III. Physikalisches Institut A, Aachen, Germany, $^{95}$Bethel University, St. Paul, USA, $^{96}$Brown University, Providence, USA, $^{97}$Nikhef National Institute for Subatomic Physics and University of Amsterdam, Amsterdam, Netherlands, $^{98}$Institute for Mathematics, Astrophysics and Particle Physics, Radboud University Nijmegen/Nikhef, Nijmegen, Netherlands, $^{99}$Universit{'a}t Zurich, Physik-Institut. Winterthurerstrasse 190, CH-8057 Zurich, Switzerland, $^{100}$Tata Institute of Fundamental Research-B, Mumbai, India, $^{101}$Excellence Cluster Universe, TUM, Boltzmannstr. 2, 85748 Garching, Germany, $^{102}$Department of Physics and Astronomy, University of Sussex, Brighton BN1 6NL, United Kingdom, $^{103}$LAL, Univ. Paris-Sud, IN2P3/CNRS, Universit\'{e} Paris-Saclay, Paris, France, $^{104}$Sezione di Roma, Istituto Nazionale di Fisica Nucleare, Roma, Italy, $^{105}$Sapienza Universit\`a di Roma, Dipartimento di Fisica, Roma, Italy, $^{106}$School of Physics and Astronomy, Tel Aviv University, Israel, $^{107}$Nagoya University, Furo-cho, Chikusa-ku, Nagoya-shi 464-6802, Japan, $^{108}$ETH Zurich - Institute for Particle Physics and Astrophysics (IPA), Zurich, Switzerland, $^{109}$I. Physikalisches Institut, RWTH Aachen University, Aachen, Germany, $^{110}$Paul Scherrer Institut, Villigen, Switzerland, $^{111}$Department of Physics, Arizona State University, Tempe, AZ 85287, USA, $^{112}$State Key Laboratory of Nuclear Physics and Technology, Peking University, Beijing, China, $^{113}$Department of Physics, Nankai University, $^{114}$Lawrence Berkeley National Laboratory ; University of California, Berkeley, USA, $^{115}$University of Wisconsin - Madison, Madison, USA, $^{116}$Indiana University, 727 E 3rd St, Bloomington, IN 47405, USA, $^{117}$Univ. Lyon, Univ. Lyon 1, CNRS/IN2P3, Institut de Physique Nucl\'eaire de Lyon, UMR5822, F-69622 Villeurbanne, France, $^{118}$Institute for High Energy Physics of National Research Centre 'Kurchatov Institute', Protvino, Russia, $^{119}$Physics Department, University of Siegen, 57072 Siegen, Germany , $^{120}$Univ. Grenoble Alpes, Univ. Savoie Mont Blanc, CNRS, IN2P3-LAPP, Annecy, France, $^{121}$Instituto Galego de Fisica de Altas Enerxias (IGFAE), Universidade de Santiago de Compostela, $^{122}$University of Notre Dame, Notre Dame, Indiana 46556, USA, $^{123}$Instituto de Fisica Corpuscular, Centro Mixto Universidad de Valencia - CSIC, Valencia, Spain, $^{124}$INFN, Sezione di Trieste, Trieste, Italy, $^{125}$Universitat Autonoma de Barcelona and IFAE, 08193 Bellaterra (Barcelona), $^{126}$Universit\"{a}t Z\"{u}rich, Zurich, Switzerland, $^{127}$Department of Particle Physics and Astrophysics, Weizmann Institute of Science, Rehovot, Israel 7610001, $^{128}$Pisa, Scuola Normale Superiore, Pisa, Italy, $^{129}$INFN Sezione di Milano, Milano, Italy, $^{130}$Universit{\`a} degli Studi di Milano, Milano, Italy, $^{131}$Johannes Gutenberg University, Mainz, Germany, $^{132}$Institute for Theoretical Particle Physics, Karlsruhe Institute of Technology (KIT), Wolfgang-Gaede-Str. 1, 76131 Karlsruhe, Germany , $^{133}$Instituto de Fisica Corpuscular (IFIC), Centro Mixto Universidad de Valencia - CSIC , $^{134}$Bogazici University, Istanbul, Turkey, $^{135}$Laborat\'{o}rio de Instrumentacao F\'{i}sica Experimental de Part\'{i}culas, Lisboa, Portugal, $^{136}$Physics Division, Lawrence Berkeley National Laboratory and University of California, Berkeley CA, United States of America, $^{137}$Department of Physics, University of California (UC), Berkeley, United States of America, $^{138}$Universidad de Oviedo, Oviedo, Spain, $^{139}$INFN Laboratori Nazionali di Frascati, Frascati, Italy, $^{140}$Center for Exploration of Energy and Matter, Indiana University, Bloomington, IN 47408, USA, $^{141}$Theory Center, Thomas Jefferson National Accelerator Facility, 12000 Jefferson Ave, Newport News VA 23606, USA, $^{142}$Fermi National Accelerator Laboratory, Batavia, USA, $^{143}$Institute of Physics, Ecole Polytechnique  F{\'e}d{\'e}rale de Lausanne (EPFL), Lausanne, Switzerland, $^{144}$European Laboratory for Particle Physics, CERN, Geneva, Switzerland, $^{145}$Department of Physics and Astronomy, Wayne State University, Detroit, MI 48201, USA, $^{146}$European Centre for Theoretical Studies in Nuclear Physics and Related Areas (ECT$^*$) ) and Fondazione Bruno Kessler, Strada delle Tabarelle 286, I-38123 Villazzano (TN), Italy, $^{147}$Sapienza Universit\`{a} di Roma and INFN, Piazzale Aldo Moro 2, I-00185, Rome, Italy, $^{148}$Universit{\`a}t Regensburg, Fakult{\`a}t f{\"u}r Physik, Universitatsstr. 31, 93053 Regensburg, Germany, $^{149}$Universit{\`a} di Pisa, Pisa, Italy, $^{150}$H.H. Wills Physics Laboratory, University of Bristol, Bristol, United Kingdom, $^{151}$Charles University, Faculty of Mathematics and Physics, Prague, Czech Republic, $^{152}$University of Chicago, Enrico Fermi Institute, 5640 S Ellis Ave, Chicago IL 60637, USA, $^{153}$Universit\'{e} Catholique de Louvain, Centre for Cosmology, Particle Physics, and Phenomenology, Louvain-la-Neuve, Belgium, $^{154}$LAL, Univ. Paris-Sud, CNRS/IN2P3, Universit{\'e} Paris-Saclay, Orsay, France, $^{155}$Departamento de Sistemas F{\'i}sicos, Qu{\'i}micos y Naturales, Universidad Pablo de Olavide, E-41013 Sevilla, Spain, $^{156}$ Universit\`{a}t Hamburg, II Institut f\"{u}r Theoretische Physik, Luruper Chaussee 149, 22761 Hamburg, Germany, $^{157}$NHETC, Dept of Physics and Astronomy, Rutgers University, 136 Frelinghuysen Rd, Piscataway, NJ 08854 USA, $^{158}$Vrije Universiteit Brussel, Brussel, Belgium, $^{159}$High Energy Theory, Brookhaven National Lab, Upton, NY 11973, USA, $^{160}$Department of Physics, Durham University, $^{161}$School of Physics and Astronomy, University of Glasgow, Glasgow, United Kingdom, $^{162}$INFN, Genoa  Istituto Nazionale di Fisica Nucleare (INFN)  Sezione di Genova  Via Dodecaneso, 33 I-16146 Genova Italy, $^{163}$University of Cyprus, Nicosia, Cyprus, $^{164}$JINR, Joint Institute for Nuclear Research, Dubna, Russia, $^{165}$University of Colorado Boulder, Boulder, USA, $^{166}$INFN Sezione di Bari, Bari, Italy, $^{167}$Center for Theoretical Physics, Massachusetts Institute of Technology   77 Massachusetts Ave, Cambridge, MA 02139, USA, $^{168}$Physik Department, Technische Universit\"at M\"unchen, James-Franck-Strasse 1, D-85748 Garching, Germany, $^{169}$Department Physik, Universit\"{a}t Siegen, Siegen, Germany, $^{170}$SKLPPC, School of Physics and Astronomy, Shanghai Jiao Tong University, Shanghai  200240, China, $^{171}$University of Pittsburgh, Department of Physics and Astronomy, 3941 O'Hara Street, Pittsburgh, PA 15260, USA, $^{172}$Massachusetts Institute of Technology, Cambridge, MA, United States, $^{173}$STFC Rutherford Appleton Laboratory, Didcot, United Kingdom, $^{174}$Institute of Particle Physics, Central China Normal University, Wuhan, Hubei, China, $^{175}$Center for High Energy Physics, Tsinghua University, Beijing, China, $^{176}$Institute of High Energy Physics, Beijing, China, $^{177}$University of Cambridge, Cavendish Laboratory, Madingley Road, Cambridge, CB3 0HE, United Kingdom, $^{178}$Department of Applied Mathematics and Theoretical Physics, University of Cambridge, Wilberforce Road, United Kingdom, $^{179}$Johannes Gutenberg Universit{\`a}t Mainz, Intitut f{\"u}r Physik, Staudinger Weg 7, 55128 Mainz, Germany, $^{180}$Institute of High Energy Physics (IHEP), Chinese Academy of Sciences (CAS), $^{181}$Centro Brasileiro de Pesquisas F{\'\i}sicas (CBPF), Rio de Janeiro, Brazil
}\normalsize

%% file: section1/section.tex
\section{Introduction}
\label{sec:Intro}

The past decade has witnessed a highly successful programme of flavour physics at the LHC, building on and greatly expanding the pioneering work at the Tevatron's CDF and D\O. The unprecedented breadth and precision of the physics results produced by the LHC's dedicated flavour physics experiment, LHCb, has been complemented by crucial measurements at ATLAS and CMS. Together, they have probed the Standard Model at energy scales  complementary to the direct LHC searches, and proven that it is possible to carry out a broad programme of precision flavour physics in such a challenging hadronic environment. This document offers a glimpse of the future -- the potential for flavour physics in the High-Luminosity phase of the Large Hadron Collider (HL-LHC) and its possible upgrade to a 27 TeV proton collider, the High-Energy LHC (HE-LHC). 
The landscape of flavour physics is considered and theoretical arguments are presented for measurements with higher precision and of qualitatively new observables.
The prospective experimental sensitivities for the HL-LHC assume 3000~fb$^{-1}$ recorded by ATLAS and CMS, and 300~fb$^{-1}$ recorded by a proposed Upgrade II of LHCb.

The main points, detailed in the subsequent sections, are:
\begin{itemize}
\item[$\bullet$]  The flavour physics programme at the LHC comprises many different probes: the weak decays of  beauty, charm, strange and top quarks, as well as of the $\tau$ lepton and the Higgs;
\item[$\bullet$] $\CP$ violation and Flavour Changing Neutral Currents (FCNCs) are sensitive probes of short-distance 
physics, within the Standard Model (SM) and beyond (BSM);
\item[$\bullet$] Flavour physics probes scales  much greater than $1$\,TeV, with the  
sensitivity often limited by statistics and not by theory;
\item[$\bullet$] For most FCNC processes, a New Physics (NP) contribution at 20\% of the SM is still allowed, so there is plenty of discovery potential;
\item[$\bullet$] Spectroscopy and flavour changing transitions serve as laboratories for a better understanding of nonperturbative Quantum Chromodynamics (QCD);
\item[$\bullet$] Some of the several tensions between flavour physics data and the SM may soon become decisive;
\item[$\bullet$]  Precision tests of the SM flavour sector will improve by orders of magnitude including Charged Lepton Flavour Violating transitions (CLFV);
\item[$\bullet$]  Flavour physics will teach us about physics at shorter distances, complementary to the high-\pt physics programme, whether NP is seen or not, and could point to the next energy scale to explore.
\end{itemize}

\subsection{Theoretical considerations}
{\it \small Authors (TH): G.\ Isidori, Z.\ Ligeti.}

As a community, we are now in a strikingly different position than we were a decade ago, before the LHC  turned on. Already before the start of the LHC it was clear from unitarity considerations that the LHC experiments
were basically guaranteed to uncover the origin of the electroweak symmetry breaking, i.e., 
the breaking of the $SU(2)_L \times U(1)_Y$ gauge symmetry to the $U(1)$ of electromagnetism. The discovery of the Higgs boson
by ATLAS and CMS in 2012 was a triumph, confirming these expectations. Since then we have learned that the properties of the Higgs boson are in increasing
agreement with the SM. Coupled with the lack of direct signals of BSM 
particles so far, this increasingly points to a mass gap between the SM particle spectrum and the BSM one. 

After completion of the first phase of the LHC programme, the field entered into a more uncertain, yet possibly more exciting exploratory era. 
We are still faced by a number of key
open questions, e.g., the need for dark matter and how to generate the baryon asymmetry. We thus do know that BSM physics must exist. However,
we do not know which experiments, at what energy scale, and probing which aspects of our understanding of nature, may provide the first unambiguous evidence for BSM phenomena.
The phenomenological successes of the SM, in conjunction with being a renormalizable quantum field
theory, means that there is no clear guidance where to search for clues on how to extend 
the SM.
\footnote{The only clearly established exception 
are neutrino masses which require non-renormalizable 
operators (or new degrees of freedom) and seem to point toward a very high scale of new physics that is not accessible in practice.
However, 
the existence of a high new scale connected to neutrino mass generation does not prevent other BSM physics to appear at lower scales.}
This calls for a diversified programme of BSM searches, with no stone left unturned. 
A deeper study of the properties of the Higgs boson is one of the pillars of this programme, and will be the central focus of the HL-LHC.  
The same programme also offers unique opportunities for tremendous improvements in 
indirect NP searches via precision studies of low-energy flavour-changing observables.
Here the expected increase in statistics may be even larger than in the Higgs sector. 
As explained below, this programme is complementary to both the high-\pt NP searches as well as 
to the indirect NP searches performed via the Higgs precision measurements. To show this, we first give 
a brief introduction to flavour physics, starting with the ``flavour puzzle'' and the general discussion of probing BSM through flavour transitions. 

\smallskip
\subsubsection{The flavour puzzle}
Flavour is the label generically used to differentiate the 12 fermions which, according to the SM, 
are the basic constituents of matter.
These particles can be grouped into 3 families, each containing two quarks and two leptons. The particles within a given family 
have different combinations of strong, weak, and electromagnetic charges. This in turn implies differing behaviors under the SM interactions. 
Across the three families the particle content is identical except for the masses. That is, the second and third family are copies of the first family, with the same SM quantum numbers for the copies of particles across generations, but with different masses. 
Ordinary matter consists of particles of the first family: the up and down quarks that form atomic nuclei, as well as the electrons and the corresponding neutrinos.  The question why there are three almost identical replicas of quarks and leptons as well as the origin of their different mass matrices are among the big open questions in fundamental physics, often referred to as the ``SM flavour puzzle''. 

Within the SM, the hierarchy of fermion masses originates from the hierarchy in the strengths of interactions between the fermions and the Higgs field, namely from the structure of the Yukawa couplings.  However, this prescription does not provide any explanation  
for the origin of the large hierarchies observed among fermion masses. Putting aside the special case of neutrinos, there are five orders of magnitudes between the mass of an electron and a top quark. Similarly, we do not know what determines the peculiar and rather different mixing structure in the quark and lepton mass matrices observed through the misalignment of mass and weak-interaction eigenstates in flavour space. 
We do know experimentally, that the Higgs field is responsible for the bulk of the heaviest quark and lepton masses: the top and bottom quarks and the tau leptons. The generation of at least some of the quark masses and mixing angles is thus connected to the Higgs sector. This
suggests a possible connection between the flavour puzzle and the electroweak hierarchy puzzle, 
another big open question pointing toward some form of new physics.

\smallskip
\subsubsection{Model-independent considerations}
The above puzzling aspects make flavour physics, i.e., the precision study of flavour-changing processes in the quark and lepton sector, 
a very interesting window on possible physics beyond the SM.  We do not know if there is an energy scale at which the  flavour structure observed assumes a simpler form, i.e., we do not know if 
 the masses and mixing angles, as observed at low energies, can be predicted in terms of a reduced number of 
more fundamental parameters in a theory valid at some high scale.  On the other hand, precision measurements of flavour-changing transitions may probe such scales, 
even if they are well above the LHC center-of-mass energy.  

This statement can be made quantitative by considering the SM as 
a low-energy effective theory that is valid up to a cut-off scale $\Lambda$, taken to be bigger than the 
electroweak scale $v = (\sqrt{2}\,G_F)^{-1/2} \approx 246$~GeV. 
According to such an assumption of heavy NP, the amplitudes describing a flavour changing transition 
of a fermion $\psi_i$ to a fermion $\psi_j$ 
can be cast into the following general form
\begin{equation}
\mathcal{A}(\psi_i \to \psi_j + X) = \mathcal{A}_{0} \left( \frac{c_{\rm SM}}{v^2} 
+ \frac{c_{\rm NP}}{\Lambda^2} \right) .
\label{eq:fl1}
\end{equation} 
Since in many cases $c_{\rm SM} \ll 1$, NP effects can have a large impact even if $\Lambda \gg v$.
For instance, in the quark sector the reason that often $c_{\rm SM} \ll 1$ stems from the facts that:
\vspace{-\topsep}
\begin{itemize}
\item[(i)] $c_{\rm SM}$ can be proportional to small 
entries of the Cabibbo-Kobayashi-Maskawa (CKM) matrix and/or to small SM Yukawa couplings; 
\item[(ii)] $c_{\rm SM}$ may include a loop factor $1/(16\pi^2)$, if the corresponding transition is forbidden at tree level, 
as is the case for flavour-changing neutral-current 
(FCNC) transitions or meson-antimeson mixing transitions. 
\end{itemize}
\vspace{-\topsep}
As a result, 
these low-energy processes can probe indirectly, via quantum effects, scales of order $v/\sqrt{c_{\rm SM}}$. These can easily 
exceed those directly reachable via production of on-shell states in current and planned accelerators.
As an explicit example, in the case of $B$\,--\,$\bar B$ mixing, $\sqrt{c_{\rm SM}} \sim |V_{td}|/(4\pi) \sim 10^{-3}$,
hence this observable can probe NP scales up to $10^3$~TeV in models with $c_{\rm NP} \sim 1$.

The precise values of the NP scale probed at present 
vary over a wide range, depending on the specific observable and the specific NP model
($c_{\rm NP}$ can span a large range, too).
However, the form of Eq.~(\ref{eq:fl1}) does allow us to predict how the bounds will 
improve with increasing datasets. For the observables that are SM dominated, are already observed, 
and whose uncertainties are dominated by statistics,
the corresponding bound on $\Lambda$ scales as $N^{1/4}$, 
where $N$ is the relative increase in the number of events.
The same scaling occurs for forbidden or highly suppressed SM processes, i.e., in the limit $c_{\rm SM} \ll c_{\rm NP}$,
if the search is not background dominated. 
Thus, with two orders of magnitude increase in statistics one can probe scales roughly 3 times higher
than at present. This is well above the increase in NP scale probed in on-shell heavy particle searches at high-\pt that can be achieved at fixed collider energy by a similar increase in statistics.

While theoretical uncertainties are often important, there are enough measurements which are known 
not to be limited by theoretical uncertainties. Improved experimental results will therefore directly translate to better NP sensitivity.  
There are also several cases of observables sensitive to NP where the theoretical uncertainties  
are mainly of parametric nature (e.g., our ability to precisely compute $c_{\rm SM}$ is dominated by the knowledge of CKM elements, 
quark masses, etc.). For such cases, we can expect significant increase in precision with higher 
statistics thanks to the improvement in the reduction of parametric uncertainties. This also highlights the importance of a broad 
flavour physics programme where the focus is not only on rare or $\CP$ violating processes ``most likely'' affected by NP
but also on core SM measurements which help to reduce the theoretical uncertainties.

\smallskip
\subsubsection{Current anomalies and historical comments}
Due to the generic sensitivity to high scales, flavour physics has historically played a major role in developing and understanding 
the Standard Model. Flavour physics measurements signalled the presence of ``new'' particles 
well before these were directly observed (this was the case for charm and top quarks from $K_L\to \mu^+\mu^-$ decays and $K$-meson mixing, and from $B$-meson mixing, respectively). 
With the completion of the SM, and the increasingly precise tests that the SM predictions have  successfully passed, one may draw the naive conclusion that the discovery potential of precision experiments has declined in the last decades. 
However, the opposite is true. First of all, a qualitative change in our understanding has been achieved during that time. Before the asymmetric $B$ factory experiments, BaBar and Belle, it was not known whether the SM accounted for the dominant or just a small part of \CP violation observed in kaon mixing.  We now know that the bulk of it is due to the SM Kobayashi-Maskawa mechanism. However, 
even after decades of progress, for most FCNC amplitudes the NP is still allowed to contribute at $\sim$\,20\% of the SM contribution.

The great improvements in precision for several flavour-changing processes achieved in the last 20 years, both at experimental and theoretical levels, represent a very important advancement of the field. 
We learned that either NP is much heavier than the electroweak scale,
or, if it is not far above the electroweak scale as required by most solutions of the hierarchy puzzle, it must have a highly 
nontrivial flavour structure that is able to mimic the strong suppression of FCNC transitions in the SM.
The latter statement has often been oversimplified, assuming that there is little hope to observe significant deviations from the SM in  
flavour physics. The anomalies recently observed in semileptonic $B$ decays clearly demonstrated a genuine discovery potential, regardless of whether or not their significance increase with improved measurements.

Recent measurements, both in charged-current and in neutral-current 
semileptonic $B$ decays, hint at a violation of one of the 
key predictions of the SM -- the universality of interactions for leptons of different generations
(in the limit where their masses can be neglected).  These anomalies represent the strongest tensions
with the SM predictions currently observed in laboratory experiments. The statistical significance of the anomalies is not high enough to 
claim a discovery but the situation is very interesting. 
More precise measurements of some of these observables, in particular the lepton flavour universality violating ratios
$R_{K^{(*)}} = \Gamma(B\to K^{(*)} \mu^+\mu^-) / \Gamma(B\to K^{(*)} e^+e^-)$ and 
$R(D^{(*)}) = \Gamma(B\to D^{(*)} \tau\bar\nu) / \Gamma(B\to D^{(*)} l\bar\nu)$, where $l=e,\, \mu$,
could establish the presence of NP even 
with modest improvements in statistics.  At the current central values for these anomalies, 
analyzing all of the Run~1 and Run~2 data could already establish a discrepancy with the SM expectation 
in a single observable with $5\,\sigma$ significance.

Whether or not these anomalies will gain significance to become unambiguous signals of physics beyond the SM, 
they have clearly exemplified the discovery potential of flavour-physics observables 
and enlarged our horizon regarding possible BSM scenarios. Before the appearance of these anomalies, 
lepton flavour universality (LFU) was an implicit assumption adopted by the vast majority of BSM scenarios proposed .
It is now better appreciated that LFU is an accidental property of the SM. It is well tested in transitions involving only 
the first two generations of quarks and leptons, while it is rather poorly tested in processes involving the third generation 
(and may indeed be violated at a detectable level in $B$ decays). A deeper scrutiny of this SM property has highlighted 
the interest in a large variety of observables, with small theoretical uncertainties, which would strongly benefit from more statistics. 
Similarly, it has often (though not always) been taken for granted that NP 
effects in tree-level dominated processes, such as those affecting $R(D^{(*)})$, are negligible, while it is now clear that 
there are many NP scenarios where this assumption does not hold. This observation has important 
 phenomenological consequences and signals the limitation of a significant fraction of current NP analyses.
Last but not least, theoretical models addressing the anomalies have highlighted the  interest of BSM constructions
containing heavy leptoquark fields -- a class of NP models that was not popular until a few years ago. 

The current central values of $R_{K^{(*)}}$ and, especially, $R(D^{(*)})$ imply that NP needs to be at a fairly low scale: below few tens of TeV in the former,  
and a few TeV in the latter case.  This can be easily understood given that the NP effects need to give ${\mathcal O}(10\%-20\%)$ corrections to the amplitudes which are one-loop and tree level in the SM, respectively. Models addressing the anomalies are therefore a perfect laboratory to explore the interplay between 
indirect NP searches from flavour observables and direct searches at high-\pt. Interestingly enough, even in the low-scale models 
addressing $R(D^{(*)})$, with or without $R_{K^{(*)}}$, there exist ample regions of parameter space 
that are able to explain the anomalies and that are at the same time consistent with the null results of NP searches performed so far at high \pt.

\smallskip
\subsubsection{Connections to lepton flavour violation}
The CLFV processes, such as $\tau \to 3 \mu$, 
are essential parts of the flavour-physics programme.  
CLFV amplitudes can also be decomposed as in Eq.~(\ref{eq:fl1}), with the advantage that in this case 
$c_{\rm SM}$ vanish.  If the SM is extended to describe neutrino masses, non-zero predictions arise but are suppressed by $m_\nu^2/m_W^2$. The predicted CLFV rates are thus many orders of magnitudes below the 
detection reach of any present or planned facility. As a consequence the searches for CLFV are very clean 
and powerful ways to search for physics beyond the SM.

Any attempt to solve the flavour puzzle with new dynamics not far from the TeV scale, such that  the observed hierarchies in the Yukawa couplings are accounted for by the new dynamics, 
naturally leads to CLFV rates not far from the present bounds. The recent LFU anomalies 
have strengthened the case further. Many models explaining these anomalies predict 
CLFV at a detectable level, in many cases just below the current bounds. 
Of noteworthy interest, triggered by the recent anomalies, are processes 
that violate both quark and lepton flavour, such as $B \to K \tau \mu$.  
There is a large variety of observables of this type that, together with purely leptonic observables, 
form a large and very promising sub-field of NP probes. Such searches can be organized 
in a large matrix, with the row and column indices determined by lepton and quark flavours, which is largely unexplored at present. For any NP model that may populate entries in this matrix, there is a large  
complementarity between the HL-LHC
experiments, Belle-II, and dedicated experiments at muon beams searching for 
$\mu\to e$ conversion, $\mu\to e\gamma$, and $\mu\to 3 e$, as well as with the flavour diagonal probes such as the measurements of the $(g-2)$ of the muon and the electron, or the searches for electric dipole moment of the electron.

\smallskip
\subsubsection{Connections with the hierarchy problem and complementarity with high-\pt searches} 
BSM models proposed to address the electroweak hierarchy problem, such as supersymmetric models or composite Higgs models, predict new particles around the TeV scale. For all these models, flavour physics imposes very stringent bounds, requiring a flavour structure not far from that in the SM. This was the main rationale underlying 
the hypothesis of Minimal flavour Violation (MFV). The MFV hypothesis is an ansatz for the flavour structure of NP that assumes that the SM Yukawa couplings are the only sources of flavour non-degeneracy even beyond the SM.
This requirement is nowadays partially relaxed by the absence of direct signals of NP in high-\pt experiments, allowing non-trivial modifications from the strict MFV.
This example illustrates nicely the importance of flavour physics in reconstructing the structure of any NP model addressing the electroweak hierarchy problem. But it also reveals its complementarity with the high-\pt experiments, where improved direct bounds relax the flavour structure requirements.

If there are new particles which couple to the SM quarks or leptons,  then, in general, there are corresponding new flavour parameters. Measuring them would be very important in order to understand the structure of NP.  This has been studied in great detail in the context of Supersymmetry (SUSY) (alignment mechanism
of the soft-breaking terms) and in composite models 
(partial-compositeness mechanism).
In the specific case of low energy supersymmetry, the squark and slepton couplings may yield measurable effects in FCNC processes and \CP violating observables and may give rise to detectable CLFV transitions.  Observable \CP violation is also possible in neutral currents and in electric dipole moments, for which the SM predictions are below the near future experimental sensitivities. The supersymmetric flavour problems, namely the 
observation that TeV-scale SUSY models with generic parameters are excluded by FCNC and \CP violation measurements, can be alleviated in several scenarios: (i) universal squark masses (e.g., gauge mediation); (ii) quark--squark alignment, (e.g., horizontal symmetry); (iii) very heavy squarks (e.g., split SUSY). All viable models incorporate some of these ingredients. Conversely, if SUSY is discovered, mapping out its flavour structure, with the help of future more precise flavour tests,
may help answer questions about even higher scales, the mechanism of SUSY breaking and the way it is communicated to the Minimal Supersymmetric Standard Model (MSSM), etc.

\smallskip
\subsubsection{Nonperturbative QCD and its role in flavour physics} Of special interest are the theoretical uncertainties due to our incomplete understanding of QCD dynamics at low energies. In order to extract information on short-distance physics from weak decays of hadrons, knowledge of nonperturbative matrix elements, encoded in decay constants and form factors, is usually needed. 
Refinements in the effective-field-theory (EFTs) approaches exploiting heavy-quark and/or low-energy perturbative expansions and, especially, major progress in lattice QCD calculations seen in the last decade, make possible a full exploitation of the BSM flavour physics programme. Some of the hadronic uncertainties have already reached the per-mille level, e.g., the theoretical precision in the calculation of nonperturbative quantities crucial for the extraction of the CKM matrix element $|V_{us}|$, and many more are at the percent level, for instance, the theory prediction for the rare FCNC decay $B_q\to\mu^+\mu^-$. 

On the other hand, there are many other transitions 
that would benefit from further theoretical breakthroughs. 
In the past, large increases in available data always triggered new theory developments, and better understanding of the domain of applicability and accuracy of existing theoretical tools.  It can be anticipated that these fruitful cross-fertilizations will continue to occur in the HL-LHC era between flavour physics experiments and theory.  While there is a substantial suite of measurements whose interpretations will not be limited by hadronic uncertainties, the experimental programme can still benefit a lot from theoretical improvements.  For many observables, lattice QCD improvements are important.  The anticipated improvements in experimental precision also pose interesting challenges for lattice QCD, to robustly address isospin violating and electromagnetic effects in flavour observables.  For many nonleptonic decays, relevant for \CP violation, lattice QCD is unlikely to make a big impact. Nevertheless, developing new methods based on effective theories and testing existing approaches can be expected to improve the theoretical understanding of many observables, further enhancing the sensitivity of the experimental programme to possible BSM phenomena.

Understanding the nonperturbative structure of QCD of course has significant scientific merit per se, independent of the searches for NP. A very active area of research that will benefit from the flavour programme at HL/HE-LHC
is hadron spectroscopy. 
A plethora of new states, many of which were unexpected or show intriguing features, have been discovered at the $B$-factories, Tevatron and the LHC. The increase of data samples at the HL-LHC will make it possible to discover many more of these states and chart their quantum numbers and properties. Accommodating them into our theoretical understanding of the nonperturbative regime of QCD will be a major challenge for the next decades. 

\smallskip
\subsubsection{Unexpected discoveries}
It goes without saying that it is impossible to predict truly unexpected future discoveries.  However, it cannot be emphasized enough that the large increase in datasets has the potential to revolutionize the field by unexpected discoveries and trigger entirely new areas of experimentation.  It should be obvious that exact and approximate conservation laws should be tested as precisely as possible, especially when the experimental sensitivity can substantially increase.  Recall that the discovery of \CP violation itself was unexpected, in an experiment whose primary goal was checking an anomalous kaon regeneration result.  
 New particles with surprising properties were in fact discovered at each of BaBar, Belle, and LHCb, respectively: the discoveries of the $D_{sJ}(2317)$ meson with a mass much below expectations, the discovery of the unexpectedly narrow charmonium-like state $X(3872)$ and the $Z(4430)$, and the discovery of pentaquarks.  
Thus, beside the ``classical'' searches for flavour-violating processes mentioned so far, both in quark and lepton sectors, other searches like those related to dark sectors in many channels or  BSM searches not yet conceived will all form important parts of the flavour physics programme in the HL-LHC era.

\smallskip
\subsection{Experimental considerations and the breadth of flavour physics} 
At the end of the HL-LHC the useful datasets will have increased by a factor of order 30--100 compared to the present ones. However, due to improvements in detector capabilities and changing running conditions, robust estimates of sensitivity improvements are complicated tasks 
discussed in details
in the next sections.  At LHCb, in most analyses, one may expect faster improvements in the results than simply scaling with collected total luminosity, due to improvements in detector capabilities in the upcoming upgrades.
At ATLAS and CMS the large number of interactions per bunch crossing during the HL-LHC will be a challenge. However, the upgraded detectors will have higher granularity and timing information to mitigate pileup effects \cite{Collaboration:2650976,ATL-PHYS-PUB-2019-005}. 
It is important to pursue as broad a programme as possible, since several key channels are expected to remain competitive with LHC. 

In many cases Belle-II and tau-charm factories such as BES III will provide competition and cross-checks of LHCb results.  However, especially in the very low rate modes, such as $B_d\to \mu^+\mu^-$, it is ATLAS and CMS and not Belle-II which are expected to best compete with LHCb.  If there are anomalies in $B_s$, and especially in $\Lambda_b$ decays, they can only be cross-checked at the LHC experiments.
 
As mentioned above, our ignorance about BSM physics requires a diversified programme that,  
even within the flavour-physics domain, calls for a large set of complementary measurements. 
To properly identify the BSM model, if deviations are observed, measuring its imprint on different observables is very important, as stressed, e.g., in Ref. \cite{Buras:2013ooa}.
These measurements cannot all be 
performed at a single facility. There is full complementarity and many potential synergies 
in case some BSM signal emerges, among different $b$-hadron decays ($B_{u,d}$, 
$B_s$, $\Lambda_b$, etc.), \CP violating and rare processes involving  
charm and kaons, as well as possible FCNC transitions with top-quark. For instance, the measurements of theoretically precisely known $s\to d\nu\nu$ FCNC transitions are expected from NA62 and KOTO, and will be directly complementary to the results from the flavour programme at the LHC.  Such measurements of different flavour transitions are important to determine the BSM flavour structure, while measurements of the same quark level transition, but with different hadronic initial and final states determine the chiral structure of the BSM model. 
\smallskip

\subsubsection{Key experimental capabilities at ATLAS and CMS}

The upgraded high-luminosity LHC (HL-LHC) will deliver to the CMS and ATLAS experiments proton-proton collisions at a center-of-mass 
energy of 14 TeV for a total integrated luminosity of about 3000\,fb$^{-1}$. 
 This goal will be achieved through a high instantaneous luminosity which implies up to 200 proton-proton interactions per bunch-crossing. In this regime, the experimental sensitivity to new physics is enhanced and complemented by flavour physics measurements, with sensitivities in specific modes 
(e.g., $B_{s,d} \to \mu\mu$, $B^0_s \to J/\psi \phi$, $B^0 \to K^{*0}\mu\mu$ ) comparable to those of dedicated experiments.
The ability of general purpose detectors to make precision heavy flavour measurements has been clearly demonstrated
by the results from Run-1 and Run-2 data.
HL-LHC can be a unique test bench for $B$ physics studies in ATLAS and CMS: $\sim$ 10$^{15}$ $b\bar{b}$ pairs will be produced for the integrated luminosity goal.

ATLAS and CMS will exploit this potential thanks to some projected Phase-2 upgrades which promise good detection capability at low momenta, good pileup effect mitigation and even, in some cases, an improved performance. 
 Examples are the new inner trackers, improvements of the muon systems, topological trigger capabilities, and the possibility to use tracking in the early stages of the trigger chain \cite{Collaboration:2650976,ATL-PHYS-PUB-2019-005}.
 
The  high integrated luminosity expected will allow ATLAS and CMS to study some rare processes at a precision never attained before. The excellent tracking and muon identification performances are highlighted by a number of  benchmark channels, $B_{s,d} \to \mu\mu$, $B^0 \to K^{*0} \mu\mu$, $B_s \to J/\psi \phi$, and $\tau \to 3\mu$, that are used for projections. 
Precision measurements at the level of $5\%$ to $10\%$ for the $B^0_s \to \mu^+\mu^-$  branching fraction,
are expected, along with the observation of the 
$B^0 \to \mu^+ \mu^-$ decay with more than 5$\sigma$, and a measurement of the $B^0_s \to \mu^+\mu^-$ effective lifetime with a $3\%$ statistical precision.
The sensitivity to the \CP-violating phase $\phi_s$ in the $B^0_s \to J/\psi \phi$ mode is estimated to be at the level of $\sim$ 5 mrad, 
i.e., a factor of $\sim$ 20 better than the corresponding Run-1 analyses values (a factor of $\sim 5 $ with respect to the current combination of $b \to \bar c c s$ measurements). 
The uncertainty on the angular variable $P_5'$ in $B^0 \to K^{*0} \mu^+ \mu^-$  as a function 
of the dimuon squared invariant mass ($q^2$ ) is expected to improve by a factor of 15 with respect to the published Run-1 measurements. 
With the HL-LHC high statistics the $B^0 \to K^{*0} \mu^+ \mu^-$ analysis can be performed in narrow bins of $q^2$ to reach a more precise determination of the angular observables.
Finally, the $\tau \to 3 \mu$ decay is expected to be probed down to $\mathcal{O} (10^{-9})$.

The lack of particle-ID detectors is bound to limit the investigation of fully hadronic final states at ATLAS and CMS. Nevertheless, some capability is retained through the early use of tracking in the trigger selection.  The $B_s \to \phi \phi \to 4K$ decay is an example of a hadronic final state that would benefit from the tracking performance at trigger level and the $\phi$ resonance signature.
Furthermore, the precision time information from the 
timing detector \cite{CMS:TPMTD} will bring new and unique capabilities to the detectors in the heavy flavour sector.
 
The heavy flavour programme at ATLAS and CMS requires dedicated low-\pt triggers, in contention for bandwidth with high-\pt measurements and searches. The physics scenario at the time of HL-LHC will motivate the optimal trigger bandwidth allocation for low-\pt studies. Indeed, considering the tenfold increase in the High Level Trigger rates and pileup mitigation, it could be conceivable to think of analysis dedicated streams to be performed with the whole 3000 fb$^{-1}$ statistics or in dedicated runs, with minimal impact on the 
high-\pt physics. 
Still unexplored options, such as 40MHz data scouting, will be also studied. Furthermore, the high-\pt searches in ATLAS and CMS will allow for a programme of measurements which are complementary to the low-\pt flavour investigations and will help to build a coherent theoretical picture. 

\subsubsection{Key experimental capabilities at LHCb}
\begin{figure}[t]
\begin{center}
\includegraphics[width=0.75\textwidth]{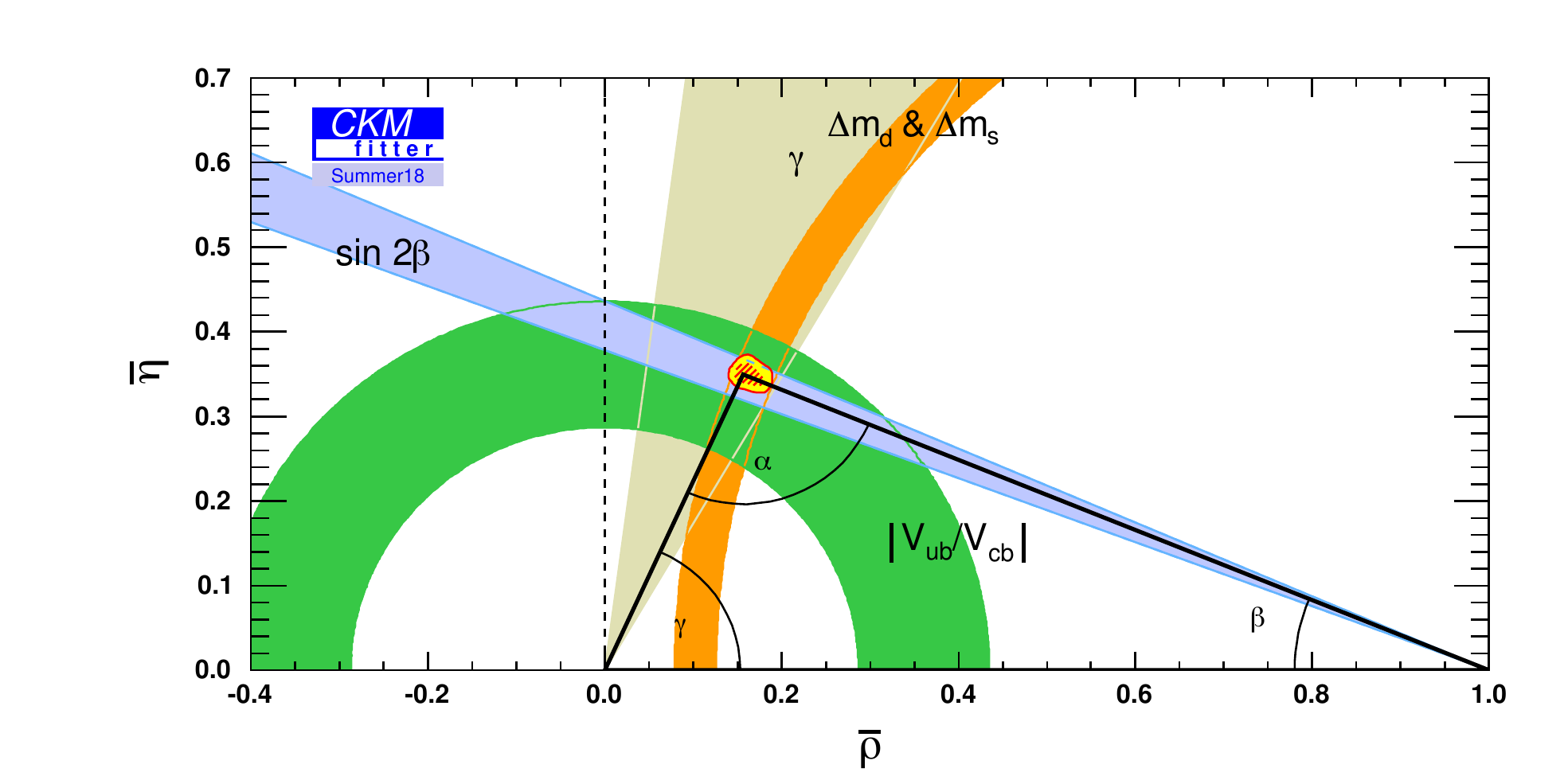}

\begin{minipage}{0.75\textwidth}
\includegraphics[width=\textwidth]{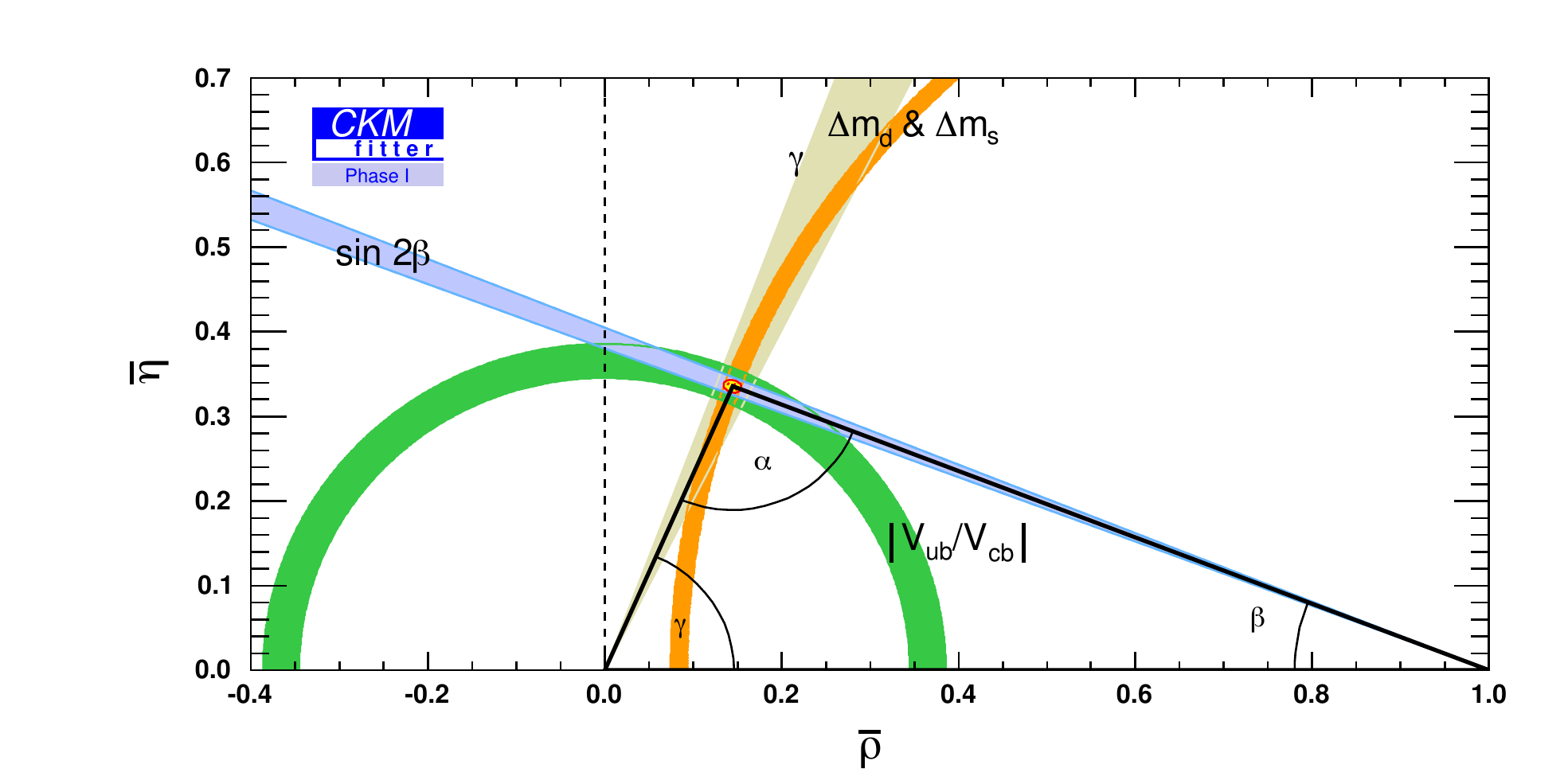}\vspace{-5.36cm}
\mbox{\hspace{7.9cm}\includegraphics[width=0.26\textwidth]{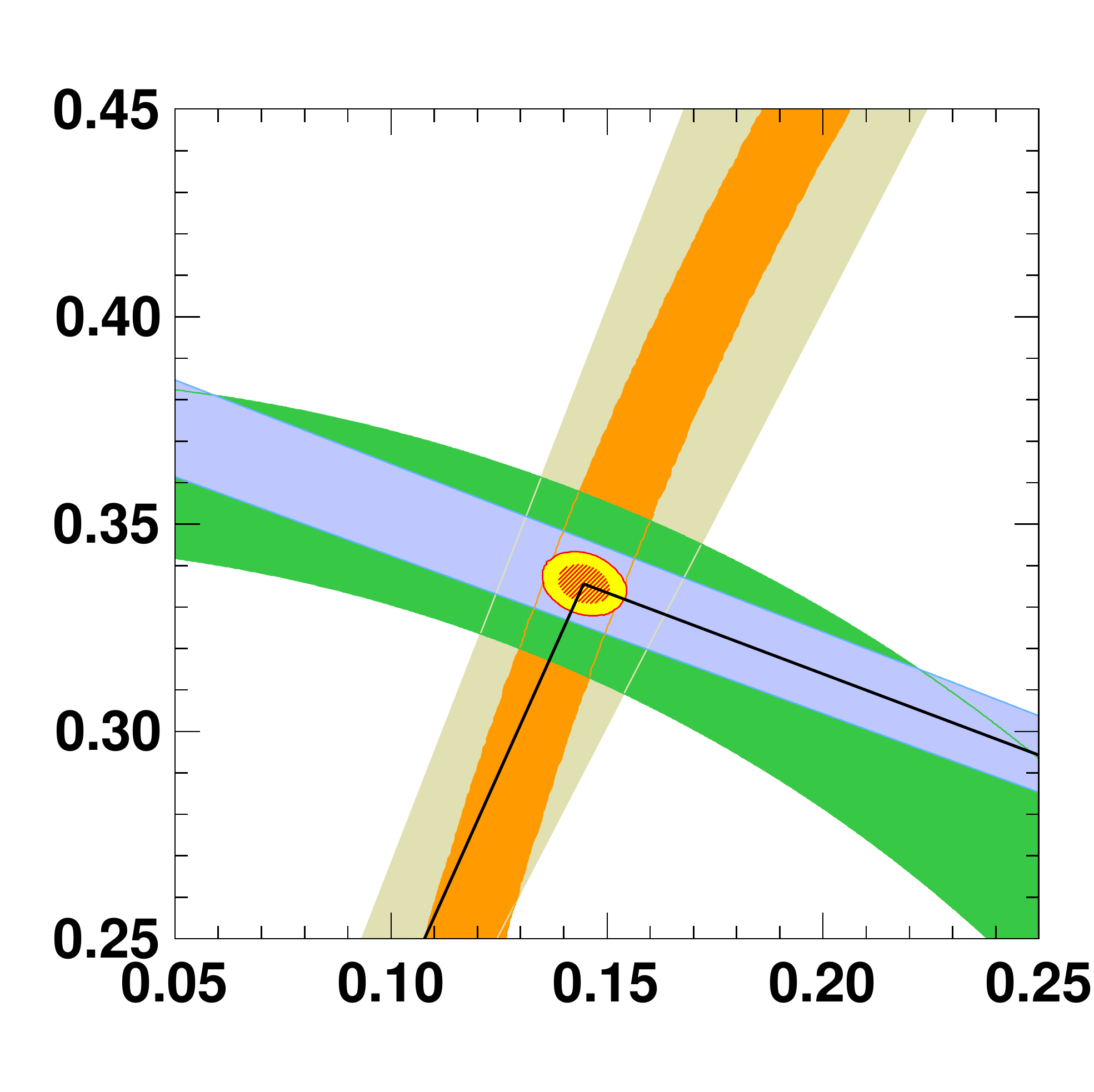}}
\vspace{2.25cm}
\end{minipage}

\begin{minipage}{0.75\textwidth}
\includegraphics[width=\textwidth]{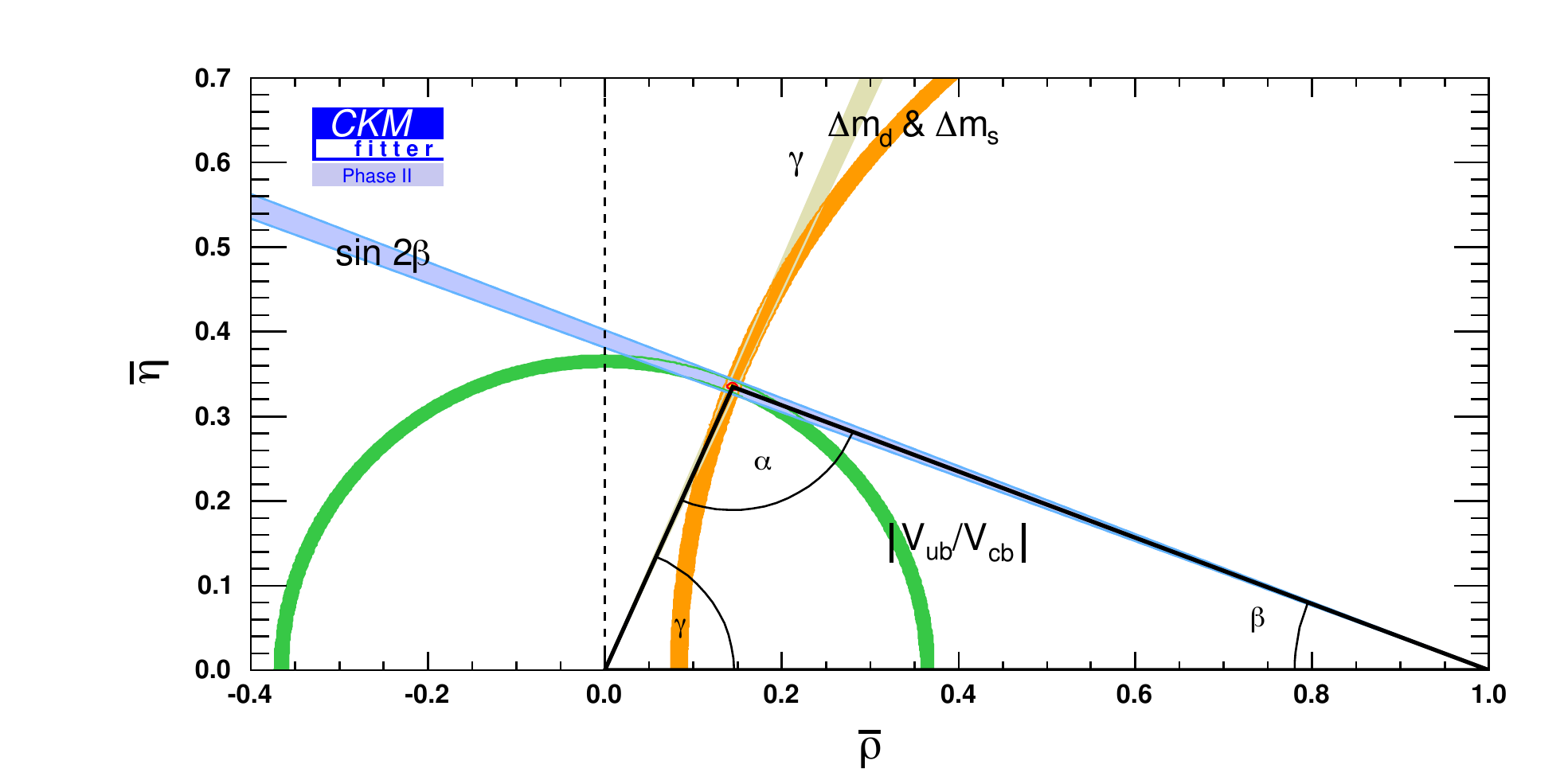}\vspace{-5.36cm}
\mbox{\hspace{7.9cm}\includegraphics[width=0.26\textwidth]{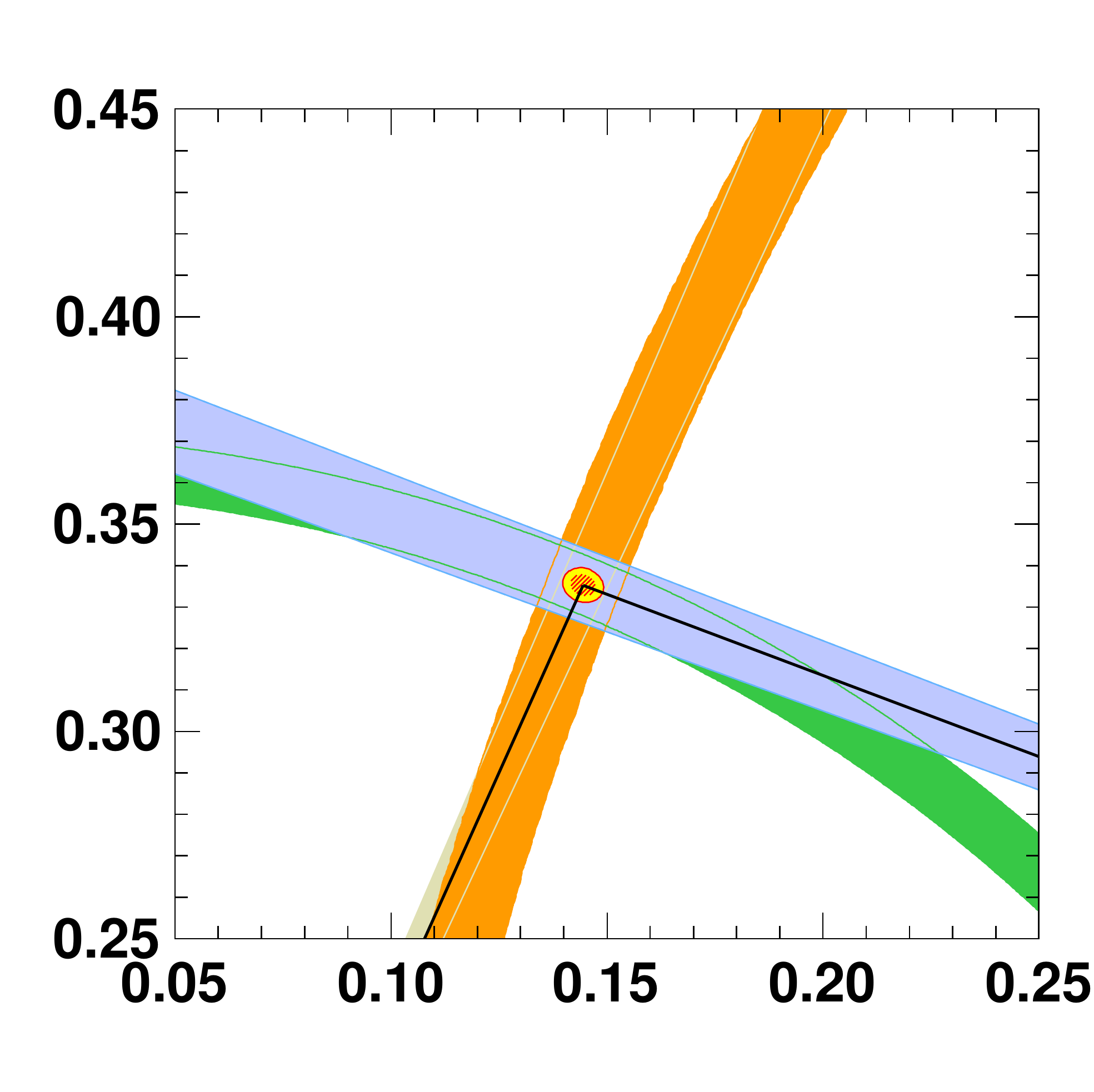}}
\vspace{2.25cm}
\end{minipage}

\caption{\small  Evolving constraints in the $\bar{\rho}-\bar{\eta}$ plane from LHCb measurements and lattice QCD calculations, alone, with current inputs (2018), and the anticipated improvements from the data accumulated by 2025  (23\,fb$^{-1}$) and 2035 (300\,fb$^{-1}$). More information on the fits may be found in Sec.~\ref{sec:metro-CKM} and~\cite{Bediaga:2018lhg}.
 }
\label{fig:UTprojection}
\end{center}
\end{figure}

The Upgrade II of LHCb will enable a very wide range of flavour observables to be determined with unprecedented precision, which will give the experiment sensitivity to NP scales several orders of magnitude above those accessible to direct searches. 
The expected uncertainties for a few key measurements  with 300\invfb are presented in Table~\ref{tab:physummary}. The future LHCb estimates are all based on extrapolations from current measurements, and take no account of detector improvements apart from an approximate factor two increase in efficiency for hadronic modes, arising from the full software trigger that will be deployed from Run~3 onwards. Three principal arguments motivate the Upgrade II of LHCb, and the full exploitation of the HL-LHC for flavour physics.
\begin{enumerate}
\item
There is a host of measurements of {\it theoretically clean} observables, such as the \CP-violating phase $\gamma$, the lepton-universality ratios $R_K$, $R_{K^\ast}$ etc., or the ratio of branching fractions $R\equiv {\cal{B}} (\Bd \rightarrow \mu^+ \mu^-)$/${\cal{B}} (\Bs \rightarrow \mu^+ \mu^-)$, where knowledge will still be statistically limited after Run~4.  The same conclusion applies for other observables such as $\phi_s$ and $\sin 2 \beta$, where strategies exist to monitor and control possible Penguin pollution. The HL-LHC and the capabilities of LHCb Upgrade II offer a unique opportunity to take another stride forward in precision for these quantities.  Advances in lattice-QCD calculations will also motivate better measurements of other critical observables, e.g. $|\Vub|/|\Vcb|$.  

The anticipated impact of the improved knowledge of Unitarity Triangle parameters can be seen in Fig.~\ref{fig:UTprojection}, which shows the evolving constraints in the $\bar{\rho}-\bar{\eta}$ plane from LHCb inputs and lattice-QCD calculations, alone.   The increased sensitivity will allow for extremely precise tests of the CKM paradigm. In particular, it will permit the tree-level observables, which provide SM benchmarks, to  be assessed against  those with loop contributions, which are more susceptible to NP.  In practice, this already very powerful ensemble of constraints will be augmented by complementary measurements from Belle-II, particularly in the case of $|\Vub|/|\Vcb|$.

\begin{table}[ptb!]
\caption{\small Summary of prospects for future measurements of selected flavour observables for LHCb.  The projected LHCb sensitivities take no account of potential detector improvements, apart from in the trigger. See subsequent chapters for definitions. \vspace*{0.1cm}}\label{tab:physummary}
\centering
\begin{tabular}{lrrr} \hline\hline
Observable  &   Current LHCb  &  LHCb 2025  &  Upgrade II  \\ \hline
\underline{\bf EW Penguins}  &  & & \\
$R_K$ $(1<q^2<6\,{{\rm GeV}^2c^4})$            & 0.1\cite{LHCb-PAPER-2014-024}  &  0.025   &   0.007  \\
$R_{K^\ast}$ $(1<q^2<6\,{{\rm GeV}^2c^4})$  & 0.1\cite{LHCb-PAPER-2017-013}  &  0.031  &   0.008 \\
$R_\phi$,  $R_{pK}$, $R_{\pi}$         & -- & 0.08, 0.06, 0.18 & 0.02, 0.02, 0.05 \\
\rule{0pt}{3ex}\underline{\bf CKM tests}  &   & &\\
$\gamma$, with $\Bs \to \Ds\Km$ & $(^{+17}_{-22})^\circ$~\cite{LHCb-PAPER-2017-047} & 4$^\circ$ & 1$^\circ$ \\
$\gamma$, all modes & $(^{+5.0}_{-5.8})^\circ$~\cite{LHCb-CONF-2018-002} & 1.5$^\circ$ & $0.35^\circ$ \\
$\sin 2 \beta$, with $\Bd \to J/\psi\KS$ & $0.04$\cite{LHCb-PAPER-2017-029} & $0.011$ & $0.003$ \\
$\phi_s$, with $\Bs \to J/\psi \phi$ & 49 mrad\cite{LHCb-PAPER-2014-059}  & 14 mrad  & 4 mrad \\
$\phi_s$, with $\Bs \to \Dsp\Dsm$   & 170 mrad\cite{LHCb-PAPER-2014-051}  & 35 mrad  & 9 mrad \\
$\phi_s^{s{\bar{s}s}}$, with $\Bs \to \phi\phi$  & 154 mrad\cite{LHCb-PAPER-2014-026} & 39 mrad & 11 mrad \\
$a^s_{\rm sl}$ & $33 \times 10^{-4}$\cite{LHCb-PAPER-2016-013}  &  $10 \times 10^{-4}$  & $3 \times 10^{-4}$\\
$|\Vub|/|\Vcb|$  & 6\%\cite{LHCb-PAPER-2015-013}  & 3\%   & 1\% \\
\rule{0pt}{3ex}\underline{\bf ${\bm{B^0_s, B^0} {\to} \bm{\mu^+}\bm{\mu^-}}$}  &  & \\
${\cal{B}} (\Bd \rightarrow \mu^+ \mu^-)$/${\cal{B}} (\Bs \rightarrow \mu^+ \mu^-)$ & 90\%\cite{LHCb-PAPER-2017-001}  & 34\%  & 10\% \\
$\tau_{\Bs \to \mu^+\mu^-}$ &22\%\cite{LHCb-PAPER-2017-001} & 8\%  & 2\% \\
$S_{\mu\mu}$ & -- &  -- & 0.2\\
\rule{0pt}{3ex}\underline{\bf\boldmath {${b \to c \ell^-\bar{\nu_l}}$} LUV studies}  &    & \\
$R(D^\ast)$ & 0.026\cite{LHCb-PAPER-2015-025,LHCb-PAPER-2017-027}  & 0.0072 & 0.002\\
$R(J/\psi)$ & 0.24\cite{LHCb-PAPER-2017-035} & 0.071 & 0.02 \\
\rule{0pt}{3ex}\underline{\bf Charm}  &   &  &\\
$\Delta A_{\CP}(KK-\pi\pi)$ &  $8.5 \times 10^{-4}$\cite{LHCb-PAPER-2015-055}  & $1.7\times 10^{-4}$ &  $3.0 \times 10^{-5}$ \\
$A_\Gamma$ ($\approx x \sin\phi$) & $2.8 \times 10^{-4}$\cite{LHCb-PAPER-2016-063} & $4.3 \times 10^{-5}$ &  $1.0 \times 10^{-5}$ \\
${x \sin \phi}$ from $\Dz \to K^+\pi^-$ & $13\times 10^{-4}$\cite{LHCb-PAPER-2017-046} & $3.2 \times 10^{-4}$ & $8.0 \times 10^{-5}$ \\
${x \sin \phi}$ from multibody decays & -- & ($K3\pi$) $4.0\times 10^{-5}$ & ($K3\pi$) $8.0 \times 10^{-6}$ \\
\hline\hline
\end{tabular}
\end{table}

The increasing precision of observables from measurements of statistically-limited FCNC processes will provide significant improvements in sensitivity to the scale of NP.  As an example, Table~\ref{tab:wilson_sum} shows the expected improvement with integrated luminosity in the knowledge of the Wilson coefficients $C_9$ (vector current) and $C^\prime_{10}$ (right-handed axial-vector current), and the corresponding 90\% exclusion limits to the NP scale $\Lambda$ under various scenarios. The reach for generic NP at tree-level in Upgrade~II is found to exceed 100~TeV. 

\item
It will be essential to {\it widen the set of observables under study} beyond those accessible at the current LHCb experiment or its first upgrade, {\it e.g.} by including additional important measurements involving  $b \to s \ellell$, $b \to d \ellell$ and $b \to c \ell^-\bar{\nu_l}$ decays.  Improving our knowledge of the flavour sector both through better measurements and through new observables will be essential in searching for and then characterising NP in the HL-LHC era.

\item
Due to its ability to reconstruct and analyze all collisions in real-time and the statistical power of the HL-LHC dataset, LHCb \upgradetwo will be able to collect a unique dataset for hadronic spectroscopy. 
This will enable not only the precise understanding of higher-excited states of mesons and baryons, but also a detailed and broad understanding of multiquark systems, containing (or not) multiple heavy quarks, and other yet-to-be-discovered exotic states of matter.
While not directly sensitive to BSM effects, these measurements will play an important role in sharpening our understanding of QCD at the energy scales relevant for flavour physics, and hence make an important contribution to the accurate interpretation of any  BSM anomalies observed. 
\end{enumerate}
The intention to operate a flavour-physics experiment at luminosities of $10^{34}\,{\rm cm^{-2}s^{-1}}$ is already an ambitious one, but the planned improvements to the detector's capabilities will extend the physics gains still further.  These gains are not included in Table~\ref{tab:physummary} as full simulations have not yet been performed,  but a summary of the expected benefits can be found in~\cite{Bediaga:2018lhg}. It is intended to take first steps towards some of these detector enhancements already in LS3, before the start of the HL-LHC, thereby improving the performance of the first LHCb upgrade, and laying the foundations for Upgrade II. Finally, it must be emphasised that the raw gain in sample sizes during the HL-LHC era will have great consequences for the physics reach, irrespective of any detector improvements.   The energy scale probed by virtual loops in flavour observables will rise by a factor of up to 1.9 with respect to the pre-HL-LHC era, with a corresponding gain in discovery potential similar to what will apply for direct searches if the beam energy is doubled, as proposed for the HE-LHC.

\begin{table}[t]
\caption{\small
Uncertainty on Wilson coefficients and 90\% exclusion limits on NP scales $\Lambda$ for different data samples.  The  $C_9$ analysis is based on the ratio of branching fractions $R_K$ and $R_{K^\ast}$ in the range $1<q^2<6\,{\rm GeV}^2/c^4$.  The $C^\prime_{10}$  analysis exploits the angular observables $S_i$ from the decay $B^0 \to K^{\ast 0} \mu^+\mu^-$ in the ranges $1 < q^2 < 6\,{\rm GeV}^2/c^4$ and  $15 < q^2 < 19\,{\rm GeV}^2/c^4$.   The limits on the scale of NP, $\Lambda_{\rm NP}$, are given for the following scenarios: tree-level generic, tree-level minimum flavour violation, loop-level generic and loop-level minimal flavour violation.  More information on the fits may be found in~\cite{Bediaga:2018lhg}.} \label{tab:wilson_sum}
  \begin{center}
  \begin{tabular}{lrrr}\hline\hline
    Integrated Luminosity & $3\invfb$ & $23\invfb$ & $300\invfb$\\ \hline
    \multicolumn{4}{c}{$R_K$ and $R_{K^*}$ measurements}\\\hline
    $\sigma(C_9)$ & 0.44 & 0.12 & 0.03\\
    $\Lambda^\textrm{tree\,generic}~[\tev]$ & 40 & 80 & 155\\
    $\Lambda^\textrm{tree\,MFV}~[\tev]$ & 8 & 16 & 31\\
    $\Lambda^\textrm{loop\,generic}~[\tev]$ & 3 & 6 & 12\\
    $\Lambda^\textrm{loop\,MFV}~[\tev]$ & 0.7 & 1.3 & 2.5\\ \hline
    \multicolumn{4}{c}{$\decay{\Bd}{\Kstarz\mumu}$ angular analysis}\\\hline   
    $\sigma^\textrm{stat}(S_i)$ & 0.034--0.058 & 0.009--0.016 & 0.003--0.004\\
    $\sigma(C_{10}^\prime)$ & 0.31 & 0.15 & 0.06 \\
    $\Lambda^\textrm{tree\,generic}~[\tev]$ & 50 & 75 & 115\\
    $\Lambda^\textrm{tree\,MFV}~[\tev]$ & 10 & 15 & 23\\
    $\Lambda^\textrm{loop\,generic}~[\tev]$ & 4 & 6 & 9\\
    $\Lambda^\textrm{loop\,MFV}~[\tev]$ & 0.8 & 1.2 & 1.9\\
    \hline\hline
    \end{tabular}
    \end{center}
\end{table}

%% file: section2/section.tex
\def\vub {\ensuremath{|V_{\uquark\bquark}|}\xspace}
\def\vcb {\ensuremath{|V_{\cquark\bquark}|}\xspace}
\def\qq {\ensuremath{q^{2}}\xspace}
\def\PMuNu{\decay{\Lb}{p \mun \neumb}}
\def\LcMuNu{\decay{\Lb}{\Lc \mun \neumb}}
\def\KMuNu{\decay{\Bs}{\Km \mup \neum}}
\def\DsMuNu{\decay{\Bs}{\Dsm \mup \neum}}
\def\Bq      {{\ensuremath{\B^0_\quark}}\xspace}
\def\Bqb     {{\ensuremath{\Bbar{}^0_\quark}}\xspace}
\def\dbtaunu {\decay{\B}{D^{(*)} \tau \nu}}
\def\dbmunu {\decay{\B}{D^{(*)} \mu \nu}}
\def\dbenu {\decay{\B}{D^{(*)} e \nu}}
\def\dmunu {\decay{\B}{D \mu \nu}}
\def\denu {\decay{\B}{D e \nu}}
\def\dsttaunu {\decay{\B}{D^{*} \tau \nu}}
\def\dstztaunu {\decay{\B}{D^{*0} \tau \nu}}
\def\dstptaunu {\decay{\B_{d}}{D^{*-} \tau^{+} \nu}}
\def\dstmunu {\decay{\B}{D^{*} \mu \nu}}
\def\dstenu {\decay{\B}{D^{*} e \nu}}
\def\dstpmunu {\decay{\B}{D^{*+} \mu \nu}}
\def\dstzmunu {\decay{\B}{D^{*0} \mu \nu}}
\def\dstlnu {\decay{\B}{D^{*} \ell \nu}}
\def\dblnu {\decay{\B}{D^{(*)} \ell \nu}}
\def\RDst {\ensuremath{\mathcal{R}(\Dstar)}\xspace}
\def\RDstp {\ensuremath{\mathcal{R}(\Dstarp)}\xspace}
\def\RD {\ensuremath{\mathcal{R}(D)}\xspace}
\def\RDb {\ensuremath{\mathcal{R}(D^{(*)})}\xspace}
\def\vub {\ensuremath{|V_{\uquark\bquark}|}\xspace}
\def\vcb {\ensuremath{|V_{\cquark\bquark}|}\xspace}
\def\qq {\ensuremath{q^{2}}\xspace}
\def\PMuNu{\decay{\Lb}{p \mun \neumb}}
\def\LcMuNu{\decay{\Lb}{\Lc \mun \neumb}}
\def\KMuNu{\decay{\Bs}{\Km \mup \neum}}
\def\DsMuNu{\decay{\Bs}{\Dsm \mup \neum}}
\def\vub {\ensuremath{|V_{ub}|}\xspace}
\def\vcb {\ensuremath{|V_{cb}|}\xspace}
\def\qq {\ensuremath{q^{2}}\xspace}
\def\PMuNu{\decay{\Lb}{p \mun \neumb}}
\def\LcMuNu{\decay{\Lb}{\Lc \mun \neumb}}
\def\KMuNu{\decay{\Bs}{\Km \mup \neum}}
\def\DsMuNu{\decay{\Bs}{\Dsm \mup \neum}}
\def        \DtoKspp        {\ensuremath{\D\to\KS\pip\pim}\xspace}
\def        \DtoKskk        {\ensuremath{\D\to\KS\Kp\Km}\xspace}
\def        \DtoKshh        {\ensuremath{\D\to\KS\hadron^+\hadron^-}\xspace}
\def        \BtoDpi         {\ensuremath{\Bpm\to\D\pipm}\xspace}
\def        \BtoDK          {\ensuremath{\Bpm\to\D\Kpm}\xspace}
\def        \BztoDstmuX     {\ensuremath{\Bz\to\Dstarpm\mu^\mp\nu_\mu X}\xspace}
\def        \hp            {\ensuremath{\hadron^+}\xspace}
\def        \hm            {\ensuremath{\hadron^-}\xspace}
\def        \BtoDKst       {\ensuremath{\Bz\to\PD\Kstarz}\xspace}
\def        \mpsq          {\ensuremath{m_{+}^{2}}\xspace}
\def        \mmsq          {\ensuremath{m_{-}^{2}}\xspace}
\def        \mpmsq          {\ensuremath{m_{\pm}^{2}}\xspace}
\def        \mmpsq          {\ensuremath{m_{\mp}^{2}}\xspace}
\def        \rb            {\ensuremath{r_B}\xspace}
\def        \db            {\ensuremath{\delta_B}\xspace}
\def        \g             {\ensuremath{\gamma}\xspace}
\def        \dpos          {\ensuremath{(\mpmsq,\mmpsq)\xspace}}
\def        \dbpos         {\ensuremath{(\mmpsq,\mpmsq)\xspace}}
\newcommand{\rbk}{\ensuremath{r_{B}^{DK}}\xspace}

\section{Testing the CKM unitarity and related observables}
{\it \small Authors (TH): J\'er\^ome Charles, Marco Ciuchini,  Olivier Deschamps, S\'ebastien Descotes-Genon, Luca Silvestrini, Vincenzo Vagnoni.}
\label{sec:metro-CKM}

In the SM, the weak charged-current transitions mix quarks of different generations, which is encoded in the
unitary Cabibbo-Kobayashi-Maskawa (CKM) matrix~\cite{Cabibbo:1963yz,Kobayashi:1973fv}.
The SM does not predict the values of the weak flavour-couplings, and so all matrix elements must be measured experimentally.
However, the unitary nature of the CKM matrix, and the assumptions of the SM, impose relations between the elements that are
often expressed graphically in the complex plane as the so-called unitarity triangle.
Overconstraining the apex of this unitarity triangle from tree- and loop-level quark mixing processes is therefore a powerful way to probe for virtual BSM effects at mass scales complementary or superior to those which can be directly searched for at the HL-LHC. \Blu{As we shall see below, in many cases such indirect probes of BSM physics will not be limited by either experimental or theoretical systematics in the HL-LHC era.}

\subsection{Structure of the CKM matrix}

 The part of the SM 
 Lagrangian which is relevant for describing quark mixing is
\begin{equation}\label{eq:chargeInteraction}
\mathcal{L}_{W^{\pm}} = -
\frac{g}{\sqrt{2}}\left(V_{\rm
CKM} \right )_{ij}  \big(\overline{u}_{i}\gamma^{\mu}\tfrac{(1-\gamma_5)}{2}d_{j}\big) W_{\mu}^{+} + \mathrm{h.c.},
\end{equation}
where $g$ is the electroweak coupling constant, and $V_{\rm
CKM}$ the unitary CKM matrix,
\begin{equation}\label{eq:ckmMatrix}
V_{\rm CKM} = \left(
\begin{array}{lcr}
V_{ud} & V_{us} & V_{ub} \\
V_{cd} & V_{cs} & V_{cb} \\
V_{td} & V_{ts} & V_{tb}
\end{array}
\right).
\end{equation}
The CKM matrix induces flavour-changing transitions inside and between generations in the charged currents at tree level ($W^\pm$ interaction). By contrast, there are no flavour-changing transitions in the neutral currents at tree level.

Experimentally, a strong hierarchy is observed  among the CKM matrix elements: transitions within the same generation are characterised by $V_{\rm CKM}$ elements of $\mathcal{O}(1)$, whereas
there is a suppression of $\mathcal{O}(10^{-1})$
between 1st and 2nd generations, $\mathcal{O}(10^{-2})$ between 2nd and 3rd and $\mathcal{O}(10^{-3})$ between 1st and 3rd.
This hierarchy is expressed by defining the four phase convention--independent
quantities,
\begin{equation}
\label{eq:lambda:A:rhoeta}
\lambda^2 =\frac{|V_{us}|^2}{|V_{ud}|^2+|V_{us}|^2}\,, \qquad
A^2\lambda^4=\frac{|V_{cb}|^2}{|V_{ud}|^2+|V_{us}|^2}\,, \qquad
\bar\rho+i\bar\eta=-\frac{V_{ud}V_{ub}^*}{V_{cd}V^*_{cb}}.
\end{equation}
The four independent quantities, $\lambda$, $A$, $\bar \rho$, $\bar \eta$,  fully determine the CKM matrix in the SM.

The CKM matrix can be expanded in powers of the small parameter
$\lambda$ (which corresponds to the Cabibbo parameter $\sin{\theta_{C}}\simeq
0.22$)~\cite{Wolfenstein:1983yz} by exploiting the unitarity of
$V_\text{CKM}$. This
expansion yields the following parametrisation, valid up to
$\mathcal{O}\big(\lambda^{6}\big)$,
\begin{equation}\label{eq:ckmWolfenstein4}
V_{\rm CKM} = \left(
\begin{array}{ccc}
1-\frac{1}{2}\lambda^{2}-\frac{1}{8}\lambda^{4} & \lambda & A\lambda^{3}\left(\bar\rho -i\bar\eta\right) \\
-\lambda +\frac{1}{2}A^{2}\lambda^{5}\left[1-2(\bar\rho +i\bar\eta)\right] & 1-\frac{1}{2}\lambda^{2}-\frac{1}{8}\lambda^{4}(1+4A^{2}) & A\lambda^{2} \\
A\lambda^{3}\left[1-(\bar\rho +i\bar\eta)\right]~ &
~-A\lambda^{2}+\frac{1}{2}A\lambda^{4}\left[1-2(\bar\rho
+i\bar\eta)\right]~ & ~1-\frac{1}{2}A^{2}\lambda^{4}
\end{array}
\right).
\end{equation}
The CKM matrix is complex, i.e., it contains a phase that cannot be rotated away, if $\bar\eta\ne 0$.  Furthermore, \CP is violated,  if
and only if $\bar\eta$ differs from zero. 

\begin{figure}[!tb]
\centering
\includegraphics[width=0.6\linewidth]{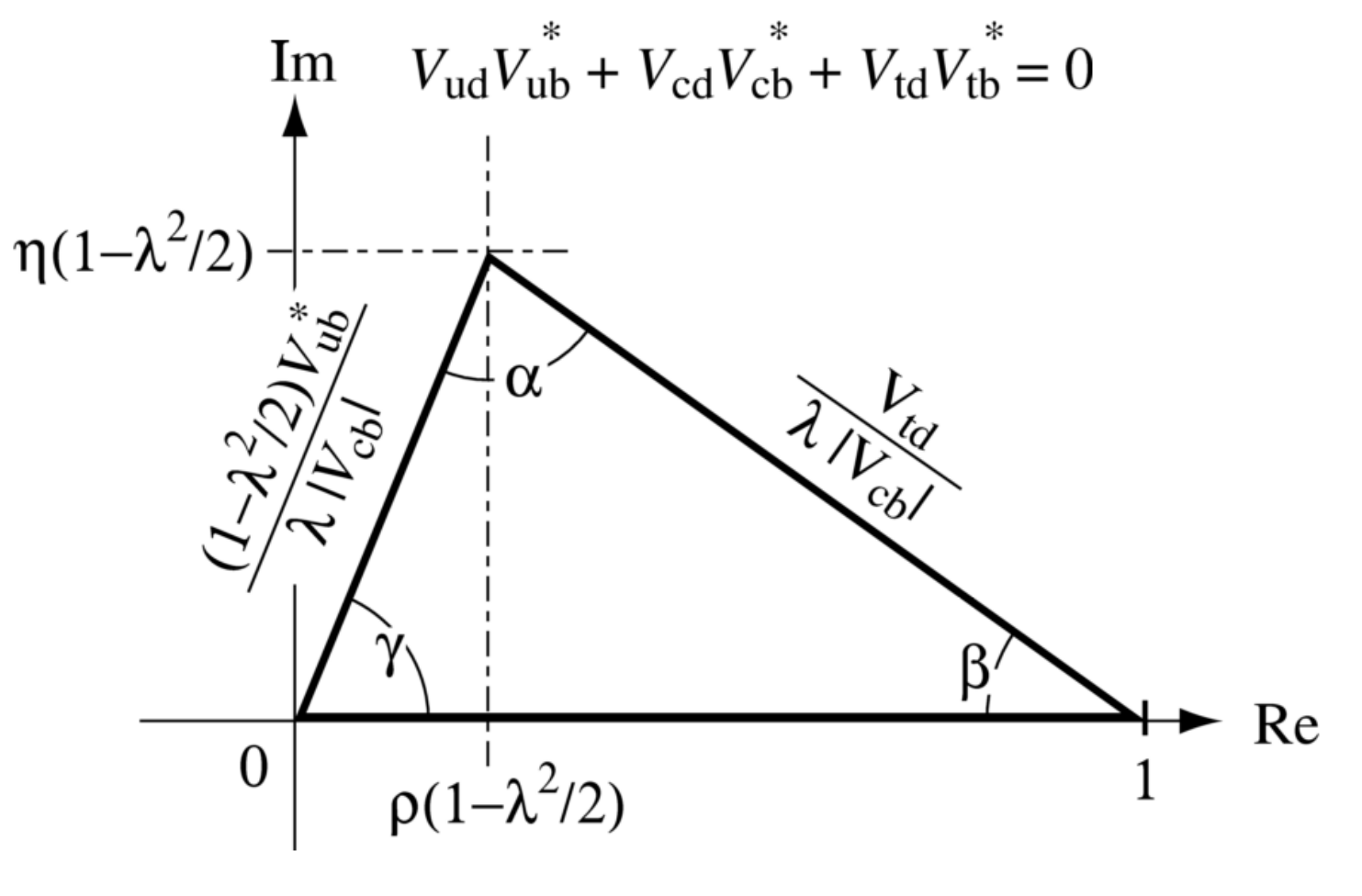} 
\caption{
The standard CKM unitarity triangle. The parameters $\rho$ and $\eta$ are defined as 
$\rho+i\eta=\big(\bar\rho+i\bar\eta\big)\big(1-\lambda^2/2+O(\lambda^4)\big)$, with $\bar \rho, \bar \eta$ defined in \eqref{eq:lambda:A:rhoeta}.
}
\label{fig:UTtriangle}
\end{figure}

Orthogonality relations can be written involving two columns or two rows of the unitary CKM matrix, and they can be represented as triangles in the complex plane. It is standard to focus on the following orthogonality relation,
\begin{equation}
\label{eq:orthog:rel}
{V_{ud}V_{ub}^{*}}+{V_{cd}V_{cb}^{*}}+{V_{td}V_{tb}^{*}} = 0,
\end{equation}
as the three products of CKM elements 
are of similar size, ${\mathcal O}(\lambda^3)$. Fig.~\ref{fig:UTtriangle} shows the standard unitarity triangle (UT), obtained from Eq.~\eqref{eq:orthog:rel} by rescaling the three terms in the orthogonality relation by $V_{cd}V_{cb}^{*}$. 

The apex of the UT is at $(\bar\rho,\bar\eta)$, while 
the angles 
are related to the CKM matrix elements as
\begin{equation}
\alpha= \arg{\left(-\frac{V_{td}V_{tb}^{*}}{V_{ud}V_{ub}^{*}}\right)}, \qquad
\beta=\arg{\left(-\frac{V_{cd}V_{cb}^{*}}{V_{td}V_{tb}^{*}}\right)}, \qquad
\gamma\equiv\arg{\left(-\frac{V_{ud}V_{ub}^{*}}{V_{cd}V_{cb}^{*}}\right)}.
\end{equation}

\subsection{Current status of the constraints}

\subsubsection{$|V_{ud}|,|V_{us}|,|V_{cd}|,|V_{cs}|$}

Accurate constraints on the first and second rows and columns of the CKM matrix come from leptonic decays, $\pi\to e\nu$, $K\to e\nu$, $K\to\mu\nu$, $\tau\to \pi\nu_\tau$,  $\tau\to K\nu_\tau$, $D\to \mu\nu$, $D_s\to \mu\nu$, and from semileptonic decays, $K\to\pi e\nu$, $D\to \pi e\nu$, $D_s\to Ke\nu$. The extraction of CKM matrix elements requires knowledge of hadronic inputs (decay constants for the leptonic decays, normalisations of the form factors at $q^2=0$ for the semileptonic decays) and electromagnetic/isospin corrections when available (i.e., for kaon and pion decays)~\cite{Antonelli:2010yf}. Another prominent input for the $|V_{ud}|$ determination comes from the consideration of the superallowed $\beta$ decays of 20 different nuclei~\cite{Hardy:2014qxa,Towner:2015woa,Hardy:2018zsb}, which provides a very accurate constraint on $|V_{ud}|$. There are also other constraints, but less powerful due to experimental uncertainties and/or theoretical systematics that are difficult to assess.

\subsubsection{$|V_{cb}|$ and $|V_{ub}|$} 

Tree-level semileptonic decays of beauty mesons and baryons allow for the extraction of $|V_{cb}|$ and $|V_{ub}|$. The current determination is dominated by the $B$-factories data on $B$ decays and by the measurement of $|V_{ub}|/|V_{cb}|$ from baryonic decays at LHCb. For $B$ decays, both inclusive and exclusive semileptonic decays have been used to extract $|V_{cb}|$ and $|V_{ub}|$. The two approaches have different sources of theoretical uncertainties: inclusive analyses rely on quark-hadron duality, involve hadronic matrix elements in subleading powers of the heavy quark expansion and, for $|V_{ub}|$, on additional hadronic quantities called shape functions; exclusive analyses require the knowledge of the relevant form factors over the entire kinematic range, a very difficult task for lattice QCD. Currently, the HFLAV averages for inclusive and exclusive determinations of $|V_{cb}|$ and $|V_{ub}|$ disagree at the $3\sigma$ level.
While recently the choice of the parameterization of the form factor dependence on the recoil for $B \to D^*$ decays has been shown to have a large impact on the extracted value of $|V_{cb}|$ \cite{Abdesselam:2018nnh,Bigi:2017njr,Grinstein:2017nlq}, the situation is still rather unclear. 

\begin{figure}[!tb]
\centering
\includegraphics[width=0.5\linewidth]{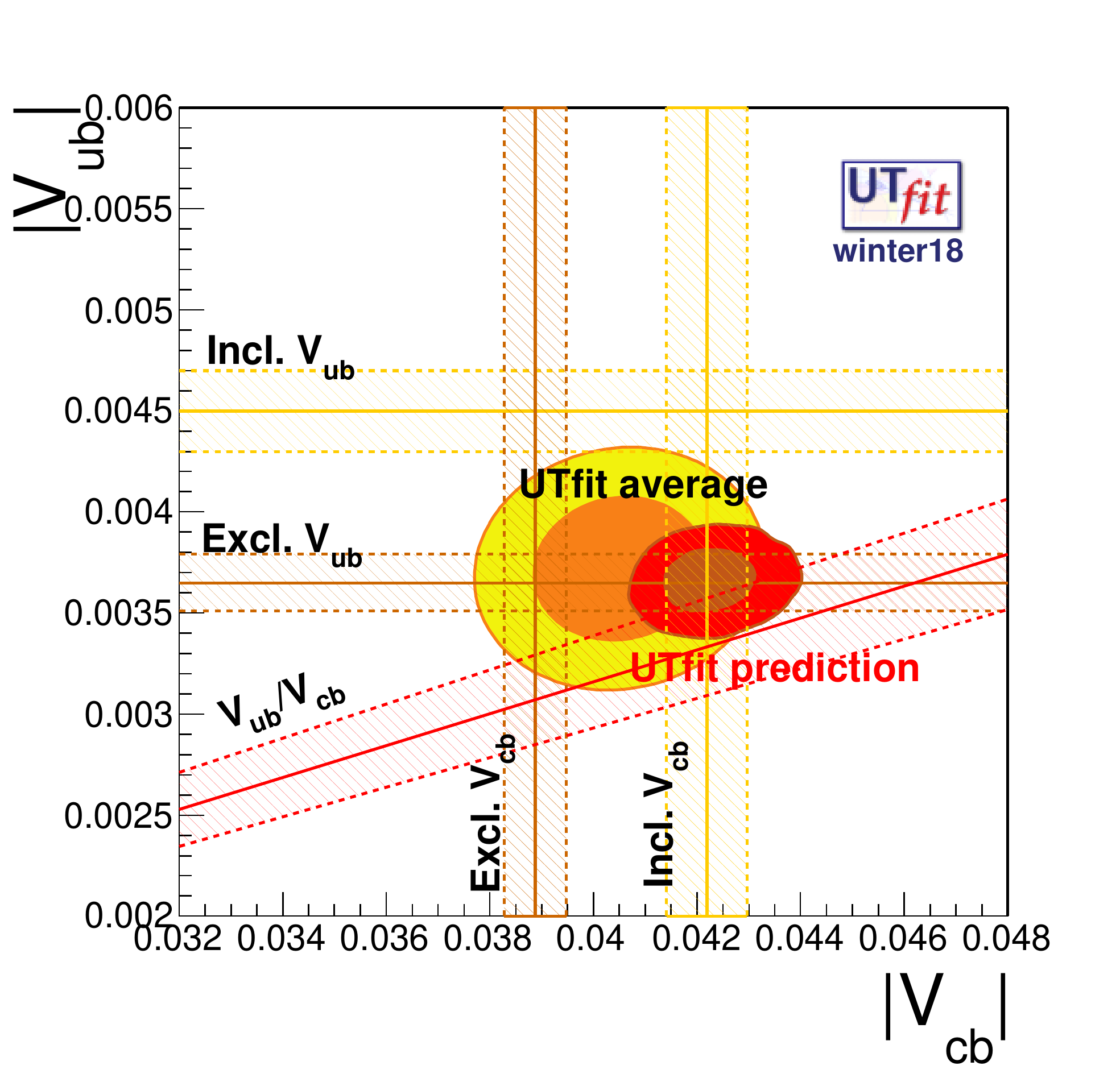} 
\caption{
Measurements of $|V_{cb}|$ and $|V_{ub}|$ from exclusive and inclusive $B$ decays (vertical and horizontal bands) and from \PMuNu and \LcMuNu (diagonal band), skeptical average and indirect determination of $|V_{cb}|$ and $|V_{ub}|$.
}
\label{fig:UTfitvubvcb}
\end{figure}

One way to deal with this, followed by the UTFit collaboration, is to assume that the uncertainty of each determination of $|V_{cb}|$ and $|V_{ub}|$ might have been underestimated, and perform a ``skeptic combination'' of all available data following the method of Ref.~\cite{DAgostini:1999niu}, which is equivalent to the PDG prescription in one dimension, and is a straightforward generalization to the two-dimensional case of $|V_{cb}|$ and $|V_{ub}|$. The result 
 is reported in Fig.~\ref{fig:UTfitvubvcb}, where in addition to the average we also plot the ``indirect determination'' obtained from all the other flavour observables.
 The numerical results of the skeptic average are $|V_{cb}| = 0.0405 \pm 0.0011$, $|V_{ub}| = 0.00374 \pm 0.00023$, with correlation $\rho = 0.09$. CKMfitter collaboration uses instead the exclusive value for $V_{cb}$ from $B\to D^* \ell \nu$  that was obtained using more model independent BGL parametrisation, and which is in agreement with both $B\to D \ell \nu$  and the inclusive extraction. This leads to the combined value $|V_{cb}|=(41.8\pm0.4({\rm exp})\pm0.6({\rm theory}))\cdot 10^{-3}$. For $V_{ub}$ the CKMfitter collaboration performs an R-fit of inclusive and exclusive determinations to obtain $|V_{ub}|=(3.98\pm0.08({\rm exp})\pm0.22({\rm LQCD}))\cdot 10^{-3}$.

Lattice QCD determinations of the form factors in a large kinematic range are under way. It is expected that they will allow Belle-II
to perform a much more parameterization-independent extraction of $V_{cb}$ from exclusive decays, hopefully reconciling it with the inclusive one. Also for $|V_{ub}|$, the much larger statistics should allow to get an insight on the origin of the discrepancy between inclusive and exclusive determinations, and provide a consistent determination of $|V_{ub}|$. 

\subsubsection{Angle $\alpha$}

The constraints on the CKM angle $\alpha$ are derived from the isospin analysis of the charmless decay modes $B\to\pi\pi$, $B\to\rho\rho$ and $B\to\rho\pi$ \cite{Gronau:1990ka,Lipkin:1991st,Snyder:1993mx}. This approach has the  interesting feature of being almost free from hadronic uncertainties. Assuming isospin symmetry and neglecting the electroweak penguin contributions, the amplitudes of the isospin $SU(2)$-conjugated modes are related. The measured branching fractions and asymmetries in the $\B^{\pm,0}\to(\pi\pi)^{\pm,0}$ and $B^{\pm,0}\to(\rho\rho)^{\pm,0}$ modes and the bilinear form factors in the Dalitz analysis of the neutral $B^{0}\to(\rho\pi)^0\to\pi^+\pi^-\pi^0$ decays thus provide enough observables to simultaneously extract the weak phase $\beta+\gamma=\pi-\alpha$ together with the hadronic tree and penguin contribution to each mode. The combination of the experimental data for the three decay modes above, mostly provided by the $B$-factories,  gives the world-average value at 68\% 
Confidence  Level (CL)\cite{Charles:2017evz}: 
$\alpha_{\rm dir} = (86.2^{+4.4}_{-4.0} \xspace\cup\xspace 178.4^{+3.9}_{-5.1})^\circ\,$.
The experimental uncertainty is so far significantly larger than the theoretical uncertainty 
due to isospin $SU(2)$ breaking, which
may affect the  $\alpha_{\rm dir}$ determination at around $2^\circ$. The solution near $90^\circ$ is in  good agreement with the indirect determination obtained by the global fit of the flavour data\cite{CKMfitterwebsite}.

A detailed analysis of the prospects for the CKM angle $\alpha$ has been performed in~\cite{Charles:2017evz}.
The global
determination of $\alpha$ is dominated by the $B\to\rho\rho$ data,
constraining $\alpha$ with about 5\% uncertainty.  Improving the measurements of the neutral modes, especially the colour-suppressed $B^0\to\rho^0\rho^0$ decay, can have  a sizeable impact. In the $B\to\pi\pi$ system any sizable improvement is  driven by the increased accuracy in the measurement of the direct $\CP$ asymmetry in the colour-suppressed decay  $B^0\to\pi^0\pi^0$, potentially reachable through Dalitz decays $\pi^0\to\gamma e^+e^-$. However, all $\alpha$ measurements at subdegree precision must address isospin-breaking effects such as electroweak penguin contributions~\cite{Charles:2017evz,Gronau:2005pq}.

\subsubsection{Angle $\beta$} 
	The CKM angle $\beta$ is measured from the time-dependent \CP asymmetry of $b\to c\bar c s$ decays, such as  
	$B\to J/\psi K_S$.  The decay amplitude,  $A_{J/\psi K_S} = V_{cs}V_{cb}^* T + V_{us}V_{ub}^* P$, is dominated by the isoscalar tree-level amplitude $T$, while the subleading term $P$ is both loop and doubly Cabibbo suppressed. Neglecting $P$, the time-dependent \CP asymmetry, $a^{\CP}_t(B\to J/\psi K_S)=\sin2\beta\,\sin(\Delta m_B\, t)$,  has a single oscillatory term, the amplitude of which gives experimental access to $\sin2\beta$ (the fourfold ambiguity in $\beta$ can be reduced through time-dependent transversity analysis of $B\to J/\psi K^*$, sensitive to $\cos2\beta$).
	The $P=0$ approximation is justified as long as the relative error on $\sin2\beta$ is larger than the correction induced by the subleading term $\sim \lambda^2 P/T$. The present relative uncertainty on $\sin2\beta$ from various charmonium states, including $J/\psi$ / $\psi(2S)$ / $\chi_{c1}$ / $\eta_c$ $K_S$, $J/\psi K_L$, $\psi(nS)K^0$, is about $2.4\%$. Double Cabibbo suppression amounts to $5\%$ and $|P/T| < 1$ is expected. Yet the precise suppression due to $P/T$ is unknown, likely ranging between some tens and a few percent, the latter estimate given by the perturbative estimate of the penguin amplitude. Thus, depending on the actual value of $P/T$, the subleading amplitude could be a limiting factor for the extraction of $\sin2\beta$ already now or, in the opposite case, be safely neglected
	down to an experimental precision of few permille. In order to place a bound on the subleading amplitude, one can use $SU(3)$ flavour (or $U$-spin) related decays where the penguin contribution is not suppressed, such as $B_s\to J/\psi K_s$ or $B\to J/\psi \pi$~\cite{Fleischer:1999nz,Ciuchini:2005mg,Faller:2008zc,Ciuchini:2011kd,Jung:2012mp,DeBruyn:2014oga}. Measuring branching ratios and (time-dependent) \CP asymmetries, it is possible to extract the correction to $\sin2\beta$ induced by the subleading term and the corresponding uncertainty, due to the experimental errors and $SU(3)$-breaking effects. The latter, being proportional to a CKM suppressed amplitude, are expected to be at the permille level.
	
	With present data, the typical shift in $\beta$ is $0.5$--$1^\circ$ with about $100\%$ uncertainty, dominated by the experimental uncertainties. Improvements in the measurement of the time-dependent asymmetry in $B\to J/\psi K_S$ should therefore be accompanied by a corresponding improvement in the measurement of the control channels in order to keep the subleading amplitude under control and extract $\sin2\beta$ with a subpercent precision.
	It is worth mentioning that a complemetary approach proposing a theoretical computation of the subleading amplitude using an OPE based on soft-collinear factorization in QCD can be found in Ref.~\cite{Frings:2015eva}.
	Assuming that the subleading amplitude is kept under control, the uncertainty on the indirect determination of $\sin2\beta$ from the UT analysis is expected to go from the present 4.5\% down to 0.6\%, providing an improved sensitivity to NP in the $B$--$\bar B$ mixing amplitude, as discussed further below.

\subsubsection{Angle $\gamma$}

The current best sensitivity is obtained from charged 
$B$ decays into $\tilde{D} K^-$, $\tilde{D}^{*} K^-$, $\tilde{D} K^{*-}$, and $\tilde{D} K^-\pi^+\pi^-$, where $\tilde{D}$
is a $D^0$ or $\overline{D}^0$ meson decaying to the same final state. The $D^0$ is produced through
the leading $b \to c$ transition, with the amplitude ${\cal A}_{b\to c}\propto \lambda^3$, and $\overline{D}^0$ in the
CKM and colour-suppressed $b \to u$ transition,  ${\cal A}_{b\to u}\propto \lambda^3 \big({\bar{\eta}^2+\bar{\rho}^2}\big)^{1/2} e^{i( \delta_B - \gamma)}$.
Since $D^0$ and $\overline{D}^0$ decay to a common final state, 
the interference between the two tree amplitudes leads to observables that 
depend on the relative weak phase $\gamma$ and results in different 
$B^+$ and $B^-$ decay rates.
The size of this interference also depends on the magnitude of the ratio
of the two amplitudes, $r_B\equiv \vert {\cal A}_{b\to u}/$ ${\cal A}_{b\to c} \vert$,
and the relative strong phase $\delta_B$.
The value of $r_B$, which determines the size of the direct \CP\ asymmetry,
is predicted~\cite{Gronau:2002mu} to lie in the range 0.05--0.2. Experimental sensitivity to $\gamma$
decreases with smaller $r_B$.

There are several different methods that exploit the above interference pattern: the Gronau-London-Wyler (GLW)
method~\cite{Gronau:1990ra,Gronau:1991dp} where the neutral $D$ meson is
reconstructed in a \CP\ eigenstate; the Atwood-Dunietz-Soni (ADS)
method~\cite{Atwood:1994zm,Atwood:1996ci,Atwood:2000ck} where the $D^0$ meson,
associated with the ${\cal A}_{b\to c}$ amplitude, is required to decay to a doubly
Cabibbo-suppressed decay (DCSD), while the $\overline{D}^0$ meson, associated with the
${\cal A}_{b\to u}$ amplitude, decays to a Cabibbo-favoured final state, such as
$K^+\pi^-$; the Giri-Grossman-Soffer-Zupan (GGSZ) method~\cite{Giri:2003ty} 
where the neutral $D$ meson is reconstructed in self-conjugate three-body 
final state such as $K_S^0 h^+ h^-$ (where $h=\pi$ or $K$). These $\gamma$ extraction techniques 
rely on clean theoretical assumptions since the decay amplitudes 
are completely tree-level dominated, to an excellent 
approximation~\cite{Brod:2013sga}.
All variants are sensitive to the same $B$ decay parameters and can therefore
be combined in a single fit to extract the common weak phase $\gamma$, as well 
as the hadronic parameters associated with each method.

In addition to the above approaches, it is also possible to measure $\gamma$ from time-dependent analyses of $B^0$ and $B^0_s$ decays, as well as from decays of $b$-baryons. These methods offer significant additional sensitivity and are discussed further in Sec.~\ref{sec:expckm}.

\subsubsection{Angle $\beta_s$}
	The angle $\beta_s=\arg\big(- V_{ts} V_{tb}^*/V_{cs} V_{cb}^*\big)$ belongs to the squeezed UT $\sum_{q_u} V_{q_u b}^*V_{q_u s}=0$ instead of the usual $\sum_{q_u} V_{q_u b}^*V_{q_u d}=0$. Given its reduced sensitivity to $\bar{\rho}$ and $\bar{\eta}$, it is not a strong constraint in the UT plane. On the other hand, the UT analysis provides a very precise indirect determination of its value that can be compared with the direct determination extracted from time-dependent transversality analysis of $B_s\to J/\psi \phi$. At present, we have $\beta_s^\mathrm{direct} = (0.60\pm 0.88)^\circ$ and  $\beta_s^\mathrm{UTA} = (1.06\pm 0.03)^\circ$. The future uncertainties of the direct and indirect determinations are expected to be $\pm 0.01$ and $\pm 0.007$ respectively, allowing for a deeper investigation of BSM contributions to $B_s$--$\bar{B}_s$ mixing amplitude.
	It is worth noting that, similarly to the measurement of phase $\beta$, the shifts of a few degrees in the value of $\beta_s$ extracted from the time-dependent analysis caused by the subleading amplitude $P$ cannot be excluded. These shifts, however, now  produce a much larger relative uncertainty, since $\beta_s$ itself is doubly Cabibbo suppressed. 
	Some control of doubly Cabibbo suppressed contributions is thus required to meaningfully extract $\beta_s$ at the SM level.
	This can be achieved by using 
	the $SU(3)_f$ related channels, with the additional caveat that $\phi$ is an (almost equal) admixture of $SU(3)_f$ octet and singlet, introducing an additional source of uncertainty.

\subsubsection{$K-\bar{K}$ mixing} 
The CP-violating parameter $\varepsilon_K$, historically the first constraint in the UT plane, is becoming limited by long-distance contributions.
In the last years, the improvements in the constraining power of $\varepsilon_K$ have come from new determination of the bag parameter, $B_K$, on the lattice. Nowadays, the theoretical error on $B_K$  is approaching percent level. The dominating uncertainty in the SM prediction of $\varepsilon_K$ comes from the long-distance contributions, the best estimate of which has an error of $2\%$~\cite{Buras:2010pza}. A breakthrough could come from developing ideas for computing these terms on the lattice, using recent techniques to cope with the non-local operators that were used to obtain predictions for a closely related quantity, $\Delta m_K$ \cite{Bai:2014cva,Wang:2018csg}.

\subsubsection{$B_d-\bar{B}_d$ and $B_s-\bar{B}_s$ mixing}

The dispersive and absorptive contributions to  $B_d$ and $B_s$ mixing are described by $2\times2$ matrices, 
$M^q=M^{q\dagger}$ and
$\Gamma^q=\Gamma^{q\dagger}$, respectively ($q=d,s$). 
They describe the quantum-mechanical evolution 
of the  $B_q-\overline{B}_q$ system. 
Their diagonalisation defines the physical eigenstates $\ket{B^q_H}$ and
$\ket{B^q_L}$ with masses
$M^q_H,\,M^q_L$ and decay rates $\Gamma^q_H,\,\Gamma^q_L$.
One can reexpress these quantities in terms of three parameters:   $|M_{12}^q|$,
$|\Gamma_{12}^q|$ and the relative phase $\phi_q=\arg(-M_{12}^q/\Gamma_{12}^q)$.
The SM prediction for the mass splitting, $\Delta m_q=M_H^q-M_L^q$,  is dominated by boxes involving top quarks, and is given by
	\begin{equation}
	\label{eqof-dmd}
	   \dm_q = \frac{G_F^2}{6 \pi^2}\eta_B m_{B_q}f_{B_q}^2 {\hat{B}}_q
		  m_W^2 S(x_t) \left|V_{tq} V_{tb}^*\right|^2,
	\end{equation}
with  the Inami-Lim function $S(x_t)$~\cite{Inami:1980fz} evaluated at $x_{t}=\overline{m}_t^2/m_W^2$, while
$\eta_B$ encodes the perturbative QCD corrections, 
originally estimated at NLO in Ref.~\cite{Buras:1990fn}.
Using up-to-date $\alpha_{s}$ and the top-quark mass gives $\eta_{B} = 0.5510 \pm 0.0022$~\cite{Lenz:2010gu}.
The decay constant $f_{B_q}$ and the so-called bag parameter, $\hat{B}_q$, parametrize the nonperturbative hadronic matrix elements.  

The translation of the measured value for $\Delta m_d$ into constraints
on the CKM parameter combination ${|V_{td}^{\phantom{*}} V_{tb}^{*}|}^{2}$
is limited at present by uncertainties in the lattice QCD calculation of the
hadronic parameters $f_{B_{d}},  \hat B_{{d}}$. The ratios $f_{B_s}/f_{B_d}$ and $\hat B_{s}/\hat B_{d}$ are much better determined, so that 
$\Delta m_d/\Delta m_s$ gives a much better constraint in the $(\bar \rho,\bar \eta)$ plane, see Fig. \ref{CKMfitterSummer2018}. 
The hadronic
parameters, $f_{B_{s}}$, $\hat B_{{s}}$, as well as $f_{B_s}/f_{B_d}$ and $\hat B_{s}/\hat B_{d}$,
are set to have their errors significantly reduced  by future lattice QCD computations, see Sec.~\ref{sec:lattice}.

\subsection{Combined constraints on the CKM parameters}\label{sec:2018ckmfit}
Due to its economical structure in terms of only four parameters 
the CKM matrix can be
determined through many different quark transitions, both the 
 $\Delta F = 1$ decays and the $\Delta F=2$ neutral-meson mixing transitions. A consistent determination of the four CKM parameters from all these processes is a fundamental test of the Kobayashi-Maskawa mechanism. 
Extracting the information on the four CKM parameters from data poses both experimental and theoretical challenges.
First of all, the SM depends on a number of other parameters, masses and couplings, which are not predicted within the SM, but rather need to also be determined experimentally. An additional difficulty relates to the presence of strong interactions, binding quarks into hadrons. This is responsible for most of the theoretical uncertainties in the determinations of CKM matrix elements.

\subsubsection{Statistical approaches}

The CKMfitter group determines the CKM parameters from a large set of flavour physics constraints using a standard $\chi^2$-like frequentist approach, in addition to a specific (Rfit) scheme to treat theoretical uncertainties~\cite{Charles:2004jd,Charles:2011va,Charles:2015gya}. The set of experimental observables, denoted $\vec {\cal O}_{exp}$, is measured in terms of likelihoods that can be used to build a \chisq-like test statistic,
$\chi^2(\vec{p})=-2\log {\mathcal{L}}(\vec {\cal O}_{exp}-\vec {\cal O}_{th}(\vec p))$,
with $\vec {\cal O}_{th}(\vec p)$ the theoretical values of the observables depending on $N$ fixed parameters $\vec p$. The absolute minimum value of the test statistic, $\chi^2_{\rm min}$, quantifies the agreement of the data with the theoretical model, once converted into a $p$-value (interpreting $\chi^2(\vec{p})$
as a random variable distributed according to a $\chi^2$ law).
It is also possible to perform the metrology of specific parameters of the model, 
by considering the  hadronic parameters $\vec\mu$ as ``nuisance parameters'' 
and 
defining the test statistic,
$ \Delta\chi^2(\alpha) = \min_{\vec\mu}[\chi^2(\alpha)] - \chi^2_{\rm min}$~\cite{Hocker:2001xe,Charles:2004jd,Charles:2016qtt}.
Here,  $\min_{\vec\mu}[\chi^2(\alpha)]$ is the value of  \chisq, minimised with respect to the nuisance parameters for a  fixed $\alpha$ value. This test statistic assesses how a given hypothesis on the true value of $\alpha$ agrees with the data, irrespective of the value of the nuisance parameters. Confidence intervals on $\alpha$ can be derived from the resulting $p$-value, which is computed assuming that $\Delta\chi^2(\alpha)$ is \chisq-distributed with one degree of freedom,
\begin{equation}\label{eq:prob}
 p(\alpha) = {\rm Prob}(\Delta\chi^2(\alpha),N_{dof}=1)\,, \qquad  {\rm Prob}\!\left(\Delta \chi^2,N_{dof}\right)=\frac{\Gamma(N_{dof}/2,\Delta\chi^2/2)}{\Gamma(N_{dof}/2)}\,,
\end{equation}
where $\Gamma(x)$ is the usual Euler factorial function, and $\Gamma(s,x)$ is the upper incomplete gamma function.
Confidence intervals at a given confidence level (CL) are obtained by selecting the values of $\alpha$ with $p$-value larger than $1-$CL.

In addition to the frequentist statistical treatment outline above, the CKMfitter collaboration relies on a specific treatment of theoretical uncertainties (e.g., systematics due to uncertainties on hadronic matrix elements not scaling with the size of the sample).
The current approach is the so-called Rfit~\cite{Hocker:2001xe,Charles:2004jd,Charles:2016qtt} model, 
where the theoretical parameter/observable is restricted to a range, without any possibility to exceed this range. Most of the systematic uncertainties come from lattice QCD. The CKMfitter collaboration follows the recommendations of the Flavour Lattice Averaging Group~\cite{Aoki:2016frl} and uses a specific procedure to perform the average of the lattice inputs based on the Rfit model combined with a linear addition of systematic uncertainties for the individual inputs~\cite{Charles:2016qtt}.

	The UTfit collaboration follows a Bayesian approach to combine the constraints in the UT plane.  Bayesian statistics allows for a unified treatment of systematic and theoretical uncertainties in a scheme of ``updating of knowledge'' from prior to posterior distributions.
Following Bayes' theorem, the unnormalized posterior probability density function (p.d.f.)\ for $\bar\rho$ and $\bar\eta$ (given the constraints) is,
	\begin{equation}
	f(\bar\rho,\bar\eta|\mathbf{\hat{c}}, \mathbf{\hat{f}}) \propto 
	{\cal L}(\mathbf{\hat{c}}\,|\,\bar\rho,\bar\eta,\mathbf{f})\, f_0(\bar\rho,\bar\eta).
	\label{eq:flat_inf}
	\end{equation}
Here $\mathbf{\hat{c}}=\{c_1,c_2,\dots,c_M\}$ is a set of measured constraints, whose theoretical expressions are given by functions $c_j(\bar\rho,\bar\eta;\mathbf{x})$ that depend on $\bar\rho$, $\bar\eta$, and a set of additional experimental and theoretical parameters ${\mathbf x}=\{x_1,x_2,\dots,x_N\}$. The $f_0(\bar\rho,\bar\eta)$ is the prior p.d.f.\ for $\bar\rho$ and $\bar\eta$, assumed to be flat on the UT plane, and is multiplied by the effective overall likelihood,
    \begin{equation}
    {\cal L}(\mathbf{\hat{c}}\,|\,\bar\rho,\bar\eta,\mathbf{f})
    = \int 
    \prod_{j=1,{\rm M}}f_j(\hat{c}_j\,|\,\bar\rho,\bar\eta,\mathbf{x})
    \prod_{i=1,{\rm N}}f_i(x_i)\, \mbox{d}{ x_i}\,.
    \label{eq:lik_int}
    \end{equation}
	The p.d.f.\ $f_i=\{f_1,f_2,\dots,f_N\}$ are the prior distributions of the parameters, while  $f_j(\hat{c}_j\,|\,\bar\rho,\bar\eta,\mathbf{x})$ are the conditional probabilities of $\hat{c}_j$ given $\bar\rho$, $\bar\eta$, and $\mathbf{x}$, that in the Gaussian approximation become
    \begin{equation}	
	f_j(\hat{c}_j\,|\,\bar\rho,\bar\eta,\mathbf{x})=\frac{1}{\sqrt{2 \pi}\,\sigma(c_j)}
	\exp\left[-\frac{(c_j(\bar\rho, \bar\eta; \mathbf{x})-
		\hat{c}_j)^2}{2\,\sigma^2(c_j)}\right]\,.
    \end{equation}
    The integration in Eq. (\ref{eq:lik_int}) is usually carried out using Monte Carlo methods. More details on the Bayesian approach of the UTfit collaboration can be found in Ref.\cite{Ciuchini:2000de}.

\subsubsection{Current combined constraints on the CKM parameters}

\begin{figure}[!tb]
  \centering
  \includegraphics[width=0.60\textwidth]{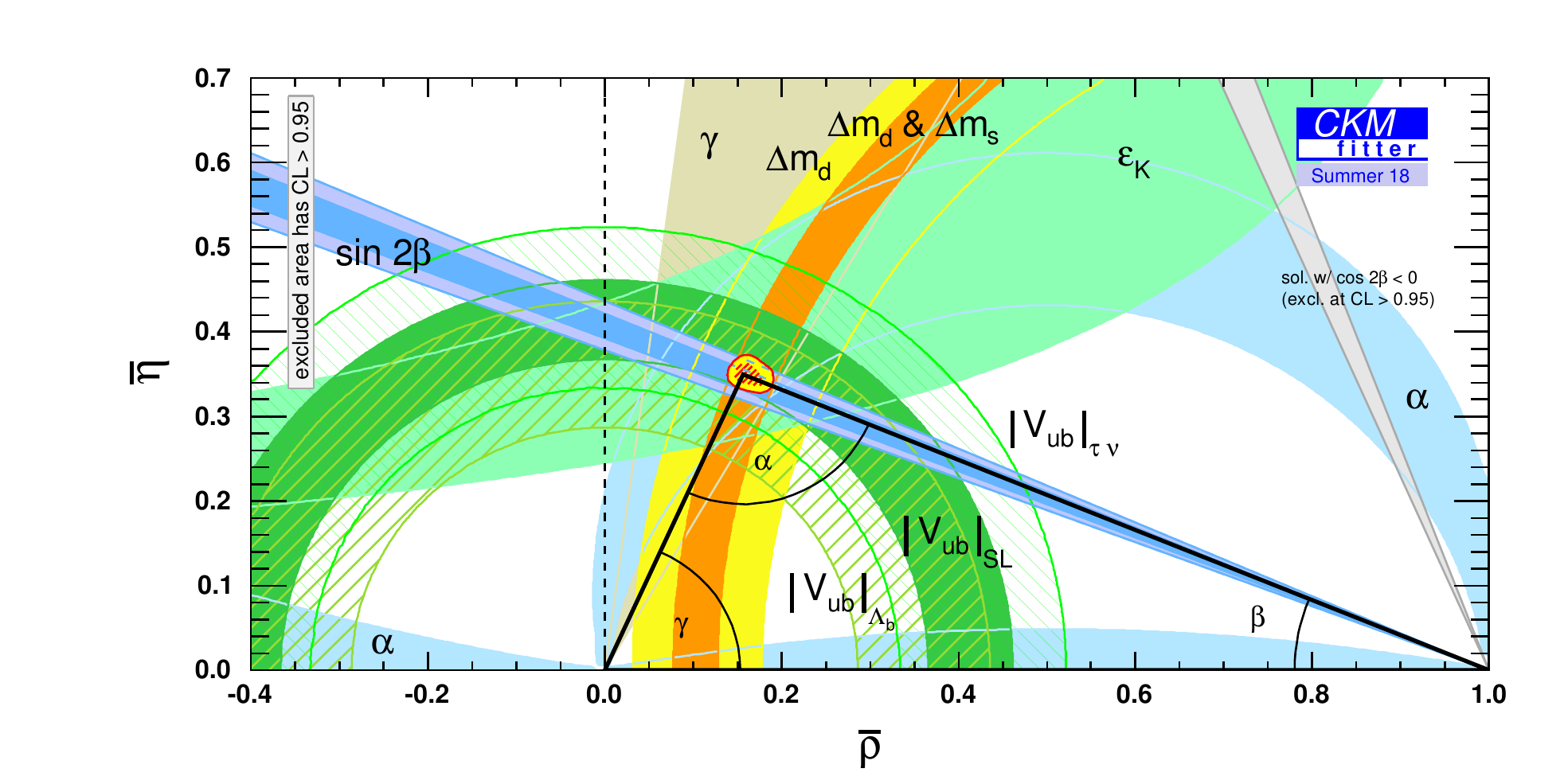}
   \includegraphics[width=0.35\textwidth]{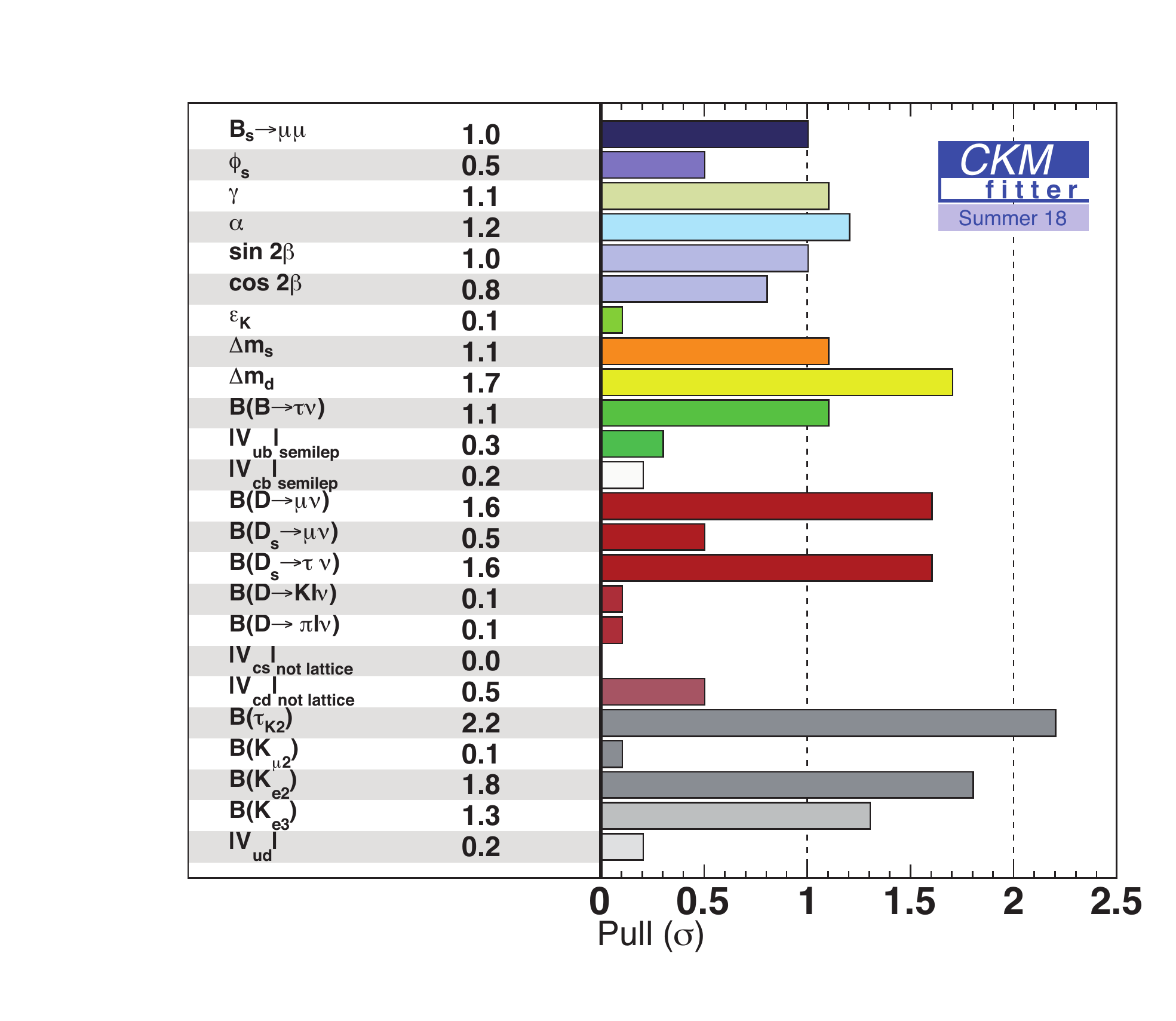}
  \caption{Results for the CKMfitter global fit of the CKM parameters as of Summer 2018. See Ref.~\cite{CKMfitterwebsite} for more detail.
    \label{CKMfitterSummer2018}
  }
\end{figure}

From the SM global fit the CKMfitter collaboration finds for 
the four CKM parameters, 
\begin{equation}
\label{eq:CKMfitter:Alambdarhoeta}
A=0.8403^{\,+0.0056}_{\,-0.0201},
\quad
\lambda=0.224747^{\,+0.000254}_{\,-0.000059},
\quad
\bar\rho=0.1577^{\,+0.0096}_{\,-0.0074},
\quad
\bar\eta=0.3493^{\,+0.0095}_{\,-0.0071}.
\end{equation}
The asymmetric errors come from the combination of several constraints containing both statistical and systematic uncertainties, so that the resulting $\chi^2$ has a rather complicated asymmetric shape. The corresponding results are shown in Fig.~\ref{CKMfitterSummer2018}.
The fit shows a good overall consistency among the various constraints. The main pulls come from the kaon and charm sector, but nothing exceeds the 2-$\sigma$ level, confirming the very good overall agreement of the various constraints involved here.

 Detailed input values as well as predictions for various CKM-related parameters and observables in the SM from the UTfit collaboration can be found in the Summer 2018 update page on the UTfit website~\cite{UTfitwebsite18}.
The CKM matrix is determined as
\begin{eqnarray}
&&\!\!\!\!\! V_{\rm CKM}=\nonumber\\ 
&&\!\!\! \left(\begin{array}{ccc} 0.97431(12) & 0.22514(55) & 0.00365(10)e^
{-i 66.8(2.0)^\circ}\\ -0.22500(54)e^{i 0.0351(10)^\circ} & 0.97344(12)e^{-i 0
.00188(5)^\circ} & 0.04241(65) \\ 0.00869(14)e^{-i 22.2(0.6)^\circ} & -
0.04124(56)e^{i 1.056(32)^\circ} & 0.999112(24)\end{array}\right)\,.
\end{eqnarray}
while the Wolfenstein parameters are,
\begin{equation}
A=0.826\pm,0.012,\quad \lambda=0.2255\pm0.0005, \quad \bar\rho=0.148\pm0.013,\quad \bar \eta=0.348\pm 0.010,
\end{equation}
in good agreement with the CKMfitter determination in \eqref{eq:CKMfitter:Alambdarhoeta}.

\subsection{Theoretical prospects}
\subsubsection{Lattice extrapolations}
Advancement in the lattice QCD determination of hadronic matrix elements, used in the extraction of CKM elements, need to go hand in hand with the improved experimental precision. The projections on the expected errors for the most important hadronic matrix elements are  discussed in detail in Sec.~\ref{sec:lattice} with the main results collected in Table \ref{table:Proj}.

The accuracy of the projections below will also require an improvement in the understanding of electromagnetic corrections, which for the moment are only partially addressed in theoretical predictions. Further details on the inclusion of QED corrections in lattice predictions are discussed in Sec.~\ref{sec:lattice}.

\subsubsection{Other theoretical issues}

The determination of $\gamma$ has negligible theoretical errors. 
In contrast, 
extractions of $\beta$ and $\beta_s$ 
from $b\to c\bar cs$ transitions receive small theoretical uncertainty from penguin diagrams. For present data the penguin contributions in $b\to c\bar cs$ are assumed to be negligible. However, they may compete with the statistical errors in the future. Techniques to estimate the penguin contributions from data on SU(3)-related channels are described in Sec.~\ref{sec:b2ccspenguin}. An independent, more theory-driven, cross-check of the SU(3) approach was suggested  
in Ref.~\cite{Frings:2015eva} -- to use Operator Product Expansion to calculate the penguin-to-tree ratio.
\Blu{ It is important to note that the penguin contributions in various $b\to c\bar c s$ channels are different, so that with increased precision of the measurements one cannot anymore average $\sin 2\beta$ from $B\to J/\psi K_S, B\to \eta_c K_S,\ldots$, without first correcting for the penguin contributions on a channel by channel basis}. 

In the determination of $\alpha$, the small penguin contributions are avoided using isospin symmetry. The associated theory error on $\alpha$ due to isospin breaking  
has been estimated to be at sub degree level~\cite{Charles:2017evz}. It is neglected at present, but will need to be taken into account in significantly larger statistics samples.

An open issue is the (in)compatibility of the inclusive and exclusive determinations of $|V_{cb}|$ and $|V_{ub}|$,
where a $\sim 3\sigma$ discrepancy 
has been observed ever since the first precise measurements.
Recently, the $|V_{cb}|$ discrepancy was shown to depend significantly on the choice of the form factor parametrisation, and on the hypotheses made about the underlying heavy quark symmetry~\cite{Grinstein:2017nlq,Bigi:2017njr}. Relaxing these choices leads to an exclusive determination of $|V_{cb}|$  that is in good agreement with the inclusive one, but with unexpectedly large corrections to the heavy quark symmetry predictions~\cite{Bernlochner:2017xyx}, unless one employs additional theoretical information, as shown in Ref. \cite{Bigi:2017jbd}. Lattice calculations should soon settle the matter. As for $|V_{ub}|$ no satisfying explanation of the discrepancy has been found so far. In the future more precise exclusive measurements by LHCb and Belle-II, and more precise inclusive analysis by Belle-II, should shed light on this topic by investigating whether or not it could be an experimental effect.

\input{section2/experimental}

\subsection{Future of global CKM fits}

\subsubsection{Summary of the projections}

As discussed above, HL-LHC will improve the determination of several flavour observables crucial for the extraction of  CKM parameters.
We consider two phases for the HL-LHC projections: in Phase 1, we assume an integrated luminosity of 23 fb$^{-1}$ for LHCb and 300 fb$^{-1}$ for CMS/ATLAS; in Phase 2 we have 300 fb$^{-1}$ for LHCb  and 3000 fb$^{-1}$ for CMS/ATLAS.
Several observables will be measured more precisely at Belle-II. For uncertainties on these observables we use the 50 ab$^{-1}$ projections in Ref.~\cite{Kou:2018nap}. 
 Since we are
interested in  the future sensitivity for Phase 1 and Phase 2, we choose the central values of future measurements to coincide
with their SM predictions using the current best-fit values of $\rhobar$ and
$\etabar$.

\begin{table}
\caption{Uncertainties on inputs for the CKMfitter and UTFit projections.}\label{tab:CKMfitterprojinputs}
\begin{center}
{
\begin{tabular}{ccccc}
\hline\hline
  &  Current  &  Phase 1  & Phase 2 & Ref.  \\
\hline\hline
$|V_{ud}|$  &  $\pm 0.00021$ & $\pm 0.00021$ & $\pm 0.00021$ & \cite{CKMfitterwebsite} \\
$|V_{us}|f_+^{K\to\pi}(0)$  & $\pm 0.0004$  & $\pm 0.0004$ & $\pm 0.0004$ & \cite{CKMfitterwebsite}\\
$|\epsilon_K|\times 10^{3}$ & $\pm 0.011$   & $\pm 0.011$  & $\pm 0.011$ & \cite{CKMfitterwebsite}\\
$\Delta m_d$ [${\rm ps}^{-1}$]   &  $\pm 0.0019$ & $\pm 0.0019$ & $\pm 0.0019$ & \cite{Amhis:2016xyh} \\
$\Delta m_s$ [${\rm ps}^{-1}$]  &  $\pm 0.021$ & $\pm 0.021$ & $\pm 0.021$ &  \cite{Amhis:2016xyh} \\
$|V_{ub}|\times 10^3$ ($b\to u\ell\bar\nu$)  &  $\pm 0.23$ & $\pm 0.04$ & $\pm 0.04$ & \cite{Kou:2018nap}\\
$|V_{cb}|\times 10^3$  ($b\to c\ell\bar\nu$) &  $\pm 0.7$ & $\pm 0.5$ & $\pm 0.5$ & \cite{Kou:2018nap}\\
$|V_{ub}/V_{cb}|$ ($\Lambda_b$) & $\pm 0.0050$ & $\pm 0.0025$ & $\pm 0.0008$ &  See above \\
\hline
$\sin 2\beta$ &  $\pm 0.017$ & $\pm 0.005$ & $\pm 0.003$ & Above \& \cite{Kou:2018nap}\\
$\alpha\ [^\circ]$  &  $\pm 4.4$ & $\pm 0.6$ & $\pm 0.6$& \cite{Kou:2018nap}\\
$\gamma\ [^\circ]$  &  $\pm 5.6$ & $\pm 1$ & $\pm 0.35$& Above \&  \cite{Kou:2018nap}\\
$\beta_s\ [{\rm rad}]$ &  $\pm 0.031$ & $\pm 0.014$ & $\pm 0.004$ & See above\\
${\cal B}(B\to\tau\nu)\times 10^4$ &  $\pm 0.21$ & $\pm 0.04$ & $\pm 0.04$ & \cite{Kou:2018nap}\\
\hline
$\bar{m}_c$ [GeV] &  $\pm 0.012$ $(0.9\%)$ & $\pm 0.005$ $(0.4 \%)$ & $\pm 0.005$ $(0.4\%)$ & See Sec. 11\\
$\bar{m}_t$ [GeV] &  $\pm 0.73$ $(0.4\%)$ & $\pm 0.35$ $(0.2\%)$& $\pm 0.35$ $(0.2\%)$ & \cite{CKMfitterwebsite}\\
$\alpha_s(m_Z)$  &  $\pm 0.0011$ $(0.9\%)$ & $\pm 0.0011$ $(0.9\%)$ & $\pm 0.0011$ $(0.9\%)$ & \cite{CKMfitterwebsite}\\
$f_+^{K\to\pi}(0)$ &  $\pm 0.0026$ $(0.3\%)$ & $\pm 0.0012$ $(0.12\%)$ & $\pm 0.0012$ $(0.12\%)$ & See Sec. 11\\
$f_K$ & $\pm 0.0006$ $(0.5\%)$ & $\pm 0.0005$ $(0.4\%)$ & $\pm 0.0005$ $(0.4\%)$ & See Sec. 11\\
$B_K$  &  $\pm 0.012$ $(1.6\%)$ & $\pm 0.005$ $(0.7\%)$ & $\pm 0.004$ $(0.5\%)$ & See Sec. 11\\
$f_{B_s}$ [GeV] &  $\pm 0.0025$ $(1.1\%)$ & $\pm 0.0011$ $(0.5\%)$ & $\pm 0.0011$ $(0.5\%)$ & See Sec. 11\\
$B_{B_s}$ &   $\pm 0.034$ $(2.8\%)$ & $\pm 0.010$ $(0.8\%)$ & $\pm 0.007$ $(0.5\%)$ & See Sec. 11\\
$f_{B_s}/f_{B_d}$  &  $\pm 0.007$ $(0.6\%)$ & $\pm 0.005$ $(0.4\%)$ & $\pm 0.005$ $(0.4\%)$ & See Sec. 11\\
$B_{B_s}/B_{B_d}$  &  $\pm 0.020$ $(1.9\%)$ & $\pm 0.005$ $(0.5\%)$ & $\pm 0.003$ $(0.3\%)$ & See Sec. 11\\
\hline\hline
\end{tabular}}
\end{center}
\end{table}

\subsubsection{CKMfitter results}

The inputs used by CKMfitter collaboration for the fits are shown in Table~\ref{tab:CKMfitterprojinputs}. For easier comparison the ``Current'' column shows present uncertainties, taking  
 central values corresponding to a perfect agreement of the various constraints in the SM. Note that this choice does change slightly the present determination of CKM parameters -- the global fit described in Sec.~\ref{sec:2018ckmfit} exhibits slight discrepancies, in particular for $|V_{us}|$, which currently increases the accuracy of the determination of the CKM parameters. In order to determine the increase in accuracy on the CKM parameters in a fair way, we therefore compare the three scenarios presented in Table~\ref{tab:CKMfitterprojinputs} with the same central values taken to have perfect agreement (rather than comparing the future projections with the CKMfitter results for the Summer 2018 update).

\begin{figure}
\begin{center}
\centering
\includegraphics[width=0.45\linewidth]{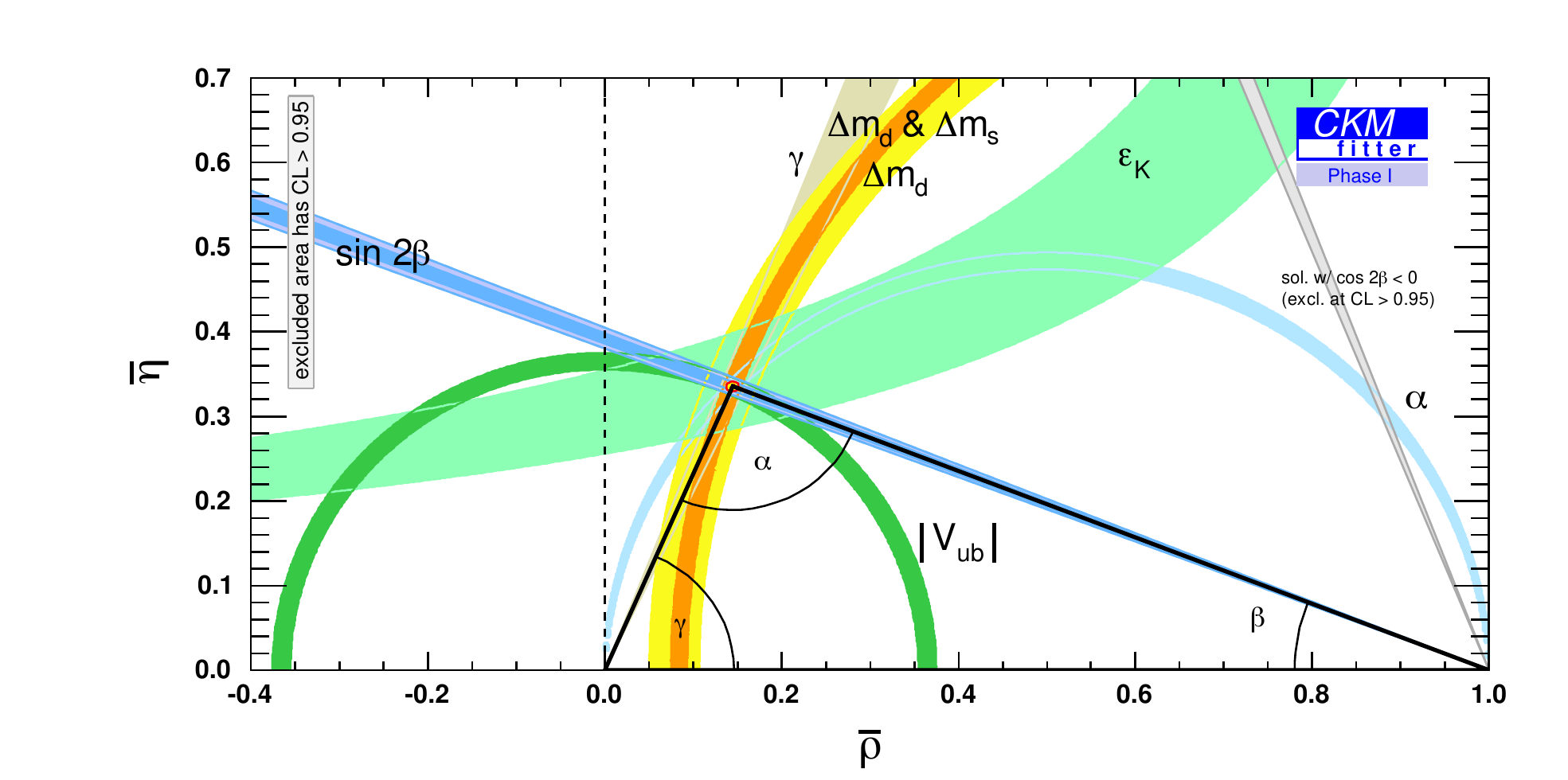} 
\includegraphics[width=0.45\linewidth]{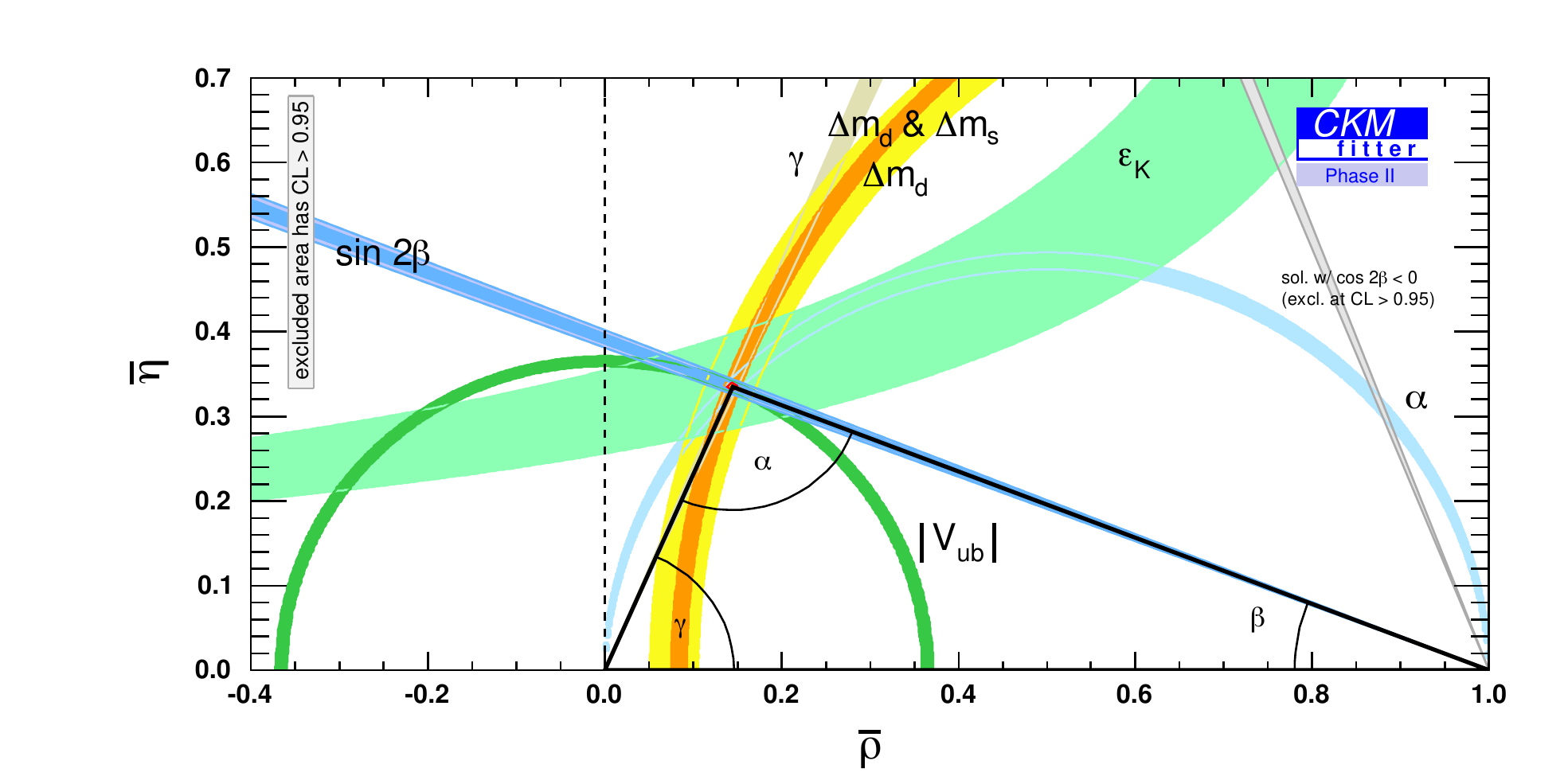}\\ 
\includegraphics[width=0.45\linewidth]{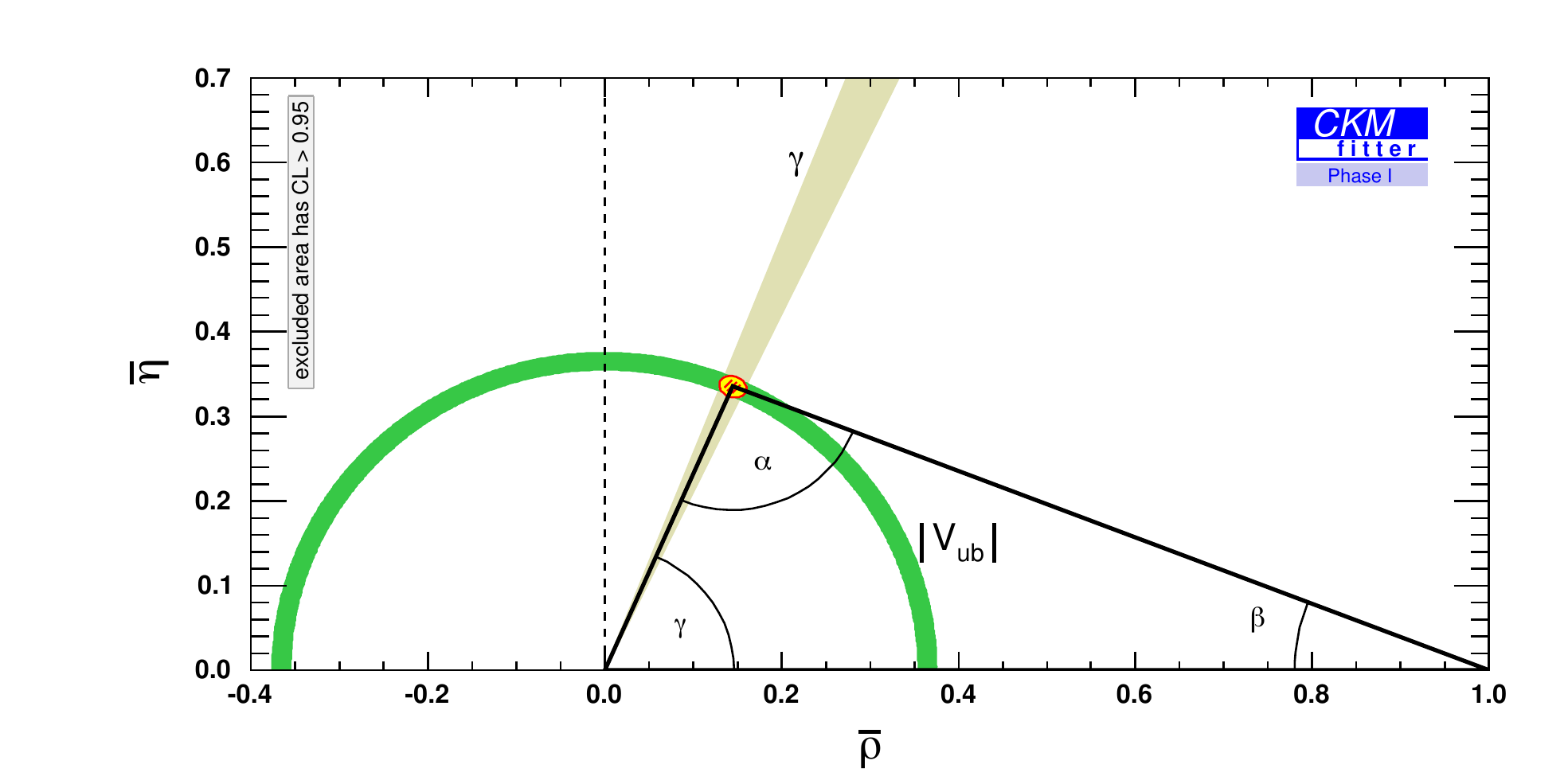} 
\includegraphics[width=0.45\linewidth]{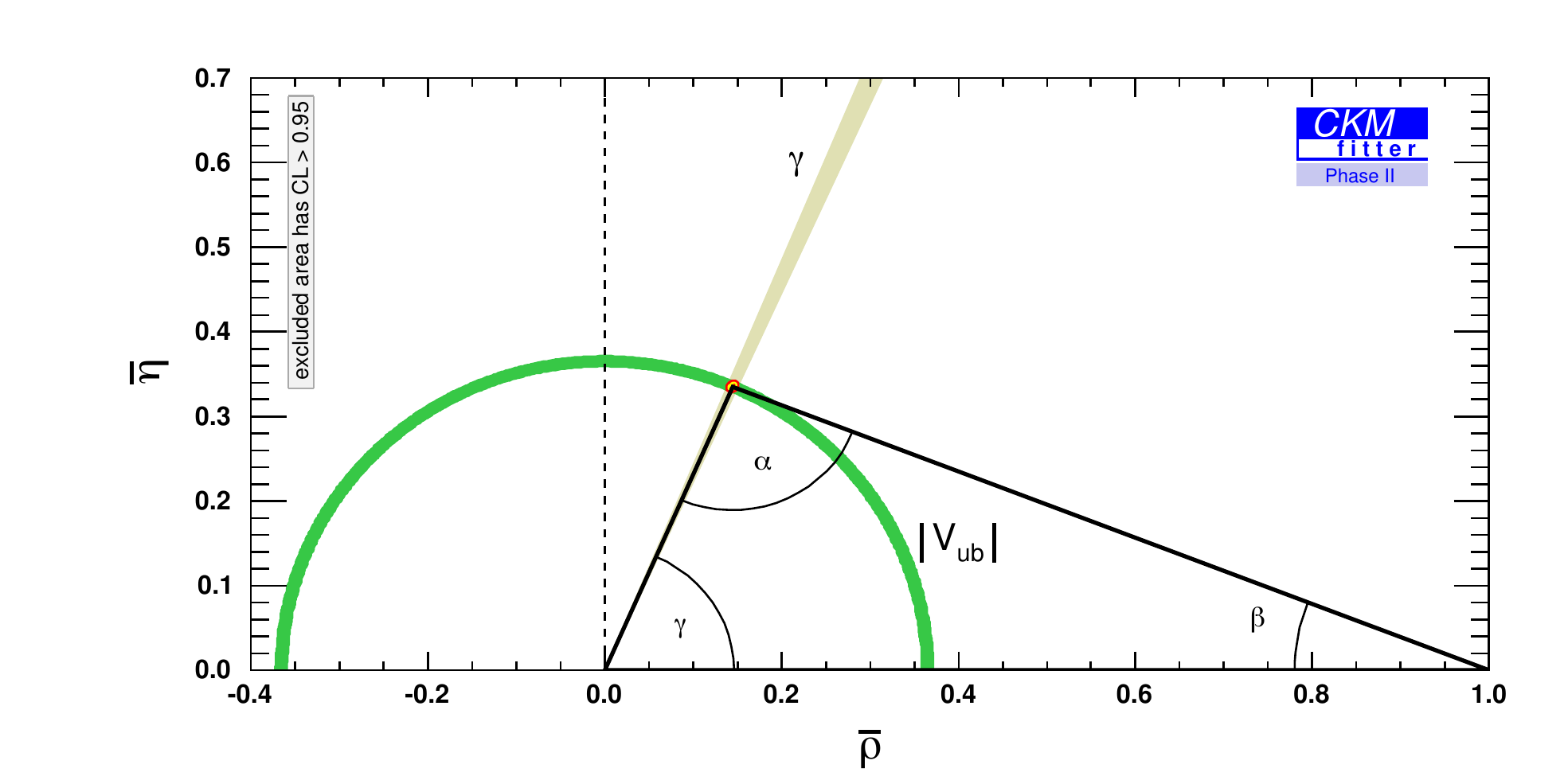} \\
\includegraphics[width=0.45\linewidth]{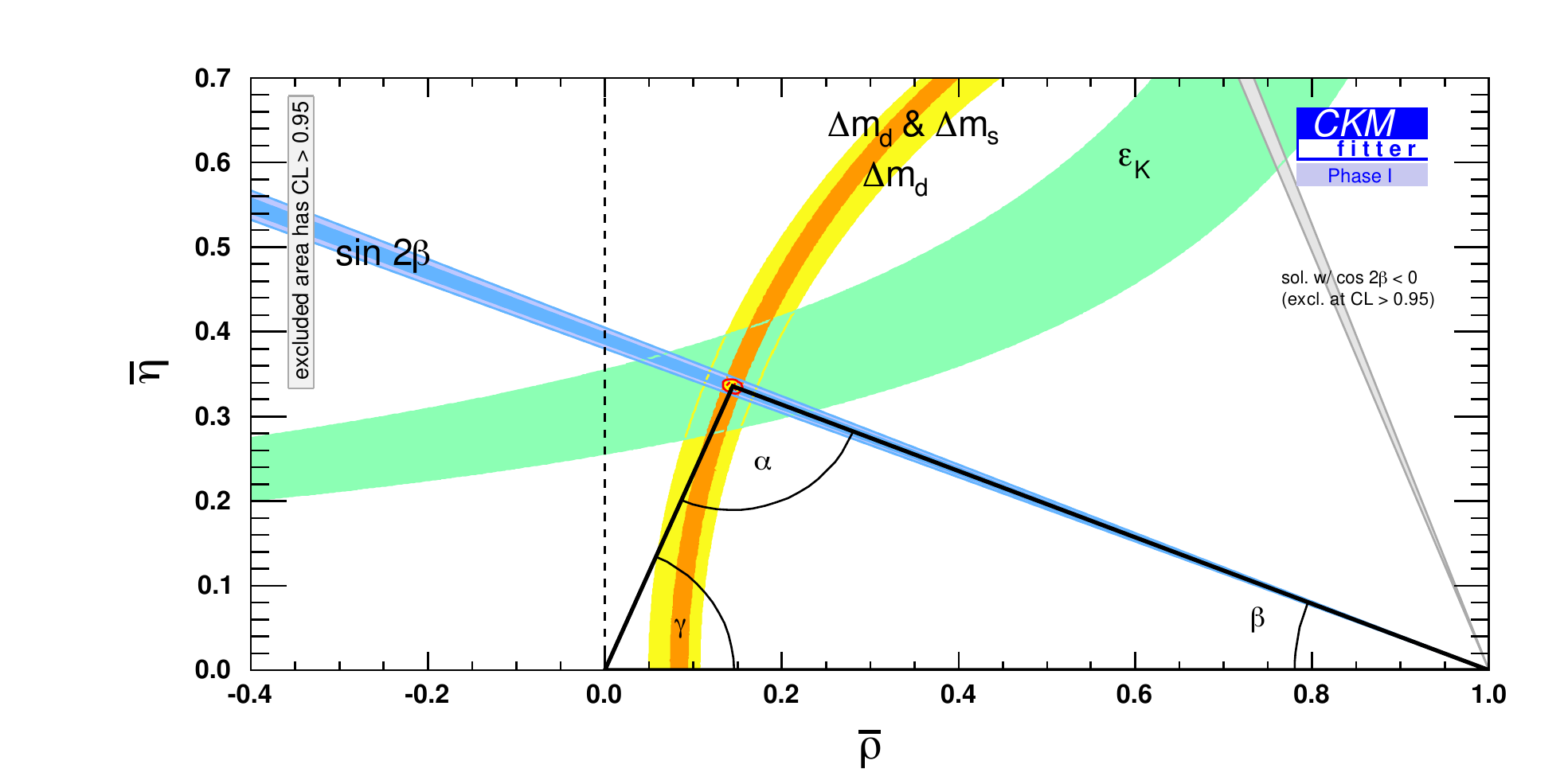} 
\includegraphics[width=0.45\linewidth]{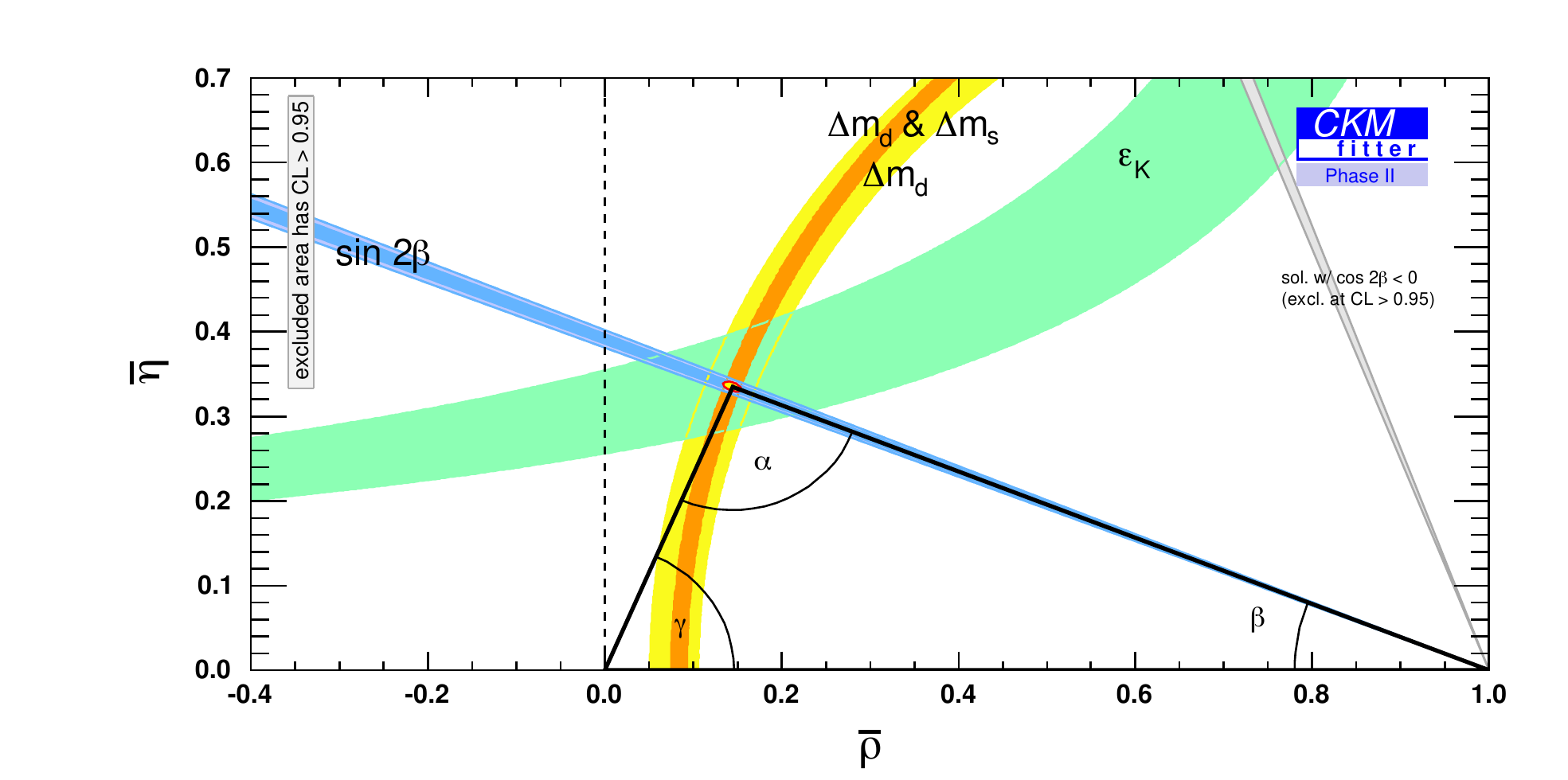} 
\caption{Constraints on the unitarity triangle from the CKMfitter global analysis: global fit (top), tree only (center), loop only (bottom),  for Phase 1 (left) and Phase 2 (right).}
\label{fig:CKMfitterglobalsmall}
\end{center}
\end{figure}

\begin{figure}
\begin{center}
\centering
\includegraphics[width=0.45\linewidth]{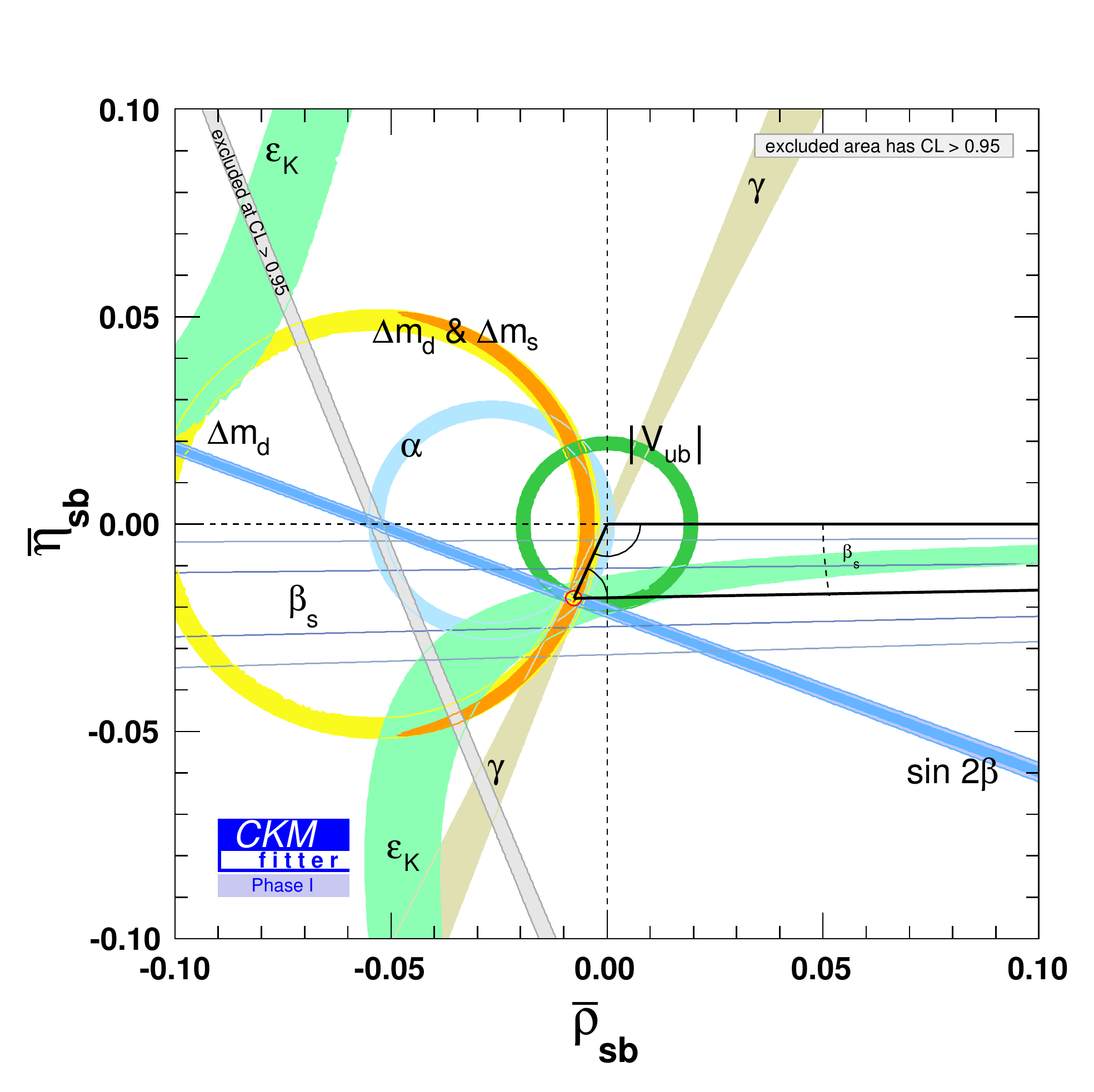} 
\includegraphics[width=0.45\linewidth]{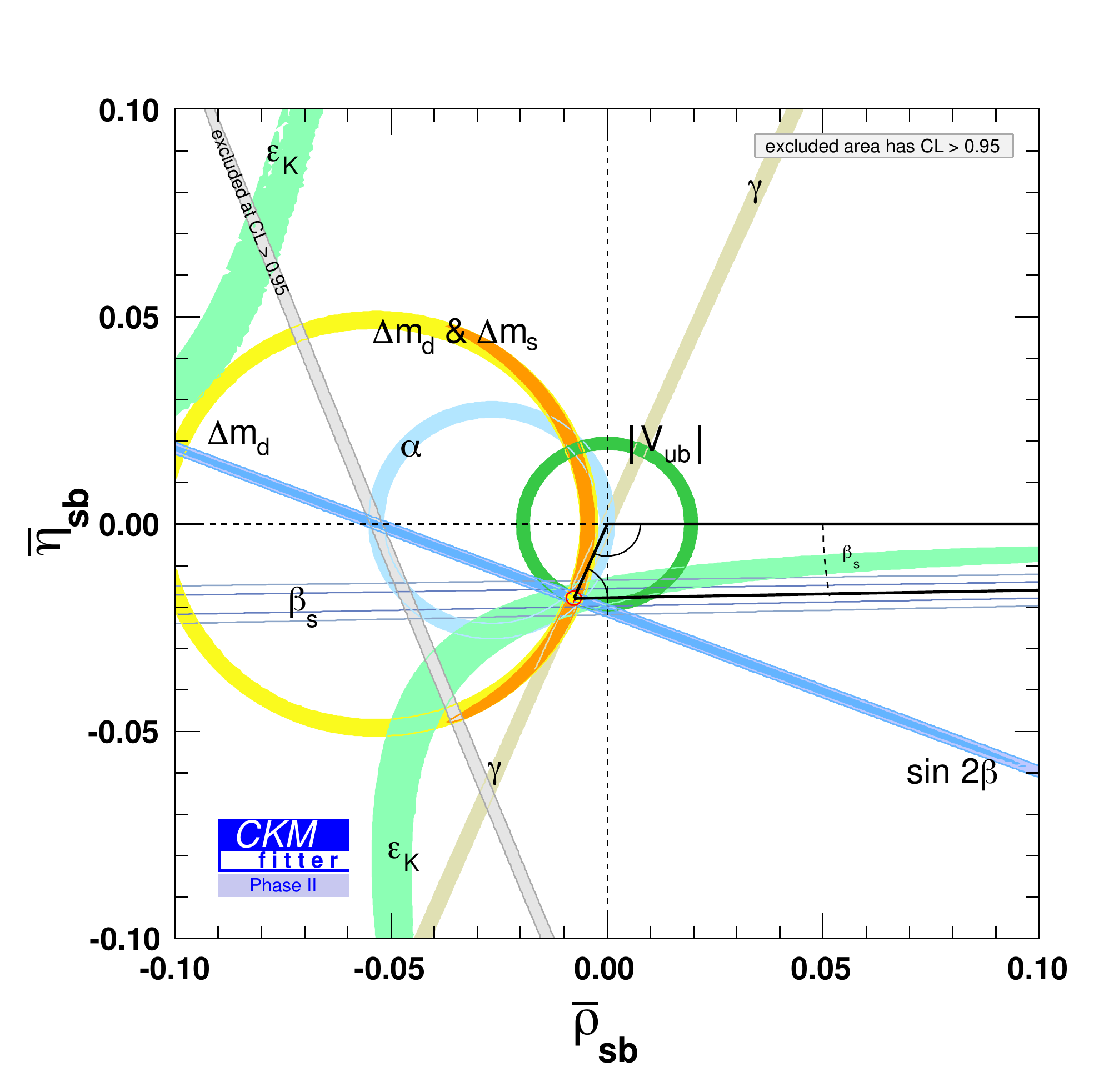} 
\caption{Constraints on the unitarity triangle for the $B_s$ meson from the CKMfitter global analysis for Phase 1 (left) and Phase 2 (right).}
\label{fig:CKMfitterglobalBs}
\end{center}
\end{figure}

From the global fit we determine the 68\% CL intervals for the 4 CKM parameters and other parameters of interest. 
In Fig.~\ref{fig:CKMfitterglobalsmall} we show the Phase 1 (left panels) and Phase 2 (right panels)  determinations in the standard UT plane for a global fit (upper), as well as when using only subsets of constraints, tree only (middle) or loop only (lower panels).
The uncertainties obtained are listed in Table~\ref{tab:CKMfitterprojresults}.
We show  the corresponding constraints for the $B_s$ meson system also in Fig.~\ref{fig:CKMfitterglobalBs}, defining the apex of the $B_s$ unitarity triangle as~\cite{Charles:2015gya}
\begin{equation}
\bar\rho_{sb}+i\bar\eta_{sb}=-\frac{V_{us}V_{ub}^*}{V_{cs}V_{cb}^*}.    
\end{equation}

\begin{table}
\caption{The 68\% CL uncertainties on the determination from the CKMfitter global fit.}\label{tab:CKMfitterprojresults}
\begin{center}
{
\begin{tabular}{cccccc}
\hline\hline
  & Summer 18 &  Current  &  Phase I  & Phase II  \\
\hline
$A$  & 0.0129 & 0.0120 & 0.0058 & 0.0057 \\
$\lambda$ & 0.0002  & 0.0007 & 0.0004 & 0.0004 \\
$\bar\rho$ & 0.0085 & 0.0085 & 0.0027& 0.0018\\
$\bar\eta$ & 0.0083 & 0.0087& 0.0024 & 0.0015\\
\hline
$|V_{ub}|$ & 0.000076 & 0.000096 & 0.000027 & 0.000023\\
$|V_{cb}|$ & 0.00073 & 0.00070 & 0.00026 & 0.00025\\
$|V_{td}|$ & 0.00017 & 0.00014 & 0.00006 & 0.00006\\
$|V_{ts}|$ & 0.00068 & 0.00054 & 0.00026 & 0.00025\\
\hline
$\sin 2\beta$ & 0.012 & 0.015 & 0.004 & 0.003\\
$\alpha\ (^\circ)$ & 1.4 & 1.4 & 0.4 & 0.3\\
$\gamma\ (^\circ)$ & 1.3 & 1.3 & 0.4 & 0.3\\
$\beta_s\ ({\rm rad})$ & 0.00042 & 0.00042 & 0.00012 & 0.00010 \\
\hline\hline
\end{tabular}}
\end{center}
\end{table}

\subsubsection{UTfit results}
The projection of the UT analysis in the HL-LHC era is obtained by performing a global fit using the same future expected values of experimental and theoretical input parameters as CKMfitter, Table \ref{tab:CKMfitterprojinputs}. In particular, the lattice uncertainties for Phase 1 and Phase 2 are the same as Table~\ref{table:Proj} in Sec.~\ref{sec:lattice}. 
For both theoretical and experimental parameters the SM expectations were taken as central values, in order to ensure the compatibility of the extrapolated constraints.

\begin{figure}
\begin{center}
\centering
\includegraphics[width=0.32\linewidth]{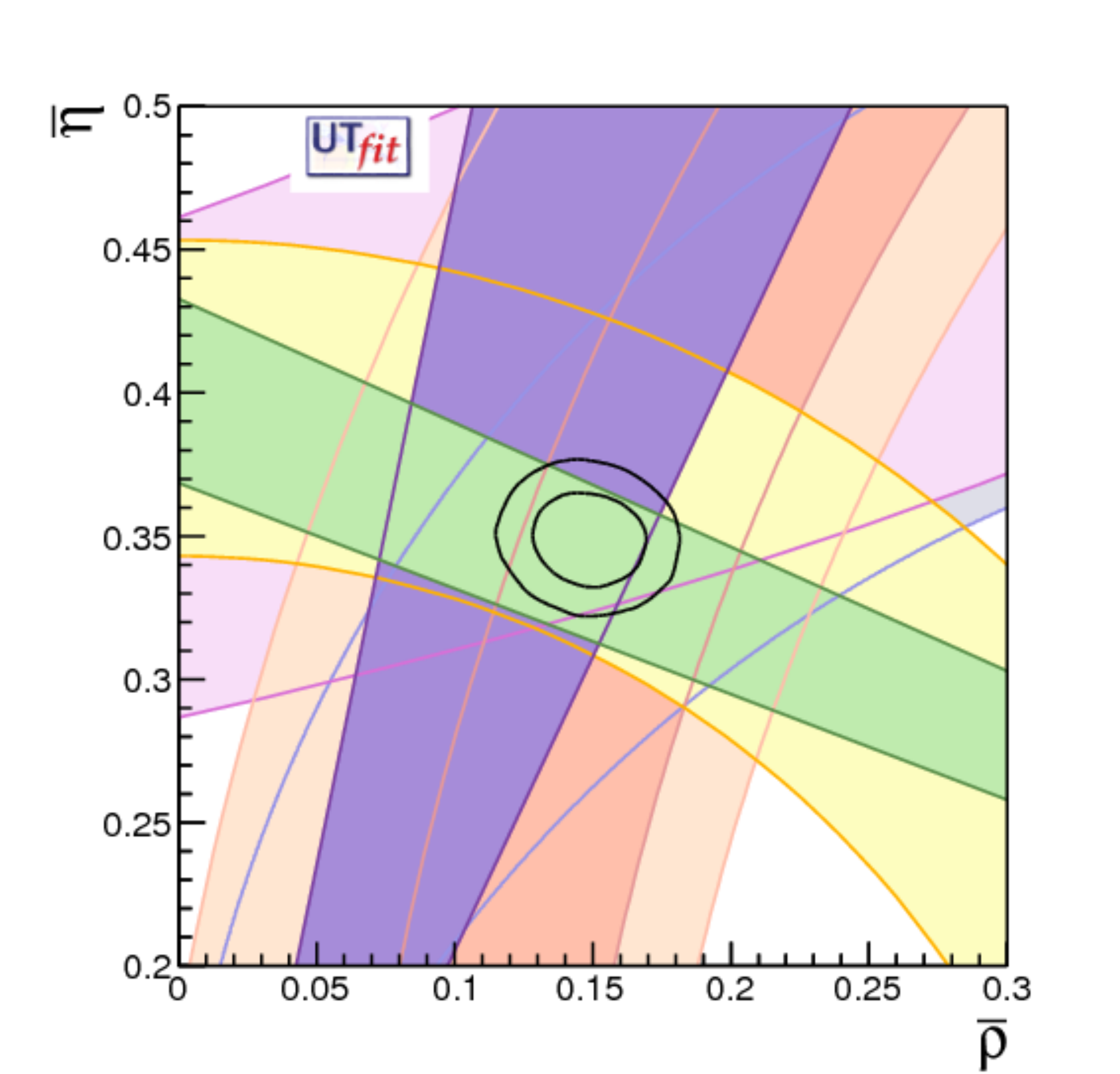} 
\includegraphics[width=0.32\linewidth]{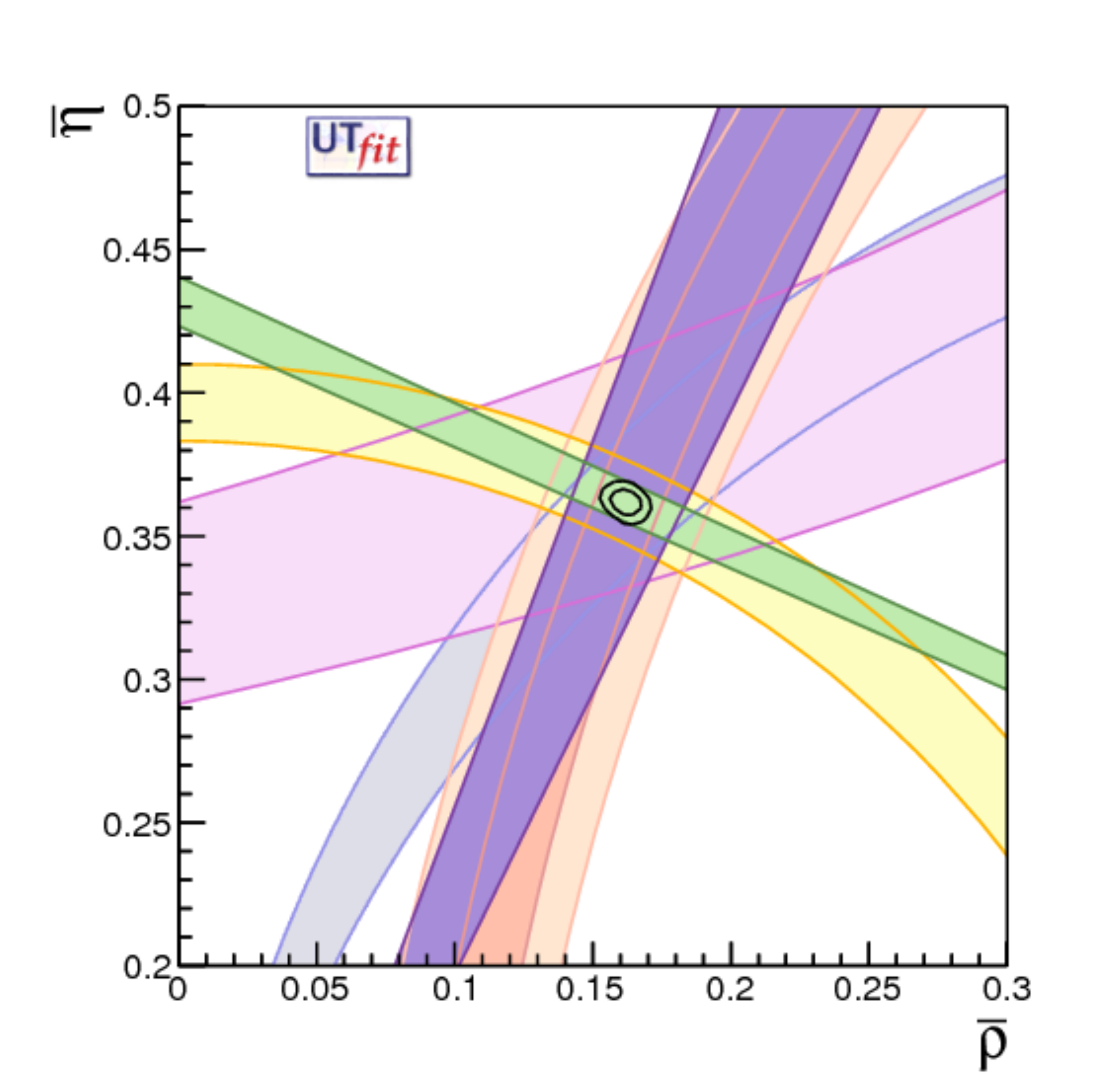}
\includegraphics[width=0.32\linewidth]{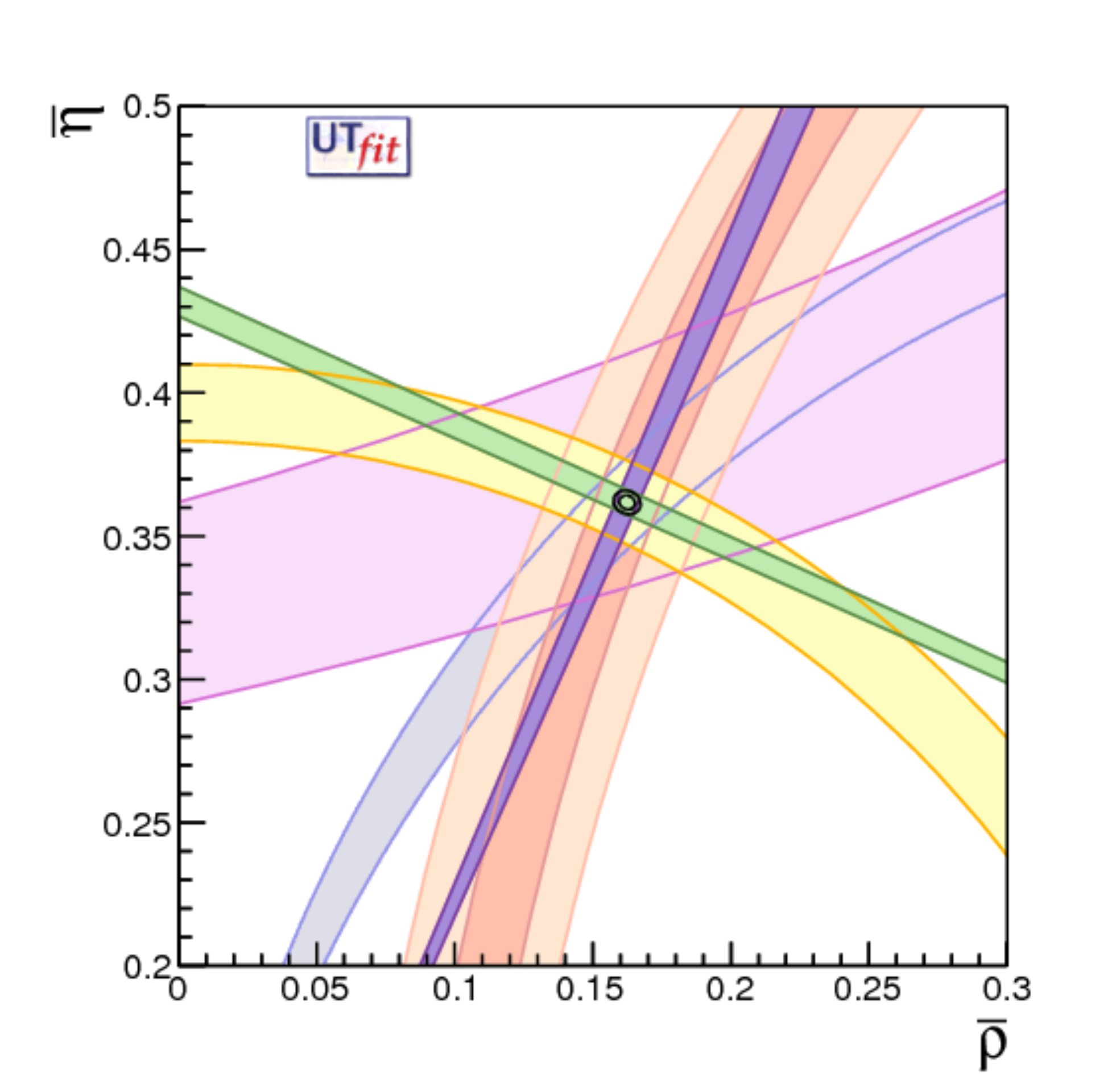} 
\caption{Present (left) and future (center: phase 1, right: phase 2) constraints in the $(\bar \rho, \bar \eta)$ plane (UTfit collaboration).}
\label{fig:UTfitFut}
\end{center}
\end{figure}

The improvement of the UT global analysis can be appreciated in Fig.~\ref{fig:UTfitFut}, where the present and future Phase 1 and Phase 2 constraints on the standard UT plane are shown next to each other, after zooming into the SM preferred region.
For a more quantitative comparison we collect in Table~\ref{tab:utfit_global} the uncertainties on the indirect determination of CKM parameters and angles, obtained from the predictive posterior p.d.f.'s (i.e., obtained without including the corresponding direct measurements in the fit). These uncertainties are reduced by a factor 3--5 for Phase 1, and are further reduced by up to a factor of 2 for Phase 2,
allowing for an increasingly improved tests of the SM, as discussed next. A similar progression of improvements is seen in the projections from CKMfitter collaboration, cf. Fig.~\ref{fig:CKMfitterglobalsmall}.

\begin{table}
\caption{Relative uncertainties on the predictions of UT parameters and angles, using current and extrapolated input values for measurements and theoretical parameters (UTfit collaboration).} 
\label{tab:utfit_global}
  \begin{center}
	\begin{tabular}{ccccccccc}
		\hline\hline 
		 & $\lambda$ & $\bar{\rho}$ & $\bar{\eta}$ & $A$ &  $\sin2\beta$ & $\gamma$ & $\alpha$ & $\beta_s$ \\ \hline
			Current &0.12\%& 9\% & 3\% & 1.5\% & 4.5\% & 3\% & 2.5\% & 3\% \\ 
			Phase 1  &0.12\%& 2\% & 0.8\% & 0.6\% & 0.9\% &  0.9\% & 0.7\% & 0.8\%\\ 
			Phase 2  &0.12\%& 1\% & 0.6\% & 0.5\% & 0.6\% & 0.8\% & 0.4\% & 0.5\%\\ \hline
 	\hline
	\end{tabular} 
\end{center}
\end{table}

\subsection{Future extrapolation of constraints on NP in $\Delta F=2$
  amplitudes}
\label{sec:DF2NP}

\begin{figure}[htb]
\begin{center}
\includegraphics[width=0.37\linewidth]{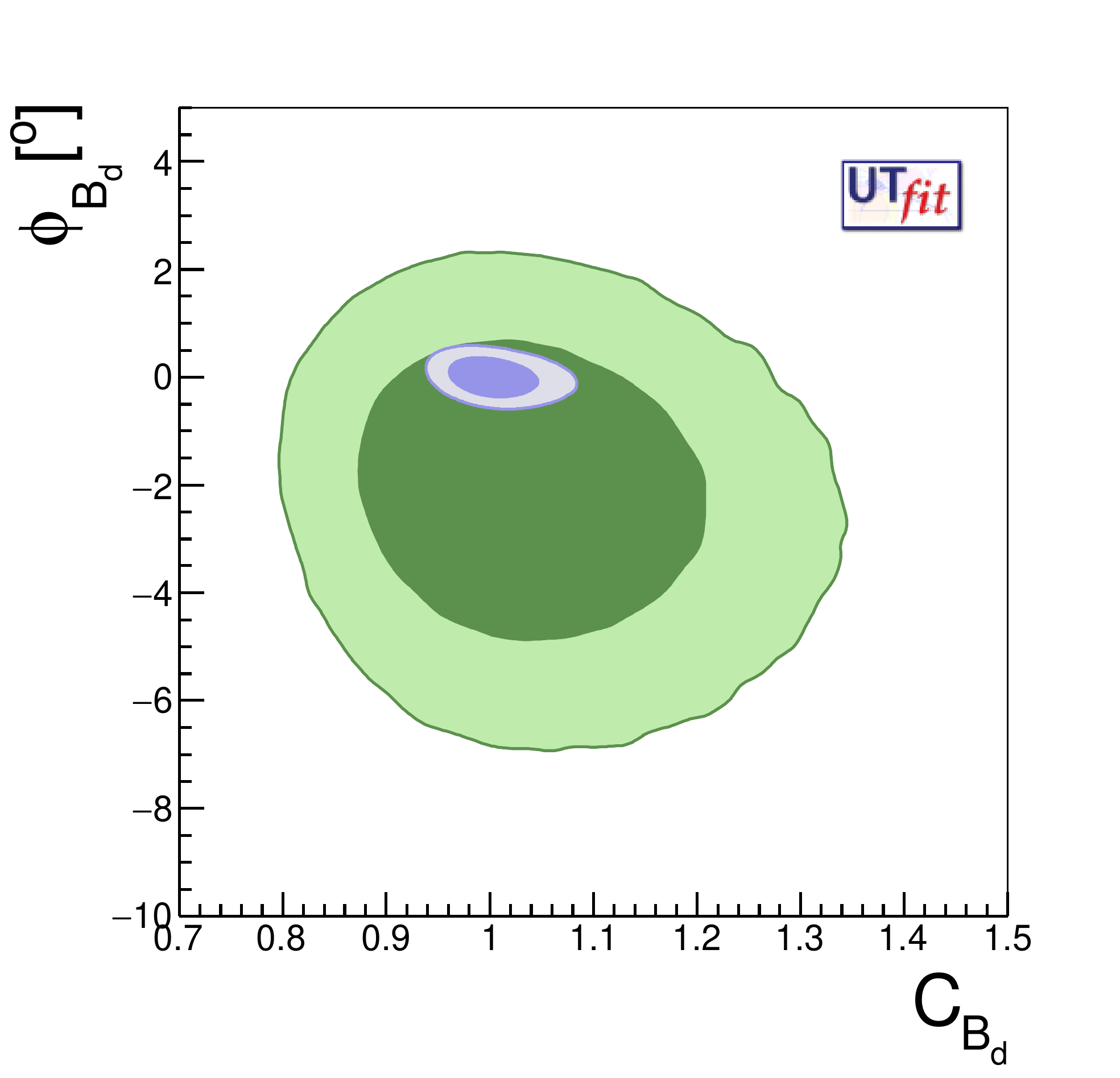}~~~~~~~~~~~~
\includegraphics[width=0.37\linewidth]{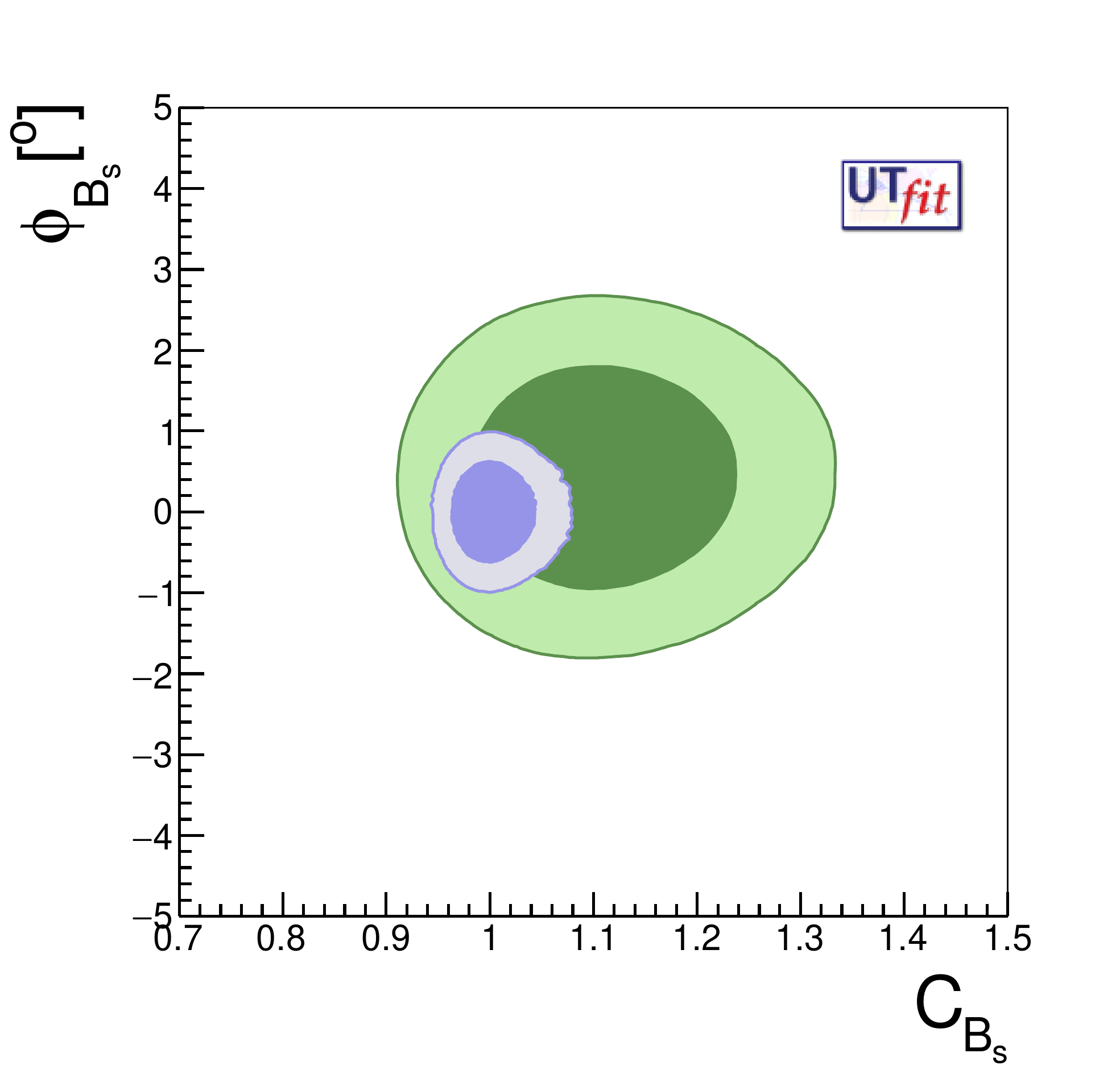} 
\caption{The present (green) and future Phase 2 (blue) constraints on NP contributions to $B_d$-$\bar{B}_d$ (left) and $B_s$-$\bar{B}_s$ (right) mixing, with $1\sigma$ ($2\sigma$) regions shown with darker (lighter) shading. }
\label{fig:UTAgen}
\end{center}
\end{figure}

The Unitarity Triangle Analysis can be generalized beyond the SM to
obtain a simultaneous determination of CKM parameters and NP
contributions to $\Delta F=2$ amplitudes \cite{Bona:2007vi,Lenz:2010gu}. Assuming
that NP is absent (or negligible) in charged current amplitudes, but
allowing for NP to be present in FCNC amplitudes, where its virtual
effects compete with loop-level SM amplitudes, we can still use the
measurements of $\vert V_{ud}\vert$, $\vert V_{us}\vert$, $\vert
V_{cb}\vert$, $\vert V_{ub}\vert$, $\gamma$ and $\alpha$ (allowing for
NP contributions in penguins, but barring order-of-magnitude
enhancements of electroweak penguins) to obtain the ``tree-level''
determination of the UT. This allows us
to obtain the SM prediction for $K$, $B_{d}$ and $B_{s}$ mixing
amplitudes. Comparing them with the experimental results we can
extract 
$C_{\varepsilon_K} = \varepsilon_K/\varepsilon_K^\mathrm{SM}$
and
\begin{equation}
  \label{eq:cbqphibq}
  C_{B_q} e^{i \phi_{B_q}} = \frac{\langle B_q \lvert H^\mathrm{SM+NP}
    \rvert \bar{B}_q \rangle}{\langle B_q \lvert H^\mathrm{SM}
    \rvert \bar{B}_q \rangle}\,.
\end{equation}
The SM point is $C_i=1, \phi_i=0$. 
Using semileptonic asymmetries it is possible to break the degeneracy
for $\gamma \leftrightarrow \gamma + 180^\circ$ present in the
tree-level determination of the CKM matrix \cite{Laplace:2002ik},
getting rid of the solution in the third quadrant. We then obtain the
results in Table~\ref{tab:UTAgen} for the projected errors on CKM parameters and on the NP parameters. Note that at present the NP contribution that are about an order of magnitude smaller than the SM are still perfectly allowed. At the end of Phase 2 we will be able to probe amplitudes that are another factor of 4 smaller than possible at present (corresponding to about a factor of 2 higher reach in the NP scale for dimension 6 NP operators).  The
corresponding two-dimensional distributions for $B_d$ and $B_s$ mixing are shown in
Fig.~\ref{fig:UTAgen}. 

\begin{table}
  \begin{center}
  \caption{Present and future uncertainties on CKM and NP parameters from the generalized UT analysis (UTfit collaboration).
  }   \label{tab:UTAgen}
	\begin{tabular}{cccccccc}
		\hline \hline
		 & $\bar{\rho}$ & $\bar{\eta}$ & $C_{\varepsilon_K}$ &  $C_{B_d}$ & $\phi_{B_d}[^\circ]$ & $C_{B_s}$ & $\phi_{B_s}[^\circ]$ \\ \hline
		Current & 0.030 & 0.028 & 0.12 & 0.11 & 1.8 & 0.09 & 0.89 \\ 
		Phase 2  & 0.0047 & 0.0040 & 0.036 & 0.030 & 0.28 & 0.026 & 0.29\\ \hline \hline
	\end{tabular} 
\end{center}
\end{table}

\begin{figure}[t]
\begin{center}
\includegraphics[width=0.49\linewidth]{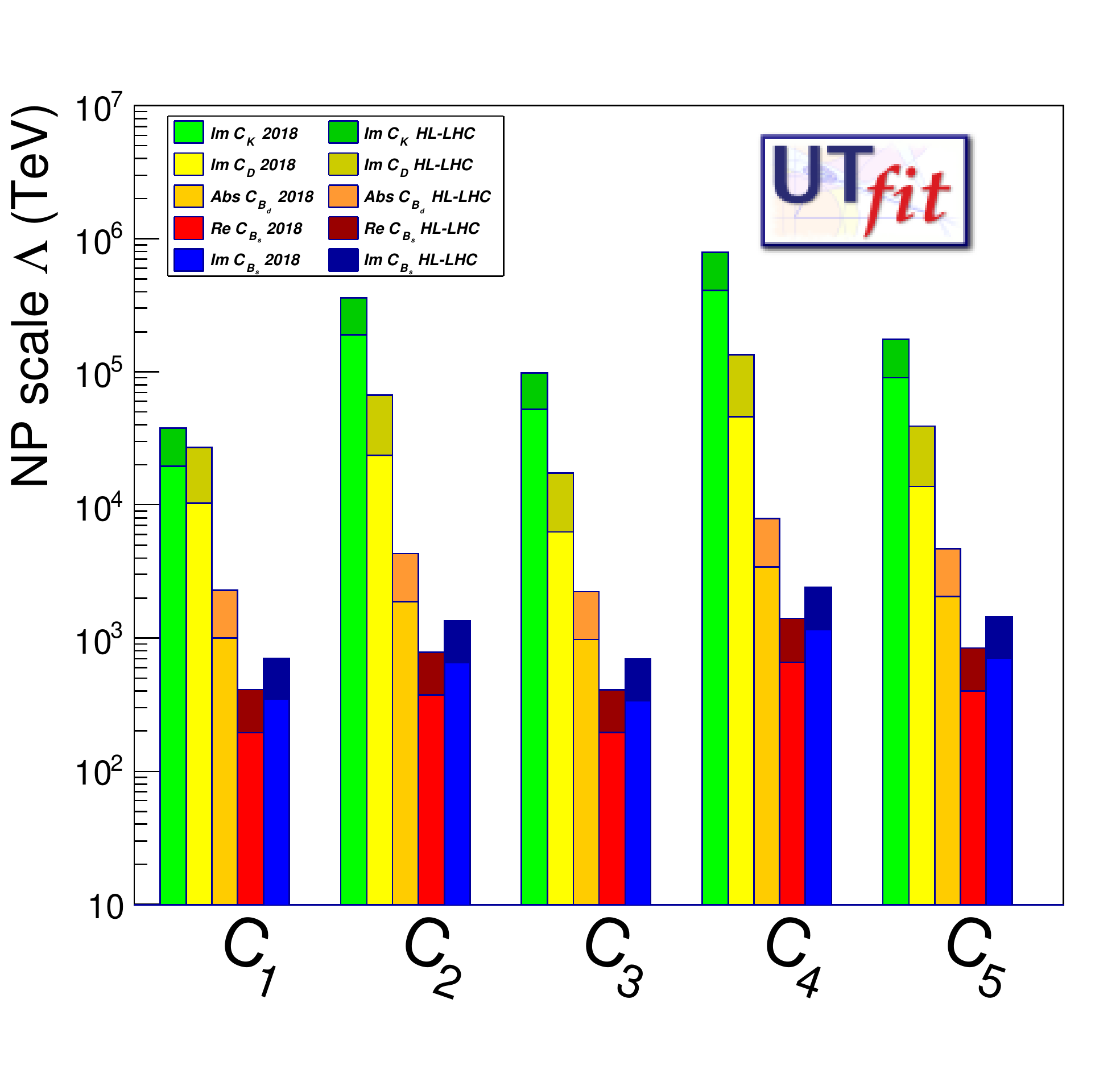}
\includegraphics[width=0.49\linewidth]{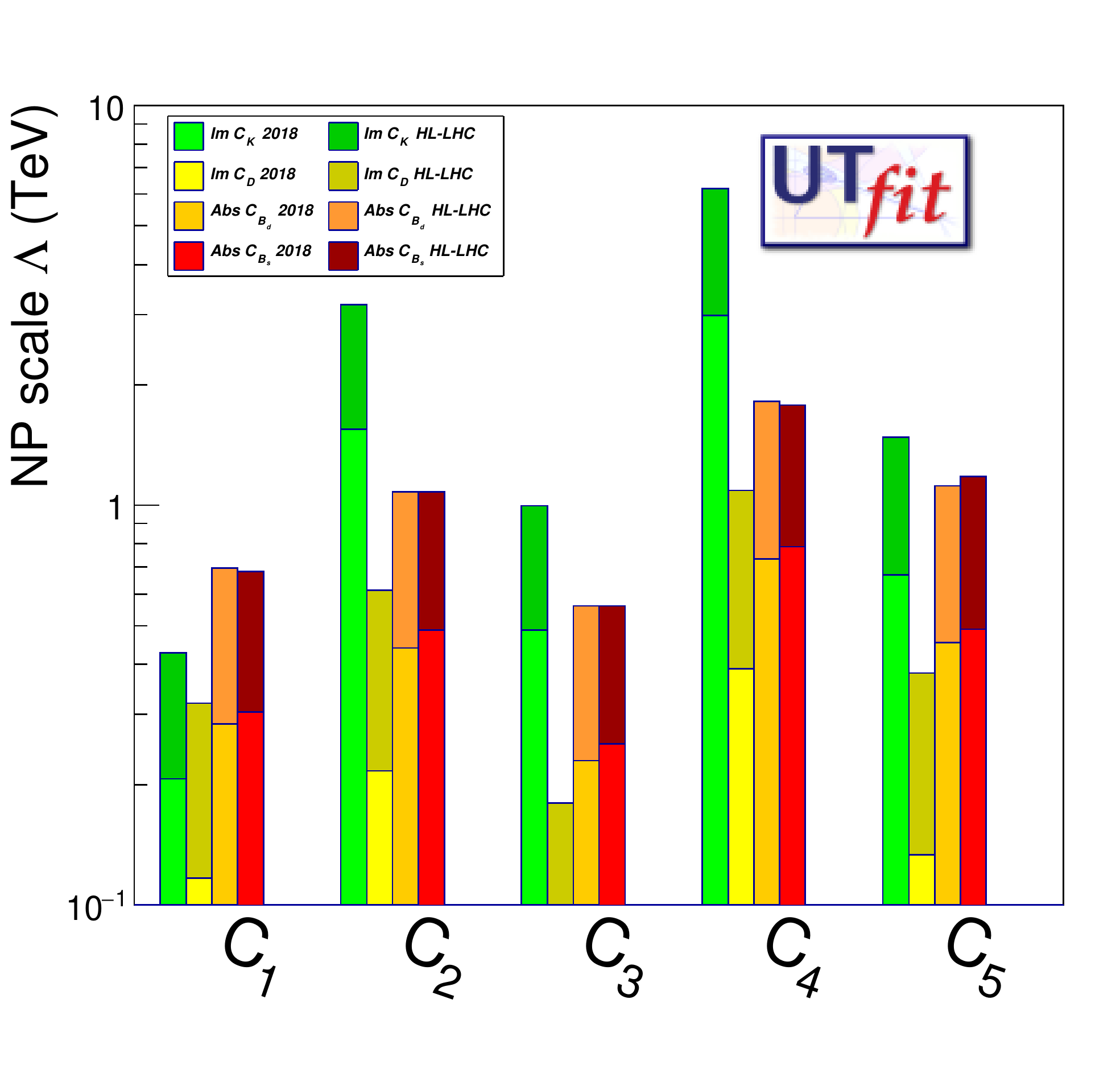}
\caption{Present (lighter) and future Phase 2 (darker) constraints on the NP scale from the UTfit NP analysis. The right panel shows constraints assuming NP is weakly coupled, has MFV structure of couplings, and enters observables only at one loop, see text for details. }
\label{fig:DF2Lambda}
\end{center}
\end{figure}

Combining the results of the generalized UT analysis with the
constraints on \CP violation in $D$ mixing from Sec. \ref{sec:CPV:mixing:Dmeson},
   we can consider the most general $\Delta
F=2$ effective Hamiltonian and place bounds on its coefficients (barring
accidental cancellations). The most general effective Hamiltonians for $\Delta F=2$ processes
beyond the SM have the following form \cite{Gabbiani:1996hi} (with $q_1q_2=sd,uc,bq$ for $M=K,D,B_q$)
\begin{equation}
  \label{eq:defheff}
  {\cal H}_{\rm eff}^{M-\bar M}=\sum_{i=1}^{5} C_i\, Q_i^{q_1q_2} +
  \sum_{i=1}^{3} \tilde{C}_i\, \tilde{Q}_i^{q_1q_2}, 
\end{equation}
where the operator basis consists of  dimension 6 operators
$Q_1^{q_iq_j} =\big(\bar q^{\alpha}_{jL} \gamma_\mu q^{\alpha}_{iL}\big)\big( \bar q^{\beta}_{jL} \gamma^\mu q^{\beta}_{iL}\big)$, 
$ Q_2^{q_iq_j}=\big(\bar q^{\alpha}_{jR}  q^{\alpha}_{iL}\big)\big( \bar q^{\beta}_{jR} q^{\beta}_{iL}\big)$, 
 $Q_3^{q_iq_j} =\big(\bar q^{\alpha}_{jR}  q^{\beta}_{iL}\big)\big(\bar q^{\beta}_{jR} q^{\alpha}_{iL}\big)$, 
$ Q_4^{q_iq_j} = \big(\bar q^{\alpha}_{jR}  q^{\alpha}_{iL}\big)\big(\bar q^{\beta}_{jL} q^{\beta}_{iR}\big)$, 
$ Q_5^{q_iq_j} = \big(\bar q^{\alpha}_{jR}  q^{\beta}_{iL}\big)\big(\bar q^{\beta}_{jL} q^{\alpha}_{iR}\big)$, 
and the operators $\tilde{Q}_{1,2,3}^{q_iq_j}$ that are obtained from the
$Q_{1,2,3}^{q_iq_j}$ by exchanging $ L \leftrightarrow R$.  Here
$q_{R,L}=P_{R,L}\,q$, with $P_{R,L}=(1 \pm \gamma_5)/2$, and $\alpha$
and $\beta$ are colour indices. Following the procedure detailed in Ref.~\cite{Bona:2007vi}, the UTFit collaboration obtained
p.d.f.'s for the Wilson coefficients based on the extrapolated UT and
$D$ mixing analyses. For self-consistency, the coefficients are computed
at a scale $\mu_H$ roughly corresponding to the bound on the NP scale
$\Lambda$ that one obtains from the analysis (see below). The
present and expected Phase 2 allowed regions at $95\%$ probability on the Wilson coefficients 
\begin{equation}
  \label{eq:CiLambda}
  C_i(\Lambda) = \frac{F_i L_i}{\Lambda^2}\,,
\end{equation}
are reported in Fig.  \ref{fig:DF2Lambda}. In the left panel in Fig.  \ref{fig:DF2Lambda} the flavour and loop factors were set to $F_i=L_i=1$, i.e., this shows the limits on the mass of NP states that contribute to meson mixing at tree level and couple with ${\mathcal O}(1)$ strength to the corresponding SM fermions. In Fig. \ref{fig:DF2Lambda} right, the flavour factor was set to $F_i=V_{t q_1}V_{t q_2}^*$, and the loop factor to $L_i=\alpha_2^2$, with $\alpha_2$ the weak structure constant. That is, the right panel of Fig. \ref{fig:DF2Lambda} shows the reach for masses of NP states that have MFV-like couplings to SM fermions and contribute only at one loop level. 

%% file: section2/experimental.tex
\subsection{Experimental prospects}
\label{sec:expckm}
\subsubsection{\vub and \vcb from semileptonic decays}
LHCb is well suited to measuring ratios of $b\to u\ell\nu$ to $b \to c\ell\nu$ decay rates,
in which the unknown $b$ production cross sections, and to some extent also efficiency corrections, cancel.
LHCb reported the first study of the \PMuNu and \LcMuNu decays  with Run~1 data, which resulted in a determination
of \vub/\vcb~\cite{LHCb-PAPER-2015-013}, exploiting precise lattice QCD calculations of the decay form factors~\cite{Detmold:2015aaa,Flynn:2015mha}.

LHCb Upgrade~II presents an exciting opportunity for new measurements of this type.
An excellent example is an analogous analysis of \KMuNu and \DsMuNu decays.
The relatively large spectator $s$ quark mass allows the form factors of these decays to 
be computed with lattice QCD to higher precision than decays of other $b$ hadrons.
There are also good prospects to extend the approach of~\cite{LHCb-PAPER-2015-013}, with a single $q^2$ bin,
to perform a differential measurement in many fine bins of $q^2$~\cite{Ciezarek:2016lqu},
which clearly demands substantially larger sample sizes.
Furthermore, there are several reasons to expect that, compared to the study of \Lb decays~\cite{LHCb-PAPER-2015-013}, far larger luminosities are required for the
ultimate precision with \Bs decays. 
Firstly, the \KMuNu signal rate is roughly one order of magnitude smaller compared to \PMuNu.
Secondly, the \KMuNu decay is subject to backgrounds from all $b$ meson species, whereas \PMuNu is primarily contaminated
by other \Lb decays.
Thirdly, the \KMuNu decay rate is further suppressed with respect to \PMuNu at the higher $q^2$ values at which the lattice QCD uncertainties are smallest.

\begin{figure}[t]
\centering
\includegraphics[width=0.6\linewidth]{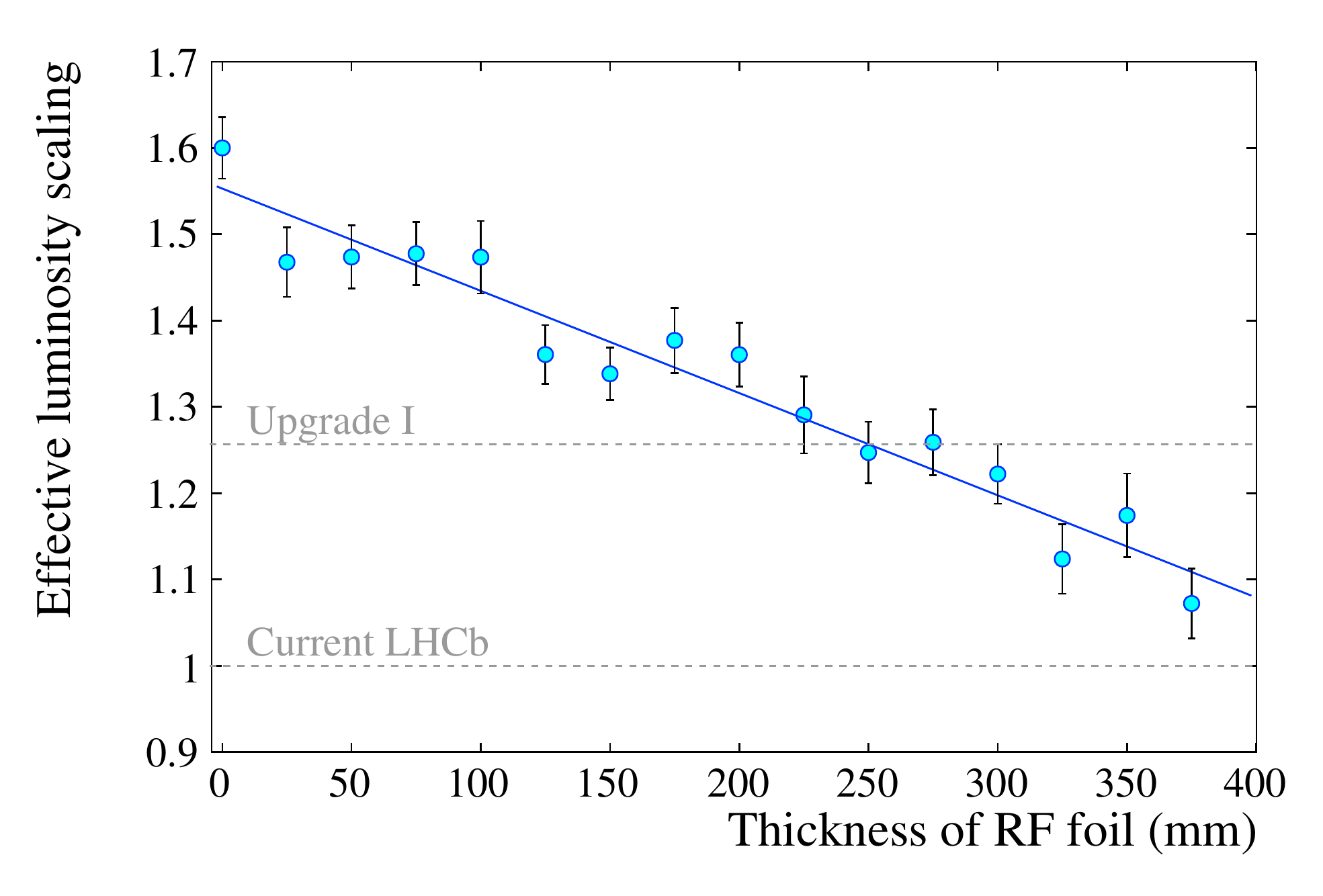} 
\caption{
The factor by which the effective luminosity in LHCb would be increased in an analysis of \KMuNu decays, as a function of the thickness of the RF foil.
This factor is defined to be with respect to the vertex resolution of the current detector.
It is determined using pseudo experiments in which the \KMuNu signal yield is extracted through a likelihood fit to the corrected mass distribution.
The other parameters affecting the vertex resolution are assumed to be
the same as in the upgraded LHCb detector, which will have a baseline foil thickness of 250 $\mu$m.
}
\label{fig:CorrectedMass}
\end{figure} 

LHCb Upgrade~II should also include several potential gains in the detector performance which are highly relevant to the reconstruction
of decays like \KMuNu.
The key variable which distinguishes the signal from background processes is the {\em corrected mass},
which depends on the reconstructed line-of-flight between the primary vertex (PV) and the \Bs decay vertex.
It is the resolution on this direction which dominates the corrected mass resolution.
The removal, or further thinning, of the RF foil is therefore a very appealing prospect, since this would reduce
the multiple scattering contribution to the corrected mass resolution.
Fig.~\ref{fig:CorrectedMass} shows the potential gain in effective luminosity that can be achieved by reducing the RF foil thickness.
This analysis only considers the effect of the improved corrected mass resolution, while further improvements
are expected in the selection efficiency, purity and $q^2$ resolution.

The dominant systematic uncertainty in the \Lb analysis~\cite{LHCb-PAPER-2015-013} (material budget
and its effect on the charged hadron reconstruction efficiency) can be tightly constrained 
with new methods, the performance of which will be greatly enhanced by any reduction in the RF foil. 
The lattice QCD form factors are most precise at large $q^2$ values, which correspond to low momentum kaons
that are not efficiently identified with the RICH approach of the current detector. 
The low momentum particle identification (PID) performance of the proposed TORCH detector would greatly enhance 
our capabilities with \KMuNu decays at high $q^2$.

The combination of these detector improvements with the LHCb Upgrade II data set is expected to reduce the experimental systematic uncertainties on \vub/\vcb to the 0.5\% level. 
External branching ratio uncertainties will also be greatly reduced at the BESIII experiment. 
For example, $\mathcal{B}(D_{s}^{+}\to K^{+} K^{-} \pi^{+})$ will be determined at the 1\% level, translating to a $\sim 0.5\%$ uncertainty on \vub/\vcb.
Combined with the differential shape information of the signals and continued improvements to the lattice 
calculations this will lead to a \vub/\vcb measurement uncertainty of less than 1\% with the LHCb Upgrade II data set.

LHCb Upgrade II will allow several currently inaccessible decays, not least those of the rarely produced \Bc mesons,
to be studied in detail.
\Blu{A prime example is the decay \decay{\Bc}{\Dz\mup\neum}, which is potentially
very clean from a theory point of view, once lattice QCD calculations of the form factor become available.}
Approximately 30,000  reconstructed candidates can be expected with the 300\invfb LHCb Upgrade II data set, which could lead to a competitive measurement of \vub
in this, currently unexplored, system.
Purely leptonic decays, such as $B^{+} \to \mu^{+}\mu^{-}\mu^{+}\nu_{\mu}$, 
will also become competitive, with similar signal yields, and can provide information on the 
$B$-meson light cone distribution amplitude, which is a crucial input to the widely used theoretical tool of QCD factorisation \cite{Beneke:1999br,Beneke:2000ry}.

The standalone determination of $|V_{cb}|$ will also be increasingly important when other measurements get more precise,
as it will become the limiting factor in many SM predictions such as the branching fraction of $\Bs\to\mumu$~\cite{Buras:2012ru}, as well as $K\to \pi \nu \bar \nu$, and for $\epsilon_K$. 
The current uncertainty is inflated due to the disagreement between measurements from inclusive and exclusive final states and currently appears to 
critically depend on the parametrisation used to fit the form factors~\cite{Grinstein:2017nlq,Bigi:2017njr}. 
LHCb has already performed a measurement of the differential rate of the decay $\Lb \to \Lc \mu \nu$, allowing a determination of the form factors
of that decay~\cite{LHCb-PAPER-2017-016}. A first determination of the absolute value of $|V_{cb}|$, exploiting theoretical predictions for the ratios of semileptonic
decay widths between different $b$ hadron species~\cite{Bigi2011}, is in progress.
The LHCb Upgrade II data set would provide large samples of exclusive $b \to c \ell\nu$ decays, with the full range of $b$ hadron species,
with which very precise shape measurements could be performed, as a crucial ingredient to reach the ultimate precision on \vcb.

\subsubsection{Time-integrated tree-level measurements of $\gamma$}

LHCb has observed and studied two-body ADS/GLW modes during Run 1 and Run 2~\cite{LHCb-PAPER-2016-003,LHCb-PAPER-2017-021}; all ADS/GLW asymmetries are statistically limited.
The systematic uncertainties are small and arise predominantly from sources that will naturally decrease with increasing data, notably knowledge of instrumentation asymmetries. Methods for measuring and correcting for the $B$-meson production asymmetry and the \Kpm/\pipm reconstruction asymmetries are established using calibration samples; such samples will continue to be collected.
The dominant systematic uncertainties for GLW decays are due to background contributions from $\Lb \to \Lc K^-$ decays and charmless decays, while the dominant uncertainty for ADS decays arises from the $B_{s}^0 \to \Dzb \Km \pip$ background.
All will be better determined with dedicated studies as the sample size increases.

LHCb is expanding the ADS/GLW technique to the other $\B \to D K$ decays, which share the same quark-level transition.
An analysis of GLW observables in $\Bpm \to D^{*0} \Kpm$ decays has been developed for the case where the $D^{*0}$ vector meson is not fully reconstructed~\cite{LHCb-PAPER-2017-021}.
This partial reconstruction has larger background uncertainties but these will improve with more data as dedicated studies of the background are performed.
Furthermore, ADS/GLW analyses have been developed and published in quasi-two-body modes, $\Bpm \to D \Kstarpm$~\cite{LHCb-PAPER-2017-030} and $\Bz \to D K^{*0}$~\cite{LHCb-PAPER-2014-028}.
As in the case of $\Bpm \to D^{(*)0} \Kpm$ decays, these modes have no limiting systematic and they will make competitive contributions with the \upgradetwo data sets.

\begin{figure}[t]
  \centering
  \includegraphics[width=0.50\textwidth]{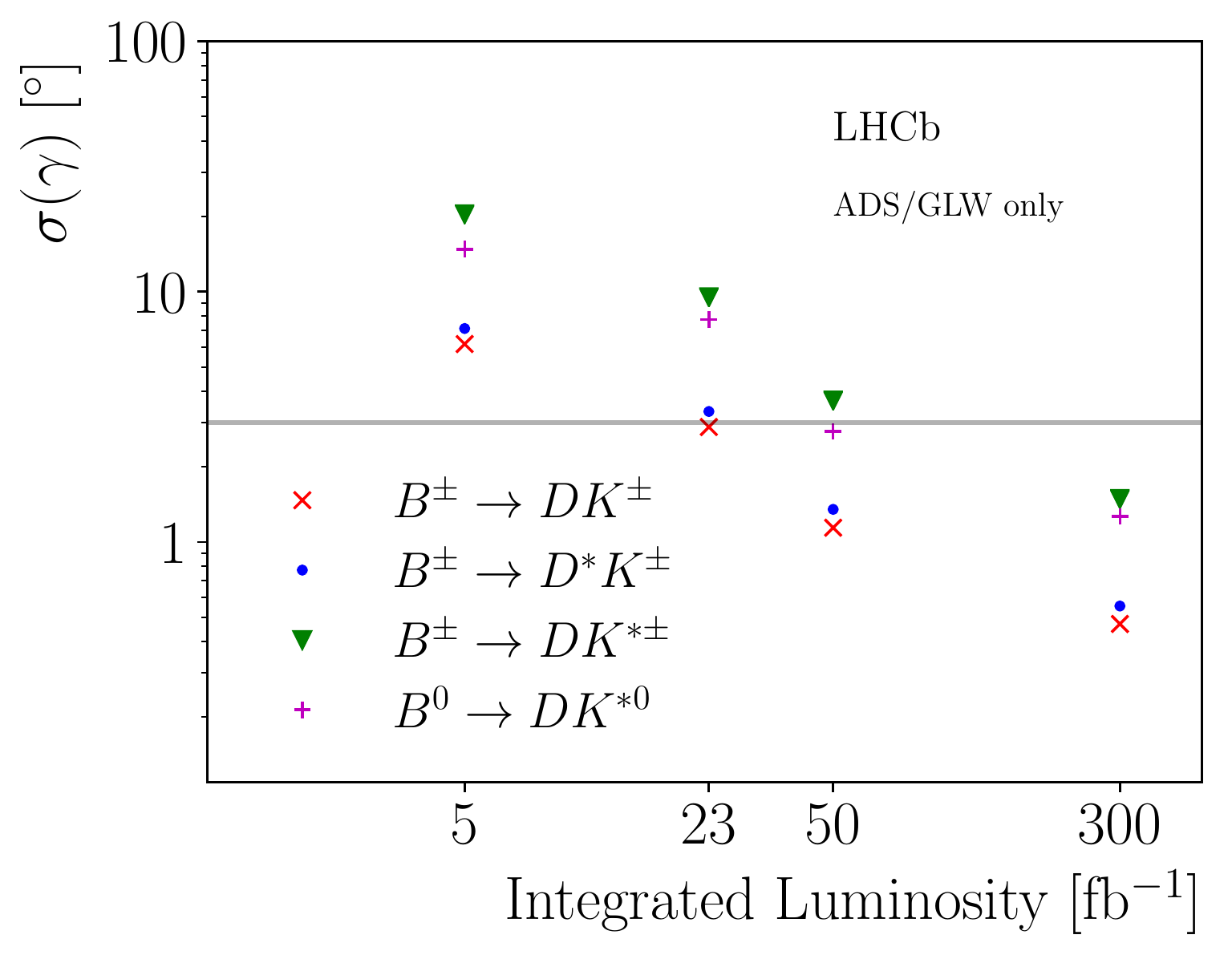}
  \caption{Extrapolation of \g sensitivity from the ADS/GLW analyses at LHCb, ignoring disfavoured ambiguities. The expected Belle-II precision on $\gamma$  at an integrated luminosity of 50 ab$^{-1}$ is shown by the horizontal grey line.
    \label{2body_gamma_projection}
  }
\end{figure}

Under the assumption that systematic uncertainties decrease in parallel with the statistical uncertainties as $\propto1/\sqrt{\mathcal{L}}$, the future precision on \g is predicted in Fig.~\ref{2body_gamma_projection}.
Fig.~\ref{2body_gamma_projection} uses central values and uncertainties in the published analyses of $B^\pm\to DK^\pm$, $B^\pm\to DK^{*\pm}$ and $B^0\to DK^{*0}$ decays.
For $B^\pm\to D^{*0}K^\pm$ both the partial and full reconstruction techniques are used in this study albeit with unpublished central values and uncertainties.

\begin{figure}[t]
  \centering
  \includegraphics[width=0.42\textwidth]{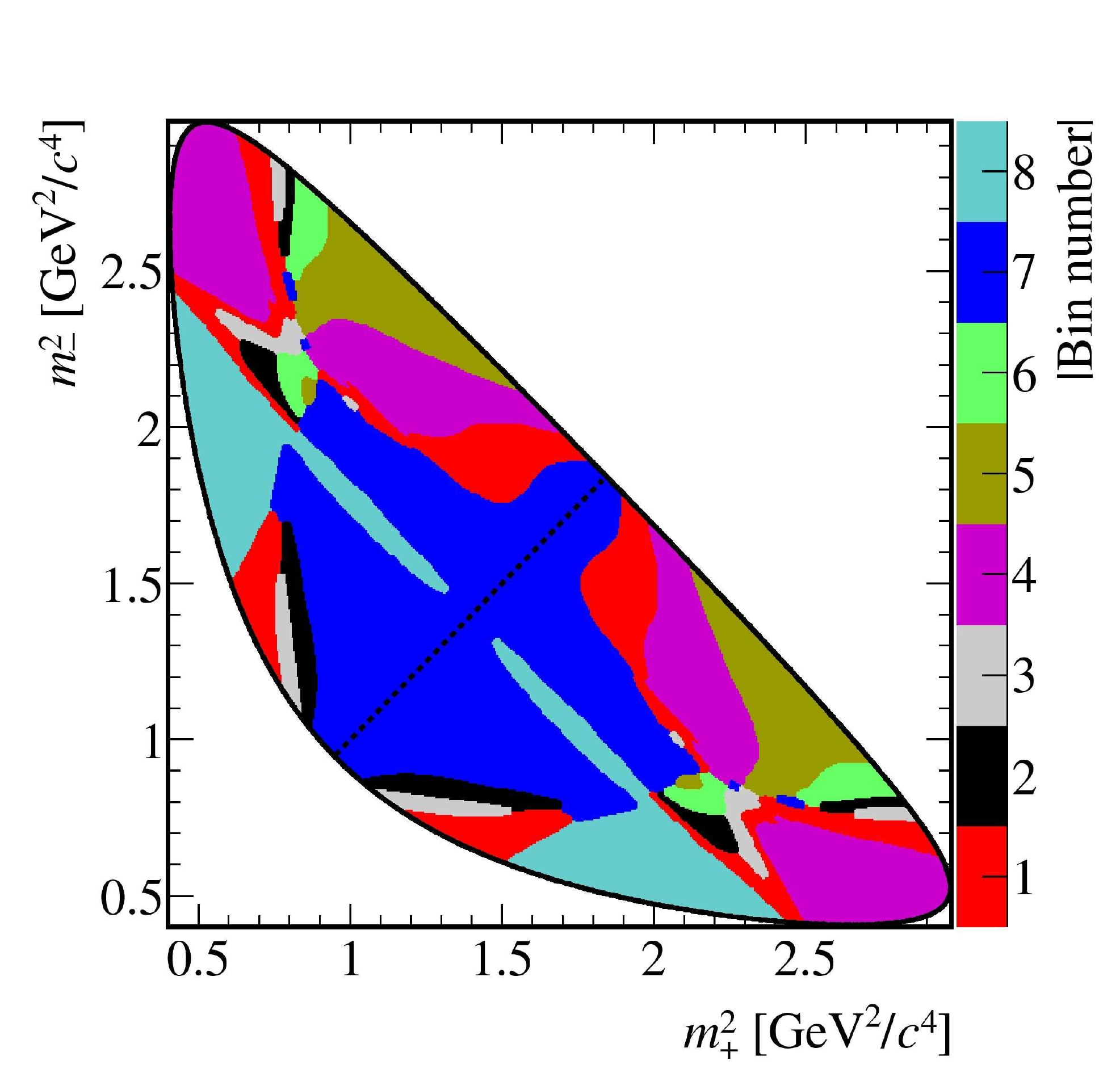}
  \includegraphics[width=0.57\textwidth]{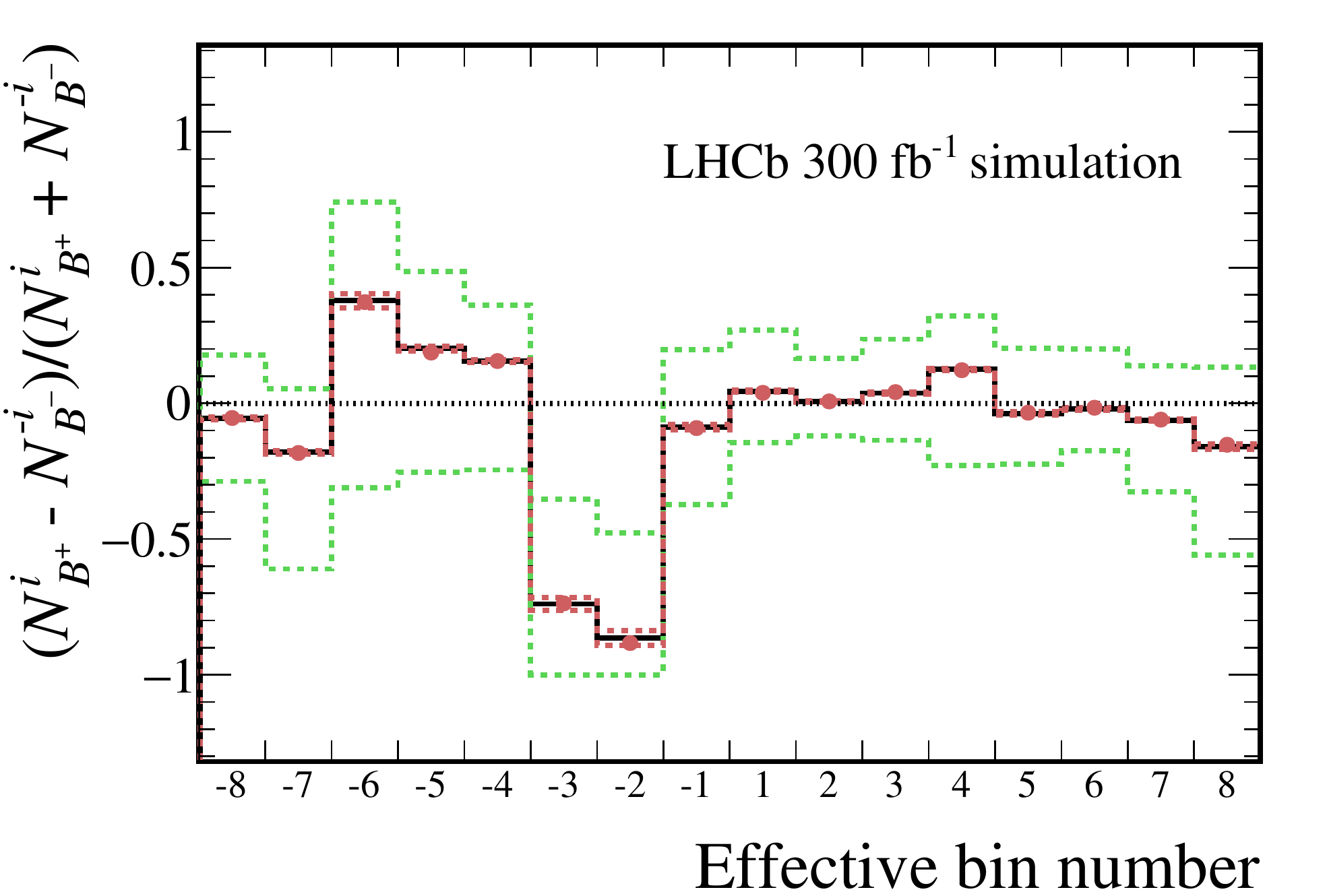}
  \caption{
    (Left) Bin definitions for \DtoKspp\ as a function of $\mmsq$ and $\mpsq$, the invariant masses squared of the $\KS h^-$ and  $\KS h^+$.
    (Right) Asymmetry between yields for $\Bpm \to D\Kpm$ decays, with \DtoKspp\ in bin $i$ (for \Bp) and $-i$ (for \Bm).
    The data points are obtained from simulation with the expected sample size at $300 \invfb$, assuming the current performance of the LHCb experiment.
    The black histogram shows the predicted asymmetry based on the current world average values of $\gamma$ and relevant hadronic parameters, the red dots show the result of a single pseudoexperiment, while the red bands show the expected uncertainties from an ensemble.
    The green bands show the corresponding uncertainties with the current LHCb data set.
    \label{fig:bins}
  }
\end{figure}

For the GGSZ family of measurements, the 
model-independent method is expected to be the baseline for LHCb \upgradetwo\ , and its
uncertainty is currently statistically dominated. Although systematic uncertainties
will already become significant compared to the statistical uncertainty in Run 3, studies
performed so far give confidence that the systematic uncertainties will generally scale with the
statistical reach well into the \upgradetwo\ period. An example of the bin definitions and expected
per-bin asymmetries in LHCb \upgradetwo\ can be seen in Fig.~\ref{fig:bins}.

\begin{figure}[t]
  \centering
  \includegraphics[width=0.49\textwidth]{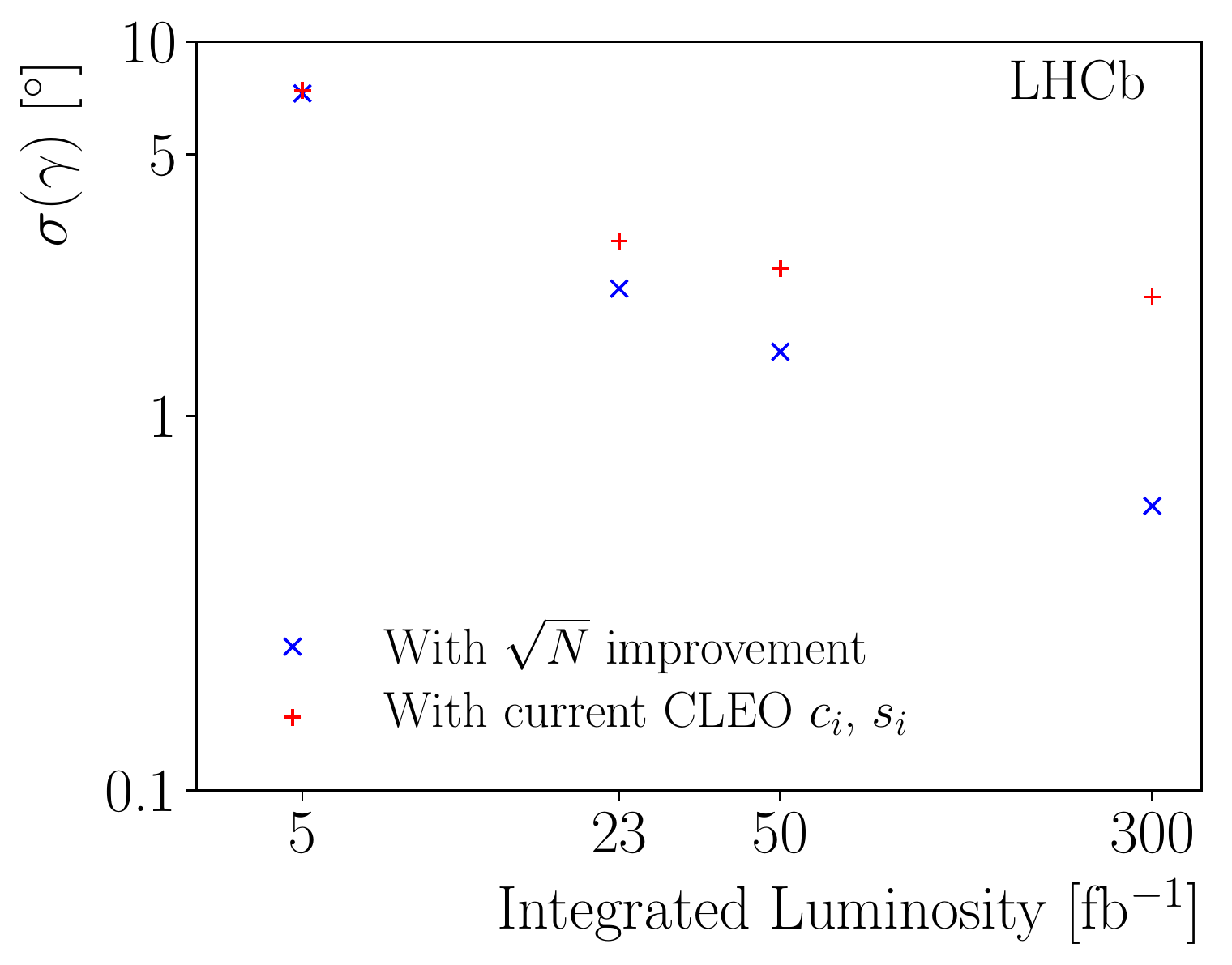}
  \caption{Expected evolution of \g sensitivity with (red crosses) current CLEO inputs and with (blue x-marks) $\sqrt{N}$ improvements on their uncertainty.
    \label{fig:41:sensitivity}
  }
\end{figure}

The largest systematic uncertainty is due to the precision of the
external strong-phase inputs coming from the CLEO-c data which currently contribute approximately $2\degrees$ to the overall uncertainty
on $\gamma$~\cite{LHCb-PUB-2016-025}.
The impact of this uncertainty on GGSZ measurements is estimated in Fig.~\ref{fig:41:sensitivity} which compares a $\sqrt{N}$ improvement with the
expected yield increase and the projected uncertainty if the current external information on $c_i$ and $s_i$ is not improved.
It can be seen that this starts to approach a limit originating from the fixed size of the quantum-correlated charm input from CLEO-c.
This contribution will naturally decrease with larger $\Bpm\to\Dz\Kpm$ samples, however, as the $B$ decays themselves
also have sensitivity to $c_i$ and $s_i$. Studies performed using pseudoexperiments with different size quantum correlated $D$ samples and LHCb
$B$ data suggest that the optimal sensitivity is only reached when the size of the input $D$ sample is at least as big as the overall $B$ sample.
More precise measurements with data already recorded by the BESIII experiment will be able to
reduce the external contribution to the uncertainty by around 50\% but analysis of future larger data sets with BESIII and at \lhcb will be vital
in order to avoid the external input compromising the ultimate sensitivity to \g.

The current second largest source of systematic uncertainty
comes from the knowledge of the distribution of $D$ decays
over the Dalitz plane in the flavour-specific $\B$ final state, with reconstruction and efficiency effects incorporated.
These are determined with a flavour-specific control decay mode \BztoDstmuX, where the \Dstarm decays
to \Dzb\pim and $X$ represents any unreconstructed particles.
The ultimate systematic uncertainty is particularly sensitive to data-simulation agreement and size of simulated samples because
of a need to model unavoidable differences in the signal and control modes.
Fast simulation techniques which are being deployed and further
developed at LHCb will therefore be crucial for keeping up with the large data samples,
while a fully software-based trigger will allow for a better alignment of the signal and control channel selections compared to today.
Uncertainties from sources such as low mass backgrounds can be expected to remain subdominant
with higher statistics, as further studies
will give better understanding to their rates and shapes.
Some additional complications will present themselves as the yields will eventually be high enough
that it will become necessary to take into account effects induced from asymmetries in the
\KS system, and eventually through mixing in the $\Dz$ system. However these are tractable problems,
and studies have already been done to understand when these effects will become important.

There are many prospects for adding orthogonal information on \g by applying the ADS/GLW and GGSZ techniques to modes that have an additional \piz.
The first use of \piz mesons in an LHCb \g analysis occurred with the Run 1 ADS/GLW analysis of $\Bpm \to D\Kpm$ decays with $D \to K\pi \piz,\ KK\piz, \pi\pi\piz$ 
final states~\cite{LHCb-PAPER-2015-014}. The $K\pi\piz$ and $\pi\pi\piz$ modes have branching fractions 3 and 10 times larger than their two-body equivalents. 
However, the \piz reconstruction+selection efficiency in these decays is low, around 3\% with the current calorimeter. 
Also the analysis is complicated by a combinatorial background arising from random $\piz$ association. 
Improvements to the \upgradetwo\ calorimeter granularity and energy 
resolution will therefore be crucial in achieving the ultimate sensitivity with these modes, especially by improving the resolution of merged \piz, their separation
for photons, and improving the \piz mass resolution.

An important mode under development for the upgrade era is $\Bpm \to D^{*0} \Kpm$ decays, with $D^{*0}\to\Dz\piz$ and $D^{*0}\to\Dz\gamma$ decays. 
These twin modes provide an excellent sensitivity to \g as an exact phase difference between the two $D^{*0}$ modes can be exploited~\cite{Bondar:2004bi}. 
For this case, the efficient distinction of \piz and \g calorimeter objects is critical as the two $D^{*0}$ modes exhibit opposite \CP asymmetries. 
The initial studies show small, but clean signals.
As long as the fully and partially reconstructed data sets are kept statistically independent, the final sensitivity from $\Bpm \to D^{*0} \Kpm$ decays will be around $0.5^\circ$ as seen in Fig.~\ref{2body_gamma_projection}.

A GGSZ-like analysis of $\Bpm\to\D[ \to K_{s}^{0} \pi\pi\piz]\Kpm$ decays has recently been proposed for \belletwo, where a sensitivity approaching that of 
the $D \to K_{s}^{0} \pi\pi$ GGSZ analysis is expected~\cite{K:2017qxf}. With improved \piz efficiency, LHCb \upgradetwo can exploit this mode competitively. 
Lastly, higher \piz efficiency will merit the analysis of $B^\pm\to DK^{*\pm}[\to \Kpm\piz]$ decays. The \piz reconstruction efficiency is typically a factor 
3-4 lower than that of the \KS so the $\KS\pipm$ mode is preferred. 
However the $B^\pm\to D\Kpm\piz$ Dalitz analysis for \g should share many advantages of the isospin-conjugate decays $\Bd\to D \Kp\pim$ analysis (discussed next) 
but with reduced \Bs feed down and large asymmetries in the ADS-like region of the Dalitz space.

Analogously to the neutral modes, a variety of high-multiplicity $B$ and $D$ modes are already being established and will play an important role in a future determination of \g.
The most developed multibody $B$ decay channel is $\Bd\to D \Kp\pim$ decays, where the \D meson is found in an ADS/GLW-like ($K\pi,\ KK,\ \pi\pi,\ K3\pi,\ ... $) or GGSZ-like ($\KS\pi\pi$,\ $\KS KK,\ ...$) final states. 
These modes are less abundant than the equivalent \Bpm modes but the fact that both the $b\to c\bar{u}s$ and $b\to u\bar{c}s$ transitions proceed by colour-suppressed amplitudes means the GLW asymmetries can be very large. Furthermore, understanding the pattern of asymmetry across the \Bd Dalitz plane quashes the trigonometric ambiguities. 
This analysis has been established with the $KK$ and $\pi\pi$ modes using Run 1 data in Ref.~\cite{LHCb-PAPER-2015-059} after the development of the $\Bd\to\Dzb\Kp\pim$ amplitude model~\cite{LHCb-PAPER-2015-017}. 
Although the statistical sensitivity to \CP violation using Run 1 data alone was not significant, the method remains promising for future analysis given the high value of $r_B$ in this mode ($\sim0.25$).
The extension to $\Bd\to D[\to\KS\pip\pim] \Kp\pim$ decays is of particular importance as it allows for a so-called ``double Dalitz'' model-independent analysis to be performed~\cite{Gershon:2016fda,Gershon:2008pe}. 
Extrapolating yields from a Run 1 $\Bd\to\Dz\Kstarz$ analysis to a dataset corresponding to an integrated luminosity of 23 \invfb, sensitivity studies indicate that a precision on $\gamma$ of $3^\circ$ can be expected~\cite{Craik:2017dpc}, thus by extension, a sub-degree precision is expected for \upgradetwo.
Another high-multiplicity mode that holds promise is $B^\pm \to D \Kpm \pip\pim$, which has been studied with Run 1 data with two-body \D decays~\cite{LHCb-PAPER-2015-020}, though due to the unknown Dalitz structure of the $B$ decay a much larger dataset is needed to fully exploit this mode. 

\begin{figure}[t]
\begin{center}
\includegraphics[width=0.5\textwidth]{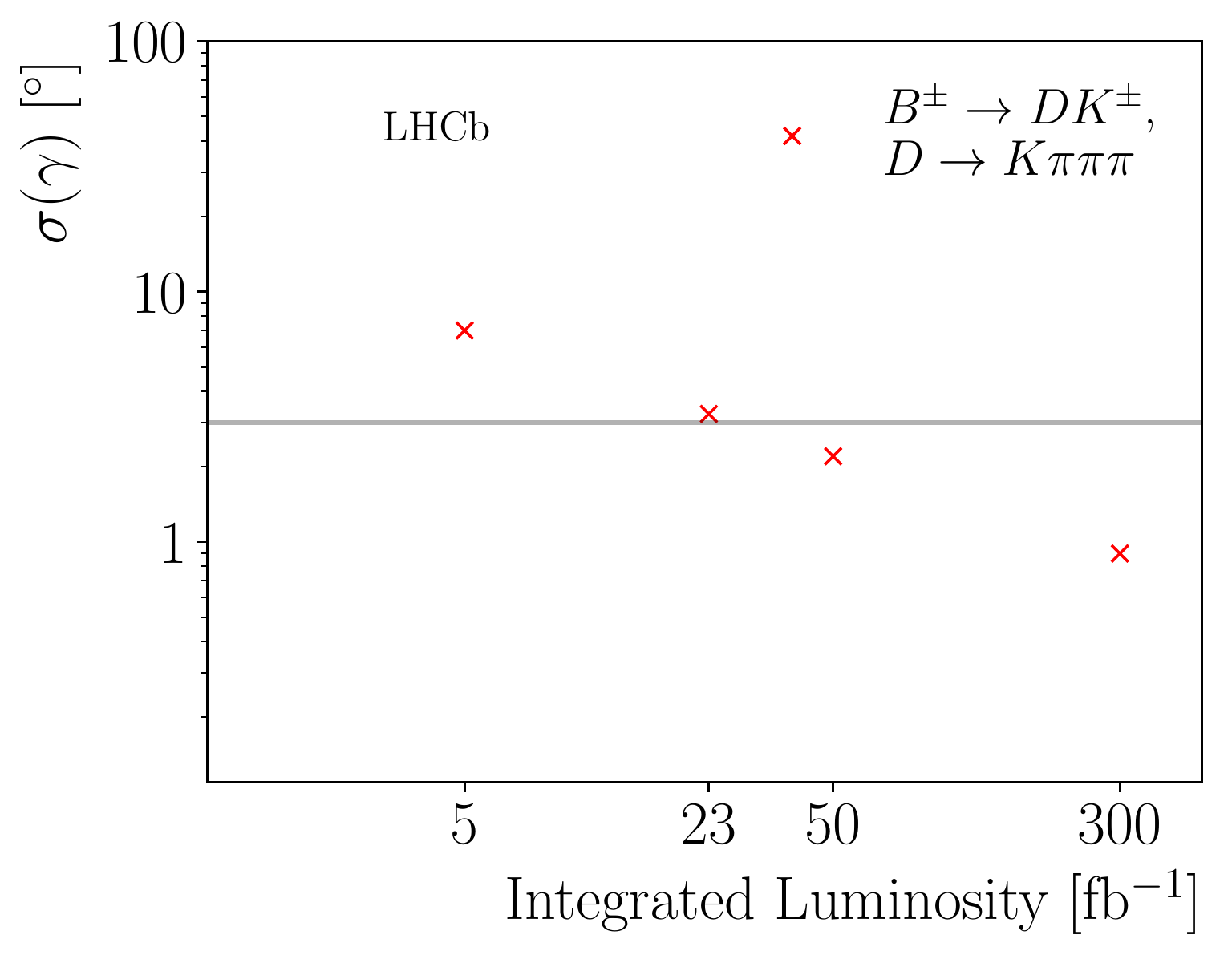}
    \captionof{figure}{Projected sensitivity to $\gamma$ from a binned analysis of $\Bpm \to D[\to\pi K\pi\pi] \Kpm$ decays. The expected Belle-II precision on $\gamma$  at an integrated luminosity of 50 ab$^{-1}$ is shown by the horizontal grey line.
    \label{4body_gamma_projection}}
  \end{center}
\end{figure}

Another exciting extension to the standard ADS technique uses $\Bpm \to D[\to\pi K\pi\pi] \Kpm$ decays where the five-dimensional (5D) Dalitz volume of the $D$ decay is split into bins. 
Excellent sensitivity to \g is achieved as long as the $D$-system parameters ($r_D,\,\delta_D,\,\kappa$) in each bin are known.  
An optimal binning scheme is soon to be defined from the amplitude analysis of the suppressed and favoured $D^0\to\pi K\pi\pi$ and $D^0\to K\pi\pi\pi$ decays~\cite{LHCb-PAPER-2017-040}.
The expected sensitivity is shown in Fig.~\ref{4body_gamma_projection}. An additional irreducible uncertainty from the \D system measurements of $<1^\circ$ is expected.

Other multibody \D decays are under development with reciprocal charm-system measurements underway: a $\Bpm \to D[\to4\pi] \Kpm$ analysis can build on the \D-system knowledge gathered in Ref.~\cite{Harnew:2017tlp}; an ADS analysis of $\Bpm \to D[\to\KS K\pi] \Kpm$ decays has been demonstrated with Run 1~\cite{LHCb-PAPER-2013-068}; and work on $\Bpm \to D[\to KK\pi\pi] \Kpm$ decays is envisaged.

\subsubsection{Time-integrated loop-level measurements of $\gamma$}

There are numerous methods that allow the extraction of 
CKM information by combining amplitude measurements made in
different decay channels, in particular the ${B^0_{(s)} \to \KS h^+h^-}$,
${B^0_{(s)} \to h^+h^-\piz}$, and ${B^+ \to h^+h^+h^-}$ families.
Some such methods exploit the interference between various $\Kstar\pion$ or
$\rho\kaon$ contributions, which can be related using isospin
symmetry~\cite{Ciuchini:2006kv,Ciuchini:2006st,Gronau:2006qn,Gronau:2007vr,Bediaga:2006jk,Charles:2017ptc}.
Constructing such relations enables the determination of the angle
$\gamma$, up to corrections for contributions from electroweak penguins.
These methods are particularly promising when using decays such as
${\Bs \to \KS \pip \pim}$ and ${\Bs \to \Km \pip \piz}$, since here the
electroweak penguin contributions are expected to be negligible.
Other methods use the whole Dalitz plot amplitude and relate numerous
decays via flavour symmetries~\cite{ReyLeLorier:2011ww,Bhattacharya:2013cla,Bhattacharya:2015uua}
or exploit interference between the charmless decays and those that proceed
via intermediate charmonium states such as
$\Bp\to\chiczero\pip$~\cite{Bediaga:1998ma,Blanco:2000gw}.

Extrapolating the observed yields in Ref.~\cite{LHCb-PAPER-2014-044} to the data sample 
expected to be collected at the end of LHCb \upgradetwo\ gives signal yields of between 
1.2 and 36 million depending on the ${B^+ \to h^+h^+h^-}$ final state.
These can be compared to an extrapolation of the yields obtained by the
\B factories~\cite{Aubert:2008bj,Hsu:2017kir} to the expected 50\invab
sample to be collected by \belletwo, which gives approximately 47,000 to
660,000 events for the same range of modes.
The far larger yields, combined with the much better signal to background
ratios, mean that LHCb will continue to dominate the precision in these modes.

Amplitude analyses cannot benefit from cancellations of systematic
effects to the same degree as the binned measurements of asymmetries performed in Ref.~\cite{LHCb-PAPER-2014-044}, and are more likely to become systematically limited.
On the other hand, the extremely large ${B^+ \to h^+h^+h^-}$ signal samples
will allow new information to be extracted.
In particular, by performing coupled-channel analyses of these decay modes,
contributions from $\pip\pim\leftrightarrow\Kp\Km$ rescattering processes
can be better understood and constrained.
The development of such new, improved models will also be of enormous
benefit to analyses of many other decay modes, such as the closely related
$\B \to \KS h^+ h^-$ and $\B \to h^+ h^- \piz$ families.
This will help to reduce the corresponding uncertainty on the CKM phases
that can be determined from those channels.

\subsubsection{Time-dependent measurements of $\gamma$} 
\label{sec:gammatd}
Using Run~1 data, \lhcb~has determined the $B_s^0 \to D_s^\mp K^\pm$ oscillation parameters, which determine $\gamma-2\beta_s$, from a sample of about 6000 signal decays~\cite{LHCb-PAPER-2017-047}. 
From the measured parameters,  a value of $\gamma$ of $(128\,^{+17}_{-22})^\circ$ (modulo $180^\circ$) is determined. 
The result is dominated by statistical uncertainties thanks to the wide use of data-driven methods to determine the decay-time acceptance and resolution, and to calibrate the flavour tagging.
The systematic uncertainties are, in decreasing order of importance, related to background from $b$-hadron decays, to uncertainty on the value of $\Delta m_s$, to the calibration of the decay-time resolution and to the flavour tagging. 
The first contribution can be significantly reduced by a tighter signal selection or by using a different fitting approach; the remaining contributions are expected to scale with the statistics accumulated due to their data-driven nature.

In the case of $\Bz \to \D^\mp \pi^\pm$ decays, the smallness of the ratio of amplitudes, $r_{D \pi}$, which limits the sensitivity to $S_f$ and $S_{\bar{f}}$, is compensated for by a large signal yield.
About 480\,000 flavour-tagged signal decays are available in the Run 1 LHCb data sample~\cite{LHCb-PAPER-2018-009}. 
Analysis of this sample gives measurements of $S_f$ and $S_{\bar{f}}$ that are more precise than those from BaBar and Belle~\cite{Aubert:2006tw,Ronga:2006hv}.
Also in this case the precision is limited by the statistical uncertainty. 
The dominant sources of systematic uncertainty, such as due to knowledge of $\Delta m_d$ and of background subtraction, are expected to be reducible with larger samples.

In $\Bz \to \D^\mp \pi^\pm$ decays there are only two observables, $S_f$ and $S_{\bar{f}}$, that depend on three unknown quantities, $r_{D \pi}$, $\delta_{D \pi}$ and $2\beta+\gamma$. External input must thus be used to obtain a constraint on $2\beta+\gamma$.
A common approach is to determine $r_{D\pi}$ from the branching fraction of $\Bd \to \Dsp \pim$ decays, assuming \grpsuthree\ symmetry,
\begin{equation}
r_{D\pi}= \tan \theta_c  \frac{f_{D^+}}{f_{D_s}}\sqrt{\frac{\mathcal{B}(\Bd \to \Dsp \pim)}{\mathcal{B}(\Bd \to \Dm \pip)}}\,,
\end{equation}
where $\tan\theta_c$ is the tangent of the Cabibbo angle and  ${f_{D^+}}/{f_{D_s}}$ is the ratio of decay constants.
Using the resulting value of $r_{D\pi}=0.0182\pm0.0012\pm0.0036$, where the second uncertainty accounts for possible nonfactorisable \grpsuthree-breaking effects~\cite{DeBruyn:2012jp}, the intervals $|\sin(2\beta + \gamma)|\in [0.77,1.00]$ and $\gamma\in [5,86]\degrees \cup [185,266]\degrees$ are obtained, at the 68\% CL. 
The uncertainties on $r_{D\pi}$ and $\beta$ have negligible impact on these intervals, as the dominant uncertainties are from the $S_f$ and $S_{\bar{f}}$ measurements. 

The expected statistical sensitivities for the \CP violation parameters in $B^0_s \to D_s^\mp K^\pm$ and $B^0 \to D^\mp \pi^\pm$ decays are shown in Table~\ref{tab:Dh_expected_errors}.
These are based on scaling of yields, and as such assume that the same detector performance as achieved in Run~I can be maintained.
In particular, the sensitivity depends strongly on the performance of the particle identification, decay time resolution and flavour tagging.  
The results can be complemented by studies of the related $B^0_s \to D_s^{*\mp} K^\pm$ and $B^0 \to D^{*\mp} \pi^\pm$ channels.
In particular, the $D^{*\mp} \pi^\pm$ mode has an all charged final state, and with a possible gain in the acceptance of slow pions from $D^*$ decays from the addition of magnet side stations, comparable precision to that for $B^0 \to D^\mp \pi^\pm$ may be possible.
 
The corresponding expected sensitivities of $\gamma$ from $\Bs \to \D_s^\mp K^\pm$ decays are about $4^\circ$, $2.5^\circ$ and $1^\circ$ after collecting 
23, 50 and 300\invfb, respectively. 
It is more challenging to estimate the constraints on $\sin(2\beta +\gamma)$ and $\gamma$ from $\Bd \to D^\mp \pi^\pm$ decays, since the precision of the external value of $r_{D\pi}$ will become the dominant source of systematic uncertainty.
Theoretical advancements on understanding the nonfactorisable \grpsuthree-breaking effects are thus required.

\begin{table}[t]
        \centering
        \caption{Expected statistical uncertainties from LHCb on parameters of  $\Bs \to \D_s^\mp K^\pm$ and $\Bd \to \D^\mp \pi^\pm$ decays.}
        \begin{tabular}{lcccccccc}
		\hline\hline
		                        & \multicolumn{4}{c}{$\Bs \to \D_s^\mp K^\pm$} & \multicolumn{4}{c}{$\Bd \to \D^\mp \pi^\pm$}  \\
		    Parameters  & Run~1 & 23\invfb & 50\invfb & 300\invfb & 23\invfb & 50\invfb & 300\invfb \\ 
		\hline
		$S_f$, $S_{\bar{f}}$ & 0.20 & 0.043 & 0.027 & 0.011  & 0.02 & 0.0041 & 0.0026 & 0.0010\\
		$A^{\Delta\Gamma}_f$, $A^{\Delta\Gamma}_{\bar{f}}$ & 0.28 & 0.065 & 0.039 & 0.016 & -- & -- &-- \\
		$C_f$ & 0.14 & 0.030 & 0.017 & 0.007 & -- & -- & -- \\
	  \hline\hline
	\end{tabular}
        \label{tab:Dh_expected_errors}
\end{table}

\lhcb has measured the \CP violation parameters in \BdTopipi and \BsToKK decays (\Cpipi, \Spipi, \CKK, \SKK and \ADGKK) using the full sample of \proton\proton collisions collected during Run~1 corresponding to $3.0\invfb$ of integrated luminosity~\cite{LHCb-PAPER-2018-006}.
The results are reported in Table~\ref{tab:b2hhStatusAndProjections}, together with the projections of the statistical uncertainties to larger samples.
The scaling of statistical uncertainties assumes the same detector performances as in Run~1, in particular regarding the flavour tagging, the decay-time resolution and the particle identification performance, which are particularly important for the determination of these observables. 
The main sources of systematic uncertainties are due to limited knowledge of: the variation of the selection efficiency as a function of the $B_q$ meson decay time, the parameters \Gs and \DGs, and the calibration of the decay-time resolution.
The evaluation of these uncertainties is based on the study of control modes, and hence they are expected to decrease in a statistical manner as the available sample size grows.

\begin{table}[t]
	\begin{center}
	\caption{Current \lhcb measurements of \Cpipi, \Spipi, \CKK, \SKK and \ADGKK using the full sample of \proton\proton collisions collected during Run~1, where the first uncertainty is statistical and the second systematic. The projection of statistical precisions for each variables are also reported.}
\label{tab:b2hhStatusAndProjections}
	\resizebox{\columnwidth}{!}{
		\renewcommand{\arraystretch}{1.2}
		\begin{tabular}{lccccc}
			\hline\hline 
			Data sample &  \Cpipi & \Spipi & \CKK & \SKK & \ADGKK \\
			\hline 
			Run 1 ($3\invfb$~\cite{LHCb-PAPER-2018-006}) & $-0.34 \pm 0.06 \pm 0.01$ & $-0.63 \pm 0.05 \pm 0.01$ & $\phantom{-}0.20 \pm 0.06 \pm 0.02$ & $\phantom{-}0.18 \pm 0.06 \pm 0.02$ & $-0.79 \pm 0.07 \pm 0.10$ \\
			\hline
			& \multicolumn{5}{c}{$\sigma$ (stat.)} \\
			\hline
			Run 1-3 ($23\invfb$)  & $0.015$ & $0.013$ & $0.015$ & $0.015$ & $0.018$ \\
			Run 1-6 ($300\invfb$) & $0.004$ & $0.004$ & $0.004$ & $0.004$ & $0.005$ \\
			\hline \hline
		\end{tabular}
		\renewcommand{\arraystretch}{1.0}
		}
	\end{center}
\end{table}

\subsubsection{Gamma combination and impact of external inputs}

\begin{figure}[t]
\centering
\includegraphics[scale=0.5]{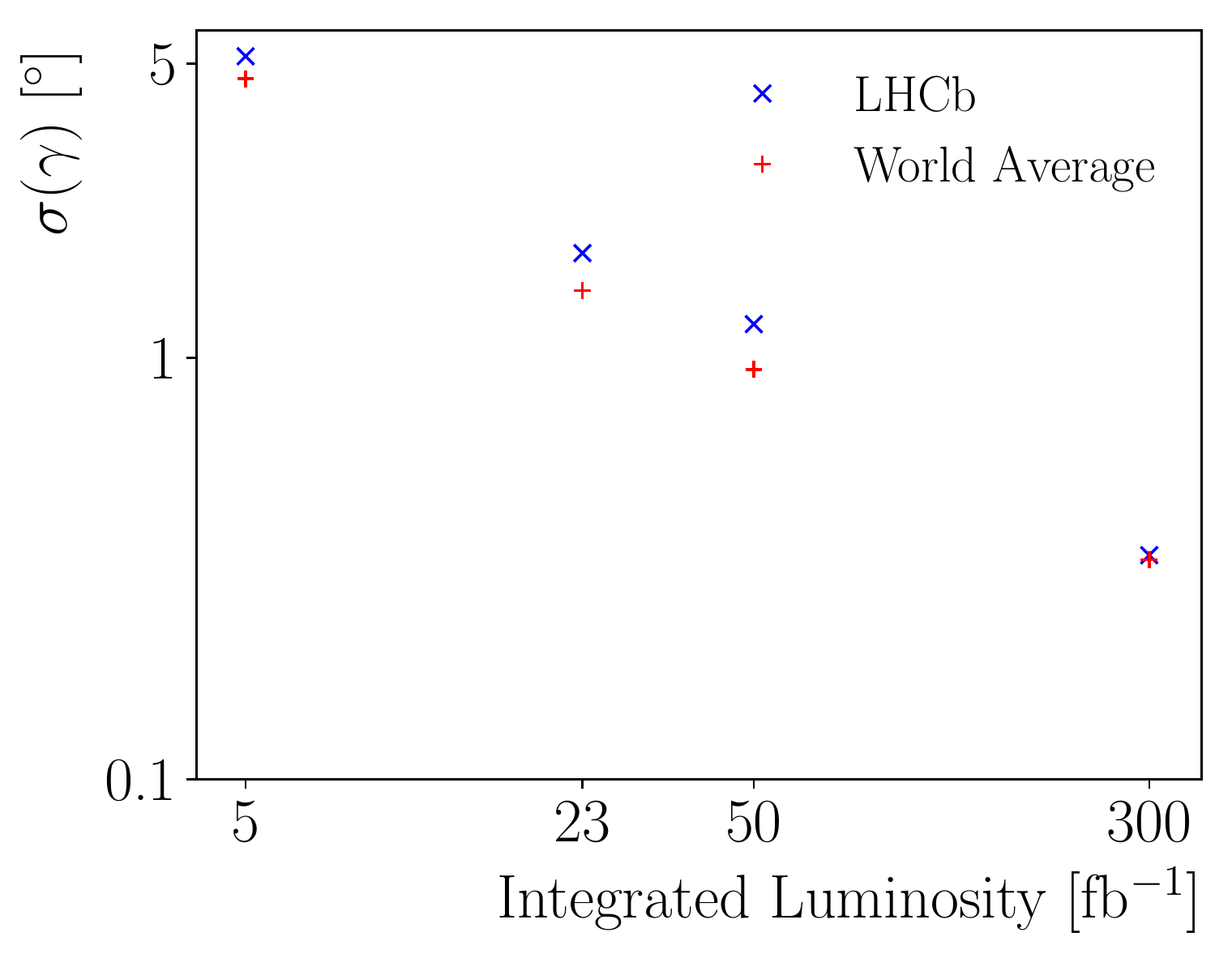}
  \caption{\small Projected sensitivity for the LHCb $\gamma$ combination with the currently used strategies and the world average using projections from \belletwo in addition.}
\label{fig:42:proj}
\end{figure}

\begin{figure}[t]
\centering
\includegraphics[width=0.48\textwidth]{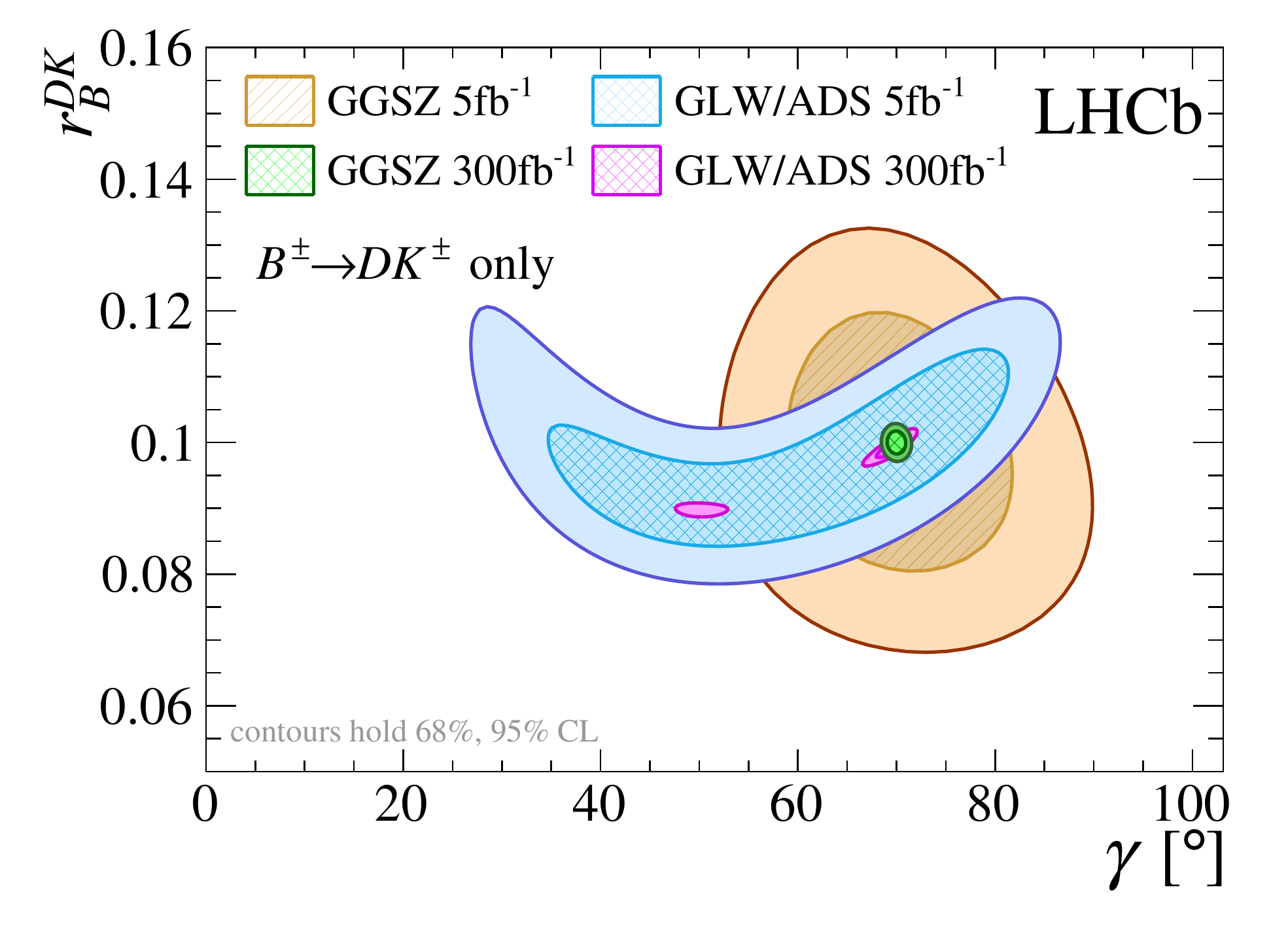}
  \caption{\small Comparison
  between the current LHCb 3-body GGSZ and 2-body GLW/ADS measurements alongside their future projections with 300\invfb in the plane of \g vs.
   \rbk (note the curtailed $y$-axis for \rbk). The scan is produced
  using a pseudo-experiment, centred at $\g=70\degrees$, $\rbk=0.1$, with $\Bpm\to\D\Kpm$ decays only.}
\label{fig:42:compare}
\end{figure}

Projections for the expected precision of the LHCb $\gamma$ combination are shown in Fig.~\ref{fig:42:proj}, estimating the uncertainty on $\gamma$ to be $1.5^\circ$ and $0.35^\circ$ with 23~\invfb and 300~\invfb data samples, respectively.
The LHCb projections assume that the statistical uncertainty scales with the data sample size
and  include the effect of the increased centre-of-mass energy, increased trigger performance and increased integrated luminosity. Most of the systematic uncertainties are driven by the size of
 the data samples and corresponding simulation samples.
The overall sensitivity is predominantly driven by the GLW modes which provide
the narrowest solutions for \g. The dominant systematic uncertainty for these modes will depend on knowledge of both the shape and rate of the background
from $\Lb\to\Lc\Km$, as well as uncertainties arising from particle identification calibration and instrumental charge asymmetries.
In order to obtain the best possible precision on \g it will be necessary to ascertain the correlation of these uncertainties between
 different GLW modes. The ultimate sensitivity to these modes from \belletwo\ will be considerably less than that at \lhcb, however detailed analysis from
both experiments will provide an important cross check. For decays with neutrals in the final state, particularly the \CP-odd GLW mode with \Dz\to\KS\piz,
the \lhcb detector has considerable disadvantages over \belletwo. 
These would be mitigated by an improved electromagnetic calorimeter for the \lhcb \upgradetwo.
The GGSZ modes are a powerful way to unambiguously resolve the multiple solutions of the GLW method and furthermore offer considerable standalone
sensitivity to \g.
Accurate knowledge of the selection efficiency across the Dalitz plane
is vital for these modes and contributes a considerable systematic uncertainty. This will naturally reduce with larger datasets as it is obtained
via semi-leptonic $\Bd\to\Dstarp\mun\neumb X$ control modes but requires large simulation samples.
More precise measurement of important external parameters, particularly $c_i$ and $s_i$ from BESIII, will be required to reduce the uncertainty associated with the model independent GGSZ method.
The uncertainties of inputs from charm threshold data collected by CLEO-c will begin to limit the sensitivity by the end of Run 2, so it is essential to work together with BESIII to provide updated measurements for the suite of charm decays and \DtoKshh in particular.
Provided that the charm inputs are improved sub-degree level precision on $\gamma$ is attainable. Understanding the correlations between different
$B$ decay modes that all use these external parameters will be vitally important as they are likely to contribute one of the largest overall systematic
uncertainties in the combination.
A comparison between the current LHCb GGSZ and GLW/ADS measurements~\cite{LHCb-PAPER-2018-017,LHCb-PAPER-2017-021} with their future projections
at 300\invfb is shown in Fig.~\ref{fig:42:compare}. The order of magnitude increase in precision is very apparent and the importance of the
combination clear, given the multiple ambiguous solutions for GLW/ADS measurements is not resolved with increased luminosity.

The GGSZ modes are considered the \emph{golden modes} at \belletwo and drive the overall uncertainty on \g which is expected to reach
$1.5^\circ$ with a data sample of $50~{\rm ab}^{-1}$. This is comparable to the sensitivity that the \lhcb \g combination will
achieve with a data sample
corresponding to approximately $23\invfb$. Subsequently input from \belletwo will still contribute towards the world average by the end
of Run~4 but \lhcb will dominate \g measurements with \upgradetwo (300\invfb) contributing entirely towards a world average precision of
$\sim0.35\degrees$.
It should be emphasised that this projection includes only the currently used strategies, and does not include improvements from other approaches.
A comparison between the projected uncertainties for \lhcb and the world average as a function of integrated luminosity is shown in Fig.~\ref{fig:42:proj}.

\subsubsection{Determinations of $\Delta m_s$, $\Delta m_d$, and interplay with $b$-hadron lifetimes}

The world-leading measurements of both $\Delta m_{\dquark}$ and $\Delta m_{\squark}$ are from LHCb~\cite{LHCb-PAPER-2015-031,LHCb-PAPER-2013-006}, and can be improved further assuming that good flavour tagging performance can be maintained.
This will not only reduce systematic uncertainties in \CP-violation measurements but also provide a strong constraint on the length of one side of the unitarity triangle, although progress here is mainly dependent on improvements in lattice QCD calculations. The decay-time-dependent angular analysis of $\Bs \to \jpsi\phi$ allows measurement of $\Delta \Gamma_{\squark}$ simultaneously with \CP-violation parameters. Therefore, improved knowledge of $\Delta \Gamma_{\squark}$ will be obtained together with measurements of $\phi_s^{c\bar{c}s}$.
Precision at the LHC is expected not to be systematically limited.
The LHCb \upgradetwo will allow to exploit measurements in various channels.
ATLAS and CMS projections in the $B_s \to \jpsi \phi$ decay mode can be found in~\ref{sec:jpsiphihllhc}. 
For theory predictions of $\Delta \Gamma_s$ see Ref. \cite{Lenz:2006hd}.

Width differences between different types of \bquark\ hadrons, such as $\Gamma_s - \Gamma_d$, are also of interest.
They test the heavy-quark expansion, used to make theoretical predictions. In addition, their precise knowledge is important to control systematic uncertainties in measurements where a decay mode of one type of \bquark\ hadron is used as a control channel in studies of a decay of another.
Detailed understanding of acceptances is necessary for such measurements, which can be achieved using topologically similar final states (see, e.g., Ref.~\cite{LHCb-PAPER-2014-003}).
These measurements are therefore expected to be significantly improved with LHCb \upgradetwo.

\subsubsection{Semileptonic asymmetries and prospects for $\Delta \Gamma_d$}

 Semileptonic decays, being flavour specific, provide a unique probe of \Bq, where $q = s,d$, meson mixing phenomena.
In particular, \CP violation in \Bq meson mixing can be expressed through the semileptonic asymmetries
\begin{equation}
\label{eq:5.3.1}
  a_{\rm sl}^{q} = \frac{\Gamma (\Bqb \to f) - \Gamma (\Bq \to \bar{f})}{\Gamma (\Bqb \to f) + \Gamma (\Bq \to \bar{f})} \approx \frac{\Delta\Gamma_q}{\Delta M_q}\tan\phi_{12}^q,
\end{equation}
where $f$ is a flavour-specific final state that is only accessible through the decay of the $B_q^0$ state.
Mixing is required to mediate the transition $\Bqb \to \Bq \to f$, and its conjugate.
Semileptonic decays of the type $\Bq \to D_q^- \mu^+ \nu_{\mu} X$ are well suited because (i) they are immune to any unknown \CP violation in decay and (ii) they have large branching ratios.

Including the effect of an unknown production asymmetry, $a_p$, the time-dependent {\em untagged} asymmetry is defined as: 

\begin{equation}
\label{eq:5.3.2}
  A_{\rm sl}^{q}(t) \equiv \frac{N - \bar{N}}{N + \bar{N}} = \frac{a_{\rm sl}^{q}}{2} - \left[ a_{p} + \frac{a_{\rm sl}^{q}}{2} \right] \cdot \left[\frac{\cos \Delta M_q t }{\cosh \Delta \Gamma_q t /2}\right],
\end{equation}
where $N$ and $\bar{N}$ denote the number of observed decays to $f$ and $\bar{f}$ final states, respectively.
A decay-time-dependent fit can disentangle the $B_d^0-\bar{B}_d^0$ production asymmetry from $a_{\rm sl}^d$~\cite{LHCb-PAPER-2014-053}.
In the $B_s^0$ case the {\em time integrated} asymmetry is employed~\cite{LHCb-PAPER-2016-013}.
Owing to the large value of $\Delta m_s$ the term containing $a_p$ is suppressed to a calculable correction of a few $\times 10^{-4}$,
after integrating over decay time.
These approaches have been applied in the measurements with the Run 2 dataset that are listed in Table~\ref{tab:5.3.1}, giving 
the world's best single measurements.
These measurements are far from any uncertainty floor in the SM predictions, and are 
sensitive to anomalous NP contributions to $\Gamma_{12}^{q}$ and $M_{12}^{q}$. \Blu{The difference $\Gamma_d-\Gamma_s$ further probes NP in penguin coefficients \cite{Keum:1998fd}.}

The following briefly reviews the dominant sources of uncertainty on the current LHCb measurements, and considers strategies to 
fully exploit the potential of the LHCb \upgradetwo.
All uncertainties are as evaluated on $a_{sl}^{q}$ (\ie all sources of raw asymmetry, and their uncertainties, are scaled by  
a factor of two as expected from Eq.~\ref{eq:5.3.2}).

\begin{figure}[t]
\centering
\includegraphics[width=0.6\linewidth]{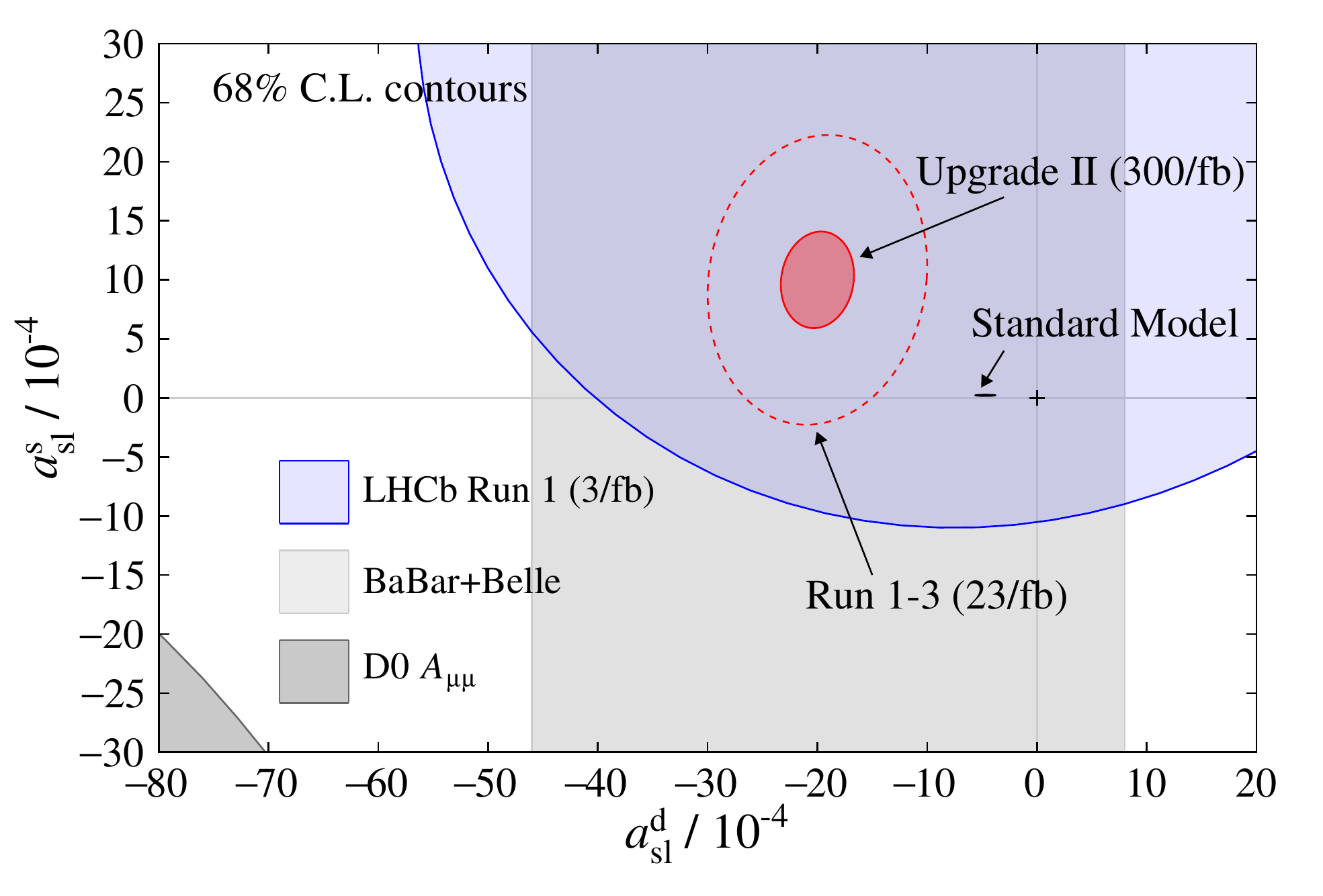}
\caption{
Current and future landscape for the semileptonic asymmetries.
The grey vertical band indicates the current \B-Factory average for $a_{\rm sl}^d$.
The blue ellipse represents the current LHCb Run 1 measurements~\cite{LHCb-PAPER-2016-013,LHCb-PAPER-2014-053}.
The red ellipse, which is arbitrarily centred, delineates the LHCb \upgradetwo projected precision.
The black ellipse shows the SM prediction, the uncertainty of which is barely visible.
\label{fig:5.3}}
\end{figure}

\begin{description}
\item[Statistical precision:] 
The statistical uncertainty on $a^s_{\rm sl}$ straightforwardly extrapolates to $2.1 \times 10^{-4}$.
In the case of $a^d_{\rm sl}$ it should be noted that stringent fiducial cuts and weights were imposed on the signal sample
to control certain tracking asymmetries that were not well known at the time. By the time of the subsequent $a^s_{\rm sl}$ measurement,
a new method with $J/\psi \to \mu^+\mu^-$ decays had been developed, and others are in the pipeline. 
A further factor of two  increase in yields is therefore assumed,
which extrapolates to an uncertainty of $1.1 \times 10^{-4}$.
\item[Detection asymmetries:] The single largest contributor is the $K^-\pi^+$ asymmetry in $a^d_{\rm sl}$.
This asymmetry was determined with a single method using $D^+$ 
decays to the $K^-\pi^+\pi^+$ and $\KS\pi^+$ final states,
with a precision of around $2.0 \times 10^{-3}$~\cite{LHCb-PAPER-2014-053}.
Thanks to trigger improvements, a factor of two increase in the effective yield of the
most limiting $\KS\pi^+$ final state~\cite{Davis:2310213,Davis:2284097} can be anticipated. 
This will extrapolate to an uncertainty of $1.1 \times 10^{-4}$. 
Improvements in the reconstruction of downstream tracks in the earliest stage of the software trigger may also allow us to exploit 
$\KS\pi^+$ final states with $\KS$ decays downstream of the LHCb vertex detector (VELO).
Further methods have since been proposed using partially reconstructed $D^{*+}$ decays and $D^0 \to K^-K^+$ decays.
The partial reconstruction method will be greatly improved by the reduction of material before the first VELO measurement point.
The systematic uncertainties in these approaches will be controlled by using ultra high statistics fast simulations to track the 
kinematic dependencies in the asymmetries. The target uncertainty is $1.0 \times 10^{-4}$, including systematic uncertainties.
The equivalent $K^+K^-$ asymmetry in the $a^s_{\rm sl}$ measurement will be smaller and more precisely controlled.
The $\mu^+\pi^-$ asymmetry will be controlled by a combination of $J/\psi \to \mu^+\mu^-$ decays, partially reconstructed $D^{*+}$ decays,
$D^0 \to h^-h^+$ decays, and high statistics fast simulations. 
\item[Background asymmetries:] These measurements are challenging because the \hbox{$\Bq \to D_q^- \mu^+ \nu_{\mu} X$} final states can be fed by
the decays of other $b$ hadron species. This dilutes the relation between the raw asymmetry and $a_{\rm sl}^q$ and leads to a cocktail of 
production asymmetry corrections. We assume that these backgrounds can be statistically subtracted by extending the signal fits to include the
$D_q^- \mu^+$ corrected mass dimension. It is assumed that the background asymmetry uncertainties can be controlled to the $1.0 \times 10^{-4}$ level.
\item[Other considerations:] We must carefully consider the impact of having unequal sample sizes in the two polarities. This can be partially compensated for by
assigning weights to one polarity~\cite{Vesterinen:1642153}. 
We note that the choice of crossing angles should be carefully considered~\cite{Dufour:2304546}.
While we do not account for them in the current estimation, we could consider using other $D^{+}_{q}$ decay modes to better align the detection asymmetries between $a_{\rm sl}^s$ and $a_{\rm sl}^d$.
For example, $D^+ \to K^-K^+\pi^+$ and $D^+ \to \KS \pi^+$ decays can be used, taking advantage of possible improvements in the trigger efficiency for \KS decays in LHCb Upgrade II.
While the former decay is singly Cabibbo suppressed, its \CP asymmetry could be measured using promptly produced $D^+$ mesons.
\end{description}
In summary the LHCb Upgrade II dataset should allow both $a^s_{\rm sl}$ and $a^d_{\rm sl}$ asymmetries to be measured to the level of a few parts in $10^{-4}$,
see Table~\ref{tab:5.3.1}. This will give unprecedented new physics sensitivity, and is still far from saturating the {\em current}
theory uncertainties in the SM predictions.
Fig.~\ref{fig:5.3} shows the prospective LHCb \upgradetwo measurement, arbitrarily centred at a value that differs from the SM
prediction at the $10^{-3}$ level. 

\begin{table}[t]
\caption{\label{tab:5.3.1}Current theoretical and experimental determinations of the semileptonic asymmetries $a^d_{\rm sl}$ and $a^s_{\rm sl}$.}
\centering
\begin{tabular}{lcc}
\hline\hline
Sample ($\mathcal{L}$) & $\delta a^s_{\rm sl}/10^{-4}$ & $\delta a^d_{\rm sl}/10^{-4}$ \\
\hline
Run 1 (3 \invfb)~\cite{LHCb-PAPER-2016-013,LHCb-PAPER-2014-053} & $33$  & $36$\\
Run 1-3 (23 \invfb) &  $10$ &  $8$ \\
Run 1-5 (300 \invfb) &  $3$ &  $2$ \\
\hline
Current theory~\cite{Lenz:2006hd,Artuso:2015swg} & $0.03$ & $0.6$\\
\hline\hline
\end{tabular}

\end{table}

The ratio $\Delta \Gamma_{\dquark}/\Gamma_{\dquark}$, is typically measured  from the difference in effective lifetimes between $\Bd$ decays to final states that are flavour-specific, namely $\jpsi \Kstarz$, and \CP-eigenstates, namely $\jpsi \KS$~\cite{Gershon:2010wx}. With this approach LHCb determined $\Delta \Gamma_{\dquark}/\Gamma_{\dquark} = -0.044\pm0.025\pm0.011$ using $1\invfb$ of data~\cite{LHCb-PAPER-2013-065}. ATLAS and CMS published competitive measurements~\cite{Aaboud:2016bro,Sirunyan:2017nbv} reaching at their best statistical and systematic uncertainties of $0.011$ and $0.009$ respectively. 
The expected statistical uncertainty for LHCb \upgradetwo, taking into account the larger centre-of-mass energy and the increase in luminosity, is $\sigma(\Delta \Gamma_{\dquark}/\Gamma_{\dquark}) \sim 0.001$.
This can be compared with the SM prediction $\Delta \Gamma^{\rm SM}_{\dquark}/\Gamma^{\rm SM}_{\dquark}=(0.00397 \pm 0.00090)$~\cite{Artuso:2015swg}. 
Thus, if systematic uncertainties can be controlled sufficiently, it will be possible to measure a significantly non-zero value of $\Delta \Gamma_{\dquark}$ even if it is not enhanced above its SM prediction.

The challenge in controlling the systematic uncertainty is to understand precisely the decay-time acceptance difference between the $\jpsi \Kstarz$ and $\jpsi \KS$ decay topologies. 
However, if the $\Bd$ vertex position is reconstructed identically, namely using only the $\jpsi$ decay products, and only $\KS$ mesons decaying inside the VELO are considered, the largest sources of systematic uncertainty should cancel almost exactly and therefore not dominate the measurement. 
Similarly, the asymmetry in production rate between $\Bd$ and $\Bdb$ is expected to be precisely measured in independent control samples, and thus will not limit the achievable precision. 
In addition to the intrinsic interest in determining $\Delta \Gamma_{\dquark}$, precise knowledge of its value will benefit many other studies of $\Bd$ decay modes, since any systematic uncertainties associated with the assumption that $\Delta \Gamma_{\dquark} = 0$ can be removed.

\subsubsection{Measurements of $\phis$ from $b\to c\bar{c}s$ transitions}

The single statistically most  sensitive measurement of $\phi_{\squark}^{c\bar{c}s}$ is given by the flavour-tagged decay-time-dependent angular analysis of $\Bs\to \jpsi(\mu^+\mu^-)\phi(\Kp\Km)$ decay.
The current world average is consistent with the SM prediction~\cite{Charles:2015gya,Artuso:2015swg}, with new physics effects of a few tens of miliradians still allowed~\ref{HFLAVPDG18}. 
As the experimental precision improves it will be essential to have good control over possible hadronic effects~\cite{Faller:2008gt,Bhattacharya:2012ph} that could mimic the BSM signal. It will also be crucial to achieve precise control of penguin pollution, both in  $\Bs\to \jpsi(\mu^+\mu^-)\phi(\Kp\Km)$ and in other $b\to c\bar{c}s$ transitions. The wide range of $b\to c\bar{c}s$ modes accessible at LHCb \upgradetwo will be vital to achieving both of these, and therefore contribute significantly more than their naive statistical sensitivity to the overall reach of this physics programme.

\subsubsubsection{Projections for $\Bs\to\jpsi(\mu^+\mu^-)\phi(\Kp\Km)$}
\label{sec:jpsiphihllhc}
Fully exploiting the statistical power of HL-LHC for measurements of 
$\Delta\Gamma_s$ and the weak phase difference $\phi_s$ will require excellent flavour tagging performance, invariant mass resolution and proper-time resolution. The planned detector improvements in all three experiments are critical to deliver these requirements despite the much higher HL-LHC pileup.
 
\begin{figure}[t]
\begin{center}
\includegraphics[width=0.6\textwidth]{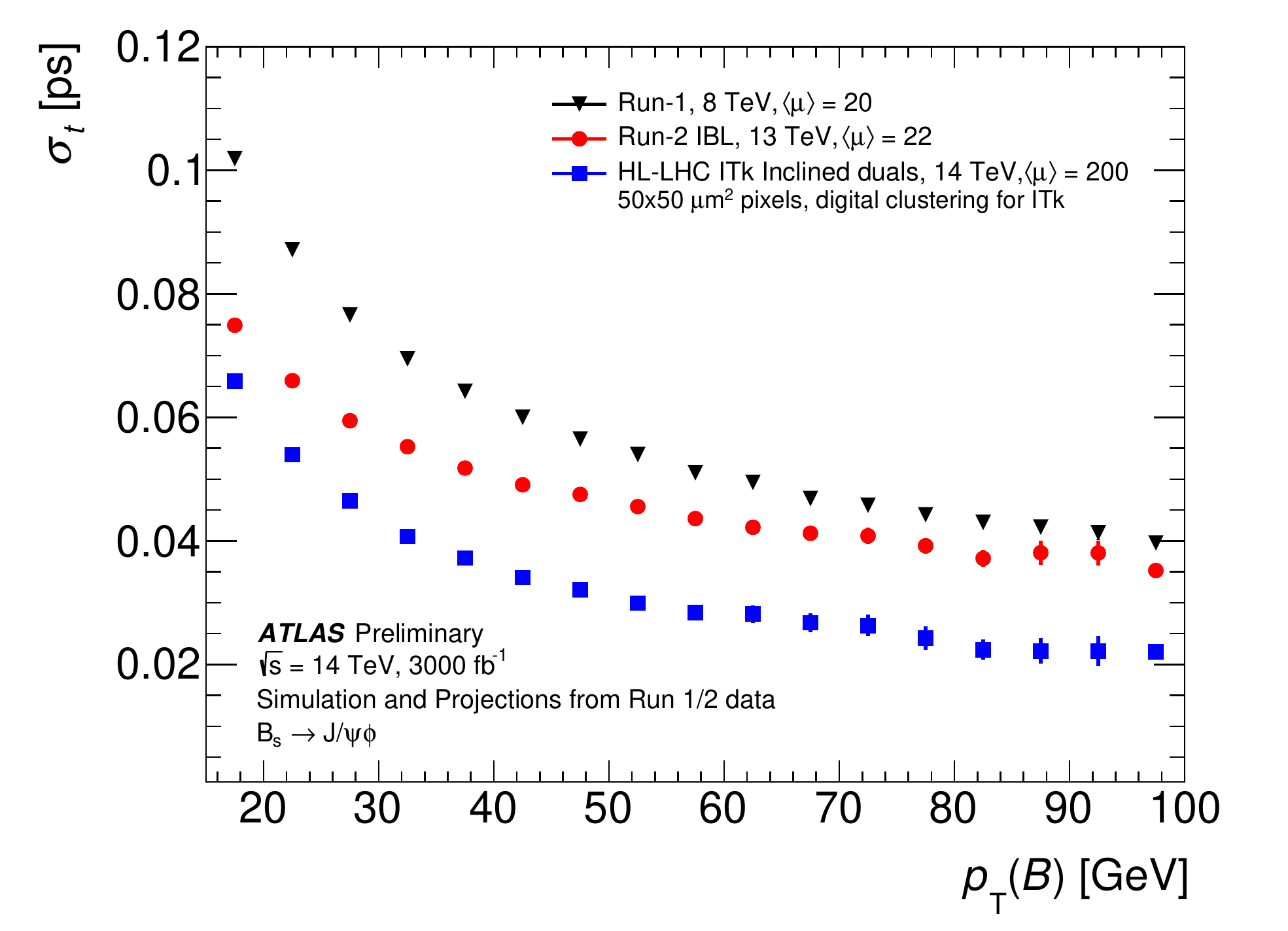}
\end{center}
\caption{The proper decay time resolution of the $B_s^0$ meson in ATLAS for the signal $B_s\to J/\psi\phi$ decay as function of the $B$ meson \pt. Per-candidate resolutions corrected for scale-factors are shown, comparing the performance in Run 1, Run 2 and upgrade HL-LHC MC simulations. All samples use 6 GeV muon \pt cuts.}
\label{jpsiphiSensitivity:Mass}
\end{figure}

\begin{figure}[t]
\begin{center}
\includegraphics[width=0.55\textwidth]{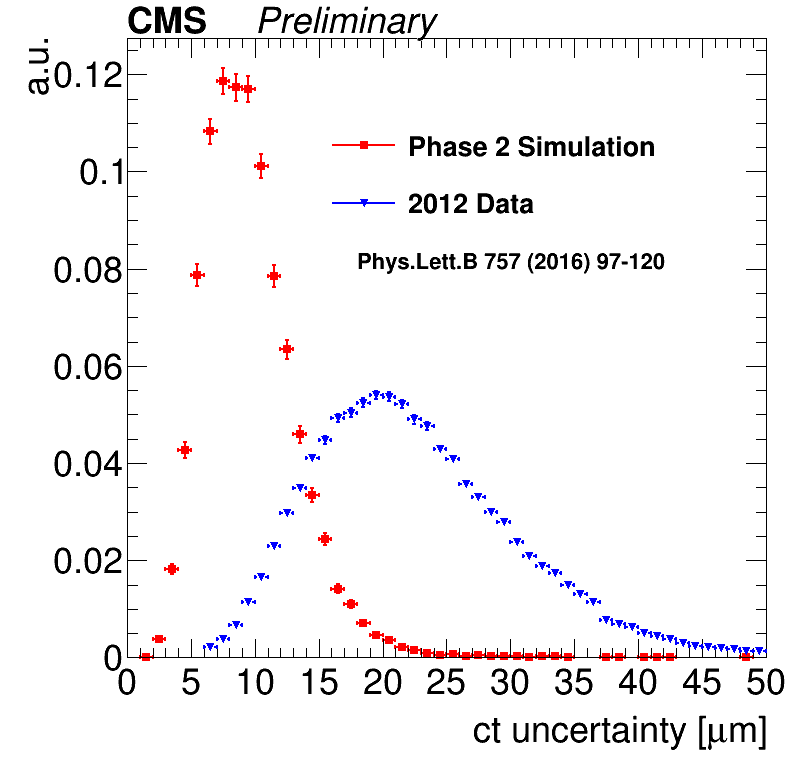}
\caption{CMS ct uncertainty distribution in 2012 data (blue) and HL-LHC Monte Carlo (red) samples (taken from \cite{CMS-PAS-FTR-18-041}).}
\label{fig:CMSreso}
\end{center}
\end{figure}

Improvements in the decay time resolution are expected in ATLAS and CMS thanks to their upgraded inner detectors.
Fig.~\ref{jpsiphiSensitivity:Mass} shows the ATLAS proper time resolution for Run-1, Run-2, and HL-LHC as a function of the B meson transverse momentum. 
Similarly Fig.~\ref{fig:CMSreso} shows the CMS \cite{ CMS-PAS-FTR-18-041} expected HL-LHC performances in the proper decay length uncertainty, obtained from a simulation of 
of an ideal Phase-2 detector response.
Besides the improvements in the tracker, which include the L1 trigger capability to reconstruct charged tracks above 2 GeV in transverse momentum with almost offline-like resolutions, the extended pseudorapidity 
coverage (up to $|\eta|$ = 4) will increase the CMS acceptance for track reconstruction \cite{Collaboration:2650976}.
The CMS Phase-2 L1 (hardware) and HLT (software) trigger performances are expected to be comparable to those in Run~2 and sustainable in terms of rates; the offline selections for this projection are thus the same as in the 2012 data analysis \cite{CMSPhis}. The same signal purity is assumed as in the previous analysis; this assumption also relies on the future presence of the timing layer \cite{CMS:TPMTD} which will mitigate the background pollution due to tracks coming from pileup vertices.

\begin{table}[t]
\caption{Summary of the ATLAS $B_s\to J/\psi\phi$ performance at the HL-LHC compared with the Run~1 measurements and projections (numbers in parenthesis are predictions). The $\epsilon$ is the flavour-tagging efficiency, $D=1-2\omega$ is the dilution factor, $\omega$ the wrong tag fraction, and $\sigma(t)$ the proper time resolution.}
\label{jpsiphitable}
\center
\begin{tabular}{lcccccc}
\hline\hline
Period & $L_{int}$ [fb$^{-1}$]  & $N_{sig}$ & $\epsilon D^2$ & $\sigma\left(t\right)$ [ps] & $\delta_{\phi_s}^{stat}$ [rad] & $\delta_{\Delta\Gamma_s}^{stat}$ [ps$^{-1}$] \\[1mm]
\hline
2011 & 4.9 & 22700 & 1.45 & 0.1 & 0.25 (0.22) & 0.021 \\
2012 & 14.3 & 73700 & 1.5 & 0.09 & 0.082 & 0.013 \\
\hline 
HL-LHC $\mu 6\mu 6$     & 3000 & $9.7\cdot 10^6$ & 1.5 & 0.05 & (0.004) & (0.0011)\\
HL-LHC $\mu 10\mu 6$   & 3000 & $5.9\cdot 10^6$ & 1.5 & 0.04 & (0.005) & (0.0014)\\
HL-LHC $\mu 10\mu 10$ & 3000 & $1.7\cdot 10^6$ & 1.5 & 0.04 & (0.009) & (0.003)\\
\hline\hline
\end{tabular}
\end{table}

\begin{figure}[t]
\begin{center}
\includegraphics[width=0.5\textwidth]{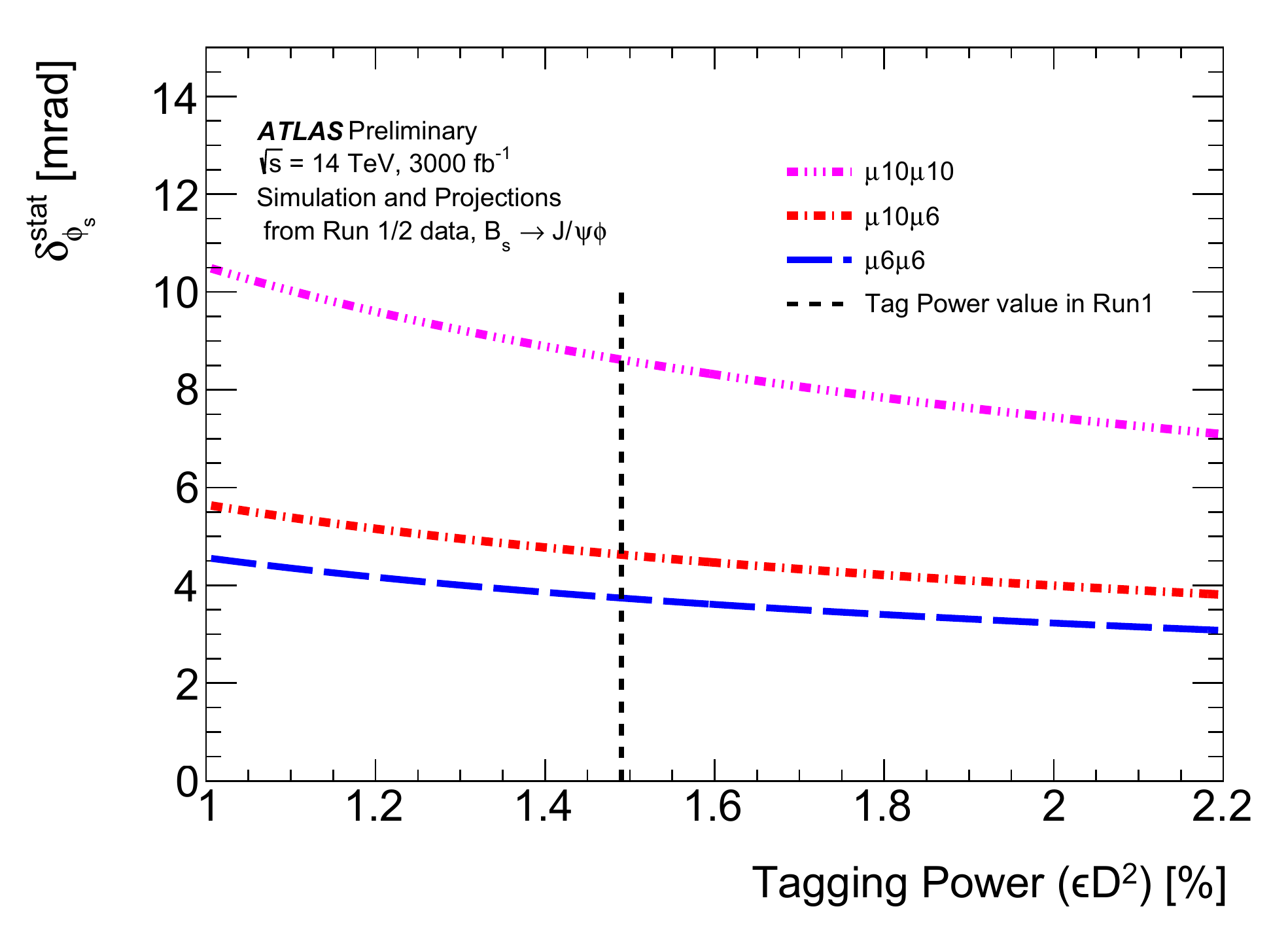}
\includegraphics[width=0.4\textwidth]{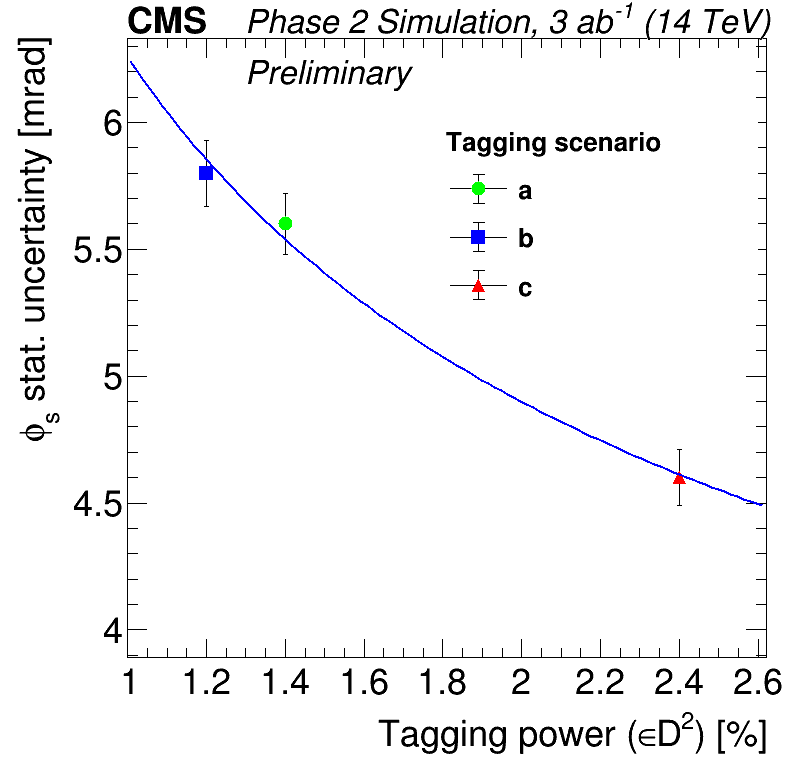}
\caption{Variation of the $\phi_s$ statistical uncertainty in ATLAS (left) and CMS (right)  as function of the tagging power ($\epsilon D^2$).
 The behaviour is shown for different ATLAS triggers and CMS flavour tagging scenarios as explained in the text. 
 The vertical dashed line in the ATLAS plot corresponds to the value assumed in the
analysis extrapolation.
A function proportional to $1/\sqrt{\epsilon D^2}$ is shown in the CMS plot to describe the $\phi_s$ uncertainty behaviour in a continuous range of the tagging power from 1.0 to 2.6 \% (The CMS plot is taken from \cite{CMS-PAS-FTR-18-041}).}
\label{fig:CMStag}
\end{center}
\end{figure}

The ATLAS sensitivity study \cite{ATL-PHYS-PUB-2018-041} follows the same approach as the previous study found in \cite{ATL-PHYS-PUB-2013-010} and is based on the extrapolation of the ATLAS 2012 analysis \cite{Aad:2016tdj}, correcting for the full-simulation based observation of the signal invariant mass and proper-time resolutions. A full simulation of the HL-LHC ATLAS tracking system is employed, including the effect of an average of 200 pile-up event per bunch crossing. An offline emulation of HL-LHC trigger responses is employed to evaluate the signal yield expectations corresponding to different dimuon transverse momentum thresholds: $(p_T^1,p_T^2)$= (6 GeV, 6 GeV), (6 GeV, 10 GeV) and (10 GeV, 10 GeV). 
Table \ref{jpsiphitable} summarises the expected sensitivities, compared to the ATLAS Run-1 measurements and Fig.~\ref{fig:CMStag} left shows the $\delta \phi_s$ distributions. 
The CMS study is based on fully simulated signal events using the same model as in the previous CMS analysis \cite{CMSPhis} and toy Monte Carlo experiments for three different tagging scenarios: $a$ where the flavour tagging performance is based on opposite-side muons and 
jet-charge, $b$ where a muon and electron  flavour tagging is used (as in the CMS 2012 data analysis), and $c$ where 
a well performing flavour tagging based on leptons, jet-charge, and same side jet-charge/kaon tagging is tested.
Fig.~\ref{fig:CMStag} right shows the CMS $\phi_s$ statistical uncertainty predictions for the different tagging scenarios.
Assuming the new tagging power to be in the range 1.2-2.4\% , and a
total of 9 million fully reconstructed $B^0_s \to J/\psi\phi(1020)$ candidates, CMS expects the $\phi_s$ statistical
uncertainty to be 5-6 mrad at the end of the HL-LHC data taking.
Both ATLAS and CMS systematic uncertainties are expected to be reduced to 1 mrad in the HL-LHC period, and thus the total $\phi_s$  uncertainty will still be statistically limited. 

\begin{figure}[t]
\begin{center}
\includegraphics[width=0.6\textwidth]{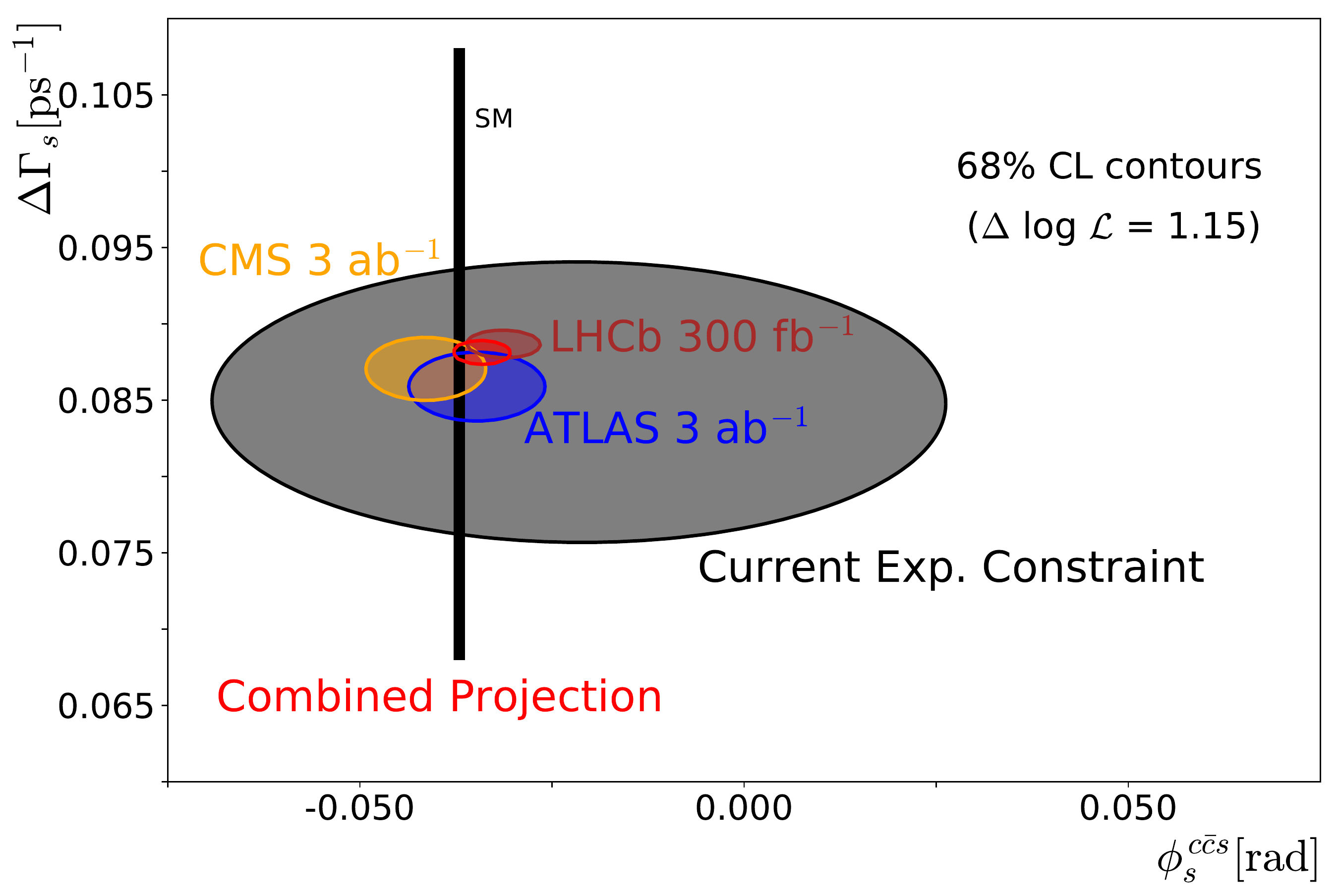}
\caption{Projected 68\% confidence-level contour in the $\Delta \Gamma_s$ vs $\varphi_s$ plane for the ATLAS, CMS (still preliminary) and LHCb sensitivity at the HL-LHC compared with the current experimental limit. For clarity of the representation each projection is centered on a random value generated with its uncertainty assuming the SM value as truth.
 The combined contour is obtained following the HFLAV approach  for averages for the PDG 2018 review \cite{HFLAVPDG18}.} 
\label{jpsiphiSensitivity:Contour}
\end{center}
\end{figure}

The $\phi_s$ measurement is usually illustrated as a constraint in the  $\phi_s$-$\Delta \Gamma_s$ plane.
Fig.~\ref{jpsiphiSensitivity:Contour} summarises the projected contours in the $\Delta \Gamma_s$ vs $\varphi_s$ plane for the ATLAS, CMS and LHCb experiments, with 
the CMS systematic uncertainty on $\Delta \Gamma_s$ assumed to be equal to the statistical uncertainty and an ATLAS estimated systematic uncertainty on the same parameter of approximately 0.0005 ps$^{-1}$. The red contour in the figure
illustrates the combined HL-LHC sensitivity, equivalent to $\sim 2$ $\mrad$ in $\varphi_s$ and $\sim 5$ ps$^{-1}$ in $\Delta \Gamma_s$. 

\begin{figure}[t]
  \centering
  \includegraphics[width=0.50\textwidth]{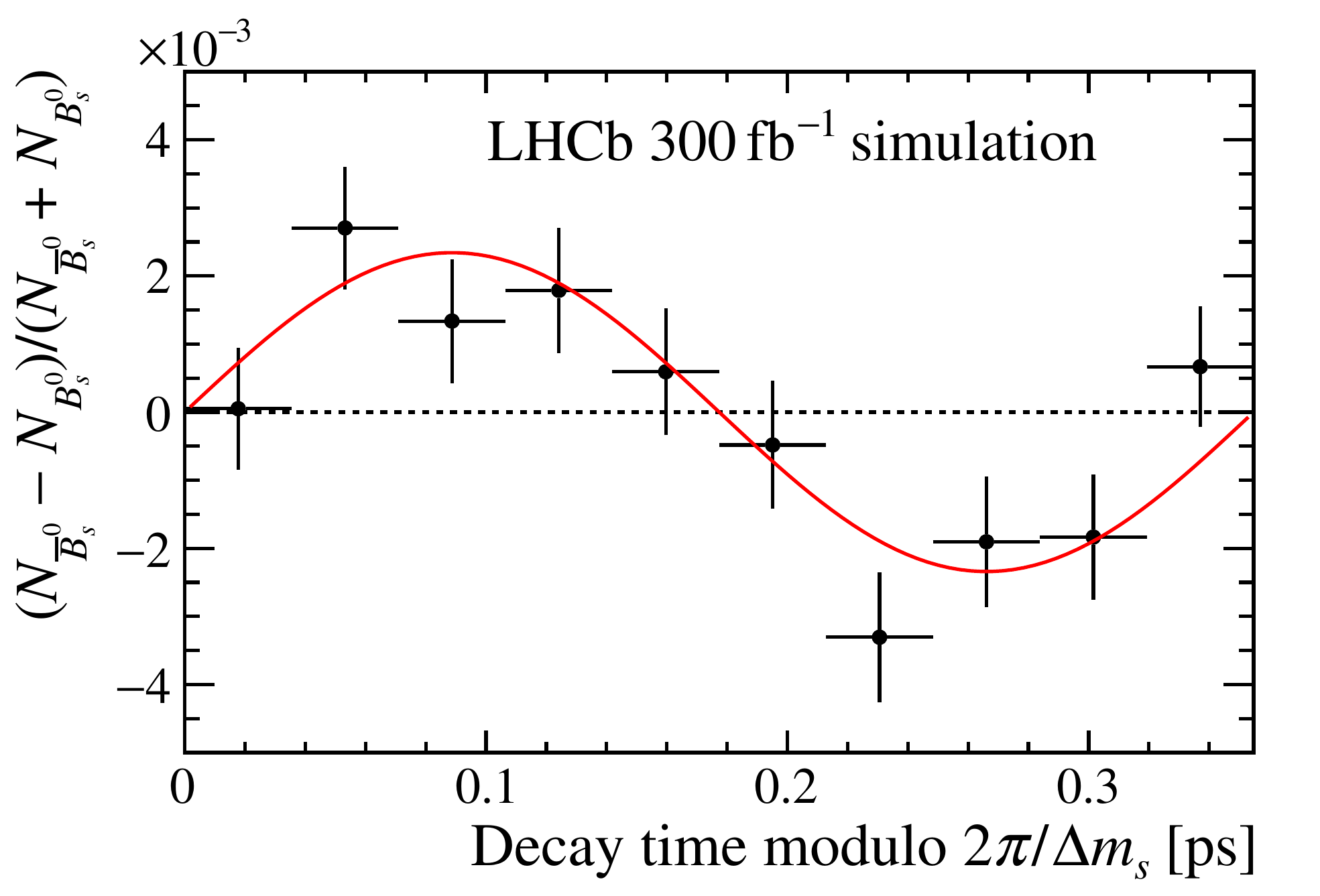}
  \includegraphics[width=0.48\textwidth]{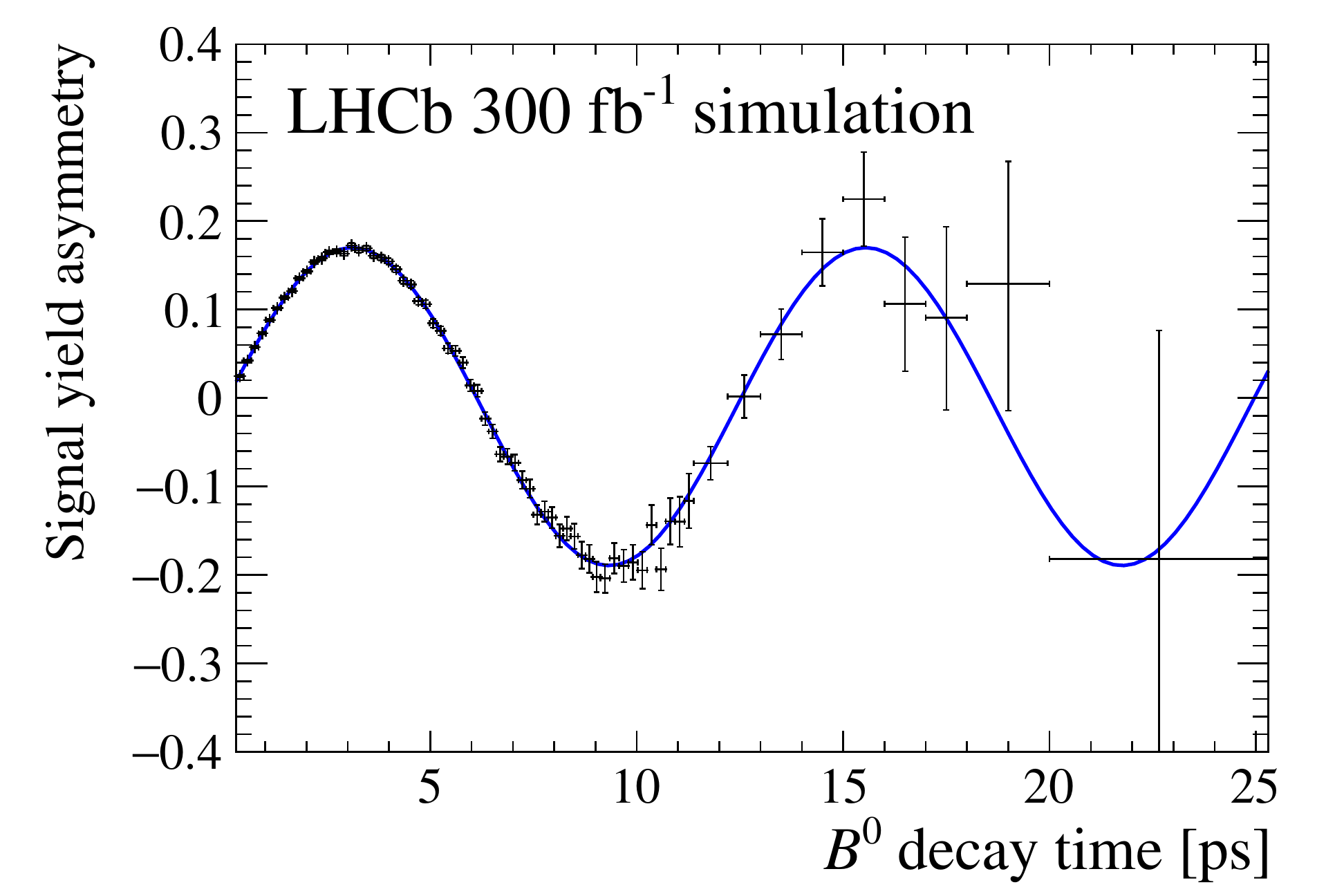}
  \caption{\small 
    Signal-yield asymmetry as a function of the $\Bds$ decay time, $(N_{\Bdsb}-N_{\Bds})/(N_{\Bdsb}+N_{\Bds})$. 
    Here, $N_{\Bdsb}$ ($N_{\Bds}$) is the number of (left) $\Bs\to\jpsi\phi$ or (right) $\Bz\to\jpsi\KS$ decays with a $\Bdsb$ ($\Bds$) flavour tag. 
    The data points are obtained from simulation with the expected sample size at $300\invfb$, and assuming the current performance of the LHCb experiment. 
    The solid curves represent the expected asymmetries for $\phi^{c\bar{c}s}_{\squark}=-36.4\mrad$~\cite{Charles:2015gya} and $\sin\phi^{c\bar{c}s}_{\dquark}=0.731$~\cite{LHCb-PAPER-2015-004}), the values used in the simulation.
    The height of the oscillation is diluted from $\sin\phi^{c\bar{c}s}_{\dquark(\squark)}$ due to mistagging, decay time resolution, and, for $\Bs\to\jpsi\phi$, the mixture of \CP-even and \CP-odd components in the final state.
  }
  \label{fig:Bx2JpsiXs_asymmetry_U2}
\end{figure}

The expected LHCb precision on $\phi_{\squark}^{c\bar{c}s}$ after \upgradetwo\ has been estimated based on the current published results, assuming that the detector performance remains the same in the HL-LHC period. Because of the data-driven nature of the LHCb analysis systematic uncertainties are expected to scale with luminosity and the overall sensitivity is expected to be $\sim 4 \mrad$.
This will be at the same level as the current precision on the indirect determination based on the CKM fit using tree-level measurements, which in turn is expected to improve with better measurements of the other CKM matrix parameters.  
Fig.~\ref{fig:Bx2JpsiXs_asymmetry_U2} left shows the signal-yield asymmetry as a function of the $\Bs$ decay time, folded at the frequency of $\Bs$ oscillations, for $\Bs\to\jpsi\phi$ decays from a simulated data set corresponding to $300\invfb$. It clearly shows that a visible \CP-violation effect will be observable.

\subsubsubsection{Projections for other $b\to c\bar{c}s$ transitions}

LHCb foresees extending the study of $b\to c\bar{c}s$  transitions to cover  multiple independent precision measurements. 
It permits not only improved precision of the average, but
a powerful consistency check of the SM.
One important way in which this can be done is by allowing independent \CP-violation effects for each polarisation state in the $\Bs\to\jpsi\phi$.
This has been done as a cross-check in the LHCb Run~I analysis~\cite{LHCb-PAPER-2014-059}, but this strategy will become the default in LHCb \upgradetwo.
Additional information can be obtained from $\Bs\to\jpsi\Kp\Km$ decays with $\Kp\Km$ invariant mass above the $\phi(1020)$ meson, where higher spin $\Kp\Km$ resonances such as $f^\prime_2(1525)$ meson contribute~\cite{LHCb-PAPER-2017-008}.  
Among other channels, competitive precision can be obtained with $\Bs\to\jpsi\pip\pim$ decays~\cite{LHCb-PAPER-2014-019}, which have been found to be dominated by the \CP-odd component.
The $\Bs\to\Dsp\Dsm$~\cite{LHCb-PAPER-2014-051} and $\Bs\to\psi(2S)\phi$~\cite{LHCb-PAPER-2016-027} modes have also been studied with LHCb, and give less precise but still important complementary results.
Other channels, which have not been exploited yet, but could be important in LHCb \upgradetwo, if good calorimeter performance can be achieved, include $\Bs\to\jpsi\phi$ with $\jpsi \to \epem$ and $\Bs\to\jpsi\eta^{(\prime)}$ with $\etapr \to \rhoz\gamma$ or $\eta\pip\pim$, and $\eta \to \pip\pim\piz$ or $\gamma\gamma$~\cite{LHCb-PAPER-2014-056,LHCb-PAPER-2016-017}.

\begin{figure}[t]
\centering
\includegraphics[width=0.47\linewidth]{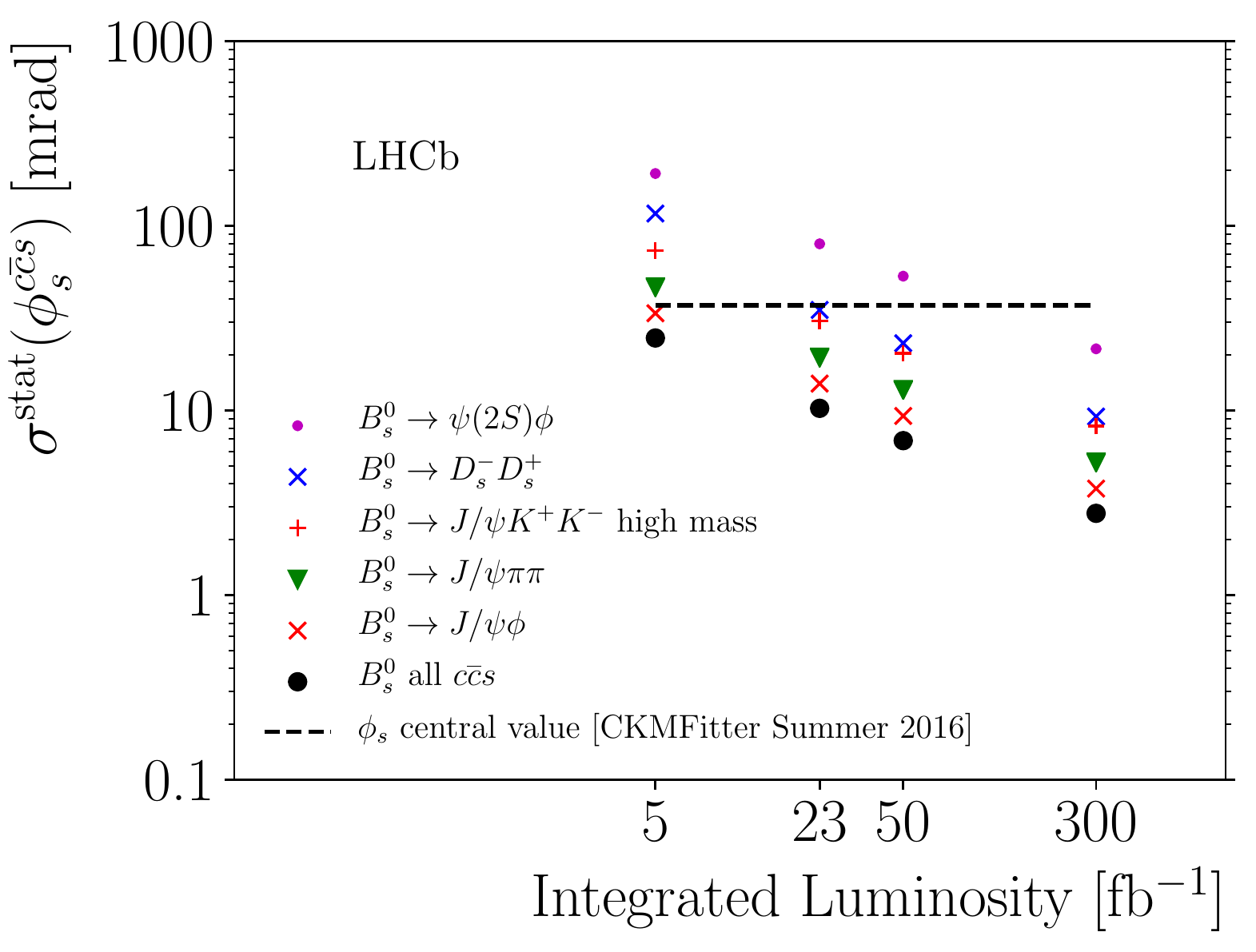}
    \caption{\small Scaling of the LHCb statistical precision on $\phi_{\squark}$ from several
    tree-dominated \Bs meson decay modes.}
\label{fig:32:hflav}
\end{figure}

The scaling of the $\phi_{\squark}^{c\bar{c}s}$ precision with integrated luminosity for individual decay modes and for their combination is shown in Fig.~\ref{fig:32:hflav}. 
These uncertainties are statistical only and are scaled from existing results, taking into account the gain in efficiency expected for $\Bs\to\Dsp\Dsm$ from the removal of the hardware trigger.
Maintaining the current performance will put stringent constraints on the design of the detector with regards to the momentum and vertex position resolution, as well as particle identification performance.
A key ingredient is the flavour tagging that is very sensitive to event and track multiplicity.
Systematic uncertainties are mainly based on the sizes of control samples, and are therefore expected to remain subdominant even with very large samples.
Therefore, it is expected that the small value of $-2\beta_{\squark}$ predicted in the SM can be measured to be significantly non-zero in several channels.
The expected precision on $\phi_{\squark}^{c\bar{c}s}$ after LHCb \upgradetwo\ will be  $\sim 3 \mrad$ from all modes combined.

\subsubsection{Measurements of $\phid$ from $b\to c\bar{c}s$ transitions}

The world average of $\sin2\beta$ is dominated by results from BaBar, Belle and LHCb using $\Bz\to\jpsi\KS$ decays.
The single most precise measurement is from Belle ($\sin\phi_{\dquark}^{c\bar{c}s} = 0.670 \pm 0.029 \pm 0.013$~\cite{Adachi:2012et}), while the LHCb result has competitive uncertainty ($0.731 \pm 0.035 \pm 0.020$~\cite{LHCb-PAPER-2015-004}). With $50\invfb$ of data, LHCb will reach a precision on $\sin\phi_{\dquark}^{c\bar{c}s}$ of about $0.006$ with $B^0 \to \jpsi \KS$ decays. The \belletwo\ experiment is expected to achieve a precision of about $0.005$ after accumulating $50 \invab$~\cite{Gaz:2017fgl}.  After \upgradetwo, LHCb will be able to reach a statistical precision below $0.003$. 
Fig.~\ref{fig:Bx2JpsiXs_asymmetry_U2}(right) shows the signal-yield asymmetry as a function of the $\Bz$ decay time for $\Bz\to\jpsi\KS$ decays from a simulated data set corresponding to $300\invfb$.

The majority of systematic uncertainties on $\sin\phi_{\dquark}^{c\bar{c}s}$ depend on the size of control samples, and are therefore not expected to be limiting. However, 
at this level of precision it will be necessary to understand possible biases on the result due to \CP violation in \Kz--\Kzb mixing, and from the difference in the nuclear cross-sections in material between \Kz and \Kzb states. Therefore, some irreducible systematic uncertainties are unavoidable. 
It is notable that the leading sources of systematic uncertainty are different for \belletwo\ and LHCb, so that having measurements from both experiments will be important.  
As for the $\phi_s^{c\bar{c}s}$ case, continued good flavour tagging performance and improved understanding of subleading contributions to the decay amplitudes will be required.

The decay $\Bzb \to \Dz \pi^+ \pi^-$, and related decays involving excited charm mesons,
offer a purely tree-level measurement of $\phi_d= 2\beta$.
BaBar and Belle have performed measurements using $B^0 \to D^{(*)}h^0$ with both $D$ decays to \CP\ eigenstates~\cite{Abdesselam:2015gha} and to the three-body $\KS \pi^+\pi^-$ final state~\cite{Adachi:2018itz,Adachi:2018jqe}, where $h^0$ is a light neutral meson such as \piz.
The combined results are  $\sin{\phi_d} = 0.71 \pm 0.09$ and $\cos{\phi_d} = 0.91 \pm 0.25$ including all uncertainties.
While $B^0 \to Dh^0$ can also be studied at LHCb, it is more attractive to measure the same quantities using the $B^0 \to D \pi^+ \pi^-$ mode (here $D$ indicates an admixture of $\Dz$ and $\Dzb$ states).
The Dalitz-plot structure of the decay $B^0 \to \Dzb \pi^+ \pi^-$ has been previously studied~\cite{Kuzmin:2006mw,delAmoSanchez:2010ad,LHCb-PAPER-2014-070}, and the models obtained from these studies could be used in a decay-time-dependent amplitude analysis using the $D \to K^+ K^-$ and $D \to \pi^+ \pi^-$ channels to determine $\phi_d$~\cite{Charles:1998vf,Latham:2008zs}. 
In practice it will be more convenient to perform a simultaneous fit including also the $\Dzb \to K^+ \pi^-$ mode, which acts as a control mode to determine the amplitude model, flavour-tagging response and decay-time acceptance.

An estimate of the achievable sensitivity has been made using pseudoexperiments. 
The expected statistical precisions on $\sin{\phi_d}$ and $\cos{\phi_d}$ are $\pm 0.06$ and $\pm 0.10$, respectively, for the LHCb Run 1 and 2 data samples combined.
Extrapolating this to $300\invfb$ gives $\pm 0.007$ for $\sin{\phi_d}$ and $\pm 0.017$ for $\cos{\phi_d}$.
Further experimental studies are needed to understand the impact of systematic uncertainties, although the use of the $\Dzb \to K^+ \pi^-$ control sample is expected to minimise effects from many potential sources of systematic bias.  
In case model uncertainties become a limiting factor, a model-independent version of the method can be used instead~\cite{Bondar:2018gpb}.
Thus, it is expected that a penguin-free measurement of $\phi_d$ can be achieved with sensitivity better than \belletwo, and comparable to that with $\Bz \to \jpsi\KS$.

\subsubsection{Measurements of penguin pollution in $b\to c\bar{c}s$}\label{sec:b2ccspenguin}

Strategies to measure penguin pollution have already been tested using $\Bd\to\jpsi\rho^0$~\cite{LHCb-PAPER-2014-058} and $\Bs\to\jpsi\Kstarzb$~\cite{LHCb-PAPER-2015-034} decays.
\Blu{These modes constrain the penguin contribution in $B^0\to J/\psi K^{*0}$.}
The best constraint on penguin pollution in $\phi_s^{c\bar{c}s}$ is currently obtained from the $\Bd\to\jpsi\rho^0$ channel, benefiting from the fact that $S_{\jpsi\rhoz}$ and $C_{\jpsi\rhoz}$ have been measured, giving a constraint on $\phid^{\jpsi\rhoz}$.
Following Ref.~\cite{LHCb-PAPER-2014-058}, the bias on \phis is 
\begin{equation}
\Delta\phi_{s}^{c\bar{c}s} \approx - \epsilon \left( \phid^{\jpsi\rhoz} - 2\beta \right), 
\label{eq:penguinCorrection}
\end{equation}
where $\epsilon =  \left|\Vus\right|^2 / (1 - \left|\Vus\right|^2) = 0.0534$, and $2\beta$ is mainly determined by $\Bz \to \jpsi \KS$. 
Since $2\beta$ is, and will continue to be, determined precisely, the sensitivity on $\Delta\phi_{s}^{c\bar{c}s}$ will be driven by the precision on $\phid^{\jpsi\rhoz}$.
Scaling the uncertainties obtained from Ref.~\cite{LHCb-PAPER-2014-058}, the expected statistical precision on $\phid^{\jpsi\rhoz}$ will be $\lesssim 1^\circ$ with $300 \invfb$. 
It is expected that systematic uncertainties, such as those from modelling the S-wave component in $\Bd \to \jpsi\pip\pim$ decays, can be kept under control, so that the uncertainty due to penguin pollution is expected to be $\lesssim 1.5\mrad$.
Thus, this is not expected to limit the sensitivity of the \phis measurement with $\Bs\to\jpsi\phi$.
However, if significant effects of penguin pollution become apparent it may complicate the combination of results from different modes, since a separate correction will be required for each.
For some modes such as $\Bs \to \jpsi f_0(980)$ this may be challenging, since the identification of the states in the \grpsuthree\ multiplet is not trivial. 
 However, this is not a problem for $\Bs \to \Dsp\Dsm$, related by U-spin subgroup to $\Bd \to \Dp\Dm$, which can be used to control penguin pollution~\cite{Fleischer:2007zn,Jung:2014jfa,Bel:2015wha,LHCb-PAPER-2016-037}. 

Similar strategies can be applied using $\Bs \to \jpsi \KS$ and $\Bd \to \jpsi \pi^0$ decays to control the penguin contributions in $\phi_d^{c\bar{c}s}$~\cite{Ciuchini:2005mg,Faller:2008zc,DeBruyn:2014oga}. 
A first analysis of $\Bs \to \jpsi \KS$ decays has been performed~\cite{LHCb-PAPER-2015-005} as a proof of concept for constraining $\Delta\phi_d^{c\bar{c}s}$ with larger datasets. 
The \CP\ violation parameters in $\Bd \to \jpsi \pi^0$ have been previously measured by BaBar and Belle~\cite{Aubert:2008bs,Lee:2007wd}, and the \belletwo\ experiment is expected to reach a sensitivity to $S_{\jpsi\piz}$ of $\sim 0.03$, which should be sufficient to keep penguin pollution under control.
LHCb can also study $\Bd \to \jpsi \pi^0$ decays, although the presence of a neutral pion in the final state makes the analysis more challenging and there is currently no public result from which to extrapolate the sensitivity.  
Improving the capabilities of LHCb's electromagnetic calorimeter (ECAL) will enhance prospects for studying this mode.

It should be stressed that the methods to constrain penguin pollution rely on \grpsuthree\ symmetry, and the approximations associated with the method and inherent in Eq.~(\ref{eq:penguinCorrection}) must also be investigated. 
This can be done by studying the full set of modes related by \grpsuthree, namely $\Bds \to \jpsi \big\{ \piz, \eta, \etapr, \Kz, \Kzb \big\}$ and $\Bpm \to \jpsi \big\{ \pipm, \Kpm \big\}$ for the vector-pseudoscalar final states and $\Bds \to \jpsi \big\{ \rhoz, \omega, \phi, \Kstarz, \Kstarzb \big\}$ and $\Bpm \to \jpsi \big\{ \rhopm, \Kstarpm \big\}$ for the vector-vector final states.
Several of these modes have not yet been measured, but with the data sample of LHCb \upgradetwo\ it should be possible to measure all branching fractions and \CP\ asymmetry parameters, allowing a full theoretical treatment and more detailed understanding of subleading contributions. \Blu{An alternative approach to controlling penguin contributions, which does not involve $SU(3)$ symmetry, can be found in Ref. \cite{Frings:2015eva}.}

\subsubsection{Measurements of $\phi_d$ and $\phi_s$ in charmless decays}
The $\Bs\to\phi\phi$ decay is forbidden at tree level in the SM and proceeds predominantly via a gluonic penguin $\bquarkbar \to \squarkbar \ssbar$ loop process.
Hence, this channel provides an excellent probe of physics beyond the SM that may contribute to the penguin diagram~\cite{Bartsch:2008ps,Beneke:2006hg,PhysRevD.80.114026}.
This mode is well suited for study at the LHC, as both $\phi$ mesons can be reconstructed through their decay to $\Kp\Km$, leading to clean signatures even in the absence of hadronic particle identification.

\subsubsubsection{CMS $B_s \to \phi \phi$ studies}
The lack of particle-ID detectors will limit the CMS investigation of fully hadronic final states. However, some capability is retained through the early use of tracking in the trigger selection.
The $B_s \to \phi\phi \to 4K$ is an example of a hadronic final state that would benefit from the tracking performance at trigger level and the $\phi$ resonance signature.
While the full study of the sensitivity to the $\phi$ measurement in this channel is still ongoing, 
an analysis was performed by CMS to see if  $B^0_s \to \phi \phi \to 4 K$ events can be 
triggered with high efficiency at L1 using only the tracks reconstructed at that level (L1 tracks)~\cite{CMSTDR:trk}.

The L1 track finder will allow identification of $B^0_s \to \phi \phi \to 4 K$ candidates at L1 by forming 
$\phi$ candidates from pairs of oppositely charged L1 tracks constrained to come from the same vertex and then  combining  pairs of  such candidates into a $B^0_s$  candidate.
The \pt of the lowest-\pt kaon lies very close to the lowest possible trigger threshold of the L1 tracking of 2 GeV, possibly causing loss of signal efficiency. 

\begin{figure}[t]
\begin{center}
  \includegraphics[width=0.495\textwidth]{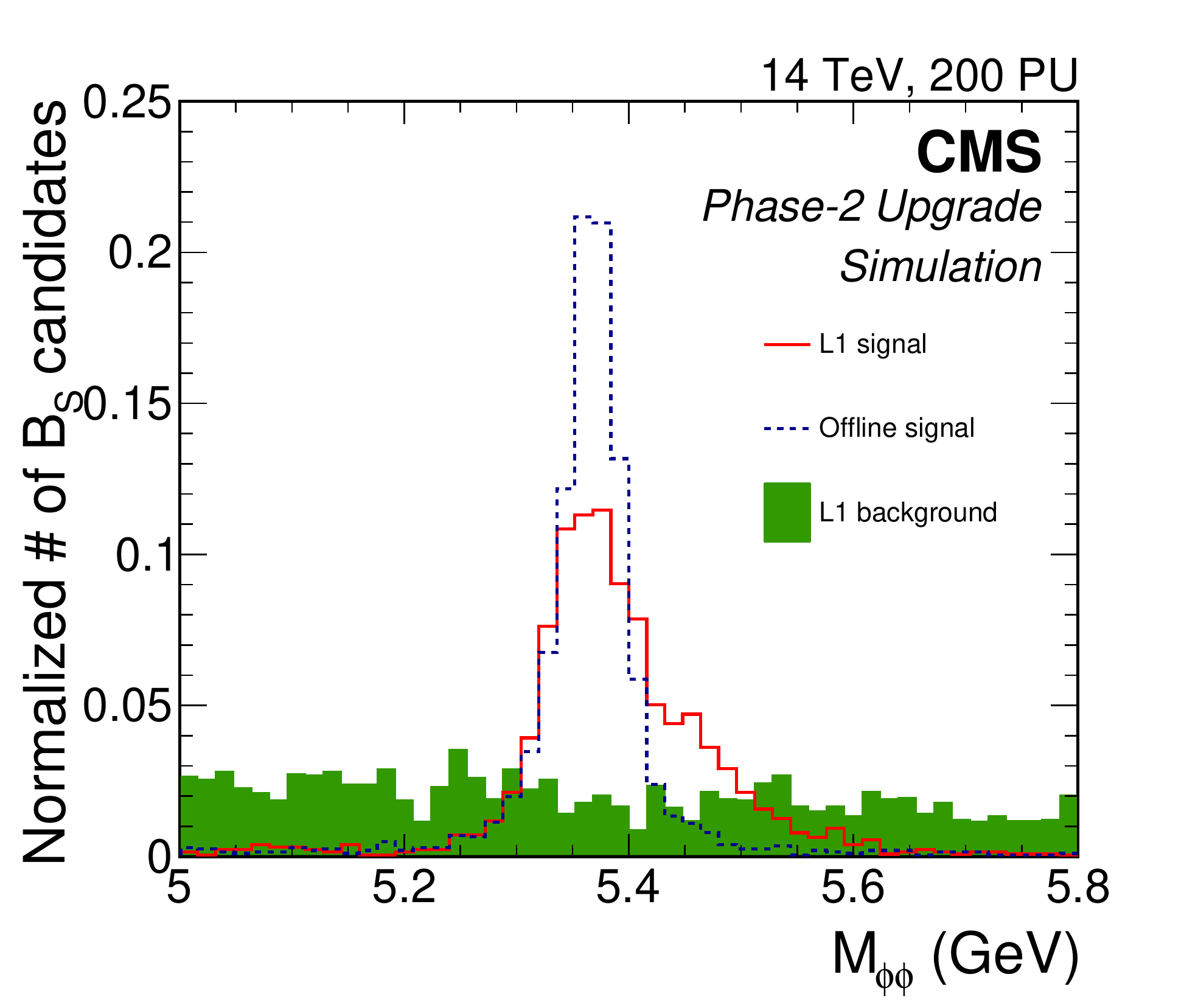}
  \caption{Normalized invariant mass distribution in CMS of all the $\phi$-pairs with $\Delta z (\phi\text{-pair}) < 1$ cm, $\Delta xy (\phi\text{-pair}) <1$ cm, $0.2 < \Delta R(\phi\text{-pair}) < 1$, $\Delta R(K^+,K^-) < 0.12$. 
The blue dashed  line corresponds to the signal events reconstructed with offline tracks.
The signal and background distributions obtained using L1 tracks  are shown as 
red solid line and green histograms, respectively. A pile up scenario of 200 interactions is assumed (taken from \cite{CMSTDR:trk}).}
    \label{fig:cms_bs2phiphi}
\end{center}
\end{figure}

Fig.~\ref{fig:cms_bs2phiphi} shows the invariant mass of the $B^0_s \to \phi \phi $ candidates for all $\phi$-pairs having separation along the beam axis ($z$) of  $\Delta z(\phi\text{-pair}) < 1$ cm, distance in the plane perpendicular to the beam axis  $\Delta xy(\phi\text{-pair}) < 1$ cm, $0.2 < \Delta R (\phi\text{-pair}) < 1$, and $\Delta R (K^+, K^-) < 0.12$,  in events with 200 pile-up (PU) interactions.
Simulations show that an efficiency of 30-35 \% can be achieved at L1 trigger level, depending on the track candidate selection,  and that the events selected at L1 would be accepted by the subsequent offline analysis with high efficiency.
For the scenario with 200 PU and a moderate signal efficiency of around 30\%, the expected L1 trigger rate is about 15 kHz, within the acceptable limit according to the present understanding of the expected detector performance.

\subsubsubsection{LHCb $B_s \to \phi \phi$ projections}

\begin{figure}[t]
\begin{center}
  \includegraphics[width=0.495\textwidth]{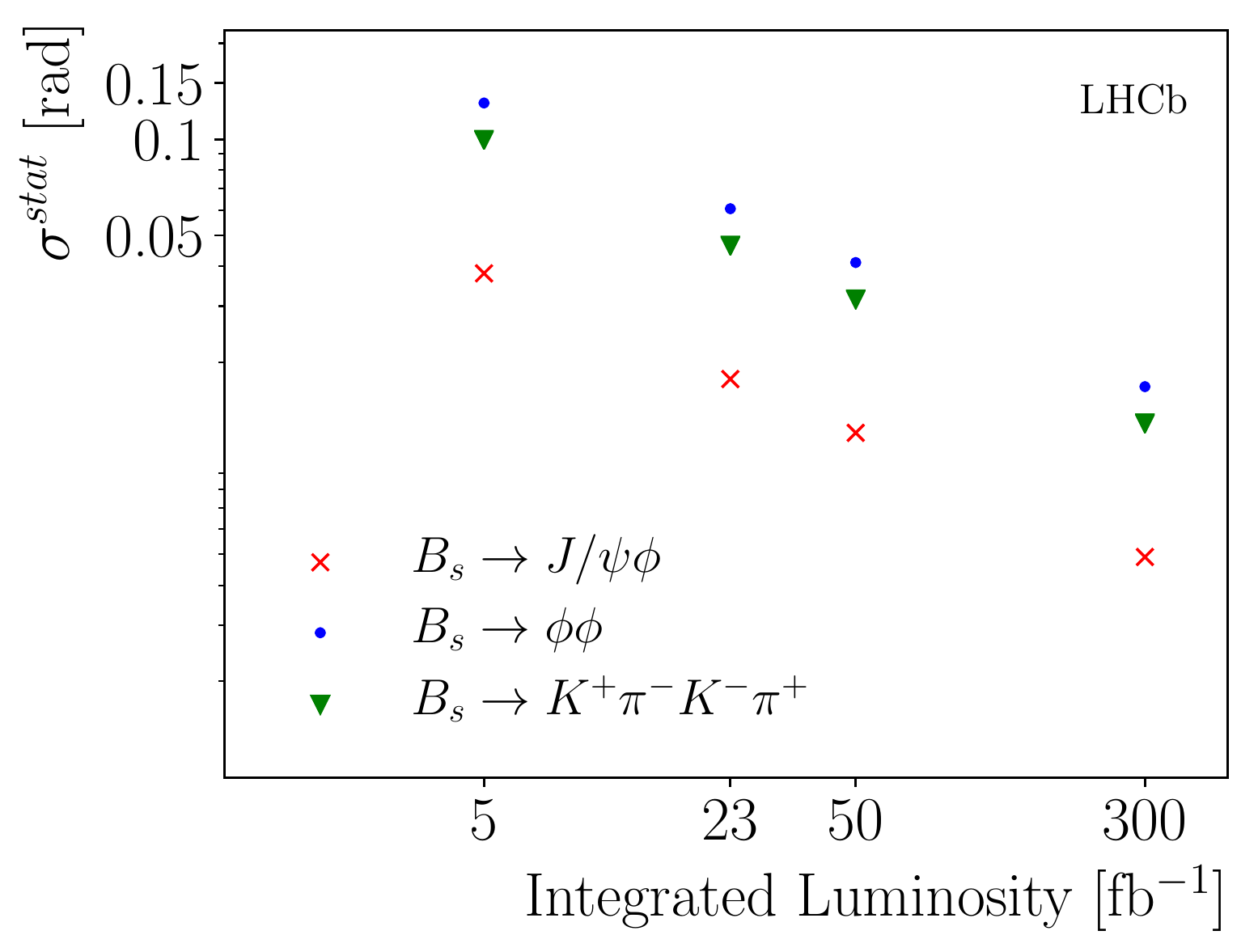}
  \caption{Comparison of $\phi_s$ statistical sensitivity at LHCb from different decay modes.}
    \label{fig:td_comp}
\end{center}
\end{figure}

The measured and extrapolated LHCb statistical sensitivities for $\phisPP$ and similar \CP-violating phases measured in other decay modes are shown in Fig.~\ref{fig:td_comp}.
A statistical uncertainty on $\phisPP$ of $0.011 \rad$ can be achieved with $300 \invfb$ of data collected at LHCb \upgradetwo.
Similarly to other measurements of \CP violation parameters from decay-time-dependent analyses, many systematic uncertainties are evaluated from control samples, and are therefore expected to scale accordingly with integrated luminosity.
Among others, there is an important uncertainty associated with knowledge of the angular acceptance, which is determined from simulation.
This therefore relies on good agreement between data and simulation, which can be validated using control channels such as $\Bd \to \phi \Kstarz$. 
Thus the determination of $\phisPP$ is expected to remain statistically limited even with the full LHCb \upgradetwo\ data sample.  

\subsubsubsection{LHCb projections for $\phi_s$ from other charmless decays}

Another way of measuring $\phi_s$ is the $B_s^0\to \Kstar(892)^0\Kstarb(892)^0$ family of decays, 
which in the SM is dominated by a gluonic penguin  $b\to d\overline{d}s$ diagram.
LHCb has recently published the first measurement of $\phi_s^{d\bar{d}s}$~\cite{LHCb-PAPER-2017-048} 
using Run~1 data. In this groundbreaking analysis, it was realised that a significant gain in sensitivity can be obtained by including the full $B_s^0\to(K^+\pi^-)(K^-\pi^+)$ phase space in the $K\pi$-mass window from 750 to $1600 \mevcc$, since the fraction of $B_s^0\to \Kstar(892)^0\Kstarb(892)^0$ in this region is only $f_{VV}=0.067 \pm 0.004 \pm 0.024$ (the other contributions are from $K\pi$ S-wave and the $K_2^*(1430)^0$ resonance).
The result, $\phi_s^{d\bar{d}s}= -0.10 \pm 0.13 \pm 0.14 \rad$, is compatible with the SM expectation.

The current result has statistical and systematic uncertainties of comparable size, but both are expected to be reducible with larger data samples.
The largest systematic uncertainty, corresponding to the treatment of the acceptance, is mostly driven by the limited size of the simulation samples --- due to the large phase space investigated in this analysis, very large simulation samples are required.
In order to produce significantly larger samples it will be necessary to exploit rapid simulation production mechanisms, since increases in available CPU power are not expected to keep pace with the size of the data samples.  
Another important systematic uncertainty due to the modelling of the $K\pi$ resonant and non-resonant components can be reduced by incorporating results of state-of-the-art studies of the $K\pi$ system, but some component of this may be irreducible.
Other systematic uncertainties are mainly based on control samples.
Therefore it is expected that the limiting systematic uncertainty will be not larger than $\sigma(\text{syst.})\sim0.03 \rad$.

The measured and extrapolated statistical sensitivities for $\phi_s^{d\bar{d}s}$ are given in \tabref{tab:BsKpiKpi_sigmastat}, both for the average over the $B_s^0\to(K^+\pi^-)(K^-\pi^+)$ system and for the exclusive $B_s^0\to \Kstar(892)^0\Kstarb(892)^0$ decay.
The sensitivities for $B_s^0\to(K^+\pi^-)(K^-\pi^+)$ are also included in Fig.~\ref{fig:td_comp}.
In the current analysis, the same weak phase is assumed for all contributions, but as the precision increases it will be possible to determine $\phi_s^{d\bar{d}s}$ separately for each, including possible polarisation dependence in the $B_s^0\to \Kstar(892)^0\Kstarb(892)^0$ decay.
The systematic uncertainty related to modelling of components is expected to be smaller when focusing on the $\Kstar(892)$ resonance, since its lineshape is well known.
Moreover, by making similar studies with the $B^0 \to (K^+\pi^-)(K^-\pi^+)$ mode, it will be possible to obtain all necessary inputs for the U-spin analysis of each component separately, leading to good control of the theoretical uncertainty on the prediction for $\phi_s^{d\bar{d}s}$.

\begin{table}[t]
\begin{center}
\caption{Statistical sensitivity on $\phi_s^{s\bar{s}s}$ and $\phi_s^{d\bar{d}s}$ at LHCb.}
  \renewcommand{\arraystretch}{1.2}
 \begin{tabular}{ccccc}
 \hline\hline
\multirow{ 2}{*}{Decay mode} & \multicolumn{4}{c}{$\sigma$(stat.) [rad]} \\
& 3~fb$^{-1}$ & 23~fb$^{-1}$ & 50~fb$^{-1}$ & 300~fb$^{-1}$ \\
\hline
$B_s^0 \to \phi\phi$ &  $0.154$  & $0.039$  & $0.026$ & $0.011$ \\
$B_s^0 \to (K^+\pi^-)(K^-\pi^+)$ (inclusive) & $0.129$  & $0.033$  & $0.022$ & $0.009$ \\
$B_s^0 \to \Kstar(892)^0\Kstarb(892)^0$ & $-$  & $0.127$  & $0.086$ & $0.035$ \\
 \hline \hline
\end{tabular}
\renewcommand{\arraystretch}{1.0}
 \label{tab:BsKpiKpi_sigmastat}
\end{center}
\end{table}

Finally, LHCb can also make measurements of $\phi_{\squark}^{\dquark\uquarkbar\uquark}$ using a decay-time-dependent flavour-tagged Dalitz-plot analysis of \decay{\Bs}{\KS\pip\pim} decays~\cite{Gershon:2014yma}. Preliminary sensitivity studies indicate that the precision achievable on $\phi_{\squark}^{\dquark\uquarkbar\uquark}$ with the full Run 1 + Run 2 dataset is approximately $0.4\rad$. 
Extrapolation to $300\invfb$ indicates potential for a precision of around $0.07\rad$. 
Similar studies for the \decay{\Bs}{\KS\Kpm\pimp} decay mode can be performed, which requires a more complicated analysis since both $\KS\Km\pip$ and $\KS\Kp\pim$ final states are accessible to both \Bs and \Bsb decays with comparable magnitude.
Although currently the precision of such measurements are dominated by the statistical uncertainty~\cite{SilvaCoutinho:2045786}, the expected yields of more than $10^6$ signal decays for $300 \invfb$ will allow a complete understanding of the \decay{\Bs}{\KS\Kpm\pimp} phase space. 

\subsubsubsection{LHCb projections for $\phi_d$ from other charmless decays}
The large yields of $\Bd \to \KS h^+h^-$ decays  available~\cite{LHCb-PAPER-2017-010} at LHCb enable a relevant reach for $\phi_d$ from these modes.
For the $\Bd \to \KS\pip\pim$ mode, a decay-time-integrated analysis has been performed resulting in the first observation of \CP violation in the $\Bd \to \Kstarp\pim$ channel~\cite{LHCb-PAPER-2017-033}; with more data this analysis can be updated to include decay-time-dependence and determine also \CP-violating parameters for the $\Bd \to \rhoz\KS$ and $f_0(980)\KS$ channels.

Studies of $\Bds \to h^+ h^- \piz$ decays will provide further sensitivity.
For example, resonant contributions of the type $\Kstarpm h^\mp$ will decay to the final state $\Kpm h^\mp \piz$ in addition to $\KS \pipm h^\mp$, and therefore a combined analysis of both can provide additional information that helps to test the SM prediction~\cite{Ciuchini:2006kv,Ciuchini:2006st,Gronau:2006qn,Gronau:2007vr}.
Large yields will be available and the improved capabilities of the \upgradetwo ECAL will allow backgrounds to be controlled. 
Two particularly important features of these decays are that background from $B \to V \gamma$ decays, where $V \to h^+h^-$, must be suppressed and that it must be possible to resolve \Bd\ and \Bs\ decays to the same final state.
Thus, it will be important to have both good $\gamma$--$\piz$ separation and good mass resolution.

\subsubsection{Measurements of $\alpha$}

The main input from LHCb to the
 isospin analysis determination of $\alpha$ from $B \to \pi\pi$ decays will be world-leading measurements of the \CP-violating parameters in \BdTopipi decay. Important input can also be expected on the \decay{\Bu}{\pip\piz} decay, using the method pioneered in Ref.~\cite{LHCb-CONF-2015-001} for \decay{\Bu}{\Kp\piz} decays.
Clearly, good performance of the electromagnetic calorimeter will be critical for such a measurement.
Progress on the \decay{\Bd}{\piz\piz} mode, which is currently limiting the precision of $\alpha$ determination from $B \to \pi\pi$ decays, will mainly come from the \belletwo\ experiment.

The situation is quite different for the $B \to \rho\rho$ system, which currently provides the strongest constraints on $\alpha$.
Although the presence of two vector particles in the final state makes the analysis more complicated in principle~\cite{Gronau:1990ka,Dunietz:1990cj,Falk:2003uq,Beneke:2006rb}, the observed dominance of longitudinal polarisation and the smaller penguin contribution compared to $B \to \pi\pi$ lead to good sensitivity on $\alpha$. 
The rarest of the three isospin-partner modes is the $\decay{\Bz}{\rho^0\rho^0}$ decay, which has a final state of four charged tracks following $\rho^0 \to \pip\pim$ decays, making it well suited for study at LHCb. 
Indeed, this decay was first observed by LHCb, and a time-integrated angular analysis on Run 1 data was performed~\cite{LHCb-PAPER-2015-006}.
With larger data samples it will be possible not only to improve the measurements of the branching fraction and longitudinal polarisation fraction, but to make precise determinations of the \CP-violating parameters. 
Consequently, the determination of $\alpha$ from $B \to \rho\rho$ decays will benefit from the additional information inherent in $S_{\rhoz\rhoz}$, compared to 
 the $B \to \pi\pi$ system for which $S_{\piz\piz}$ is barely measurable (\belletwo\ plans to measure $S_{\piz\piz}$ using  Dalitz decay  of $\piz$, but only rather limited sensitivity is possible).

First measurements of $\Bz \rightarrow \rho \pi$ have been published by BaBar and Belle~\cite{Lees:2013nwa,Kusaka:2007dv,Kusaka:2007mj}, but do not yet provide strong constraints on $\alpha$.
LHCb has not yet published any result on this channel, but it is expected that large yields will be available in LHCb \upgradetwo, and that it should be possible to control backgrounds with good understanding of $\piz$ reconstruction.  
It is worth noting that a significant proportion of the photons from neutral pions produced at the \B decay vertex convert into $\ep\en$ pairs in the LHCb detector material, with approximately half of these conversions occurring before the magnet.
Tracks from these converted photons provide additional information with which to constrain the \B decay vertex and the neutral pion momentum, resulting in improved resolution and background rejection.

\subsubsection{Measurements of \CP violation in baryon decays\footnote{We thank I. Bigi for stressing the importance of baryon decays.}}
In contrast with the study of \CP violation in beauty-meson decays, the sector of beauty baryons remains
almost unexplored. Previous to the \lhc era, only a measurement of direct \CP asymmetries
in \decay{\Lb}{\proton\Km} and \decay{\Lb}{\proton\pim} decays was available with $\order(0.1)$
precision~\cite{Aaltonen:2014vra}. Thanks to the large production cross-section of beauty baryons in \proton\proton collisions at the \lhc, the \lhcb experiment is the only experiment capable of expanding our knowledge
in this sector, as these decays are not accessible at the \ep\en KEK collider. Hence, even though a handful
of \CP asymmetries of \Lb decays have been measured so far by
\lhcb~\cite{LHCb-PAPER-2013-061,LHCb-PAPER-2016-004,LHCb-PAPER-2017-034,LHCb-PAPER-2016-059},
the landscape of \CP violation in the decays of beauty baryons is expected to change rapidly in the
next few years. 

The unprecedented number of beauty baryons available with the data sample expected to be collected in the \lhcb \upgradetwo phase, will allow a precision measurement programme of \CP violation observables in \bquark-baryon decays to be pursued, analogously to \bquark-meson decays.
A very interesting example is the study of decays governed by \decay{\bquark}{\uquark} and \decay{\bquark}{\cquark} tree-level transitions, like the decays \decay{\Lb}{\Dz\Lambdares} and \decay{\Lb}{\Dz\proton\Km}. These decays can be used to measure the angle $\gamma$ of the unitarity triangle~\cite{Dunietz:1992ti,Fayyazuddin:1998aa,Giri:2001ju} in a similar way to what can be done with \decay{\Bd}{\D\Kp\pim} decays. The \lhcb experiment reported the first observation of the \decay{\Lb}{\Dz(\Km\pip)\proton\Km}, based on a signal yield of $163 \pm 18$ using a sample of \proton\proton collisions corresponding to 1\invfb of integrated luminosity at a centre-of-mass energy of 7\tev~\cite{LHCb-PAPER-2013-056}. Extrapolating to \upgradetwo\ approximately 95000 signal decays are expected. However, extrapolating the sensitivity to $\gamma$ is not easy, since it strongly depends on the values of the hadronic parameters involved in the process. In addition, even though the determination of $\gamma$ from the analysis of these decays is expected to be theoretically very clean, the possible polarisation of \Lb baryons produced in \proton\proton collisions has to be taken into account and might represent a limiting factor for high-precision measurements.

Another very interesting sector is that of beauty baryons decaying to final states without a charm quark. These decays receive relevant contributions from \decay{\bquark}{\dquark,\,\squark} loop-level transitions, where new physics beyond the SM may appear. Also in this case, similar quantities to those measured with \B-meson decays are available. For example, statistical precisions of $\order(10^{-3})$ and $\order(10^{-2})$ are expected for the \CP asymmetries of \decay{\Lb}{\proton h^-} and \decay{\Lb}{\Lz h^+h^-} decays (with $h = \kaon,\,\pion$), respectively. Very large signal yields are also expected in several multibody final states of \Lb and \Xib decays: about $10^6$ \decay{\Lb}{\proton\pim\pip\pim} and \decay{\Lb}{\proton\Km\Kp\Km} decays, and about $10^5$ \decay{\Xibz}{\proton\Km\pip\Km} decays~\cite{LHCb-PAPER-2016-030,LHCb-PAPER-2018-001}.
Such a signal yield will allow very precise measurements of \CP-violating quantities to be made over the phase space of these decays, characterised by a rich set of resonances. Unfortunately, as for the charmless decays of \B mesons, the interpretation of these quantities in terms of CKM parameters is still unclear from the theoretical point of view. Hence, more theoretical work is crucial to exploit the full potential of beauty baryons.

Experimentally, the main issues are the determination of particle-antiparticle production asymmetries and detection asymmetries that could mimic \CP-violation effects. This task is generally more difficult for heavy baryons, with respect to \B mesons, since methods used for measuring meson production asymmetries~\cite{LHCb-PAPER-2016-062} cannot be applied. In addition, different interactions of baryons and antibaryons with the detector material are difficult to calibrate. Nonetheless, several quantities can be measured in \bquark-baryon decays that are sensitive to different manifestations of \CP violation and are largely unaffected by experimental effects. A few examples are the difference of \CP-violating asymmetries of particles decaying to a similar final state, $\Delta A_{\CP}$~\cite{LHCb-PAPER-2017-044}, triple-product asymmetries (TPA)~\cite{LHCb-PAPER-2016-030} and energy-test (ET)~\cite{LHCb-PAPER-2014-054}. It is important to note that TPA and ET are important tools for discovery of \CP violation in multibody decays, while an amplitude analysis is required  to study the source of \CP violation.

%% file: section3/section.tex
\section{Charm-quark probes of new physics}
{\it \small Authors (TH): J. Brod,  S. Fajfer, A. Kagan, A. Lenz, L. Silvestrini.}

In the SM, the FCNC processes involving charmed hadrons are
suppressed compared to those involving strange or beauty hadrons since they are proportional to 
the small breaking of the GIM mechanism by the bottom quark mass. This also means that 
 the contributions from long-distance physics, due to intermediate $d$, $s$ quarks, are relatively more important, complicating the predictions. 
Moreover, due to small off-diagonal CKM matrix elements the third generation approximately factorizes from the first two generations, 
leading to additional suppression of the \CP violating effects in charmed hadrons.
The charmed
hadrons can then be used as sensitive probes of new physics in the up-quark
sector, to the extent that theoretical uncertainties can be brought
under control, e.g., by constructing null tests, or circumvented by using experimental data.

\subsection{Charm mixing}
\label{sec:Dmixingth}
Weak interactions mix $D^0$ and $\bar D^0$ mesons, so that the mass eigenstates are
$|D_{1,2}\rangle = p |D^0 \rangle \pm q |\dzbar\rangle$. By convention $|D_{2}\rangle$
is CP-even in the absence of \CP violation.  The mass and width differences, 
$\Delta M = m_2 - m_1$, and $\Delta \Gamma = \Gamma_2 - \Gamma_1$,  are parametrized as 
\begin{eqnarray}\label{xyexp}
x \equiv \Delta M^D / \Gamma^D = 0.46 \% \pm 0.13 \% \, ,
&&
y \equiv \Delta \Gamma^D / (2 \Gamma^D) = 0.62 \% \pm 0.07 \%.
\end{eqnarray}
Here $\Gamma$ is the total decay rate of the neutral $D$ mesons, while the numerical values are from the fits to the experimental measurements \cite{Amhis:2016xyh}. As follows from Eq. \eqref{xyexp}, it appears that in $D$ system $x \sim y$ or $\Gamma_{12} \sim M_{12}$.
This is to be contrasted with the $B$ system, where $|\Gamma_{12}/M_{12}| \ll 1$ holds.
We also define the ``theoretical'' mixing parameters, 
\beq 
x_{12}\, \equiv\, \frac{2|M^D_{12}|}{\Gamma^D}\,
,~~ y_{12}\, \equiv\,\frac{|\Gamma^D_{12}|}{\Gamma^D}\,,~~\phi_{12} \,\equiv\, {\rm arg}\left(\frac{M^D_{12}}{\Gamma^D_{12} }\right)\,,
\label{x12yi12phi12}
\eeq
where $M_{12}^D $ and $\Gamma_{12}^D$ are the dispersive and absorptive contributions to the $\DDbar$ mixing amplitude, $\langle  D^0 | H | \dzbar \rangle \ =\  M^D_{12} - \frac{i}{2} \Gamma^D_{12}$. 
The phase $\phi_{12}$ gives rise to \CP violation in mixing, cf. Sec.~\ref{sec:CPVDmixth}. Its magnitude is currently bounded to lie below $\sim 100$ mrad at 95\% CL~\cite{LHCbimplications,Amhis:2016xyh}.
These parameters are related to $x$ and $y$ as,
\beq (x- i y)^2 = x_{12}^2 - y_{12}^2  - 2 \,i\, x_{12} y_{12} \cos\phi_{12}\,,\label{xyx12y12}\eeq
so that, up to negligible corrections quadratic in $\sin\phi_{12}$ (in general, $|y| \le y_{12}$ \cite{Nierste:2009wg,Jubb:2016mvq}),
\beq \begin{split} |x|=x_{12} ,~~|y| = y_{12}\,.\label{x12vsx}
\end{split}
\eeq

It is convenient to begin the discussion of the SM mixing amplitudes 
with their $U$-spin flavor symmetry decomposition \cite{Falk:2001hx,Gronau:2012kq}.
Employing CKM unitarity, $\lambda_d + \lambda_s +\lambda_b = 0$, with  $\lambda_x = V_{cx} V^*_{ux}$, 
$\Gamma^D_{12}$ can be written as 
\beq
 \Gamma^D_{12 } = \frac{(\lambda_s - \lambda_d )^2}{4 }\, \Gamma_2 + \frac{(\lambda_s - \lambda_d ){\lambda_b} }{ 2 }\, \Gamma_1 + 
\frac{{ \lambda_b^2} }{4} \, \Gamma_0 \,,  \label{Uspindecomp}\eeq
where
\beq \begin{split} \label{Uspinamps}
\Gamma_2  &= \Gamma_{ss} + \Gamma_{dd} - 2 \Gamma_{sd} ~\sim~ (\bar s s - \bar d d )^2 ={\mathcal O}(\epsilon^2)  \,,\cr
\Gamma_1 & = \Gamma_{ss} - \Gamma_{dd}~ \sim~ (\bar s s - \bar d d ) (\bar s s + \bar d d) ={\mathcal O}(\epsilon)\,,\cr
\Gamma_0 & = \Gamma_{ss} +\Gamma_{dd}+2 \Gamma_{sd} ~ \sim~ (\bar s s+ \bar d d )^2 ={\mathcal O}(1)  \,,
\end{split}
\eeq
with $\epsilon\sim0.2$ denoting the $U$-spin breaking parameter. 
The $\Gamma_{2,1,0}$ are the $\Delta U_3 =0$ elements of the $U$-spin 5-plet, triplet, and singlet, respectively.
The individual $\Gamma_{ij}$ are identified, at the quark level, with box diagrams containing on-shell internal $i$ and $j$ quarks. They possess flavor structure $\Gamma_{ij}\sim  (\bar i i) (\bar j j) (\bar u c)^2$, where
the external $u $ and $c$ quarks are irrelevant for the $U$-spin decomposition.
The $U$-spin decomposition of $M^D_{12}$ is analogous to Eq.~\eqref{Uspindecomp}, with $\Gamma\to M$ replacement everywhere. 
At the level of quark box diagrams, $M_{1,0}$ also receive contributions containing internal $b$ quarks.
The small value of $\lambda_b\sim {\mathcal O}(10^{-4})$ implies that we can neglect the $\Delta U= 1, 0$ contributions to the mass and width differences, even though the $\Delta U=2$ piece is higher order in $\epsilon$.

Evaluation of the SM mixing amplitudes is very challenging,
because the charm quark mass lies at an intermediate scale between the masses of the light quarks and the bottom quark.
Broadly speaking, there are two approaches: (i) an inclusive one employing the operator product expansion (OPE), which expands in powers of $\Lambda_{\rm QCD}/m_c$, as in the heavy quark expansion (HQE), and assumes that local quark-hadron duality holds \cite{Georgi:1992as,Ohl:1992sr,Bigi:2000wn,Bobrowski:2010xg}; and (ii) 
an exclusive one in which $y$ is estimated by summing over contributions of exclusive states, and $x$ is estimated via a dispersion relation which relates
it to $y$.
In the first approach, 
the HQE applied to $\Gamma_{12}^D$, combined with the relevant non-perturbative dimension six operator matrix elements evaluated in 
\cite{Carrasco:2014uya,Carrasco:2015pra,Bazavov:2017weg,Kirk:2017juj}, yields
contributions of the individual $\Gamma_{ij}$ 
to $y$ that are five times larger than the experimental value \cite{Lenz:2016fcv}.
This corresponds to $\Gamma_{ij} \sim \Gamma^D$, which is not surprising, given that the HQE can accommodate the charm meson lifetimes
\cite{Lenz:2013aua,Kirk:2017juj}.
However, the result for $\Gamma_2$, cf. Eqs. \eqref{Uspindecomp}, \eqref{Uspinamps}, yields a value of $y$
lying about four orders 
of magnitude below experiment, due to large GIM-cancellations between the $\Gamma_{ij} $ contributions.
Evidently, the inclusive approach is not well suited for analyzing the $U$-spin breaking responsible for $\DDbar$ mixing, i.e., the charm quark is not sufficiently massive, and $(m_s - m_d )/\Lambda_{\rm QCD}$ is not sufficiently small.
First estimates 
of the dimension nine contribution in the HQE \cite{Bobrowski:2012jf} indicate an enhancement compared to the leading dimension six terms, but do not alter this conclusion.  The HQE result $\theta_c^2 \,\Gamma_{ij}\sim 5 \,y $ would require 
large $U$-spin violation, e.g. ${\mathcal O}(\epsilon^2 ) = 20\% $ in $\Gamma_2$, cf. \eqref{Uspinamps}, which could be attributed to long-distance duality violation \cite{Jubb:2016mvq}.  One possibility for directly addressing the origin of $U$-spin violation in $\DDbar$ mixing is the second (exclusive) approach mentioned above.

The starting point for the exclusive approach\cite{Falk:2001hx,Falk:2004wg,Cheng:2010rv,Gronau:2012kq,Jiang:2017zwr} 
is a sum over the decay modes contributing to the absorptive and dispersive mixing amplitudes, see e.g.~\cite{Nir:1992uv},
\beq
\Gamma_{12}^D =  \sum_f \rho_f                                            A_f^* \, \bar A_f  \, ;
~~~~~M_{12}^D  =   \langle {D}^0 | H^{\Delta D=2} |\dzbar \rangle 
+ P \sum_f  \frac{    A_f^* \, \bar A_f  }
                                          {m_{D^0}^2 - E_f^2} \, ,
\label{Dmix_exclusive}
\eeq
where $A_f =\langle f | H_{\Delta c=1} | D^0\rangle $ and $\bar A_f =\langle f | H_{\Delta c=1} | \dzbar \rangle $ are the $D^0 \to f$ and $\dzbar \to f$ decay amplitudes, respectively, 
$\rho_f$ is the 
density of the state $f$, and $P$ is the principal value.   Unfortunately, the charm quark mass is not sufficiently light
for $D$ meson decays to be dominated by a few final states.  Moreover, the strong phase differences entering $\Gamma_{12}^D$,  and the off-shell decay amplitudes
entering $M_{12}^D$ 
are not calculable from first principles.  
Thus, simplified treatments of $SU(3)_F$ flavor symmetry breaking have been utilized.
A rough $U$-spin based estimate for $y$ is simply obtained from the first term in \eqref{Uspindecomp}, 
\beq y \,= \, \sin^2 \theta_c \times  \Gamma_2/ \Gamma^D\, \sim \,\sin^2\theta_C \,\times \, \epsilon^2  \sim (0.2 -5)\%,\eeq
where we have taken $\Gamma_2 \sim \Gamma^D \epsilon^2 $ (which can be motivated by the HQE result $\Gamma_{ij} \sim \Gamma^D$); and 
$\epsilon \sim 0.2 - 1$, corresponding to variation from nominal to maximal $U$-spin breaking.
The authors of \cite{Falk:2001hx,Falk:2004wg} only took $SU(3)_F$-breaking phase space effects into account in the exclusive sum.  They found that 
$y\lsim 1\%$ could naturally be realized, where a value at the high end would require contributions from higher multiplicity final states, due to the larger $SU(3)_F$ breaking effects near threshold (consistent with the large $U$-spin breaking required from duality violations in the OPE/HQE approach).    
This conclusion was subsequently supported in \cite{Cheng:2010rv,Jiang:2017zwr}, which added experimental branching ratio inputs together with factorization based models for dynamical $SU(3)_F$ breaking effects, e.g., in the form factors and strong phases.
Rough dispersion relation estimates in \cite{Falk:2004wg,Cheng:2010rv } suggested that $|x/y| \sim 0.1 -1$.

We conclude that the estimates of $x$ and $y$ in the SM are consistent with their measured values, cf. \eqref{xyexp}.  
Unfortunately, the large theoretical uncertainties eliminate the window for NP in these quantities. 
The situation is markedly different for the CPV mixing observables, due to their large suppression in the SM, as discussed below.
On a very long time-scale, direct lattice calculations might be able to predict the
SM values of $x$ and $y$ by building on the methods described in \cite{Hansen:2012tf}.

\subsection{CP violation in $\DDbar$ mixing}
\label{sec:CPVDmixth}

In the SM, \CP violation (CPV) in $\DDbar$ mixing is highly suppressed, entering at 
{ ${\mathcal O}(|V_{cb} V_{ub} / V_{cs} V_{us}| ) \sim 10^{-3}$}. 
This raises several questions, which we briefly address, based on work to appear in \cite{GKLPPS}:
What is the resulting theoretical uncertainty on the indirect CPV 
observables? How large is the current window for New Physics (NP)?  
What is an appropriate parametrization for indirect CPV effects, 
given the expected sensitivity in the LHCb/Belle-II era?

There are two types of CPV due to mixing; both are referred to 
as ``indirect'' CPV.  The first is CPV due to interference between the dispersive and absorptive mixing amplitudes (``CPVMIX''), which arises when $\phi_{12}\ne 0$.  CPVMIX can be directly measured via 
the semileptonic \CP asymmetry 
\beq \label{ASL}
A_{\rm SL} \ \equiv\  
\frac{\Gamma(D^0\to K^+\ell^-\nu) - \Gamma(\dzbar\to K^-\ell^+\nu)}
{\Gamma(D^0\to K^+\ell^-\nu) + \Gamma(\dzbar\to K^-\ell^+\nu)} \ =\ 
 \frac{|q/p|^4 -1}{|q/p|^4 +1}\ = \frac{2 x_{12} y_{12}}{x_{12}^2 + y_{12}^2 } \sin\phi_{12}\,.
\eeq
The second type of indirect CPV is due to interference between a direct decay 
amplitude and a ``mixed'' amplitude followed by decay (``CPVINT''), 
i.e., interference between $D^0 \to f$ and $D^0 \to \dzbar \to f $. 
For decays to a \CP eigenstate final state, there are two CPVINT observables
\cite{Branco:1999fs,Bigi:2000yz}, 
\beq 
\lambda^M_{f} \equiv \,\frac{M_{12}}{|M_{12} |}\,
\frac{A_f } { \overline A_f} = \,\eta_{f}^{CP}\,\left|\frac{A_f }{ \overline A_f }\right| \, e^{ i  \phi_f^M }\,,~~~
\lambda^\Gamma_{f} \equiv \,\frac{\Gamma_{12} }{|\Gamma_{12} |}\,
\frac{A_f } {\overline A_f} = \,\eta_{f}^{CP}\,\left|\frac{A_f } {\overline A_f}\right| \, e^{ i  \phi_f^\Gamma }
   \,,\label{lambdaMf}\eeq   
parametrizing the interference for a dispersive and absorptive mixing amplitudes, respectively.
The $\phi^{M}_f$ and $\phi^{\Gamma}_f$ are CPV weak phases, with $\phi_{12} = \phi_{f}^M - \phi_{f}^\Gamma$,  while $\eta_f^{CP} = + (-)$ for \CP even (odd) final states. 
In general, $\phi^{M}_f$ and $\phi^{\Gamma}_f$ 
are final-state specific due 
to non-universal weak and strong phases entering the CKM suppressed SM 
(and potential NP) contributions to the subleading decay amplitudes.

Non-vanishing $\phi^{M}_f$ and $\phi^\Gamma_f$
yield {\it time-dependent} \CP asymmetries. For example, in singly Cabibbo suppressed (SCS) decays to 
CP-eigenstates, $f=K^+K^-,\,\pi^+ \pi^-,...$, the effective decay widths, $\hat{\Gamma}$, for $D^0$ and
$\dzbar$ decays (the time-dependence of these decays can, to good approximation, be parametrized in exponential form $\propto  e^{- \hat \Gamma \,\tau}$, where $\tau \equiv \Gamma_D t $ ) will differ,
\beq 
\Delta Y_f \ \equiv\ 
\frac{\hat \Gamma^{}_{\dzbar \to f} - \hat \Gamma^{}_{D^0 \to f}} {2 \Gamma_D}
\ = \,- x_{12} \sin\phi_f^M  + a_f^d \, y_{12} \,.
\label{eqn:scs_asymm1}
\eeq
The first and second terms on the RHS are the dispersive CPVINT and direct CPV contributions, respectively, 
where the direct \CP asymmetry is defined as $a_f^d = 1 - \left|\bar A_f / A_f \right|$. 
They can, in principle, be disentangled
via measurements of the corresponding time-integrated \CP asymmetries \eqref{ACPhh}, which satisfy $A_{CP} (D^0 \to h^+ h^-) = \Delta Y_{h^+ h^-} \,\langle t\rangle/ \tau^{}_D   +\ a_{h^+ h^-}^{d}$.
Examples of time-dependent \CP asymmetries in decays to non-CP eigenstates include
the SCS final states $f=K^* K$ or $f=\rho\pi$, and the Cabibbo favored/doubly Cabibbo suppressed
(CF/DCS) final states $f=K^\pm \pi^\mp$.  These asymmetries generally depend on both $\phi_f^M$ and $\phi_f^\Gamma$, unlike decays to \CP eigenstates, due to the additional strong phases \cite{GKLPPS}.

The dispersive and absorptive observables are simply related \cite{GKLPPS} 
to the more familiar parametrization of indirect CPV, see, e.g., \cite{Nir:1992uv}. The latter consists 
of $ |q/p|-1 $, 
and  
\beq{\lambda_f} \equiv \frac{q}{p} \frac{ \bar A_f }{A_f} = - \eta_f^{ CP} \left|\lambda_f \right| e^{i \,\phi_{\lambda_f}},\label{lambdaf}\eeq
for \CP eigenstate final states, and pairs of observables ${\lambda_f} $, $\lambda_{\bar f} $ 
for non-CP eigenstate final states.
The same number of independent parameters is employed in each case (recall that $\phi_{12} = \phi_f^M - \phi_f^\Gamma$).

Single mrad precision for $\phi_{12}$ could become a realistic target at LHCb, see below and Sec.~ \ref{sec:expcharmmix}, to be 
compared with the current $\sim 50-100$ mrad bounds 
[95\% CL] quoted by HFLAV and UTfit Collaboration~\cite{Amhis:2016xyh,LHCbimplications}. 
Thus, we must estimate the final state dependence in $\phi_f^M$, $\phi_f^\Gamma$
due to the subleading decay amplitudes, and consider how to best parameterize this.
This is accomplished via the $U$-spin flavor symmetry decomposition of the $\DDbar$ mixing amplitudes given in Eq. \eqref{Uspindecomp}.
We define three theoretical CPV phases 
\beq \label{theorphasesdef}
{ \phi_2^\Gamma }\equiv  {\rm arg}\left(\frac{\Gamma_{12}} {\Gamma_{12}^{\Delta U=2} }\right),~~{ \phi_2^M }\equiv  {\rm arg}\left(\frac{M_{12} }{\Gamma_{12}^{\Delta U=2} }\right),~~\phi_2 \equiv  {\rm arg}\left(\frac{q} { p}{\Gamma_{12}^{\Delta U=2} } \right)\,,
\eeq
which are the theoretical analogs of the final state dependent phases $\phi_f^M$, $\phi_f^\Gamma$, and $\phi_{\lambda_f}
$, respectively.
The $U$-spin breaking hierarchy $\Gamma_1/ \Gamma_2 = {\mathcal O}(1/\epsilon)$ 
yields
the estimate
\beq \begin{split} \label{theorphases}
{ \phi_2^\Gamma }&  \approx {\rm Im}\left(  \frac{2 \lambda_b }{ \lambda_s - \lambda_d}  \frac{\Gamma_1 }{\Gamma_2} \right)
\sim \left| \frac{\lambda_b }{ \theta_c } \right| \sin \gamma  \times \frac{1} { \epsilon},
\end{split}
\eeq
and similarly for $\phi_2^M$ (the $O(\lambda_b^2)$ contributions are negligible).
Taking the nominal value $\epsilon \sim 0.2$ for $U$-spin breaking, 
we obtain the rough 
SM estimates
\beq  \phi_{12} \sim \phi_2^\Gamma \sim \phi_2^M \sim 3 \times 10^{-3} 
\,.\label{phiGphiMnaive}\eeq
Comparison with the current 95\% CL bounds on $\phi_{12}$ 
implies that there is an ${\mathcal O}(10)$ window for NP in indirect CPV.
An alternative expression for $\phi_2^\Gamma$ follows from \eqref{theorphases} via the relation $\Gamma_2  \cong y \,\Gamma^D$, 
 \beq |\phi_2^\Gamma| =  \left|\frac{\sin\gamma\,\, \lambda_b\,\lambda_s 
} { y }\right| \,\frac{|\Gamma_1 |} { \Gamma^D } \approx 0.005 \, \frac{|\Gamma_1 |} { \Gamma^D } \sim 0.005\, \epsilon  \,,\label{refined}\eeq
where in the last relation we have taken $\Gamma_1 \sim \Gamma^D \epsilon$ (recall that the HQE yields $\Gamma_{ij} \sim \Gamma^D$).
In principle, $\Gamma_1$ can be estimated via the exclusive approach as more data on SCS $D^0$ decay branching ratios and direct \CP asymmetries becomes available.

The misalignments between $\phi_f^M$, $\phi_f^\Gamma$, $\phi_{\lambda_f}$ in \eqref{lambdaMf}, \eqref{lambdaf}, and their theoretical counterparts satisfy  
\beq {{ \delta \phi_f } \equiv \phi_{f}^\Gamma - \phi_{2}^\Gamma = \phi_{f}^M - \phi_{2}^M  = \phi_2 -\phi_{\lambda_f}} \,.\eeq
We can characterize the magnitude of the misalignment $\delta \phi_f$ in the SM as follows:
(i) For CF/DCS decays it is precisely known and negligible, i.e., $\delta \phi_f = {\mathcal O}(\lambda_b^2 / \theta_c^2 )$;
(ii) In SCS decays, $\delta \phi_f$ is related to direct CPV as $\delta \phi_f = a_f^d  \,\cot \delta $
 via the $U$-spin decomposition of the decay amplitudes \cite{Brod:2012ud}, where
a strong phase $\delta = {\mathcal O}(1)$ is expected due to 
large rescattering at the charm mass scale.  Thus, for $f = \pi^+ \pi^-,\, K^+ K^-$,
the experimental bounds $a_f^d \lsim {\mathcal O}(10^{-3})$ imply that $\delta \phi_f \lsim {\mathcal O}(10^{-3})$; 
(iii) In SCS decays, $\delta \phi_f = {\mathcal O}( \lambda_b \sin\gamma /\theta_c ) \times \cot\delta$, i.e., it is ${\mathcal O}(1)$ in $SU(3)_F$ breaking.  Thus,
\eqref{theorphases} yields ${\delta \phi_f  /\phi_{2}^\Gamma } = {\mathcal O}({ \epsilon})$,
implying an order of magnitude suppression of the misalignment. 
An exception to property (iii) arises in $D^0 \to K_s K_s$, where the leading ``tree-level" decay amplitude enters at $O(\epsilon)$ \cite{Nierste:2015zra}, thus yielding ${\delta \phi_{K_s K_s} /\phi_{2}^\Gamma } = {\mathcal O}(1)$. Note that $\delta \phi_{K^{*0} K_S} $ could also be enhanced due to  a suppression of the leading amplitudes \cite{Nierste:2017cua}.

In the HL-LHC era, 
a single pair of dispersive and absorptive phases, identified with $\phi_2^M$ and  $\phi_2^\Gamma$, respectively, should suffice to parametrize all indirect CPV effects. We refer to this fortunate circumstance as {\it approximate universality}.  Moreover, approximate universality generalizes beyond the SM under the following conservative assumptions about NP 
contributions:
(i) they can be neglected in CF/DCS decays (a highly exotic NP flavor structure would otherwise be required in order to evade the $\epsilon_K$ constraint~\cite{Bergmann:1999pm}); (ii) in SCS decays they are of similar size or  
smaller than the SM QCD penguin amplitudes, as already hinted at by the experimental bounds on the direct \CP asymmetries $a^d_{K^+K^-},\,a^d_{\pi^+\pi^-}$. These assumptions can be tested by future direct CPV measurements.

Under approximate universality, 
$\phi_f^M\to \phi_2^M$ and $\phi_f^\Gamma\to\phi_2^\Gamma$
in expressions for time-dependent \CP asymmetries. 
A global fit to the CPV data for any two of the three phases, 
$\phi_2^M$, $\phi_2^\Gamma$, $\phi_{12}$, is equivalent to the traditional
two-parameter fit for 
$|q/p|$ and $ \phi$
, where $ \phi$ is identified with $\phi_2$~\cite{GKLPPS}.
The relations 
\beq
\left| \frac{q} { p}\right| -1   \approx  \frac{|x| |y| } { x^2 + y^2 } \sin{ \phi_{12}} ,~~~~~
\tan 2 (\phi_2 + { \phi^\Gamma})  \approx  - \frac{x_{12}^2 } { x_{12}^2 + y_{12}^2  }   
\sin 2{ \phi_{12}} \,,
\label{beyondsuperweak}
\eeq
together with $\phi_{12} = \phi_2^M - \phi_2^\Gamma$, allow one to translate between 
$(\phi_2, |q/p|)$ and $(\phi_2^M, \phi_2^\Gamma)$. 
To illustrate the potential reach of LHCb in $\phi_2^M$ and $\phi_2^\Gamma$ from prompt charm production at 300 fb$^{-1}$,
the projected statistical errors on $\phi$, $|q/p|$, $x$, $y$ from $D^0 \to K_s \pi^+ \pi^-$ (combined with the Belle 
error correlation matrix \cite{Peng:2014oda}), and on $\Delta Y_f = -A_\Gamma$,  
given in the last rows of Tables \ref{tab:D0KSPiPi_yields} and \ref{tab:Agamma_yields}, and assuming the central values $|q/p|=1$, $\phi=0$, $x=0.57\%$, and $y=0.7\%$, yield 
\beq \sigma (\phi_2^M ) = 2~\text{mrad},~~~ \sigma (\phi_2^\Gamma ) = 5~\text{mrad}\,.\eeq 
The projected statistical errors are  smaller for $D^0 \to K^+ \pi^-\pi^-\pi^+$ than for $D^0 \to K_s \pi^+ \pi^-$, cf. Tab.~\ref{tab:WS-D2K3pi}.
Thus, if the systematic errors are not prohibitively large, cf. Sec.~\ref{sec:expcharmmix},
LHCb could probe indirect CPV in the SM.

Finally, we remark, that if NP predominantly couples to left-handed quark currents, there are strong correlations between NP contributions to $D-\bar D$ and $K-\bar K$ mixing \cite{Blum:2009sk,Kagan:2009gb}.  In such a case, the combination of measurements in these two systems is particularly powerful.

\subsection{Direct \CP violating probes}

Direct \CP asymmetries,
\begin{equation}\label{ACPhh}
A_{\CP}(\Dz\to h^-h^+)\equiv \frac{\Gamma(D^0\rightarrow h^-h^+)-\Gamma(\Dzb\rightarrow h^-h^+)}{\Gamma(D^0\rightarrow h^-h^+)+\Gamma(\Dzb\rightarrow h^-h^+)},
\end{equation}
with $h$ a light meson, are suppressed in the SM but could be enhanced by NP.
A prominent test of direct CPV in charm is the
observable 
\begin{equation}\label{eq:Delta:ACP}
\Delta A_{\CP}=A_{\CP}(\Km\Kp)-A_{\CP}(\pim\pip),
\end{equation}
which measures the difference between direct CPV in SCS modes $D^0 \to K^+ K^-$ and $D^0 \to \pi^+ \pi^-$. This is despite that the
SM prediction is hard to obtain.  However, assuming nominal breaking of
the $SU(3)$ flavor symmetry, $\epsilon\approx f_K/f_\pi - 1 \approx
{\mathcal O}(20\,\%)$, one can infer from the observed branching ratios of the
CF decay$D^0 \to \pi^+ K^-$ and DCS decay $D^0
\to \pi^- K^+$ a consistent picture involving large
 matrix elements for $U$-spin breaking penguin~\cite{Brod:2012ud,
  Savage:1991wu, Pirtskhalava:2011va, Feldmann:2012js, Franco:2012ck,
  Hiller:2012xm} (see also \cite{Atwood:2012ac}). These can account, in the presence of large strong
phases,  for values of $\Delta A_{\CP}
\lesssim 0.2\%$~\cite{Brod:2011re}, in accordance with the current
measurements. Light cone QCD sum rule calculations, on the other hand, predict a much smaller quantity, $\Delta A_{CP}=(0.020 \pm 0.003)\%$ \cite{Khodjamirian:2017zdu}.  LHCb is expected to probe far into this region after
Upgrade~II (see Sec.~\ref{sec:directCP:exp}).

SCS $D$ decay modes are sensitives probes of \CP violation in and
beyond the SM~\cite{Grossman:2006jg}. Other decay modes can also be
sensitive to \CP asymmetries of the same order, both in the SM and in
NP extensions that modify the QCD penguin operators. Besides the modes
$D^+ \to K^+ \bar K^0$ and $D_s^+ \to \pi^+ K^0$, obtained from the
above via exchange of the spectator quark, the mode $D^0 \to K_S K_S$
is particularly interesting, because a large \CP asymmetry $\lesssim
1\,\%$ can be expected~\cite{Brod:2011re, Nierste:2015zra}. The
related $D \to K K^*$ modes have smaller asymmetries, but this can be
compensated by the higher experimental
efficiency~\cite{Nierste:2017cua}. Such quasi two-body decays can 
interfere, if they contribute to the same three- or four-body final state, which needs 
to be taken into account in the analysis (for theory discussion, see, e.g., \cite{Ryd:2009uf}, for
experimental prospects at LHCb \upgradetwo see Sec.~\ref{sec:directCP:exp}).

Another interesting class of observables are the semileptonic $D$
decays, $D \to P \ell^+ \ell^-$ and $D \to P_1 P_2 \ell^+ \ell^-$,
where $P, P_1, P_2$ are light pseudoscalars. In analogy to the
well-studied corresponding semileptonic $B$ decay modes, interesting
null tests of \CP violation can be constructed~\cite{deBoer:2015boa,
  deBoer:2017que, deBoer:2018buv}. In certain regions of phase space,
the \CP asymmetries can be largely enhanced via interferences with
resonances, which makes these decays an interesting class of
observables for LHCb~\cite{LHCb-PAPER-2018-020}.

\Blu{A set of observables related to direct \CP asymmetry are  ${\widehat{T}}$-violating triple products \cite{Datta:2003mj} (see also Sec.~\ref{sec:directCP:exp}). To be nonzero they require a source of CPV, and at least three independent momenta or polarization vectors in the final state, such as in $D\to VV$ decays or decays to four pseudoscalars.  Unlike direct \CP asymmetry, the  ${\widehat{T}}$-violating triple products  can be nonzero even in the case of vanishing strong phase differences. }

\subsection{Null tests from isospin sum rules}
\label{sec:isospin:null:test}
SM predictions for hadronic $D$ decays are notoriously difficult. In
some cases it is possible to obtain strong indications for the
presence of NP by relating various modes using the
approximate flavor symmetry $SU(3)$ (invariance under interchange of
up, down, and strange quarks) or the more precise isospin (invariance under interchange of
up and down quarks). 

Probably the simplest example is the $D^+ \to \pi^+
\pi^0$ decay~\cite{Brod:2011re}. The final state has isospin $I=2$ which
cannot be reached from the $I = 1/2$ initial state via the $\Delta I =
1/2$ QCD penguin operators, predicting very suppressed direct CPV in the SM. An important question in this context is
the size of isospin-breaking effects. Isospin-breaking due to QED and
the difference of up- and down-quark mass is \CP conserving and can be
safely neglected. The electroweak penguin contribution is relatively
suppressed by $\alpha/\alpha_s$ in the SM. Thus, enhanced direct CPV in $D^+ \to \pi^+ \pi^0$ would signal the presence of
isospin-violating NP.

Another example is the sum of rate differences~\cite{Grossman:2012eb}
\begin{equation}
\begin{split}
  |A_{\pi^+ \pi^-}|^2 - |\bar A_{\pi^+ \pi^-}|^2 & + |A_{\pi^0
    \pi^0}|^2 - |\bar A_{\pi^0 \pi^0}|^2 \\ & - \frac{3}{2} \big(
  |A_{\pi^+ \pi^0}|^2 - |\bar A_{\pi^- \pi^0}|^2 \big) = 3 \big(
  |A_{1/2}|^2 - |\bar A_{1/2}|^2 \big),
\end{split}
\end{equation}
which depends only on $\Delta I = 1/2$ amplitudes. There are two
possibilities. If the sum is non-zero, there are $\Delta I = 1/2$
contributions to \CP violation; they can be due to SM or NP. If the sum
is zero, but the individual asymmetries are non-zero, the CP
asymmetries are likely dominated by $\Delta I = 3/2$ NP
contributions. More sum rules, involving also vector meson final
states, can be devised for the decay modes $D \to \rho \pi$, $D \to
K^{(*)} \bar K^{(*)} \pi (\rho)$, $D_s^+ \to K^* \pi
(\rho)$~\cite{Grossman:2012eb}. For an exhaustive list of sum rules
based on the flavor $SU(3)$ or its subgroups, 
see Refs.~\cite{Grossman:2012ry,
  Grossman:2013lya, Muller:2015rna}. \Blu{For decays with only charged particles in the final state a significant improvement in precision is projected at LHCb \upgradetwo. The final states with neutral pions are more challenging (for experimental prospects at LHCb \upgradetwo see Sec.~\ref{sec:directCP:exp}). However, precise information on these \CP asymmetries is expected from Belle-II. }

\subsection{Radiative and leptonic charm decays}
\Blu{In the down-type quark sector the GIM suppression is less effective because of the large top quark mass, so that the diagrams with the top-quark running in the loop dominate. In the \emph{charm} sector the GIM suppression is more efficient at least at the perturbative level, since none of the down-type quarks are heavy. The exact suppression for a particular FCNC decay depends on how well the GIM suppression is carried over to the non-perturbative contributions (``long distance'' physics). The result is that the SM branching ratios for radiative and leptonic charm decays, while suppressed, are in general not too small to be out of reach. The branching ratios for semi-leptonic FCNC decays, such as  $D\to P \ell\ell$, $D\to V\ell \ell$, and $D\to PP \ell \ell$ are at the level of $10^{-7}-10^{-6}$, while for the radiative decays, $D\to V \gamma$, they are  at the level of  $10^{-4}-10^{-5}$. The branching ratios for these decays are resonance dominated, and this holds true even in the presence of new physics. However, one can still search for new physics effects by using appropriate 
 observables such as \CP asymmetries, polarization asymmetries, and angular observables, as well as enhancements of differential rates away from resonance regions.}

\subsubsection{Radiative Decays}

The effective weak Lagrangian for the exclusive $c\to u \gamma$  FCNC transitions $D\to V\gamma$ is, see, e.g., \cite{deBoer:2017que},
\begin{align}
\label{eq:Leff:e1}
 \mathcal L_\text{eff}^\text{weak}=\frac{4G_F}{\sqrt 2}\biggr(\sum_{q\in\{d,s\}}V_{cq}^*V_{uq}\sum_{i=1}^2C_iO_i^{(q)}+\sum_{i=3}^6C_iO_i+\sum_{i=7}^8\left(C_i O_i+C_i'O_i'\right)\biggr)\,,
\end{align}
with the operators ${\cal O}_{2(1)}=(\bar u_L\gamma_\mu(T^a)q_L)(\overline q_L\gamma^\mu(T^a)c_L)$, ${ \cal O}_7^{(\prime)}={e\, }{m_c}/{16\pi^2}(\bar u_{L(R)}\sigma^{\mu\nu}c_{R(L)})F_{\mu\nu}$ and $ { \cal O}_8^{(\prime)}={g_s\,m_c}/{16\pi^2}(\bar u_{L(R)}\sigma^{\mu\nu}T^ac_{R(L)})G^a_{\mu\nu}$. 
The SM effective Wilson coefficients $C_i^\text{(eff)}$, which absorb the universal long-distance
effects from quark loops in perturbation theory, are known at  two-loop level in QCD \cite{deBoer:2017que,deBoer:2016dcg,Greub:1996wn,Fajfer:2002gp}.
The authors of Ref. \cite{deBoer:2017que}  improved the SM prediction for the branching ratios of $D\to V\gamma$, by  including power corrections and updating the hybrid model predictions. The hybrid model combines   the heavy quark effective theory and chiral perturbation theory using experimentally measured parameters \cite{Fajfer:1997bh,Fajfer:1998dv}. 
They also  included corrections to the perturbative Wilson coefficients by employing  a QCD based approach, worked out for $B$ physics as reviewed in  \cite{deBoer:2017que}. Updated values for the  SM Wilson coefficients 
at leading order in $\alpha_s$ are given in  \cite{deBoer:2017que,deBoer:2015boa,deBoer:2016dcg}. 
The GIM mechanism  suppresses strongly  $C_8^{\text{eff}}$ in the SM. 
The experimental branching ratio  for the Cabibbo allowed decay is $\BR(D^0\to K^*(892) \gamma) = (4.1\pm 0.7)\times 10^{-4}$ and for the Cabibbo suppressed  decays 
$\BR(D^0\to\phi\gamma) = (2.74\pm 0.19)\times 10^{-5}$,    
$\BR( D^0\to \rho^0 \gamma) = (1.76\pm 0.31)\times 10^{-5}$ \cite{Tanabashi:2018oca}.
The hybrid model 
predicts in the SM 
$\BR(D^0 \to \rho^0 \gamma)_{\rm SM}= (0.041-1.17)\times 10^{-5}$, $\BR(D^0\to\phi\gamma)_{\rm SM} = (0.24-2.8))\times 10^{-5}$,   $\BR(D^0\to\bar K^{*0} \gamma)_{\rm SM} = (0.26-4.6)\times 10^{-4}$    \cite{deBoer:2017rzd,deBoer:2017que}. The NP 
scenarios discussed in Ref.   \cite{deBoer:2017rzd} can contribute at the loop level to $C_{7,8}^\text{eff}$, and cannot significantly modify the branching ratios.
 However, the NP induced CPV asymmetry was found to still be modest, e.g.,  $\sim {\mathcal O}(10\%)$ for $D^0 \to \rho^0 \gamma$. In the case of baryonic mode $\Lambda_c \to p \gamma$ the rate is estimated to be $\sim {\cal O} (10^{-5})$. The forward-backward asymmetry of photon momentum relative to $\Lambda_c$ boost probes the handedness of $c \to u\gamma$ transitions. It can be $0.2$ in the SM, and somewhat smaller  within scenarios of NP discussed in \cite{deBoer:2017que,deBoer:2017rzd}. The probes of photon polarization include time-dependent analysis in $D^0, \bar \D^0 \to V \gamma$, where $V=\rho^-, K^0, \phi$, or an up-down asymmetry in $D_{(s)} \to K_1 (\to K \pi \pi) \gamma$~\cite{deBoer:2018zhz}.

\subsubsection{Rare leptonic  decays of charm} 
To describe NP effects the effective Lagrangian for the $c\to u \bar \ell \ell $ has to be extended by the following operators
\begin{equation}
  \label{eq:operators}
  \begin{split}
  \qquad \quad {\cal O}_{9(10)} &= \frac{e^2}{(4\pi)^2}\, (\bar u \gamma^\mu P_L c)
  (\bar\ell \gamma_\mu (\gamma_5)\ell) \,, \qquad \quad  
   {\cal O}_{S(P)}  = \frac{e^2}{(4\pi)^2}\, (\bar u P_R c)
  (\bar\ell (\gamma_5) \ell)\,, 
\\
    \qquad \quad  {\cal O}_{T} &= \frac{e^2}{(4\pi)^2}\, (\bar u \sigma_{\mu\nu} c)
  (\bar\ell \sigma^{\mu\nu}  \ell)\,,
 \qquad \qquad\quad   {\cal O}_{T5} = \frac{e^2}{(4\pi)^2}\, (\bar u \sigma_{\mu\nu} c)
  (\bar\ell \sigma^{\mu\nu} \gamma_5  \ell).
\end{split}
\end{equation}
Among these only $C_9$ is nonzero in the SM. For each ${\cal O}_i$ one can introduce ${\cal O}_i^\prime$ with the corresponding Wilson coefficient $C_i^\prime$ by replacing the chirality operator $P_L= 1/2(1-\gamma^5)$ by $P_R= 1/2(1+\gamma^5)$ \cite{Fajfer:2015mia}. 
In Ref. \cite{deBoer:2015boa} the authors obtained (N)NLO QCD SM Wilson coefficients at $\mu_c=m_c$, 
$C_{7} \simeq( -0.0011-0.0041 i)$ and $C_9 \simeq  -0.021 X_{ds}$, where $X_{ds}=V_{cd}^*V_{ud}L(m_d^2,q^2)+V_{cs}^*V_{us}L(m_s^2,q^2)$,  with $L(m^2, q^2)$ defined in eq. (B1) of \cite{deBoer:2015boa}. 
 In the range of $m_c/\sqrt2\le\mu_c\le\sqrt2m_c$ the  effective Wilson coefficients  were found to be $(-0.0014-0.0054i)\le C_7\le(-0.00087-0.0033i)$ and 
 $-0.060X_{ds}(\mu_c=\sqrt2m_c)\le C_9\le0.030 X_{ds}(\mu_c=m_c/\sqrt2)$. In the small 
 $q^2 \gtrsim 1 \, \mbox{GeV}^2$ region  $|C_{9}| \lesssim 5\cdot10^{-4}$.  

One can consider contributions of this effective Lagrangian in the exclusive decay channels. For the $D$ meson di-leptonic decays the best upper bound to date is obtained by the LHCb collaboration at  the $90\%$ CL 
~\cite{LHCb-PAPER-2013-013}
$\BR(D^0 \to \mu^+ \mu^-) < 6.2 \times 10^{-9}$.
In the decay
$D^+ \to \pi^+ \mu^+ \mu^-$ the LHCb experiment 
determined bounds on the branching ratio in the two kinematic regions of the di-lepton mass, 
 chosen to be either below or
above the dominant resonant contributions. The measured
total branching ratio, obtained by extrapolating spectra over the
non-resonant region, is~\cite{LHCb-PAPER-2012-051}
$\BR(D^+ \to \pi^+ \mu^+ \mu^-) < 8.3 \times 10^{-8}$,
while the separate branching fractions in the low- and high-$q^2$ bins are bounded to be below 
$\BR(\pi^+ \mu^+ \mu^-)_{I}<  2.5\times 10^{-8}$ for region I, ${q^2 \in [0.0625,0.276]\,{\rm GeV}^2} $, and
    $\BR(\pi^+ \mu^+ \mu^-)_{II}  <  2.9 \times 10^{-8}$ for region II, ${q^2 \in [1.56,4.00]{\rm GeV}^2}$~\cite{LHCb-PAPER-2012-051}. 
These can be used to put bounds on Wilson coefficients. Allowing for NP contributions to only one 
 Wilson coefficient at a time gives the upper bounds listed in Table~\ref{tab:BR}, where $\tilde C_i = V_{ub} V_{cb}^* C_i$\cite{Fajfer:2015mia}.
The bound on $\BR(D^0 \to \mu^+ \mu^-)$ 
gives the most stringent bounds on 
$C_{S,P,10}$ Wilson coefficients.

For baryonic $c\to u \ell^+ \ell^-$ transitions the relevant  form factors are known from lattice QCD calculations for the 
 $\Lambda_c \to p \ell^+\ell^-$ decay \cite{Meinel:2017ggx}. The dominant contributions to the branching ratio come from resonant regions of $\Lambda_c  \to p \rho, (\omega, \phi)$ with $\rho$, $\omega$ and $\phi$ decaying to $\mu^+ \mu^-$. 
 This permits to  investigate the impact of NP on the differential branching ratio, the  fraction of longitudinally polarized di-muons and the forward-backward asymmetry. 
 The upper 90\% CL bound $\BR(\Lambda_c \to p  \mu^+ \mu^-)_{exp}  <7.7 \times 10^{-8}$  \cite{LHCb-PAPER-2017-039}, obtained by excluding the $\pm40$ MeV intervals around resonances still allows NP contribution in $C_9$ and $ C_{10}$. 
The $\Lambda_c \to p \mu^+ \mu^-$ forward-backward asymmetry appears as a result of  a nonzero $C_{10}$ Wilson coefficient, generated by the NP.  Therefore  this observable  provides a clean null test of the SM as suggested in Ref. \cite{Meinel:2017ggx}.

\begin{table}[t]
\caption{ 
Maximal experimentally allowed magnitudes of the Wilson coefficients,
  $|\tilde C_i| = |V_{ub}  V_{cb}^* C_i|$, obtained from 
  non-resonant part of $D^+ \to \pi^+ \mu^+ \mu^-$ decay, low $q^2$ region I: $q^2 \in [0.0625,0.276]\,{\rm GeV}^2$; high $q^2$ region: $q^2 \in
  [1.56,4.00] {\rm GeV}^2$); and from $\BR (D^0 \to \mu^+ \mu^-) < 7.6 \times 10^{-9}$ at 95\% C.L.~\cite{LHCb-PAPER-2013-013}.
  The last row applies to the case $\tilde C_9= \pm\tilde C_{10}$. All the bounds assume real
  $C_{i}$, and also apply to coefficients with flipped chirality, $\tilde C_j^\prime$.
  }
\label{tab:BR}
\begin{center}
\begin{tabular}{cr@{.}lr@{.}lr@{.}l} 
\hline\hline\\[-4mm]
\multicolumn{1}{c}{}&\multicolumn{6}{c}{$|\tilde C_i|_\mathrm{max}$}\\
 \multicolumn{1}{c}{}  & \multicolumn{2}{c}{$\BR(\pi \mu \mu)_{I}$} &  \multicolumn{2}{c}{$\BR(\pi \mu \mu)_{II}$} &   \multicolumn{2}{c}{$\BR(D^0\to \mu \mu)$} \\ 
\hline\\[-4mm] 
$\tilde C_7$ & \hspace{0.8cm} 2&4 \hspace{0.8cm} & \hspace{0.8cm} 1&6 \hspace{0.8cm} & \multicolumn{2}{c}{-} \\
$\tilde C_9$ & 2&1 & 1&3 & \multicolumn{2}{c}{-}  \\
$\tilde C_{10}$ & 1&4  & 0&92 & \hspace{.7cm}0&63\\
$\tilde C_S$ & 4&5 & 0&38& 0&049\\
$\tilde C_P$ & 3&6 & 0&37 & 0&049\\
$\tilde C_T$ & 4&1 & 0&76& \multicolumn{2}{c}{-}  \\
$\tilde C_{T5}$ & 4&4& 0&74 & \multicolumn{2}{c}{-}  \\
$\tilde C_9= \pm \tilde C_{10}$ & 1&3& 0&81&  0&63\\
 \hline\hline
\end{tabular}
\end{center}
\end{table}

The above bounds allow for appreciable  NP contributions. An example are leptoquark mediators \cite{Dorsner:2016wpm}, which are well motivated as the NP explanations of  
the $B$-meson anomalies, see, e.g., \cite{Buttazzo:2017ixm}.
Refs. \cite{deBoer:2015boa} and \cite{Fajfer:2015mia} showed that leptoquark exchanges do not affect much 
the branching ratios, but can lead to \CP asymmetries in $D\to \pi l^+ l^-$ and $D_s \to K l^+ l^-$  of a few percent \cite{Fajfer:2012nr,deBoer:2015boa}. Namely, such \CP asymmetries are 
defined close to the $\phi$  resonance that couples to the lepton pair and they  can be generated by imaginary parts of the $C_{7}$ Wilson coefficients in the effective Lagrangian  for $c \to  u l^+l^−$ processes. 
In the case of any NP scenario one cannot consider any charm rare decays without considering bounds from the $D^0- \bar D^0 $ oscillation as pointed out in \cite{Golowich:2009ii}. In some particular NP scenarios, bounds on NP are stronger if from  the $D^0- \bar D^0 $ oscillation. 

 In addition, if the NP is realized by the left-handed doublets of the weak isospin, then NP present in $B$ physics also inevitably appears in charm physics, accompanied with the appropriate CKM matrix elements.  
Tests of lepton flavour universality are of particular interest, such as the observable  
$R_{hh} =\BR(D\to hh \mu^+ \mu^-)/\BR(D\to hh e^+ e^-)$.  The ratio $R_{hh}$ is theoretically clean, but only when both numerator and denominator have the same kinematic cuts, and when these are well above the muon threshold \cite{deBoer:2018buv,Fajfer:2015mia}. 
LHCb has already measured $D\to \pi^+ \pi^-  \mu^+ \mu^-$ and $D\to K^+ K^-  \mu^+ \mu^-$ \cite{LHCb-PAPER-2017-019}, and BESIII set upper limits on the electron modes \cite{Ablikim:2018gro}, but with different cuts on $q^2$.

\subsection{Inputs for $B$ physics}
If discovered, the presence of direct \CP violation in $D$ decays can
affect the extraction of the CKM angle $\gamma$ from the
``tree-level'' decays $B \to DK$~\cite{Gronau:1990ra, Gronau:1991dp,
  Atwood:1996ci, Giri:2003ty}. If one includes the $B \to D \pi$
modes, the effect can be of order one~\cite{Martone:2012nj}. One of
the advantages of obtaining $\gamma$ from tree decays is that all
hadronic parameters can be fit from data. This remains essentially
true even if direct CPV is present in the decays of the final-state
$D$ mesons~\cite{Martone:2012nj, Wang:2012ie, Bhattacharya:2013vc}, as
long as one includes all direct \CP asymmetries in the fit -- there are
still more observables than parameters in the fit. However, one can
show that there remains an ambiguity: A shift in the angle $\gamma$
can be compensated by a corresponding, unobservable shift in the
contributing strong phases. This shift symmetry can be broken by
assuming the absence of CPV in one of the $D$ decay modes (for
instance, in the SM this is the case for CF and DCS decay
modes). Alternatively, one could consider ratios of $B \to f_D K$ and
$B \to f_D \pi$ modes where the strong phase
cancels, or use information on the relative
strong phases, e.g., by measuring entangled decays at $D$
factories~\cite{Martone:2012nj}.

\input{section3/experimental}

%% file: section3/experimental.tex
\subsection{Experimental prospects}%
We now summarize the status of experimental measurements and their prospects for the future. We generally follow the notation of Sec.~\ref{sec:CPVDmixth}. In the case of time-dependent CPV, we also define the parameters $x'$
and $y'$ which depend linearly on the mixing parameters,
$x' \equiv x\cos\delta+y\sin\delta$ and
$y' \equiv y\cos\delta-x\sin\delta$. Here $\delta$ is the strong phase difference between the favoured and doubly Cabibbo suppressed final states. 
\label{sec:expcharmmix}
\subsubsection{Mixing and time-dependent CPV in two-body decays} 
The mixing and CPV parameters in \Dz--\Dzb oscillations can be accessed by comparing the decay-time-dependent ratio of \decay{\Dz}{K^+\pi^-} to \decay{\Dz}{K^-\pi^+} rates with the corresponding ratio for the charge-conjugate processes.

The latest measurement from LHCb~\cite{LHCb-PAPER-2017-046} uses Run~1
and early Run~2 (2015--2016) data, corresponding to a total sample of
about $\mathcal{L}=5\invfb$ of integrated luminosity.  Assuming \CP
conservation, the mixing parameters are measured to be
$x_{K\pi}'^2=(3.9 \pm 2.7) \times10^{-5}$,
$y_{K\pi}'= (5.28 \pm 0.52) \times 10^{-3}$, and
$R_D^{K\pi} = (3.454 \pm 0.031)\times10^{-3}$.  Studying \Dz and \Dzb decays
separately shows no evidence for \CP violation and provides the
current most stringent bounds on the parameters $A_D^{K\pi}$ and $|q/p|$ from
a single measurement, $A_D^{K\pi} =(-0.1\pm9.1)\times10^{-3}$ and
$1.00< |q/p| <1.35$ at the $68.3\%$ confidence level.  

\begin{table}[tb]
\centering
\caption{Extrapolated signal yields, and statistical precision on the mixing and \CP-violation parameters, from the analysis of promptly produced  DCS \decay{\Dstarp}{\Dz(\to K^+\pi^-)\pi^+} decays.
Signal yields of promptly produced CF \decay{\Dstarp}{\Dz(\to K^-\pi^+)\pi^+} decays are typically $250$ times larger.\label{tab:WS-D2Kpi-yields}}
\begin{tabular}{lcccccc} 
\hline\hline\\[-4mm]
Sample ($\mathcal{L}$)  & Yield ($\times10^6$) & $\sigma(x_{K\pi}'^2)$ & $\sigma(y_{K\pi}')$ & $\sigma(A_D)$ & $\sigma(|q/p|)$ & $\sigma(\phi)$ \\
\hline
Run 1--2 (9\invfb)   &  1.8   & $1.5\times10^{-5}$ & $2.9\times10^{-4}$ & 0.51\% & 0.12 & 10\degrees \\
Run 1--3 (23\invfb)  &   10   & $6.4\times10^{-6}$ & $1.2\times10^{-4}$ & 0.22\% & 0.05 &  4\degrees \\
Run 1--4 (50\invfb)  &   25   & $3.9\times10^{-6}$ & $7.6\times10^{-5}$ & 0.14\% & 0.03 &  3\degrees \\
Run 1--5 (300\invfb) &  170   & $1.5\times10^{-6}$ & $2.9\times10^{-5}$ & 0.05\% & 0.01 &  1\degrees \\
\hline\hline
\end{tabular}
\end{table}

In Table~\ref{tab:WS-D2Kpi-yields} the signal yields and the statistical precision from Ref.~\cite{LHCb-PAPER-2017-046} are extrapolated to the end of Run 2
and to the end of \upgradetwo, assuming that the central values of the measurements stay the same. This assumption is particularly important for the \CP-violation parameters, as their precision may depend on the measured values.

Systematic uncertainties are estimated using control samples of data and none of them are foreseen to have irreducible contributions that exceed the ultimate statistical precision, if the detector performance (particularly in terms of vertexing/tracking and particle identification capabilities) is kept at least in line with what is currently achieved at LHCb.

\subsubsection{Mixing and time-dependent CPV in $\Dz\to\KS\pip\pim$} 
The self-conjugate decay \decay{D^0}{\KS \pi^+ \pi^-} includes both CF and  DCS, 
as well as $\CP$-eigenstate processes reconstructed in the same final state.
This allows for the relative strong phase between different contributions to be determined from data, and, in turn, enables both the mixing parameters $x$ and $y$, as well as the \CP-violation parameters $|q/p|$ and $\phi$ to be directly measured without need for external input.
As a result, this channel provides the dominant constraint on the parameter $x$ in the global fits.

The mixing and CPV parameters modulate the time-dependence of the complex amplitudes, and these amplitudes themselves vary over the two-dimensional final state phase-space. 
As such, the measurement relies both on the precise understanding of the detector acceptance as a function of phase-space and decay time, and on the accurate description of the evolution of the underlying decay amplitudes over the Dalitz plane. Both model-dependent and model-independent approaches using quantum-correlated $D\bar{D}$ pairs from $\psi(3770)$ decays can be applied.

Previous measurements from the CLEO~\cite{Asner:2005sz}, BaBar~\cite{delAmoSanchez:2010xz}, and Belle~\cite{Peng:2014oda} collaborations have used the model-dependent approach, with the Belle measurement having the best precision to date, $x = (0.56 ^{+0.20}_{-0.23})\%$, $y = (0.30 ^{+0.16}_{-0.17})\%$ (assuming \CP symmetry), and $|q/p| = 0.90 ^{+0.18}_{-0.16}$, $\phi = (-6 \pm 12)\degrees$.
The one published LHCb result was based on 1~fb$^{-1}$ of Run 1 data~\cite{LHCb-PAPER-2015-042}, and used a model-independent approach with strong phases taken from the CLEO measurement~\cite{Libby:2010nu} to determine $x = (-0.86 \pm 0.56)\%$, $y = (0.03 \pm 0.48)\%$.
This analysis used around $2\times10^5$ \decay{D^{*+}}{D^0 \pi^+}, \decay{D^0}{\KS \pi^+ \pi^-} decays from 2011, which suffered from low \KS trigger efficiencies that were significantly increased for 2012 and beyond, and will benefit further from software trigger innovations at the LHCb in the upgrade era.

At LHCb these decays can be reconstructed either through semileptonic decays, for instance \decay{B^-}{D^0 \mu^- \bar{\nu}_{\mu}}, where the muon charge is used to tag the initial $D^0$ flavour, or through prompt charm production, where the charge of the slow pion in the decay \decay{D^{*+}}{D^0 \pi^+} tags the initial flavour.
The two channels have complementary properties and both will be important components of future mixing and \CP violation analyses at LHCb.

The prompt charm yields are significantly larger than for the semileptonic channel, due to the increased production cross-section. However, for the semileptonic channel the triggering on signal candidates is much more efficient, and introduces fewer non-uniformities in the acceptance.
The estimated future yields are presented in Table~\ref{tab:D0KSPiPi_yields}.
Also shown are projected statistical precisions on the four mixing and CPV parameters, which have been extrapolated from complete analyses of the Run 1 data for both the semileptonic and prompt cases. 

\begin{table}[tb]
 \caption{Extrapolated signal yields at LHCb, together with statistical precision on the mixing and \CP violation parameters, for the analysis of the decay \decay{D^0}{\KS \pi^+ \pi^-}.
Candidates tagged by semileptonic $B$ decay (SL) and those from prompt charm meson production are shown separately.}
 \label{tab:D0KSPiPi_yields}
\begin{center}
  \renewcommand{\arraystretch}{1.2}
 \begin{tabular}{lcccccc} 
 \hline\hline
 Sample (lumi $\mathcal{L}$)              & Tag    & Yield & $\sigma(x)$ & $\sigma(y)$ & $\sigma(|q/p|)$ & $\sigma(\phi)$ \\ \hline
\multirow{2}{*}{Run 1--2 (9~fb$^{-1})$}   & SL     & 10M   & 0.07\%      & 0.05\%      & 0.07            & 4.6\degrees    \\
                                          & Prompt & 36M   & 0.05\%      & 0.05\%      & 0.04            & 1.8\degrees    \\ %
\multirow{2}{*}{Run 1--3 (23~fb$^{-1})$}  & SL     & 33M   & 0.036\%     & 0.030\%     & 0.036           & 2.5\degrees    \\
                                          & Prompt & 200M  & 0.020\%     & 0.020\%     & 0.017           & 0.77\degrees   \\ %
\multirow{2}{*}{Run 1--4 (50~fb$^{-1})$} & SL     & 78M   & 0.024\%     & 0.019\%     & 0.024           & 1.7\degrees    \\
                                         & Prompt & 520M  & 0.012\%     & 0.013\%     & 0.011           & 0.48\degrees   \\
\multirow{2}{*}{Run 1--5 (300~fb$^{-1})$} & SL     & 490M  & 0.009\%     & 0.008\%     & 0.009           & 0.69\degrees   \\
                                          & Prompt & 3500M & 0.005\%     & 0.005\%     & 0.004           & 0.18\degrees   \\
 \hline\hline
\end{tabular}
\renewcommand{\arraystretch}{1.0}
\end{center}
\end{table}

For this channel the dominant systematic uncertainties on mixing parameters  come from two main sources.
First is the precision with which the non-uniformities in detector acceptance  can be determined versus as a function of phase space and decay time. Second is the knowledge of the strong-phase variation across the Dalitz plane.
For the LHCb Run 2 analysis, both contributions are significantly smaller than the statistical precision. In the longer term new approaches will be necessary to further reduce these systematic uncertainties.
Trigger and event selection techniques should be adapted to emphasise uniform acceptance, a task made easier by the removal of the calorimeter-based hardware trigger. New techniques, such as the bin-flip method~\cite{DiCanto:2018tsd}, 
can further reduce dependence on the non-uniform acceptance, although at the cost of
degraded statistical precision on the mixing and \CP-violation parameters.
In the model-dependent approach many of the model systematic uncertainties may reduce or vanish with increased integrated luminosity, as currently fixed parameters are incorporated into the data fit, and the data become increasingly capable of rejecting unsuitable models provided that there is suitable evolution in the model descriptions.
For the model-independent approach, the uncertainty from external inputs (currently from CLEO-c, later with 50\% reduction from BESIII) will also reduce with luminosity as the LHCb data starts to provide constraining power.
There are no systematic uncertainties which are known to have irreducible contributions that exceed the ultimate statistical precision.

For the \CP violation parameters additional sources of systematic uncertainty come from the knowledge of detector-induced asymmetries.
In particular, there is a known asymmetry between $K^0$ and $\bar{K}^0$ in their interactions with material.
The limitation here will be the precision with which the material traversed by each $\KS$ meson can be determined.
The LHCb \upgradetwo detector will be constructed to minimise material, and to allow precise evaluation of the remaining contributions. In summary, this channel has comparable power on \CP violating parameters, but with a simpler two-dimensional phase space and complementary detector systematic uncertainties, as the four-body decays that we discuss next. 

\subsubsection{Mixing and time-dependent CPV in four-body decays} 
Like \decay{\Dz}{\Km\pip} and \decay{\Dz}{\Kp\pim}, the decays
\decay{\Dz}{\Km\pip\pim\pip} and \decay{\Dz}{\Kp\pim\pim\pip} are a
pair of CF and DCS decays with high sensitivity to charm mixing.
However, the rich amplitude structure across the five dimensional phase
space of the latter decays offers unique opportunities (and
challenges) in these four-body modes.

In the phase-space integrated analysis using $3\mathrm{fb}^{-1}$ of data,
LHCb made the first observation of mixing in this decay mode, and
measured quantities $R_D^{K3\pi} = (3.21 \pm 0.014)\cdot 10^{-3}$,
as well as
$R^{K3\pi}_{\mathrm{coher}}y'_{K3\pi} = (0.3\pm 1.8)\cdot 10^{-3}$, and
$\frac{1}{4}(x^2 + y^2) = (4.8\pm 1.8) \cdot
10^{-5}$~\cite{LHCb-PAPER-2015-057}. The coherence factor,
$R^{K3\pi}_{\mathrm{coher}}$, 
measures the effect of integrating
over the entire four-body phase
space~\cite{Atwood:2003mj,Evans:2016tlp}. %

The unique power of multibody decays lies to a large extent in the
fact that the strong phase difference between the interfering \Dz and
\Dzb amplitudes varies across the phase space. This can be fully
exploited only by moving away from the phase-space-integrated approach to the analyses of phase space distributions, either in
bins or unbinned.
Such
a ``phase space resolved'' approach allows a direct measurement of
$x_{K\pi\pi\pi}'$ and $y_{K\pi\pi\pi}'$ (rather than only the
$x'^2$ and $y'$ as in the 2-body case), and,
crucially, provides high sensitivity to the \CP\ violating variables $\phi$ and
$|q/p|$.

On the other hand, the same phase variations that make multibody decays so
powerful, are also a major challenge, as they need to be known precisely in
order to cleanly extract the mixing and \CP violation parameters of
interest. In principle, the relevant phases can be inferred from an
amplitude model such as that obtained from $3\mathrm{fb}^{-1}$ of LHCb
data~\cite{LHCb-PAPER-2017-040}. Such models may introduce  theoretical uncertainties that are unacceptably large for
the precision era of \lhcb \upgradetwo,  unless there are significant innovations
in the theoretical description of four-body amplitudes. The alternative is to use 
model-independent approaches.
These use 
quantum-correlated events at the charm threshold to infer
the required phase information in a model-unbiased way. BESIII is
working closely with \lhcb~\cite{Malde:2223391} to provide the necessary model-independent
inputs for \decay{\Dz}{\Kp\pim\pim\pip} across different regions of
phase space for measurements of the $\gamma$ angle as well as charm mixing and
\CP violation measurements.

Sensitivity studies with model-dependent approaches give a
useful indication of the precision that can be achieved. A recent such
study, Ref.~\cite{Muller:2297069}, uses LHCb's latest
\decay{\Dz}{\Kp\pim\pim\pip} amplitude
model~\cite{LHCb-PAPER-2017-040}.  Table~\ref{tab:WS-D2K3pi} gives the
yields and sensitivities scaled from the study in~\cite{Muller:2297069},  illustrating the
impressive sensitivity of this decay mode.
The study is based on
promptly produced \Dstarp mesons, decaying in the flavour-conserving
 \decay{\Dstarp}{\Dz\pip} channel.  Several systematic uncertainties
require improvements in the analysis method in order to scale with
increasing sample sizes. However, given the huge potential of this
channel, sufficient effort is expected to be dedicated to this challenge, such
that adequate methods can be developed, and that the necessary input from
threshold measurements is both generated at BESIII and exploited
optimally at \lhcb. Indeed, once these are in place, this channel has
the potential for probing \CP violation at the $\order(10^{-5})$ level, given the
current world average value of $x$. 
\begin{table}[tb]
\caption{Extrapolated signal yields for LHCb, and sensitivity to the mixing and
  \CP-violation parameters, from the analysis of
  \decay{\Dz}{\Kp\pim\pim\pip} decays (statistical uncertainties only).\label{tab:WS-D2K3pi}}
\centering
\begin{tabular}{lccccc} 
\hline\hline\\[-4mm]
Sample ($\mathcal{L}$)  & Yield ($\times10^6$) & $\sigma(x_{K\pi\pi\pi}')$ & $\sigma(y_{K\pi\pi\pi}')$ & $\sigma(|q/p|)$ & $\sigma(\phi)$ \\
\hline
Run 1-2 (9\invfb)       &  0.22                & $2.3\times10^{-4}$       & $2.3\times10^{-4}$       & 0.020           & 1.2\degrees \\
Run 1-3 (23\invfb)      &  1.29                & $0.9\times10^{-4}$
                                                                           & $0.9\times10^{-4}$       & 0.008           & 0.5\degrees \\
Run 1-4 (50\invfb)      &  3.36                & $0.6\times10^{-4}$       & $0.6\times10^{-4}$       & 0.005           & 0.3\degrees \\
Run 1-5 (300\invfb)     & 22.5\phantom{00}     & $0.2\times10^{-4}$       & $0.2\times10^{-4}$       & 0.002           & 0.1\degrees \\
\hline\hline
\end{tabular}
\end{table}

\subsubsection{Measurement of $A_\Gamma$}
\label{sec:AGamma}
The parameter \agamma is related to indirect \CP violation ($\simeq -A_\CP^\mathrm{indir}$) and is defined as
\begin{align}
\agamma &\equiv \frac{\hat{\Gamma}(\decay{\Dz}{ \hadron^+ \hadron^-)} - %
 			\hat{\Gamma}(\decay{\Dzb}{ \hadron^+ \hadron^-)}}%
 			{\hat{\Gamma}(\decay{\Dz}{ \hadron^+ \hadron^-)}+%
 			\hat{\Gamma}(\decay{\Dzb}{ \hadron^+ \hadron^-)}} \label{eq:agamma_def} %
 \end{align}
 Neglecting contributions from subleading amplitudes%
, \agamma is independent of the final state $f$. 

The large yields available in the SCS modes, $f=\pi^+\pi^-$ or $f=K^+K^-$, together with tagging from the \Dstarpm decay, allow for a precise measurement of $A_\Gamma$, provided the systematic uncertainties can be controlled with high degree of precision.
Tagging based on semileptonic decays of a parent bottom hadron is also possible and has been used in a published LHCb measurement~\cite{LHCb-PAPER-2014-069}, but contributes significantly lower yields. 

Most potential systematic effects are essentially constant in $t$ and
therefore cause little uncertainties in the observed decay time
evolution of the asymmetry.  However, second-order effects and
detector--induced correlation between momentum and proper decay time
are sufficient to produce spurious asymmetries, that must be
appropriately corrected. In addition, contamination from secondary
decays is a first-order effect in time that must be suppressed, and
its residual bias accounted for.  Both corrections are dependent on
the availability of a large number of CF $\Dz\to\Km\pip$  decays
as calibration, and can be expected to scale with statistics;
collection of this sample with the same trigger as for the signal
modes is therefore a crucial tool for performing this measurement with high precision in the future.

The Run~1 LHCb measurement of this quantity gave consistent results in the two $h^+h^-$ modes, averaging  $\agamma= (- 0.13 \pm 0.28 \pm 0.10)\times 10^{-3}$ \cite{LHCb-PAPER-2016-063}, which is still statistically dominated. For the reasons mentioned above, this precision is at the threshold of becoming physically interesting, making it a worthy target to pursue with more data. It seems highly unlikely that any experiment built in the foreseeable future will be able to do this, except for an upgrade of LHCb to higher luminosity. 

Table~\ref{tab:Agamma_yields} shows expected yields and precisions attainable in  \lhcb \upgradetwo, under the same assumptions on efficiencies adopted in the previous sections; this must include provisions for acquiring and storing $5\times 10^{10}$ CF decays. The ultimate combined precision is $1\times 10^{-5}$.

\begin{table}[t]
\caption{Extrapolated signal yields at LHCb, and statistical precision on
  indirect \CP violation from \agamma. 
}
 \label{tab:Agamma_yields}
\begin{center}
  \renewcommand{\arraystretch}{1.2}
 \begin{tabular}{lccccc} 
 \hline\hline
 Sample ($\mathcal{L}$)   & Tag   & Yield $K^+K^-$ & $\sigma(\agamma)_{K^+K^-}$  & Yield  $\pi^+\pi^-$  & $\sigma(\agamma)_{\pi^+\pi^-}$ \\[1mm] \hline
Run 1--2 (9~fb$^{-1})$    & Prompt   & 60M    & 0.013\% & 18M & 0.024\%       \\ %
Run 1--3 (23~fb$^{-1})$  & Prompt   & 310M   & 0.0056\% &92M& 0.0104 \%       \\ %
Run 1--4 (50~fb$^{-1})$  & Prompt   & 793M   & 0.0035\% &236M& 0.0065 \%       \\ %
Run 1--5 (300~fb$^{-1})$ & Prompt  & 5.3G   & 0.0014\% & 1.6G & 0.0025 \%         \\
 \hline\hline
\end{tabular}
\renewcommand{\arraystretch}{1.0}
\end{center}
\end{table}

\subsubsection{Combined mixing and time dependent CPV sensitivity}
\label{sec:CPV:mixing:Dmeson}
\begin{figure}
  \centering
  \includegraphics[width=1.0\textwidth]{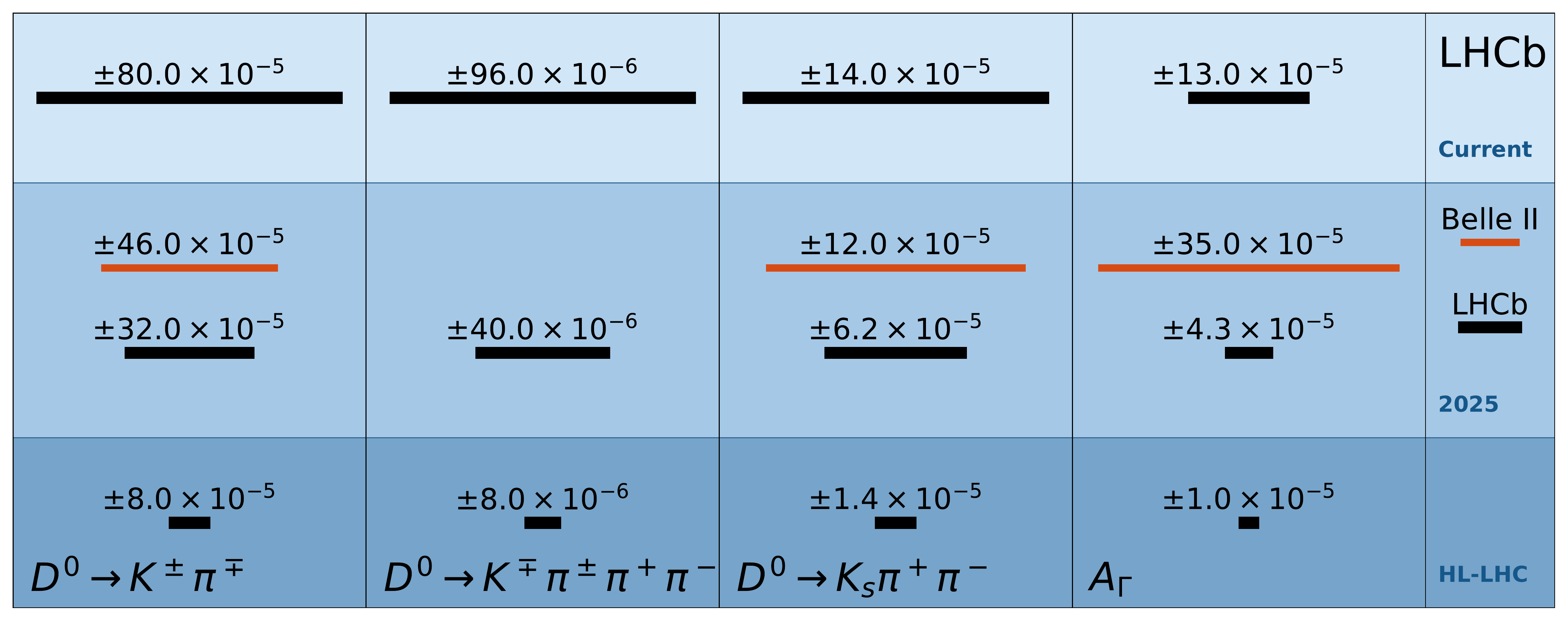}
  \caption{The predicted constraints on the indirect \CP violation
    asymmetry in
    charm from the decay channels indicated in the labels at the
    bottom of the columns. Predictions are shown in LS2
    (2020) from LHCb, LS3 (2025) from LHCb, at the end of Belle~II (2025), and at the end
    of the HL-LHC LHCb \upgradetwo\ programme. } 
\label{fig:charm_indirect_channels}
\end{figure}

The projected precisions of the analyses presented in
the previous sections are shown in
Fig.~\ref{fig:charm_indirect_channels}, and are compared with the expected
precisions at Belle~II. The expected LHCb constraints on $\phi$ are
translated into asymmetry constraints ($A_{\CP}^{ind.} \approx x \sin(\phi)$) by multiplying by the current
HFLAV average of $x$ and neglecting the uncertainty on this under the
assumption that $x$ will be comparatively well determined in the
future. This comparison neglects additional constraining power from $|q/p|$.
The relative values of these asymmetry constraints with those from
\agamma is indicative only.

\begin{figure}
  \centering
  \includegraphics[width=0.6\textwidth]{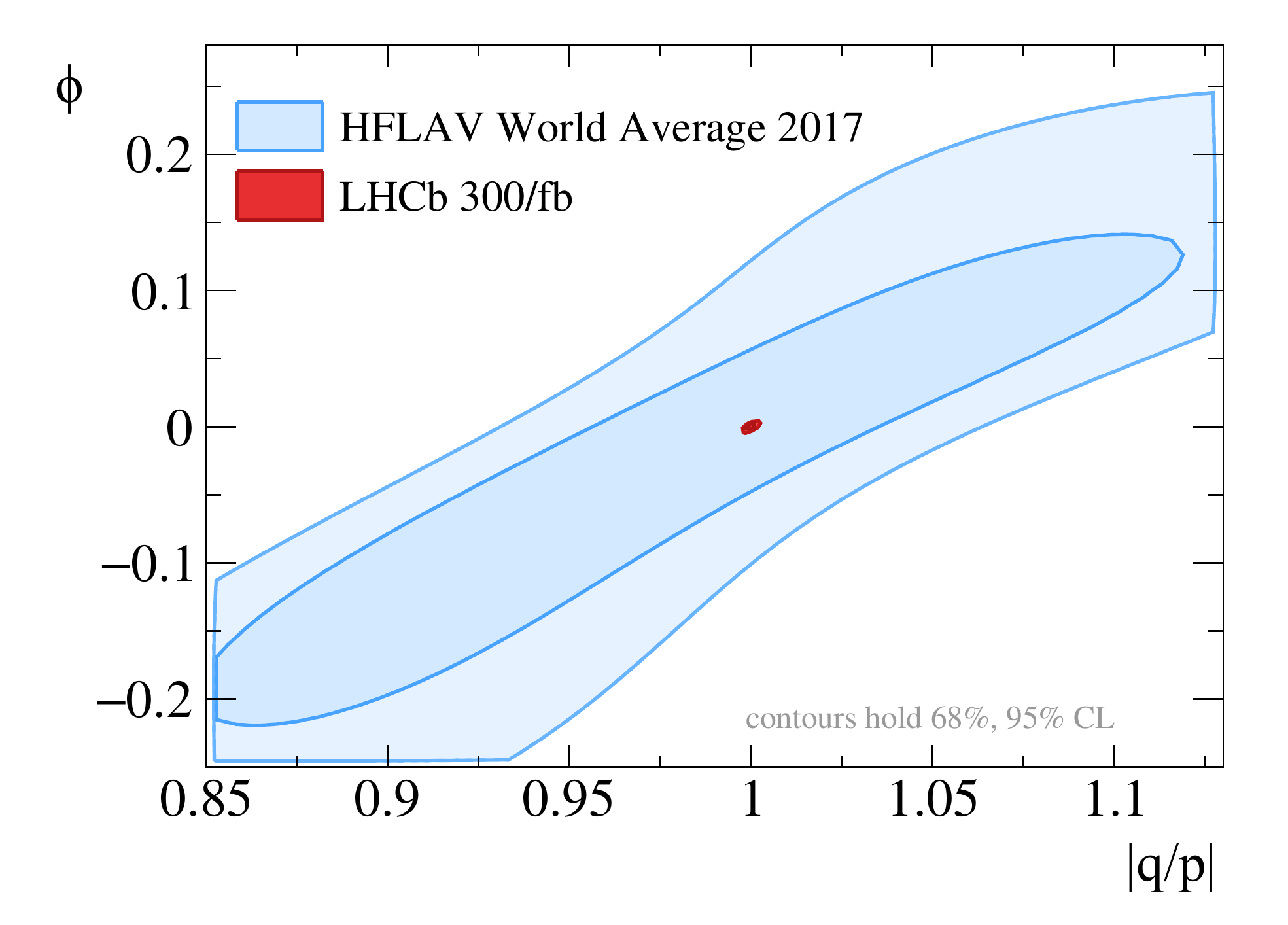}
  \caption{The estimated constraints for LHCb \upgradetwo\ on $\phi$, $|q/p|$ from the combination of the analyses (red), see main text for details, compared to the current world-average precision (light blue). The notation on the axes or lowercase $a$ corresponds to the uppercase $A$ used in the section text.}  
\label{fig:charm_indirect}
\end{figure}

The analyses presented in the previous sections are also combined to
establish the sensitivity to the \CP-violating parameters $|q/p|$ and
$\phi$. The combination is performed using the method described in Ref.~\cite{LHCb-PAPER-2016-032}.
At an integrated luminosity of $300\invfb$ the sensitivity to $|q/p|$ is expected to be $0.001$ and that to $\phi$ to be $0.1^\circ$.
This remarkable sensitivity is contrasted in Fig.~\ref{fig:charm_indirect} with the HFLAV world average
as of 2017. We can conclude that the
LHCb \upgradetwo will have impressive power to characterise NP
contributions to \CP violation and is the only foreseen facility with strong potential of
probing the SM contribution.

\subsubsection{Direct \CP violation}\label{sec:directCP:exp}

The SCS decays $D^0 \rightarrow \Km\Kp$ and $D^0 \rightarrow \pim\pip$ 
play a critical role in the measurement of time-integrated direct \CP violation
through time-integrated \CP asymmetry in the $h^-h^+$ decay rates, Eq. \eqref{ACPhh}.
The sensitivity to direct \CP violation is enhanced through a measurement of the difference in \CP asymmetries between \Dz\to\Km\Kp and \Dz\to\pim\pip decays, $\Delta A_{\CP}$, Eq.~\eqref{eq:Delta:ACP}.
The individual asymmetries $A_{\CP}(\Km\Kp)$ and $A_{\CP}(\pim\pip)$ can also be measured. 

A measurement of the time-integrated \CP asymmetry in $D^0 \rightarrow \Km\Kp$ has been performed in \lhcb\ with 3\invfb 
collected at centre-of-mass energies of 7 and 8\tev. The flavour of the charm meson at production is determined from the charge of the pion in
$\Dstarp\to\Dz\pi^+$ decays, or via the charge of the muon in semileptonic \bquark-hadron decays, $\Bb\to\Dz\mun\neumb X$. 
The analysis strategy so far relies on the $D^+\rightarrow K^+\pi^+\pi^- $, $D^+\rightarrow K_s^0\pi^+$ and $\Dstarp\to\Dz (\to K^-\pi^+)\pi^+$ decays as control samples~\cite{LHCb-PAPER-2016-035}. In this case, due to the weighting procedures aiming to fully cancel the production and reconstruction asymmetries, the effective prompt signal yield for $A_{\CP}(\Km\Kp)$ is reduced. The expected signal yields and the corresponding statistical precision in LHCb \upgradetwo are summarised in Table~\ref{tab:expectedyields}.

\begin{table}[t]
\caption{\small Extrapolated signal yields at LHCb and statistical precision on direct \CP violation observables for the promptly produced samples.}
\centering
\begin{tabular}{l c  c  c  c  c }
\hline\hline\\[-4mm]
Sample ($\mathcal{L}$) & Tag & Yield & Yield  & $\sigma(\Delta A_{\CP})$ &  $\sigma(A_{\CP}(hh))$ \\
 &  & \Dz\to\Km\Kp & \Dz\to\pim\pip & [$\%$] &  [\%] \\
\hline
Run 1-2 (9 \invfb)  & Prompt      &  52M &   17M & $0.03$  & $0.07$ \\   
Run 1-3 (23 \invfb) & Prompt      & 280M &   94M & $0.013$  & $0.03$ \\   
Run 1-4 (50 \invfb) & Prompt     &  1G  &  305M & $0.01$ & $0.03$ \\ 
Run 1-5  (300 \invfb) & Prompt    & 4.9G &  1.6G & $0.003$ & $0.007$ \\ 
\hline\hline
\end{tabular}
\label{tab:expectedyields}
\end{table}
 
\begin{figure}
  \centering
  \includegraphics[width=0.7\textwidth]{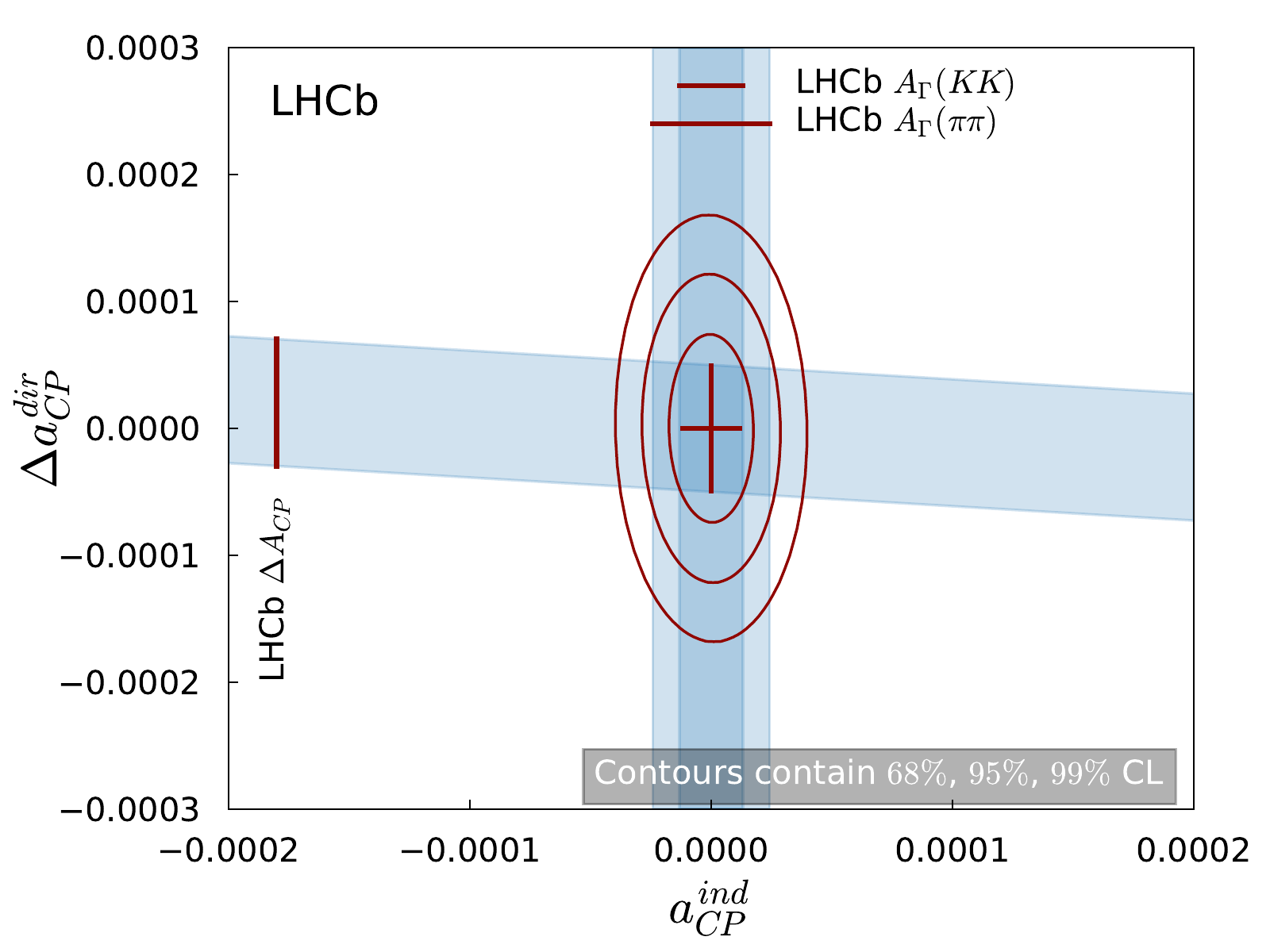}
  \caption{The estimated constraints for LHCb \upgradetwo\ on indirect and direct charm \CP violation from the analysis of two-body \CP eigenstates. The current world-average precision~\cite{HFLAV16} is $\pm2.6 \times 10^{-4}$ for indirect and $\pm 18 \times 10^{-4}$ for direct \CP violation and thus larger than the full scale of this plot.} 
\label{fig:agammadeltaacp}
\end{figure}

The observable $\Delta A_{\CP}$  is robust against systematic uncertainties. The main sources of systematic uncertainties are inaccuracies in the fit model, the weighting procedure, the contamination of the prompt sample with secondary $\Dz$ mesons and the presence of peaking backgrounds.
There are no systematic uncertainties with expected irreducible contributions above the ultimate statistical precision.  This channel is already entering the upper range of the physically interesting sensitivities, and will likely continue to provide the world's best sensitivity to direct \CP violation in charm at LHCb \upgradetwo. The power of these two-body \CP eigenstates at LHCb \upgradetwo is illustrated in Fig.~\ref{fig:agammadeltaacp}, which shows the indirect (see Sect.~\ref{sec:AGamma}) and direct \CP constraints that will come from these modes.

There are a number of other two-body modes of strong physics interest for which \upgradetwo will also make important contributions. These include the decay modes $\Dz\to\KS\KS$ (0.28\%), 
$\Dz\to\KS\bar{K^{*0}}$ (0.21\%), $\Dz\to\KS{K^{*0}}$ (0.15\%), $\Dsp\to\KS\pip$ $(3.2\times 10^{-4})$, $\Dp\to\KS\Kp$ $(1.2\times 10^{-4})$, $\Dp\to \phi \pip$ $(6\times 10^{-5})$, $\Dp \to \eta'\pip$ $(3.2\times 10^{-5})$, $\Dsp \to \eta'\pip$ $(3.2\times 10^{-4})$, where the projected statistical only \CP asymmetry sensitivities are given in brackets after the decay mode. The first three modes mentioned are notable as they receive sizeable contributions from exchange amplitudes at tree-level and could have a relatively enhanced contribution from penguin annihilation diagrams which are sensitive to NP. Consequently, 
they could be potential \CP violation discovery channels~\cite{Nierste:2015zra},\cite{Nierste:2017cua}.

Searches for direct \CP violation in the phase space of SCS $D^+\to h_1h_2h_3$ 
decays, hereafter referred to as $D\to 3h$, are complementary to that of $D^{(0,+)}\to h_1h_2$ 
($h_i = \pi, K$). In charged $D$ systems, only \CP violation in the decay is possible. The main observable 
is the \CP asymmetry, which, in the case of two-body decays, is a single number. In contrast,  $D\to 3h$ 
decays allow to study the variation of the \CP asymmetry across the two-dimensional phase space 
(usually represented by the Dalitz plot). 

The estimated  signal yields in future LHCb upgrades  are summarised in Table~\ref{tab:Dhhh-yields}, based on
an extrapolation of the Run 2 yields per unit luminosity. The
estimated sensitivities to observation of \CP violation, using $D^+ \to \pi^-\pi^+\pi^+$ as an example, are presented 
in Table~\ref{tab:Dhhh-sensitivity}.

\begin{table}[tb]
\centering
\caption{Extrapolated signal yields at LHCb, in units of 10$^6$, for the SCS decays $D^+ \to K^-K^+\pi^+$, $D^+ \to \pi^-\pi^+\pi^+$,
and for the DCS decays $D^+ \to K^-K^+K^+$, $D^+ \to \pi^-K^+\pi^+$. \label{tab:Dhhh-yields}}
\begin{tabular}{lcccc}
\hline \hline
Sample ($\mathcal{L}$)  & $ K^-K^+\pi^+$  & $\pi^-\pi^+\pi^+$ &  $K^-K^+K^+$ & $ \pi^-K^+\pi^+$ \\
\hline
Run 1--2 (9\invfb)     &   200       & 100     & 14       & 8   \\
Run 1--4 (23\invfb)   &   1,000    & 500  & 70     & 40   \\
Run 1--4 (50\invfb)   &   2,600    & 1,300  & 182     & 104   \\
Run 1--6 (300\invfb) &  17,420   &  8,710 & 1,219  &  697  \\
\hline \hline
\end{tabular}
\end{table}

\begin{table}[bt]
\centering
\caption{Sensitivities to several illustrative \CP-violation scenarios in $D^+ \to \pi^-\pi^+\pi^+$ decay. Simulated $D^+$ and $D^-$
Dalitz plots are generated with relative changes in the phase of the $R\pi^{\pm}$ amplitude, $R=\rho^0(770)$, 
$f_0(500)$ or $f_2(1270)$. The values of the phase difference are given in degrees and correspond to 
a 5$\sigma$  \CP-violation effect. Simulations are performed with $3\invfb$ and extrapolated to the 
expected luminosities.}
\begin{tabular}{lc c c c}
\hline\hline
  resonant channel           & $9\invfb$   & $23\invfb$     & $50\invfb$     & $300\invfb$\\
\hline
$f_0(500)\pi$                    & 0.30 &  0.13   &  0.083   &  0.032 \\
$\rho^0(770\pi$                & 0.50 &  0.22   &  0.14     &  0.054\\
$f_2(1270)\pi$                  & 1.0  &  0.45   &   0.28    & 0.11\\
\hline\hline
\end{tabular}
\label{tab:Dhhh-sensitivity}
\end{table}

The SM generated \CP violation could be observed in the SCS decays, such as \decay{\Dz}{\pip\pim\pip\pim} and \decay{\Dz}{\Kp\Km\pip\pim}, while NP is needed to justify any observation of \CP violation in the DCS decays, such as \decay{\Dz}{\Kp\pim\pip\pim}.
Many techniques can be adopted to search for \CP violation, all of them exploiting the rich resonant structure of the decays.
The methods used so far  at \lhcb are based on ${\widehat{T}}$-odd asymmetries and the energy test, while studies are ongoing to measure model-dependent \CP asymmetries in the decay amplitudes.

The study of ${\widehat{T}}$-odd asymmetries exploits potential $P$-odd \CP violation from the interference between different amplitude structures in the decay, as described in Ref.~\cite{Durieux:2015zwa}.  This uses a triple product $C_T=\vec{p}_A\cdot (\vec{p}_B
\times \vec{p}_C)$ constructed from the momenta of three of the final state particles $\vec{p}_A,\vec{p}_B,\vec{p}_C $.
\lhcb has studied ${\widehat{T}}$-odd asymmetries using 3\invfb data from the Run~1 dataset, and obtained a sensitivity of $2.9\times10^{-3}$ with very small systematic uncertainties~\cite{LHCB-PAPER-2014-046}.
The peculiarity of this measurement is the absence of instrumental asymmetries, since it is given by the difference of two asymmetries measured separately on \Dz and \Dzb decays, $a_{\CP}=(A_T-\bar{A}_T)/2$, where
\begin{align}
  A_T = \frac{(\Gamma(\Dz,C_T>0)-\Gamma(\Dz,C_T<0))}{(\Gamma(\Dz,C_T>0)+\Gamma(\Dz,C_T<0))},\quad \bar{A}_T = \frac{(\Gamma(\Dzb,\bar{C}_T>0)-\Gamma(\Dzb,\bar{C}_T<0))}{(\Gamma(\Dzb,\bar{C}_T>0)+\Gamma(\Dzb,\bar{C}_T<0))}.
\end{align}
One can therefore expect the errors to scale with luminosity to reach a sensitivity down to $2.9\times10^{-5}$ ($9.4\times10^{-5}$) for $\Dz\to\pip\pim\pip\pim$ ($\Dz\to\Kp\Km\pip\pim$) decays, as detailed in Table~\ref{tab:CS-D24h-yields}.
\begin{table}[tb]
\centering
\caption{ Extrapolated signal yields, and statistical precision on ${\widehat{T}}$-odd \CP-violation observables at LHCb.\label{tab:CS-D24h-yields}}
\begin{tabular}{lcccc}
\hline\hline
  & \multicolumn{2}{c}{$\Dz\to\pip\pim\pip\pim$} & \multicolumn{2}{c}{$\Dz\to\Kp\Km\pip\pim$} \\
Sample ($\mathcal{L}$)  & Yield ($\times10^6$) & $\sigma(a_{\CP}^{\widehat{T}\text{-odd}})$  & Yield ($\times10^6$) & $\sigma(a_{\CP}^{\widehat{T}\text{-odd}})$ \\
\hline
Run 1--2 (9\invfb)   & 13.5   & $2.4\times10^{-4}$ &  4.7   & $5.4\times10^{-4}$ \\
Run 1--3 (23\invfb)  &   69   & $1.1\times10^{-4}$ &   12   & $3.4\times10^{-4}$ \\
Run 1--5 (300\invfb) &  900   & $2.9\times10^{-5}$ &  156   & $9.4\times10^{-5}$ \\
\hline\hline
\end{tabular}
\end{table}

The energy test method is insensitive to global asymmetries. However, it is expected that it will become sensitive to variations in phase space of production and detection asymmetries.
These can be controlled in data by application of the method to CF decays, such as \decay{\Dz}{\Km\pip\pip\pim}.
Assuming scaling with the square-root of the ratio of sample sizes, the same p-values can be expected for the \CP asymmetries given in Table~\ref{tab:charm_623_energy_test}.

\begin{table}[bt]
\centering
\caption{Overview of sensitivities to various \CP-violation scenarios for \decay{\Dz}{\pip\pim\pip\pim} decays as extrapolated from Ref.~\cite{LHCb-PAPER-2016-044}. The relative changes in magnitude and phase of the amplitude of the resonance $R$ to which sensitivity is expected are given in $\%$ and $^\circ$, respectively. The $P$-wave $\rho^0(770)$ is a $P$-odd component. The phase change in this resonance is tested with the $P$-odd \CP-violation test. Results for all the other scenarios are given with the standard $P$-even test.}
\begin{tabular}{lc c c}
\hline\hline
$R$ (partial wave)        & $9\invfb$   & $23\invfb$ & $300\invfb$\\
\hline
\decay{a_1}{\rhoz\pi} (S) & $1.4\%$     & $0.6\%$      & $0.17\%$\\
\decay{a_1}{\rhoz\pi} (S) & $0.8^\circ$ & $0.35^\circ$ & $0.10^\circ$\\
\rhoz\rhoz (D)            & $1.4\%$     & $0.6\%$      & $0.17\%$\\
\hline
\rhoz\rhoz (P)            & $0.8^\circ$ & $0.35^\circ$ & $0.10^\circ$\\
\hline\hline
\end{tabular}
\label{tab:charm_623_energy_test}
\end{table}

Charm decays with neutrals in the final state can help to shed light on the SM or beyond-SM origin of possible \CP-violation signals by testing correlations between \CP asymmetries measured in various flavour-\grpsuthree or isospin related decays, see Sec.~\ref{sec:isospin:null:test} and Refs. \cite{Bhattacharya:2012ah,Pirtskhalava:2011va,Feldmann:2012js}. These modes are, however, particularly challenging in hadronic collisions, where the calorimeter background for low energy clusters is high, while the trigger retention rate needs to be kept low to allow for affordable rates. 

Nevertheless, good performances are achieved when considering decays with at least two charged particles in the final states, such as \decay{\Dz}{\pi^+\pi^-\pi^0}, since the charged particles help to identify the displaced decay vertex of the charm meson. In only 2\invfb of data, collected during 2012, LHCb has reconstructed about 660,000 \decay{\Dz}{\pi^+\pi^-\pi^0} decays~\cite{LHCb-PAPER-2014-054}, \ie about five times more than Babar from its full data set~\cite{TheBABAR:2016gom}, with comparable purity. Preliminary estimates for Run 2 data, give about 500\,000 signal decays per \invfb, making future \CP-violation searches in this channel very promising. Similarly, large samples of \decay{\D_{(\squark)}^+}{\eta^{(\prime)}\pi^+} decays, with \decay{\eta^{(\prime)}}{\pi^+\pi^-\gamma}, or \decay{\Dp}{\pi^+\pi^0} decays, with \decay{\pi^0}{e^+e^-\gamma}, are already possible with the current detector. The \decay{\D_{(\squark)}^+}{\eta^{\prime}\pi^{+}} mode, as an example, has been used by LHCb during Run 1 to perform the most precise measurement of \CP asymmetries in these channels to date, with uncertainties below the 1\% level~\cite{LHCb-PAPER-2016-041}.

More challenging final states consisting only of neutral particles, such as $\pi^0 \pi^0$ or $\eta \eta$, can still be reconstructed with \decay{\pi^0}{\gamma\gamma} or \decay{\eta}{\gamma\gamma} candidates made of photons which, after interacting with the detector material, have converted into an $e^+e^-$ pair. Such conversions must occur before the tracking system to have electron tracks reconstructed. Although the reconstruction efficiency of these ``early'' converted photons in the current detector reaches only a few percent of the calorimetric photon efficiency, their purity is much higher. This approach may become interesting only with the large data sets that are expected to be collected by the end of \upgradetwo. The $\eta$ decays can also be reconstructed through the $\pi^+\pi^-\gamma$ final state.

Unlike the \Dz decays which are usually tagged with a soft pion from \Dstarp decays, there is no easy tagging of the \Dp modes, which thus often suffer from a high combinatorial background. Employing a $\pi^0$ tag using \decay{\Dstarp}{\Dp\pi^0} decays could facilitate future studies of \Dp decays, in particular those with challenging and/or high multiplicity final states.

The study of these modes may be challenging in Run~3 due to the cluster pile-up at higher luminosities and radiation damage of the current calorimeter. The new calorimeter proposed for LHCb \upgradetwo would have an improved granularity. It would therefore improve the efficiency for \decay{\pi^0}{\gamma\gamma} decays and, in particular, could make the $\pi^0$ tag feasible.

\subsubsection{Rare leptonic and radiative charm decays}
The most experimentally accessible very rare charm decay is \dmumu. The world's best limit on \brdmumu was obtained by LHCb with $0.9\invfb$ of 2011 data~\cite{LHCb-PAPER-2013-013}, resulting in
\begin{equation}
\brdmumu< 6.2 \times 10^{-9}\text{ at 90\% CL. }
\end{equation}
Extrapolating the current detector performance, the expected limit is about $5.9 \times 10^{-10}$ with $23~\invfb$ and $1.8 \times 10^{-10}$ with $300~\invfb$ of integrated luminosity,
covering a large part of the unambiguous space to search for NP without being affected by long distance uncertainties in the SM predictions.

\begin{figure}[tb]
\includegraphics[width=0.5\textwidth]{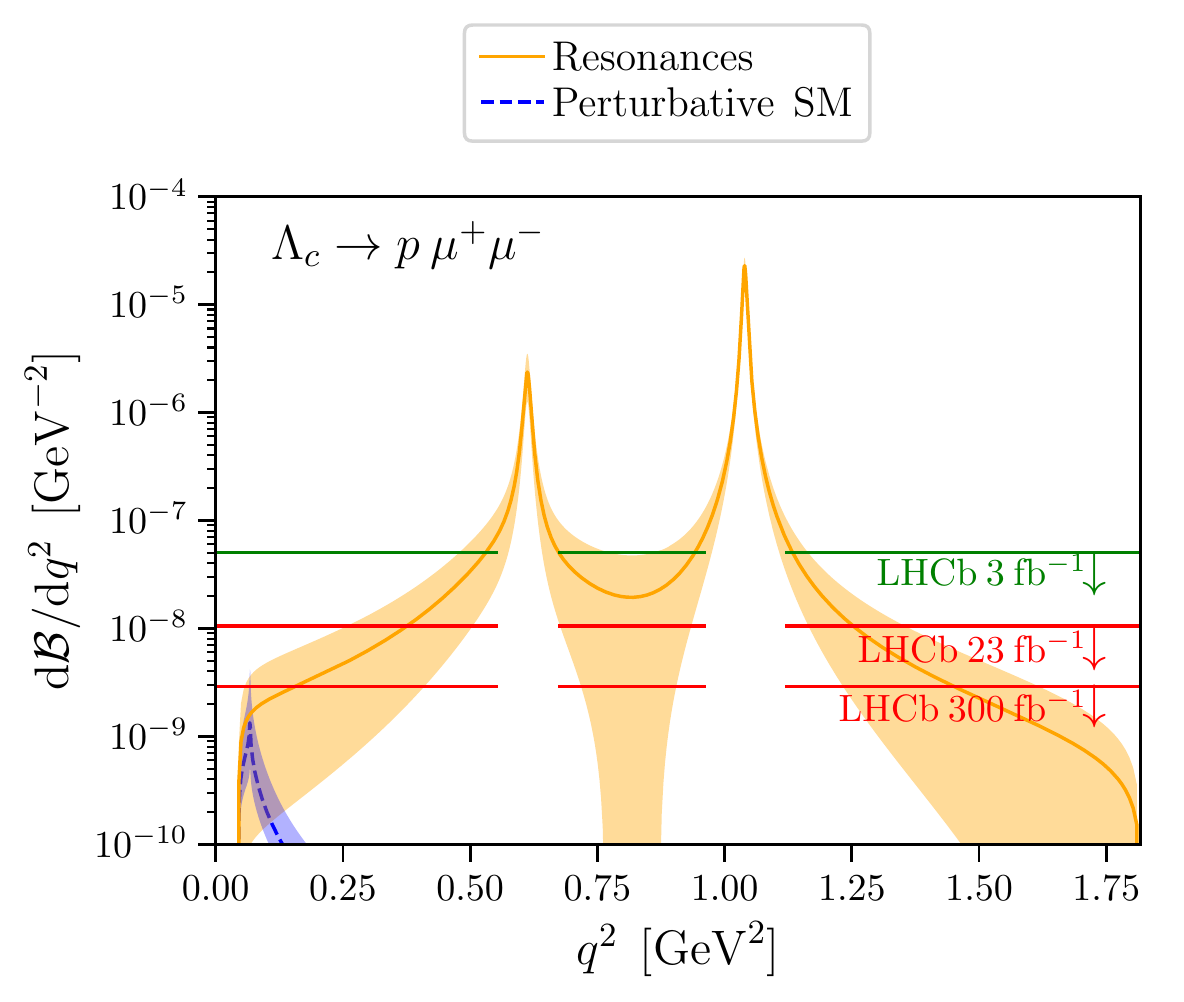}
\includegraphics[width=0.5\textwidth]{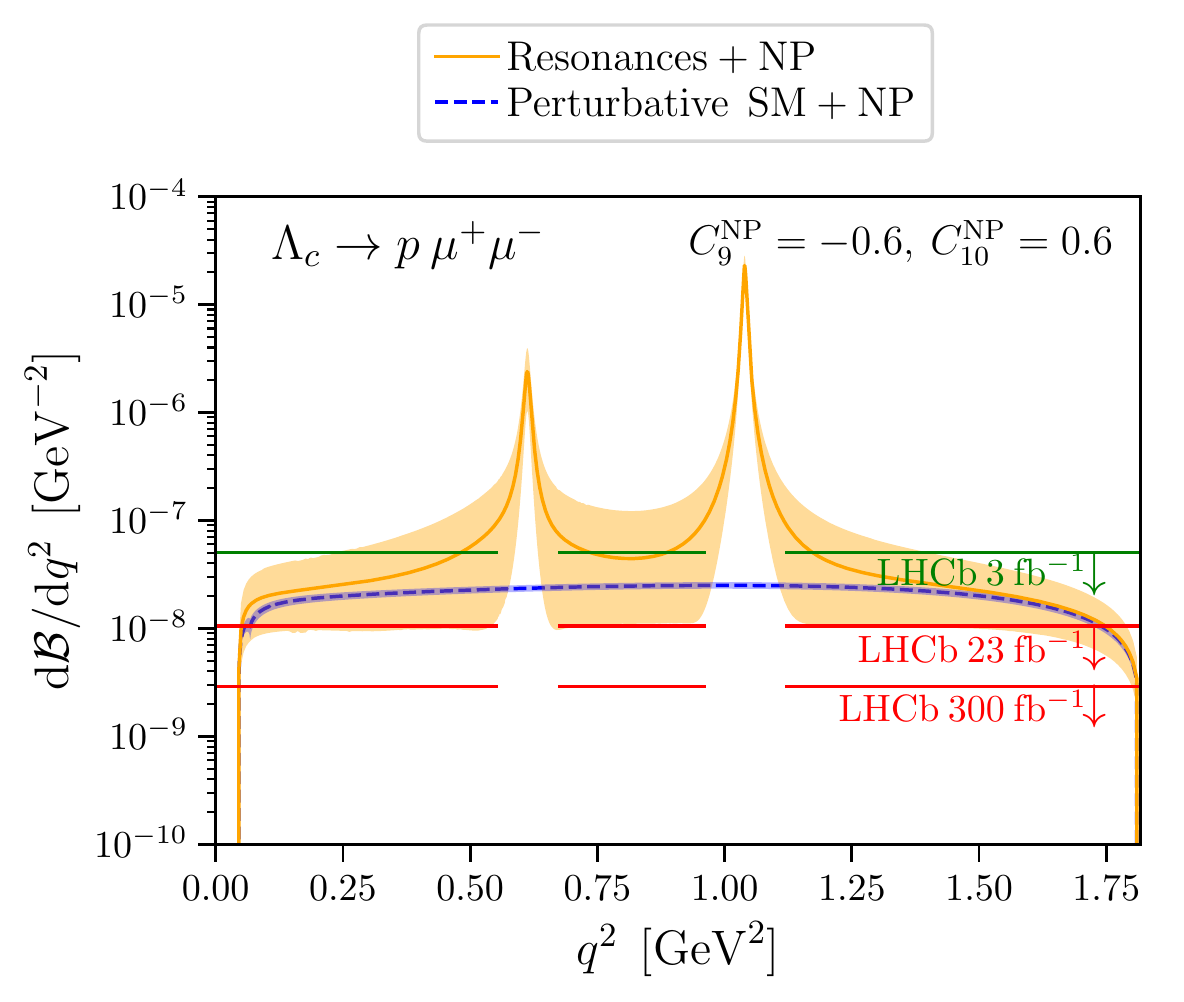}
\caption{Differential branching fraction for the decay $\Lc \to p \mumu$ as a function of $q^2$ (left) in the SM and (right) in a NP scenario with ${\cal C}_9^{\rm NP}=-0.6$ and ${\cal C}_{10}^{\rm NP}=-0.6$~\cite{Meinel:2017ggx}. The current LHCb limit is marked with green. The extrapolated upper limit with the $23$ and $300~\invfb$ data sets is marked in red.\label{fig:LC2PMUMU}}
\end{figure}

The next class of rare charm decays which are particularly suited for experiments at hadron colliders 
are three-body decays with a pair of leptons in the final state, such as $D^{+}_{(s)}/ \Lc \to h^+ \ellell$ and $D^0 \to h^+h^- \ellell$. In some NP scenarios the short distance contributions can be enhanced by several orders of magnitude allowing NP to manifest as an enhancement of the branching fraction. An example of such a model is shown in Fig.~\ref{fig:LC2PMUMU}, where the Wilson coefficients were assumed to obtain NP contributions, $C_9^{\rm NP} =-0.6$ and $C_{10}^{\rm NP} =0.6$ (cf. Eqs. \eqref{eq:Leff:e1}, \eqref{eq:operators}).
Outside of the resonance regions the short distance contributions are 
comparable to the long distance effects. The chosen NP benchmark point leads to branching ratios just below the current LHCb limit~\cite{LHCb-PAPER-2017-039} %
\begin{equation}
\BF( \Lc \to p \mumu) < 5.9 \times 10^{-8}~{\rm at}~90\%~{\rm CL}.
\end{equation}
In the \upgradetwo LHCb is expected to improve the limit to
\begin{equation}
\BF( \Lc \to p \mumu) < 4.4 \times 10^{-9}~{\rm at}~90\%~{\rm CL}.
\end{equation}

\begin{figure}[tb]
\includegraphics[width=0.33\textwidth]{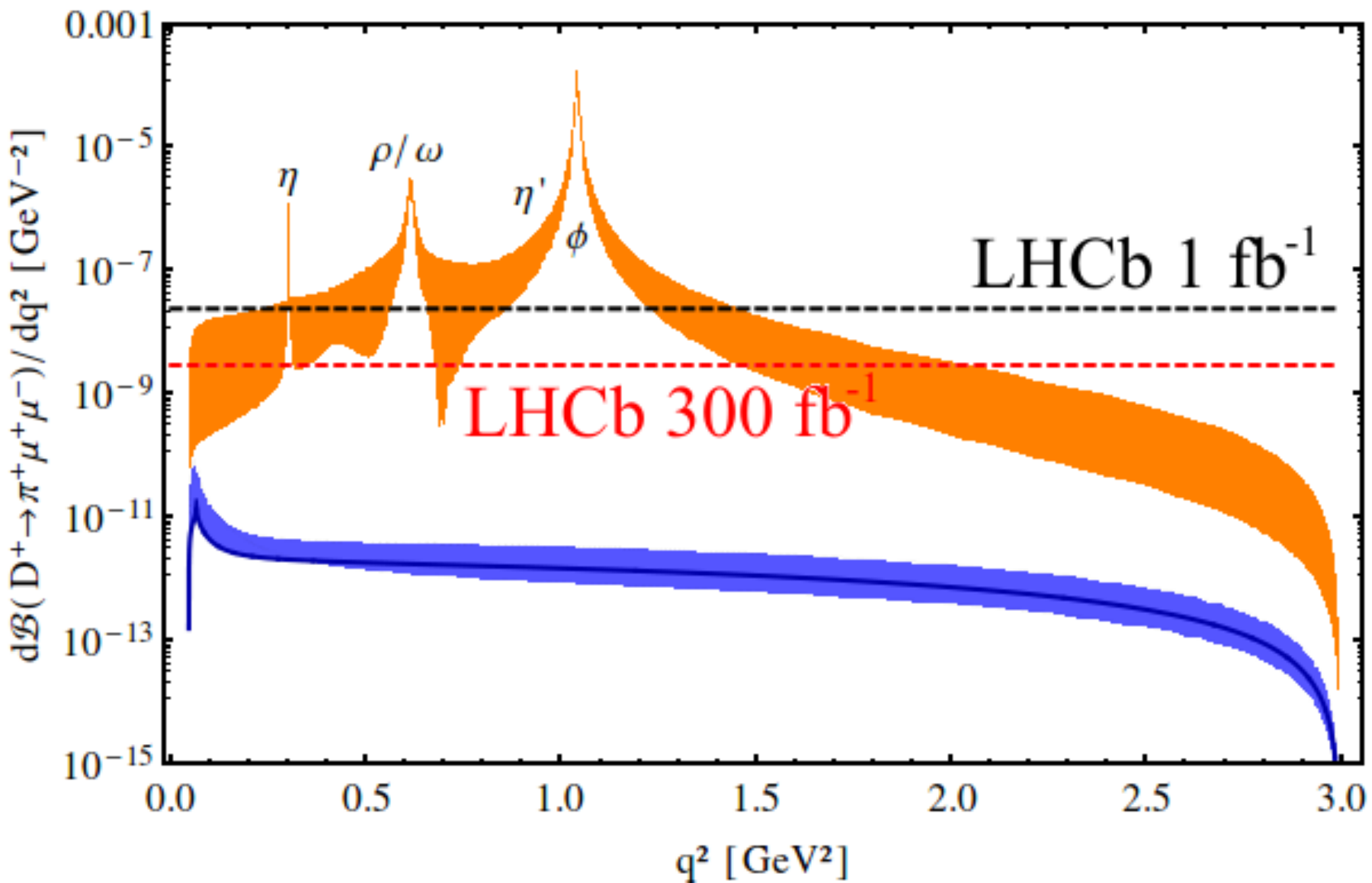}
\includegraphics[width=0.66\textwidth,clip=true,trim=0mm 60mm 0mm 0mm]{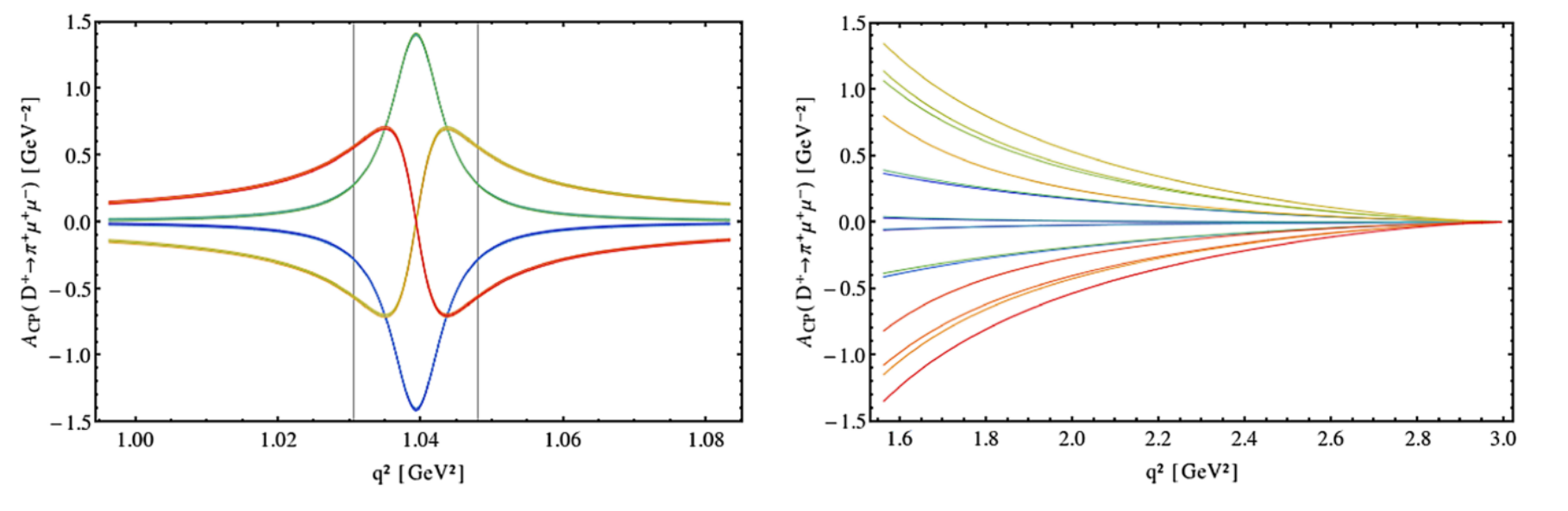}
\caption{
    (Left) Differential branching fraction of $\D^+ \to \pi^+ \mumu$ as a function of $q^2$ in the SM. The current LHCb limit is marked with a dashed black line. (Middle and right) Predictions for $A_{\CP}$ in a particular NP scenario of Ref.~\cite{deBoer:2015boa} with different strong phases $\delta_{\rho,\phi}$.\label{fig:D2PIMUMU}
}
\end{figure}

An order of magnitude improvement is expected for the limit on $\BF(D^{+}_{(s)}\to\pi^+\mu^+\mu^-)$ decays, which is currently at $7.3\times10^{-8}$
at 90\% CL \cite{LHCb-PAPER-2012-051}. The expected upper limits are about $1.3\times10^{-8}$ with $23 \invfb$ and $0.37\times10^{-8}$  with $300 \invfb$.
In addition, LHCb will have the ability to measure angular observables such as the forward-background asymmetry, $A_{\rm FB}$, or time integrated $A_{\CP}$, 
which will provide additional handles to separate the long distance from the short distance contributions, and for which some theoretical predictions in NP scenarios already exist \cite{deBoer:2015boa},
as shown in Fig.~\ref{fig:D2PIMUMU}.

Last but not least with a sample corresponding to an integrated luminosity of $300~\invfb$ LHCb will be able to perform searches for the LFV decays $D^{+}_{(s)}/ \Lc\to h^+ \ell^+ \ell^{\prime -}$
and perform tests of lepton universality via the ratios ${\cal B} (\decay{D^{+}_{(s)}/\Lc}{h^+ \mu^+ \mu^{-}})/{\cal B}(\decay{D^{+}_{(s)}/\Lc}{h^+ e^+ e^{-}})$. 

The decays $D^0\to h^+h^-\ellell$ have richer dynamics than the 
two- and three-body decays, allowing for a variety of differential distributions to be investigated.
Due to the huge charm production cross-section at the LHC, and LHCb's ability to trigger on low \pt\ dimuons, 
LHCb has unique physics reach in studying these decays. 
In fact, significant progress has already been made with the observation of the CF decay $\decay{\Dz}{\Km\pip\mun\mup}$ (with the dimuon mass in the $\rho/\omega$ region) with a branching fraction $(4.17\pm0.42)\times10^{-6}$~\cite{LHCb-PAPER-2015-043}  and the SCS decays $\decay{\Dz}{\pip\pim\mup\mun}$ and $\decay{\Dz}{\Kp\Km\mup\mun}$ with branching fractions of $(9.64\pm1.20)\times10^{-7}$ and $(1.54\pm0.33)\times10^{-7}$, respectively~\cite{LHCb-PAPER-2017-019}. For the latter, also the differential branching fraction as a function of the dimuon mass squared, $q^2$, was measured.
Furthermore, LHCb has performed the first measurement of \CP- and angular asymmetries in $\decay{\Dz}{\pip\pim\mumu}$ and $\decay{\Dz}{\Kp\Km\mumu}$ decays using Run~2 data. 
This resulted in the first determination of the forward-backward asymmetry $A_{\rm FB}$, the triple-product asymmetry $A_{2\phi}$ and the \CP-asymmetry $A_{\rm CP}$ with uncertainties at the percent-level~\cite{LHCb-PAPER-2018-020}. 
Moreover, the implementation of triggers for dielectron modes 
opens the possibility of measuring branching-fraction ratios between di-muon and di-electron modes.
Since the main limit for these studies comes from the available statistics, LHCb will already have excellent prospects in Run~3 and 4. 
Projected signal yields for the muonic modes of ${\mathcal O}(10^4)$ will allow more sensitive studies of angular asymmetry and the first amplitude analyses to attempt to disentangle short distance and long distance components.
However,  the full potential for these decays will be exploited only with the $300$\invfb LHCb upgrade.

\subsubsection{Prospects with charm baryons}

In general, charmed baryon decays offer a rich laboratory within which to study 
matter-antimatter asymmetries.
The most easily accessible modes for \lhcb are multibody decays containing one 
proton and several kaons or pions, such as the CF $\decay{\Lc}{\proton\Km\pip}$ decay and the SCS 
$\decay{\Lc}{\proton\Km\Kp}$ and $\proton\pim\pip$ decays.
Unlike the analogous final states of the spinless neutral and charged \PD mesons, such 
charmed baryon final states have at least five degrees of freedom, allowing for 
a complex variation of the strength of \CP violation across the phase space.
A complete description of such a space is experimentally challenging.
This is compounded by the relatively short lifetime of the \Lc baryon with 
respect to those of the \PD mesons, which decreases the power of selections 
based on the displacement of the charm vertex, increasing the prompt 
combinatorial background, which mainly comprises low-momentum pions and kaons.
Selections of charm baryons then heavily rely on good discrimination between 
proton, kaon, and pion hypotheses in the reconstruction of charged tracks, and 
on a precise secondary vertex reconstruction.
The latter is also necessary when reconstructing charm baryons originating from 
semileptonic \bquark-hadron decays, which has the advantage of both a simple 
trigger path (a high-\pt, displaced muon) and a cleaner experimental signature 
due to the large \bquark-hadron lifetime.
The presence of the proton in the final state, whilst a useful handle for 
selections, poses experimental challenges for \CP-violation measurements in 
addition to those present for measurements with \PD mesons, as the 
proton-antiproton detection asymmetry must be accounted for.
This has not yet been measured at \lhcb due to the lack of a suitable control 
mode.

The most precise measurement of \CP violation in charm baryons was made 
recently by the \lhcb collaboration using Run 1 data, corresponding to 
$3$\invfb of integrated luminosity~\cite{LHCb-PAPER-2017-044}.
The difference between the phase-space-averaged \CP asymmetries in 
$\decay{\Lc}{\proton\Km\Kp}$ and $\decay{\Lc}{\proton\pim\pip}$ decays was 
found to be consistent with \CP symmetry to a statistical precision of 
$0.9\,\%$.
This difference is largely insensitive to the proton detection asymmetry, but 
masks like-sign \CP asymmetries between the two modes.
Further studies must then gain a precise understanding of the proton detection 
asymmetry, in addition to measuring the variation of \CP violation across the 
decay phase space.

Although there is little literature on the subject, the magnitude of direct
\CP violation in charm baryon decays is expected to be similar to that for 
charmed mesons~\cite{Bigi:2012ev}, and so the first step in furthering our 
understanding is to reach a precision of $0.5\times10^{-4}$. This can be met by LHCb
given the $300$\invfb of integrated luminosity collected by the end of Run 5.
The acquisition of more data is vital in enabling studies of states heavier 
than the \Lc baryon, as their alternate compositions may permit considerably different dynamics. In this respect it is interesting to note that the \Xiccpp baryon was discovered~\cite{LHCb-PAPER-2018-026} by LHCb using the $\Lc\Km\pip\pip$ final state with 
data taken in 2016, corresponding to an integrated luminosity of $1.7$\invfb.
A signal yield of $313 \pm 33$ was determined, which can be 
extrapolated to around $100,000$ such decays obtainable with data corresponding to
$300$\invfb, allowing asymmetry measurements with a precision of $0.4$\,\%. The flexible trigger and real-time analysis concept of LHCb's \upgradetwo will however enable not only this measurement but a detailed mapping of the entire landscape of multiply-heavy charmed hadrons.

%% file: section4/section.tex
\section[Strange-quark probes of new physics]{Strange-quark probes of new physics}
{\it\small Authors (TH): Giancarlo D'Ambrosio, Diego Guadagnoli, Teppei Kitahara, Diego Martinez Santos.}

\noindent Although not specifically built for the study of strange hadrons, LHCb's forward geometry allows for a rich physics programme with strange hadrons. LHCb \upgradetwo phase will substantially enlarge in scope this programme. Its relevance is evident from the fact that kaon physics provides among the most stringent constraints on 
new-physics interactions. 
As with the charm and beauty counterparts, the HL-LHC will be a uniquely powerful factory for strange baryons which the LHCb \upgradetwo is well placed to exploit.

\subsection{The (HL) LHC as a strangeness factory}

The capabilities of LHCb for strangeness decays are due to the large strange-hadron production rates at the LHC, combined with the specific features of LHCb in comparison with the other LHC detectors. These are the higher efficiency for low transverse-momentum particles and the better invariant-mass and vertex resolutions. A simple estimate with {\tt Pythia 8.226} yields a $K^0$ cross section in 14 TeV $pp$ collisions of about 0.6 barn, as well as another 0.6 barn for $K^{\pm}$. These are roughly 1000 times larger than the production cross sections for $B$ mesons. Hyperon cross sections are also large, attaining in {\tt Pythia} 0.14, 0.04, 0.01 barn for $\Lambda^0$, $\Sigma^+$, and $\Xi^-$, respectively (including their antiparticles), while the cross section for $\Omega^-$ is similar to that of $B$'s. 

The LHCb capabilities for strange decays were first demonstrated in Ref.~\cite{LHCb-PAPER-2012-023}, which achieved the world's best result in $\mathcal{B}(K_S^0\to\mu^+\mu^-)$ even though the trigger efficiency on well reconstructed decays was only $\sim 1\%$ (to be compared to $\sim 90\%$ for $B_s^0\to\mu^+\mu^-$). In Run 2, dedicated trigger lines have been implemented, selecting muons down to 80 MeV in transverse momentum and increasing the trigger efficiency by one order of magnitude for strangeness decays to dimuons ~\cite{Dettori:2297352}. 
\Blu{The main limitation is} the hardware trigger (L0). In Run 3 the LHCb  trigger will be fully software-based, which can in principle allow for efficiencies as high as those achieved for $B$'s. The main challenge for LHCb \upgradetwo will be to fully exploit the software trigger, and enable high trigger efficiency at low transverse momentum using the displacement of the decay products. In addition, using downstream reconstruction in the trigger may boost the LHCb capabilities for charged kaon decays by up to a factor 5~\cite{kaon_paper}.

Among the different strange hadron species, the LHCb detector layout is particularly suitable for $K^0_S$ and hyperons, which have relatively shorter lifetimes compared to $K^{\pm}$ or $K^0_L$, which mostly decay outside the detector. An approximate acceptance ratio for $K_S^0:K^{\pm}:K_L^0$ is estimated as $1 : 0.01 : 3\times10^{-3}$ ~\cite{kaon_paper} for full tracking and $1: 0.02: 0.01$ for downstream tracking (i.e, no usage of VELO information).
Hence, measurements with $K^\pm$ decays will also be possible, with potential overlap with NA62.
LHCb  \upgradetwo
will reach sensitivities for $\mathcal{B}(K_S^0\to\mu^+\mu^-)$ below the $10^{-11}$ level if it keeps the performance of the current detector~\cite{dmsFPCP}, taking into account that the full software trigger will allow for very high trigger efficiencies. Similarly, sensitivities at the $10^{-10}$ level are expected for $\mathcal{B}(K_S^0\to\pi^0\mu^+\mu^-)$~\cite{Chobanova:2195218}. 

The same analysis strategy as $\mathcal{B}(K_S^0\to\pi^0\mu^+\mu^-)$ can be applied to other decays, such as $K_S^0\to\gamma\mu^+\mu^-$, although the sensitivity will be worse due to poorer mass resolution~\cite{kaon_paper}. Other kaon decays that can be studied at LHCb include $K^+\rightarrow\pi\mu\mu$ (both with opposite-sign and same-sign muon pairs), $K_S^0\rightarrow4\mu$, or decays involving electrons, especially interesting in order to search for lepton flavor violation (LFV) and lepton-flavor universality violation (LFUV). The 
LHCb \upgradetwo will also have an abundant enough sample of $\Sigma^+\to p \mu^+\mu^-$ to do precise study of the differential decay rate. The $\Sigma^+\to p e^+e^-$ can be studied using dedicated triggers. Semileptonic hyperon decays are reconstructed in LHCb~\cite{kaon_paper} and kinematic constraints can be used to reconstruct the mass peak of the hyperon. Since the expected yields for these decays
can be very large, the main challenge is the fight against peaking backgrounds, such as $\Lambda\to p\pi^-$ or $\Xi^-\to\Lambda\pi^-$. 
LHCb \upgradetwo will also be able to update existing limits on LFV kaon decays~\cite{Borsato:2018tcz}. A longer list of decays that LHCb will be able to probe can be found in Ref.~\cite{kaon_paper}. 

\subsection{$K_S^0\to \mu^+ \mu^- $ and $K_L^0\to \mu^+ \mu^- $ decays}

LHCb \upgradeone and \upgradetwo are expected to be able to probe short-distance physics 
using the $K^0\to \mu^+ \mu^- $ decay. 
In the SM, $K_S \to \mu^+ \mu^-$ is significantly dominated by $P$-wave $\CP$-conserving long-distance (LD) contribution, while $S$-wave $\CP$-violating short-distance (SD) contributions from $Z$-penguin and $W$-box are small \cite{Ecker:1991ru,Isidori:2003ts,DAmbrosio:2017klp}:
\begin{align}
\mathcal{B}(K_S \to \mu^+ \mu^-)_{\rm SM} &= \left[ \left(4.99   \pm 1.50\right)_{\rm LD} + \left( 0.19 \pm 0.02 \right)_{\rm SD} \right] \times 10^{-12}.
\label{eq:KSmumu:SM}
\end{align}
The large uncertainty comes from the LD contribution which  has been computed in ChPT \cite{Ecker:1991ru}. This uncertainty 
 is expected to be reduced by a dispersive treatment of $K_S \to \gamma^{\ast} \gamma^{\ast}$ \cite{Colangelo:2016ruc}, where  $K_S \to \gamma \gamma$, $K_S \to  \mu^+ \mu^- \gamma $, $K_S \to \mu^+ \mu^- e^+ e^-$  and $K_S \to \mu^+ \mu^- \mu^+ \mu^-$ are measurable in LHCb experiment and KLOE-2 experiment at DA$\Phi$NE \cite{AmelinoCamelia:2010me}.
  \Blu{
  It is important to note that, in contrast to the $K_L\to\mu^+\mu^-$ decay:
\begin{itemize}
\item[$\bullet$]
The decay $K_S\to\mu^+\mu^-$ provides another sensitive probe of {\em imaginary} parts of SD couplings and consequently is very sensitive to new sources of  \CP  violation. This is not the case of $K_L\to\mu^+\mu^-$ which is governed by {\em real} couplings.
\item[$\bullet$]
As seen in Eq.~(\ref{eq:KSmumu:SM}) the %
LD
and 
SD
contributions to the total rate are added incoherently \cite{Ecker:1991ru, Isidori:2003ts}, which represents a big theoretical advantage over $K_L\to\mu^+\mu^-$, where LD and SD amplitudes interfere.
\end{itemize}

This means that in BSMs in which the SD contribution is significantly enhanced, the LD-uncertainty in $K_S\to\mu^+\mu^-$ ceases to be important, and theoretically clean tests of new-physics scenarios are possible. In particular, being the SD contribution dominant, correlations of $K_S\to\mu^+\mu^-$ with $\varepsilon' /\varepsilon$ and also $K_L\to\pi^0\nu\bar\nu$ are present within many new-physics models.

Indeed, within concrete BSMs, $\mathcal{B}(K_S \to \mu^+ \mu^-)$ can be modified substantially, for example $\mathcal{B}\sim \mathcal{O}(10^{-10})$ in the leptoquark models \cite{Bobeth:2017ecx} and $\mathcal{B}\sim \mathcal{O}(10^{-11})$, or even saturate the current experimental bound in certain MSSM parameter space \cite{Chobanova:2017rkj}, albeit somewhat fine-tuned. Already the present upper bound from LHCb can have some impact on the allowed  parameter range of certain models. This shows that future improvement of this bound can have important impact on various BSM scenarios.
}

\begin{figure}[t]
\begin{center}
\vspace{-0.2cm}
\includegraphics[scale=1.0]{\main/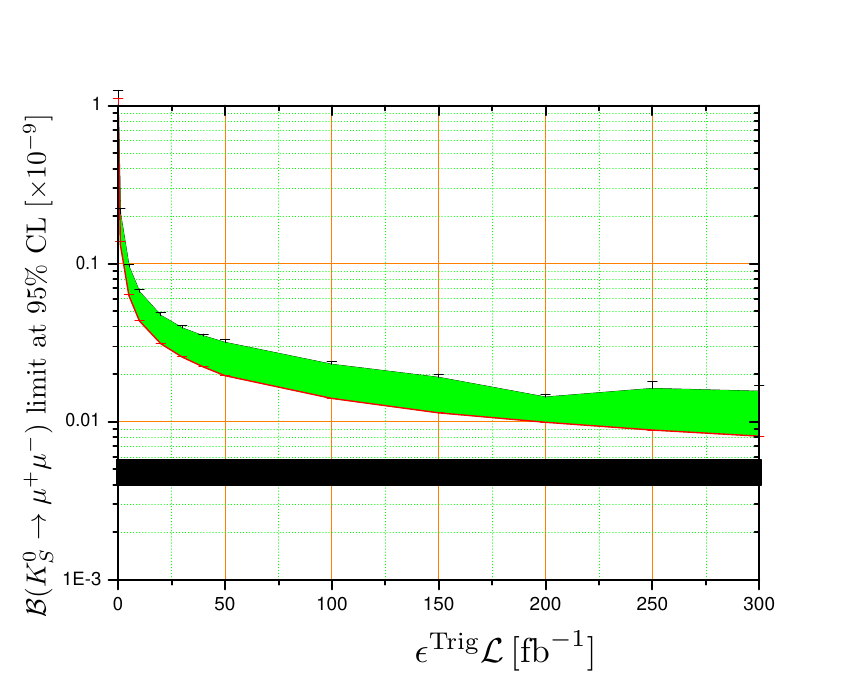}
\vspace{-0.1cm}
\caption{Projected LHCb reach for $\mathcal{B}(K_S^0\to\mu^+\mu^-)$ as a function of the integrated luminosity times trigger efficiency (which is expected to be ${\mathcal O}(1)$ in LHCb \upgradetwo. The black band is the SM prediction. Figure adapted from Ref. \cite{dmsFPCP}, which is based on the data used in~\cite{LHCb-PAPER-2017-009}.}
\label{fig:KmmSens}
\vspace{-1cm}
\end{center}
\end{figure}

The LHCb full Run1 analysis has set the  upper limit for $K_S \to \mu^+ \mu^- $ \cite{LHCb-PAPER-2017-009}, 
\begin{align}
\mathcal{B}(K_S \to \mu^+ \mu^-)_{\rm{LHCb~Run1}}< 0.8 \,(1.0)\times 10^{-9}  {\rm ~at~}   90 \%  \,(95\%)\,   {\rm CL},
\label{eq:KSmumu:exp}
\end{align}
which is $2$ orders of magnitude larger than the SM sensitivity.
With LHCb upgrades the sensitivity is significantly improved. Using the upgraded software trigger,  the LHCb experiment is aiming to reach the SM sensitivity, as shown in Fig.~\ref{fig:KmmSens}.

A crucial aspect of the $K^0 \to \mu^+ \mu^- $ decay is a flavor-tagged measurement which can probe $\CP$-violating SD contributions directly. Its numerical effect is $\mathcal{O}(1)$ compared to the prediction in Eq.~\eqref{eq:KSmumu:SM} even in the SM \cite{DAmbrosio:2017klp}.
While $K_L$ decays typically outside the LHCb fiducial volume, for $K_S$
the interference between $K_L$ and $K_S$ affects the number of signal events as 
$ \Gamma_{\rm{int.}} \propto 
\mathcal{A}(K_L \to \mu^+ {\mu}^-)  
\mathcal{A}(K_S \to \mu^+ {\mu}^- )^{\ast}
$ 
when the flavor tagging, ${K^0}$ or $\overline{K}{}^0$  at $t=0$, is performed. 
An effective branching ratio into  $\mu^+ \mu^-$,
which includes the interference correction and would correspond to experimental event numbers 
after the removal of $K_L \to \mu^+ \mu^- $ background,
is given by \cite{DAmbrosio:2017klp},
\begin{equation}
\begin{split}
\mathcal{B} (K_S \to \mu^+ \mu^-)_{\rm eff}=& 
\tau_S  \left[ \int^{t_{{\rm max}}}_{t_{{\rm min}}} d t e^{- \frac{t}{\tau_S}}\varepsilon (t)\right]^{-1}
 \int^{t_{{\rm max}}}_{t_{{\rm min}}} d t 
 \Biggl\{  \Gamma (K_{S} \to \mu^+ \mu^- )  e^{-  \frac{t}{\tau_S}}   \\
& +   \frac{ D f_K^2 m_K^3 \beta_{\mu}}{ {8} \pi} \textrm{Re}\left[  i \left( A_S A_L - \beta_{\mu}^2 {B_S^{\ast}} B_L \right) e^{ - i \Delta m_K t}  \right] e^{- \frac{ t }{ 2 \tau_S} \left( 1 + \frac{ \tau_S}{\tau_L} \right)} \Biggr\} \varepsilon (t), 
\label{eq:effBR}
\end{split}
\end{equation}
with  
$
 \Gamma (K_{S} \to \mu^+ \mu^- ) =   f_K^2 m_K^3 \beta_{\mu}
  \left( A_{S}^2 + \beta_{\mu}^2 | B_{S} |^2\right) / (16 \pi),
$
where  final-state muon polarizations are summed over, ${t_{{\rm min}}}$ to ${t_{{\rm max}}}$ corresponds to a range of detector for $K_S$ tagging,  $\varepsilon(t)$ is the decay-time acceptance of the detector,  $
\beta_{\mu} =\big(1 - 4 m_{\mu}^2 / m_K^2 \big)^{1/2}
$, 
and $f_K = ( 155.9 \pm 0.4 )$~MeV~\cite{PDG2018}.
The $K_L  \to \mu^+\mu^-$ background can be subtracted by a combination of the simultaneous measurement of $K_S \to \pi^+ \pi^- $ and the knowledge of the observed value of $\mathcal{B}(K_L \to \mu^+\mu^-)$  \cite{DAmbrosio:2017klp}. 
The dilution factor $D$ is a measure of the initial  $K^0$--$\overline{K}{}^0$  asymmetry,
\begin{align}
D= \big( K^0 - \overline{K}{}^0 \big)/ \big( K^0 + \overline{K}{}^0\big).
 \label{eq:DforK}
\end{align}
The $A_{S,L}$ and $B_{S,L}$ are the $S$-wave and $P$-wave contributions in $K_{S,L} \to \mu^+ \mu^-$ transitions, respectively. The expressions for them are given in Ref.~\cite{Chobanova:2017rkj} using the general $\Delta S= 1 $ effective Hamiltonian. Note that $A_S$ and $B_L$ are real, while $B_S$ and $A_L$ are complex.

The interference contribution is proportional to the dilution factor, $D$, which requires flavor tagging.
This can be done by detecting the accompanying 
  $K^-$  in the process  $p p \to K^0 K^{-} X  $,  $\Lambda^0$ in the process $p p \to K^0 \Lambda^0 X  $, or  $\pi^+$ in the process  $ p p \to K^{\ast +} X  \to K^0 \pi^+ X$ \cite{DAmbrosio:2017klp}.

\begin{figure}[t]
\centering 
  \includegraphics[width=0.46\textwidth]
{\main/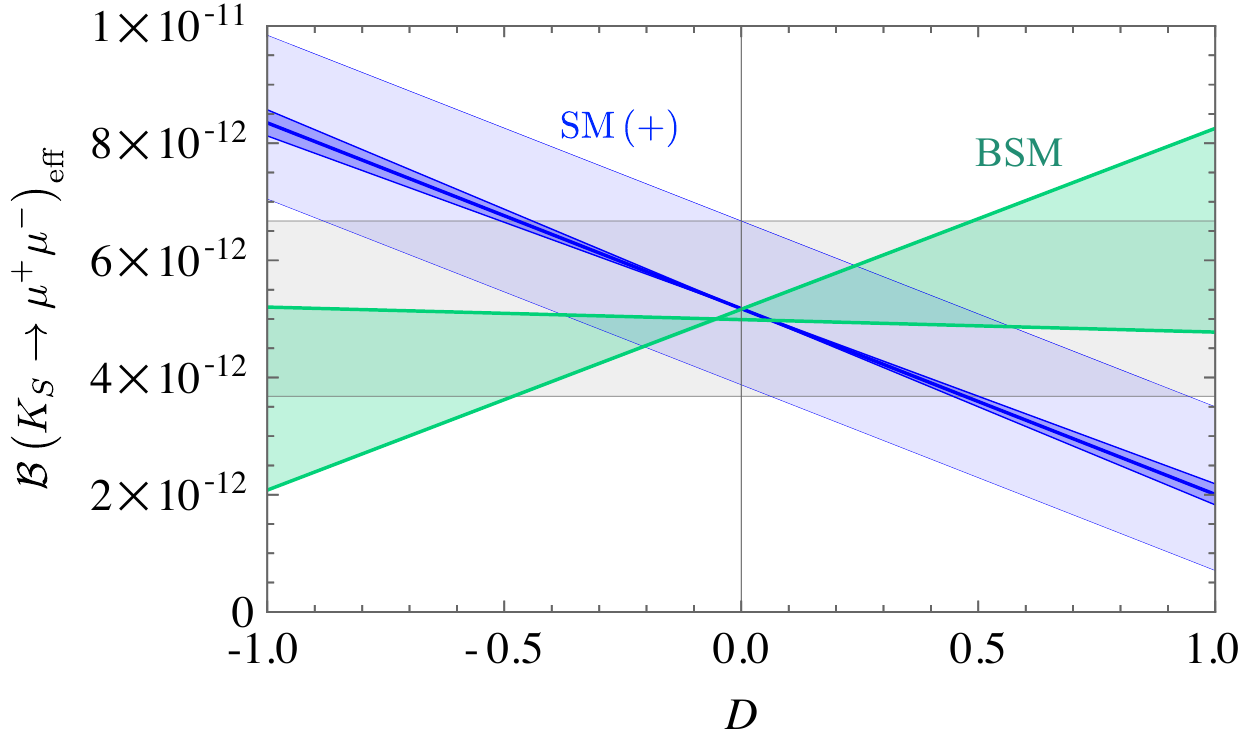}
 ~~  \includegraphics[width=0.42 \textwidth]{\main/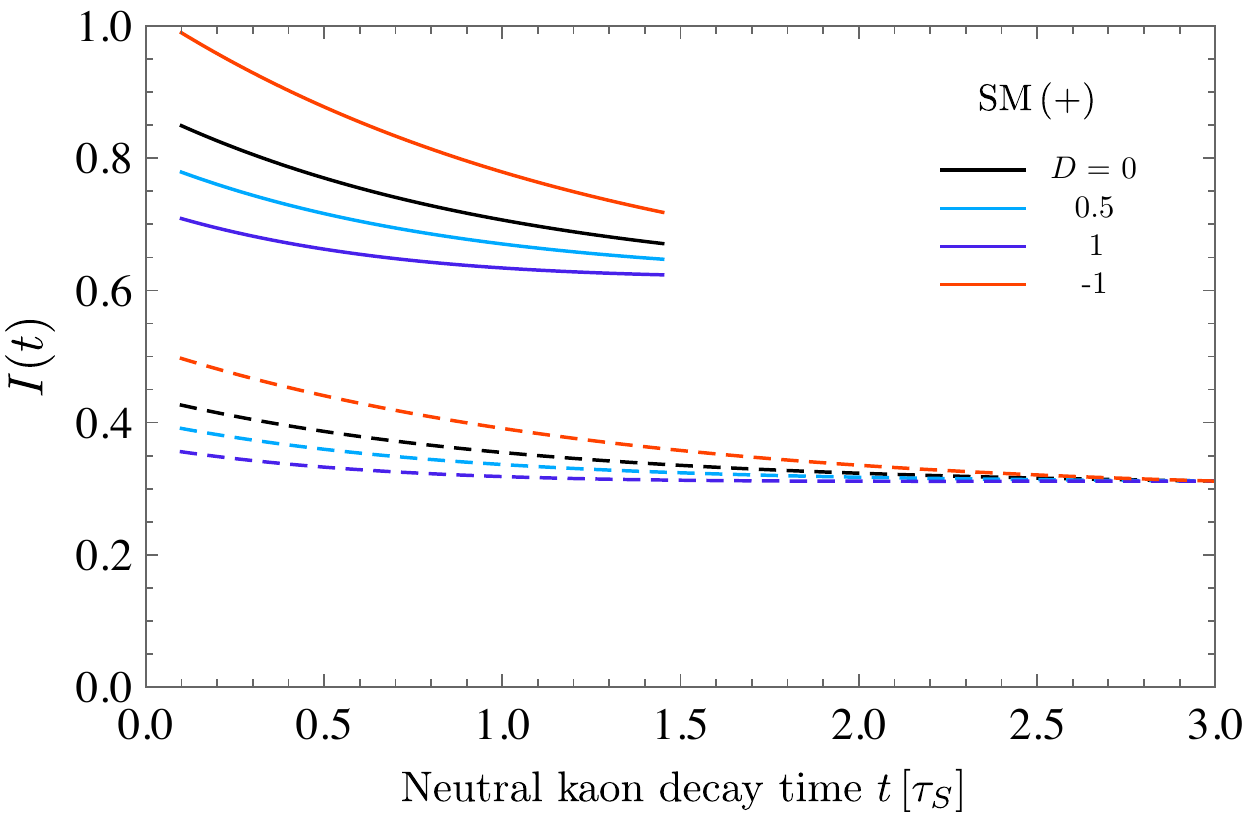}
    \caption{ 
Left panel:  The effective branching ratios 
with the blue (gray) band denoting the 
SM prediction with (without) taking into account the dilution $D$. \Blu{The green band corresponds to a BSM scenario entering through electroweak penguins studied in Ref.~\cite{DAmbrosio:2017klp}.} 
Right panel: The $K \to \mu^+ \mu^- $  time distributions in the SM for several choices of $D$.  
    The $D=0$ decay intensity is normalized, either over the interval $0.1 \tau_S$ to $1.45 \tau_S$ (solid lines) or $0.1 \tau_S$ to $3 \tau_S$ (dashed lines).
    Positive sign of $A^{\mu}_{L \gamma \gamma}$ is assumed in Eq.~\eqref{eq:ALgammagamma}.
    A detailed explanation and results for the negative sign of $A^{\mu}_{L \gamma \gamma}$ are given in Ref.~\cite{DAmbrosio:2017klp}.
}
\label{fig:time_distribution}
\end{figure}

In the SM, the effective branching ratio in Eq.~\eqref{eq:effBR} can be reduced to \cite{DAmbrosio:2017klp, Chobanova:2017rkj}
\begin{align}
A_S A_L - \beta_{\mu}^2 {B_S^{\ast}} B_L =   
\frac{4 G_F^2 M_W^2  m_{\mu}^2 }{ m_K^2 \pi^2 }  
\underbrace{\textrm{Im}C_{A,\textrm{SM}}}_{\rm SD\,(CPV)} \biggr( \underbrace{{A}^{\mu}_{L\gamma\gamma}}_{\rm LD\,(CPC)} - \frac{\pi^2}{G_F^2 M_W^2}  \underbrace{\textrm{Re} C_{A,\textrm{SM}} }_{\rm SD\,(CPC)}\biggr).
\end{align}
The Wilson coefficient $C_A$ is defined by, $\mathcal{H}_{\rm eff} = - C_A  ( \overline{s} \gamma^{\mu} P_L d )(\overline{\ell} \gamma_{\mu} \gamma_5 \ell)+ \textrm{h.c.} $, and
\begin{align}
C_{A, {\rm SM}} &= - \frac{\left[ \alpha_2(M_Z)\right]^2}{2 M_W^2}    \left(V_{ts}^{\ast} V_{td}  Y_t + V_{cs}^{\ast} V_{cd} Y_{c} \right), 
\end{align}
where $\alpha_2 = g^2 / (4 \pi)$, $Y_t = 0.950  \pm  0.049 $ and $Y_c = ( 2.95 \pm 0.46 ) \times 10^{-4}$ \cite{Gorbahn:2006bm}.
The large $\CP$-conserving LD two-photon contribution to $K_L$ is \cite{Isidori:2003ts, Mescia:2006jd}
\begin{align}
  {A}^{\mu}_{L\gamma \gamma} = \pm 2.01(1)\times 10^{-4} \times \left( 0.71(101)  - i \,5.21 \right),
  \label{eq:ALgammagamma}
\end{align}
where the sign is theoretically and experimentally unknown. 
The large imaginary (absorptive) component in $  {A}^{\mu}_{L\gamma \gamma}$ can amplify the small $\CP$-violating SD contribution in $K_S \to \mu^+ \mu^-$.
Fig.~\ref{fig:time_distribution} shows the effective branching ratio and time distributions in the SM, where
the interference \Blu{of the} $\CP$-violating contribution can affect $K_S \to \mu^+ \mu^-$ decay up to $\mathcal{O}(60)\%$. \Blu{This quantity is also sensitive to NP contributions to the electroweak penguin, that contributes to $\varepsilon^\prime/\varepsilon$ (direct $\CP$ violation in $K_L^0 \to \pi \pi$). To illustrate this correlation we plot with a green band in Fig.~\ref{fig:time_distribution} (left) the effect in the effective lifetime (as a function of $D$) of a $Z$ penguin BSM contribution that would explain the experimental value of $\varepsilon^\prime/\varepsilon$ assuming the SM predictions obtained in Refs.~\cite{Buras:2015yba,Kitahara:2016nld}. These are based on the current lattice result \cite{Bai:2015nea} or in Ref.~\cite{Buras:2015xba}, although it is important to note that there are SM calculations of $\varepsilon^\prime/\varepsilon$ consistent with the experimental data~\cite{Pallante:1999qf,Gisbert:2017vvj}.}
Some studies in new physics models are also given in Refs.~\cite{Chobanova:2017rkj,Endo:2017ums}.

Using the effective branching ratio in Eq.~\eqref{eq:effBR}, one can define the flavor-tagging asymmetry in $K_S \to \mu^+ \mu^-$ by \cite{Chobanova:2017rkj}
\begin{align}
 A_{CP} (K_S \to \mu^+ \mu^-)_{D,D'} = 
\frac{\mathcal{B} (K_S \to \mu^+ \mu^-)_{\rm eff}(D)  - \mathcal{B} (K_S \to \mu^+ \mu^-)_{\rm eff}(D')}
{\mathcal{B} (K_S \to \mu^+ \mu^-)_{\rm eff}(D)  + \mathcal{B} (K_S \to \mu^+ \mu^-)_{\rm eff}(D')},
\end{align}
where $D'$ is obtained by requiring an opposite flavor tagging.
For instance, when a positive dilution factor is achieved by collecting the di-muon signals accompanying $K^-$,
the negative dilution factor, that is expressed as $D'$, can be obtained by collecting the signals accompanying $K^+$. 
This asymmetry is a theoretically clean quantity that emerges from a genuine direct $\CP$ violation in general new-physics models. 
In the SM, $ A_{CP} (K_S \to \mu^+ \mu^-)^{\textrm{SM}}_{D,-D} = \mathcal{O}(0.7)\times D$ is predicted in the case of $D' = -D$ 
and it is sensitive to a new  \CP -violating phase beyond the SM  \cite{Chobanova:2017rkj}. 

In a similar way, the $\CP$ asymmetry of $B_{d,s} \to \mu^+ \mu^- $ has been studied  \cite{DeBruyn:2012wk,Buras:2013uqa}. However, for the $B_s$ system the situation differs substantially from $K_S$, since 
the LD contributions are negligible and the life-time difference between the two mass eigenstates is small compared to $K_{S,L}$. 
The $\CP$ asymmetry in $B_{d,s} \to \mu^+ \mu^- $ vanishes in the SM, 
but is also sensitive to a new  \CP -violating phase.

Before closing this section, we briefly comment on $K_L \to e^+ e^-$.
Both $K_L \to e^+ e^-$ and $K_L \to \mu^+ \mu^-  $ are dominated by LD two-photon contributions \cite{GomezDumm:1998gw,Cirigliano:2011ny}.
The branching ratio for $K_L \to e^+ e^- $ is dominated by the double logarithm contribution,  $\propto \log ^2 (m_e/m_K)$. 
The subdominant   local term is fixed,  up  to a two-fold ambiguity, from the
measured $\mathcal{B}(K_L \to \mu^+ \mu^- )$;  so that $\mathcal{B}(K_L \to e^+ e^- )/\mathcal{B}(K_L \to \gamma\gamma )=(1.552\pm 0.014)\times 10^{-8}$ or 
$\mathcal{B}(K_L \to e^+ e^- )/\mathcal{B}(K_L \to \gamma\gamma )=(1.406\pm 0.013)\times 10^{-8}$ is predicted  \cite{Cirigliano:2011ny}.
The measured value $(1.65\pm0.91)\times 10^{-8}$ \cite{PDG2018} is not yet precise enough to resolve the ambiguity.

\subsection{$K_S \to  \mu^+ \mu^- \gamma $, ~$K_S \to \mu^+ \mu^- e^+ e^-$  ~and ~$K_S \to \mu^+ \mu^- \mu^+ \mu^-$}

To improve the LD determination of $\mathcal{B}(K_S \to \mu^+ \mu^-)$, it is important  to measure $K_S \to  \mu^+ \mu^- \gamma $, $K_S \to \mu^+ \mu^- e^+ e^-$  and $K_S \to \mu^+ \mu^- \mu^+ \mu^-$.
These channels are at reach for  LHCb \upgradetwo
and thus may give necessary LD  information needed for a better control of $K_{L}\rightarrow \mu^+ \mu^-$. 
These four body decays have a peculiar feature: similarly to $K_{S,L}\rightarrow \pi ^+ \pi ^- e^+ e^- $ \cite{Cirigliano:2011ny},
the two different helicity amplitudes interfere. 
One can then measure the sign of $ K_{L} \rightarrow \gamma ^\ast \gamma ^\ast \rightarrow  \ell^+ \ell^- \ell^+ \ell^-$  by studying  the time interference between $K_{S}$ and $ K_{L}$, which has a decay length	$2 \Gamma_S$ \cite{DAmbrosio:2013qmd}.

The 
ChPT prediction for $\mathcal{B}(K_S \to \gamma\gamma)$~\cite{DAmbrosio:1986zin}, which has been experimentally confirmed, allows one to make a  prediction for $\mathcal{B}(K_S \to  \mu ^+  \mu ^-  \gamma)=7.25\times 10^{-10}$; 
this value is increased by vector meson dominance (VMD) and unitarity corrections to
$\mathcal{B}(K_S \to \mu ^+  \mu^- \gamma  )=(1.45\pm 0.27)\times 10^{-9}$ \cite{Colangelo:2016ruc}.

\subsection{$K_S\to \pi^0 \ell^+ \ell^- $ and $K^\pm \to \pi^\pm \ell^+ \ell^- $}

At low dilepton mass squared, $q^2$,
the dominant contribution to $K^\pm (K_{S}) \rightarrow \pi ^{\pm} 
(\pi ^{0})\ell ^{+}\ell ^{-}$ 
is due to a single virtual-photon exchange.
The resulting amplitude involves a vector form factor $V_i(z)$  ($i=\pm,S$),
which  can be decomposed in the general form up to ${\mathcal O}(p^6)$ in the chiral expansion as~\cite{DAmbrosio:1998gur,DAmbrosio:2018ytt}, 
\begin{equation}
  V_i(z) = a_i+ b_i z + V_i ^{\pi\pi}(z)\,, \quad z = q^2 / m_K^2\,,
  \quad \textrm{for}\quad i=\pm,S.
  \label{V_dec}
\end{equation}
Here the low energy constants (LECs), $a_i$ and $b_i$, parametrize the polynomial part of the amplitude, while the rescattering contribution $V_i^{\pi\pi}$ can be determined from fits to $K\to\pi\pi$ and $K\to\pi\pi\pi$ data (no $\Delta I=1/2$ contribution to $V_S^{\pi\pi}$). \Blu{The chiral loops, encoded in $V_i^{\pi \pi}$, have been computed, model-independently and
at leading order, in the framework of chiral perturbation theory~\cite{DAmbrosio:1998gur, DAmbrosio:2018ytt}.}
Chiral symmetry alone does not constrain the values of the LECs,
so instead, we consider the differential decay rate $d \Gamma / dz \propto |V_+(z)|^2$ 
as a means to extract $a_+$ and $b_+$ from experiment. 
The resulting fit to the decay spectra from all available high-statistics experiments is given in Table~\ref{tab:a+b+}.
The experimental size of the $b_{+}/a_{+}$ ratio 
exceeds the naive dimensional analysis estimate, calling  
for a large VMD contribution. 

\begin{table}[t]
\renewcommand{\arraystretch}{1.3}
\centering
\caption{Fitted values of coefficients entering the vector form factor in Eq.~\eqref{V_dec}.}
\label{tab:a+b+}
\begin{tabular}{cccr}
\hline\hline
Channel & $a_+$ & $b_+$ & Reference \\
\hline
$ee$ & $-0.587\pm 0.010$ & $-0.655\pm 0.044$ & E865~\cite{Appel:1999yq}\\
$ee$ & $-0.578\pm 0.016$ & $-0.779\pm 0.066$ & NA48/2~\cite{Batley:2009aa}\\
$\mu\mu$ & $-0.575\pm 0.039$ & $-0.813\pm 0.145$ & NA48/2~\cite{Batley:2011zz}\\
\hline\hline
\end{tabular}
\end{table}

\begin{figure}[t]
\begin{center}
\vspace{-0.2cm}
\includegraphics[scale=0.8]{\main/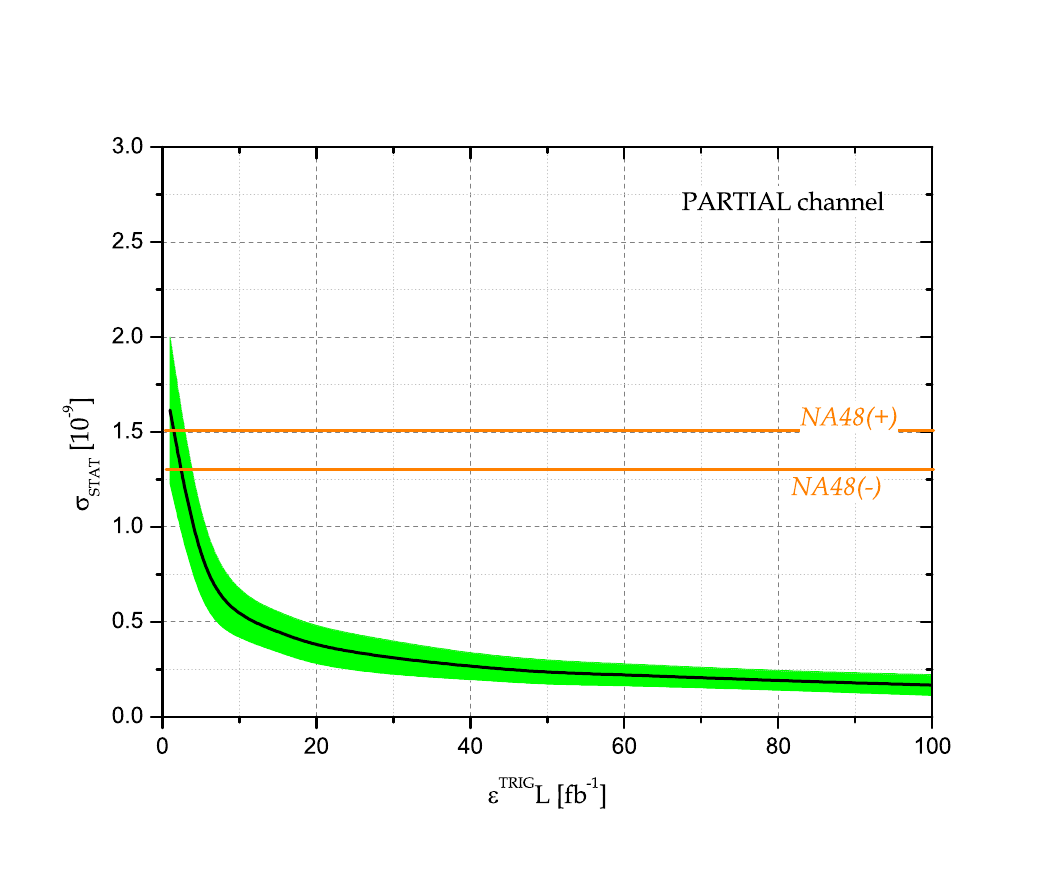}
\vspace{-0.1cm}
\caption{Projected LHCb reach for $\mathcal{B}(K_S^0\to\pi^0\mu^+\mu^-)$ as a function of the integrated luminosity times trigger efficiency (which is expected to be ${\mathcal O}(1)$ in LHCb \upgradetwo. Figure adapted from Ref. \cite{Chobanova:2195218}.}
\label{fig:KpzSens}
\vspace{-1cm}
\end{center}
\end{figure}

The branching ratios of $K_S \to \pi^0 \ell^+ \ell^-$, on the other hand, are  approximately 
\begin{equation}
 \mathcal{B}(K_S \rightarrow \pi^0 e^+ e^-)   
   \approx     5 \times 10^{-9} \cdot a_S^2, \qquad  
   \mathcal{B}(K_S \rightarrow \pi^0 \mu ^+  \mu ^-)  
   \approx     1.2  \times 10^{-9}  \cdot  a_S^2,
\label{eq:BRKS}
\end{equation}
and   NA48, assuming just a VMD form factor, finds respectively \cite{Batley:2003mu,Batley:2004wg} 
 \begin{equation}
  | a_S|_{e e } = 1.06 ^{+0.26} _{-0.21} \pm 0.07, 
 \qquad
 |a_S|_{\mu \mu } = 1.54 ^{+0.40} _{-0.32} \pm 0.06.
\end{equation}
The uncertainty of $|a_S|_{\mu \mu }$ gives the dominant theoretical uncertainty in $K_L \to \pi^0 \mu^+ \mu^-$.
Therefore,
this measurement is a crucial piece of information required to establish the relative roles of indirect $\CP$ violation vs. direct $\CP$ violation  in $ K_L \rightarrow \pi^0 \mu ^+  \mu ^-$, in order to probe SD effects \cite{Mescia:2006jd}.
The LHCb \upgradetwo can reach a precision in $\mathcal{B}(K_S^0\to\pi^0\mu^+\mu^-)$  at the $10^{-10}$ level (see Fig. \ref{fig:KpzSens}), and, through an analysis of the differential decay rate, a $10\%$ statistical precision on the form factor term $|a_S|$ with free $b_S$~\cite{kaon_paper}.

\subsubsection{Lepton flavor universality violation in $K^\pm \to \pi^\pm \ell^+\ell^-$}
The coefficients $a_+$ and $b_+$  can be used to test for LFUV. 
 If lepton flavour universality applies, each of the two coefficients, averaged over different experiments, have to be equal for the $ee$ and $\mu\mu$ channels. Within errors this is indeed the case, see Table \ref{tab:a+b+}.  Since the SM interactions are lepton flavour universal, deviations from zero in differences, such as $a_+^{\mu\mu} - a_+^{ee}$, would  be a sign of NP. 
\Blu{Such a test is} especially interesting in view of the $B$-physics anomalies (see Sec.~\ref{secBanom}), with rare kaon decays providing a complementary role in testing the NP explanations.  Our analysis~\cite{Crivellin:2016vjc} is based on the observation that at low energy scales, $\mu \ll m_{t,b,c}$, 
the strangeness-changing transitions are described in terms of the effective Lagrangian~\cite{Cirigliano:2011ny}
\begin{equation}
\label{Lagr_K}
  {\cal L}_\text{eff}^{|\Delta S|=1} = 
  - \frac{G_F}{\sqrt{2}} V_{ud}V_{us}^* \sum_{i} C_i(\mu) Q_i(\mu) + \text{h.c.} \,,
\end{equation}
which contains semileptonic operators
\begin{align}\label{eq:Q11-12}
Q_{7V} =   \left[\bar s \gamma^\mu (1 - \gamma_5) d \right]
\sum_{\ell=e,\mu} \left[\bar{\ell} \gamma_\mu  \ell \right], \quad \mbox{and} \quad
Q_{7A} = \left[\bar s \gamma^\mu (1 - \gamma_5) d \right]
\sum_{\ell=e,\mu}  \left[\bar{\ell} \gamma_\mu \gamma_5 \ell \right], 
\end{align}
that are the $s\to d$ analogues of the $b\to s$ operators, $Q_{9,10}^B$.  In the framework of minimal flavour violation (MFV), the Wilson coefficients of the two sectors are correlated. We use this feature to convert knowledge of $C_{7V,7A}$ into bounds on $C_{9,10}^B$.  

 To convert the allowed range on $a_+^\text{NP}$ into a corresponding range in the Wilson coefficients $C_{7V}^{\ell\ell}$, we make use of the ${\mathcal O}(p^2)$ chiral realization of the $SU(3)_L$ current
\begin{equation}
  \bar{s}\gamma^\mu(1-\gamma_5)d \leftrightarrow i f_\pi^2 (U\partial^\mu U^\dagger)_{23}\,, 
  \qquad U = U(\pi,K,\eta)\,, 
  \label{bosonize}
\end{equation}
to obtain 
\begin{equation}
 a_+^\text{NP}=\frac{2\pi\sqrt{2}}{\alpha}V_{ud}V_{us}^*C_{7V}^\text{NP}\,.
 \label{aNP}
\end{equation}
Contributions due to NP in $K^+ \to \pi^+ \ell^+\ell^-$ can then be probed by considering the difference between the two channels
\begin{equation}
\label{limit_Kp}
 C_{7V}^{\mu\mu}-C_{7V}^{ee}=\alpha\frac{a_+^{\mu\mu}-a_+^{ee}}{2\pi\sqrt{2}V_{ud}V_{us}^*}\,.
\end{equation}
Assuming MFV, this can be converted into a constraint on the NP contribution to $C_9^B$,
\begin{equation}
\label{C_charged}
 C_{9}^{B,\mu\mu}-C_{9}^{B,ee}=-\frac{a_+^{\mu\mu}-a_+^{ee}}{\sqrt{2}V_{td}V_{ts}^*}\approx -19\pm 79\,,
\end{equation}
where we have averaged over the two electron experiments listed in Table~\ref{tab:a+b+}.

The determination of $a_+^{\mu\mu}-a_+^{ee}$ needs to be improved by an ${\mathcal O}(10)$ factor in order to probe the parameter space relevant for the $B$-anomalies, which require Wilson coefficients $C_{9,10}^B={\mathcal O}(1)$~\cite{Descotes-Genon:2015uva}. Improvements of this size may be possible at NA62, especially for the experimentally cleaner dimuon mode which currently has the larger uncertainty.

\subsection{$K_S\to \pi^+ \pi^- e^+ e^-$}
The $  K_S \rightarrow \pi^+ \pi^- e^+ e^- $ decays 
can be interesting, if one can test  beyond the dominant bremsstrahlung contribution, 
performing  ChPT tests. %
In principle, 
$\CP$ violation is also of interest for NP searches.
So far,  NA48   has, using 676 events, obtained a measurement
$\mathcal{B}(  K_S \rightarrow \pi^+ \pi^- e^+ e^- )=(4.79\pm 0.15)\times 10^{-5}$, 	 		
\cite{PDG2018} 
to be compared with the theoretical prediction \cite{Cappiello:2017ilv}
\begin{align}
\mathcal{B}(K_S\to \pi^+\pi^-e^+e^-)=\underbrace{4.74\cdot 10^{-5}}_{\text{Brems.}}+\underbrace{4.39\cdot 10^{-8}}_{\text{Int.}}+\underbrace{1.33\cdot 10^{-10}}_{\text{DE}}~.
\end{align}
Similarly, one can predict for the dimuon channel, 
\begin{align}
\mathcal{B}(K_S\to \pi^+\pi^-\mu^+\mu^-)=\underbrace{4.17\cdot 10^{-14}}_{\text{Brems.}}+\underbrace{4.98\cdot 10^{-15}}_{\text{Int.}}+\underbrace{2.17\cdot 10^{-16}}_{\text{DE}}~.
\end{align}
The LHCb upgrade expects a yield of up to 5$\times10^{-4}$ events per fb-1 ~\cite{MarinBenito:2193358}.

\subsection{LFV modes}

Modes with LFV, such as $K \to (n \pi) \mu^\pm e^\mp$, provide null tests of the SM. The interest on such processes has been renewed because they can receive sizable contributions from BSM addressing the hints for LFUV in $B \to K^{(\ast)} \ell^+ \ell^-$. Both types of processes can arise from NP contributions to the product of the two neutral currents, composed of the down-type quarks and charged leptons. The only difference between the two is the strength of the flavour couplings involved.
Using general EFT arguments, the amount of LFUV hinted at in $B \to K^{(\ast)} \ell^+ \ell^-$ decays, generically imply $B \to K^{(\ast)}$ LFV rates of the order of $10^{-8}$ \cite{Glashow:2014iga}. (More quantitative estimates require introduction of flavour models \cite{Guadagnoli:2015nra,Boucenna:2015raa,Celis:2015ara,Alonso:2015sja,Gripaios:2015gra,Crivellin:2016vjc,Becirevic:2016zri,Becirevic:2016oho,Hiller:2016kry,Becirevic:2017jtw,Buttazzo:2017ixm,King:2017anf,Bordone:2018nbg}.) 

\begin{figure}[t]
\begin{center}
\vspace{-0.2cm}
\includegraphics[scale=0.5]{\main/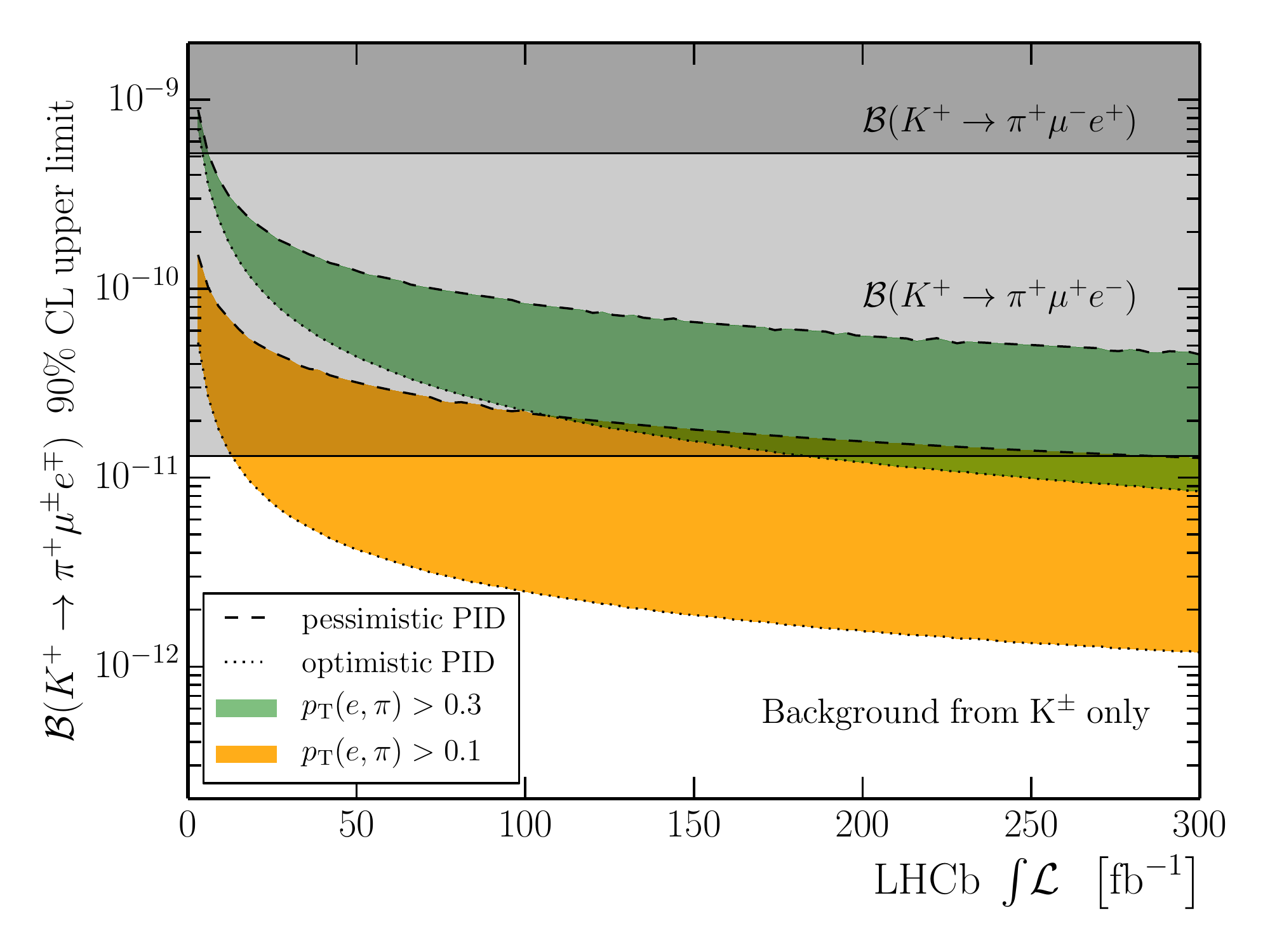}
\vspace{-0.1cm}
\caption{Expected reach for $K^+\to \pi^+\mu^\pm e^\mp$ as a function of the integrated luminosity with 13~TeV $pp$ collisions based on a fast simulation of LHCb. Different scenarios in terms of PID performance and $p_{\rm T}$ thresholds of the $\pi^+$ and $e^{\pm}$ candidates are shown. Combinatorial background is neglected in the study. Figure adapted from Ref. \cite{Borsato:2018tcz}.}
\label{fig:K_LFV}
\vspace{-1cm}
\end{center}
\end{figure}

As discussed in Ref. \cite{Borsato:2018tcz}, such arguments can be extended to $K \to (\pi) \mu^\pm e^\mp$, with fairly general assumptions on different flavour couplings involved. Expected rates can be as large as $10^{-10} - 10^{-13}$ for the $K_L \to \mu^\pm e^\mp$ mode and a factor of $\sim 100$ smaller for the $K^+ \to \pi^+ \mu^\pm e^\mp$ one. Taking into account the suppression mechanisms at play, such ``large'' rates are a non-trivial finding. The relatively wide predicted ranges are due to the inherent model dependence, especially in the choice of the leptonic coupling and of the overall scale of the new interaction, typically between $5$ and $15$ TeV \cite{Borsato:2018tcz}. On the experimental side, the limits on $K \to (\pi) \mu^\pm e^\mp$ modes are, somewhat surprisingly, decades-old
\begin{eqnarray}
\label{eq:limits}
\begin{tabular}{clcl}
$\mathcal B(K_L \to e^\pm \mu^\mp) < 4.7 \times 10^{-12}$ & \hspace{-0.3cm}\cite{Ambrose:1998us}~, &~~~~
$\mathcal B(K_L \to \pi^0 e^\pm \mu^\mp) < 7.6 \times 10^{-11}$ & \hspace{-0.3cm}\cite{Abouzaid:2007aa}~,\\
[0.1cm]
$\mathcal B(K^+ \to \pi^+ e^- \mu^+) < 1.3 \times 10^{-11}$ & \hspace{-0.3cm}\cite{Sher:2005sp}~, &~~~~
$\mathcal B(K^+ \to \pi^+ e^+ \mu^-) < 5.2 \times 10^{-10}$ & \hspace{-0.3cm}\cite{Appel:2000tc}~.\\
\end{tabular}
\end{eqnarray}
These modes can be profitably pursued at the upgraded LHCb, which will benefit from huge yields. %
Ref. \cite{Borsato:2018tcz} presented a feasibility study for the modes in Eq. (\ref{eq:limits}), taking $K^+ \to  \pi^+ \mu^\pm e^\mp$ as a benchmark. The expected reach is displayed in Fig. \ref{fig:K_LFV} as a function of the integrated luminosity and for different scenarios of detector performance. This shows that LHCb \upgradetwo could update some of the existing limits, a result that per se already makes these searches promising. 
Even more interestingly, LHCb \upgradetwo 
 could probe part of the parameter space for LFV kaon decays suggested by the $B$-physics anomalies. This conclusion in turn calls attention to other running and upcoming facilities, including Belle-II, NA62~\cite{NA62:2312430}, as well as the newly proposed TauFV~\cite{PBCTalk}. %
Dedicated sensitivity studies for these facilities are required in order to make more quantitative statements. Yet, the experimental outlook for all these modes is certainly promising.

\subsection{Hyperons at HL-LHC}

The LHCb can contribute significantly to the strangeness-physics programme with measurements of hyperon decays. There is vast room for improvement in this sector as most of the data for the standard modes, both in nonleptonic and semileptonic decays, is about 40 years old, and many of the rare decays sensitive to SD physics have not even been searched for. 
An exploratory study of the LHCb's potential in hyperon physics was presented in~\cite{Junior:2018odx}; here we summarize the main conclusions. 

Current experimental data on the semileptonic hyperon decays $\Lambda\to p\mu^-\bar\nu$, $\Xi^-\to\Lambda\mu^-\bar\nu$ and $\Xi^-\to\Sigma^0\mu^-\bar\nu$ is quite poor, with relative uncertainties in the range of 20\%-100\%. These decay modes can be partially reconstructed at the LHCb, where the kinematic distributions allow one to discriminate from the peaking backgrounds of $\Lambda\to p\pi^-$ and $\Xi^-\to\Lambda\pi^-$. Besides testing lepton-universality by comparing with the semi-electronic modes, these decays are sensitive to BSM scalar and tensor currents~\cite{Chang:2014iba}. 
If percent precision is achieved in the measurement of semi-muonic branching fractions, they could contribute to a better determination of the CKM matrix element $|V_{us}|$ from hyperon decays~\cite{Cabibbo:2003cu,Cabibbo:2003ea,Mateu:2005wi,Sasaki:2017jue}. 

A golden mode among the rare hyperon decays is $\Sigma^+\to p \mu^+\mu^-$, to which LHCb recently contributed with an evidence for the decay at 4.1$\sigma$ and a di-muon invariant distribution consistent with the SM~\cite{LHCb-PAPER-2017-049}, thus challenging the HyperCP anomaly~\cite{Park:2005eka}. With dedicated triggers and the upgraded LHCb detector in the HL phase about a thousand events per year of data-taking could be measured, which will allow one to measure angular distributions or direct $\CP$ asymmetries that have been shown to be sensitive to SD physics~\cite{Meinel:2017ggx,He:2018yzu}. The equivalent channel with lepton number violation, $\Sigma\to\bar{p}\mu^+\mu^+$ can also be searched for, with potential sensitivities at the $10^{-9}$ level. Other $\Delta S=1$ semileptonic rare hyperon decays, except those of the $\Omega^-$ (which will be produced at a rate similar to that of a $B$-meson), must have electrons in the final state due to phase space. Sensitivity to $\mathcal B(\Sigma^+\to p e^+e^-)\sim 10^{-6}$ should be accessible at LHCb, whereas other modes like $\Lambda\to p \pi^- e^+e^-$ suffer from low electron reconstruction efficiency. More exotic modes, probing baryon-number violation, lepton-flavor violation, or $\Delta S=2$, can also be measured with projected sensitivities improving by orders of magnitude the current limits~\cite{Junior:2018odx}. 
Radiative decays such as $\Lambda \to p \pi \gamma$ are also accessible to LHCb.

%% file: section5/section.tex
\section[Tau leptons]{Tau leptons}
{\small\it Authors (TH):  Vincenzo Cirigliano, Martin Gonzalez-Alonso, Adam Falkowski, Emilie Passemar.}
\label{sec:5}

The physics of the tau lepton is an exceptionally broad topic, both experimentally and theoretically. A large variety of processes involving taus have been used in the last decades to learn about fundamental physics~\cite{Pich:2013lsa}. 
In some cases the results obtained in $e^+e^-$ machines~\cite{Schael:2005am} are very hard to improve in the LHC environment, due to large backgrounds, even if the number of taus produced is much larger. This is the case, {\it e.g.}, for the study of basic tau properties (mass, lifetime, etc.) or the standard tau decay channels. However, there are also many processes involving taus that are relevant for HL/HE-LHC. Here we focus on those that are more directly connected to flavour and that are not covered by other sections (for instance $h \to \tau \tau$, $h\to\tau\ell$, Sec.~\ref{sec:9} or heavy meson decays to taus, Sec.~\ref{sec:7} ).

To simplify the discussion we will assume 
that the scale of New Physics  (NP) relevant for the processes under consideration is much heavier than the energy scales probed by the  LHC. We can then use the SMEFT framework~\cite{Buchmuller:1985jz}. We first discuss lepton flavour conserving observables, where the SM contribution has to be calculated with some accuracy in order to probe NP. In contrast, the SM contribution is negligible for lepton flavour violating processes, where the challenges are purely experimental.
As a result, the NP scales that are probed are much higher  in the latter case.

\subsection{Lepton-flavour-conserving processes}

Collider phenomenology is usually very different for the so-called vertex corrections and contact interactions. For vertex corrections the NP contributions mimic the structure of the SM gauge couplings, $Z\tau\tau$ and $W\tau\nu$. Their study requires, in most cases, very precise measurements, which are challenging in a hadron collider. On the other hand, four-fermion contact interactions (generated for example by the tree-level exchange of a heavy mediator) give a contribution to high-energy observables qualitatively different from the SM one. For instance, a contribution from an off-shell mediator grows with the partonic center of mass energy,
and thus one does not require very high precision in order for the measurements to probe interesting NP scales~\cite{Cirigliano:2012ab,deBlas:2013qqa}. For these contact interactions, we focus on flavour-diagonal couplings involving first-generation quarks and third-generation leptons.

\subsubsection{High-energy tails}
LHC measurements of differential distributions in the Drell-Yan lepton production can be sensitive to new effective interactions between leptons and quarks~\cite{Cirigliano:2012ab,deBlas:2013qqa}. 
As an example, consider that the following dimension-6 interaction is added to the SM Lagrangian,
\begin{equation}
\label{eq:olq}
{\cal L} \supset {C_{lq}^{(3)} \over \Lambda^2} (\bar L_3 \gamma_\mu  \sigma^i L_3)  (Q_1 \gamma^\mu \sigma^i Q_1), 
\end{equation}  
where $\sigma^i$ are the Pauli matrices, $L_3 = (\nu^\tau_L,\tau_L)$ is the 3rd generation lepton doublet, and $Q_1 = (u_L,d_L)$ is the 1st generation quark doublet. This interaction term can be interpreted as an effective field theory (EFT) description of a more fundamental theory, for example an exchange of an $SU(2)$ triplet of vector bosons with masses $m_V = \Lambda \gg v$ that couples to $L_3$ and $Q_1$ with coupling strength $g_{V} \approx  2 |C_{lq}^{(3)}|^{1/2}$.  
Even when $m_V$ is too heavy to be directly produced at the LHC, the effective interaction in Eq.~(\ref{eq:olq}) may produce observable effects, and therefore provide
information about the theory that completes the SM. 

At the LHC, contact interactions between quarks and 3rd generation leptons can be probed via the processes $p p \to \tau^+ \tau^- X$ and $p p \to \tau \nu X$~\cite{Faroughy:2016osc,Cirigliano:2018dyk}. 
Here we focus on the latter, taking as a template the existing Run-2 ATLAS analysis in Ref.~\cite{Aaboud:2018vgh}. 
This measured  the  $\tau \nu$ transverse mass
distribution, where the transverse mass is given by
$m_T =\sqrt{ 2 \pt E_T (1-\cos \phi) }$, with $\vec \pt$  the transverse momentum of the (hadronic) tau candidate,  $\vec E_T$ the missing transverse momentum, and $\phi$ the azimuthal angle between the two. 
The effects of the interaction in Eq.~(\ref{eq:olq}) are most pronounced at higher $m_T$.
To estimate the current and future sensitivity, we pick $m_T > 1.8$~TeV as our signal region (a more elaborate analysis with a wider, binned, $m_T$ range would lead to stronger bounds~\cite{Cirigliano:2018dyk}).
Using the {\tt Madgraph}\cite{Alwall:2014hca}/{\tt Pythia~8}~\cite{Sjostrand:2014zea}/{\tt Delphes} \cite{deFavereau:2013fsa} simulation chain and imposing the selection cuts we can estimate the expected number of events in the presence of the interaction in Eq.~(\ref{eq:olq}): 
\begin{equation}
N_{p p \to \tau \tau}^{m_T > 1.8~\tev} \approx  \left ({ {\cal L} \over 36.1~{\rm fb}^{-1} } \right )
 \left [ 0.36  + 4.0 \times 10^3 {C_{lq}^{(3)} v^2 \over \Lambda^2} + 3.6 \times 10^5  \left ( {C_{lq}^{(3)} v^2 \over \Lambda^2}  \right )^2 \right ], 
\end{equation}
 where $ {\cal L}$ is the integrated luminosity, $v \approx 246$~GeV, and we will ignore the theoretical error of this estimation. 
The ATLAS collaboration observed $0$ events in  ${\cal L} = 36.1~{\rm fb}^{-1}$~\cite{Aaboud:2018vgh},  from which  one can derive the 68\%~CL limit on the Wilson coefficient in Eq.~(\ref{eq:olq}):  $- (5.3~\tev)^{-2} \leq  C_{lq}^{(3)} / \Lambda^2 \leq (7.7~\tev)^{-2}$.  
For the HL-LHC with ${\cal L} = 3000~{\rm fb}^{-1}$,  assuming the observed number of events will be exactly equal to the SM prediction, the expected limit becomes
\beq
- {1 \over (17.5~\tev)^2} \leq  {C_{lq}^{(3)} \over \Lambda^2} \leq {1 \over (20.5~\tev)^2} \qquad @~68\%~{\rm CL}. 
\eeq 
Compared to the present LHC bound, this represents an $\cO(10)$ stronger bound on $C_{lq}^{(3)}/\Lambda^2$, or $\cO(3)$ times improvement in the reach for the mass scale of NP. 
Assuming maximally strongly coupled NP, $g_V \sim 4 \pi$, the HL-LHC can probe particles even as heavy as $m_V \sim 100$~TeV. 

There are 3 more independent dimension-6 operators that can be probed by the $p p \to \tau \nu$ process: 
 $O_{l e d q}  =  (\bar L_3 \tau_R)  (\bar d_R  Q_1)$,  $O_{l e q u} =  (\bar L_3 \tau_R) (Q_1 u_R)$, and  
 $O^{(3)}_{l e q u}  =   (\bar L_3  \sigma_{\mu \nu} \tau_R)  (\bar Q_1 \sigma^{\mu \nu} u_R)$~\cite{Cirigliano:2012ab,Grzadkowski:2010es}. 
We estimate that the HL-LHC will be sensitive to $C / \Lambda^2 \sim (10~\tev)^{-2}$ through a one-bin analysis like the one presented above. 
The slightly smaller sensitivity than for the operator in Eq.~(\ref{eq:olq}) follows from the fact that these do not interfere with the SM amplitudes, and thus their effect enters only at quadratic order in $C/\Lambda^2$.~\footnote{Contributions from SMEFT dim-8 operators are assumed to be subleading with respect to dim-6 squared terms. Although not true in general, this is indeed the case for strongly coupled UV completions~\cite{Contino:2016jqw}.}

We expect that our simple analysis provides a good qualitative estimate of the HL-LHC reach.  
However, a more sophisticated analysis using the full information about the $m_T$ distribution (such as in Ref.~\cite{Cirigliano:2018dyk}), and possibly also about the $\tau$ polarization, should lead to a further $\cO(1)$ increase of sensitivity. 
Ideally, the optimal analysis should be able to distinguish between different dimension-6 operators (except between $O_{l e d q}$ and  $O_{l e q u}$, which give exactly the same contribution to the $p p \to \tau \nu$ cross section), and provide constraints on the Wilson coefficients in the situation when all the independent operators are simultaneously present.  
Furthermore, the process $p p \to \tau^+ \tau^-$ probes a large set of dimension-6 operators: 
$O_{l q} =  (\bar L_3 \gamma_\mu  L_3)  (Q_1 \gamma^\mu Q_1)$, 
$O_{l u} =  (\bar L_3 \gamma_\mu  L_3)    (\bar u_R \gamma^\mu u_R)$,   
$O_{l d} =  (\bar L_3 \gamma_\mu  L_3)    (\bar d_R \gamma^\mu d_R)$,   
$O_{e q} =  (\bar \tau_R \gamma_\mu  \tau_R)    (\bar Q_1 \gamma^\mu Q_1)$,    
$O_{e u} =  (\bar \tau_R \gamma_\mu  \tau_R)  (\bar u_R \gamma^\mu u_R)$, and
$O_{e d} =  (\bar \tau_R \gamma_\mu  \tau_R)$  $(\bar d_R \gamma^\mu d_R)$, in addition to the operators discussed above. 
We expect comparable sensitivity of  $p p \to \tau^+ \tau^-$ to $C/\Lambda^2$ as in the case of $p p \to \tau \nu$~\cite{Cirigliano:2012ab}. 
It is unlikely that the LHC alone can discriminate between all these operators; to this end combining with low energy precision measurements of $\tau$ decays will be necessary. 
Finally, we mention that $\tau \tau$ and $\tau \nu$ production also probes analogous operators with heavier ($s$, $c$, $b$) quarks instead of the 1st generation ones \cite{Faroughy:2016osc,Altmannshofer:2017poe,Greljo:2018tzh}. 
The presence of such operators in the Lagrangian can be motivated by the anomalies observed by BaBar and LHCb in the $B \to D^{(*)} \tau \nu$ decays~\cite{Lees:2013uzd,Huschle:2015rga,Sato:2016svk} (see Sec~\ref{sec:7}).

\subsubsection{Beyond tails: lepton flavour universality}
It has been pointed out recently that there is an interesting complementarity between the  high-energy searches and low-energy precision studies, presented above, and hadronic tau decays~\cite{Cirigliano:2018dyk}. The latter are equally sensitive to vertex and contact interactions, and thus effectively become model-independent probes of LFU violations in the $W\tau\nu$ vertex once the strong LHC bounds described above are taken into account. This is interesting because the only low-energy model-independent measurement of this effect, which was carried out at LEP2, found a $\sim2\sigma$ tension with LFU~\cite{Patrignani:2016xqp,Filipuzzi:2012mg}. Including also hadronic tau decays and LHC bounds on contact interactions, the result is improved by a factor of two, but the (dis)agreement with LFU remains at the $\sim2\sigma$ level: $\delta g_L^{W\tau}-\delta g_L^{We}=0.0134(74)$~\cite{Cirigliano:2018dyk}.

Finally, it is interesting to mention the possibility of accessing LFU violations in the vertex corrections at the (HL/HE) LHC. In the ratio of $W\to\tau\nu$ and $W\to\ell\nu$ many experimental and QCD uncertainties  cancel, which make a per-cent level extraction a bit less complicated. In fact, past studies carried out by the D0 collaboration~\cite{Abbott:1999pk} showed that such precision is indeed possible in a collider environment.

\subsection{Lepton Flavour Violation}
\label{LFV}
Lepton Flavour Violating (LFV)   processes  involving charged leptons  are  very interesting because 
their observation would be a clear indication of physics beyond the SM. 
While lepton family number is an accidental symmetry of the SM, 
we know it must be broken in order to account for neutrino masses and mixings. 
If the only low-energy manifestations of LFV are neutrino masses and mixings (which corresponds to 
a very high scale for LFV),  then charged LFV amplitudes are suppressed by the GIM mechanism 
and the predicted rates are un-observably  small, e.g., $\mathcal B  (\mu \to e\gamma) \sim 10^{-52}$, $\mathcal B (\tau \to \mu\gamma) \sim 10^{-45}$ and $\mathcal B (\tau \to 3 \mu) \sim 10^{-54}$~\cite{Cheng:1977nv, Lee:1977tib, Petcov:1976ff, Hernandez-Tome:2018fbq}). 
However, if the breaking of the lepton family symmetry happens not too much above
the electroweak symmetry breaking scale, as borne out in many NP scenarios,  
then one can expect charged LFV BRs  quite close to existing limits, 
and therefore within reach of ongoing searches. In fact in some cases experimental limits are already excluding regions of parameter space 
in specific weak scale NP models.  While less severe than in the quark sector,  one has a ``flavour problem'' in the lepton sector as well.

A rich literature exists on this topic,  including studies in  supersymmetric extensions of the 
SM,  little Higgs models, low-scale seesaw models, 
leptoquark models,   $Z'$ models,  left-right symmetric models,  and extended Higgs models. 
For  recent reviews on both theoretical and experimental aspects we refer the reader to Refs.~\cite{Calibbi:2017uvl,Bernstein:2013hba}. 
Current limits on BRs in  $\mu$-$e$ transitions are at the level of  $10^{-13}$  
(e.g.  $\mathcal B(\mu^+ \to e^+ \gamma) < 4.2
\times 10^{-13}$ (90\% CL) \cite{TheMEG:2016wtm}), while  
$\tau$-$\mu$ and $\tau$-$e$ BRs are bound at the $10^{-8}$ level~~\cite{Amhis:2016xyh}. 
As discussed below,  improvements in LFV $\tau$ decays
 will offer the opportunity to explore  
\textit{(i)} correlations with  $\mu$-$e$ transitions, probing underlying sources of flavour breaking; 
\textit{(ii)} correlations among different LFV $\tau$ decays, probing the nature of the underlying 
mechanism. 
In parallel to the ambitious programme constraining LFV for tau lepton, a similar programme 
exists in the muonic sector improving the limit on $\mu \to e \gamma$ with MEGII
~\cite{Baldini:2013ke, Baldini:2018nnn}, $\mu \to 3 e$ with Mu3e~\cite{Blondel:2013ia,Berger:2014vba},  
and $\mu \to e$ conversion on Aluminium target
with Mu2e~\cite{Bartoszek:2014mya} at Fermilab and COMET~\cite{COMET} at JPARC and on silicon-carbide and graphite target with DeeMe at JPARC~\cite{DeeMe,Nguyen:2015vkk}. LFV studies can also be performed in $B$ and $D$ decays, as well as the decays of $c\bar c$ and $b\bar b$ quarkonia, though the effective NP scale reach is lower, see, e.g., \cite{Hazard:2017udp,Hazard:2016fnc}.
Probing the relative strength of $\mu \to e$ and $\tau \to \mu$ LFV transitions will shed light on the underlying sources of family symmetry breaking in the lepton sector and their link to the quark sector in grand unified scenarios.

\Blu{Finally, LFV decay modes of the $Z^0$ are also of considerable interest. Their BRs in the SM are again negligible, $\BR(Z \to e^\pm \mu^\mp)  \sim \BR(Z \to e^\pm \tau^\mp) \sim 10^{-54}$,  $\BR(Z \to \mu^\pm \tau^\mp) \sim 10^{-60}$ and the LHC can obtain competitive bounds in some of these channels~\cite{Davidson:2012wn}.  
The current upper limits are due to LEP measurements (at 95\% CL)~\cite{Decamp:1991uy,Adriani:1993sy,Akers:1995gz,Abreu:1996mj},
\begin{equation}
\BR(Z \to e^\pm \mu^\mp) < 1.7\times 10^{-6},
\quad \BR(Z \to e^\pm \tau^\mp)< 9.8\times 10^{-6}, 
\quad\BR(Z \to \mu^\pm \tau^\mp) < 1.2 \times 10^{-5},
\end{equation} 
while the ATLAS Collaboration has recently obtained the following 95\% CL upper limits: $\BR(Z \to e^\pm \tau^\mp)< 5.8  \times 10^{-5}$~\cite{Aaboud:2018cxn}, and $\BR(Z \to \mu^\pm \tau^\mp) <1.3  \times 10^{-5}$~\cite{Aaboud:2018cxn}. 
The HL/HE LHC will therefore provide the best limit on LFV $Z$ decays. 
The detection of a signal in the $Z^0 \to \ell \ell^\prime$ channel, in combination with the
information from charged lepton LFV decays,  
would also allow one to learn about features of the underlying LFV dynamics. 
An explicit example is provided by the Inverse Seesaw (ISS) and ``3+1''  effective models 
which add one or more sterile neutrinos to the particle content of the SM~\cite{Abada:2014nwa} (see also, e.g., Ref.~\cite{DeRomeri:2016gum,Bhattacharya:2018ryy,Cai:2018cog}).}

\subsubsection{Lepton Flavour Violation in $\tau$ decays}
\label{LFVTAU}
\begin{figure}[ht]
\begin{center}
\includegraphics[width=1.\textwidth]{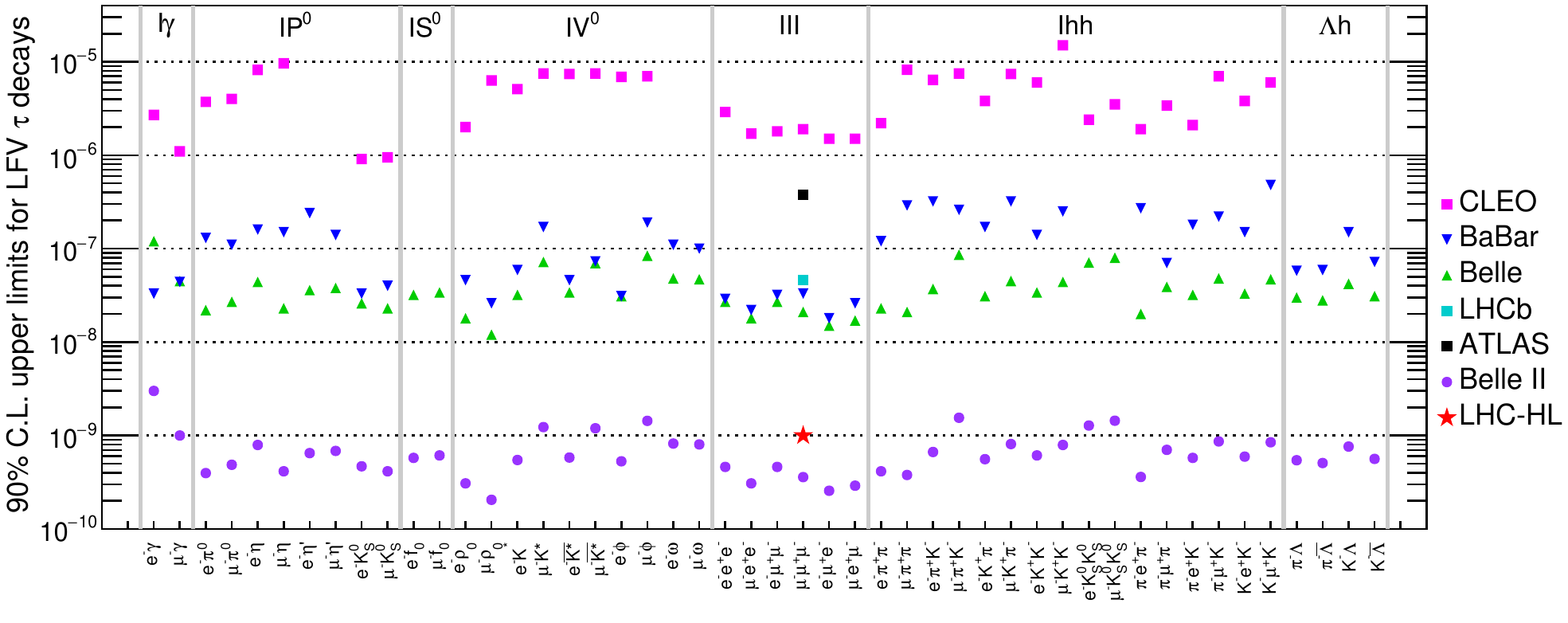}
\caption{Bounds on Tau Lepton Flavour %
Data from the existing experiments are compiled by HFLAV~\cite{Amhis:2016xyh}; projections of the Belle-II bounds were performed by the Belle-II collaboration assuming 50 ab$^{-1}$ of integrated luminosity~\cite{Kou:2018nap}.}
\label{fig:Tau_LFV}
\end{center}
\end{figure}

Tau decays offer a  rich  landscape to search for CLFV. 
The  $\tau$  lepton is heavy enough to decay into hadrons. Until now already 48 LFV modes have been bounded at the level of $10^{-8}$~\cite{Amhis:2016xyh}, 
as can be seen in 
Fig.~\ref{fig:Tau_LFV}.  

The $B$ factories, BaBar and Belle, have improved by more than an order of magnitude
~\cite{Aubert:2006cz,Aubert:2007kx,Aubert:2009ag,Aubert:2009ys,Aubert:2009ap,Lees:2010ez,Miyazaki:2005ng,Hayasaka:2007vc,Miyazaki:2007jp,
Miyazaki:2008mw,Miyazaki:2010qb,Hayasaka:2010np,Miyazaki:2011xe} 
the previous CLEO bounds~\cite{Bowcock:1989mq,Bonvicini:1997bw,Chen:2002ug} for a significant 
number of modes. 
Some of the modes, for instance, $\tau \to \ell \omega$, have been bounded for the first time~\cite{Aubert:2007kx}. 

Table~\ref{tab:tauto3mu} shows a list of limits obtained for the $\tau \to 3\mu$ channel by different experiments. 
The strongest limits come from the $B$-factories, with a competitive limit obtained by LHCb~\cite{LHCb-PAPER-2014-052}. Table~\ref{tab:tauto3mu} also contains the recent measurement by ATLAS~\cite{Aad:2016wce}, as well as 
the expected limit from the Belle-II experiment at the SuperKEKB collider, which will improve current limits by almost two orders of magnitude~\cite{Kou:2018nap}. Finally, Table~\ref{tab:tauto3mu} also summarizes the expected limits from the HL-LHC that we discuss in more detail below.
\begin{table}[t]
\centering
\caption{Actual and expected limits on BR$(\tau\to 3\mu)$ for different experiments and facilities. The ATLAS projections are given for the medium background scenario, see main text for further details. }%
\label{tab:tauto3mu}
\begin{tabular}{ p{4cm}  l  l }
 \hline\hline
 BR$(\tau\to 3\mu)$		&	Ref.						&	Comments	\\ 
 (90\% CL limit)		&								&				\\ \hline
$3.8\times 10^{-7}$		&	ATLAS~\cite{Aad:2016wce}	&	Actual limit (Run 1)	\\
$4.6\times 10^{-8}$		&	LHCb~\cite{LHCb-PAPER-2014-052}	&	Actual limit (Run 1)	\\
$3.3\times 10^{-8}$		&	BaBar~\cite{Lees:2010ez}	&	Actual limit	\\
$2.1\times 10^{-8}$		&	Belle~\cite{Hayasaka:2010np}&	Actual limit	\\ \hline
$3.7\times 10^{-9}$		&	CMS HF-channel at HL-LHC		&	Expected limit (3000 fb$^{-1}$)	\\
$6\times 10^{-9}$		&	ATLAS W-channel at HL-LHC		&	Expected limit (3000 fb$^{-1}$)	\\
$2.3\times 10^{-9}$		&	ATLAS HF-channel at HL-LHC		&	Expected limit (3000 fb$^{-1}$)	\\
${\cal O}(10^{-9})$		&	LHCb at HL-LHC		&	Expected limit (300 fb$^{-1}$)	\\
$3.3\times 10^{-10}$	&	Belle-II~\cite{Kou:2018nap}	&	Expected limit	(50 ab$^{-1}$)\\
\hline\hline
\end{tabular}
\end{table}

The physics reach and model-discriminating power of LFV tau decays is most efficiently analyzed above the electroweak scale using SMEFT, and in a corresponding 
low-energy EFT when below the weak scale~\cite{Celis:2014asa}.
Several  classes of  dimension-six operators contribute  to LFV tau decays at the low-scale, with effective couplings denoted by $C_i/\Lambda^2$.  
Loop-induced dipole  operators mediate radiative decays $\tau \to \ell \gamma$ as well as  purely leptonic $\tau \to 3 \ell$  and semi-leptonic 
decays. Four-fermion -- both four-lepton and semi-leptonic -- operators with different Dirac structures 
can be induced at tree-level or loop-level, and contribute to $\tau \to 3 \ell$ and $\tau \to \ell  \, +  \, {\rm hadrons}$.  
As a typical example,  we note that current limits on $\tau \to \mu \gamma$ probe  scales on the order of $\Lambda / \sqrt{C_{\rm Dipole}} \sim 500$~TeV. 
Besides probing  high scales,  LFV $\tau$ decays offer two main handles to discriminate among underlying models of NP, i.e., to identify 
which operators are present at low energy and what is their relative strength:
(i) correlations among different LFV  $\tau$ decay rates~\cite{Celis:2014asa}; 
(ii)  differential distributions in higher multiplicity decays, such as the $\pi \pi$ invariant mass in 
$\tau \to \mu \pi \pi$ ~\cite{Celis:2014asa} 
and the Dalitz plot in $\tau \to 3 \mu$~\cite{Dassinger:2007ru,Matsuzaki:2007hh}.

In addition to Belle-II, 
HL/HE LHC will be able 
to search for the ``background-free''   $\tau \to 3 \mu$.  
A detailed summary of projected sensitivities is given below. This is a particularly crucial discovery mode when LFV is introduced at tree-level,
for example by the exchange of $Z'$ or doubly charged Higgs bosons.
Efforts should also go into understanding the 
backgrounds and improving the sensitivity in semi-leptonic three-body decays $\tau  \to \mu \pi^+ \pi^-$ 
and $\tau \to \mu K \bar{K}$, which have a particularly high discovery potential in Higgs-mediated~\cite{Paradisi:2005tk,Celis:2013xja,Petrov:2013vka} or leptoquark-mediated LFV.
If LFV is discovered, these modes also have significant model-diagnosing power because of their Dalitz structure, and therefore also probe a wider class of models than the dominant one-loop-induced LFV process $\tau \to \mu \gamma$. Of particular interest is the sensitivity of $\tau \to \mu \pi \pi$ to extended Higgs sectors and to non-standard Yukawa couplings of the SM Higgs to light quarks and leptons~\cite{Celis:2013xja,Celis:2014asa}. This channel also allows a particularly robust theoretical interpretation thanks to advances in the calculation of all the relevant hadronic form factors~\cite{Celis:2013xja}.

\subsubsection{HL-LHC experimental prospects}
    \def\mupm   {{\ensuremath{\mu^\pm}}\xspace}
    \def\mump   {{\ensuremath{\mu^\mp}}\xspace}
    \def\taupm   {{\ensuremath{\tau^\pm}}\xspace}
    \def\taump   {{\ensuremath{\tau^\mp}}\xspace}
    \def\taummm{\decay{\taupm}{\mupm \mu^+ \mu^-}}
    \def\Dsetamunu{\decay{\Ds}{\eta(\mu^+\mu^-\gamma)\mu^+\nu_{\mu}}}
The LHC proton collisions at 13 TeV produce $\tau$ leptons with a cross-section five orders of magnitude 
larger than at \belletwo. As a result, during the HL-LHC running period, about $10^ {15}$~$\tau$~leptons will be produced in 4$\pi$. Most will be
produced in the decay of heavy flavour hadrons, specifically $D_s$ meson decays.
This high production cross-section compensates for the higher background levels 
and lower integrated luminosity, in particular for the $\tau \to 3\mu$ golden mode.
Background events arise dominantly from badly reconstructed heavy flavour decays like \Dsetamunu, lepton fakes from hadrons ($c\bar{c}/b\bar{b}\to X\mu\mu$), and pile-up. A particular challenge for this production channel is the soft momentum spectrum of the $\tau$ decay products, which places stringent requirements on both the trigger and offline reconstructions of all the HL-LHC experiments.

$W$~and $Z$~bosons offer a complementary source of $\tau$  leptons. Their production cross sections are considerably smaller than those for $B$~and $D$~mesons, but $\tau$~leptons from $W$~and $Z$~afford much cleaner experimental signatures with far better signal-to-background ratios for CMS and ATLAS; the LHCb forward geometry is less well suited to exploting these decays. For instance, in a $\tau \to 3 \mu$ search relying on $W \to \tau \nu$ decays as a source of $\tau$~leptons, one can benefit from $\tau$~leptons having relatively large transverse momenta and being isolated, from large missing transverse momentum $p_{\mathrm{T}}^{\mathrm{miss}}$ in an event, and from the transverse mass of the $\tau$-$p_{\mathrm{T}}^{\mathrm{miss}}$ system being close to the $W$~mass. 

\begin{figure}[t]
\begin{center}
\includegraphics[width=0.495\textwidth]{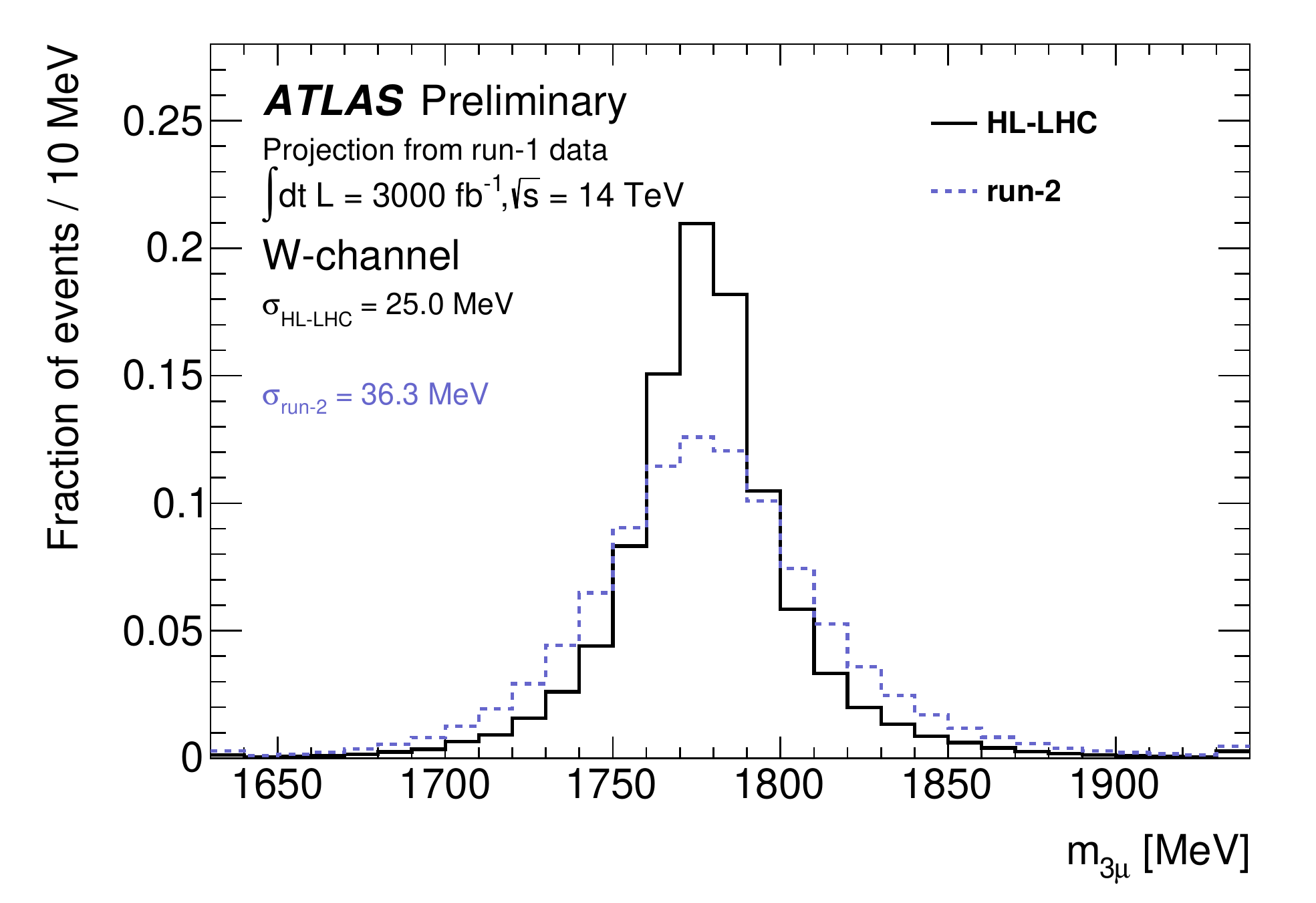}
\includegraphics[width=0.495\textwidth]{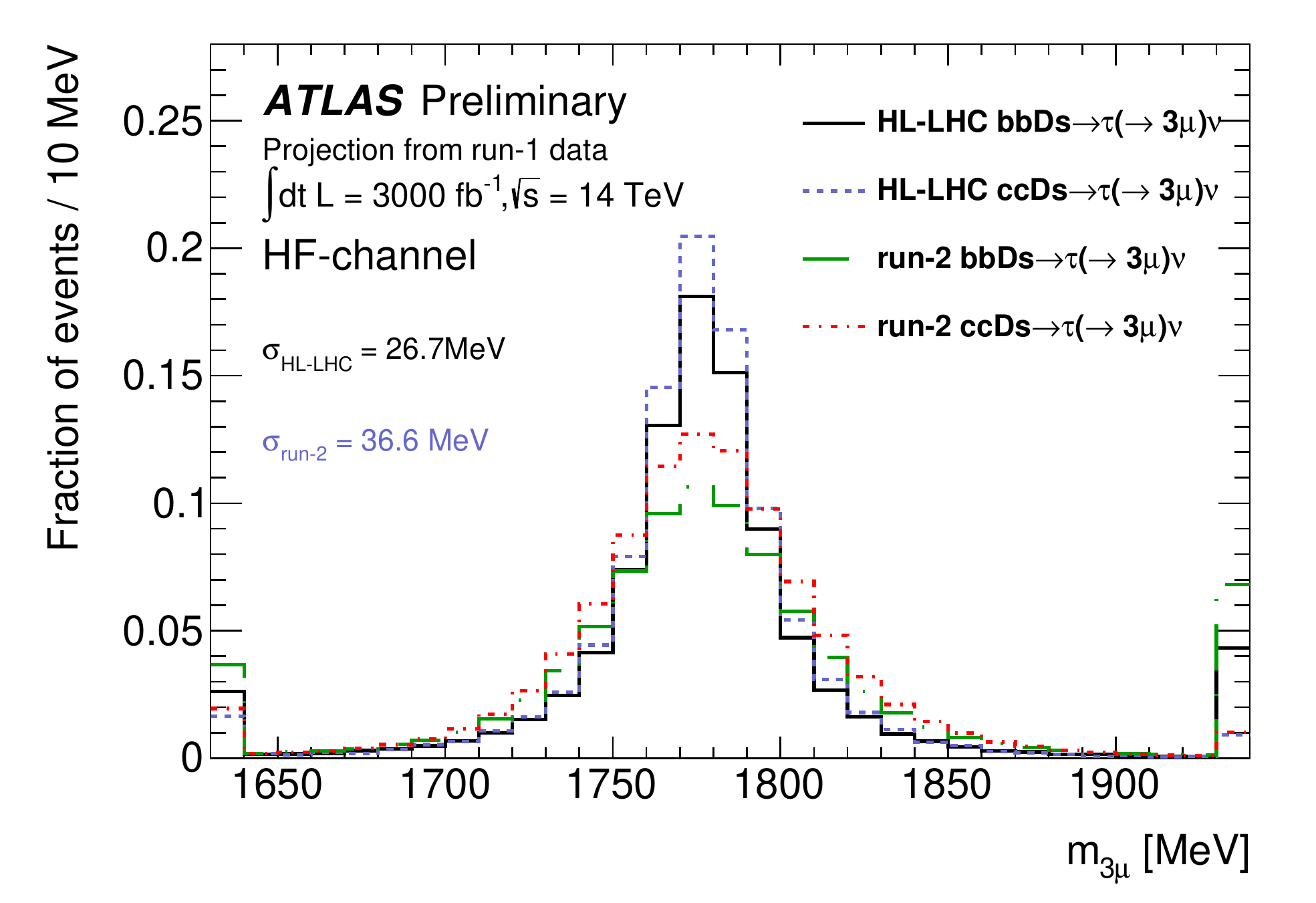}
  \includegraphics[width=0.495\textwidth]{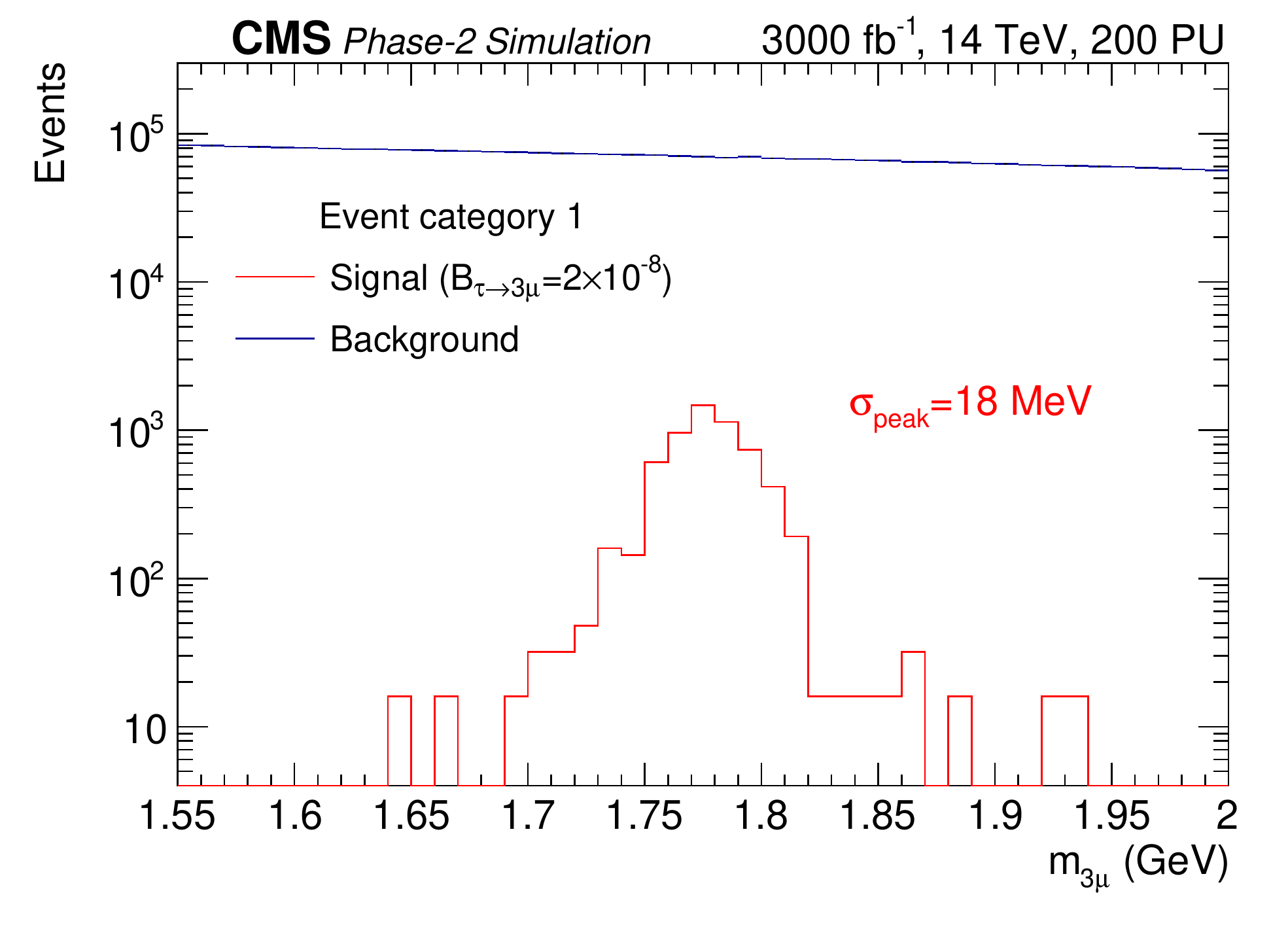}
  \includegraphics[width=0.495\textwidth]{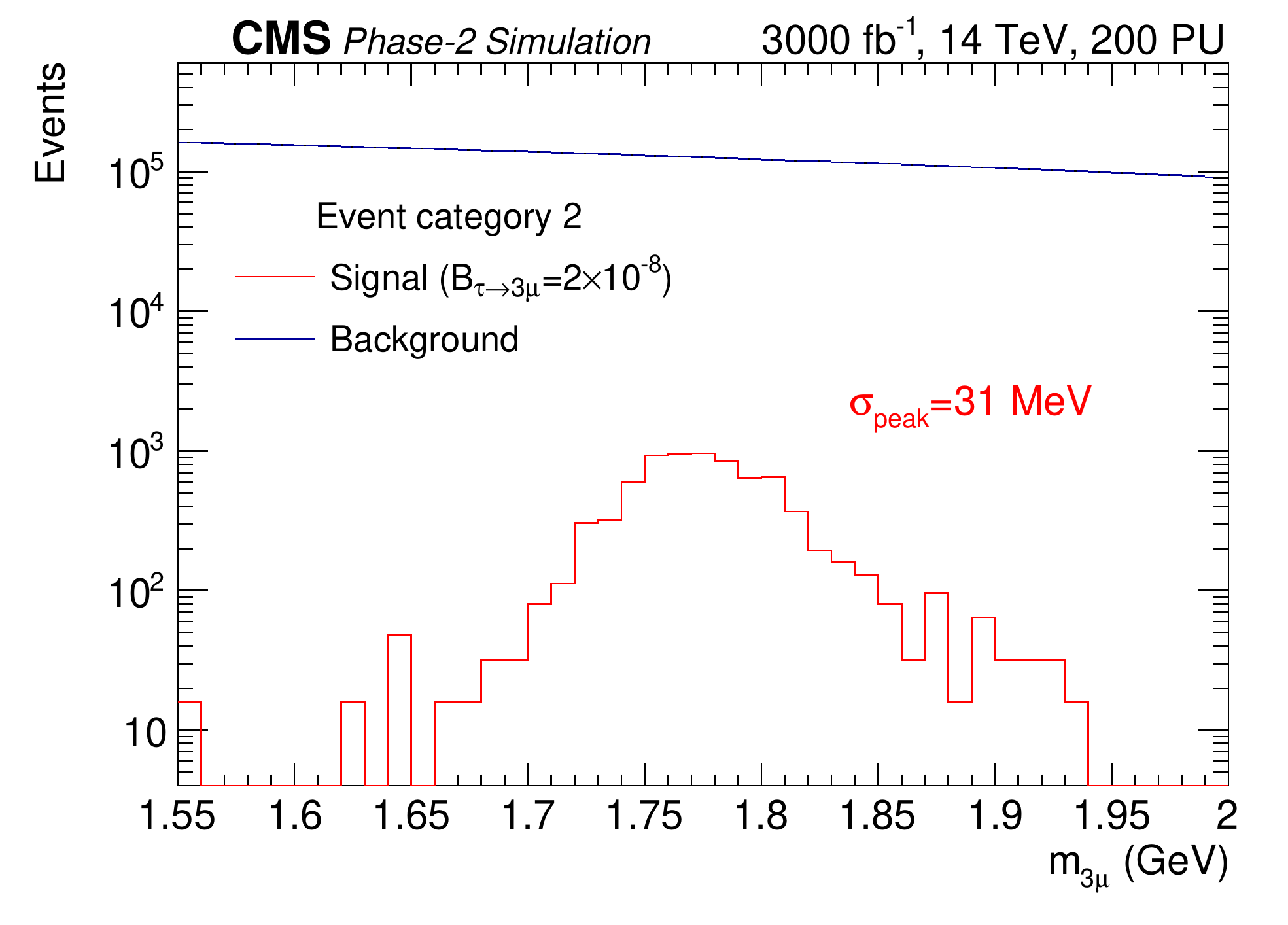}
\caption{(Top) Comparison of ATLAS tau mass resolutions in the $W$- (left) and HF-channels (right) in Run-2 and under HL-LHC detector conditions.  Widths are estimated from a double-Gaussian fit. (Bottom) CMS trimuon invariant mass $m_{3\mu}$  for the $\tau \to 3\mu$
signal (red) and background (blue) after all event selection cuts, for 
Category 1 (left) and Category 2 (right) events, as defined in the text.
 The signal is shown for  $\mathcal{B}(\tau \to 3\mu) = 2\times10^{-8}$. CMS results refer to the HF channel; plots  are taken from \cite{CMSTDR:muon} .}
\label{fig:mass}
\end{center}
\end{figure}

Detector improvements planned for the HL-LHC period will significantly enhance the capabilities of all three experiments in this area. In the case of ATLAS, the installation of a new tracking system will improve the vertex and momentum resolution. The trigger system upgrade will include additional capabilities, ultimately improving the online selection, and allowing to maintain a low muon triggering threshold. 
The CMS upgraded  muon system, whose coverage is extended from $|\eta|$ = 2.4 to ~2.8,  increases  
the signal fiducial acceptance by a factor of two and, also, enhances the capability to trigger on and reconstruct low momentum muons \cite{CMSTDR:muon}.
The additional events with muons at high $|\eta|$ have worse trimuon mass resolution. Hence, two event categories are introduced: 
Category 1 for events with all three muons reconstructed only with the Phase-1 detectors, and Category 2 for
events with at least one muon 
reconstructed by the new triple Gas Electron Multiplier (GEM) detectors, which  will be installed in the first station of the upgraded muon system. The impact of the proposed ATLAS detector upgrades and CMS event categories is shown in Fig.~\ref{fig:mass}. In the case of LHCb, the deployment of a fully software trigger will remove one of the key sources of inefficiency in the current analysis. In addition, the proposed calorimeter improvements during the LHCb \upgradetwo  will play an important role in suppressing backgrounds such as \Dsetamunu. 

\begin{figure}[h]
\begin{center}
\includegraphics[width=0.47\textwidth]{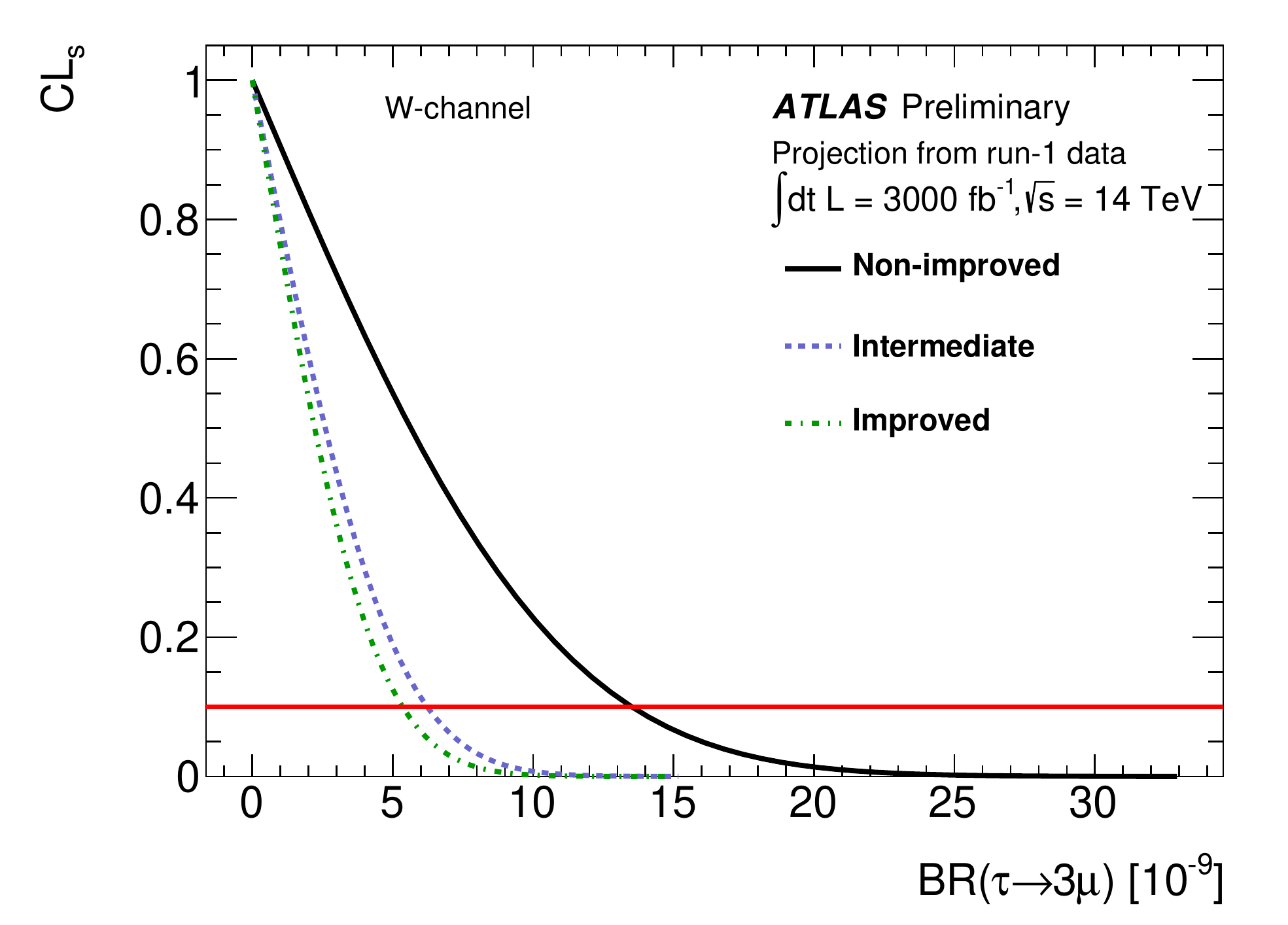}
\includegraphics[width=0.47\textwidth]{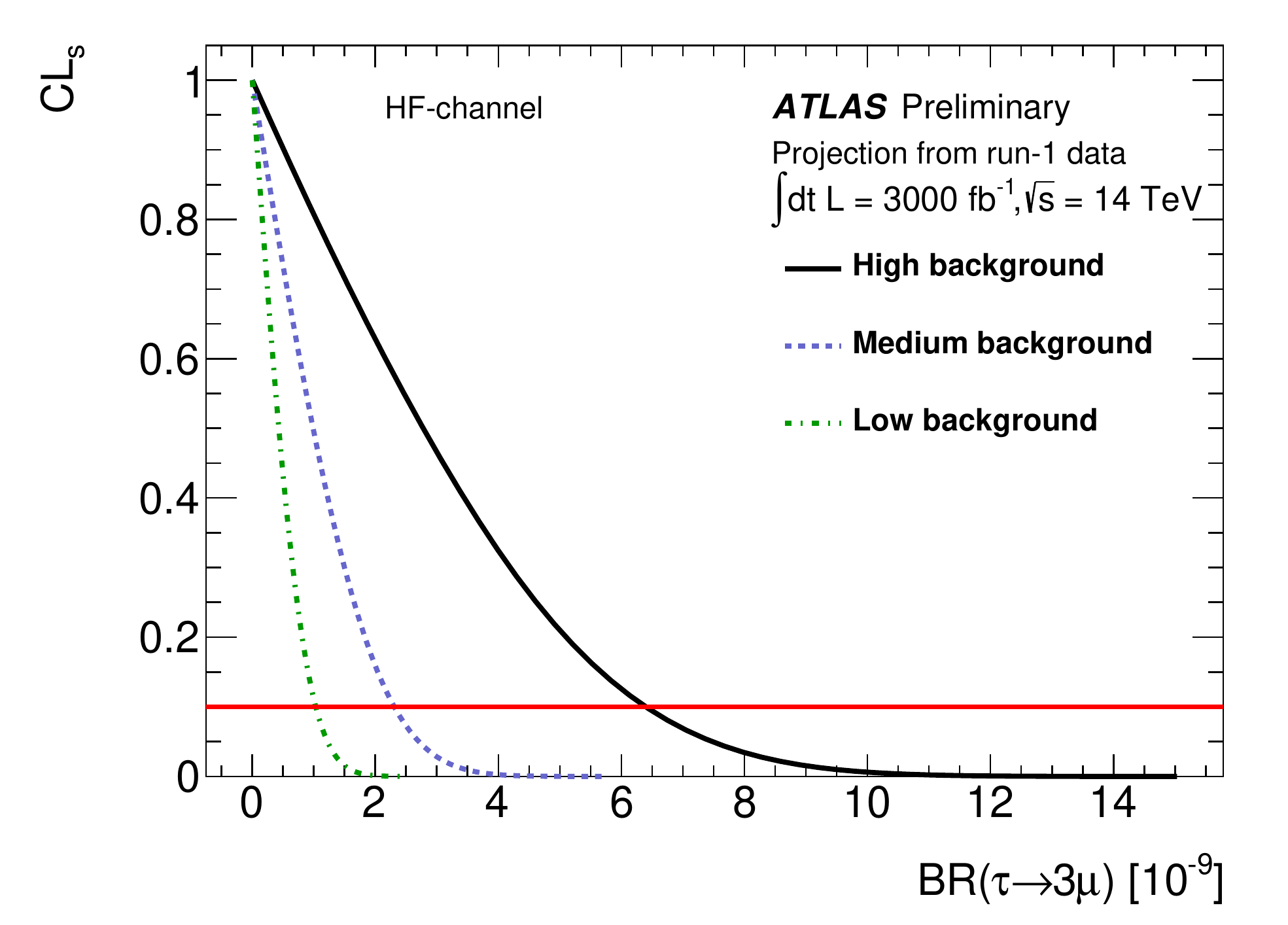}
\caption{ATLAS extrapolated CL$_{\mathrm{s}}$ versus the $\tau\rightarrow 3\mu$ branching fraction, BR($\tau\rightarrow 3\mu)$ for each of the discussed scenarios in the $W$-channel (left) and HF-channel (right). The horizontal red line denotes the 90\% CL. The limit is obtained from the intersection of the CLs and this line.}
\label{fig:limits}
\end{center}
\end{figure}

\begin{table}[t]
\caption{Summary of the ATLAS expected 90\% CL upper limit on $\BR(\tau\to3\mu)$ for an assumed luminosity of 3 ab$^{-1}$ of $pp$ collisions at $\sqrt{s} = 14$ TeV in the $W$ and Heavy Flavour (HF) channels, for different signal and background yield scenarios.}
\label{tab:limitsatlas}
\center
\begin{tabular}{lcc}
\hline\hline
Scenario & $W$-channel                      & HF-channel  \\
               & 90\% CL UL [$10^{-9}$]  & 90\% CL UL [$10^{-9}$]  \\
\hline
ATLAS High      & 5.4 & 1 \\
ATLAS Medium & 6.2   & 2.3 \\
ATLAS Low       & 13.5   & 6.4    \\
\hline\hline
\end{tabular}
\end{table}

\begin{table}[t]
\begin{center}
\caption{(Top) The expected numbers of signal and background events in the mass window 1.55 -2.0 GeV for CMS. An integrated luminosity of 3000 fb$^{-1}$ and a signal  $\mathcal{B}(\tau \to 3 \mu$) = 2  $\times 10^{-8}$ is assumed. (Bottom) The search sensitivities for the combined categories.}
\label{tab:limitscms}
\begin{tabular}{l c c}   
\hline\hline
& Category 1 & Category 2 \\ \hline 
Number of background events & 2.4 $\times 10^6$ & 2.6 $\times 10^6$ \\
Number of signal events & 4580 & 3640 \\
Trimuon mass resolution & 18 MeV& 31 MeV \\
$\mathcal{B}(\tau \to 3 \mu$) limit per event category & 4.3 $\times 10^{-9}$ & 7.0 $\times 10{^{-9}}$ \\ \hline
$\mathcal{B}(\tau \to 3 \mu$) 90\% C.L. limit &  \multicolumn{2}{c}{$3.7 \times 10{^{-9}}$}
 \\
$\mathcal{B}(\tau \to 3 \mu$) for 3-$\sigma$ evidence   &  \multicolumn{2}{c}{$6.7 \times 10{^{-9}}$}
 \\
$\mathcal{B}(\tau \to 3 \mu$) for 5-$\sigma$ observation   &  \multicolumn{2}{c}{$1.1 \times 10{^{-8}}$} \\ 
\hline\hline
\end{tabular}
\end{center}
\end{table}

The extrapolated sensitivities for ATLAS and CMS are shown in Fig.~\ref{fig:limits} and Tables~\ref{tab:limitsatlas}~and~\ref{tab:limitscms}.
The ATLAS sensitivities \cite{ATL-PHYS-PUB-2018-032} are extrapolated based on the Run~1 
measurement~\cite{Aad:2016wce} taking into account the expected detector 
improvements. Three scenarios for the acceptance, 
efficiency and background yields are considered.
Systematic uncertainties are extrapolated from the Run-1 measurement 
scaling down by the increased statistics with preserved constant terms for the 
reconstruction efficiency. A 15\% systematic uncertainty dominated by the background 
estimation is derived. Varying the systematic uncertainty by 5\% translates into a 
10\% change of the expected upper limit. Limits of up to $\BR(\tau\rightarrow 3\mu) = 
1.03 (5.36) \times 10^{-9}$ are derived in the HF-($W$-)channel.
In the case of CMS, the absence of tails in the signal trimuon mass distribution, shown in Fig.~\ref{fig:mass}, demonstrates the robustness of the reconstruction even at the high pileup (PU=200) of HL-LHC.
The projected exclusion sensitivity in the absence of a signal is $\BR( \tau \to 3 \mu) < 3.7 \times 10^{-9}$  at 90\% CL, while the expected 5 $\sigma$-observation sensitivity is $\BR(\tau \to 3 \mu) = 1.1  \times 10^{-8}$. CMS studies focusing on $\tau$~leptons originating from $W$~boson decays are ongoing.

Extrapolations based on the current LHCb $\tau \to 3\mu$ result~\cite{LHCb-PAPER-2014-052} and assuming no detector or trigger improvements show that, similarly to \belletwo, CMS, and ATLAS, \upgradetwo LHCb would also be able to probe branching ratios down to ${\cal O}(10^{-9})$. This will allow \upgradetwo LHCb to independently confirm any earlier \belletwo discovery, or to
significantly improve the combined limit.

%% file: section6/section.tex
\section{Hadron spectroscopy and QCD exotica}
{\it \small Authors (TH): Angelo Esposito, Feng-Kun Guo, Juan Nieves, Alessandro Pilloni, Antonio Polosa, Sasa Prelovsek.}

\Blu{Determination of hadron spectrum is key to understanding the strong interactions, in particular the mechanisms underlying color confinement.}
While \Blu{BESIII is the leading experiment in the charm region, and} \belletwo will make a major impact 
in the study of excited mesons and lighter exotic hadrons, only the HL-LHC experiments 
have the potential to comprehensively study the full range of possible excited
hadronic states. Such a comprehensive understanding and characterisation of quark
structures would not only, in some sense, greatly enhance the taxonomy of the SM particles, but
would also sharpen our understanding of QCD in a particularly difficult energy
regime. Moreover, while not directly probing BSM physics, spectroscopy
measurements will continue to make essential contributions to the interpretation
of any observed BSM effects in the flavour sector. In addition, spectroscopy can provide tools to probe BSM effects by using a new observed state as a tagger\cite{LHCb-PAPER-2017-023}, \cite{LHCb-PAPER-2016-039}.

While the \upgradetwo of LHCb will
by design offer unparalleled capabilities in this area, the legacy of HL-LHC for
spectroscopy will be much stronger, if ATLAS and CMS are also able to continue to
pursue this work in the HL-LHC period. The planned hardware tracking
triggers and much higher data rates sent to their software triggers are promising
in this respect, and we encourage further study of what capabilities they might bring to this area of physics. 
\Blu{In particular, heavy-ion experiments might have an advantage for doubly-heavy exotics, if they can improve their particle identification.}

\Blu{Gauge theories play a major role in BSM and some of them might be strongly coupled. 
In this sense, QCD serves as a prototype of a strongly-coupled gauge theory. By studying hadron spectroscopy and the QCD exotica we learn about the range of possibilities of analogous states in BSM theories.}

\input{section6/theory}
\input{section6/experimental}

%% file: section6/theory.tex
\subsection{Open questions in spectroscopy}

A large number of  ``exotic'' experimental discoveries, which did not fit the expectations of the (until then) very successful valence quark model, as well as the unprecedented statistical precision obtained by LHCb, BESIII, and other experiments, have led to a recent renaissance of hadron spectroscopy. For recent reviews, see~\cite{
Chen:2016qju,Hosaka:2016pey,Lebed:2016hpi,Esposito:2016noz,Guo:2017jvc,Ali:2017jda,Olsen:2017bmm,Karliner:2017qhf,Yuan:2018inv,Kou:2018nap,PDG2018}.
Understanding strong interactions at low energies requires  an explanation of how the emergent hadron spectrum is  organized,  which can shed light on the confinement property of QCD.  

To answer this question, a joint effort by experimental and theoretical communities is needed, making a combined use of all available tools, including lattice QCD results and effective field theory (EFT) methods, as well as phenomenological tools. The spectra of weakly decaying 
hadrons, as post-dicted or predicted by lattice QCD, generally agree with the observed ones. Valuable conclusions 
can be drawn from the lattice  also  for several   strongly decaying resonances,  where the  challenge increases with the number of open strong decay channels.  From the phenomenological side, quark potential models, inspired by QCD, describe mesons as bound states of a valence quark-antiquark pair, and baryons as bound states of three valence quarks. %
One would expect these models~\cite{Eichten:1978tg, Eichten:1979ms,Godfrey:1985xj} to work  particularly well in describing the heavy quarkonium sector. \Blu{They indeed do a very good job for 
the spectrum of charmonia and bottomonia below their respective open-flavor thresholds. This success can be connected with QCD by using nonrelativistic EFTs such as the ones described in Refs.~\cite{Caswell:1985ui, Bodwin:1994jh, Brambilla:2004jw, Pineda:2011dg}.} 

However,  other color configurations are  allowed,  such as  mesons with two quarks and two anti-quarks, and baryons with four quarks and one anti-quark, dubbed tetraquarks and pentaquarks, respectively.\footnote{In fact, such configurations were already suggested in Gell-Mann's seminal paper on the quark model~\cite{GellMann:1964nj}. Multiquark configurations are also suggested, but not listed, in Zweig's original paper~\cite{Zweig:1964jf}. The name of ``pentaquark'' and a pioneering dynamical model first appeared in Ref.~\cite{Lipkin:1987xs}.} These are examples of the so-called exotic hadrons \Blu{which also include glueballs and hybrid hadrons with gluonic excitations}.  As already mentioned, their search has been an important theme in high-energy experiments, although plagued by several ambiguous claims. The key questions are: \Blu{Can new states be unambiguously extracted from the experimental data?} Are exotic hadrons really allowed by QCD? If yes, what kinds are allowed? Why are they so scarce? Can lattice QCD predict their spectrum? Is it possible to construct QCD-based phenomenological models for (at least) some of them?

The situation dramatically changed in 2003,  when the charm-strange mesons $D_{s0}^*(2317)$, $D_{s1}(2460)$ and the charmonium-like $X(3872)$ were discovered by the BaBar, CLEO and Belle Collaborations, respectively.  None of them is in line with the predictions from potential quark models. For instance, their masses are lower than the predictions by Godfrey and Isgur by about 160~MeV, 70~MeV and 80~MeV, respectively~\cite{Godfrey:1985xj}.  As time passed, more and more unknown resonant structures were reported from various high-energy experiments, mostly BaBar, Belle, BESIII and LHCb.  Many new structures, not compatible with predictions, have been found in the charmonium mass region. These are traditionally called $XYZ$ states.\footnote{Note that the naming scheme was changed in the 2018 edition of the Reviews of Particle Physics (RPP) by the Particle Data Group~\cite{PDG2018}. For charmonium(-like) states, the isoscalar states with quantum numbers $J^{--}$, $J^{++}$ are called $\psi_{J}$, $\chi_{cJ}$, respectively, and the isovector states with $J^{--}$, $J^{+-}$  are called $R_{cJ}$, $Z_{cJ}$, respectively.} 
In particular, the isovector $Z$ states are explicitly exotic, because they have nonzero isospin and at the same time contain a heavy $Q\bar Q$ pair. Similarly, LHCb~\cite{Aaij:2015tga} reported two structures in the $J/\psi\, p$ spectrum.   If these were produced  by real QCD baryon resonances ($P_c$),  these states would contain at least five valence quarks/antiquarks.  So far, there has been no consensus about the nature of these resonances, and  some of the related experimental peaks might even receive large contributions from kinematical effects, making their interpretation as real states ambiguous. To resolve these issues, high statistics data and the search for signatures of these states in several different processes are highly desirable.

Next, we introduce briefly a few main types of models beyond the naive quark model for these intriguing exotic hadron candidates.

\subsubsection{Tetraquarks} \label{tetra}

Two quarks and two antiquarks can lead to color singlets in different ways, which are difficult to choose from first principles. One possibility is for the constituents to be bound in a compact tetraquark (see, e.g.,~\cite{Maiani:2004vq,Maiani:2014aja}). 
For experimental reasons, 
the most studied systems are of the form $Q\bar Q q\bar q^{\,\prime}$, where $Q=c,b$ and $q^{(\prime)}$ are light quarks. Their spectrum is well described in terms of diquarks, with a spin-spin interaction between the constituents given by $H_I=2\kappa_{Qq}\big(\bm{S}_Q\cdot \bm S_q+\bm S_{\bar Q}\cdot \bm S_{\bar q^{\,\prime}}\big)$, with no coupling between the constituents of different diquarks~\cite{Maiani:2014aja}. The chromomagnetic couplings $\kappa$ are extracted from observed masses. It is found that two quasi-degenerate $1^{++}$ and $1^{+-}$ states are expected (identified with the $X(3872)$ and $Z_c(3900)$), together with a higher $1^{+-}$ state (the  $Z_c^\prime (4020)$). The same pattern should be replicated in the beauty system, where the two $Z_b(1^{+-})$'s have been discovered but the $X_b(1^{++})$ is still missing. 
The $Z_c$'s and $Z_b$'s are expected to fill in triplets of states, as confirmed experimentally.  The $X$ is an exception: only the neutral state is observed. This could be due to its accidental vicinity to threshold~\cite{Maiani:2017kyi}. \Blu{Moreover, the remarkable isospin violation observed in its decays can be explained by the fact that, being $\alpha_s(2m_c)$ small, the mixing between the mass eigenstates is suppressed~\cite{Rossi:2004yr,Maiani:2004vq}.} The $X^{\pm}$ components are forced to decay only into charmonia, and the computed rates lie below the current exclusion limits set experimentally. Decay modes such as, for example, $X^\pm\to J/\psi\rho^\pm$ should be eagerly sought in data.

Radial excitations are also possible, the charged $Z(4430)$ being the most remarkable example~\cite{Maiani:2008zz,Maiani:2007wz,Maiani:2004vq}. Its mass and decay modes fit perfectly a picture where it is the first radial excitation of the $Z_c$.
Lastly, one also has orbital excitations, allowing for a $1^{-}$ spectroscopy.  Prominent candidates are the observed  $Y(1^-)$ states. An example is the $Y(4630)-Y(4660)$ system (identified as a single particle in~\cite{Cotugno:2009ys}). Its mass fits the diquarkonium spectroscopy and its strong preference to decay into two baryons is easily explained by the breaking of the QCD string between the diquarks. For reviews see~\cite{Esposito:2014rxa,Chen:2016qju,Lebed:2016hpi,Esposito:2016noz,Ali:2017jda}.
Remarkably, the $X$, $Z^{(\prime)}_c$ and $Z_b^{(\prime)}$ resonances are found experimentally $\delta\lesssim 10$~MeV {\it above} their related meson-meson thresholds ($D^0\bar D^{*0}$ for the $X$, etc.). 
Their measured widths follow a $\Gamma=A\sqrt{\delta}$ law, with the same coefficient $A$ for both charm and beauty states, as first observed in~\cite{Esposito:2016itg}.

The above feature, the suppression of the interactions across different diquarks and the preference for decaying into open-flavour mesons, can be qualitatively explained with a separation in space of the two diquarks due to a (tiny) potential barrier between the two~\cite{Jaffe:2003sg, Selem:2006nd, Maiani:2017kyi,Esposito:2018cwh}. \Blu{With this conjecture, one can also understand why the discrepancy between the decay into quarkonia and open flavor is more prominent in the bottom sector than in the charm one.} This would also make these exotic hadrons slightly larger than the standard ones ($\approx 1.3$~fm)~\cite{Esposito:2018cwh}. 
The idea of the diquark--antidiquark pair being slightly separated in space has also been presented in~\cite{Brodsky:2014xia}, where it was shown that considering diquarks as dynamical and moving with respect to each other leads to an interesting way of computing tetraquark decays. The same picture has also been employed to study the spectrum of tetraquarks and possible selection rules for their decays~\cite{Lebed:2017min}, as well as to study pentaquarks~\cite{Lebed:2015tna}.
Among other proposals to calculate the tetraquark spectrum, we recall the Born-Oppenheimer approximation, where the tetraquark is described as a heavy $Q\bar Q$ pair in color octet, immersed into an effective potential generated by the light pair, again in color octet. Under the assumption that the effective potential is the same as the one calculated on the lattice for heavy hybrids, the spectrum is derived~\cite{Braaten:2013boa}.
Lastly, it was also proposed that the scaling of the tetraquark exclusive production cross section with the center-of-mass energy might reveal important information on the composite nature of these states~\cite{Brodsky:2015wza,Lebed:2018jcr}. It follows that the study of production of exotic states is crucial.

\subsubsection{Hadronic molecules}

These are complex systems for which the dominant components are two or more interacting hadrons. For a review of theoretical and phenomenological aspects of hadronic molecules, with a focus on the new heavy-flavour hadrons, we refer to~\cite{Guo:2017jvc}. For a review focusing on the $X(3872)$, see~\cite{Kalashnikova:2018vkv}.

The simplest  hadronic molecule  would be a loosely bound state of two hadrons ($H_1, H_2$) with a sizable extension.  Such an object  would be the analog to a neutron-proton pair bound to form a deuteron.  If the physical state is located slightly below an $S$-wave threshold of two hadrons $H_1$ and $H_2$, the scattering length ($a$) and effective range ($r$)  of the continuum $H_1+H_2\to H_1+H_2$  scattering amplitude would be approximately given by
\begin{equation}
a=- 2 \frac{1-\lambda^2}{2-\lambda^2}\,\frac1\gamma + {\cal O} (\gamma/\beta), \qquad
r =- \frac{\lambda^2}{1-\lambda^2}\,\frac1\gamma + {\cal O} (\gamma/\beta), 
\end{equation}
where $\gamma = \sqrt{2\mu E_B}$ ,  with $\mu$ the $H_1, H_2$ reduced mass and $E_B>0 $  the binding energy of the physical state (i.e., the difference between the $H_1, H_2$ threshold and its mass),  $\beta$ the inverse range of forces  and $\lambda^2$ the wave function renormalization constant, which for a pure molecule vanishes. 
There is also a relation for the effective coupling of the physical state to $H_1$ and $H_2$, given by (using the non-relativistic normalization)
\begin{equation}
  g_\text{NR,eff}^2 = \frac{2\pi\gamma}{\mu} (1-\lambda^2) + {\cal O}(\gamma/\beta) .
\end{equation}
The $(1-\lambda^2)$ factor gives the molecular probability, i.e., the probability of finding the $H_1H_2$ component in the physical state (the Weinberg compositeness criterion~\cite{Weinberg:1965zz}).  Thus, for a pure molecule ($\lambda^2=0$),  one finds that the scattering length takes its maximum value,  $a=-1/\gamma$,  and in addition $r={\cal O} (1/\beta)$, while for a compact state $(\lambda^2=1)$ one gets $ a={\cal O} (1/\beta)$  and $r\to -\infty$. 
In addition, the effective coupling takes the maximal value for a pure molecule  and it vanishes when there is no $H_1H_2$ component in the physical state.
These differences produce distinctive  signatures in the line shapes of near-threshold states (see, e.g., the discussion in~\cite{Guo:2017jvc,  Sekihara:2014kya}) and not-so-near-threshold states~\cite{Molina:2016pbg}.	In the case of coupled-channel dynamics, one might also find states  that can decay into some of the open channels. These may still qualify as hadronic molecules, were they to couple strongly to one of the channels considered in the dynamical space. 
A detailed review on weak decays of heavy hadrons leading to molecular states in the final state is given in~\cite{Oset:2016lyh}.

The hadron-hadron interactions are usually taken from EFTs  based on exact or 
approximate QCD symmetries, which are employed to construct amplitudes satisfying unitarity in coupled channels.  The %
undetermined low-energy constants (LECs) are fitted to data or lattice QCD results, both  in the perturbative expansion (implicit in  the EFT) and  in the unitarization procedure.    

\Blu{
\subsubsubsection{Effect of meson-meson channels on naive quark-antiquark bound states}

Many of the hadronic structures discovered since 2003 appear close to hadron-hadron thresholds, and thus their dynamics can be strongly dictated by the nearby multiquark hadron-hadron channels. In fact, the discovery of the $XYZ$ particles is opening the door to systematically explore higher Fock components of the hadron wave function~\cite{Eichten:2005ga,Kalashnikova:2005ui,Baru:2010ww}. When the hadron-hadron channels dominate, one gets states that can be approximately by hadronic molecules introduced above. Within quark models, the hadron-hadron effects can be calculated as pioneered in Refs.~\cite{Eichten:1978tg,Eichten:1979ms,Tornqvist:1979hx}.

Within the quark model picture, the coupling between the quark-antiquark and meson-meson sectors requires that the hadronic state is written as~\cite{Ortega:2012rs, Segovia:2016xqb}
\begin{equation} 
| \Psi \rangle = \sum_\alpha c_\alpha | \psi_\alpha \rangle + \sum_\beta \chi_\beta(P) |\phi_A \phi_B \beta \rangle,
\label{eq:funonda}
\end{equation}
where $|\psi_\alpha\rangle$ are eigenstates of the quark-antiquark Hamiltonian~\cite{Vijande:2004he, Segovia:2013wma}, $|\phi_A \phi_B \beta \rangle$ is the two-meson state with $\beta$ quantum numbers and $\chi_\beta(P)$ is the relative wave function between the two mesons. In principle one should couple with an infinite number of hadron-hadron channels. However, it has been argued in Refs.~\cite{Swanson:2005rc, Barnes:2007xu}  that the only relevant thresholds are those close to the naive states, while the rest have only little effect, which can be absorbed in the quark model parameters. The reason is that the energy nonanalyticity at thresholds does not cause any nontrivial behavior, if the thresholds are far away.

The meson-meson interaction can be derived from the quark-antiquark one using the resonating group method (RGM)~\cite{Tang:1978zz, Entem:2000mq}. The coupling between the quark-antiquark and meson-meson sectors requires the creation of a light quark-antiquark pair, and thus the operator associated with this should describe also the open-flavor meson strong decays of heavy quarkonia. The most simple decay model is the well-known $^3P_0$ model~\cite{Micu:1968mk, LeYaouanc:1972vsx, LeYaouanc:1973ldf}, in which the transition potential between quark-antiquark and meson-meson sectors can be defined as 
\begin{equation}
\langle \phi_{A} \phi_{B} \beta | T | \psi_\alpha \rangle = P \, h_{\beta \alpha}(P) \,\delta^{(3)}(\vec P_{\rm cm})\,,
\label{eq:Vab}
\end{equation}
where $P$ denotes the relative momentum of the two-meson state.

Using Eq.~(\ref{eq:funonda}) and the transition potential in Eq.~(\ref{eq:Vab}), one gets the coupled equations
\begin{align}
c_\alpha M_\alpha +  \sum_\beta \int h_{\alpha\beta}(P) \chi_\beta(P)P^2 dP &= E
\, c_\alpha\,, \label{eq:Ec-Res1} \\
\sum_{\beta}\int H_{\beta'\beta}(P',P)\chi_{\beta}(P) P^2 dP + \sum_\alpha h_{\beta'\alpha}(P') \, c_\alpha &= E \, \chi_{\beta'}(P') \,, \label{eq:Ec-Res2}
\end{align}
where $M_\alpha$ are the masses of the bare quark-antiquark mesons and $H_{\beta'\beta}$ is the RGM Hamiltonian for the two-meson states. Solving the coupled-channel equations, Eqs.~\eqref{eq:Ec-Res1} and~\eqref{eq:Ec-Res2}, as indicated, e.g., in Ref.~\cite{Ortega:2016hde}, allows to study hadronic states as poles of the $S$-matrix in any possible Riemann sheet.
}

\subsection{Hadrons with a single heavy quark}

\noindent
\subsubsection{Mesons}

  The lowest positive-parity states include the very narrow charm-strange mesons $D_{s0}^*(2317)$, $D_{s1}(2460)$, and the very broad charm-nonstrange mesons $D_0^*(2400)$ and $D_1(2430)$~\cite{PDG2018}. Their bottom partners are still waiting for discovery,  likely by LHC experiments.
  \Blu{Bottom partners of the first two might be also accessible at Belle-II, though being at the upper edge of its energy reach.}

\noindent
\paragraph{Phenomenology}\label{sec:spectro:mesons:pheno}
There were attempts trying to interpret the $D_{s0}^*(2317)$ and $D_{s1}(2460)$ as $c\bar s$ mesons, see, e.g.,~\cite{Godfrey:2003kg,Colangelo:2003vg,Mehen:2005hc,Lakhina:2006fy}, or as the positive-parity chiral partners of the ground-state $D_s$ and $D_s^*$~\cite{Bardeen:2003kt,Nowak:2003ra}. In these cases, the isospin-breaking hadronic decay widths are predicted to be of the order of only 10~keV. Predictions for the radiative decays can be found in~\cite{Godfrey:2003kg,Bardeen:2003kt}. An updated calculation of the excited $c\bar q$ ($q=u,d,s$) meson spectrum and the decay properties in the relativized quark model can be found in~\cite{Godfrey:2015dva}.

The $D_{s0}^*(2317)$ and $D_{s1}(2460)$ states could be compact tetraquarks, namely $[cq][\bar s\bar q^{\,\prime}]$ states,
where the  two quarks and  the two antiquarks are coupled to form a color singlet, see, e.g.,~\cite{Browder:2003fk,Maiani:2004vq}. \Blu{Having a single heavy quark, the mechanism driving isospin violation described in~\cite{Rossi:2004yr,Maiani:2004vq} is expected not to be as effective as for the $X(3872)$, making the $D_{sJ}$ states isosinglet. The only kinematically allowed hadronic decay channel is the isovector $D_s^{(*)} \pi$ state. Hence, the lack of isospin violation, together with the predilection for tetraquarks to decay into baryons, explains the narrowness of the states.}  
Using the mass of the $X(3872)$  to fix the chromomagnetic coupling in the tetraquark model,  one can indeed accommodate both of them. Of course, this predicts other similar states with quantum numbers $(0,1,2)^+$, all expected to decay into $D_s^{(*)+} \pi^0$. The $D^{(*)} K$ mode should be open for the heavier members in the multiplet.  %

Another possibility is to describe   the $D_{s0}^*(2317)$ and  $D_{s1}(2460)$ as $DK$ and $D^*K$ molecules~\cite{Barnes:2003dj,vanBeveren:2003kd}.
Dynamically they can be generated from charmed-meson--light-meson scattering in unitarized heavy-meson chiral effective theory (UHMChPT)~\cite{Kolomeitsev:2003ac,Guo:2006fu,Guo:2006rp,Gamermann:2006nm},  using as input the scattering lengths calculated on the lattice~\cite{Liu:2012zya}.  Such scheme predicts  a (strangeness, isospin) $(S,I)=(1,0)$ state  with a  mass $2315^{+18}_{-28}$~MeV, in agreement with that of the  $D_{s0}^*(2317)$ resonance.  Applying the Weinberg's compositeness condition allows one to estimate the size of the molecular component to be about 70\%~\cite{Liu:2012zya}.  Such a picture is supported by the lattice energy levels reported in~\cite{Bali:2017pdv, Lang:2014yfa}, as discussed in \cite{Torres:2014vna, Albaladejo:2018mhb}. 

A decisive measurement is that of the width of the $D_{s0}^*(2317)$, which is expected to be around 100~keV in the molecular picture and smaller in other models~\cite{Colangelo:2003vg, Godfrey:2003kg}. So far, only an upper limit has been provided. It would be difficult to measure such a tiny width at LHC experiments for two reasons: because it requires a very high mass resolution, and because there is a neutral pion in the dominant decay mode $D_s^+\pi^0$. However, it could be possible to measure the width of its spin partner $D_{s1}(2460)$, which carries similar information, through the decay mode $D_s\pi^+\pi^-$ and the Dalitz decay $D_{s1}(2460)\to D_s \gamma(\to\mu^+\mu^-)$. Dalitz decays have already been probed successfully for the $\chi_c$ states~\cite{Aaij:2017vck}. Both decay modes benefit from a very good mass resolution ($\sim1$~MeV), given the small $Q$ values of the reactions. Though a measurement of the width with a precision of 100~keV might be 
optimistic, 
LHCb should at least be able to improve on the current upper limit ($<3.5$~MeV). The proposed decay modes suffer from tiny branching fractions, but the large integrated luminosity in the HL-HE era would help cope with that. 

 In the $(S,I)=(0,1/2)$ sector,  UHMChPT suggests the presence of two broad $D_0^*$ states at about $2.10-i\,0.10$~GeV and $2.45-i\,0.13$~GeV~\cite{Albaladejo:2016lbb}, which would have masses different from the $D_0^*(2400)$ given in the RPP~\cite{PDG2018},  determined from fitting to the $D\pi$ mass distributions using a single Breit--Wigner function. 
Ref.~\cite{Du:2017zvv} showed that these {amplitudes} can reproduce well the $B^-\to D^+\pi^-\pi^-$ process measured by the LHCb Collaboration~\cite{Aaij:2016fma}. The combination of angular moments, $\langle P_1\rangle- \frac{14}{9}\langle P_3\rangle$, is particularly sensitive to the $D\pi$ $S$-wave, and a {higher-statistics} measurement in the energy range between 2.4 and 2.5~GeV will provide invaluable information on the $D\pi$--$D\eta$--$D_s\bar K$ coupled-channel dynamics. 
{Ref.~\cite{Du:2017zvv} also finds} two broad $D_1$ states at $2.25-i\,0.11$~GeV and $2.56-i\,0.20$~GeV, in addition to the relatively well-understood narrow $D_1(2420)$. A precise study of $\langle P_1\rangle- \frac{14}{9}\langle P_3\rangle$  for the $D^*\pi$ $S$-wave is needed in the $\bar B\to D^{*}\pi\pi$ channel,
as well as in $\bar B\to D^{(*)}\eta\pi$, $\bar B\to D^{(*)}_s\bar K\pi$, $\bar B\to D^{(*)}_s\bar K\bar K$, $\bar B_s\to D^{(*)}\bar K\pi$, $\bar B_s\to D^{(*)}_s\eta\pi$, etc. In particular, signals of the predicted higher $D_0^*$ and $D_1$ states, coupled dominantly to $D_s^{(*)}\bar K$, could be near-threshold enhancements in the $D_s^{(*)}\bar K$ spectrum. Such enhancements exist in low-statistics data by BaBar~\cite{Aubert:2007xma} and Belle~\cite{Wiechczynski:2009rg,Wiechczynski:2014kxh}, and need to be investigated at LHCb using data sets with much higher statistics. 

Heavy-quark flavour symmetry allows one to predict the bottom partners of charmed mesons. The lowest $B_{s0}^*$ and $B_{s1}$ are predicted to be at about 5.72~GeV and 5.77~GeV~\cite{Albaladejo:2016lbb,Du:2017zvv}, consistent with lattice results~\cite{Lang:2015hza}. A good channel to search for both of them is $B_s^*\gamma$~\cite{Bardeen:2003kt,Cleven:2014oka}.
They can also decay into the isospin-breaking hadronic channels $B_s^{(*)}\pi^0$, whose widths are expected to be smaller than those of their charm partners because the isospin splitting between $B^0$ and $B^\pm$ is an order of magnitude smaller than that between $D^0$ and $D^\pm$. The axial state $B_{s1}$ can decay into $B_s\gamma$ as well. Both the $B_s\gamma$ and $B_s^*\gamma$ decay modes should appear in the $B_s\gamma$ spectrum with similar efficiencies (similarly to the $B_{s2}^*(5840) \to B^*K$ decay, which peaks in the $BK$ spectrum as well~\cite{Aaij:2012uva}).
{Large integrated luminosity will cope with the low efficiency for detecting soft photons, and the LHCb experiment will have sensitivity to observe such states for the first time.} %
The predicted poles for the $B_0^*$ mesons are at about $5.54-i\,0.11$~GeV and $5.85-i\,0.04$~GeV, while those for the $B_1$ mesons they are at about $5.58-i\,0.12$~GeV and $5.91-i\,0.04$~GeV~\cite{Albaladejo:2016lbb,Du:2017zvv}. The bottom meson spectrum in quark models is different. For an updated calculation of the excited $b\bar q$ ($q=u,d,s$) meson spectrum and the decay properties in the relativized quark model can be found in~\cite{Godfrey:2016nwn}. And the predictions of bottom mesons in the parity-doubling model can be found in~\cite{Bardeen:2003kt}. These resonances can be searched for in $B\pi$ and, for higher excited states, in $B_s\bar K$ final states.

\Blu{The $D_{sJ}(2860)$ discovered by BaBar~\cite{Aubert:2006mh} was split into a $1^-$ state, the $D^*_{s1}(2860)$, and a $3^-$ state, the $D_{s3}^*(2860)$ by the LHCb measurement with higher statistics~\cite{Aaij:2014xza}. Yet, the ratio of two branching fractions $\mathcal{B}(D^*K)/\mathcal{B}(DK)=1.10\pm0.24$~\cite{Aubert:2009ah} is assigned to the spin-1 $D^*_{s1}(2860)$ in the RPP2018~\cite{PDG2018}. This value is much larger than the expectations in Ref.~\cite{Colangelo:2006rq}, based on heavy quark spin symmetry (HQSS) and the leading order chiral Lagrangian, for the $D$-wave $c\bar s$ mesons, which are the only available quark-model option in that mass region. The predictions in~\cite{Colangelo:2006rq} are 0.06 and 0.39 for the $1^-$ and $3^-$ $D$-wave states, respectively. However, the quark-model calculation in Ref.~\cite{Segovia:2015dia} gives larger values, 0.72 and 0.68 for the $1^-$ and $3^-$ $D$-wave states, respectively. On the other hand, in the picture for the $D^*_{s1}(2860)$ as mainly a $D_1(2420)K$ bound state, the ratio is predicted to be 1.23 as a natural consequence of HQSS~\cite{Guo:2011dd}. Heavy quark symmetries allow to predict several related mesons in this picture: a $2^-$ spin partner at about 2.91~GeV, decaying into $D^*K$ and $D_s^*\eta$; the bottom partners at about 6.15~GeV and 6.17~GeV, decaying into $B^{(*)}\bar K$ and $B_s^{(*)}\eta$~\cite{Guo:2011dd}.
}

\noindent
\paragraph{Lattice QCD}

The most extensive spectrum  of higher-lying $D$ and $D_s$ mesons  on the lattice \cite{Cheung:2016bym} was obtained in the single-hadron approximation, in which the decays of resonances and the effects of thresholds are not taken into account.  This lattice study predicts a large number  of states with $J\leq 4$~\cite{Cheung:2016bym}, most of which have not been discovered yet. Some of them might contain substantial gluonic components (hybrid mesons).   

A proper   treatment of  a strongly decaying resonance $R\to H_1H_2$,
with one open channel, 
 requires simulations of single-channel  $H_1H_2$ scattering. This has been  accomplished  recently by several lattice collaborations for a series of resonances. Mostly, resonances composed of $u,d,s$ were considered.    The infinite-volume scattering matrix $T(E)$ is extracted from the energies of $H_1H_2$ eigenstates on the finite lattice via the rigorous L\"uscher's formalism~\cite{Luscher:1990ux}. The resulting scattering matrix $T(E)$ renders resonance masses, $M$, and decay widths, $\Gamma$, via Breit--Wigner-type fits. A related strategy is to analytically continue   $T(E_c)$ to the complex-energy plane, where the pole positions, $E_c\simeq M-\tfrac{i}{2}\Gamma$, are related to the resonance parameters.   The resonances  that strongly decay   to several final states require simulation of coupled-channel scattering, which is much more challenging. The Hadron Spectrum Collaboration managed to extract the coupled-channel scattering matrix   for a few selected channels, while most channels  are awaiting future simulations.

Strong decays of resonances containing a single heavy quark were considered only for  the   low-lying charmed resonances  with $J^P=0^+$, $1^\pm $,  $2^+$~\cite{Mohler:2012na,Moir:2016srx}.
The $N_f=2$   simulation of $D\pi$ scattering~\cite{Mohler:2012na} finds a broad $D_0^*$, in rough agreement with experiment.    As commented above,  a reanalysis of  $D\pi$--$D\eta$--$D_s\bar K$  coupled channels~\cite{Moir:2016srx} for $N_f=2+1$ suggests two $D_0^*$ states, with masses located around $2.10$ and $2.45$ GeV~\cite{Albaladejo:2016lbb}.

The strongly stable states that lie closely below the $H_1H_2$ threshold might be sensitive to threshold effects. A proper way to treat these is to simulate $H_1H_2$ scattering. The mass of a shallow bound state corresponds to an  energy $E<m_1+m_2$, of which the scattering matrix, $T(E)$, has a pole.  In this way, the effects of   $D^{*}K$ thresholds were found to push the masses  of the   $D_{s0}^*$ and $D_{s1}^*$ down, bringing them close to the experimental values \cite{Bali:2017pdv,Lang:2014yfa}.  Analogously, the yet-undiscovered strongly stable $B_{s0}^*$ and $B_{s1}^0$ were predicted at  $5.71$ and  $5.75$ GeV,  respectively~\cite{Lang:2015hza}. For the decay modes that can be used in searching, we refer to the discussion in \ref{sec:spectro:mesons:pheno}.
\subsubsection{Baryons}

The most extensive spectrum of yet-undiscovered singly charmed baryons was predicted in 2013 \cite{Padmanath:2013bla}. It predicted five $\Omega_c$ baryons in the region 3.0-3.2 GeV, in impressive agreement with the 2017 LHCb discovery~\cite{Aaij:2017nav}. 
This work  predicted also up to ten $\Lambda_c,~\Sigma_c,~\Omega_c,~\Xi_c$ states in each  channel  with $J\leq 7/2$,  where resonances are treated in a simplified single-hadron approach.   A follow-up precision lattice study of the five discovered $\Omega_c$ resonances  confirms their most likely quantum numbers \cite{Padmanath:2017lng}.  

The experimental discovery of the five $\Omega_c$ states has triggered extensive theoretical activity,
with some of the quark models 
revisited in view of the new result~\cite{Karliner:2017kfm, Wang:2017hej, Chen:2017gnu, Cheng:2017ove, Wang:2017vnc, Yang:2017rpg}. 
The role of diquarks in the $\Omega_c$ spectrum was discussed in~\cite{Karliner:2017kfm,Ali:2017wsf, Kim:2017jpx},  while the odd-parity molecular interpretation for two or three of the states seen by LHCb was proposed in Refs.~\cite{Nieves:2017jjx, Montana:2017kjw,  Debastiani:2017ewu}.  The
meson-baryon interactions used in the  molecular schemes, derived in \cite{Nieves:2017jjx,  Debastiani:2017ewu},  are consistent with both chiral and heavy-quark spin symmetries, and lead to  successful descriptions of the observed lowest-lying odd parity $\Lambda_c(2595)$, $\Lambda_c(2625)$~\cite{GarciaRecio:2008dp, Romanets:2012hm, Liang:2014kra, Liang:2014eba} and $\Lambda_b(5912)$,  $\Lambda_b(5920)$~\cite{GarciaRecio:2012db} resonances.  Some of the $\Omega_c$ states observed by LHCb could thus be  spin-flavour symmetry partners of these  $\Lambda^*_{Q=c,b}$ baryons.  However,  the masses and decay widths of $\Lambda^*_{Q=c,b}$,  at least the charmed ones,  can also be accommodated  within  usual constituent quark models (see for instance \cite{Yoshida:2015tia, Nagahiro:2016nsx }), and thus the importance of the molecular components in their structure has not been settled yet. 
\Blu{Information obtained from the $\Lambda_b\to \bar \nu_\ell \ell\Lambda_c(2595)$, $\Lambda_b\to \bar \nu_\ell \ell\Lambda_c(2625)$, and related reactions will put some constraints on these models, as shown in~\cite{Liang:2016exm}.}

Molecular schemes also predict partners of the $\Omega_c$ baryons in the bottom sector (see for instance, Ref.~\cite{Liang:2017ejq}). 
LHCb recently reported~\cite{Aaij:2018yqz}  a peak in both the  $\Lambda_b^0 K^-$ and $\Xi^0_b\pi^-$ invariant mass spectra that  might  correspond to either a radially or orbitally excited quark-model  $\Xi_b^-$(6627) resonance with quark content $bds$, or to  a hadron molecule. In the latter, it would be  dynamically generated from the coupled-channel interaction between Goldstone bosons and the lowest even-parity bottom baryons~\cite{GarciaRecio:2012db,Lu:2014ina,Huang:2018bed,Yu:2018yxl}.  For a review of recent observations and phenomenological models of open-flavour heavy hadrons, we refer to~\cite{Chen:2016spr}.

\subsection{Hadrons containing $\bar cc$, $\bar bb$ or $\bar cb$}
 
The current experimental status for the hadrons containing a heavy quark-antiquark pair can be found in  several recent reviews~\cite{
Chen:2016qju,Hosaka:2016pey,Lebed:2016hpi,Esposito:2016noz,Guo:2017jvc,Ali:2017jda,Olsen:2017bmm,Karliner:2017qhf,Yuan:2018inv,Kou:2018nap,PDG2018},  and will not be repeated here. In the following, we discuss  isospin-scalar quarkonium(-like) states and, separately,  
exotic charged states, and focus on a few selected important issues.
\noindent
\subsubsection{Quarkonia}

\paragraph{Phenomenology}
The $J^{PC}=1^{++}$ $X(3872)$ state has several salient features: the central value of its mass coincides 
with the $D^0\bar D^*$ threshold within a small uncertainty of 180~keV~\cite{PDG2018}; despite the tiny phase space, its decay branching fraction into $D^0\bar D^0\pi^0$ is larger than 40\%~\cite{PDG2018}; it decays into  $\omega J/\psi$ ($I=0$) and $\rho J/\psi$ ($I=1$) with similar partial decay widths. 

This state could be interpreted in terms of a $[cq][\bar c\bar q]$ compact tetraquark. In this case, the isospin violation is explained by the smallness of $\alpha_s(m_c)$~\cite{Rossi:2004yr,Maiani:2004vq}, which suppresses the mixing between the almost degenerate mass eigenstates $X_u=[cu][\bar c\bar u]$ and $X_d=[cd][\bar c\bar d]$.
Isospin symmetry predicts a degenerate charged partner $X^+$\Blu{-- experimental search for which has so far been unsuccessful}. Similarly, degenerate isoscalar and isovector states with signature $0^{++}$, $0^{++\prime}$ and $2^{++}$ are expected, with masses around $M(0^{++})\simeq3.8$~GeV and $M(0^{++\prime})\simeq M(2^{++})\simeq4$~GeV~\cite{Maiani:2014aja}.  Were the $Z_c(4050)$ a scalar or a tensor, it would have been a suitable candidate for one of the above isovector states (see e.g.~\cite{Olsen:2015zcy}). If confirmed, another possible candidate  for the heavier scalar state could be the resonance observed by LHCb in $\eta_c\pi$~\cite{Aaij:2018bla}; see also Ref.~\cite{Braaten:2013boa}. 
Given the expected mass,  the charged partner predicted in the tetraquark picture, $X^+$, might only decay into a charmonium and a light meson. %
 \Blu{As discussed in Sec.~\ref{tetra}, a potential barrier between the diquarks has been conjectured~\cite{Selem:2006nd, Jaffe:2003sg, Maiani:2017kyi, Esposito:2018cwh}. This would solve the issue about the elusive $X^+$: the decay into $J/\psi$ plus hadrons would be largely suppressed by the tunneling factor for a heavy quark, and could make the hadronic decay comparable to the electromagnetic one, as is indeed observed in data. Although physically and phenomenologically motivated, such a picture awaits experimental confirmation. Specifically, the charged partners of the $X$ could be searched in the $J/\psi\rho^\pm$ channel.}

The properties of the $X(3872)$, on the other hand,  indicate that it couples strongly to $D\bar D^*$. This leads to the proposal that it could have a large  $D\bar D^*$ molecular component~\cite{Tornqvist:2004qy,Swanson:2003tb,AlFiky:2005jd}. 
\Blu{The isospin violation in this case stems from the difference of masses between the neutral $D^0\bar D^{*0}$ and charge $D^+D^{*-}$ in the loops~\cite{Gamermann:2009fv,Hanhart:2011tn,Li:2012cs}.}
The line shapes of the $X(3872)$ in both $J/\psi\pi^+\pi^-$ and $D^0\bar D^{*0}$ modes are crucial to reveal its nature and binding mechanism~\cite{Hanhart:2007yq,Kalashnikova:2009gt,Artoisenet:2010va,Baru:2011rs,Kang:2016jxw} (for a review, see~\cite{Kalashnikova:2018vkv}). Improved measurements at the LHC experiments are foreseen. In the hadronic molecular picture,  the $2^{++}$ heavy-quark spin partner of the $X$, dubbed $X_2$, would decay into $D\bar D^{*}$ and $D\bar D$ in a $D$ wave,  and it is expected to be narrow~\cite{Albaladejo:2015dsa,Baru:2016iwj}. However,
it has  been suggested~\cite{Cincioglu:2016fkm} that the mixing of the $D^*\bar D^{*}$ molecule with bare charmonium  $\chi_{c2}(2P)$ might destabilize the $X_2$~\cite{Ortega:2017qmg}, making it hardly visible. 
Better experimental information on the $2^{++}$ spectrum around 4~GeV is hence of great importance, and the $X_2$ can be searched for in $D\bar D$ and $J/\psi\omega$. So far, there is only one observed state,  $\chi_{c2}(3930)$,  compatible with a standard charmonium assignment, and its mass is lower than what is normally expected for the $X_2$. No evidence for an additional $2^{++}$ state that could be identified as the $X_2$  has been observed so far. Hence, the existence of the  $\chi_{c1}(2P)$ and/or the $X_2$, in addition to the $X(3872)$ and $\chi_{c2}(2P)$ states, respectively, are still open questions which need to be addressed and that will definitely shed light into the dynamics of the mysterious $X(3872)$~\cite{Cincioglu:2016fkm}.

The $0^{++}$ spectrum is also still unclear: the spin of the narrow $X(3915)$ is not fixed, and the broad $X(3860)$~\cite{Guo:2012tv,Chilikin:2017evr} needs confirmation. Since the $(0,1,2)^{++}$ and $1^{+-}$ heavy quarkonia differ from each other only by the quark polarization, it is necessary to consider systematically physical states with these quantum numbers ~\cite{Cincioglu:2016fkm,Zhou:2017dwj}.  Thus, searching for an isoscalar $1^{+-}$ state around 3.9~GeV is also of high interest~\cite{Lebed:2017yme}. Its important decay modes include $D\bar D^*$, $J/\psi\eta$ and $J/\psi\pi\pi$. Hints for such a state have been seen by COMPASS~\cite{Aghasyan:2017utv}, albeit with low statistics.

\Blu{
The $X(3872)$ could emerge as a dynamically generated mixed state of a $DD^\ast$ molecule and the $\chi_{c1}(2P)$ as shown in~\cite{Ortega:2009hj}, see also Refs.~\cite{Kalashnikova:2005ui,Kalashnikova:2009gt,Danilkin:2009hr,Takizawa:2012hy,Meng:2013gga,Takeuchi:2014rsa,Zhou:2017dwj,Zhou:2017txt}. The $c\bar{c}$ mixture is less than $10\%$ but it is important to bind the molecular state. The proposed structure would allow to understand the origin of its prompt production rate and to describe correctly its isospin violating decays and radiative transitions. The original
$c\bar{c}(2^3P_1)$ state is dressed by meson-meson coupled-channel effects and goes up in the spectrum allowing its identification with the $X(3940)$. Within the same coupled-channels scheme, Ref.~\cite{Ortega:2017qmg} analyzes the $J^{PC}=0^{++}$ and $2^{++}$ charmonium sectors. The two hadron states found in the $0^{++}$ channel were assigned to the $Y(3940)$ and to the recently identified resonance $X(3860)$~\cite{Guo:2012tv,Chilikin:2017evr}. The bound state with quantum numbers $2^{++}$ found in~\cite{Ortega:2017qmg} is interesting because, within the uncertainties of the model, its mass and decay properties suggest that the $X(3915)$ and $X(3930)$ meson-like structures could correspond to the same state, as also claimed in Ref.~\cite{Zhou:2015uva}.

}

It is also crucial to look for the analogue of the $X(3872)$ in the bottom sector. Such a state could decay, for example, into $B\bar B\gamma$, 
 $\chi_{bJ}\pi\pi$, $\Upsilon\pi\pi\pi$ and $\Upsilon\gamma$~\cite{Guo:2013sya,Guo:2014sca,Karliner:2014lta}. 
 The $X_b\to\Upsilon\rho$ channel has already been investigated, with no observation~\cite{Chatrchyan:2013mea,Aad:2014ama}. \Blu{It should be noticed, however, that this channel should be highly suppressed due to $G$-parity violation. There is an important difference between the $X(3872)$ and the $X_b$: The mass difference between the charged and neutral charmed mesons is one order of magnitude larger than that for the bottom mesons, because of the different interference between the QCD and QED contributions~\cite{Guo:2008ns}, and thus the $X_b$ should be a well-defined isoscalar state~\cite{Guo:2013sya}. The null results of Refs.~\cite{Chatrchyan:2013mea,Aad:2014ama} are perfectly consistent with the existence of an isoscalar $X_b$~\cite{Guo:2014sca,Karliner:2014lta}.}  Note that in the tetraquark model one would expect isospin violation, while the opposite is true in a molecular scheme. The isospin-conserving channel $X_b \to \Upsilon \omega$ has found no evidence for the state either~\cite{He:2014sqj}. The spin-2 bottomonium-like $X_{b2}$ and $B_c$-like state $X_{bc2}$~\cite{Guo:2013sya} can instead be searched for in $D$-wave $B\bar B$ and $DB$ final states, respectively.

Vector $1^{--}$ states have also been observed at $e^+e^-$ colliders (see,  e.g.,~\cite{Aubert:2005rm,Liu:2013dau,Ablikim:2013dyn,Wang:2014hta,Lees:2012pv}). The number of observed vector charmonium-like states far exceeds the predictions in the $c\bar c$ potential quark models, and thus some of them must have an exotic origin. In the tetraquark model these are diquarkonia with orbital angular momentum $L=1$, and their spectrum easily matches the  experimental observations~\cite{Maiani:2014aja} (see also~\cite{Wang:2018ntv}), which also leads to a distinctive spectrum 
with a low-mass $3^{--}$ state~\cite{Cleven:2015era}. Complexity in understanding these structures also comes from the many $S$-wave thresholds of hadron pairs, such as the $\bar D D_1(2420)$ for the $Y(4260)$~\cite{Wang:2013cya} and $\psi'f_0(980)$ for the $Y(4660)$~\cite{Guo:2008zg}. \Blu{Effects of open charm channels for the $Y(4260)$ are considered in a quark model framework in Ref.~\cite{Lu:2017yhl}. The $Y(4260)$ is also widely regarded as a candidate for a hybrid charmonium~\cite{Zhu:2005hp,Kalashnikova:2008qr,Berwein:2015vca,Chen:2016ejo,Oncala:2017hop}. The hadro-charmonium model for the $Y(4260)$ and $Y(4360)$~\cite{Li:2013ssa} can hardly explain why the $Y(4260)$ was seen in several different channels including $J/\psi \pi^+\pi^-, h_c\pi^+\pi^-, \chi_{c0}\omega$ and $D^0D^{*-}\pi^++c.c.$ (the data are nicely summarized in Ref.~\cite{Yuan:2018inv} and a combined fit in these channels leads to a lower mass around 4.22~GeV for the $Y(4260)$~\cite{Gao:2017sqa}.} The search for these states in, for example, prompt production and/or $B$ decays can help to understand their nature.

\paragraph{Lattice QCD}

The lattice spectrum of bottomonia presented in~\cite{Wurtz:2015mqa} contains almost all the observed $\bar bb$ states up to the $\bar BB$ threshold. It also predicts a plethora of undiscovered states where the $\bar bb$ pair carries an orbital angular momentum $L=2,3,4$, and total angular momentum $J\leq 4$.   Fourteen $B_c$ mesons with $J\leq 3$   are predicted up to the $BD$ threshold~\cite{Wurtz:2015mqa}. Only two of them have been discovered so far.  

All the charmonia below the $D\bar D$ threshold have been experimentally discovered, while the treatment of strongly decaying resonances is much more challenging. The most extensive spectrum obtained in the simplified  single-hadron approach predicts several excited $\bar cc$ states, as well as $\bar cc g$ hybrids up to $4.7$ GeV, carrying   $J\leq 4$~\cite{Cheung:2016bym}, also including exotic $1^{-+},~0^{+-}, ~2^{+-}$ quantum numbers. Most of them have not been observed yet.  Only one exploratory study considered the resonant nature of charmonia above the open-charm threshold \cite{Lang:2015sba}, and underlines the need to  experimentally explore further $\bar DD$ in $S$ and $D$ waves.    A neutral  is found  as a state slightly below $D\bar D^*$~\cite{Prelovsek:2013cra}. \Blu{A neutral $X(3872)$ is found slightly below $D\bar D^*$ threshold~\cite{Prelovsek:2013cra} and corresponds to a bound-state pole in the $D\bar D^*$ scattering matrix. The energy eigenstate related to $X(3872)$ appears in the simulation only if $D\bar D^∗$ as well as $c\bar c$  interpolating fields are employed. The $X(3872)$ does not appear in absence of $c\bar c$ interpolators. Although the overlaps are scheme and scale dependent, and no theoretically strict conclusion can be driven from them, this might nevertheless suggest that the $c\bar c$ Fock component is most likely more essential for creating the $X(3872)$ than the diquark-antidiquark one.}  No other   $1^{++}$ state is found between threshold and  $4.1$~GeV, in agreement with the experiment. Similarly, no indication for the isospin-1 partner of $X(3872)$  is found   \cite{Padmanath:2015era}. Note that these simulations were performed  in the $m_u = m_d$  limit.

\noindent
\subsubsection{Charged exotic states}

\paragraph{Phenomenology}

\Blu{Several charged strutures have been reported in the charmonium region, such as the $Z_c(3900)$, $Z_c(4020)$, $Z_c(4200)$, $Z_c(4430)$ and others; only two were observed in the bottomonium region, i.e., the $Z_b(10610)$ and $Z_b(10650)$. The $Z_c(3900)$, $Z_c(4020)$ and the two $Z_b$ states (the lower and higher sates will be referred to as $Z_{c(b)}^{}$ and $Z_{c(b)}'$, respectively), share some similarities: Their quantum numbers are $J^{PC}=1^{+-}$; they are rather close to the corresponding $D^{(*)}\bar D^*$ and $B^{(*)}\bar B^*$ thresholds, respectively; despite of the little phase space, they decay dominantly into the open-heavy-flavor channels (notice that being isospin-vector states, their decays into a heavy quarkonium and a pion do not violate the OZI rule). 
They also have noticeable differences: The $Z_c(3900)$ and $Z_c(4020)$ were discovered in the $J/\psi\pi$ and $h_c\pi$, respectively, while both $Z_b$ states decay into the $\Upsilon(1S,2S,3S)\pi$ and $h_b(1P,2P)\pi$ with similar rates.}

The $Z_c^{(\prime)}$ and $Z_b^{(\prime)}$ are 
predicted by the diquarkonium model as part of the $[cq][\bar c\bar q]$ spectrum~\cite{Ali:2011ug,Maiani:2014aja}. A possible hint on their internal structure could be provided by the $Z_c^{(\prime)}\to\eta_c\rho$ decay, for which the tetraquark and molecular models predict statistically different results~\cite{Esposito:2014hsa,Agaev:2016dev,Agaev:2017tzv,Xiao:2018kfx}. Similar analyses could also be performed for $Z_b^{(\prime)}\to\eta_b\rho$. Searching for the decay $Z_c^\prime\to J/\psi\pi$ with higher statistics is also important to confirm/disprove a possible tension between the molecular model and the experimental data~\cite{Esposito:2014hsa}. Analyses as in~\cite{Pilloni:2016obd} can give a more robust extraction of the resonant parameters, thus helping in distinguishing between the possible interpretations.
The $Z(4430)$ is also easily accommodated in the tetraquark model, as the first radial excitation of the $Z_c$~\cite{Maiani:2014aja}. Such a state has been discovered in a $\psi(2S)\pi$ final state, but the search for it  in other channels is also important (e.g., $J/\psi\pi$).

From a molecular perspective, once again although the $Z_c^{(\prime)}$ and $Z_b^{(\prime)}$ are close to the $D^{(*)}\bar D^*$ and $B^{(*)}\bar B^*$ thresholds, respectively, \Blu{the predominance of their decays into the open-flavor channels is quite natural due to the strong couplings.}  The pole locations of the near-threhold $Z_c$ and $Z_b$ states are not precisely determined. It was shown in~\cite{Albaladejo:2015lob,Pilloni:2016obd} that the BESIII data on the $J/\psi\pi$ and $D\bar D^*$ distributions are consistent with either a virtual state or a resonance, which is also consistent~\cite{Albaladejo:2016jsg} with the energy levels calculated on the lattice~\cite{Prelovsek:2014swa}. The data for the lighter $Z_b$ are also consistent with a virtual state~\cite{Guo:2016bjq,Wang:2018jlv},  although the state lies so close to threshold, that it is hard to draw final conclusions based on  the present data. Line shapes near the open-flavour thresholds in both the open- and hidden-flavour channels need to be extracted precisely, in order to pin down the positions of the poles.

It is worth noticing that a threshold cusp with a nearby pole is not enough to produce peaks as narrow as the ones observed  for the $Z_c$~\cite{Guo:2014iya}.
There are suggestions that nearby triangle singularities play an important role in producing the $Z_{c(b)}$ structures associated with a pion in $e^+e^-$ collisions~\cite{Wang:2013cya,Wang:2013hga,Szczepaniak:2015eza,Pilloni:2016obd,Gong:2016jzb,Bondar:2016pox}. In order to disentangle such effects, the $Z_{c(b)}$ states need to be searched for in other processes. Recently, the D0 Collaboration reported evidence for the $Z_c$ in semi-inclusive $b$-flavoured hadron decays~\cite{Abazov:2018cyu}, which is waiting for confirmation from the LHC experiments. 

In the molecular picture the $Z_b$ states may have more isospin-vector partners with quantum numbers $(0,1,2)^{++}$, denoted as $W_{bJ}$ states~\cite{Voloshin:2011qa}. The searches for an isospin-scalar $X_b$ in $J/\psi\pi^+\pi^-$, by CMS~\cite{Chatrchyan:2013mea} and ATLAS~\cite{Aad:2014ama}, should be reinterpreted as negative results for seeking the $W_{bJ}$ states. Improved experimental information would be of great importance for understanding the charged bottomonium-like structures.

\paragraph{Lattice QCD}

Lattice QCD does not find candidates for $\bar QQ\bar du$ ($Q=c,b$) states below the strong-decay thresholds $m_{\bar QQ}+m_\pi$. The study of the $Z_c$ and $Z_b$ resonances above threshold is much more challenging as it requires the simulation of coupled-channel scattering.  For this reason, lattice  results for the charged $Z_c$ and $Z_b$  states using L\"uscher's approach are not (yet) available. 

  The  indication for the lowest-lying $Z_c(3900)^+$   was found in a lattice study based on the somewhat simpler HALQCD method~\cite{Ikeda:2016zwx}, where the peak arises due to a threshold cusp and not to a resonance pole.   This seems to be in accordance with the absence of additional energy eigenstates in \cite{Prelovsek:2014swa,Cheung:2017tnt}, but more work based on the more rigorous L\"uscher's method is needed. An indication for the $Z_b$ below the $B\bar B^*$ threshold was found using static $b$ quarks, but $\Upsilon \pi$ states  with non-zero pion momentum must be incorporated before drawing any firm conclusion~\cite{Peters:2017hon}.

\subsubsection{Pentaquarks}

\paragraph{Phenomenology}

The observation of  two peaks in the $J/\psi p$  spectrum  in $\Lambda^0_b \to J/\psi K^-p$ decays by LHCb in 2015~\cite{Aaij:2015tga} triggered a large activity in the field.  If they are interpreted as  actual  QCD resonances,  they will be  consistent with pentaquark states with opposite parities. This 
 seems to indicate that diquarks play a role in their structure~\cite{Maiani:2015vwa,Lebed:2015tna}; it is 
 hard to accommodate orbitally excited states in the framework of a bound state of color neutral hadrons. Moreover, the diquark-diquark-antiquark model correctly reproduces the observed mass splitting between the two pentaquarks~\cite{Maiani:2015vwa}. The SU(3) flavour symmetry then opens up a vast spectroscopy~\cite{Maiani:2015vwa,Maiani:2015iaa,Ali:2016dkf,Ali:2017ebb}, for which an experimental study is of great interest. Such pentaquarks could be found in reactions as, for example, $\Xi_b(5749)\to K\,P_c\to K(J/\psi\Sigma(1385))$, $\Omega_b(6049)\to \phi \,P_c\to\phi(J/\psi\Omega(1672))$ or $\Omega_b(6049)\to \phi \,P_c\to K(J/\psi\Xi(1387))$~\cite{Maiani:2015vwa}. More pentaquarks could also be observed in the $J/\psi p$ or $\eta_c p$ final states~\cite{Ali:2017ebb}. Predictions of hidden-charm pentaquarks as hadronic molecules in the right mass region have been made~\cite{Wu:2010jy,Wang:2011rga,Yang:2011wz,Wu:2012md} a few year before the discovery, \Blu{which was also preceded by a prediction of a $\Sigma_c\bar D^*$ hadronic molecule with properties consistent with the observed $P_c(4450)$~\cite{Karliner:2015ina}}.
After the discovery, there exists also molecular  ($\Sigma_c^{(*)}{\bar D}^{(*)}$) interpretations for the narrower of the peaks (see for instance~\cite{Roca:2015dva, Roca:2016tdh,Karliner:2015ina}), which imply the existence of additional  pentaquark states, not yet observed~\cite{Xiao:2013yca}.   

One complexity in interpreting the experimental observation comes from triangle singularities,  which are indeed present in this mass region~\cite{Guo:2015umn,Liu:2015fea}. In particular, a triangle singularity can produce a narrow peak around  $\simeq4.45$~GeV, if coupled to the $\chi_{c1}p$ in an $S$ wave~\cite{Bayar:2016ftu}. \Blu{The $3/2^-$ and $5/2^+$  quantum numbers, the most preferred in the original LHCb analysis~\cite{Aaij:2015tga}, for the narrow $P_c(4450)$, require $\chi_{c1}p$ in $P$ and $D$ waves, respectively. Yet, the option of $3/2^+$, which allows for an $S$-wave coupling to $\chi_{c1}p$, belongs to one of the preferred sets  in an analysis considering more $\Lambda^*$ hyperons decaying into the final state $pK^-$~\cite{Jurik:2016bdm}.} There are ways to distinguish triangle singularities from real resonances: $i)$ quantum numbers, since triangle singularities produce narrow peaks only for $S$-wave couplings; $ii)$ the near-threshold $\chi_{c1}p$ distribution for the process $\Lambda_b\to \chi_{c1}p K$,  since a triangle singularity would not lead to a near-threshold peak in the $\chi_{c1}p$ invariant mass distribution~\cite{Guo:2015umn} if the inelasticity between the channels is not large; $iii)$ searching for the $P_c$ in processes having different kinematics, as in photoproduction~\cite{Kubarovsky:2015aaa,Blin:2016dlf,Karliner:2015voa}.
Hidden-charm pentaquarks also need to be searched for in $\Lambda_c\bar D^{(*)}$ since their branching fractions could be much larger than that of $J/\psi p$~\cite{Shen:2016tzq,Lin:2017mtz,Eides:2018lqg}. 
There exist model predictions of hidden-bottom pentaquarks~\cite{Wu:2010rv,Wu:2017weo,Yamaguchi:2017zmn,Shen:2017ayv,Lin:2018kcc,Yang:2018oqd}, which can only be searched for at LHC currently. Important modes include $\Upsilon p$ and $\Lambda_b B^{(*)}$.

 \subsection{Doubly-heavy hadrons}

\subsubsection{Baryons}

\paragraph{Phenomenology}

    The quark-model predictions for the spectrum of the states can be found in~\cite{Gershtein:2000nx,Kiselev:2001fw,Ebert:2002ig,Kiselev:2002iy, Albertus:2006ya,Padmanath:2015jea,Karliner:2014gca}. \Blu{The most precise quark-model prediction for the mass of the ground-state $\Xi_{cc}$ appears in Ref.~\cite{Karliner:2014gca}, $(3627\pm12)$~MeV, and a compilation of many other predictions can also be found in that reference.} Decay properties of the doubly-charmed baryons are discussed in~\cite{Karliner:2014gca,Yu:2017zst,Wang:2017mqp,Xiao:2017udy,Cui:2017udv,Xiao:2017dly,Li:2017pxa}. It is worth noticing that the recent LHCb determination~\cite{Aaij:2017ueg}  of the $\Xi_{cc}^{++}$ (ground state) mass ($3621.40 \pm  0.72 \pm  0.27 \pm  0.14$) MeV solves a longstanding discrepancy between quark-model and lattice QCD predictions, and the previous value ($3519 \pm 1$ MeV) reported by the SELEX Collaboration~\cite{Mattson:2002vu}. 
    \Blu{Several experiments attempted to verify the SELEX result, but did not see anything in the corresponding channel. There is consensus that the SELEX result is wrong.\footnote{For  discussions of reasonable values for the isospin splitting between doubly-charmed baryons within QCD$+$QED, see~\cite{Brodsky:2011zs,Karliner:2017gml}.}}
    
    The next doubly-charmed baryons above the $1/2^+$ and $3/2^+$ states are the $1/2^-$ ones.  There are two possible origins: either the $[cc]$ diquark is in a $P$-wave state, or the diquark and the remaining quark have an $L=1$ orbital angular momentum relatively to each other. These two different possibilities mix to form the physical $1/2^-$ states. As a result, there will likely be two $1/2^-$ $\Omega_{cc}$ states below the $\Xi_{cc} \bar K$ threshold, and three  $1/2^-$ $\Xi_{cc}$ isospin-doublets below 4.2~GeV. Among the latter, the lowest one (about 200~MeV heavier than the observed ground state $\Xi_{cc}$) is narrow and can be searched for in the $\Xi_{cc}^{++}\pi^-$ channel~\cite{Yan:2018zdt}. \Blu{The douly-charmed baryons of molecular type without considering the mixture with the $P$-wave $cc$ excitation have been studied in~\cite{Guo:2011dd,Guo:2017vcf,Dias:2018qhp}.}

\paragraph{Lattice QCD}

 The mass of the only discovered doubly-heavy baryon  $\Xi_{cc}^{++}=ccu$ is in very good agreement with the lattice predictions for the $1/2^+$ state.     Other low-lying baryons, containing at least two heavy quarks, have been predicted by a number of simulations, which broadly agree on their masses. The most extensive spectra of $\Xi_{cc}=ccq$ and $\Omega_{cc}=ccs$ baryons were predicted  using the single-hadron approximation in \cite{Padmanath:2015jea}, where up to ten of them were found in each channel with $J\leq 7/2$. Finding the lowest-lying $3/2^+$ state $\Xi_{cc}$ may be challenging for LHCb, as it is predicted to be only  80-100 MeV above the  discovered $\Xi_{cc}^{++}$. The next candidate for discovery may well be $\Omega_{cc}$~\cite{Karliner:2018hos}, predicted at  $3712\pm 10 \pm 12$ MeV \cite{Mathur:2018rwu},  with decay modes such as $\Xi_c^+\bar K^0$, $\Xi_c^+K^-\pi^+$, $\Xi_c^0 K^- \pi^+\pi^+$   and $\Omega_c \pi^+$. 
 The recent predictions for the masses of hadrons containing both charm and bottom quarks are discussed, for example, in~\cite{Mathur:2018epb}. Their production and decays are discussed, for example, in~\cite{Karliner:2018hos,Karliner:2014gca}.

   \noindent
\subsubsection{Tetraquarks}

\paragraph{Phenomenology}

The discovery of the $\Xi_{cc}^{++}$ revived the study of doubly-heavy tetraquarks. Using heavy-quark symmetry arguments, it is generally expected~\cite{Eichten:2017ffp,Czarnecki:2017vco} that the doubly-bottom ground state $\bar b\bar b ud$ is stable so that it can only decay weakly, in line with the lattice results. \Blu{The same conclusion can be reached starting from the observed mass of the $\Xi_{cc}$ and extrapolating to the bottom case~\cite{Karliner:2017qjm}.}
The stability of the ground state $\bar b\bar b ud$ has also been pointed out long ago~\cite{Ader:1981db,Manohar:1992nd}. 
Excited states are discussed in Ref.~\cite{Mehen:2017nrh}.  The doubly-charmed states are likely above the open-charm thresholds, and may form resonances~\cite{Esposito:2013fma,Karliner:2017qjm,Eichten:2017ffp}. The results of various models are collected in Ref.~\cite{Luo:2017eub}. 
Being explicitly exotic, the search for doubly-heavy tetraquarks is of high interest. The production of doubly-heavy tetraquarks at the LHC has been discussed in Refs.~\cite{Yuqi:2011gm,Ali:2018xfq}, and the results show that it is promising to discover them at the LHC experiments CMS, ATLAS and LHCb. The weakly decaying ground-state doubly-beauty hadrons could be searched for by displaced $B_c^-$ mesons as an inclusive signature~\cite{Gershon:2018gda}. \Blu{This provides a competitive channel that, being inclusive, requires a much smaller integrated luminosity.}

One of the most appealing features of these states is the presence of a doubly-charged particle in their spectrum~\cite{Esposito:2013fma}. If it were to be discovered, this resonance would be \Blu{a good candidate for} a compact tetraquark: Coulomb repulsion prevents molecules of this kind from binding. \Blu{However, a sufficiently large strong interaction could overcome the Coulomb repulsion. It is known that the proton-proton system fails to be bound by little, and $^3$He is already bound.}

\paragraph{Lattice QCD}
 
A strongly stable exotic state $\bar b\bar b du$ with $I(J^P)=0(1^+)$   is found up to $0.2$ GeV below the $B\bar B^*$ threshold in several independent lattice approaches, for example \cite{Bicudo:2012qt,Francis:2016hui}. The search for this exotic tetraquark  should be of prime interest at LHC.  Examples of fully reconstructible modes for its weak
decays are   $B^+ \bar D^0$ and $J/\psi B^+ K^0$,  with $\bar D^0\to K^+\pi^-$, $B^+\to \bar D^0\pi^+$ and $K^0\to \pi^+\pi^- K_s$ \cite{Francis:2016hui}. 
   The   $1^+$ state $\bar b\bar bqs$  is also found below the $BB_s$ threshold~\cite{Francis:2016hui} and  it could be   searched for in the weak decays to $J/\psi B_sK^+$, $J/\psi B^+\phi$ for $q=u$, and $B^+D_s^-$, $B_s\bar D^0$, $J/\psi B^0\phi$, $J/\psi B_s K^0$ for $q=d$.   The  systems $\bar c\bar cud$, $\bar c\bar cus$, $\bar b\bar bss$, $\bar b\bar bcc$, $\bar b\bar bbb$, on the other hand, were not found to have compelling signals of bound states below strong decay thresholds  in the channels  explored (see for example \cite{Bicudo:2015vta,Hughes:2017xie}).

\subsection{All-heavy states}

States with all heavy constituents were predicted in Refs.~\cite{Heller:1985cb,Berezhnoy:2011xn,Wu:2016vtq,Chen:2016jxd,Karliner:2016zzc,Bai:2016int,Wang:2017jtz,Richard:2017vry,Anwar:2017toa,Vega-Morales:2017pmm,Eichten:2017ual,Esposito:2018cwh}. The widths of the ground-state $cc\bar c\bar c$ and $bb\bar b\bar b$ are expected to be of the order of a few tens of MeV due to the decays through annihilation of a quark-antiquark pair~\cite{Chao:1980dv,Anwar:2017toa}.
The experimental search for a $bb\bar b\bar b$ state has been of particular interest recently. Indeed, the presence/absence of such a state would be very informative about the nature of the exotic mesons, and different models predict that, with sufficient luminosity, it should be observed~\cite{Eichten:2017ual,Esposito:2018cwh}.

The only lattice simulation of the  $\bar b\bar bbb$ system \cite{Hughes:2017xie} found no evidence for  bound states with a mass below the lowest   bottomonium-pair thresholds, i.e., no $0^{++}$, $1^{+-}$ and $2^{++}$ states    below $\eta_b\eta_b$,   $\Upsilon\Upsilon$ and $\eta_b\Upsilon$ thresholds, respectively. The $b$ quarks were treated using improved non-relativistic QCD (NRQCD), and all four   Wick contractions were taken into account (omitting bottom annihilation). If only part of the Wick contractions (for example ``direct'') are taken into account, the system seems to show false binding  \cite{Hughes:2017xie}, which  could be the reason why some model approaches (effectively treating only part of the Wick contractions) find false bound states below thresholds.     No indication was found also of  $\bar b\bar b cc$ bound states with $J^P=0^+,~1^+,~2^+$ below the   thresholds \cite{Bicudo:2015vta}, where  $cc$ carry spin $1$; this was explored using static $b$ quarks using the Born--Oppenheimer approach. The existence of resonances with two heavy quarks and two heavy antiquarks ($Q=b,c$) above strong-decay thresholds has not been explored on the lattice.   
\Blu{Finally, the most extensive lattice QCD  spectra of $ccc$ baryons was calculated in~\cite{Padmanath:2013zfa},
where up to ten states were predicted for each quantum number $J^P$ ($J\leq7/2$ and $P=\pm$). 
Effects of strong decays and thresholds are omitted in this study.}

\subsection{Probes from prompt production in $pp$}
The sizable prompt-production cross section of the $X(3872)$ in high-energy hadron collisions has triggered lots of debates in the literature on whether it is compatible or not with a pure molecular nature.
The argument, first formulated  in~\cite{Bignamini:2009sk}, relies on the observation that in ${\mathcal O}(\text{TeV})$ collisions, it is unlikely to produce a pair of $D^0$ and $\bar D^{*0}$ --- the alleged constituents of the $X(3872)$ --- with a relative momentum $k_\text{rel}$ small enough for the binding to occur. The identification of the support of $k_\text{rel}$ is crucial, since by phase-space arguments the cross section scales with the cube of the maximum $k_\text{rel}$ allowed. Using the indetermination relations and the upper bound on the binding energy of the $X$, Ref.~\cite{Bignamini:2009sk} estimated $k_\text{rel} \lesssim 30~\text{MeV}$. Using Monte Carlo generators, this turns into a production cross section that is two orders of magnitude smaller then the one measured by CDF. 
The choice $k_\text{rel}\sim 30~\text{MeV}$ was criticized in~\cite{Artoisenet:2009wk}: the inclusion of final state interactions (FSI) generating the bound state would require a relative momentum of the scale of the mediator, i.e., the pion mass. Integrating up to $k_\text{rel}\sim 300~\text{MeV}$ permits to reach the experimental cross section. However,
implementation of FSI in an environment polluted by hundreds of other particles is controversial~\cite{Bignamini:2009fn,Artoisenet:2010uu,Esposito:2013ada,Guerrieri:2014gfa}. The choice of the cutoff has  been the object of a new debate recently~\cite{Albaladejo:2017blx,Esposito:2017qef}, \Blu{and it was shown that the maximum $k_\text{rel}$ should be at least 4.5~times larger than the binding momentum~\cite{Braaten:2018eov} to fulfill the inequality derived in Ref.~\cite{Bignamini:2009sk}. Notice that being allowed by quantum numbers, the hadronic molecular interpretation of the $X(3872)$ necessarily involves mixing with the nearby $\chi_c(2P)$ (see, e.g, ~\cite{Meng:2013gga,Karliner:2014lta}), and the production can occur through the charmonium component.}
In Ref.~\cite{Meng:2013gga},  the $X(3872)$ is considered to be a superposition of a molecular state and of the ordinary $\chi_{c1}(2P)$. The former does not contribute to the production cross section, whereas the latter is estimated in NRQCD. An alternative estimate using NRQCD, mostly agnostic about the nature of the $X$, was anticipated in~\cite{Artoisenet:2009wk}. Note that, if the $X$ were to be a compact tetraquark, the high production cross section would not be an issue.

In~\cite{Esposito:2015fsa} the production cross section of the $X(3872)$ was compared, after high-\pt extrapolation, with the one of deuteron and other light nuclei. The cross section of the $X(3872)$ overshoots the others by a few orders of magnitude, thus undermining the idea that they have the same nature. Refs.~\cite{Albaladejo:2017blx,Wang:2017gay} proposed that the production of exotic states depends on their short-range nature, namely on the number of quarks in the minimal Fock component of the state, regardless of their molecular or compact nature, so that prompt production could give little information about the nature of these states. This would imply that some form of mixing with more compact states is required to reproduce the cross section. Prompt production, together with other processes sensitive to the short-range structure of the exotics, would thus probe the details of the mixing with compact charmonia and tetraquarks.  
Finally, under some  assumptions for the FSI, also predictions for production of the $Z_{c,b}$~\cite{Guo:2013ufa}, $X_b$~\cite{Guo:2014sca}, and $D_{sJ}$~\cite{Guo:2014ppa} have been made.

\subsection{Summary of interesting processes and states}

Here we summarize the interesting states and structures, with an emphasis on candidates of heavy-flavor exotic hadrons, in Table~\ref{tab:interesting}.

\begin{table}[tbp]
\centering
\caption{\label{tab:interesting}
Interesting resonant structures,  decay modes or processes and relevant observables. The last column contains some comments when necessary.}
\begin{tabularx}{\linewidth}{p{2.4cm} p{3.4cm} p{3.7cm} X}
\hline\hline
  States/structures  & Channels & Observables  & Further comments \\ \hline
  $D_{s0}^*(2317)$ & $D_s^{}\pi^0, D_s^{*}\gamma$ & Width upper limit; branching fractions & $\pi^0$ is difficult \\
  $D_{s1}^*(2460)$ & $D_s^{*+}\pi^0, D_s^+\pi^+\pi^-$, $D_s^{(*)+}\gamma (\to \mu^+\mu^-)$ & see above & May use the Dalitz decay to probe photons \\
Broad $D_0^*$, $D_1$ structures & $\bar B\to D^{(*)}\pi^-\pi^-$, $ D^{(*)}_s\bar K\pi$, $D^{(*)}_s\bar K\bar K$, $\bar B_s\to D^{(*)}\bar K\pi$  & $D^{(*)}\pi$ angular moments; $D_s^{(*)}\bar K$ invariant mass distribution & $\langle P_1\rangle - \frac{14}{9}\langle P_3\rangle$ is particularly sensitive to the $D^{(*)}\pi$ $S$-wave; possible enhancement above $D_s\bar K$ threshold  \\ \hline
$B_{s0}^*$ (?)  &  $B_s\pi^0,B_s^*\gamma$ &  & $M\sim5.72$~GeV; not seen yet\\
$B_{s1}$ (?)  & $B_s^*\pi^0, B_s\pi^+\pi^-$, $B_s^{(*)}\gamma$ & & $M\sim5.77$~GeV, lower than $B_{s1}(5810)$; not seen yet\\ \hline
Excited single-heavy baryons & & & A whole SU(3) family; \Blu{determination of spin and parity} \\ \hline
$X(3872)$ & $D^0\bar D^0\pi^0, D \bar D\gamma,$ $J/\psi \pi^+\pi^-$, $J/\psi3\pi$, $J/\psi\gamma$, $\psi'\gamma$ & Line shapes; decay width; production rates & \\ 
$X_2$ (?) & $D\bar D$, $D\bar D^*+c.c.$, $J/\psi\omega$ &  & $J^{PC}=2^{++}$, $M\sim 4$~GeV, $\Gamma\lesssim50$~MeV; existence unknown \\
$\chi_{c1}(2P)$ (?) & $D\bar D^*+c.c., J/\psi \omega$ & & $M\sim3.9$~GeV, broad; existence unknown\\
$h_c(2P)$ (?) & $D\bar D^*+c.c.$, $J/\psi \eta$, $\eta_c\omega$ & & $M\sim3.9$~GeV, broad; not seen yet \\
\hline
$X_b$ (?) & $\Upsilon\omega$, $\chi_{bJ}\pi^+\pi^-$, $B\bar B\gamma$, $\Upsilon\gamma$ & & Bottom analogue of $X(3872)$; existence unknown\\
$X_{b2}$ (?) & $B\bar B$, $\Upsilon\omega$ & & Bottom analogue of $X_2$; existence unknown\\\hline
$Z_c$ structures &  $(c\bar c)\pi^\pm$, $(D^{(*)}\bar D^{(*)})^\pm$ & Line shapes; production rates; Argand plots & Sensitivity to kinematics \\ 
$Z_{cs}$ (?)  & $(c\bar c)K$, $D^{(*)}_s\bar D^{(*)}$ &  & Existence unknown \\
\hline
$Z_b$ structures &  $(b\bar b)\pi^\pm$, $(B^{(*)}\bar B^{(*)})^\pm$ & Line shapes & Not seen at LHC yet \\ 
$W_{bJ}$ (?) & $\Upsilon \pi^+\pi^-, \Upsilon\gamma$ &  & $I^G(J^{PC})=1^-(J^{++})$, possible spin partners of $Z_b$ states; existence unknown
\\ \hline
$P_c$ and relatives & $J/\psi p$, $\chi_{cJ}p$, $\Lambda_c\bar D^{(*)}$,  $\Sigma_c\bar D^{(*)}$  &  & Hidden-charm pentaquarks \\  \hline
Doubly-heavy baryons & & & Displaced $B_c$ as an inclusive signature of weakly decaying double-beauty hadrons  \\ 
$\bar b\bar b ud$ (?); $\bar b\bar b qs$ ($q=u,d$)  (?) & $B^+\bar D^0$, $J/\psi B^+ K^0$; $B\bar D_s$, $B_s \bar D$, $J/\psi B \phi$, $J/\psi B_s K$  & & Ground states likely stable against strong decays; not seen\\\hline
$cc\bar c\bar c$ (?); $bb\bar b\bar b$ (?) & $H_Q\bar H_Q+$anything, $J/\psi (\Upsilon) \mu^+\mu^-$, $\mu^+\mu^-,4\mu$  & 
 & Widths: tens of MeV if below double-$(Q\bar Q)$ thresholds; $H_Q$ denotes any heavy hadron; existence unknown \\ 
\hline\hline
\end{tabularx}
\end{table}

%% file: section6/experimental.tex
\subsection{Experimental prospects}

The LHCb \upgradetwo\ detector will have a large impact on sensitivity in searches for heavy states. Aside from the much larger integrated luminosity, many of the 
detector improvements planned for LHCb \upgradetwo may have significant benefits for spectroscopy studies. For example, the potential removal of the VELO RF foils, together with the improved particle identification provide by the TORCH, will enhance the reconstruction efficiency for multibody \B decays, such as \decay{\Bcp}{\Dsp \Dz \Dzb}; the selection of short-lived particles ($\eg$ \Bc, \Xicc, \Omegacc, $\Xires_{\bquark\cquark}$, etc) will also benefit from an improved vertex resolution; the Magnet Side Stations will help in studying di-pion transitions such as \decay{\PX(3872)}{\chicone \pip \pim} or \decay{\B_\cquark^{**+}}{\Bc \pip \pim}; improved \piz and \etaz mass resolutions will increase the sensitivity in searching for the $C$-odd and charged partners of the $\PX(3872)$ meson by \decay{\PX(3872)^{C\rm{-odd}}}{\jpsi \etaz} and \decay{\PX(3872)^\pm}{\jpsi \piz \pipm}. A summary of the expected yields in certain important modes, and a comparison with \belletwo, is given in Table~\ref{tab:sec9_summary}.
Below
we 
give further details on the prospects for specific studies and analyses.

\begin{table}[tb]
\centering
\caption{\label{tab:yields}Expected data samples at \lhcb \upgradetwo and \belletwo for key decay modes for the spectroscopy of heavy flavoured hadrons. The expected yields at \belletwo are estimated by assuming similar efficiencies as at \belle.}
\begin{tabular}{lcccc}
\hline\hline
                      & \multicolumn{3}{c}{\lhcb} & \belletwo\\
Decay mode & $23$\invfb & $50$\invfb & $300$\invfb & $50$\invab\\
\hline
\decay{\Bp}{\PX(3872) (\to \jpsi \pip \pim)\Kp} &$14$k &30k&$180$k&$11$k\\
\decay{\Bp}{\PX(3872) (\to \psitwos \g)\Kp}     &$500$ &$1$k& $7$k & $4$k \\
\decay{\Bz}{\psitwos \Km \pip}                  &$340$k &$700$k&$4$M& $140$k\\
\decay{\Bcp}{\Dsp \Dz \Dzb}                     & $10$ &$20$& $100$ & ---\\
\decay{\Lb}{\jpsi \proton \Km}                  &$340$k &$700$k&$4$M& ---\\
\decay{\Xibm}{\jpsi \Lz \Km}                    &$4$k &$10$k&$55$k& ---\\
\decay{\Xiccpp}{\Lc\Km\pip\pip}                 &$7$k &$15$k&$90$k & $<6$k\\
\decay{\Xires_{\bquark\cquark}^+}{\jpsi \Xires_\cquark^+} & $50$ &$100$& $600$ &---\\
\hline\hline
\end{tabular}
\label{tab:sec9_summary}
\end{table}

\subsubsection{Taxonomy of tetraquarks and pentaquarks}

To advance our understanding of the $\PX(3872)$ state, it will be very important to learn even more about its decay pattern.
In particular, if it really has a strong $\Pchi_{\cquark 1}(2P)$ component, it should have \pip\pim transitions to the $\Pchi_{\cquark1}(1P)$ state.
Unfortunately at \lhcb, the reconstruction efficiency for the dominant $\Pchi_{\cquark 1}(1P)$ decay  to \g\jpsi decay is low, making this prediction hard to test.
 The very large data set of LHCb \upgradetwo  will allow one to overcome this problem, and will be essential
in detecting or refuting such transitions.
Studies of the $\PX(3872)$ lineshape by a simultaneous fit to all detected channels are important for pinning down the location
of its resonant pole and determining its natural width; both are very important inputs in helping with the understanding of the state.
Therefore a very large data set will be essential for the statistical precision of such studies and reconstruction of decays to \Dz\Dstarzb, which are relevant given the proximity of the $\PX(3872)$ mass to the \Dz\Dstarzb threshold.

Searching for prompt production of any known exotic hadron candidates at the \lhc remains an important task,
since its detection would signify a compact component, either conventional quarkonium,  or a tightly bound tetraquark or pentaquark.
To date, the $\PX(3872)$ is the only exotic hadron candidate with $\PQ\Qbar$ content that has been confirmed to be produced promptly. 
However, it will be important to quantify the upper limits in negative searches to allow more rigorous phenomenological analysis.

\begin{figure}[t]
 \centering
\includegraphics[width=0.5\textwidth]{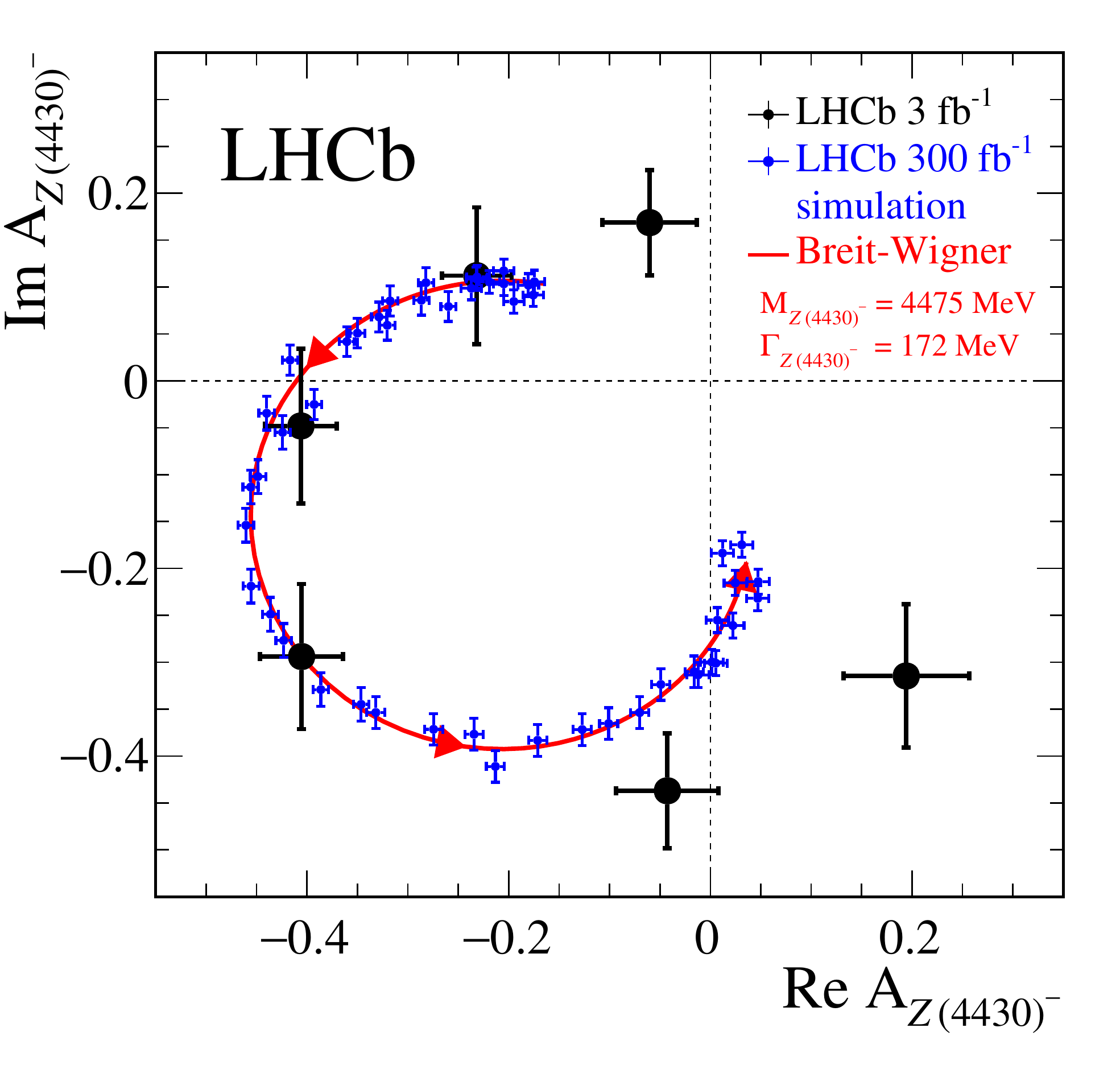}
\caption {Argand diagram of the $\PZ(4430)^-$ amplitude (${\rm A}_{\PZ(4430)^-}$) in bins of $m^2_{\psitwos \pim}$ from a fit to the $\decay{\Bz}{\psitwos \Kp \pim}$ decays. The black points are the results based on Run 1 data~\cite{LHCb-PAPER-2014-014} while the blue points correspond to an extrapolation to an integrated luminosity of 300\invfb expected at the \lhcb \upgradetwo. The red curve is the prediction from the Breit-Wigner formula with a resonance mass (width) of 4475 (172)\mev. Units are arbitrary.}
 \label{fig:argand}
\end{figure}

Many puzzling charged exotic meson candidates (\eg $\PZ(4430)^+$) decaying to \jpsi, \psitwos or $\Pchi_{\cquark 1}$ plus a charged pion have been observed in \B decays.
Some of them are broad, and none can be satisfactorily explained by any of the available phenomenological models.
The hidden-charm mesons, observed in the $\jpsi\Pphi$ decay~\cite{LHCb-PAPER-2016-018,Chatrchyan:2013dma,Aaltonen:2009tz}, also belong to this category.
The determination of their properties, or even the claim of their existence, relies on an advanced amplitude analysis,
which allows the exotic contributions to be separated from the typically dominant non-exotic components.
Further investigation of these $\PQ\Qbar\quark\quarkbar$ structures
will require much larger data samples and refinement of theoretical approaches to parametrisations of hadronic amplitudes.
 Similar comments apply to improvements in the determination of the properties of the pentaquark candidates
$\PP_\cquark(4380)^+$ and $\PP_\cquark(4450)^+$ and to the spectroscopy of excited \Lz baryons in $\decay{\Lb}{\jpsi \proton \kaon}$ decays. The large data set collected during the \lhcb \upgradetwo would allow one to test further the resonant character of the $\PP_\cquark(4380)^+$, $\PP_\cquark(4450)^+$ and $\PZ(4430)^+$ states (Fig.~\ref{fig:argand}), while improvements in calorimetry would help in searching for new decay modes (\eg $\decay{\PP_\cquark^+}{\Pchi_{\cquark 1,2} (\to \jpsi \g) \proton}$) by amplitude analyses of $\decay{\Lb}{\Pchi_{\cquark 1,2}\proton\Km}$ decays~\cite{LHCb-PAPER-2017-011,Guo:2015umn}. %

\subsubsection{Searches for further tetra- and pentaquarks}

Though the true nature of the $\PX(3872)$ meson is still unclear, both the molecular~\cite{Nieves:2012tt}
and tetraquark~\cite{Maiani:2004vq} models predict that a $C$-odd partner ($\PX(3872)^{C\rm{-odd}}$)
and charged partners ($\PX(3872)^\pm$) may exist and decay to $\jpsi \etaz / \g \Pchi_{\cquark J}$ and \jpsi \piz \pipm respectively.

\begin{figure}[t]
 \centering
\includegraphics[width=0.32\textwidth]{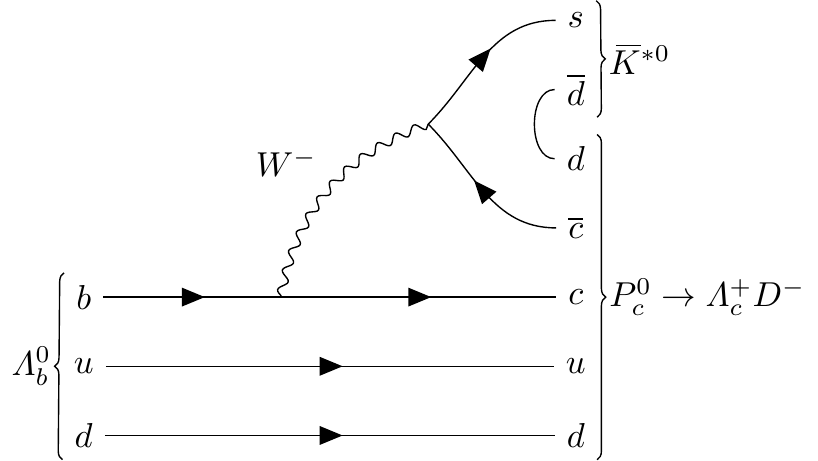}
\includegraphics[width=0.32\textwidth]{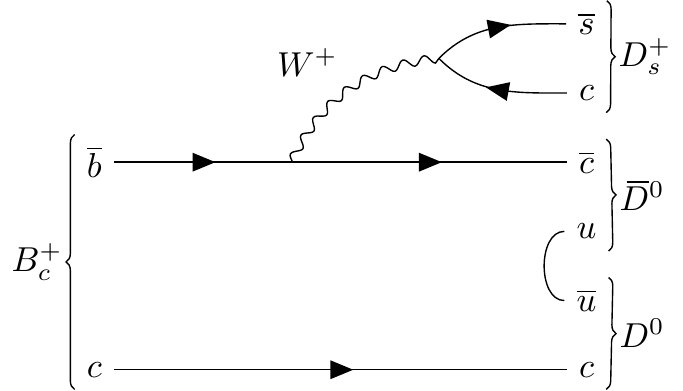}
\includegraphics[width=0.32\textwidth]{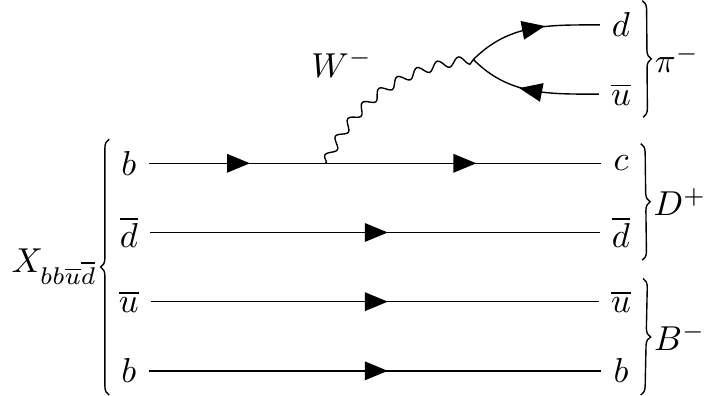}
\caption {Schematic Feynman diagrams for decays of $\Lambda_b^0\to \bar{K}^{*0}\Lambda_c^+D^-$, $B_c^+\to D^0\bar D^0 D_s^+$, and $X_{bb\bar u\bar d}\to B^- D^+ \pi^-$ (from left to right).}
 \label{fig:feynman}
\end{figure}

Similarly, the existence of the $P_\cquark(4380)^+$ and $P_\cquark(4450)^+$ pentaquark states raises the question
of whether there is a large pentaquark multiplet. The observed states have an isospin
3-component of $I_3=+\frac{1}{2}$. If they are part of an isospin doublet with $I=\frac{1}{2}$, 
there should be a neutral $I_3=-\frac{1}{2}$ state decaying to \jpsi \neutron~\cite{Lebed:2015tna}. However this final state does
not lend itself well to observation. Instead, the search for the neutral pentaquark candidate
 can be carried out using decays into pairs of open charm, in particular in the process
\decay{\Lb}{\Lc\Dm\Kstarzb}, where the neutral pentaquark states would appear as resonances
in the \Lc\Dm subsystem (Fig.~\ref{fig:feynman} left). Such decays can be very well reconstructed, but the total reconstruction
efficiency suffers from the large number of tracks and the small branching fractions of $\Lc$ and $\Dm$ reconstructable final states; the total reconstruction
efficiency is about a factor 50
smaller than the efficiency for the \decay{\Lb}{ \jpsi \proton \Km} channel.
If there could exist pentaquarks of 
an isospin quadruplet, then there is the interesting possibility to find doubly charged
pentaquarks decaying into $\Sigmares_\cquark^{++} \Dzb$. Channels such as these
require very large data sets to offset the low efficiency. The magnet side stations will also improve the reconstruction efficiency of such decay modes with several tracks in the final states. 

The relative coupling of the pentaquark states to their decays into the double open-charm channels will
depend on their internal structure and the spin structure of the respective decay. For that reason
it is important to study decays involving \Dstarp resonances as well (\eg \decay{P_\cquark^+}{\Dstarm \Sigmares_\cquark^{++}})
 to investigate the internal structure of pentaquarks~\cite{Lin:2017mtz}. Since
these decays require the reconstruction of slow pions from the \Dstarp decays, the proposed tracking stations inside the magnet, enhancing the acceptance for low-momentum particles, will be highly beneficial for this study.

Invoking $\rm{SU}(3)$ flavour symmetry,
 one would expect the existence of pentaquarks with strangeness, which would
decay into channels like \jpsi \Lz or \Lc\Dsm. Such states and their decays into these channels were investigated in~\cite{Wu:2010jy,Wu:2010vk}. To explore the potential of
the former case, the decay \decay{\Xires_\bquark^-}{ \jpsi \Lz\Km} has been studied using Run~1 data at LHCb, with about
300 signal decays observed~\cite{LHCB-PAPER-2016-053}.
Complementary information can be achieved by a study of the \decay{\Lb}{ \jpsi \Lz \Pphi} decays.
An increase of the available integrated luminosity by a factor of 100 would allow detailed amplitude analyses to be performed for 
 these final states, with a similar sensitivity as was the case for the pentaquark discovery channel.

The history of $\PX(3872)$ studies illustrates well the difficulty of distinguishing between exotic and conventional explanations for a hidden-charm state.
 Therefore, it is appealing to search for  states with  uncontroversial exotic signatures.
A good candidate in this category would be a $\mathcal T_{\cquark\cquark}$ doubly charmed tetraquark~\cite{Moinester:1995fk,DelFabbro:2004ta,Carames:2011zz,Yuqi:2011gm,Hyodo:2012pm,Esposito:2013fma,Ikeda:2013vwa,Guerrieri:2014nxa,Maciula:2016wci,Richard:2016eis,Esposito:2016noz,Luo:2017eub,Eichten:2017ffp,Hyodo:2017hue,Cheung:2017tnt,Wang:2017dtg,Yan:2018gik}, being a meson with constituent-quark content $\cquark\cquark\quarkbar\quarkbar^{\prime}$, where the light quarks \quark and $\quark^{\prime}$ could be \uquark, \dquark or \squark.
If the masses of the doubly charmed tetraquarks are below their corresponding open-charm thresholds, they would manifest as 
weakly decaying hadrons with properties including masses, lifetimes and decay modes not too different from the recently observed \Xiccpp baryons~\cite{LHCb-PAPER-2017-018},
and, as for the $\Xires_{\cquark\cquark}$ baryons, the most promising searches are in prompt production. 
Instead, if the masses of $\mathcal T_{\cquark\cquark}$ states are  above the open-charm threshold and their widths are broad, it will be very challenging to observe these states via prompt production. 
Instead, \Bcp decays to open-charm mesons can offer unique opportunity to test for their existence. 
In Run 5 the \Bc mesons will be copiously produced at the \lhc, because of the large production cross-sections of \bbbar and \ccbar pairs and of the enormous data sample.
Similarly to the amplitude analysis of the $\decay{\Lb}{\jpsi\proton\Km}$ decay, which led to the observation of the $P_\cquark^+$ pentaquark candidates~\cite{LHCb-PAPER-2015-029}, 
studying the angular distributions of the multi-body final states of the \Bc meson
has the potential of indicating new states, \eg $\mathcal T_{\cquark\cquark}$, inaccessible through decays of lighter hadrons.
It also allows for the determination of the spin-parity quantum numbers of any state that is observed. 
A good example is to study the $\mathcal T_{\cquark\cquark}^+$ state in the decay mode $\decay{\Bcp}{\Dsp \Dz \Dzb}$ (Fig.~\ref{fig:feynman} middle)
through the decay chain $\decay{\Bc}{\mathcal T_{\cquark\cquark}^+ \Dzb}$ and 
$\decay{\mathcal T_{\cquark\cquark}^+}{\Dsp \Dz}$, as discussed in Ref.~\cite{Esposito:2013fma}.

The decay $\decay{\Bc}{\Dsp \Dz \Dzb}$ has not been observed with the Run 1 data,
and predictions on the branching fractions of \Bcp decays are subject to very large 
uncertainties. Estimates of the integrated luminosity needed to perform a full amplitude analysis
are therefore imprecise, and can only be formulated through considerations of other decay modes such as $\decay{\Bc}{\jpsi \Dsp}$.
The signal yield of $\decay{\Bc}{\jpsi \Dsp (\to \Pphi \pip)}$ decays observed in Run 1 data is $30 \pm 6$~\cite{LHCb-PAPER-2013-010}.
Considering the branching fraction of the decay of the additional charm hadron and the lower efficiency due to the higher track multiplicity,
the estimated number of signal of $\decay{\Bc}{\Dsp \Dz \Dzb}$ decays is $\mathcal O(10^2)$ 
in a future dataset corresponding to an integrated luminosity of $300$\invfb, collected with $\mathcal O(100\%)$ trigger efficiency~\cite{LHCb-PAPER-2013-010}.
Since the \Dz and \Dsp mesons are pseudoscalars, the amplitude analysis simplifies, 
 and can provide conclusive results already with a few hundred decays.

Finally, strongly decaying doubly charmed tetraquarks with a narrow decay width, as predicted by pure tetraquark models with 
spin-parity quantum numbers of $0^+$, $1^+$ and $2^+$, can also be searched for in prompt production.
The expected yields can be estimated by the associated production of open-charm mesons measured with a fraction of the Run 1 data~\cite{LHCb-PAPER-2012-003}. 
With a data sample of $300$\invfb, the yield for \Dp\Dp (\Dp\Dsp) associated production is around $750$k ($150$k), which is a very promising sample in which to search for narrow $\mathcal T_{\cquark\cquark}$ states.

If the coincidence of the $\Pchi_{\cquark 1}(2P)$ charmonium state with
the \Dz \Dstarzb threshold is responsible for the $\PX(3872)$ state,
 there is likely no bottomonium analogue of it, since  the $\Pchi_{\bquark 1}(3P)$
 state was detected well below the $\B\Bbar{}^{*}$ threshold,
and the $\Pchi_{\bquark 1}(4P)$ state is predicted to be too far above it.
However, if molecular forces dominate its dynamics, there could be an isosinglet state below this threshold
decaying to \omegaz\OneS, where \omegaz could be reconstructed via the decay to \pip \pim \piz.
Unfortunately, its prompt production would likely be very small unless
driven by tightly bound tetraquark dynamics or by the $b\bar b$ Fock-state component of the same quantum numbers~\cite{Guo:2014sca}. 
The improved $\piz$ reconstruction in the LHCb \upgradetwo will help for these searches.

The prompt production at the \lhc remains the best hope for unambiguously establishing the existence   
of a stable, weakly decaying $\bquark\bquark\uquarkbar\dquarkbar$ tetraquark predicted by both lattice QCD and phenomenological models. 
However, the inclusive reconstruction efficiencies for such states are tiny due to the small branching fractions of \B and \D meson decays to low-multiplicity final states (Fig.~\ref{fig:feynman}c).
Recently, there have  been also several predictions for an exotic state with quark composition $\bquark\bquarkbar\bquark\bquarkbar$~\cite{Heller:1985cb,Berezhnoy:2011xn,Wu:2016vtq,Chen:2016jxd,Karliner:2016zzc,Bai:2016int,Wang:2017jtz,Richard:2017vry,Anwar:2017toa,Vega-Morales:2017pmm,Eichten:2017ual} with a mass below the $2m_{\Peta_\bquark}$ threshold,
which implies that it can decay to  $\PUpsilon \mumu$. However, lattice QCD calculations do not find evidence for such a state in the hadron spectrum~\cite{Hughes:2017xie}. Given the presence of four muons in the final state, \lhcb will have good sensitivity for observing the first exotic state composed of more than two heavy quarks~\cite{LHCb-PAPER-2018-027}.

Motivated by the discovery of the hidden-charm pentaquarks, theorists have extended
the respective models for multiplet systems to include beauty quarks. In Ref.~\cite{Wu:2017weo} 
$\PQ\Qbar\quark\quark\quark$ ground states were investigated in an effective-Hamiltonian
framework assuming a colour-magnetic interaction
between colour-octet \quark\quark\quark and $\PQ\Qbar$ subsystems. Several resonant states are predicted.
Such beautiful pentaquarks could be searched for in the
$\PUpsilon\proton$, $\PUpsilon\Lz$,  $\Bcpm \proton$ and $\Bcpm \Lz$ mass spectra.
In analogy with the popular $\Sigmares_\cquark \Lcbar$ molecular model,
Refs.~\cite{Yamaguchi:2017zmn} and~\cite{Shen:2017ayv} investigate similar dynamics in the hidden-bottom sector and predict
a large number of exotic resonances. Indeed, in the hidden-beauty sector the theory calculations are
found to be even more stable than for hidden charm, motivating searches for resonances close to the
$\B^*\Sigmares_\bquark, \B\Sigmares_\bquark^*, \B^*\Sigmares_\bquark^*$ and $\B \Lz^{*0}_\bquark, \B^*\Lb$ thresholds.

Another possibility is the existence of pentaquarks with open beauty and quark contents such as
\bquarkbar\dquark\uquark\uquark\dquark, \bquark\uquarkbar\uquark\dquark\dquark,
\bquark\dquarkbar\uquark\uquark\dquark and \bquarkbar\squark\uquark\uquark\dquark~\cite{Stewart:2004pd, Oh:1994np}.
If those states lie below the respective baryon-meson thresholds containing beauty, then
they could be stable against strong decay and would predominantly decay through the weak transition
\decay{\bquark}{\cquark\cquarkbar\squark}. A search using a $3\invfb$ data set 
in four decay channels $\jpsi\proton\hadron^+\hadron^-$ (\hadron = \kaon, \pion)  has been performed in Ref.~\cite{LHCb-PAPER-2017-043}.
No signals were found and $90\%$ CL limits were set on the production cross section times branching fraction
relative to the \Lb in the \jpsi\proton\Km mode. The obtained limits are of the order of $10^{-3}$, which
does not yet rule out the estimates for the production of such an object provided in Ref.~\cite{Stewart:2004pd}.
Similar searches in channels with open-charm hadrons in the final state again lead
to large multiplicities and the respective small reconstruction efficiencies, but could
profit from favoured branching fractions. Investigations of a large number of channels
will maximise sensitivity for weakly decaying exotic hadrons. It has  also been proposed to search for  excited \Omegab states~\cite{Liang:2017ejq}
in analogy to the recently discovered excited \Omegac states~\cite{LHCb-PAPER-2017-002}.
Such open-beauty excited states could be searched for in decays to the \Xib\kaon final state.

\subsubsection{Study of doubly-heavy baryons}

  The discovery of the \Xiccpp baryon has opened an exciting new line of research that
  \lhcb is avidly pursuing.
  Measurements of the lifetime and relative production cross-section of \Xiccpp,
  searches for additional decay modes, and searches for its isospin partner
  \Xiccp and their strange counterpart \Omegacc are underway.

  A signal yield of  $313 \pm 33$ \decay{\Xiccpp}{\Lc\Km\pip\pip} decays
  was observed by LHCb in $1.7\invfb$ of Run 2 data~\cite{LHCb-PAPER-2017-018}.
  Improvements in the trigger for the \lhcb \upgradetwo detector are projected to
  increase selection efficiencies by a factor two for most charm decays, with
  decays to high-multiplicity final states, such as those from the cascade
  decays of doubly charmed baryons, potentially benefiting much
  more~\cite{LHCb-TDR-012,LHCb-PAPER-2012-031}.
	Thus the  Run 5 sample  will contain more than 90\,000 decays of this mode.
  The branching fraction for \decay{\Xiccpp}{\Lc\Km\pip\pip} is
  theoretically estimated to be up to 10\%, making it one of the most frequent
  nonleptonic decay modes. Several other lower multiplicity modes with
  $\mathcal{O}(1\%)$ predicted branching fractions will yield samples of 
  comparable size~\cite{LHCb-PAPER-2018-026, Wang:2017mqp,Gutsche:2017hux,Sharma:2017txj}.

  The efficiency with which \lhcb can disentangle weak decays of doubly charmed
  baryons from prompt backgrounds depends on the lifetime of the
  baryon~\cite{LHCb-PAPER-2013-049}.
  Although the predicted lifetimes for the \Xiccp, \Xiccpp, and \Omegacc baryons span
  almost an order of magnitude, the relative lifetimes of \Xiccp and \Omegacc
  are expected to be approximately $1/3$ that of the
  \Xiccpp baryon~\cite{Kiselev:2001fw,Karliner:2014gca,Fleck:1989mb,Guberina:1999mx,Kiselev:1998sy,Chang:2007xa,Berezhnoy:2016wix,LHCb-PAPER-2018-019}.
  Assuming a relative efficiency of 0.25 with respect to \Xiccpp due to the
  shorter lifetimes and an additional production suppression of
  $\sigma(\Omegacc)/\sigma(\Xiccpp) \sim 0.2$ for
  \Omegacc~\cite{Kiselev:2001fw}, \lhcb will have Run 5 yields of around 25\,000 for \Xiccp and 4500 for \Omegacc in each of several decay modes.

  \lhcb will be the primary experiment for studies of the physics of doubly
  charmed baryons for the foreseeable future, and its potential will not be
  exhausted by the end of Run 5.
  With the data collected in Run 2, \lhcb should observe all three weakly decaying
  doubly charmed baryons and characterise their physical properties.
  Run 5  will supply precision measurements of doubly differential
  cross sections that will provide insight into production
  mechanisms of doubly heavy baryons. In addition 
  Run 5   will allow the spectroscopy of excited states and bring studies of the
  rich decay structure of doubly charmed hadrons into the domain of precision
  physics.

The production cross section of the $\Xires_{\bquark\cquark}$ baryons within the \lhcb acceptance is expected to be about 77\nb~\cite{Zhang:2011hi}.
This value is about 1/6 of the expected production cross-section of a \Bcp meson~\cite{Chang:2003cr,Gao:2010zzc}. It should be noted that the relative \Lb production rate is \pt-dependent. In the typical \pt range in the 
\lhcb acceptance, a ratio of production rates, 
$\sigma(\decay{\proton\proton}{\Lb \PX})/\sigma(\decay{\proton\proton}{\Bzb X})\sim0.5$~\cite{LHCb-PAPER-2011-018,LHCb-PAPER-2014-004} is measured. It is therefore
conceivable that the $\Xires_{bc}$ production rates are also larger than predicted by the above calculations.

It is quite challenging to observe and study $\Xires_{\bquark\cquark}$ and $\Omegares_{\bquark\cquark}^0$ baryons  because of the low production rates, the small product of branching fractions
and the selection efficiencies for reconstructing all of the final-state particles. To collect a large sample of $\Xires_{\bquark\cquark}$
baryons will require the higher integrated luminosity and detector enhancements planned for in the \lhcb \upgradetwo.  
  Using the notation that $\PX_\cquark$ is a charmed baryon
containing a single charm quark, some of the most promising decay modes to detect $\Xires_{\bquark\cquark}$ and $\Omegares_{\bquark\cquark}^0$ baryons are: {\em (i)}  $\jpsi \PX_\cquark$ modes \jpsi\Xicp, \jpsi\Xicz, \jpsi\Lc, \jpsi\Lc\Km; 
 {\em (ii)} $\Xires_{\cquark\cquark}$ modes $\Xires_{\cquark\cquark}\pim$; {\em (iii)} doubly charmed modes  \Dz \Lc, \Dz \Lc \pim, \Dz\Dz \proton; {\em (iv)} penguin-topology modes \Lc\Km, \Xicp\pim; {\em (v)} the $\Xires_\bquark$ , \Bd or \Lb modes \Xib\pip, \Lb\pip, \Bz \proton, using fully reconstructed or semileptonic \Bz, \Lb or \Xib decays~\cite{LHCb-PAPER-2018-032};  {\em (vi)} the decays due to $W$-exchange between \bquark-\cquark quarks, that is not helicity suppressed, and which can give rise to a final state with just one charmed particle, $\eg$ \Lc\Km.

To put this in context, the \lhcb collaboration observed $30\pm6$ $\decay{\Bcp}{\jpsi \Dsp (\to\Pphi\pip)}$ decays with 3\invfb data at 7 and 8\tev~\cite{LHCb-PAPER-2013-010}. 
With looser selections about $100$ signal decays can be obtained with reasonably good signal-to-background ratio.
The $\decay{\Xires^+_{\bquark\cquark}}{\jpsi\Xicp}$ decay is kinematically very similar.
Assuming
${f_{\Xires_{\bquark\cquark}^+}}/{f_{\Bcp}}\sim0.2$, with
${\BR(\decay{\Xires_{\bquark\cquark}^+}{\jpsi\Xicp})}/{\BR(\decay{\Bcp}{\jpsi\Dsp})}\sim~1$, while
${\BR(\decay{\Xicp}{p\Km\pip})}/{\BR(\decay{\Dsp}{\Kp\Km\pip})}\sim0.1$, and
$\epsilon_{\Xires_{\bquark\cquark}^+}/\epsilon_{\Bcp}\sim~0.5$,
a signal yield of about 600 $\decay{\Xires_{\bquark\cquark}^+}{\jpsi\Xicp}$ decays
is expected in Run 5, albeit with sizeable uncertainties.
Other modes could also provide sizeable signal samples.
It is likely \lhcb will observe the $\Xires_{\bquark\cquark}$ baryons in Run 3/4, and further probe the spectrum of other doubly heavy baryons with the large samples accessible in the proposed \upgradetwo. 

\subsubsection{Precision measurements of quarkonia}

The correct interpretation of the experimental polarisation results
for $S$-wave quarkonia requires a rigorous analysis of the feed-down contributions from
higher excited states ~\cite{LHCb-PAPER-2011-030,LHCb-PAPER-2014-031}.
The direct measurement of the polarisation for $\Pchi_\cquark$ and $\Pchi_\bquark$ states is necessary to decrease this 
model dependence. Since $P$-wave states are practically free from the feed-down from higher excited states,
any $\Pchi_\cquark$ and $\Pchi_{\bquark}$ polarisation measurements could be interpreted in a robust manner
 without additional model assumptions.

The recent discovery of the $\decay{\Pchi_{\cquark1,2}}{\jpsi\mumu}$ decays~\cite{LHCb-PAPER-2017-036} opens the possibility
to perform a detailed study of $\Pchi_{\cquark}$ production, allowing almost background-free measurements even for a
very low transverse momentum of $\Pchi_\cquark$~candidates. 
Due to the excellent mass resolution, the vector state $\Pchi_{\cquark1}$  and the tensor state $\Pchi_{\cquark2}$ are well separated,
eliminating the possible systematic uncertainty  caused by the large overlap of
these states in the $\decay{\Pchi_{\cquark}(\Pchi_\bquark)}{\jpsi(\PUpsilon)\g}$ decay\cite{
  LHCb-PAPER-2011-019,
  LHCb-PAPER-2011-030,
  LHCb-PAPER-2012-015,
  LHCb-PAPER-2013-028,
  LHCb-PAPER-2014-031,
  LHCb-PAPER-2014-040}.
An integrated luminosity of 300\invfb will allow one to probe the high-multipole contributions 
to the  $\decay{\Pchi_{\cquark}}{\jpsi\mumu}$ amplitude, namely
the magnetic-dipole contribution for $\Pchi_{\cquark1}$~decays and
the magnetic-dipole and electric-octupole contributions for $\Pchi_{\cquark2}$ decays. 
Use of the Run 5  data set is also necessary  to measure the $(\pt,y)$ dependence of $\Pchi_\cquark$ polarisation parameters. 
 In addition, the effect of the form factor in
the~decays $\decay{\Pchi_{\cquark}}{\jpsi\mumu}$~\cite{Faessler:1999de,Luchinsky:2017pby}
could be probed with the precision of several percent from the shape of the $m(\mumu)$ spectra.

Studies of double quarkonia production allows independent tests for the  quarkonia production mechanism,
and in particular for the role of the colour octet. So far the \lhcb collaboration has  analysed double-\jpsi production in 7\tev and 13\tev data with relatively small
datasets~\cite{LHCb-PAPER-2011-013,LHCb-PAPER-2016-057}.
Using $280$\invpb of data collected at $\sqs=13\tev$,
$(1.05\pm0.05)\times10^3$ signal \jpsi\jpsi events are observed.
However, even with the larger sample of \jpsi\jpsi events now available, 
it is not possible to distinguish between 
 different theory descriptions of the single-parton scattering (SPS) mechanism~\cite{Sun:2014gca,Likhoded:2016zmk, Lansberg:2013qka, Lansberg:2014swa,Lansberg:2015lva,Shao:2012iz,Shao:2015vga, Baranov:2011zz,Baranov:1993qv}, nor to separate the
  contributions from the SPS and double-parton  scattering (DPS) mechanisms~\cite{Bansal:2014paa,Belyaev:2017sws}. 
Though the larger samples collected during  Runs~3 and 4  will allow some progress on these questions, the measurement of the correlation of \jpsi polarisation parameters will only be possible with the \upgradetwo data set.

In addition, while it is likely that  $\PUpsilon\PUpsilon$ and $\jpsi\PUpsilon$ production will be observed in the near future (assuming the dominance of the DPS mechanism),
the determination of the relative SPS and DPS contributions, as well as the discrimination between different SPS theory models, will require precision measurements only possible with  Run 5  data.

%% file: section7/section.tex
\section{Bottom-quark probes of new physics and prospects for $B$-anomalies \label{secBanom}}
\label{sec:7}

A key contribution of the HL-LHC data sets will be a comprehensive global picture of rare and (semi)-leptonic quark transitions. This has value whether or not the current ``flavour anomalies'' are confirmed with Run~2 or Run~3 LHC data. If the improvements on statistically limited observables, $R_{K^{(*)}}$ and $R(D^{(*)})$,  
confirm the anomalies, a  global analysis of all related channels may 
discriminate between different 
BSM explanations. Such a global analysis will also help to constrain the mass scale of the BSM particles and mediators, which, together with the direct HL-LHC searches, will further restrict the range of possible explanations for the anomalies. On the other hand, if the signifcance of the anomalies decreases, or they even disappear with Run 2 and Run 3 data, we should still search 
for small BSM
effects on top of the dominant SM amplitudes. In this case, global analyses of the data would lead to improved bounds on the BSM affecting these flavor observables. The statistical power of the HL-LHC datasets will also enable a unique reach for other BSM signatures such as Lepton-Flavour Violation (LFV) and Baryon-Number Violation (BNV), particularly in the baryon and heavy-meson sectors.

We would like to stress that, in the HL-LHC era, many rare beauty decays will become abundant and enter the precision-measurement regime. Therefore, it will  become increasingly important that the LHC collaborations and Belle II publish results in such a way that systematic uncertainties can be treated in a coherent manner and their correlations taken into account. 
A good example of this is the treatment of the ratio of
hadronization fractions $f_s/f_d$ in any LHC $B_q\to \ell\ell$ combination, discussed 
in Sec.~\ref{sec:7_exp}. It will also be important to correctly treat systematic uncertainties due to the use of common software packages, e.g., the use of {\tt PHOTOS} by  the LHC experiments and Belle~II, in future lepton universality measurements.

While much of the power of the HL-LHC and Belle~II datasets will lie in the breadth of precisely measured rare and (semi-)leptonic decays, practical considerations mean that experiments will continue to publish analyses of individual decay modes as each is completed. Systematically publishing experimental likelihoods, efficiency maps, and resolution unfoldings, as already done for some of the most important analyses, should be strongly encouraged even now. 
This facilitates both the combination of results between experimental collaborations, and the inclusion of the results in global fits and in tests of phenomenological models~\cite{Hurth:2017sqw,Blake:2017fyh,Chrzaszcz:2018yza}.

\subsection{Phenomenology of $b\to s \ell\ell$ decays}
{\it\small Authors (TH): Wolfgang Altmannshofer,  David Straub, Javier Virto.}

\input{section7/bsll.tex}

\subsection{Phenomenology of $b\to c \ell \nu$ decays}
{\it\small Authors (TH): Marat Freytsis, Martin Jung, Dean Robinson, Stefan Schacht.}

\label{sec:7_RDs}
\input{section7/bctaunu.tex}

\input{section7/experimental}

%% file: section7/bsll.tex
In this section we discuss the status of the interpretation of $b\to s$ transitions in and beyond the SM, the prospects for future sensitivities at the LHC, and the complementarity with  Belle-II. All current and planned measurements of processes involving $b\to s$ transitions are related to weak decays of $b$-hadrons. At present, most of the data is on $B$ decays, with a smaller fraction on $B_s$ decays. LHCb has already provided a few measurements on $\Lambda_b$ decays, with data sets that will increase dramatically in future runs, including a significant output on decays of other $b$-hadrons such as $\Omega_b$
or $\Xi_b$~\cite{Aaij:2244311} (see Sec.~\ref{sec:7_exp}). 

Theoretically, the decays of $b$-hadrons are best described within the Weak Effective Theory (WET), where flavor-changing transitions are mediated by ``effective'' dimension-six
operators with Wilson coefficients (WCs) that encapsulate all SM and heavy NP effects. Thus all observables can be calculated in full generality, model-independently in terms of the WCs and hadronic matrix elements. The relevant effective Lagrangian for $b\to s$ transitions at low-energies in the SM is~\cite{Buchalla:1995vs}
\begin{equation}
\label{eq:Heff}
  {\cal{H}}_{\rm eff}^{\rm SM}=
  \frac{4 G_F}{\sqrt{2}}\sum_{p=u,c}\lambda_{ps}\left(C_1 O_1^p+C_2 O_2^p +\sum_{i=3}^{10} C_i  O_i\right),
\end{equation}
with $\lambda_{ps}=V_{pb}^{} V_{p s}^\ast$. As defined in Ref.~\cite{Buchalla:1995vs}, the $O_{1,2}^p$,  $O_{3,\ldots,6}$, and $O_{8}$ are the so-called ``current-current'', ``QCD-penguin'' and ``chromo-dipole'' operators, respectively, and they contribute to the $b\to s\gamma$ and $b\to s\ell\ell$ transitions via an electromagnetic interaction. The $O_{7\gamma}$ and $O_{9,10}$ are the ``electromagnetic dipole'' and the ``semileptonic'' operators, respectively.
The BSM can enter through the WCs $C_1,\ldots,C_{10}$, or the chirally-flipped versions, $P_{L(R)}\to P_{R(L)}$, giving $O_{7\gamma,\ldots, 10}'$, or through scalar and tensor semileptonic operators~\cite{Bobeth:2007dw}. Furthermore, beyond the SM, semileptonic operators can induce lepton-flavor universality violation (LFUV) or charged-lepton flavor violation (CLFV).
For the purpose of this Section, the relevant operators for the interpretation of $b\to s\gamma$ and $b\to s\ell\ell^\prime$ data are
\begin{equation}
\begin{split}
&O_{7\gamma}^{(\prime)}=\frac{e\,m_b}{16\pi^2}(\bar s\sigma_{\mu\nu}P_{R(L)} b)F^{\mu\nu},
\\
&O_9^{\ell\ell^\prime(\prime)} =\frac{\alpha_{\rm em}}{4\pi}
(\bar{s} \gamma_{\mu} P_{L(R)} b)(\bar{\ell} \gamma^\mu \ell^\prime),\quad
\,\,
O_{10}^{\ell\ell^\prime(\prime)} =\frac{\alpha_{\rm em}}{4\pi}
(\bar{s} \gamma_{\mu} P_{L(R)} b)( \bar{\ell} \gamma^\mu \gamma_5 \ell^\prime),\label{eq:O10}
\end{split}
\end{equation}
where we have added leptonic flavor labels ($C_i^\ell\equiv C_i^{\ell\ell}$). 

By calculating and measuring a large set of independent observables, one can perform global fits to all the relevant WCs, and by comparing with their SM values,
extract information on NP model-independently. 
This programme has been carried out since the start of the LHC, and culminated in the current $b\to s\ell\ell$ anomalies~\cite{1304.6325,1307.5683,1308.1501,Beaujean:2013soa,Hurth:2013ssa,Altmannshofer:2017fio,Capdevila:2017bsm,Altmannshofer:2017yso,Hurth:2017hxg,Ciuchini:2017mik,Geng:2017svp}.
The future runs of the LHC
 will allow either to establish the anomalies or to refine our understanding of these transitions.

\subsubsection{Observables and Hadronic Matrix Elements}

\subsubsubsection{$B_q\to\ell^+\ell^-$}

From a theoretical perspective the purely leptonic decay, $B_{s}\to \ell^+ \ell^-$, is the cleanest exclusive $b\to s\ell\ell$ process. Up to QED corrections~\cite{1708.09152}, all QCD effects are contained in a decay constant,  which is precisely and reliably computed using lattice QCD, giving $f_{B_s}=230.7(1.3)$~MeV~\cite{Bazavov:2017lyh}.~\footnote{The average of lattice $N_f=2+1$ results from FLAG 2016~\cite{Aoki:2016frl} and other recent $N_f=2+1+1$ lattice calculations of the decay constant~\cite{Bussone:2016iua,Hughes:2017spc} give similar central values but with larger errors.} Going beyond this accuracy in $f_{B_s}$ is difficult (see Sec.~\ref{sec:lattice}), and the current theoretical error is less than 5\%~\cite{1708.09152,Bazavov:2017lyh},
\begin{align}\label{eq:Bsmumu_th}
\mathcal B(B_s\to\mu^+\mu^-)_{\rm SM}= (3.64\pm0.11)\cdot10^{-9},    
\end{align}
which is dominated by the uncertainties of the relevant CKM parameters. The latest measurements by LHCb and ATLAS~\cite{LHCb-PAPER-2017-001,Aaboud:2018mst}, have an error of~$\sim 25\%$ (see Sec.~\ref{sec:7_exp} for HL-LHC prospects). Beyond the SM, the decay $B_{s}\to \mu^+ \mu^-$ gives very strong constraints on the scalar and pseudoscalar operators~\cite{Alonso:2014csa,Arbey:2018ics}, and also on $C^{\mu(\prime)}_{10}$, which has an impact on the fits and the $b\to s\mu\mu$ anomalies. Searches for the CLFV channels $B_s\to \tau\mu$ and $B_s\to\mu e$ are important, because an observation would be an unambiguous signal of NP, which can be connected to the LFUV signals in $R_{K^{(*)}}$~\cite{Glashow:2014iga,Alonso:2015sja,Calibbi:2015kma,Boucenna:2015raa,Crivellin:2015era}. Besides the branching fractions, an effective lifetime observable \Blu{and the \CP-violating observable $S_{\mu\mu}$ are accessible by exploiting the nonvanishing width difference, $\Delta \Gamma_s$,} in the $B_s$ system~\cite{DeBruyn:2012wk,DeBruyn:2012wj}. This provides complementary constraints on the WCs~\cite{Altmannshofer:2017wqy}, and a precise measurement will be possible at the LHCb. 

These measurements are simultaneously sensitive to the decay $B_d\to\mu^+\mu^-$, although its branching fraction is further suppressed by a CKM factor~\cite{Bobeth:2013uxa,Bazavov:2017lyh},
\begin{align}\label{eq:Bdmumu_th}
\mathcal B(B_d\to\mu^+\mu^-)_{\rm SM}= (1.00\pm 0.03)\cdot10^{-10}. 
\end{align}
The ratio $\mathcal B(B_s\to\mu^+\mu^-)/\mathcal B(B_d\to\mu^+\mu^-)$ can be predicted more accurately in the SM \Blu{and models with Minimal-Flavor Violation (MFV) than the single branching ratios, testing the flavor structure of the short-distance dynamics. In particular, it is related to the ratio $\Delta m_s/\Delta m_d$ \cite{Buras:2003td} with small uncertainties from hadronic inputs and none from CKM ones. A similar comment applies to the ratios $\mathcal B(B_q\to\mu^+\mu^-)/\Delta m_q$~\cite{Buras:2003td}. It is important to emphasize that the measurements of $\mathcal B(B_d\to\mu^+\mu^-)$ that can be done at the LHC are of utmost importance as they cannot be done in any other facility in the coming decade.}

Finally, the $m_{\mu^+\mu^-}$ spectrum is sensitive to the decay $B_s\to\mu^+\mu^-\gamma$ (where the photon is not detected)~\cite{Dettori:2016zff}. Theoretically, this mode is interesting because the extra photon lifts the chiral suppression of the leptonic mode and gives access to the WC $C_9^\mu$. However, it is also challenging to predict because of long-distance hadronic contributions~\cite{Guadagnoli:2017quo}.

\subsubsubsection{$B_q\to M \ell^+\ell^-$}

The most prominent semileptonic decay modes are $B\to K^{(*)}\mu^+\mu^-$ and $B_s\to \phi \mu^+\mu^-$. These are sensitive to all the WCs and they  currently dominate the global fits to $b\to s\ell\ell$ data because of the experimental precision achieved
and the large number of observables they give access to. This allows one to constrain some independent combinations of WCs and hadronic parameters. Since there is no experimental information related to the polarization of the final-state leptons,
the measurable observables arise from the kinematic differential distributions of the final-state momenta. These are customarily written as  angular distributions with coefficients that depend on the dilepton invariant mass squared $q^2$, measured in specific bins of $q^2$.

In the case of the three-body mode $B\to K\mu^+\mu^-$, there are three
observables: the differential branching fraction, $d{\cal B}/dq^2$, the forward-backward asymmetry, $A_{\rm FB}(q^2)$,
and the ``flat term'', $F_H(q^2)$~\cite{Bobeth:2007dw}.
The kinematic distribution of the four-body decay, $B\to V(\to M_1M_2) \ell^+\ell^-$, contains many more independent angular coefficients, up to 11 in the most general case \Blu{(plus the total rate)}, called $I_i(q^2)$ or $J_i(q^2)$~\cite{Melikhov:1998cd,0811.1214,1202.4266}.
Normalizing these by the total differential rate $d\Gamma/dq^2$ and symmetrizing or antisymmetrizing with respect to the two charge-conjugate modes leads to the
observables $S_i$, $A_i$~\cite{0811.1214}.
A subset of these observables can also be constructed for
$B_s\to \phi \mu^+\mu^-$~\cite{Bobeth:2008ij}. It is convenient to define certain combinations of angular
observables where form-factor uncertainties largely cancel in the heavy-quark limit, called ``optimized observables''. An independent set of these, optimized at low-$q^2$,
is given by the $P_i^{(\prime)}$ basis~\cite{1202.4266,1207.2753,1303.5794}. Optimized observables at large-$q^2$ also exist~\cite{1006.5013,1303.5794}.

The  observables in exclusive semileptonic decays, $B\to M \ell^+ \ell^-$,  
are specified
by transversity (or helicity) amplitudes. They depend on two types of hadronic matrix elements: ``local'' (form factors) and ``non-local'' (see, e.g.,~\cite{Jager:2012uw,Jager:2014rwa,1707.07305}).
Local form factors for $B\to K$, $B\to K^*$ and $B_s\to \phi$ transitions can be calculated at low-$q^2$ in two different versions of light-cone sum rules (LCSRs)~\cite{hep-ph/0611193,1503.05534,Khodjamirian:2017fxg,Gubernari:2018wyi}, and at large-$q^2$ using Lattice QCD~\cite{1509.06235,1310.3722} (LQCD). A comparison between both determinations can be done by parametrizing
the $q^2$ dependence via the $z$-expansion~\cite{Meiman:1963,Boyd:1994tt,0807.2722}, which is based on the analytic structure of the matrix elements. Future prospects for the theoretical
precision of 
the form factors rely on improvements in LQCD calculations. Note that both LQCD and LCSRs work in the narrow-width limit for  
vector mesons.
A calculation beyond this approximation is possible within the LCSRs~\cite{1701.01633,1709.00173}, and points to
a correction
of up to $10\%$. There are  prospects for treating hadronic resonances on the lattice~\cite{1704.05439}, \Blu{and calculating directly the $B\to K\pi$ form factors should play an important role in the next decade.}

Non-local effects are significantly more difficult to estimate~\cite{hep-ph/0106067,1006.4945,1101.5118}. Data-driven methods might be able to reduce the uncertainties on these hadronic contributions and will benefit significantly from the high statistics collected by LHCb in the HL phase.
All of these methods are based on precise measurements of the $q^2$ spectra in conjunction with a theoretically motivated parametrization of the $q^2$ dependence
of the amplitudes and a theory benchmark that allows one to separate short-distance contributions from long-distance contributions.

At low $q^2$, the theory input is based on the light-cone Operator Product Expansion (OPE)
at very low (or negative) $q^2$~\cite{hep-ph/0106067,1006.4945},
an expansion that breaks down at the perturbative $c\bar c$ threshold $q^2\simeq 4m_c^2$.
Parametrizations of the $q^2$ dependence are based on a Taylor expansion in powers of $q^2$~\cite{Jager:2014rwa,Ciuchini:2017mik} or on  dispersion relations~\cite{1006.4945} such as in the 
$z$-expansion~\cite{1707.07305}. The latter two parametrizations implement analyticity constraints and use
extra information, such as data on (or in between) the $B\to \psi K^*$ decays, with $\psi=\{J/\psi, \psi(2S)\}$. Short- and long-distance effects are disentangled by the experimental input from $B\to \psi K^*$, the fixed $q^2$ dependence of the NP contribution, and by the theory constraints at negative $q^2$. The experimental prospects for this data-driven approach were studied in~\cite{1805.06378}, showing that future LHC data could provide a higher level of control over the long-distance contribution at low $q^2$.

At high $q^2$, the theory input is based on the low-recoil OPE~\cite{hep-ph/0404250,1006.5013,1101.5118}. This method relies
on the theoretical assumption that resonant effects from ``above-threshold'' charmonia average out within sufficiently broad $q^2$ bins (``quark-hadron duality''). Thus, only a single bin
in the whole high-$q^2$ region is typically considered in the global fits. The $q^2$ spectrum can be used to give estimates and test models of the intrinsically nonperturbative ``duality-violating'' effects. Currently, these analyses are carried out within the
``Kr\"{u}ger-Sehgal'' (naive factorization) approach~\cite{hep-ph/9603237}, which allows one to use data on the $R(s)$ ratio in $e^+e^-$ annihilation~\cite{1101.5118,1406.0566,1606.00775}. Ref.~\cite{1606.00775} uses all currently available data on $B\to K^*\mu^+\mu^-$
at low recoil and finds agreement with the OPE within $\sim 20\%$ in all the bins. Notably, future precision data from the LHC with the
expected fine binning will be essential in refining these data-driven methods and to disentangle NP contributions. As in the low-$q^2$ case, combined fits to hadronic parameters and NP are also beneficial~\cite{1606.00775}.

Hadronic uncertainties largely cancel in the SM in lepton-universality ratios such as  $R_{K^{(*)}}$~\cite{hep-ph/0310219},
\Blu{\begin{equation}
  R_{K^{(*)}} = \frac{\mathcal{B}(B \to K^{(*)}\mu^+\mu^-)}{\mathcal{B}(B \to K^{(*)}e^+e^-)}.
\end{equation}}
The SM predictions are thus limited only by the size of the electromagnetic corrections~\cite{Bordone:2016gaq}. Current LHCb measurements show tensions with the SM in $R_{K}$~\cite{1406.6482} and in two bins of $R_{K^*}$ ~\cite{1705.05802} at approximately $\sim 2.5\,\sigma$ each. Much higher precision will be achieved with future data at LHCb~\cite{Aaij:2244311}.
and, independently, with Belle-II, which can confirm $R_K$ at $5\,\sigma$ with $20\,\text{ab}^{-1}$~\cite{Kou:2018nap} at the current experimental central value. In the presence of LFUV contributions the predictions are less precise and ``optimized'' observables based on angular analyses of muonic and electronic modes can improve the sensitivity to different BSM scenarios~\cite{1605.03156,1612.05014}. In fact, LFUV and CLFV can be connected, and decays such as $B\to K^{(*)}\tau\mu$ and $B\to K^{(*)}\mu e$ become clear targets for the HL-LHC. \Blu{Semitauonic decays $B\to K^{(*)}\tau^+\tau^-$ are very challenging at the LHC but theoretically interesting because their branching fractions can receive enhancements of several orders of magnitude in BSM ~\cite{Alonso:2015sja,Capdevila:2017iqn} (see also~\cite{Kamenik:2017ghi} and \cite{Das:2018iap}).}
Assuming that the LFUV anomalies persist and the BSM contributions are in the muonic WCs, as the global fits to $b\to s \mu\mu$ currently suggest, one can use future precise and fine-binned measurements in $b\to se^+e^-$ exclusive modes to fit the hadronic parameters directly~\cite{1805.06401}. 

Other measurements with potential impact on the $b\to s\ell\ell$ fits will be possible at the HL-LHC.  The LHC experiments have 
a unique opportunity to  measure the $\Lambda_b\to\Lambda\mu^+\mu^-$ decays~\cite{LHCb-PAPER-2016-059}, probing the $b\to s\mu^+\mu^-$ transition in a baryonic system~\cite{Boer:2014kda,Das:2018sms,Blake:2017une}. 
For $B_s\to\phi \mu^+\mu^-$, a flavor-tagged time-dependent analysis allows one to access independent observables~\cite{1502.05509}
sensitive to BSM.
The expected sensitivity to the $B_s$ and $\Lambda_b$ decay measurements will make it possible to extend the 
global fit programme, outlined above,
 to these modes.
  Finally, the study of exclusive $b\to d\mu^+\mu^-$ transitions will reach the 
  level of precision we have now in $b\to s\mu^+\mu^-$ (see Sec.~\ref{sec:7_exp}).
This will allow one to extend the programme of the global fits to $b\to d\mu^+\mu^-$ transitions, 
setting constraints in a different flavor sector or, if  the $b\to s\ell\ell$ anomalies remain, to give further insights on the BSM flavour structure. The $b\to d\mu^+\mu^-$ modes are theoretically more challenging than their $b\to s\ell\ell$ counterparts, since new large long-distance contributions appear at low $q^2$, e.g., in the form of light resonances~\cite{Hambrock:2015wka}. 

\subsubsubsection{Radiative decays: $B_s\to \phi\gamma$ and related modes}

Radiative decays are obvious probes of the electromagnetic dipole operators.  Strategies to determine with these decays the helicity of the photon (and therefore the presence of BSM WC $C_{7\gamma}^{\prime}$  
have intensively been investigated~\cite{Atwood:1997zr,
Gronau:2001ng,Gronau:2002rz,Ball:2006eu,
Kou:2010kn,Becirevic:2012dx}. Approaches based on
the $\CP$-averaged exclusive decay rates are prone to hadronic uncertainties
and depend quadratically on $C_{7\gamma}^{\prime}$~\cite{Kou:2010kn,DescotesGenon:2011yn,Becirevic:2012dx}.
In contrast, interference between helicity amplitudes of the $B^0_{(s)}$ and $\bar B^0_{(s)}$ decays is directly sensitive to the photon polarization. This can be measured by constructing time-dependent $\CP$-asymmetries in $B\to V\gamma$ (the $S_{V\gamma}$ observable), or those induced by the width-differences $\Delta\Gamma_q$, 
shown to be clean null tests of the SM~\cite{Atwood:1997zr,Muheim:2008vu}. Experimentally, $S_{K^*\gamma}$ has been measured in the $B$-factories~\cite{Aubert:2005bu,Ushiroda:2006fi}, while a measurement of the width-difference effects has been achieved at LHCb~\cite{LHCb-PAPER-2016-034}. Prospects in the HL-LHC include reaching a few-percent precision in this observable, which will provide a strong constraint on the chirality of the electromagnetic dipole operators.  

The $B\to K^{(*)}\ell^+\ell^-$ decays are also sensitive to the interference of the two helicities  through the angular observables $P_1$ (also called $A_{\rm T}^{(2)}$) and $P_3^{CP}$ close to the photon pole $q^2\simeq0$, \Blu{ region where these observables are particularly clean from hadronic uncertainties~\cite{Jager:2014rwa}. Thus, the electronic mode is especially suited for these measurements, providing a theoretically clean window to right-handed currents beyond the SM.} This has been  measured by LHCb with a $\sim20\%$ precision with data of Run 1~\cite{1501.03038} and will be improved to a few-percent precision in the HL-LHC.

\subsubsubsection{Interplay with Belle-II: Inclusive $B$ decays and $B_q\to M \nu \bar \nu$}
Apart from contributing to the above-mentioned measurements~\cite{Kou:2018nap}, Belle-II will also measure decays which are very challenging in the LHC environment. This comprises inclusive decays and the $b\to s\nu\bar\nu$ transitions. The inclusive observables used in the current fits are
${\cal B}(B\to X_s \gamma)$ and ${\cal B}(B\to X_s \ell^+\ell^-)$. 
Belle-II will improve these, including precise measurements of the forward-backward asymmetry in $B\to X_s \ell^+\ell^-$, which will impact the fits~\cite{Kou:2018nap}. In particular, Belle-II measurements of $B\to X_s \mu^+\mu^-$ will be able
to test the LHCb anomalies independently by 2024~\cite{Kou:2018nap}. Theoretically, these inclusive rates can be calculated perturbatively in terms of the
partonic decay of the $b$ quark up to small non-perturbative effects~\cite{1003.5012,1705.10366}. These include local power corrections, non-local shape function-like effects, and additional contributions to the absence of an OPE for some operator insertion. \Blu{The latter effects represent irreducible uncertainties which cannot  be removed by relaxing the experimentally necessary cuts in the hadronic mass spectrum~\cite{1705.10366}.}
Calculations of both rates have been done with high accuracy~\cite{1503.01789,1503.04849}. 
The exclusive decays $B\to K^{(*)} \nu \bar \nu$ and related modes~\cite{1409.4557}, on the other hand, depend only on local form factors. Belle-II is expected to provide measurements of these modes with an uncertainty of about $10\%$,
assuming the rates are SM-like~\cite{Kou:2018nap}. These decays do not probe directly the WCs entering the $b\to s\nu\bar\nu$ transitions. However, they
can be correlated through $SU(2)_L$ gauge invariance in the SMEFT if the BSM is realized above the EW scale, and direct correlations are obtained in specific models of NP (see e.g.~\cite{Buras:2014fpa,Alonso:2015sja}).

\subsubsection{Model-Independent Fits}
\label{sec7:bsll:model:indep:fits}

Existing measurements show hints of deviations from the SM expectations in three
different classes of measurements:
in $B\to K^*\mu^+\mu^-$ angular observables \cite{LHCb-PAPER-2015-051},
in branching fractions of exclusive $b\to s\mu^+\mu^-$ decays (in particular
$B_s\to\phi\mu^+\mu^-$ \cite{LHCb-PAPER-2015-023}),
and in $\mu$-$e$ universality (LFU) tests \cite{LHCb-PAPER-2014-024,LHCb-PAPER-2017-013}.
None of the individual measurements is in tension with the SM by more than
$4\sigma$. However, a global significance of the tensions can be defined in a specific
framework of NP, such as the model-independent framework provided by the weak effective Hamiltonian.

Several groups have performed global fits of the WCs to existing $b\to s\ell\ell$ data
(see \cite{Altmannshofer:2017fio,Capdevila:2017bsm,Altmannshofer:2017yso,Hurth:2017hxg,Ciuchini:2017mik,Geng:2017svp,Arbey:2018ics}
for recent fits).
Three classes of fits can be distinguished: fits to $b\to s\mu\mu$ data only,
fits to $\mu$-$e$ LFU ratios only, and combined fits assuming no NP
in $b\to see$ transitions. All these fits agree -- up to differences that can be
attributed to different theoretical inputs or  different selection
of observables -- and arrive at two basic conclusions.
First, there is a tension in $b\to s\mu\mu$ data alone, and could be
explained by a NP shift in the WC $C_9^\mu$~\cite{1307.5683} (maybe combined with $C_{10}^\mu$), or by a not well understood
hadronic effect.
Second, the LHCb measurements of $R_{K^{(*)}}$, if combined,are already in tension with the SM and lepton-flavor universality at 4$\sigma$, \Blu{assuming there is no experimental correlation between $R_K$ and $R_{K^*}$.}  This cannot be explained by hadronic effects. Assuming that it is due to NP coupling only to muons, one finds that it is consistent with the NP contribution needed to accommodate the $b\to s\mu\mu$ anomaly~\cite{Alonso:2014csa}.
Singling out the WC $C_9^\mu$ in the muonic transition and performing
a global fit to all the data, the log-likelihood ratio between the best-fit
point and the SM hypothesis corresponds to a deviation ranging from $4\sigma$ to more than $6\sigma$, depending on the theoretical assumptions~\cite{Altmannshofer:2017fio,Capdevila:2017bsm,Altmannshofer:2017yso,Hurth:2017hxg,Ciuchini:2017mik,Geng:2017svp}. 

Clearly, future experimental efforts that can clarify the origin of the above tensions 
are of 
utmost
importance. If 
LFU is indeed violated in 
$b\to s\ell\ell$ transitions, then 
$R_K$ and
$R_{K^*}$ are theoretically 
clean smoking guns that allow one to establish a possible deviation from the SM. 
At the same time, global fits to all the relevant observables will remain 
relevant 
for several reasons: {\em (i)}
 if the hints for LFUV disappear with more statistics, LFU new physics effects~\cite{Geng:2017svp,Jager:2017gal,Alguero:2018nvb} might still hide in  observables that are theoretically less clean ; {\em (ii)}
 to identify the nature of NP
 (and not just its presence), the values of the BSM Wilson coefficients
 have to be determined; \Blu{{\em (iii)} in particular, if LFUV persists, to understand whether the NP effects are due to the muons, the electrons, or to which part of admixture one needs to perform lepton specific measurements and corresponding global fits; }and {\em (iv)}
 they allow one to simultaneously determine poorly known hadronic effects from the data.

\begin{figure}
    \centering
    \includegraphics[width=0.5\textwidth]{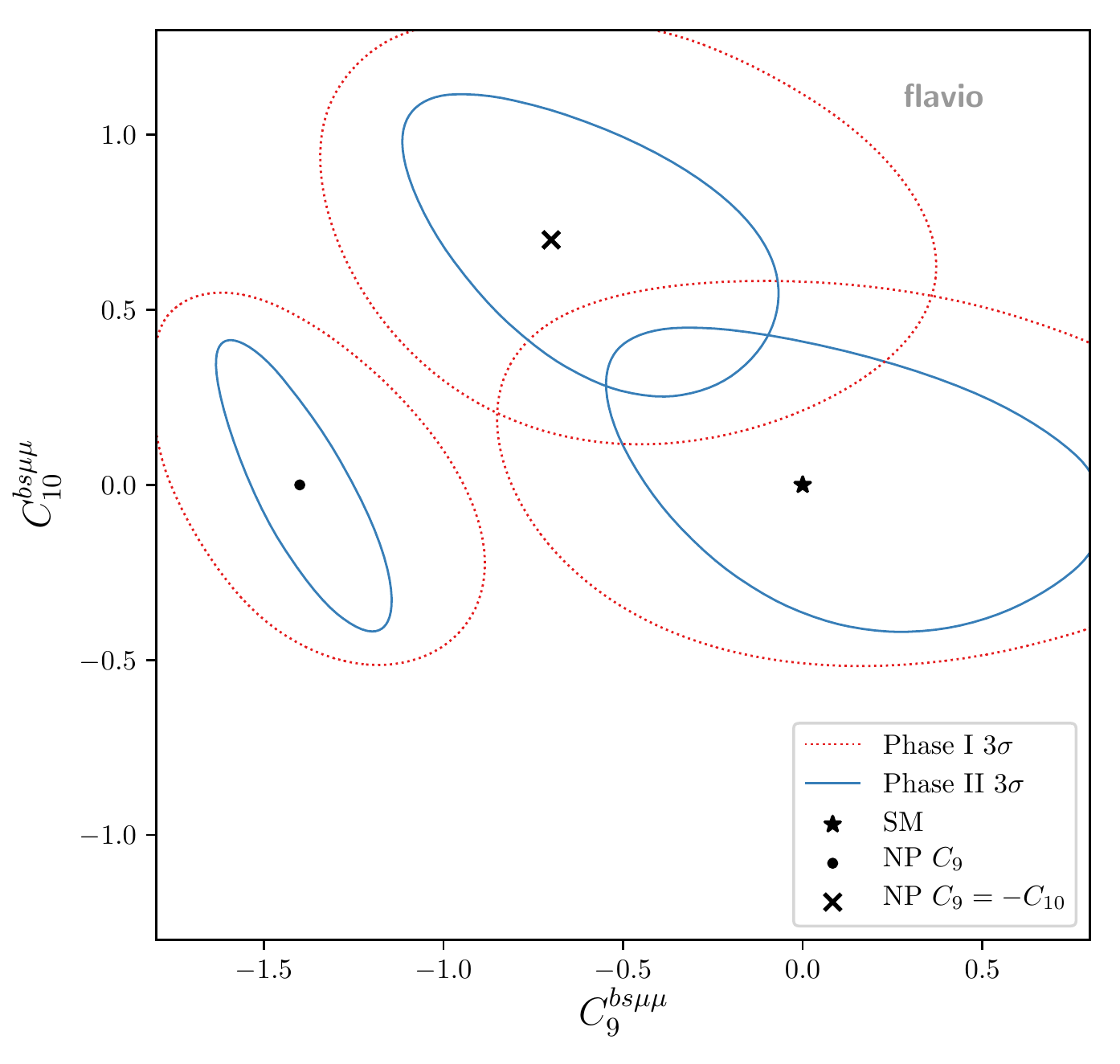}
    \caption{Potential sensitivity to the SM and to NP scenarios motivated by the anomalies of LHCb, ATLAS and CMS combined after the HL-LHC phase. These scenarios are $C_9=-1.4$ (vector current) and $C_9=-C_{10}=-0.7$ (pure left-handed current). The observables included are the branching fraction of $B_s\to\mu^+\mu^-$ and the angular observables of the decay $B^0 \to K^{\ast 0} \mu^+\mu^-$ in the low-$q^2$ region (e.g., $P_5^\prime$). To produce the $P_5^\prime$ expectation for ATLAS and CMS in Phase I, the result from a CMS projection was scaled by $1/\sqrt{2}$, assuming that the two experiments have the same sensitivity and the uncertainties are uncorrelated. \Blu{This plot has been done using the {\tt flavio} software package~\cite{Straub:2018kue}.}}
    \label{fig:WC}
\end{figure}

Extrapolations of global fits to the LHC data after the HL phase, in the  $(C_9^\mu$, $C_{10}^\mu)$ plane,
are shown in Fig.~\ref{fig:WC}. 
Further improvements beyond those taken into account in  projections in Fig.~\ref{fig:WC} are:
{\em (i)}  $B\to K^*e^+e^-$ angular analysis can be included, where for LFUV NP a simultaneous amplitude analysis of $B\to K^*\mu^+\mu^-$ and  $B\to K^*e^+e^-$ decays is more powerful than separate analyses~\cite{1805.06401};
{\em (ii)} combined fits to semi-leptonic and non-leptonic decays as well as the
precise dilepton invariant mass spectra will allow  one to better disentangle
long- and short-distance effects in $B\to K^*\mu^+\mu^-$~\cite{1805.06378}; {\em (iii)} 
unbinned determination of Wilson coefficients~\cite{Hurth:2017sqw} will fully exploit the experimental
potential.

\subsubsection{Models of NP for $b\to s\ell\ell$}
\label{sec:7:bsll:NP}
Finally, we briefly review the NP models that can explain the $b\to s\ell \ell$ anomalies, assuming these become statistically significant. The model-independent analysis gives for the NP scale \Blu{of a tree-level mediator with couplings of $\mathcal O(1)$},
\begin{equation}
 \Lambda_\text{NP} = \frac{4\pi}{e} \frac{1}{\sqrt{|V_{tb} V_{ts}^*|}} \frac{1}{\sqrt{|\Delta C_9^\mu|}} \frac{v}{\sqrt{2}} \simeq \frac{35~\text{TeV}}{\sqrt{|\Delta C_9^\mu|}}~,
\end{equation}
where $\Delta C_i^\ell$ denotes the NP contribution to $C_i^\ell$.
The actual mass of the NP degrees of freedom responsible for the anomalies can be much smaller if the NP couplings are small and/or if the NP contributions arise at loop level, instead of at tree level.

\begin{figure}[t]
\begin{center} 
\includegraphics[width=0.23\textwidth]{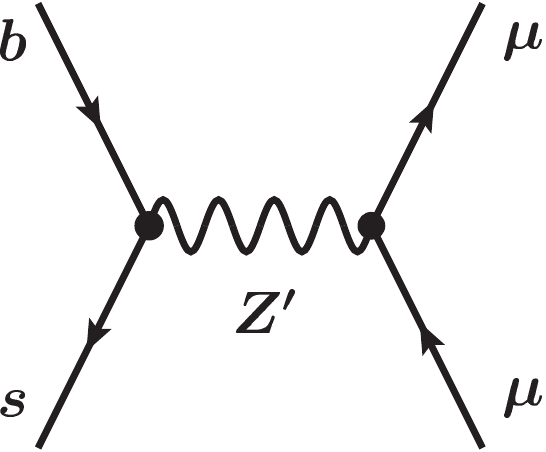} ~~~~~~~~
\includegraphics[width=0.23\textwidth]{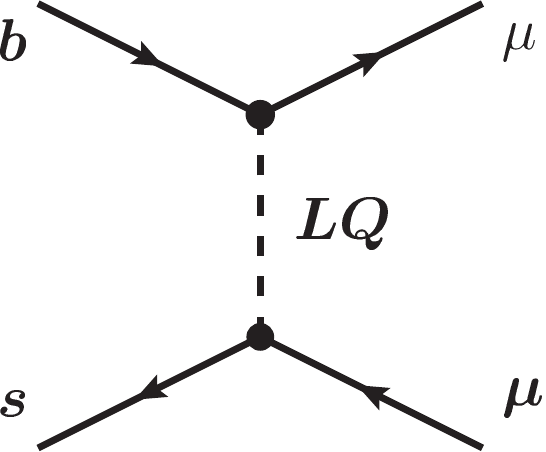} ~~~~~~~~
\includegraphics[width=0.26\textwidth]{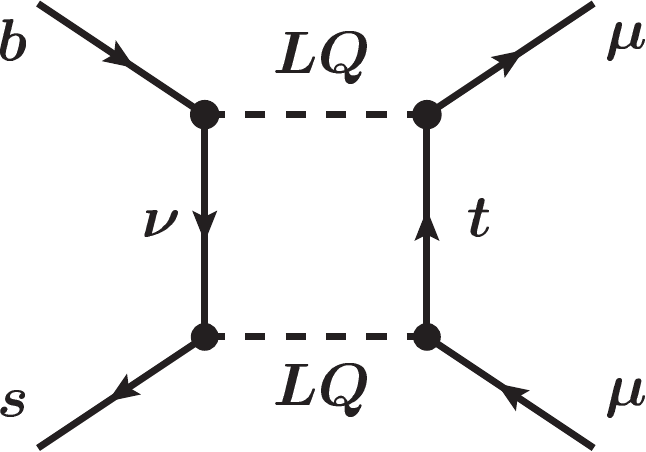}
\caption{Examples of NP contributions to $b \to s \mu \mu$ transitions. Left: tree-level $Z^\prime$ contribution. Center: tree-level leptoquark contribution. Right: One-loop leptoquark contribution.}
\label{fig:bsll_diagrams}
\end{center}
\end{figure}

At tree level there are two types of NP particles that can contribute to $b \to s \ell \ell$ transitions: $Z^\prime$ and leptoquarks (see left and center diagrams in Fig.~\ref{fig:bsll_diagrams}). Loop-level contributions of leptoquarks have been studied extensively in the literature (see right diagram in Fig.~\ref{fig:bsll_diagrams}). Other loop-level models have been put forward, see, 
e.g., 
Refs.~\cite{Aebischer:2015fzz,Belanger:2015nma,Gripaios:2015gra,Arnan:2016cpy,Kamenik:2017tnu,Crivellin:2018yvo}, but we will not discuss them in detail.

\subsubsubsection{$Z^\prime$ models}
Models with a $Z^\prime$ that has flavour-violating couplings to quarks and that couples non-universally to leptons can explain the observed anomalies in $b \to s \ell \ell$. In a ``simplified model'' approach, treating the $Z^\prime$ mass and its couplings to the SM fermions as free parameters, irreducible constraints arise from $B_s$ mixing, neutrino-trident production and LEP bounds on four-lepton contact interactions.

Combining these constraints gives the following maximal values for the $Z^\prime$ contributions to the relevant Wilson coefficients $\Delta C_{9,10}^{e,\mu}$ in the case of {\em (i)} left-handed lepton couplings~(\ref{eq:C910_LH}), or {\em (ii)} in the case of vectorial lepton couplings~(\ref{eq:C910_vec})~\cite{Altmannshofer:2014rta},
\begin{eqnarray} \label{eq:C910_LH}
 && |\Delta C_9^\mu| = |\Delta C_{10}^\mu| \lesssim 5.4 ~,\quad |\Delta C_9^e| = |\Delta C_{10}^e| \lesssim 0.64 ~, \\  \label{eq:C910_vec}
 && |\Delta C_9^\mu| \lesssim 9.3 ~,\quad |\Delta C_9^e| \lesssim 0.72 ~,\quad \Delta C_{10}^\mu = \Delta C_{10}^e = 0 ~.
\end{eqnarray}
The $Z^\prime$ coupling to muons 
can comfortably explain the anomalies in $R_K$ and $R_{K^*}$ and the 
other anomalies in $b \to s \mu \mu$ transitions.
Addressing $R_K$ and $R_{K^*}$ through a $Z^\prime$ coupling to electrons
is  possible only in the parameter region close to the current upper bounds from $B_s$ meson mixing and
four-lepton contact interactions. \Blu{The sensitivity to NP of  $B_s$-mixing~\cite{DiLuzio:2017fdq,DiLuzio:2018wch} will become stronger at the time scale of the HL/HE-LHC due to improved lattice predictions of hadronic matrix elements (see Secs.~\ref{sec:metro-CKM} and \ref{sec:lattice}).}
The $Z^\prime$ mass can be at most several TeV; otherwise an explanation of the anomalies requires couplings to leptons that are non-perturbatively large. If the $Z^\prime$ has very weak couplings to light quarks and electrons, its production cross section at colliders is tiny. Therefore the $Z^\prime$ could in principle be light, for example at the electroweak scale, or in certain models even much lighter~\cite{Datta:2017pfz,Sala:2017ihs,Altmannshofer:2017bsz}. 

A popular class of UV-complete $Z'$ models is based on gauging the difference of muon-number and tau-number, $L_\mu - L_\tau$. Once physics is introduced that generates flavor-violating couplings of the $Z'$ to quarks, the observed tensions in $b \to s \ell \ell$ decays can be explained~\cite{Altmannshofer:2014cfa,Crivellin:2015mga,Altmannshofer:2015mqa,Fuyuto:2015gmk,Altmannshofer:2016jzy,Baek:2017sew,Chen:2017usq,Altmannshofer:2016oaq}. $L_\mu - L_\tau$ models predict
\begin{equation}
 \Delta C_9^e = 0 ~,\quad \Delta C_9^\mu = - \Delta C_9^\tau ~,\quad \Delta C_{10}^e = \Delta C_{10}^\mu = \Delta C_{10}^\tau = 0~.
\end{equation}
Besides $L_\mu - L_\tau$, various other combinations of gauged flavor symmetries have been used to construct $Z'$ models that can address the $b \to s \ell \ell$ anomalies~\cite{Crivellin:2015lwa,Celis:2015ara,Falkowski:2015zwa,Boucenna:2016wpr,Boucenna:2016qad,Celis:2016ayl,Crivellin:2016ejn,Alonso:2017bff,Ellis:2017nrp,Alonso:2017uky,Bonilla:2017lsq,Babu:2017olk,Bian:2017rpg,Tang:2017gkz,Cline:2017ihf}.
Also, scenarios where the $Z'$ couples to both quarks and leptons indirectly through the mixing with heavy vectorlike fermions have been considered~\cite{Sierra:2015fma,King:2017anf,Fuyuto:2017sys}.

In models with partial compositeness the $Z'$ can be identified with a heavy neutral spin-1 resonance of the composite sector, typically denoted as $\rho$. Such a resonance generically features flavor-violating couplings to quarks. A large degree of compositeness of the left-handed muons is required to explain the $B$-decay anomalies~\cite{Niehoff:2015bfa,Carmona:2015ena,Megias:2016bde,Carmona:2017fsn,Megias:2017ove,Sannino:2017utc}.
The generic expectation in models with partial compositeness is that the $\rho$ couplings are strongest to SM fermions of the third generation, reflecting the mass hierarchy of the SM fermions that is related to their degree of compositeness.
Assuming the dominance of couplings to left-handed leptons, these models 
suggest the pattern 
\begin{equation}
 \Delta C_9^e \simeq -\Delta C_{10}^e \ll \Delta C_9^\mu \simeq -\Delta C_{10}^\mu \ll \Delta C_9^\tau \simeq -\Delta C_{10}^\tau ~.
\end{equation}
Models in which the $Z^\prime$ is part of a $SU(2)_L$ triplet have been suggested as a simultaneous explanation of the $b \to s \ell \ell$ anomalies and hints for LFUV in semileptonic charged-current decays, $R(D)$ and $R({D^*})$~\cite{Bhattacharya:2014wla,Greljo:2015mma,Bhattacharya:2016mcc,Boucenna:2016wpr,Boucenna:2016qad,Buttazzo:2017ixm,Kumar:2018kmr}. 

Different $Z'$ models predict different patterns of NP effects in $b \to s \ell \ell$ and the related $b \to s \nu \nu$ transitions. Future measurements of these transitions at LHCb and Belle II will therefore allow one to narrow down viable $Z^\prime$ models.
For example, the $Z^\prime$ models based on the gauged $L_\mu - L_\tau$ symmetry predict effects in the semileptonic $b \to s \mu^+ \mu^-$ and $b \to s \tau^+ \tau^-$ transitions of opposite sign, while $b \to s e^+ e^-$ transitions remain SM-like.
In these models, the purely leptonic $B_s \to \mu^+\mu^-$ and $B_s \to \tau^+\tau^-$ decays, as well as the neutrino modes $B \to K^{(*)} \nu \bar \nu$, are predicted to be SM-like~\cite{Altmannshofer:2014cfa}.

A markedly different pattern arises in the $Z^\prime$ scenarios based on dominant couplings to left-handed fermions of the third generation. In those models the $b \to s \tau^+ \tau^-$ and $B \to K^{(*)} \nu \bar \nu$ rates are typically enhanced compared to the SM predictions by a factor of a few. The $B_s \to \mu^+\mu^-$ rate is predicted to be suppressed by approximately $25\%$ compared to the SM prediction. Finally, in contrast to the $L_\mu - L_\tau$ models, rare lepton flavor-violating decays like $B \to K^{(*)} \tau \mu$ are predicted at levels of $O(10^{-8})$~\cite{Glashow:2014iga} which might be in reach at the HL-/HE-LHC.\\

\subsubsubsection{Leptoquark models}
There are seven quantum number assignments for leptoquarks that allow tree-level couplings to down-type quarks and charged leptons of the SM.
These are~\cite{Dorsner:2016wpm} $S_3 = (\bar 3,3,1/3)$, $R_2 = (3,2,7/6)$, $\tilde R_2 = (3,2,1/6)$, $\tilde S_1 = (\bar 3,1,4/3)$, $U_3 = (3,3,2/3)$, $V_2 = (\bar 3,2,5/3)$ and $U_1 = (3,1,-1/3)$.
Among these, the couplings of $R_2$, $\tilde R_2$, $\tilde S_1$, and $V_2$ necessarily involve right-handed currents and therefore cannot explain the anomalies.
Thus one is left with the triplet scalar $S_3$, the singlet vector $U_1$, and the triplet vector $U_3$. They all contribute at tree level to the operator $(\bar s \gamma_\alpha P_L b)(\bar \mu \gamma^\alpha P_L \mu)$ and can explain the $b \to s \ell \ell$ anomalies.

In a simplified model approach the constraints on the leptoquark models are very weak.
While the leptoquarks $S_3$, $U_1$, and $U_3$ contribute to $B_s$-mixing, they only do so at the 1-loop level. 
Correspondingly, the bounds on the leptoquark couplings from $B_s$ mixing allow leptoquark masses as high as several 10's of TeV.
Lower bounds on the leptoquark masses come from direct searches at hadron colliders. All leptoquarks are charged under color 
and 
can be pair-produced through strong interactions in $pp$ collisions. Bounds from direct searches at the LHC are \Blu{currently above 1~TeV in all different channels~\cite{Diaz:2017lit}}. 

\Blu{There is an additional leptoquark that has been identified as a possible explanation of the rare $B$-decay anomalies. With the appropriate couplings, the scalar-doublet leptoquark $R_2 = (3,2,7/6)$ can contribute to $b \to s \ell \ell$ transitions through 1-loop box diagrams~\cite{Becirevic:2017jtw} (for an earlier attempt with the scalar singlet leptoquark $S_1$ see~\cite{Bauer:2015knc}). The $R_2$ model largely avoids constraints from the neutrino modes $B \to K^{(*)} \nu \bar \nu$, and $B_s$ mixing as well as from lepton universality in $b \to c \mu \nu$ and $b \to c e \nu$ transitions. If $R_2$ is responsible for the anomalies, it is expected to be very close to the current sensitivity of direct searches. }

Most studies treat leptoquarks in the simplified-model approach~\cite{Hiller:2014yaa,Bauer:2015knc,Fajfer:2015ycq,Alonso:2015sja,Becirevic:2016oho,Bhattacharya:2016mcc,Hiller:2017bzc,Chen:2017hir,Crivellin:2017zlb,Kumar:2018kmr}.
Going beyond simplified models, one finds that the leptoquark couplings 
required to explain the anomalies 
are likely not of Minimal-Flavor Violation type~\cite{Aloni:2017ixa}, and constitute new sources of flavor violation beyond the SM Yukawa couplings.
Both scalar and vector leptoquarks could arise, for example, in composite models~\cite{Gripaios:2014tna,Barbieri:2016las}. The vector leptoquarks could also be the gauge bosons of an enlarged gauge group that is broken not far above the TeV scale~\cite{Das:2016vkr,Fornal:2018dqn,Assad:2017iib,Bordone:2017bld,DiLuzio:2017vat,Bordone:2018nbg,Blanke:2018sro,Calibbi:2017qbu}.
The scalar-singlet leptoquark that contributes to $b \to s \ell \ell$ transitions at the loop level can be identified with the right-handed sbottom in the Minimal Supersymmetric Standard Model with $R$-parity violation~\cite{Das:2017kfo,Earl:2018snx}.

Some leptoquark scenarios are also able to address simultaneously the $b \to s \ell \ell$ anomalies as well as the hints for LFUV in $R({D^{(*)}})$~\cite{Bauer:2015knc,Bhattacharya:2016mcc,Barbieri:2016las,Das:2016vkr,Chen:2017hir,Buttazzo:2017ixm,Bordone:2017bld,DiLuzio:2017vat,Crivellin:2017zlb,Bordone:2018nbg,Blanke:2018sro,Kumar:2018kmr,Becirevic:2018afm}.
However, many of the models that attempt a simultaneous explanation are strongly constrained by measurements of the $\tau^+\tau^-$ invariant mass spectrum at the LHC~\cite{Faroughy:2016osc}; by existing bounds on $B \to K \nu\nu$ and $B \to K^* \nu\nu$ from BaBar and Belle; by existing bounds on LFV tau decays like $\tau \to 3\mu$; from precision measurements of the leptonic couplings of the $Z$ at LEP~\cite{Feruglio:2016gvd,Feruglio:2017rjo,Cornella:2018tfd}, and by lepton-universality tests in leptonic tau decays $\tau\to\ell \nu_\tau\bar\nu_\ell$~\cite{Feruglio:2016gvd,Feruglio:2017rjo,Cornella:2018tfd}.

%% file: section7/bctaunu.tex
\subsubsection{$B\rightarrow D^{(*)}l\nu$ form factors and anatomy of $R(D^{(*)})$}
Signs of LFUV have not only been seen in loop-suppressed flavor-changing 
neutral currents discussed above, but also in tree-level decays, namely, in tensions with the SM predictions for the ratios
\begin{equation}
  R(D^{(*)}) = \frac{\mathcal{B}(B \to D^{(*)}\tau\nu)}{\mathcal{B}(B \to D^{(*)}l\nu)},\,\hspace{0.5cm}(l=\mu,~e).
\end{equation}
Assuming for the moment the SM particle content only, the $b\to c\ell\nu_{\ell^\prime}$ transitions at scales near $m_b$ can be described by an $SU(3) \times U(1)$-invariant effective Hamiltonian,
\begin{equation}\label{eq:Heffbcellnu}
  \mathcal H_\text{eff}^{b\to c\tau\nu} = \frac{4 G_F}{\sqrt{2}} V_{cb}
    \sum_{\ell\ell^\prime}\Big(\Blu{(\delta_{\ell\ell^\prime}+C_L^{\ell\ell^\prime})}O_{V_L}^{\ell\ell^\prime} + \sum_{i} C_i^{\ell\ell'} O_i^{\ell\ell'} + \text{h.c.}\Big),
\end{equation}
where $\ell,\ell'=e,\mu,\tau$ denotes the charged-lepton and neutrino flavor, respectively, and the sum over $i$ runs over the following operators,
\begin{equation}
\label{eq:ops}
\begin{split}
  O_{V_{L,R}}^{\ell\ell'} &= (\bar c_{L,R} \gamma^\mu b_{L,R})(\bar \ell_L \gamma_\mu \nu_{\ell' L}), 
  \qquad O_{S_{L,R}}^{\ell\ell'} = (\bar c_{R,L} b_{L,R})(\bar \ell_R \nu_{\ell' L}), 
  \\
  O_T^{\ell\ell'} &= (\bar c_R \sigma^{\mu\nu} b_L)(\bar \ell_R \sigma_{\mu\nu}\nu_{\ell' L}).
\end{split}
\end{equation}
The NP coefficients $C_i^{\ell\ell'}$ depend, in general, on both charged-lepton and neutrino flavor. These operators arise from more fundamental interactions at a higher scale $\Lambda$ which, in accordance with available LHC data, can be taken to be larger than the electroweak scale $v˜$. The operators of Eq.~(\ref{eq:ops}) should then be embedded into $SU(2)_L\times U(1)$ invariant operators composed of the full SM field content. For $C_{V_R}$ there is no contribution violating lepton-universality at dimension-6, leading to a parametric suppression of at least $v^2/\Lambda^2$ for tree-level ultraviolet (UV) completions. Linearly realized electroweak symmetry breaking, captured by the Standard Model effective field theory (SMEFT)~\cite{Buchmuller:1985jz,Grzadkowski:2010es}, also yields the prediction $C_{V_R}^{\ell\ell'} \equiv C_{V_R}\delta_{\ell\ell'}$~\cite{Cirigliano:2009wk,Cata:2015lta}. In this case, sizable contributions to $C_{V_R}$ are excluded by $b\to c(e,\mu)\nu$~\cite{Jung:2018lfu}. While deviations from this prediction are possible in a non-linear realization of electroweak symmetry breaking~\cite{Cata:2015lta}, at least one of the above-named sources of suppression always remains and right-handed currents do not play a role.

The relevant hadronic matrix elements of the SM operator $O_{V_L}^{\ell\ell}$, i.e., $ \bra{D} \bar{c} \gamma^\mu b \ket{\bar{B}}$, $\bra{D^*} \bar{c} \gamma^\mu b \ket{\bar{B}}$, and 
$\bra{D^*} \bar{c} \gamma^\mu \gamma_5 b \ket{\bar{B}}$, are parametrized using one, two, and three form factors, respectively~\cite{Caprini:1997mu}.
The helicity form factors obey unitarity bounds which can be written in an elegant way using the parametrization of Boyd-Grinstein-Lebed (BGL) \cite{Boyd:1997kz}. The contributions of the helicity form factors $S_1$ and $P_1$ \Blu{(as defined in~\cite{Caprini:1997mu})} to the branching ratios of $B\rightarrow Dl\nu$ and $B\rightarrow D^*l\nu$ are
suppressed by the mass of the final-state lepton. In view of the lack of experimental information on these form factors we need  input from theory. 
In the case of $B \to D$, we are in the fortunate position that lattice results
for both $V_1$ and $S_1$ at $w\geq 1$ exist~\cite{Lattice:2015rga,Na:2015kha,Aoki:2016frl}, where $w = (m_B^2 + m_{D^{(*)}}^2-q^2)/(2 m_B m_{D^{(*)}})$.
This is the reason for the very good agreement of all SM predictions in this case, see Table~\ref{tab:RD-and-RDstar-anomaly}. \Blu{Recently, also soft photon corrections to $R(D)$ have been discussed~\cite{deBoer:2018ipi}.}
For $B \to D^*$, we have only one data point from lattice, namely $A_1(w=1)$~\cite{Bailey:2014tva,Harrison:2017fmw} so that it is necessary to use 
Heavy Quark Effective Theory (HQET)~\cite{Bernlochner:2017jka,Caprini:1997mu,Luke:1990eg,Neubert:1991xw,Neubert:1993mb,Neubert:1992wq,Neubert:1992pn,Ligeti:1993hw}
to relate the $B \to D$ and $B \to D^*$ form factors order-by-order in the heavy quark expansion in terms of Isgur--Wise functions
in order to obtain a prediction for $P_1$ and hence $R(D^*)$. 

At NLO in the heavy-quark expansion, {i.e.}, expanding to linear order in $\alpha_s/\pi$ and $1/m_{c,b}$,
there is sufficient differential information in the $B \to D^{(*)}l\nu$ decays to fit to the four Isgur-Wise functions
that arise at this order \cite{Bernlochner:2017jka,Jaiswal:2017rve}. 
In the literature the theoretical error from using NLO HQET results is under discussion, leading to different 
results for the error of the SM prediction. 
HQET can also be used to obtain a stronger version of the unitarity bounds
(see Refs.~\cite{Boyd:1997kz,Bigi:2017jbd} for details).
Additional model-dependent theoretical input at maximal $w$ can be provided by Light Cone Sum Rules (LCSR)~\cite{Faller:2008tr,Gubernari:2018wyi},
or at zero recoil by QCD sum rules~\cite{Neubert:1992wq,Neubert:1992pn,Neubert:1993mb,Ligeti:1993hw}.
We give a summary of theoretical predictions for $R(D^{(*)})$ in Table~\ref{tab:RD-and-RDstar-anomaly}.

Since lattice data is presently available only at zero recoil for $B \to D^*$
and the kinematic suppression requires $d\Gamma[B \to D^*l\nu]/dw$ to vanish at $w=1$,  $|V_{cb}|$ can be obtained from $B \to D^*l\nu$ only by extrapolating $d\Gamma[B \to D^*l\nu]/dw$ back to zero recoil. 
This procedure can be highly sensitive to the chosen form-factor parameterization and other theoretical inputs;
for recent analyses, see 
Refs.~\cite{Abdesselam:2017kjf,Bigi:2017njr,Bigi:2017jbd,Grinstein:2017nlq,Bernlochner:2017jka,Bernlochner:2017xyx,Jaiswal:2017rve,Harrison:2017fmw,Schacht:2017vfd}. 
Lattice data beyond zero recoil is expected soon (see \cite{Aviles-Casco:2017nge} for preliminary results)
from both domain wall and AsqTad ensemble approaches. 
Combined with abundant future data for $B \to D^*l\nu$ from HL-LHC, extractions of $|V_{cb}|$ that are less sensitive to theoretical inputs
will become possible, thereby either resolving or more concretely establishing tensions between exclusive and inclusive measurements of $|V_{cb}|$.

With future experimental data and lattice QCD results (see Sec.~\ref{sec:lattice} and Table~\ref{table:Proj} therein) the SM 
prediction of $R(D^*)$ will improve considerably. For an estimate, one may note that the dependence of $d\Gamma/dw$ on the $P_1$ form factor arises 
only incoherently, via a contribution of the form $m_l^2|P_1(w)|^2$. This $P_1$ term contributes approximately 10\% of the total 
integrated $B \to D^*\tau\nu$ rate, which suggests that the dependence of $R(D^*)$ on this form factor should be limited. 
Assuming a 1\% future precision for $P_1$, using the 50 ab$^{-1}$ which Belle~II will presumably gather by 2025, and taking into 
account the expected improvement of the form factors $A_{1,5}$ and $V_4$ from the lattice, it is hence reasonable to assume that
an error of 0.001 for the SM value of $R(D^*)$ might be achieved.

\begin{table*}[t]
\caption{
Current SM $R(D^{(*)})$ theory predictions and their deviation from experiment 
$R(D)^{\mathrm{exp}} = 0.407(39)(24)$ \cite{Amhis:2016xyh,Lees:2012xj,Lees:2013uzd,Huschle:2015rga}
and
$R(D^*) = 0.306(13)(7)$ \cite{Amhis:2016xyh,Lees:2012xj,Lees:2013uzd,Huschle:2015rga,Sato:2016svk,LHCb-PAPER-2015-025,Hirose:2016wfn,Hirose:2017dxl,LHCb-PAPER-2017-017,LHCb-PAPER-2017-027}
(HFLAV 2018 summer update). 
For older SM predictions see Refs.~\cite{Fajfer:2012vx,Celis:2012dk,Tanaka:2012nw}.
Table adapted and extended from Ref.~\cite{Schacht:2017vfd}.
\label{tab:RD-and-RDstar-anomaly}} 
\begin{center}
\begin{tabular}{ccc}
  \hline\hline
  Ref.                       & $R(D)$      &  Exp. deviation \\
  \hline
  \cite{Bigi:2016mdz}        & $0.299(3)$  & $2.4\sigma$ \\ 
  \cite{Bernlochner:2017jka} & $0.299(3)$  & $2.4\sigma$ \\
  \cite{Jaiswal:2017rve}     & $0.302(3)$  & $2.3\sigma$ \\
  \hline\hline 
\end{tabular}
\qquad
\begin{tabular}{ccc}
  \hline\hline
  Ref.                       & $R(D^*)$   &  Exp. deviation \\
  \hline 
  \cite{Bernlochner:2017jka} & $0.257(3)$ & $3.3\sigma$ \\
  \cite{Bigi:2017jbd}        & $0.260(8)$ & $2.7\sigma$ \\ 
  \cite{Jaiswal:2017rve}     & $0.257(5)$ & $3.1\sigma$ \\
  \hline\hline
\end{tabular}
\end{center}
\end{table*}

\subsubsection{Excited states and other $b$ hadrons: $R(D^{**})$, $R(J/\psi)$, $R(\Lambda_c^{(*)})$ and $R(X_c)$}

Measurements of $|V_{cb}|$ and lepton universality can also be probed via $B$ decays to the $D^{**}$ excited states,
as well as decays of strange or charmed $b$ hadrons, including $B_s \to D^{(*)}_s$, $B_c \to J/\psi$,
and the baryonic $\Lambda_b \to \Lambda_c^{(*)}$ transitions. In comparison to the $B \to D^{(*)}\ell \nu$ decay modes, these modes may 
variously exhibit higher sensitivities to specific NP operators or, in some cases,
may be theoretically cleaner than the decays to the $D^{(*)}$ ground states. 
These modes can also be important downfeed or crossfeed backgrounds to the $B \to D^{(*)}\ell \nu$ decays and,
to the extent they are affected by the same NP operators, must also be understood and measured carefully.
The HL-LHC is the only experiment planned that will yield significant samples of these heavier $b$-hadrons, 
with precision analyses anticipated from the LHCb experiment.
In this subsection we present the motivations and theoretical prospects for the measurement of each of these \Blu{exclusive decay modes, 
as well as for measurement of inclusive semileptonic $B$ decays.}

The $D^{**}$ excited states comprise four different charmed hadrons: The $D_0^*$, $D_1^*$, $D_1$, $D_2^*$.
In the language of HQET, these furnish two doublets, $\{D_0^*\,,D_1^*\}$ and $\{D_1\,,D_2^*\}$,
with spin-parity $s_{\ell}^{\pi_\ell} = \frac{1}{2}^+$ and $\frac{3}{2}^+$, respectively.
(In the heavy-quark limit, spin-parity is a conserved quantum number.)
The $\frac{3}{2}^+$ states are narrow, with $\Gamma \sim 30$--$50$\,MeV,
because their hadronic decays to $D^{(*)}\pi$ either proceed via a $D$-wave or violate heavy quark-symmetry,
while the $\frac{1}{2}^+$ states are quite broad.  Although isolating these excited-state decays will likely be simpler at $e^+e^-$ $B$ factories, 
which can more easily reconstruct $\pi^0$'s and photons, analyses of $B \to D^{**}\ell\nu$ decays are also feasible at LHCb \upgradetwo, 
especially for the narrow $\frac{3}{2}^+$ states subsequently decaying to charged hadrons.

The crucial attractive feature of the $B \to D^{**}$ transitions is that various leading-order contributions to the form factors
vanish in the heavy-quark limit at zero recoil ($w=1$), so that $\mathcal{O}(\alpha_s)$ and $\mathcal{O}(1/m_{c,b})$ corrections
become important~\cite{Leibovich:1997tu,Leibovich:1997em,Bernlochner:2012bc,Bernlochner:2016bci}. 
The richer structure of the subleading form-factor contributions has the consequence that sensitivity to various NP currents
can be much larger than in the ground state decays~\cite{Bernlochner:2016bci, Bernlochner:2017jxt}. 
For example, including only a NP tensor operator, $O^{\ell\ell^\prime}_T$, one finds the ratios
$R(X)/R(X)_{\text{SM}} \simeq \{1.5, 1.3\}$ for $X = \{D, D^*\}$ at the best fit to the $R(D^{(*)})$ data. 
However, the same Wilson coefficients result in $R(X)/R(X)_{\text{SM}}$ greater than $4.0$ or less than $0.5$ for $X=\{D_0^*, D_1^*, D_1, D_2^*\}$. \Blu{Interference between different $B\to D^{**}$ transitions also offers the possibility to probe for new CP-violating phases in the NP operators~\cite{Aloni:2018ipm}.}
The current SM predictions for all four modes, from fits to Belle data including NLO HQET contributions, are~\cite{Bernlochner:2017jxt}
\begin{equation}
\label{eq:RDDs}
  R(D_0^*) = 0.08 \pm 0.03,\quad  R(D_1^*) = 0.05 \pm 0.02, \quad
  R(D_1) =  0.10 \pm 0.02,\quad  R(D_2^*) = 0.07 \pm 0.01\,.
\end{equation}

Decays of $B$ mesons to these excited states have total SM branching ratios comparable to the $B \to D^{(*)}\ell \nu$ decays themselves.
Combined with the possible large enhancement of the semitauonic modes by NP contributions,
this means that the subsequent $D^{**} \to D^{(*)}X$ decays can then induce important downfeed  backgrounds to the $D^{(*)}$ measurements.
Analyses of $B \to D^{(*)}\ell \nu$ will therefore typically have to fold in contributions from these excited states.
Moreover, anticipated analyses for the inclusive $B \to D\pi\ell\nu$ decays provide an opportunity to probe these excited-state decays
and their associated larger NP sensitivities collectively with the ground-state decays.
In this context, rather than being thought of as a background, these contributions should be more properly thought of as additional sources of (NP) signal.
The large data sets from HL-LHC will then provide a very sensitive set of channels for probing NP contributions to $b \to c \ell \nu$.

The $B_s$ and $B_c$ mesons have production ratios $\sigma(B_s)/\sigma(b\bar{b}) \sim 10\%$~\cite{LHCb-PAPER-2013-004}
and $\sigma(B_c)/\sigma(b\bar{b}) \sim 0.2\%$~\cite{LHCb-PAPER-2013-044} in Run 1 LHC.
A much smaller sample of $B_s$ mesons may also be produced at $B$ factories running on the $\Upsilon(5S)$ resonance.
However, for the $B_c$ the only significant sample of mesons will be produced at HL-LHC.
At first glance, the theoretical structure of $B_s \to D_s^{(*,**)}\ell\nu$ can be mapped directly from $B \to D^{(*,**)}\ell\nu$
via the approximate $SU(3)$ flavor symmetry.
Some crucial differences are that $\mathcal{B}(D_s^{*} \to D_s\gamma) \simeq 94\%$, which will be difficult to see at LHCb \upgradetwo,
while the four $D_s^{**}$ excited states are all narrow, and therefore may be easier to resolve.

The leptonic $B_c \to (J/\psi \to \mu\mu) \ell \nu$ decay mode is reasonably clean experimentally, with measurements for $R(J/\psi)$ already available from LHCb Run 1 data,
$  R(J/\psi) = 0.71 \pm 0.17\,\text{(stat)}\, \pm 0.18\,\text{(sys)}$~\cite{LHCb-PAPER-2017-035},
albeit with large uncertainties at present. A central difficulty in probing this mode lies in the large theoretical uncertainties for the $B_c \to J/\psi$ form-factor parameterizations. Predictions for the form factors are typically hadronic-model-dependent, making use of either perturbative QCD, the constituent-quark model, the (non)relativitistic quark model, or QCD sum rules~\cite{Anisimov:1998uk,Kiselev:1999sc,Ivanov:2000aj,Kiselev:2002vz,Hernandez:2006gt,Ivanov:2006ni,Wen-Fei:2013uea,Qiao:2012vt,Rui:2016opu}. The LHCb results have motivated several studies of the form factors~\cite{Dutta:2017xmj,Tran:2018kuv,Issadykov:2018myx,Watanabe:2017mip}. A recent, more model-independent result combines preliminary lattice QCD results with dispersive bounds and zero-recoil heavy-quark relations, leading to the prediction
$  0.20 \le R(J/\psi) \le 0.39$~\cite{Cohen:2018dgz}, 
at $95\%$ CL, implying a mild tension at the $1.3\sigma$ level with the data.

Abundant samples of $\Lambda_b$'s will be produced only at (HL-)LHC, with a production cross-section $\sigma(\Lambda_b)/\sigma(b\bar{b}) \sim 10\%$~\cite{LHCb-PAPER-2011-018}. 
From an HQET point of view, the $\Lambda_b \to \Lambda_c$ transitions are theoretically cleaner than the $B \to D^{(*)}$ decays,
because the ``brown muck'' dressing the heavy quark lies in the $s_{\ell}^{\pi_\ell}= 0^+$ ground state.
A consequence of this is a relatively simpler form-factor structure, where not only the $\mathcal{O}(\alpha_s)$
but also the $\mathcal{O}(1/m_{c,b})$ and $\mathcal{O}(\alpha_s/m_{c,b})$ subleading contributions
are fully fixed by the leading-order HQET structure, reducing the number of free parameters in the form-factor fits.
These modes are therefore promising, clean candidates for testing the behavior of the HQET expansion itself,
by e.g., assessing the impact of $\mathcal{O}(1/m_c^2)$ contributions.
Fitting the subsubleading $\mathcal{O}(\alpha_s, \alpha_s/m_{c,b}, 1/m_c^2)$ HQET structure to existing LHCb data~\cite{LHCb-PAPER-2017-016}
and lattice form factor results~\cite{Detmold:2015aaa} implies such terms are of the expected size~\cite{Bernlochner:2018kxh}. 
More $\Lambda_b \to \Lambda_c \ell \nu$ data from LHCb will improve the precision of these results, allowing access to other subleading terms.
Moreover, with precision lattice calculations of the form factors (see, e.g., Ref.~\cite{Detmold:2015aaa}), 
additional data for these modes may permit precision measurement of $|V_{cb}|$ in an environment with reduced theoretical uncertainties.

\Blu{The heavy quark expansion is also applicable to $\Lambda_b$ transitions into excited
    $\Lambda_c^* = \Lambda_c(2595)$, $\Lambda_c(2625)$ states \cite{Leibovich:1997az}.
    The complexity of the HQET description lies between that of $\Lambda_b\to\Lambda_c$
    and $B\to D^{(*)}$ transitions, with two unknown functions up to $\mathcal{O}(1/m_{c,b})$.
   It was recently demonstrated that simultaneous binned likelihood fits to the rich angular distributions
   of $\Lambda_b\to \Lambda_c^*\mu\bar\nu$ decays can determine these two functions
   and produce data-driven predictions of $R(\Lambda_c^*)$~\cite{Boer:2018vpx}. The projected
   precision due to parametric effects reaches $\sim 2\%$ for the LHCb Upgrade 1 data set.
   Due to the spin structure of the $\Lambda_c^*$ states, these decays provide a
   complementary LFU probe with a different set of systematic uncertainties
   when compared to $R(D^{(*)})$ and $R(\Lambda_c)$. More data from LHCb will similarly improve the precision of these results.}

\Blu{
Also of interest is the measurement of the inclusive process $B \to X_c \ell \nu$, where $X_c$ can be a multibody charmed state of arbitrary invariant mass, and the associated ratio $R(X_c)$. In the heavy quark limit $m_b \to \infty$, the amplitude for this process corresponds simply to that of the free quark decay $b \to c \ell \nu$. Corrections to the heavy quark limit are obtained via an OPE in terms of local heavy-quark operators: SM predictions exist at $\mathcal{O}(1/m_b^2)$ including two-loop QCD corrections~\cite{Ligeti:2014kia,Freytsis:2015qca}, yielding
\begin{equation}
    R(X_c) = 0.223 \pm 0.004\,,
\end{equation}
although recent analyses~\cite{Mannel:2017jfk,Bhattacharya:2018kig} have shown that $1/m_b^3$ corrections can decrease $R(X_c)$ by as much as 5\%. Combining this with the inclusive measurement $\BR(B \to X_c e \nu) = (10.65 \pm 0.16)\%$~\cite{Amhis:2016xyh} implies $\text{Br}(B \to X_c \tau \nu) = (2.38 \pm 0.05)\%$. This is in good agreement with the LEP measurement $\BR(b \to X \tau \nu)= (2.41 \pm 0.23) \%$~\cite{PDG2018}, where $X$ can be any multibody state. However, the present measurements for $\BR(B \to D \tau \nu) + \BR(B \to D^* \tau \nu) = (2.71 \pm 0.18)\%$~\cite{Freytsis:2015qca}, when further combined with the SM predictions for $\BR(B \to D^{**}\tau\nu)$~\cite{Bernlochner:2016bci, Bernlochner:2017jxt}, implies $\BR(B \to X_c \tau \nu)  > (2.8 \pm 0.2)\%$, already well in excess of the SM prediction for the semitauonic inclusive process. An indirect $R(X_c)$ anomaly then arises independent from the details of the form factor parameterizations involved in $R(D^{(*)})$. Given these tensions, and since measurement of the inclusive process would involve both different theory uncertainties and different systematics compared to the exclusive modes, direct measurement of $B \to X_c\tau\nu$ is of high interest in further understanding the $b \to c \tau \nu$ anomalies. 
}

\subsubsection{Models of NP for $b\to c\tau\nu$}
\label{sec7:ModelsNP:bctaunu}
\begin{figure}
\includegraphics[width=0.5\textwidth]{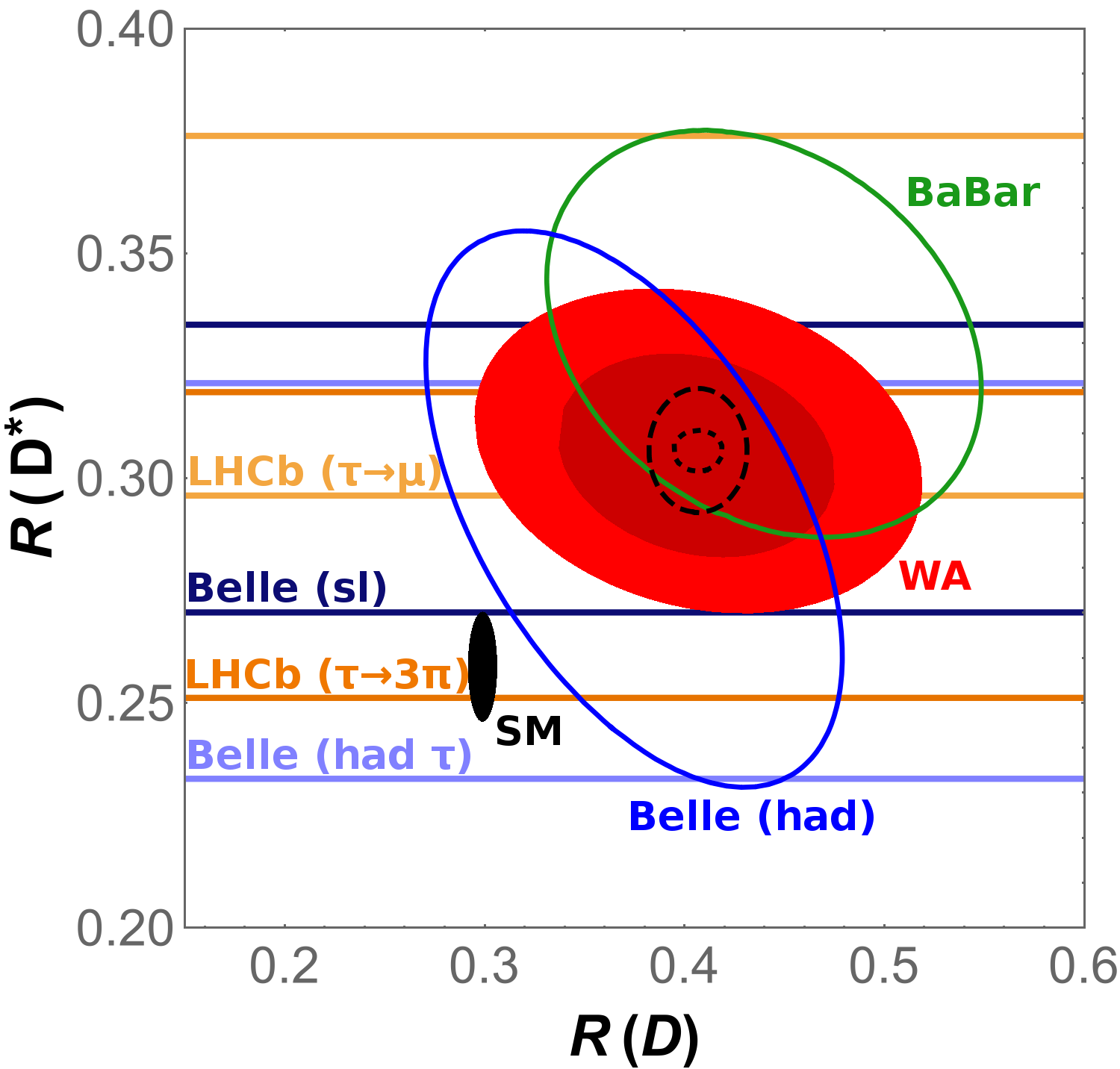}
\qquad\parbox[b]{0.5\textwidth}{\includegraphics[width=7cm,height=4cm]{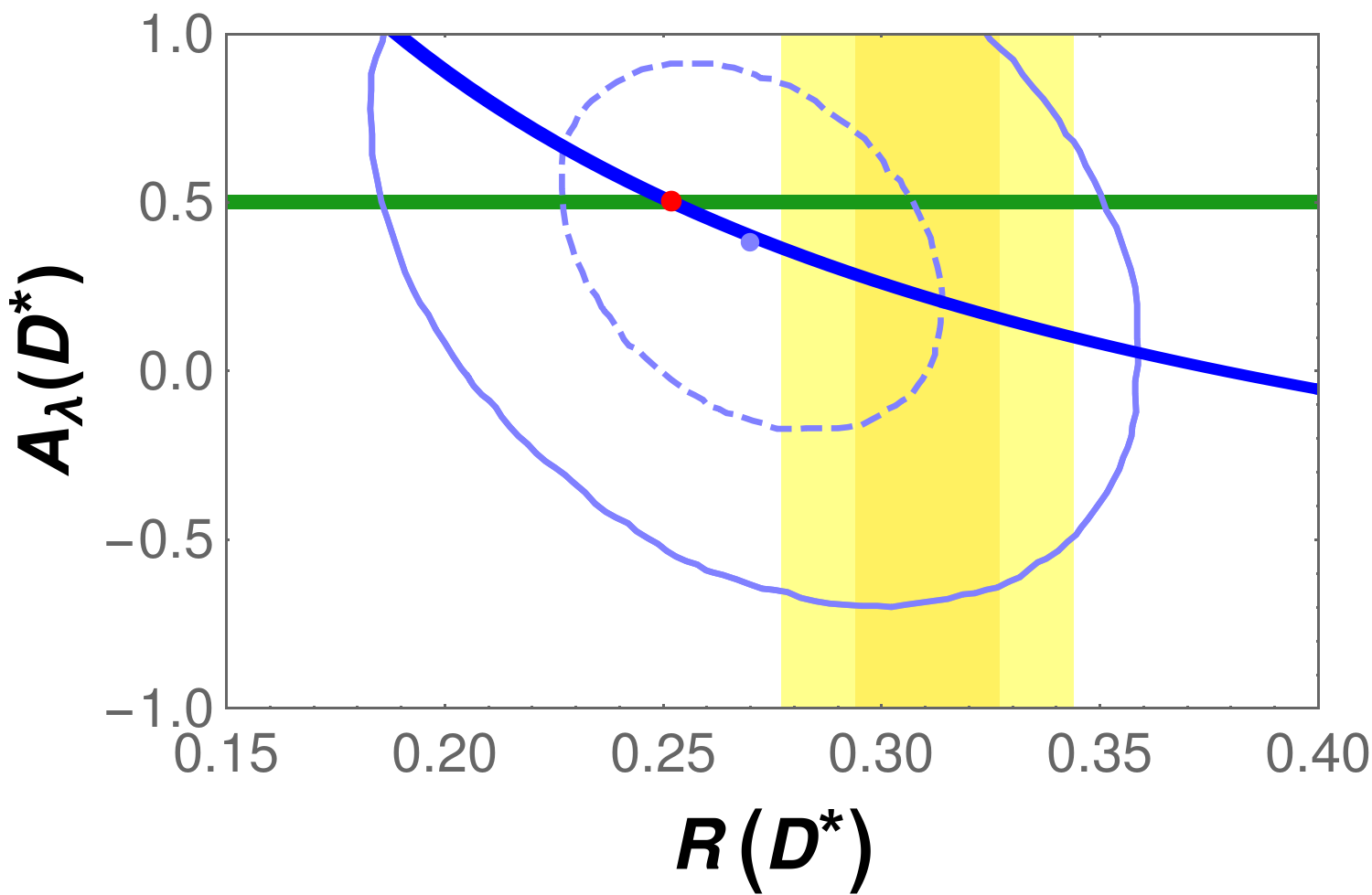}\\
\includegraphics[width=7cm,height=4cm]{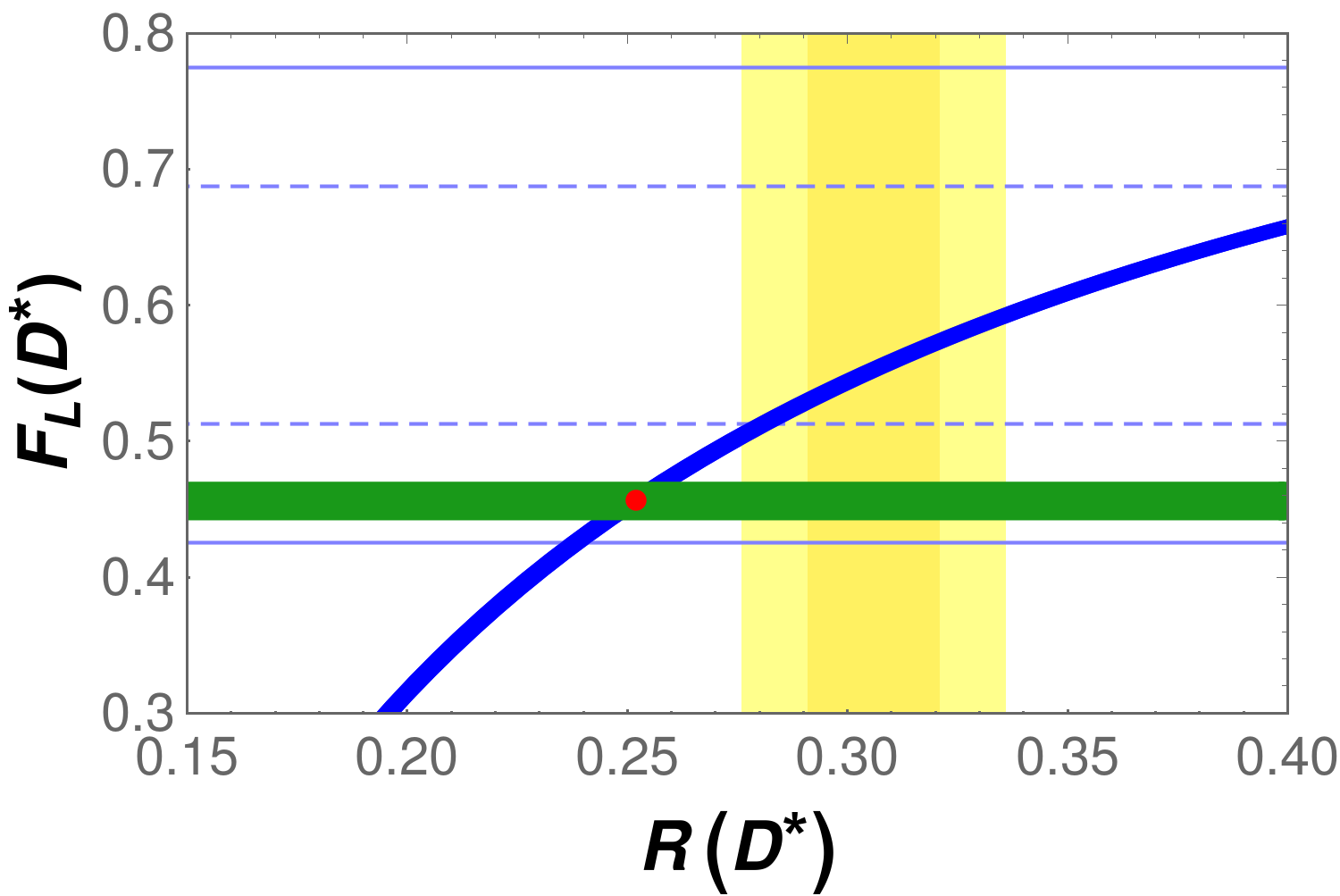}}
\caption{\label{fig:projections}
Left: Present status of the $R(D)$-$R(D^*)$ anomaly, showing the individual measurements (68\% CL contours), the world average (68\% and 95\% CL filled ellipses), the SM prediction (95\% CL filled ellipse) as well as the projections for LHCb measurements by 2025 (dashed contour), and \upgradetwo (dotted contour, both at 68\% CL, assuming the present central value).
Right: Correlations between $R(D^*)$ and the $\tau$-polarization asymmetry $A_\lambda(D^*)[=-P_\tau(D^*)]$ (upper panel) and the longitudinal fraction $F_L(D^*)$ (lower panel) for NP scenarios with only a left-handed vector coupling (green) or only scalar couplings (blue), together with the first measurements~\cite{Hirose:2016wfn,Hirose:2017dxl} (light blue) and the experimental average for $R(D^*)$ (yellow bands), excluding \cite{Hirose:2016wfn} in the upper panel. Updated and adapted from Ref.~\cite{Celis:2016azn}.}
\end{figure}

A possible NP origin of the deviations in $R(D^{(*)})$ requires a large 
contribution with respect to the SM. 
Defining $\hat{R}(X)=R(X)/R(X)|_{\rm SM}$, we have with present data $\hat{R}(D)=1.36 \pm 0.15$ and $\hat{R}(D^*)=1.19 \pm 0.06$. This points to $\gtrsim 10\%$ NP contribution to the amplitude  when the SM and NP contributions interfere, and 
$\gtrsim 40\%$ if they do not. 
\Blu{The scale of NP for the $R(D^{(*)})$ anomaly is then
\begin{equation}
 \Lambda_\text{NP} = \frac{1}{\sqrt{|V_{cb}|}} \frac{1}{\sqrt{|\Delta C_i|}} \frac{v}{\sqrt{2}} \sim \frac{1~\text{TeV}}{\sqrt{|\Delta C_i|}}~,
\end{equation}
and the contributions are expected to enter at} tree level. Possible mediators, assuming they couple only to SM fields, were classified in~\cite{Freytsis:2015qca}. If the present central values $R(D^{(*)})$ are sustained, 
this anomaly will be established by the time of the measurements at HL/HE-LHC, cf. Fig.~\ref{fig:projections} left. The focus will hence shift to model differentiation in $b\to c\tau\nu$ and the analysis of the lepton- and quark-flavor structure of the NP contributions.
A completely general NP analysis will require theoretical care. For instance, form-factor determinations from $b\to c \ell \nu$ decays could also be sensitive to NP contributions, subject to the constraint that extractions of $|V_{cb}|$ from these  decays can be made consistent with all other global data on CKM unitarity. Consequently, experimentally determined form-factor parameters may require a simultaneous fit to the deviations or additional determinations of form-factor ratios, which, however, are expected to be available at the required precision by the start of HL/HE-LHC. See Sec.~\ref{sec:lattice} for corresponding prospects from lattice QCD.

The tree-level mediators that can explain the $R(D)$-$R(D^*)$ anomaly
are a $W'$~\cite{Greljo:2015mma,Boucenna:2016wpr,Boucenna:2016qad,Megias:2017ove} generating $C_{V_L}$, a charged color-neutral scalar~\cite{Crivellin:2012ye,Celis:2012dk,Crivellin:2015hha,Celis:2016azn,Chen:2017eby,Iguro:2017ysu,Chen:2018hqy,Li:2018rax} generating $C_{S_{L,R}}$ in Eq.~(\ref{eq:Heffbcellnu}), and leptoquarks~\cite{Fajfer:2012jt,Deshpande:2012rr,Sakaki:2013bfa,Duraisamy:2014sna,Calibbi:2015kma,Fajfer:2015ycq,Barbieri:2015yvd,Alonso:2015sja,Bauer:2015knc,Das:2016vkr,Deshpand:2016cpw,Sahoo:2016pet,Dumont:2016xpj,Li:2016vvp,Becirevic:2016yqi,Barbieri:2016las,DiLuzio:2017vat,Chen:2017hir,Bordone:2017bld,Altmannshofer:2017poe,Becirevic:2018afm,Iguro:2018vqb} generating various couplings, mostly $C_{V_L}$ or $C_{S_L} \sim C_T$. For comparisons between these models,  see for instance~\cite{Tanaka:2012nw,Freytsis:2015qca,Bhattacharya:2016zcw,Ivanov:2017mrj,Alok:2017qsi,Bifani:2018zmi,Blanke:2018yud}.
Allowing for additional light particles opens up the possibility to address the anomalies with contributions involving right-handed neutrinos~\cite{He:2012zp,He:2017bft,Becirevic:2016yqi,Greljo:2018ogz,Asadi:2018wea,Robinson:2018gza,Azatov:2018kzb,Heeck:2018ntp}, since the neutrino is not detected. 

\Blu{In order to differentiate between the different combinations of NP operators that these models produce at low energies,} information beyond $R(D^{(*)})$ is needed \cite{Bhattacharya:2018kig,Bhattacharya:2015ida}. Presently, the available additional observables in $b\to c\tau\nu$ transitions are: {\em (i)} differential distributions in $q^2$~\cite{Lees:2013uzd,Huschle:2015rga} already excluding some fine-tuned scenarios despite their large uncertainties; {\em (ii)} the first measurements of the $\tau$ polarization asymmetry and the longitudinal fraction in $B\to D^*\tau\nu$~\cite{Hirose:2016wfn,Hirose:2017dxl}; {\em (iii)} the measurement of $R(J/\psi)$~\cite{LHCb-PAPER-2017-035}, which is up to $2\sigma$ above the SM prediction, as well as that of any NP model (see the previous subsection); {\em (iv)} the measurement of the inclusive rate $b\to X\tau\nu$ at \Blu{LEP~\cite{Abbaneo:2001bv,PDG2018}, yielding $\hat R(X_c)=1.01\pm0.10$}, in slight tension with the measurements for $R(D^{(*)})$~\cite{Ligeti:2014kia,Freytsis:2015qca,Celis:2016azn,Mannel:2017jfk}; {\it (v)} the (indirect) bound on $B_c \to \tau\nu$ from the $B_c$ lifetime~\cite{Li:2016vvp,Alonso:2016oyd,Akeroyd:2017mhr}, providing a strong constraint on models with only scalar couplings and disfavoring them as a solution for $R(D^*)$ for values close to the present central value.

Additional indirect constraints apply within UV-complete NP models:
high-\pt searches for signatures related to potential mediators of these transitions often provide strong constraints via, e.g., $b\bar{b} \to \tau\tau$~\cite{Faroughy:2016osc},   $b\bar{c} \to \tau\nu$~\cite{Greljo:2018tzh} or $h\to\tau\tau$~\cite{Feruglio:2018fxo}, while radiative corrections
can result in constraints from lepton universality in $\tau$ decays~\cite{Feruglio:2017rjo}, lepton-flavor violating decays~\cite{Feruglio:2017rjo}, charged-lepton magnetic moments~\cite{Feruglio:2018fxo} and electric dipole moments~\cite{Dekens:2018bci}. Current data on $\Upsilon(1S) \to \tau\tau$ decays also constrain most mediator models, and a future programme of measuring both $\Upsilon$ and $\psi$ decays can have sensitivity to all NP UV completions~\cite{Aloni:2017eny}. The current bounds on $b \to s \nu\bar{\nu}$, now only $\mathcal{O}(1)$ above the SM~\cite{Lees:2013kla,Grygier:2017tzo}, can also put severe constraints on particular NP models~\cite{Blake:2016olu,Cai:2017wry}. \Blu{Finally, it is interesting to note that contributions to $b\to s\ell\ell$ and $b\to s\gamma$ are also generically produced at loop level~\cite{Crivellin:2018yvo}.} 

At the HL/HE-LHC qualitative progress in identifying potential NP in $b\to c\tau\nu$ can be understood chiefly in terms of two classes of observables: 
\begin{itemize}
\item[$\bullet$] Precision results for $R(X)$: $R(D^{(*)})$ can establish NP and give basic model differentiation. $R(\Lambda_c^{(*)})$ is a measurement with independent systematics that improves model discrimination since it is sensitive to a different combination of NP parameters. The same applies to inclusive $R(X_c)$, to be 
measured
by Belle~II. Other modes, like $R(D_s^{(*)})$ or $R(J/\psi)$, will add to these and provide cross-checks with independent systematic uncertainties.
\item[$\bullet$] Differential-rate measurements in $B\to X\tau\nu$: Differential measurements in $q^2$ as well as in the helicity angles and the $\tau$-polarization are powerful discriminators between the SM and NP, as well as between different NP models \cite{Korner:1987kd,Hagiwara:1989gza,Tanaka:1994ay,Chen:2005gr,Chen:2006nua,Nierste:2008qe,Tanaka:2010se,Fajfer:2012vx,Datta:2012qk,Sakaki:2012ft,Celis:2012dk,Duraisamy:2013kcw,Alonso:2016gym,Ligeti:2016npd,Celis:2016azn,Ivanov:2017mrj,Alonso:2017ktd,Jung:2018lfu}.
For instance, the low-recoil region in $B\to D\tau\nu$ is very sensitive to scalar contributions, tensor contributions change the polarization in the high-recoil region in $B\to D^*$ in a unique manner, and left-handed vector contributions leave normalized quantities unchanged while affecting the total rates sizably. Examples for the discriminating power of such measurements are given in Fig.~\ref{fig:projections} (right), where correlations in NP models are shown together with the first measurements of two proposed quantities by Belle~\cite{Hirose:2016wfn,Hirose:2017dxl}. First studies regarding the reach of LHCb for such observables are presented in Sec.~\ref{sec:bclnu:exp}. 
Already semi-integrated quantities like the forward-backward asymmetry or the observables shown in Fig.~\ref{fig:projections} can be very powerful in distinguishing different NP scenarios.
\end{itemize}

At high \pt, $b c \tau \nu$ operators can also be directly probed by the $pp \to \tau X$+MET (missing transverse energy) signature at the LHC, inclusively~\cite{Greljo:2018tzh}, or with a $b$-tag in the final state, with model discrimination possible at the beyond the $3\sigma$ level in at the HL-LHC~\cite{Altmannshofer:2017yso}. This channel would allow one to probe all the NP scenarios addressing the $R(D^{(*)})$ anomalies in the HL-LHC phase (see Sec.~\ref{sec:highpT}).  Analysis of NP effects in $b \to c \tau \nu$ will be made more powerful and self-consistent by the development of dedicated NP reweighting tools such as \texttt{Hammer}~\cite{Duell:2016maj}. These tools will permit experimental collaborations to efficiently reweight their very large simulated datasets to arbitrary NP models and thus fit for WCs as part of experimental analyses. 
Observation of NP in $b\to c\tau\nu$ would warrant precise measurements at HL-/HE-LHC of related modes, 
$b\to c (e,\mu)\nu$, $b\to u\tau\nu$ and $t\to b\tau\nu$ transitions. 

\FloatBarrier

%% file: section7/experimental.tex
\subsection{Experimental perspectives}%
\label{sec:7_exp}
\subsubsection{Measurements of $B_q\to ll$ from LHCb/ATLAS/CMS}
\begin{figure}[!tb]
\centering
\hbox{\hspace{-0.2cm}}
\includegraphics[width=0.35\textwidth]{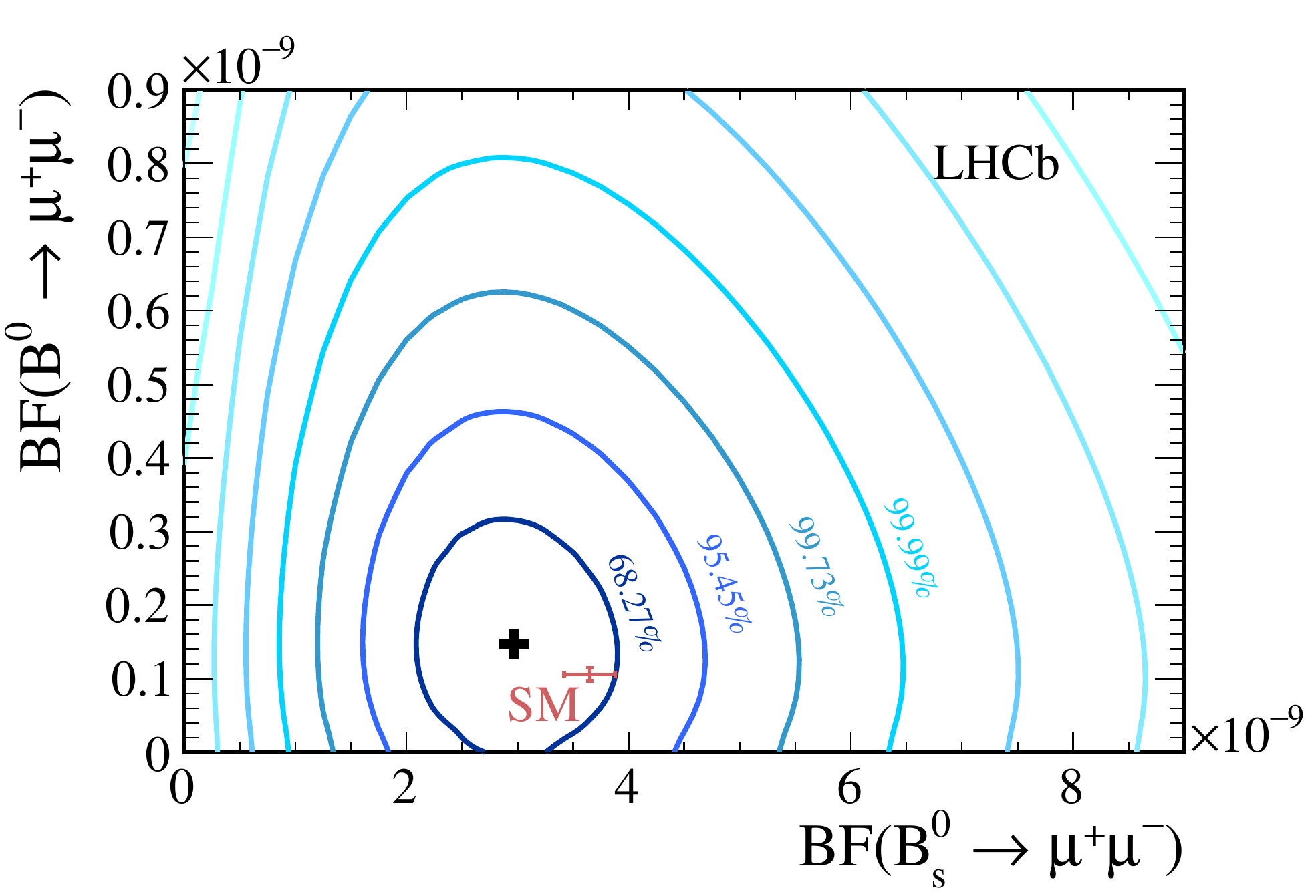}
\hbox{\hspace{-0.7cm}}
\includegraphics[width=0.315\textwidth]{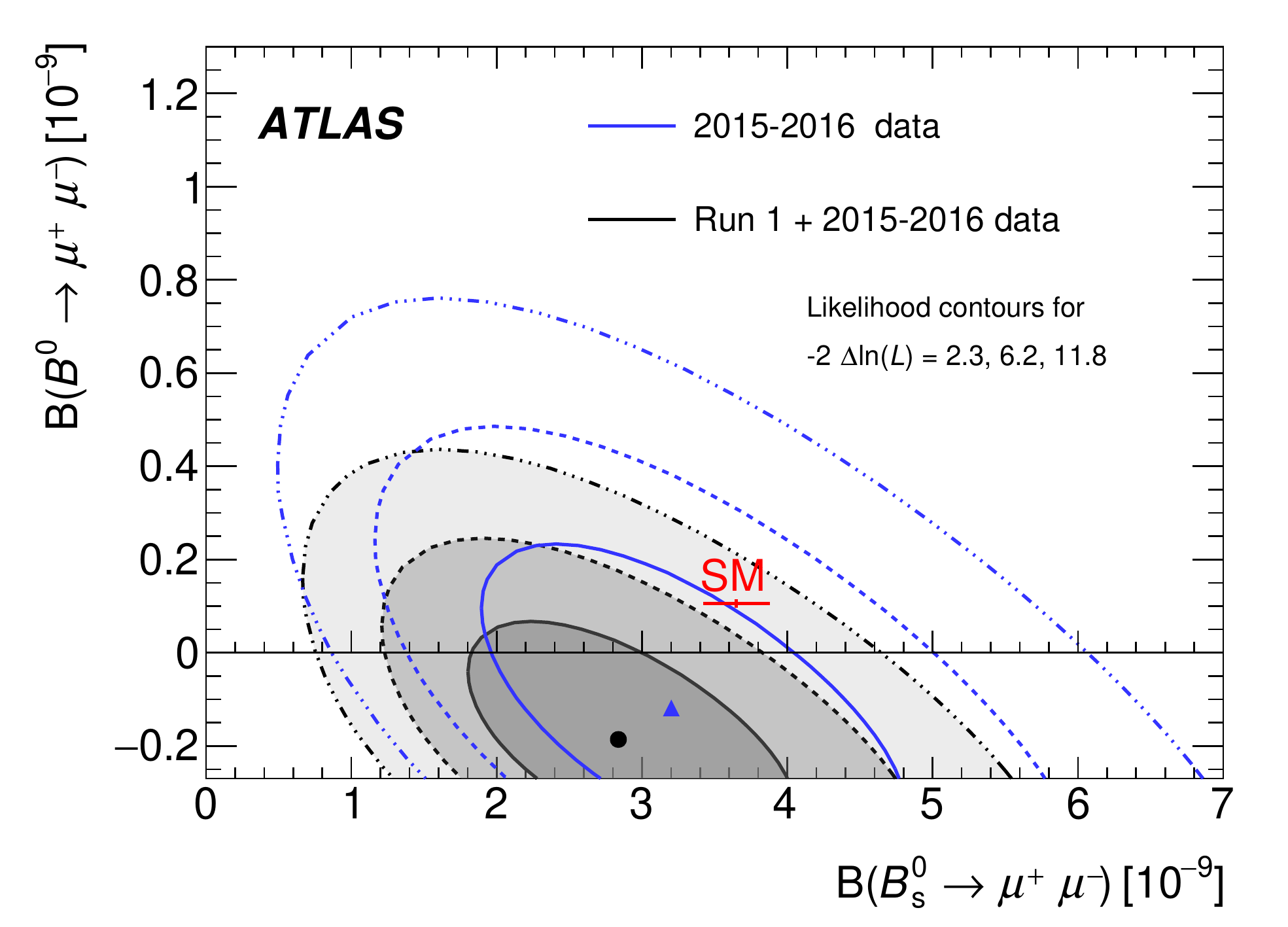}
\hbox{\hspace{-0.35cm}}
\includegraphics[width=0.375\textwidth]{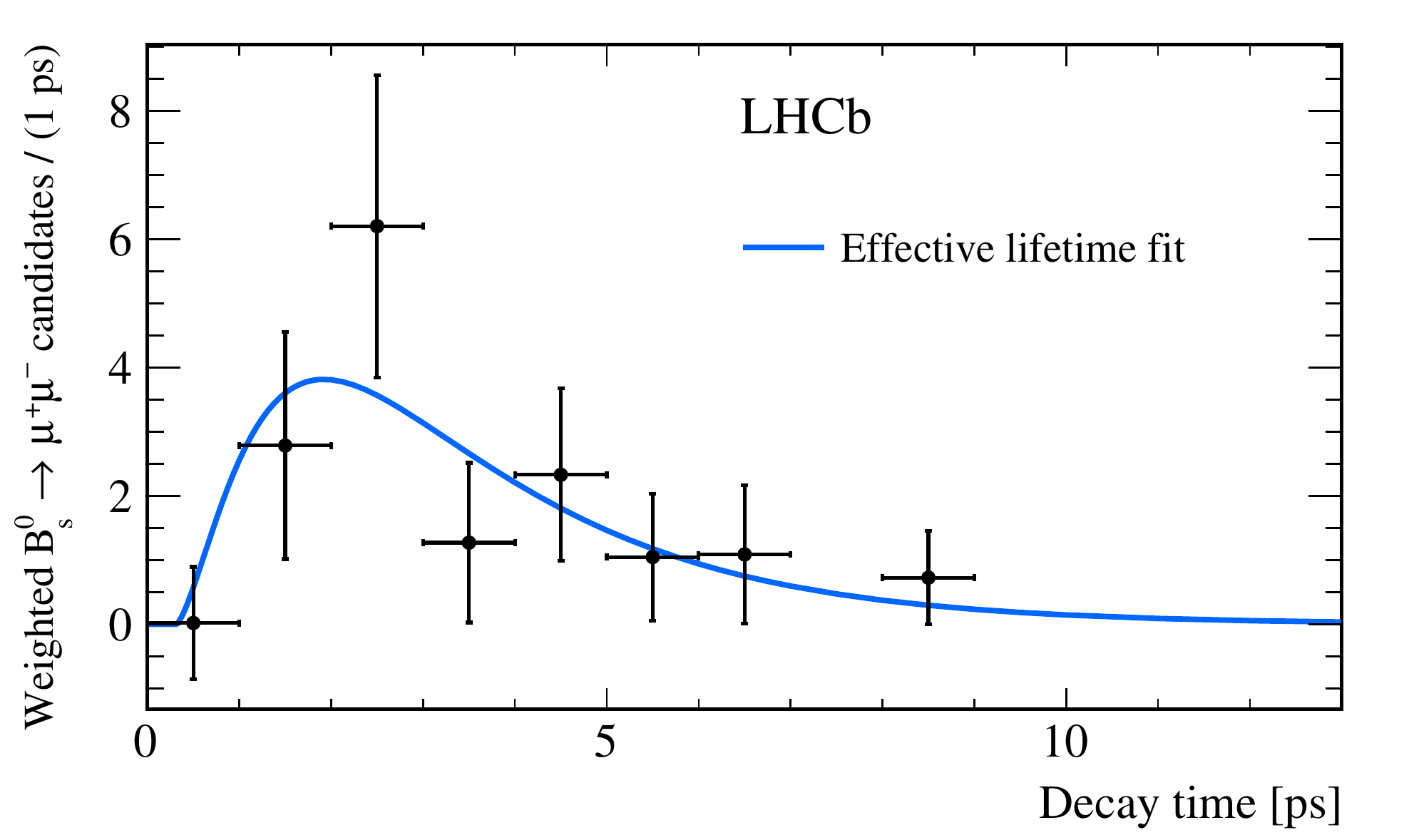}
\caption{A two-dimensional representation of the LHCb (left) and ATLAS (middle) branching fraction measurements for \Bdmm and \Bsmm. The central values are indicated with the marker. The profile likelihood contours for 1,2,3\dots Gaussian $\sigma$ are shown as blue contours (left) and grey shaded areas (middle). The red cross labelled SM reports the Standard Model predictions. Right: background-subtracted \Bsmm decay-time distribution with the LHCb fit result superimposed.}
\label{fig:btomumu}
\end{figure}

Following the observation of $\decay{\Bs}{\mumu}$ by the CMS and LHCb collaborations~\cite{CMS:2014xfa}, the most stringent constraints on $\decay{\Bs}{\mumu}$ and $\decay{\Bd}{\mumu}$ have been set by the LHCb~\cite{LHCb-PAPER-2017-001} and ATLAS~\cite{ATLAS-CONF-2018-046,Aaboud:2018mst} collaborations, cf., Fig.~\ref{fig:btomumu}.
Both sets of measurements are compatible with the SM predictions.%

\begin{figure}[t]
   \begin{center}
    \includegraphics[width=0.4\textwidth]{\main/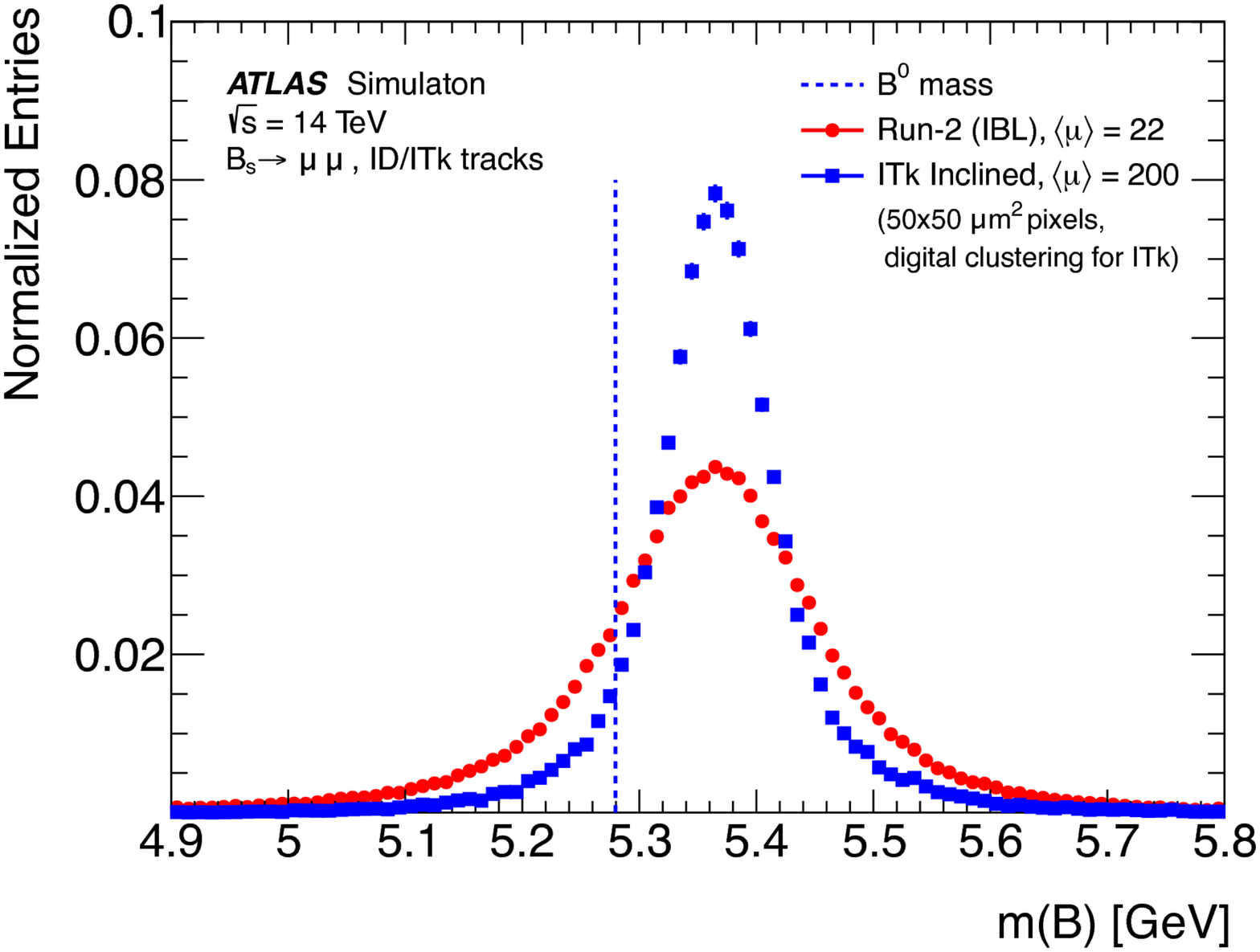}\hbox{\hspace{-0.5cm}}
     \includegraphics[width=0.3\textwidth]{\main/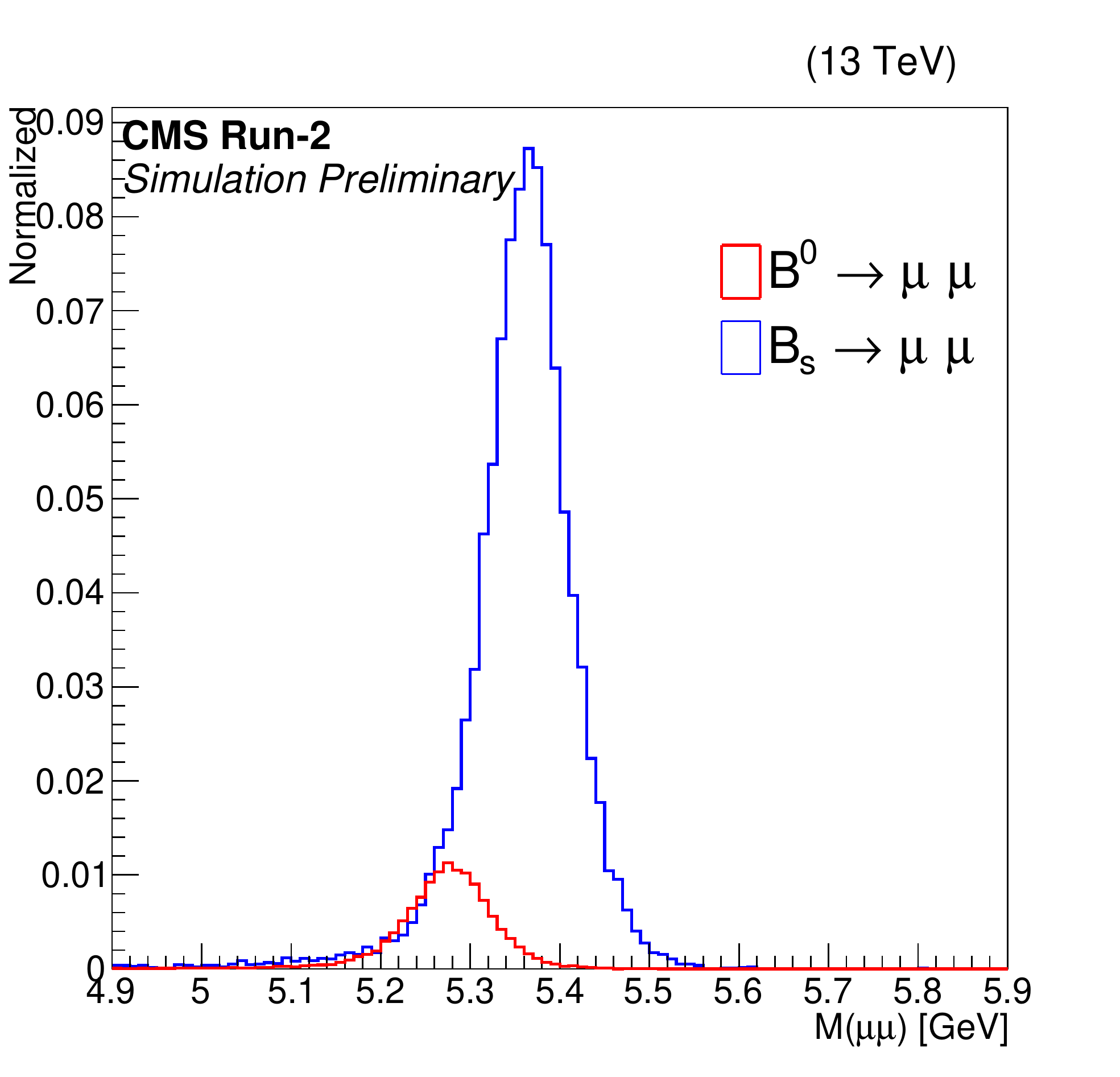}\hbox{\hspace{-0.2cm}}
     \includegraphics[width=0.3\textwidth]{\main/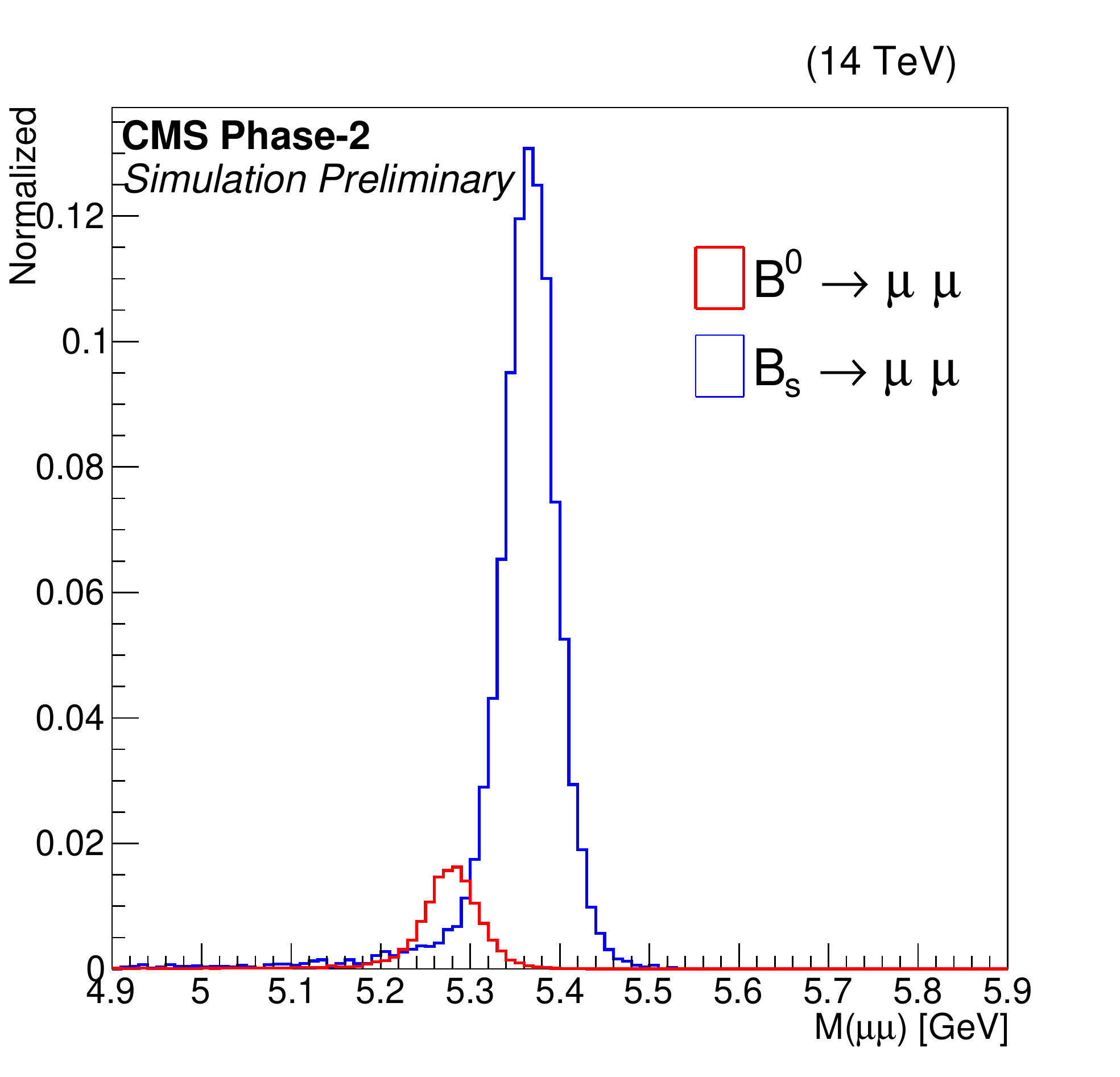}
      \caption{Left: comparison of the ATLAS invariant mass spectra for simulated $B^0_s \to \mu^+\mu^-$ events with the current (Run-2) and upgraded (ITk) detectors. A vertical line at the $B_d$ mass value is drawn to visualize the $B_s$-$B_d$ signals separation. Middle and right plots:  CMS $B^0_s$ and $B^0$ invariant mass distributions in the Run-2 and Phase-2 scenario respectively. The $B^0_s$ distribution is normalized to unity and the $B^0$  distribution is normalized according to the SM expectation.
 CMS plots are taken from \cite{CMS-PAS-FTR-18-013}.}
     \label{fig:GPDMass}\end{center}\end{figure}

HL-LHC will offer a compelling opportunity to extend the ATLAS and CMS  $B^0_s \to \mu^+\mu^-$ and  $B^0\to \mu^+\mu^-$ studies to the expected integrated high luminosity.
Flexible trigger systems and inner tracker improvements will allow both experiments to maintain efficient low-\pt dimuon triggers and achieve good mass resolution, which are the key ingredients for the $B_{s,d} \to\mu^+\mu^-$ analysis.
Fig.~\ref{fig:GPDMass} demonstrates the ATLAS and CMS $B \to \mu^+\mu^-$ invariant mass reconstruction capabilities in the HL-LHC era.
 The pseudorapidity |$\eta_f$| of the most forward muon (of the candidate) is used to visualize the CMS results and compare the performance of Phase-2 against Run-2. 
 In Fig.~\ref{fig:GPDMass} the signal mass distributions for |$\eta_f$| < 1.4 are overlaid. The CMS improved separation between $B^0 \to \mu^+\mu^-$ and $B^0_s \to \mu^+\mu^-$ in Phase-2 is evident; this will help to separate the $B^0$ signal from the tails of the $B^0_s$ signal,which now becomes a background for the $B^0$ measurement.
 
The LHCb detector is already well optimised for this decay, and planned improvements to the tracking and muon detector shielding in LHCb's upgrades will ensure that the muon reconstruction and identification performance does not degrade with increasing pileup.   %
 \setlength{\extrarowheight}{4pt}    
Both ATLAS and CMS make HL-LHC extrapolations with a PU scenario of 200 interactions per bunch-crossing, which was found not to have strong impact on the analysis performance. 
While the uncertainty on ${\cal B}(\decay{\Bd}{\mumu})$  remains statistically limited for HL-LHC projections (300\invfb/3\invab), the projected uncertainty on ${\cal B}(\decay{\Bs}{\mumu})$ depends on the assumptions made for the systematic uncertainties. 
The current systematic uncertainty is dominated by sources external to the analysis, such as the relative uncertainty associated with the $b$-quark fragmentation probability ratio, $f_s/f_d$~\cite{fsfd}, followed by the branching fractions of the normalisation modes and less significant systematics arising from internal analysis effects (e.g., $2\%$ each from particle identification and track reconstruction in the case of LHCb, individual  efficiencies in the case of ATLAS and CMS).

\begin{figure}[htb]
  \centering  
  \includegraphics[width=0.5\textwidth]{\main/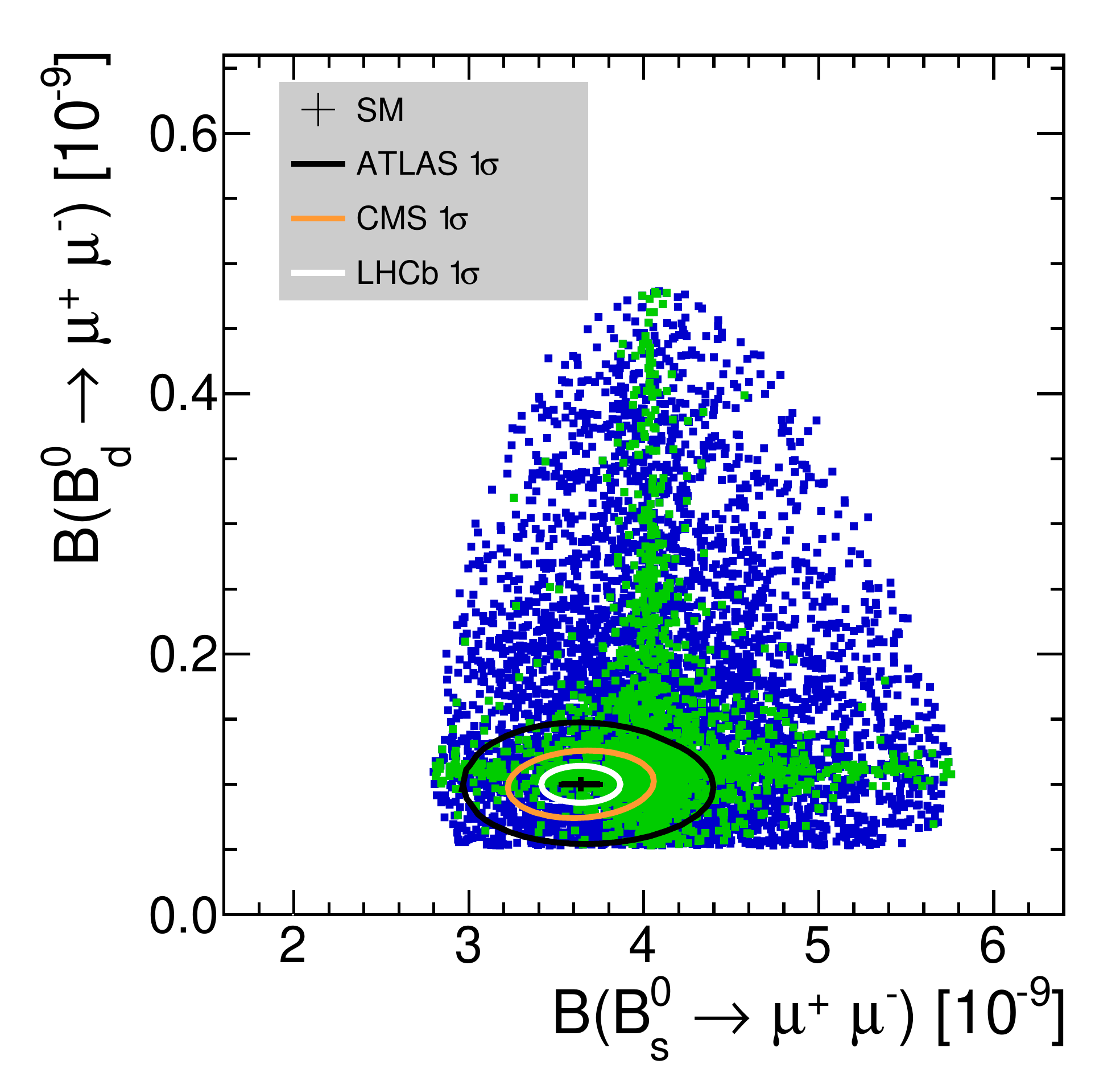}\hbox{\hspace{-0.2cm}}
  \caption{$B^0_s \to \mu^+ \mu^-$ and $B^0 \to \mu^+ \mu^-$ branching ratios as computed using new sources of flavour-changing neutral currents, as discussed in Ref.~\cite{Dutta:2015dla}. The green points are the subset consistent with other measurements. The black cross point is the SM prediction, while the coloured contours show the expected 1-sigma HL-LHC sensitivites of ATLAS, CMS, and LHCb.}
  \label{fig:HLCombinedReachFromVava}
\end{figure}

Systematic uncertainties are treated slightly differently in the projections of the three experiments.
ATLAS conservatively assumes in the HL-LHC projections that the $f_s/f_d$ and the normalization modes branching fraction uncertainties will be at the same level as previously used, i.e., $5.8\%$ and $3\%$, while 
 CMS and LHCb project them to be $3.5\%$ and $1.4\%$, respectively, based on reasonable assumptions about additional Belle~II inputs and improvements in the  knowledge of form-factor ratios and branching fraction measurements.
 A more complete discussion of sensitivity projections and systematics therein from ATLAS and CMS 
are presented in Refs
 \cite{ATL-PHYS-PUB-2018-005} and \cite{CMS-PAS-FTR-18-013}. 
For LHCb, the remaining experimental systematic uncertainties are already at the $\sim 3$~percent level, and because they rely on data-driven corrections from calibration samples they can be expected to be reduced to the $\sim 1.4$~percent level in \upgradetwo.
The resulting HL-LHC projected statistical and systematic uncertainties for the three experiments are summarized in table~\ref{LHC_Bmumu_Reach}; for the comparisons of the ATLAS, CMS and LHCb reaches (such as the one carried out in Fig.~\ref{fig:HLCombinedReachFromVava}) , the reference scenarios considered are respectively the ``3$\ab^{-1}$ Intermediate'', ``3 $\ab^{-1}$'' and ``300 $\fb^{-1}$''.

\begin{table}
  \centering
  \caption{Projected ATLAS, CMS and LHCb uncertainty on $\cal{B}$$(B^0_s \to \mu^+ \mu^-)$ and $\cal{B}$$(B^0 \to \mu^+ \mu^-)$.The HL-LHC scenario corresponds to an integrated luminosity of 300 fb$^{-1}$ for LHCb and 3 ab$^{-1}$ for ATLAS and CMS. 
    For each extrapolation the  total (statistical+systematic) uncertainties are reported. 
  }
  \label{LHC_Bmumu_Reach}
  \renewcommand{\arraystretch}{1.3}
  \begin{tabular}{ l c c c }
    \hline\hline
    & & $\BR(B^0_s \to \mu^+ \mu^-)$ & $\BR(B^0 \to \mu^+ \mu^-)$ \\
   Experiment & Scenario & stat + syst $\%$ & stat + syst $\%$ \\
    \hline
    LHCb   & 23\invfb          &   8.2  &  33       \\
    LHCb   & 300\invfb         &   4.4  &  9.4      \\
     CMS   & 300\invfb         &   12   &  46  \\
     CMS   & 3 \invab          &   7    &  16  \\
    ATLAS & Run 2              & 22.7 & 135\\
    ATLAS & 3 \invab Conservative  & 15.1 & 51\\
    ATLAS & 3 \invab Intermediate   &  12.9 & 29\\
    ATLAS & 3 \invab High-yield       & 12.6 & 26\\
    \hline\hline
  \end{tabular}
\end{table}

At the end of the Upgrade II data taking period, LHCb assumes to achieve an overall uncertainty on $\BR(\decay{\Bs}{\mumu})$ of about $4.4\%$, which
would imply an uncertainty on $\BR(\decay{\Bs}{\mumu})$ to be approximately $0.30\times 10^{-9}$ with 23\invfb and $0.16\times 10^{-9}$ with $300\invfb$.  The LHCb reach on the ratio of branching fractions ${\cal B}(\decay{\Bd}{\mumu})/{\cal B}(\decay{\Bs}{\mumu})$ is expected to remain limited by statistics and decrease from $90\%$ for the current measurement to $\sim 34\%$ with $23\invfb$ and $\sim 10\%$ with $300\invfb$. 

\begin{figure}[t]
   \begin{center}
     \includegraphics[width=0.45\textwidth]{\main/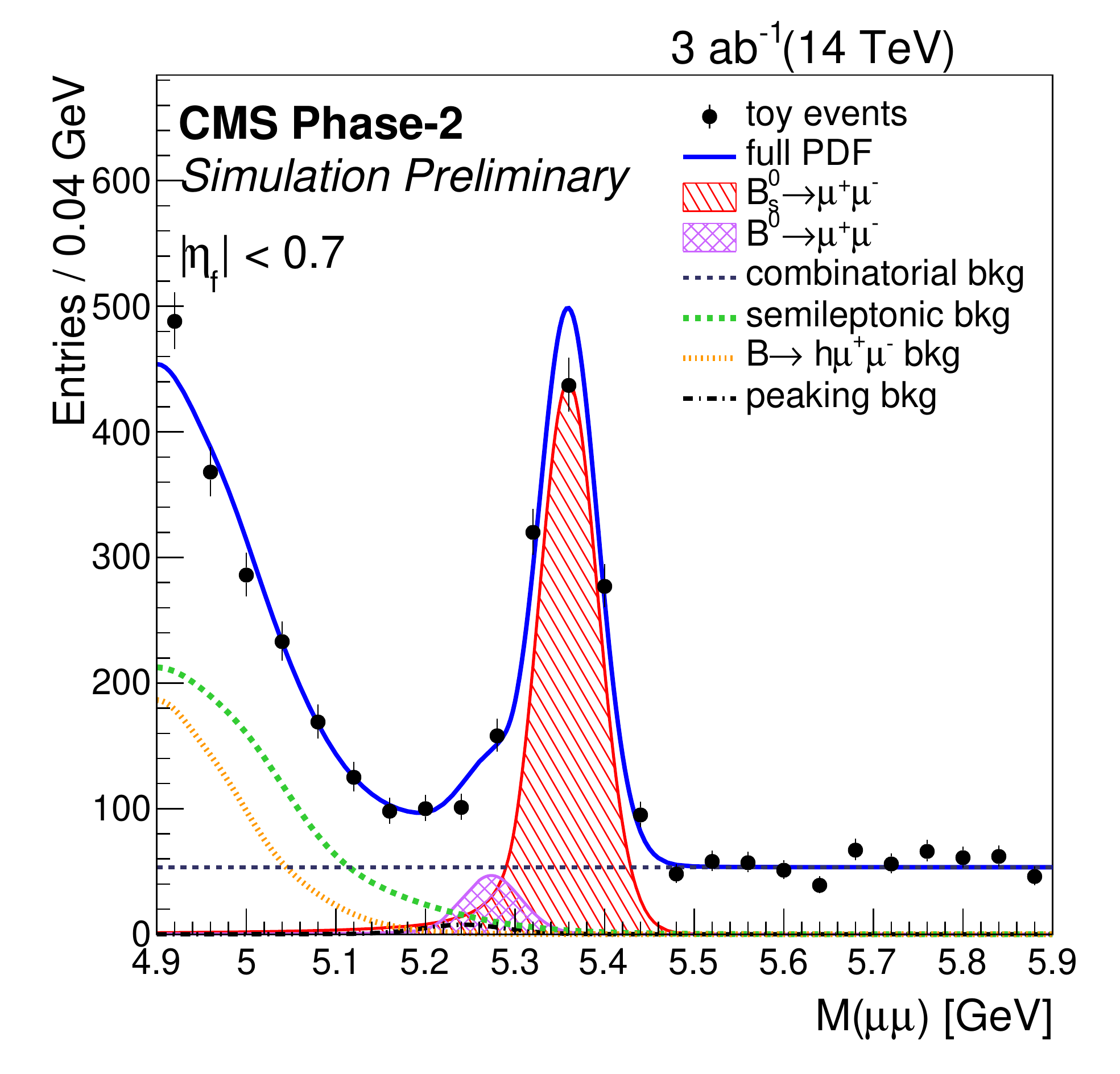}\hbox{\hspace{-0.2cm}}
     \includegraphics[width=0.45\textwidth]{\main/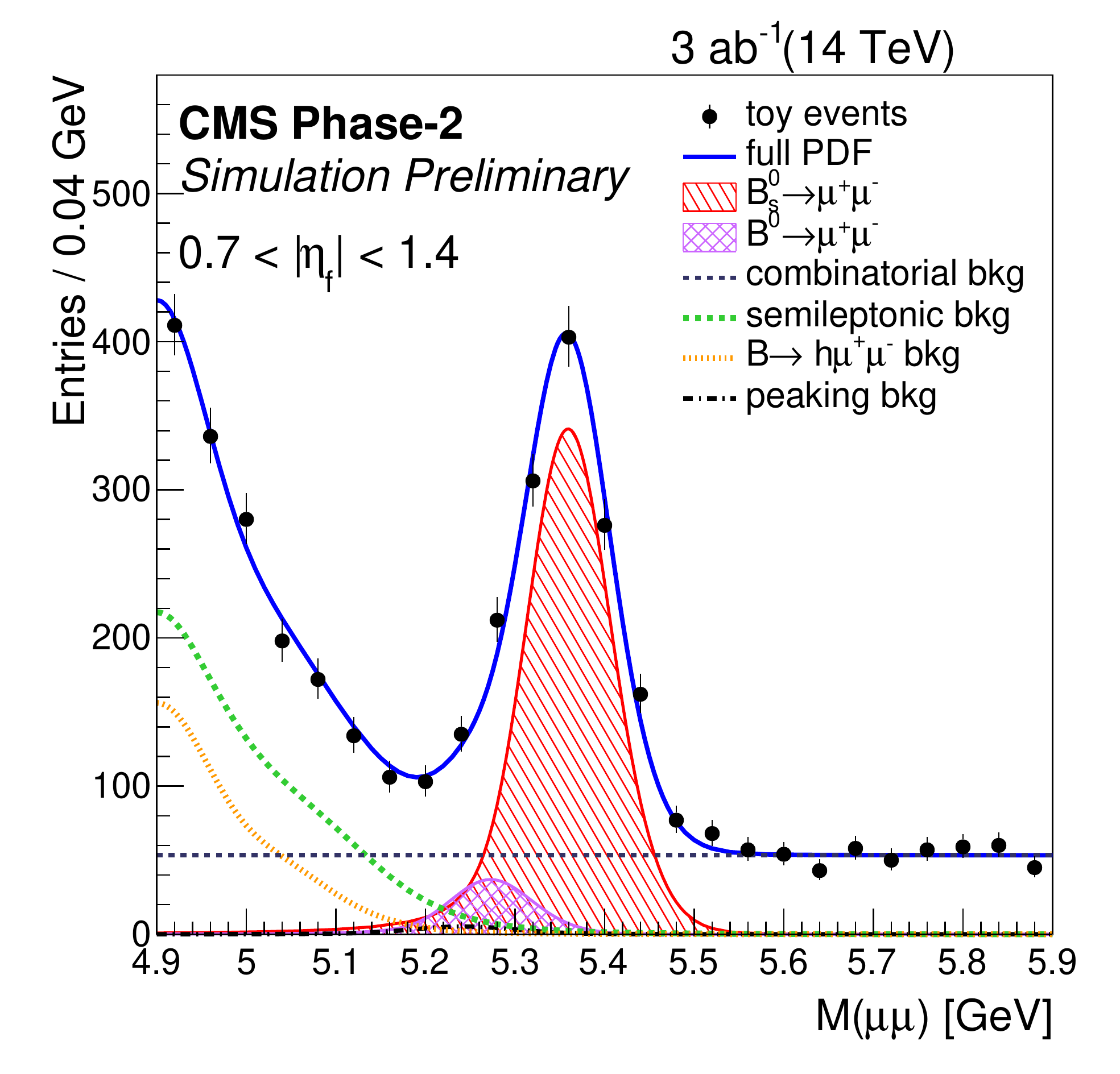}
      \caption{
      Projected dimuon invariant mass distributions with overlaid fit results for the $B_{d,s} \to \mu^+\mu^-$ analyses by CMS, with an integrated luminosity of 3000 fb$^{-1}$.  Events where the most forward muon lies in the  barrel (left) and forward (right)  regions of the detector display different mass resolutions and are categorized accordingly in the analysis. The SM relative $B^0_s$ and $B^0$ contributions are here assumed. (Plots are taken from \cite{CMS-PAS-FTR-18-013}).}
     \label{fig:CMSFitMass}\end{center}\end{figure}

The CMS projections are obtained by
 extrapolating the Run-2 analysis performances to the HL-LHC scenario. Trigger efficiencies comparable to those of Run-2, with manageable rates, are expected to be attained \cite{CMS-PAS-FTR-14-015}
 and are here assumed. The effect of the increased pileup on the signal selection efficiency was also found to be manageable. The inner tracker of the CMS HL-LHC detector is estimated to provide a 40-50\% improvement relative to Run-2 on the dimuon mass resolution. This results in an improved separation of the $B^0_s$ and $B^0$ signals, lowering the signal cross-feed contamination that is specially crucial for the $B^0$ observation, and a reduction of the level of the semileptonic background in the signal region (see Fig.~\ref{fig:GPDMass}). With the full Phase-2 integrated luminosity of 3000  fb$^{−1}$, CMS expects to measure the 
 $B^0_s \to \mu^+ \mu^-$  branching fraction at the level of 7\%.
  precision, and observe 
  the $B^0 \to \mu^+\mu^-$ decay with a significance in excess of  $5\sigma$. 
Fig.~\ref{fig:CMSFitMass} shows the invariant mass fit projections
for the $B_{s,d} \to \mu^+\mu^-$ analyses for an integrated luminosity of 3000 fb$^{-1}$. 
 \setlength{\extrarowheight}{4pt}     

\begin{figure}[t]
  \centering  
  \includegraphics[width=0.33\textwidth]{\main/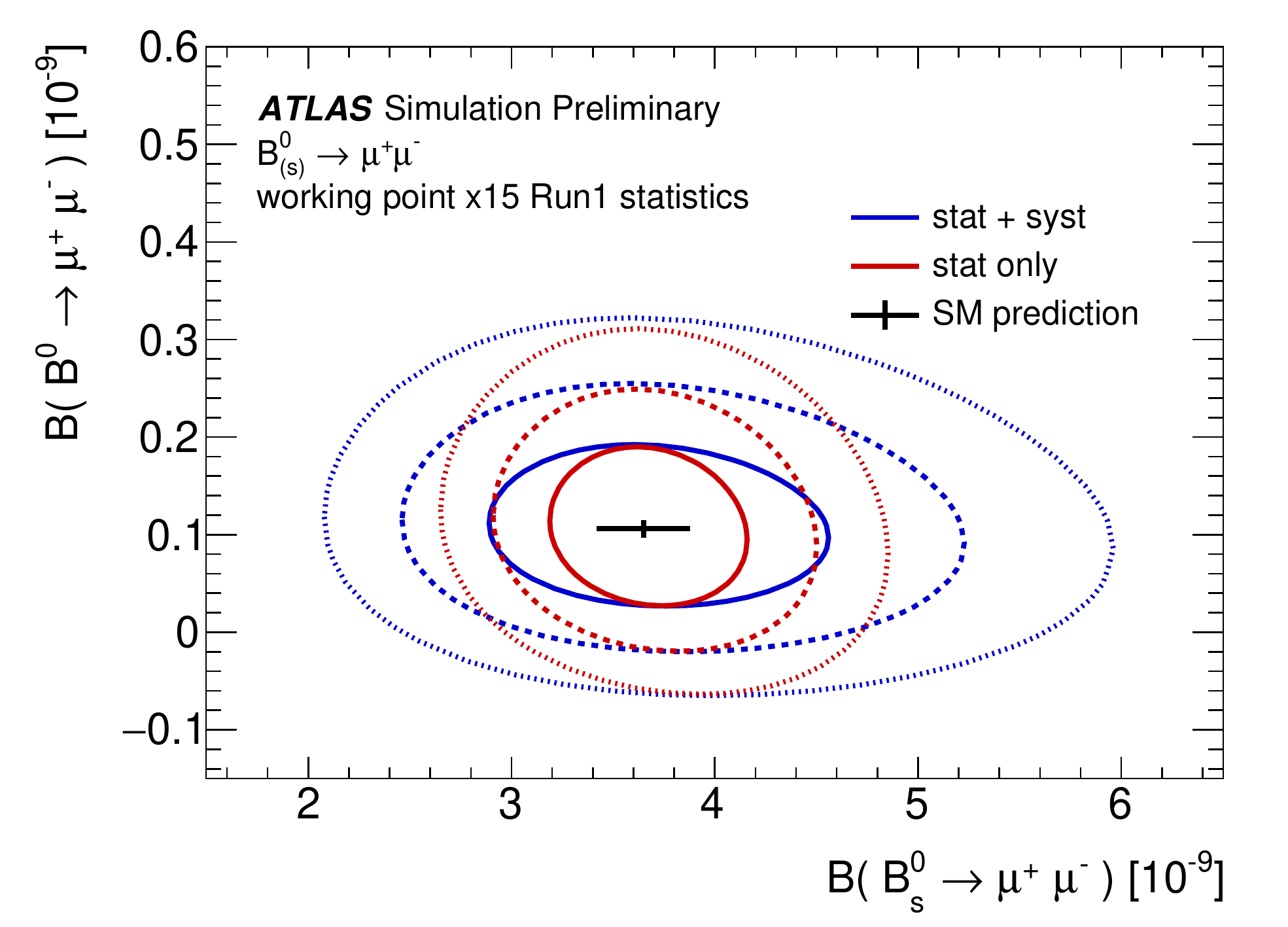}\hbox{\hspace{-0.2cm}}
  \includegraphics[width=0.33\textwidth]{\main/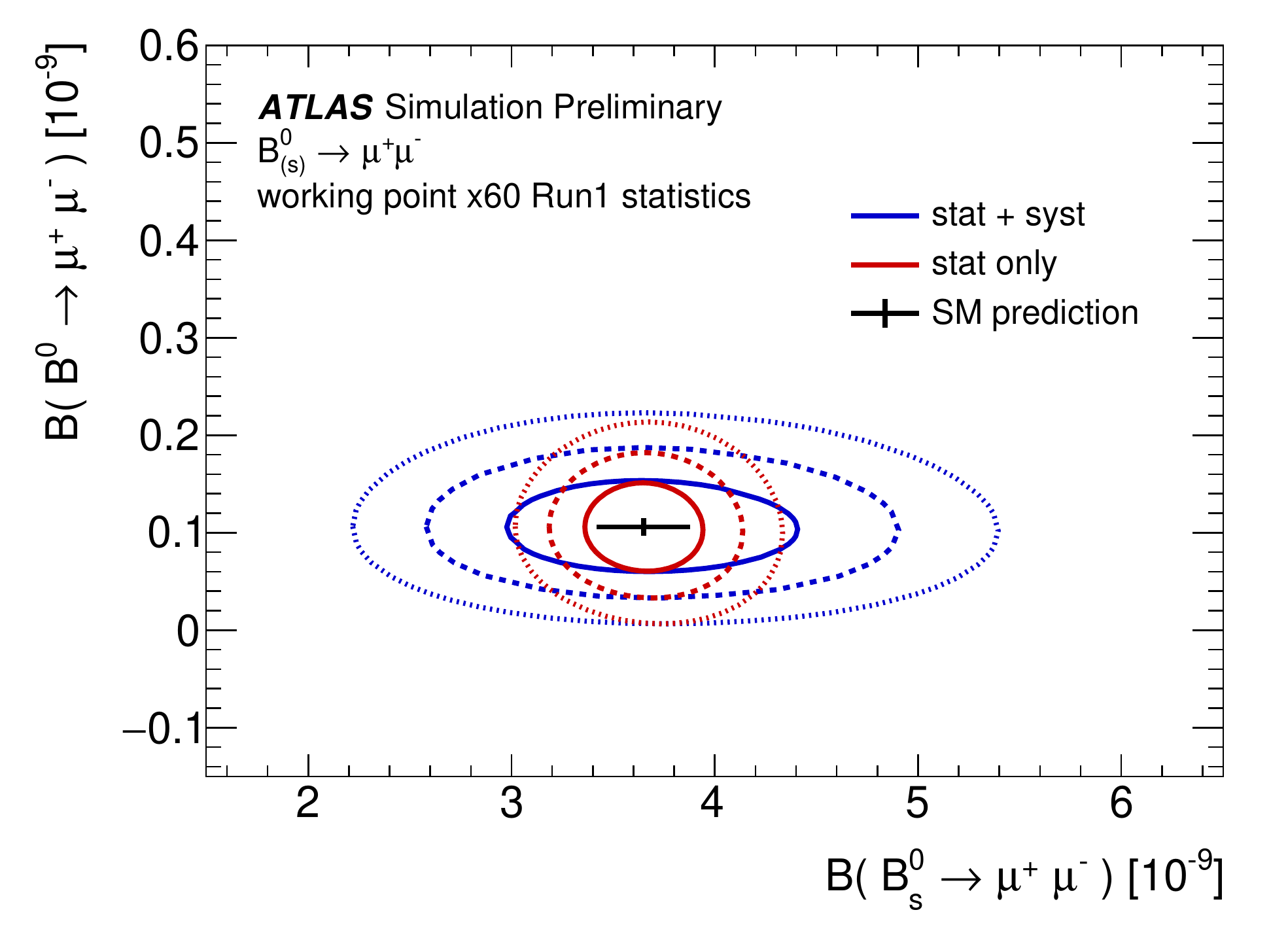}\hbox{\hspace{-0.2cm}}
  \includegraphics[width=0.33\textwidth]{\main/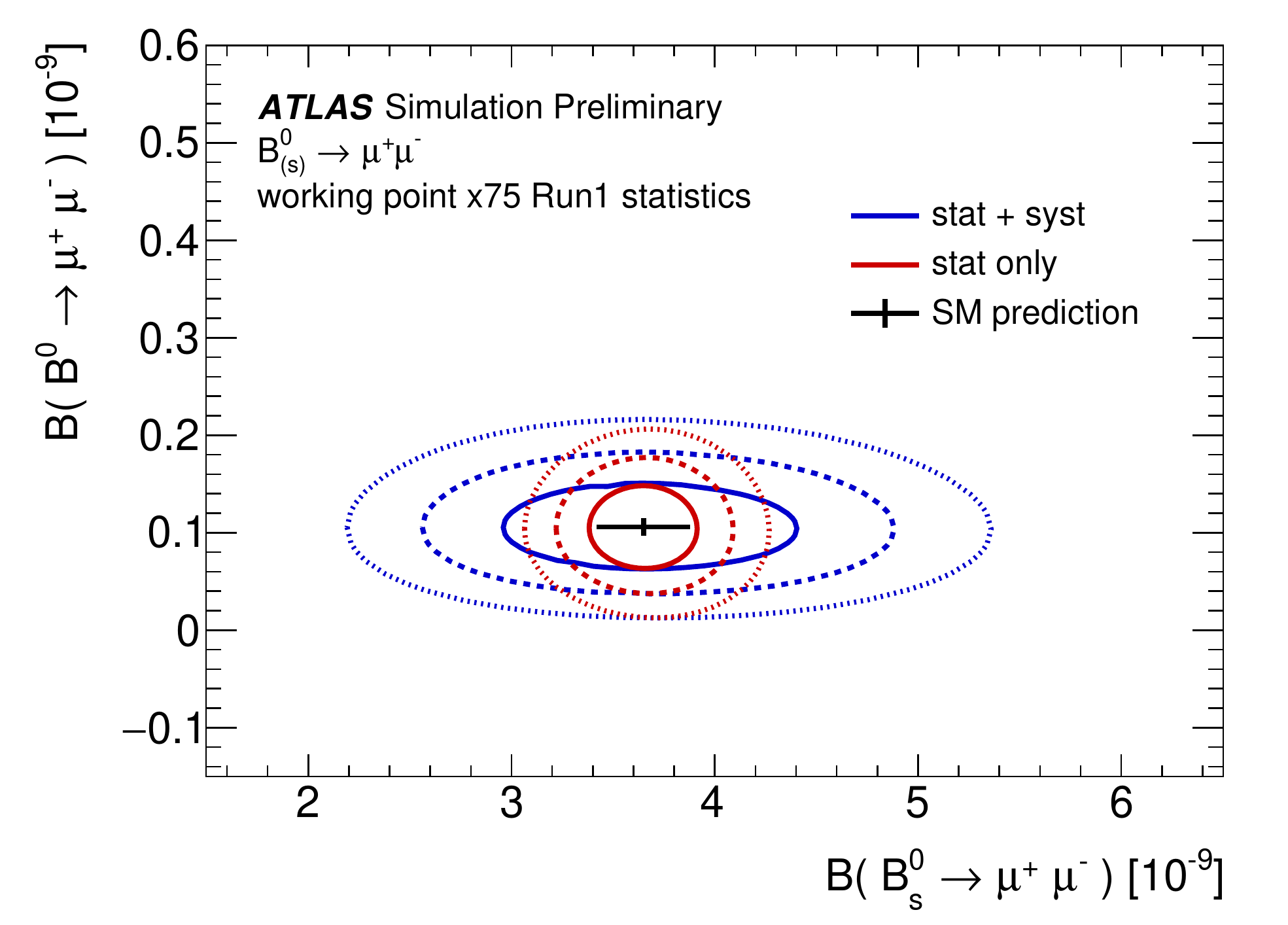}
  \caption{ATLAS projected 68.3\% (solid), 95.5\% (dashed) and 99.7\% (dotted) confidence level profiled likelihood ratio contours for the ``conservative'' (top), ``intermediate'' (middle) and ``high-yield'' HL-LHC extrapolations.
    Red contours do not include the systematic uncertainties, which are then included
    in the blue ellipsoids. The black point shows the SM theoretical prediction and its uncertainty \cite{Bobeth}.}
  \label{fig:ATLAS_HLLHCextrapolation}
\end{figure}

The ATLAS projections  \cite{ATL-PHYS-PUB-2018-005} are extrapolated from the ATLAS Run~1 analysis \cite{Aaboud:2016ire}. Assumptions include the training of a multivariate classifier capable of similar background rejection and signal purities and an analysis selection with comparable pile-up immunity as Run~1. The study takes into account the scaling of $B$ production cross-section and integrated luminosity relative to Run~1, and explores different triggering scenarios corresponding to different dimuon transverse momentum thresholds, ($p_T^{\mu_1}$,$p_T^{\mu_2}$): $(6 \,\text{GeV},6 \,\text{GeV})$, $(6 \, \text{GeV},10 \,\text{GeV})$ and $(10 \,\text{GeV},10 \,\text{GeV})$. For each of these scenarios the sensitivity is categorized on the basis of the signal statistics expected relative
to the Run~1 analysis (x15, x60 and x75 respectively, in $3\ab^{-1}$ of HL-LHC ATLAS data), yielding the projected $68.3\%$, $95.5\%$ and $99.7\%$ likelihood contours in Fig.~\ref{fig:ATLAS_HLLHCextrapolation}.

Fig.~\ref{fig:HLCombinedReachFromVava} compares the projected experimental sensitivities of ATLAS, CMS, and LHCb with the BR predictions from a particular class of BSM models~\cite{Dutta:2015dla}.   
All estimates use the quoted SM predictions as central values for the branching fractions. The estimated experimental sensitivity at HL-LHC is close to the uncertainty of the current SM prediction from theory, which is dominated by the uncertainty on the $\Bs$ decay constant, $f_{B_s}$, determined from lattice QCD calculations, and the CKM matrix elements. Both are expected to improve in precision in the future. The power of the HL-LHC data set to discriminate not only between the SM and BSM models, but also within the parameter space of those BSM models, is clear. %

With the HL-LHC data set, precise measurements of additional observables are possible, namely the effective lifetime, $\tau_{\mu\mu}^{\rm eff}$, and the time-dependent \CP\ asymmetry of \Bsmm decays. Both quantities are sensitive to possible new contributions from the scalar and pseudo-scalar sector in a way complementary to the branching ratio measurement\cite{DeBruyn:2012wk}. The effective lifetime is related to the mean \Bs lifetime $\tau_{B_s}$ through the relation 
\begin{equation}
\tau_{\mu\mu}^{\rm eff}=\frac{\tau_{B_s}}{1-y^2_s}\frac{1+2A^{\mu\mu}_{\Delta\Gamma}y_s+y^2_s}{1+A^{\mu\mu}_{\Delta\Gamma}y_s}\,, 
\end{equation}
where $y_s=\tau_{B_s}\Delta\Gamma_s/2$, and $\Delta\Gamma_s=\Gamma_{B^0_{sL}}-\Gamma_{B^0_{sH}}$. In the SM,  $A^{\mu\mu}_{\Delta\Gamma}=1$, with only the heavy mass eigenstate decaying to \mumu. In BSM scenarios it
 can take any value between $-1$ and 1. LHCb has performed the first measurement of the \Bsmm effective lifetime using a dataset of 4.4\invfb, resulting in $\tau_{\mu\mu}^{\rm eff}=2.04\pm 0.44\pm 0.05 \ps$~\cite{LHCb-PAPER-2017-001} (Fig.~\ref{fig:btomumu}, right). The relative uncertainty on $\tau_{\mu\mu}^{\rm eff}$ is expected to decrease to approximately $8\%$  with 23\invfb and $2\%$ with 300\invfb, being statistically limited. 

\begin{figure}[t]
\begin{center}
\includegraphics[width=0.4\textwidth]{\main/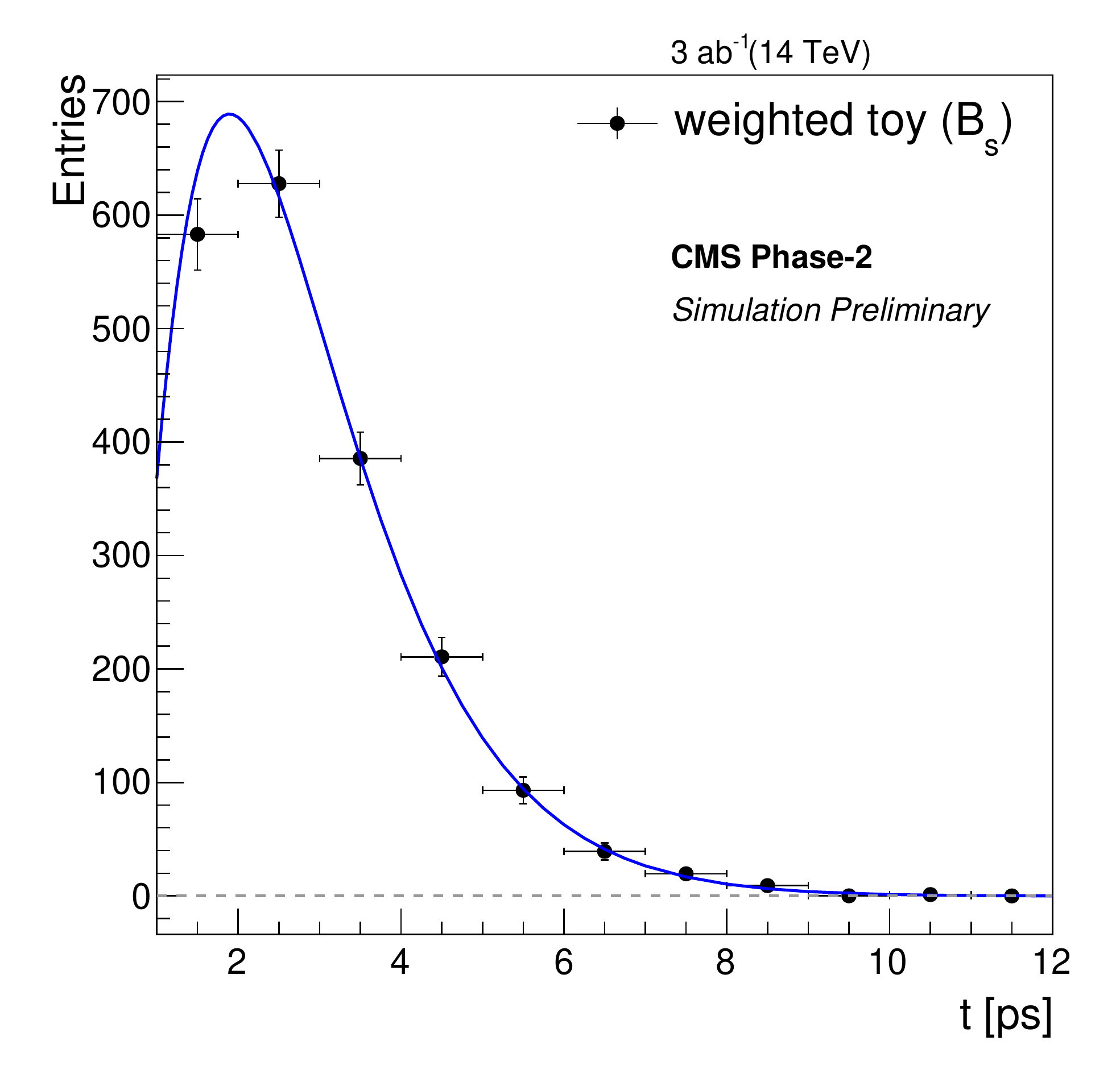}
\caption{Projection of a  background-subtracted proper decay time distribution, with fit result overlaid, with the CMS experiment and 3000 fb$^{-1}$. The estimated uncertainty in the $B^0_s \to \mu ^+ \mu^-$ effective lifetime is obtained by fitting an ensemble of corresponding pseudo-experiments. (The plot is taken from \cite{CMS-PAS-FTR-18-013}).}
 \label{fig:CMS_lifetime}\end{center}\end{figure}

The CMS sensitivity for a measurement of the
 $B^0_s \to \mu ^+ \mu^-$ effective 
lifetime is estimated using an ensemble of  pseudo-experiments generated with parameters reflecting the projected Phase-2 conditions. 
The signal lifetime distribution for each pseudo-experiment is obtained using the sPlot technique\cite{splot} to separate out the background, and then fitted with a model consisting of an exponential function, convolved with a Gaussian function that describes the expected decay time resolution, and multiplied by an efficiency function that accounts for reconstruction effects. The outcome of such a pseudo-experiment is shown in Fig.~\ref{fig:CMS_lifetime}.
The effective lifetime is expected to be measured with a statistical precision of 3\% at 3000 fb$^{-1}$.

\setlength{\extrarowheight}{4pt}     

While the current experimental uncertainty is larger than for $\tau_{B^0_{sH}}-\tau_{B^0_{sL}}$,
a $2-3\%$ uncertainty on $\tau_{\mu\mu}^{\rm eff}$ would allow to set stringent constraints on $A^{\mu\mu}_{\Delta\Gamma}$ and in particular would allow to break the degeneracy between any possible contribution from new scalar and pseudoscalar mediators.

Assuming a tagging power of about $3.7\%$~\cite{LHCb-PAPER-2014-059}, a dataset of 300\invfb allows LHCb to reconstruct a pure sample of more than 100 flavour-tagged \Bsmm decays (effective yield) and measure their time-dependent \CP asymmetry. From the relation  
\begin{equation}
\frac{\Gamma(B^0_s(t)\to\mu^+\mu^-)-\Gamma(\bar{B}^0_s\to\mu^+\mu^-)}{\Gamma(B^0_s(t)\to\mu^+\mu^-)+\Gamma(\bar{B}^0_s\to\mu^+\mu^-)}=\frac{S_{\mu\mu}\sin(\Delta m_st)}{\cosh(y_st/\tau_{B_s})+A^{\mu\mu}_{\Delta\Gamma}\sinh(y_st/\tau_{B_s})}\,, 
\end{equation}
where $t$ is the signal proper time and $\Delta m_s$ is the mass difference of the heavy and light \Bs mass eigenstates, $S_{\mu\mu}$ can be measured with an uncertainty of about 0.2. The signal yield expected in a 23\invfb dataset, on the other hand,  is too low to allow a meaningful constraint to be set on $S_{\mu\mu}$. A nonzero value for $S_{\mu\mu}$ would automatically indicate evidence of \CP-violating phases beyond the SM.

Being sensitive to a wider set of effective operators ($O_7$, $O_9$ and $O_{10}$)~\cite{Kruger:2002gf}, the \bsmumugamma decay offers 
an interesting counterpart to \bsmumu. 
The theoretical branching fraction is expected to be one order of magnitude 
larger than the \bsmumu one~\cite{Melikhov:2004mk}, owing to the removal of the helicity suppression, 
when integrated over the full $q^2$ spectrum. 
However, the presence of the photon makes the direct reconstruction 
challenging at LHCb.
No limit exist today on the \bsmumugamma channel, 
while the \bdmumugamma is limited at $1\times 10^{-7}$ at 90\% CL by the 
\babar experiment~\cite{Aubert:2007up}.

Given the experimental difficulty, two complementary techniques are employed  
for the study of the \bsmumugamma decay at LHCb. The first is a full reconstruction
which is more sensitive at low and mid $q^2$, where the photon energy is higher, 
and the second, recently proposed in Ref.~\cite{Dettori:2016zff}, without photon reconstruction
but only sensitive at high $q^2$. 

The only non-negligible partially reconstructed background is the not yet 
measured $B_d \to \mu^{+} \mu^{-} \pi^{0} \xspace$, whose branching fraction is 
theoretically estimated to be of the same order of magnitude as the signal.
The main difficulty of the measurement is therefore the combinatorial 
background, because the uncertainty on the photon momentum enlarges the signal 
width and blurs its kinematics.
Based on current reconstruction efficiencies, the expected sensitivity at the 
end of Run~3 (Upgrade II) is $\sim 9 \sigma$ ($\sim 22 \sigma$).
The use of $B_s \to J/\psi \eta \xspace$ and $B_d \to K\pi\gamma \xspace$ as 
normalisation channels reduces the systematic uncertainties due to the selection.

The partially reconstructed method consists of studying the \bsmumugamma decay 
as a shoulder on the left of the \bsmumu peak in the dimuon mass distribution. 
The SM contribution as background has been considered negligible so far.
Conversely, large branching fractions could be easily excluded~\cite{Dettori:2016zff}
when considering this as an additional component. 
The SM branching fraction for this region would be around $2 \times 10^{-10}$, 
implying a first observation would be possible with Run~3 and certainly with Run~4 data, 
while extremely tight limits could already be determined with Run~2. 

\subsubsection{Measurements of $b\to s\ell\ell$ from LHCb/ATLAS/CMS}

\subsubsubsection{Yield and systematics evolution}

With the large data set that will be collected at the end of Run~5,   it will
be possible to make a precise determination of the angular observables in narrow
bins of \qsq or using a \qsq-unbinned approach~\cite{Hurth:2017sqw,Chrzaszcz:2018yza}; LHCb foresees to achieve around 440\,000 fully reconstructed \decay{\Bz}{\Kstarz\mumu} decays, CMS around 700\,000 excluding the $q^2$ range 
 overlapping with the resonant decays \cite{CMS-PAS-FTR-18-033}. 
 
 The current ATLAS, CMS and LHCb measurements are statistically limited.
 For LHCb 
 the major systematics are those which affect the calculation of the angular acceptance. Most of the systematic uncertainties are expected to reduce significantly with more integrated luminosity due to larger control samples. Both the LHCb statistical and systematic uncertainties are therefore scaled with integrated luminosity when obtaining the projections. In this context it is interesting to note that \upgradetwo will provide signal yields that are of the order of the current tree-level control modes $B^0\to \jpsi \Kstarz$ and $B^0_s\to \jpsi\phi$. %
 In the most recent measurement of $\phis$ from $B^0_s\to \jpsi\phi$, the systematic uncertainties on the decay amplitudes due to modeling of the angular acceptance were already at the 0.001-0.002 level. Therefore, even without considering further improvements from the larger control samples available in \upgradetwo, these uncertainties will not systematically limit the \upgradetwo analysis.
 
 The CMS sensitivity for the measurement of the $P_5'$ parameter at HL-LHC is extrapolated \cite{CMS-PAS-FTR-18-033} 
  from the CMS Run-1 results \cite{Sirunyan:2017dhj} under some assumptions: effects of improvements in the analysis strategy (e.g., different selection criteria or fits) are not considered and the trigger thresholds and efficiency are assumed to remain the same. The latter is likely to be a conservative assumption since the availability of tracking information at the first level trigger may result in a higher efficiency than in Run-1. The extrapolation method also assumes that the signal-to-background is the same; indeed,
  the main 
  source of background is from other b-hadron decays, whose cross-section scales the same as the signal.
 Samples of simulated events are used to evaluate three relevant aspects of the analysis, namely mass resolution, CP mistagged rate, and the effect of pileup, to justify the extrapolation method; no degradation in the projected analysis performance was found.
  For each $q^2$ bin, the expected $B^0 \to K^{*0}\mu^+\mu^-$ signal yields are obtained from a sample of simulated signal events generated with the Phase-2 conditions, including an average of 200 pileup, and scaled to the integrated luminosity of 3000 fb $^{-1}$.

The estimated  statistical uncertainty on the ${P'_5}$  parameter at 3000 fb$^{-1}$ is obtained by scaling the statistical uncertainty measured in Run-1 
by the square root of the ratio between the yields observed in Run-1 and the Phase-2 simulation.
 The evolution of the systematic uncertainties is also extrapolated from the Run-1 analysis.
 Improved understanding of theory and the experimental apparatus is expected to reflect in a factor of 2 reduction in many uncertainties in the Phase-2 scenario. 
The uncertainties which depend on the available amount of data
are scaled the same as the statistical uncertainties. 
The uncertainty related to the limited number of simulation events is neglected, under the assumption that sufficiently large simulation samples will be available by the time the HL-LHC becomes operational.\\

The ATLAS study extrapolates the signal and background yields based on the Run 1 analysis data observations \cite{Aaboud:2018krd}, accounting for 3000 fb$^{-1}$ of integrated luminosity and a $\times 1.7$ increase in the $b$ production cross-section. Monte-Carlo simulations are employed to estimate the experimental precision achievable, including the improved 4-prong invariant mass resolution projected with the ATLAS upgraded tracking system \cite{ATLAS:2285585}.
Particular care is taken in applying likely trigger selections, emulating trigger thresholds and accounting for the $q^2$-dependency of the trigger efficiency. 
Three different sets of di-muon trigger threshold scenarios are considered: two 6 GeV muons, the combinaton of 6 GeV and a 10 GeV muon, or two muons of 10 GeV.
Although not dominant in the result, 
systematic uncertainties are carefully extrapolated as well: signal and background fit model accuracies are expected to improve in proportion to the available data statistics, detector acceptance and mistagging accuracy are driven by MC statistics and thus assumed to produce negligible effects, the inclusion of S-wave contributions to the data fit is extrapolated to improve the effect of this systematic by a factor 5, and detector alignment and B-field systematic uncertainties are expected to improve approximately by a factor 4 with larger calibration samples and the use of new techniques \cite{ATL-PHYS-PUB-2018-003}. 

 The precision of the future 
 $b\to s\ell\ell$ branching fraction measurements from the LHC experiments will be limited also by the knowledge of the \decay{\B}{\jpsi X} decay modes that are used to normalise the observed signals. 
The knowledge of these branching fractions will be improved by the \belletwo collaboration but will inevitably limit the precision of the absolute branching fractions of rare $\bquark\to\squark \ellell$ processes. 
The comparison between the predicted and measured branching fractions will in any case be limited by the theoretical knowledge of the form factors.
 
 \subsubsubsection{Sensitivity projections}

In order to estimate the sensitivity to BSM effects in $b\to s\ell\ell$ decays, a number of benchmark NP scenarios are considered, see Table~\ref{tab:penguins:NPscenarios}. 
Scenarios~I and II are inspired by the current discrepancies. The first scenario is the one that best explains the present $b\to s\mu\mu$ decay data. The second is the best explaining the rare semileptonic measurements within a purely left-handed scenario. This requirement is theoretically well motivated and arises in models designed to simultaneously explain the discrepancies seen in both tree-level semitauonic and loop-level semileptonic decays. The third and fourth scenarios assume that the current discrepancies are not confirmed but there is instead a small contribution from right-handed currents that would not be visible with the current level of experimental precision. These scenarios will serve to illustrate the power of the large LHCb \upgradetwo data set to distinguish between different NP models. This power relies critically on the ability to exploit multiple related decay channels. 

\begin{table}[!tb]
   \centering
\caption{
    Wilson coefficients in benchmark NP scenarios. The first scenario is inspired by the present discrepancies in the
    rare decays, including the angular distributions of the decay \decay{\Bz}{\Kstarz\mumu} and the measurements of the branching fraction ratios $R_K$ and $R_{K^*}$. 
    The second
    scenario is inspired by the possibility of explaining the
    rare decays discrepancies and those measured in the observables
    $R(D^{(*)})$. The third and fourth scenarios assume small nonzero
    right-handed couplings.} 
            \label{tab:penguins:NPscenarios}
   \begin{tabular}{crrrr}
       \hline\hline
            scenario & $C_9^{\rm NP}$ & $C_{10}^{\rm NP}$ & $C_{9}^{\prime}$ & $C_{10}^{\prime}$ \\
       \hline
       I & $-1.4$ & 0 & 0 & 0 \\
       II & $-0.7$ & 0.7 & 0 & 0\\
       III & 0 & 0& 0.3 & 0.3\\
       IV & 0 & 0& 0.3 & $-0.3$\\
       \hline\hline
   \end{tabular}
\end{table}

Unlike the systematics limited branching fractions, a more precise comparison between theory and experiment can be achieved by studying isospin and \CP asymmetries, which will be experimentally probed at percent-level precision with the LHCb \upgradetwo data set. This will also enable new decay modes to be studied, for example higher spin \Kstar states and modes with larger numbers of decay products. It is also possible to reduce theoretical and experimental uncertainties by comparing regions in angular phase-space of $\bquark\to\squark \ell\ell$ decays. The angular distribution of \decay{\B}{V \ellell} decays, where $V$ is a vector meson, can be expressed in terms of eight \qsq-dependent angular coefficients that depend on the Wilson coefficients and the form factors. 
Measurements of angular observables in \decay{\Bz}{\Kstarz\mumu} decays show a
discrepancy with respect to the SM predictions~\cite{LHCb-PAPER-2013-037,LHCb-PAPER-2015-051,LHCb-PAPER-2014-006,LHCb-PAPER-2013-019,Aubert:2006vb,Lees:2015ymt,Wei:2009zv,Aaltonen:2011ja,Chatrchyan:2013cda,Khachatryan:2015isa,Aaboud:2018krd,Sirunyan:2017dhj,Beneke:2004dp,Egede:2015kha,Kruger:2005ep,Lyon:2014hpa,Hurth:2013ssa,Beaujean:2013soa,Altmannshofer:2013foa,Descotes-Genon:2013wba}. 
This discrepancy is largest in the so-called optimized observable $P'_{5}$~\cite{LHCb-PAPER-2015-051}. 
The decay \decay{\Bs}{\phi\mumu} can also be described by the same angular formalism as the \decay{\Bz}{\Kstarz\mumu} decay.
However, in this case the \Bs and \Bsb mesons decay to a common final state and it is not possible to determine the full set of observables   without tagging the initial flavour of the \Bs. 

\begin{figure}[!tb]
\centering
\includegraphics[width=0.45\linewidth]{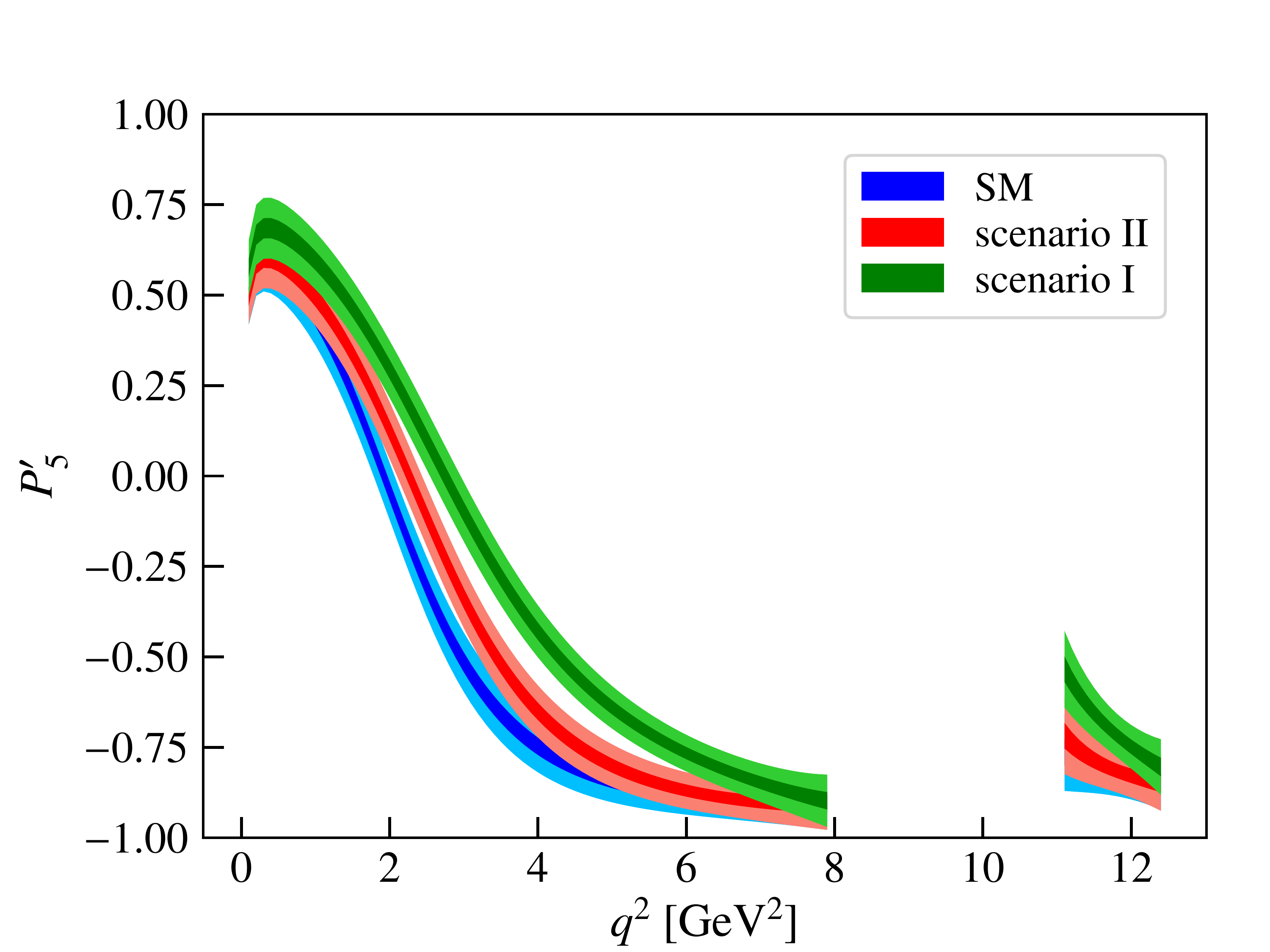}
\includegraphics[width=0.45\linewidth]{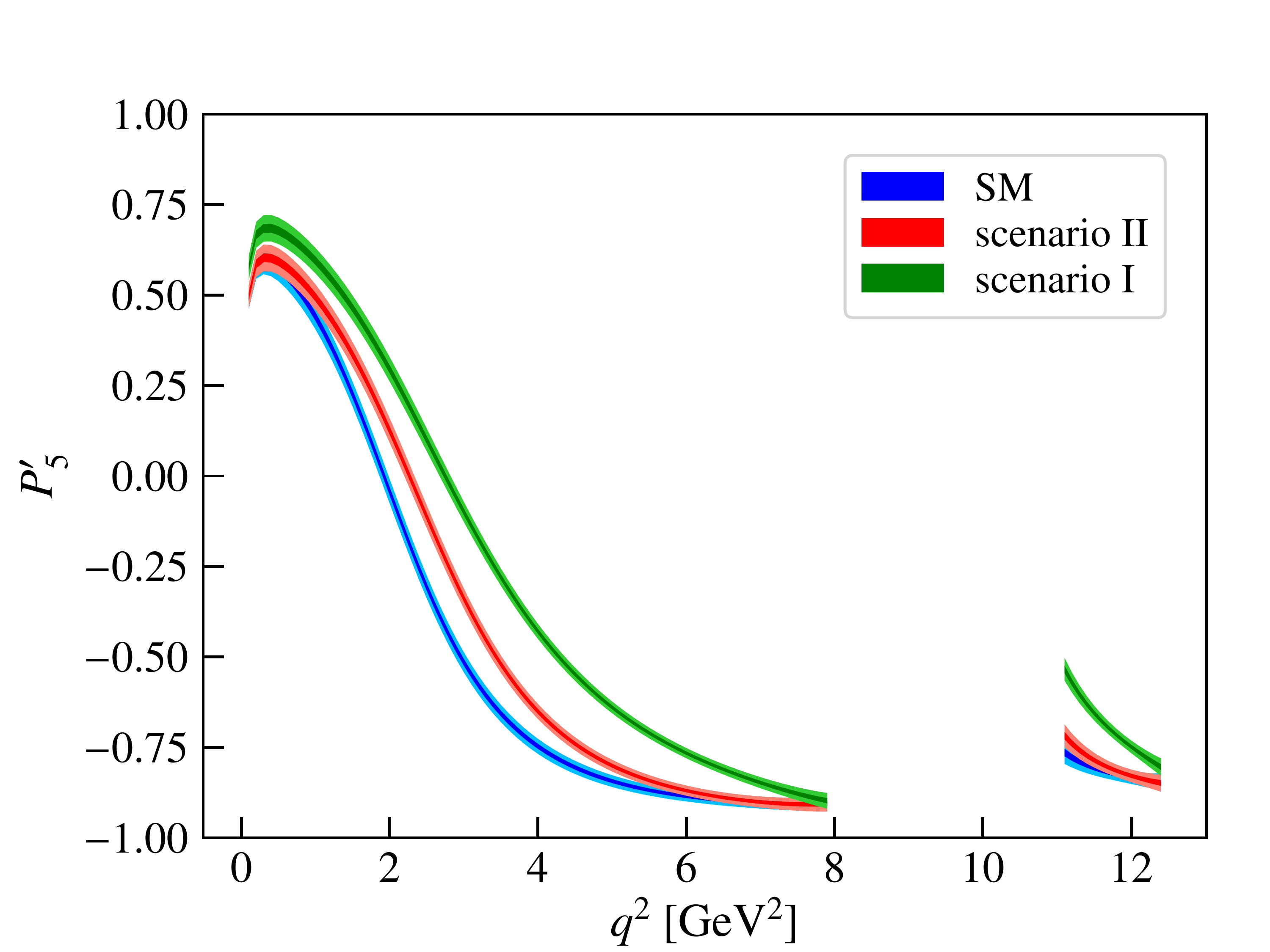}
\caption{Total experimental sensitivity, including systematics, at LHCb to the $P_5^{\prime}$ angular observable in the SM,
  Scenarios~I and II for  the Run~3 (left) and  the \upgradetwo (right) data sets. 
  The sensitivity is computed assuming that the charm-loop contribution is determined from the data. %
}
\label{fig:penguin:P5primeUnbinned}
\end{figure}

\begin{figure}[!tb]
\centering
\includegraphics[width=0.4\linewidth]{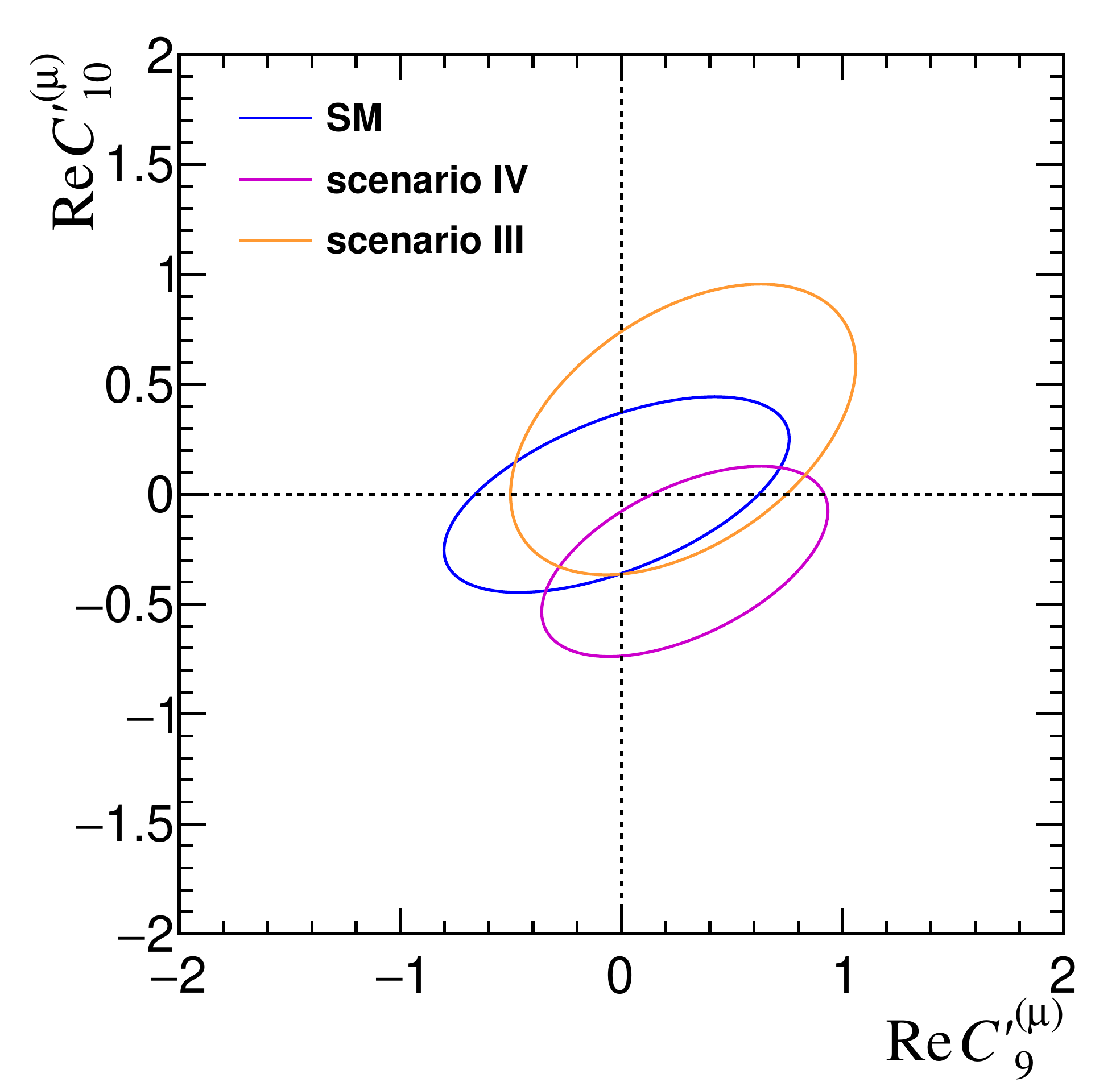}
\includegraphics[width=0.4\linewidth]{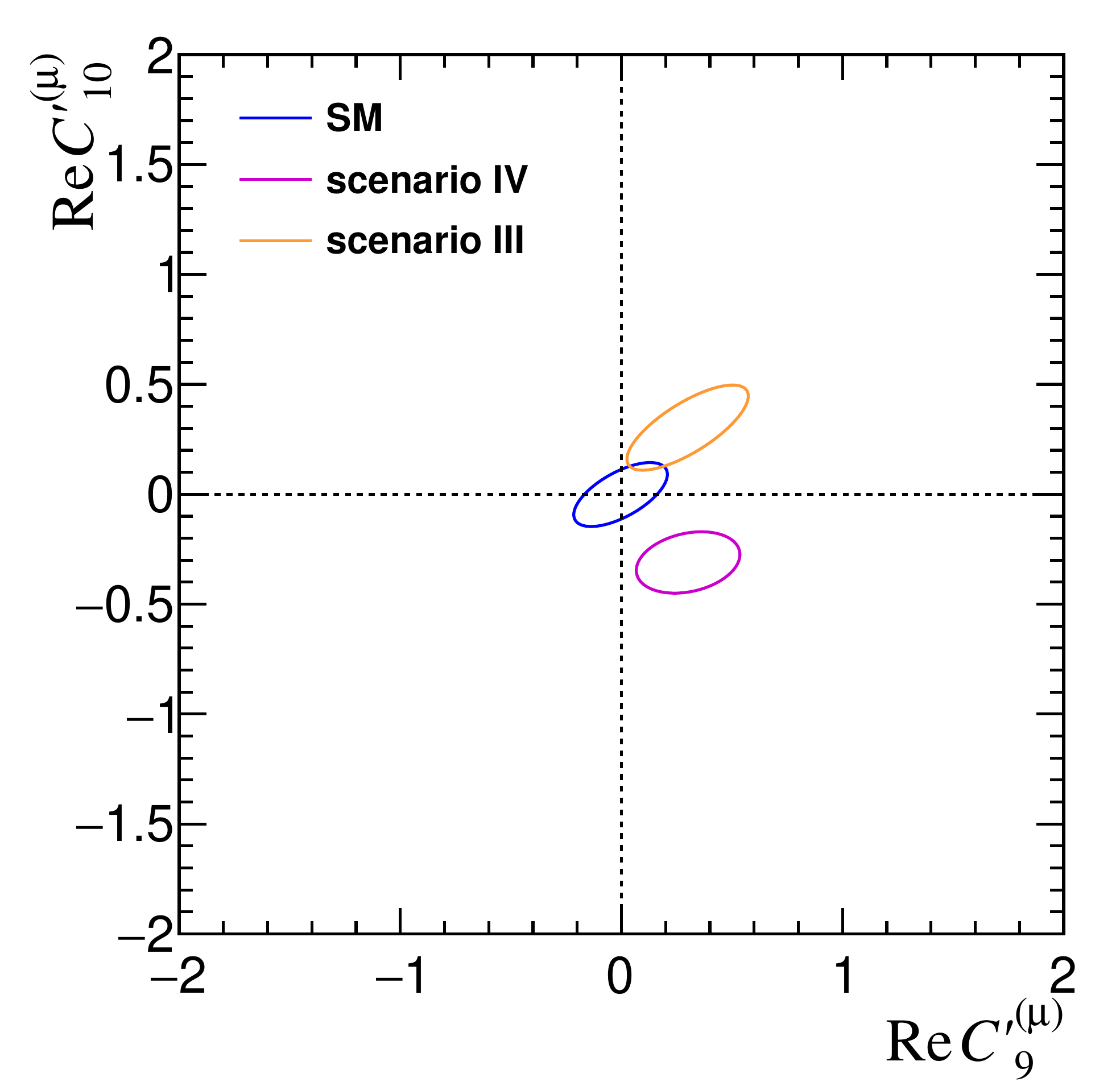}
\caption{Expected sensitivity to the Wilson coefficients $C_9^{\prime}$ and
  $C_{10}^{\prime}$ from the future LHCb analysis of the  \decay{\Bz}{\Kstarz\mumu} decay. 
  The ellipses correspond to 3\,$\sigma$ contours for
  the SM, Scenario~III and Scenario~IV for  the Run~3 (left) and  the \upgradetwo (right) data sets. 
  }
  \label{fig:penguin:WCKstmm}
\end{figure}

The expected precision of an unbinned LHCb-only determination of $P_5^{\prime}$ in the SM and in Scenarios~I and II is illustrated in Fig.~\ref{fig:penguin:P5primeUnbinned}, where we have followed the theoretical approaches in Refs.~\cite{Hurth:2017sqw,Chrzaszcz:2018yza} for the predictions. 
\upgradetwo will enable these scenarios to be clearly separated from the SM and from each other.
By combining information from all of the angular observables in the decay, it will also be possible for LHCb to distinguish models with much smaller NP contributions. 
Fig.~\ref{fig:penguin:WCKstmm} shows the expected $3\,\sigma$ sensitivity to NP in the Wilson
coefficients $C_{9,10}^{\prime}$ assuming the central values for the SM, Scenario~III and Scenario~IV.  
These scenarios are also clearly distinguishable with the precision that will be available with the \upgradetwo data set.

\begin{figure}[t]
  \begin{center}
  \includegraphics[width=0.6\textwidth]{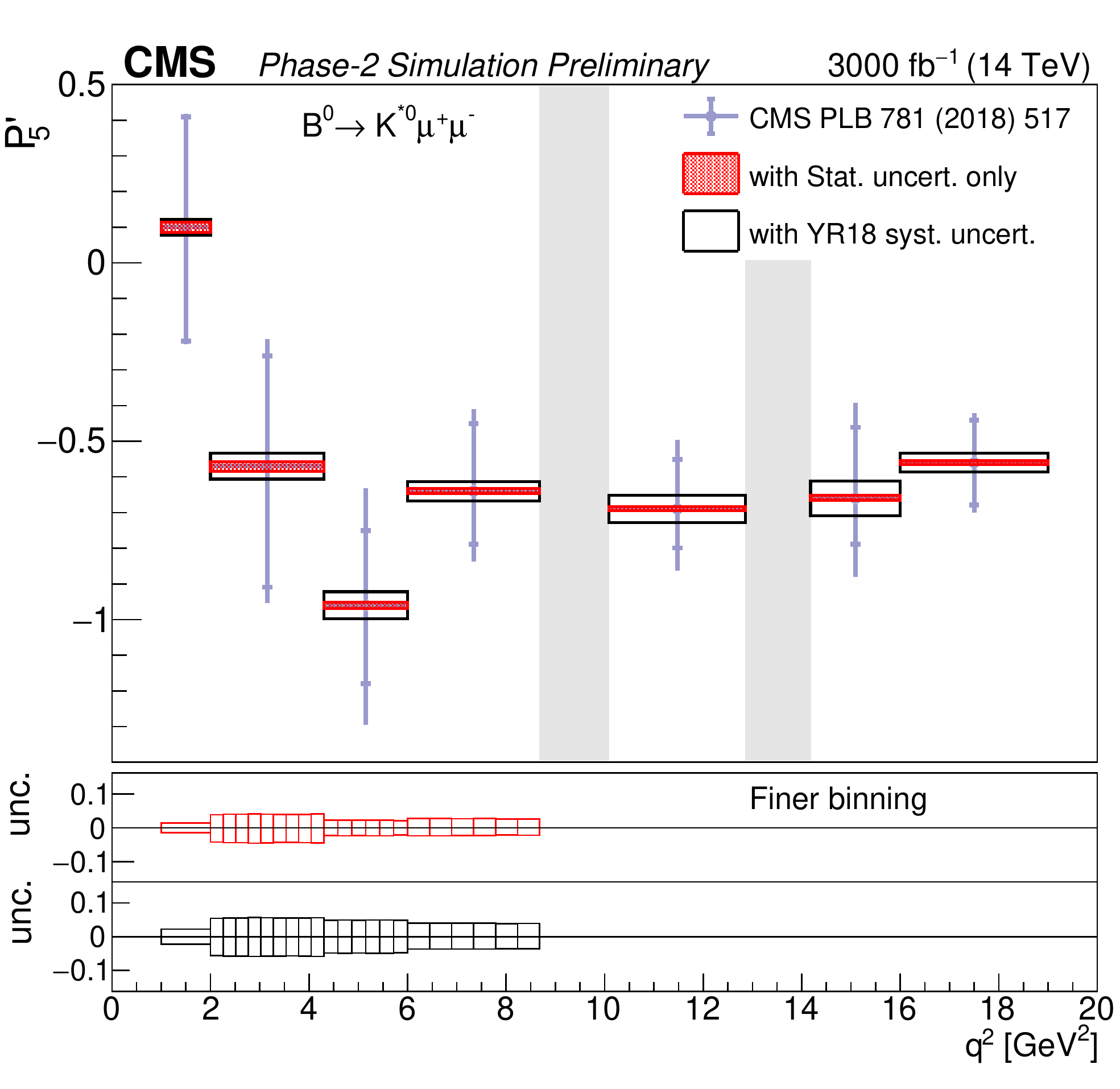}
 \caption{Projected statistical (hatched regions) and total (open box) uncertainties on the CMS $P'_5$ parameter versus $q^2$ in the Phase-2 scenario at 3000 fb$^{-1}$. The CMS Run-1 measurement of $P'_5$ is also shown by circles with inner vertical bars representing the statistical uncertainties and outer vertical bars representing the total uncertainties.
The vertical shaded regions correspond to the $J/\Psi$ and $\psi'$ resonances.
The two lower pads represent the  statistical (upper pad) and total (lower pad)  uncertainties with the finer $q^2$ binning. (The plot is taken from \cite{CMS-PAS-FTR-18-033}).} 
 \label{fig:CMSP5presults}
 \end{center}
 \end{figure}
 
The CMS projected statistical uncertainties and total uncertainties for the measurement of $P'_5$ versus $q^2$ for an integrated luminosity of 3000 fb$^{-1}$ are shown in Fig.~\ref{fig:CMSP5presults}, along with the Run~1 results. 

The $P_5'$ total uncertainties in the $q^2$ bins are estimated to improve up to a factor of 15 in the  3000 fb$^{-1}$ scenario \cite{CMS-PAS-FTR-18-033}, compared to those quoted in the Run-1 analysis.
The extrapolation to 300 fb$^{-1}$ integrated luminosity is also estimated by CMS \cite{CMS-PAS-FTR-18-033} and provides an improvement of up to a factor of 7 with respect to the Run-1 analysis.
These results are among the inputs of the global fit to the HL-LHC experimental projections shown in Fig.~\ref{fig:WC}, which
indicates the potential sensitivity to the SM and NP scenarios in the $C_9$- $C_{10}$ Wilson-coefficient plane.

The foreseen Phase-2 total integrated luminosity offers the opportunity to perform the angular analysis in narrower $q^2$ bins, in order to measure the $P'_5$  shape as a function of $q^2$ with finer granularity. The $q^2$ region below the $J/\psi$ mass(-squared), which is more sensitive to possible new physics effects, is considered. Each Run-1 $q^2$ bin is split into smaller and equal-size bins to achieve a statistical uncertainty of the order of the total systematic uncertainty in the same bin.
With respect to the Phase-2 systematic uncertainties with wider bins, the systematic uncertainties that were scaled the same as the statistical uncertainties are adjusted to account for less data in each finer bin while the other uncertainties are unchanged.
The corresponding  statistical and total uncertainties on $P'_5$  are shown in the lower two pads of Fig.~\ref{fig:CMSP5presults}.

The analysis in narrow $q^2$ bins provides a better determination of the $P_5'$ parameter shape which will allow for testing  theoretical predictions. 
The CMS available projection is for the single $P_5'$ angular parameter; with the foreseen HL-LHC statistics, CMS will have the capability to perform a full angular analysis of the \decay{\Bz}{\Kstarz\mumu} decay mode.

The ATLAS \cite{ATL-PHYS-PUB-2019-003} projected statistical and systematic uncertainties are provided in the same $q^2$ bins employed in the Run 1 analysis, and for the same set of angular parameters already published. Table \ref{tab:ATLASExtrapolation} reports the combined statistical and systematic uncertainty expectations for the different trigger scenarios considered, with the projected uncertainties for all the angular parameters compared to current theoretical predictions in figure \ref{fig:ATLASExtrapolation}. Reference \cite{ATL-PHYS-PUB-2019-003} includes full detail on the breakdown of statistical and systematic contributions.
The precision in measuring e.g. the $P'_5$ parameter is expected to improve by a factor between 5 and 9 relative to the Run 1 result, depending on the trigger scenario considered.
\newcommand{\Pany}{\ensuremath{P_i^{(\prime)}}\xspace}
\newcommand{\Pone}{\ensuremath{P^{}_1}\xspace}
\newcommand{\Pfour}{\ensuremath{P_4^\prime}\xspace}
\newcommand{\Pfive}{\ensuremath{P_5^\prime}\xspace}
\newcommand{\Psix}{\ensuremath{P_6^\prime}\xspace}
\newcommand{\Peight}{\ensuremath{P_8^\prime}\xspace}
\newcommand{\totFL}{\ensuremath{\delta^\mathrm{tot}_{F_L}}\xspace}
\newcommand{\totPone}{\ensuremath{\delta^\mathrm{tot}_{P^{}_1}}\xspace}
\newcommand{\totPfour}{\ensuremath{\delta^\mathrm{tot}_{P_4^\prime}}\xspace}
\newcommand{\totPfive}{\ensuremath{\delta^\mathrm{tot}_{P_5^\prime}}\xspace}
\newcommand{\totPsix}{\ensuremath{\delta^\mathrm{tot}_{P_6^\prime}}\xspace}
\newcommand{\totPeight}{\ensuremath{\delta^\mathrm{tot}_{P_8^\prime}}\xspace}

\newcommand{\trigSixSix}{\ensuremath{\upmu 6 \upmu 6}\xspace}
\newcommand{\trigTenSix}{\ensuremath{\upmu 10 \upmu 6}\xspace}
\newcommand{\trigTenTen}{\ensuremath{\upmu 10 \upmu 10}\xspace}
\newcommand{\trigXX}{\ensuremath{\upmu\mathrm{X}\upmu\mathrm{X}}\xspace}

\begin{figure}[htb]
  \begin{center}
      \includegraphics[width=0.42\columnwidth]{\main/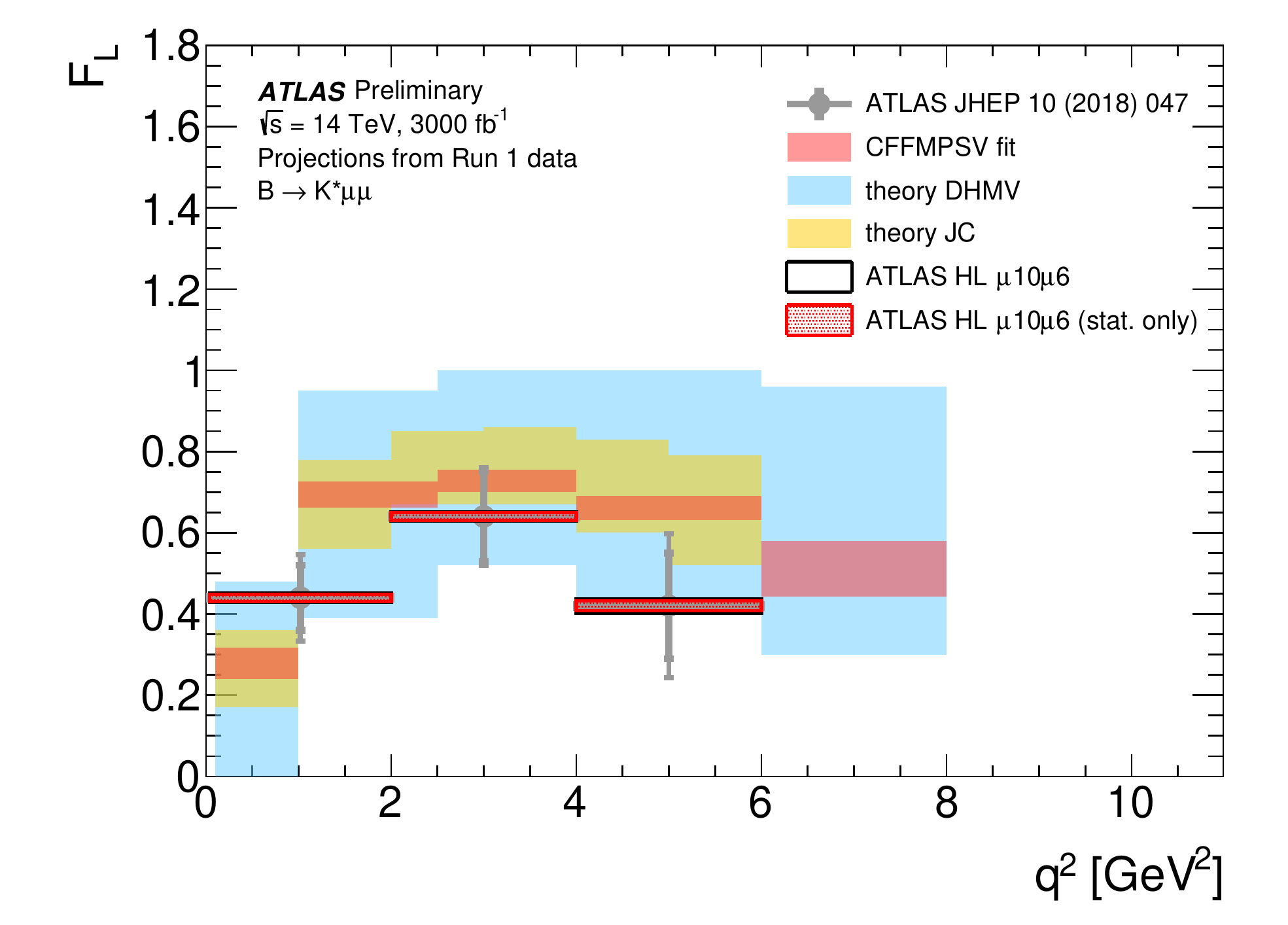}
      \includegraphics[width=0.42\columnwidth]{\main/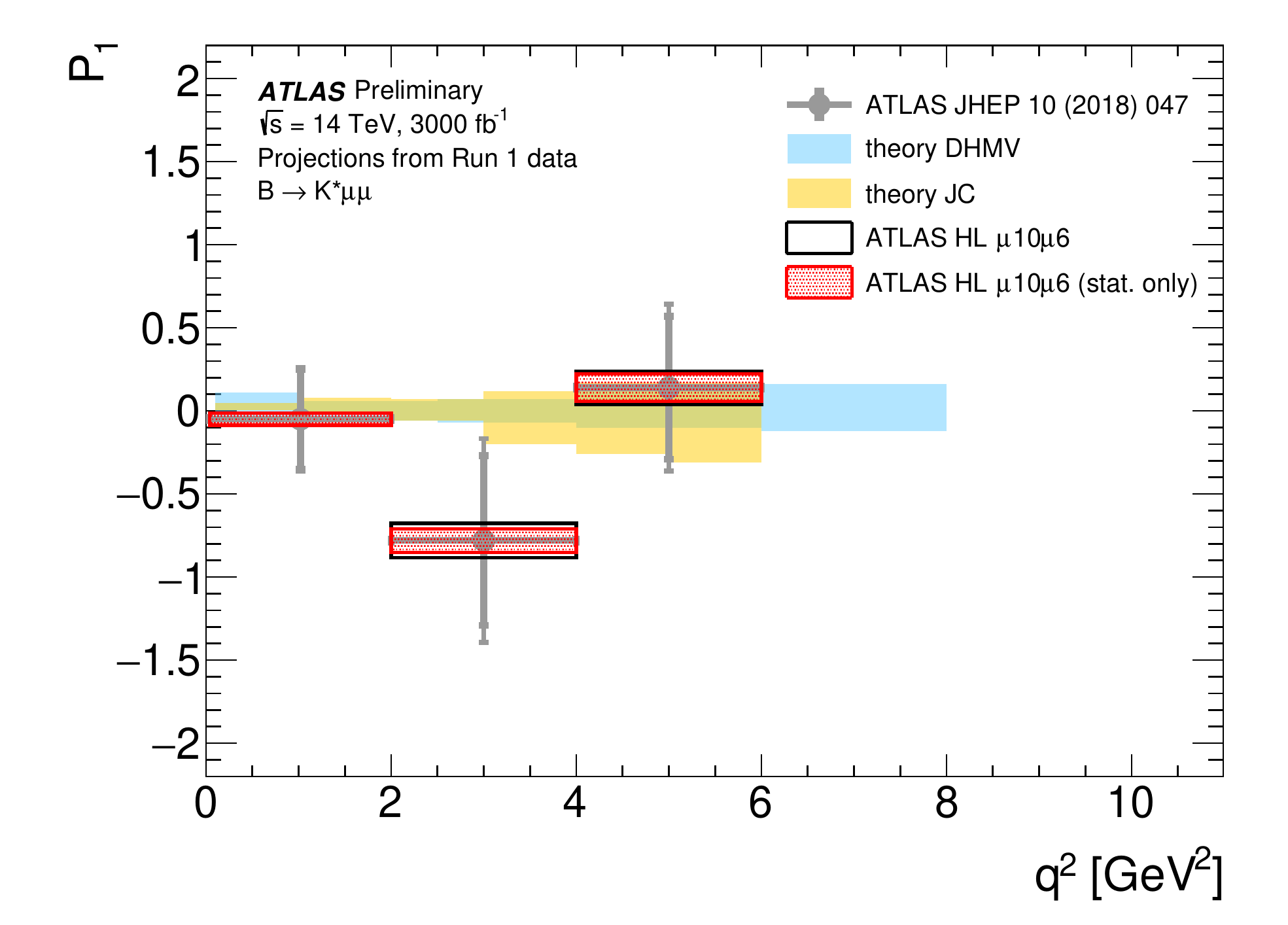}
      \includegraphics[width=0.42\columnwidth]{\main/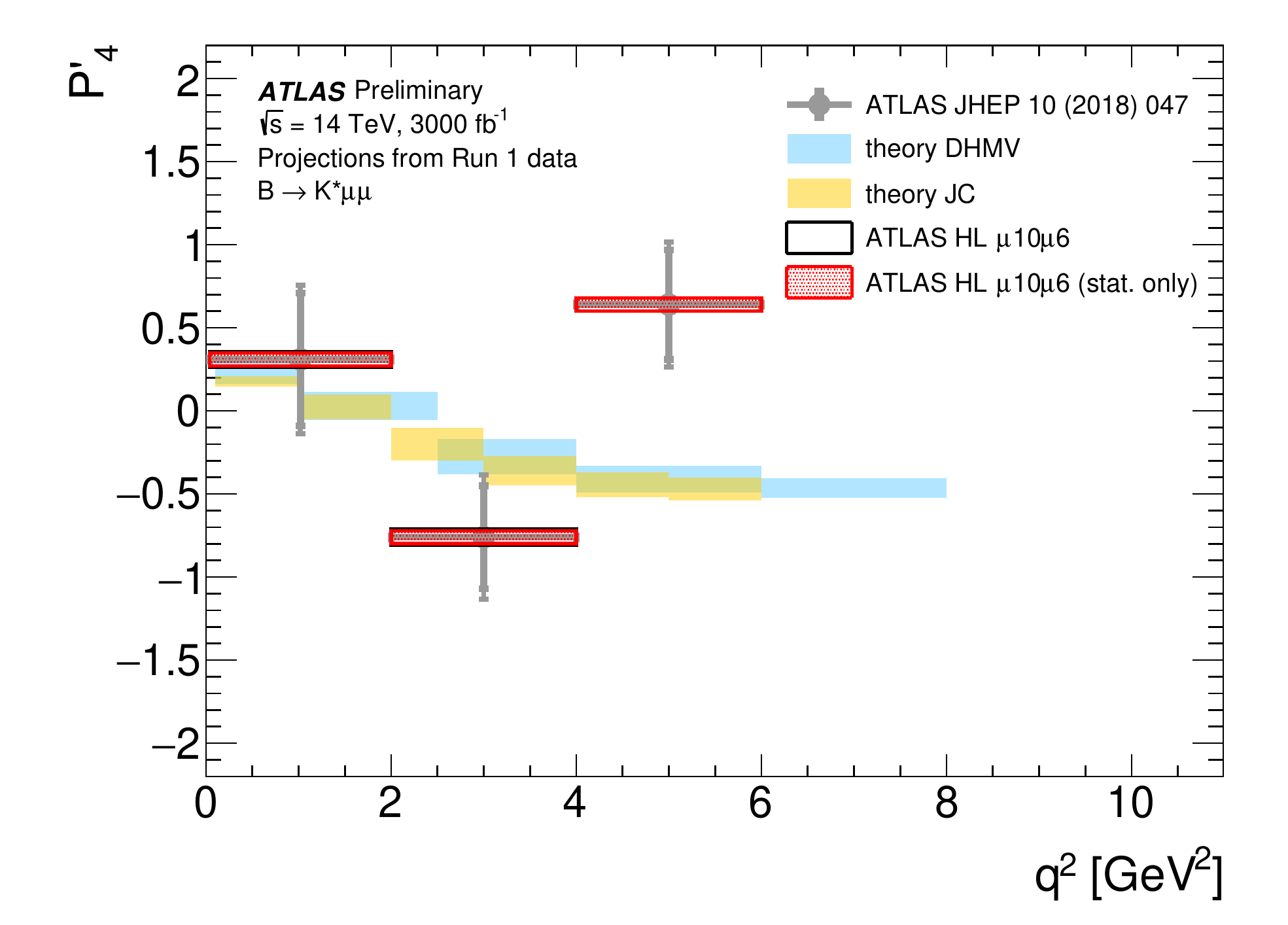}
      \includegraphics[width=0.42\columnwidth]{\main/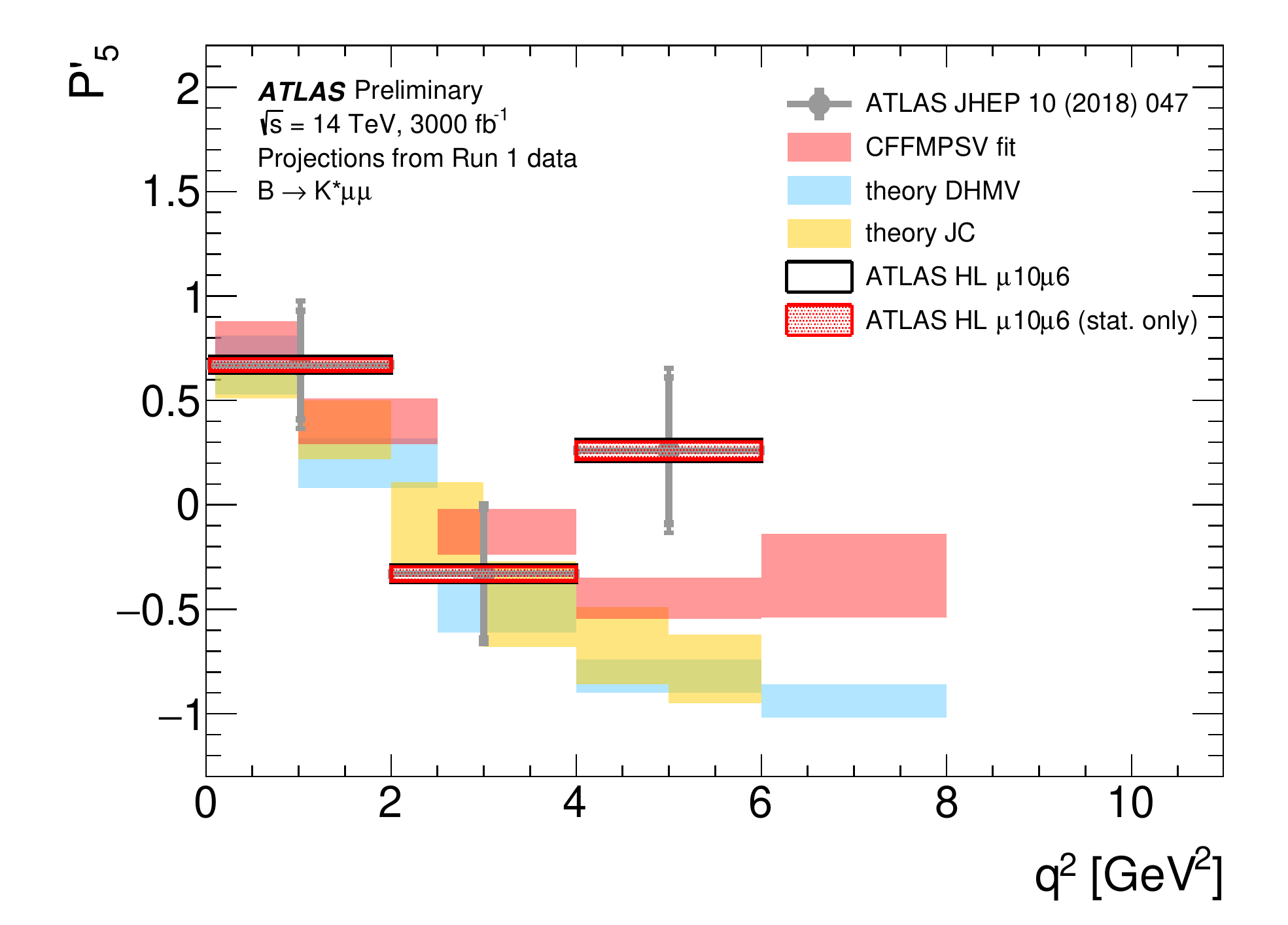}
      \includegraphics[width=0.42\columnwidth]{\main/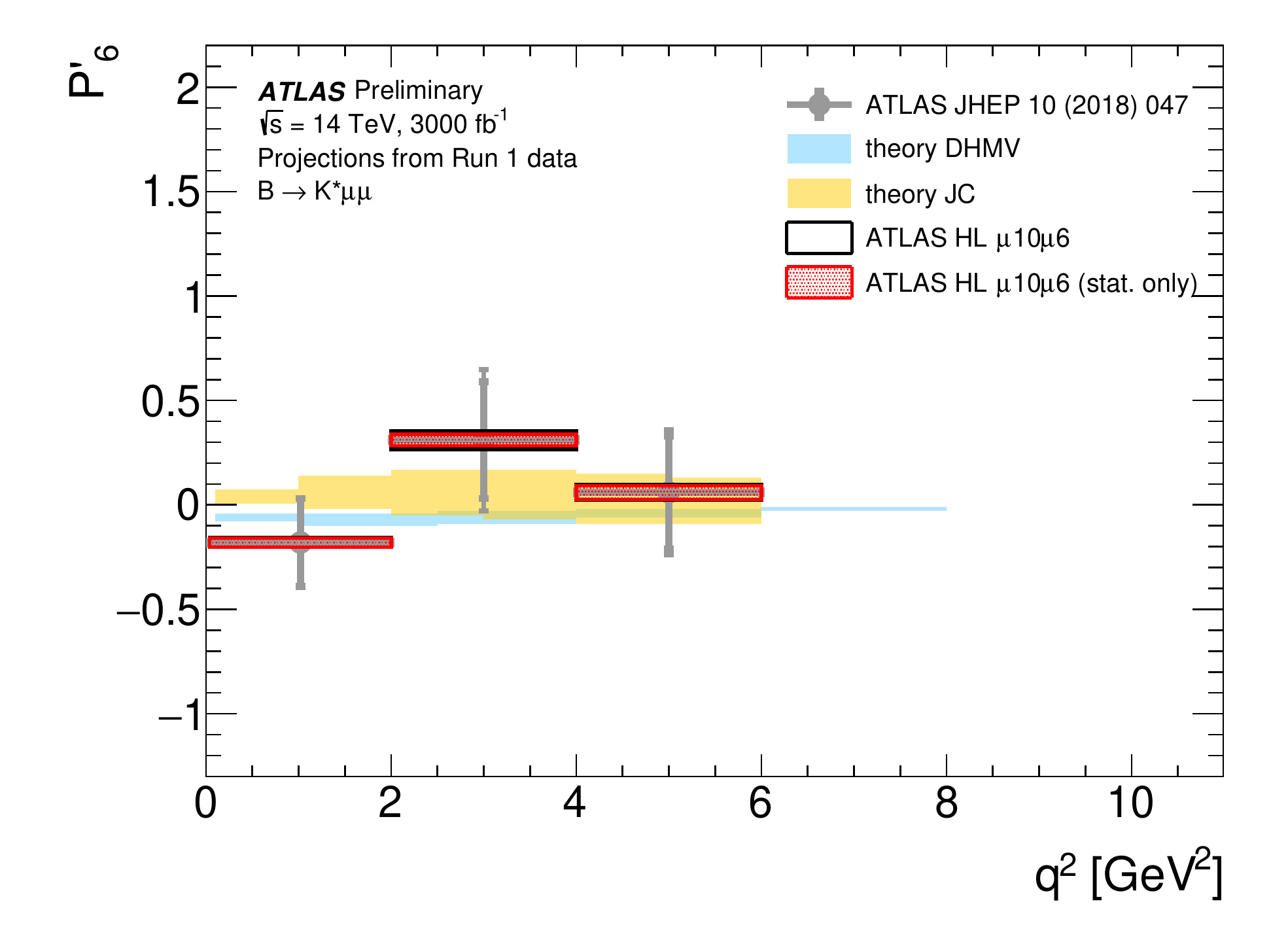}
      \includegraphics[width=0.42\columnwidth]{\main/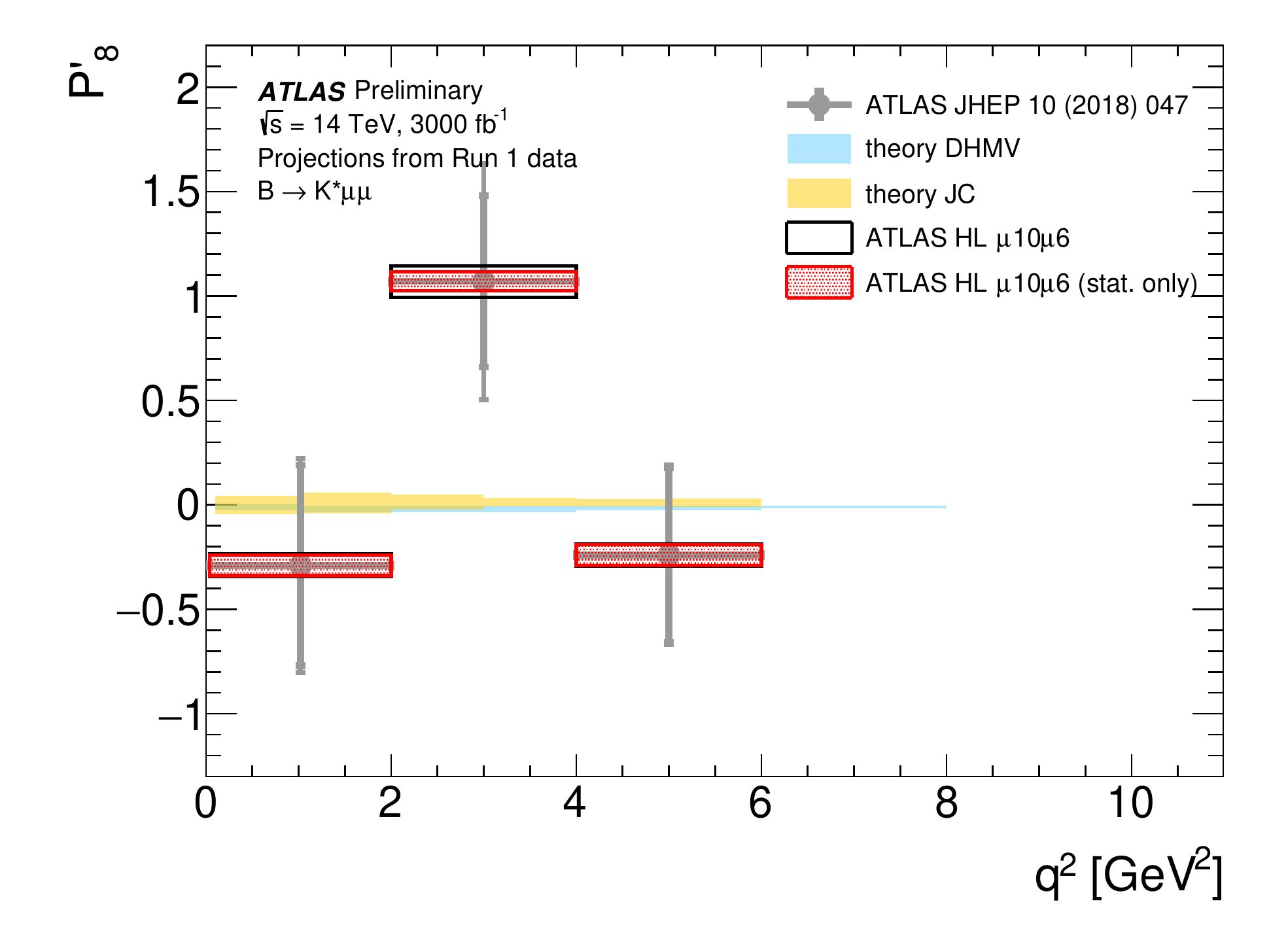}
 \caption{Projected ATLAS HL-LHC measurement precision in the \FL, \Pone, \Pfour, \Pfive, \Psix and \Peight parameters for the intermediate \trigTenSix trigger scenario compared to the ATLAS Run~1 measurement. Alongside theory predictions (CFFMPSV~\cite{Ciuchini:2015qxb}, DHMV~\cite{Descotes-Genon:2014uoa}, JC~\cite{Jager:2012uw}) are also shown. Both the projected statistical and the total (statistical and systematic) uncertainties are shown. While the HL-LHC toy-MC were generated with the DHMV central values of the \FL and \Pany parameters, in these plots the central values are moved to the ATLAS Run 1 measurement for better visualization of the improvement in the precision.} 
 \label{fig:ATLASExtrapolation}
 \end{center}
 \end{figure}
 
 \begin{table}[!tb]
  \begin{center}
    \caption{Combined systematic and statistical uncertainties on the \FL and \Pany parameters from the 2012 ATLAS data measurement
           and projected to the HL-LHC phase for the three trigger scenarios discussed in the text.}
            \label{tab:ATLASExtrapolation}
    \begin{tabular}{lrcccccc}
      \hline
      \hline
      \rule{0pt}{15pt}LHC phase &  $q^2[\GeV^2]$ & \totFL  & \totPone  & \totPfour  & \totPfive  & \totPsix  & \totPeight \\[5pt]
      \hline
      Run~1                     & $[0.04, 2.0]$    & 0.11    & 0.31      & 0.45       & 0.31       & 0.21      & 0.51 \\
                                & $[2.0, 4.0]$     & 0.12    & 0.61      & 0.37       & 0.34       & 0.34      & 0.57 \\
                                & $[4.0, 6.0]$     & 0.18    & 0.50      & 0.38       & 0.39       & 0.30      & 0.43 \\
      \hline
      HL-LHC \trigSixSix        & $[0.04, 2.0]$    & 0.010   & 0.027     & 0.037      & 0.037      & 0.019     & 0.046 \\
                                & $[2.0, 4.0]$     & 0.008   & 0.093     & 0.040      & 0.038      & 0.040     & 0.070 \\
                                & $[4.0, 6.0]$     & 0.016   & 0.083     & 0.032      & 0.047      & 0.033     & 0.041 \\
      \hline
      HL-LHC \trigTenSix        & $[0.04, 2.0]$    & 0.011   & 0.037     & 0.046      & 0.040      & 0.023     & 0.055 \\
                                & $[2.0, 4.0]$     & 0.011   & 0.103     & 0.047      & 0.042      & 0.044     & 0.075 \\
                                & $[4.0, 6.0]$     & 0.018   & 0.100     & 0.040      & 0.053      & 0.038     & 0.052 \\
      \hline
      HL-LHC \trigTenTen        & $[0.04, 2.0]$    & 0.018   & 0.065     & 0.076      & 0.059      & 0.041     & 0.093 \\
                                & $[2.0, 4.0]$     & 0.017   & 0.15      & 0.074      & 0.068      & 0.059     & 0.100 \\
                                & $[4.0, 6.0]$     & 0.026   & 0.17      & 0.074      & 0.082      & 0.063     & 0.090 \\
      \hline
      \hline
    \end{tabular}
  \end{center}
   \centering
   \begin{tabular}{crrrr}
   \end{tabular}
\end{table}

Furthermore, with large data sets expected at the HL-LHC it will  be possible for the LHC experiments to probe \decay{\B}{V \ellell} SM contributions, under the premise that a genuine NP contribution is expected to have no $q^2$ dependence, while, \eg a 
charm loop contribution is expected to grow when approaching the pole of
the charmonia resonances.  
A measurement using Breit-Wigner functions to parametrise the
resonances, and their interference with the
short-distance contributions to the decay, was proposed in Ref.~\cite{Blake:2017fyh}. 
A similar technique has already been applied by LHCb to the Run-1 data for the \decay{\Bp}{\Kp\mumu} decay~\cite{LHCb-PAPER-2016-045}. 
An alternative approach using additional phenomenological inputs has also been proposed~\cite{Bobeth:2017vxj}.
Such a combination of phenomenological and experimental
methods may improve our knowledge of the charm-loop contribution and form factors, which would allow $C_9^\mu$ and $C_{10}^\mu$ to be determined with great precision in $\bquark\to\squark\mu\mu$ transitions.

\subsubsection{Measurements of $b\to dll$}

Thanks to LHCb's particle identification capabilities, the \upgradetwo data set will provide a unique opportunity to make precise measurements of $\bquark\to\dquark \ell\ell$ processes.
Using the Run~1 and 2 data sets, LHCb data have been used to observe the decays \decay{\Bp}{\pip\mumu}~\cite{LHCb-PAPER-2012-020,LHCb-PAPER-2015-035} and \decay{\Lb}{p\pim\mumu}~\cite{LHCb-PAPER-2016-049}, and to find evidence for the decays \decay{\Bz}{\pip\pim\mumu} (in a $\pip\pim$ mass region that  is expected to be dominated by \decay{\Bz}{\rhoz\mumu}) and \decay{\Bs}{\Kstarzb\mumu}~\cite{LHCb-PAPER-2018-004} with branching fractions at the $\mathcal{O}(10^{-8})$ level.
The existing data samples comprise $\mathcal{O}(10)$ decays in these decay modes. 
The upgrade will provide samples of thousands, or tens of thousands of such decays.
The ability to measure the properties of these processes depends heavily on the PID performance of the LHCb subdetectors. 
In the case of the \decay{\Bs}{\Kstarzb\mumu} decay, excellent mass resolution is also critical to separate \Bs and \Bz decays. 

The ratio of branching fractions between the CKM-suppressed $b \to d\ell\ell$ transitions and their CKM-favoured  $\bquark \to \squark \ellell$ counterparts, together with theoretical input on the ratio of the relevant form factors, enables the ratio of CKM elements  $|V_{td}|/|V_{ts}|$ to be determined. The precision on $|V_{td}|/|V_{ts}|$ from such decays is dominated at present by the statistical uncertainty on the experimental measurements of \decay{\Bp}{\pip\mumu}, and is much less precise than the determination from mixing measurements. The theoretical uncertainty at high-$q^2$ is at the level of $4\%$ %
and is expected to improve with further progress on the form factors from lattice QCD~\cite{Du:2015tda}. Around 17\,000 \decay{\Bp}{\pip\mumu} decays are expected in the full 300\invfb dataset, allowing an experimental precision better than $2\%$.

The current set of measurements of $\bquark\to\squark \ell\ell$ processes have demonstrated the importance of angular measurements in the precision determination of Wilson coefficients. With the LHCb \upgradetwo dataset, where a sample of 4300  \decay{\Bs}{\Kstarzb\mumu} decays is expected,  it will be possible to make a full angular analysis of a $\bquark\to\dquark \ellell$ transition. The \decay{\Bs}{\Kstarzb\mumu} decay is both self-tagging and has a final state involving only charged particles. The LHCb \upgradetwo data set will allow the angular observables in this decay to be measured with better precision than the existing measurements of the \decay{\Bz}{\Kstarz\mumu} angular distribution. 

The LHCb \upgradetwo dataset will also give substantial numbers of \decay{\B^{0,+}}{\rho^{0,+}\mumu}  and \decay{\Lb}{N\mumu} decays. Although the  \decay{\Bz}{\rhoz\mumu} decay does not give the flavour of the initial \B meson, untagged measurements will give sensitivity to a subset of the interesting angular observables. 
Analysis of the \decay{\Lb}{N^{*}\mumu} decay will require statistical separation of  overlapping $p\pim$ resonances with different $J^P$ by performing an amplitude analysis of the final-state particles. 

The combination of information from $\BF(\decay{\Bz}{\mumu})$, the differential branching fraction of the \decay{\Bp}{\pip\mumu} decay, and angular measurements, notably of \decay{\Bs}{\Kstarzb\mumu}, will indicate whether NP effects are present in $\bquark\to\dquark$ transitions at the level of $20\%$ of the SM amplitude with more than $5\sigma$ significance. 

\subsubsection{LFU tests in $b\to (s,d)\ell\ell$}

\begin{figure}[!t] 
\centering
\includegraphics[width=0.75\textwidth]{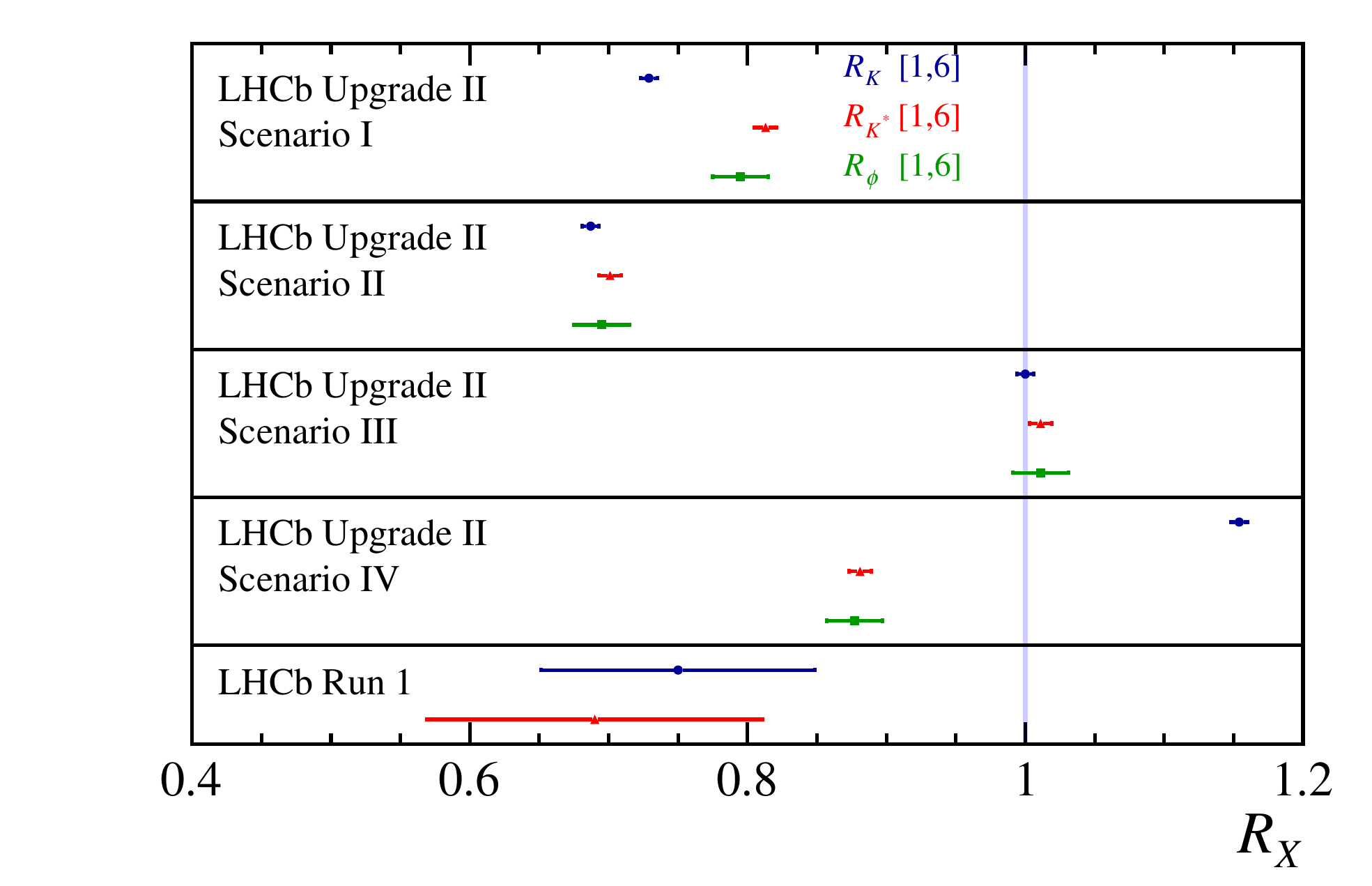}
\caption{
Projected sensitivity for the $R_{K}$, $R_{K^*}$ and $R_{\phi}$ measurements in different NP scenarios with the Upgrade~II data set. 
The existing Run~1 measurements of $R_{K}$ and $R_{K^*}$ are shown for comparison. 
}
\label{fig:penguin:RXscenarios}
\end{figure}

The Run~1 \lhcb data have been used to perform the most precise measurements of $R_{K}$ and $R_{\Kstar}$ to-date~\cite{LHCb-PAPER-2017-013,LHCb-PAPER-2014-024} (see Fig.~\ref{fig:penguin:RXscenarios}).
These measurements are compatible with the SM at the level of 2.1--2.6 standard deviations. Assuming the current detector performance, approximately 46\,000 \decay{\Bp}{\Kp\epem} and 20\,000 \decay{\Bz}{\Kstarz\epem} candidates are expected in the range $1.1 < \qsq < 6.0\gevgevcccc$ in the \upgradetwo data set.
The ultimate precision on $R_{K}$ and $R_{\Kstar}$ will be better than 1\%.
The importance of the \upgradetwo data set in distinguishing between different NP scenarios is highlighted in Fig.~\ref{fig:penguin:RXscenarios}. 
With this data set all four NP scenarios could be distinguished at more than 5\,$\sigma$ significance.

The \upgradetwo data set will also enable the measurement of other $R_{X}$ ratios \eg $R_{\phi}, R_{p\kaon}$ and the ratios in CKM suppressed decays. 
For example, with 300\invfb, it will be possible to determine $R_{\pi}=\BR(\decay{\Bp}{\pip\mumu})/\BR(\decay{\Bp}{\pip\epem})$ with a few percent statistical precision.
A summary of the expected performance for a number of different $R_X$ ratios is indicated in Table~\ref{tab:penguin:LU_extrapolations}.

\begin{table}[!tb]
\caption{
Estimated yields of $\bquark\to\squark\ep\en$ and $\bquark\to\dquark\ep\en$ processes and the statistical uncertainty on $R_{X}$ in the range $1.1<\qsq<6.0\gevgevcccc$ extrapolated from the Run~1 data. 
A linear dependence of the \bbbar production cross section on the $pp$ centre-of-mass energy and unchanged Run~1 detector performance are assumed. 
Where modes have yet to be observed, a scaled estimate from the corresponding muon mode is used.
}
\label{tab:penguin:LU_extrapolations}
\centering
\begin{tabular}{lrrrrr}
\hline\hline
Yield  & Run~1 result & 9\invfb & 23\invfb & 50\invfb & 300\invfb \\
\hline
 \decay{\Bu}{\Kp\epem} 		& $254 \pm 29$\cite{LHCb-PAPER-2014-024} & 1\,120 & 3\,300 & 7\,500 & 46\,000 \\
 \decay{\Bd}{\Kstarz\epem}	& $111 \pm 14$\cite{LHCb-PAPER-2017-013} & 490 & 1\,400 & 3\,300 & 20\,000 \\
 \decay{\Bs}{\phi\epem}		& -- & 80 & 230 & 530 & 3\,300 \\
 \decay{\Lb}{\proton\kaon\epem} & -- & 120 & 360 & 820 & 5\,000 \\
\decay{\Bp}{\pip\epem}		& -- & 20 & 70 & 150 & 900 \\ 
\hline
$R_X$ precision &  Run~1 result & 9\invfb & 23\invfb & 50\invfb & 300\invfb \\
\hline
$R_{\kaon}$					& $0.745 \pm 0.090\pm 0.036$\cite{LHCb-PAPER-2014-024} & 0.043 & 0.025 & 0.017 & 0.007 \\
$R_{\Kstarz}$					& $0.69 \pm 0.11\pm 0.05$\cite{LHCb-PAPER-2017-013} & 0.052 & 0.031 & 0.020 & 0.008 \\
$R_{\phi}$					& -- & 0.130 & 0.076 & 0.050 & 0.020 \\
$R_{\proton\kaon}$  & -- & 0.105 & 0.061 & 0.041 & 0.016 \\
$R_{\pi}$						& -- & 0.302 & 0.176 & 0.117 & 0.047 \\ 
\hline\hline
\end{tabular}
\end{table}

\begin{figure}[t!]
\centering
\includegraphics[width=0.49\textwidth]{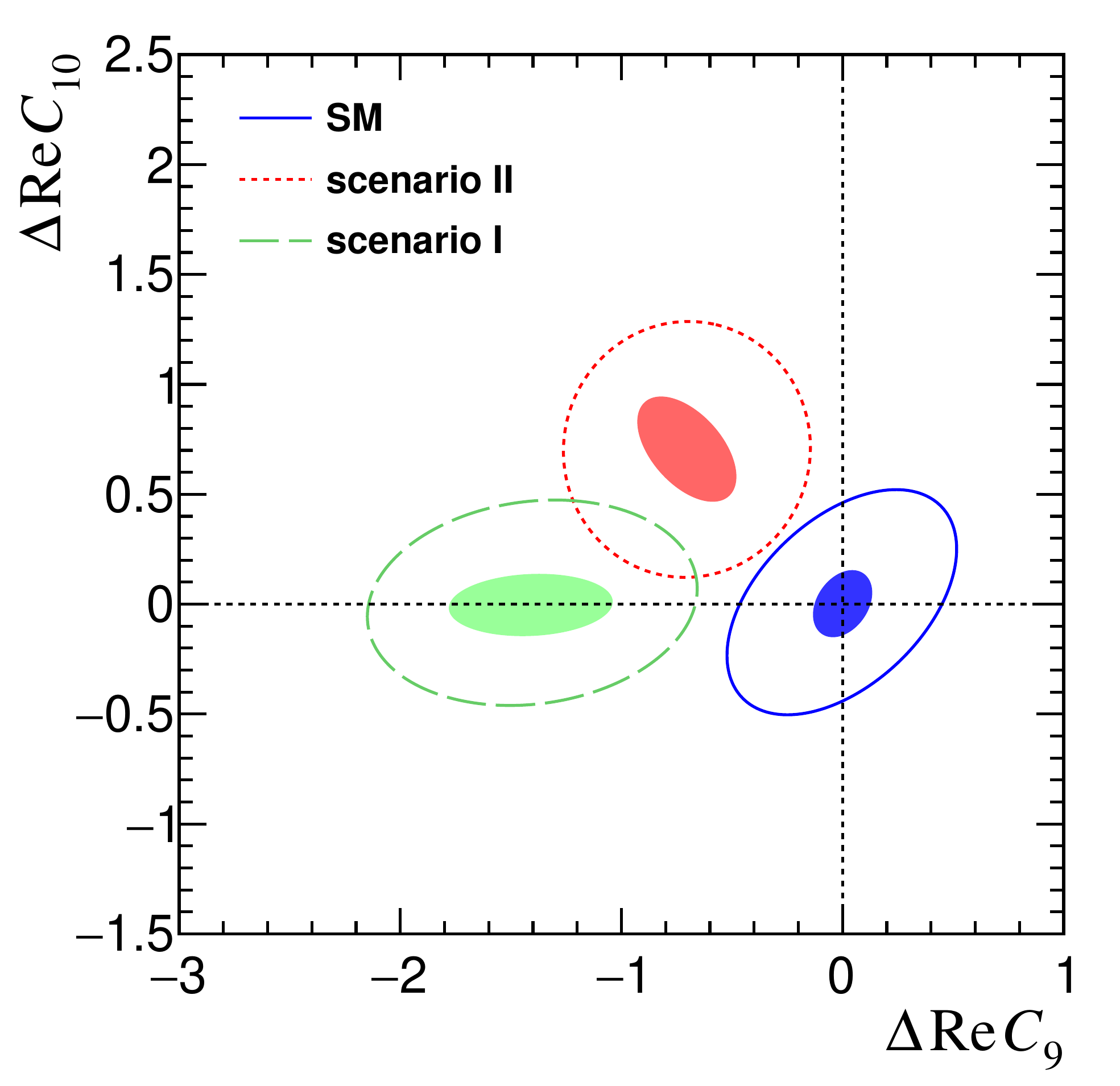}
\caption{
  Constraints on the difference in the $C_9$ and $C_{10}$ Wilson coefficients from angular analyses of the electron and muon modes with the Run~3 and \upgradetwo data sets.
  The $3\sigma$ regions for the Run~3 data sample are shown for the SM (solid blue), a vector-axial-vector new physics contribution (red dotted) and for a purely vector new physics contribution (green dashed). 
  The shaded regions denote the corresponding constraints for the \upgradetwo data set.
}
\label{fig:penguin:DeltaC}
\end{figure}

In addition to improvements in the $R_{X}$ measurements, the enlarged \upgradetwo data set will give access to new observables. For example, the data will allow precise comparisons of the angular distribution of dielectron and dimuon final-states. 
Differences between angular observables in \decay{\B}{X\mumu} and \decay{\B}{X\ep\en} decays are theoretically pristine~\cite{Capdevila:2016ivx,Serra:2016ivr}
and are sensitive to different combinations of Wilson coefficients compared to the $R_X$ measurements. 
Fig.~\ref{fig:penguin:DeltaC} shows that an upgraded \lhcb detector will enable such decays to be used to discriminate between different NP models, for example separating between Scenarios~I and II~\cite{Mauri:2018vbg}. 
Excellent NP sensitivity can be achieved irrespective of the assumptions made about the hadronic contributions to the decays.

In the existing \lhcb detector, electron modes have an approximately factor five lower efficiency than the corresponding muon modes, owing to the tendency for the electrons to lose a significant fraction of their energy through bremsstrahlung in the detector. 
This loss impacts on the ability to reconstruct, trigger and select the electron modes.
The precision with which observables can be extracted therefore depends primarily on the electron modes and not the muon modes.
In order for $R_{X}$ measurements to benefit from the large \upgradetwo data samples, it will be necessary to reduce systematic uncertainties to the percent level.
These uncertainties are controlled by taking a double ratio between $R_{X}$ and the decays \decay{\B}{\jpsi X}, where the \jpsi decays to \mumu and $\ep\en$.
This approach is expected to work well, even with very large data sets.  

Other sources of systematic uncertainty can be mitigated through design choices for the upgraded detector.
The recovery of bremsstrahlung photons is inhibited by the ability to find the relevant photons in the ECAL (over significant backgrounds) and by the energy resolution.
A reduced amount of material before the magnet would reduce the amount of bremsstrahlung and hence would increase the electron reconstruction efficiency and improve the electron momentum resolution. 
Higher transverse granularity would aid signal selection and help reduce the backgrounds. 
With a large number of primary $pp$ collisions, the combinatorial background will increase and will need to be controlled with the use of timing information.
However, the Run~1 data set indicates that it may be possible to tolerate a significant (\ie\ larger than a factor two) increase in combinatorial backgrounds without destroying the signal selection ability.

\subsubsection{Time dependent angular analyses in $b\to (s,d)ll$}

Time dependent analyses of rare decays into \CP-eigenstates can deliver orthogonal experimental information to time-integrated observables. 
So far, no time-dependent measurement of the $\decay{\Bs}{\phi\mumu}$ decay has been performed due to the limited signal yield of $432\pm 24$ in the Run~1 data sample~\cite{LHCb-PAPER-2015-023}.
However, the larger data samples available in \upgradetwo will enable time-dependent studies. 
The framework describing $\Bbar$ and $\B\to V\ellell$ transitions to a common final-state is discussed in  Ref.~\cite{Descotes-Genon:2015hea}, 
where several observables are discussed that can be accessed with and without flavour tagging.  
Two observables called $s_8$ and $s_9$, which are only accessible through a time-dependent flavour-tagged analysis, are of particular interest. 
These observables  are proportional to the mixing term $\sin\left(\Delta m_s t\right)$ and provide information that is not available through flavour specific decays.
Assuming a time resolution of around $45\fs$ and an effective tagging power of $5\%$ 
results in an effective signal yield of  $2000$ decays for the \upgradetwo data set. 

As a first step towards a full time-dependent analysis, the effective lifetime of the decay $\decay{\Bs}{\phi\mumu}$ can be studied. 
The untagged time-dependent decay rate is given by
\begin{align}
  \frac{\deriv\Gamma}{\deriv t} &\propto  e^{-\Gamma_s}\left[\cosh\left(\frac{\Delta\Gamma_st}{2}\right)+A^{\Delta\Gamma}\sinh\left(\frac{\Delta\Gamma_st}{2}\right)\right].
\end{align}
The observable $A^{\Delta\Gamma}$ can be related to the angular observables $F_{\rm L}$ and $S_3$ via $A^{\Delta\Gamma}=2S_3-F_{\rm L}$. 
Due to the significant lifetime difference $\Delta\Gamma_s$ in the \Bs\ system, even an untagged analysis can probe right-handed currents. 
For the combined low- and high-\qsq regions, preliminary studies suggest a statistical sensitivity to $A^{\Delta\Gamma}$ of $0.05$ can be achieved with a $300\invfb$ data set. 

With the \upgradetwo data set it will also be possible to perform a time-dependent angular analysis of the $\bquark \to \dquark$ process \decay{\Bd}{\rho^0\mumu}. 
This process differs from \decay{\Bs}{\phi\mumu} in two important regards: it is CKM suppressed and therefore has a smaller SM branching fraction; and $\Delta \Gamma_d \approx 0$, removing sensitivity to $A^{\Delta\Gamma}$. 
The uncertainties on the angular observables are expected to be of the order of $0.1$ for this case. 

The time-dependent angular analyses will still be statistically limited even with $300\invfb$. 
It will be important to maintain good decay-time resolution and the performance of the particle identification will be crucial to control backgrounds, as well as to improve flavour tagging performance. 

\subsubsection{Measurements of $b\to s\gamma$}

The time dependent \CP asymmetry of \decay{\B}{f_{\CP}\gamma} arises from the interference between decay amplitudes with and without $\Bds-\Bdsb$ mixing and is predicted to be small in the SM\ \cite{Atwood:1997zr,Ball:2006cva,Matsumori:2005ax}.
As a consequence, a large asymmetry due to interference between the \B mixing and decay diagrams can only be present if the two photon helicities contribute to both \B and \Bb decays.
From the time dependent decay rate
\begin{align}
    \Gamma(\decay{\Bds(\Bdsb)}{f_{\CP}\gamma})(t) \sim e^{-\Gamma_s t}
    &\big[
        \cosh\left(\frac{\Delta\Gamma_{(s)}}{2}\right)
        - \mathcal{A}^{\Delta}\sinh\left(\frac{\Delta\Gamma_{(s)}}{2}\right) \pm \nonumber\\
        & \pm \mathcal{C}_\CP\cos\left(\Delta m_{(s)} t\right)
        \mp \mathcal{S}_\CP\sin\left(\Delta m_{(s)} t\right)
    \big],
\end{align}
where $A^\Delta$, $\mathcal{C}_\CP$ and $\mathcal{S}_\CP$ depend on the photon polarisation~\cite{Muheim:2008vu}.
Two strategies can be devised:
one studying the decay rate independently of the flavour of the \B meson, which allows $A^\Delta$ to be accessed, and one tagging the flavour of the \B meson, which accesses $\mathcal{S}_\CP$ and $\mathcal{C}_\CP$.
The first strategy has been exploited at \lhcb to study the $4000$ \decay{\Bs}{\phi\gamma} candidates collected in Run 1 to obtain $A^\Delta=-0.98^{+0.46}_{-0.52}\stat ^{+0.23}_{-0.20}\syst$~\cite{LHCb-PAPER-2016-034}, compatible at two standard deviations from the prediction of $A^\Delta_{\text{SM}}=0.047^{+0.029}_{-0.025}$.

With $\sim60$k signal candidates expected with $50\,\invfb$, the full analysis, including flavour tagging information, will improve the statistical uncertainty on $A^\Delta$ to $\sim0.07$, and will need a careful control of the systematic 
uncertainties.
The analysis performed with $\sim800$k signal decays expected with $300\,\invfb$, with a statistical uncertainty to $\sim0.02$, requires some of the possible improvements in \piz reconstruction of the \upgradetwo detector 
to be able to use the full statistical power of the data.

In addition to studying the \Bs system, \lhcb can study the time-dependent decay rate of $\Bz\to\KS\pip\pim\gamma$ decays, which permits access of the photon polarisation through the $\mathcal{S}_\CP$ term.
With $\mathcal{O}(1000)$ signal events in Run 1, around $35$k and $200$k are expected at the end of Run~4 and \upgradetwo, respectively
($1.75$k and $10$k when considering the flavour tagging efficiency), opening the doors to a very competitive measurement of $\mathcal{S}_\CP$ in the \Bz system.

Another way to study the photon polarisation is through the angular correlations among the three-body decay products of a kaonic resonance in \decay{\B}{K_{\mathrm{res}}(\to K\pi\pi)\gamma},
which allows the direct measurement of the photon polarisation parameter in the effective radiative weak Hamiltonian~\cite{Gronau:2002rz}. 
As a first step towards the photon polarisation measurement, \lhcb observed nonzero photon polarisation for the first time by studying the photon angular distribution in bins of $\Kp\pim\pip$ invariant mass\ \cite{LHCb-PAPER-2014-001}, but the determination of the value of this polarisation could not be performed due to the lack of knowledge of the hadronic system.
To overcome this problem, a method to measure the photon polarisation using a full amplitude analysis of \decay{\B}{K\pi\pi\gamma} decays is currently under development~\cite{Bellee:2018}, with an expected statistical sensitivity on the photon polarisation parameter of $\sim5\%$ in the charged mode with the Run 1 dataset.
The extrapolation of the precision to $300\,\invfb$  results in a statistical precision better than $1\%$, and hence control of the systematic uncertainties will be crucial.

The polarisation of the photon emitted in $b\to s\gamma$ transitions can also be accessed via semileptonic $b\to s\ell\ell$ transitions, for example in the decay $B^0\to\Kstarz\ell^+\ell^-$. Indeed, as mentioned in 
previous sections, at very low $q^2$ these decays are dominated by the electromagnetic dipole operator ${\cal O}_7^{(\prime)}$. Namely, the longitudinal polarisation fraction ($\FL$) is expected to be below $20\%$ for $q^2<0.2\gevgevcccc$.
In this $q^2$ region, the angle $\phi$ between the planes defined by the dilepton system and the $\Kstarz\to K^+\pi^-$ decay is sensitive to the $b\to s\gamma$ photon helicity.

While the $\Kstarz\mumu$ final state is experimentally easier to select and measure at LHCb, the $\Kstarz\epem$ final state allows $q^2$ values below $4m_\mu^2$ to be probed, 
where the sensitivity to the photon helicity is maximal.
Compared to the radiative channels used for polarisation measurements, the $B^0\to\Kstarz\epem$ final state is fully charged and gives better mass resolution and therefore better separation from partially reconstructed backgrounds.

The sensitivity of this decay channel at LHCb was demonstrated by an angular analysis performed with Run 1 data~\cite{LHCb-PAPER-2014-066}. The angular observables most sensitive to the photon polarisation at low $q^2$ are $A_{\rm T}^{(2)}$ and $A_{\rm T}^{\rm Im}$, as defined in Ref.~\cite{Becirevic:2011bp,LHCb-PAPER-2014-066}. 
Indeed, in the limit $q^2\to 0$, these observables can be expressed by the following functions of $C_{7\gamma}^{(\prime)}$ (assuming NP contributions to be much smaller than $|C_{7\gamma}^{\rm SM}|$):
\begin{equation}
  \label{eq:qSqToZero} 
  A_{\rm T}^{(2)} (\qsq \to 0) \simeq 2 \frac{\Real (C_7^{' \ast}) } {| C_7 |}  \quad {\rm and} \quad  A_{\rm T}^{\rm Im} (\qsq \to 0) \simeq 2 \frac{\Imag (C_7^{' \ast}) } {| C_7 |}.
\end{equation}
In order to maximise the sensitivity to the photon polarisation, 
the angular analysis should be performed as close as possible 
to the low $q^2$ endpoint. However, the events at extremely low $q^2$ have worse $\phi$ resolution (because the two electrons are almost collinear) and are polluted by $B^0\to\Kstarz\gamma$ decays with the $\gamma$ converting in the VELO material. In the Run 1 analysis~\cite{LHCb-PAPER-2014-066} the minimum required $m(\epem)$ was set at 20\mevcc, but this should
be reduced as the \upgradetwo VELO detector will have a significantly lower material budget (multiple scattering is the main effect worsening the $\phi$ resolution). 
Similarly, the background from $\gamma$ conversions will be reduced with a lighter RF-foil or with the complete removal of it in \upgradetwo~\cite{Aaij:2244311}.

Using the signal yield as given in Table~\ref{tab:penguin:LU_extrapolations} leads to the following statistical sensitivities to $A_{\rm T}^{(2)}$ and $A_{\rm T}^{\rm Im}$: 12\% with 8\invfb, 7\% with 23\invfb and 2\% with 300\invfb.
The theoretical uncertainty induced when this observable is translated into a photon polarisation measurement is currently at the level of $2\%$ but should improve by the time of the \upgradetwo analyses.
The current measurements performed with Run 1 data have a systematic uncertainty of order $5\%$ coming mainly from the modelling of the angular acceptance and from the uncertainty on the angular shape of the combinatorial background. The acceptance is independent of $\phi$ at low $q^2$ and its
modelling can be improved with larger simulation samples and using the proxy channel $\Bd\to\Kstarz\jpsi(\to\epem)$. 

Weak radiative decays of \bquark baryons are largely unexplored, with the best limits coming from CDF: $\mathcal{B}(\decay{\Lb}{\Lz\gamma}) < 1.3\times10^{-3}$ at $90\%\,$CL~\cite{Acosta:2002fh}.
They offer a unique sensitivity to the photon polarisation through the study of their angular distributions, and will constitute one of the main topics in the radiative decays programme in the \lhcb \upgradetwo.

With predicted branching fractions of $O(10^{-5}-10^{-6})$, the first challenge for \lhcb will be their observation, as the production of long-lived particles in their decay, in addition to the photon, means in most cases that the $b$-baryon secondary vertex cannot be reconstructed. This makes their separation from background considerably more difficult than in the case of regular radiative \bquark decays.

The most abundant of these decays is \decay{\Lb}{\Lz(\to p\pim)\gamma}, which is sensitive to the photon polarisation mainly\footnote{In the following, we assume that the \Lb (and any other beauty baryon) polarisation is zero~\cite{LHCb-PAPER-2012-057}, removing part of the photon polarisation dependence.} through the distribution of the angle between the proton and the \Lz momentum in the rest frame of the \Lz ($\theta_p$), 
\begin{equation}
    \frac{\operatorname{d}\Gamma}{\operatorname{d}\cos\theta_p} \propto 1 - \alpha_\gamma \alpha_{p,1/2}\cos\theta_p, 
\end{equation}
where $\alpha_\gamma$ is the asymmetry between left- and right-handed amplitudes and $\alpha_{p,1/2}=0.642\pm0.013$~\cite{PDG2018} is the \decay{\Lz}{p\pim} decay parameter.
Using specialised trigger lines for this mode $15-150$ signal events are expected using the Run 2 dataset. Preliminary studies show that a statistical sensitivity to $\alpha_\gamma$ of $(20-25)\%$ is expected with these data, which would be reduced to $\sim15\%$ with $23\,\invfb$ and below $4\%$ with $300\,\invfb$.
In the \lhcb \upgradetwo, the addition of timing information in the calorimeter will be important to be able to study this combinatorial-background dominated decay;
additionally, improved downstream reconstruction would allow the use of downstream \Lz decays, which make up more than $2/3$ of the total signal.

The \decay{\Xib}{\Xi^-(\to\Lz(\to p\pim)\pim)\gamma} decay presents a richer angular distribution, with dependence to the photon polarisation in both the \Lz angle ($\theta_\Lz$) and proton angle ($\theta_p$),
\begin{equation}
    \frac{\operatorname{d}\Gamma}{\operatorname{d}\cos\theta_\Lz\cos\theta_p} \propto 1 - \alpha_\gamma \alpha_\Xi \cos\theta_\Lz + \alpha_{p,1/2}\cos\theta_p\left(\alpha_\Xi-\alpha_\gamma\cos\theta_\Lz\right),
\end{equation}
but the lower $\sigma(\decay{pp}{\Xib})$, combined with a lower reconstruction efficiency due to the presence of one extra track, results in an order of magnitude fewer events than in the \Lb case, making the increase of 
statistics from the \upgradetwo even more relevant. With a similar sensitivity to the photon polarisation to that of \decay{\Lb}{\Lz(\to p\pim)\gamma}, \decay{\Xib}{\Xi^-\gamma} decays will allow this parameter to be probed with a precision of $40\%$ and $10\%$ with $23$ and $300\,\invfb$, respectively.

\subsubsection{Measurements of $b\to c\ell\nu$ including $B_c$ and $b$-baryon prospects}
\label{sec:bclnu:exp}

\begin{figure}[!tb]
\centering
\includegraphics[width=0.7\linewidth]{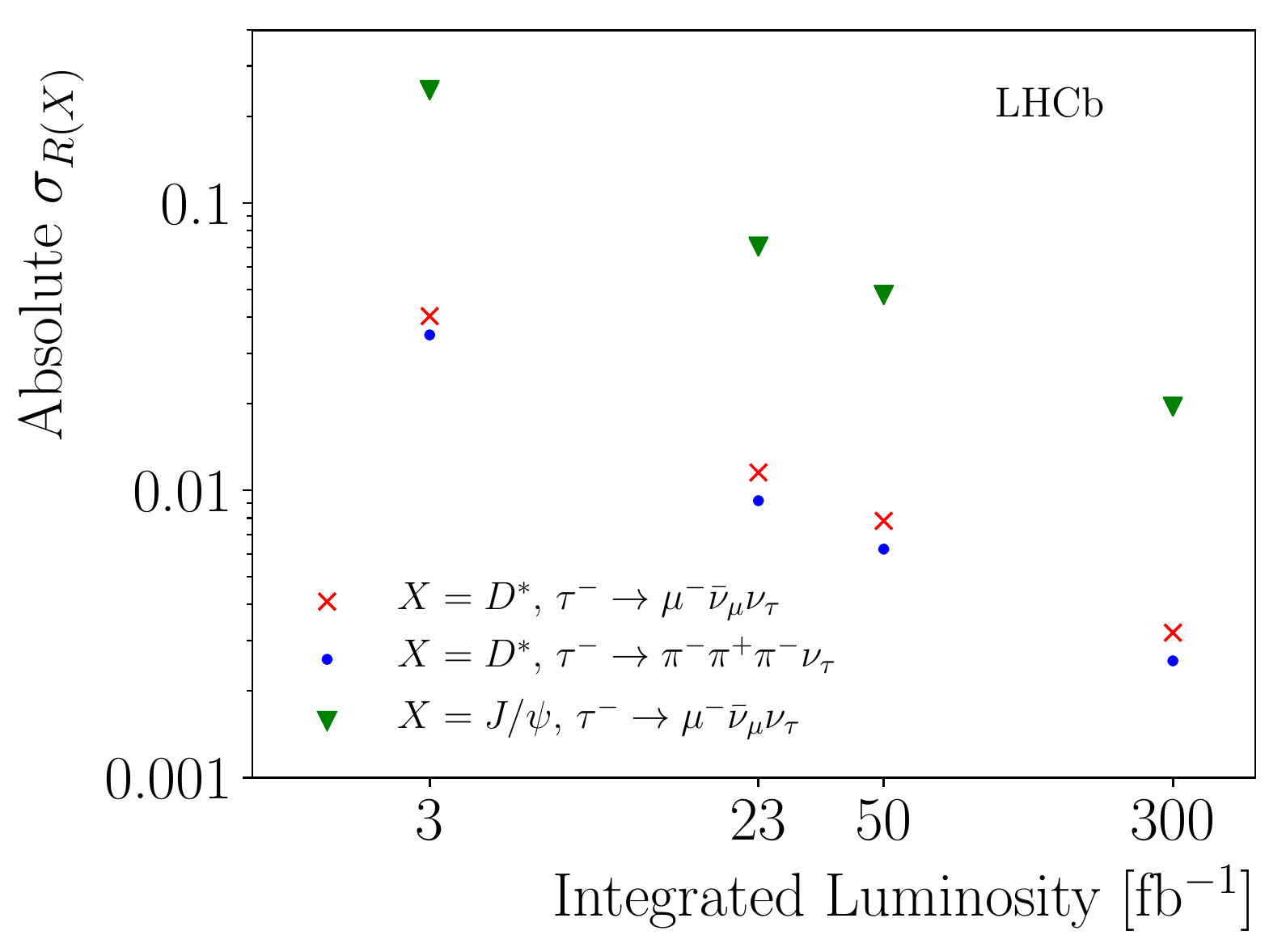}
\caption{
The projected absolute uncertainties on $\mathcal{R}(D^{\ast})$ and $\mathcal{R}(J/\psi)$ from the current sensitivities (at 3\invfb) to 23\invfb, 50\invfb, and 300\invfb.
}
\label{fig:RXproj}
\end{figure} 

LHCb has made measurements of $R(D^{(*)})$ using both muonic ($\tau^{+} \to \mup \nu \nu$) and hadronic ($\tau^{+} \to \pip \pim \pip \nu$) decays of the tau lepton~\cite{LHCb-PAPER-2015-025,LHCb-PAPER-2017-017,LHCb-PAPER-2017-027}.
Due to the presence of multiple neutrinos these decays are extremely challenging to measure.
The measurements rely on isolation techniques to suppress partially reconstructed backgrounds, 
\B meson flight information to constrain the kinematics of the unreconstructed neutrinos, and a multidimensional template fit to determine the signal yield. 
Fig.~\ref{fig:RXproj} shows how the absolute uncertainties on the LHCb muonic and hadronic $\mathcal{R}(D^{\ast})$ measurements are projected to evolve with respect to the current status.
The major uncertainties are the statistical uncertainty from the fit, the uncertainties on the background modelling and the limited size of simulated samples.
A major effort is already underway to commission fast simulation tools.
The background modelling is driven by a strategy of dedicated control samples in the data, and so this uncertainty will continue to improve with larger data samples.
From Run 3 onward it is assumed that, taking advantage of the full software HLT, the hadronic analysis can normalise directly to the $B^0 \to D^{\ast -}\mu^+\nu_{\mu}$  decay,
thus eliminating the uncertainty from external measurements of $\mathcal{B}(B^0 \to D^{\ast -} \pi^+\pi^-\pi^+)$.
It is assumed that all other sources of systematic uncertainty will scale as $\sqrt{\mathcal{L}}$.
With these assumptions, an absolute uncertainty on $\mathcal{R}(D^{\ast})$ of $0.003$ will be achievable for the muonic and hadronic modes with the 300\invfb Upgrade II dataset.

\begin{figure}[!tb]
\centering
\includegraphics[width=0.45\linewidth]{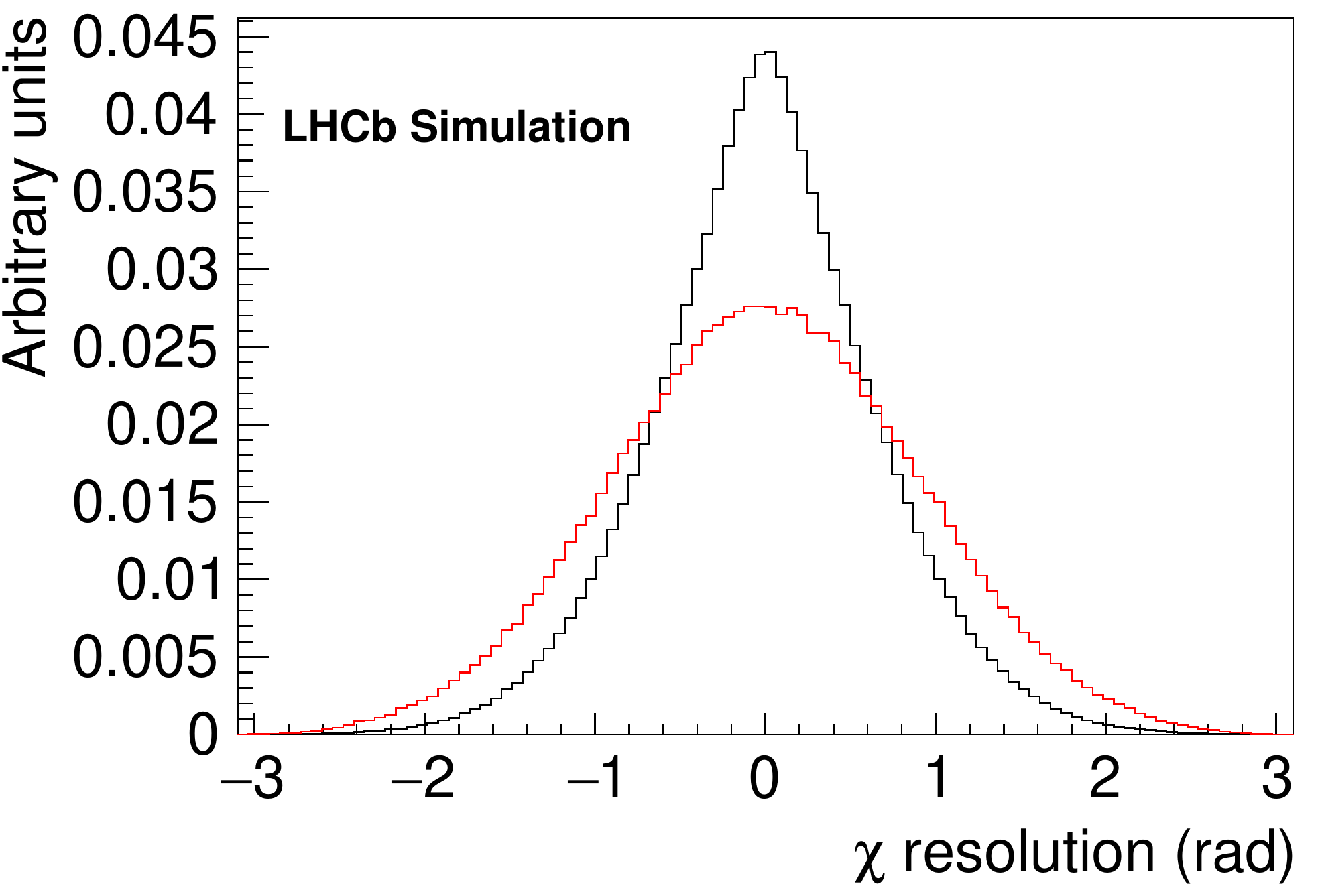}
\includegraphics[width=0.45\linewidth]{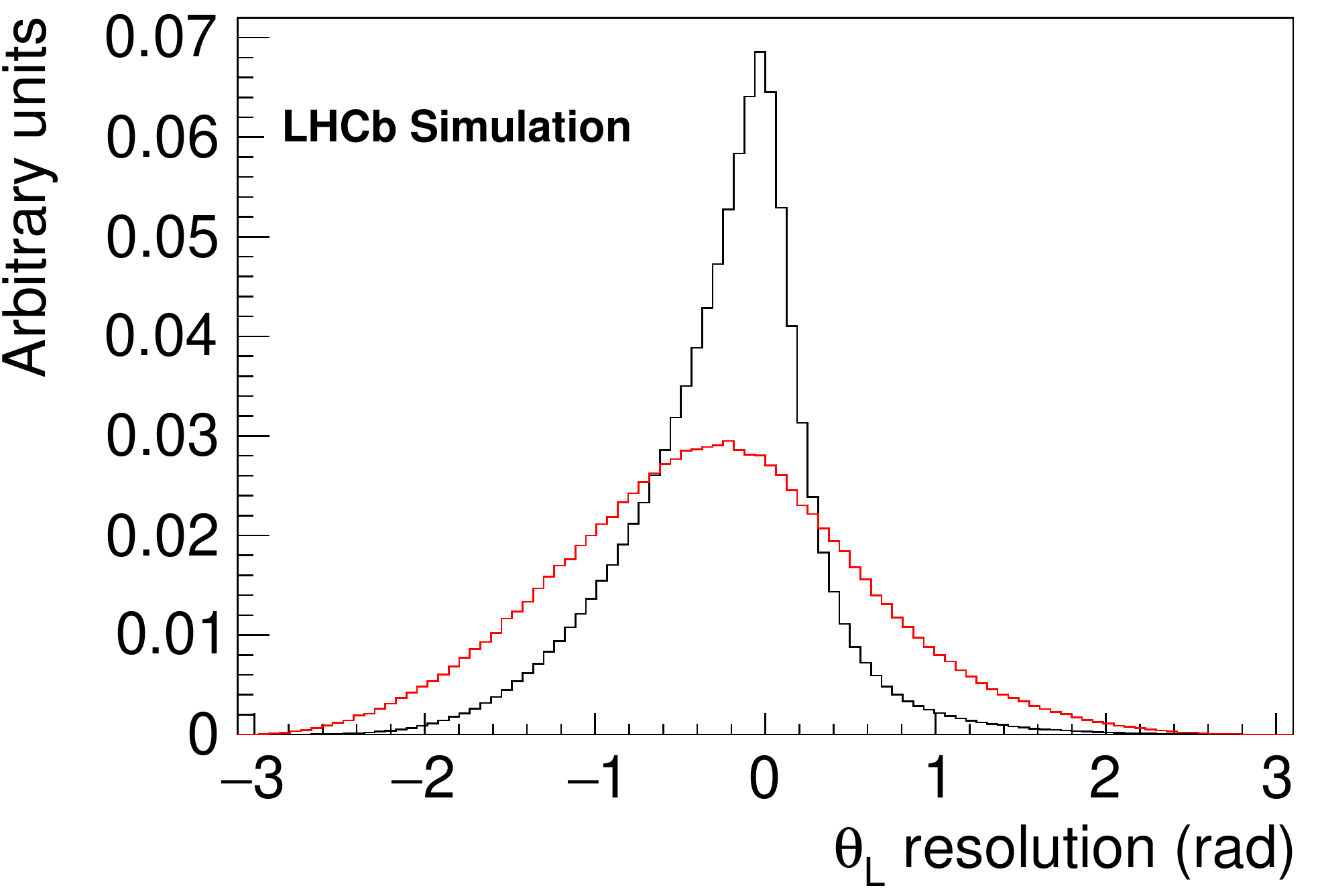}
\includegraphics[width=0.45\linewidth]{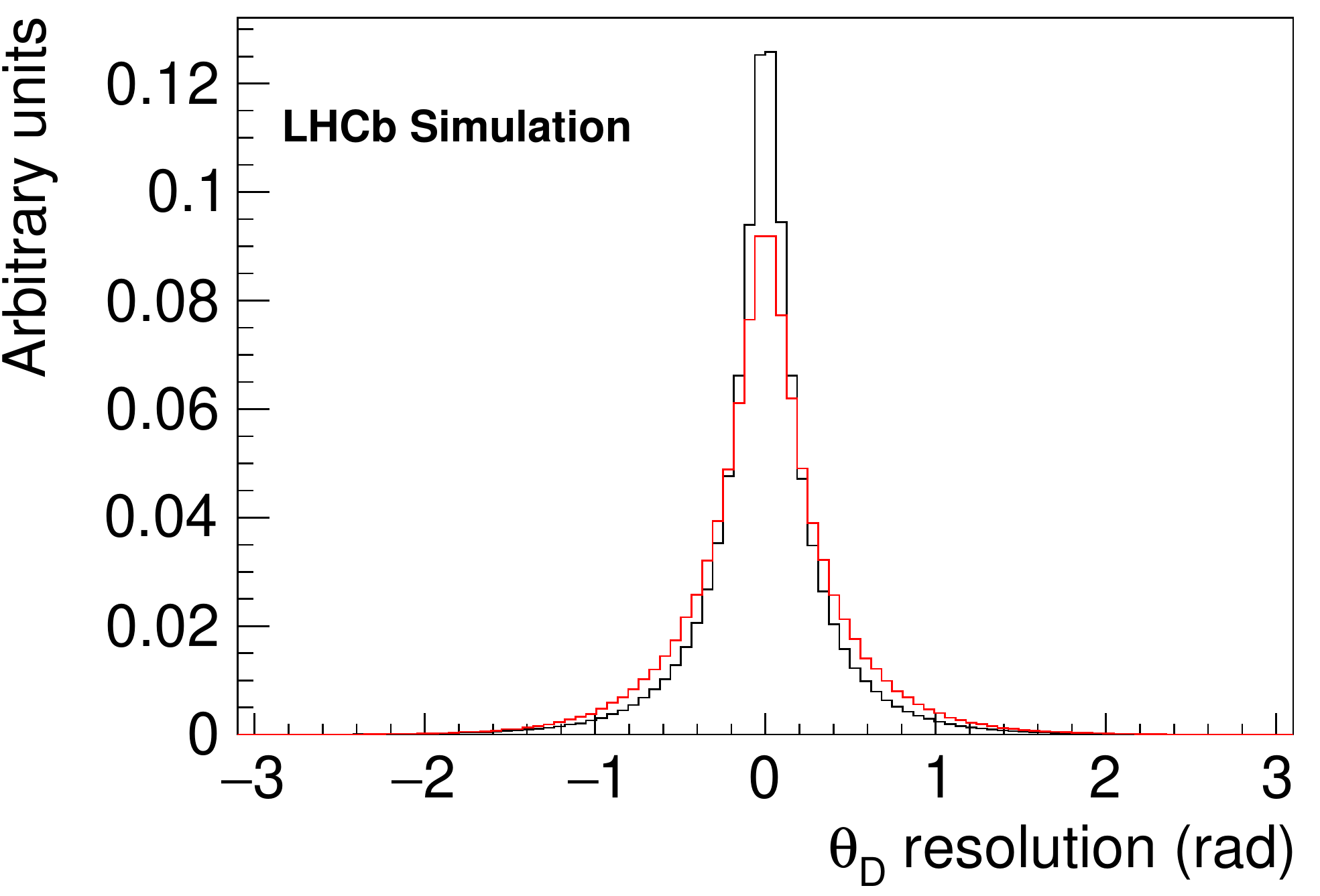}
\caption{
Angular resolution for simulated \dstmunu (black) and \dsttaunu  (red) decays, with $\tau^{+} \to \mup \nu \nu$. 
This demonstrates our ability to resolve the full angular distribution, with some level of statistical dilution.
}
\label{fig:RDAngles}
\end{figure}

On the timescale of \upgradetwo, interest will shift toward new observables  beyond the branching fraction ratio~\cite{Becirevic:2016hea}.
The kinematics of the \dsttaunu decays is fully described by the dilepton mass, and three angles which are denoted $\chi$, $\theta_L$ and $\theta_D$. 
LHCb is capable of resolving these three angles, as can be seen in~\figref{fig:RDAngles}.
However, the broad resolutions demand very large samples to extract the underlying physics. The decay distributions within this kinematic space are governed by the underlying spin structure, and precise measurements of these distributions will allow the different helicity amplitudes to be disentangled.
This can be used both to constrain the Lorentz structure of any potential NP contribution, and to measure the hadronic parameters
governing the \dsttaunu decay, serving as an essential baseline for SM and non-SM studies.
The helicity-suppressed amplitude which presently dominates the theoretical uncertainty on $R(D^{(*)})$ is too strongly suppressed in the 
\dbmunu decays to be measurable, however this can be accessed in the \dbtaunu decay directly.
If any potential NP contributions are assumed not to contribute via the helicity-suppressed amplitude then the combined measurements of \dbmunu and \dbtaunu decays will allow for a fully data-driven prediction for
$R(D^{(*)})$ under the assumption of lepton universality, eliminating the need for any theory input relating to hadronic form factors. 
However, these measurements have yet to be demonstrated with existing data.
This exciting programme of differential measurements needs to be developed on Run~1 and 2 data before any statement is made about the precise sensitivity, 
but it offers unparalleled potential to fully characterise both the SM and non-SM contributions to the $b \to c \tau \nu$ transition.

As measurements in \RDst become more statistically precise, it will become increasingly more 
urgent to provide supplementary measurements in other \bquark-hadron species with different background 
structure and different sources of systematic uncertainties. For example, the \decay{\Bsb}{\Dsp \taum\neub} and $\Bsb\to\Dssp\taum\neub$ decays
will allow supplementary 
measurements at high yields, and do not suffer as badly from cross-feed backgrounds from other 
mesons, unlike, for example, \decay{\Bzb }{ \Dstarp \taum\neub}, where the \Bp and \Bs both contribute to 
the $\Dstarp \mu X$ or $\Dstarp \threepim X$ final states. 
Furthermore, the comparison of decays with different spins of the $b$ and $c$ hadrons can enhance our sensitivity 
to different NP scenarios~\cite{Azatov:2018knx,LHCb-PAPER-2017-016}.
No published measurements exist for the $\Bs$ case yet, but based on known relative efficiencies and assuming 
the statistical power of this mode tracks \RDorDst, we expect less than $6\%$ relative uncertainty 
after Run~3, and $2.5\%$ with the \upgradetwo data, where limiting systematic uncertainties 
are currently expected to arise from corrections to 
simulated pointing and vertex resolutions, from knowledge of particle identification efficiencies, 
and from knowledge of the backgrounds from random combinations of charm and muons. It is conceivable 
that new techniques and control samples could further increase the precision of these measurements. 

Methods 
are currently under development for separating the \decay{\Bsb}{\Dssp\ell^{-}\neub} and \decay{\Bsb}{\Dsp\ell^{-}\neub} modes, and given the
relative slow pion (\decay{\Dstarp }{ \Dz\pip}) and soft photon (\decay{\Dssp }{ \Dsp\gamma}) efficiencies, the precision in \decay{\Bsb}{ \Dsp \tau \nu} decays can be expected to exceed that in
\decay{\Bsb }{ \Dssp \tau \nu}, the reverse of the situation for $\RDorDst$.
An upgraded ECAL would extend the breadth and sensitivity of $\mathcal{R}(D^{*(*)+}_{s})$ measurements possible
in the \upgradetwo\ scenario above and beyond the possible benefits of improved neutral isolation in 
$\RD$ or $\mathcal{R}(\Dsp)$ measurements.

Of particular interest are the semitauonic decays of
\bquark baryons and of \Bcp mesons.
The former provides probes of entirely new Lorentz structures of NP operators
which pseudoscalar to pseudoscalar or vector transitions simply do not access.
The value of probing this supplementary space of couplings has already been demonstrated by
LHCb with its Run~1
measurement of $|\Vub|$ via the decay \decay{\Lb}{\proton\mun\neub}, which places strong constraints on 
right-handed currents sometimes invoked to explain the inclusive-exclusive tensions in that quantity. By the end of Run~3, it is expected that the relative uncertainty for $\mathcal{R}(\Lc)$ will reach below $4\%$, and $2.5\%$ by the end of \upgradetwo. 
A further exciting prospect is the study of $b \to u\mu \nu$ decays, which have been beyond experimental reach thus far.
For example the decay $B^+ \to p\bar{p}\mu\nu$ offers a clean experimental signature. 
Our capabilities with this decay could benefit from the enhanced low momentum proton identification with the TORCH subdetector.

Meanwhile, the \decay{\Bcp }{ \jpsi \taum \neub} decay
is an entirely unique state among the flavoured mesons as the bound state of two distinct 
flavors of heavy quark, and, through its abundant decays to charmonium final states, provides a 
highly efficient signature for triggering and reconstruction at high instantaneous luminosities. 
Measurements of \decay{\Bcp }{ \jpsi \ell \neub} decays involve a trade-off between the approximately 100 times 
smaller production cross-section for \Bcp verses the extremely efficient \decay{\jpsi}{\mup\mun} signature in the LHCb 
trigger. For illustration, in Run~1, LHCb reconstructed 
and selected 19\,000 \decay{\Bcp }{ \jpsi \mun \neub} decays, compared with 360\,000 \decay{\Bzb}{\Dstarp\mun\neub}. This resulted in a measurement of $\mathcal{R}(\jpsi)=0.71\pm 0.17\pm 0.18$ \cite{LHCb-PAPER-2017-035}. As a result of the smaller production cross-section, the muonic measurements have large backgrounds from \decay{h}{ \mu} misidentification from the relatively abundant \decay{\B }{ \jpsi X_h} decays, where $X_h$ is any collection of hadrons, and so they
are very sensitive to the performance of the muon system and PID algorithms in the future. Here it is assumed that 
it will be possible to achieve similar performance to Run 1 in the upgraded system. 

To project the sensitivity for \decay{\Bcp }{ \jpsi \taum \neub} based on Ref.~\cite{LHCb-PAPER-2017-035}, it is assumed that 
all the systematic uncertainties can be reduced with the size of the input data except for those that were assumed not to
scale with data for the previous predictions. For these, we assume that they can be reduced down until they reach 
the same absolute size as the corresponding systematic uncertainties in the Run~1 muonic $\RDst$ analysis. In addition, it is
assumed that sometime in the 2020s lattice QCD calculations of the form factors for this process will allow 
the systematic uncertainty due to signal form factors to be reduced by an additional factor of two. This results in
a projected absolute uncertainty for the muonic mode of $0.07$ at the end of Run~3 
and $0.02$ by the end of \upgradetwo,  as can be seen in Fig.~\ref{fig:RXproj}.
Measurements in the hadronic mode can be expected to reach similar sensitivities.

\subsubsection{Searches for LFV, LNV, BNV and interplay with tests of LU}

The LHCb collaboration has recently published~\cite{LHCb-PAPER-2017-031} the world's best limits on the branching fractions of the \BsToEMu
and \BdToEMu decays using the first 3\invfb collected in 2011 and 2012 at 7 and 8\tev respectively. The acceptance of the \BsToEMu decays can be affected by the relative contribution of the two \Bs mass eigenstates to the total decay amplitude, due to their large lifetime difference. Therefore, the upper limit on the branching fraction of \BsToEMu decays is evaluated in two extreme hypotheses: where the amplitude is completely dominated by the heavy eigenstate or by the light eigenstate. The results are  $\BF(\BsToEMu) < 6.3\,(5.4) \times 10^{-9}$ and $\BF(\BsToEMu) < 7.2\,(6.0) \times 10^{-9}$ at $95\%\,(90\%)$ CL, respectively. The limit for the branching fraction of the \Bd mode is  $\BF(\BdToEMu) < 1.3\,(1.0) \times 10^{-9}$ at $95\%\,(90\%)$ CL.

Assuming similar performances in background rejection and signal retention as in the current analysis, at the end of Run~4 the LHCb
experiment will be able to probe branching fractions of \BsToEMu and \BdToEMu decays down to $8\times 10^{-10}$
and $2\times 10^{-10}$, respectively. The additional statistics accumulated during the \upgradetwo data taking period will push down
these limits to $3\times 10^{-10}$ and $9\times 10^{-11}$ respectively, close to the interesting region where NP effects may appear. The \upgradetwo improvement in electron  reconstruction will be very important in attaining, or exceeding, this goal. 

An upper limit on the \BdToTauMu channel has been already set by $\babar$: $\BR\left( \BdToTauMu\right)< 2.2\times 10^{-5} $ at $90\%$~CL~\cite{Aubert:2008cu}. The first search for the \BsToTauMu channel is in progress in LHCb and the results are expected soon on data recorded in 2011 and 2012 using the \tauppp and \tauppppi0 decay modes.
Given the presence of a neutrino that escapes detection this kind of analysis is much more complicated than those investigating electron or muon final states. A specific reconstruction technique is used in order to infer the energy of the $\nu$, taking advantage of the known $\tau$ vertex position given by the $3\pi$ reconstructed vertex. 
This way, the complete kinematics of the process can be solved up to a two-fold ambiguity. LHCb expects to reach sensitivities of a few times $10^{-5}$ with the Run~1 and 2 data sets. Extrapolating the current measurements to the \upgradetwo LHCb could reach  $\BR\left( \BdToTauMu\right)< 3\times 10^{-6} $ at $90\%$~CL. The mass reconstruction technique depends heavily on the uncertainty on the primary and the $\tau$ decay vertices, hence improvement in the tracking system in \upgradetwo, including a removal or reduction in material of the VELO RF foil, will be very valuable.

In many generic NP models with LFUV, CLFV decays of $b$-hadrons can be linked with the anomalies recently measured in $b\rightarrow s\ell\ell$ decays~\cite{LHCb-PAPER-2014-024,LHCb-PAPER-2017-013,LHCb-PAPER-2015-051}. If NP indeed allows for CLFV then the branching fractions of $B\to K \ell \ell^{'}$  or $\Lb \to \Lz \ell \ell^{'}$ will be enhanced with respect to their purely leptonic counterparts, since the helicity suppression is lower. Furthermore, if observed, they would allow the measurement of more observables with respect to the lepton flavour violating decays discussed in the previous sections, thanks to their multi-body final states and, in the case of $\Lb$, to the non-zero initial spin. 

The current  limits set by the \bfactories on the branching fractions of $B \to Ke \mu$ and $B \to K\tau \mu$ decays are $< 13\times 10^{-8}$~\cite{Aubert:2006vb} and $< 4.8\cdot 10^{-5}$~\cite{Lees:2012zz}  at 90\% confidence level, respectively.  

At LHCb, searches for $\decay{\Bu}{\Kp e^\pm\mu^\mp}$, $\decay{\Bd}{\Kstarz\tau^\pm\mu^\mp}$, $\decay{\Bu}{\Kp\tau^\pm\mu^\mp}$ and $\decay{\Lb}{\Lz e^\pm\mu^\mp}$ are ongoing. These searches are complementary, as charged lepton flavour violation couplings among different families are expected to be different.
The analyses involving $\tau$ leptons reconstruct candidates via the $\decay{\taum}{\pim\pim\pip\nu_{\tau}}$ channel,
which allows the reconstruction of the $\tau$ decay vertex.\footnote{It should be noted that searches for $\decay{\Bu}{\Kp\tau^\pm\mu^\mp}$ from $B_{s2}^*$ without $\tau$ reconstruction can give complementary information.} 
All these decays contain at least one muon, which is used to efficiently trigger on the event. Usually, since these decays involve combinations of leptons that are not allowed in the SM, the backgrounds can be kept well under control,  leaving very clean samples only polluted by candidates formed by the random combinations of tracks. This combinatorial effect is higher for the channel with a $\tau$ in the final state decaying into three charged pions.  The other relevant background  comes from chains of semileptonic decays, where two or more neutrinos are emitted and therefore combinations of leptons of different flavours are possible. These decays have typically a low reconstructed invariant mass, due to the energy carried away by the neutrinos, and so they do not significantly pollute the signal region. 

The expected upper limits at LHCb using the first $9 \invfb$ of data taken are
$\mathcal{O}(10^{-9})$ and $\mathcal{O}(10^{-6})$ for the $\decay{\Bu}{\Kp e^\pm\mu^\mp}$ and $\decay{\Bd}{\Kstarz\tau^\pm\mu^\mp}$ decays respectively, at 90\% confidence level.
The limit for $\decay{\Bu}{\Kp\tau^\pm\mu^\mp}$ is expected to be similar to $\decay{\Bd}{\Kstarz\tau^\pm\mu^\mp}$.
The sensitivity of these analyses scales almost linearly with luminosity  for $\decay{\Bu}{\Kp e^\pm\mu^\mp}$, 
and with the square root of the luminosity for $\decay{\Bd}{\Kstarz\tau^\pm\mu^\mp}$.
In both cases, the expected limits using the \upgradetwo data are in the region of interest of the models currently developed for explaining the $B$ anomalies, so they will provide strong constraints on the NP scenarios with CLFV 

Experimentally, Lepton-Number Violating (LNV) and Baryon-Number Violating (BNV) measurements are null searches, so sensitivity is assumed
to scale linearly with luminosity $\mathcal L$ when the background
is negligible and as $\sqrt \mathcal L$ if the background is significant. LHCb has already published searches in certain channels, and others are
in progress:
\begin{itemize}
  \item[$\bullet$] Searches for LNV in various $B$-meson decays of the form
    $B \to X \mu^+ \mu^+$, where $X$ is a system of one or more hadrons.
    The principal motivation is the sensitivity to contributions from Majorana neutrinos~\cite{Atre:2009rg},
    which may be on-shell or off-shell, depending on the decay mode.
    The published results consist of
      searches for $\decay{\Bp}{\Km \mu^+ \mu^+}$, $\decay{\Bp}{\pim \mu^+ \mu^+}$ and $\decay{\Bu}{D^+_{(s)}\mun\mun}$~\cite{LHCb-PAPER-2011-009, LHCb-PAPER-2011-038, LHCb-PAPER-2013-064}. A limit of $\mathcal{B}(B^+ \to \pi^- \mu^+ \mu^+) < 4 \times 10^{-9}$ is set at the 95\% confidence level, along with more detailed limits as a function of the Majorana neutrino mass. Since the combinatorial background was found to be low but not negligible with the Run~1 data, we estimate that the limit can be improved by a factor of ten with the full \upgradetwo dataset.
  \item[$\bullet$] Search for BNV in \Xibz oscillations~\cite{LHCb-PAPER-2017-023}.
    Six-fermion, flavour-diagonal operators, involving two fermions from each
    generation, could give rise to BNV without violating the nucleon stability
    limit~\cite{Smith:2011rp,Durieux:2012gj}.
    Since the \Xibz~($bsd$) has one valence quark from each generation, it could
    couple directly to such an operator and oscillate to \Xibzbar.
    The published search used the Run~1 data and set a lower limit on the oscillation
    period of 80\ps. Since events are tagged by decays of
    the \XibPrimeMinus and \XibStarMinus resonances, with the former being particularly
    clean, and since the analysis also uses the decay-time distribution of events,
    the sensitivity is expected to scale linearly. 
    Although the decay mode used in the published analysis is hadronic
    ($\Xibz \to \Xicp \pim$), future work could also benefit from the
    lower-purity but higher-yield semileptonic mode $\Xibz \to \Xicp \mun \neumb$.
  \item[$\bullet$] $\Lc \to \antiproton \mu^+ \mu^+$.
    This channel has previously been investigated at the $e^+ e^-$ \bfactories.
    The current upper limit, obtained by \babar~\cite{Lees:2011hb}, is
    $\mathcal{B}(\Lc \to \antiproton \mup \mup) < 9.4 \times 10^{-6}$
    at the 90\% confidence level.
    With Run~1 and~2 data alone, it should be possible to reduce this
    to $1 \times 10^{-6}$. Further progress depends on the background
    level, but an additional factor of 5--10 with the full \upgradetwo statistics
    is likely.
  \item[$\bullet$] $\Lc \to \mu^+ \mu^- \mu^+$.
    Experimentally, this is a particularly promising decay mode:
    the final state with three muons is very clean, and there are
    no known sources of peaking background.
    This search could be added for little extra effort to the
    $\taum \to \mup \mun \mun$ search described in the
    preceding section.
\end{itemize}

%% file: section8/section.tex
\section{The top quark and flavour physics}
\label{sec:top:quark}

Among the SM fermions the top quark stands out. It has a large mass and an ${\mathcal O}(1)$ coupling to the Higgs, quite distinct from any other SM fermion. Studying top quark  properties may shed light on the resolution of the SM flavour puzzle, or at least as to why one and only one Yukawa coupling is large. The large Higgs-top coupling is also the reason for the weak scale hierarchy problem to be so acute -- the quadratic divergent corrections to the Higgs mass are driven almost completely by this coupling. BSM models addressing the hierarchy problem may thus well leave an imprint in the top quark properties and decays.  For instance, the FCNC top decays, $t\to c\gamma, cZ, cg$, are null tests of the SM and are used as BSM probes.

Top quark also directly enters the flavour phenomenology. Loops with the top quark are 
responsible for the largest short-distance contributions to the down-quark FCNCs. 
The SM predictions are thus controlled by the flavour couplings of the top -- with $B$ and $K$ transitions determining the CKM matrix elements $V_{tb}, V_{ts}, V_{td}$ through these virtual effects. Determining $V_{tx}$ directly from high \pt transitions, as well as the structure of the $Wtb$ vertices, can then serve as independent consistency checks of the SM. 

\vtwo{Top quarks can also be used as a clean source of tagged $B$ mesons. Ref. \cite{Gedalia:2012sx}, for instance, suggested to use top-pair events, where at least one of the tops decays semi-leptonically, to measure CP violation in heavy flavor mixing and decays. This measurement was then performed for the first time at by ATLAS in Ref. \cite{Aaboud:2016bmk}.}

The LHC is already a top factory and the currently available statistics has allowed ATLAS and CMS to perform a vast campaign of top related measurements. However, the larger number of top quarks at HL-LHC and HE-LHC will open new possibilities for precise measurements of top-quark properties and for significant 
improvements probing NP, such as the rare FCNC decays. 

\subsection{Global effective-field-theory interpretation of top-quark FCNCs} \label{sec:top:FCNC} 
{\it\small Authors (TH): Gauthier Durieux, Teppei Kitahara, Cen Zhang.}
{
\newcommand{\hc}[1]{#1}%

\newcommand{\ifb}{\ensuremath{\text{fb}^{-1}}}
\newcommand{\iab}{\ensuremath{\text{ab}^{-1}}}

\newcommand{\FDF}{(\varphi^\dagger i\!\!\overleftrightarrow{D}_\mu\varphi)}
\newcommand{\FDFI}{(\varphi^\dagger i\!\!\overleftrightarrow{D}^I_\mu\varphi)}

\newcommand*{\tmp}[4]{\ensuremath{%
	{#4%
	\ifx\empty#3\empty\ifx\empty#1\empty\else^{#1}\fi\else^{#1(#3)}\fi%
	\ifx\empty#2\empty\else_{#2}\fi}}}%
\newcommand{\qq }[4][]{\tmp{#2}{#3}{#4}{#1{O}}}
\newcommand*{\ccc}[4][]{\tmp{#2}{#3}{#4}{#1{c}}}

\newcommand{\ReIm}{{}_{\Re}^{[\Im]}\!}
\newcommand*{\sw}{s_W}
\newcommand*{\cw}{c_W}

\subsubsection{Effective operators}

Starting from an Effective Field Theory (EFT) with full $SU(3)_C\times SU(2)_L\times U(1)_Y$ gauge symmetry and matter content of the SM,
one can show that odd-dimensional operators all violate baryon or lepton 
numbers~\cite{Degrande:2012wf}. Imposing the conservation of these quantum numbers, the leading higher dimensional operators of the SM arise at dimension six.
We follow the top-quark EFT conventions set by the LHC TOP WG in 
Ref.~\cite{AguilarSaavedra:2018nen}. The LHC TOP WG employs linear
combinations of Warsaw-basis operators~\cite{Grzadkowski:2010es} which appear in
interactions with physical fields after electroweak symmetry breaking.

The operators contributing to top-quark FCNC processes fall into
several categories. We consider operators involving
exactly two quarks, as well as those involving two quarks and two leptons.
Operators containing four quarks only start contributing at next-to-leading order in QCD in most of the measurements we consider ($pp\to tj$ is the exception).
The corresponding Warsaw-basis operators are~\cite{AguilarSaavedra:2018nen} 
\begin{equation}
	\begin{aligned}
	\hc{\qq{}{u\varphi}{ij}}
	&=\bar{q}_i u_j\tilde\varphi\: (\varphi^{\dagger}\varphi)
	,\\
	\qq{1}{\varphi q}{ij}
	&=\FDF (\bar{q}_i\gamma^\mu q_j)
	,\\
	\qq{3}{\varphi q}{ij}
	&=\FDFI (\bar{q}_i\gamma^\mu\tau^I q_j)
	,\\
	\qq{}{\varphi u}{ij}
	&=\FDF (\bar{u}_i\gamma^\mu u_j)
	,\\
	\hc{\qq{}{\varphi ud}{ij}}
	&=(\tilde\varphi^\dagger iD_\mu\varphi)
	  (\bar{u}_i\gamma^\mu d_j)
	,\\
	\hc{\qq{}{uW}{ij}}
	&=(\bar{q}_i\sigma^{\mu\nu}\tau^Iu_j)\:\tilde{\varphi}W_{\mu\nu}^I
	,\\
	\hc{\qq{}{dW}{ij}}
	&=(\bar{q}_i\sigma^{\mu\nu}\tau^Id_j)\:{\varphi} W_{\mu\nu}^I
	,\\
	\hc{\qq{}{uB}{ij}}
	&=(\bar{q}_i\sigma^{\mu\nu} u_j)\quad\:\tilde{\varphi}B_{\mu\nu}
	,\\
	\hc{\qq{}{uG}{ij}}
	&=(\bar{q}_i\sigma^{\mu\nu}T^Au_j)\:\tilde{\varphi}G_{\mu\nu}^A
\end{aligned}\quad
\begin{aligned}
	\qq{1}{lq}{ijkl}
	&=(\bar l_i\gamma^\mu l_j)
	  (\bar q_k\gamma^\mu q_l)
	,\\
	\qq{3}{lq}{ijkl}
	&=(\bar l_i\gamma^\mu \tau^I l_j)
	  (\bar q_k\gamma^\mu \tau^I q_l)
	,\\
	\qq{}{lu}{ijkl}
	&=(\bar l_i\gamma^\mu l_j)
	  (\bar u_k\gamma^\mu u_l)
	,\\
	\qq{}{eq}{ijkl}
	&=(\bar e_i\gamma^\mu e_j)
	  (\bar q_k\gamma^\mu q_l)
	,\\
	\qq{}{eu}{ijkl}
	&=(\bar e_i\gamma^\mu e_j)
	  (\bar u_k\gamma^\mu u_l)
	,\\
	\hc{\qq{1}{lequ}{ijkl}}
	&=(\bar l_i e_j)\;\varepsilon\;
	  (\bar q_k u_l)
	,\\
	\hc{\qq{3}{lequ}{ijkl}}
	&=(\bar l_i \sigma^{\mu\nu} e_j)\;\varepsilon\;
	  (\bar q_k \sigma_{\mu\nu} u_l)
	,\\
	\hc{\qq{}{ledq}{ijkl}}
	&=(\bar l_i e_j)
	  (\bar d_k q_l)
  \end{aligned}
\end{equation}
The operators $\qq{}{\varphi ud}{}$, $\qq{}{dW}{}$, and $\qq{}{ledq}{}$,  only contribute to charged top-quark currents (not considering SM electroweak corrections) and are therefore not relevant for our purposes.
The EFT degrees of freedom appearing in top-quark FCNC processes were defined in Appendices~E.1-2 of Ref.\,\cite{AguilarSaavedra:2018nen}. They are:%
{%
\renewcommand*{\cc }[4][]{\tmp{#2}{#3}{#4}{#1{C}}}%
\begin{align}
\ccc{[I]}{t\varphi}{3a} &\equiv \ReIm\{\cc{}{u\varphi}{3a}\}, &\ccc{[I]}{uA}{3a} &\equiv \ReIm\{\cw\cc{}{uB}{3a}+\sw\cc{}{uW}{3a}\},
\\
\ccc{[I]}{t\varphi}{a3} &\equiv \ReIm\{\cc{}{u\varphi}{a3}\},&\ccc{[I]}{uA}{a3} &\equiv \ReIm\{\cw\cc{}{uB}{a3}+\sw\cc{}{uW}{a3}\},
\\
\ccc{-[I]}{\varphi q}{3+a}  &\equiv \ReIm\{\cc{1}{\varphi q}{3a}-\cc{3}{\varphi q}{3a}\},
&\ccc{[I]}{uZ}{3a} &\equiv \ReIm\{-\sw\cc{}{uB}{3a}+\cw\cc{}{uW}{3a}\},
\\
\ccc{[I]}{\varphi u}{3+a}  &\equiv \ReIm\{\cc{}{\varphi u}{3a}\},
&\ccc{[I]}{uZ}{a3} &\equiv \ReIm\{-\sw\cc{}{uB}{a3}+\cw\cc{}{uW}{a3}\},
\end{align}
as well as 
$\ccc{[I]}{uG}{3a} \equiv \ReIm\{\cc{}{uG}{3a}\}$, 
$\ccc{[I]}{uG}{a3} \equiv \ReIm\{\cc{}{uG}{a3}\},$ and 
\begin{align}
\ccc{-[I]}{lq}{\ell,3+a} &\equiv \ReIm\{\cc{-}{lq}{\ell\ell 3a}\}, &\ccc{S[I]}{lequ}{\ell,3a} &\equiv \ReIm\{\cc{1}{lequ}{\ell\ell 3a}\},
\\
\ccc{[I]}{eq}{\ell,3+a} &\equiv \ReIm\{\cc{}{eq}{\ell\ell 3a}\},& \ccc{S[I]}{lequ}{\ell,a3} &\equiv \ReIm\{\cc{1}{lequ}{\ell\ell a3}\},\\
\ccc{[I]}{lu}{\ell,3+a} &\equiv \ReIm\{\cc{}{lu}{\ell\ell 3a}\},& \ccc{T[I]}{lequ}{\ell,3a} &\equiv \ReIm\{\cc{3}{lequ}{\ell\ell 3a}\},\\
\ccc{[I]}{eu}{\ell,3+a} &\equiv \ReIm\{\cc{}{eu}{\ell\ell 3a}\},& \ccc{T[I]}{lequ}{\ell,a3} &\equiv \ReIm\{\cc{3}{lequ}{\ell\ell a3}\}.
\end{align}
}%

\begin{figure}[tb]\centering
\includegraphics[width=.6\linewidth]{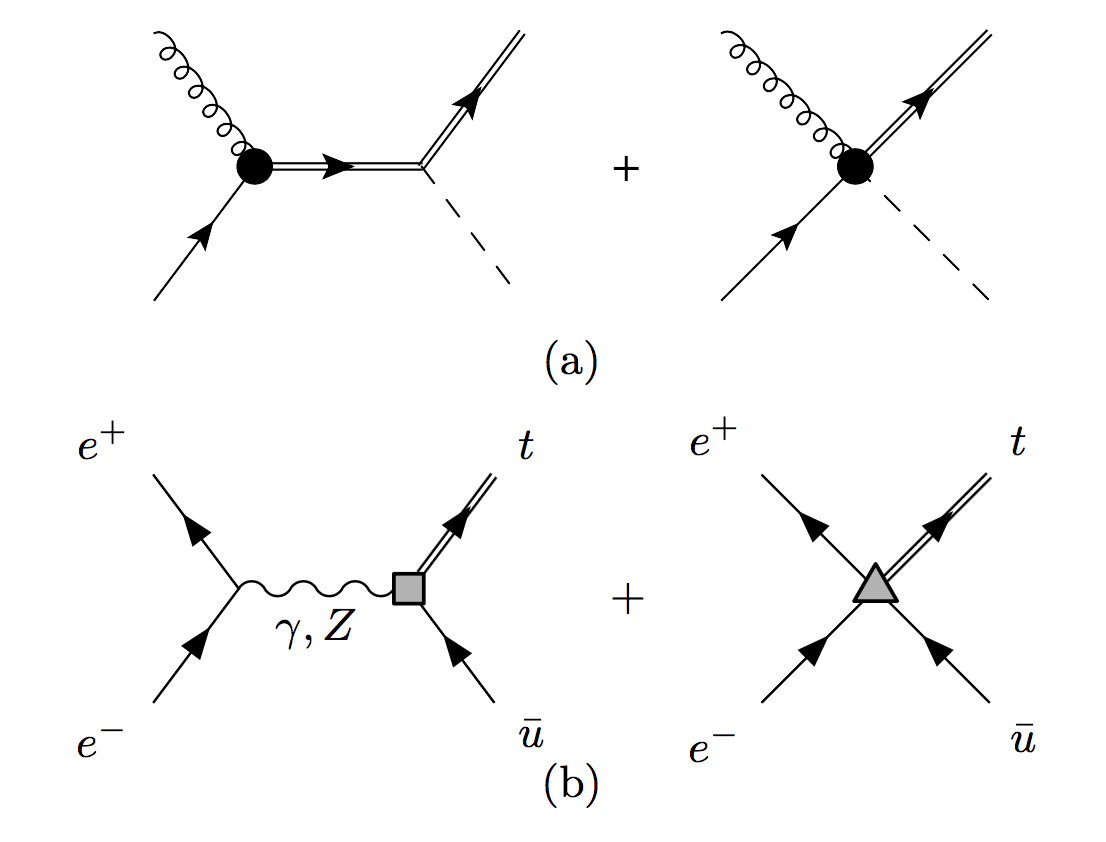}
\caption{Effective operators give rise to four-point contact interactions that
are overlooked in the approach with anomalous couplings,  although they contribute to
FCNC processes at the same order in $1/\Lambda^2$ as the three-point ones.
Representative diagrams are shown for $ug\to th$ production (or radiative $t\to
hug$ decay), and $e^+e^-\to t\,\bar u$ (or $t\to u\,e^+e^-$ decay). Figure taken from Ref.~\cite{Durieux:2014xla}.}
\label{fig:missing_four_point_operators}
\end{figure}

Compared to the anomalous coupling parametrization, the EFT approach has two
features that are worth emphasizing. First, it includes four-fermion operators,
which have  been unduly neglected in most
experimental analyses (apart from Ref.~\cite{Achard:2002vv}).  
Second, the EFT approach captures the correlations between interaction terms that derive from electroweak gauge invariance. For instance, the $\bar
t\sigma^{\mu\nu}T^Aq \; h\; G_{\mu\nu}^A$ and $\bar t\sigma^{\mu\nu}T^Aq\;
G_{\mu\nu}^A$ interactions arise from the same
$O_{uG}$ operator and their coefficients are thus related. 
In Fig.~\ref{fig:missing_four_point_operators}, we show examples of four-point interactions contributing to single top-quark FCNC production.
Correlations also arise from the fact that left-handed down- and up-type quarks
belong to a single
gauge-eigenstate doublet. Operator coefficients measurable in $B$-meson physics
are thus related to those relevant to top-quark physics (see Secs.~\ref{sec:B:K:top} and \ref{sec:8:2:Indirect}).

\subsubsection{Theory predictions}
Because the LHC is a hadron collider, theory predictions at LO are
often not sufficient when an accurate interpretation of observables in terms of
theory parameters is needed. Typical NLO QCD
corrections in top-decay processes~\cite {Kidonakis:2003sc, Zhang:2008yn,
Drobnak:2010wh, Drobnak:2010by, Zhang:2010bm, Zhang:2014rja}
amount to approximately $10\%$, while in production processes they can reach
between about $30\%$ and $80\%$~\cite{Liu:2005dp, Gao:2009rf, Zhang:2011gh,
Li:2011ek, Wang:2012gp}. 
Theory predictions for top-quark FCNC processes are in general available at NLO
accuracy in QCD.  Complete results at NLO in QCD for top-quark FCNC decays
through two-quark and two-quark--two-lepton operators can be found in
Ref.~\cite{Zhang:2014rja}.
Single top-quark production associated with a neutral gauge boson, $\gamma$, $Z$, or
the scalar boson $h$ have also been studied.  Two-quark operators have been
implemented in the {\tt FeynRules}/{\tt MadGraph5\_aMC@NLO} simulation
chain~\cite{Alloul:2013bka,Degrande:2011ua,Alwall:2014hca}, allowing for
automated NLO QCD predictions matched to parton shower. 
Details on this implementation have been presented in Ref.~\cite{Degrande:2014tta}.
Two-quark-two-lepton operators are now also
available in {\tt MadGraph5\_aMC@NLO} (see Ref.~\cite{Durieux:2018tev}). Finally, the direct top-quark
production with decay process, $pp\to bW^+$, involves additional technical
difficulties due to the intermediate top-quark resonance. It is now being
studied, and the corresponding NLO generator matched to parton shower will be
available in the future \cite{directtop}.

\begin{table}[t]\centering
\caption{%
Summary of the existing $95\%$~C.L.\ limits on top-quark FCNC branching fractions obtained at the LHC.
A summary plot is available at \protect\url{https://twiki.cern.ch/twiki/bin/view/LHCPhysics/LHCTopWGSummaryPlots\#table21}.
 Ref.\,\cite{Khachatryan:2015att} has set limits in a fiducial volume at the particle level for $pp\to t\gamma$ (not displayed in this table).
Numbers in bold are used as inputs to the global EFT analysis.}
\begin{tabular*}{\textwidth}{@{\extracolsep{\fill}}c@{}r@{\,}*{4}{c@{\,}}c}
\hline\hline
Mode	& $\br^{95\%\text{CL}}$
	& Ref.
	& exp.
	& $\sqrt{s}$
	& $\mathcal{L}$
	& remarks
\\
\hline
$t\to q Z$
\\
\noalign{\vskip1mm}
$u$ 	& $\bf 1.7\times 10^{-4}$
	& \cite{Aaboud:2018nyl}		& ATLAS	& $13\,\tev$	& $36.1\,\ifb$
	& decay, $|m_{\ell\ell}-m_Z|<15\,\gev$
\\
$c$	& $\bf 2.4\times 10^{-4}$
\\
$u$	& $2.4\times 10^{-4}$
	& \cite{CMS:2017twu}		& CMS	& $13\,\tev$	& $35.9\,\ifb$
	& production plus decay
\\
$c$	& $4.5\times 10^{-4}$
\\
$u$	& $2.2\times 10^{-4}$
	& \cite{Sirunyan:2017kkr}	& CMS	& $8\,\tev$	& $19.7\,\ifb$
	& production, $76 < m_{\ell\ell} < 106\,\gev$
\\
$c$	& $4.9\times 10^{-4}$
\\[2mm]
\hline
$t\to qg$
\\\noalign{\vskip1mm}
$u$	& $0.40\times 10^{-4}$
	& \cite{Aad:2015gea}		& ATLAS	& $8\,\tev$	& $20.3\,\ifb$
	& $\sigma(pp\to t)\times \br(t\to bW) < 3.4\,\pb$
\\
$c$	& $\bf 2.0\times 10^{-4}$
\\
$u$	&  $\bf 0.20\times 10^{-4}$
	& \cite{Khachatryan:2016sib}	& CMS	& $7,8\,\tev$	& $5.0,17.9\,\ifb$
	& in $pp\to tj$
\\
$c$	& $4.1\times 10^{-4}$
\\[2mm]
\hline
$t\to q\gamma$
\\\noalign{\vskip1mm}
$u$	& $\bf 1.3\times 10^{-4}$
	& \cite{Khachatryan:2015att}	& CMS	& $8\,\tev$	& $19.8\,\ifb$
	& $\sigma(pp\to t\gamma)\times \br(t\to bl\nu)<26\,\fb$
\\
$c$	& $\bf 17\times 10^{-4}$
	&&&&
	& $\sigma(pp\to t\gamma)\times \br(t\to bl\nu)<37\,\fb$
\\[2mm]
\hline
$t\to qh$
\\\noalign{\vskip1mm}
$u$	& $\bf 19\times 10^{-4}$
	& \cite{Aaboud:2018pob}		& ATLAS	& $13\,\tev$	& $36.1\,\ifb$
	& multilepton channel
\\
$c$	& $\bf 16\times 10^{-4}$
	&
\\
$u$	& $55\times 10^{-4}$
	& \cite{Khachatryan:2016atv}	& CMS	& $8\,\tev$	& $19.7\,\ifb$
	& multilepton, $\gamma\gamma$, $b\bar b$
\\
$c$	& $40\times 10^{-4}$
	&
\\
$u$	& $47\times 10^{-4}$
	& \cite{Sirunyan:2017uae}	& CMS	& $13\,\tev$	& $35.9\,\ifb$
	& $b\bar{b}$
\\
$c$	& $47\times 10^{-4}$
	&
\\[1mm]\hline\hline
\end{tabular*}
\label{tab:current_limits}
\end{table}

\subsubsection{Existing limits}
Table~\ref{tab:current_limits} lists the existing limits on FCNC processes.
We follow Ref.\,\cite{Durieux:2014xla} and interpret them in
a global EFT analysis. Several additional remarks are in order:
\begin{itemize}

\item[$\bullet$] For $t\to q\ell\ell$ we use the predictions at NLO in QCD provided in Ref.\,\cite{Durieux:2014xla} for the $m_{\ell\ell}\in[78,102]\,\gev$ range although the most stringent constraints by ATLAS are set using $|m_{\ell\ell}-m_Z|<15\,\gev$. The CMS bounds obtained by combining production and decay process cannot be reinterpreted to include four-fermion operators.

\item[$\bullet$] For single top-quark production through the $tqg$ interaction, we use the best constraints: in the up-quark channel by CMS and in the charm-quark channel by ATLAS. We naively combine them using the numerical value at NLO in QCD for the $t\to jj$ branching fraction provided in Sec.~V.B of Ref.\,\cite{Durieux:2014xla}.

\item[$\bullet$] For single top-quark production in association with a photon, we note the very interesting fiducial limit provided by Ref.\,\cite{Khachatryan:2015att} which allows for an accurate reinterpretation. 
However, we use the simplified approach of Ref.\,\cite{Durieux:2014xla} based on the limit quoted on the total cross section and use the numerical values computed at NLO in QCD for $pp\to t\gamma+\bar{t}\gamma$ with a $30\,\gev$ cut on the photon \pt although a $50\,\gev$ cut was applied in Ref.\,\cite{Khachatryan:2015att}.

\item[$\bullet$] For $t\to h j$ decay we use the most stringent limits set by ATLAS in the multilepton channel. The dependence on all operator coefficients, except $C_{t\phi}$ and $C_{tG}$, is assumed to be negligible.

\item[$\bullet$] Limits on $e^+e^-\to tj+\bar{t}j$ obtained at LEP~II~\cite{Aleph:2001dzz} still dominate constraints on four-fermion operators involving electrons, while $t\to q\ell\ell$ at hadron colliders are the only measurements constraining four-fermion operators featuring a pair of muons. However, 
the latter limits are not explicitly shown below.
We use the limit from the highest LEP II centre-of-mass energy, $\sqrt{s}=207\,\gev$, which is the most sensitive to four-fermion operators, $\sigma(e^+e^-\to tj+\bar tj)<170\,\fb$.
\end{itemize}

The global analysis based on existing data gives $95\%\,$C.L.\ limits
on the EFT Wilsons coefficients in the notation of Ref.\,\cite{AguilarSaavedra:2018nen}, shown 
in Fig.~\ref{fig:limits_current} (left panel). 
As explained in Ref.\,\cite{Durieux:2014xla}, 
no statistical combination is
attempted, i.e.,~limits from different measurements are only overlaid.
 Fig.~\ref{fig:limits_pu_eu_current} (left panel) show two-dimensional constraints in the $c_{\varphi q,\varphi
u}^-$,$c_{eq,eu}$ plane. 
This illustrates the complementarity between the LHC and the LEP~II measurements;
the former gives better constraints on two-fermion operators, the latter
on four-fermion operators.

\begin{figure}
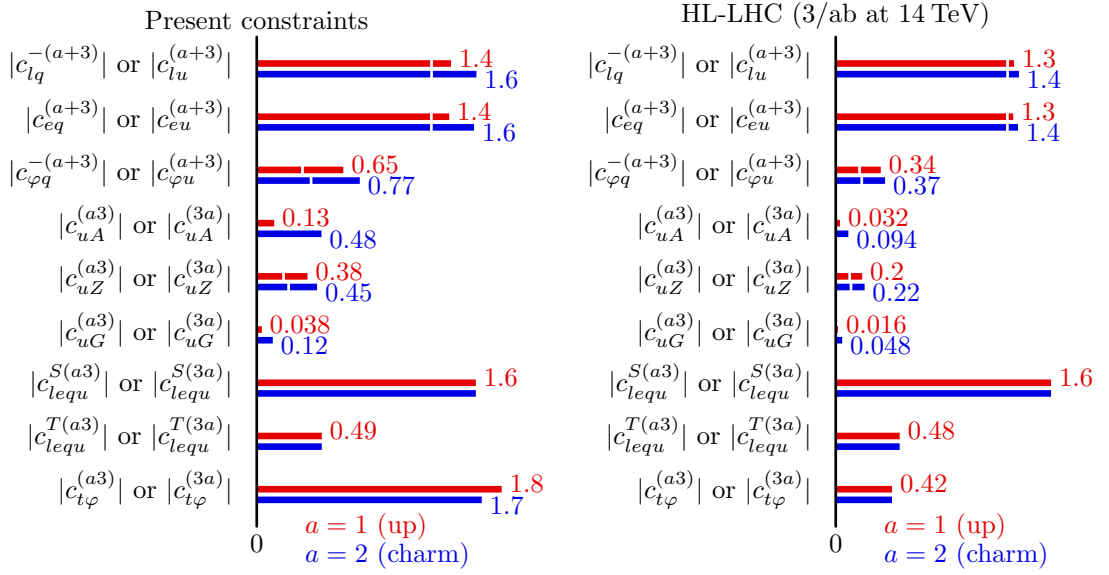
\centering
\hfil\includegraphics[scale=1]{section8/figs/limits.mps}%
\hfil\includegraphics[scale=1]{section8/figs/limits_prospects.mps}
\hfil
\caption{Current (left) and projected HL-LHC (right) $95\%\,$C.L.\ limits on top-quark FCNC operator coefficients in the conventions of Ref.\,\cite{AguilarSaavedra:2018nen}. Red and blue bars denote  top-up and top-charm FCNCs, respectively. White marks indicate individual limits, obtained under the unrealistic assumption that all the other operator coefficients vanish.}
\label{fig:limits_current}
\label{fig:limits_prospects}
\end{figure}

\begin{figure}
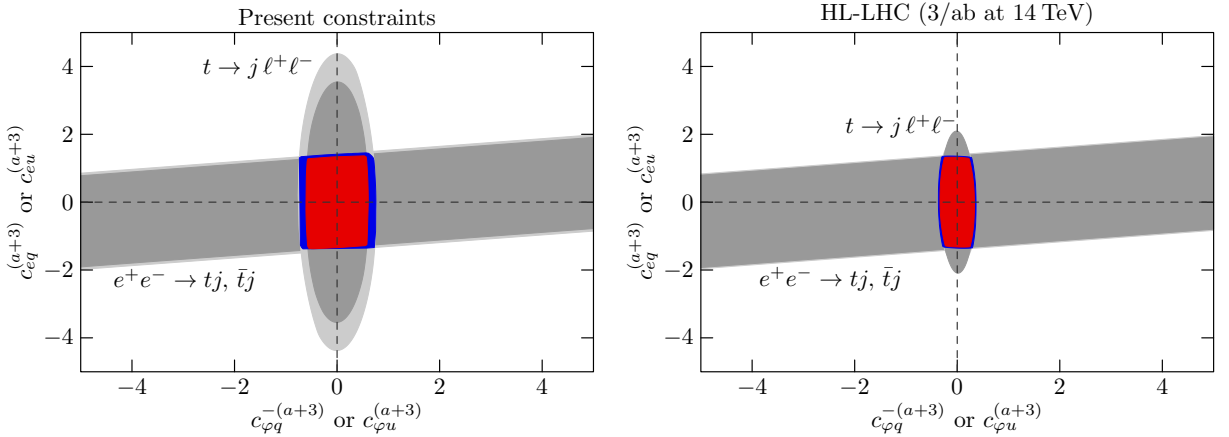
\centering
\includegraphics[width=.49\textwidth]{section8/figs/limits_pu_eu.mps}\hfill
\includegraphics[width=.49\textwidth]{section8/figs/limits_pu_eu_prospects.mps}
\caption{Current (left) and prospective HL-LHC (right) $95\%\,$C.L.\ limits on top-quark FCNC operator coefficients in a two-dimensional plane formed by two- ($x$ axis) and four-fermion ($y$ axis) operator coefficients. Other parameters are marginalized over, within the constraints obtained when all measurements are included. Red and blue regions are the combined constraints for top-up and top-charm FCNCs. The impact of $t\to j\ell^+\ell^-$ and $e^+e^-\to tj,\bar{t}j$ measurements is displayed separately in dark and light gray colors for top-up and top-charm FCNCs, respectively.}
\label{fig:limits_pu_eu_current}
\label{fig:limits_pu_eu_prospects}
\end{figure}

\subsubsection{Future limits}
We use the prospects presented in Sec.~\ref{sec:8:EXP:FCNC} to estimate the future reach of global constraints for the HL-LHC scenario.  As previously, we assume that the limits quoted on the $\br(t\to qZ)$ branching fraction are derived using the dilepton decays of the $Z$ boson, in a $m_{\ell\ell}\in[78,102]\,\gev$ window for the dilepton invariant mass. This determines the sensitivity to four-fermion operators.
The limits on the $tq\gamma$ interaction were derived by a combination of production and decay processes~\cite{CMS:2017gvo}. Since the only prospect provided is on the $\BR(t\to q\gamma)$ branching fraction, we approximate it as though it is from the measurement of the decay process only. This assumption affects primarily the dependence of bounds on $tqg$ interactions.

The results of a global fit based on the HL-LHC prospects are displayed
in Fig.~\ref{fig:limits_prospects} (right panel). Comparing with the left panel,
constraints on two-fermion operators are typically improved by a factor of a few,
while those on four-fermion operators are only marginally improved, mostly
indirectly through the improvement of the limits on other operator
coefficients.  The two dimensional plane of
Fig.~\ref{fig:limits_pu_eu_prospects} (right panel) shows that LEP~II
constraints on $e^+e^-\to tj,\bar{t}j$ production will start having little impact, even
on the four-fermion operators, after the HL-LHC phase.
}

\subsubsection{Complementarity with $B$-meson and kaon rare processes}
\label{sec:B:K:top}
The NP effective operators which include the top quark also contribute to
the low-scale effective Hamiltonian for meson decays, 
and hence, precision measurements of the $B$-meson and kaon rare processes provide complementary constraints to the NP top-quark operators~\cite{Fox:2007in,Drobnak:2011aa,Brod:2014hsa}.

In the framework of SMEFT, 
the $SU(2)_L$ gauge symmetry between $t_L$ and $b_L$ provides 
a direct constraint to the NP top-quark operators arising from the tree-level matching onto the $B$ physics Hamiltonian \cite{Drobnak:2011aa,Aebischer:2015fzz}.
For instance, in the flavour basis, the following operators
\begin{align}
O_{\ell q}^{1(ij32)} &= (\bar{\ell}_i \gamma_{\mu} P_L \ell_j)(\bar{q}_3 \gamma^{\mu} P_L q_2) = 
(\bar{\ell}_i \gamma_{\mu} P_L \ell _j)(\bar{t} \gamma^{\mu} P_L c) + (\bar{\ell}_i \gamma_{\mu} P_L \ell _j)(\bar{b} \gamma^{\mu} P_L s)
\,,\\
O_{\ell e q u}^{1( i j 3 2)} &= ( \bar{\ell}_i P_R e_j ) \varepsilon (\bar{q}_3 P_R c)  =
- ( \bar{\ell}_i{}^{+} P_R e_j ) (\bar{t} P_R c)+( \bar{\nu}_i P_R e_j ) (\bar{b} P_R c)  \,,
\end{align}
are
constrained from $B_s \to \ell^+ \ell^-$, $B \to K^{(\ast)} \overline{\nu} \nu$, $b \to s \ell^+ \ell^-$, and $b \to c\ell^- \bar{\nu}$ observables.
Each of the constraints significantly depends on the lepton-flavour dependence
(\textit{e.g.}, see Ref.~\cite{Fajfer:2015ycq}).

Also,
the NP operators which include single top quark, 
\textit{e.g.},  
\begin{align}
O^{1(32)}_{\varphi q} = (\varphi^{\dag} \overleftrightarrow{i D}_{\mu} \varphi)(\bar{t} \gamma^{\mu} P_L c) + (\varphi^{\dag} \overleftrightarrow{i D}_{\mu} \varphi)(\bar{b} \gamma^{\mu} P_L s)\,,
\end{align}
and two top quarks, 
\textit{e.g.},  
\begin{align}
O_{qd}^{1 (33kl)} = (\bar{t}\gamma_{\mu} P_L t )(\bar{d}^k \gamma^{\mu} P_R d^l) 
+ (\bar{b}\gamma_{\mu} P_L b)(\bar{d}^k \gamma^{\mu} P_R d^l) \quad \textrm{with} ~ k\neq l \,,
\end{align}
can contribute to the low-scale effective Hamiltonian 
through the one-loop matching conditions at the electroweak symmetry breaking  scale by integrating out the top quark, $W$, $Z$ and the SM Higgs boson.
These one-loop contributions are enhanced by the top-quark mass.
Although such a two top-quark operator does not contribute to 
the single top-quark production mentioned in this section, 
once a UV completion is considered, 
single and two top-quarks operators could be related. 
The one-loop matching conditions onto $\Delta F= 1$ processes are given in Ref.~\cite{Aebischer:2015fzz}, while 
the conditions onto  $\Delta F= 2$ ones are
 given in Ref.~\cite{Endo:2018gdn}.
Besides,
 one-loop matching conditions to $\Delta F = 0$, \textit{e.g.}, $h \to \tau^+ \tau^-$ and the  leptonic dipole moments, are investigated in Ref.~\cite{Feruglio:2018fxo}.

\subsubsection{Experimental perspectives}
\label{sec:8:EXP:FCNC}

\subsubsubsection{Signal modeling}

The generation of signal events at ATLAS is done at NLO with {\tt MadGraph5\_aMC@NLO}~\cite{Alwall:2014hca, Mangano:2006rw} and the effective field theory framework developed in the TopFCNC model is used~\cite{Degrande:2014tta,Durieux:2014xla}. In the case of $gqt$ coupling, the MEtop generator is used instead~\cite{Coimbra:2012ys}. At CMS signal events are simulated at LO with {\tt MadGraph5\_aMC@NLO} with the effective lagrangian implemented by means of the {\tt FEYNRULES} package, except in the simulation of signal events for $gqt$ and $\gamma qt$ couplings where {\tt CompHEP}~\cite{Boos:2004kh} and {\tt PROTOS 2.0}~\cite{AguilarSaavedra:2010rx} are used, respectively. In both experiments {\tt Pythia 8} is used to simulate the parton showering and hadronization. The generation of signal events is done under the assumption of only one non-vanishing FCNC coupling at a time. 


\begin{table}
\centering
\caption{Summary of the projected reach for the 95\% C.L. limits on the branching ratio for anomalous flavor changing top couplings.}
\begin{tabular}{lcccc}
\hline\hline
$\mathcal{B}$ limit at 95\%C.L. & 3 ab$^{-1}$, 14 TeV & 15ab$^{-1}$, 27 TeV & Ref. \\
\hline
$t\to gu $ & $3.8\times 10^{-6}$ & $5.6\times 10^{-7}$ & \cite{CMS:2018kwi}\\
$t\to gc $ & $32.1\times 10^{-6}$ &  $19.1\times 10^{-7}$ & \cite{CMS:2018kwi}\\
$t\to Zq$ & $2.4-5.8\times 10^{-5}$ &  & \cite{ATL-PHYS-PUB-2016-019}\\
$t\to \gamma u $ & $8.6\times 10^{-6}$ & & \cite{Collaboration:2293646} \\
$t\to \gamma c $ & $7.4\times 10^{-5}$ & & \cite{Collaboration:2293646} \\
$t\to Hq $ & 10$^{-4}$ & & \cite{ATL-PHYS-PUB-2016-019} \\
\hline\hline
\end{tabular}
\end{table}

\subsubsubsection{Top-gluon}

\begin{figure}[h]
\includegraphics[scale=0.4]{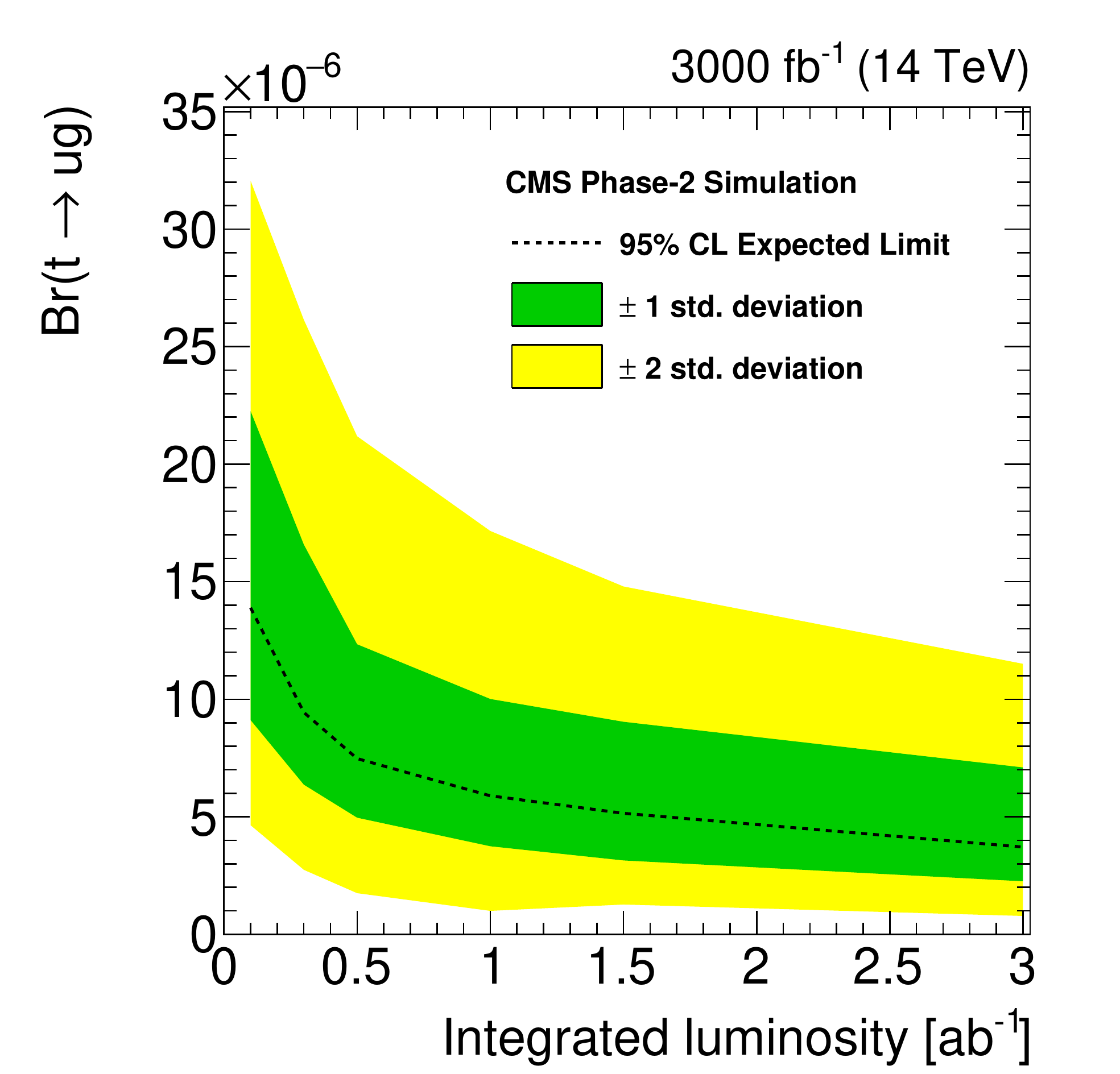}
\vspace{0.3cm}
\includegraphics[scale=0.4]{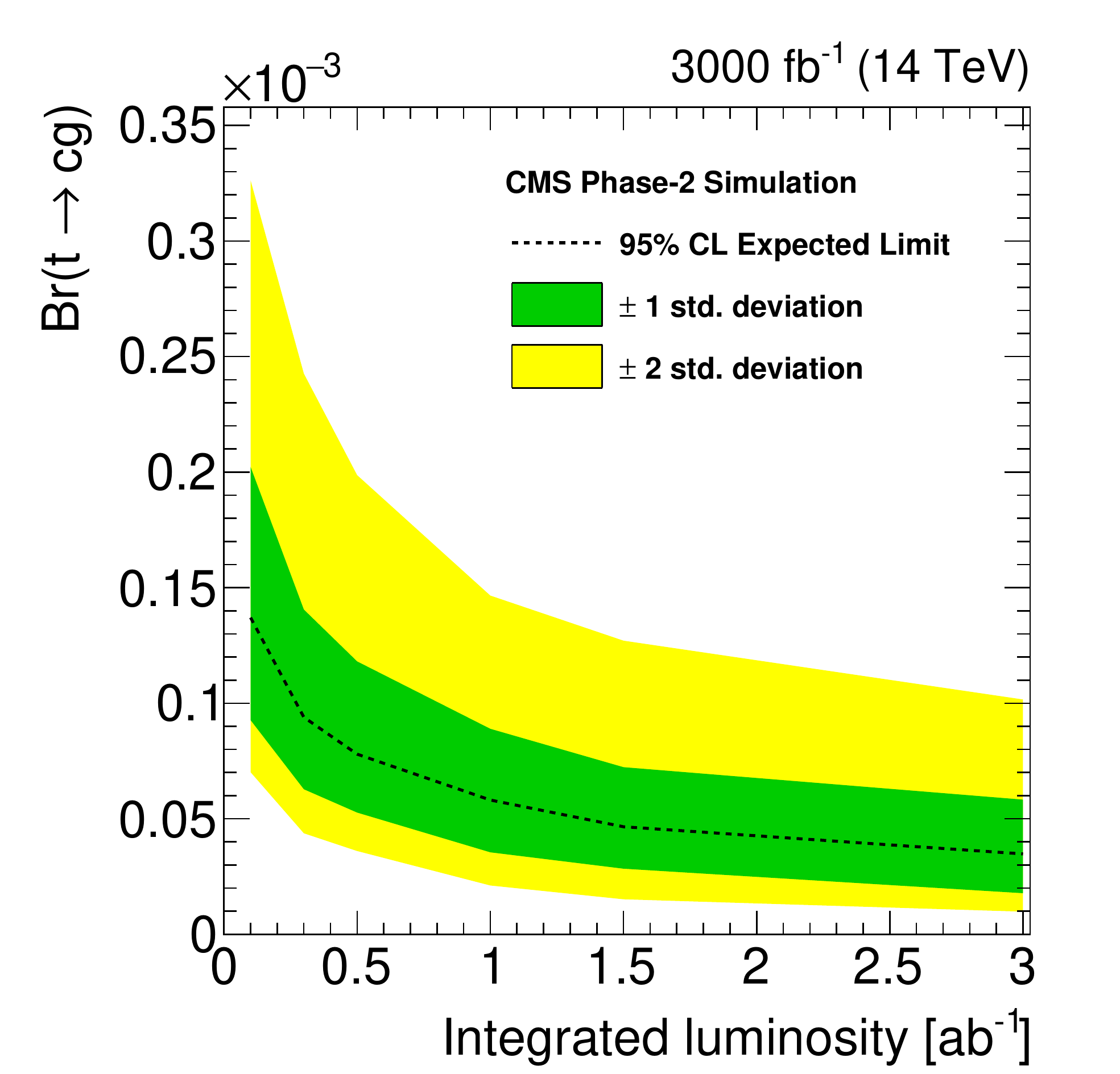}
\hfil
\caption{The expected exclusion limits at 95\% C.L. on the branching fractions for the $t\to u g$ (left panel) and $t\to c g$ (right panel) FCNC processes as a function of integrated luminosity.}
\label{fig:topglue_lumi}
\end{figure}

The $gqt$ FCNC process was studied by ATLAS~\cite{Aad:2015gea} and CMS~\cite{Khachatryan:2016sib} in single top quark events. The event signature includes the
requirement of one isolated lepton and the presence of a significant amount of transverse
missing energy ($E^{miss}_{T}$). The analysis at CMS requires exactly one $b$ and one non-$b$ jet to be present in the final state with the dominant background arising from the $t\bar{t}$+jets
production, while the analysis at ATLAS vetoes any additional jets resulting in the dominant source of background associated with the $W$+jets production. A neural network-based
technique is used to separate signal from background events.
The observed~(expected) 95\% C.L. upper limits in the CMS analysis are $\mathcal{B}(t \to gu) < 2.0~(2.8) \times 10^{-5}$ and
$\mathcal{B}(t \to gc) < 4.1~(2.8) \times 10^{-4}$, while the resultant limits in case of ATLAS are $\mathcal{B}(t \to gu) < 4.0~(3.5) \times 10^{-5}$ 
and $\mathcal{B}(t \to gc) < 2.0~(1.8) \times 10^{-4}$. The projected limits for 3 ab$^{-1}$ are $\mathcal{B}(t \to gu) < 3.8 \times 10^{-6}$ and $\mathcal{B}(t \to gc) < 32.1 \times 10^{-6}$~\cite{CMS:2018kwi}. 

\begin{figure}[t]
\begin{center}{
\includegraphics[width=8cm]{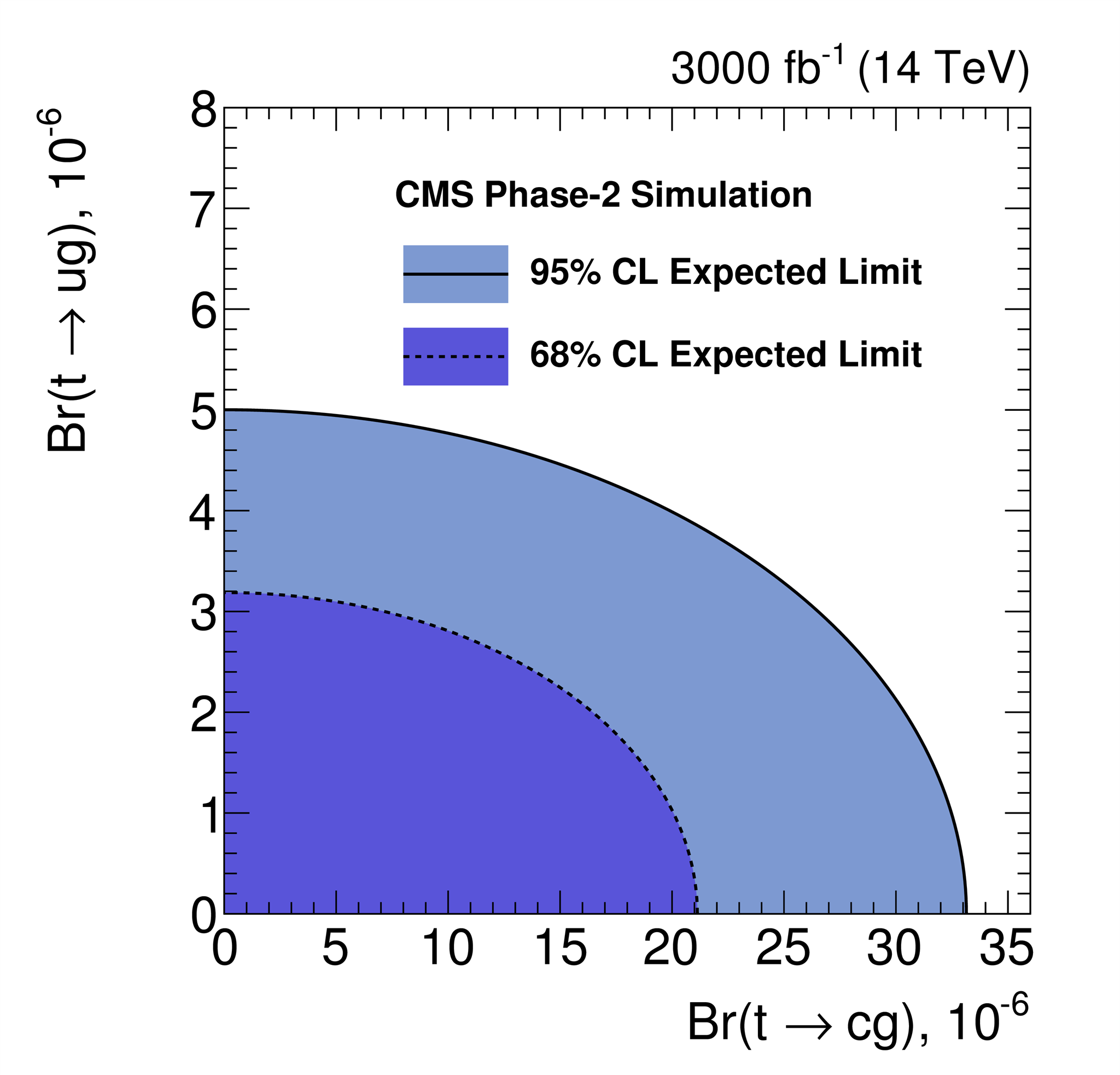}
\caption{{Two-dimensional expected limits on the $t\to u g$ vs and $t\to c g$ branching fractions at 68\% and 95\% C.L. for an integrated luminosity of 3000 fb$^{-1}$}} \label{fig:topglue}}
\end{center}
\end{figure}

The dependence of the 
$\mathcal{B}(t \to ug)$ and $\mathcal{B}(t \to cg)$
exclusion upper limits on integrated luminosity is shown in Fig.~\ref{fig:topglue_lumi}
 with 1 and 2 $\sigma$ bands corresponding to 68 \% and 95 \% C.L intervals of distributions of the limits .
In addition the two-dimensional contour that reflects the possible simultaneous presence of both FCNC processes. is shown in Fig.~\ref{fig:topglue}.

\subsubsubsection{Top-Z}


%
%
ATLAS studied the sensitivity to the $tqZ$ interaction, by performing an analysis~\cite{ATL-PHYS-PUB-2019-001} based on simulated samples and following both the strategy of its 13~TeV data analysis on the same subject~\cite{Aaboud:2018nyl} and of the general recommendations for this HL-LHC study. The study is performed in the three charged lepton final state of $t\bar t$ events, in which one of the top quarks decays to $qZ$, $(q=u,c)$ and the other one decays to $bW$ ($t\bar t\to bWqZ\to b\ell\nu q\ell\ell$). The kinematics of the events are reconstructed through a $\chi^2$ minimisation and dedicated control regions are used to normalize the main backgrounds and constrain systematic uncertainties. The main uncertainties, in both the background and signal estimations, are expected to come from theoretical normalization uncertainties and uncertainties in the modeling of background processes in the simulation. Different scenarios for the systematic uncertainties are considered, ranging from the full estimations obtained with the 13~TeV data analysis, to the ones expected with improvements in theoretical predictions, which should be half of the former ones. A binned likelihood function $L(\mu, \theta)$ is used to do the statistical analysis and extract the signal normalisation. An improvement by a factor of five is expected in relation to the current 13~TeV data analysis results. Obtained branching ratio limits are at the level of 4 to 5 $\times 10^{-5}$ depending on the considered scenarios for the systematic uncertainties.

\subsubsubsection{Top-gamma}

The $t\gamma q$ anomalous interactions have been probed by CMS at 8 TeV in
events with single top quarks produced in association with a photon~\cite{Khachatryan:2015att} and the resulting exclusion limits are $\mathcal{B}(t \to \gamma u) < 1.3~(1.9) \times 10^{-4}$ and $\mathcal{B}(t \to \gamma c) < 2.0~(1.7) \times 10^{-3}$. 

\begin{figure}[!h!]
	\begin{minipage}[t]{\linewidth}
		\begin{center}
        \includegraphics[width=0.49\textwidth]{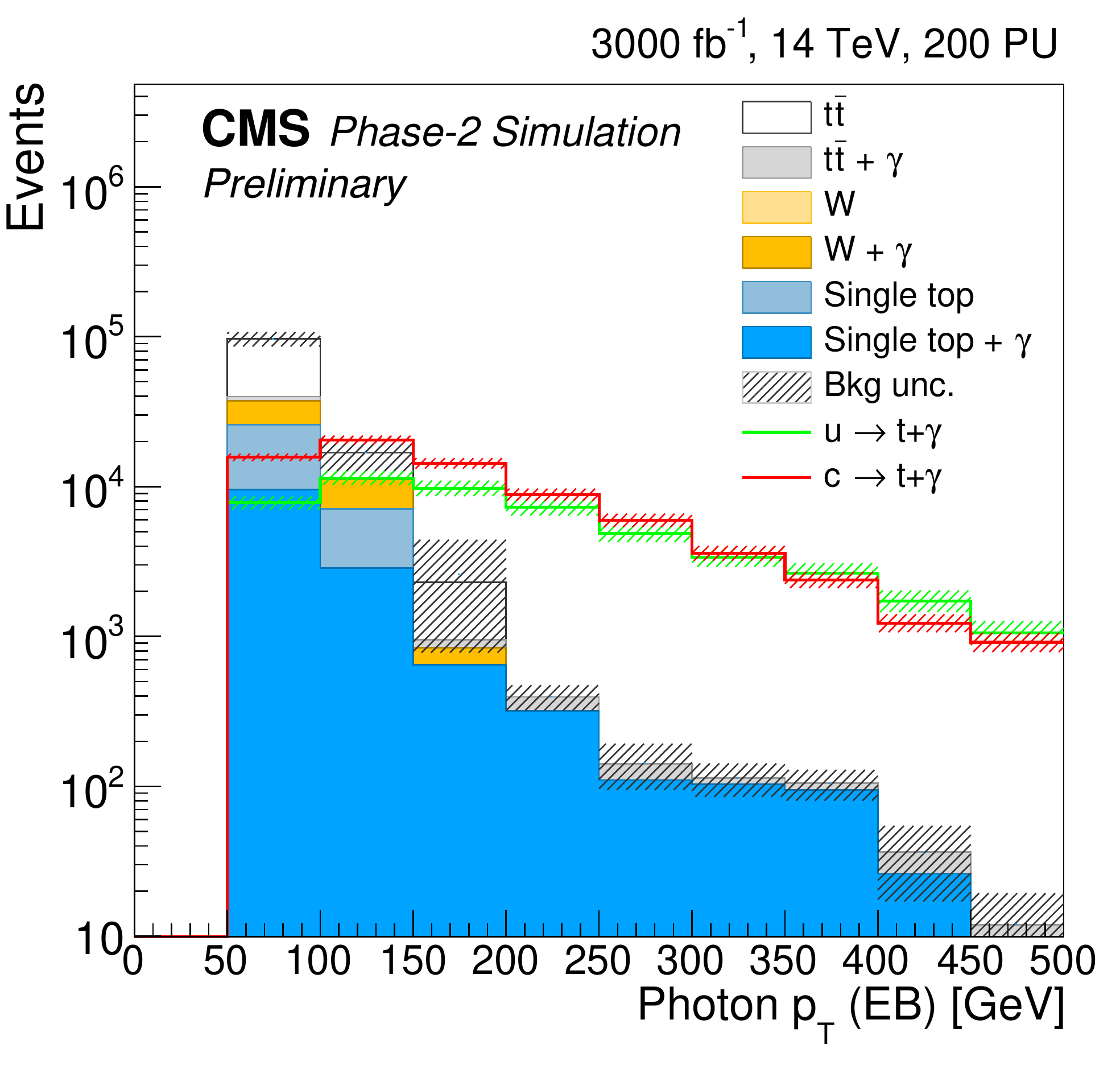}
         \includegraphics[width=0.49\textwidth]{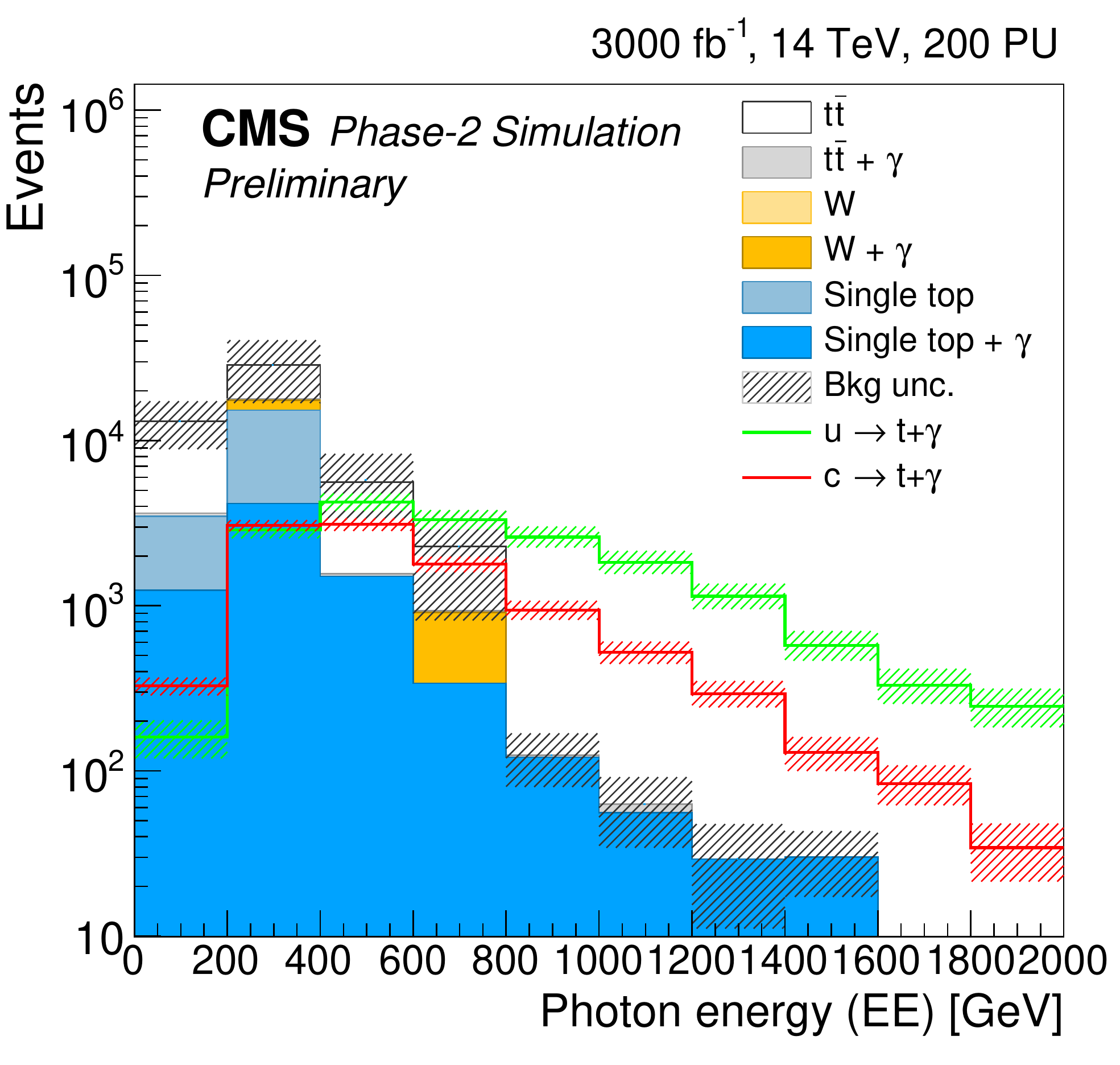}
		\end{center}
	\end{minipage}
	\caption{Transverse momentum of photon candidates for the central $\eta$ region (left) and
energy of photon candidates in the forward region (right).}
	\label{fig:tqgamma}
\end{figure}

In this section, the sensitivity of the upgraded CMS detector to $tq\gamma$ FCNC transitions is estimated for integrated luminosities of 300 and 3000 fb$^{-1}$ using single top quark production via $q\rightarrow q\gamma$, with $q$ being a $u$ or a charm quark~\cite{Collaboration:2293646}. This analysis focuses on subsequent SM decays of the top quark in a $W$ boson and bottom quark, with the $W$ boson decays leptonically to a muon or electron and a neutrino. 
The final state signature is the presence of a single muon or electron, large missing transverse momentum, a $b$ jet, and an isolated high energy photon, with a broad $\eta$ spectrum. 
The photon properties themselves provide good separation with respect to the dominant background processes from $W+$jets, and single top or top quark pair production in association with photons. 
For the discrimination of signal and background events, and to set the limits on the FCNC couplings, the events are split into two categories depending on the pseudo-rapidity of the photon (central region with $|\eta_{\gamma}| < 1.4$ and forward region with $1.6<|\eta_{\gamma}| < 2.8$). In the central (forward) region the photon \pt (energy) is used as a discriminating distribution: the low \pt(energy) is background dominated, while the high \pt(energy) region is populated by signal events. The distributions are shown in Fig.~\ref{fig:tqgamma}.

The limits on the cross section for the single top quark production via $tq\gamma$ are obtained considering systematic uncertainties from variations of the renormalization and factorization scale, $b$-tagging and jet energy scale corrections and their effects as propagated to missing transverse energy, lepton efficiency and luminosity. 

These studies yield the following upper limite on the branching ratios at 95\%C.L.:  $\mathcal{B}(t \to \gamma u) < 8.6 \times 10^{-6}$, $\mathcal{B}(t \to \gamma c) < 7.4 \times 10^{-5}$.

%
\subsubsubsection{Top-Higgs}

The $tHq$ interactions are studied by ATLAS in top quark
pair events with $t \rightarrow qH, H \rightarrow \gamma\gamma$~\cite{Aaboud:2017mfd} and $H \rightarrow WW$~\cite{Aaboud:2018pob} at 13
TeV. The former analysis explores the final state with two isolated
photons. For leptonic top quark decays the selection criteria includes
the requirement of one isolated lepton, exactly one b jet, and at least
one non-b jet. In case of hadronic top quark decays the analysis
selects events with no isolated leptons, at least one b jet, and at
least three additional non-b jets. The dominant background processes
are associated with the production of non-resonant $\gamma\gamma$+jets, $t\bar{t}$+jets and
$W$+$\gamma\gamma$ events. 
The resultant limits are $\mathcal{B}(t \to Hu) < 2.4~(1.7) \times
10^{-3}$ and $\mathcal{B}(t \to Hc) < 2.2~(1.6) \times 10^{-3}$.
The search for FCNC in $H \rightarrow WW$ includes the analysis of multilepton final states with either two same-sign or three leptons. The dominant backgrounds arising from the $ttW$, $ttZ$ and non-prompt lepton production are suppressed with a BDT.
The obtained limits are $\mathcal{B}(t \to Hu) < 1.9~(1.5) \times
10^{-3}$ and $\mathcal{B}(t \to Hc) < 1.6~(1.5) \times 10^{-3}$.
The $tHq$ anomalous couplings are probed by CMS in $H \rightarrow
b\bar{b}$ channel in top quark pair events, as well as in single top
associated production with a Higgs boson, at 13
TeV~\cite{Sirunyan:2017uae}. The event selection includes the
requirement of one isolated lepton, at least two b jets, and at least
one additional non-b jet. The dominant $t\bar{t}$ background is
suppressed with a BDT discriminant to set the exclusion limits of
$\mathcal{B}(t \to Hu) < 4.7~(3.4) \times 10^{-3}$ and $\mathcal{B}(t
\to Hc) < 4.7~(4.4) \times 10^{-3}$. Preliminary projections suggest $\mathcal{B}(t \to Hq) < \mathcal{O}(10^{-4})$~\cite{ATL-PHYS-PUB-2016-019, ATL-PHYS-PUB-2013-012}.

\subsection{Anomalous $Wtb$ vertices and \CP-violation effects from $T$-odd kinematic distributions}
{\it \small Authors (TH): Frederic Deliot, Wouter Dekens.}

Beyond the SM, contributions to the $Wtb$ vertex have been widely studied in the literature  \cite{Grzadkowski:2008mf,Drobnak:2010ej,AguilarSaavedra:2008zc,GonzalezSprinberg:2011kx,Drobnak:2011aa,Brod:2014hsa,Cao:2015doa,Hioki:2015env,Schulze:2016qas,Kamenik:2011dk,Zhang:2012cd,deBlas:2015aea,Buckley:2015nca,Buckley:2015lku,Bylund:2016phk, Castro:2016jjv,Deliot:2017byp}.
Assuming that new physics is much heavier than the electroweak scale, the BSM corrections to the $Wtb$ vertex can be described using an EFT. The first contributions appear at dimension six and, in the conventions of \cite{AguilarSaavedra:2018nen}, take the following form
\begin{equation}\label{eq:Wtb}
\mathcal L_{tb} =-\bar t\left[ \frac{g}{\sqrt{2}}\,\gamma^\mu \big(V_{tb}\big(1+\tfrac{v^2}{2}c_{\varphi q}^{-}\big)P_L+\tfrac{v^2}{2} C_{\varphi tb}P_R\big)W_\mu^+-v\sigma^{\mu\nu}W_{\mu\nu}^+\left(C_{bW}P_R+C_{tW}^*P_L\right)\right]b+{\rm h.c.}\,\,,
\end{equation}
where $C_i = (c_{i}+ic_i^{ [I]})/\Lambda^2$. In writing \eqref{eq:Wtb} we follow Ref.~\cite{Grzadkowski:2008mf} and \Blu{enforce $C_{\varphi q}^{1(33)}+C_{\varphi q}^{3(33)} = 0$ to avoid tree-level FCNC decays of the $Z$ (one also obtains a strong constraint from $Z\to b b$ decays~\cite{Zhang:2012cd})}, such that only the real combination $c_{\varphi q}^-=C_{\varphi q}^{1(33)}-C_{\varphi q}^{3(33)}$ appears.
The Wilson coefficients in Eq.~\eqref{eq:Wtb} are in terms of the often-used anomalous couplings \cite{AguilarSaavedra:2008zc,AguilarSaavedra:2006fy} given by
\begin{equation}
V_L = V_{tb}^*\big(1+\tfrac{v^2}{2}c_{\varphi q}^{-}\big),\qquad V_R = \tfrac{v^2}{2} C_{\varphi tb}^*,\qquad g_L = -\sqrt{2}v^2 C_{bW}^*,\qquad g_R = -\sqrt{2}v^2 C_{tW}.
\end{equation}
Note that within SMEFT gauge invariance links the above vertices to additional interactions. The full form of the relevant SMEFT operators is given in Sec.~\ref{sec:top:FCNC}.

The above interactions give tree-level contributions to single-top production and to the observables in $t\to W^+b$ decay, in particular, to the $W$-boson helicity fractions. Apart from these processes that are ``directly'' sensitive to the $Wtb$ vertices, there are ``indirect'' observables that receive contributions from the same operators through loop diagrams. Examples are $h\to \bar bb$, $B\to X_s\gamma$, and searches for electric dipole moments (EDMs).
Below we  discuss briefly both the direct and indirect limits on the operators in Eq.\ \eqref{eq:Wtb}, as well as the projections for the HL-LHC (see also Working-Group 1 Section on this topic).

\subsubsection{Direct probes}
\begin{table}
\center
\caption{The current  and projected $95\%$ C.L.\ constraints from direct observables on the real and imaginary parts of the Wilson coefficients that contribute to the $Wtb$ vertex, assuming $\Lambda = 1$ TeV and $c_{\varphi q}^-=0$. 
}
\vspace{.2cm}
\begin{tabular}{llccc}
	\hline\hline
	& Coeff. &$c^{}_{tW}$& $c^{}_{bW}$& $c^{}_{\varphi tb}$ \\\hline
Current&	Re$(\ldots)$ &$[-0.70,\,0.82]$ &$[-2.2,\,2.2]$ &$[-8.9,\,10.9]$  	\\
	& Im$(\ldots)$ &$[-1.5,\,2.2]$ &$[-2.1,\,2.2]$ &$[-9.9,\,9.9]$  
	\\ 
	$3000~{\rm fb}^{-1}$ &	Re$(\ldots)$ &$[-0.23,\,0.58]$&$[-2.2,\,2.0]$&$[-9.3,\,10.6]$\\
 & Im$(\ldots)$ &$[-1.2,\,1.3]$&$[-2.2,\,2.1]$&$[-9.9,\,9.9]$\\
	\hline\hline
\end{tabular}
\label{tabdirectWtb}
\end{table}
The $Wtb$ interactions can be probed directly in single-top production and through the $W$ boson helicity fractions in top decays. In the case of polarized top quarks it is possible to construct additional $T$-odd observables that are sensitive to \CP-violating phases in the $Wtb$ couplings. Here we discuss briefly  these observables and the resulting (projected) constraints.

\smallskip
{\bf Single top production~~}
The single-top production cross section has been measured in the $t$ and $s$ channels at the LHC at $\sqrt{s} = 7$, $8$, and $13$ TeV \cite{Aad:2014fwa,Aaboud:2016ymp,Aaboud:2017pdi,Chatrchyan:2012ep,Khachatryan:2014iya,Sirunyan:2016cdg}. In principle, these cross sections  receive contributions from all $Wtb$ couplings. The measurements can be compared with the SM prediction \cite{Brucherseifer:2014ama} at NNLO in QCD, while BSM contributions have been evaluated at NLO \cite{Zhang:2016omx,Cirigliano:2016nyn,Alioli:2017ce,deBeurs:2018pvs}.
In SMEFT the single-top cross sections can receive contributions from other operators, in particular, from several four-quark operators, see, e.g., Refs.~\cite{Buckley:2015lku,Buckley:2015nca}. Thus, in order to perform an exhaustive global analysis one would have to include these effects as well. %

\smallskip
{\bf Top quark decays~~}
The helicity fractions of the $W$ boson in top decays are mostly sensitive to $c_{tW}$, $c_{bW}$, and $c_{\varphi tb}$, and have been measured both at the Tevatron and the LHC \cite{Aaltonen:2012rz,Aad:2012ky,Chatrchyan:2013jna,Aad:2015yem,Khachatryan:2014vma,Aaboud:2016hsq}. In addition, the phase $\delta^-$ between the amplitudes of longitudinally and transversely polarized $W$ bosons, recoiling against a left-handed $b$ quark, carries information on the imaginary part of $C_{tW}$ \cite{Aad:2015yem,Boudreau:2013yna}.
The SM predictions for the helicity fractions are known to NNLO in QCD \cite{Czarnecki:2010gb}, while the BSM contributions have been computed to NLO \cite{Drobnak:2010ej,Zhang:2014rja}.  

As mentioned above, in the case of polarized top-quark decays, it becomes possible to construct additional observables that are sensitive to both the real and imaginary parts of the $Wtb$ couplings. In particular, both the asymmetries constructed in Ref.\ \cite{Aguilar-Saavedra:2015yza,AguilarSaavedra:2010nx} and the triple-differential measurements of Ref.\ \cite{Aaboud:2017yqf} are sensitive to $c_{tW}^{[I]}$ (see also Ref.~\cite{Schulze:2016qas}). As a result, the limits on $c_{tW}^{[I]}$ already improve noticeably when including current experimental measurements of these angular asymmetries \cite{Deliot:2017byp}.

After taking into account the current experimental results on single-top production, the helicity fractions, and angular asymmetries one obtains the current (projected) limits in the left (right) panel of Table \ref{tabdirectWtb} \cite{Deliot:2017byp,Deliot:2018jts}. For an analysis focused on single-top production see Ref.~\cite{CMS-PAS-TOP-17-017}.
These limits were obtained by assuming that $c_{\varphi q}^-=0$.~\footnote{This coupling is harder to constrain as it is degenerate with a shift in $V_{tb}$, see Eq.\ \eqref{eq:Wtb}. This degeneracy can be broken by using CKM unitarity tests, or by considering additional interactions, such as the $Wts$, $Wtd$, and $Ztt$ vertices, that are linked to $c_{\varphi q}^-$ by gauge invariance. A more naive constraint, obtained by setting $V_{tb}=1$, leads to $|c_{\varphi q}^-|\lesssim 1$ \cite{Alioli:2017ce}.}
 Comparing the two panels one sees that the current and projected limits on the $c_{bW}$ and $c_{\varphi tb}$ couplings are very similar, while the projected limits on the real and imaginary parts of $c_{tW}$ are roughly a factor of $2$ stronger than the current constraints.

Top decays can also be used to probe physics beyond that of the top sector. For example, Ref. \cite{Harrison:2018bqi} recently suggested that the process $t\to bW\to b\bar b c$ can be used to measure $V_{cb}$ at the $m_W$ scale, instead of $V_{cb}(m_b)$ which is probed in $B$ decays.

\subsubsection{Indirect probes}
\label{sec:8:2:Indirect}
\begin{table}
\center
\caption{Indicative constraints on the SMEFT operators contributing to the $Wtb$ vertex from indirect observables. A single Wilson coefficient is taken to be nonzero at a time, and we set $\Lambda = 1$ TeV.}
\vspace{.2cm}
\begin{tabular}{ccclll}
\hline\hline
Coeff. &$h\to \bar bb$ & EWPT&$B\to X_s\gamma$ &EDMs\\\hline
$c^{-}_{\varphi q}$&$-$&$-$&$[-2.0,2.2]$ & $-$ \\
$c^{}_{\varphi tb}$&$[-0.3,4.3]$ & $-$&$[-4.6,4.9]\cdot 10^{-2}$&$-$\\
$c^{}_{tW}$&$-$&$[-1.1,0.7]$&$[-0.22,0.89]$&  $-$\\
 $c^{}_{bW}$ &$-$&$-$& $[-13,3.5]\cdot 10^{-3}$& $-$\\
$c^{[I]}_{\varphi tb}$&$[-7.3,7.3]$ &$-$&$[-0.20,0.20]$&$[-0.019,0.019]$ \\
$c^{[I]}_{tW}$&$-$&$-$& $[-2.4, 4.5]$ & $[-1.0,1.0]\cdot 10^{-3}$\\
 $c^{[I]}_{bW}$ &$-$&$-$& $[-4.3, 2.3] \cdot 10^{-2}$& $[-5.5,5.5] \cdot 10^{-4}$\\
 \hline\hline
\end{tabular}\label{tabindirectWtb}
\end{table}
The anomalous $Wtb$ interactions contribute to other processes through loop diagrams.  Although this can give rise to stringent constraints, their interpretation requires some care. As indirect observables receive contributions from additional operators apart from the $Wtb$ couplings, cancellations between different Wilson coefficients are possible. 
We will assume this is not the case and derive limits for the case that a single dimension-six operator is generated at the BSM scale.

\smallskip
{\bf Electric dipole moments.~}
Electric dipole moments are probes of \CP violation and, therefore, receive contributions from the imaginary parts of the $Wtb$ couplings. 
The most stringent experimental limits have been obtained on the EDMs of the neutron, mercury, and the ThO molecule, the latter of which can be interpreted as a limit on the electron EDM for our purposes.
To obtain the contributions to EDMs one first has to evolve the $Wtb$ operators to low energies, $\mu\sim 2$ GeV. QCD becomes non-perturbative below this scale, and one has to match to Chiral perturbation theory, which describes the \CP-odd interactions in terms of hadrons, photons, and electrons. These interactions can then be used to calculate the EDMs of nucleons, atoms, and molecules.

Among the operators in Eq.\ \eqref{eq:Wtb}, the $c_{\varphi tb}^{[I]}$ and $c_{bW}^{[I]}$ couplings are mainly constrained by the neutron EDM, while $c_{ tW}^{[I]}$ contributes to the electron EDM. The $c_{\varphi tb}^{[I]}$ and $c_{bW}^{[I]}$ first generate the bottom-quark chromo EDM,  $O_{dG}^{(33)}$ (in the notation of \cite{Grzadkowski:2008mf}), through one-loop diagrams \cite{Dekens:2013zca,Alonso:2013hga,Alioli:2017ce}, which subsequently induces the Weinberg operator, $O_{\tilde G}$, after integrating out the bottom quark \cite{Dicus:1989va,Weinberg:1989dx,BraatenPRL,Boyd:1990bx}. The hadronic matrix element of the Weinberg operator contributing to the neutron EDM is poorly known. Combining naive-dimensional-analysis and sum-rule estimates \cite{Pospelov_Weinberg,Weinberg:1989dx,deVries:2010ah}, one has $ |d_n|  = 6|e\,(50\, {\rm MeV})\, C_{\tilde G}(1\, {\rm GeV})|$, with an $\mathcal O(100\%)$ uncertainty.

The contributions to the electron EDM also arise from a two-step process: $c_{ tW}^{[I]}$ first induces \CP-odd operators of the form $X_{\mu\nu }\tilde X^{\mu\nu }\varphi^\dagger\varphi$, with $X_{\mu\nu}$ a $SU(2)_L$ or $U(1)_Y$ field strength, as well as semileptonic interactions of the form, $(\bar e_L \sigma_{\mu\nu}\,e_R) \, (\bar t_L\sigma^{\mu\nu}t_R)$, through the renormalization group equations \cite{Cirigliano:2016nyn,Cirigliano:2016njn,Fuyuto:2017xup}. These additional operators then induce the electron EDM at one loop. 
Using the above contributions and the current experimental limits \cite{Afach:2015sja,Baker:2006ts,Andreev:2018ayy}, we obtain the constraints in Table \ref{tabindirectWtb}. \Blu{The limits from the neutron EDM are expected to improve by $1$ to $2$ orders of magnitude in the next generation of experiments \cite{Chupp:2017rkp}, while proposals exist to improve the electron EDM limits by several orders of magnitude \cite{Kozyryev:2017cwq,Vutha:2017pej,PhysRevLett.120.123201}.}

\smallskip
{\bf Rare $B$ decays.~}
Unlike EDMs, measurements of $B\to X_s\gamma$ are sensitive to both the real and imaginary parts of the couplings. Although all four of the $Wtb$ vertices give rise to  flavour-changing $b\to s$ transitions through one-loop diagrams \cite{Grzadkowski:2008mf,Brod:2014hsa,Aebischer:2015fzz}, the largest effects are due to $C_{\varphi tb}$ and $C_{bW}$. Both of these couplings induce contributions proportional to $m_t$ instead of $m_b$ that appears in SM (as well as for $c^-_{\varphi q}$ and $C_{tW}$), leading to a relative enhancement of $m_t/m_b$. Here we consider the constraints from measurements of the $B\to X_s\gamma$ branching ratio and the \CP asymmetry \cite{Amhis:2016xyh}, for which we use the theoretical expressions of \cite{Hurth:2003dk} and \cite{Benzke:2010tq}, respectively. This leads to the constraints  in Table \ref{tabindirectWtb}, \Blu{which are expected to improve by a factor of a few in the future. In particular, the uncertainty on the branching ratio is expected to decrease by a factor of $2$ to $3$ at Belle II, while the improvement is projected to be a factor of $5$ for the CP asymmetry \cite{Kou:2018nap}.}

\smallskip
{\bf Electroweak precision tests.~}
The $Wtb$ operators also modify the self energies of the SM gauge bosons through one-loop diagrams, which are often parametrized by the $S$, $T$, and $U$ parameters \cite{Peskin:1990zt,Peskin:1991sw,Barbieri:2004qk}.
Taking into account the RGE contributions, only $c_{tW}$ induces the $S$ parameter by mixing with the $O_{\varphi WB}$ operator ($c_{bW}$ contributions are proportional to $m_b$), which leads to the limits in Table \ref{tabindirectWtb}. Here we assumed only a single dimension-six operator is present at $\mu = \Lambda$, but one can include the couplings of the operators that induce $S$ and $T$ at tree level, $O_{\varphi WB}$ and $O_{HD}$, and marginalize over them. This can be done because electroweak precision observables carry more information than is captured by the $S$, $T$, and $U$ parameters alone. This approach is discussed Ref.\ \cite{Greiner:2011tt,Zhang:2012cd} and leads to weaker limits, $c_{tW}\in [-1.6,0.8]$, $c_{bW}\in [-2,24]$, $c_{\varphi q}^-\in [-0.7,4.7]$ for $\Lambda= 1\,{\rm TeV}$.

\smallskip
{\bf Higgs decays.~}
Finally, the $Wtb$ interactions contribute to the process $h\to \bar bb$ by inducing corrections to the SM bottom-quark Yukawa coupling \cite{Alioli:2017ce}. In particular, the $C_{\varphi tb}$ coupling generates a contribution to the Yukawa coupling that scales as $y_b\sim {y_t} v^2C_{\varphi tb}/{(4\pi)^2}$. Thus, although this contribution only appears at one loop, the suppression is offset by the appearance of the top-quark Yukawa instead of that of the bottom quark. Using the combined ATLAS and CMS analysis \cite{Khachatryan:2016vau} of the Higgs decays to $\gamma\gamma$, $WW$, $ZZ$, $\tau\tau$, $\mu\mu$, and $\bar bb$ signal strengths we obtain the limits in Table \ref{tabindirectWtb}.

\subsection{Determinations of $V_{tx}$} 
{\it\small Authors (TH): Mariel Estevez, Darius Faroughy, Jernej Kamenik. }

The LHC as a top-quark factory has the ability to directly probe the $V_{tx}$ matrix elements. In particular, measuring the fractions of $b$-tagged jets in leptonic top decays $t\to Wj$ allows one to set a limit on the ratio 
\begin{equation}
\mathcal {R}=\frac{\mathrm{Br}(t\to bW)}{\sum_{x=d,s,b}\mathrm{Br}(t\to Wx)}>0.995
\end{equation}
 at 95\% CL \cite{Khachatryan:2014nda}. This result, when combined with measurements of $t$-channel single-top production \cite{Aaboud:2016ymp}, provides a direct determination of $|V_{tb}|=1.07\pm0.09$ in the limit $|V_{tb}|\gg|V_{ts}|,\,|V_{td}|$. Given that these measurements are already dominated by systematic effects, in particular the knowledge of the $b$-tagging efficiencies and theoretical uncertainties in single-top production, significant improvement in precision of $V_{tb}$ measurement at the HL-LHC or HE-LHC would arguably require novel strategies.  

\subsubsection{Measuring $|V_{td}|$ at HL-LHC and HE-LHC}

A possible experimental strategy to probe the $|V_{td}|$ matrix element directly at the HL(HE)-LHC is using single top production associated with a $W$ boson, $pp \rightarrow tW$. The idea is to exploit the production cross-section enhancement, as well as boosts of the top quarks coming from initial state valence $d$-partons. The $d$-quark is a valence constituent of the proton and there is an imbalance with the $\bar{d}$-quark that motivates to explore charge asymmetries as possible $V_{td}$-sensitive observables.
The main backgrounds, contrary to $t$-channel single top production, are charge symmetric or have very small charge asymmetries. 
The $dg \to tW$ associated production process is interesting because of its sizeable charge asymmetry in proton collisions and also because its kinematics predicts a characteristic angular distribution. We expect a relatively large incoming momentum on average from the valence $d$-quark. Consequently, a forward $W^-$ is preferred in the lab frame, which is supposed to produce a forward $\ell ^-$ in signal events. The main two backgrounds to the $\ell ^+ \ell ^- b E_T^{\rm{miss}}$ final state are the dileptonic $t \bar{t}$ production (missing one of the $b$-jets from top decays) and $g b \to tW$ associated production, proportional to $|V_{tb}|^2$. Both backgrounds have very small charge asymmetries. In order to increase the sensitivity and enhance the signal cuts can be imposed that reduce the cross-sections of the backgrounds, see Ref.~\cite{Alvarez:2017ybk} for details. 
The most important difference between signal and background comes from the $\eta (\ell ^-)$ distribution, where the signal clearly prefers forward negatively charged leptons. The asymmetry
\begin{equation}
A(\eta,p_T) = \frac{N^+ - N^-}{N^+ + N^-},
\label{A}
\end{equation}
where
$
N^\pm = N\left( \Delta |\eta(\ell)| \gtrless 0 \,\& \, \Delta p_T(\ell) \gtrless 0 \right),
$
is a $|V_{td}|$ sensitive observable. Each process contributes with
$N^+_i - N^-_i = \sigma_i \cdot {\cal A}_i \cdot \epsilon_i \cdot A_i(\eta,p_T)$,
where the factors on the right hand side are the cross-section, acceptance, selection efficiency and asymmetry, respectively. 
To quantify the versatility of the proposed charge asymmetry we study the prospective experimental reach in $r\equiv |V_{td}/V_{td}^{\rm SM}|$ by computing the difference of $A(\eta, p_T)$ to its SM expectation in units of the uncertainty. Based on existing experimental studies of charge asymmetries in top production~\cite{Khachatryan:2016ysn} we include an estimate for the systematic uncertainty of $\Delta _{sys}=0.2 \%$, and 
define the significance as
\begin{equation}
\mbox{significance} = {\left| A(\eta,p_T) - A(\eta,p_T)^{\rm SM}\right| }/{\sqrt{(N^+ + N^-)^{-1} + \Delta_{\rm syst}^2}} .
\end{equation}

\begin{figure}[t]
\centering
\includegraphics[width=0.45\textwidth]{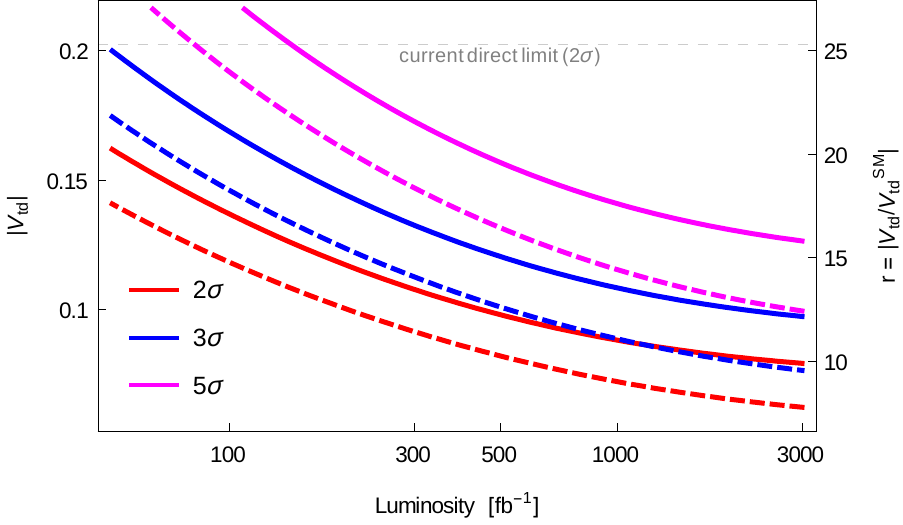}
\hspace{7mm}
\includegraphics[width=0.45\textwidth]{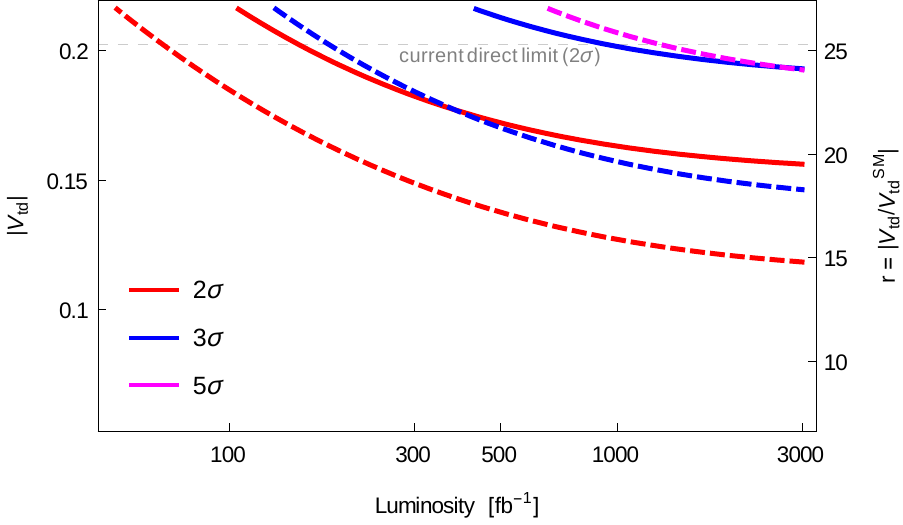}
\caption{Contour lines for the projected $2\sigma$, $3\sigma$ and $5\sigma$ upper bounds on $|V_{td}|$ (and $r\equiv|V_{td}/V_{td}^{\rm SM}|$) as functions of the LHC luminosity at $13$ TeV (left)  and $27$ TeV (right). Dashed lines follow if  the dominant $t\bar t$ background is reduced by half, see text for details.} 
\label{figure:Vtd}
\end{figure}

Fig.~\ref{figure:Vtd} shows the contours of expected experimental significance for $A(\eta,p_T)$ (and $r$) as functions of luminosity, for $13$ TeV and $27$ TeV LHC. As a rough guidance we also show with dashed lines the significance for the case that the dominant $t \bar{t}$ background were further reduced by a factor of $2$, e.g., by using multivariate discrimination techniques as already done in existing single top analyses~\cite{Chatrchyan:2014tua, Aad:2015eto, Aaboud:2016lpj}.
Values of $r<10$ could be directly accessible at the LHC@13TeV, improving the existing best direct constraint~\cite{Khachatryan:2014nda} by roughly a factor of three. The current direct bound on $|V_{td}|$ can be surpassed with the Run 2 dataset, while with $3000\, {\rm fb}^{-1}$ it could be possible to probe $|V_{td}|\sim |V_{ts}^{\rm SM}| \simeq 0.04$. 

The dominant $t \bar{t}$ background is mostly generated via gluon fusion, which is charge symmetric. There are subleading contributions from processes like $u \bar{u} \rightarrow t \bar{t}$ which are charge asymmetric, but only enter at higher orders in QCD. This is the main reason the dominant $t \bar{t}$ background has a strongly suppressed charge asymmetry. At higher collision energies, the probability of finding energetic enough gluons in the proton increases faster than that of valence quarks. Consequently, the fraction of $t\bar t$ events from the quark-antiquark initial state is reduced~\cite{Khachatryan:2016ysn}.  This leads to a shrinking charge asymmetry with growing collision energy. %
Unfortunately, the same happens to the signal, this time because at higher energies the asymmetry between $d$ and $\bar{d}$ partons inside the proton is reduced. 
The net effect is a severely diminished significance at $27$\,TeV compared to 13\,TeV for comparable luminosities. 
\subsubsection{Measuring Cabibbo-suppressed decays of the top quark at HL-LHC and HE-LHC}

	With the large $t\bar t$ statistics at the LHC, one might attempt a direct measurement of Cabibbo-suppressed decays of the top quark, $t\to (s,d)\,W^\pm$, in leptonic $t\bar t$ events. Since there is no practical way to distinguish between strange and down quark jets at the detector level (without dedictated PID systems), one measures 
\begin{equation}
\rho\equiv\sqrt{\big(\BR(t\to sW)+\BR(t\to dW)\big)/\BR(t\to bW)}\,.
\end{equation}
This gives
a direct information on $({|V_{ts}|^2+|V_{td}|^2})^{1/2}$, but not on $V_{ts}$ and $V_{td}$ separately. 
 In the SM, $\rho\approx |V_{ts}|\approx0.04$. 

To perform the measurement it is necessary to discriminate between heavy-flavoured jets, e.g., $b$-jets from the $t\to b W$ background, gluons from ISR/FSR contamination, and the signal -- the light-quark jets ($q$-jets) from the Cabibbo-suppressed top decays. This can be achieved through a $q$-tagger based on existing techniques used for $b$-tagging and quark/gluon jet discrimination~\cite{Gallicchio:2011xq}. There are several known useful observables which can be used in a $q$-tagger. In the following we use: (i) the multiplicity $N_{\mathrm{SV}}$ of secondary vertices (SV) in the jet within a fiducial volume of the tracker, (ii) the fraction of longitudinal momentum of the jet carried by the hardest prompt charged track 
$ z_{\mathrm{max}}\equiv\mathrm{max}\big[{\vec{p}_{x}\cdot\vec{p}_{\text{jet}}}/{|\vec{p}_\text{jet}|^2}\big]_{x\in\mathrm{jet}}$,
and (iii) the 2-point energy correlation function 
$U^\beta_1=\sum_{i,j\in \mathrm{jet}} z_T^i z_T^j\,(R_{ij})^\beta$, $z_T^i\equiv p_T^i /p_T^{\mathrm{jet}}$,
where $R_{ij}^2=(\eta_i-\eta_j)^2+(\phi_i-\phi_j)^2$ and $\beta$ is a free real parameter for which quark/gluon discrimination is optimized at $\beta=0.2$~\cite{Larkoski:2013eya,Moult:2016cvt}, see
Ref.\cite{thePaper} for more details. 
\begin{table}[t!]
\begin{center}
 {
\centering
\caption{ Tagging and mis-tagging efficiencies for the $q$-tagging working point used in the $V_{tq}$ analysis.}
\label{tab:effs} 
\begin{tabular}{cccccccc}
\hline\hline 
(t) type & Cuts &\  $\epsilon^{\mathrm t}_{q}$\ &\ $\epsilon^{\mathrm t}_{b}$\ &\ $\epsilon^{\mathrm t}_{c}$\ &\ $\epsilon^{\mathrm t}_{g}$
\\ 
\hline
$q$-tagger &$N_{\mathrm SV}=0$ \& $z_{\mathrm max}>0.3$    &  $0.18$ & $0.0031$ & $0.038$ & $ 0.049$ \\
$b$-tagger & $N_{\mathrm SV}>3$     		                          &  $0.0091$ & $0.64$ & $0.09$ & $0.016$ \\ 
\hline\hline
\end{tabular}
}
\end{center}
\end{table}

\begin{figure}[t]
\begin{center}{
\includegraphics[width=10cm]{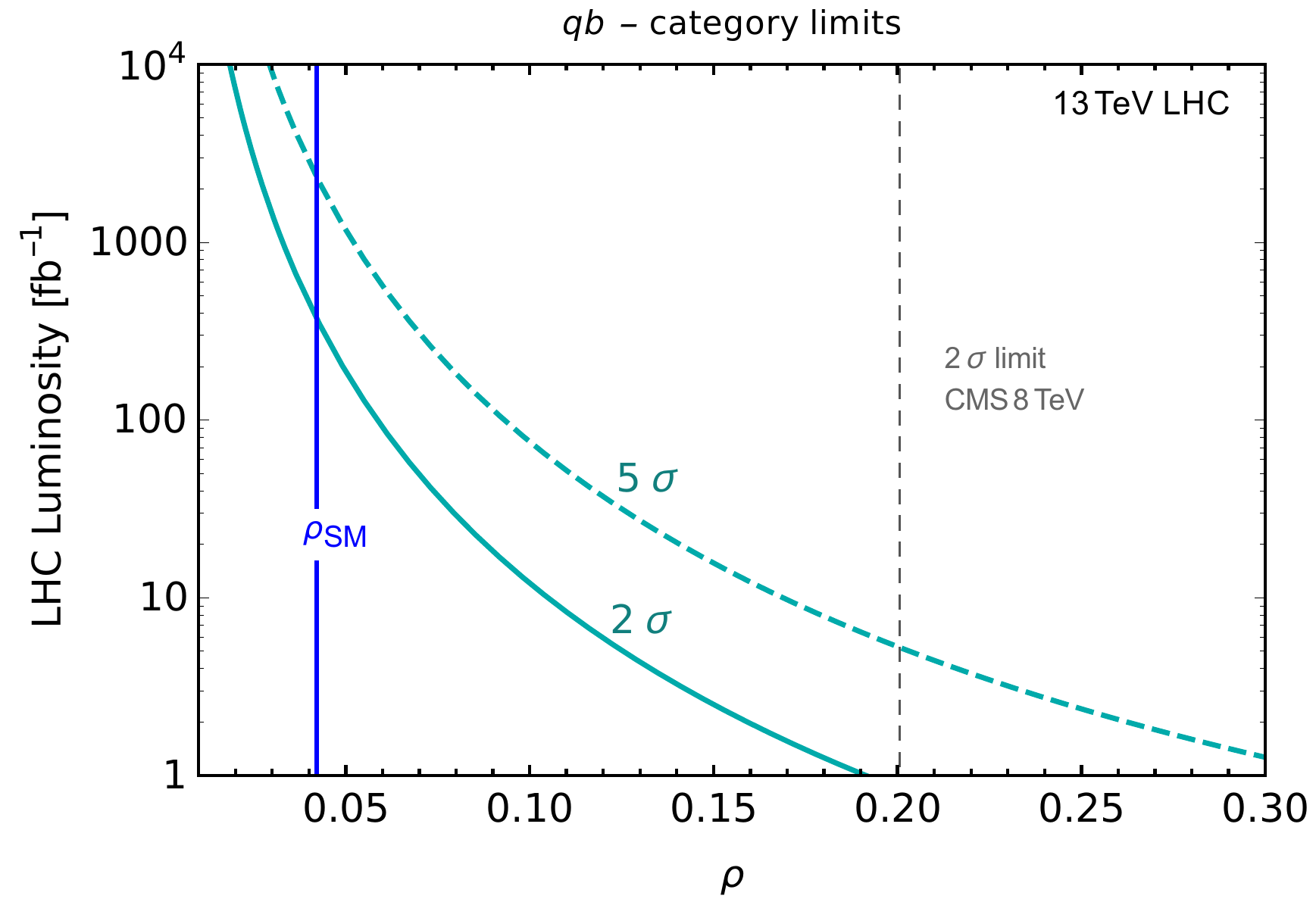} 
\caption{{Illustration of statistical limits on the ratio $\rho \simeq \big(|V_{ts}|^2 + |V_{td}|^2\big)^{1/2}/|V_{tb}|$ from an analysis of Cabbibo-suppressed top decays in leptonic $t\bar t$ events at 13TeV LHC. See text for details.}} \label{fig:master}}
\end{center}
\end{figure}

For the projections we use the reference working point 
shown in Table~\ref{tab:effs}.	
	We bin preselected events into one of the six tagged dijet categories $\{\jmath\jmath,\jmath b,\jmath q, qb,qq,bb\}$ where $q$, $b$ and $\jmath$ represent $q\,$-jets, $b\,$-jets and non-tagged jets (a jet failing both taggers), respectively. 
	Fig.~\ref{fig:master} shows the resulting upper limits on $\rho$ taking into account only statistical uncertainties in the $qb$ category for the signal significance $S/\sqrt{B}$ at $2\sigma$ (solid boundary) and $5\sigma$ (dashed boundary) as a function of the LHC luminosity, compared to the current best limit by CMS shown by the gray dashed curve. The results suggest that the HL-LHC could find evidence of Cabbibo-suppressed top decays and determine $V_{ts}$ CKM element directly, though the precise precision depends crucially on how well systematic uncertainties can be controlled. %
	The uncertainties in $q$- and $b$-tagging efficiencies could be controlled using $Zj$ production. Then a fit of the categorized dijet data $\{\jmath\jmath,\jmath b,\jmath q, qb,qq,bb\}$ in inclusive dilepton events to a probabilistic model taking into account the tagging efficiencies could be performed, similar to that performed in \cite{Silva:2010qt,Khachatryan:2014nda} for the extraction of $V_{tb}$. Especially relevant for the HE-LHC would be to measure $t\to (s,d)W$ from boosted semi-leptonic $t\bar t$ events. Events with one top-tagged fat jet, one narrow jet and one lepton (both daughter candidates of the leptonic top) can be categorized by the flavour content of the narrow jet. A preliminary analysis~\cite{thePaper} suggests comparable sensitivity to the leptonic $t\bar t$ dataset already at the 13TeV LHC.

\end{document}

%% file: section9/section.tex
\section{The Higgs boson and flavour physics}
{\it\small Authors (TH): F. Bishara, R. Harnik,  A. Martin, M. Schlafer, Y. Soreq, E. Stamou, F. Yu.}
\label{sec:9}
In the SM the Higgs couplings to the fermions are the origin of the flavour structure. In the SM the  
Yukawa couplings, $y_f$, are \CP conserving and proportional to the fermion masses, $m_f$, with a common proportionality factor, 
\begin{align}
\label{eq:SM:Yukawa:proportionality}
    y^{\rm SM}_f = \sqrt{2} m_f / v \, ,
\end{align}
while the tree-level flavour changing couplings are zero. 
Currently, only the third generation Yukawa couplings have been measured and found to be in agreement with the SM predictions, see Refs.~\cite{Aaboud:2017jvq,Aaboud:2018urx,Sirunyan:2018shy,Aad:2015vsa,Aaboud:2018pen,Sirunyan:2018kst,Aaboud:2018zhk}.
For the Higgs couplings to the first two generations, only upper bounds exist at present~\cite{Aaboud:2018fhh,Aaboud:2017ojs,Perez:2015aoa,Altmannshofer:2015qra,Kagan:2014ila}.

To parametrize the deviations from the SM, 
it is useful to introduce
a generalized $\kappa$ framework (the summation is over fermion type $f=u,d,\ell$ and generations $i,j=1,2,3$)
\begin{equation}
\label{eq:L_eff}
\begin{split}
	\mathcal{L}_{\rm eff} &= -\kappa_{f_i} \frac{m_{f_i}}{v} h \bar f_i f_i + i \tilde \kappa_{f_i} \frac{m_{f_i}}{v} h \bar f_i \gamma_5 f_i  
	- \left[\left( \kappa_{f_i f_j} + i \tilde \kappa_{f_i f_j} \right) h \bar f_L^i f_R^j +{\rm h.c.}\right]_{i\neq j}.
\end{split}
\end{equation}
Experimentally, we want to test a number of SM predictions: {\em (i)} proportionality, $y_f\propto m_f$; {\em (ii)} the factor of proportionality, $\kappa_{f_i}=1$; {\em (iii)} diagonality (no off-diagonal flavour violation), $\kappa_{{f_i}{f_j}}=0$; {\em (iv)} reality (no \CP violation), $\tilde
\kappa_{f_i}=\tilde \kappa_{{f_i}{f_j}}=0$ \cite{Nir:2016zkd}.
Different Higgs Yukawa couplings are probed both directly and indirectly. 
Direct methods are: for top Yukawa the $t\bar{t}h$ production~\cite{Aaboud:2017jvq,Aaboud:2018urx,Sirunyan:2018shy}); for bottom and charm, $pp\to Vh, h\to b\bar{b}, c\bar{c} $~\cite{Aaboud:2018zhk,Aaboud:2018fhh,Sirunyan:2018kst}; for leptons $pp\to h\to \ell^+ \ell^-$~\cite{Aaboud:2018pen,Aaboud:2017ojs,Aad:2015vsa}; for light quarks the exclusive $h\to V\gamma$ decays~\cite{Aad:2015sda,Khachatryan:2015lga,Aaboud:2018txb,Aaboud:2017xnb}. The 
kinematical distributions~\cite{Soreq:2016rae,Bishara:2016jga} and
global fits to all the Higgs data also provide bounds on different Yukawa couplings.

The Higgs production and decay signal strengths from the CMS collaboration~\cite{Sirunyan:2018koj}, and from ATLAS for $h\to c\bar{c}$~\cite{Aaboud:2018fhh} from a global fit, which includes the direct observation of $t\bar t h$ production, 
gives for the signal strengths
 $\mu_{tth}= 1.18^{+0.30}_{-0.27}, \mu_{bb}=1.12^{+0.29}_{-0.29}, 
    \mu_{cc} < 105,  \mu_{\tau\tau} = 1.20^{+0.26}_{-0.24}, \mu_{\mu\mu}=0.68^{+1.25}_{-1.24}$.
In terms of modifications of the flavour-diagonal \CP-conserving Yukawas,  the best fit values are, see also \cite{Khachatryan:2016vau,Sirunyan:2018koj},
\begin{equation}
\label{eq:kappalimits}
    \kappa_t = 1.11^{+0.12}_{-0.10}, \qquad \kappa_b = -1.10^{+0.33}_{-0.23},\qquad
    \kappa_\tau = 1.01^{+0.16}_{-0.20}, \qquad \kappa_\mu = 0.79^{+0.58}_{-0.79}.
\end{equation}
The $u$, $d$, $s$, and charm Yukawa couplings can be constrained from a global fit to the Higgs data and the precision electroweak measurements at LEP. Floating all the couplings in the fit gives~ \cite{Perez:2015aoa,Kagan:2014ila},
\begin{align}
\kappa_u&<3.4 \cdot 10^{3},  &\kappa_d&<1.7 \cdot 10^3, &\kappa_s&< 42, & \kappa_c \lesssim 6.2.
\end{align}
The upper bound on $\BR(h\to e^+e^-)$ gives for the electron Yukawa $|\kappa_e|<611$~\cite{Khachatryan:2014aep,Altmannshofer:2015qra}.  
The upper bounds on $\kappa_{c,s,d,u}$ roughly
correspond to the size of the SM bottom Yukawa coupling and are thus much bigger than the corresponding SM Yukawa couplings. The upper bounds can be saturated only if one allows for large cancellations between the
contribution to fermion masses from the Higgs vev and an equally large contribution from NP, but
with an opposite sign. 
In models
of NP motivated by the hierarchy problem, the effects of NP are
generically well below these bounds.

 In the rest of this section we first briefly discuss the expected deviations in a set of NP models, and give the prospects for HL-/HE-LHC (see also~\cite{Perez:2015lra,Brivio:2015fxa,Koenig:2015pha,Aad:2015sda,Bodwin:2014bpa,Bodwin:2013gca}). An expanded version of the discussion is available in write-up of WG2, Sec.~7. 

\subsection{New Physics benchmarks for modified Higgs couplings}

The expected sizes of effective Yukawa couplings,
$\kappa_f$, $\tilde \kappa_f$ and $\kappa_{ff'}, \tilde \kappa{ff'}$, in popular models of weak scale NP models, some of them motivated by the hierarchy problem are shown in  
Tables~\ref{tab:upyukawa} and \ref{tab:upFVyukawa}, 
adapted from
\cite{Bishara:2015cha,Dery:2014kxa,Dery:2013aba,Dery:2013rta,Bauer:2015kzy}.
The predictions are shown for the Standard Model, multi-Higgs-doublet models
(MHDM) with natural flavour conservation (NFC)~\cite{Glashow:1976nt,
  Paschos:1976ay}, a ``flavourful'' two-Higgs-doublet model beyond NFC
(F2HDM)~\cite{Altmannshofer:2015esa, Altmannshofer:2016zrn, 
Altmannshofer:2017uvs, Altmannshofer:2018bch} the MSSM at tree level, 
a single Higgs doublet with
a Froggat-Nielsen mechanism (FN)~\cite{Froggatt:1978nt}, the
Giudice-Lebedev model of quark masses modified to 2HDM
(GL2)~\cite{Giudice:2008uua}, NP models with minimal flavour violation
(MFV)~\cite{D'Ambrosio:2002ex}, Randall-Sundrum models
(RS)~\cite{Randall:1999ee}, and models with a composite Higgs where
Higgs is a pseudo-Nambu-Goldstone boson (pNGB)~\cite{Dugan:1984hq,
  Georgi:1984ef, Kaplan:1983sm, Kaplan:1983fs}. Tables~\ref{tab:upyukawa} and \ref{tab:upFVyukawa} only show predictions for up-quark sector, while the results for down-quark and lepton sectors can be found in (see also Sec.~7 of WG2 write-up).

\begin{table}[t]
\begin{center}
\begin{tabular}{l  c  c  c c  }
\hline\hline
Model	& $\kappa_t$ & $\kappa_{c (u)}/\kappa_t$  & $\tilde \kappa_t/\kappa_t$ & $\tilde \kappa_{c (u)}/\kappa_t$ \\ %
\hline
SM	& 1	& 1 & 0 & 0 \\
MFV &$1+\frac{\Re(a_uv^2+2b_u m_t^2)}{\Lambda^2}$
&$1-\frac{2\Re(b_u)m_t^2}{\Lambda^2}$
&$\frac{\Im(a_uv^2+2b_u m_t^2)}{\Lambda^2}$ & $\frac{\Im(a_u
  v^2)}{\Lambda^2} $ \\
NFC & $V_{hu}\,v/v_u$	& 1 &  0 &0 \\
F2HDM & $\cos\alpha/\sin\beta$	& $-\tan\alpha/\tan\beta$ & $\mcO\left(\frac{m_c}{m_t}\frac{\cos(\beta-\alpha)}{\cos\alpha\cos\beta}\right)$ & $\mcO\left(\frac{m_{c(u)}^2}{m_t^2} \frac{\cos(\beta-\alpha)}{\cos\alpha\cos\beta}\right)$ \\
MSSM	& $\cos\alpha/\sin\beta$	&1 &0  &0\\
FN & $1+\mcO\left(\frac{v^2}{\Lambda^2}\right)$ &
	$1+\mcO\left(\frac{v^2}{\Lambda^2}\right)$ &
	$\mcO\left(\frac{v^2}{\Lambda^2}\right)$ &
	$\mcO\left(\frac{v^2}{\Lambda^2}\right)$ \\
GL2 	& $\cos\alpha/\sin\beta$& $\simeq 3(7)$ & 0 & 0 \\
RS &$1-{\mathcal O}\Big(\frac{ v^2}{m_{KK}^2}\bar Y^2\Big)$&$1+{\mathcal O}\Big(\frac{ v^2}{m_{KK}^2}\bar Y^2\Big)$ &${\mathcal O}\Big(\frac{ v^2}{m_{KK}^2}\bar Y^2\Big)$ &${\mathcal O}\Big(\frac{ v^2}{m_{KK}^2}\bar Y^2\Big)$ \\
pNGB & $1+{\mathcal O}\Big(\frac{ v^2}{f^2}\Big)+{\mathcal O}\Big(y_*^2 \lambda^2 \frac{ v^2}{M_*^2}\Big)$ & $1+{\mathcal O}\Big(y_*^2 \lambda^2 \frac{ v^2}{M_*^2}\Big)$ & ${\mathcal O}\Big(y_*^2 \lambda^2 \frac{ v^2}{M_*^2}\Big)$ & ${\mathcal O}\Big(y_*^2 \lambda^2 \frac{ v^2}{M_*^2}\Big)$ \\
\hline\hline
\end{tabular}
\caption{Predictions for the flavour-diagonal up-type Yukawa couplings
  in a sample of NP models (see text for details).
}
\label{tab:upyukawa}
\end{center}
\end{table}

\begin{table}[t]
\begin{center}
\begin{tabular}{l  c  c  c }
\hline\hline
Model	& $\kappa_{ct (tc)}/\kappa_t$ & $\kappa_{ut (tu)}/\kappa_t$  & $\kappa_{uc (cu)}/\kappa_t$ \\ 
\hline
MFV &$ \frac{\Re\big( c_u m_b^2 V_{cb}^{(*)}\big)}{\Lambda^2}\frac{\sqrt2 m_{t(c)}}{v} $~&~ $ \frac{\Re\big( c_u m_b^2 V_{ub}^{(*)}\big)}{\Lambda^2} \frac{\sqrt2 m_{t(u)}}{v}$~&~ $ \frac{\Re\big( c_u m_b^2 V_{ub(cb)}V_{cb(ub)}^{*}\big)}{\Lambda^2} \frac{\sqrt2 m_{c(u)}}{v}$\\\vspace{0.15cm}
F2HDM & $\mcO\left(\frac{m_c}{m_t}\frac{\cos(\beta-\alpha)}{\cos\alpha\cos\beta}\right)$ & $\mcO\left(\frac{m_u}{m_t}\frac{\cos(\beta-\alpha)}{\cos\alpha\cos\beta}\right)$ & $\mcO\left(\frac{m_c m_u}{m_t^2}\frac{\cos(\beta-\alpha)}{\cos\alpha\cos\beta}\right)$ \\\vspace{0.15cm}
FN &  $\mathcal{O}\left(\frac{v m_{t(c)}}{\Lambda^2} |V_{cb}|^{\pm 1}\right)$ &
	$\mathcal{O}\left(\frac{v m_{t(u)}}{\Lambda^2} |V_{ub}|^{\pm 1}\right)$ &
	$\mathcal{O}\left(\frac{v m_{c(u)}}{\Lambda^2} |V_{us}|^{\pm 1}\right)$\\\vspace{0.15cm}
GL2	& $\epsilon (\epsilon^2)$ & $\epsilon (\epsilon^2)$ & $\epsilon^3$ \\\vspace{0.15cm}
RS & $\sim \lambda^{(-)2} \frac{m_{t(c)}}{v} \bar Y^2\frac{v^2}{m_{KK}^2} $&$\sim \lambda^{(-)3} \frac{m_{t(u)}}{v} \bar Y^2\frac{v^2}{m_{KK}^2} $&$\sim \lambda^{(-)1} \frac{m_{c(u)}}{v} \bar Y^2\frac{v^2}{m_{KK}^2} $ \\\vspace{0.15cm}
pNGB & ${\mathcal O}(y_*^2 \frac{m_t}{v}\frac{\lambda_{L (R),2} \lambda_{L(R),3}m_W^2}{M_*^2})$ & ${\mathcal O}(y_*^2 \frac{m_t}{v}\frac{\lambda_{L (R),1} \lambda_{L(R),3}m_W^2}{M_*^2})$  & ${\mathcal O}(y_*^2 \frac{m_c}{v}\frac{\lambda_{L (R),1} \lambda_{L(R),2}m_W^2}{M_*^2})$ \\
\hline\hline
\end{tabular}
\caption{Same as Table \ref{tab:upyukawa} but for flavour-violating up-type Yukawa couplings. In the SM,
  NFC and the tree-level MSSM the Higgs Yukawa couplings are flavour
  diagonal. The CP-violating $\tilde \kappa_{ff'}$ are obtained by replacing the real part, ${\Re}$, with the imaginary part, ${\Im}$. All the other models predict a zero contribution to these flavour changing couplings.
}
\label{tab:upFVyukawa}
\end{center}
\end{table}

In Tables~\ref{tab:upyukawa} and \ref{tab:upFVyukawa}, $v=246$ GeV is the electroweak vev, while $\Lambda$ is the typical NP scale. For instance, if SM is corrected by dimension six operators with Minimal Flavour Violation (MFV), then the up-quark couplings receive a contribution $ Y_u^\prime\bar{Q}_L H^c u_R/{\Lambda^2}$, so that the Yukawa coupling is $y_u=Y_u +3 Y_u' {v^2}/({2 \Lambda^2})$, with $Y_u^{\prime}= a_uY_u +
        b^{\phantom\dagger}_uY^{\phantom\dagger}_uY_u^\dagger
        Y^{\phantom\dagger}_u + c^{\phantom\dagger}_u
        Y^{\phantom\dagger}_dY_d^\dagger
        Y^{\phantom\dagger}_u+\cdots$, where $Y_{u,d}$ are the SM Yukawas. The $v^2/\Lambda^2$ contributions correct both the diagonal Yukawa couplings, and lead to off-diagonal, flavour violating, couplings. 
        
 Not all NP models lead to flavour-violating Yukawa couplings. For instance, in multi-Higgs-doublet models with natural flavour conservation by assumption only one doublet, $H_u$, couples to the up-type quarks, only one Higgs doublet, $H_d$, couples to the down-type quarks, and only one doublet, $H_\ell$ couples to leptons (it is possible that any of these coincide, as in the SM where $H=H_u=H_d=H_\ell$)~\cite{Glashow:1976nt, Paschos:1976ay}. This only modifies diagonal Yukawa couplings, while off-diagonal remain to be zero, as in the SM. Similar result applies to the MSSM tree-level Higgs potential  where $h_{u,d}$ mix into the Higgs mass-eigenstates $h$ and
$H$ as $h_u=\cos \alpha h+\sin\alpha H$, $h_d = -\sin\alpha h +\cos\alpha H$,
where $h$ is the observed SM-like Higgs, and the vevs are $v_u=\sin\beta\, v$, $v_d=\cos\beta$. Flavourful two-Higgs-doublet model~\cite{Altmannshofer:2015esa}, on the other hand,  introduces mass suppressed off-diagonal and \CP violating contributions, a direct consequence of the fact that one Higgs doublet couples only to top, bottom and tau, and a second Higgs doublet couples to the remaining fermions (see also~\cite{Botella:2016krk,Ghosh:2015gpa,Das:1995df,Blechman:2010cs}). Off-diagonal and \CP violating Yukawa couplings are typical of any model of flavour with new degrees of freedom that are light enough, such as if the Higgs mixes with the flavon from the Froggatt-Nielsen (FN) mechanism~\cite{Froggatt:1978nt}, or if FN mechanism gives the structure of both dimension 4 and dimension 6 operators in SMEFT. Another example is the model of quark masses introduced by Giudice and Lebedev~\cite{Giudice:2008uua}, where the quark masses, apart from the top mass, are small, because they arise from higher dimensional operators. 

In Randall-Sundrum warped extra-dimensional models, that address simultaneously the hierarchy problem and the hierarchy of the SM fermion masses ~\cite{Randall:1999ee, Gherghetta:2000qt, Grossman:1999ra, Huber:2000ie,
	Huber:2003tu}, the corrections to the Yukawa couplings are suppressed by the masses of Kaluza-Klein (KK) modes,  $m_{KK}$. If Higgs is a
pseudo-Goldstone boson arising from the spontaneous breaking of a
global symmetry in a strongly coupled sector, coupling to the composite
sector with a typical coupling $y_*$ \cite{Dugan:1984hq,
  Georgi:1984ef, Kaplan:1983sm, Kaplan:1983fs} (for a review, see~\cite{Panico:2015jxa}), the corrections to the Yukawa couplings are suppressed by the mass of composite resonance with a typical mass $M_*\sim \Lambda$.
  
  In conclusion, we see that the NP effects in Higgs couplings to the SM fermions are either
suppressed by $1/\Lambda^2$, where $\Lambda$ is the NP scale, or are
proportional to the mixing angles with the extra scalars. This means that in the decoupling limit, $\Lambda \to \infty$ and/or small mixing angles, all the NP effects vanish. In the decoupling limit the Higgs couplings coincide with  the SM predictions, $\kappa_f=1$, while $\tilde \kappa_f=0, \kappa_{ff'}=0, \tilde \kappa_{f f'}=0$.

\subsection{Probing charm and light quark Yukawa couplings}

The inclusive method of probing the charm-quark Yukawa is in many ways complementary 
to searches for exclusive decays (see discussion of Sec.~\ref{sec:exclusiveHiggs}) or 
searches for deviations in Higgs distributions (see Sec.~\ref{sec:HiggsDist}).
For example, in the inclusive approach an underlying assumption is 
that the Higgs coupling to $WW$ and $ZZ$ ---entering Higgs production--- is SM-like,
while the interpretation of Higgs distributions assumes no additional new physics contribution that affects them in a significant way.
An important difference between the inclusive and the exclusive approach is that the latter relies on interference with the SM $H\to \gamma\gamma$ amplitude while the former does not.
Therefore, in principle the exclusive approach may be sensitive to the sign and 
 \CP  properties of the coupling to which the inclusive approach is insensitive to.
At the same time, measurements of exclusive decays of the Higgs are challenging due 
to the small probability of fragmenting into the specific final state and 
large QCD backgrounds, which is why the inclusive approach appears to be the most promising one to probe deviations in the magnitude of the Higgs to charm coupling.

\begin{figure}[t]
	\centering
	\includegraphics[width=0.49\textwidth]{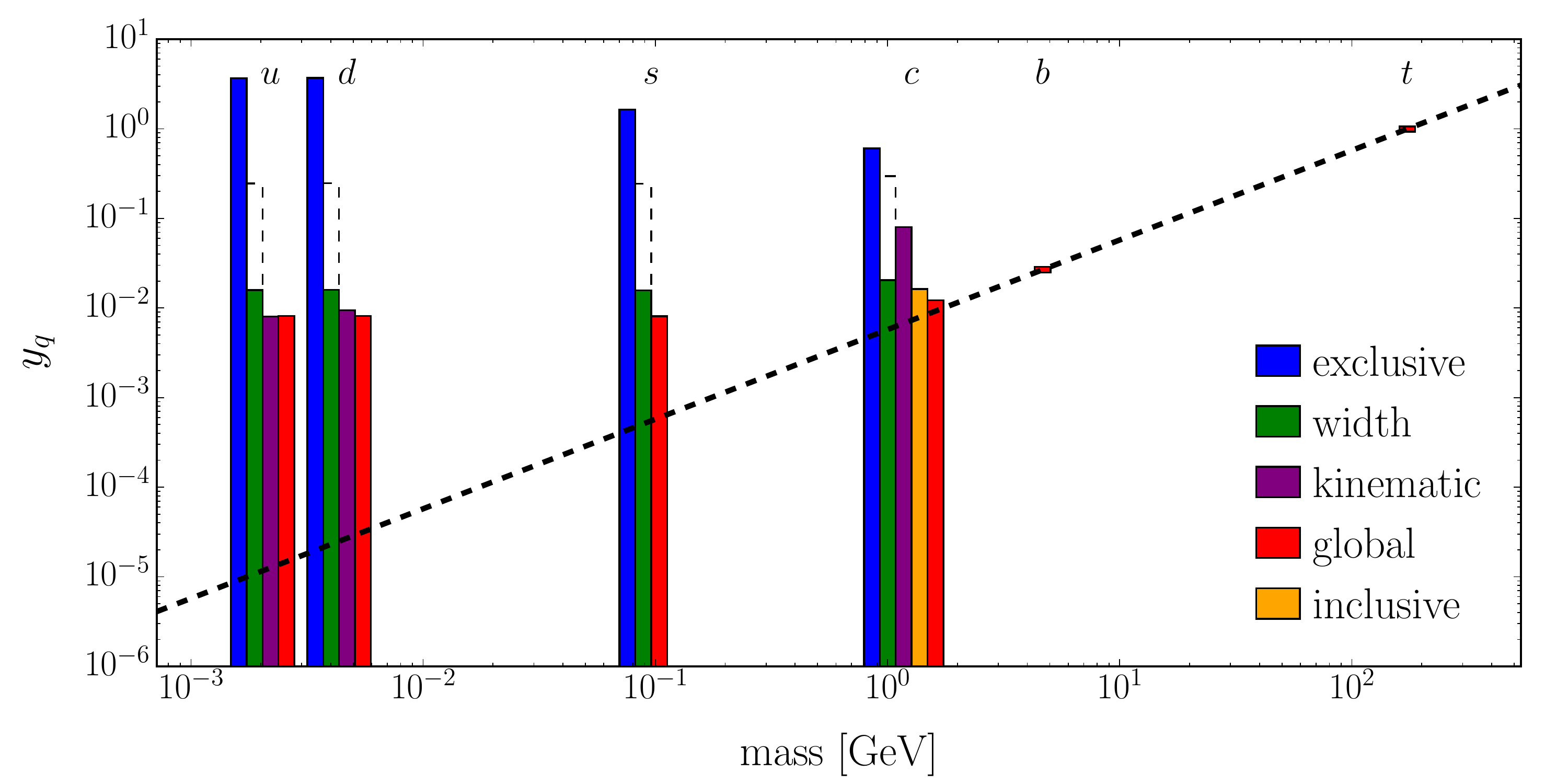}
	\includegraphics[width=0.49\textwidth]{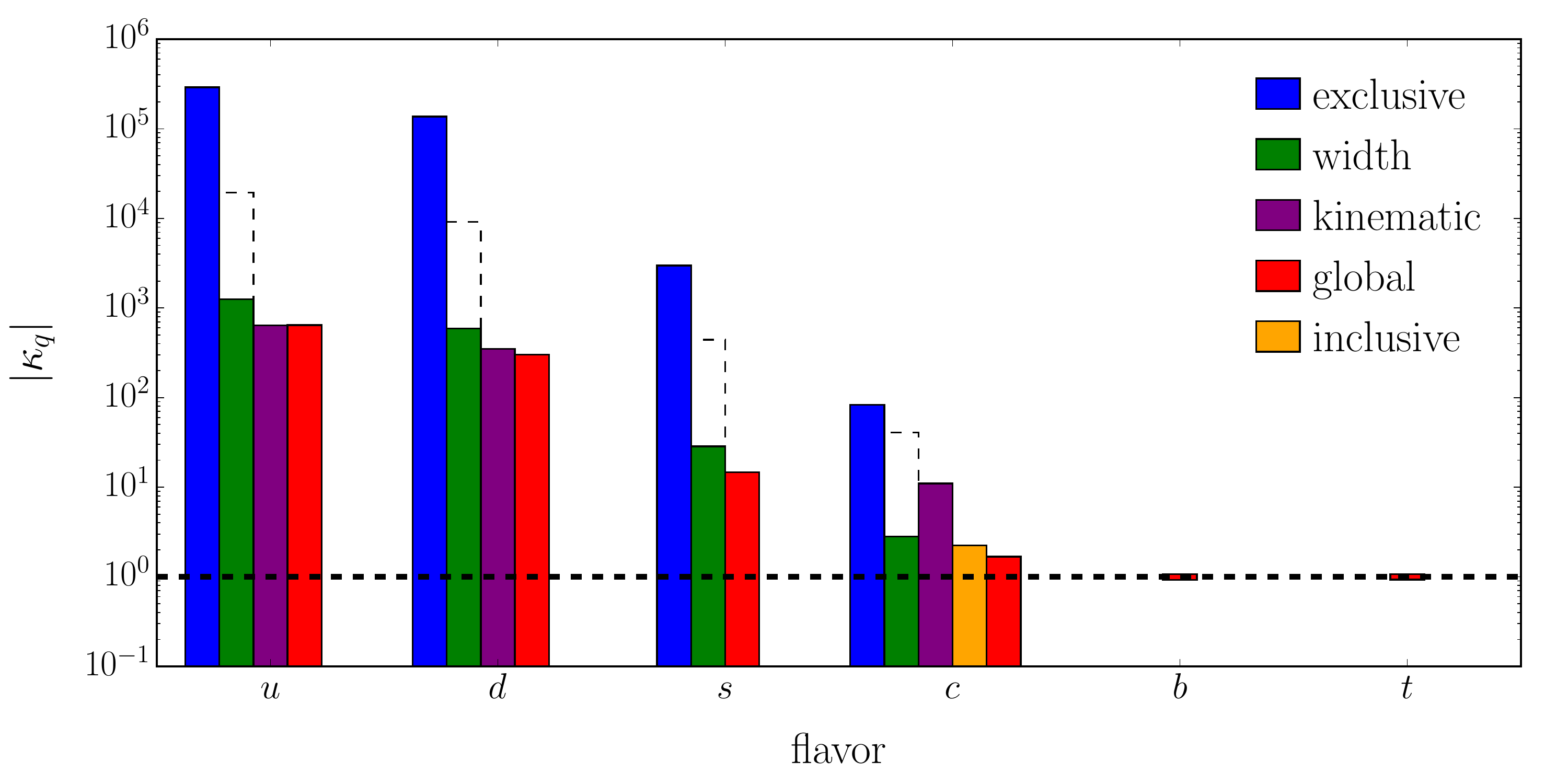}
	\caption{\vtwo{Summary of the HL-LHC projections for measurements of Higgs Yukawa couplings to quarks discussed in the main text:  from exclusive decays (blue), from Higgs kinematic distributions (purple) and from inclusive constraints (yellow), from off-shell bounds on total decay width (green), from the  $h\to \gamma \gamma$ and  $ZZ$ line shapes (dashed), and from global fits (red).}	
	\label{fig:higgs:forecast}}
\end{figure}

\vtwo{
The summary of the projections that are discussed in the following sections, is given in Fig. \ref{fig:higgs:forecast}. Shown are the expected HL-LHC constraints from exclusive decays (blue), from Higgs kinematic distributions (purple) and from inclusive $c$-tagging measurements (yellow), as well as the constraints from the combined $h\to \gamma \gamma$ and  $h\to ZZ$ line shapes  (dashed lines), from the total width using off-shell methods (green), and from the global fits to Higgs data (red).  For the global fit one assumes $\BR(h\to {\rm BSM})<5\%$, while the constraints from the total width assume that one will be able to constrain $\Gamma < 1.2\Gamma_{\rm SM}$ (green bars), and that the resolution for the $\gamma \gamma/ZZ$ line shape will be 200MeV at the HL-LHC. Note that the off-shell constraints are model dependent. 
}

\begin{figure}[t]
	\centering
	\includegraphics[width=0.49\textwidth]{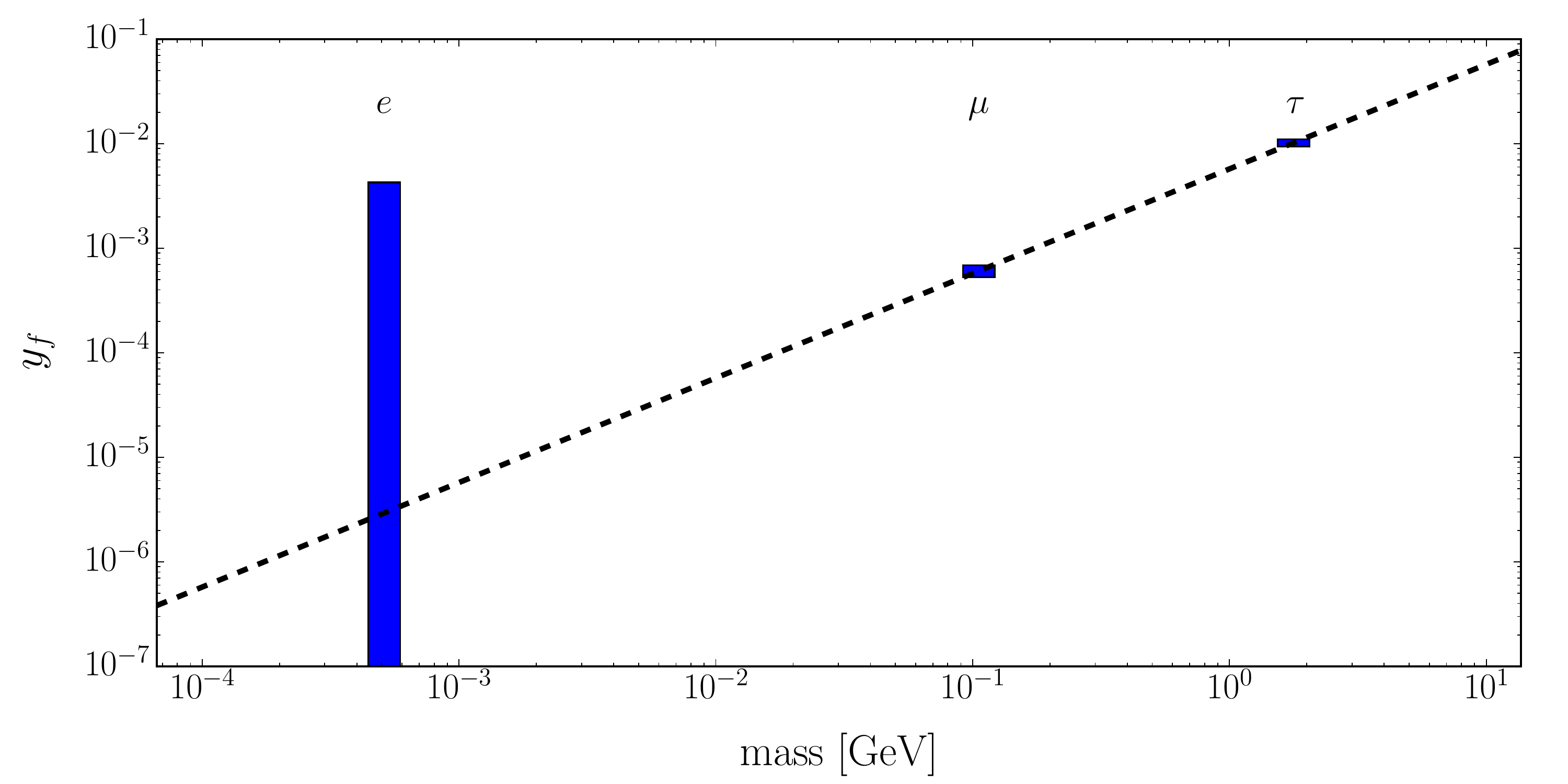}
	\includegraphics[width=0.49\textwidth]{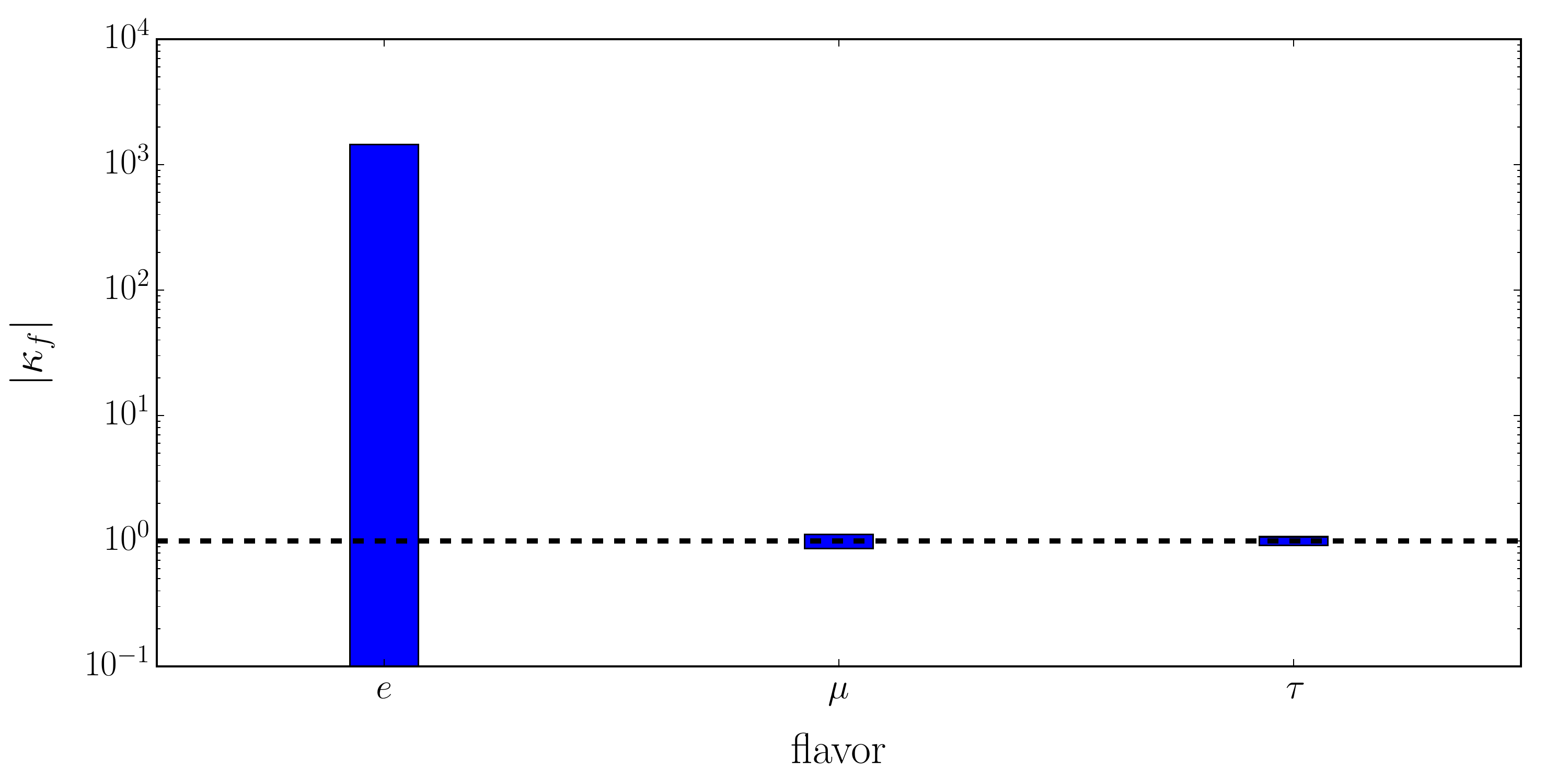}
	\caption{\vtwo{Summary of the HL-LHC projections for measurements of Higgs Yukawa couplings to leptons from searches for $h\to\ell^+\ell^-$ decays.}	
	\label{fig:higgs:forecast}}
\end{figure}

\subsubsection{Inclusive charm quark tagging}

The most straight-forward way of inclusively probing charm quark Yukawa is to 
expand the $h\to b\bar b$ search with a search for 
$pp \to (Z/W\to\ell\ell/\nu) (h\to c\bar c)$ \cite{Perez:2015aoa}. Another possibility is to search for $pp\to h c$ production~\cite{Brivio:2015fxa}. The search for $h\to c \bar c$ from $pp\to Z h$ at $\sqrt{s}=13$~TeV was recently performed on a $36.1$~fb$^{-1}$ sample by ATLAS \cite{Aaboud:2018fhh}. The 
$c$-tagging algorithms are similar to $b$-tagging ones, with the most relevant quantities the 
displaced vertices due to a finite lifetime of $c$-hadrons.
Prior to its use in Higgs physics, $c$-tagging was used early on in Run I of the LHC by ATLAS and CMS in searches for supersymmetry~\cite{Aad:2014nra,Aad:2015gna}.
Its usefulness in relations to Higgs physics was first discussed in Ref.~\cite{Delaunay:2013pja} and subsequently
used in Ref.~\cite{Perez:2015aoa} to recast ATLAS and CMS Run~I analyses for $h\to b\bar b$ to provide the first direct LHC constraint on charm Yukawa.

The efficiency of jet flavour tagging algorithms in associating a jet to a specific quark is 
correlated with the confidence to reject other hypotheses, e.g., production from light-quarks.
Given the rather similar lifetimes of $b$ and $c$ hadrons, there is always
a non-negligible contamination of the $c$-jet sample with
$b$ jets~\cite{Perez:2015aoa}. The ATLAS analysis \cite{Aaboud:2018fhh} used a working point  with an  efficiency of approximately $41\%$ to tag $c$-jets and rejection factors of roughly $4$ and $20$ for $b$- and light-quark-jets, respectively.
An inclusive $h\to c\bar c$ analysis
must thus either assume a SM value 
for the bottom Yukawa (as in Ref.~\cite{Aaboud:2018fhh}), or 
break the degeneracy between $y_b$ and $y_c$, e.g., by using more than one tagging working point with different ratios of $c$-tagging to $b$-tagging efficiencies~\cite{Perez:2015aoa,Perez:2015lra}.  

\begin{figure}[t]
	\centering
	\includegraphics[width=0.45\textwidth]{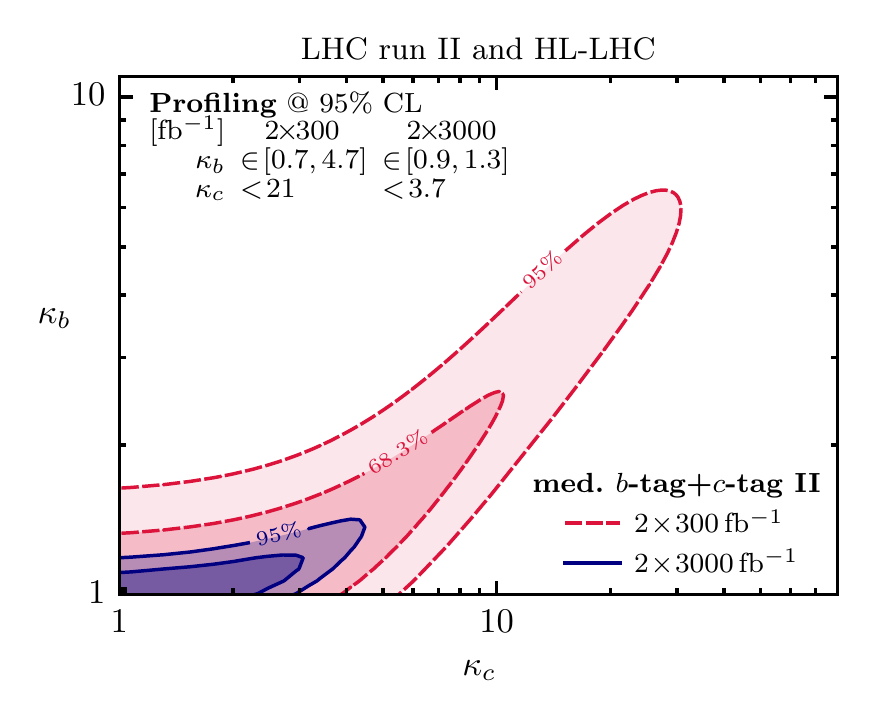}
	\includegraphics[width=0.45\textwidth]{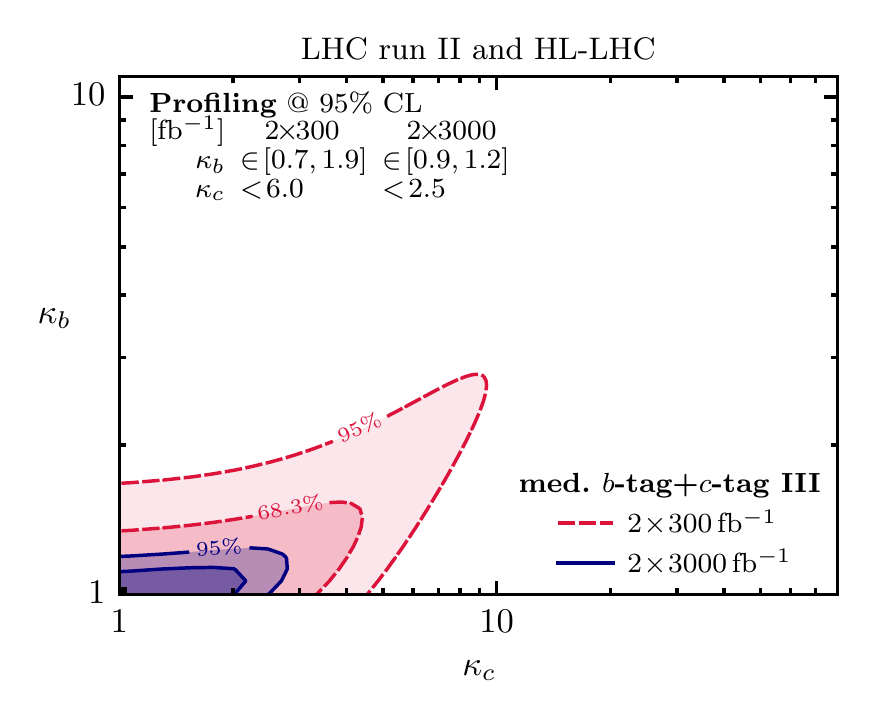}
	\caption{Projections for measuring charm Yukawa modifications from an inclusive 
		$H\to c\bar c$ search at $\sqrt{s}=14$\,TeV using two different 
		$c$-taggers (left and right panel) \cite{Perez:2015lra}.
		In red the $95\%$ CL region employing an integrated luminosity 
		of $2\times 300$~fb$^{-1}$ and in blue the region 
		employing $2\times 3000$~fb$^{-1}$.
	\label{fig:inclusiveforecast}}
\end{figure}

The prospects for measuring $pp\to Zh(\to c\bar c)$ at the HL-LHC were obtained in Ref. \cite{ATL-PHYS-PUB-2018-016}, by rescaling the 
Run 2 analysis \cite{Aaboud:2018fhh} 
to an integrated luminosity of $3000$~fb$^{-1}$.
Possibilities to reduce the systematic uncertainties were discussed as well.
Assuming $\kappa_b=1$,
an upper bound  on the signal strength 
 of $\mu_{Zh(c\bar c)} <6.3$ at $95\%$ CL can be set.
In Ref.~\cite{Perez:2015lra} instead, 
the prospects for measuring $h\to b\bar b$ at $\sqrt{s}=14$\,TeV 
\cite{ATLAS-collaboration:2012iza} were recast 
 to obtain the projection for an inclusive $h\to c \bar c$ measurement. Both $\kappa_c$ and $\kappa_b$ 
 are treated as free variables. 
Fig.~\ref{fig:inclusiveforecast} shows results for two flavour tagging working points: a $c$-tagging efficiency of $30\%$ ($c$-tag II, left panel) or $50\%$ ($c$-tag III, right panel), while 
in both cases the $b$(light)-jet rejection is $5$($200$). 
 The %
$\kappa_b$ direction is profiled away, giving the HL-LHC sensitivity of 
$\kappa_c \simeq 21 (6)$  at $95\%$ CL with
$c$-tag II ($c$-tag III) and $2\times 3000$\,fb$^{-1}$ at $\sqrt{s}=14$\,TeV,
indicated by the blue regions in Fig.~\ref{fig:inclusiveforecast}.

Even though the LHCb experiment operates at lower luminosity  compared to ATLAS and CMS, it has unique capabilities for discrimination between $b$- and $c$-jets thanks to its excellent vertex reconstruction system~\cite{LHCb-PAPER-2015-016}. With the secondary vertex tagging (SV-tagging) LHCb achieved an identification efficiency of 60$\%$ on $b$-jets, of 25$\%$ on $c$-jets and a light jets (light quarks or gluons) mis-identification probability of less than 0.2$\%$. Further discrimination between light and heavy jets and between $b$- and $c$-jets is achieved by exploiting the secondary vertex kinematic properties, using Boosted Decision Tree techniques (BDTs). For instance, an additional cut on the BDT that separates $b$- from $c$-jets removes 90$\%$ of $h\rightarrow b \bar{b}$ while retaining 62$\%$ of $h\rightarrow c \bar{c}$ events~\cite{LHCb:2016yxg}. 

\begin{figure}[t]
	\centering
	\includegraphics[width=0.3\linewidth]{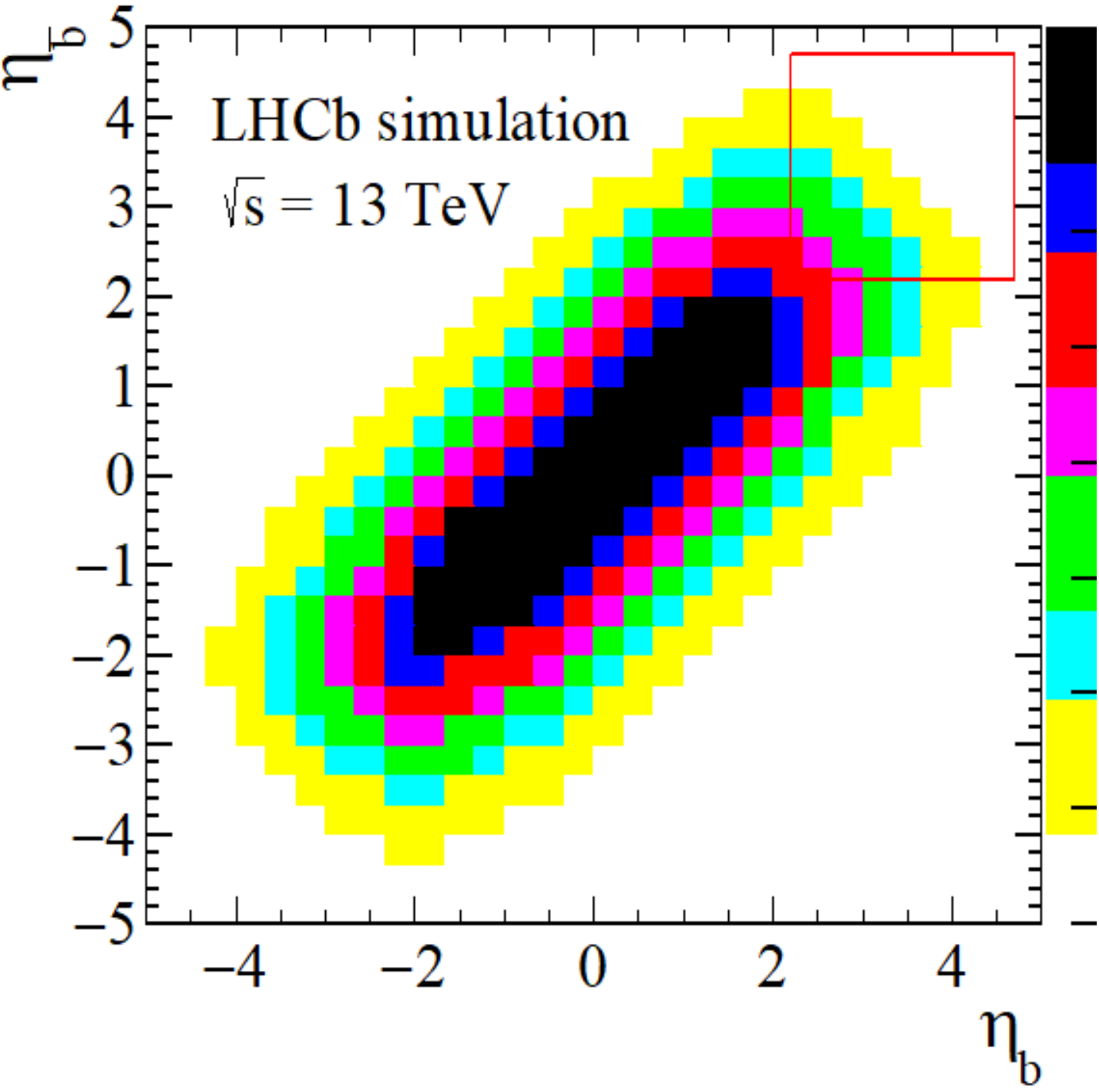}~~~~~~~~~
	\includegraphics[width=0.65\linewidth]{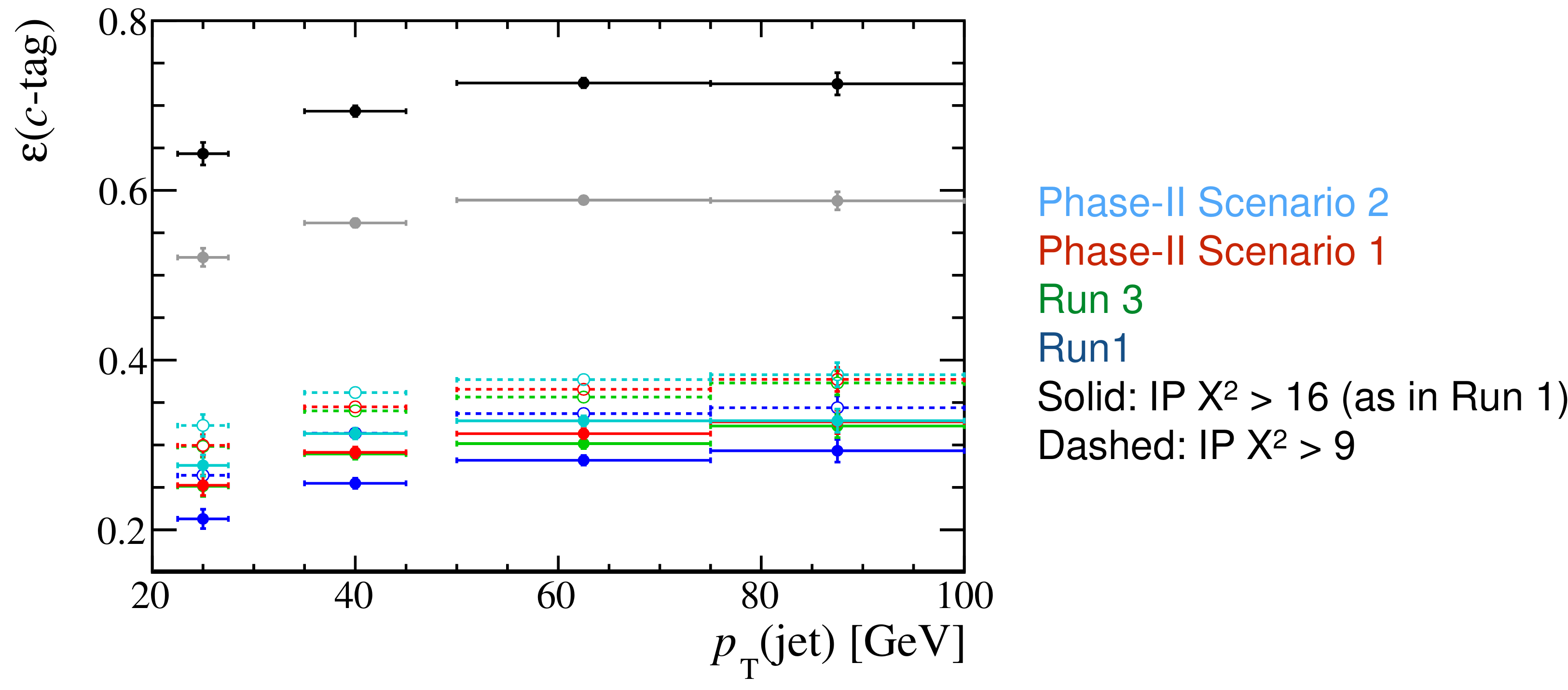}
	\caption{Left: 2D histogram showing the coverage of the LHCb acceptance for the $b\bar{b}$ pair produced by the Higgs decay in associated production with a $W$ or a $Z$ boson. Right: LHCb $c$-jet SV-tagging efficiency for different scenarios in the HL-LHC conditions.}
\label{fig:hbbetas}
\end{figure}

The LHCb acceptance covers $\sim 5$\% of the associated production of $W/Z+h$ at 13 TeV.  Fig.~\ref{fig:hbbetas} (left) shows the coverage of  LHCb for the $b\bar{b}$ pair produced in the decay of the  Higgs boson in association with a vector boson. When the two $b$-jets are within the acceptance, the lepton from $W/Z$ tends to be in acceptance as well ($\sim 60$\% of the time). Due to the forward geometry, Lorentz-boosted Higgs bosons are  likely to be properly reconstructed.

LHCb set  upper limits on the $V+h(\rightarrow b \bar{b})$ and $V+h(\rightarrow c \bar{c})$ production~\cite{LHCb:2016yxg} with data from LHC Run 1.
Without any improvements in the analysis or detector, the extrapolation of this  to 300fb$^{-1}$ at 14\,TeV leads to a sensitivity of $\mu_{Zh(cc)}\lesssim 50\,$.  Detector improvements are expected in future upgrades, in particular in impact parameter resolution which directly affects the $c$-tagging performance. 
If the detector improvement is taken into account, the $c$-jet tagging efficiency with the SV-tagging is expected to  improve as shown in Fig.~\ref{fig:hbbetas} (right). 
Further improvement is expected from the electron reconstruction due to upgraded versions of the electromagnetic calorimeter. Electrons are used in the identification of the vector bosons associated with the Higgs.
With these improvements, the expected limit can be pushed down to $\mu_{Zh(cc)}\lesssim 5-10$ which corresponds to a limit of 2-3 times the Standard Model prediction on the charm Yukawa coupling.
This extrapolation does not include improvements in analysis techniques: for instance  Deep Learning methods can be applied to exploit correlations in  jets substructure properties to reduce the backgrounds.

\subsubsection{Strange quark tagging}
\label{sec:strange-tagging}

The main idea behind the strange tagger described in~\Bref{Duarte-Campderros:2018ouv} is that
strange quarks---more than other partons---hadronize to prompt kaons which carry a large fraction of the jet momentum. %
Although the current focus at LHC is
mainly on charm and bottom tagging, recognizing strange jets has been attempted before at
DELPHI~\cite{Boudinov:1998fao} and SLD~\cite{Kalelkar:2000ig}, albeit in $Z$ decays.

\begin{figure}[t]
  \centering
  \includegraphics[width=0.5\textwidth]{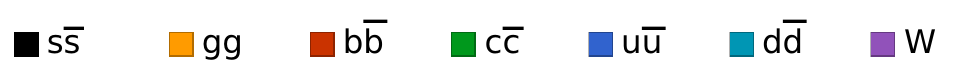}\\
  \includegraphics[width=0.45\textwidth]{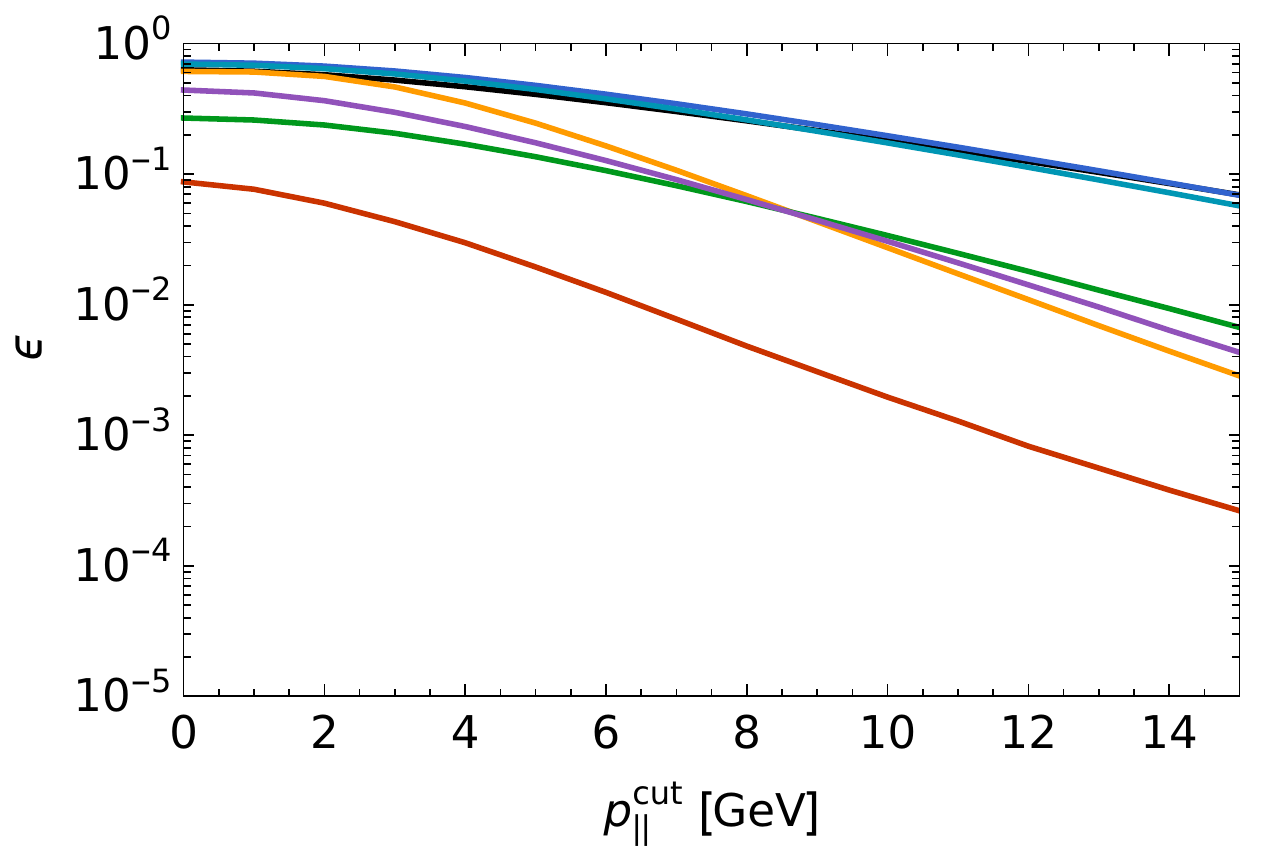}
  \includegraphics[width=0.45\textwidth]{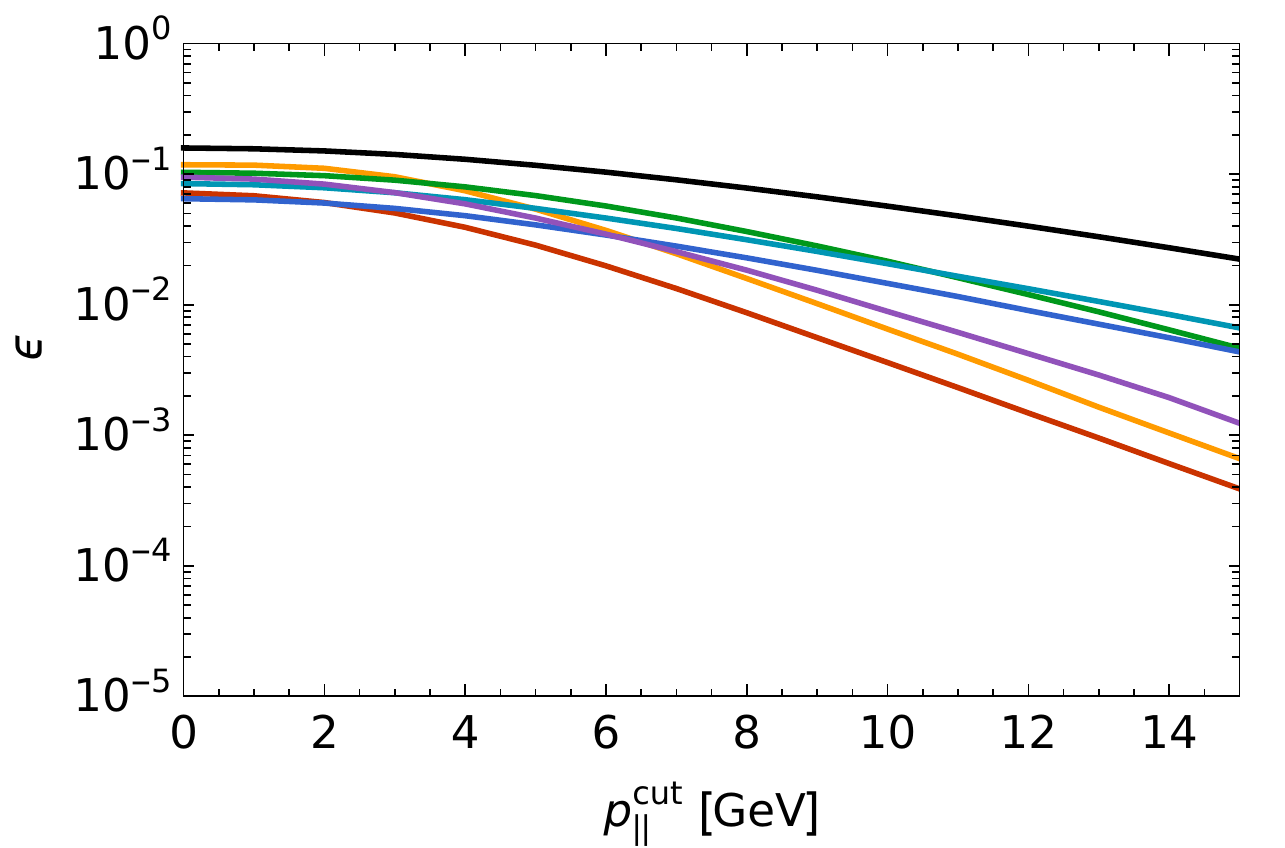}
  \caption{Efficiencies as function of the cut on $p_{||}$ and for $d_0<$\Unit{14}{$\mu$m} to
    reconstruct the different Higgs decay channels and $W$ decays as $s\bar s$ event by the
    described tagger. The left plot shows the CC channel, the right the CN channel.}
  \label{fig:stagger-efficiencies}
\end{figure}

Fig.~\ref{fig:stagger-efficiencies} shows results for a strange tagger
based on an analysis of event samples of Higgs and $W$ events generated with
\texttt{PYTHIA} 8.219~\cite{Sjostrand:2006za, Sjostrand:2014zea}. In each of the two hemispheres of the resonance decay, the charged pions and kaons stemming from the resonance are selected with an
assumed efficiency of 95\%. Similarly, $K_S$ are identified with an efficiency of 85\% if they decay via $K_S\to \pi^+\pi^-$
within \Unit{80}{cm} of the interaction point, which 
allows one to reconstruct
the decaying neutral kaon. Among the two lists of kaon candidates---one per hemisphere---one kaon in
each list is chosen for further analysis such that the scalar sum of their momenta is maximized
while rejecting charged same-sign pairs. The events are separated into the categories
charged-charged~(CC), charged-neutral~(CN) and neutral-neutral~(NN) with relative abundances of about CC:CN:NN$\approx 9:6:1$.

All selected candidates are required to carry a large momentum $p_{||}$ along the hemisphere
axis. This cut reduces the background from gluon jets as gluons radiate more than quarks
and therefore tend to spread their energy among more final state particles. In addition, charged kaons need to be produced promptly, in order to reject heavy flavour jets. The latter requirement is implemented by a cut on the impact parameter $d_0$ after the truth value has been smeared by the detector resolution.

The efficiencies obtained in the CC and CN channel for a cut of $d_0<$\Unit{14}{$\mu$m} are shown in
Fig.~\ref{fig:stagger-efficiencies}. While there is clearly still ample room for improvement, this
simple tagger  already shows  a good suppression of the bottom, charm and gluon
background  by orders of magnitude. Due to missing particle identification at ATLAS and CMS, the efficiencies for first-generation jets and
strange jets are degenerate in the CC channel. However, in the CN channel, due to the required
$K_S$, a suppression of pions is achieved that breaks this degeneracy. This is particularly
interesting in light of the HL-LHC, where a large background from first generation jets is expected.

\subsubsection{Exclusive Higgs decays}
\label{sec:exclusiveHiggs}

Exclusive Higgs decays to a vector meson, $V$, and a photon, $h\to V\gamma$, can directly probe  bottom, charm~\cite{Bodwin:2013gca,Bodwin:2014bpa}, strange, down and up~\cite{Kagan:2014ila} quark Yukawas.
On the experimental side, both ATLAS and CMS reported first upper bounds on $h\to \Upsilon\gamma, J/\psi\gamma$~\cite{Aad:2015sda,Khachatryan:2015lga}, $h\to\phi\gamma$ and $h\to\rho\gamma$~\cite{Aaboud:2016rug,Aaboud:2017xnb}, sensitive to diagonal Yukawa couplings. 
The $h\to V\gamma$ decays
receive two contributions, from $h\to \gamma\gamma^*$ decay followed by a $\gamma^*\to V$ fragmentation, and from a numerically smaller amplitude that involves the direct coupling of quarks to the Higgs~\cite{Keung:1983ac,Bodwin:2013gca,Bodwin:2014bpa,Kagan:2014ila}. The sensitivity to the quark Yukawa couplings thus comes mostly from the interference of the two amplitudes.

Normalizing to the $ h\to ZZ^*\to4\ell$ channel the total Higgs width cancels~\cite{Perez:2015aoa,Koenig:2015pha}, 
\begin{equation}
	\cR_{V\gamma,ZZ^*} 
\equiv	\frac{\mu_{V\gamma} }{\mu_{ZZ^*}}  \frac{ \BR^{\rm SM}_{h\to V\gamma}}{ \BR^{\rm SM}_{h\to ZZ^*}}  
	\simeq  
	\frac{\Gamma_{h\to V\gamma}}{ \Gamma_{h\to ZZ^*}},
\end{equation}
where $ \BR^{\rm SM}_{h\to ZZ^*}$ is the SM branching ratio for $h\to ZZ^*\to 4\ell$.
In the last equality we also assumed perfect cancellation of the production mechanisms (this is entirely correct, if $h\to \gamma\gamma$ is used as normalization channel, but at present this leads to slightly worse bounds on light quark Yukawas). Using predictions from Ref.~\cite{Koenig:2015pha} gives the currently allowed ranges for light quark Yukawa couplings, collected in Table \ref{tab:hexclusive} (here $\kappa_Z$ and $\kappa_{\gamma\gamma}^{\rm eff}$ parametrize deviations of $h\to ZZ, \gamma\gamma$ amplitudes relative to their SM values).

\begin{table}[t]
\begin{center}
\begin{tabular}{llcc}
\hline\hline
mode & ~~~~~~~~~~~~$\BR_{h\to V\gamma}<$ & $\cR_{V\gamma,ZZ^*}<$ &Yukawa range  \\
\hline
$J/\psi\,\gamma$  &
$1.5\cdot 10^{-3}$~\cite{Aad:2015sda,Khachatryan:2015lga} & 
$9.3$ &
$-295\kappa_Z + 16\kappa^{\rm eff}_{\gamma\gamma} < \kappa_c < 295\kappa_Z + 16\kappa^{\rm eff}_{\gamma\gamma} $
\\
$\phi\,\gamma$ &
$4.8 \cdot 10^{-4}$~\cite{Aaboud:2016rug,Aaboud:2017xnb} & 
$3.2$ &
$-140\kappa_Z + 10\kappa^{\rm eff}_{\gamma\gamma} < \bar{\kappa}_s < 140\kappa_Z + 10\kappa^{\rm eff}_{\gamma\gamma} $
\\
$\rho\,\gamma$ &
$8.8 \cdot 10^{-4}$~\cite{Aaboud:2017xnb} & 
$5.8$ &
$-285\kappa_Z + 42\kappa^{\rm eff}_{\gamma\gamma} < 2\bar{\kappa}_u + \bar{\kappa}_d < 285\kappa_Z + 42\kappa^{\rm eff}_{\gamma\gamma} $\\
\hline\hline
\end{tabular}
\end{center}
\caption{The current 95\%\,C.L. upper bounds, assuming SM Higgs production, on different exclusive $h\to V\gamma$ decays, and the interpretation in terms of the Higgs Yukawa couplings. Note that $\bar{\kappa}_q = y_q/y^{\rm SM}_b\,$. %
}
\label{tab:hexclusive}
\end{table}%

For prospects to probe light quark Yukawa at HL-/HE-LHC 
we follow 
Ref.~\cite{Perez:2015lra}.
Rescaling with luminosity and the increased production cross sections both the signal and backgrounds, while ignoring any changes to the analysis that may change the ratios of the two, gives the projected sensitivities 
listed in Table \ref{tab:exclusiveproj}.
The estimates in Table~\ref{tab:exclusiveproj} are in agreement with the ATLAS projection for $h\to J/\psi\gamma$~\cite{ATL-PHYS-PUB-2015-043}, which quotes $\cR_{J/\psi\gamma,ZZ^*}<0.34^{+0.14}_{-0.1}$.
We see that only large enhancements, with Yukawa coupling of light quarks well above the values of bottom Yukawa will be probed. To be phenomenologically viable, these would require large cancellations in the contributions to the light quark masses (and a mechanism to avoid indirect bounds from global fits to the Higgs data).
\begin{table}[t]
\begin{center}
\begin{tabular}{cccc}
\hline\hline
mode & collider energy & $\cR_{V\gamma,ZZ^*}<$ &Yukawa range ($\kappa_V=\kappa^{\rm eff}_{\gamma\gamma}=1$)   \\
\hline
$J/\psi\,\gamma$ &
14\,TeV &  $0.47\sqrt{L_3}$ &
$16 - 67 L_3^{1/4}  < \kappa_c < 16 + 67 L_3^{1/4} $
\\
&27\,TeV &  $0.28\sqrt{L_3}$ &
$16 - 52 L_3^{1/4}  < \kappa_c < 16 + 52 L_3^{1/4} $\\
\hline
$\phi\,\gamma$ &
14\,TeV & 
$0.33\sqrt{L_3}$ &
$11 - 46 L_3^{1/4} < \bar{\kappa}_s < 11 + 46 L_3^{1/4} $\\
&
27\,TeV & 
$0.20\sqrt{L_3}$ &
$11 - 35 L_3^{1/4} < \bar{\kappa}_s < 11 + 35 L_3^{1/4} $\\
\hline
$\rho\,\gamma$ &
14\,TeV & 
$0.60\sqrt{L_3}$ &
$44 - 93 L_3^{1/4} < 2\bar{\kappa}_u + \bar{\kappa}_d < 44 + 93 L_3^{1/4} $\\
&
27\,TeV & 
$0.36\sqrt{L_3}$ &
$44 - 72 L_3^{1/4} < 2\bar{\kappa}_u + \bar{\kappa}_d < 44 + 72 L_3^{1/4} $\\
\hline\hline
\end{tabular}
\end{center}
\caption{The projections of bounds on Yukawa couplings for HL-/HE-LHC as functions of integrated luminosity, $L_3\equiv (3/{\rm ab})/\cL\,$. Note that $\bar{\kappa}_q = y_q/y^{\rm SM}_b\,$. }
\label{tab:exclusiveproj}
\end{table}%
Higgs exclusive decays can in principle also probe off-diagonal couplings by measuring modes such as $h\to B^*_s \gamma$~\cite{Kagan:2014ila}. 
However, the Higgs flavour violating couplings are strongly constrained by meson mixing~\cite{Blankenburg:2012ex,Harnik:2012pb}, so that the expected rates are too small to be observed. 
For a detailed discussion of $h \to MZ,MW$ channels see~\cite{Alte:2016yuw}.

\subsubsection{Yukawa constraints from Higgs distributions}
\label{sec:HiggsDist}

\subsubsubsection{Higgs \pt and rapidity distributions}

\begin{figure}[t]
\begin{center}
\includegraphics[width=0.28\textwidth]{\main/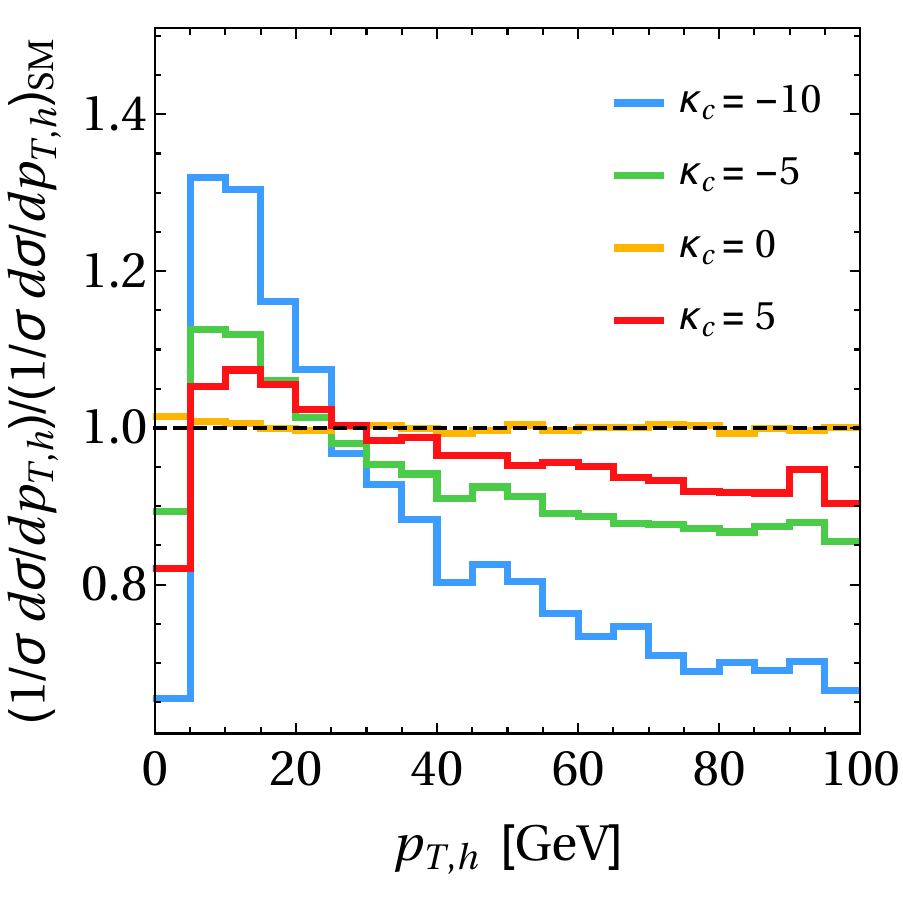}~~~~
\includegraphics[width=0.34\textwidth]{\main/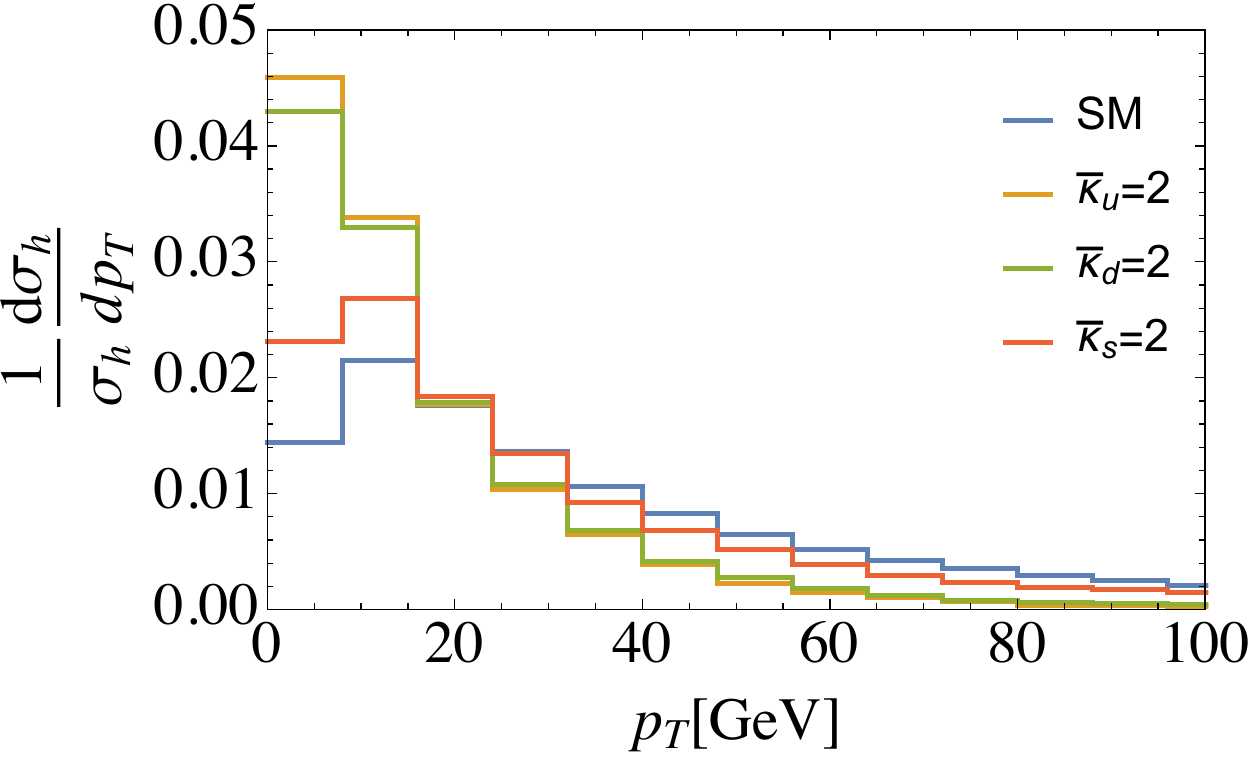}~~~~
\includegraphics[width=0.34\textwidth]{\main/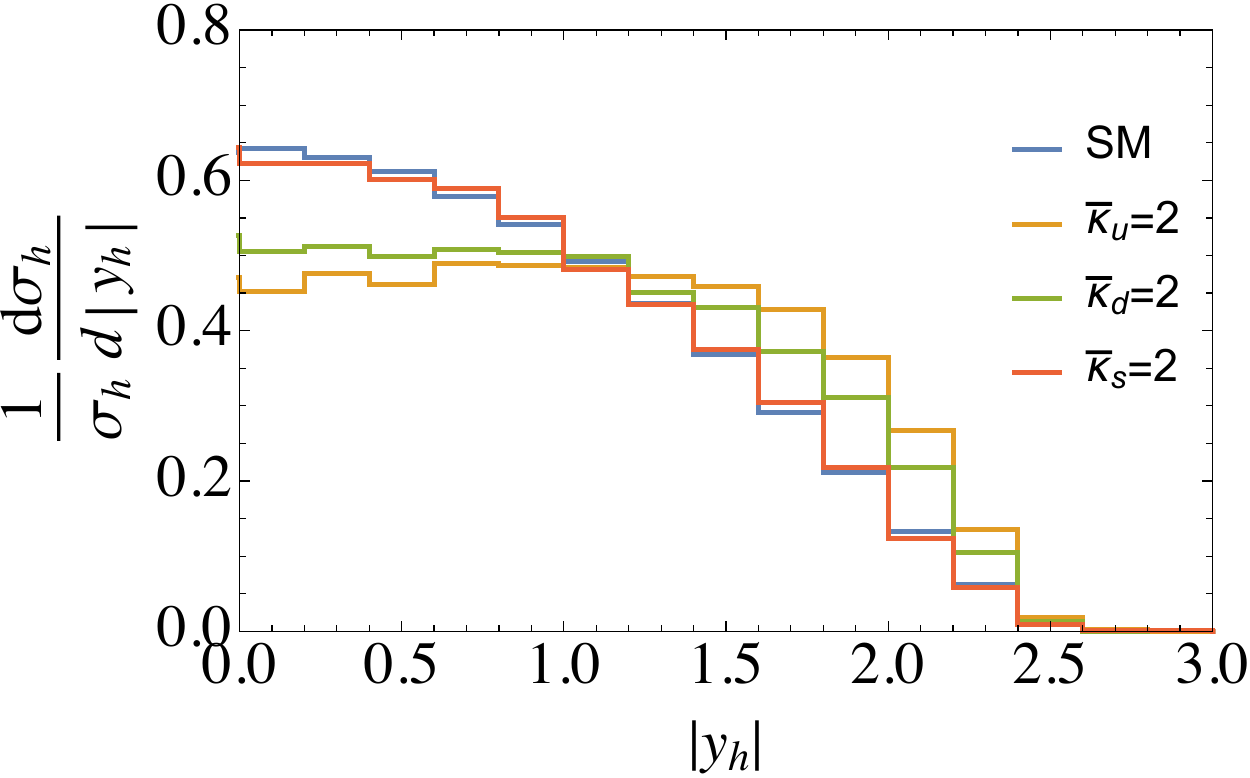}
\caption{Normalized Higgs \pt and rapidity distributions for modified charm (left) and $u$, $d$ and $s$ Yukawas (middle and right). Taken from~\cite{Bishara:2016jga} and~\cite{Soreq:2016rae}. 
}
\label{fig:HiggsDist}
\end{center}
\end{figure}

In general, the Higgs \pt distribution probes
whether NP particles are running in the $gg\to h +(g)$ loops 
\cite{Arnesen:2008fb,Biekotter:2016ecg,Brehmer:2015rna,Dawson:2015gka,Schlaffer:2014osa,Grojean:2013nya,Langenegger:2015lra,Bramante:2014hua,Buschmann:2014twa,Azatov:2013xha,Banfi:2013yoa,Buschmann:2014sia}. 
However,  the soft part of the \pt spectrum is also an indirect probe of light quark Yukawas~\cite{Soreq:2016rae,Bishara:2016jga}. If 
Higgs, unlike in the SM, is produced from the $u \bar u$ or $d \bar d$ fusion,
then {\em (i)} the Sudakov peak is shifted to $p_T\sim 5$\,GeV from $\sim10$\,GeV in the SM~\cite{Collins:1984kg,Soreq:2016rae}, and  {\em (ii)} the rapidity distributions are more forward, i.e., shifted toward larger $\eta$~\cite{Soreq:2016rae}, see Fig. \ref{fig:HiggsDist}.
Enhanced $s$ or $c$ Yukawa couplings also lead to softer Higgs \pt spectrum. The dominant effect for charm quark is due to one loop contributions  to $gg\to hj$ process that are enhanced by double logarithms~\cite{Baur:1989cm}, while for strange quark it is due to $s\bar s$ fusion production of the Higgs~\cite{Soreq:2016rae}. 
Fig.~\ref{fig:HiggsDist} shows the corresponding normalized distributions, for which many theoretical and experimental uncertainties cancel.

\begin{figure}[t]
\begin{center}
\includegraphics[width=0.35\textwidth]{\main/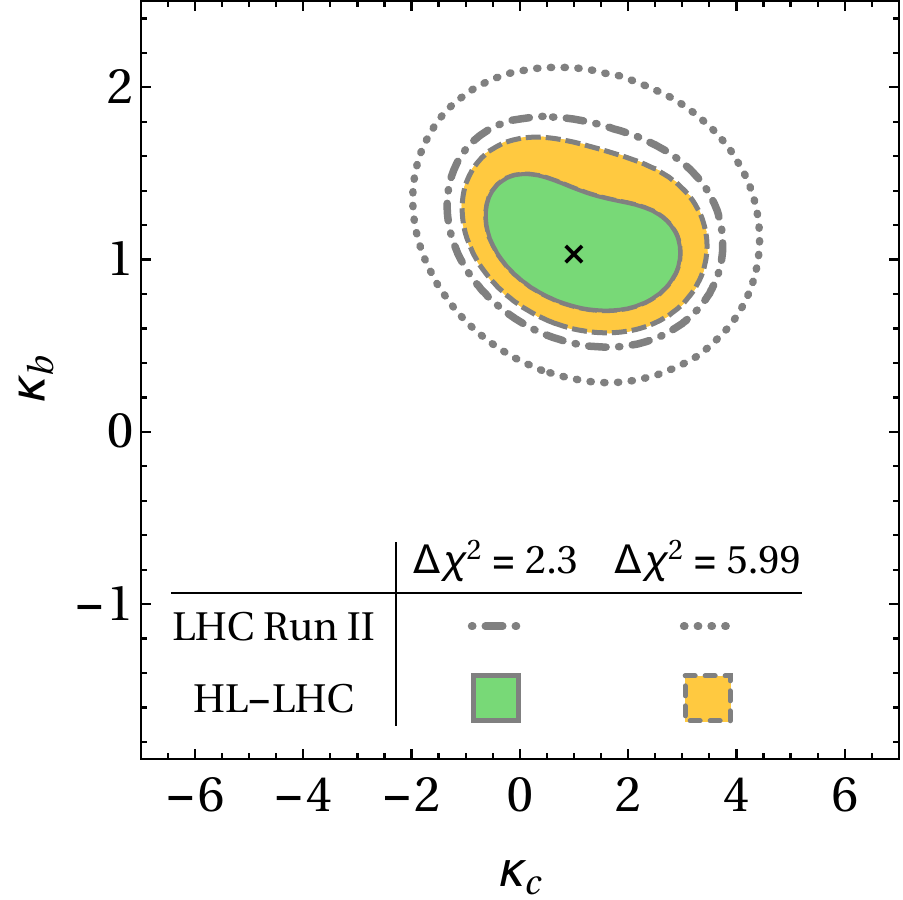}~~~~~~~~~~~~~~~
\includegraphics[width=0.37\textwidth]{\main/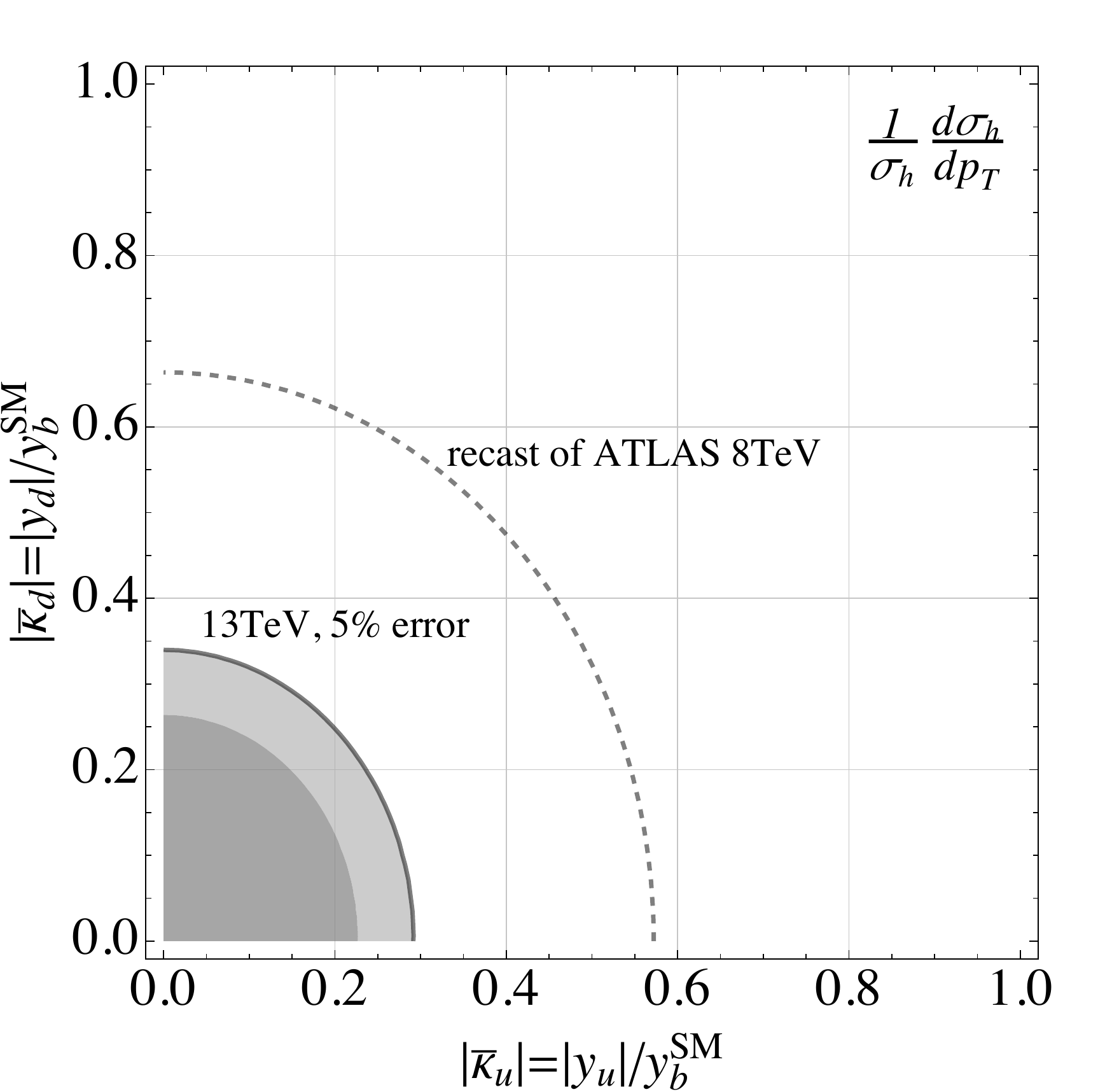}
\caption{The sensitivity of  Higgs \pt distributions to the light quark Yukawa couplings at 13\,TeV LHC Run2 and HL-LHC: to charm and bottom~\cite{Bishara:2016jga} (left) and to up and down~\cite{Soreq:2016rae} (right, assuming total $5\%$ systematic+statistical uncertainty).  
}
\label{fig:HiggsDistFuture}
\end{center}
\end{figure}

The 8\,TeV ATLAS results on Higgs \pt distributions~\cite{Aad:2015lha} were converted to  the following $95\,\%$ CL bounds in Ref.~\cite{Soreq:2016rae}
\begin{align}
	\bar{\kappa}_u = y_u/y^{\rm SM}_b < 0.46 \, , \qquad\qquad
	\bar{\kappa}_d = y_d/y^{\rm SM}_b< 0.54 \, ,
\end{align}
stronger than the corresponding bounds from fits to the inclusive Higgs production cross sections. 
The sensitivities expected at HL-LHC are shown in Fig.~\ref{fig:HiggsDistFuture} (right), assuming a $5\%$  total uncertainty in each \pt bin. CMS interpreted the $35.9\,$fb$^{-1}$ measurement of 13\,TeV Higgs \pt spectrum in terms of  95\,\% CL  bounds on $c$ and $b$ Yukawa couplings~\cite{Sirunyan:2018sgc}
\begin{align}
    -4.9 (-33) < \kappa_c < 4.8 (38),  \qquad
    -1.1(-8.5) < \kappa_b < 1.1(18),
\end{align}
assuming the branching fractions depend only on $\kappa_c$ or $\kappa_b$ (in addition the total decay width is allowed to float freely). 
These bounds on the $c$ and $b$ Yukawa are  weaker than the bounds from the global fit of the 8\,TeV Higgs data along with the electroweak precision data allowing all Higgs coupling to float~\cite{Perez:2015aoa} and from the direct measurement of $h\to b\bar{b}$ by using $b$-tagging. The projected HL-LHC sensitivity for $\kappa_{c,b}$ from higgs $p_T$ distributions, while assuming SM branching ratios, is shown in Fig. \ref{fig:HiggsDistFuture} (left).
 \vtwo{For the case that tha bounds depends only on the $\kappa_c$ and $\kappa_b$, the bound on the $c$ Yukawa is stronger than the bound from the global fit of the 8\,\UTeV Higgs data along with the electroweak precision data, allowing all Higgs couplings to float~\cite{Perez:2015aoa}. However, it relies on strong assumptions that the Higgs couplings (besides $c$ or $b$) are SM like, and it is mostly sensitive to the cross section and not to the angular shape as the latter bound.   }  

\begin{figure}[t]
  \begin{center}
  \includegraphics[width=0.4\linewidth]{\main/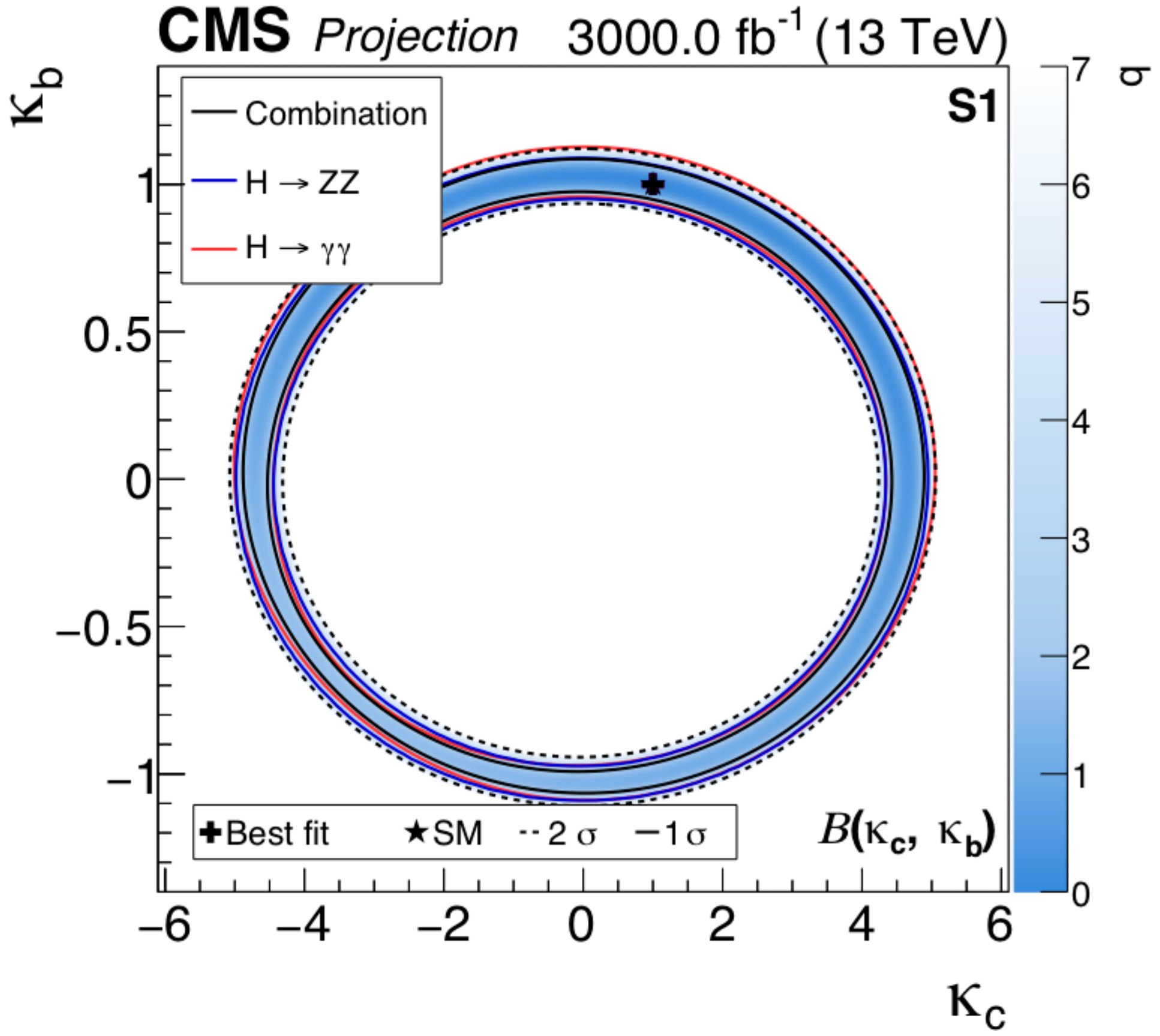}~~~~~~~~~~
   \includegraphics[width=0.4\linewidth]{\main/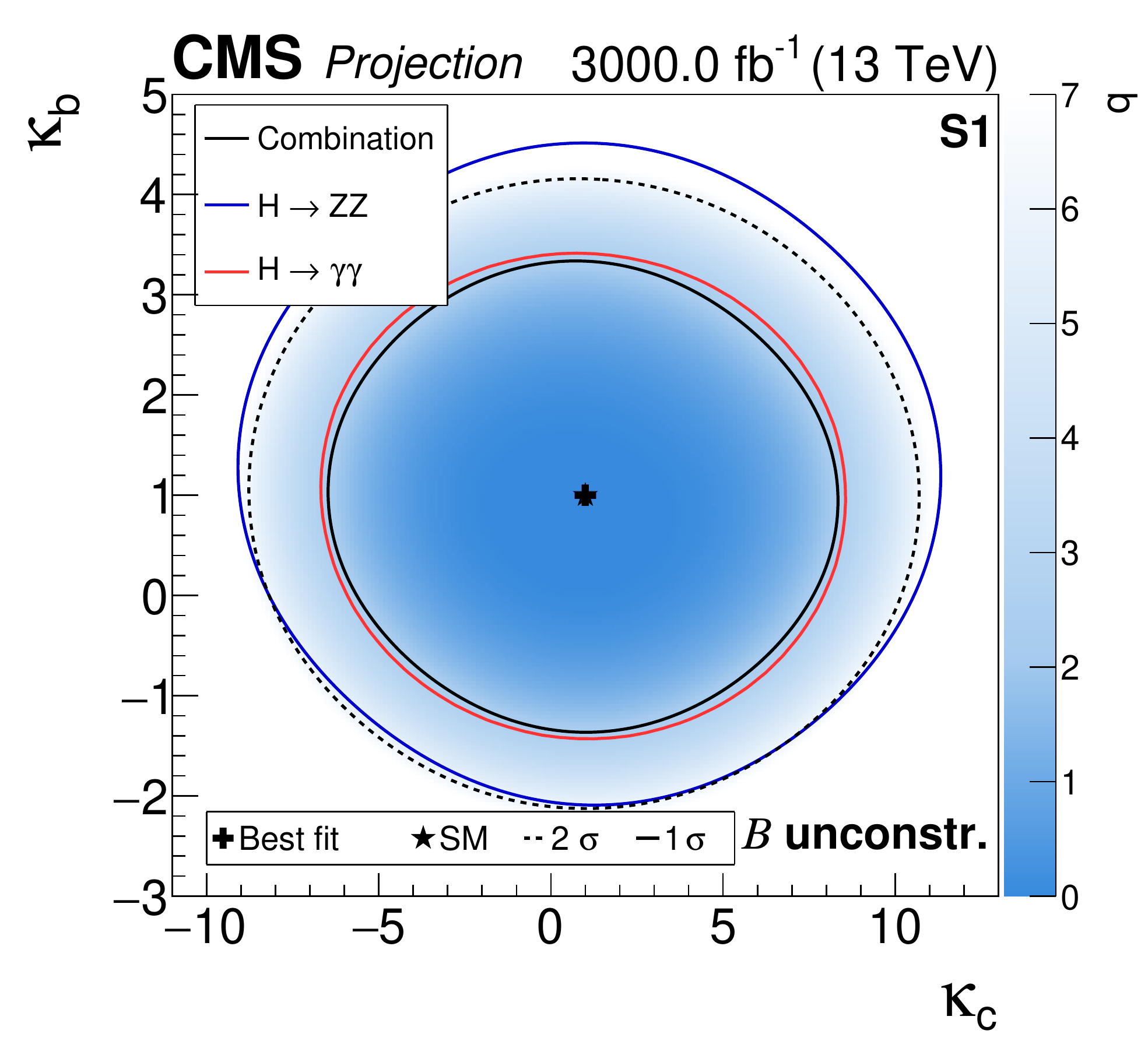} 
\caption{
 Left: projected simultaneous fit of $\kappab$ and $\kappac$ 
 from $3{\rm ab}^{-1}$ 13 TeV data, assuming current systematics (Scenario 1).
The one standard deviation contours are drawn for the $\hgg$ channel (the $\hzz$, the combination of $\hgg$ and $\hzz$) with solid red (blue, black).
For the combination the two standard deviation contour is drawn as a black dashed line. The negative log-likelihood value are given on the coloured axis.   Right: same as left, but with the branching fractions implemented as nuisance parameters with no prior constraint, i.e., the total Higgs width is floated freely. (Taken from \cite{CMS-PAS-FTR-18-011}.).
        }
    \label{fig:kbkc_couplingdependentBRs}
  \end{center}
\end{figure}

\Blu{ Fig.~\ref{fig:kbkc_couplingdependentBRs} shows expected one sigma $\kappac, \kappab$ contours from $3000$fb$^{-1}$ global fits,
obtained by extrapolating the measured constraints in Ref.~\cite{CMS:2018hhg}. The fit uses expected differential distributions, obtained by extrapolating the $\sqrt{s}=13\,$TeV \pt distributions in the $\hgg$~\cite{Sirunyan:2018kta} and 
$\hzz^{(*)}\to 4\ell$~\cite{Sirunyan:2017exp}
($\ell = e$ or $\muon$) decay channels, as well as 
 a search for $\hbb$ at large \pt~\cite{Sirunyan:2017dgc},
 which enhances the sensitivity  to $\kappat$.
 The simultaneous extended maximum likelihood fit to the diphoton mass, four-lepton mass, and soft-drop mass, $\msd$,~\cite{Dasgupta:2013ihk,Larkoski:2014wba} spectra in all the analysis categories of the $\hgg$, $\hzz$, and $\hbb$ channels, gives the results in Fig.~\ref{fig:kbkc_couplingdependentBRs}  (left) when only $\kappa_{c}$ and $\kappa_b$ are varied, while all the other couplings are set to the SM value, and in Fig.~\ref{fig:kbkc_couplingdependentBRs} (right), if in addition the Higgs total decay width is allowed to float.  In the fit for Fig.~\ref{fig:kbkc_couplingdependentBRs}  (left) the largest sensitivity to $\kappa_c, \kappa_b$ comes from the total cross sections times branching ratios, while for Fig.~\ref{fig:kbkc_couplingdependentBRs}  (right) it is due to normalized differential distributions $(1/\sigma)(d\sigma/dp_T)$. 
In Fig.~\ref{fig:kbkc_couplingdependentBRs} the systematic uncertainties are conservatively kept at current level (dubbed Scenario 1).
}
There is only a minor change in the projected constraints on $\kappa_c, \kappa_b$ in case of reduced systematic uncertainties (Scenario 2) compared to Scenario 1.

\subsubsubsection{$W^\pm h$ charge asymmetry}

The $W^\pm h$ charge asymmetry, 
\begin{equation}
	A 
= 	\frac{ \sigma (W^+ h) - \sigma (W^- h)}
	{\sigma (W^+ h) + \sigma (W^- h) },
\end{equation}
  is a  production-based probe of light quark Yukawa
couplings~\cite{Yu:2016rvv}.  
In the SM, the inclusive HE-LHC charge asymmetry is expected to be $17.3\%$, while the HL-LHC charge asymmetry is expected to be $21.6\%$.  
The dominant $W^\pm h$ production mode is due to Higgs boson
radiating from $W^\pm$ intermediate lines, if the Yukawa-mediated
diagrams are negligible.  If the quark Yukawas are not SM-like,
however, the charge asymmetry can either increase or decrease,
depending on the overall weight of the relevant PDFs.  In particular,
the charge asymmetry will increase if the down or up quark Yukawa
couplings are large, reflecting the increased asymmetry of $u \bar{d}$
vs.~$\bar{u} d$ PDFs; the charge asymmetry will decrease if the
strange or charm Yukawa couplings are large, reflecting the symmetric
nature of $c \bar{s}$ vs.~$\bar{c} s$ PDFs.  The subleading correction
from the Cabibbo angle-suppressed PDF contributions determines the
asymptotic behavior for extremely large Yukawa enhancements.

\begin{figure}[t]
  \begin{center}
 \includegraphics[width=0.6\textwidth, angle=0]{\main/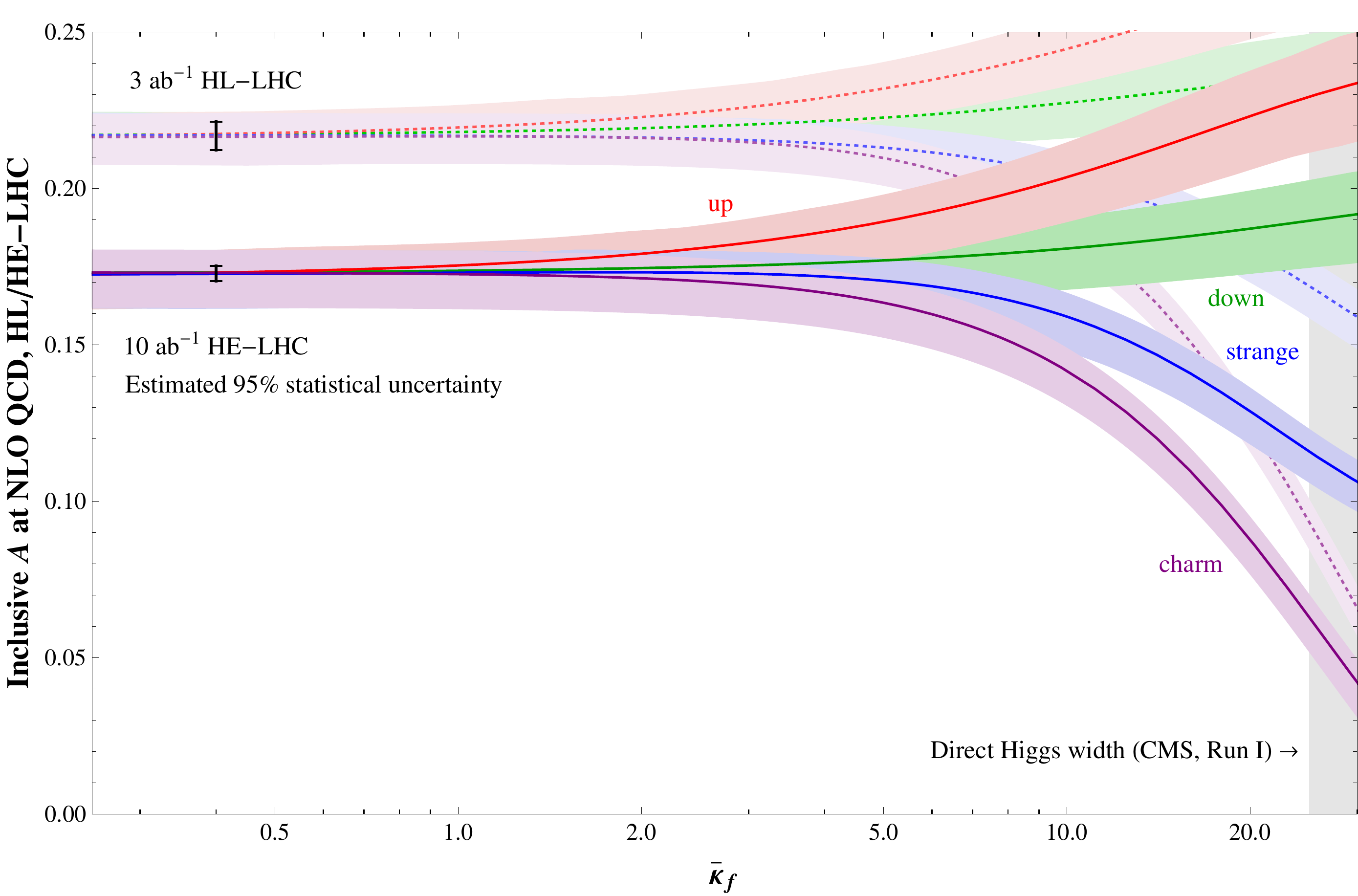}
 \caption{Inclusive charge asymmetry for $W^\pm h$ production at the
   27 TeV HE-LHC (solid colored bands), and 14 TeV HL-LHC (dotted
   colored bands), calculated at NLO QCD from MadGraph\_aMC@NLO using
   NNPDF 2.3 as a function of individual Yukawa rescaling factors
   $\bar{\kappa}_f$ for $f = u$ (red), $d$ (green), $s$ (blue), and
   $c$ (purple).  Shaded bands correspond to scale uncertainties at
   $1\sigma$ from individual $\sigma(W^+ h)$ and $\sigma(W^- h)$
   production, which are conservatively taken to be fully
   uncorrelated.  The expected statistical errors from this
   measurement using 10 ab$^{-1}$ of HE-LHC data and 3 ab$^{-1}$ of
   HL-LHC data are also shown.}
  \label{fig:asymmetry}
  \end{center}
\end{figure}

The effect of individual $d$, $u$, $s$, or $c$ quark Yukawa
enhancements on the inclusive charge asymmetry is shown in
Fig.~\ref{fig:asymmetry}, in units of $\bar{\kappa}_f = y_f /
y_{b,\text{SM}}$, evaluated at the Higgs mass scale.  Since $W^\pm h$
production probes lower Bjorken-$x$ at the HE-LHC compared to the
HL-LHC, the expected SM charge asymmetry is lower at the higher energy
collider.  The bands denote the change in the charge
asymmetry from varying the renormalization and factorization
scales within a factor of 2. The error bars denote 
the expected
$0.45\%$ ($0.25\%$) statistical sensitivity to the charge asymmetry at HL-LHC (HE-LHC) in the $W^\pm h \to
\ell^\pm \ell^\pm jj \nu \nu$ final state~\cite{Yu:2016rvv}.  The HE-LHC sensitivity was estimated
by simply rescaling with the appropriate luminosity ratio, since we expect the increases in both signal and
background electroweak rates to largely cancel.  The
constraint from the CMS Run I direct Higgs width upper bound is also shown~\cite{Yu:2016rvv}.  
If the signal strengths are fixed to the
SM expectation and the central prediction is used, the HE-LHC charge
asymmetry measurement could constrain $\bar{\kappa}_f \lesssim 2-3$
for up and charm quarks, and $\bar{\kappa}_f \lesssim 7$ for down or
strange quarks.

\subsection{LFV decays of the Higgs}

The flavour violating Yukawa couplings are well constrained by the low-energy
flavour-changing neutral current measurements
\cite{Harnik:2012pb,Blankenburg:2012ex,Gorbahn:2014sha}. 
\Blu{
For instance, CMS bounds $\kappa_{\mu e}^2+\kappa_{e\mu}^2<(5.4\times 10^{-4})^2$  from the $h\to e\mu$ search \cite{Khachatryan:2016rke}, compared to indirect bound from $\mu\to e\gamma$, which is $\kappa_{\mu e}^2+\kappa_{e\mu}^2<(3.6\times 10^{-6})^2$ \cite{Harnik:2012pb}.
}
A notable
exception are the flavour-violating couplings involving a tau lepton, where the
strongest constraints on $\kappa_{\tau\mu}, \kappa_{\mu\tau},
\kappa_{\tau e}, \kappa_{e \tau}$ are from direct searches for flavour-violating Higgs decays at
the LHC \cite{Sirunyan:2017xzt,Aad:2016blu}.
Currently, the CMS 13\,TeV search with 35.9\,fb$^{-1}$~\cite{Sirunyan:2017xzt} gives the strongest constraint 
\Blu{
\begin{align}
    \sqrt{\kappa_{\mu\tau}^2 + \kappa^2_{\tau\mu}} < 1.43 \times 10^{-3}\, , \qquad
    \sqrt{\kappa_{e\tau}^2 + \kappa^2_{\tau e}} < 2.26 \times 10^{-3}\, ,
\end{align}
}
obtained from  95\,\%~CL upper limits $\BR(h\to \mu\tau)<0.25\%$ and $\BR(h\to e\tau)<0.61\%$, respectively. One can also directly measure the difference between the branching ratios of $h\to\tau e$ and $h\to\tau\mu$, as proposed in~\cite{Bressler:2014jta}.
Assuming naively that both systematics and statistical error scale with square root of the luminosity, one can expect that the sensitivity of $3000\,$fb$^{-1}$ HL-LHC will be  around the half per-mil level for both the  $h\to e\tau$ and $h \to \mu\tau$ branching ratios.

The LHC can also set bounds on rare FCNC top decays involving a Higgs~\cite{Aaboud:2017mfd,Khachatryan:2016atv,Aad:2015pja,Aad:2014dya}. The strongest current bounds are ${|\kappa_{ct}|^2+|\kappa_{tc}|^2}<(0.06)^2$ 
\Blu{and ${|\kappa_{ut}|^2+|\kappa_{tu}|^2}<(0.07)^2$ at 95\,\%~CL}.

\subsection{CP violating Yukawa couplings}
The CP-violating flavour-diagonal Yukawa couplings, $\tilde \kappa_{f_i}$,
are well constrained from bounds on the electric dipole moments~(EDMs)
\cite{Brod:2013cka,Chien:2015xha,Altmannshofer:2015qra,Egana-Ugrinovic:2018fpy,Brod:2018pli} under the assumption of no 
cancellation with other contributions to EDMs.
For the electron Yukawa, the latest ACME measurement~\cite{Baron:2013eja,Andreev:2018ayy} results in an upper bound of $\tilde\kappa_e<1.9\times 10^{-3}$~\cite{Altmannshofer:2015qra}. For the bottom and charm Yukawas the strongest limits come from the neutron EDM~\cite{Brod:2018pli}. Using the NLO QCD theoretical prediction, this translates into the upper bounds $\tilde\kappa_b<5$ and $\tilde\kappa_c<21$ when theory errors are taken into account.
For the light quark CPV Yukawas, measurements of the Mercury EDM place strong bounds on the up and down Yukawas of $\tilde\kappa_u<0.06$ and $\tilde\kappa_d<0.03$~\cite{Brod:2018xyz} (no theory errors, $90\%$ CL), while the neutron EDM measurement gives a weaker constraint on the strange quark Yukawa of $\tilde\kappa_s<2.2$~\cite{Brod:2018xyz} (no theory errors, $90\%$ CL).

The top and $\tau$ Yukawa phases can be directly probed at HL-LHC, as we discuss below. For constraints from EDMs and other phases see~\cite{Brod:2013cka,Chien:2015xha,Altmannshofer:2015qra,Brod:2018pli,Baron:2013eja,Andreev:2018ayy,Altmannshofer:2015qra}. 

\subsubsection{$t\bar{t} h$}

CP violation in the top quark-Higgs coupling is strongly constrained by EDM measurements~\cite{Brod:2013cka}, if the light quark Yukawa couplings and $hWW$ couplings have their SM values. If this is not the case, the indirect constraints on the phase of the top Yukawa coupling can be substantially relaxed.
    Assuming the EDM constraints can be avoided, the \CP structure of the top quark Yukawa can be probed directly in $pp \to t\bar t h$. Many simple observables, such as $m_{t\bar t h}$ and $p_{T,h}$ are sensitive to the \CP structure, but require reconstructing the top quarks and Higgs.

Recently, several $t\bar t h$ observables have been proposed  that access the \CP structure without requiring full event reconstruction. These include the azimuthal angle between the two leptons in a fully leptonic $t/\bar{t}$ decay with the additional requirement that the $p_{T,h} > 200\, \text{GeV}$~\cite{Buckley:2015vsa}, and the angle between the leptons, in a fully leptonic $t/\bar t$ system, projected onto the plane perpendicular to the $h$ momentum~\cite{Boudjema:2015nda}. These observables only require that the Higgs is reconstructed and are inspired by the sensitivity of $\Delta \phi_{\ell^+\ell^-}$ to top/anti-top spin correlations in $pp \to t\bar t$~\cite{Mahlon:1995zn}. The sensitivity of both of these observables improves at higher Higgs boost, and therefore higher energy, making them promising targets for the HE-LHC, though no dedicated studies have been carried out to date.

\begin{figure}[t]
\centering
 \includegraphics[width=0.32 \textwidth]{\main/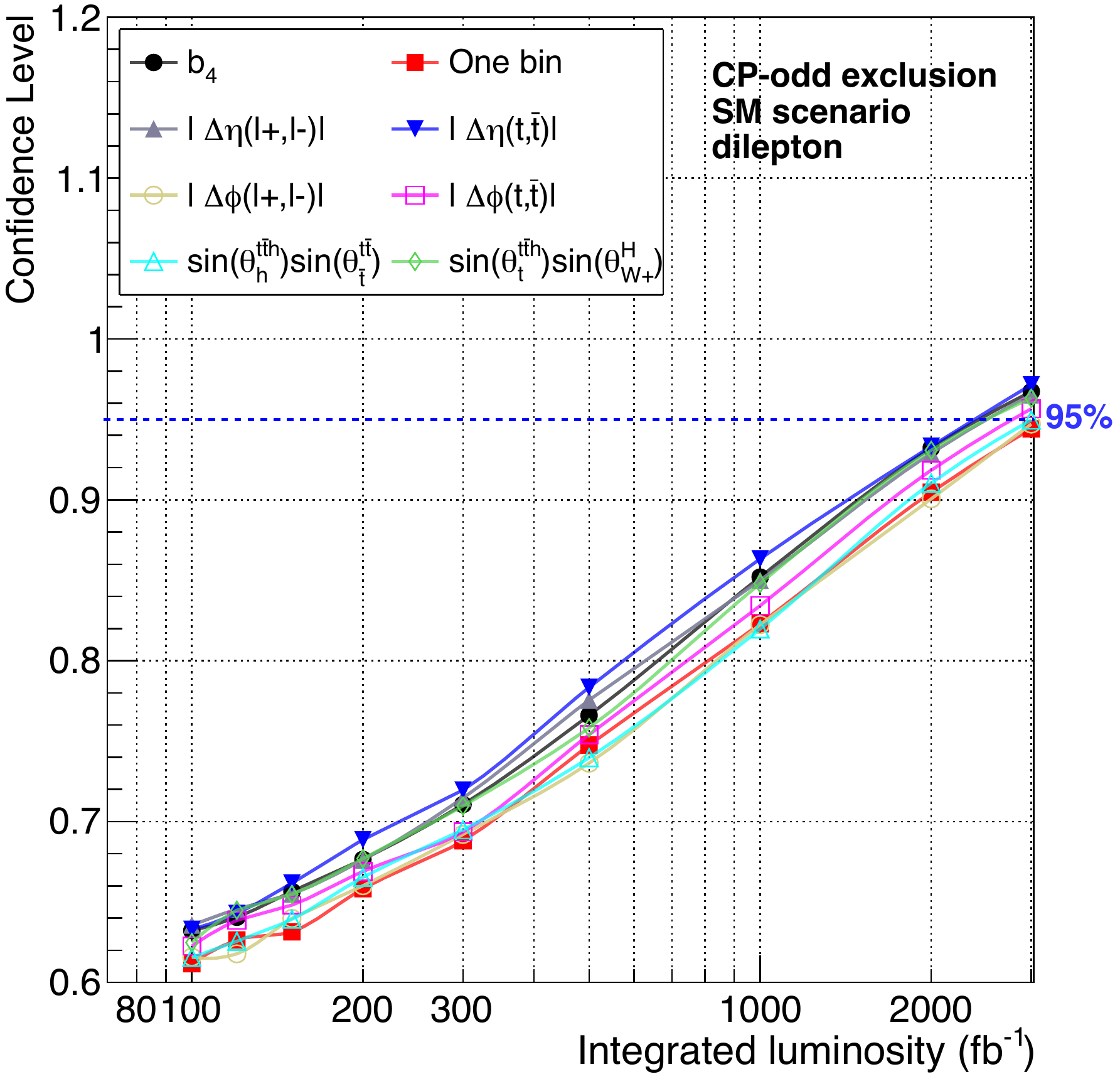}
~
 \includegraphics[width=0.32\textwidth]{\main/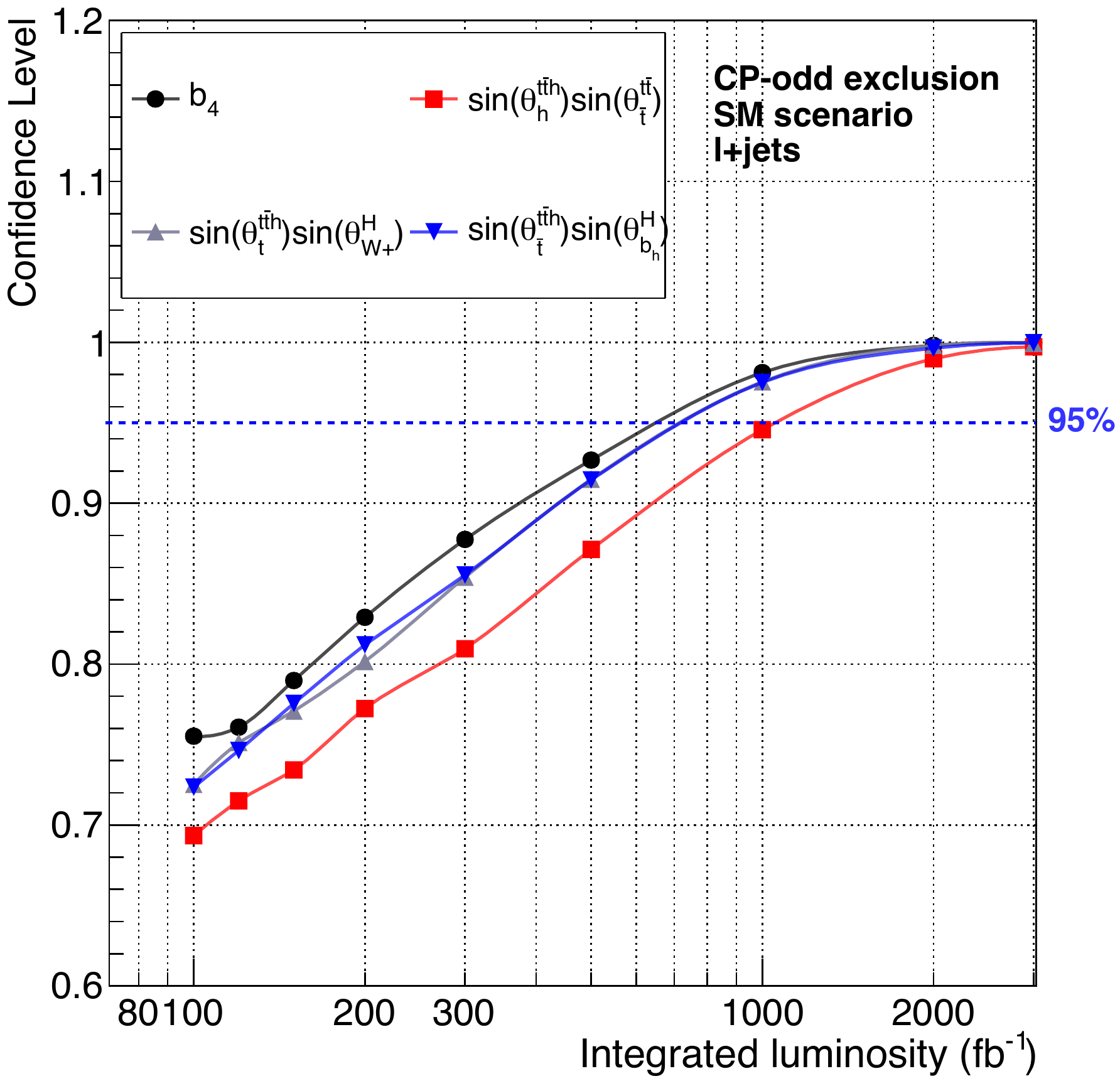}
~
 \includegraphics[width=0.32\textwidth]{\main/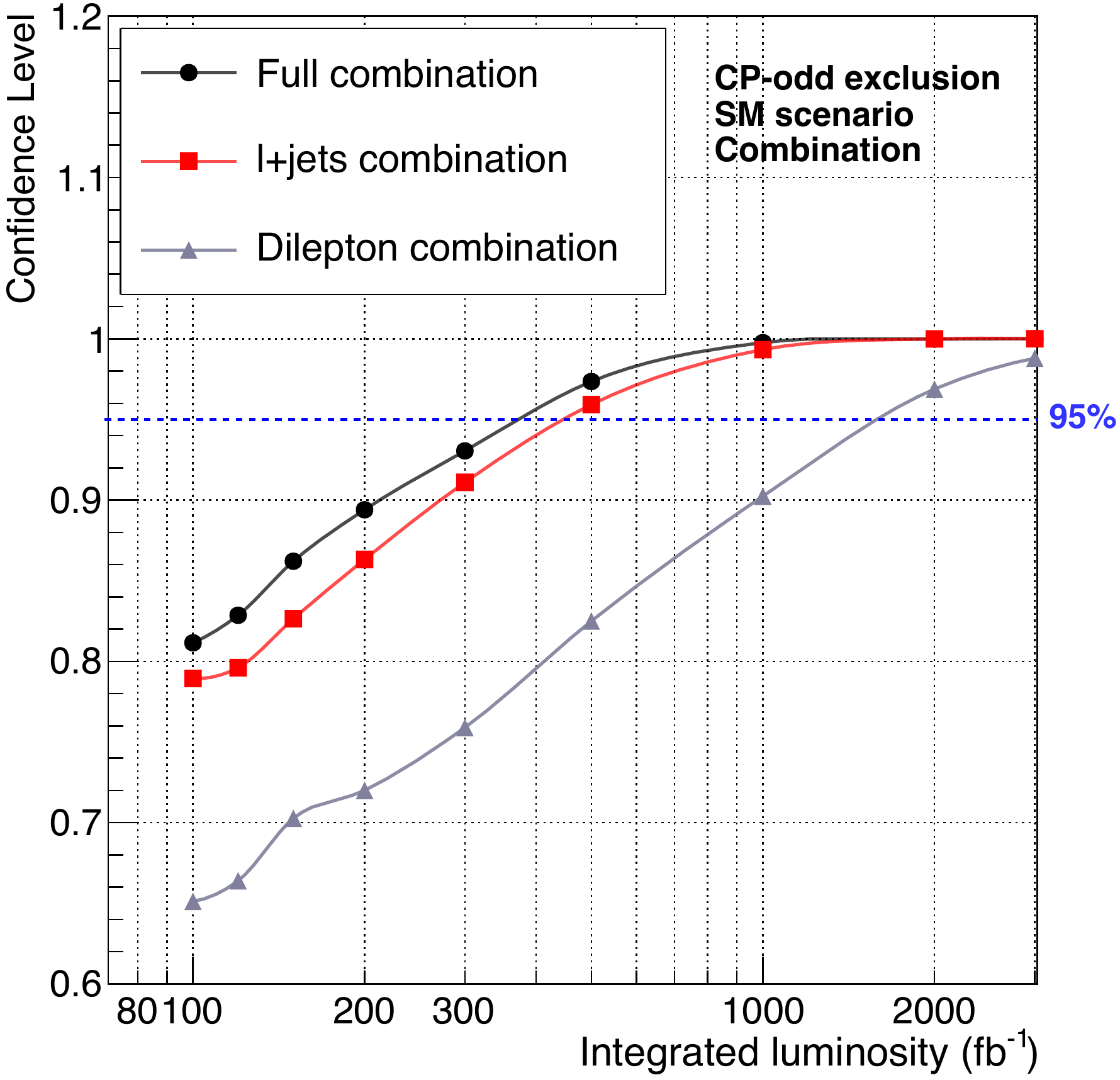}
 \caption{Expected CL, assuming the SM, for exclusion of the pure \CP-odd scenario, $\kappa_t=0, \tilde \kappa_t=1$, as a function of the integrated luminosity. Left: using the $t\bar{t}h$ ($h\rightarrow b\bar{b}$) dileptonic analysis only, middle: using the $t\bar{t}h$ ($h\rightarrow b\bar{b}$)  semileptonic analysis only, right: combining observables in each individual channel and combining both channels, treating  the observables as uncorrelated.  A likelihood ratio computed from the binned distribution of the corresponding discriminant observable was used as test statistic.}
 \label{combination}
\end{figure}

Fig.~\ref{combination} shows the expected CL, assuming the SM, for exclusion of the pure \CP-odd scenario, $\kappa_t=0, \tilde \kappa_t=1$, as a function of the integrated luminosity. Samples of $t\bar t h$($h\rightarrow b\bar{b}$) events were generated at the LHC for $\sqrt{s}=13$~TeV, with {\tt MadGraph5\_aMC@NLO}~\cite{Alwall:2014hca}
using the \texttt{HC\_NLO\_X0} model~\cite{Artoisenet:2013puc}, as were  all relevant SM background processes.
The analyses of the $t\bar t h$ ($h\rightarrow b\bar b$) events were carried out in the dileptonic and semileptonic decay channels of the $t\bar t$ system. {\tt Delphes}~\cite{deFavereau:2013fsa} was used for a parametrised detector simulation and both analyses used kinematic fits to fully reconstruct the $t\bar t h$ system. The results were extrapolated, as a function of luminosity, up to the 3000~fb$^{-1}$.

Fig.~\ref{combination} left (middle) shows results using the dileptonic (semileptonic) analysis only. The CL were obtained from a signal-enriched region (with at least 3 $b$-tagged jets) in which a likelihood ratio was computed from binned distributions of various discriminant observables~\cite{AmorDosSantos:2017ayi, Demartin:2014fia}. Only statistical uncertainties were considered. Fig.~\ref{combination} right shows CL obtained from the combination of different observables in each channel i.e., $\Delta\eta(\ell^+,\ell^-)$, $\Delta\phi(t,\bar t)$ and $\sin(\theta^{t\bar tH}_{t})\sin(\theta^{H}_{W^+})$ in the dileptonic channel and, $b_4$ and $\sin(\theta^{t\bar tH}_{\bar t})\sin(\theta^{H}_{b_H})$ in the semileptonic channel. The combination of the two channels is also shown for comparison. The observables were treated as uncorrelated. 

The main conclusions of these studies can be summarized in what follows: {\it i)} many angular observables 
are available with the potential of discriminating between values of $\kappa_t, \tilde \kappa_t$ 
 in the top quark Yukawa coupling; {\it ii)} the sensitivity of the semileptonic final state of $t\bar{t}h$($h\rightarrow b\bar{b}$) is roughly a factor 3 better than that of the dileptonic channel alone (Fig.~\ref{combination} left); {\it iii)} the combination of the two channels (semi- and dileptonic) is roughly a factor 5 more sensitive than the dileptonic channel, providing a powerful test of the top quark-Higgs interactions in the fermionic sector.

\subsubsection{$\tau\bar{\tau} h$}

The most promising direct probe of \CP violation in fermionic Higgs
decays is the $\tau^+ \tau^-$ decay channel, which benefits from a
relatively large $\tau$ Yukawa, resulting in a SM branching fraction of
$6.3\%$. Measuring the \CP violating phase in the tau Yukawa requires a measurement of the linear polarizations of both $\tau$ leptons and the azimuthal angle between them. This can be done by analyzing tau substructure, namely the angular distribution of the various components of the tau decay products.

The main $\tau$ decay modes studied include $\tau^\pm \to
\rho^\pm (770) \nu$, $\rho^\pm \to \pi^\pm \pi^0$~\cite{Bower:2002zx,
  Desch:2003mw, Desch:2003rw, Harnik:2013aja, Askew:2015mda,
  Jozefowicz:2016kvz} and $\tau^\pm \to \pi^\pm
\nu$~\cite{Berge:2008wi, Berge:2008dr, Berge:2011ij}.  Assuming CPT
symmetry, collider observables for \CP violation must be built from
differential distributions based on triple products of three-vectors.
In the first case, $h \to \pi^\pm \pi^0 \pi^\mp \pi^0 \nu \nu$,
angular distributions built only from the outgoing charged and neutral
pions are used to determine the \CP properties of the initial $\tau$
Yukawa coupling.  In the second case, $h \to \pi^\pm \pi^\mp \nu \nu$,
there are not enough reconstructible independent momenta to construct an observable sensitive to \CP violation, requiring additional kinematic information such as the $\tau$ decay impact parameter.

In the kinematic limit when each outgoing neutrino is taken to be
collinear with its corresponding reconstructed $\rho^\pm$ meson, the
acoplanarity angle, denoted $\Phi$, between the two decay planes
spanned by the $\rho^\pm \to \pi^\pm \pi^0$ decay products is exactly
analogous to the familiar acoplanarity angle from $h \to 4 \ell$
\CP-property studies.  Hence, by measuring the $\tau$ decay products in
the single-prong final state, suppressing the irreducible $Z \to
\tau^+ \tau^-$ and reducible QCD backgrounds, and reconstructing the
acoplanarity angle of $\rho^+$ vs.~$\rho^-$, the differential
distribution in $\Phi$ gives a sinusoidal shape whose maxima and
minima correspond to the \CP-phase in the $\tau$ Yukawa coupling, $\varphi_\tau=\arctan(\tilde \kappa_\tau/\kappa_\tau)$.  

An optimal observable using the colinear approximation was derived in~\cite{Harnik:2013aja}. Assuming 70\% efficiency for tagging hadronic $\tau$ final states, and
neglecting detector effects, the estimated sensitivity for the
CP-violating phase $\varphi_\tau$ using 3 ab$^{-1}$ at
the HL-LHC is 8.0$^\circ$.  A more sophisticated
analysis~\cite{Askew:2015mda} found that detector resolution effects
on the missing transverse energy distribution degrade the expected
sensitivity considerably, and as such, about 1 ab$^{-1}$ is required
to distinguish a pure scalar coupling ($\kappa_\tau=1, \tilde \kappa_\tau=0$) from a pure
pseudoscalar coupling ($\kappa_\tau=0, \tilde \kappa_\tau=1$).

At the HE-LHC, the increased signal cross section for Higgs production
is counterbalanced by the increased background rates, and so the main
expectation is that improvements in sensitivity will be driven by the
increased luminosity and more optimized experimental methodology.
Rescaling with the appropriate luminosity factors, the optimistic
sensitivity to the $\tau$ Yukawa phase $\varphi_\tau$ from acoplanarity studies is
4-5$^\circ$, while the more conservative estimate is roughly an order
of magnitude worse.

%% file: section10/section.tex
\section{The high \pt flavour physics programme} 
\label{sec:highpT}

Flavour and high \pt searches are intertwined in several ways. On the one hand the stringent bounds from low-energy constraints put severe bounds on the NP models that contain states with TeV masses with couplings to quarks.  On the other hand, TeV scale New Physics is suggested by several solutions of long standing problems of the SM, e.g., the hierarchy problem. Quite often the low energy constraints are avoided by assuming Minimal Flavour Violation (MFV), where the only flavour breaking, even in the NP sector, is due to the SM Yukawa matrices. However, more general flavour structures for the NP states are still allowed. In fact such non-MFV couplings can have interesting consequences. In general we can group the NP models into two broad classes: \textit{(i)} models that address outstanding problems of the SM, such at the SM flavour puzzle, the origin of dark matter, or the hierarchy problem, which may have non trivial flavour structure,  and \textit{(ii)} models designed to explain the $b\to s\ell\ell$ and $b\to c\tau \nu$ flavour anomalies, that almost inevitably have quite a distinct flavour structure.  The NP mediators potentially responsible for the anomalies, $Z'$, $W'$ or leptoquarks, could be found at high \pt searches in the HL- or HE-LHC.

In the rest of this section we first briefly review the models that address the SM flavour puzzle and have states that could be probed at the HL-/HE-LHC, and their implications for high \pt searches. The second part of this section is devoted to  high \pt implications of $B$ physics anomalies. 

\subsection{Models of flavour and TeV Physics}

\subsubsection{\bf Randall-Sundrum models of flavour}
Models of flavour based on warped extra dimension~\cite{Randall:1999ee} attempt to simultaneously solve
 the hierarchy problem as well as the SM flavour 
problem~\cite{Grossman:1999ra,Gherghetta:2000qt}. 
In the Randall-Sundrum (RS) models the 5-dimensional space-time has 
anti-de Sitter geometry (AdS$_5$), truncated by flat 4D boundaries, 
the Planck (UV) and the TeV (IR) branes.  
This setup gives a warped metric in the bulk~\cite{Randall:1999ee}  
$ds^2 = \exp(-2 k r_c |\phi| ) \eta_{\mu\nu} dx^\mu dx^\nu - r_c^2 d\phi^2$,
where $k$ is the 5D curvature scale, $r_c$ the radius of compactification and  
$\phi\in [-\pi,\pi]$ the coordinate along the $5$-th dimension. 
The warp factor, $\exp({-2 k r_c |\phi|})$, leads to different length scales 
in different 4D slices along the $\phi$ direction, 
which provides a solution to the hierarchy problem.  
In particular, the Higgs field is assumed to be localized
near the TeV-brane so that the metric ``warps'' 
$\langle{H}\rangle_5 \sim M_5\sim M_P \sim 10^{19}$~GeV down to the weak scale, 
$\langle{H}\rangle_4 = \exp({-kr_c \pi}) \langle{H}\rangle_5$.  
For $kr_c \approx 12$ then $\langle{H}\rangle_{\rm SM} \equiv \langle{H}\rangle_4 \sim 1$~TeV.

The hierarchies among the quark masses can be realized by localizing the Higgs on the IR brane, while the fermions have different profiles in the $5$-th dimension. 
The first and second generation zero mode fermions are localized
close to the UV-brane and have small overlaps with the Higgs, giving small effective 4D SM Yukawa interactions, and thus small quark masses after electroweak symmetry breaking.
The top quark, on the other hand, is localized near the TeV brane 
resulting in a large top Yukawa coupling.

This configuration has a built in automatic suppression of FCNCs, which are suppressed by the 
same zero mode overlaps that gives the hierarchy of masses~\cite{Grossman:1999ra,Gherghetta:2000qt}. This feature of the RS framework plays a similar role as
the SM Glashow-Iliopoulos-Maiani (GIM) mechanism, and was dubbed RS-GIM in  \cite{Agashe:2004cp,Agashe:2004ay}.
Similarly to the SM GIM, the RS GIM is violated by the large top quark mass. 
In particular, $(t,b)_L$ needs to be localized near the TeV brane, 
otherwise the 5D Yukawa coupling becomes too large and makes the theory
strongly coupled at the scale of the first Kaluza-Klein (KK) excitation. In general this 
leads to sizeable corrections to electroweak precision observables, such as the $Z b_L b_L$ couplings. 
Such problems can be largely ameliorated by enlarging the bulk symmetry such that it contains a custodial $SU(2)_L\times SU(2)_R$ symmetry \cite{Agashe:2003zs}, which for instance lowers the KK scale bounds from EW precision tests from 5 TeV to about 2TeV \cite{Agashe:2013kxa}. The consequences for flavour phenomenology have been worked out in a series of papers, see, e.g., \cite{Blanke:2008yr,Albrecht:2009xr,Casagrande:2010si,Casagrande:2008hr}, with $K-\bar K$ mixing for instance requiring the KK scale to be above 8 TeV \cite{Agashe:2013kxa}. With flavour alignment the scale of KK modes could be substantially lowered \cite{Csaki:2009wc} and could be reachable by HL/HE-LHC.

The KK gluon resonances cannot be produced from gluons \cite{Allanach:2009vz}, so that the LHC production is restricted to the quark-antiquark fusion, even though this is suppressed by the flavor dependent zero mode overlaps. This means that the LHC cross section for the first KK gluon resonances are small, suppressed also by the quark-anti-quark parton density functions (PDFs). The dominant decay mode is into $t\bar t$ final state, due to the large zero mode overlaps \cite{Agashe:2006hk}. Using the benchmark RS model from \cite{Lillie:2007yh}, the most recent CMS analysis for $t\bar t$ resonance searches, using both hadronic and leptonic tops, sets a lower bound of 4.55 TeV on the mass of the KK gluon  \cite{Sirunyan:2018ryr}. The projected reach for 33TeV and 100TeV $pp$ colliders can be found in \cite{Agashe:2013kxa}.

\subsubsection{Partial compositeness}
Partial compositeness as the origin of the flavour hierarchies in composite Higgs models \cite{Kaplan:1991dc} is the holographic dual to the RS models of flavour. While the Higgs is the lightest state of the composite sector, usually a pseudo-Nambu Goldstone boson from global symmetry breaking, the SM fermions and gauge bosons are elementary (for a review, see, e.g, \cite{Panico:2015jxa}). The elementary fermions, $Q,U,D$, are coupled to the composite sector through linear mixing with the composite operators, ${\mathcal O}_Q$, ${\mathcal O}_U$, ${\mathcal O}_D$,
\begin{equation}
{\cal L}\supset \epsilon_Q \bar Q_L {\mathcal O}_Q +\epsilon_U \bar U_R{\mathcal O}_U+\epsilon_D \bar D_R {\mathcal O}_D.
\end{equation}
The mixing parameters $\epsilon_a$ exhibit exponentially large hierarchies because of large, yet still ${\mathcal O}(1)$, differences in anomalous dimensions of the corresponding composite operators. This is the analog of the zero mode overlaps in the RS models. The SM Yukawa are given by $(Y_{U(D)})_{ij}\sim \epsilon_{Q}^i \epsilon_{U(D)}^j$. For 
$\epsilon_Q^1\ll \epsilon_Q^2\ll \epsilon_Q^3\sim 1$, $\epsilon_U^1\ll \epsilon_U^2\ll \epsilon_U^3\sim 1$, $\epsilon_D^1\ll \epsilon_D^2\ll \epsilon_D^3\ll 1$ one can obtain the SM structure of quark masses and CKM mixings. 

The composite Higgs models are described by the compositeness scale $f$ and the mass of the first composite resonances, $M_*\sim g_* f$, with $g_*$ the typical strength of the resonances in the composite sector. The searches at the HL-/HE-LHC consist of Higgs coupling measurements, including deviations in Higgs Yukawa couplings,
and searches for composite resonances preferably coupled to third generation fermions, electroweak gauge bosons, or the Higgs.
Flavour observables put strong bounds on $M_*$, if the flavour structure is assumed to be generic. Such bounds can be relaxed in the case of approximate flavour symmetries, see, e.g., Ref. \cite{Panico:2015jxa} for a review.

\subsubsection{Low scale gauge flavour  symmetries}
The SM has in the limit of vanishing Yukawa couplings a large global symmetry. In the quark sector this is $G_F=SU(3)_Q\times SU(3)_U\times SU(3)_D$. Ref. \cite{Grinstein:2010ve} showed that the SM flavor symmetry group $G_F$ is anomaly free, if one adds a set of fermions that are vector-like under the SM gauged group, but chiral under the $G_F$. This means that $G_F$ can be gauged. It is broken by a set of scalar fields that have hiearchical vevs and lead to hierarchy of SM quark masses. This also implies a hierarchy  for the masses of the flavoured gauge bosons, with gauge bosons that more strongly couple to third generation being lighter, while the flavoured gauge bosons that couple more strongly to the first two generations are significantly heavier. This pattern in the spectrum of flavoured gauge bosons then avoids too large contributions to FCNCs \cite{Buras:2011wi}.  

At the LHC one searches for the lightest flavoured gauge bosons, with ${\mathcal O}({\rm TeV})$ masses, which couple mostly to $b$ quarks and $t$ quarks, but could also have non-negligible couplings to the first two generations. The di-jet and $t\bar t$ resonance searches are thus sensitive probes. A signal could also come from production of the lightest vectorlike fermions, $t'$, or $b'$ \cite{Grinstein:2010ve}. For several further benchmarks see, e.g., Ref. \cite{Bishara:2015mha}, where also a connection with dark matter was explored.

\subsubsection{2HDM and low scale flavour models}
\label{sec:2HDM:flavour}
{\it\small Authors (TH): Martin Bauer, Marcela Carena and Adri\'an Carmona.}\\
In 2 Higgs Doublet Models (2HDMs), the two Higgs doublets, $H_1$ and $H_2$ are usually assumed not to carry flavour quantum numbers. The collider phenomenology, on the other hand, changes substantially, if they do. 
 This is an interesting possibility that could solve the SM flavour puzzle via the Froggatt-Nielsen (FN) mechanism where the flavon is replaced by the $H_1 H_2\equiv H_1^T (i\sigma_2) H_2$ operator. In this way, the NP scale $\Lambda$ where the higher dimensional FN operators are generated is tied to the electroweak scale, leading to much stronger phenomenological consequences. Let us assume for concreteness a type-I like 2DHM with the following Yukawa Lagrangian in the quark sector \cite{Bauer:2015kzy,Bauer:2015fxa}
\begin{equation}
\label{eq:yuk1}
\mathcal{L}_Y\supset  y^u_{ij} \left(\frac{H_1 H_2}{\Lambda^2}\right)^{n_{u_{ij}}}\bar{q}_L^{i}H_1 u_{R}^j+y^d_{ij} \left(\frac{H_1^{\dagger} H_2^{\dagger}}{\Lambda^2}\right)^{n_{d_{ij}}}\bar{q}_L^{i}\tilde{H}_1 d_{R}^j
+\mathrm{h.c.}\,,
\end{equation}
where $\tilde{H}_1\equiv i\sigma_2 H^{\ast}_1$ as usual, and the charges $n_{u,d,e}$ are a combination of the $U(1)$ charges of $H_1$, $(H_1H_2)$ and the different SM fermion fields (for an alternative discussion, where $H_1, H_2$ carry flavour charges, but the Yukawa interactions are taken to be renormalizable, see \cite{Dery:2016fyj}). For simplicity, we set the flavour charges of $H_1$ and $H_2$ to $0$ and $1$, respectively, such that  
$n_{u_{ij}}=a_{q_i}-a_{u_j}, n_{d_{ij}}=-a_{q_i}+a_{d_j}$,
if we denote by $a_{q_i},a_{u_i}, \ldots,$ the $U(1)$ charges of the SM fermions. In general, the fermion masses are given by
\begin{equation}\label{eq:epsilon}
m_\psi=y_{\psi} \varepsilon^{n_\psi} \frac{v}{\sqrt{2}},  \qquad \varepsilon = \frac{v_1 v_2}{2\Lambda^2}=\frac{t_\beta}{1+t_\beta^2}\frac{v^2}{2\Lambda^2},
\end{equation}
with the vacuum expectation values $\langle H_{1, 2}\rangle=v_{1, 2}$ and $t_\beta \equiv v_1/v_2$. For the right assignment of flavour charges one is able to accommodate the observed hierarchy of SM fermion masses and mixing angles.  This framework also leads to enhanced diagonal Yukawa couplings between the Higgs and the SM fermions, while FCNCs are suppressed. If we denote by $h$ and $H$ the two neutral scalar mass eigenstates, with $h$ the observed $125$ GeV Higgs, the couplings between the scalars $\varphi=h,H$ and SM fermions $\psi_{L_i, R_i}= P_{L,R} \psi_i$ in the mass eigenbasis read 
\begin{equation}\label{eq:newlagrangian}
\mathcal{L}= g_{\varphi \psi_{L_i} \psi_{R_j} }\, \varphi \,\bar \psi_{L_i} \psi_{R_j}+\mathrm{h.c.}
\end{equation}
with $i$, such that $u_i=u, c,t$,\,  $d_i=d,s,b$ and $e_i=e,\mu,\tau$. This induces 
flavour-diagonal couplings 
\begin{align}\label{eq:diagocoup}
g_{\varphi \psi_{L_i}\psi_{R_i}}= \kappa^\varphi_{\psi_i}\, \frac{m_{\psi_i}}{v} =\left(g^{\varphi}_{\psi_i}(\alpha,\beta)+n_{\psi_i}\, f^\varphi(\alpha, \beta)\right)\frac{m_{\psi_i}}{v},
\end{align}
as well as flavour off-diagonal couplings
 \begin{align}\label{eq:foffc}
g_{\varphi \psi_{L_i}\psi_{R_j}}&=  f^\varphi(\alpha, \beta)\left(\mathcal{A}_{ij}\frac{m_{\psi_j}}{v}-\frac{m_{\psi_i}}{v}\mathcal{B}_{ij} \right)\,.
\end{align}
The flavour universal functions in \eqref{eq:diagocoup} and \eqref{eq:foffc} are
$g^h_{\psi_i}={c_{\beta-\alpha}}/{t_\beta}+s_{\beta-\alpha}$, $ g^H_{\psi_i}=c_{\beta-\alpha}-{s_{\beta-\alpha}}/{t_\beta}$,
and
$ f^h(\alpha,\beta)=c_{\beta-\alpha}\big({1}/{t_\beta}-t_\beta\big)+2s_{\beta-\alpha}$, $f^H(\alpha,\beta)=-s_{\beta-\alpha}\big({1}/{t_\beta}-t_\beta\big)+2c_{\beta-\alpha}$,
where $c_{x}\equiv \cos x$, $s_x\equiv \sin x$. The entries in matrices $\mathcal{A}$ and $\mathcal{B}$ are proportional to the flavour charges of the corresponding fermions that define the coefficients in \eqref{eq:yuk1}. Unless all flavour charges for a given type of fermions are equal, the off-diagonal elements in matrices $\mathcal{A}$ and $\mathcal{B}$ lead to FCNCs which are chirally suppressed by powers of the ratio~$\varepsilon$, see \cite{Bauer:2017cov} for more details and explicit examples for scalings of matrix elements in $\mathcal{A}$ and $\mathcal{B}$.

\subsubsection{\bf A Clockwork solution to the flavour puzzle}
{\it\small Author (TH): Adri\'an Carmona.}\\
The clockwork mechanism, introduced in \cite{Choi:2015fiu, Kaplan:2015fuy} and later generalized to a broader context in Ref.~\cite{Giudice:2016yja}, allows one to obtain large hierarchies in couplings or mass scales. Ref. \cite{Alonso:2018bcg} showed that it can also be used to generate the observed hierarchy of quark masses and mixing angles with anarchic Yukawa couplings, providing a solution to the flavour puzzle. 

In the clockwork solution to the flavour puzzle, each SM chiral fermion $\psi$ is accompanied with a $N_\psi$-node chain of vector-like fermions, $\psi_{L,j},\psi_{R,j}$, with masses $m$, and
a series of nearest neighbour mass terms, $qm\bar{\psi}_{L,j}\psi_{R,j-1}$, between the nodes, where  $j=1,\ldots,N_\psi$. The mass spectrum of each chain has one chiral zero mode, the would-be SM fermion, 
and $N_\psi$  heavy Dirac fermion mass-eigenstates -- the gears.
For $q\gg 1$,  the spectrum of the gears is compressed in a $2m$ band around $qm$, with $(M_{N_\psi}-M_1)\ll M_1$. The massless zero-mode interacts with the SM Higgs, which is on the $0$-th node, through a set of Yukawa interactions described by ${\mathcal O}(1)$ Yukawa matrices, $Y_{U,D}$. 
The component of the massless mode on the $0$-th node is, 
in the large $q$ limit, given by $1/q^{N_\psi}$. This suppression is the origin of the SM Yukawa hierarchies, 
\begin{align}
\label{eqa:YuSM}
\left(Y_u^{\rm SM}\right)_{ij}&\sim q^{-N_{Q(i)}}_{Q(i)}\left(Y_U\right)_{ij}q^{-N_{u(j)}}_{u(j)},  \quad \left(Y_d^{\rm SM}\right)_{ij} \sim q^{-N_{Q(i)}}_{Q(i)}\left(Y_D\right)_{ij}q^{-N_{d(j)}}_{d(j)}.
\end{align}
The hierarchy of quark masses can be then naturally obtained for anarchical $Y_U$ and $Y_D$ Yukawa matrices if $q^{-N_{Q(i)}}\ll q^{-N_{Q(j)}}$, $q^{-N_{u(i)}}_{u(i)}\ll q^{-N_{u(j)}}_{u(j)}$, $q^{-N_{d(i)}}_{d(i)}\ll q^{-N_{d(j)}}_{d(j)}$, for $i<j$ (in the benchmark below we take $q_i$ to be universal and equal to $q$).

\begin{figure}
\begin{center}
\includegraphics[width=0.4\textwidth]{\main/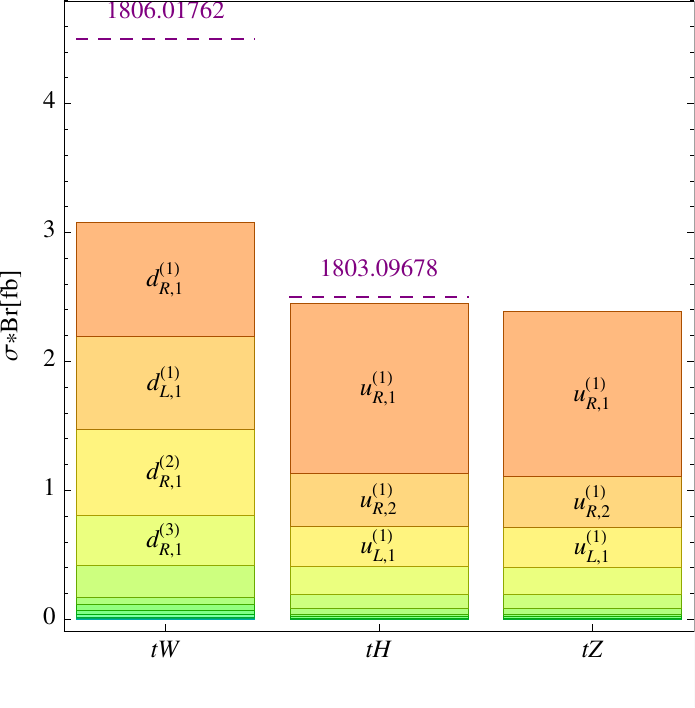}
	\includegraphics[width=0.59\textwidth]{\main/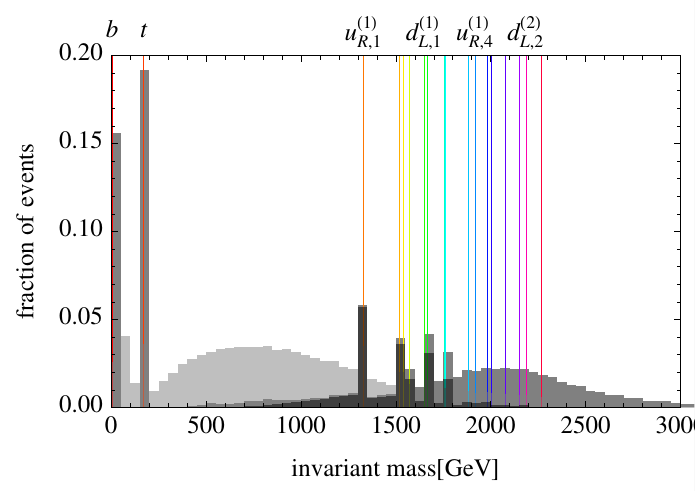}
\caption{\textit{Left:} The total gear pair production cross-sections in the final states $tW+X$, $tH+X$ and $tZ+X$ for the benchmark model from Ref. \cite{Alonso:2018bcg}, with contributions from individual gears shown stacked.
The currently most stringent upper bounds~\cite{Aaboud:2018uek,Aaboud:2018xuw} are denoted with dashed lines (for the $tZ+X$~\cite{Aaboud:2018xuw} final state the bound is too weak to be shown). 	
\textit{Right:} Invariant mass spectrum of individual pseudojets clustered using the hemisphere algorithm applied to partonic gear pair production and decay at the 13 TeV LHC. The original hemisphere clustering results are shown in light gray, and the modified hemisphere clustering results in mid-gray (dark gray if in addition 
the masses of the two pseudojets are required to differ by less than 30\%)
}
	\label{fig:both}
\end{center}
\end{figure}

The clockwork models of flavour are endowed with a powerful flavour protection against FCNCs, very similar to the RS models.  The FCNCs with light quarks on the external legs are suppressed by the same small overlaps of the zero-modes, giving rise to hierarchies between the SM quark masses. This clockwork-GIM  mechanism, along with the constraints on $Y_{U,D}$ arising from the stability of the Higgs potential,   suffices to alleviate the flavour constraints to the level that TeV scale gear masses are compatible with experimental bounds~\cite{Alonso:2018bcg}.

TeV scale gears
can be searched for at the LHC, where they are produced through QCD pair production.
The collider signatures
depend on the gear decay patterns. The gears decay predominantly through their coupling to the Higgs doublet into gears from a different-chirality chain.
The lightest gears decay directly to SM fermions, mostly $t$ and $b$, via the emission of $W,Z$ or $h$, as do heavier gears for which these are the only kinematically allowed channels. 
The main existing collider constraints 
are from searches for pair production of vector-like quarks, in final states involving third generation SM quarks. Ref.~\cite{Alonso:2018bcg} found the two
35 fb${}^{-1}$ 13 TeV ATLAS 
searches for vector-like quarks decaying into 
$tW$ final states~\cite{Aaboud:2018uek}, as well as the analogous search employing the 
$tZ$ and $tH$ final states~\cite{Aaboud:2018xuw}, to be currently most sensitive, see 
Fig.~\ref{fig:both}.

The dense spectrum of gears and the potentially complex pattern of gear decays poses an experimental challenge.
In the conventional vector-like quark searches the clockwork signal will appear as an excess of events with high transverse energies or $H_T$, but without a dominant single peak in the invariant mass of any particular final state, such as $tH$ or $tW$. Ref.~\cite{Alonso:2018bcg} proposed a novel reconstruction strategy targeting pair production of heavy quarks with a-priori unknown but potentially long decay chains that result in a single heavy flavoured quark, $t$ or $b$, plus any number of massive weak or Higgs bosons per decay chain. The proposed search strategy uses a modified hemisphere clustering algorithm, with $t-$ and $b-$tagged jets as seeds for clustering into exactly two pseudo-jets (the invariant mass of these is shown as mid-gray distribution in Fig.~\ref{fig:both} right). The original hemisphere clustering uses instead the jets with highest invariant mass as seeds and shows no sharp features (light gray). Requiring that the masses of the two pseudo-jets differ by less then 30\% gives the dark grey distribution, with clearly visible gears (in the exploratory study of~\cite{Alonso:2018bcg} tops, $b$-quarks, $W$, $Z$ and the Higgs were not decayed).

\subsection{Flavour implications for high \pt new physics searches} 
In this subsection we collect several signatures of flavour models or models where nontrivial flavour structure is relevant for high \pt searches: the FCNC top decays to exotica, the (model dependent) implications for di-Higgs production, and the set of signatures that are related to neutrino mass models. 

\subsubsection{Top decays to exotica} 
{\it\small Authors (TH): S. Banerjee, M. Chala, M. Spannowsky.} 

The FCNC mediated processes are rare within the SM. However, LHC is a top factory and significant number of events are expected even for top decays with very small branching ratios. In light of this, studies of top FCNC decays to SM particles have garnered a strong interest in the community~\cite{Agashe:2009di,Mele:1998ag,Greljo:2014dka,Azatov:2014lha,Khachatryan:2015att,Botella:2015hoa,Bardhan:2016txk,Badziak:2017wxn,Khachatryan:2016sib,Gabrielli:2016cut,CMS-PAS-TOP-17-017,Aaboud:2018nyl,Aaboud:2017mfd,Aaboud:2018pob,Papaefstathiou:2017xuv,Abbas:2015cua}. The FCNC top decays to SM particles, $t\to q Z, q \gamma, qg, q H$, $q=u,c$, and the related constraints from FCNC production processes, were discussed in Sections \ref{sec:top:FCNC} and \ref{sec:8:EXP:FCNC}.

In the presence of light NP, other exotic top FCNC decays are possible. We highlight one such possibility, where the NP spectrum contains a light 
scalar singlet, $S$, with mass $m_S$ below the top quark mass.
Such a scalar
particle is predicted in a number of well-motivated NP models, e.g., in the NMSSM~\cite{Ellwanger:2009dp} and in
non-minimal composite Higgs models~\cite{Dimopoulos:1981xc,Kaplan:1983fs,Kaplan:1983sm,Panico:2015jxa}. Moreover, quite often the induced $t\to cS, uS$ FCNC decays are easier to probe than, for instance the ones involving the SM Higgs, $t\to ch, uh$~\cite{Zhang:2013xya}. 
The reason is three-fold; \textit{(i)} The top FCNCs mediated by 
$S$ are usually suppressed by one less power of the heavy physics scale; \textit{(ii)} $S$ may have a larger decay width into cleaner final states, such as $\ell^+\ell^-$, $b\overline{b}$ 
or $\gamma\gamma$; \textit{(iii)} $S$ can be much lighter than the Higgs, reducing the 
phase space suppression. 
Note that very light $S$, i.e., with $m_S < m_h/2\sim 62.5$ GeV, need not be
excluded by the LHC constraints on the Higgs width, $\Gamma(h\rightarrow S S) \lesssim 10$ MeV ~\cite{Khachatryan:2016ctc}. Indeed, 
for a quartic coupling $\lambda_{HS} S^2 |H|^2$, 
this bound is avoided for $\lambda_{HS} < 0.05$.

\begin{figure}[t]
\includegraphics[width=0.45\textwidth]{\main/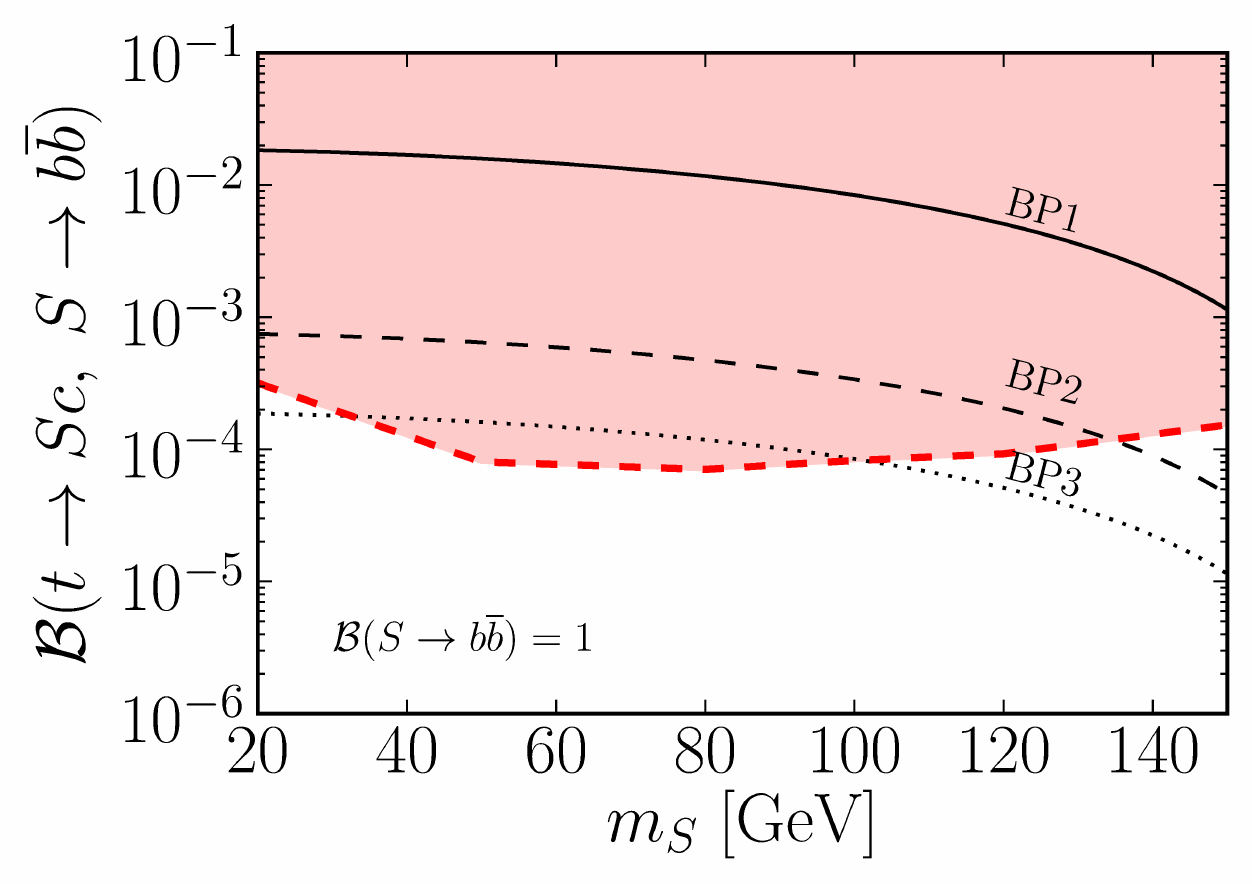}~~~
\includegraphics[width=0.45\textwidth]{\main/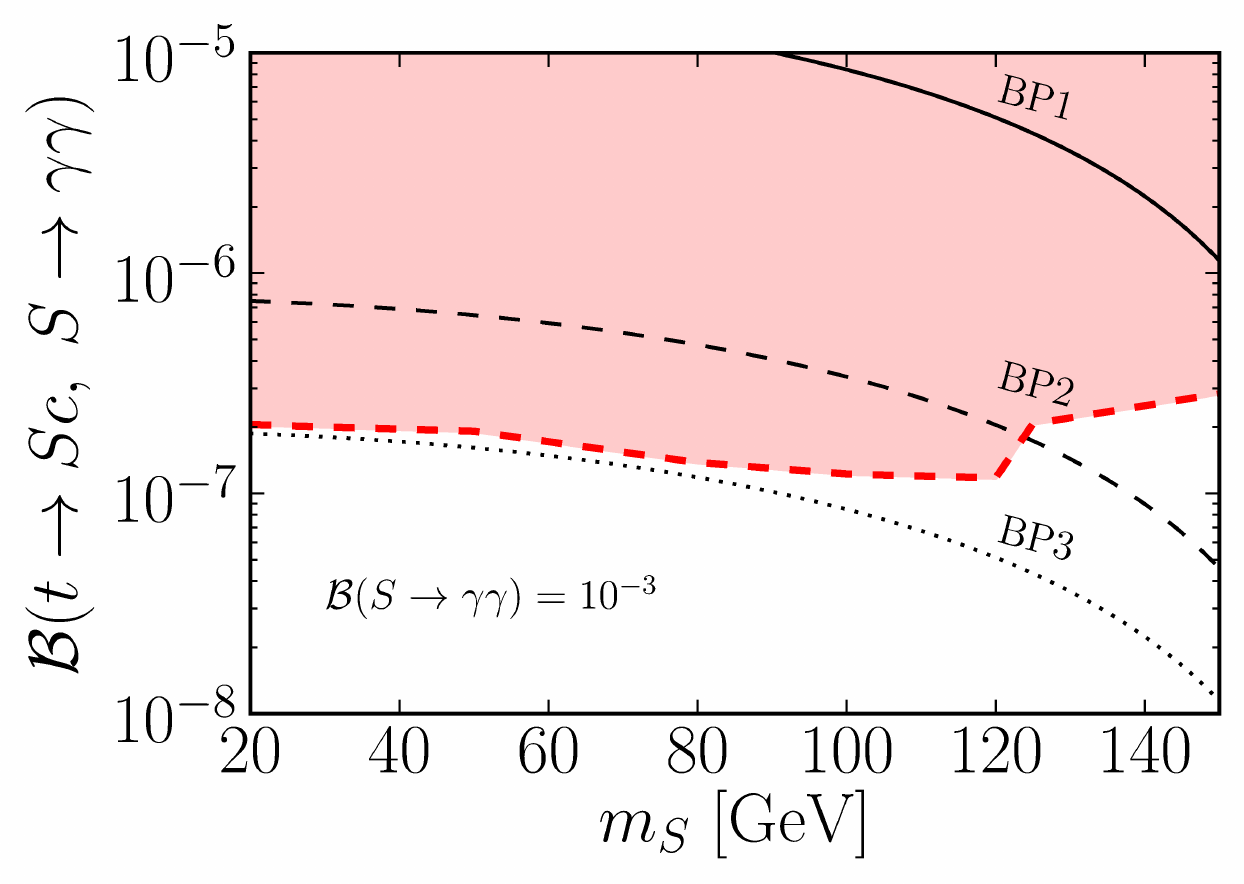}
 \caption{\label{fig:sbb2} Branching ratios that can be tested in the $b\overline{b}$ (left) and $\gamma\gamma$ (right) channels at 14 TeV HL-LHC with $3000$ fb$^{-1}$ (95\% CL upper limit is denoted by red dashed line). The three benchmark points, Eq. \eqref{eq:benchmarks:tqS}, are denoted with black lines. }
\end{figure}

There are no direct experimental limits on $t \to q S$ from colliders. The indirect constraints from 1-loop box diagrams in 
$D^0-\bar{D}^0$ oscillations constrain the products of two $S$ Yukawas, $\tilde Y_{ut} \tilde Y_{ct(tc)}$, and $\tilde Y_{tu} \tilde Y_{ct(tc)}$, to be small~\cite{Bona:2007vi,Harnik:2012pb,Agashe:2013hma}. The $S\bar t c$ or $S\bar t u$ couplings can still be sizeable, but not both at the same time. Inspired by the CMS $t\to hc$ search~\cite{Sirunyan:2017uae}, Ref.~\cite{Banerjee:2018fsx} developed a dedicated analysis for $t\to qS$, by varying $m_S$, the mass of $S$. The projections at the 14 TeV LHC for  $S \to b\bar{b}$ and $S \to \gamma \gamma$ decays  are shown in Fig.~\ref{fig:sbb2} .
Assuming that $S$ is the only light NP state, its couplings to the SM quarks are induced by dimension 5 effective operator (not displaying generational indices)
\begin{equation}\label{eq:lag}
 \mathcal{L} = -{\bar{q}_L}{\tilde{Y}}\frac{S}{f}  \tilde{H}{u_R} + \text{h.c.}\supset \tilde{g} \frac{m_t}{f} \bar{t}_L S c_R + \text{h.c.},
\end{equation}
where $f$ is the NP scale, and on the r.h.s. we introduced a new flavour violating coupling $\tilde g$. The three Benchmark Points (BP) shown in Figs. \ref{fig:sbb2} 
are
\begin{equation}
\label{eq:benchmarks:tqS}
\begin{split}
 \text{BP1(2,3)}:~~\tilde{g} = 1.0(1.0,0.1),~f=2(10,2)~\text{TeV}~~\Longrightarrow~~\BR(t\rightarrow Sc)\sim 10^{-3(4,5)}-10^{-2(3,4)}.
\end{split}
\end{equation}
If the flavour conserving couplings of $S$ to the SM fermions, $\psi$, are ${c_{\psi} m_\psi}
S(\bar{\psi}\psi)/f$, and to the photons ${c_{\gamma}\alpha} S F_{\mu\nu}\tilde{F}^{\mu\nu}/({4\pi f})$, then
$\mathcal{B}(S\rightarrow\gamma\gamma)/\mathcal{B}(S\rightarrow \overline{\psi}\psi) \sim 
({\alpha}/{\pi})^2 (m_S/m_\psi)^2$, 
taking $c_{\psi}\sim c_{\gamma}\sim \mathcal{O}(1)$.
 The suppression of $S\to \gamma\gamma$ can be partially 
compensated by scaling with $m_S$, so that $\mathcal{B}(S\to \gamma\gamma)$ can possibly be 
significantly larger than $\mathcal{B}(h\to \gamma\gamma)$. Searches should thus use both $S\to b\bar b$ and $S\to \gamma\gamma$. The details on how to reduce the backgrounds can be found in~\cite{Banerjee:2018fsx}.
Current searches for $S \rightarrow b\overline{b}$ in the gluon fusion 
channel~\cite{Sirunyan:2018ikr} constrain
only values of $c_\psi$ above $\sim 10$ for $f \sim 1$ TeV. Our analysis works
instead for very small values of $c_\psi$ provided the branching ratio 
is sizable.

For projections at future colliders we find that the increase in cross-section 
for the background at $\sqrt{s} = 27$ TeV ($100$ TeV) when compared to $\sqrt{s} = 14$ TeV is similar to 
that for the signal, and is $\sim 4(40)$. Assuming an integrated luminosity of $10$ ab$^{-1}$, 
we expect an increase in significance by a factor of $\sim 3.7$ ($\sim 11.5$). Similar results hold for the 
$b\overline{b}$ channel.

\subsubsection{Implications for di-Higgs production}
\begin{figure}
 \begin{center}
 \includegraphics[width=.52\textwidth]{\main/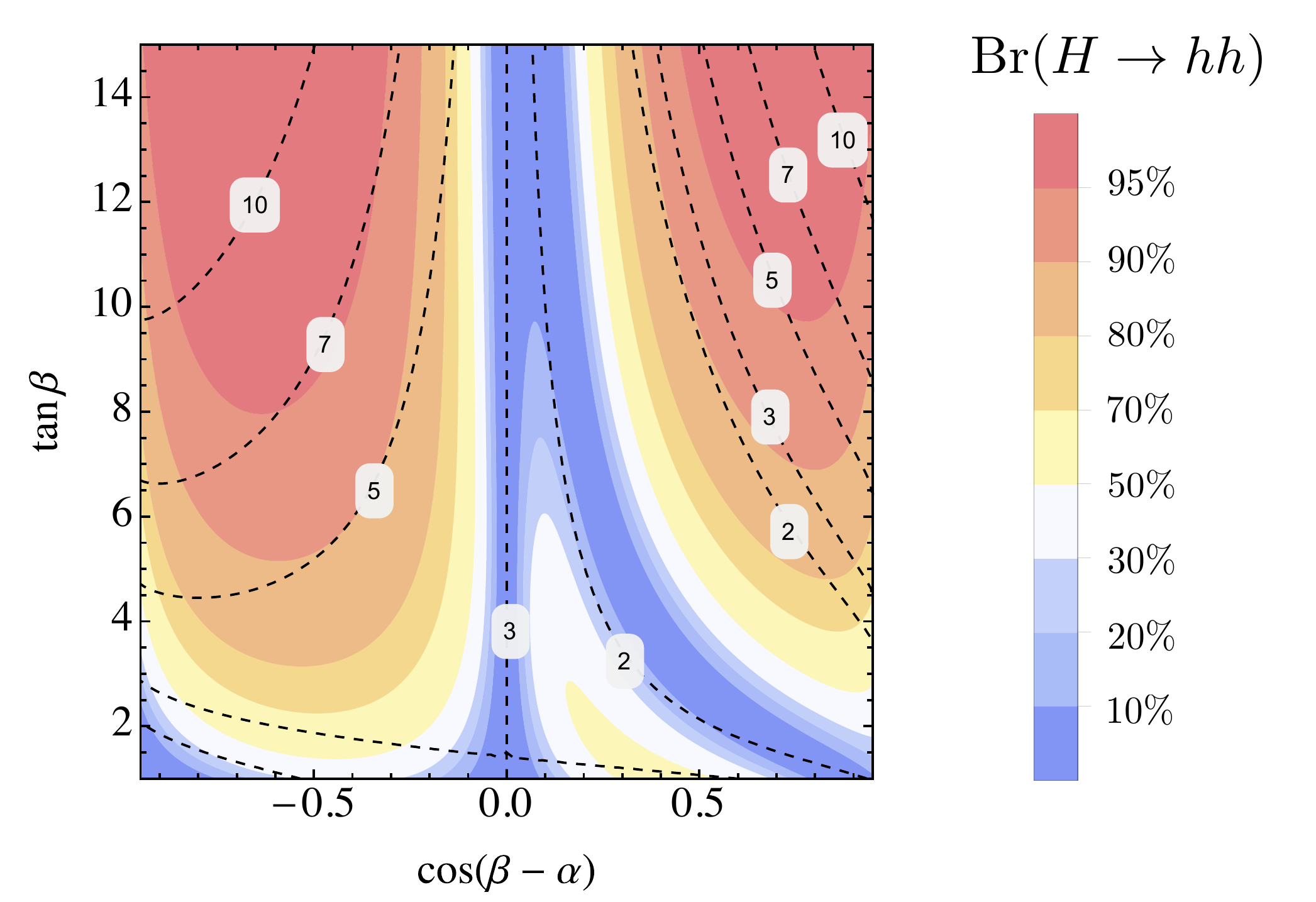}~~~~~
 \includegraphics[width=.44\textwidth]{\main/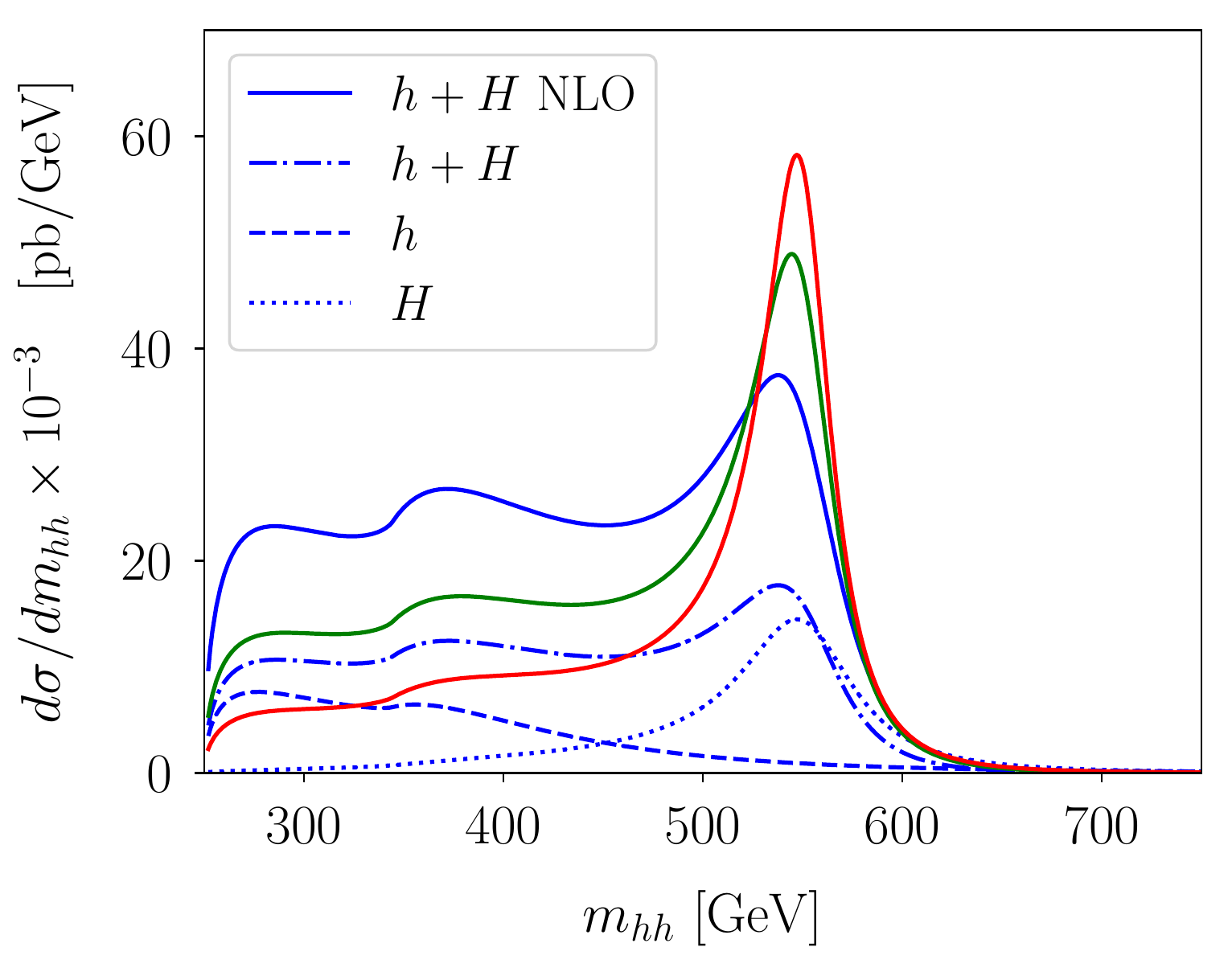}
 \caption{\label{fig:BRvskappa} Left: $\text{Br}(H \to hh)$ as a function of $\cos(\beta-\alpha)$ and $ \tan\beta$ for $M_H=M_{H^\pm}=550$ GeV and $M_A=450$ GeV. The dashed contours correspond to constant $|\kappa_{\psi}^h|$ (we set $n_{\psi}=1$). Right: Invariant mass distribution for the different contributions to the $pp\to hh$ signal with $c_{\beta-\alpha}=-0.45$ and $\kappa^h_{\psi}=5$ (blue),  $\kappa_{\psi}^h=4$ (green) and $\kappa_{\psi}^h=3$ (red) at $\sqrt{s}=27 $ TeV, respectively.}
 \end{center}
  \vspace{-.6cm}
 \end{figure}

{\it\small Authors (TH): Martin Bauer, Marcela Carena and Adri\'an Carmona.}\\
Interestingly, in some models the flavour structure can feed back into nontrivial constraints on the scalar potential. This was demonstrated in the 2HDM model with FN charges, introduced in Sec.~\ref{sec:2HDM:flavour}.
The scalar couplings to gauge bosons are the same as in the normal type-I 2HDM while the scalar coupling between the heavy Higgs $H$ and two SM Higgs scalars $h$, as well as the triple Higgs coupling can be expressed as \cite{Boudjema:2001ii,Gunion:2002zf}
\begin{align}
\label{eq:mainn1}
g_{Hhh}&=\frac{c_{\beta-\alpha}}{v}\!\left[\big(1\!-\!f^h(\alpha,\beta)s_{\beta-\alpha}\big)\big(3M_A^2\!-\!2m_h^2\!-\!M_H^2\big)\!-\!M_A^2\right],\\
g_{hhh}&= -\frac{3}{v}\!\left[f^h(\alpha,\beta)c_{\beta-\alpha}^2(m_h^2-M_A^2)+m_h^2s_{\beta-\alpha}\right],\label{eq:mainn2}
\end{align}
where $M_A$ is the pseudoscalar mass. The $U(1)$ flavour symmetry restricts the number of allowed terms in the scalar potential forbidding, e.g., terms proportional to $H_1 H_2$. Interestingly, one can rewrite such self scalar interactions with the help of function $f^h(\alpha,\beta)$, since it is related to the combination $H_1 H_2^{\dagger}$ appearing in both the scalar potential and the higher dimensional operators generating different Yukawa couplings. Therefore, the parameter space for which $f^h(\alpha,\beta)\gg 1$ and  $c_{\beta-\alpha}\neq 0$ leads to maximally enhanced diagonal couplings of the SM Higgs to fermions \eqref{eq:diagocoup} as well as to enhanced trilinear couplings \eqref{eq:mainn1} and \eqref{eq:mainn2}. For maximally enhanced Yukawa couplings, the mass of the heavy Higgs $H$ cannot be taken arbitrarily large and resonant Higgs pair production has to be present. This correlation between the enhancement of the Higgs Yukawa couplings $\kappa^h_{\psi}$ and $\text{Br}(H \to hh)$ is illustrated for $M_H=M_A=M_{H^\pm}=500$ GeV in Fig. \ref{fig:BRvskappa} (left) where we plot the dependence of $\text{Br}(H \to hh)$ on $c_{\beta-\alpha} $ and $ t_\beta$ \cite{Bauer:2017cov}.  The dashed contours correspond to constant values of $|\kappa_{\psi}^h|$ for $n_{\psi}=1$. The correlation does not depend on the factor $n_\psi$, although  $n_\psi > 1$ leads to a larger enhancement. The two exceptions for which this correlation breaks down are the limits $c_{\beta-\alpha}\approx 0$ (disfavored in the flavour model) and $c_{\beta-\alpha}\approx \pm 1$ (disfavoured by SM Higgs couplings strength measurements). Depending on the structure of the Yukawa couplings, the value of $\kappa^h_\psi$ in Fig.~\ref{fig:BRvskappa} (left) can be larger or smaller than the value $n=1$ chosen to illustrate the relation between $g_{Hhh}$ and $\kappa^h_\psi$. Current experimental limits constrain this structure. For example, since $\kappa^h_\mu < 2.1$ \cite{ATLAS:2018kbw}, either $n=0$ for the muon, or one is constrained to the $\kappa^h_\mu < 2.1$ parameter space in Fig.~\ref{fig:BRvskappa} (left).

There is  a non-trivial interplay between resonant and non-resonant contributions to $pp\to hh$, as shown in Fig. \ref{fig:BRvskappa} (right), for $\sqrt{s}=27$ TeV, setting $M_A=450$ GeV and $M_{H}=M_{H^{\pm}}=550$ GeV,  $c_{\beta-\alpha}=-0.45$ and three different values of $\kappa_{\psi}^{h}=3, 4$ and $5$. When the enhancement in the Higgs Yukawa couplings is large enough, the interference between non-resonant and resonant contributions turns the broad peak into a shoulder in the $d\sigma/ dm_{hh}$ distribution for the total cross section, as shown for the case $\kappa_{\psi}^h=5$ by the blue line.

\subsubsection{Neutrino Mass Models at the HL/HE LHC}
{\it\small Authors (TH): T. Han, T. Li, X. Marcano, S. Pascoli, R. Ruiz, C. Weiland.}

The questions pertaining to neutrino masses: whether  or not  neutrinos are Majorana particles, the origin of smallness of the neutrino masses,  as well as the reason for  large mixing angles, remain some of the most pressing open issues in particle physics today.
A set of potential solutions is provided by the seesaw models. These postulate new particles that couple to SM fields via
mixing/Yukawa couplings, SM gauge currents, and/or new gauge symmetries.
If accessible, a plethora of rich physics can be studied in considerable detail at hadron colliders. 
This would complement low energy and oscillation probes of neutrinos~\cite{Atre:2009rg,Cai:2017mow}.
In the following, we summarize the discovery potential of seesaw models at hadron colliders with collision energies of $\sqrt{s} = 14$ and 27 TeV.
For a more comprehensive reviews on the sensitivity of colliders to neutrino mass models, see \cite{Atre:2009rg,Cai:2017mow,Weiland:2013wha,Ruiz:2015gsa,Marcano:2017ucg} and references therein.

\paragraph*{The Type I Seesaw and Variants}\label{sec:lhcSeesawtype1}
In Type I seesaw the light neutrino masses and mixing are generated from couplings of SM leptons to new fermionic gauge singlets with Majorana masses. 
For low-scale seesaw models with only fermionic singlets, 
lepton number has to be nearly conserved and light neutrino masses are proportional to small lepton number violating (LNV) parameters~\cite{Kersten:2007vk,Moffat:2017feq}.
For high-scale seesaws  with only fermionic singlets, light neutrino masses are inversely proportional to large LNV mass scales,
and again lepton number is approximately conserved at low energies.
Thus LNV processes are suppressed in type I seesaw models
(unless additional particles
are introduced to decouple the light neutrino mass generation from heavy neutrino production).

\begin{figure}[t]
\begin{center}
\includegraphics[width=0.7\textwidth]{\main/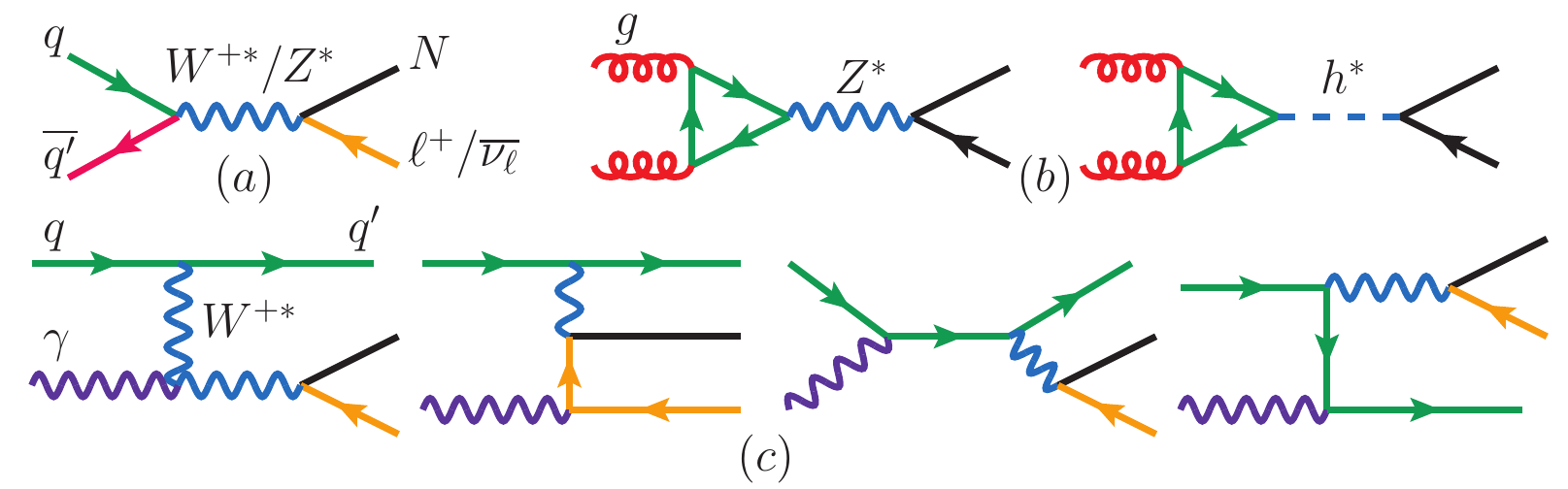}
\\[2mm]
\includegraphics[width=0.48\textwidth]{\main/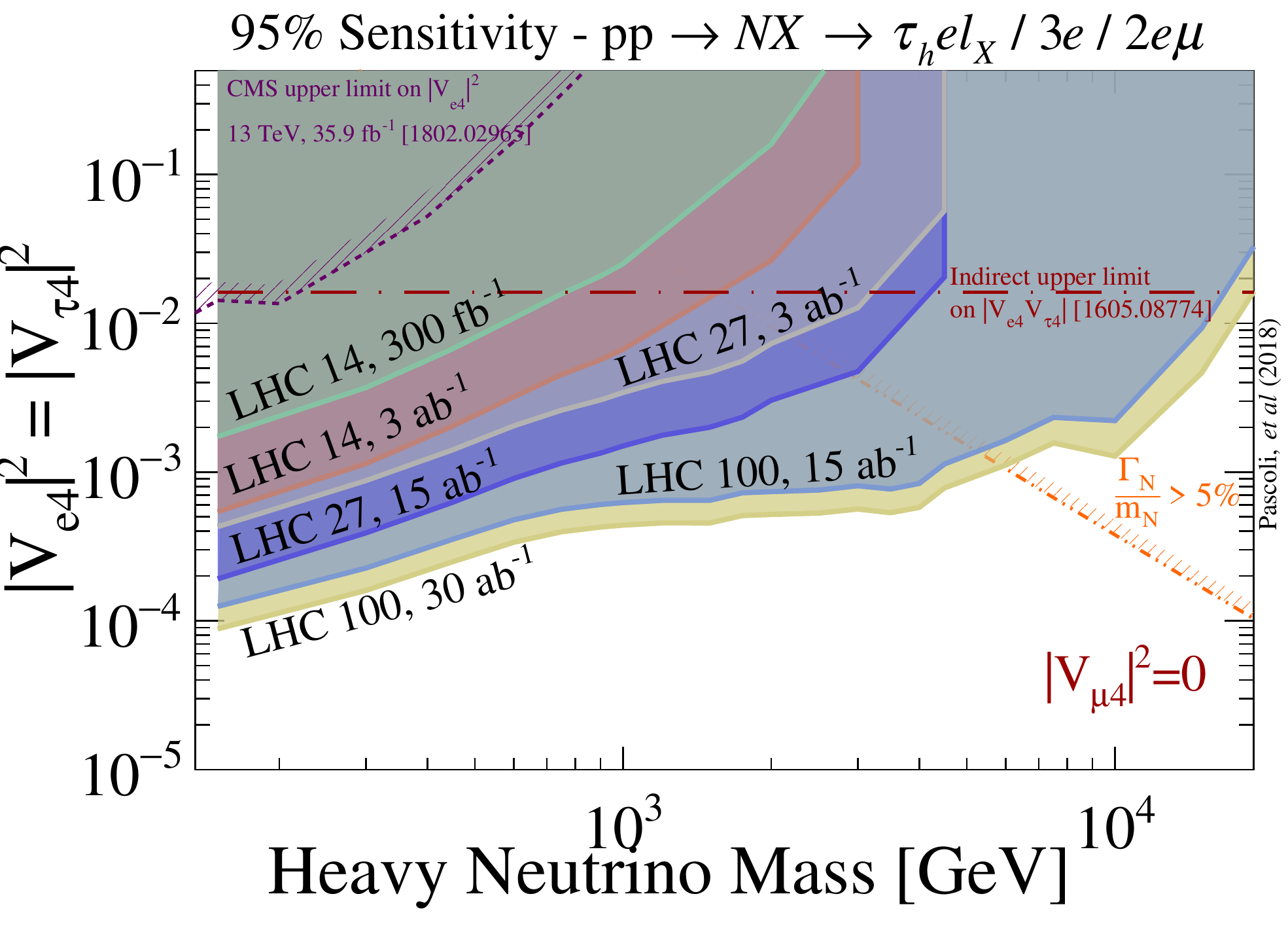}~~~~~
\includegraphics[width=0.48\textwidth]{\main/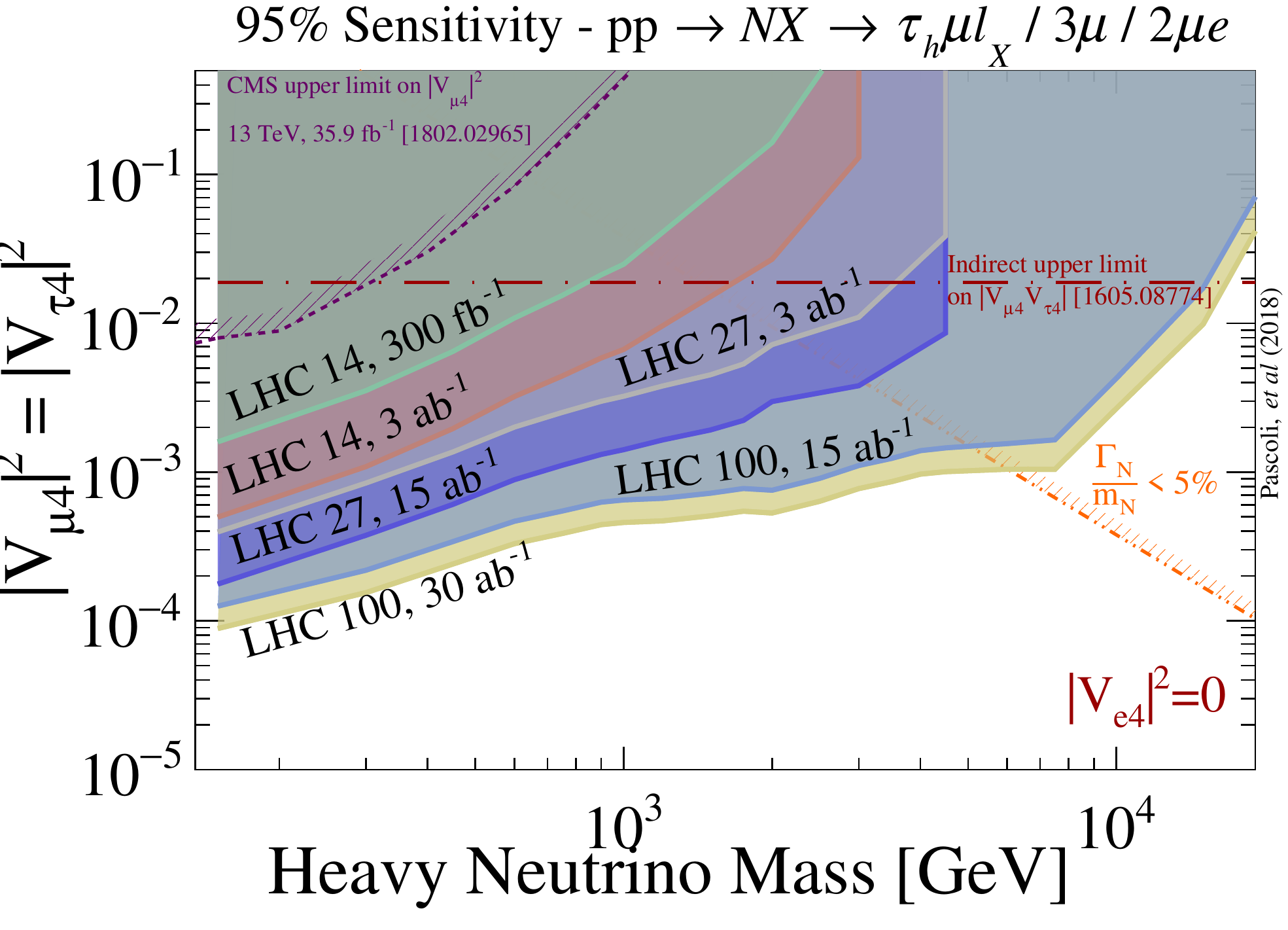}
\end{center}
\caption{
Upper:
Born-level diagrams for heavy neutrino, $N$, production via (a) Drell-Yan (DY), (b) gluon fusion (GF), and (c) vector boson fusion (VBF). 
Lower: for the benchmark mixing hypotheses
 $\vert V_{e4}\vert = \vert V_{\tau4} \vert$ with $\vert V_{\mu4}\vert = 0$ (left panel) and
 $\vert V_{\mu4}\vert = \vert V_{\tau4} \vert$ with $\vert V_{e4}\vert = 0$ (right panel),
the projected sensitivity at $\sqrt{s}=14,~27$ and $100$ TeV using the tri-lepton + dynamic jet veto analysis of Ref.~\cite{Pascoli:2018rsg}.
}
\label{fig:lhcSeesaw_ISS}
\end{figure}

If kinematically accessible, heavy neutrinos $N$ can be produced in hadron collisions through neutral current and charged current processes,
as shown in Fig. \ref{fig:lhcSeesaw_ISS} (upper).
The expected suppression of LNV processes in type I seesaw models 
motivates the study of lepton number conserving (LNC) processes, such as 
the heavy neutrino $N$ production via DY and VBF, with subsequent decays to only leptons, 
\begin{equation}
 p p \to \ell_N N +X \to \ell_N \ell_W W +X \to \ell_N \ell_W \ell_\nu \nu +X,
 \label{eq:pp3lX}
\end{equation}
giving the trilepton final state, $\ell^\pm_i \ell^\mp_j \ell^\pm_k + \textrm{MET}$.
Projections from a new tri-lepton search strategy recently proposed in Ref.~\cite{Pascoli:2018rsg}, 
based on a dynamical jet veto selection cut, are shown in Fig.~\ref{fig:lhcSeesaw_ISS}, assuming the benchmark mixing hypotheses
 $\vert V_{e4}\vert = \vert V_{\tau4} \vert$ with $\vert V_{\mu4}\vert = 0$ (left panel) and
 $\vert V_{\mu4}\vert = \vert V_{\tau4} \vert$ with $\vert V_{e4}\vert = 0$ (right panel), for $\sqrt{s}=14,~27$ and $100$ TeV.
For benchmark luminosities, the colliders can probe active-sterile mixing as small as (approximately) $\vert V_{\ell 4}\vert^2 \sim 5\times10^{-4} - 9\times10^{-5}$
and masses as heavy as (approximately) $1.5-15$ TeV for $\vert V_{\ell 4}\vert^2 \sim 10^{-2}$.

Another possibility is to search for lepton flavor violating (LFV) final states such as
\begin{equation}
 q ~\overline{q}' \to  ~N ~\ell^\pm_1 \to ~\ell^\pm_1 ~\ell^\mp_2 ~ W^\mp  \to ~\ell^\pm_1 ~\ell^\mp_2 ~j ~j.
 \label{eq:LFVfinalstate}
\end{equation}
This was, e.g., studied in Ref~\cite{Arganda:2015ija} in the context of the inverse seesaw (ISS), a low-scale variant of the type I seesaw.
Due to strong experimental limits on $\mu\to e\gamma$ by
MEG~\cite{Adam:2013mnn}, the event rates involving taus are more promising
than those for $e^\pm \mu^\mp jj$.
After  $\mathcal L=3~\rm{ab}^{-1}$ of data taking, more than 100 LFV
events of $\tau^\pm \mu^\mp jj$ type could be produced for neutrino masses below 700 (1000)
GeV for $pp$ collisions at 14 (27)~TeV.

\begin{figure}[t]
\begin{center}
\includegraphics[width=0.45\textwidth]{\main/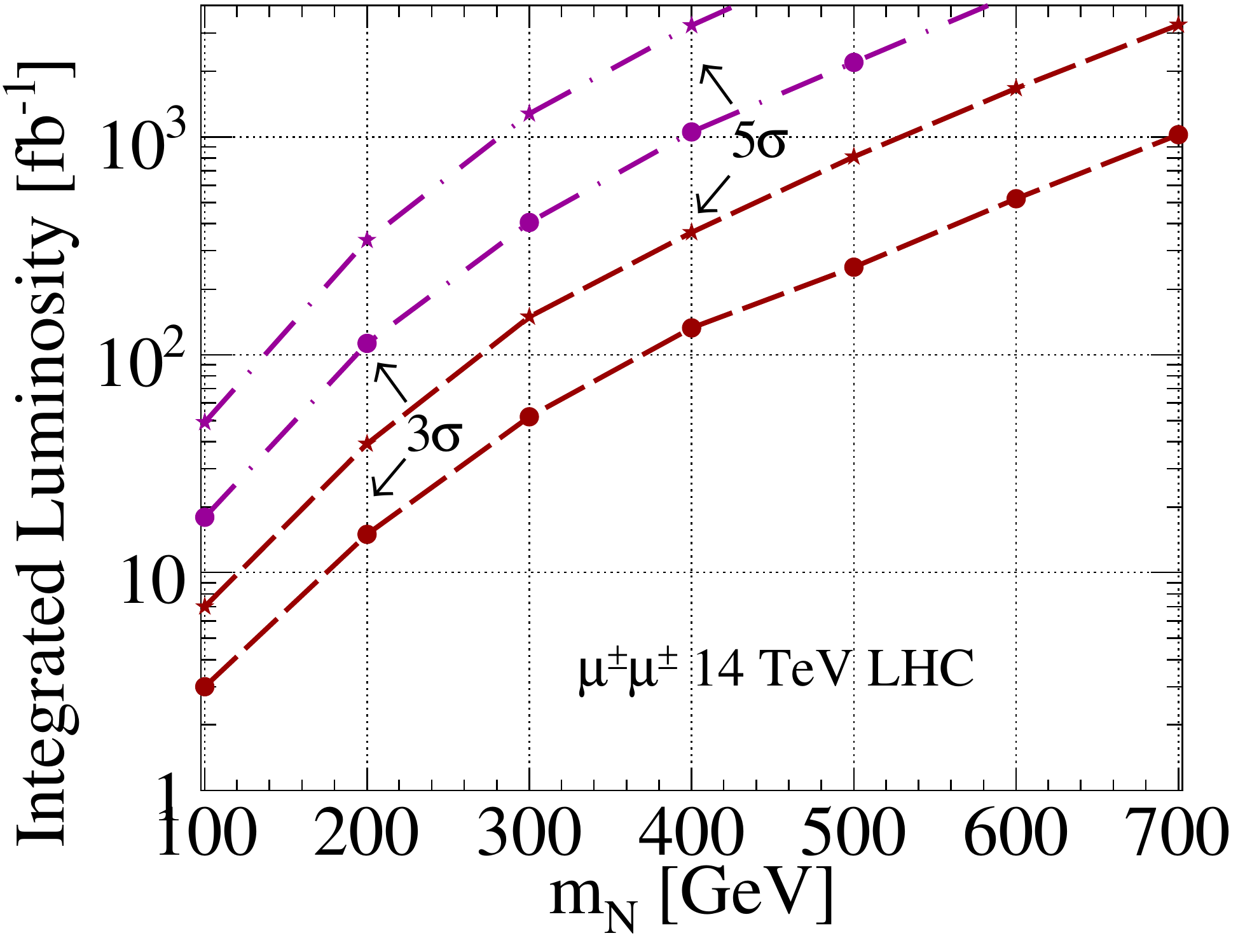}~~~~		
\includegraphics[width=0.45\textwidth]{\main/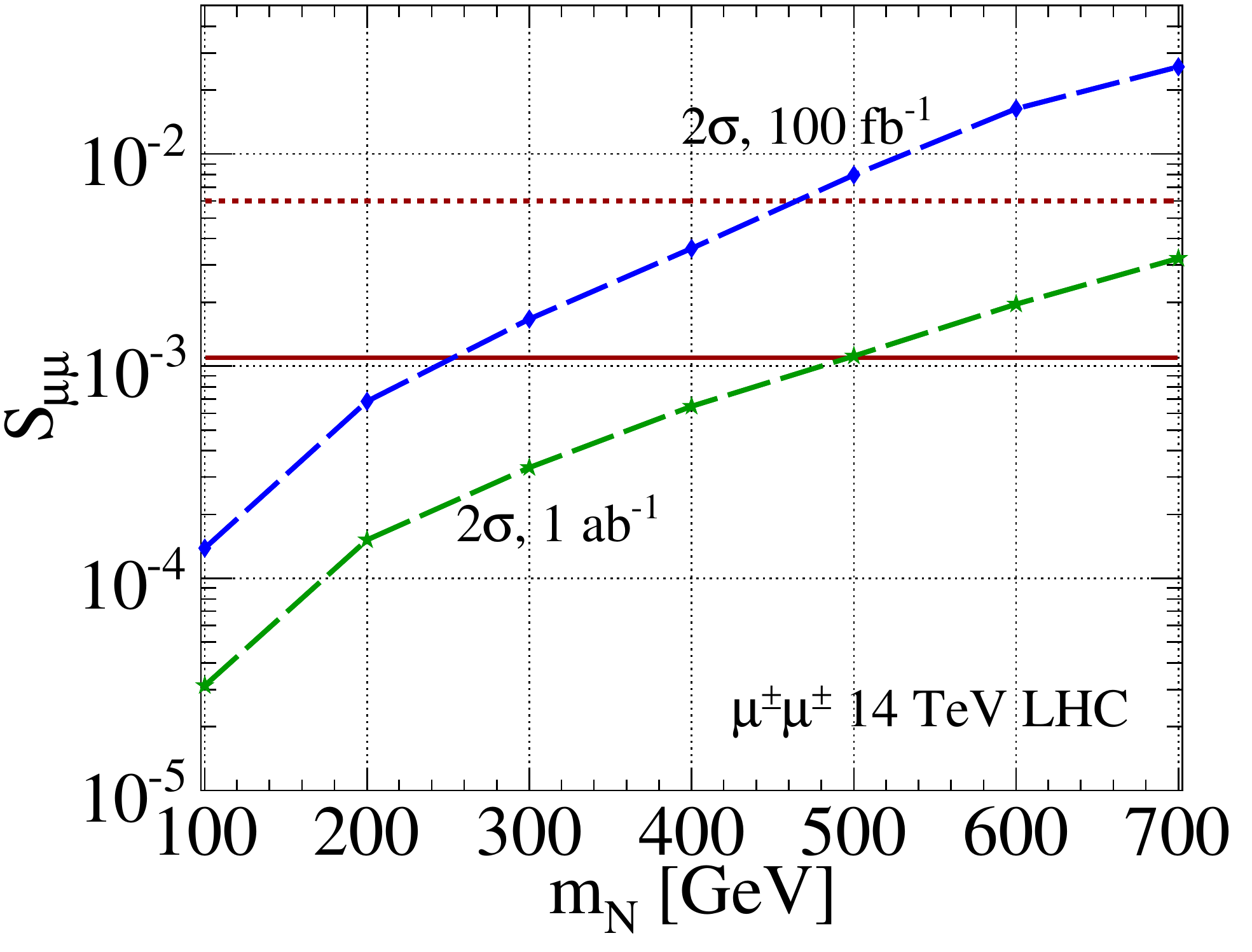}		
\end{center}
\caption{
Left: required luminosity for $3~(5) \sigma$ evidence (discovery) using the LNV final state $\mu^\pm \mu^\pm jj$, as a function of the heavy neutrino mass, $m_N$, assuming optimistic (brown) 
and pessimistic (purple) mixing scenarios~\cite{Alva:2014gxa}. Right:
sensitivity to $N-\mu$ mixing~\cite{Alva:2014gxa} with the optimistic (pessimistic) mixing scenario is given by the horizontal dashed (full) line.
}
\label{fig:SeesawIDiscoveryPotential}
\end{figure}

In the presence of additional particles that can decouple the heavy neutrino production 
from the light neutrino mass generation, e.g., new but far off-shell gauge bosons~\cite{Ruiz:2017nip},
the Majorana nature of the heavy neutrinos can lead to striking LNV collider signatures,
such as the well-studied same-sign dilepton and jets process~\cite{Keung:1983uu}
\begin{equation}
 pp \rightarrow  ~N ~\ell^\pm_1 \rightarrow ~\ell^\pm_1 ~\ell^\pm_2 ~ W^\mp  \rightarrow ~\ell^\pm_1 ~\ell^\pm_2 +nj.
 \label{eq:LNVfinalstate}
\end{equation}
Assuming that a low-scale type I seesaw is responsible for the heavy neutrino production, 
Fig. \ref{fig:SeesawIDiscoveryPotential} displays the discovery potential and active-heavy
mixing sensitivity of the $\mu^\pm\mu^\pm$ channel~\cite{Alva:2014gxa}.
Assuming the  pessimistic/conservative mixing scenario of $S_{\mu\mu} = 1.1\times10^{-3}$~\cite{Alva:2014gxa}, 
the HL-LHC with 3 ab$^{-1}$ 
would be able to discover a heavy neutrino with a mass of $m_N\simeq400\,\mathrm{GeV}$ 
and is sensitive to masses up to $550\,\mathrm{GeV}$ at $3\sigma$. 
Using only 1 ab$^{-1}$, the HL-LHC can improve on the preexisting mixing constraints summarized in the pessimistic scenario 
for neutrino masses up to $500\,\mathrm{GeV}$.

\paragraph*{Heavy Neutrinos and the Left-Right Symmetric Model}\label{sec:lhcSeesawlrsmCollider}

\begin{figure}[!t]
\begin{center}
\includegraphics[scale=1,width=.4\textwidth]{\main/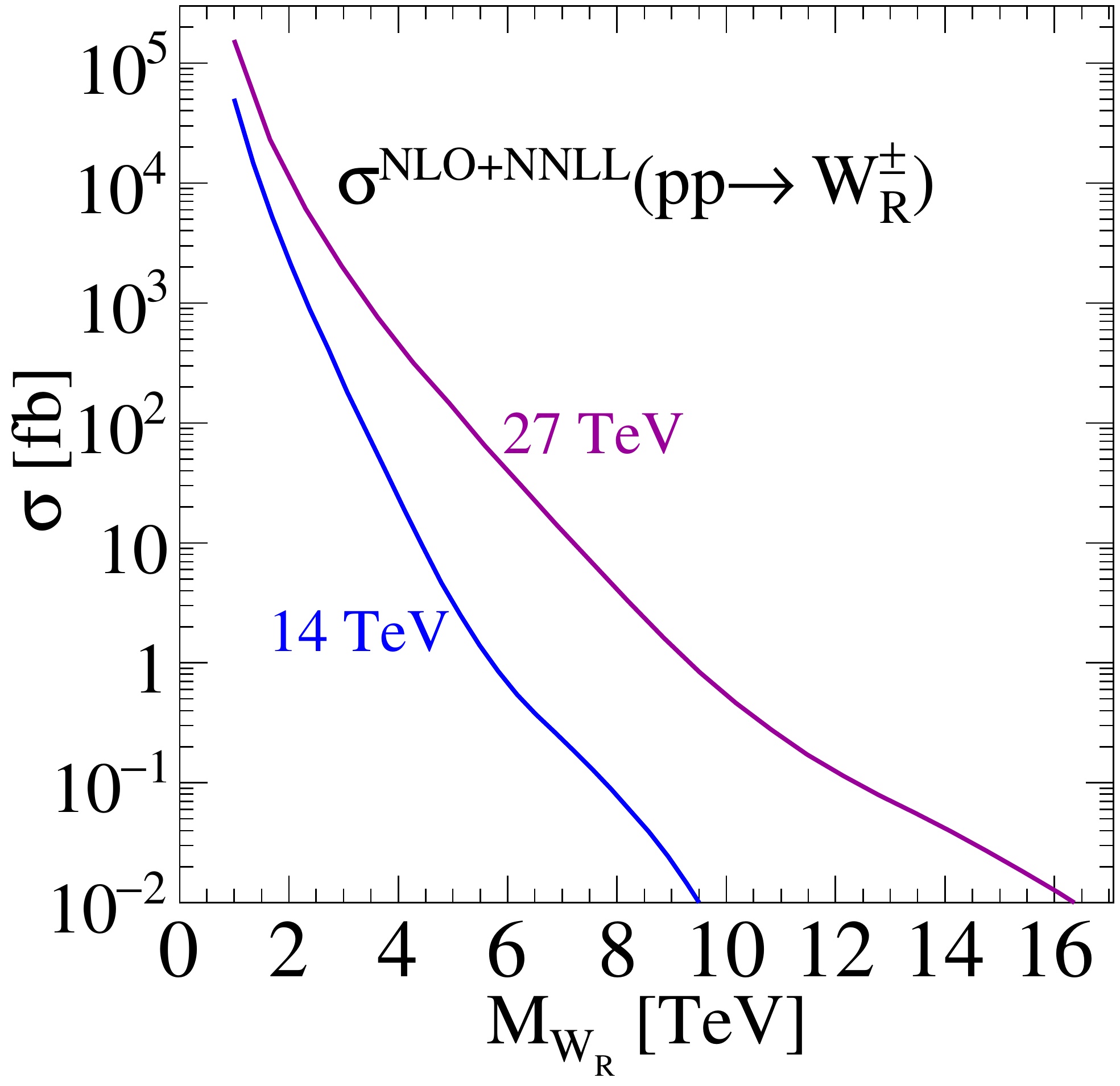}~~~~~~~~~~ 	
\includegraphics[scale=1,width=.4\textwidth]{\main/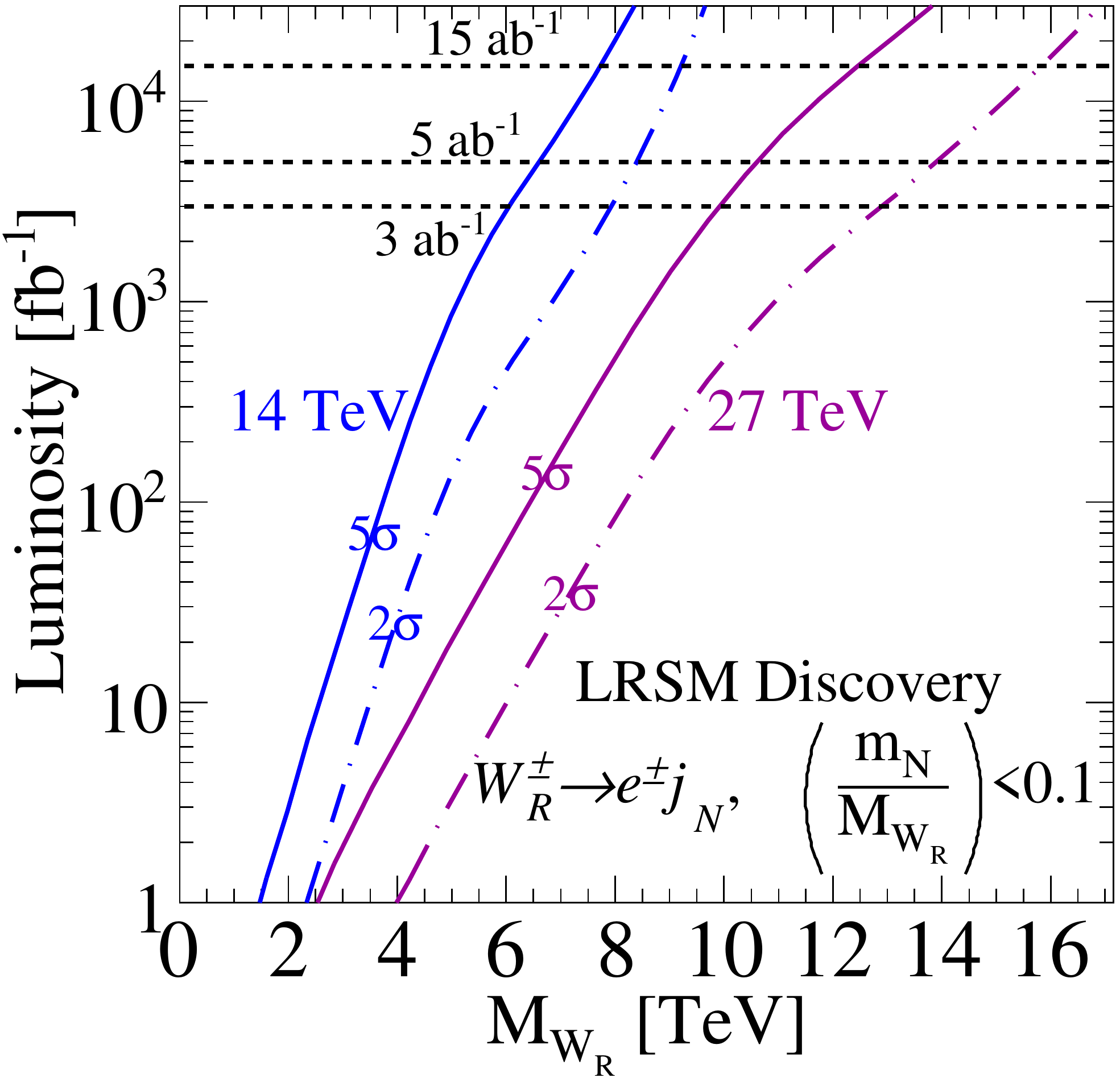} 
\end{center}
\caption{
Left: The total $pp\to W_R$  cross section at NLO+NNLL(Threshold).
Right: $5(2)\sigma$ discovery potential (sensitivity) via $W_R$ decaying to an electron and neutrino jet $(j_N)$,
as a function of $W_R$ mass, at $\sqrt{s} = 14$ and 27 TeV~\cite{Mitra:2016kov}.
}
\label{fig:lrsm_HLvsHE_LHC}
\end{figure}

The Left-Right Symmetric Model (LRSM) 
addresses the origin of both tiny neutrino masses via a Type I+II seesaw hybrid mechanism 
as well as the SM $V-A$ chiral structure through spontaneous breaking of the $SU(2)_L\times SU(2)_R$ symmetry. 
The model predicts new heavy gauge bosons $(W_R^\pm,~Z_R')$,
heavy Majorana neutrinos $(N)$, and a plethora of neutral and electrically charged scalars $(H_i^0,~H_j^\pm,~H_k^{\pm\pm})$.
The LRSM gauge couplings are fixed to the SM Weak coupling constant, up to (small) RG-running corrections.
As a result, the Drell-Yan production mechanisms for $W_R$ and $Z_R$ result in large rates at hadron colliders.
 
One of the most promising discovery channels is the production of heavy Majorana neutrinos
from resonant $W_R$, with $N$ decaying via a lepton number-violating final state. At the partonic level, the process is 
~\cite{Keung:1983uu} (for details see \cite{Cai:2017mow} and references therein)
\begin{equation}
 q_1 \overline{q}_2 \to W_R \to N ~\ell^\pm_i \to \ell_i^\pm \ell_j^\pm W_R^{\mp *} \to \ell_i^\pm \ell_j^\pm q_1' \overline{q_2'}.
 \label{eq:seesawLRSMDY}
\end{equation}
Due to the ability to fully reconstruct the final state of Eq.~\eqref{eq:seesawLRSMDY}, many properties of $W_R$ and $N$ can be extracted,
including a complete determination of $W_R$ chiral couplings to quarks independent of leptons~\cite{Han:2012vk}.
Beyond the canonical $pp \to W_R \to N\ell \to 2\ell + 2j$ channel, it may be the case that the heavy neutrino is hierarchically lighter than the 
right-handed (RH) gauge bosons. 
Notably, for $(m_N/M_{W_R})\lesssim0.1$, $N$ is sufficiently Lorentz boosted that its decay products, particularly the charged lepton, 
are too collimated to be resolved experimentally~\cite{Ferrari:2000sp,Mitra:2016kov}.
Instead, one can consider the $(\ell_j^\pm q_1' \overline{q_2'})$-system as a single object, a \textit{neutrino jet}~\cite{Mitra:2016kov,Mattelaer:2016ynf}.
The hadronic process is then
 $ p p \to W_R \to N\ell_i^\pm \to j_N ~\ell_i^\pm$,
and inherits much of the desired properties of \eqref{eq:seesawLRSMDY}, 
such as the simultaneous presence of high-\pt charged leptons and lack of MET~\cite{Mitra:2016kov,Mattelaer:2016ynf},
resulting in a very strong discovery potential.
Fig. \ref{fig:lrsm_HLvsHE_LHC} (right) shows the requisite integrated luminosity for $5(2)\sigma$ discovery at $\sqrt{s}=14$ and 27 TeV.

\paragraph*{Type II Scalars}\label{sec:lhcSeesawtype2}
\begin{figure}[!t]
\begin{center}
\includegraphics[scale=1,width=.45\textwidth]{\main/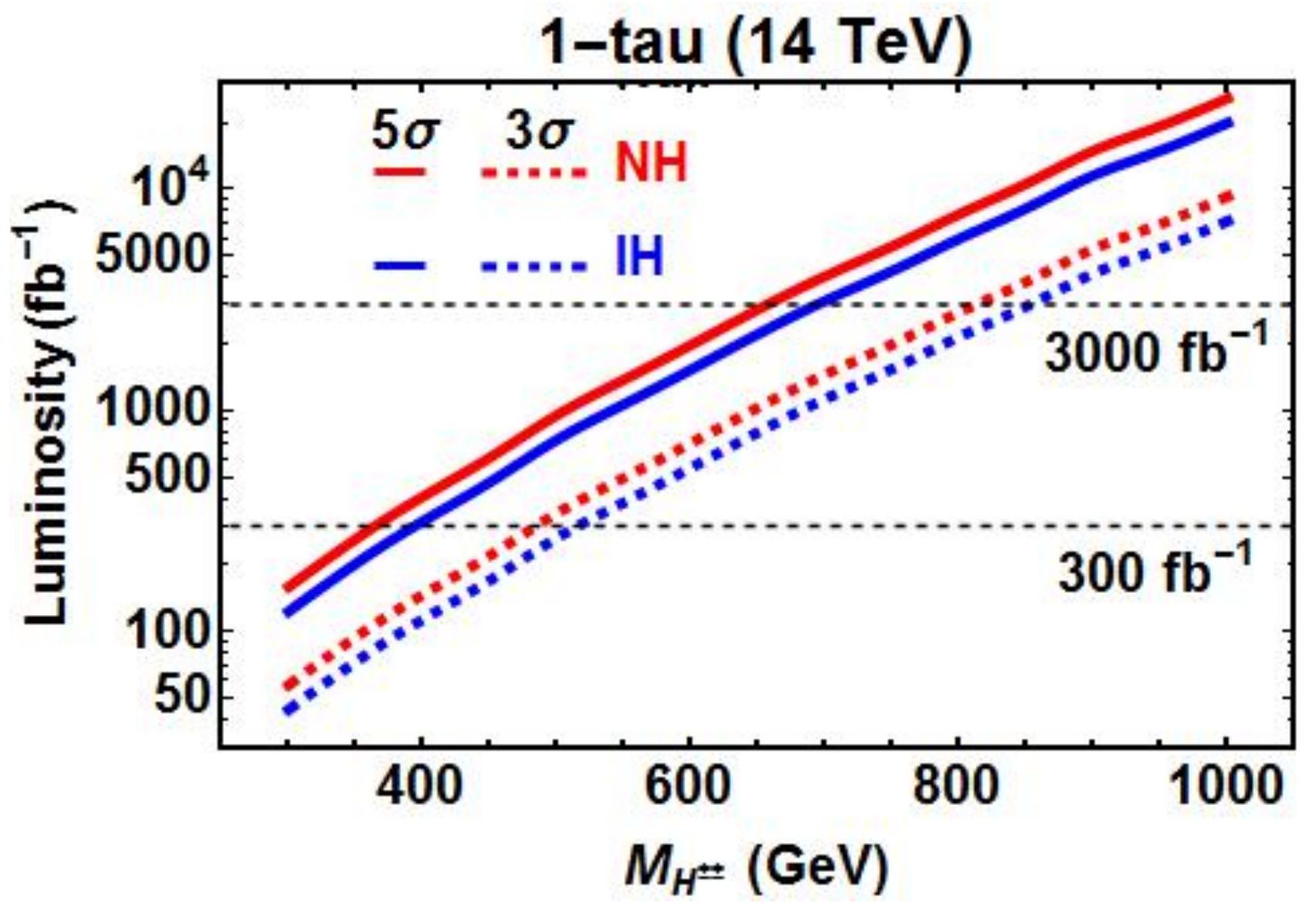}~~~~~~ 
\includegraphics[scale=1,width=.45\textwidth]{\main/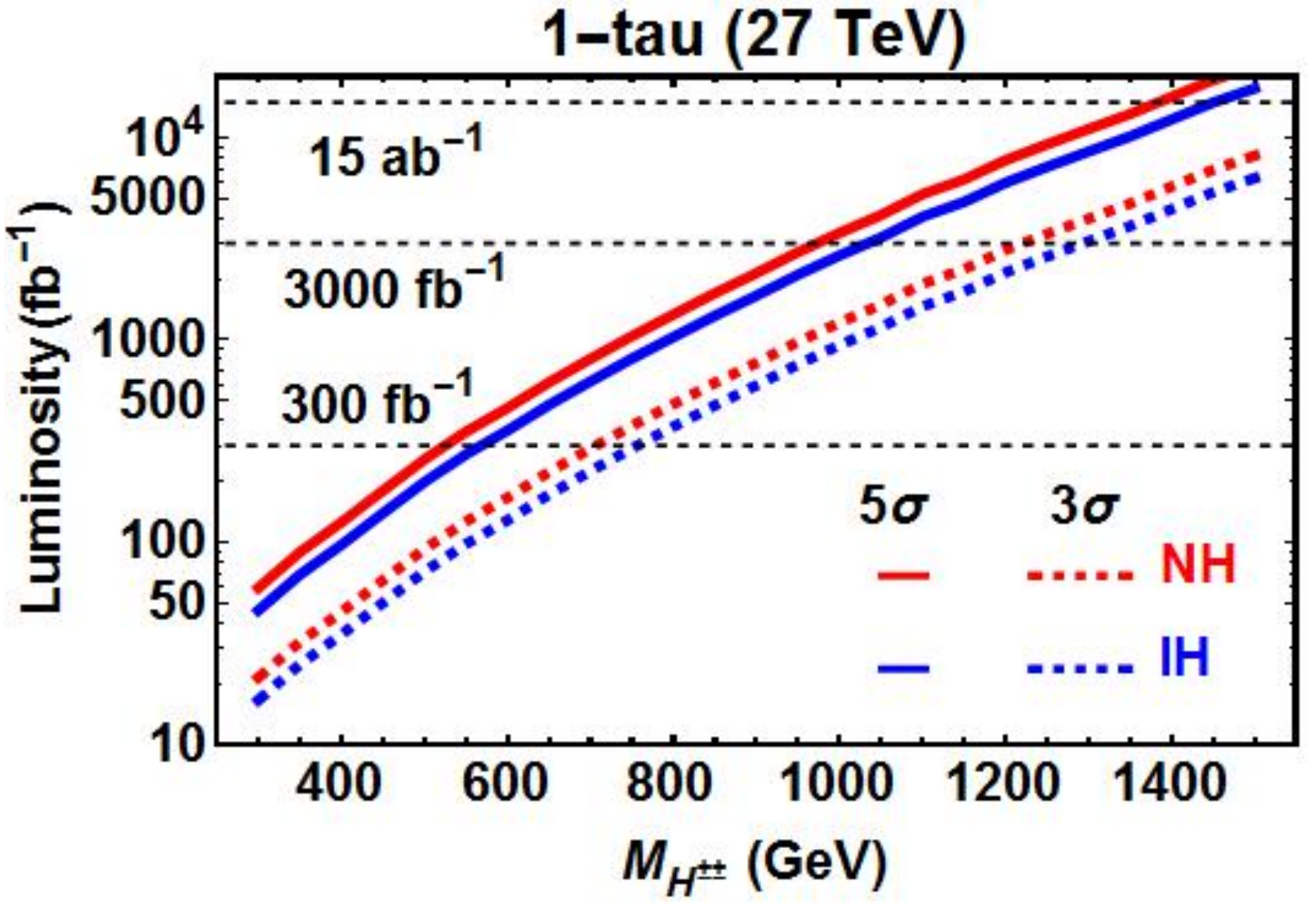}	
\end{center}
\caption{
Requisite luminosity  for $5(3)\sigma$ discovery (evidence) as a function of $M_{H^{\pm\pm}}$
for the process $pp\to H^{++}H^{--}\to \tau_h \ell^\pm \ell^\mp \ell^\mp$,
where $\tau^\pm\to \pi^\pm \nu $, 
for the NH and IH at $\sqrt{s}=14, 27$ TeV. 
}
\label{fig:lhcSeesawtypeiiScalars}
\end{figure}

Type II seesaw introduces a new scalar SU$(2)_L$ triplet that couples to SM leptons.
The light neutrinos obtain Majorana masses through  SU$(2)_L$ triplet vev, so that
type II scenario notably does not have sterile neutrinos.
The most appealing production mechanisms at hadron colliders of triplet Higgs bosons are
\begin{eqnarray}
pp\to Z^\ast/\gamma^\ast \to H^{++} H^{--}, \ \ \ pp\to W^\ast\to H^{\pm\pm}H^\mp,
\end{eqnarray}
followed, 
by lepton flavor- and lepton number-violating decays to the SM charged leptons.
In Type II scenarios, $H^{\pm\pm}$ decays to $\tau^\pm \tau^\pm$ and $\mu^\pm \mu^\pm$ pairs are comparable or greater than the
$e^\pm e^\pm$ channel by two orders of magnitude. 
Moreover, the $\tau \mu$ channel is typically dominant in decays involving different lepton flavors~\cite{Perez:2008ha,Li:2018jns}.
If such a seesaw is realized in nature, tau polarizations can help to determine the chiral property of triplet scalars.
One can discriminate between different heavy scalar mediated neutrino mass mechanisms, e.g., between Type II seesaw and Zee-Babu model, 
by studying the distributions of tau lepton decay products~\cite{Sugiyama:2012yw,Li:2018jns}. 
Due to the low $\tau_h$ identification efficiencies, 
future colliders with high energy and/or luminosity enables one to investigate and search for doubly charged Higgs decaying to $\tau_h$ pairs.
Accounting for constraints from neutrino oscillation data on the doubly charged Higgs branching ratios,
as well as  tau polarization effects~\cite{Li:2018jns},
Figs.~\ref{fig:lhcSeesawtypeiiScalars} displays
the 3$\sigma$ and 5$\sigma$ significance in the plane of integrated luminosity versus doubly charged Higgs mass 
for $pp\to H^{++}H^{--}\to \tau^\pm\ell^\pm\ell^\mp\ell^\mp$ at 
$\sqrt{s} = 14, 27$ TeV, for single $\tau$ channel with $\tau\to \pi {\nu}$, both for normal (NH) and inverted hierarchy (IH).

\paragraph*{Type III Leptons}\label{sec:lhcSeesawtype3}

\begin{figure}[!t]
\begin{center}
 \includegraphics[width=0.8\textwidth]{\main/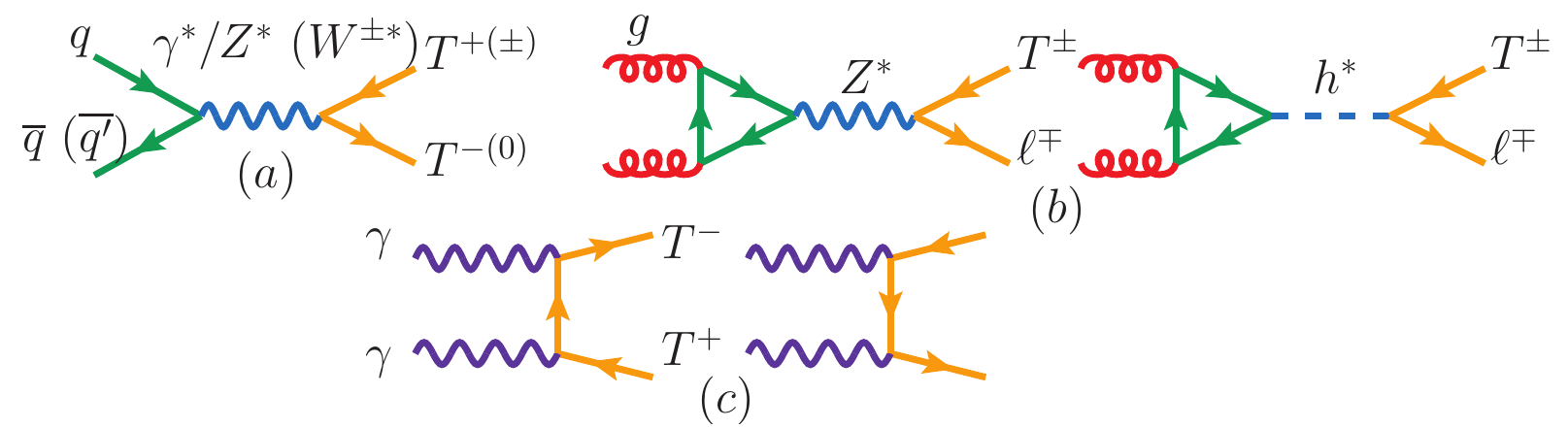}	
 \\[5mm]
  \includegraphics[scale=1,width=.40\textwidth]{\main/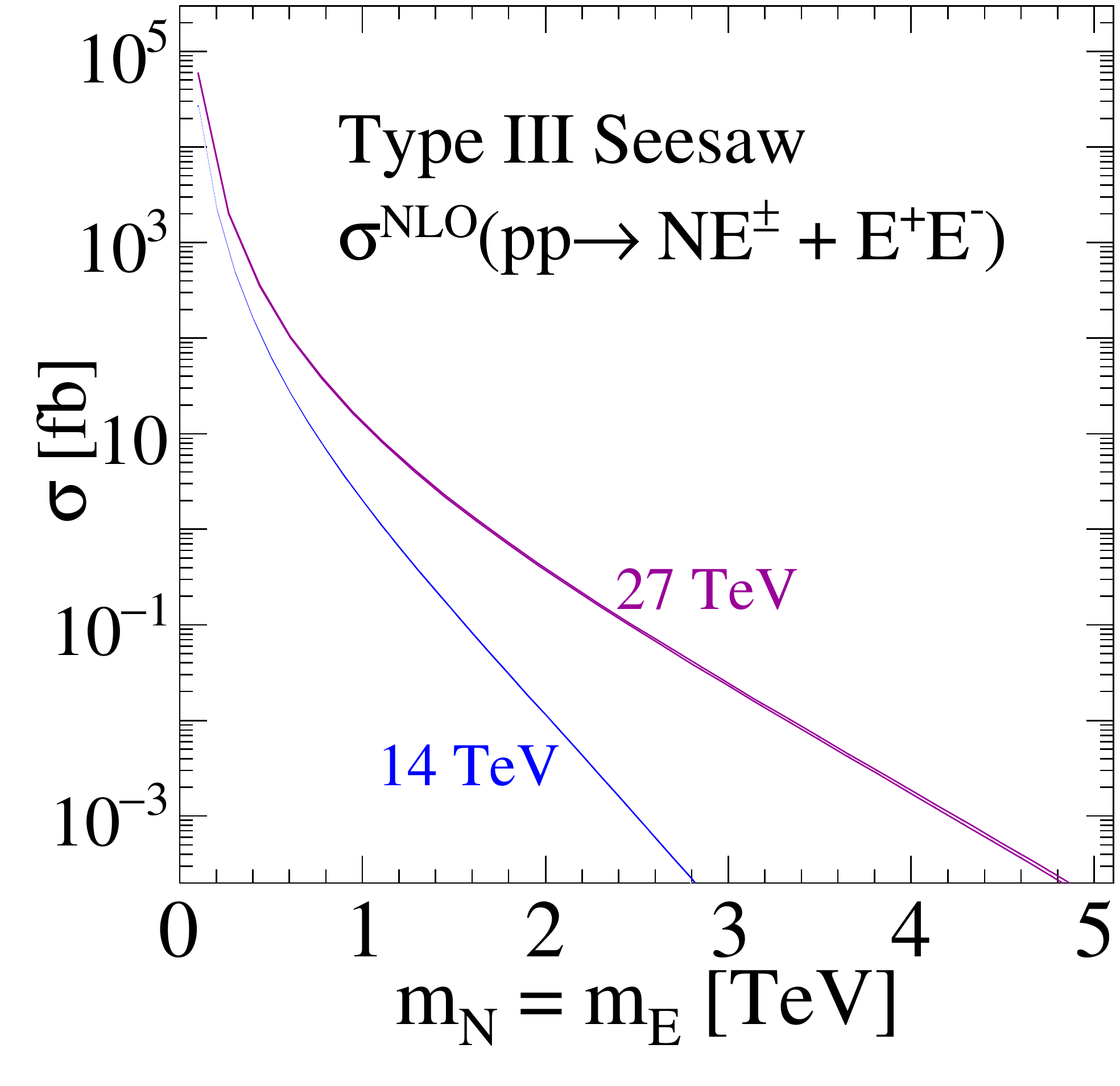}~~~~~~~~~~~~~~~~ 
\includegraphics[scale=1,width=.40\textwidth]{\main/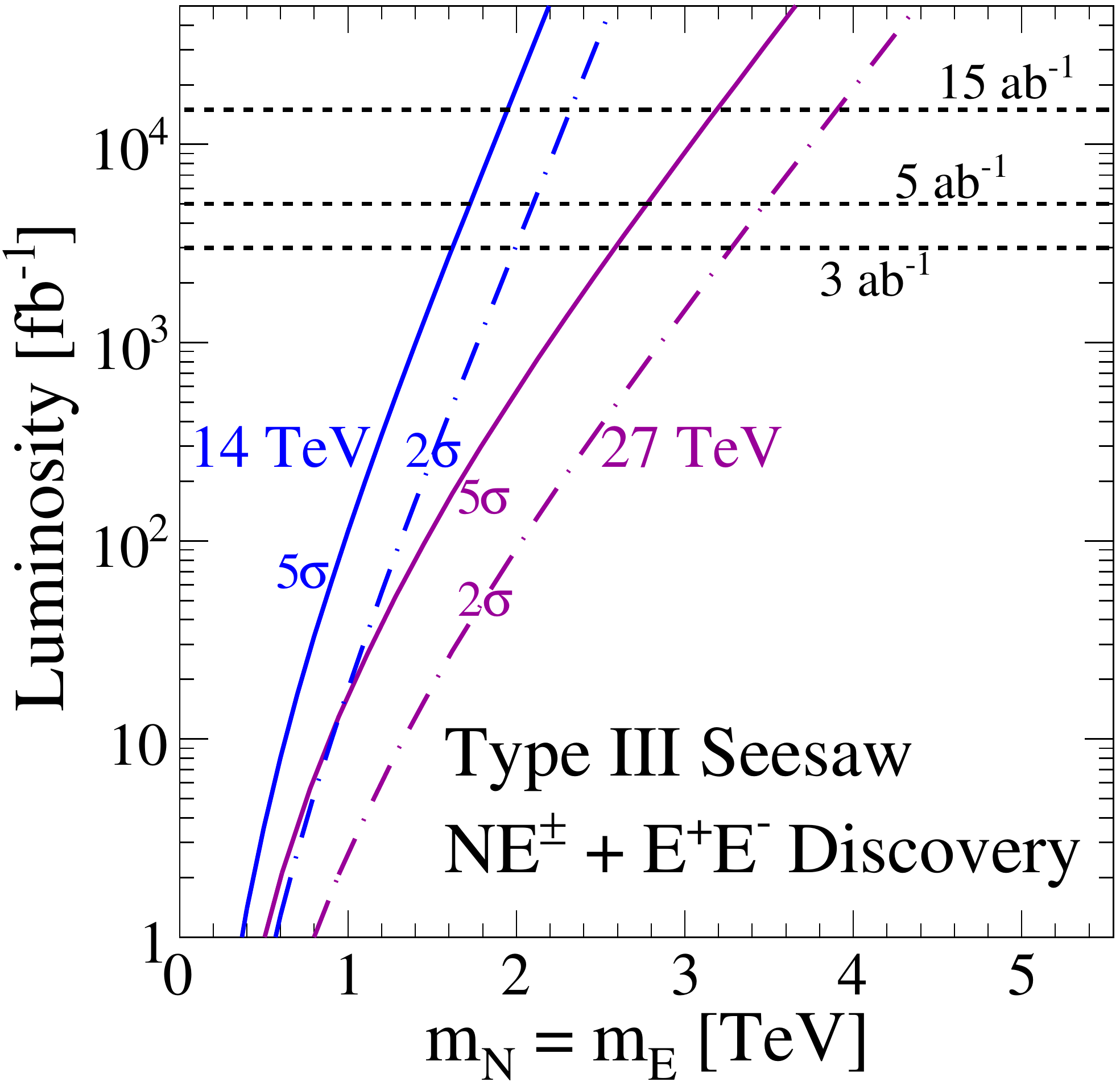} 
\end{center}
\caption{
Upper:
Born level production of Type III leptons via (a) Drell-Yan, (b) gluon fusion, and (c) photon fusion; from \cite{Cai:2017mow}.
Lower left:
  the inclusive production cross section for  $pp\to NE^\pm + E^+E^-$, at NLO in QCD~\cite{Ruiz:2015zca} for $\sqrt{s}=14$ and 27 TeV, as a function of heavy triplet lepton mass. Lower right:
  the required integrated luminosity for 5(2)$\sigma$ discovery~(sensitivity) to $N E^\pm + E^+E^-$, 
 based on the analyses of~\cite{Arhrib:2009mz,Li:2009mw}.
}
\label{fig:lhcSeesawtype3}
\end{figure}

Low-scale Type III seesaws introduce heavy electrically charged $(E^\pm)$ and neutral $(N)$ leptons, 
part of SU$(2)_L$ triplet,
that couple to both SM charged and neutral leptons through mixing/Yukawa couplings.
Triplet leptons couple appreciably to EW gauge bosons,  
and thus do not have suppressed production cross section, contrary to seesaw scenarios with gauge singlet fermions.
Up to small (and potentially negligible) mixing effects 
the triplet lepton 
pair production cross sections are fully determined, see Fig. \ref{fig:lhcSeesawtype3} (upper) for relevant tree level diagrams for the production of heavy charged leptons.
Drell-Yan processes are the dominant production channel of triplet leptons when kinematically accessible~\cite{Cai:2017mow}.
Fig. \ref{fig:lhcSeesawtype3} (lower left) shows 
the summed cross sections for the Drell-Yan processes,
$ pp\to \gamma^*/Z^* \to E^+E^-$, and  
$ pp\to W^{\pm *} \to E^\pm N$,
 at NLO in QCD, following \cite{Ruiz:2015zca}, as a function of triplet masses (assuming $m_N = m_E$), at $\sqrt{s}=14$ and 27 TeV.

Another consequence of the triplet leptons coupling to all EW bosons is the adherence to the Goldstone Equivalence Theorem.
This implies that triplet leptons with masses well above the EW scale
will preferentially decay to longitudinal polarized $W$ and $Z$ bosons as well as to the Higgs bosons. 
For decays of EW boson to jets or charged lepton pairs, triplet lepton can be fully reconstructed from their final-state 
enabling their properties to be studied in detail.
For fully reconstructible final-states,
\begin{eqnarray}
 N E^\pm	\rightarrow ~\ell\ell' + WZ/Wh		&\rightarrow& ~\ell\ell' ~+~ nj+mb,
 \\
 E^+ E^- 	\rightarrow ~\ell\ell' + ZZ/Zh/hh 	&\rightarrow& ~\ell\ell' ~+~ nj+mb,
\end{eqnarray}
which correspond approximately to the branching fractions $\BR(NE)\approx0.115$ and $\BR(EE)\approx0.116$,
search strategies such as those considered in \cite{Arhrib:2009mz,Li:2009mw} can be enacted.
Assuming a fixed detector acceptance and efficiency of $\mathcal{A}=0.75$,
which is in line to those obtained by \cite{Arhrib:2009mz,Li:2009mw},
Fig. \ref{fig:lhcSeesawtype3}(lower right) shows as a function of triplet mass 
the requisite luminosity for a 5$\sigma$ discovery (solid) and 2$\sigma$ evidence (dash-dot)
of triplet leptons at $\sqrt{s}=14$ and 27 TeV.
With $\mathcal{L}=3-5$ ab$^{-1}$, the 14 TeV HL-LHC can discover states as heavy as $m_{N},m_{E}=1.6-1.8$ TeV.
For the same amount of data, the 27 TeV HE-LHC can discover heavy leptons  $m_{N},m_{E}=2.6-2.8$ TeV;
with $\mathcal{L}=15$ ab$^{-1}$, one can discover~(probe) roughly $m_{N},m_{E}=3.2~(3.8)$ TeV.

\subsection{Implications of TeV scale flavour models for electroweak baryogenesis} 
{\it\small Authors (TH): Oleksii Matsedonskyi, Geraldine Servant.}

In most solutions to the SM flavour puzzle, Yukawa couplings have a dynamical origin, which means that they potentially impact the cosmological evolution.
 In Refs.~\cite{Baldes:2016rqn,Baldes:2016gaf,vonHarling:2016vhf,Bruggisser:2017lhc,Bruggisser:2018mus,Bruggisser:2018mrt,Baldes:2018nel,Servant:2018xcs},  such connections between flavour and cosmology, in particular with the  electroweak baryogenesis,  have been investigated in detail.

Electroweak baryogenesis (EWBG) is a framework where the
matter-antimatter asymmetry of the universe was created during the electroweak phase transition. It 
 relies on a charge transport mechanism in the vicinity of bubble walls during a first-order electroweak (EW) phase transition. EWBG uses EW scale physics only and is therefore testable experimentally. It requires an extension of the Higgs sector, giving a first-order EW phase transition as well as new sources of \CP violation. Typically, there are stringent constraints on EWBG models from bounds on Electric Dipole Moments.
 In the following,  we consider situations where the source of \CP violation has changed with time, which is a natural way to evade constraints. 
 The main motivation is to link EWBG to low-scale flavour models.  If the physics responsible for the structure of the Yukawa couplings is linked to EW symmetry breaking, we can expect the Yukawa couplings to vary at the same time as the Higgs is acquiring a VEV, in particular if the flavour structure is controlled by a new scalar field which couples to the Higgs.
 This is  precisely what can happen in Composite Higgs (CH) models ~\cite{Kaplan:1983fs,Panico:2015jxa}, on which we focus in the following.

The  Composite Higgs (CH) models assume that the Higgs boson arises as a bound state of a new strong interaction, confining around the TeV scale $f$. Other composite resonances are naturally heavier than the Higgs, due to an approximate Goldstone symmetry suppressing the Higgs mass. The rest of the SM fields do not belong to the strong sector and are elementary. The nature of the EW phase transition in CH models can be substantially different with respect to the SM. One of the reasons is that the CH models naturally feature new scalar resonances that participate in the EW phase transition and change its properties (see Refs.~\cite{Espinosa:2011eu,Chala:2016ykx}). On the other hand, the EW transition may become strongly first order even without the help of such additional states, if the EW transition happens simultaneously with the deconfinement-confinement phase transition of the new strong sector (see Refs.~\cite{Bruggisser:2018mus,Bruggisser:2018mrt} and also Refs.~\cite{vonHarling:2016vhf,Megias:2018sxv} for the dual description in the warped 5D space).

The flavour structure of the CH models is intimately tied with the viability of EWBG. The requirement of having a sizeable top quark Yukawa coupling, while suppressing the unwanted flavour-violating effects, suggests that the new sector is nearly scale-invariant
for a large range of energies above the confinement scale. As a result, one may expect that the transition dynamics is mostly determined by a single light field -- the dilaton $\chi$ (see e.g Ref.~\cite{Coradeschi:2013gda}). Depending on its mass, whose size can be related to the separation of the UV flavour scale and the EW scale, the EW phase transition may happen separately from the confinement, or simultaneously with it~\cite{Bruggisser:2018mus,Bruggisser:2018mrt}. %

Moreover, the mechanism for generating the SM flavour hierarchy in CH models may also be a source of \CP asymmetry during the EW phase transition. The hierarchy of SM Yukawas $\lambda_q$ 
is generated by the renormalization group running of the couplings $y_q$ between the elementary fermions and the strong sector operators, 
\begin{equation}
\lambda_q \propto y_q^2, \quad\text{with} \quad y_q=y_q^{\text{UV}} \left( \frac{\mu}{\Lambda_{\text{UV}}} \right)^{\gamma_{y_q}},
\end{equation}
where $\mu\sim \chi$ is the confinement scale, $\Lambda_{\text{UV}}$ is some large scale at which all the mixings are generated with a similar size, $y_q^{\text{UV}}$, and $\gamma_{y_q}$ is the anomalous dimension of the operator responsible for the mixing.
 This means that the size of the Yukawa couplings changes with the evolution of the confinement scale during the confinement phase transition. Such a change of the Yukawa interactions may efficiently source the CP-violation required for the baryogenesis~\cite{Bruggisser:2017lhc}.

\begin{figure}[!t]\centering
\includegraphics[scale=.44]{\main/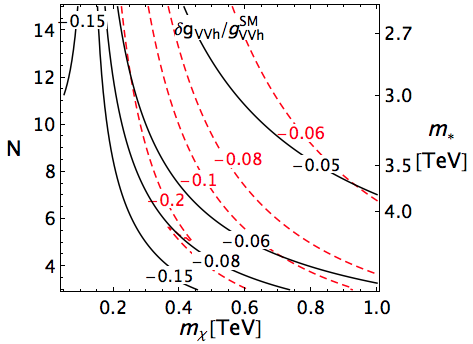}
\hspace{0.2cm}
\includegraphics[scale=.44]{\main/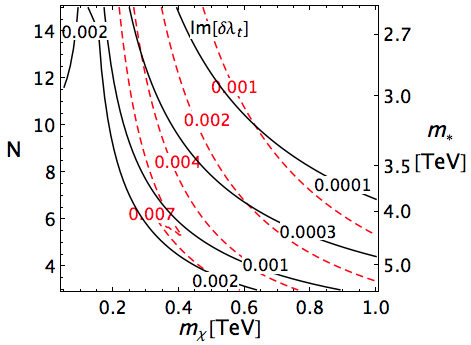}
\caption{Relative deviation of the Higgs couplings to $W$ and $Z$ bosons (left panel) and the imaginary part of the correction to the top quark Yukawa coupling (right panel), as functions of the dilaton mass, $m_\chi$, the number of colours, $N$, in the new confining sector, and generic mass of composite states, $m_*$. Solid black (red dashed) contours correspond to the glueball-like (meson-like) dilaton. The current  and near future experimental sensitivities of electron EDM experiments  to the imaginary part of the top yukawa correspond respectively to approximately $2\times 10^{-3}$ \cite{Andreev:2018ayy,Brod:2013cka}
and $2\times 10^{-4}$~\cite{Kumar:2013qya}.
}
\label{fig:ewbg_observables}
\end{figure}

For both types of transitions  mentioned, one can expect to observe deviations of the Higgs couplings from the SM predictions. These deviations are a result of contributions generic to CH models, as well as those linked to the features of the phase transition and new sources of CP-violation. For concreteness we focus on the  more minimal example, the combined electroweak and strong sector phase transition. 
The potential of the Higgs boson can be parametrized in terms of trigonometric functions of $h/f$, with the overall size of the potential controlled by the mixings between the elementary and composite fermions~\cite{Matsedonskyi:2012ym,Panico:2012uw},
\begin{equation}\label{eq:ewbg_vh}
V = c_1 \sin^2 \frac h f + c_2 \sin^4 \frac h f,
\end{equation}
where $c_1 \sim c_2 \sim \sum_q ({3 y_q^2}/{(4 \pi)^2}) g_\star^2 f^4$. The dependence of the $y_q$ mixings on the dilaton field is responsible for the mass mixing between the dilaton and the Higgs field, which we parametrise by an angle $\delta$.
To generate sufficient amount of \CP violation, such mass mixing needs to be sizeable. Let us consider its effect on the quark Yukawa coupling:
\begin{equation}\label{eq:ewbg_yuk}
{\cal L}_{\rm Yukawa} = \lambda(\chi) \left(\chi \sin \frac h f \right) \bar q_L q_R 
= \bar q_L q_R h \left(\lambda(f) \frac {\chi}{f} + \beta_{\lambda} \frac {\chi - f }{f} \right) + \dots, 
\end{equation}
where we performed an expansion in $\chi$ around its present day value, $f$. Similar, but CP-preserving, modifications are also generated in the couplings of the Higgs boson to the SM gauge fields and the Higgs self-interactions.
The complex phase of the Yukawa beta-function, $\beta_\lambda$, in \eqref{eq:ewbg_yuk} has to be different from that of the quark mass $\lambda(f) h$, as the Yukawa phase changes with energy. 
Choosing the mass parameter to be real, the CP-violating interaction resides in the term $\propto \beta_\lambda$. Rotating the $h$ and $\chi$ fields to the mass basis, the CP-even and CP-odd corrections to the Higgs Yukawa  interaction are
\begin{equation}
\text{Re} [\delta \lambda] \sim \text{Re} [\beta_\lambda] \delta (v/f) + \lambda (\delta^2/2+\delta(v/f)) ,\quad
\text{Im} [\delta \lambda] \sim \text{Im} [\beta_\lambda]  \delta (v/f).
\end{equation} 
In Fig.~(\ref{fig:ewbg_observables}) we show the values of the CP-violating top quark Yukawa modification and the deviations of the Higgs couplings to the $W$ and $Z$ bosons.
Such couplings can be tested directly at the LHC, and also in the measurements of electric dipole moments.  
The strength of the phase transition can be tested in gravitational waves signals at the future space-based observatory LISA~\cite{Caprini:2015zlo}.

\input{section10/High_pT.tex}

%% file: section10/High_pT.tex
\subsection{Phenomenology of high \pt searches in the context of flavour anomalies}

{\small\it Authors (TH): Alejandro Celis, Admir Greljo, Lukas Mittnacht, Marco Nardecchia, Tevong You.}

Precision measurements of flavour transitions at low energies, such as flavour changing $B$, $D$ and $K$ decays, are sensitive probes of hypothetical dynamics at high energy scales.  These can provide the first evidence of new phenomena beyond the SM, even before direct discovery of new particles at high energy colliders. Indeed, the current anomalies observed in $B$-meson decays, in particular, the charged current one in $b\to c\tau \nu$ transitions, and neutral current one in $b\to s\ell^+\ell^-$,  may be the first hint of new dynamics which is still waiting to be discovered at high-\pt. When considering models that can accommodate the anomalies, it is crucial to analyse the constraints derived from high-\pt searches at the LHC, since these can often rule out significant regions of model parameter space. Below we review these constrains, and assess the impact of the High Luminosity and High Energy LHC upgrades (see also the discussion in Sections \ref{sec:7:bsll:NP},\ref{sec7:bsll:model:indep:fits}, and \ref{sec7:ModelsNP:bctaunu}).

\subsubsection{EFT analysis} 

If the dominant NP effects give rise to dimension-six SMEFT operators, the low-energy flavour measurements are sensitive to $C/\Lambda^2$, with $C$ the dimensionless NP Wilson coefficient and $\Lambda$ the NP scale.    The size of the Wilson coefficient is model dependent, and thus so is the NP scale required to explain the $R_{D^{(*)}}$and $R_{K^{(*)}}$ anomalies.    Perturbative unitarity sets an upper bound on the energy scale below which new dynamics need to appear~\cite{DiLuzio:2017chi}.    
The conservative bounds on the scale of unitarity violation are   $\Lambda_U = 9.2$ TeV and $84$ TeV for $R_{D^{(*)}}$and $R_{K^{(*)}}$, respectively, obtained when the flavour structure of NP operators is exactly aligned with what is needed to explain the anomalies.       More realistic frameworks for flavour structure, such as MFV, $U(2)$ flavour models, or partial compositeness, give rise to NP effective operators with largest effects for the third generation. This results in stronger unitarity bounds,
$\Lambda_U = 1.9$ TeV and $17$ TeV for $R_{D^{(*)}}$ and $R_{K^{(*)}}$, respectively.  
These results mean that: \textit{(i)} the mediators responsible for the $b\to c \tau \nu$ charged current anomalies are expected to be in the energy range of the LHC, \textit{(ii)}  the mediators responsible for the $b\to s \ell \ell$ neutral current anomalies could well be above the energy range of the LHC. However, in realistic flavour models also these mediators typically fall within the (HE-)LHC reach.       

If the neutrinos in $b \to c \tau \nu$ are part of a left-handed doublet, the NP responsible for $R_{D^{(*)}}$ anomaly
generically implies a sizeable signal in $p p \to \tau^+ \tau^-$ production at high-\pt. For realistic flavour structures, in which $b \to c$ transition is $\mathcal{O}(V_{cb})$ suppressed compared to $b \to b$, one expects rather large $b b \to \tau \tau$ NP amplitude. 
Schematically, $\Delta R_{D^{(*)}} \sim C_{b b \tau \tau} (1 + \lambda_{bs} / {V_{cb}})$, where $C_{b b \tau \tau}$ is the size of effective dim-6 interactions controlling $b b \to \tau \tau$, and $\lambda_{bs}$ is a dimensionless parameter controlling the size of flavour violation. Recasting ATLAS 13 TeV, $3.2$~fb$^{-1}$ search for $\tau^+ \tau^-$ \cite{Aaboud:2016cre}, Ref.~\cite{Faroughy:2016osc} showed that $\lambda_{bs} = 0$ scenario is already in slight tension with data. For $\lambda_{bs} \sim 5$, which is moderately large, but still compatible with FCNC constraints, HL (or even HE) upgrade of the LHC would be needed to cover the relevant parameter space implied by the anomaly (see Fig.~[5] in~\cite{Buttazzo:2017ixm}). 
For large $\lambda_{bs}$ the limits from $p p \to \tau^+ \tau^-$ become comparable with direct the limits on $p p \to \tau \nu$ from the bottom-charm fusion. The limits on the EFT coefficients from $p p \to \tau \nu$ were derived in Ref.~\cite{Greljo:2018tzh}, and the future LHC projections are  promising. The main virtue of this channel is that the same four-fermion interaction is compared in $b \to c \tau \nu$ at low energies and $b c \to \tau \nu$ at high-\pt. 
Since the effective NP scale in $R(D^{(*)})$ anomaly is low, the above EFT analyses are only indicative. 
For more quantitative statements we review below bounds on explicit models.

The hints of NP in $R_{K^{(*)}}$ require  a $(b s) (\ell \ell)$ interaction.
Correlated effects in high-\pt tails of $p p \to \mu^+ \mu^- (e^+ e^-)$ distributions are expected, if the numerators (denominators) of LFU ratios $R_{K^{(*)}}$ are affected. 
Ref.~\cite{Greljo:2017vvb} recast the  13~TeV $36.1$~fb$^{-1}$ ATLAS search \cite{ATLAS:2017wce} (see also \cite{Aaboud:2017buh}),
to set limits on a number of semi-leptonic four-fermion operators, and derive projections for HL-LHC (see Table 1 in~\cite{Greljo:2017vvb}). These show that direct limits on the $(b s) (\ell \ell)$ operator from the tails of distributions will never be competitive with those implied by the rare $B$-decays~\cite{Greljo:2017vvb,Afik:2018nlr}. On the other hand, flavour conserving operators, $(q q) (\ell \ell)$, are efficiently constrained by the high \pt tails of the distributions. The flavour structure of an underlining NP could thus be probed by constraining ratios $\lambda^q_{b s} = C_{b s} / C_{qq}$ with $C_{b s}$  fixed by the $R_{K^{(*)}}$ anomaly. For example, in models with MFV flavour structure, so that $\lambda^{u,d}_{b s} \sim V_{c b}$, the present high-\pt dilepton data is already in slight tension with the anomaly~\cite{Greljo:2017vvb}. Instead, if  couplings to valence quarks are suppressed, e.g., if NP dominantly couples to the 3rd family SM fermions, then $\lambda^b_{b s} \sim V_{c b}$. Such NP will hardly be probed even at the HL-LHC, and  it is possible that NP responsible for the neutral current anomaly might stay undetected in the high-\pt tails at HL-LHC and even at HE-LHC. Future data will cover a significant part of viable parameter space, though not completely, so that discovery is possible, but not guaranteed.

\subsubsection{Constraints on simplified models for $b\to c \tau \nu$}

Since the $b\to c \tau \nu$ decay is a tree-level process in the SM that receives no drastic suppression, models that can explain these anomalies necessarily require a mediator that contributes at tree-level:

\noindent  $\bullet$ \, \textit{SM-like $W^{\prime}$}:  A SM-like $W^{\prime}$ boson, coupling to left-handed fermions, would explain the approximately equal enhancements observed in $R(D)$ and $R(D^{*})$.     A possible realization is a color-neutral real $SU(2)_L$ triplet of massive vector bosons~\cite{Greljo:2015mma}.  However, typical models encounter problems with current LHC data since they result in large contributions to $pp \to \tau^+ \tau^-$ cross-section, mediated by the neutral partner of the $W^{\prime}$~\cite{Greljo:2015mma,Boucenna:2016qad,Faroughy:2016osc}.    For $M_{W^{\prime}} \gtrsim 500$~GeV, solving the $R(D^{(*)})$ anomaly within the vector triplet model while being consistent with $\tau^+ \tau^-$ resonance searches at the LHC is only possible if the related $Z^{\prime}$ has a large total decay width~\cite{Faroughy:2016osc}. Focusing on the $W^{\prime}$, Ref.~\cite{Abdullah:2018ets} analyzed the production of this mediator via $gg$ and $gc$ fusion, decaying to
$\tau \nu_{\tau}$. Ref.~\cite{Abdullah:2018ets} concluded that a dedicated search using that $b$-jet is present in the final state would be effective in reducing the SM background compared to an inclusive analysis that relies on $\tau$-tagging and $E_T^{\rm{miss}}$. Nonetheless, relevant limits will be set by an inclusive search in the future~\cite{Greljo:2018tzh}.

\vspace{0.2cm}

\noindent  $\bullet$  \,  \textit{Right-handed $W^{\prime}$:} Refs.~\cite{Greljo:2018ogz,Asadi:2018wea} recently proposed that $W^{\prime}$ could mediate a right-handed interaction, with a light sterile right-handed neutrino carrying the missing energy in the $B$ decay. In this case, it is possible to completely decorelate FCNC constraints from $R(D^{(*)})$. The most constraining process in this case is instead $p p \to \tau \nu$. Ref.~\cite{Greljo:2018tzh} performed a recast of the latest ATLAS and CMS searches at 13 TeV and about 36 fb$^{-1}$ to constrain most of the relevant parameter space for the anomaly.

\vspace{0.2cm}

\noindent  $\bullet$\, \textit{Charged Higgs $H^{\pm}$:}       Models that introduce a charged Higgs, for instance a two-Higgs-doublet model, also contain additional neutral scalars.  Their masses are constrained by electroweak precision measurements to be close to that of the charged Higgs.  Accommodating the $R(D^{(*)})$ anomalies with a charged Higgs typically implies large new physics contributions to $pp \to \tau^+ \tau^-$ via the neutral scalar exchanges, so that current LHC data can challenge this option~\cite{Faroughy:2016osc}.     Note that a charged Higgs also presents an important tension between the current measurement of $R(D^*)$ and the measured lifetime of the $B_c$ meson~\cite{Li:2016vvp,Alonso:2016oyd,Celis:2016azn,Akeroyd:2017mhr}.

\begin{figure}[t]
\centering
\includegraphics[width=0.39 \textwidth]{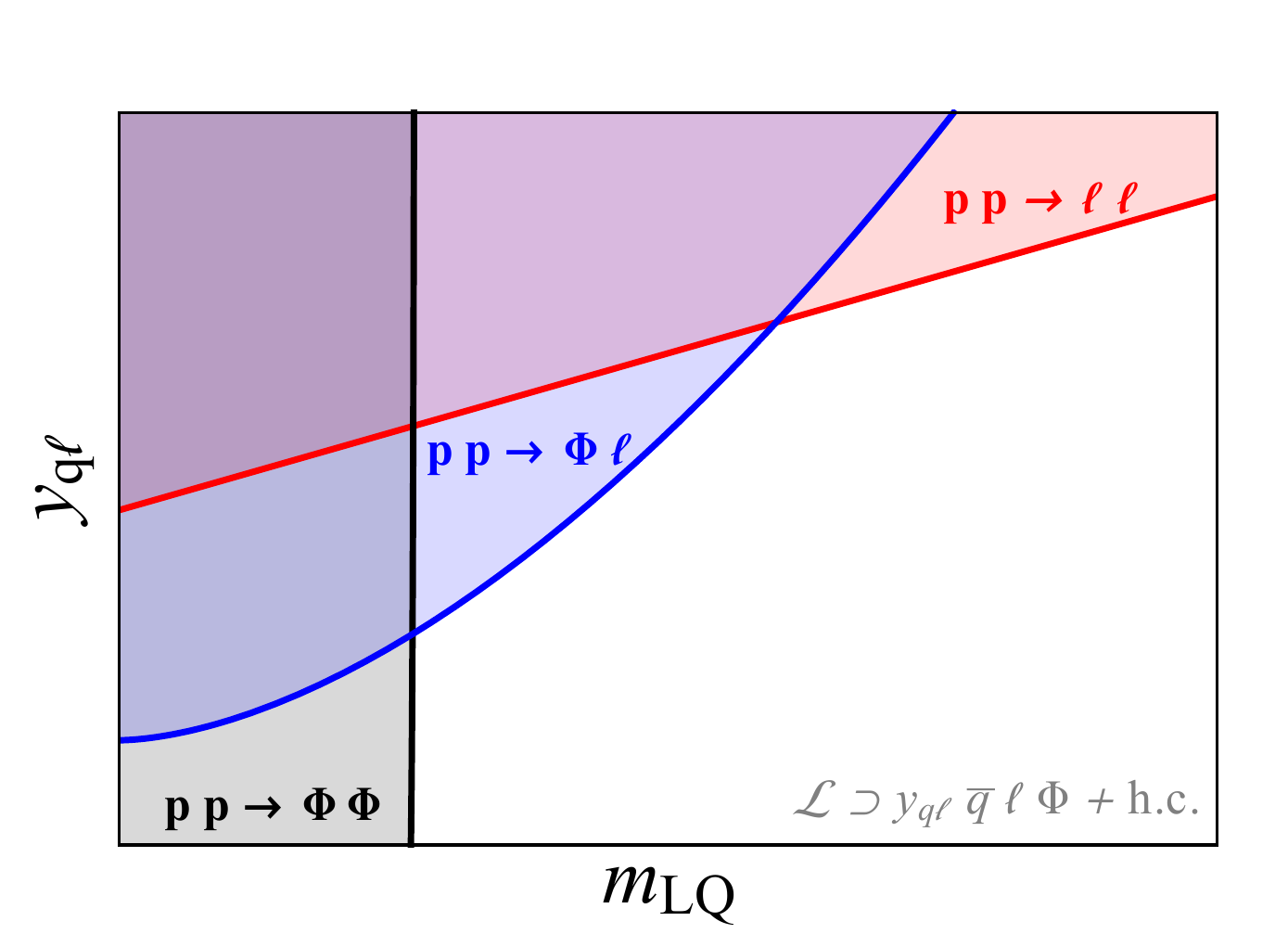}
\caption{Schematic of the LHC bounds on  LQ showing complementarity  in constraining the  $(m_{LQ},y_{q_{\ell}})$ parameters. The three cases are:
 pair production $\sigma \propto y_{q_{\ell}}^0 $, single production $\sigma \propto y_{q_{\ell}}^2 $ and Drell-Yan $\sigma \propto y_{q_{\ell}}^4$ (from~\cite{Dorsner:2018ynv}).}
\label{fig:LQstrategy}
\end{figure}

\begin{figure}[t]
\centering
\includegraphics[width=0.46 \textwidth]{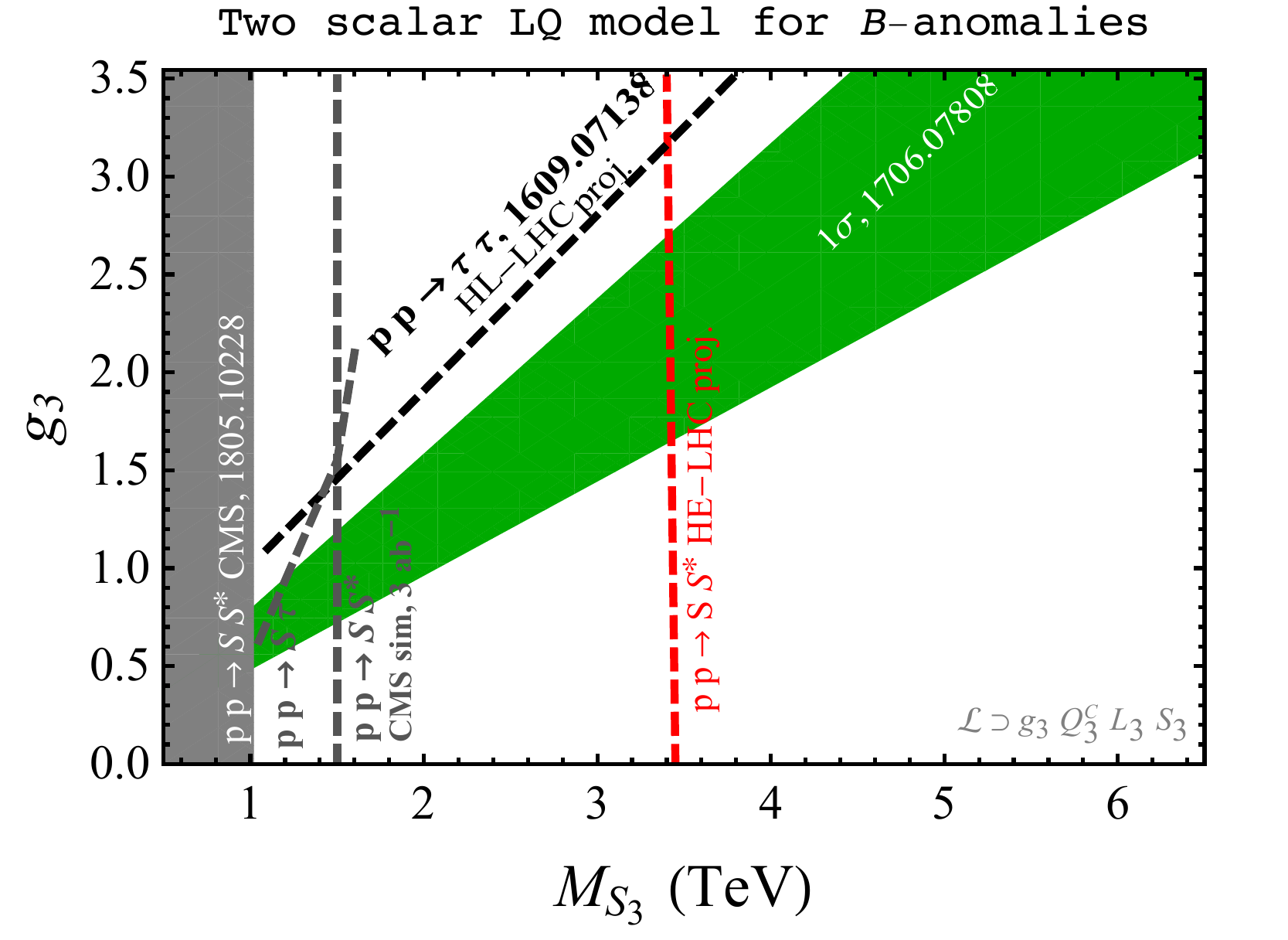}~~~
\includegraphics[width=0.46 \textwidth]{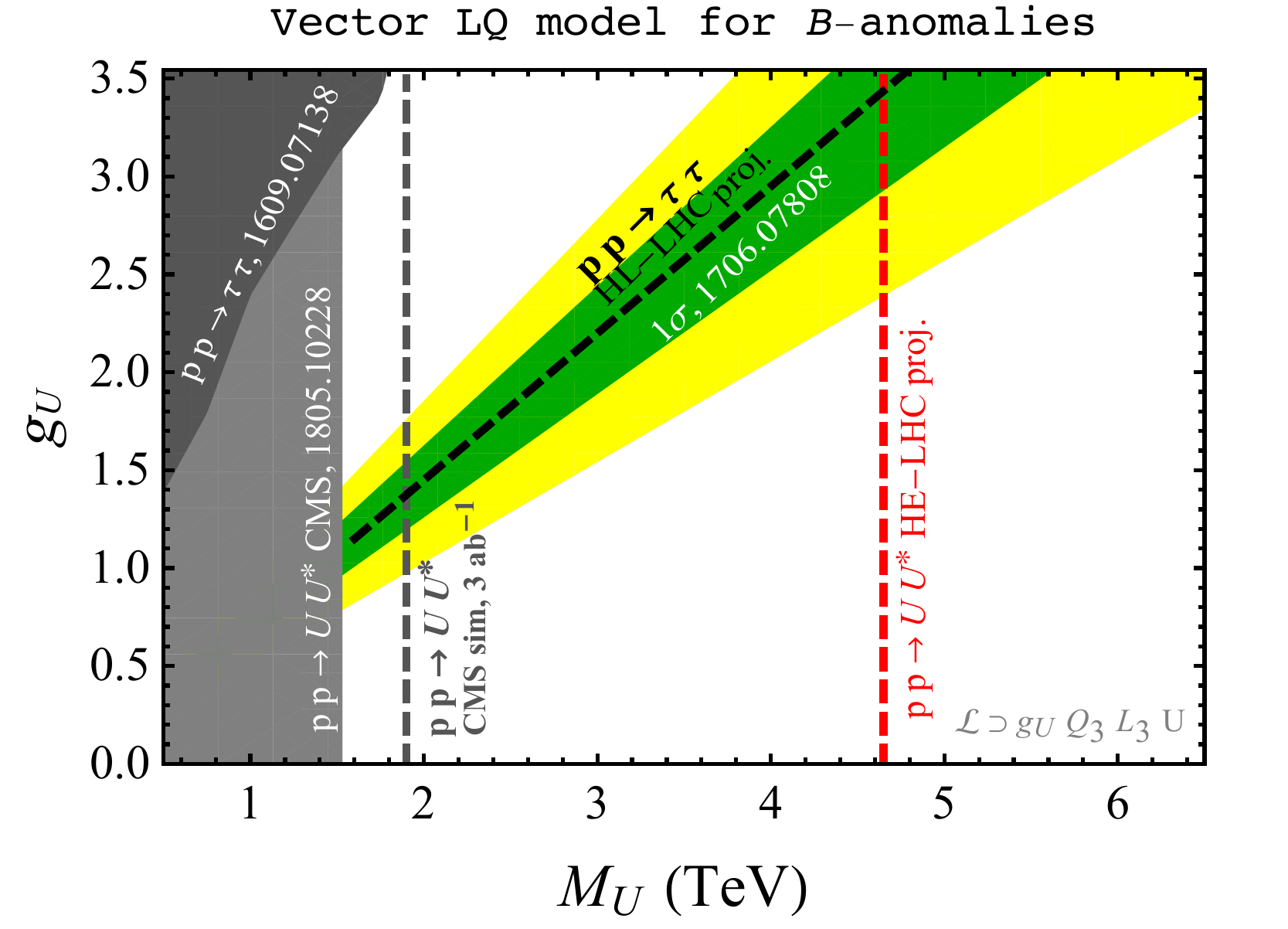}
\caption{ \small \Blu{Present constraints and HL(HE)-LHC projections in the leptoquark mass versus coupling  plane for the scalar leptoquark $S_3$ (left), and vector leptoquark $U_1$ (right). The grey and dark grey solid regions are the current exclusions. The grey and black dashed lines are the projected reach for HL-LHC (pair and single leptoquark production prospects are based on the CMS simulation from Section \ref{sec:exprospects}). The red dashed lines are the projected reach at HE-LHC (see Section \ref{sec:HE_LHC_LQ}). The green and yellow bands are the $1\sigma$ and $2\sigma$ preferred regions from the fit to $B$ physics anomalies. The second coupling required to fit the anomaly does not enter in the leading high-\pt diagrams but it is relevant for fixing the preferred region shown in green, for more details see Ref.~\cite{Buttazzo:2017ixm}. 
}
}
\label{fig:LQmediators}
\end{figure}

\vspace{0.2cm}

\noindent  $\bullet$ \, \textit{Leptoquarks:} 

The observed anomalies in charged and neutral currents appear in semileptonic decays of the $B$-mesons. This implies that the putative NP has to couple to both quarks and leptons at the fundamental level. A natural BSM option is to consider mediators that couple simultaneously quarks and leptons at the tree level. Such states are commonly referred as leptoquarks. Decay and production mechanisms of the LQ are directly linked to the physics required to explain the anomalous data.
\begin{itemize}
\item {\bf Leptoquark decays:} the fit to the $R(D^{*})$ observables suggest a rather light leptoquark (at the TeV scale) that couples predominately to the third generation fermions of the SM. A series of constraints from flavour physics, in particular the absence of BSM effects in kaon and charm mixing observables, reinforces this picture. 

\item {\bf Leptoquark production mechanism:}  The size of the couplings required to explain the anomaly is typically very large, roughly $y_{q\ell} \approx m_{LQ}$/ (1 TeV). Depending of the actual sizes of the leptoquark couplings and its mass we can distinguish three regimes that are relevant for the phenomenology at the LHC:
 	\begin{enumerate}
	\item LQ pair production due to strong interactions,
	\item Single LQ production plus lepton via a single insertion of the LQ coupling, and
	\item Non-resonant production of di-lepton through $t$-channel exchange of the leptoquark.
	\end{enumerate}
Interestingly all three regimes provide complementary bounds in the  $(m_{LQ},y_{q_{\ell}})$ plane, see Fig.~\ref{fig:LQstrategy}.

\end{itemize}

Several simplified models with leptoquark as a mediator were shown to be consistent with the low-energy data.
 A vector leptoquark with $SU(3)_c\times SU(2)_L\times U(1)_Y$ SM quantum numbers $U_\mu \sim ({\bf 3}, {\bf 1}, 2/3)$ was identified as the only single mediator model which can simultaneously fit the two anomalies (see e.g.~\cite{Buttazzo:2017ixm} for a recent fit including leading RGE effects). 
 In order to substantially cover the relevant parameter space, one needs future HL- (HE-) LHC, see Fig.~\ref{fig:LQmediators} (right) (see also Fig.~5 of~\cite{Buttazzo:2017ixm} for details on the present LHC constraints). A similar statement applies to an alternative model featuring two scalar leptoquarks, $S_1, S_3$ \cite{Crivellin:2017zlb}. The pair of plots in Fig. \ref{fig:LQmediators} summaries the current exclusion and the discovery reach for the HE and HL-LHC in the LQ coupling versus mass plane. 
 
 \Blu{Leptoquarks states are emerging as the most convincing mediators for the explanations of the flavour anomalies. It is then important to explore all the possible signatures at the HL- and HE-LHC.
 The experimental programme should focus not only on the final states containing quarks and leptons of the third generation, but also on the whole list of decay channels including the off-diagonal ones ($b \mu$, $s \tau, \dots$). The completeness of this approach would allow to shed light on the flavour structure of the putative New Physics. 
 
 Another aspect to be emphasized regarding leptoquark models is that the UV complete models often require extra fields. The accompanying particles would leave more important signatures at high $p_T$ than the leptoquarks, which is particularly true for vector leptoquark extensions (see for example \cite{DiLuzio:2017vat,DiLuzio:2018zxy})}

\subsubsection{Constraints on simplified models for $b\to s ll$}

The $b\to s ll$ transition is  both loop and CKM suppressed in the SM. The explanations of the $b\to s ll$ anomalies can thus have both tree level and loop level mediators.
Loop-level explanations typically involve lighter particles.
Tree-level mediators can also be light, if sufficiently weakly coupled. However, they can also be much heavier---possibly beyond the reach of the LHC.

\begin{figure}[t]
\centering
\includegraphics[width=0.45 \textwidth]{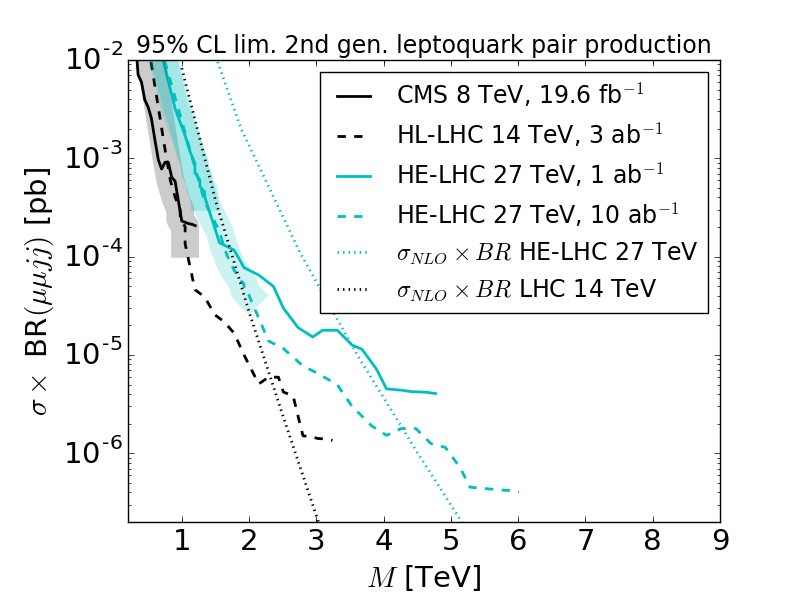}
\caption{Current and projected 95\% CL limits on $\mu\mu jj$ final state at CMS (solid black) and HL-LHC (cyan) with 1 (10) ab$^{-1}$ in solid (dashed) lines. The pair production cross-section for scalar leptoquarks is shown in dotted lines for 14 (27) TeV in black (cyan) (Fig. taken from ~\cite{Allanach:2017bta}). 
}
\label{fig:LQexclusion}
\end{figure}

\noindent $\bullet$ \textit{Tree-level mediators:}\\
For $b \to sll$ anomaly there are two possible tree-level UV-completions, the $Z'$ vector boson and leptoquarks, either scalar or vector (see Fig.~\ref{fig:bsll_diagrams} in Sec.~\ref{sec:7:bsll:NP}).
For leptoquarks, Fig.~\ref{fig:LQexclusion} shows the current 95\% CL limits from 8 TeV CMS with 19.6 fb$^{-1}$  in the $\mu\mu jj$ final state (solid black line), as well as the HL-LHC (dashed black line) and 1 (10) ab$^{-1}$ HE-LHC extrapolated limits (solid (dashed) cyan line). 
Dotted lines give the cross-sections times branching ratio at the corresponding collider energy for pair production of scalar leptoquarks, calculated at NLO using the code of Ref.~\cite{Kramer:2004df}. We see that the sensitivity to a leptoquark with only the minimal $b-\mu$ and $s-\mu$ couplings reaches around 2.5 and 4.5 TeV at the HL-LHC and HE-LHC, respectively. This pessimistic estimate is a lower bound that will typically be improved in realistic models with additional flavour couplings. Moreover, the reach can be extended by single production searches~\cite{Allanach:2017bta}, albeit in a more model-dependent way than pair production. The cross section predictions for vector leptoquark are more model-dependent and are not shown in Fig.~\ref{fig:LQexclusion}. For $\mathcal{O}(1)$ couplings the corresponding limits are typically stronger than for scalar leptoquarks. \vtwo{Searches in other decay channels should also be considered, such at $t\bar t+$MET for which HL-LHC projections can be found in Ref.~\cite{Vignaroli:2018lpq}.}

\begin{figure}[t]
\centering
\includegraphics[width=0.39 \textwidth]{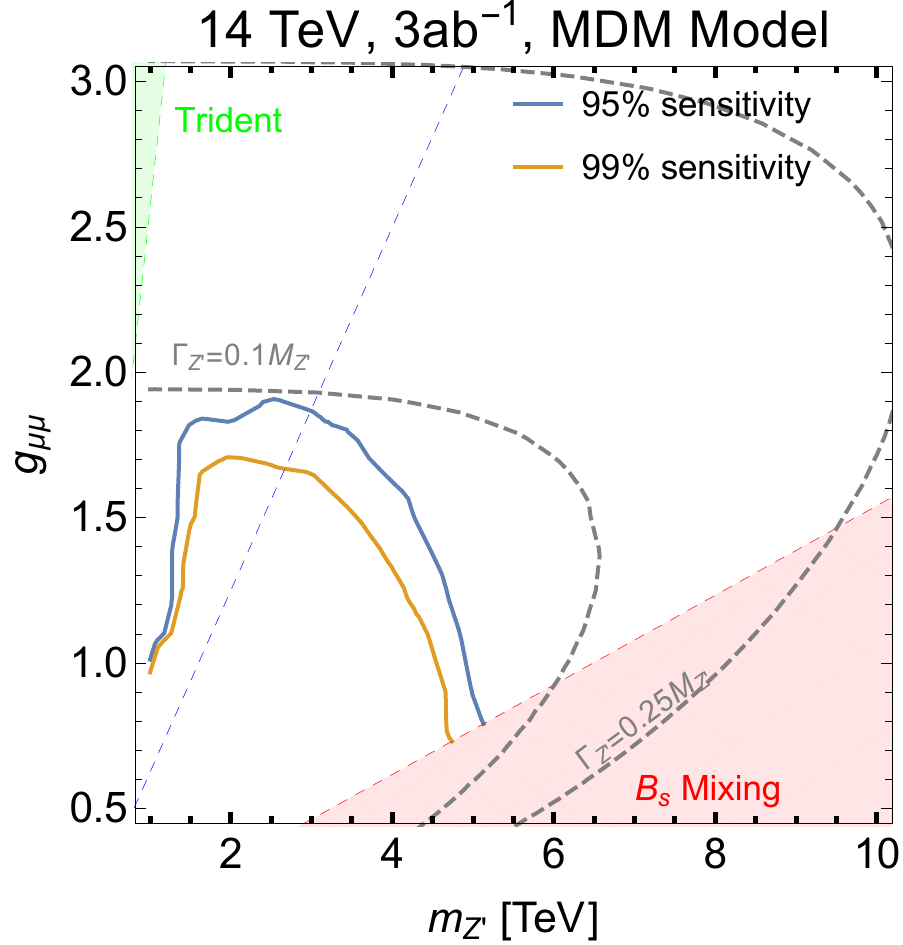}
\caption{ HL-LHC 95\% (blue) and 99\% (orange) CL sensitivity contours to $Z^\prime$ in the ``mixed-down'' model for $g_{\mu\mu}$ vs $Z^\prime$ mass in TeV. The dashed grey contours give the $Z'$ width as a fraction of mass. The green and red regions are excluded by trident neutrino production and $B_s$ mixing, respectively. The dashed blue line is the stronger $B_s$ mixing constraint from Ref.~\cite{Bazavov:2016nty} (Fig. taken from~\cite{Allanach:2018odd}). 
}
\label{fig:Zprime14TeV}
\end{figure}

For the $Z^\prime$ mediator the minimal couplings in the mass eigenstate basis are obtained by unitary transformations from the gauge eigenstate basis, which necessarily induces other couplings.  Ref.~\cite{Allanach:2018odd} defined the ``mixed-up'' model (MUM) and ``mixed-down'' model (MDM) such that the minimal couplings are obtained via CKM rotations in either the up or down sectors respectively. For MUM there is no sensitivity at the HL-LHC.  The predicted sensitivity at the HL-LHC for the MDM is shown in Fig.~\ref{fig:Zprime14TeV} as functions of $Z^\prime$ muon coupling $g_{\mu\mu}$ and the $Z^\prime$ mass, setting the $Z'$ coupling to $b$ and $s$ quarks such that it solves the $b\to s\ell\ell$ anomaly. The solid blue (orange) contours give the 95\% and 99\% CL sensitivity. The red and green regions are excluded by $B_s$ mixing~\cite{Arnan:2016cpy} and neutrino trident production~\cite{Falkowski:2017pss,Falkowski:2018dsl}, respectively. The more stringent $B_s$ mixing constraint from Ref.~\cite{Bazavov:2016nty} is denoted by the dashed blue line; see, however, Ref.~\cite{Kumar:2018kmr} for further discussion regarding the implications of this bound. The dashed grey contours denote the width as a fraction of the mass. We see that the HL-LHC will only be sensitive to $Z^\prime$ with narrow width, up to masses of 5 TeV. 

\begin{figure}[t]
\centering
\includegraphics[width=0.39 \textwidth]{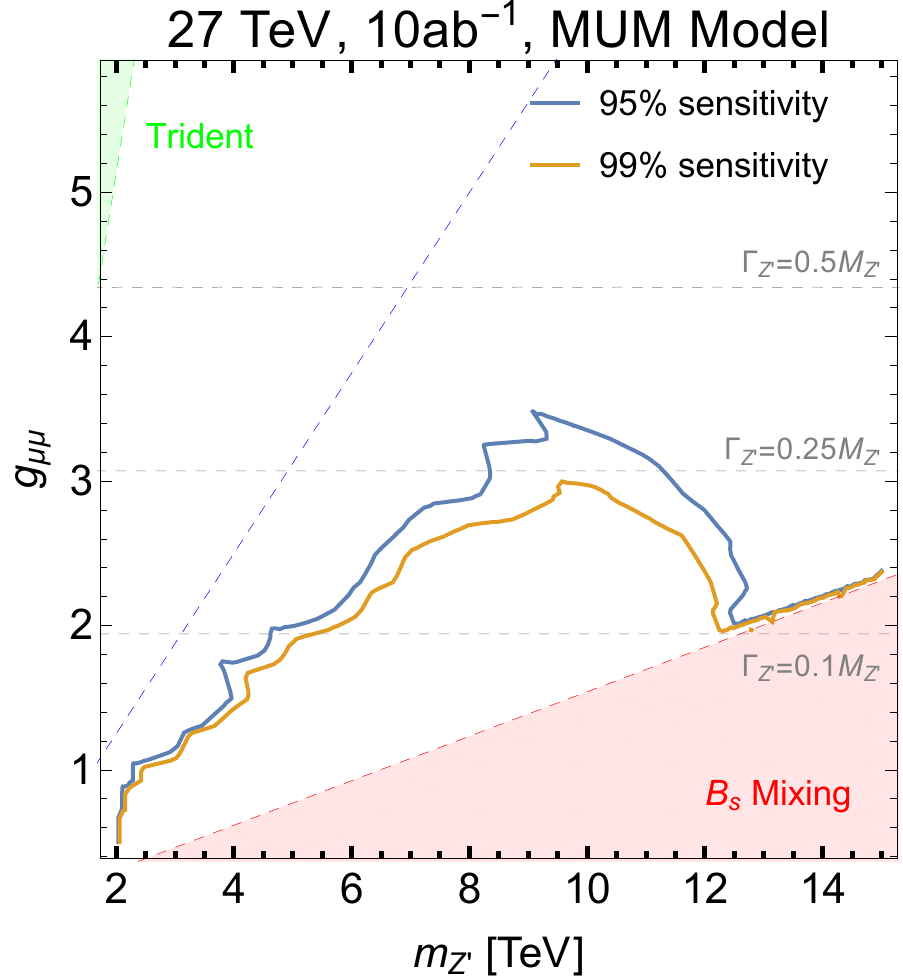}~~~~~~~~~
\includegraphics[width=0.39 \textwidth]{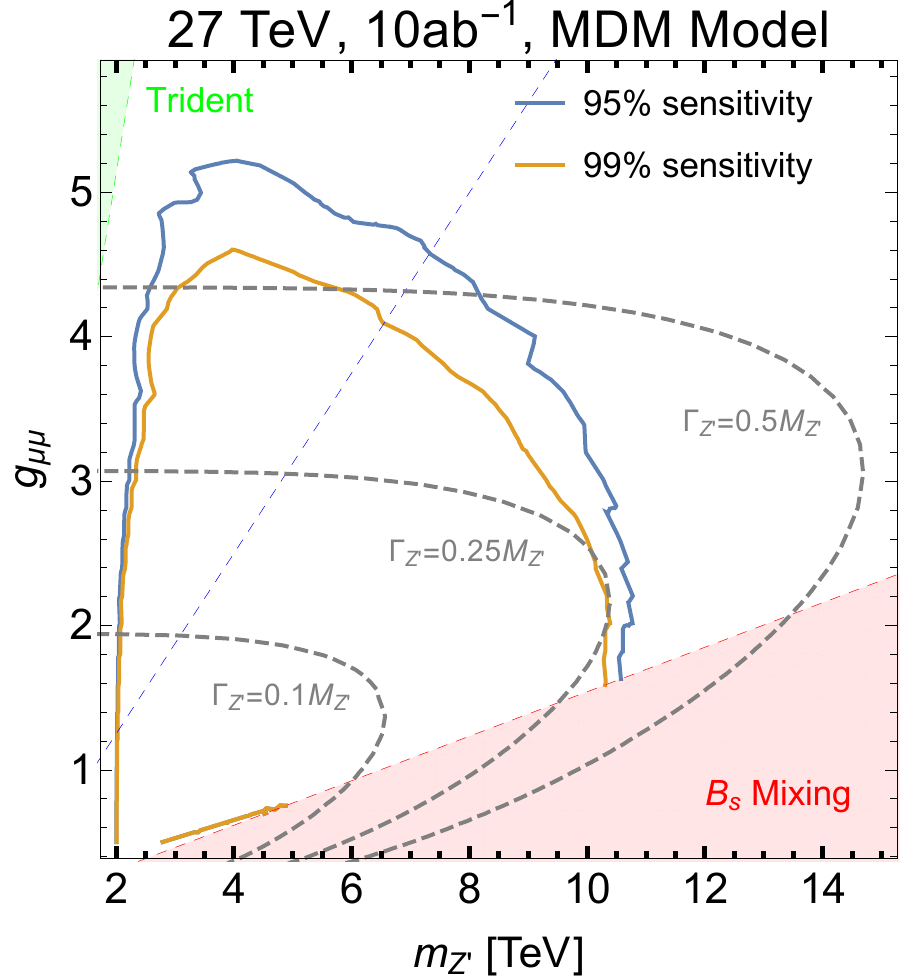}
\caption{ HE-LHC 95\% (blue) and 99\% (orange) CL sensitivity contours to $Z^{\prime}$ in the ``mixed-up'' (left) and ``mixed-down'' (right) model in the parameter space of $g_{\mu\mu}$ vs $Z^{\prime}$ mass in TeV. The dashed grey contours are the width as a fraction of mass. The green and red regions are excluded by trident neutrino production and $B_s$ mixing. The dashed blue line is the stronger $B_s$ mixing constraint from Ref.~\cite{Bazavov:2016nty} (Figs. taken from~\cite{Allanach:2018odd}). 
}
\label{fig:Zprime27TeV}
\end{figure}

At the HE-LHC, the reach for 10 ab$^{-1}$ is shown in Fig.~\ref{fig:Zprime27TeV} for the MUM and MDM on the left and right, respectively. In this case the sensitivity may reach a $Z^{\prime}$ with wider widths up to 0.25 and 0.5 of its mass, while the mass extends out to 10 to 12 TeV. We stress that this is a pessimistic estimate of the projected sensitivity, particular to the two minimal models; more realistic scenarios will typically be easier to discover.

\vspace{0.2cm}

\noindent  $\bullet$ \textit{Explanations at the one-loop level:}\\
It is possible to accomodate the $b \to s \ell^+ \ell^-$ anomalies even if mediators only enter at one-loop.  One possibility is the mediators coupling to right-handed top quarks and to muons~\cite{Celis:2017doq,Becirevic:2017jtw,Kamenik:2017tnu,Fox:2018ldq,Camargo-Molina:2018cwu}.   Given the loop and CKM  suppression of the NP contribution to the $b \to s \ell^+ \ell^-$ amplitude, these models can explain the $b \to s \ell^+ \ell^-$ anomalies for a light mediator, with mass around $\mathcal{O}(1)$~TeV or lighter.       Constraints from the LHC and future projections for the HL-LHC were derived in~\cite{Camargo-Molina:2018cwu} by recasting di-muon resonance, $pp\to t\bar tt \bar t$ and SUSY searches.  Two scenarios were considered: \textit{i)} a scalar LQ $R_2(3, 2, 7/6)$ combined with a vector LQ $\widetilde U_{1 \alpha}(3, 1, 5/3)$, \textit{ii)} a vector boson $Z^{\prime}$ in the singlet representation of the SM gauge group.    Ref.~\cite{Fox:2018ldq} also analyzed the HL-LHC projections for the $Z^{\prime}$.     The constraints from the LHC already rule out part of the relevant parameter space and the HL-LHC will be able to cover much of the remaining regions.    Dedicated searches in the $pp\to t\bar tt \bar t$ channel and a dedicated search for $t \mu$ resonances in $t \bar t \mu^+ \mu^-$ final state can improve the sensitivity to these models~\cite{Camargo-Molina:2018cwu}. 

\subsubsection{Conclusions regarding high \pt probes of flavour anomalies}

The anomalous results in $B$-meson decays cannot be considered yet as a convincing evidence of New Physics. On the other hand, the number (and quality) of observables that are not in complete agreement with the SM prediction is growing with time in a coherent way. If true, the implications for HEP will be profound. 
The conclusions we draw are the following:
\begin{itemize}
\item[$\bullet$] A conservative argument based on perturbative unitarity \cite{DiLuzio:2017chi} sets an upper bound on the New Physics scale to be 9 TeV for charged current anomalies and 80 TeV for neutral current ones. 
The analysis of explicit models show that the high-luminosity programme has a clear potential to probe a large portion of the possible BSM options. 

\item[$\bullet$] The explanation of the anomalies in $b \to c \tau \nu$ transitions requires non trivial model building. In particular, it is not possible to simply isolate the physics that mediate the flavour anomalous transitions. In complete models other signature have to be considered. Typically it would be very difficult to escape direct detection at HL/HE-LHC.

\item[$\bullet$] Even though the naive scale associated with the $b \to s \ell \ell$ anomalies   is much higher than the energy accessible at HL/HE-LHC, in motivated models the flavour suppressions and weak couplings guarantee a large coverage of the parameter space for leptoquark and $Z'$ mediators \cite{Allanach:2018odd}.
\end{itemize}

More generally, the probes of lepton flavour universality such as the ratios of inclusive $\tau^+\tau^-$ vs. $\mu^+\mu^-$ (or $\mu^+\mu^-$ vs. $e^+e^-$) mass distributions, are important, theoretically clean, tests of the SM and are well motivated observables both at HL- and HE-LHC, whether or not the current $B$-meson anomalies become statistically significant.

As a final remark, it is important to remember that the anomalies are not yet experimentally established. Among others, this also means that the statements on whether or not  the high $p_T$ LHC constraints rule out certain $R(D^{(*)})$ explanations assumes that the actual values of $R(D^{(*)})$ are given by their current global averages. If future measurements decrease the global average, the high $p_T$ constraints can in some cases be greatly relaxed so that HL- and/or HE-LHC may be essential even for these, at present tightly constrained, cases.

\input{section10/experimental.tex}

\subsubsection{High $p_T$ searches at HE-LHC relevant for $B$ physics anomalies}
\label{sec:HE_LHC_LQ}
\begin{figure}[t]
	\centering
	\includegraphics[width=0.5\textwidth]{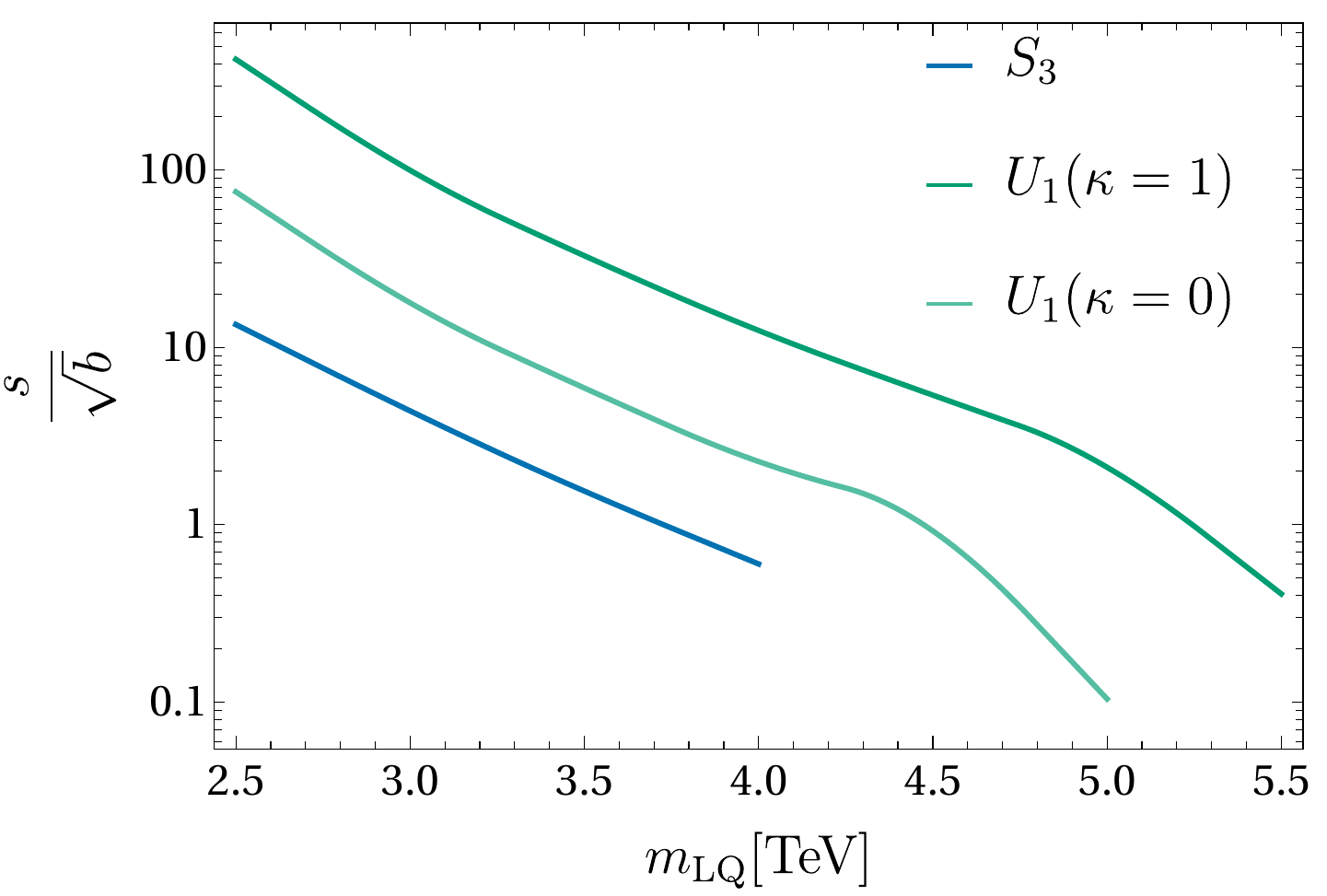}
	\caption{Expected sensitivity for pair production of scalar (\Blu{$S_3$}) and vector ($U_1$) leptoquark at 27 TeV $p p$ collider with an integrated luminosity of $15~\text{ab}^{-1}$.} \label{fig:lukas}
\end{figure}

We show next the sensitivity of the 27 TeV $pp$ collider with $15$~ab$^{-1}$ integrated luminosity to probe pair production of the scalar and vector leptoquarks decaying to $(b \tau)$ final state. The investigated events contain either one electron or muon, one hadronically decaying tau lepton, and  at least two jets. The signal events and the dominant background events ($t\bar{t}$) were generated with {\tt MadGraph5\_aMC@NLO} at leading-order. {\tt Pythia6} was used to shower and hadronise events and {\tt DELPHES3} was used to simulate the detector response. The scalar leptoquark ($S_3$) and the vector leptoquark ($U_1$) UFO model files were taken from~\cite{Dorsner:2018ynv}.

To verify the procedure,  the $t\bar{t}$ background and the scalar leptoquark signal were simulated at 13 TeV and compared to the predicted shapes in the $S_T$ distribution from the existing CMS analysis~\cite{Sirunyan:2017yrk}. After verifying the 13 TeV analysis,  the signal and the dominant background events at $27$ TeV were simulated. From these samples, all events satisfying the particle content requirements and applied the lower cut in the $S_T$ variable were selected. The cut threshold was chosen to maximize $S/\sqrt{B}$ while requiring at least 2 expected signal events at an integrated luminosity of $15~\text{ab}^{-1}$. In the case of the vector leptoquark two options were considered, the Yang-Mills ($\kappa = 1$), and the minimal coupling ($\kappa = 0$) scenario for the couplings to the gluon field strength, ${\cal L}\supset -ig \kappa U_{1\mu}^\dagger T^a U_{1\nu} G^a_{\mu\nu}$~\cite{Dorsner:2018ynv}. From the simulations of the scalar and vector leptoquark events, the ratio of the cross-sections was obtained, and assuming similar kinematics,  also the sensitivity for the vector leptoquark $U_1$.

As shown in Fig.~\ref{fig:lukas}, the HE-LHC collider will be able to probe pair produced third generation scalar leptoquark (decaying exclusively to $b \tau$ final state) up to mass of $\sim 4$~TeV and vector leptoquark up to $\sim 4.5$~TeV and $\sim 5.2$~TeV for the minimal coupling and Yang-Mills scenarios, respectively. While this result is obtained by a rather crude analysis, it shows the impressive reach of the future high-energy $pp$-collider. In particular, the HE-LHC will cut deep into the relevant perturbative parameter space for $b \to c \tau \nu$ anomaly. As a final comment, this is a rather conservative estimate of the sensitivity to leptoquark models solving $R(D^{*})$ anomaly, since a dedicated single leptoquark production search is expected to yield even stronger bounds~\cite{Buttazzo:2017ixm,Dorsner:2018ynv}.

%% file: section10/experimental.tex
\subsubsection{Experimental prospects for high $p_T$ searches at HL-LHC relevant for $B$ anomalies}
\label{sec:exprospects}

We give next  the experimental prospects for leptoquark searches, with leptoquarks decaying to the final states relevant for $B$ physics anomalies. 

\subsubsubsection{Prospects for leptoquark searches assuming $t$+$\tau$ and/or $t$+$\mu$ decays}
\label{sec:LQ:t+tau}
The reach of searches for pair production of leptoquarks (LQs) with decays to $t+\mu$ and $t+\tau$ is studied for the HL-LHC with target integrated luminosities of 
$\mathcal{L}_{\text{int}}^{\text{target}} = 300$ and $3000\,\mathrm{fb}^{-1}$~\cite{CMS-PAS-FTR-18-008}. The studies are based on published CMS results of the $t$+$\mu$~\cite{Sirunyan:2018ruf} and $t$+$\tau$~\cite{Sirunyan:2018nkj} LQ decay channels which use data of proton-proton collisions at $\sqrt{s}=13\,\mathrm{TeV}$ corresponding to $\mathcal{L}_{\text{int}} = 35.9\,\mathrm{fb}^{-1}$ recorded in 2016. While the analysis strategies are kept unchanged with respect to the ones in Refs.~\cite{Sirunyan:2018nkj,Sirunyan:2018ruf}, different total integrated luminosities, the higher center-of-mass energy of 14\,TeV, and different scenarios of systematic uncertainties are considered. In the first scenario (denoted ''w/ YR18 syst. uncert.''), the relative experimental systematic uncertainties are scaled by a factor of $1 / \sqrt{f}$, with 
$f=\mathcal{L}_{\text{int}}^{\text{target}}/35.9\,\mathrm{fb}^{-1}$, 
until they reach a defined lower limit based on estimates of the achievable accuracy with the upgraded detector~\cite{Collaboration:2650976}. The relative theoretical systematic uncertainties are halved. In the second scenario (denoted ''w/ stat. uncert. only''), no systematic uncertainties are considered. The relative statistical uncertainties in both scenarios are scaled by $1 / \sqrt{f}$.

\begin{figure}[t]
\centering
\includegraphics[width=0.49\textwidth]{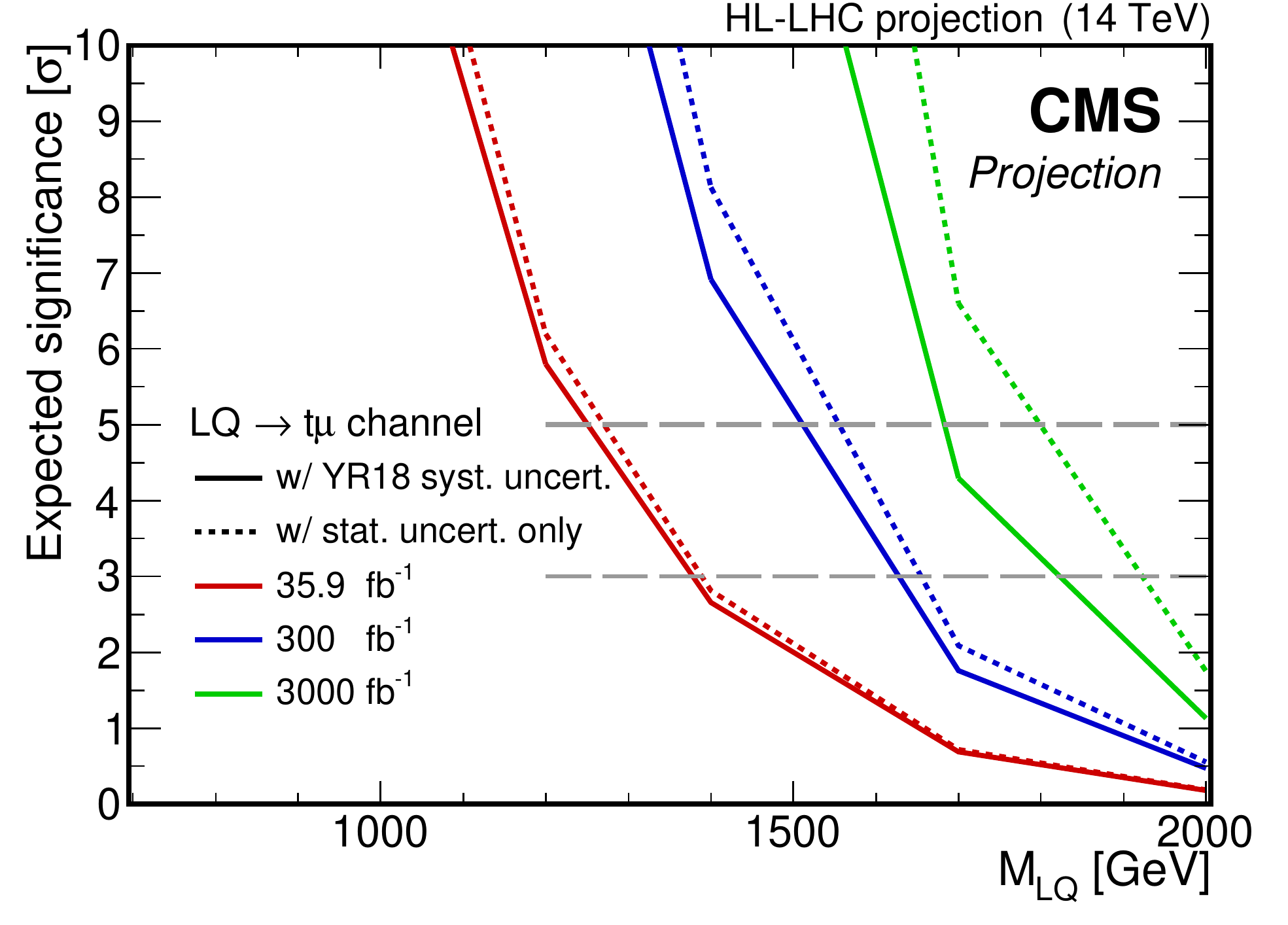}
\includegraphics[width=0.49\textwidth]{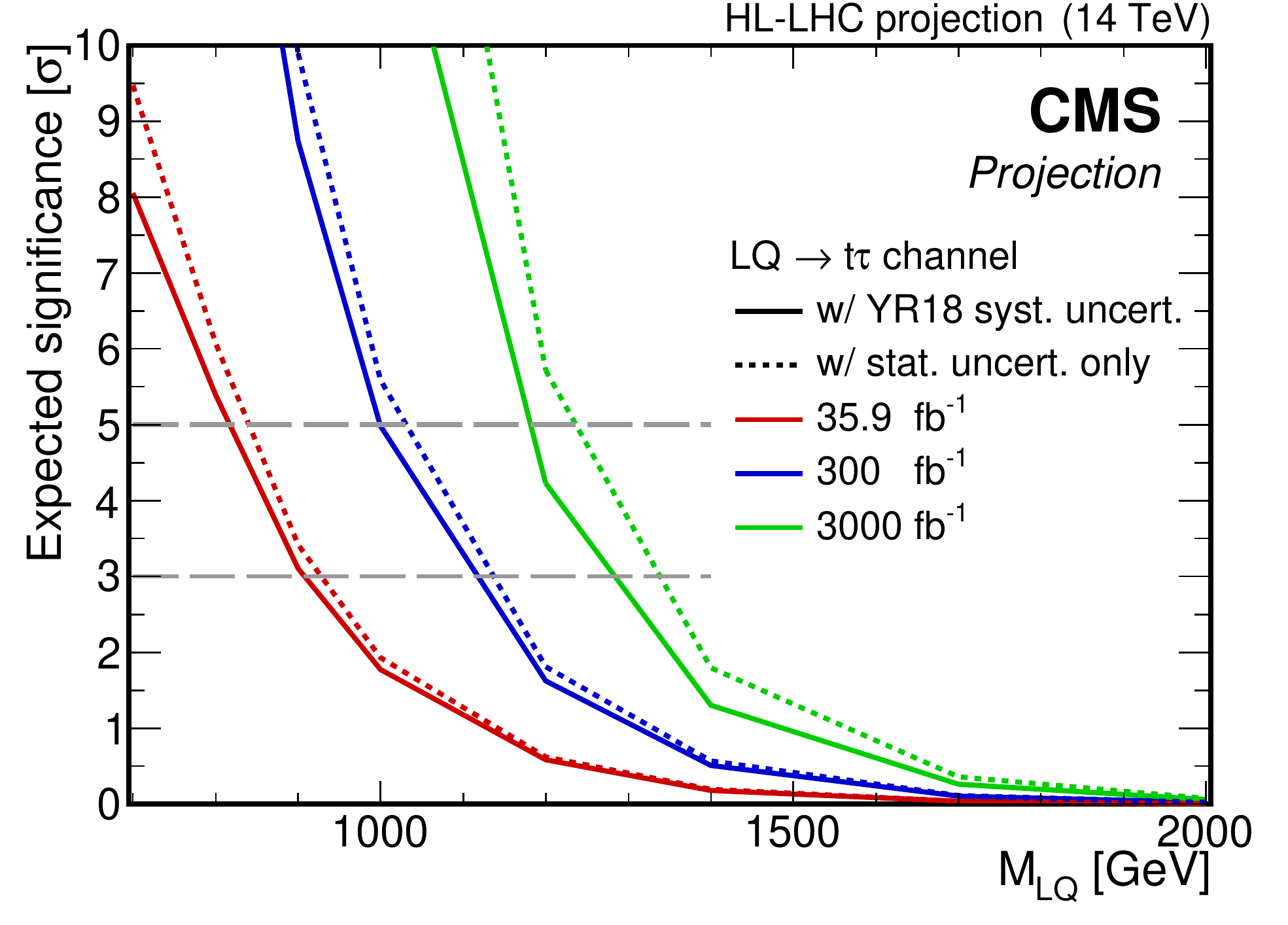}
\caption{Expected significances for an LQ decaying exclusively to top quarks and muons (left) or top quarks and $\tau$ leptons (right).}
\label{fig:LQpairs:significances}
\end{figure}

Figure \ref{fig:LQpairs:significances} presents the expected signal significances of the analyses as a function of the LQ mass for different assumed integrated luminosities in the ''w/ YR18 syst. uncert.'' and ''w/ stat. uncert. only'' scenarios. Increasing the target integrated luminosity to $\mathcal{L}_{\text{int}}^{\text{target}} = 3000\,\mathrm{fb}^{-1}$ greatly increases the discovery potential of both analyses. The LQ mass corresponding to a discovery at $5\sigma$ significance with a dataset corresponding to $3000\,\mathrm{fb}^{-1}$ increases by more than 500\,GeV compared to the situation at $\mathcal{L}_{\text{int}}^{\text{target}} = 35.9\,\mathrm{fb}^{-1}$, from about 1200\,GeV to roughly 1700\,GeV, in the $\text{LQ}\rightarrow t\mu$ decay channel. For LQs decaying exclusively to top quarks and $\tau$ leptons, a gain of 400\,GeV is expected, pushing the LQ mass in reach for a $5\sigma$ discovery from 800\,GeV to 1200\,GeV. 

\begin{figure}[t]
\centering
\includegraphics[width=0.49\textwidth]{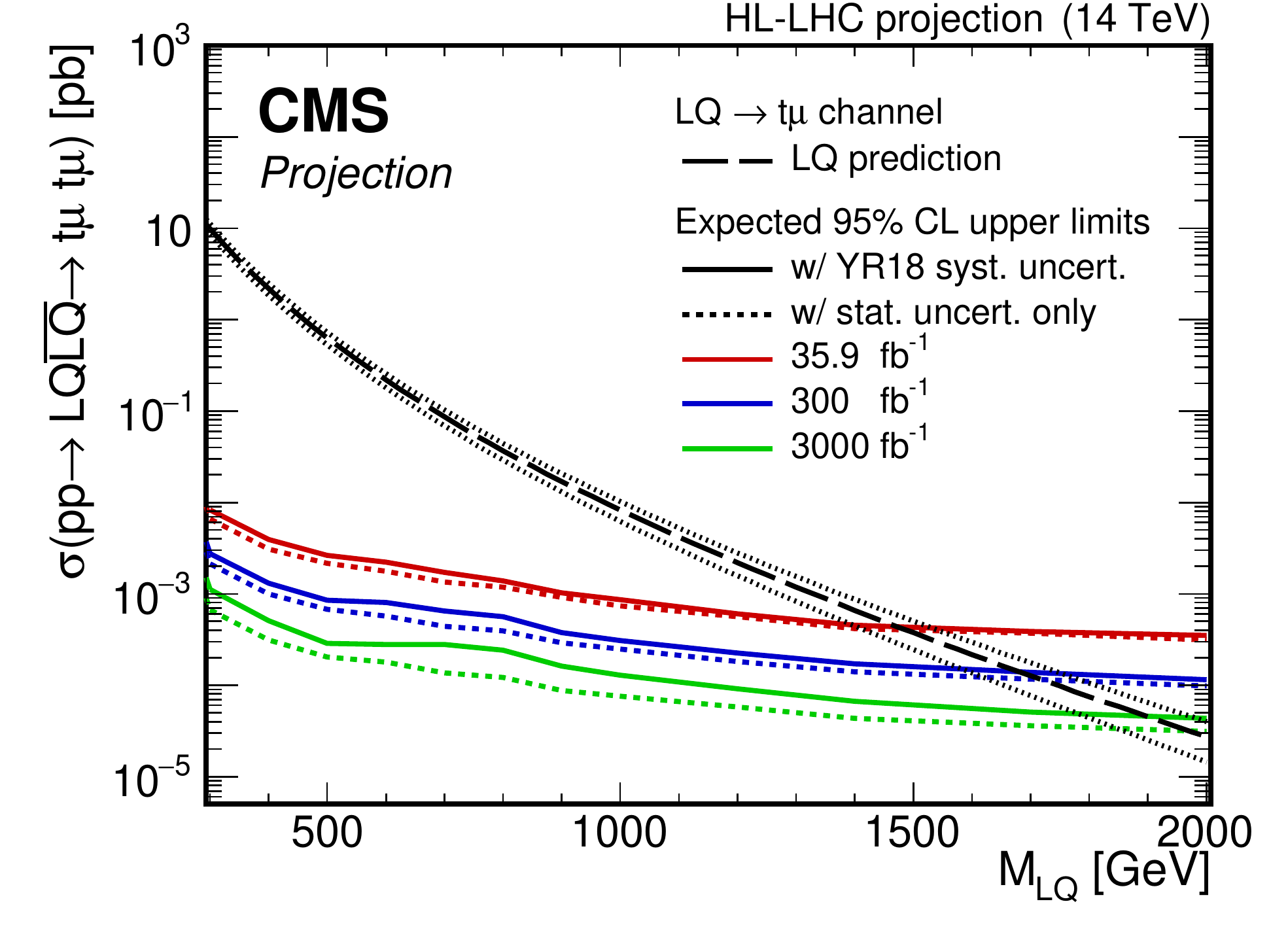}
\includegraphics[width=0.49\textwidth]{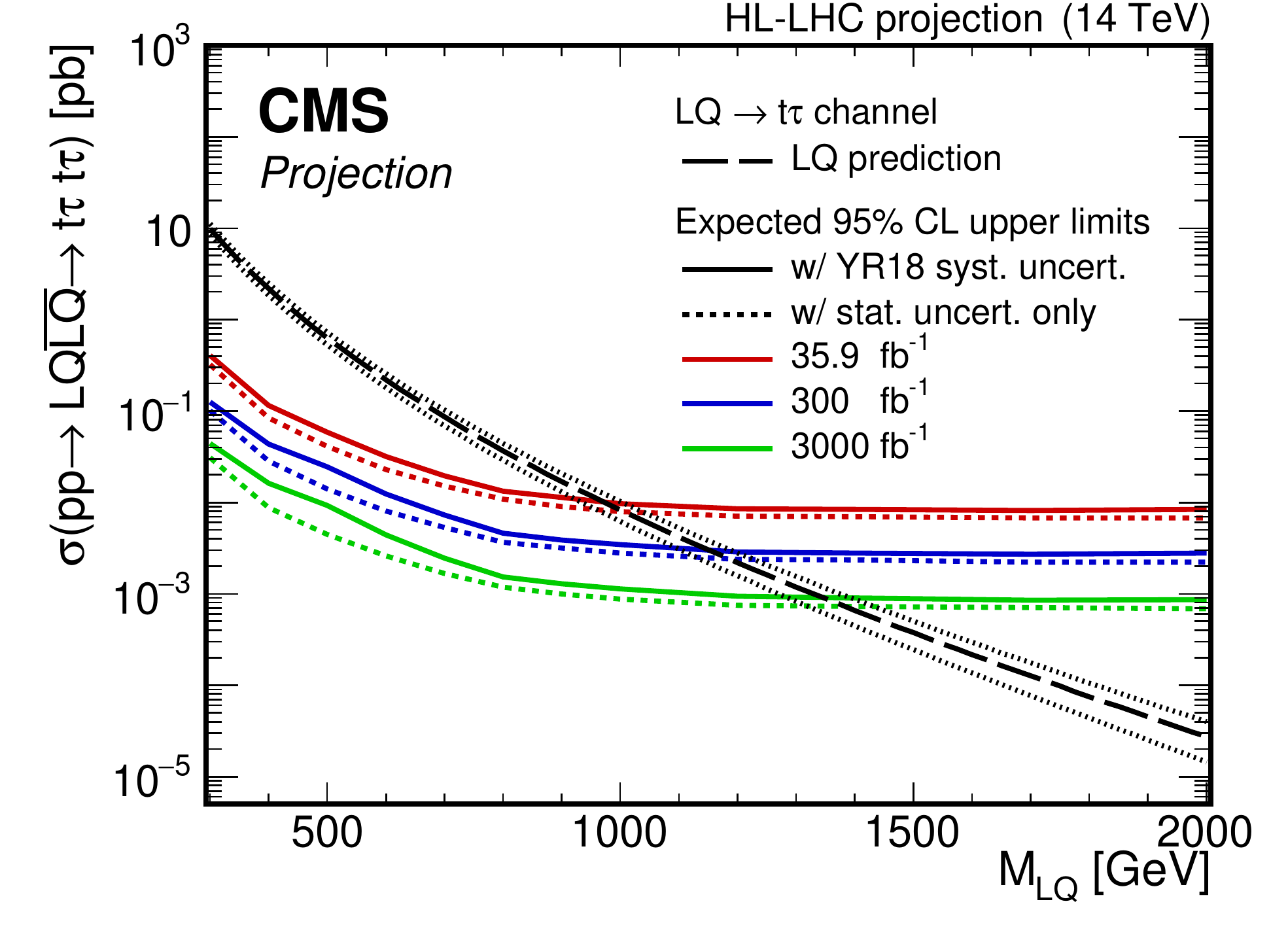}
\caption{Expected upper limits on the LQ pair production cross section at the 95\% CL for an LQ decaying exclusively to top quarks and muons (left) or $\tau$ leptons (right).}
\label{fig:LQpairs:limits1d}
\end{figure}

In Fig.~\ref{fig:LQpairs:limits1d}, the expected projected exclusion limits on the LQ pair production cross section are shown. Leptoquarks decaying only to top quarks and muons are expected to be excluded below masses of 1900\,GeV for $3000\,\mathrm{fb}^{-1}$, which is a gain of 500\,GeV compared to the limit of 1420\,GeV obtained in the published analysis of the 2016 dataset~\cite{Sirunyan:2018ruf}. The mass exclusion limit for LQs decaying exclusively to top quarks and $\tau$ leptons are expected to be increased by 500\,GeV, from 900\,GeV to approximately 1400\,GeV.

\begin{figure}[t]
\centering
\includegraphics[width = 0.49\textwidth]{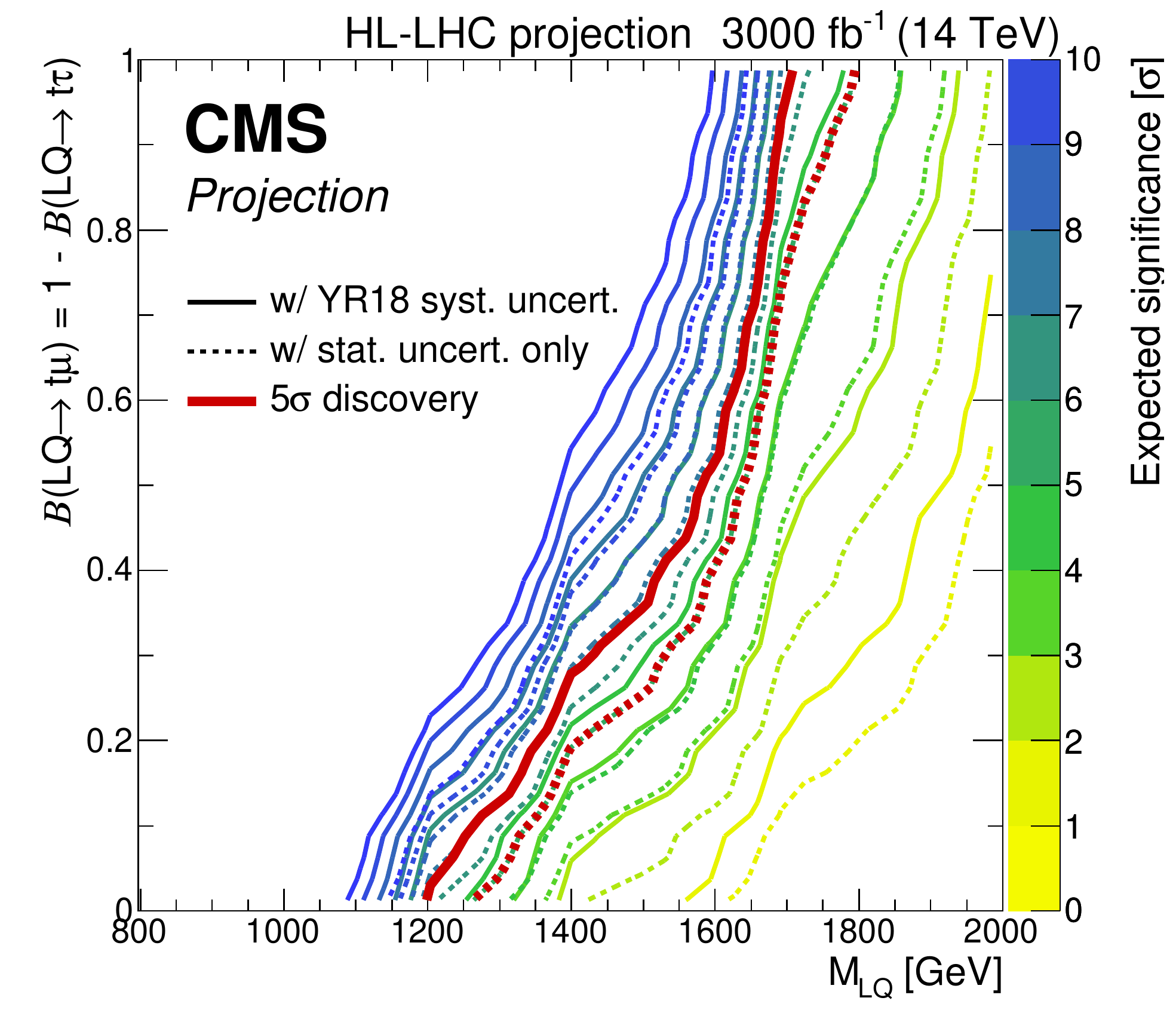}
\includegraphics[width = 0.49\textwidth]{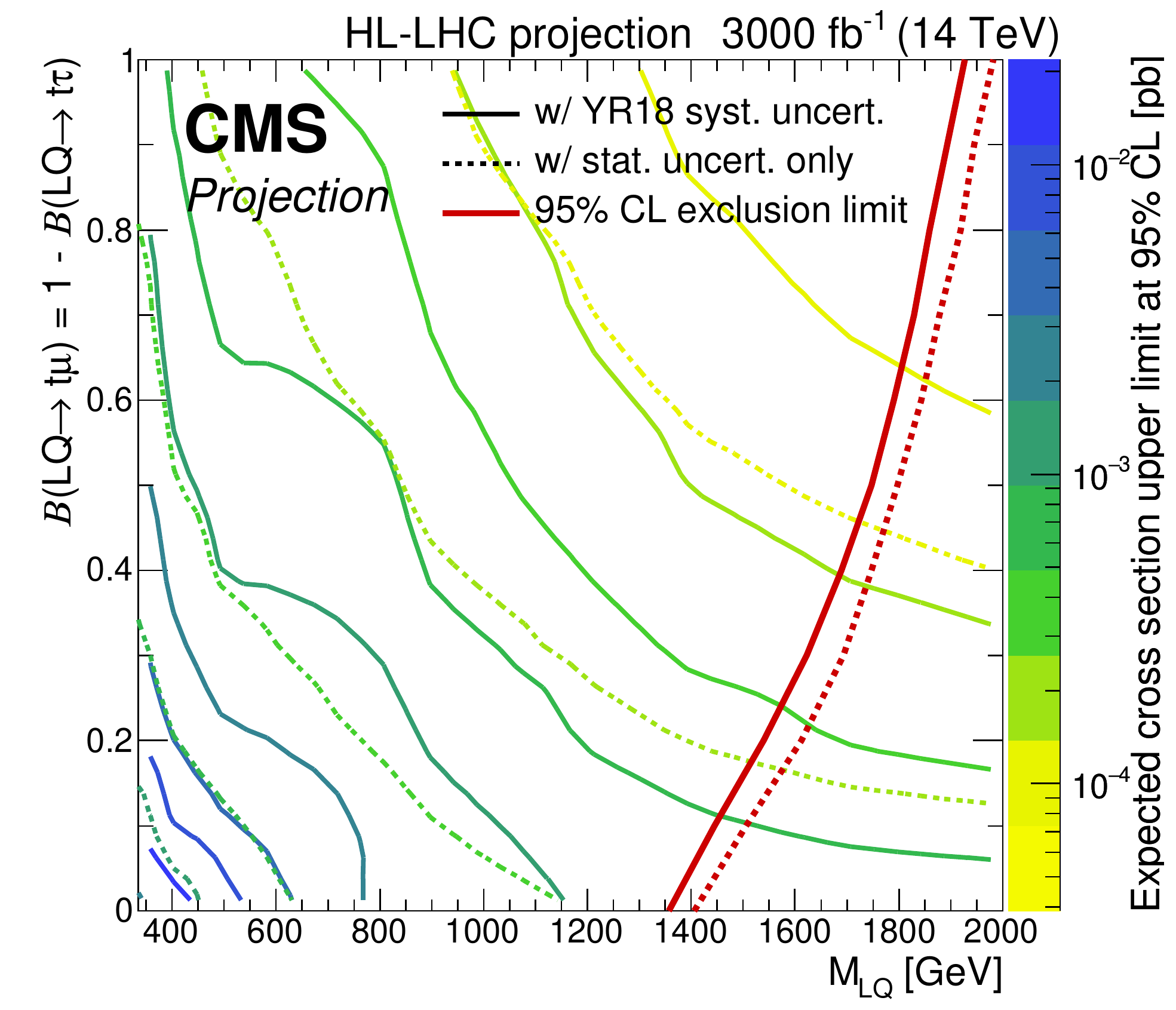}
\caption{Expected significances (left) and expected upper limits on the LQ pair-production cross section at the 95\% CL (right) as a function of the LQ mass and the branching fraction. Color-coded lines represent lines of a constant expected significance or cross section limit, respectively. The red lines indicate the 5$\sigma$ discovery level (left) and the mass exclusion limit (right).}
\label{fig:LQpairs:limits2d}
\end{figure}

Figure~\ref{fig:LQpairs:limits2d} shows the expected signal significances and upper exclusion limits on the pair production cross section of scalar LQs allowed to decay to top quarks and muons or $\tau$ leptons at the 95\% CL as a function of the LQ mass and a variable branching fraction $\mathcal{B}(\mathrm{LQ}\rightarrow t\mu) = 1 - \mathcal{B}(\mathrm{LQ}\rightarrow t\tau)$ for an integrated luminosity of $3000\,\mathrm{fb}^{-1}$ in the two different scenarios. For all values of $\mathcal{B}$, LQ masses up to approximately 1200\,GeV and 1400\,GeV are expected to be in reach for a discovery at the 5$\sigma$ level and a 95\% CL exclusion, respectively.

\subsubsubsection{CMS Search for leptoquarks decaying to $\tau$ and $b$}\label{sec:leptoqbtau}

\begin{figure}[t]
\centering
\includegraphics[width=0.25\textwidth]{\main/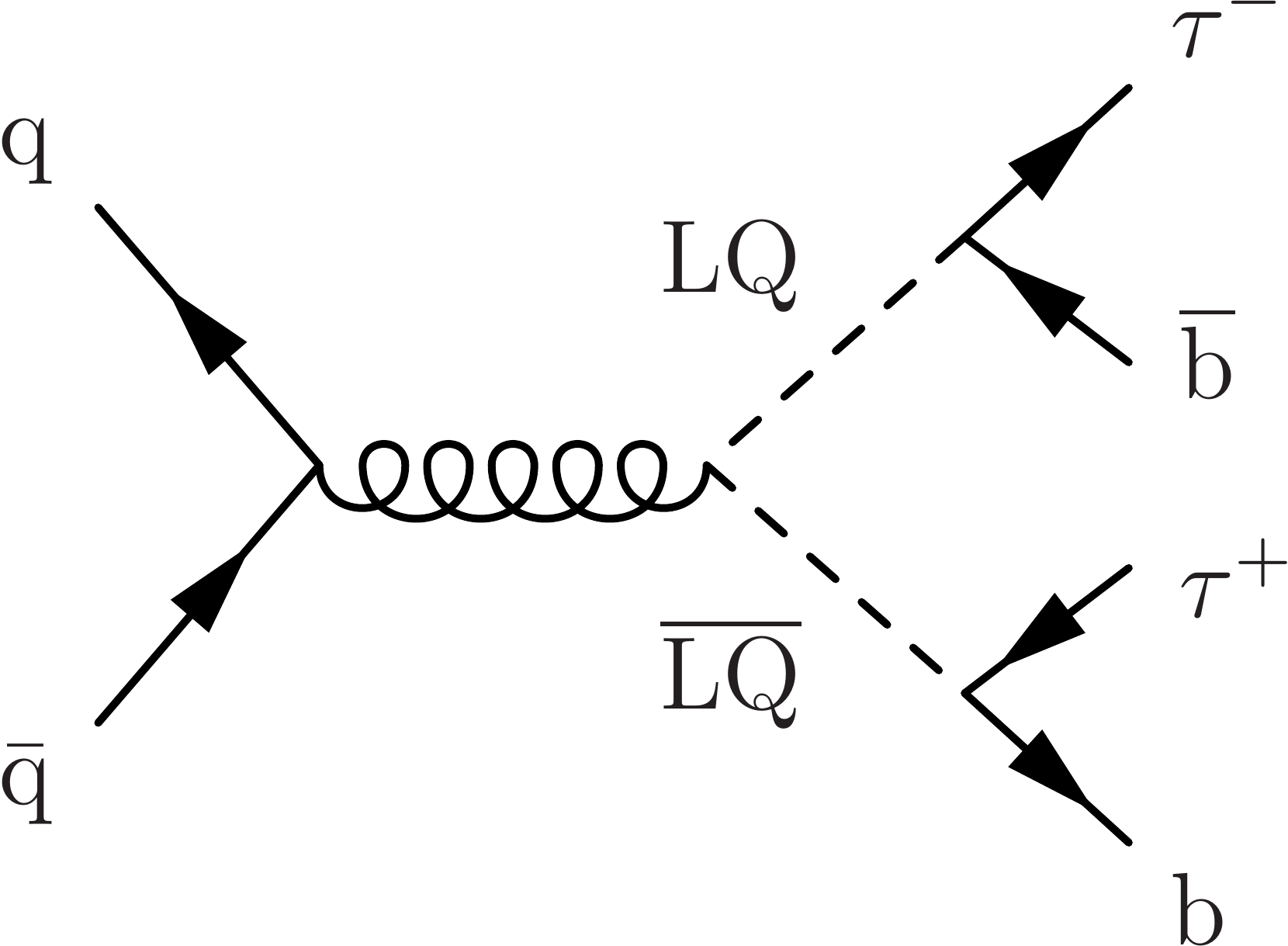}
\hspace{0.2cm}
\includegraphics[width=0.25\textwidth]{\main/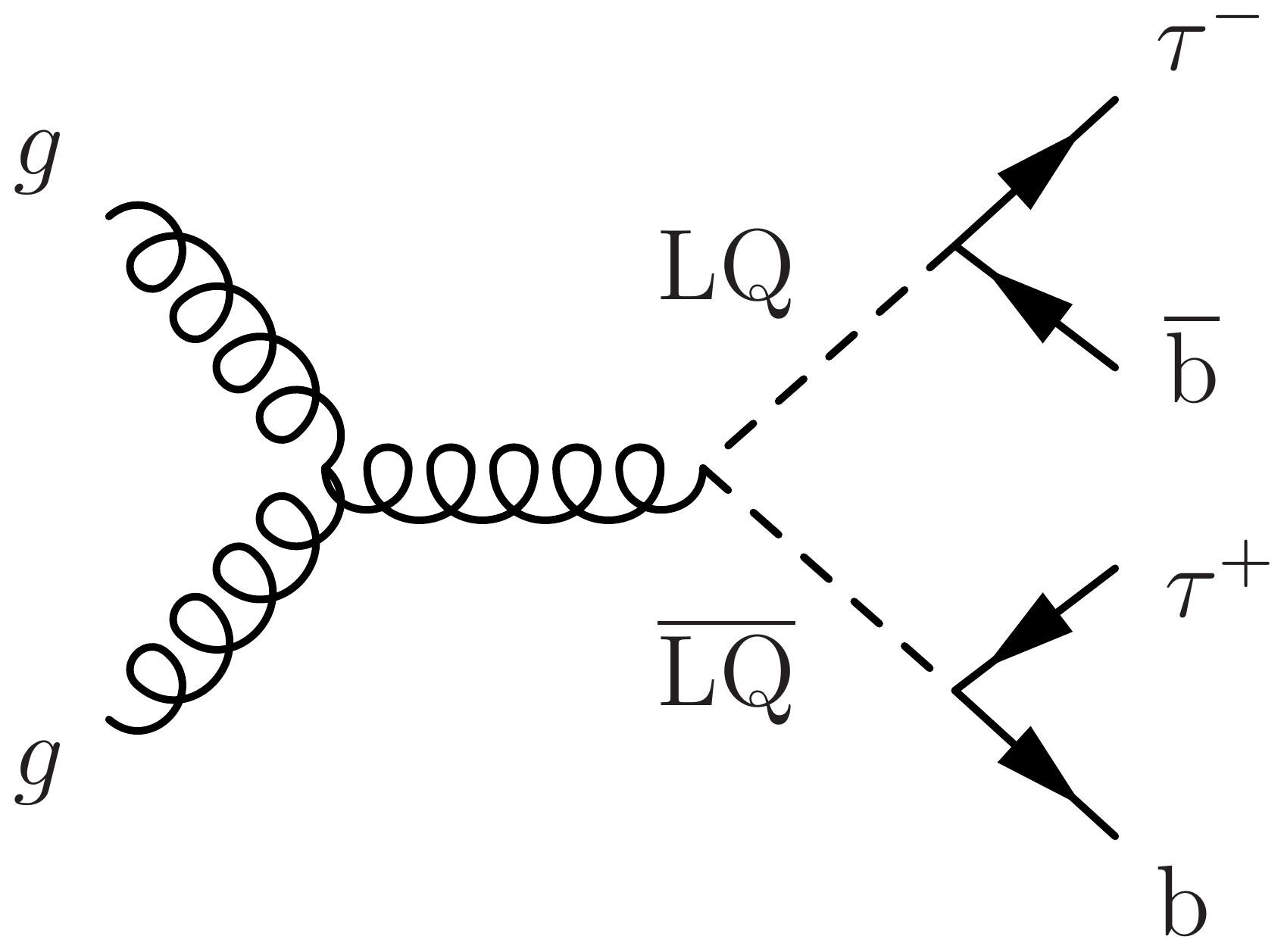}
\hspace{0.2cm}
\includegraphics[width=0.25\textwidth]{\main/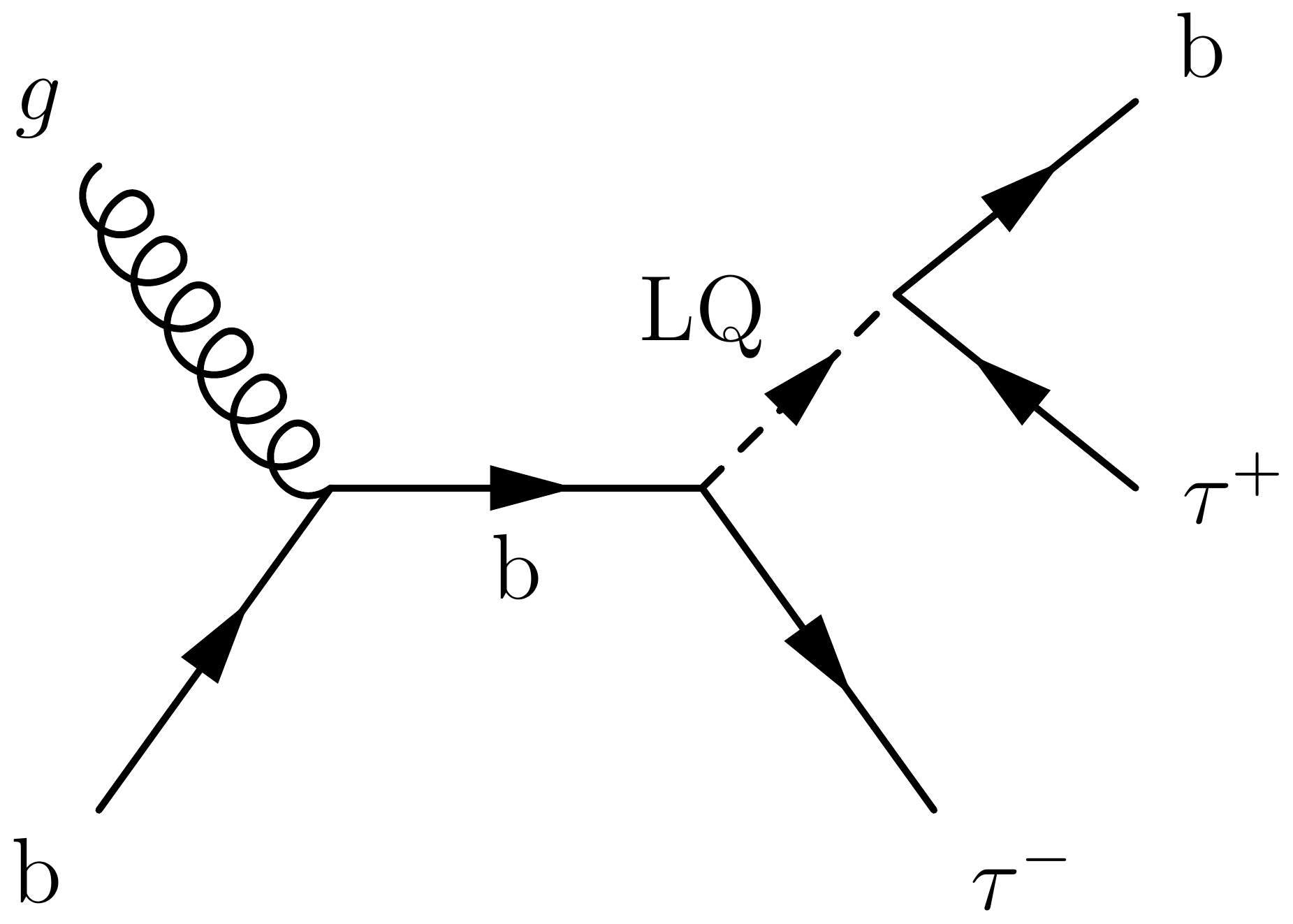}
\caption{Leading order Feynman diagrams for the production of a third-generation LQ in the single production s-channel (left) and the pair production channel via gluon fusion (center) and quark fusion (right).}
\label{LQ3-diagram}
\end{figure}

This analysis from CMS~\cite{CMS-PAS-FTR-18-028} presents future discovery and exclusion prospects for singly and pair produced 
third-generation scalar LQs, each decaying to $\tauh$ and a bottom quark. Here, $\tauh$ denotes a hadronically decaying 
$\tau$ lepton. 
The relevant Feynman diagrams of the signal processes at leading order (LO) are shown in Fig.~\ref{LQ3-diagram}.

\vspace{-5pt}

The analysis uses DELPHES~\cite{deFavereau:2013fsa} event samples of simulated pp collisions at a center-of-mass energy of 14 TeV, corresponding to integrated luminosities of 300 and 3000\fbinv. 
The matrix elements of LQ signals for both single and pair LQ production are generated at LO using version 2.6.0 of
$ {\tt MadGraph5\_aMC@NLO}$~\cite{Alwall:2014hca} for $m_{LQ}=$500, 1000, 1500, and 2000 \GeV. The branching fraction $\beta$ of the LQ to a charged
lepton and a quark, in this case LQ$\to \tau b$, is assumed to be $\beta=1$. The unknown Yukawa coupling $\lambda$ of the LQ to a
$\tau$ lepton and a bottom quark is set to $\lambda=1$. The width $\Gamma$ is calculated using $\Gamma = m_{LQ} \lambda^2 /
(16\pi)$~\cite{Plehn:1997az}, and is less than 10\% of the LQ mass for most of the considered search range. The signal samples are
normalized to the cross section calculated at LO, multiplied by a $K$ factor to account for higher order contributions~\cite{Dorsner:2018ynv}.

Similar event selections are used in both the singly and pair produced LQ searches, except for the requirement on the number of jets. 
In both channels, two reconstructed $\tauh$ with opposite-sign charge are required, each with transverse momentum $p_{T,\tau}>50$\GeV\ and a maximum pseudorapidity  ${|\eta_{\tau}|}<2.3$. 
In the search for single production, the presence of at least one reconstructed jet with $p_T > 50\GeV$ is required, while at least
two are required in the search for pair production. Jets are reconstructed with FASTJET~\cite{Cacciari:2011ma}, using the anti-\kt algorithm~\cite{Cacciari:2008gp}, with a distance parameter of 0.4. 

To reduce background due to Drell-Yan (particularly Z$\to\tau\tau$) events, the invariant mass of the two selected $\tauh$,
$m_{\tau\tau}$, is required to be $>95$\GeV.
In addition, at least one of the previously selected jets is required to be b-tagged to reduce QCD multijet backgrounds. 
Finally, an event is rejected if it contains identified and isolated electrons (muons), with $p_T>10$\GeV, $|\eta|<$2.4 (2.5).
The acceptance of the signal events is 4.9\%  (11\%) for single (pair) production, where the branching ratio of two $\tau$ leptons decaying hadronically is included in the numerator of the acceptance.

\begin{figure}[t]
\centering
\includegraphics[width=0.48\textwidth]{\main/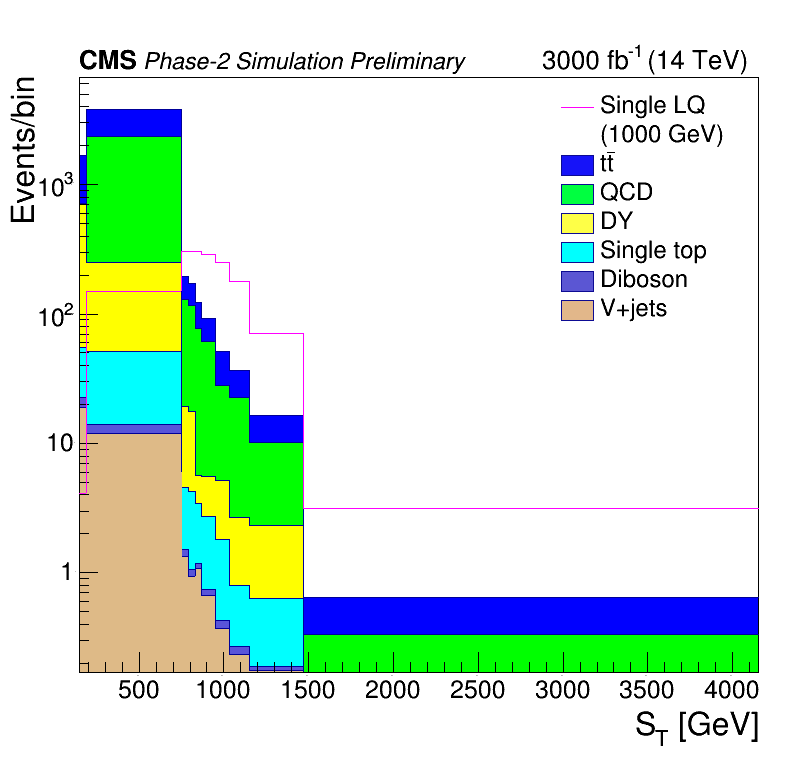}
\includegraphics[width=0.48\textwidth]{\main/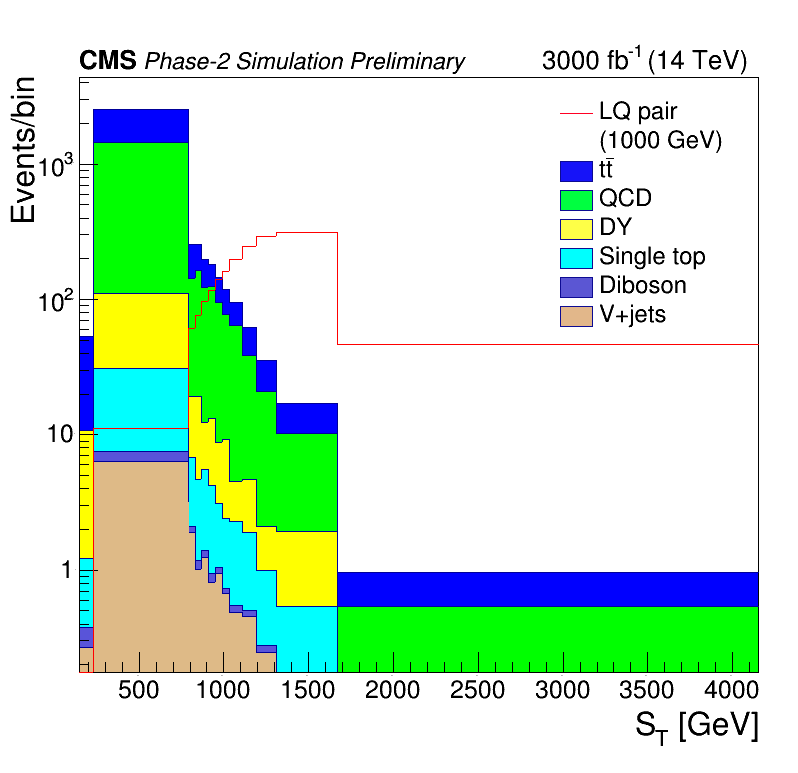}
\caption{Left: scalar sum of the \pt\ of the two selected $\tau$ leptons and the highest-$p_T$ jet in the single LQ selection region. Right: scalar sum of the $p_T$ of the two selected $\tau$ leptons and the two highest-$p_T$ jets in the LQ pair search region. The considered backgrounds are shown as stacked histograms, while empty histograms for signals for the single LQ and LQ pair channels (for $m_{LQ}=1000$\GeV) are overlaid to illustrate the sensitivity. Both signal and backround are normalised to a luminosity of 3000 $\fbinv$.} 
\label{fig:scalarptsums}
\end{figure}

Signal extraction is based on a binned maximum likelihood fit to the distribution of the scalar $p_T$ sum S$_{\text{T}}$, which 
is defined as the sum of the transverse momenta of the two $\tauh$ and either the highest-$p_T$ jet in the 
case of single LQ production, or the two highest-$p_T$ jets in the case of LQ pair production. These distributions are shown 
in Fig.~\ref{fig:scalarptsums} for the HL-LHC 3000 $\fbinv$ scenario.

Systematic uncertainties are 
calculated by scaling the current experimental uncertainties. For uncertainties limited by statistics, including the uncertainty 
on the DY (3.3\%) and QCD (3.3\%) cross sections, a scale factor of $1/\sqrt{L}$ is applied, for an integrated luminosity ratio $L$. 
For uncertainties 
coming from theoretical calculations, a scale factor of 1/2 is applied with respect to current uncertainties, as is the case for 
the uncertainties on the cross section for top (2.8\%) or diboson (3\%) events. Other experimental systematic uncertainties are 
scaled by the 
square root of the integrated luminosity ratio until the uncertainty reaches a minimum value, including uncertainties on the integrated 
luminosity (1\%), 
$\tau$ identification (5\%) and b-tagging/misidentification (1\%-5\%).

\begin{figure}[t]
\vspace{-5pt}
\centering
\includegraphics[width=0.49\textwidth]{\main/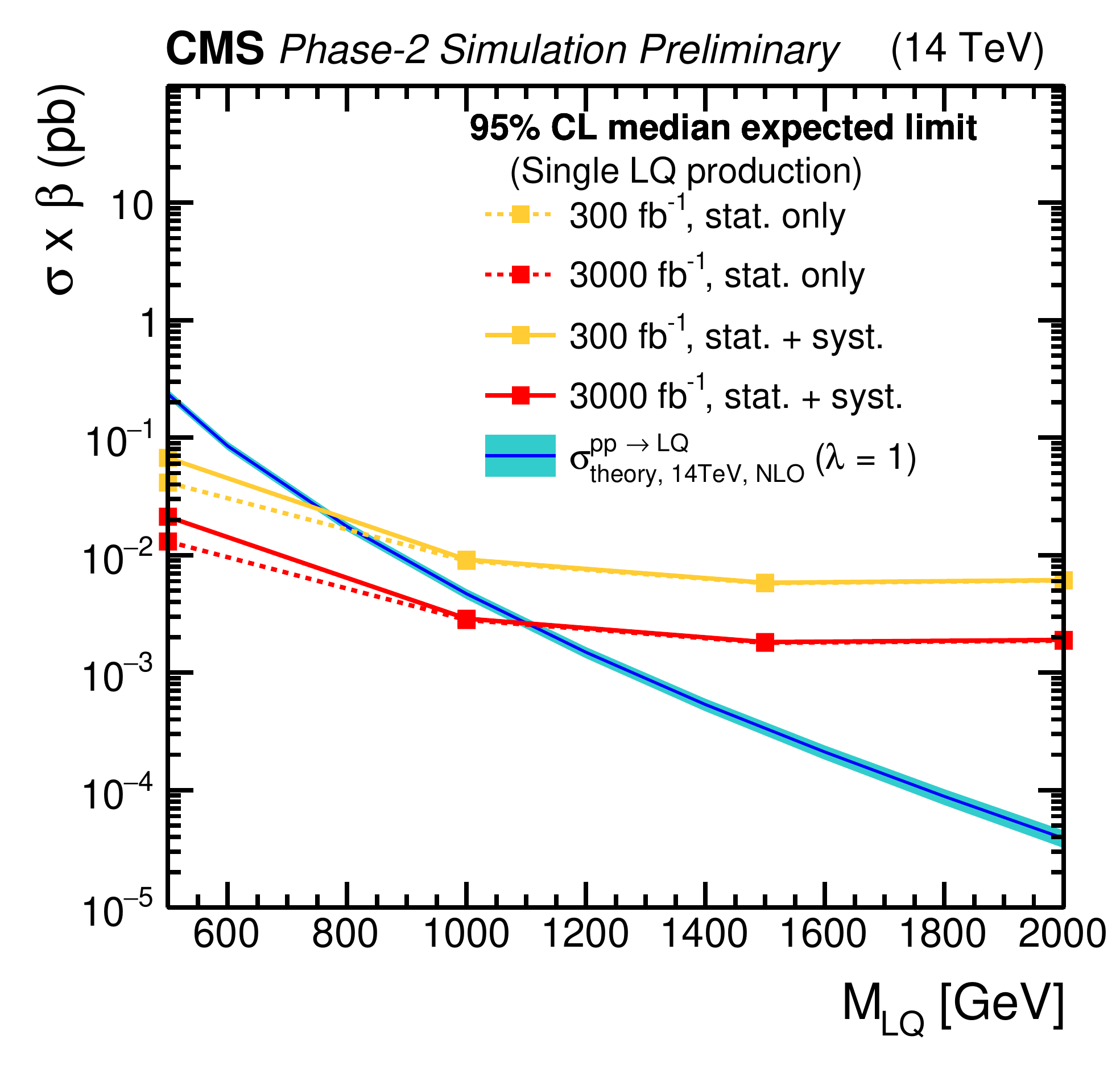}
\includegraphics[width=0.49\textwidth]{\main/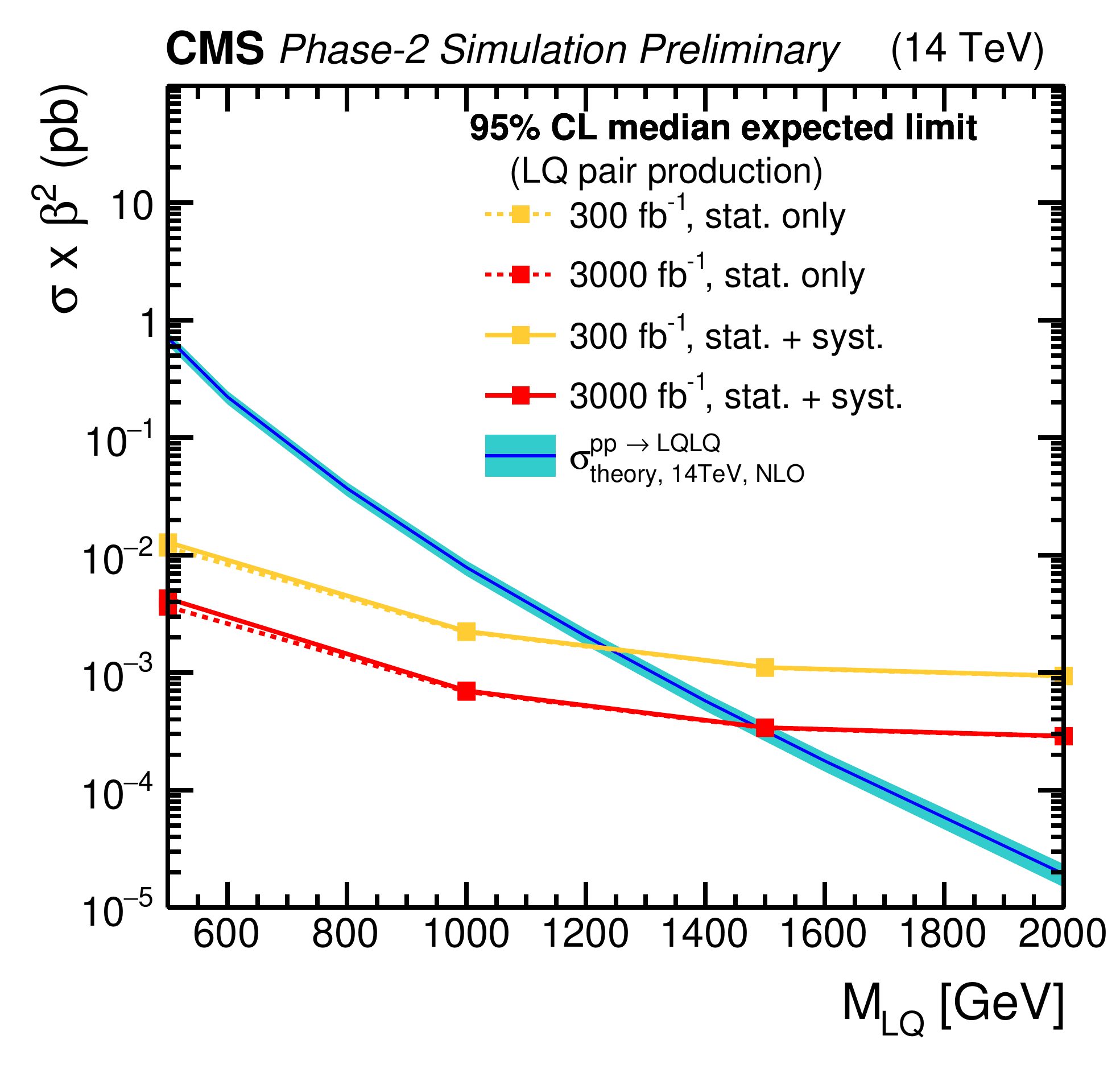}
\caption{Expected limits at 95\% CL on the product of the cross section $\sigma$ and the branching fraction $\beta$, as a function
of the LQ mass, for the two high luminosity projections, 300 \fbinv (red) and 3000 \fbinv (orange), for both the stat. only  (dashed
lines) and the stat.+syst. scenarios (solid lines). This is shown in conjunction with the theoretical predictions at NLO~\cite{Dorsner:2018ynv} in cyan. Projections are calculated for both the single LQ (left) and LQ pair production (right).}
\label{fig:LQ3-projection}
\end{figure}

Figure~\ref{fig:LQ3-projection} shows an upper limit at 95\% CL on the cross section times branching fraction $\beta$ as a 
function of $m_{LQ}$ by using the asymptotic CLs modified frequentist 
criterion~\cite{Cowan:2010js,Junk:1999kv,Read:2002hq, ATLAS:2011tau}.
Upper limits are calculated considering 
two different scenarios. The first one, hereafter abbreviated as "stat. only" considers only statistical uncertainties, to 
observe how the results are affected by the increase of the integrated luminosity. The second scenario, hereafter 
abbreviated as "stat.+syst.," also includes the estimate of the systematic uncertainties at the HL-LHC. 
For the single LQ production search, the theoretical prediction for the cross section assumes $\lambda = 1$ and $\beta = 1$. 

Comparing the limits with theoretical predictions assuming unit Yukawa coupling $\lambda = 1$, third-generation scalar 
leptoquarks are expected to be excluded at 95\% confidence level for LQ masses below 732 (1249) GeV for a luminosity 
of 300 \fbinv, and below 1130 (1518) GeV for 3000 \fbinv in the single (pair) production channel, considering both 
statistical and systematic uncertainties. 

\begin{figure}[t]
\centering
\includegraphics[width=0.5\textwidth]{\main/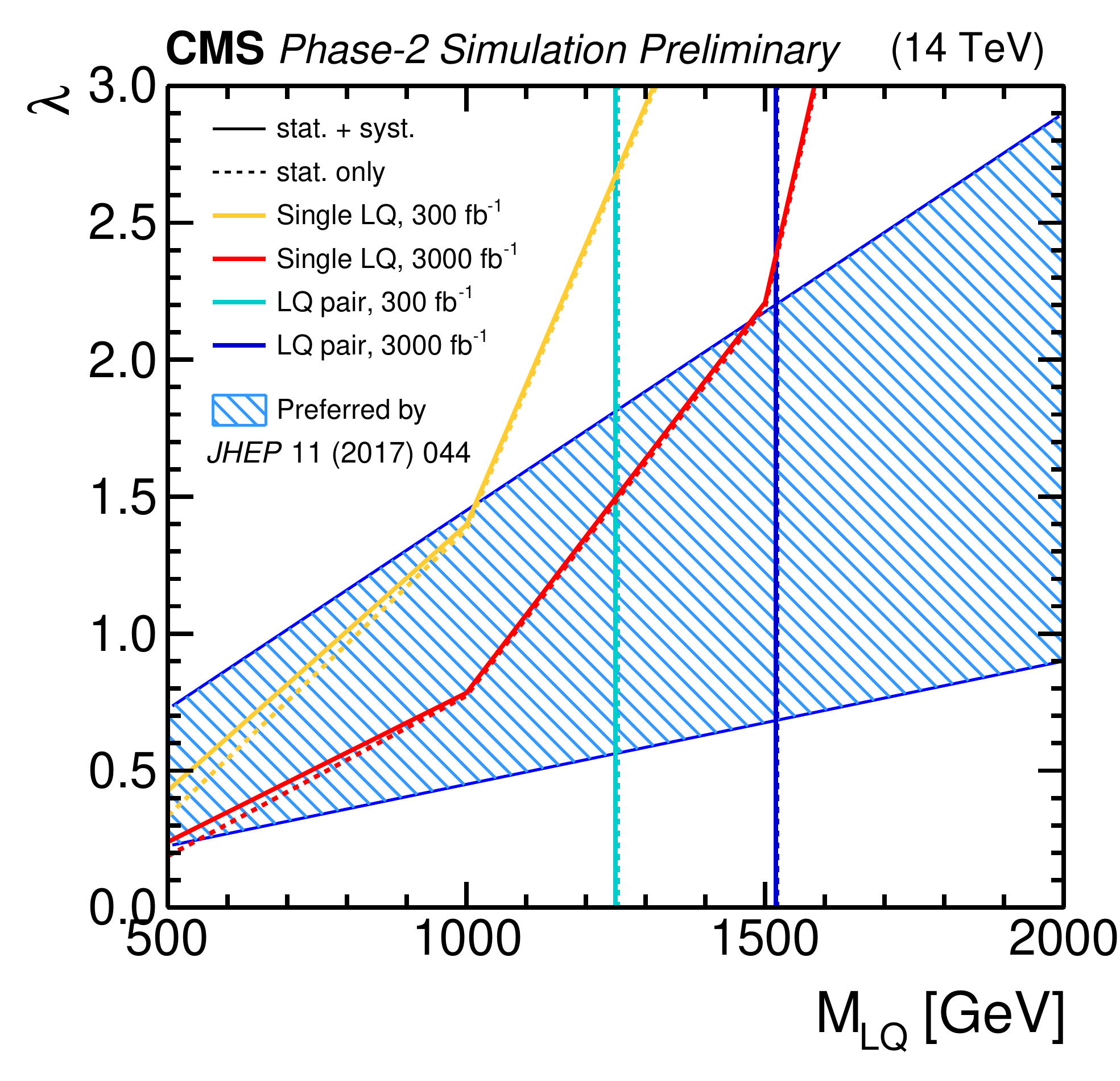}
\caption{Expected exclusion limits at 95\% CL on the Yukawa coupling $\lambda$ at the LQ-lepton-quark vertex, as a function of the
LQ mass. A unit branching fraction $\beta$ of the LQ to a $\tau$ lepton and a bottom quark is assumed. Future projections for 300
and 3000 \fbinv are shown for both the stat. only and stat.+syst. scenarios, shown as dashed and
filled lines respectively, and for both the single LQ and LQ pair production, where the latter corresponds to the vertical line
(since it does not depend on $\lambda$). The left hand side of the lines  represents the exclusion region for each of the
projections, whereas the region with diagonal blue hatching shows the parameter space preferred by one of the models proposed to
explain anomalies observed in $\PB$ physics~\cite{Buttazzo:2017ixm}.}
\label{fig:lambda_mass}
\end{figure}

Since the single-LQ signal cross section scales with $\lambda^2$, it is straightforward to recast the results presented in 
Fig.~\ref{fig:LQ3-projection} in terms of expected upper limits on $m_{LQ}$  as a function of $\lambda$, as shown in 
Fig.~\ref{fig:lambda_mass}. 
The blue band shows the parameter space (95\% CL) for the scalar LQ preferred by the $\PB$ physics 
anomalies: $\lambda = (0.95 \pm 0.50) m_{LQ} (\TeV)$~\cite{Buttazzo:2017ixm}. 
For the 300 (3000) \fbinv luminosity scenario, the leptoquark pair production channel is more sensitive if  
$\lambda<2.7$ $(2.3)$, while the single leptoquark production is dominant otherwise.
 
\begin{figure}[t]
\vspace{-5pt}
\centering
\includegraphics[width=0.49\textwidth]{\main/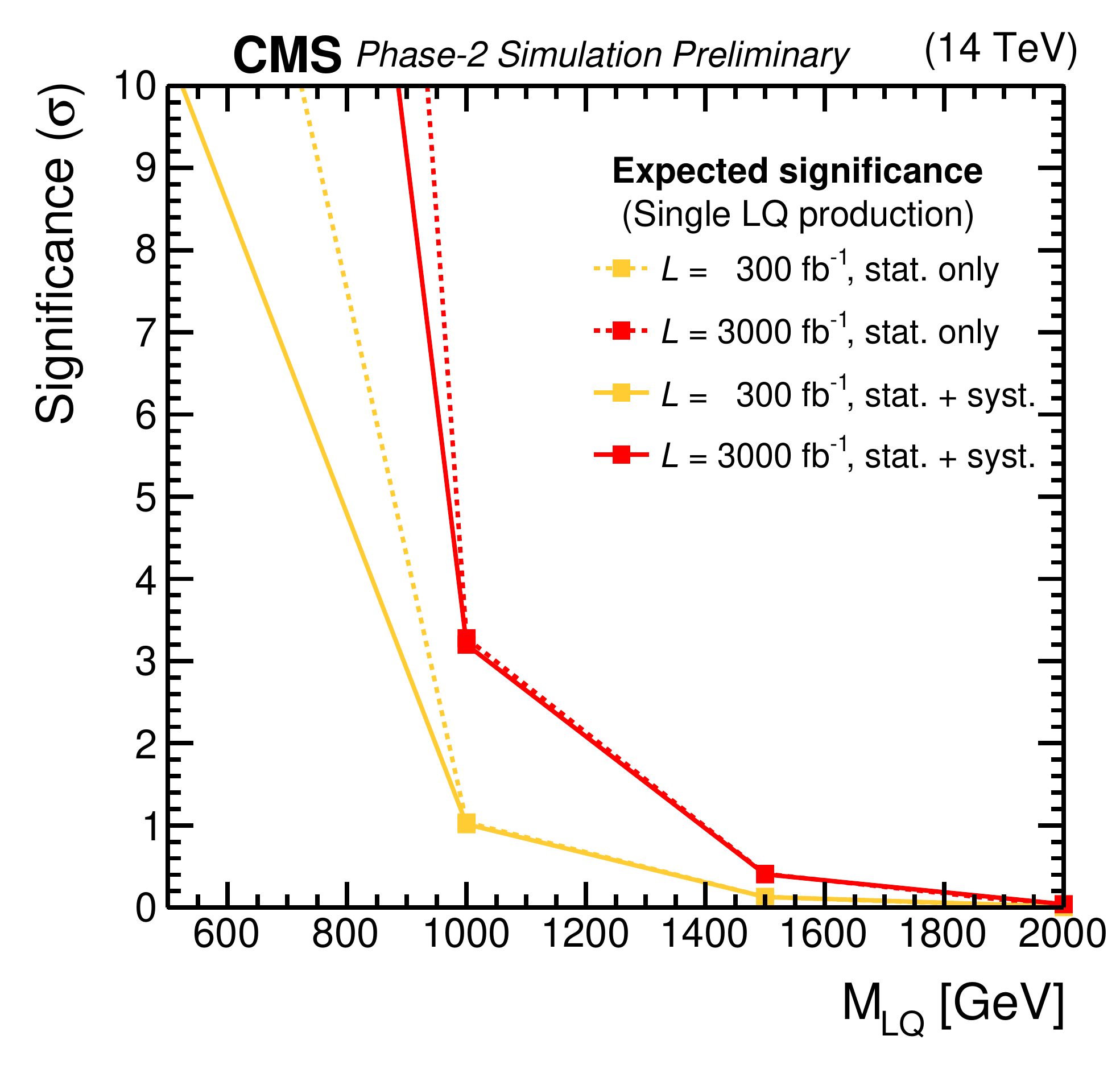}
\includegraphics[width=0.49\textwidth]{\main/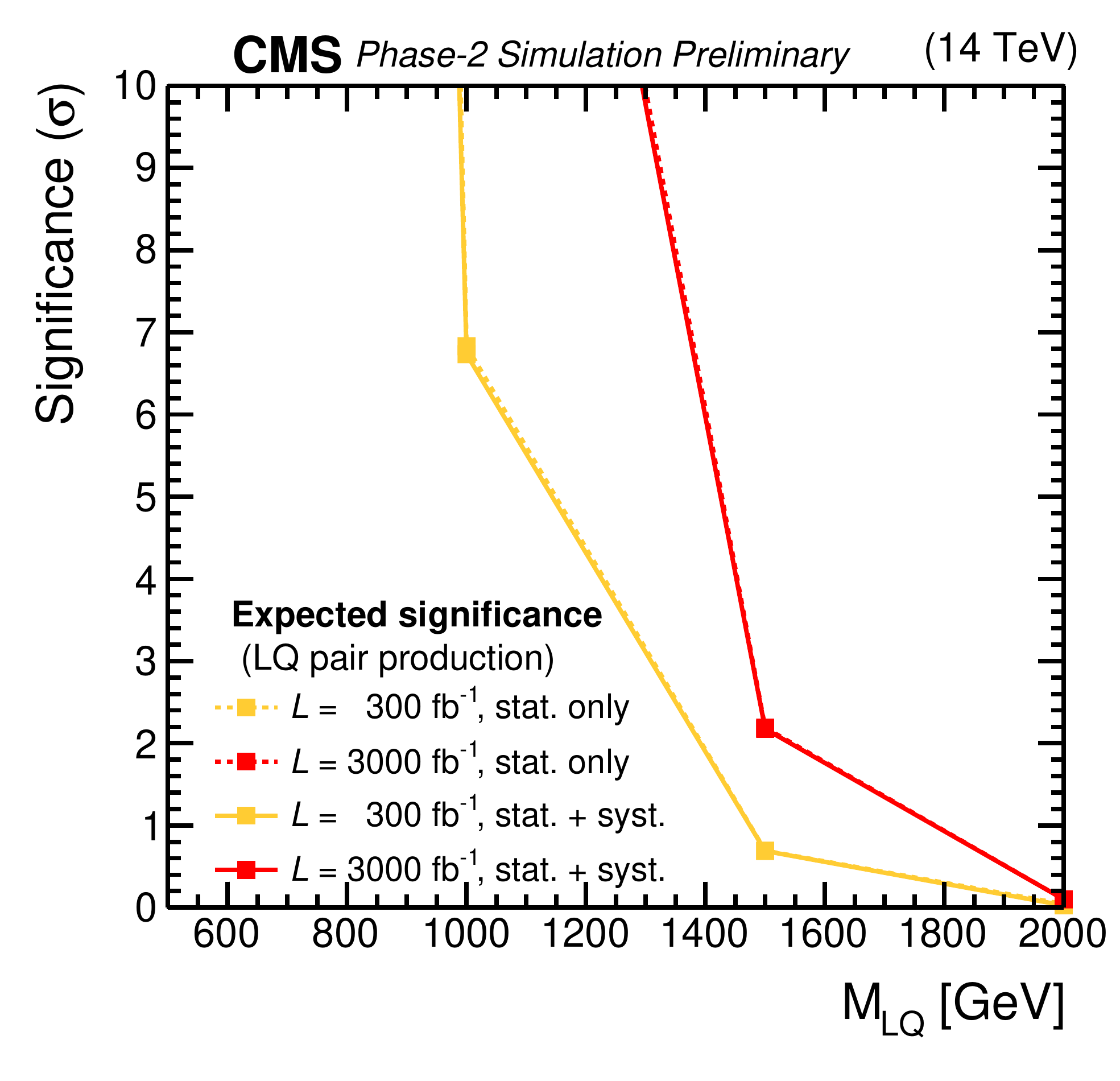}
\caption{Expected local significance of a signal-like excess as a function of the LQ mass, for the two high luminosity projections,
300 \fbinv (red) and 3000 \fbinv (orange), assuming the theoretical prediction for the LQ cross section at NLO~\cite{Dorsner:2018ynv}, calculated with $\lambda = 1$ and $\beta = 1$. Projections are calculated for both single LQ (left) and LQ pair production (right).}
\label{fig:LQ3-discovery}
\end{figure}

Using the predicted cross section~\cite{Dorsner:2018ynv} of the signal, it is also possible to estimate the maximal LQ mass expected to be in reach for a 5$\sigma$ discovery. 
Figure~\ref{fig:LQ3-discovery} shows the expected local significance of a signal-like excess as a function of the LQ mass hypothesis.

In summary, this study shows that future LQ searches under higher luminosity conditions are promising, as they are expected 
to greatly increase the reach of the search. They also show that the pair production channel is  expected to be the most 
sensitive. A significance of 5$\sigma$ is within reach for LQ masses below 800 (1200) FeV for the single (pair) production 
channels in the 300\fbinv scenario and 1000 (1500) GeV for the 3000 \fbinv scenario.

\subsubsubsection{CMS Searches for $W^\prime \to \tau + \met$}

New heavy gauge bosons are predicted by various SM extensions. The charged version of such heavy gauge bosons
is generally referred to as \PWpr.
The decay $\PWpr \rightarrow \tau \nu$ yields a single hadronically decaying tau
($\tau_h$) as the only detectable object and missing energy due to the neutrinos.
Hadronically decaying tau leptons are selected since the corresponding branching fraction,
about 60\%, is the largest among all $\tau$ decays.
Tau-jets are experimentally distinctive because of their low charged hadron
multiplicity, unlike QCD multi-jets,
which have high charged hadron multiplicity, or other leptonic \PWpr boson decays, which yield no jet. 
This Phase-2 study~\cite{CMS-PAS-FTR-18-030} follows closely the recently published Run~2 result~\cite{Sirunyan:2018lbg},
using hadronically decaying tau leptons.

\begin{figure}[htp]
\begin{center}

\includegraphics[width=.45\textwidth]{\main/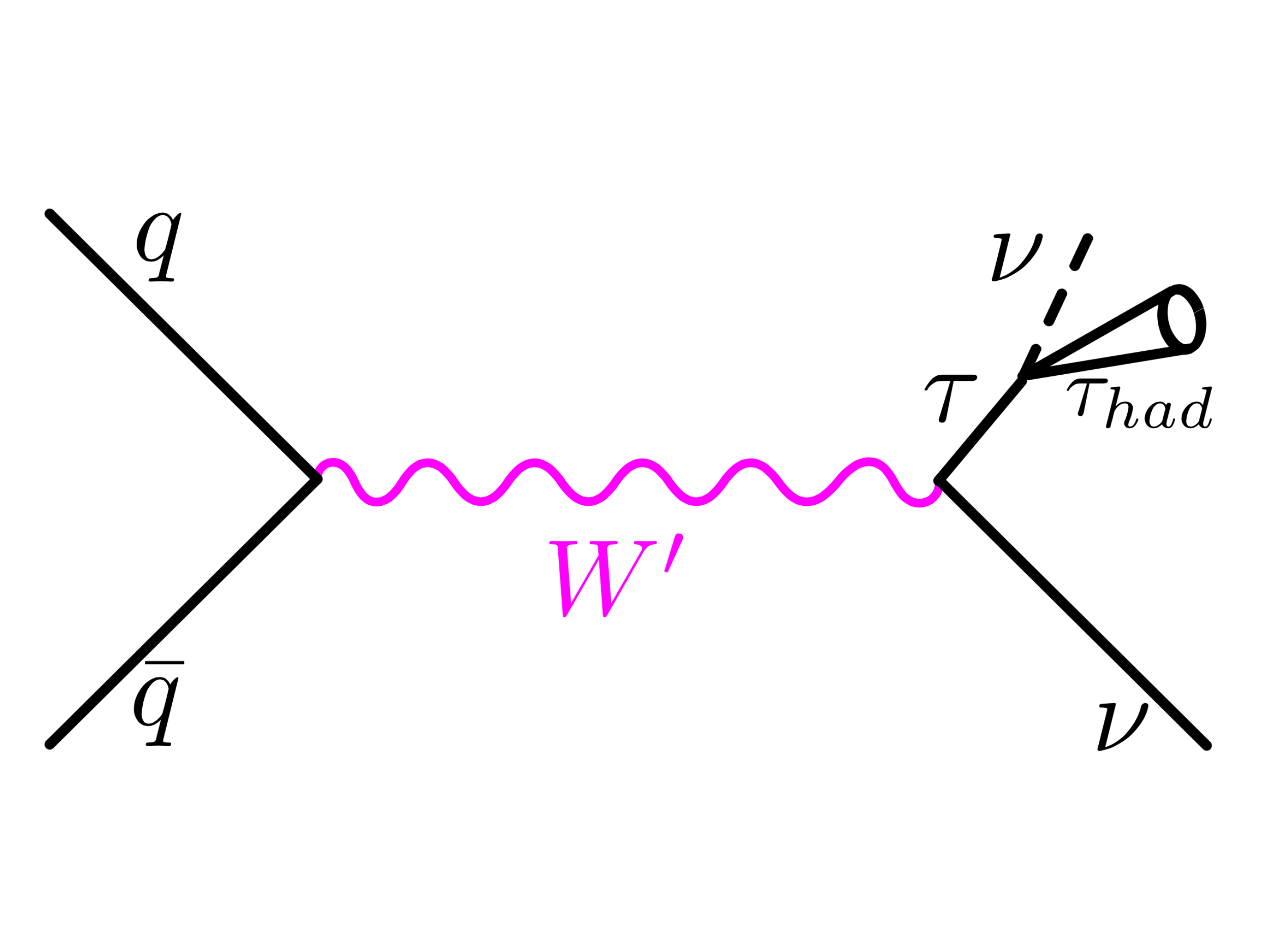}
\includegraphics[width=.49\textwidth]{\main/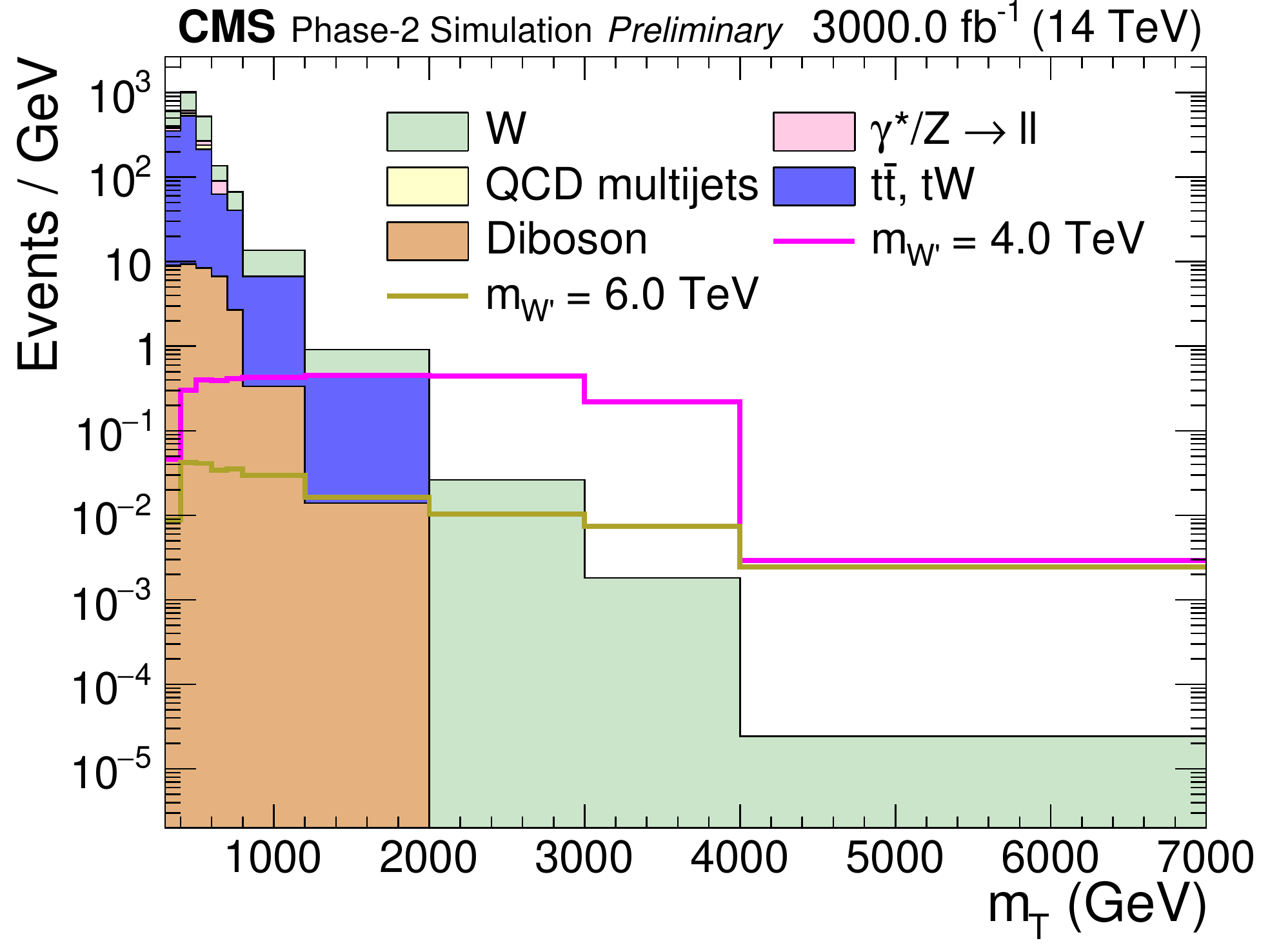}

\caption{Left: Illustration of the studied channel $W^\prime \rightarrow \tau \nu$ with the
subsequent hadronic decay of the tau ($\tau_h$).
Right: The discriminating variable, $M_T$, after all selection criteria for the HL-LHC conditions of
\intLumi and 200~PU.
The relevant SM backgrounds are shown according to the labels in
the legend. Signal examples for \PWpr boson masses of 1\TeV, 5\TeV and 6\TeV are scaled to their SSM
LO cross section and \intLumi.}
\label{fig:feynmanWprime}
\end{center}
\end{figure}

The signature of a \PWpr boson (see Fig.~\ref{fig:feynmanWprime}), is considered
similar to a high-mass W boson. 
It could be observed in the distribution of the
transverse mass ($M_T$) of the transverse momentum of the $\tau$ ($p_T^{\tau}$)
and the missing transverse momentum (\ptmiss):
$M_T = \sqrt{2 p_T^{\tau}p_T^{miss}(1-cos\Delta\phi(\tau,p_T^{miss}))}$.
Unlike the leptonic search channels, the signal shape of $W^\prime$ bosons with hadronically
decaying tau leptons does
not show a Jacobian peak structure because of the presence of two neutrinos in the final
state. Despite the multi-particle final state,
the decay appears as a typical two-body decay; the axis of the
hadronic tau jet is back to
back with \ptmiss and the magnitude of both is comparable such that their ratio is about unity.

The results are interpreted in the context of the sequential standard model (SSM)
~\cite{Altarelli:1989ff} in terms of \PWpr mass and coupling strength. 
A model-independent cross section limit allows interpretations in other models.
The signal was simulated in LO and the detector performance simulated with \delphes{}.
The \PWpr boson coupling strength, $g_{\PWpr}$, is given in terms of the SM weak coupling strength $g_{\PW} = e/\sin^2\theta_W\approx 0.65$.
Here, $\theta_{\PW}$ is the weak mixing angle.
If the \PWpr boson is a heavier copy of the SM \PW\ boson, their coupling ratio is $g_{\PWpr}/g_{\PW}=1$ and the 
SSM \PWpr boson theoretical cross sections, signal shapes, and widths apply.
However, different couplings are possible. Because of the dependence of the width of a particle on its couplings
the consequent effect on the transverse mass distribution, a limit can also be set on the coupling strength.

The dominant background appears in the off-shell tail of the $M_T$ distribution of
the SM \PW\ boson.
Subleading background contributions arise from $t\bar{t}$ and QCD multijet events. The number of background events
is reduced by the event selection.
These backgrounds primarily arise as a consequence of jets misidentified as $\tau_h$ candidates
and populate the lower transverse masses while the signal exhibits an excess of events at high $M_T$.
Events with one hadronically decaying $\tau$
and \ptmiss are selected if the ratio of $\pt^{\Pgt}$ to \ptmiss satisfies $0.7 < \pt^{\Pgt}/\ptmiss <
1.3$ and the
angle $\Delta\phi(\vec{p_T}^{\Pgt}, \vec{p_T}_{miss})$ is greater than 2.4 radians.

\begin{figure}[hbtp]\centering
    \includegraphics[width=0.49\textwidth]{\main/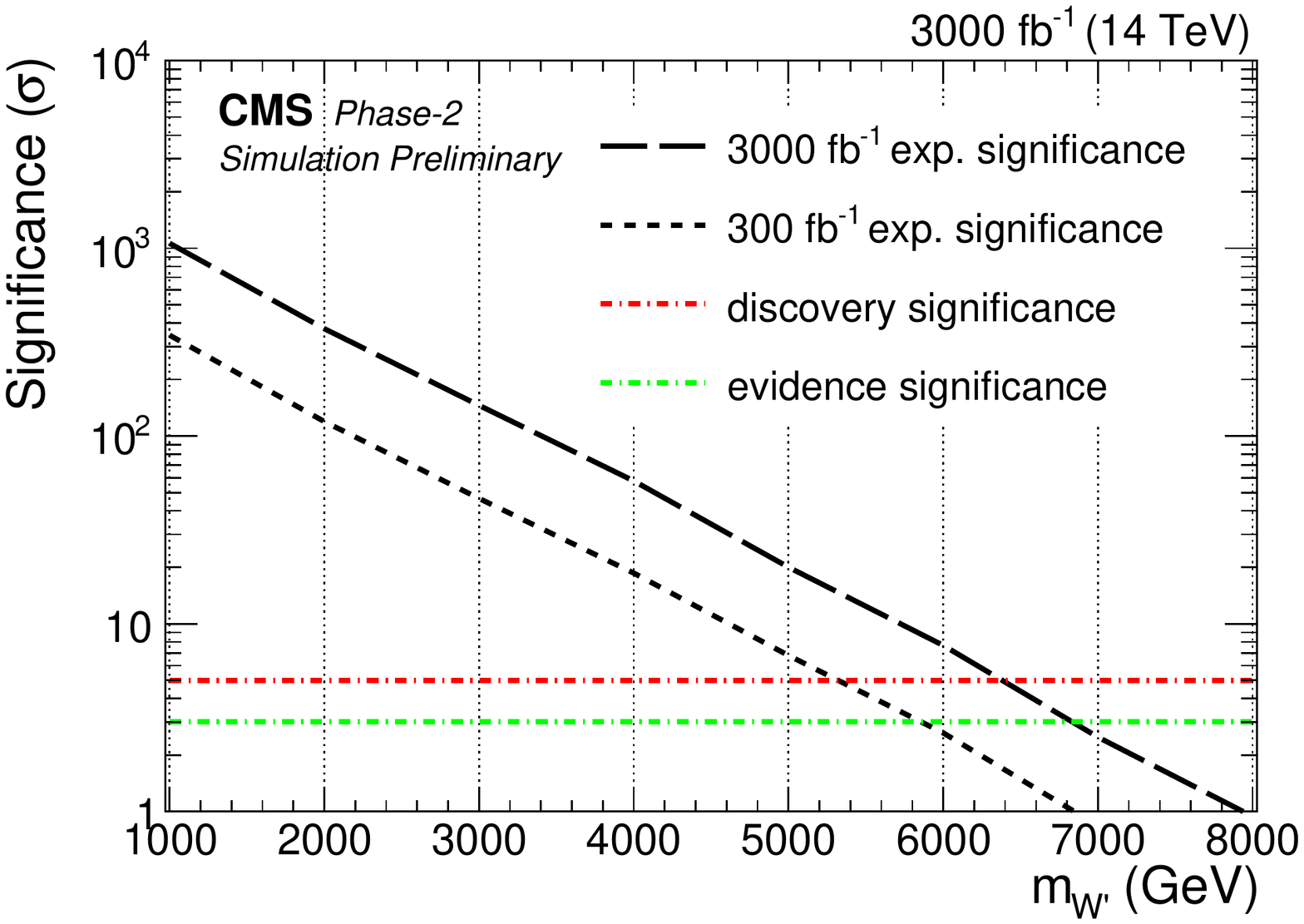}
    \includegraphics[width=0.49\textwidth]{\main/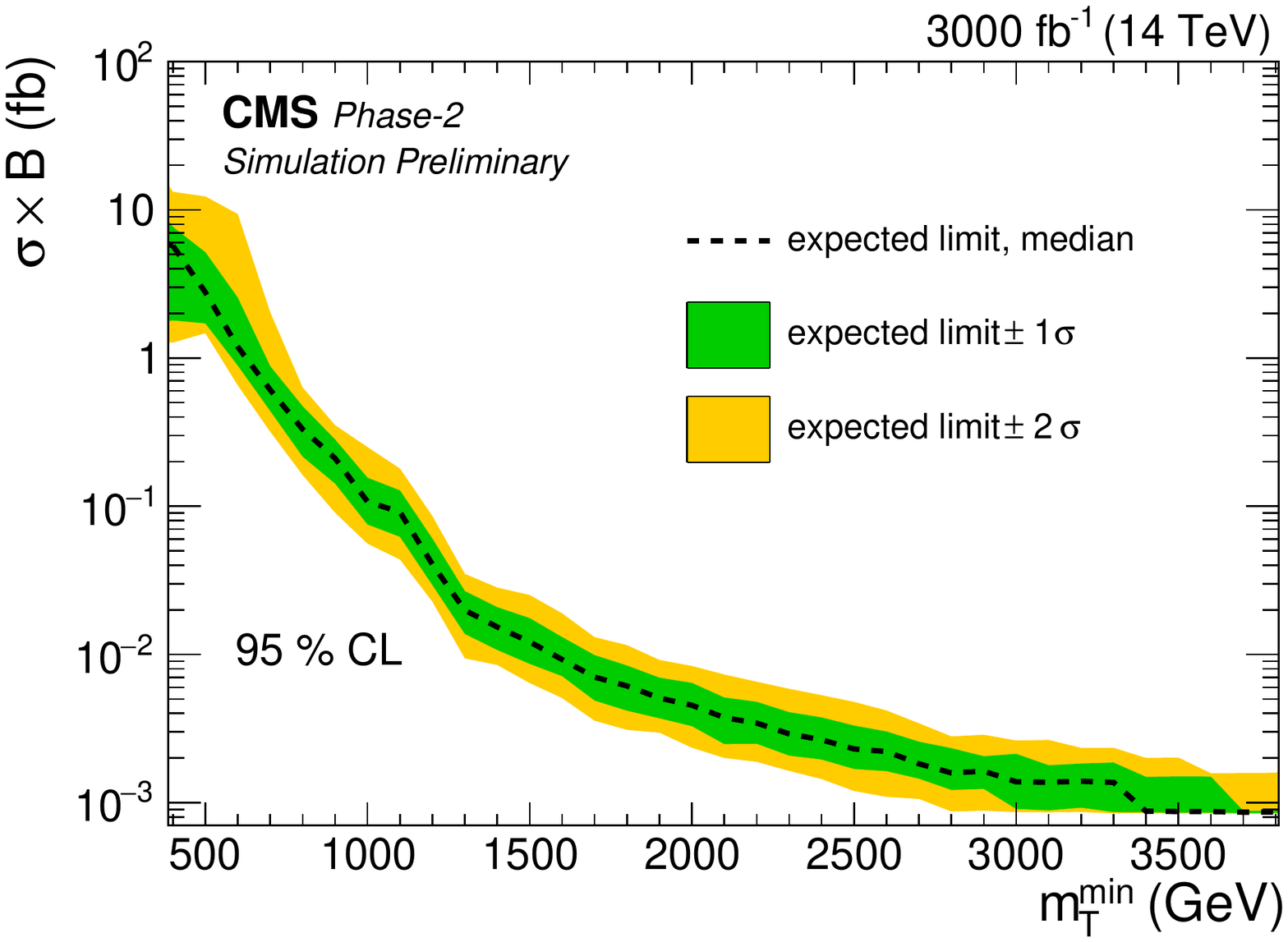}
\caption{Left: Discovery significance for SSM \PWpr to tau leptons.
Right: Model-independent cross section limit. For this, a single-bin limit is
calculated for increasing $M_T^{\rm min}$
while keeping the signal yield constant in order to avoid including any signal shape
information on this limit calculation.}
    \label{fig:WprimeResults}
\end{figure}

The physics sensitivity is studied based on the $M_T$ distribution in Fig.~\ref{fig:feynmanWprime}-right.
Signal events are expected to be particularly prominent at the upper end of the $M_T$
distribution, where the expected SM background is low.
So far, there are no indications for the existence of a SSM \PWpr\ boson~\cite{Sirunyan:2018lbg}. With the
high luminosity during Phase-2, the \PWpr mass reach for potential observation increases
to 6.9\TeV\ and 6.4\TeV\
for 3~$\sigma$ evidence and 5~$\sigma$ discovery, respectively, as shown in Fig.~\ref{fig:WprimeResults}-left. Alternatively, in case of no observation, one can exclude SSM \PWpr boson masses up to 7.0\TeV
with \intLumi.
These are multi-bin limits taking into account the full $M_T$ shape.

While the SSM model assumes SM-like couplings of the fermions, the couplings could well be weaker if
further decays occur. The HL-LHC has good sensitivity to study these couplings.
The sensitivity to weaker couplings extends significantly.
A model-independent cross section limit for new physics with $\tau$+MET
in the final state is depicted in Fig.~\ref{fig:WprimeResults}-right, calculated as a single-bin limit
by counting the number of events above a sliding threshold $M_T^{\rm min}$.

%% file: section11/section.tex
\section{Lattice QCD in the HL/HE-LHC era}\label{sec:lattice}
{\it\small Authors (TH): Michele Della Morte, Elvira G\'amiz, Enrico Lunghi.} 

We discuss the ten-year projections described in 
Sect.~\ref{sec:metro-CKM}
for the lattice inputs on hadronic parameters, by presenting the current status
and reviewing the main sources of uncertainty for different quantities.
A naive application of Moore's law (computing power doubling 
each two years) would give a reduction in the errors
by a factor around 3 by 2025. However, such an extrapolation is not always appropriate since the dominant 
uncertainties are often systematic.
For this reason attempting to extrapolate even farther in the future is subject to
very large uncertainties. The lattice approach is systematically improvable by
construction, however in order to almost completely remove the main systematic
affecting current computations one would have to extrapolate the performance of
present algorithms to unexplored regions of parameters. Any such extrapolation
would be quite unreliable. It is anyway reasonable to expect that by 2030 the
uncertainties related to the inclusion of electromagnetic effects, where relevant,
will be removed.

The last FLAG review~\cite{Aoki:2016frl}, or
 its online version~\footnote{{\tt{http://flag.unibe.ch/}}}, still provides
an almost up-to-date picture of the precision reached so far.
For example, the target accuracy on
$f_{B_s}$ and the ratio $f_{B_s}/f_{B_d}$ of about 0.4\% has already been achieved.
The latest $N_f=2+1+1$ results by the FNAL/MILC Collaboration 
in~\cite{Bazavov:2017lyh} quote
very similar errors for those two quantities.
Electromagnetic corrections (together with the known long- and short-distance 
electroweak contributions) in that case are directly subtracted from
the experimental decay rate quoted by PDG, such that the decay constant, which
is purely a QCD quantity, is the only hadronic parameter needed for
the extraction of $|V_{ub}|$ from leptonic $B$-decays.
Alternatively, one could consider the full transition rate on the lattice,
as proposed in~\cite{Carrasco:2015xwa}, and further discussed in~\cite{Patella:2017fgk}
concerning several subtleties arising in a straightforward application of the
method to heavy-meson decays.

In addition to the inclusion of isospin breaking corrections, the most important
limiting factor in the achievable precision of future lattice computations of decay constants
is probably going to be the scale setting, i.e., the conversion from
lattice units to MeV. 
This is quite obviously important for dimensionful
quantities, but it is also critical for some dimensionless ones, e.g., 
the hadronic contribution to the anomalous magnetic moment of the muon,
where one needs to convert the muon mass to lattice units (see~\cite{DellaMorte:2017dyu} for a detailed discussion of this point).
The current precision in the knowledge of the lattice spacing
is at the half-percent level. Going beyond that is challenging, as one needs
a quantity which is both very well known experimentally and precisely computable
on the lattice, with both small statistical as well as systematic uncertainties.
Future strategies may involve combining/averaging several quantities for the scale 
setting.~\footnote{We acknowledge S. Gottlieb and R. Van de Water for a discussion on this issue.}
On the other hand, ratios of decay constants can be obtained very precisely  from lattice simulations,
especially for light mesons, in which case discretization effects are not as severe as
for heavy-light mesons.
For example, the last calculations of the ratio of decay constants
$f_{K^\pm}/f_{\pi^\pm}$ has reached a 0.15\% error~\cite{Bazavov:2017lyh}.

We turn next to the bag parameter $B_K$, relevant for the theoretical prediction
of  $\varepsilon_K$, which encodes indirect CP-violation in the neutral kaon system.
The global estimate from FLAG for the $N_f=2+1$ theory has a 1.3\% error.  
Again, the error is systematic-dominated. In particular, one of the largest uncertainties
comes from the conversion between the lattice renormalization scheme (e.g., Schr\"odinger Functional (SF) 
or MOM) and $\overline{\rm MS}$. This is typically done at one-loop and to improve on that
one needs either a higher order computation~\footnote{See Ref.~\cite{Giusti:2018guw}
for recent work in this direction.}, or to follow the non-perturbative running up to
as large a scale as possible. In the SF scheme, but only in the two-flavor case, the matching scale 
has been pushed to the elctroweak scale, where truncation errors can be safely neglected~\cite{Dimopoulos:2007ht}.
The FLAG averages of the $B_K$ parameter have been very stable throughout different reviews. The main reason is,
at least for the 2+1 flavor setup, that the final value is dominated by a single computation from 2011~\cite{Durr:2011ap}.
It should therefore be possible to significantly reduce the error in the next few years.
In fact, with current methods and existing configurations several collaborations can probably reach that precision.

The main hadronic
uncertainty in the theoretical prediction of $\varepsilon_K$ is currently due to long-distance contributions 
not captured by the short-distance parameter $B_K$.
An approach for computing those non-perturbatively has been put forward in~\cite{Christ:2012se}
and preliminary results have been reported at the last Lattice conferences, with the most recent update
in~\cite{Bai:2016gzv}. Although the parameters in the lattice simulations are not physical,
producing rather heavy pions (329 MeV), and only one lattice resolution with $a>$ 0.1 fm has
been considered, the long distance correction to $\varepsilon_K$
is found to be rather large, amounting to about 8(5)\% of the experimental value.
For a more accurate estimate, the calculation is being repeated with physical kinematics
and finer lattice spacings~\footnote{See {\tt{https://indico.fnal.gov/event/15949/session/3/contribution/151/material/slides/0.pdf}}.}. 
It is worth pointing out that other approaches (e.g., in~\cite{Buras:2010pza}) produce values corresponding to a 4\% contribution at most.
Finally, it has been emphasized in~\cite{Bailey:2018feb} that the theoretical 
estimate of $\varepsilon_K$
based on lattice inputs strongly depends on the value of $V_{cb}$. Its current uncertainty, due
to the tension between inclusive and exclusive determinations, represents then the largest source of systematic error. 

In the $B$-sector, theoretical predictions of $B_q-\bar B_q$ mixing observables in the SM and beyond
depend on hadronic matrix elements of local four-fermion operators. 
Historically, these matrix elements have been parametrized in terms of the
so-called bag parameters. In the SM, the mass difference for a
$B_q^0$ meson depends on a single matrix element, proportional to
$f_{B_q}^2\hat B_{B_q}$, or, equivalently, on a single bag parameter, $\hat B_{B_q}$. 

There has been steady progress  in the last decade in lattice determinations of these matrix elements, 
with errors $\sim 4-5\%$ for $\hat B_{B_q}$~\cite{Aoki:2016frl}. However,
there is still a lot of room for improvement. First of all, current calculations
are not done on the last generation ensembles with $N_f=2+1+1$,   
physical light-quark masses, the smallest lattice spacings, and/or the most
improved lattice actions. Performing simulations on those ensembles will
reduce, and in some cases eliminate, the dominant errors in $B$-mixing: statistics,
continuum and chiral extrapolations, no inclusion of charm quarks on the sea,
and heavy-quark discretization. In addition, current calculations with $N_f=2+1$
flavors of sea quarks rely on effective field theories for the description of
the $b$ quark. Using a fully relativistic description instead, will further
reduce the heavy-quark discretization uncertainty and will also allow more precise
renormalization techniques, the other large source of uncertainty. 
For a particular lattice calculation, either the decay constants $f_B$
and $f_{B_s}$ are obtained within the same analysis in a correlated way,
or external inputs are used to get the bag parameters from the extracted matrix
elements. The recent and projected progress on the
determination of these decay constants will thus partly contribute to the reduction in the bag parameters uncertainty. 

All the improvements described above could considerably reduce the error in
$\hat B_{B_s}$ to a 0.8\% level. 
For the ratio of bag parameters, $B_{B_s}/B_{B_d}$, the error could be reduced
from the current 2\% to 0.5\% or even less. A similar reduction
can be achieved for the bag parameters describing the BSM contributions to the mixing. 
On-going calculations of hadronic matrix elements of NLO operators~\cite{Davies:2017jbi}
will also contribute to a substantial reduction in the uncertainty of $\Delta\Gamma_q$
in the next years.

Matrix elements describing the short-distance contribution to $D$-meson mixing in
the SM and beyond could benefit from a similar error reduction. However, in contrast to $B$-mixing, $D$-mixing is dominated by long-distance contributions and thus
reducing the error of the short-distance contribution is not so pressing.

Next we review the status and the prospects for lattice computations
of form factors for a number of semileptonic decays.
 The vector form factor at zero momentum transfer, $f_+^{K^0\pi^-}(q^2=0)$, needed to extract
$|V_{us}|$ from experimental measurements of $K\to\pi\ell\nu$ decay widths, 
is among the most accurate quantities obtained on the lattice. 
The most recent calculation of this form factor, not included in the last
FLAG review, has reduced the error to 0.20\%~\cite{Bazavov:2018kjg}, the 
same level as the experimental error~\cite{Moulson:2017ive}. 
The main source of uncertainty on the lattice side is
now statistics, which could be reduced by several lattice collaborations, both with
$N_f=2+1$ and $N_f=2+1+1$, extending their simulations to existing and planned
ensembles. Those ensembles include set of configurations with smaller lattice spacings
than those that are currently being used for calculations of $f_+^{K^0\pi^-}(q^2=0)$,
which will further reduce the total uncertainty. Other important sources of error
on current calculations arise from the scale setting and the 
uncertainty in the input parameters: quark masses 
and ChPT low energy constants. Better determinations of those parameters, 
which are needed in the analyses of many observables, are underway or expected. 
With those improvements the error in $f_+^{K^0\pi^-}(q^2=0)$, including
strong isospin-breaking corrections, will be soon enterely dominated by statistics
and could achieve the 0.12\% level in ten years. 

The uncertainty of the experimental average $|V_{us}|f_+^{K^0\pi^-}$, $0.19\%$, includes
errors of the estimated long-distance electromagnetic effects and
the difference between strong isospin-breaking effects in charged and neutral modes.
Those estimates, which use phenomenology and ChPT techniques, will become dominant
sources of error in the future. There are proposals to extend the study of full
leptonic transition rates to semileptonic decays~\footnote{See talk by Chris Sachrajda at
  Lattice 2018, https://indico.fnal.gov/event/15949/session/3/contribution/163}.  
Future calculations on the lattice thus should be able to not only reduce the error
on the QCD form factor, but also the electromagnetic and strong isospin-breaking
corrections contributing to the experimental uncertainty.

Lattice calculations are not limited to $q^2=0$ determination of $f_+^{K^0\pi^-}$. There exist results
for the momentum depedence of both the vector and scalar form
factors~\cite{Carrasco:2016kpy}. The $q^2$ dependence of $f_+^{K\pi}(q^2)$ obtained in
Ref.~\cite{Carrasco:2016kpy} agrees very well with experimental data, although the
determination of the constants entering in the dispersive parametrization of the
energy dependence, usually adopted in experiment, is still less precise when using lattice data.
However, in the future, with the improvements discussed above, lattice could provide
the most precise evaluation of the phase-space integral for kaon semileptonic decays. 

We discuss next three groups of $b$-hadron form factors of crucial phenomenological importance: heavy meson to stable pseudoscalar meson transitions, $B\to (\pi,K,D)$ and $B_s \to (K, D_s)$; heavy meson to unstable vector meson transitions, $B\to (D^*, K^*))$; and heavy baryon to baryon transitions, $\Lambda_b \to (p, \Lambda_c)$.
Form factors in the first group have been calculated by several collaborations, since the presence of a stable final state particle simplifies the analysis. Form factors in the second group are complicated by the presence of an unstable final state meson. While there is a solid theoretical foundation to treat such situation on the lattice~\cite{Briceno:2014uqa}, it should be noted that the huge hierarchy $\Gamma_{D^*}/M_{D^*} \sim 4 \times 10^{-5} \ll \Gamma_{K^*}/M_{K^*} \sim 6\times  10^{-2}$ reduces the impact of the final state meson decay for the $D^*$ case. 
In all cases, lattice calculations have higher accuracy for kinematical configurations in which the final state hadron has low recoil( large $q^2$) while experimental measurements tend to have better efficiency at large recoil (small $q^2$). 

Moreover, it is important to stress the role of form factors parametrizations which must be used to extrapolate the lattice results at low-$q^2$, for example the $B\to D^{(*)}$ form factors, or when combining lattice and experimental results in order to extract CKM parameters, e.g., $|V_{cb}|$ and $|V_{ub}|$. 
As mentioned above, lattice and experiments present differential (binned) distributions which tend to have higher accuracy
at low and high recoil, respectively. Their combination covers the whole kinematic spectrum, thus reducing drastically the
sensitivity to any given form factor parametrization. 
In situations in which experiments and/or lattice collaborations perform one sided extrapolations, the impact of the chosen parametrization can be very large (see, for instance, the extraction of $|V_{cb}|$ from $B\to D^*\ell\nu$ discussed in Refs.~\cite{Bigi:2017njr, Grinstein:2017nlq}).

The state of the art for the form factor determinations is 
represented by the averages in the latest FLAG review~\cite{Aoki:2016frl}. In the following we estimate the theoretical accuracy on the form factors that can be achieved over the next eight years and their impact on the extraction of $|V_{ub}|$ and $|V_{cb}|$. As a general rule, we estimate the overall drop in uncertainty to be a factor of 3 both in its statistical and systematic components. 
Thus, for many calculations the total extrapolated lattice uncertainty drops below the 1\% level, at which point QED effects need to be included. This can be done on the lattice either in the quenched QED limit or by explicitly generating mixed QCD-QED configurations. The inclusion of QED corrections to form factors is complicated  by IR safety issues which can, nonetheless,  be addressed in lattice calculations~\cite{Carrasco:2015xwa}. It is likely, but not guaranteed, that QED corrections to the form factors we discuss below will be calculated in the time frame we consider. We consider two scenarios according to whether QED corrections have or have not been calculated; in the latter case we add in quadrature an extra 1\% uncertainty. 
\vspace{-0.35cm}
\begin{itemize}
\item
{ $B\to\pi$ and $|V_{ub}|^{}$.}
The overall point-by-point uncertainty should reduce to the sub-percent level due the adoption of Highly Improved Staggered Quark (HISQ) for heavy quarks on the finer lattices and physical light quark masses. Currently the uncertainty on the extraction of $|V_{ub}|$ from a simultaneous fit of lattice form factors~\cite{Lattice:2015tia, Flynn:2015mha} and binned measurements of the semileptonic branching ratio is 3.7\%. The theoretical and experimental contributions to this error can be estimated at the 2.9\% and 2.3\% level, respectively. A reduction of the lattice uncertainties along the lines mentioned above would naively reduce the former uncertainty to about 1\%. As a cautionary note we point out that these estimates are sensitive to the information pertaining to the shape of the form factor, which is controlled by the correlation in the synthetic data and by the accuracy of future experimental results. In conclusion, the theory uncertainty on $|V_{ub}|$ is expected to decrease from 2.9\% to 1\% (1.4\% if QED corrections are not calculated).
\item
{ $B\to K$.}
These form factors have been calculated in Refs.~\cite{Bouchard:2013pna, Bailey:2015dka}. The FLAG average of the various form factors has uncertainties on the first coefficient in the BCL $z$-expansion of about 2\%. Future improvements can push this uncertainty to the 0.7\% (1.2\%) level according to the assumption on the inclusion of QED corrections.
\item
{ $B\to D$ and $|V_{cb}|^{}$.}
A simultaneous fit of the lattice $B\to D$ form factor~\cite{Na:2015kha, Lattice:2015rga}, and experimental data on the semileptonic $B\to D\ell\nu$ branching ratio yields an uncertainty on $|V_{cb}|$ at the 2.5\% level. The theoretical and experimental contributions to this error can be estimated as 1.4\% and 2.0\%, respectively. The lattice contribution to the uncertainty is essentially controlled by the 1.5\% error on the coefficients $a_0^+$ of the BCL $z$-expansion. Note that in the FLAG fit the zero-recoil value has an uncertainty of only 0.7\%: ${\cal G}^{B\to D}(1) = 1.059 (7)$. Leaving aside the issue that this uncertainty is lower than the conservative ballpark of the missing QED corrections, in order to use this very precise result one needs the corresponding zero-recoil experimental branching ratio which has an uncertainty of about 3\%. Clearly the zero-recoil extraction of $|V_{cb}|$ is inferior to the simultaneous fit method.  The uncertainty on the form factors is expected to drop by about a factor of 3 and to reach the 0.5\% (1.1\%) level. Correspondingly, the theory uncertainty on $|V_{cb}|$ is expected to decrease from 1.4\% to 0.3\% (1\%).
\item
{ $B_s\to K$,  $B_s \to D_s$ and $|V_{ub}|/|V_{cb}|$.}
These form factors have been calculated in Refs.~\cite{Bouchard:2014ypa, Flynn:2015mha}. 
The FLAG average of the various form factors has uncertainties on the first coefficient in the BCL $z$-expansion of 
about 4\%. Future improvements can push this uncertainty to the 1.3\% (1.7\%) level according to the assumption on the 
inclusion of QED corrections. Very recently the authors of Ref.~\cite{Monahan:2018lzv}
presented a calculation of the ratio $f_+^{(B\to K)} (0) / f_+^{(B_s\to D_s)} (0)$ with a total uncertainty of 13\%,
which is expected to reduce to about 4\% in the time frame we consider. This would also be the expected theory
uncertainty on the extraction of the CKM ratio $|V_{ub}|/|V_{cb}|$ from future measurements. Note that Ref.~\cite{Monahan:2018lzv}
presents full information on the $q^2$ distribution of the two form factors and that once differential experimental results become available the
above mentioned theory uncertainty will further decrease.
\item
{ $B\to D^*$ and $|V_{cb}|^{}$.}
There are two calculations of the form factor at zero-recoil~\cite{Bailey:2014tva,Harrison:2017fmw}. 
The most precise one in Ref.~\cite{Bailey:2014tva}
quotes an uncertainty of 1.4\% on ${\cal F}^{B\to D^*}(1)$. This uncertainty includes  an estimate of the size of missing QED effects at the 0.5\% level. Assuming a factor 3 reduction on lattice uncertainties one projects uncertainties on this form factor at the 0.4\% level (the estimate increases to 0.7\%, (1.1\%) in case QED corrections are not calculated, and are estimated to be 0.5\% (1\%) in size). We should point out that the Fermilab/MILC collaboration is about to publish a calculation of the form factor at non-zero recoil. This will allow simultaneous theory/experiment fits which are expected to further reduce the uncertainty on $|V_{cb}|$. 
\item
{ $B\to K^*$.}
The $B\to K^*$ form factors have been calculated in Refs.~\cite{Horgan:2013hoa, Horgan:2013pva}. These calculations are performed for a stable $K^*$ and, as we mentioned, receive potentially large corrections from the relatively large $K^*$ width. Currently an effort towards implementing the proper decay chain $B\to K^* \to K \pi$ along the lines described in Refs.~\cite{Luscher:1986pf, Luscher:1990ux, Luscher:1991cf, Lage:2009zv, Bernard:2010fp, Doring:2011vk, Hansen:2012tf, Briceno:2012yi, Dudek:2014qha, Briceno:2014uqa} is undergoing. This type of calculation will remove the uncontrolled uncertainty due to the assumption of a stable final state vector meson. Unfortunately, it is too soon to present an estimate on what the expected form factor uncertainty is going to be. Once the first complete result will become available, it will be possible to estimate the expected progression in error reduction as for all the other quantities we have considered.
\item 
{ $\Lambda_b\to p$, $\Lambda_b\to \Lambda_c$ and $|V_{ub}|/|V_{cb}|$.}
Currently there is only one calculation of these form factors~\cite{Detmold:2015aaa} which allows an extraction of the ratio $|V_{ub}|/|V_{cb}|$ with a theory uncertainty of 4.9\%. Over the next few years the total lattice uncertainty on these form factors is expected to reduce by a factor of 2~\cite{Meinel:2018MITP}. We estimate the improvement over the following decade to be another factor of 2~\footnote{We thank S. Meinel for a discussion on this point.}. This implies that in the time frame we are considering we expect a reduction in the theory uncertainty on $|V_{ub}|/|V_{cb}|$ to the 1.2\% (1.6\%) level.
\end{itemize}
An important development from the phenomenological point of view is related to
the computation of long-distance effects for rare decays as $K \to \pi \ell^+ \ell^-$.
A lattice approach has been put forward in~\cite{Christ:2015aha} and exploratory
results have appeared in~\cite{Christ:2016mmq}. The extension to $b \to s$
transitions however is by no means straightforward.

\vspace{-0.2cm}
In order to make it easily accessible, we collect all the information presented in this Section, including our projections for 2025, in Table~\ref{table:Proj}: 
\vspace{-0.3cm}
\begin{itemize}  
\item In the second and third columns we quote the current published best averages of lattice results with references.
\item In the fourth column we quote the error corresponding to the published value in the second column and,
  in parenthesis, the error coming from a couple of recent calculations, significantly reducing current errors, but that either are not published yet ($f_+(0)^{K\to\pi}$) or are not included in the FLAG-2016 averages (decay constants). The latest will be included in the next release of the FLAG averages in 2019.
\item Significant reduction of current errors in the decay constants are very
  unlikely, so we did not quote any numbers for these quantities. 
\item For semileptonic decays:
  \vspace{-0.13cm}
  \begin{itemize}
  \item $X\to Y$ for $|V_{ab}|$ means the theory error in the extraction of $|V_{ab}|$ from  that exclusive mode.
  \item Fifth column: The two errors correspond to assuming that isospin breaking corrections are calculated by that
    time (first number) or that they are still estimated phenomenologically (second number in parenthesis).
\end{itemize}
\end{itemize}

\begin{center}
\begin{sidewaystable}
\begin{center}
\begin{tabular}{cccccc}
\hline\hline
Quantity & Published averages & Reference & error (to be published/not in FLAG-2016)
& Phase I & Phase II\\
\hline
$f_{K^\pm}$ & $155.7(7)$~MeV & $N_f=2+1$~\cite{Aoki:2016frl} & 0.4\%  & 0.4\% & 0.4\%\\
$f_{K^\pm}/f_{\pi^{\pm}}$ & 1.193(3) &  $N_f=2+1+1$~\cite{Aoki:2016frl} & 0.25\%(0.15\%, symmet.~\cite{Bazavov:2017lyh})& 0.15\% & 0.15\% \\
$f_+^{K\to\pi}(0)$ & $0.9706(27)$ & $N_f=2+1+1$~\cite{Aoki:2016frl} & 0.28\% (0.20\%~\cite{Bazavov:2018kjg}) & 0.12\% & 0.12\%\\
$B_K$  & $0.7625(97)$ &  $N_f=2+1$~\cite{Aoki:2016frl} & 1.3\% & 0.7\% & 0.5\% \\
\hline
$f_{D_s}$ & 248.83(1.27) &  $N_f=2+1+1$~\cite{Aoki:2016frl} & 0.5\%(0.16\%~\cite{Bazavov:2017lyh}) & 0.16\% & 0.16\%\\
$f_{D_s}/f_{D^+}$ & 1.1716(32) &  $N_f=2+1+1$~\cite{Aoki:2016frl} & 0.27\%(0.14\%~\cite{Bazavov:2017lyh}) & 0.14\% & 0.14\%\\
$f_{B_s}$ & 228.4(3.7) &  $N_f=2+1$~\cite{Aoki:2016frl} & 1.6\%(0.56\%~\cite{Bazavov:2017lyh}) & 0.5\% & 0.5\% \\
$f_{B_s}/f_{B^+}$ & 1.205(7)& $N_f=2+1+1$~\cite{Aoki:2016frl} & 0.6\%(0.4\%~\cite{Bazavov:2017lyh}) & 0.4\% & 0.4\%\\
$B_{B_s}$ & 1.32(5)/1.35(6) & $N_f=2/N_f=2+1$~\cite{Aoki:2016frl} & $\sim 4$\% & 0.8\% & 0.5\%\\
$B_{B_s}/B_{B_d}$ & 1.007(21)/1.032(28) & $N_f=2/N_f=2+1$~\cite{Aoki:2016frl} & 2.1\%/2.7\% & 0.5\% & 0.3\%\\
$\xi$ & 1.206(17) & $N_f=2+1$~\cite{Aoki:2016frl} & 1.4\% & 0.3\% & 0.3\%\\
$\overline m_c (\overline m_c)$ & 1.275(8)~GeV &  $N_f=2+1$ ~\cite{Aoki:2016frl} & 0.6\% & 0.4\% & 0.4\%\\
\hline
$B\to \pi$ for $|V_{ub}|_{{\rm theor}}$ &  & $N_f=2+1$~\cite{Aoki:2016frl} & 2.9\% & 1\%(1.4\%) & 1\%\\
$B\to D$ for $|V_{cb}|_{{\rm theor}}$ &  & $N_f=2+1$~\cite{Aoki:2016frl} & 1.4\% & 0.3\%(1\%) & 0.3\%\\
(first param. BCL $z$-exp.) &  & $N_f=2+1$~\cite{Aoki:2016frl} & 1.5\% & 0.5\%(1.1\%) & 0.5\%\\
\begin{tabular}{c}$B\to D^*$ for $|V_{cb}|_{{\rm theor}}$\\$h_{A_1}^{B\to D^*}(\omega=1)$\\\end{tabular} &  & $N_f=2+1$~\cite{Aoki:2016frl} & 1.4\% & 0.4\%(0.7\%) & 0.4\% \\  
$P_1^{B\to D^*}(\omega=1)$ & & No LQCD available & & 1-1.5\% & 1\%\\
\begin{tabular}{c}$\Lambda_b\to p(\Lambda_c)$\\for $|V_{ub}/V_{cb}|_{{\rm theor}}$\\\end{tabular} & &
\cite{Detmold:2015aaa}  & 4.9\% & 1.2\%(1.6\%) & 1.2\%\\
\hline
$B\to K$ &  & $N_f=2+1$~\cite{Aoki:2016frl} & 2\% & 0.7\%(1.2\%) & 0.7\%\\
(first param. BCL $z$-exp.) & & & & & \\ 
$B_s\to K$ &  & $N_f=2+1$~\cite{Aoki:2016frl} & 4\% & 1.3\%(1.7\%) & 1.3\%\\
(first param. BCL $z$-exp.) & & & & & \\ 
\hline
\hline
\end{tabular}
\end{center}
\caption{Current estimates and projections for lattice QCD determinations of hadronic inputs. In the fourth column we quote the error corresponding to the published value in the second column and,
  in parenthesis, the error for forthcoming calculations that either are not published yet or are not included in the FLAG-2016 averages. See text for further explanations.}
\label{table:Proj}
\end{sidewaystable}
\end{center}

%% file: section12/section.tex
\section{Conclusions}
Flavour physics has many faces. This becomes already obvious in the Standard Model (SM), where the origin of the flavour structure are the Higgs Yukawa couplings, but which exhibits itself in the flavour non-diagonal couplings of $W$ boson, once we rotate to the quark mass eigenbases. As a result, we can search for signs of New Physics (NP) using flavour probes in many sectors: the Higgs couplings to the SM fermions, be it deviations in the flavour diagonal Yukawa couplings or in searches for flavour violating couplings, or in the flavour changing transitions of top, bottom, charm and strange quarks or of leptons. 

HL-LHC and HE-LHC experiments are set to make great strides in almost all of these probes, as exhibited in great detail in this document. The high luminosity programs at ATLAS and CMS are unmatched in their ability to push the precision frontier  in measuring Higgs couplings and searching for rare top decays, at the same time expanding their flavour-physics contribution. LHCb with its planned \upgradetwo can quite impressively cover a very wide range of low-energy probes in bottom, charm and strange flavor transitions. 

These low-energy measurements will bloom in competition with the upgraded Belle-II super $B$ factory, which is coming on line at this very moment.  There is also large complementarity between the two programs: The LHC hadronic environment gives access to a larger set of hadronic states such as heavy baryons and a larger sample of $B_s$ mesons. The HL/HE-LHC
upgrades will push well past the benchmarks set by Belle-II for many measurements in the near future.

In turn, this complementarity is very timely, given the tantalizing hints for NP surfacing in measurements of semi-tauonic and rare semi-muonic $B$-decays which suggest that Lepton Universality may be broken by new dynamics in the TeV range. The flavor anomalies thus lead to a different type of complementarity, between the low-energy flavor probes and the direct searches of NP at high \pt in ATLAS and CMS. This interplay is especially prolific in the context of model building, where the anomalies can be fitted into the more ambitious BSM program. Therefore, the HL/HE-LHC programme has the potential to shape the NP to come, in case the anomalies are confirmed in the coming years.    

Finally, it is important to note that the projected advancements in experiments are set to be accompanied by improvements in theory most notably in the calculations of the hadronic matrix elements by lattice QCD simulations. These matrix elements are required for the precision flavour-physics program, to convert the measurements in constraints or to lead to unambiguous discoveries of NP through the effects of their virtual corrections. In addition, the capabilities of LHCb to chart the spectrum and properties of resonances and exotic states in QCD, will lead to a better understanding of the nonperturbative phenomena. The combined improvements in the theoretical predictions, along with HL/HE-LHC experimental achievements are set to enhance the current reach on the NP physics mass scale by a factor as large as four. This represents a significant advance, with real discovery potential.

%% file: acknowledgments.tex
\section{Acknowledgements}

We would like to thank the LHC experimental Collaborations and the WLCG for their essential support. We are especially grateful for the efforts by the computing, generator and validation groups who were instrumental for the creation of large simulation samples. We thank the detector upgrade groups as well as the physics and performance groups for their input. 
We acknowledge support from CERN and from the national agencies:
CAPES, CNPq, FAPERJ and FINEP (Brazil); 
MOST and NSFC (China); 
CNRS/IN2P3 (France); 
BMBF, DFG and MPG (Germany); 
INFN (Italy); 
NWO (Netherlands); 
MNiSW and NCN (Poland); 
MEN/IFA (Romania); 
MSHE (Russia); 
MinECo (Spain); 
SNSF and SER (Switzerland); 
NASU (Ukraine); 
STFC (United Kingdom); 
NSF (USA).
We acknowledge the computing resources that are provided by CERN, IN2P3
(France), KIT and DESY (Germany), INFN (Italy), SURF (Netherlands),
PIC (Spain), GridPP (United Kingdom), RRCKI and Yandex
LLC (Russia), CSCS (Switzerland), IFIN-HH (Romania), CBPF (Brazil),
PL-GRID (Poland) and OSC (USA).
We are indebted to the communities behind the multiple open-source
software packages on which we depend.
Individual authors have received support from
AvH Foundation (Germany);
EPLANET, Marie Sk\l{}odowska-Curie Actions and ERC (European Union);
ANR, Labex P2IO and OCEVU, and R\'{e}gion Auvergne-Rh\^{o}ne-Alpes (France);
Key Research Program of Frontier Sciences of CAS, CAS PIFI, and the Thousand Talents Program (China);
RFBR, RSF and Yandex LLC (Russia);
GVA, XuntaGal and GENCAT (Spain);
the Royal Society
and the Leverhulme Trust (United Kingdom);
Laboratory Directed Research and Development program of LANL (USA).
Not least, we thank the many colleagues who have provided useful comments on the analyses. We thank I.~Bigi, S.~Hudan, A.~Khodjamirian, E.~Kou, L.~Maiani, J.~Portol\'es and K.~Trabelsi for useful comments. J. Mart\'in Camalich acknowledges support from the Spanish MINECO through the ``Ram\'on y Cajal'' program RYC-2016-20672. 
J.~Zupan acknowledges support in part by the DOE grant de-sc0011784.
W.~Altmannshofer acknowledges the National Science Foundation under Grant No. NSF-1912719. 
F.~Bishara acknowledges the Marie Sk\l{}odowska-Curie Individual Fellowship of the European Commission's Horizon 2020 Programme under contract number 745954 Tau-SYNERGIES.
A.~Carmona acknowledges the Cluster of Excellence Precision Physics, Fundamental Interactions and Structure of Matter (PRISMA-EXC1098) and grant 05H18UMCA1 of the German Federal Ministry for Education and Research (BMBF).
M.~Chala acknowledges the Royal Society under the Newton International Fellowship programme.
A.~Crivellin acknowledges the Ambizione grant of the Swiss National Science Foundation (PZ00P2 154834).
W.~Dekens acknowledges the US Department of Energy Grant No. DE-SC0009919.
S. Descotes-Genon acknowledges the EU Horizon 2020 program from the grants No. 690575, No. 674896 and No. 692194. 
A.~Esposito acknowledges the Swiss National Science Foundation under contract 200020-169696 and through the National Center of Competence in Research SwissMAP
R.~Fleischer acknowledges the Dutch National Organisation for Scientific Research (NWO).
E.~ G\'amiz acknowledges the Spanish State Research Agency (FPA2016-78220-C3-3-P) and Junta de Andaluc\'{\i}a (FQM 101).
L.-S.~Geng acknowledges the National Science Foundation of China, Grant Nos. 11522539 and 11735003.
B.~Grinstein acknowledges the US Depatment of Energy, Grant DE-SC0009919.
F.-K. Guo acknowledges the National Natural Science Foundation of China (NSFC) (Grant Nos. 11621131001, 11747601, and 11835015), the CAS Key Research Program of Frontier Sciences (Grant No. QYZDB-SSWSYS013),
the CAS Key Research Program (Grant No. XDPB09), the CAS Center for Excellence in Particle Physics.
S.~J\"ager acknowledges the UK STFC Consolidated Grant ST/P000819/1.
R.~Lebed acknowledges the U.S. National Science Foundation, Grant No. PHY-1803912.
A.~Lenz acknowledges the STFC through the IPPP grant.
J.~Matias has been supported by the Catalan ICREA Academia Program and acknowledges the FPA2014-55613-P and FPA2017-86989-P and 2017 SGR 1069. 
O. Matsedonskyi acknowledges the IASH postdoctoral fellowship for foreign researchers.
J. Nieves acknowledges the Spanish State Research Agency (FIS2017-84038-C2-1-P and SEV-2014-0398). 
Y.~Nir acknowledges the ISF, BSF, I-CORE, Minerva.
A.~A.~Petrov acknowledges the U.S. Department of Energy, DE-SC0007983.
T.~Pich acknowledges the Spanish State Research Agency (FPA2017-84445-P) and Generalitat Valenciana [Prometeo/2017/053].
S.~Prelovsek acknowledges the Slovenian Research Agency (No.~J1-8137 and No.~P1-0035) and DFG Grant SFB/TRR 55.
S.~Schacht acknowledges the DFG Forschungsstipendium under contract no.~SCHA~2125/1-1.
D.~Shih acknowledges the US Department of Energy Grant SC0010008.
L.~Silvestrini acknowledges the European Research Council (ERC) under the European Union's Horizon 2020 research and innovation program (grant agreement n$^o$ 772369).
D.~van Dyk acknowledges the Deutsche Forschungsgemeinschaft (DFG) within the Emmy Noether programme under grant DY 130/1-1 and through the DFG Collaborative Research Center 110 ``Symmetries and the Emergence of Structure in QCD''.
J.~Virto acknowledges the European Union's Horizon 2020 research and innovation programme under the Marie Sklodowska-Curie grant agreement No 700525, `NIOBE'.
W.~Wang acknowledges the National Science Foundation of China, Grant Nos. 11575110, 11655002, 11735010.

%% file: section13/section.tex
\section{Details on experimental extrapolations}
\subsection{Analysis methods and objects definitions}

Different approaches have been used by the experiments and in theoretical prospect studies, hereafter named projections, to assess the sensitivity in searching for new physics at the HL-LHC and HE-LHC.
For some of the projections, a mix of the approaches described below is used, in order to deliver the most realistic result.
The total integrated luminosity for the HL-LHC dataset is assumed to be $3000$~fb$^{-1}$ at a center-of-mass energy of $14$~TeV. For HE-LHC studies the dataset is assumed to be $15$~ab$^{-1}$ at a center-of-mass of $27$~TeV.
The effect of systematic uncertainties is taken into account based on the studies performed for the existing analyses and using common guidelines for projecting the expected improvements that are foreseen thanks to the large dataset and upgraded detectors, as described in Section~\ref{sec:methods:syst}.

{\bf Detailed-simulations} are used to assess the performance of reconstructed objects in the upgraded detectors and HL-LHC conditions, as described in Sections~\ref{sec:methods:perf},\ref{sec:methods:perf_LHCb}.
For some of the projections, such simulations are directly interfaced to different event generators, parton showering (PS) and hadronisation generators. Monte Carlo (MC) generated events are used for standard model (SM) and beyond-the-standard-model (BSM) processes, and are employed in the various projections to estimate the expected contributions of each process.

{\bf Extrapolations} of existing results rely on the existent statistical frameworks to estimate the expected sensitivity for the HL-LHC dataset.
The increased center-of-mass energy and the performance of the upgraded detectors are taken into account for most of the extrapolations using scale factors on the individual processes contributing to the signal regions. Such scale factors are derived from the expected cross sections and from detailed simulation studies.

{\bf Fast-simulations} are employed for some of the projections in order to produce a large number of Monte Carlo events and estimate their reconstruction efficiency for the upgraded detectors. The upgraded CMS detector performance is taken into account encoding the expected performance of the upgraded detector in {\tt Delphes}~\cite{deFavereau:2013fsa}, including the effects of pile-up interactions. Theoretical contributions use {\tt Delphes}~\cite{deFavereau:2013fsa} with the commonly accepted HL-LHC card corresponding to the upgraded ATLAS and CMS detectors.

{\bf Parametric-simulations} are used for some of the projections to allow a full re-optimization of the analysis selections that profit from the larger available datasets.
Particle-level definitions are used for electrons, photons, muons, taus, jets and missing transverse momentum. These are constructed from stable particles of the MC event record with a lifetime larger than $0.3 \times 10^{-10}$~s within the observable pseudorapidity range. Jets are reconstructed using the anti-$k_t$ algorithm~\cite{Cacciari:2008gp} implemented in the {\tt Fastjet}~\cite{fastjet} library, with a radius parameter of 0.4. All stable final-state particles are used to reconstruct the jets, except the neutrinos, leptons and photons associated to $W$ or $Z$ boson or $\tau$ lepton decays. The effects of an upgraded ATLAS detector are taken into account by applying energy smearing, efficiencies and fake rates to generator level quantities, following parameterisations based on detector performance studies with the detailed simulations. The effect of the high pileup at the HL-LHC is incorporated by overlaying pileup jets onto the hard-scatter events. Jets from pileup are randomly selected as jets to be considered for analysis with $\sim 2\%$ efficiency, based on studies of pile-up jet rejection and current experience.

\subsubsection{ATLAS and CMS performance}
\label{sec:methods:perf}

The expected performance of the upgraded ATLAS and CMS detectors has been studied in detail in the context of the Technical Design Reports
and subsequent studies; the assumptions used for this report and a more detailed description are available in Ref.~\cite{ATLAS_PERF_Note,Collaboration:2650976}. For CMS, the object performance in the central region assumes a barrel calorimeter aging corresponding to an integrated luminosity of $1000$~fb$^{-1}$.

The triggering system for both experiments will be replaced and its impact on the triggering abilities of each experiment assessed;
new capabilities will be added, and, despite the more challenging conditions, most of the trigger thresholds for common objects are expected
to either remain similar to the current ones or to even decrease~\cite{ATLAS_TDAQ_TDR,CMSL1interim}.
The inner detector is expected to be completely replaced by both experiments, notably extending its coverage to $|\eta|<4.0$.
The performance for reconstructing charged particles has been studied in detail in Ref.~\cite{ATLAS_Pixel_TDR,ATLAS_Strip_TDR,CMSTDR:trk}.
Electrons and photons are reconstructed from energy deposits in the electromagnetic calorimeter and information from the inner tracker\cite{ATLAS_LAr_TDR,CMS_Barrel_TDR,CMS_HGCAL_TDR,CMS:TPMTD}.
Several identification working points have been studied and are employed by the projection studies as most appropriate.
Muons are reconstructed combining muon spectrometer and inner tracker information~\cite{ATLAS_Muon_TDR,CMSTDR:muon}.

Jets are reconstructed by clustering energy deposits in the electromagnetic and hadronic calorimeters\cite{ATLAS_Tile_TDR,ATLAS_LAr_TDR,CMS_Barrel_TDR} using the anti-$k_{T}$ algorithm\cite{Cacciari:2008gp}.
$B$-jets are identified via $b$-tagging algorithms. B-tagging is performed if the jet is within the tracker acceptance ($|\eta|<4.0$).
Multivariate techniques are employed in order to identify $b-$jets and $c-$jets, and were fully re-optimized for the upgraded detectors~\cite{ATLAS_Pixel_TDR,CMSTDR:trk}.
An 70\% $b-$jet efficiency working point is used, unless otherwise noted.
High $p_T$ boosted jets are reconstructed using large-radius anti-$k_{T}$ jets with a distance parameter of 0.8. Various jet substructure variables are employed to identify boosted $W/Z/$Higgs boson and top quark jets with good discrimination against generic QCD jets. 

Missing transverse energy is reconstructed following similar algorithms as employed in the current data taking.
Its performance has been evaluated for standard processes, such as top pair production~\cite{ATLAS_Pixel_TDR,Contardo:2020886}.
The addition of new precise-timing detectors and its effect on object reconstruction has also been studied in Ref.~\cite{ATLAS_TP_HGTD,CMS:TPMTD}, although its results are only taken into account in a small subset of the projections in this report.

\subsubsection{LHCb}
\label{sec:methods:perf_LHCb}
The LHCb upgrades are shifted with respect to those of ATLAS and CMS. A first upgrade will happen at the end of Run 2 of the LHC, to run at a luminosity five times larger  ($2\times 10^{33}\text{cm}^{-2}\text{s}^{-1}$) in LHC Run 3 compared to those in Runs 1 and 2, while maintaining or improving the current detector performance. This first upgrade phase (named \mbox{Upgrade~I}) will be followed by by the so-called \mbox{Upgrade~II} phase (planned at the end of Run 4) to run at an even more challenging luminosity of $\sim 2\times 10^{34}\text{cm}^{-2}\text{s}^{-1}$.

The LHCb MC simulation used in this document mainly relies on the {\tt Pythia8} generator~\cite{Sjostrand:2007gs} with a specific LHCb configuration~\cite{LHCb-PROC-2010-056}, using the CTEQ6 leading-order set of parton density functions~\cite{cteq6}. The interaction of the generated particles with the detector, and its response, are implemented using the {\tt Geant4} toolkit~\cite{Allison:2006ve,Agostinelli:2002hh}, as described in Ref.~\cite{LHCb-PROC-2011-006}. 

The reconstruction of jets is done using a particle flow algorithm, with the output of this clustered using
the anti-kT algorithm as implemented in {\tt Fastjet}, with a distance parameter of
0.5. Requirements are placed on the candidate jet in order to reduce the background
formed by particles which are either incorrectly reconstructed or produced in additional pp interactions in the same event.
Concerning the increased pile-up, different assumptions are made, but in general the effect is assumed to be similar to the one in Run 2.

\subsection{Treatment of systematic uncertainties}
\label{sec:methods:syst}
It is a significant challenge to predict the expected systematic uncertainties of physics results at the end of HL-LHC running.
It is reasonable to anticipate improvements to techniques of determining systematic uncertainties over an additional decade of data-taking.
To estimate the expected performance, experts in the various physics objects and detector systems from ATLAS and CMS have looked at current limitations to
systematic uncertainties in detail to determine which contributions are limited by statistics and where there are more fundamental limitations.
Predictions were made taking into account the increased integrated luminosity and expected potential gains in technique.
These recommendations were then harmonized between the experiments to take advantage of a wider array of expert opinions and to allow the experiments to make sensitivity predictions on equal footing~\cite{ATLAS_PERF_Note,Collaboration:2650976}. For theorists' contributions, a simplified approach is often adopted, loosely inspired by the improvements predicted by experiments. 

General guide-lining principles were defined in assessing the expected systematic uncertainties.
Theoretical uncertainties are assumed to be reduced by a factor of two with respect to the current knowledge, thanks to both
higher-order calculation as well as reduced PDF uncertainties~\cite{Khalek:2018mdn}.
All the uncertainties related to the limited number of simulated events are neglected, under the assumption that sufficiently large simulation samples will be available by the time the HL-LHC becomes operational. For all scenarios, the intrinsic statistical uncertainty in the measurement is reduced by a factor $1/\sqrt{\text{L}}$, where $\text{L}$ is the projection integrated luminosity divided by that of the reference Run~2 analysis.
Systematics driven by intrinsic detector limitations are left unchanged, or revised according to detailed simulation studies of the upgraded detector.
Uncertainties on methods are kept at the same value as in the latest public results available, assuming that the harsher HL-LHC conditions will be compensated by method improvements.

The uncertainty in the integrated luminosity of the data sample is expected to be reduced down to $1\%$ by a better understanding of the calibration methods and
their stability employed in its determination, and making use of the new capabilities of the upgraded detectors.

In addition to the above scenario (often referred to as ``YR18 systematics uncertainties'' scenario), results are often
compared to the case where the current level of understanding of systematic uncertainties is assumed (``Run 2 systematic uncertainties'')
or to the case of statistical-only uncertainties.

%% file: report.bbl
\newcommand{\noopsort}[1]{} \newcommand{\printfirst}[2]{#1}
  \newcommand{\singleletter}[1]{#1} \newcommand{\switchargs}[2]{#2#1}
\providecommand{\href}[2]{#2}\begingroup\raggedright\begin{thebibliography}{1000}

\bibitem{Collaboration:2650976}
{The CMS Collaboration}, {\em {Expected performance of the physics objects with
  the upgraded CMS detector at the HL-LHC}\/},   CERN-CMS-NOTE-2018-006, CERN,
  Geneva, Dec, 2018.
\newblock \url{https://cds.cern.ch/record/2650976}.

\bibitem{ATL-PHYS-PUB-2019-005}
{ATLAS Collaboration}, {\em {Expected performance of the ATLAS detector at the
  High-Luminosity LHC}\/},   ATL-PHYS-PUB-2019-005, CERN, Geneva, Jan, 2019.
\newblock \url{http://cds.cern.ch/record/2655304}.

\bibitem{Buras:2013ooa}
A.~J. Buras and J.~Girrbach, {\em {Towards the Identification of New Physics
  through Quark Flavour Violating Processes}\/},
  \href{http://dx.doi.org/10.1088/0034-4885/77/8/086201}{Rept. Prog. Phys. {\bf
  77} (2014)  086201},
\href{http://arxiv.org/abs/1306.3775}{{\tt arXiv:1306.3775 [hep-ph]}}.

\bibitem{CMS:TPMTD}
{The CMS Collaboration}, {\em { TECHNICAL PROPOSAL FOR A MIP TIMING DETECTOR IN
  THE CMS EXPERIMENT PHASE 2 UPGRADE }\/},   CERN-LHCC-2017-027 ; LHCC-P-009,
  CERN, Geneva, 2017.
\newblock \url{https://cds.cern.ch/record/2296612}.

\bibitem{LHCb-PAPER-2014-024}
{LHCb collaboration}, R.~Aaij et al., {\em {Test of lepton universality using
  $\Bp\to \Kp\ellell$ decays}\/},  Phys. Rev. Lett. {\bf 113} (2014)  151601,
  \href{http://arxiv.org/abs/1406.6482}{{\tt arXiv:1406.6482 [hep-ex]}}.

\bibitem{LHCb-PAPER-2017-013}
{LHCb collaboration}, R.~Aaij et al., {\em {Test of lepton universality with
  $\Bz\to K^{*0} \ell^+ \ell^-$ decays}\/},  JHEP {\bf 08} (2017)  055,
  \href{http://arxiv.org/abs/1705.05802}{{\tt arXiv:1705.05802 [hep-ex]}}.

\bibitem{LHCb-PAPER-2017-047}
{LHCb collaboration}, R.~Aaij et al., {\em {Measurement of \CP asymmetry in
  $\Bs\to D_s^\mp K^\pm$ decays}\/},  JHEP {\bf 03} (2018)  059,
  \href{http://arxiv.org/abs/1712.07428}{{\tt arXiv:1712.07428 [hep-ex]}}.

\bibitem{LHCb-CONF-2018-002}
{{LHCb collaboration}}, {\em {Update of the LHCb combination of the CKM angle
  $\gamma$ using $B \to DK$ decays}\/},  {May}, {2018}.

\bibitem{LHCb-PAPER-2017-029}
{LHCb collaboration}, R.~Aaij et al., {\em {Measurement of \CP violation in
  $\Bz\to \jpsi K_S^0$ and $\Bz\to \psitwos K_S^0$ decays}\/},  JHEP {\bf 11}
  (2017)  170, \href{http://arxiv.org/abs/1709.03944}{{\tt arXiv:1709.03944
  [hep-ex]}}.

\bibitem{LHCb-PAPER-2014-059}
{LHCb collaboration}, R.~Aaij et al., {\em {Precision measurement of \CP
  violation in $\Bs\to \jpsi\Kp\Km$ decays}\/},  Phys. Rev. Lett. {\bf 114}
  (2015)  041801, \href{http://arxiv.org/abs/1411.3104}{{\tt arXiv:1411.3104
  [hep-ex]}}.

\bibitem{LHCb-PAPER-2014-051}
{LHCb collaboration}, R.~Aaij et al., {\em {Measurement of the \CP-violating
  phase $\phis$ in $\Bsb\to \Dsp\Dsm$ decays}\/},  Phys. Rev. Lett. {\bf 113}
  (2014)  211801, \href{http://arxiv.org/abs/1409.4619}{{\tt arXiv:1409.4619
  [hep-ex]}}.

\bibitem{LHCb-PAPER-2014-026}
{LHCb collaboration}, R.~Aaij et al., {\em {Measurement of \CP violation in
  $\Bs\to \phiz\phiz$ decays}\/},  Phys. Rev. {\bf D90} (2014)  052011,
  \href{http://arxiv.org/abs/1407.2222}{{\tt arXiv:1407.2222 [hep-ex]}}.

\bibitem{LHCb-PAPER-2016-013}
{LHCb collaboration}, R.~Aaij et al., {\em {Measurement of the \CP asymmetry in
  $\Bs$--$\Bsb$ mixing}\/},  Phys. Rev. Lett. {\bf 117} (2016)  061803,
  \href{http://arxiv.org/abs/1605.09768}{{\tt arXiv:1605.09768 [hep-ex]}}.

\bibitem{LHCb-PAPER-2015-013}
{LHCb collaboration}, R.~Aaij et al., {\em {Determination of the quark coupling
  strength $|\Vub|$ using baryonic decays}\/},  Nature Physics {\bf 11} (2015)
  743, \href{http://arxiv.org/abs/1504.01568}{{\tt arXiv:1504.01568 [hep-ex]}}.

\bibitem{LHCb-PAPER-2017-001}
{LHCb collaboration}, R.~Aaij et al., {\em {Measurement of the $\Bs\to
  \mup\mun$ branching fraction and effective lifetime and search for $\Bd\to
  \mup\mun$ decays}\/},  Phys. Rev. Lett. {\bf 118} (2017)  191801,
  \href{http://arxiv.org/abs/1703.05747}{{\tt arXiv:1703.05747 [hep-ex]}}.

\bibitem{LHCb-PAPER-2015-025}
{LHCb collaboration}, R.~Aaij et al., {\em {Measurement of the ratio of
  branching fractions $\BF(\Bzb\to \Dstarp\taum\neutb)/\BF(\Bzb\to
  \Dstarp\mun\neumb)$}\/},  Phys. Rev. Lett. {\bf 115} (2015)  111803,
  \href{http://arxiv.org/abs/1506.08614}{{\tt arXiv:1506.08614 [hep-ex]}}.

\bibitem{LHCb-PAPER-2017-027}
{LHCb collaboration}, R.~Aaij et al., {\em {Test of lepton flavor universality
  by the measurement of the $\Bz\to D^{\ast-} \tau^+ \nu_{\tau}$ branching
  fraction using three-prong $\tau$ decays}\/},  Phys. Rev. {\bf D97} (2018)
  072013, \href{http://arxiv.org/abs/1711.02505}{{\tt arXiv:1711.02505
  [hep-ex]}}.

\bibitem{LHCb-PAPER-2017-035}
{LHCb collaboration}, R.~Aaij et al., {\em {Measurement of the ratio of
  branching fractions $\mathcal{B}(B^+_c\to
  \jpsi\tau^+\nu_{\tau})$/$\mathcal{B}(B^+_c\to \jpsi\mu^+\nu_{\mu}$)}\/},
  Phys. Rev. Lett. {\bf 120} (2018)  121801,
  \href{http://arxiv.org/abs/1711.05623}{{\tt arXiv:1711.05623 [hep-ex]}}.

\bibitem{LHCb-PAPER-2015-055}
{LHCb collaboration}, R.~Aaij et al., {\em {Measurement of the difference of
  time-integrated \CP asymmetries in $\Dz\to \Km\Kp$ and $\Dz\to \pim\pip$
  decays}\/},  Phys. Rev. Lett. {\bf 116} (2016)  191601,
  \href{http://arxiv.org/abs/1602.03160}{{\tt arXiv:1602.03160 [hep-ex]}}.

\bibitem{LHCb-PAPER-2016-063}
{LHCb collaboration}, R.~Aaij et al., {\em {Measurement of the \CP violation
  parameter $A_\Gamma$ in $\Dz\to \Kp\Km$ and $\Dz\to \pip\pim$ decays}\/},
  Phys. Rev. Lett. {\bf 118} (2017)  261803,
  \href{http://arxiv.org/abs/1702.06490}{{\tt arXiv:1702.06490 [hep-ex]}}.

\bibitem{LHCb-PAPER-2017-046}
{LHCb collaboration}, R.~Aaij et al., {\em {Updated determination of
  $\Dz$-$\Dbar^0$ mixing and \CP violation parameters with $\Dz\to K^+\pi^-$
  decays}\/},  Phys. Rev. {\bf D97} (2018)  031101,
  \href{http://arxiv.org/abs/1712.03220}{{\tt arXiv:1712.03220 [hep-ex]}}.

\bibitem{Bediaga:2018lhg}
{LHCb Collaboration}, R.~Aaij et al., {\em {Physics case for an LHCb Upgrade II
  - Opportunities in flavour physics, and beyond, in the HL-LHC era}\/},
\href{http://arxiv.org/abs/1808.08865}{{\tt arXiv:1808.08865}}.

\bibitem{Cabibbo:1963yz}
N.~Cabibbo, {\em {Unitary Symmetry and Leptonic Decays}\/},
  \href{http://dx.doi.org/10.1103/PhysRevLett.10.531}{Phys. Rev. Lett. {\bf 10}
  (1963)  531--533}.

\bibitem{Kobayashi:1973fv}
M.~Kobayashi and T.~Maskawa, {\em {CP Violation in the Renormalizable Theory of
  Weak Interaction}\/},
\href{http://dx.doi.org/10.1143/PTP.49.652}{Prog. Theor. Phys. {\bf 49} (1973)
  652--657}.

\bibitem{Wolfenstein:1983yz}
L.~Wolfenstein, {\em {Parametrization of the Kobayashi-Maskawa Matrix}\/},
\href{http://dx.doi.org/10.1103/PhysRevLett.51.1945}{Phys. Rev. Lett. {\bf 51}
  (1983)  1945}.

\bibitem{Antonelli:2010yf}
{FlaviaNet Working Group on Kaon Decays Collaboration}, M.~Antonelli et al.,
  {\em {An Evaluation of $|V_{us}|$ and precise tests of the Standard Model
  from world data on leptonic and semileptonic kaon decays}\/},
  \href{http://dx.doi.org/10.1140/epjc/s10052-010-1406-3}{Eur. Phys. J. {\bf
  C69} (2010)  399--424},
\href{http://arxiv.org/abs/1005.2323}{{\tt arXiv:1005.2323 [hep-ph]}}.

\bibitem{Hardy:2014qxa}
J.~C. Hardy and I.~S. Towner, {\em {Superallowed $0^+\to 0^+$ nuclear decays:
  2014 critical survey, with precise results for $V_{ud}$ and CKM
  unitarity}\/},  \href{http://dx.doi.org/10.1103/PhysRevC.91.025501}{Phys.
  Rev. {\bf C91} (2015) no.~2, 025501},
\href{http://arxiv.org/abs/1411.5987}{{\tt arXiv:1411.5987 [nucl-ex]}}.

\bibitem{Towner:2015woa}
I.~S. Towner and J.~C. Hardy, {\em {Theoretical corrections and world data for
  the superallowed ft values in the $\beta$ decays of $^{42}Ti, ^{46}Cr,
  ^{50}Fe$ and $^{54}Ni$}\/},
  \href{http://dx.doi.org/10.1103/PhysRevC.92.055505}{Phys. Rev. {\bf C92}
  (2015) no.~5, 055505},
\href{http://arxiv.org/abs/1510.03793}{{\tt arXiv:1510.03793 [nucl-th]}}.

\bibitem{Hardy:2018zsb}
J.~C. Hardy and I.~S. Towner, {\em {Nuclear beta decays and CKM unitarity}\/},
  in {\em {13th Conference on the Intersections of Particle and Nuclear Physics
  (CIPANP 2018) Palm Springs, California, USA, May 29-June 3, 2018}}.
\newblock 2018.
\newblock
\href{http://arxiv.org/abs/1807.01146}{{\tt arXiv:1807.01146 [nucl-ex]}}.
\newblock

\bibitem{Abdesselam:2018nnh}
{Belle Collaboration}, A.~Abdesselam et al., {\em {Measurement of CKM Matrix
  Element $|V_{cb}|$ from $\bar{B} \to D^{*+} \ell^{-} \bar{\nu}_\ell$}\/},
\href{http://arxiv.org/abs/1809.03290}{{\tt arXiv:1809.03290 [hep-ex]}}.

\bibitem{Bigi:2017njr}
D.~Bigi, P.~Gambino, and S.~Schacht, {\em {A fresh look at the determination of
  $|V_{cb}|$ from $B\to D^{*} \ell \nu$}\/},
  \href{http://dx.doi.org/10.1016/j.physletb.2017.04.022}{Phys. Lett. {\bf
  B769} (2017)  441--445},
\href{http://arxiv.org/abs/1703.06124}{{\tt arXiv:1703.06124 [hep-ph]}}.

\bibitem{Grinstein:2017nlq}
B.~Grinstein and A.~Kobach, {\em {Model-independent extraction of $|V_{cb}|$
  from $\bar{B}\rightarrow D^* \ell \overline{\nu}$}\/},
  \href{http://dx.doi.org/10.1016/j.physletb.2017.05.078}{Phys. Lett. {\bf
  B771} (2017)  359--364},
\href{http://arxiv.org/abs/1703.08170}{{\tt arXiv:1703.08170 [hep-ph]}}.

\bibitem{DAgostini:1999niu}
G.~D'Agostini, {\em {Sceptical combination of experimental results: General
  considerations and application to epsilon-prime / epsilon}\/},  Submitted to:
  Phys. Rev. D (1999)  ,
\href{http://arxiv.org/abs/hep-ex/9910036}{{\tt arXiv:hep-ex/9910036
  [hep-ex]}}.

\bibitem{Gronau:1990ka}
M.~Gronau and D.~London, {\em {Isospin analysis of CP asymmetries in B
  decays}\/},
\href{http://dx.doi.org/10.1103/PhysRevLett.65.3381}{Phys. Rev. Lett. {\bf 65}
  (1990)  3381--3384}.

\bibitem{Lipkin:1991st}
H.~J. Lipkin, Y.~Nir, H.~R. Quinn, and A.~Snyder, {\em {Penguin trapping with
  isospin analysis and CP asymmetries in B decays}\/},
\href{http://dx.doi.org/10.1103/PhysRevD.44.1454}{Phys. Rev. {\bf D44} (1991)
  1454--1460}.

\bibitem{Snyder:1993mx}
A.~E. Snyder and H.~R. Quinn, {\em {Measuring CP asymmetry in B ---> rho pi
  decays without ambiguities}\/},
\href{http://dx.doi.org/10.1103/PhysRevD.48.2139}{Phys. Rev. {\bf D48} (1993)
  2139--2144}.

\bibitem{Charles:2017evz}
J.~Charles, O.~Deschamps, S.~Descotes-Genon, and V.~Niess,
  \href{http://dx.doi.org/10.1140/epjc/s10052-017-5126-9}{{\em {Isospin
  analysis of charmless \B-meson decays}\/}, Eur. Phys. J. {\bf C77} (Aug,
  2017)  574}, \href{http://arxiv.org/abs/1705.02981}{{\tt arXiv:1705.02981
  [hep-ph]}}. \url{https://doi.org/10.1140/epjc/s10052-017-5126-9}.

\bibitem{CKMfitterwebsite}
{CKMfitter collaboration}, {\em Summer 2018 update available on
  http://ckmfitter.in2p3.fr/\/}, .

\bibitem{Gronau:2005pq}
M.~Gronau and J.~Zupan, {\em {Isospin-breaking effects on alpha extracted in
  $B\to \pi \pi, \rho \rho, \rho \pi$}\/},
  \href{http://dx.doi.org/10.1103/PhysRevD.71.074017}{Phys. Rev. {\bf D71}
  (2005)  074017},
\href{http://arxiv.org/abs/hep-ph/0502139}{{\tt arXiv:hep-ph/0502139
  [hep-ph]}}.

\bibitem{Fleischer:1999nz}
R.~Fleischer, {\em {Extracting $\gamma$ from $B(s/d) \to J/\psi K_{S}$ and
  $B(d/s) \to D^+(d/s) D^-(d/s)$}\/},
  \href{http://dx.doi.org/10.1007/s100529900099}{Eur. Phys. J. {\bf C10} (1999)
   299--306},
\href{http://arxiv.org/abs/hep-ph/9903455}{{\tt arXiv:hep-ph/9903455
  [hep-ph]}}.

\bibitem{Ciuchini:2005mg}
M.~Ciuchini, M.~Pierini, and L.~Silvestrini, {\em {The effect of penguins in
  the $\Bd \to \jpsi \Kz$ \CP\ asymmetry}\/},
  \href{http://dx.doi.org/10.1103/PhysRevLett.95.221804}{Phys. Rev. Lett. {\bf
  95} (2005)  221804},
\href{http://arxiv.org/abs/hep-ph/0507290}{{\tt arXiv:hep-ph/0507290
  [hep-ph]}}.

\bibitem{Faller:2008zc}
S.~Faller, M.~Jung, R.~Fleischer, and T.~Mannel, {\em {The golden modes $\Bd
  \to J/\psi K_{\rm S,L}$ in the era of precision flavour physics}\/},
  \href{http://dx.doi.org/10.1103/PhysRevD.79.014030}{Phys. Rev. {\bf D79}
  (2009)  014030},
\href{http://arxiv.org/abs/0809.0842}{{\tt arXiv:0809.0842 [hep-ph]}}.

\bibitem{Ciuchini:2011kd}
M.~Ciuchini, M.~Pierini, and L.~Silvestrini, {\em {Theoretical uncertainty in
  sin $2\beta$: An Update}\/},  in {\em {CKM unitarity triangle. Proceedings,
  6th International Workshop, CKM 2010, Warwick, UK, September 6-10, 2010}}.
\newblock 2011.
\newblock
\href{http://arxiv.org/abs/1102.0392}{{\tt arXiv:1102.0392 [hep-ph]}}.
\newblock

\bibitem{Jung:2012mp}
M.~Jung, {\em {Determining weak phases from $B\to J/\psi P$ decays}\/},
  \href{http://dx.doi.org/10.1103/PhysRevD.86.053008}{Phys. Rev. {\bf D86}
  (2012)  053008},
\href{http://arxiv.org/abs/1206.2050}{{\tt arXiv:1206.2050 [hep-ph]}}.

\bibitem{DeBruyn:2014oga}
K.~De~Bruyn and R.~Fleischer, {\em {A roadmap to control penguin effects in
  $B^0_d\to J/\psi K_{\rm S}^0$ and $B^0_s\to J/\psi \phi$}\/},
  \href{http://dx.doi.org/10.1007/JHEP03(2015)145}{JHEP {\bf 1503} (2015)
  145},
\href{http://arxiv.org/abs/1412.6834}{{\tt arXiv:1412.6834 [hep-ph]}}.

\bibitem{Frings:2015eva}
P.~Frings, U.~Nierste, and M.~Wiebusch, {\em {Penguin contributions to \CP
  phases in $B_{d,s}$ decays to charmonium}\/},
\href{http://arxiv.org/abs/1503.00859}{{\tt arXiv:1503.00859 [hep-ph]}}.

\bibitem{Gronau:2002mu}
M.~Gronau, {\em {Improving bounds on γ in $B^\pm \to DK^\pm$ and $B^{\pm,0}
  → DX_s^{\pm,0}$}\/},
  \href{http://dx.doi.org/10.1016/S0370-2693(03)00192-8}{Phys. Lett. {\bf B557}
  (2003)  198--206},
\href{http://arxiv.org/abs/hep-ph/0211282}{{\tt arXiv:hep-ph/0211282
  [hep-ph]}}.

\bibitem{Gronau:1990ra}
M.~Gronau and D.~London, {\em {How to determine all the angles of the unitarity
  triangle from $B_{(d)}^0 \to D K_{(s)}$ and $B_{(s)}^0 \to D^0 \Phi$}\/},
\href{http://dx.doi.org/10.1016/0370-2693(91)91756-L}{Phys. Lett. {\bf B253}
  (1991)  483--488}.

\bibitem{Gronau:1991dp}
M.~Gronau and D.~Wyler, {\em {On determining a weak phase from $CP$ asymmetries
  in charged $B$ decays}\/},
\href{http://dx.doi.org/10.1016/0370-2693(91)90034-N}{Phys. Lett. {\bf B265}
  (1991)  172--176}.

\bibitem{Atwood:1994zm}
D.~Atwood, G.~Eilam, M.~Gronau, and A.~Soni, {\em {Enhancement of CP violation
  in $B^\pm -\to K_i^\pm D^0$ by resonant effects}\/},
  \href{http://dx.doi.org/10.1016/0370-2693(94)01317-6,
  10.1016/0370-2693(95)80017-R}{Phys. Lett. {\bf B341} (1995)  372--378},
\href{http://arxiv.org/abs/hep-ph/9409229}{{\tt arXiv:hep-ph/9409229
  [hep-ph]}}.

\bibitem{Atwood:1996ci}
D.~Atwood, I.~Dunietz, and A.~Soni, {\em {Enhanced CP violation with $B \to K
  \Dz (\Dzb)$ modes and extraction of the CKM angle \g}\/},
  \href{http://dx.doi.org/10.1103/PhysRevLett.78.3257}{Phys.Rev.Lett. {\bf 78}
  (1997)  3257},
\href{http://arxiv.org/abs/hep-ph/9612433}{{\tt arXiv:hep-ph/9612433
  [hep-ph]}}.

\bibitem{Atwood:2000ck}
D.~Atwood, I.~Dunietz, and A.~Soni, {\em {Improved methods for observing CP
  violation in $B^\pm \to K D$ and measuring the CKM phase gamma}\/},
  \href{http://dx.doi.org/10.1103/PhysRevD.63.036005}{Phys. Rev. {\bf D63}
  (2001)  036005},
\href{http://arxiv.org/abs/hep-ph/0008090}{{\tt arXiv:hep-ph/0008090
  [hep-ph]}}.

\bibitem{Giri:2003ty}
A.~Giri, Y.~Grossman, A.~Soffer, and J.~Zupan, {\em {Determining gamma using
  $B^\pm \to DK^\pm$ with multibody $D$ decays}\/},
  \href{http://dx.doi.org/10.1103/PhysRevD.68.054018}{Phys. Rev. {\bf D68}
  (2003)  054018},
\href{http://arxiv.org/abs/hep-ph/0303187}{{\tt arXiv:hep-ph/0303187
  [hep-ph]}}.

\bibitem{Brod:2013sga}
J.~Brod and J.~Zupan, {\em {The ultimate theoretical error on $\gamma$ from $B
  \to DK$ decays}\/},  \href{http://dx.doi.org/10.1007/JHEP01(2014)051}{JHEP
  {\bf 01} (2014)  051},
\href{http://arxiv.org/abs/1308.5663}{{\tt arXiv:1308.5663 [hep-ph]}}.

\bibitem{Buras:2010pza}
A.~J. Buras, D.~Guadagnoli, and G.~Isidori, {\em {On $\epsilon_K$ Beyond Lowest
  Order in the Operator Product Expansion}\/},
  \href{http://dx.doi.org/10.1016/j.physletb.2010.04.017}{Phys. Lett. {\bf
  B688} (2010)  309--313},
\href{http://arxiv.org/abs/1002.3612}{{\tt arXiv:1002.3612 [hep-ph]}}.

\bibitem{Bai:2014cva}
Z.~Bai, N.~H. Christ, T.~Izubuchi, C.~T. Sachrajda, A.~Soni, and J.~Yu, {\em
  {$K_L-K_S$ Mass Difference from Lattice QCD}\/},
  \href{http://dx.doi.org/10.1103/PhysRevLett.113.112003}{Phys. Rev. Lett. {\bf
  113} (2014)  112003},
\href{http://arxiv.org/abs/1406.0916}{{\tt arXiv:1406.0916 [hep-lat]}}.

\bibitem{Wang:2018csg}
B.~Wang, {\em {Results for the mass difference between the long- and short-
  lived K mesons for physical quark masses}\/},
\newblock 2018.
\newblock
\href{http://arxiv.org/abs/1812.05302}{{\tt arXiv:1812.05302 [hep-lat]}}.
\newblock

\bibitem{Inami:1980fz}
T.~Inami and C.~S. Lim, {\em {Effects of Superheavy Quarks and Leptons in
  Low-Energy Weak Processes $K_{L} \to \mu \bar\mu$, $K^+\to \pi^+ \nu\bar \nu$
  and $K^0 - \bar K^0$}\/},  \href{http://dx.doi.org/10.1143/PTP.65.297}{Prog.
  Theor. Phys. {\bf 65} (1981)  297}.
[Erratum: Prog. Theor. Phys.65,1772(1981)].

\bibitem{Buras:1990fn}
A.~J. Buras, M.~Jamin, and P.~H. Weisz, {\em {Leading and Next-to-leading {QCD}
  Corrections to $\epsilon$ Parameter and $B^0 - \bar{B}^0$ Mixing in the
  Presence of a Heavy Top Quark}\/},
\href{http://dx.doi.org/10.1016/0550-3213(90)90373-L}{Nucl. Phys. {\bf B347}
  (1990)  491--536}.

\bibitem{Lenz:2010gu}
A.~Lenz, U.~Nierste, J.~Charles, S.~Descotes-Genon, A.~Jantsch, C.~Kaufhold,
  H.~Lacker, S.~Monteil, V.~Niess, and S.~T'Jampens, {\em {Anatomy of New
  Physics in $B - \bar{B}$ mixing}\/},
  \href{http://dx.doi.org/10.1103/PhysRevD.83.036004}{Phys. Rev. {\bf D83}
  (2011)  036004},
\href{http://arxiv.org/abs/1008.1593}{{\tt arXiv:1008.1593 [hep-ph]}}.

\bibitem{Charles:2004jd}
{CKMfitter Group Collaboration}, J.~Charles, A.~Hocker, H.~Lacker, S.~Laplace,
  F.~R. Le~Diberder, J.~Malcles, J.~Ocariz, M.~Pivk, and L.~Roos, {\em {CP
  violation and the CKM matrix: Assessing the impact of the asymmetric $B$
  factories}\/},  \href{http://dx.doi.org/10.1140/epjc/s2005-02169-1}{Eur.
  Phys. J. {\bf C41} (2005) no.~1, 1--131},
\href{http://arxiv.org/abs/hep-ph/0406184}{{\tt arXiv:hep-ph/0406184
  [hep-ph]}}.

\bibitem{Charles:2011va}
J.~Charles et al., {\em {Predictions of selected flavor observables within the
  Standard Model}\/},
  \href{http://dx.doi.org/10.1103/PhysRevD.84.033005}{Phys. Rev. {\bf D84}
  (2011)  033005}, \href{http://arxiv.org/abs/1106.4041}{{\tt arXiv:1106.4041
  [hep-ph]}}.

\bibitem{Charles:2015gya}
J.~Charles et al., {\em {Current status of the Standard Model CKM fit and
  constraints on $\Delta F=2$ New Physics}\/},
  \href{http://dx.doi.org/10.1103/PhysRevD.91.073007}{Phys. Rev. {\bf D91}
  (2015)  073007},
\href{http://arxiv.org/abs/1501.05013}{{\tt arXiv:1501.05013 [hep-ph]}}.

\bibitem{Hocker:2001xe}
A.~Hocker, H.~Lacker, S.~Laplace, and F.~Le~Diberder, {\em {A New approach to a
  global fit of the CKM matrix}\/},
  \href{http://dx.doi.org/10.1007/s100520100729}{Eur. Phys. J. {\bf C21} (2001)
   225--259},
\href{http://arxiv.org/abs/hep-ph/0104062}{{\tt arXiv:hep-ph/0104062
  [hep-ph]}}.

\bibitem{Charles:2016qtt}
J.~Charles, S.~Descotes-Genon, V.~Niess, and L.~Vale~Silva, {\em {Modeling
  theoretical uncertainties in phenomenological analyses for particle
  physics}\/},  \href{http://dx.doi.org/10.1140/epjc/s10052-017-4767-z}{Eur.
  Phys. J. {\bf C77} (2017) no.~4, 214},
\href{http://arxiv.org/abs/1611.04768}{{\tt arXiv:1611.04768 [hep-ph]}}.

\bibitem{Aoki:2016frl}
S.~Aoki et al., {\em {Review of lattice results concerning low-energy particle
  physics}\/},  Eur. Phys. J. {\bf C77} (2017) no.~2, 112,
  \href{http://arxiv.org/abs/1607.00299}{{\tt arXiv:1607.00299 [hep-lat]}}.

\bibitem{Ciuchini:2000de}
M.~Ciuchini, G.~D'Agostini, E.~Franco, V.~Lubicz, G.~Martinelli, F.~Parodi,
  P.~Roudeau, and A.~Stocchi, {\em {2000 CKM triangle analysis: A Critical
  review with updated experimental inputs and theoretical parameters}\/},
  \href{http://dx.doi.org/10.1088/1126-6708/2001/07/013}{JHEP {\bf 07} (2001)
  013},
\href{http://arxiv.org/abs/hep-ph/0012308}{{\tt arXiv:hep-ph/0012308
  [hep-ph]}}.

\bibitem{UTfitwebsite18}
{UTfit collaboration}, {\em Summer 2018 update available at
  http://www.utfit.org/UTfit/ResultsSummer2018\/}, .

\bibitem{Bernlochner:2017xyx}
F.~U. Bernlochner, Z.~Ligeti, M.~Papucci, and D.~J. Robinson, {\em {Tensions
  and correlations in $|V_{cb}|$ determinations}\/},
  \href{http://dx.doi.org/10.1103/PhysRevD.96.091503}{Phys. Rev. {\bf D96}
  (2017) no.~9, 091503},
\href{http://arxiv.org/abs/1708.07134}{{\tt arXiv:1708.07134 [hep-ph]}}.

\bibitem{Bigi:2017jbd}
D.~Bigi, P.~Gambino, and S.~Schacht, {\em {$R(D^*)$, $|V_{cb}|$, and the Heavy
  Quark Symmetry relations between form factors}\/},  JHEP {\bf 11} (2017)
  061, \href{http://arxiv.org/abs/1707.09509}{{\tt arXiv:1707.09509 [hep-ph]}}.

\bibitem{Detmold:2015aaa}
W.~Detmold, C.~Lehner, and S.~Meinel, {\em {$\Lambdares_b \to p \ell^-
  \bar{\nu}_\ell$ and $\Lambdares_b \to \Lambdares_c \ell^- \bar{\nu}_\ell$
  form factors from lattice QCD with relativistic heavy quarks}\/},
  \href{http://dx.doi.org/10.1103/PhysRevD.92.034503}{Phys. Rev. {\bf D92}
  (2015)  034503},
\href{http://arxiv.org/abs/1503.01421}{{\tt arXiv:1503.01421 [hep-lat]}}.

\bibitem{Flynn:2015mha}
J.~M. Flynn, T.~Izubuchi, T.~Kawanai, C.~Lehner, A.~Soni, R.~S. Van~de Water,
  and O.~Witzel, {\em {$B \to \pi \ell \nu$ and $B_s \to K \ell \nu$ form
  factors and $|V_{ub}|$ from 2+1-flavor lattice QCD with domain-wall light
  quarks and relativistic heavy quarks}\/},
  \href{http://dx.doi.org/10.1103/PhysRevD.91.074510}{Phys. Rev. {\bf D91}
  (2015)  074510},
\href{http://arxiv.org/abs/1501.05373}{{\tt arXiv:1501.05373 [hep-lat]}}.

\bibitem{Ciezarek:2016lqu}
G.~Ciezarek, A.~Lupato, M.~Rotondo, and M.~Vesterinen, {\em {Reconstruction of
  semileptonically decaying beauty hadrons produced in high energy pp
  collisions}\/},  \href{http://dx.doi.org/10.1007/JHEP02(2017)021}{JHEP {\bf
  02} (2017)  021},
\href{http://arxiv.org/abs/1611.08522}{{\tt arXiv:1611.08522 [hep-ex]}}.

\bibitem{Beneke:1999br}
M.~Beneke, G.~Buchalla, M.~Neubert, and C.~T. Sachrajda, {\em {QCD
  factorization for B ---> pi pi decays: Strong phases and CP violation in the
  heavy quark limit}\/},
  \href{http://dx.doi.org/10.1103/PhysRevLett.83.1914}{Phys. Rev. Lett. {\bf
  83} (1999)  1914--1917},
\href{http://arxiv.org/abs/hep-ph/9905312}{{\tt arXiv:hep-ph/9905312
  [hep-ph]}}.

\bibitem{Beneke:2000ry}
M.~Beneke, G.~Buchalla, M.~Neubert, and C.~T. Sachrajda, {\em {QCD
  factorization for exclusive, nonleptonic B meson decays: General arguments
  and the case of heavy light final states}\/},
  \href{http://dx.doi.org/10.1016/S0550-3213(00)00559-9}{Nucl. Phys. {\bf B591}
  (2000)  313--418},
\href{http://arxiv.org/abs/hep-ph/0006124}{{\tt arXiv:hep-ph/0006124
  [hep-ph]}}.

\bibitem{Buras:2012ru}
A.~J. Buras, J.~Girrbach, D.~Guadagnoli, and G.~Isidori, {\em {On the Standard
  Model prediction for $\BR(\B_{s,d}\to\mu^+\mu^-)$}\/},
  \href{http://dx.doi.org/10.1140/epjc/s10052-012-2172-1}{Eur. Phys. J. {\bf
  C72} (2012)  2172},
\href{http://arxiv.org/abs/1208.0934}{{\tt arXiv:1208.0934 [hep-ph]}}.

\bibitem{LHCb-PAPER-2017-016}
{LHCb collaboration}, R.~Aaij et al., {\em {Measurement of the shape of the
  $\Lb\to \Lc\mu^{-}\neumb$ differential decay rate}\/},  Phys. Rev. {\bf D96}
  (2017)  112005, \href{http://arxiv.org/abs/1709.01920}{{\tt arXiv:1709.01920
  [hep-ex]}}.

\bibitem{Bigi2011}
I.~I. Bigi, T.~Mannel, and N.~Uraltsev,
  \href{http://dx.doi.org/10.1007/JHEP09(2011)012}{{\em Semileptonic width
  ratios among beauty hadrons\/}, Journal of High Energy Physics {\bf 2011}
  (Sep, 2011)  12}. \url{https://doi.org/10.1007/JHEP09(2011)012}.

\bibitem{LHCb-PAPER-2016-003}
{LHCb collaboration}, R.~Aaij et al., {\em {Measurement of \CP observables in
  $\Bpm\to \D \Kpm$ and $\Bpm\to \D\pipm$ with two- and four-body $\D$
  decays}\/},  Phys. Lett. {\bf B760} (2016)  117,
  \href{http://arxiv.org/abs/1603.08993}{{\tt arXiv:1603.08993 [hep-ex]}}.

\bibitem{LHCb-PAPER-2017-021}
{LHCb collaboration}, R.~Aaij et al., {\em {Measurement of \CP observables in
  $\Bpm\to D^{(\ast)}K^\pm$ and $\Bpm\to D^{(\ast)}\pi^\pm$ decays}\/},  Phys.
  Lett. {\bf B777} (2017)  16, \href{http://arxiv.org/abs/1708.06370}{{\tt
  arXiv:1708.06370 [hep-ex]}}.

\bibitem{LHCb-PAPER-2017-030}
{LHCb collaboration}, R.~Aaij et al., {\em {Measurement of \CP observables in
  $\Bpm\to DK^{\ast \pm}$ decays using two- and four-body $D$-meson final
  states}\/},  JHEP {\bf 11} (2017)  156,
  \href{http://arxiv.org/abs/1709.05855}{{\tt arXiv:1709.05855 [hep-ex]}}.

\bibitem{LHCb-PAPER-2014-028}
{LHCb collaboration}, R.~Aaij et al., {\em {Measurement of \CP violation
  parameters in $\Bz\to \D\Kstarz$ decays}\/},  Phys. Rev. {\bf D90} (2014)
  112002, \href{http://arxiv.org/abs/1407.8136}{{\tt arXiv:1407.8136
  [hep-ex]}}.

\bibitem{LHCb-PUB-2016-025}
S.~S. Malde, {\em {Synergy of BESIII and LHCb physics programmes}\/},  Oct,
  2016.
\newblock \url{https://cds.cern.ch/record/2223391}.

\bibitem{LHCb-PAPER-2015-014}
{LHCb collaboration}, R.~Aaij et al., {\em {A study of \CP violation in
  $\Bmp\to \D h^\mp$ $(h=K,\pi)$ with the modes $\D\to \Kmp\pipm\piz$, $\D\to
  \pip\pim\piz$ and $\D\to \Kp\Km\piz$}\/},  Phys. Rev. {\bf D91} (2015)
  112014, \href{http://arxiv.org/abs/1504.05442}{{\tt arXiv:1504.05442
  [hep-ex]}}.

\bibitem{Bondar:2004bi}
A.~Bondar and T.~Gershon, {\em {On $\phi_3$ measurements using $\Bm\to D^* K^-$
  decays}\/},  \href{http://dx.doi.org/10.1103/PhysRevD.70.091503}{Phys. Rev.
  {\bf D70} (2004)  091503},
\href{http://arxiv.org/abs/hep-ph/0409281}{{\tt arXiv:hep-ph/0409281
  [hep-ph]}}.

\bibitem{K:2017qxf}
P.~K. Resmi, J.~Libby, S.~Malde, and G.~Wilkinson, {\em {Quantum-correlated
  measurements of $D\to K^{0}_{\rm S}\pi^{+}\pi^{-}\pi^{0}$ decays and
  consequences for the determination of the CKM angle $\gamma$}\/},
  \href{http://dx.doi.org/10.1007/JHEP01(2018)082}{JHEP {\bf 01} (2018)  082},
\href{http://arxiv.org/abs/1710.10086}{{\tt arXiv:1710.10086 [hep-ex]}}.

\bibitem{LHCb-PAPER-2015-059}
{LHCb collaboration}, R.~Aaij et al., {\em {Constraints on the unitarity
  triangle angle $\gamma$ from Dalitz plot analysis of $\Bz\to \D\Kp\pim$
  decays}\/},  Phys. Rev. {\bf D93} (2016)  112018,
  \href{http://arxiv.org/abs/1602.03455}{{\tt arXiv:1602.03455 [hep-ex]}}.

\bibitem{LHCb-PAPER-2015-017}
{LHCb collaboration}, R.~Aaij et al., {\em {Amplitude analysis of $\Bz\to
  \Dzb\Kp\pim$ decays}\/},  Phys. Rev. {\bf D92} (2015)  012012,
  \href{http://arxiv.org/abs/1505.01505}{{\tt arXiv:1505.01505 [hep-ex]}}.

\bibitem{Gershon:2016fda}
T.~Gershon and V.~V. Gligorov, {\em {$CP$ violation in the $B$ system}\/},
  \href{http://dx.doi.org/10.1088/1361-6633/aa5514}{Rept. Prog. Phys. {\bf 80}
  (2017)  046201},
\href{http://arxiv.org/abs/1607.06746}{{\tt arXiv:1607.06746 [hep-ex]}}.

\bibitem{Gershon:2008pe}
T.~Gershon, {\em {On the measurement of the Unitarity Triangle angle $\gamma$
  from $B^0 \to D K^{\ast 0}$ decays}\/},
  \href{http://dx.doi.org/10.1103/PhysRevD.79.051301}{Phys. Rev. {\bf D79}
  (2009)  051301},
\href{http://arxiv.org/abs/0810.2706}{{\tt arXiv:0810.2706 [hep-ph]}}.

\bibitem{Craik:2017dpc}
D.~Craik, T.~Gershon, and A.~Poluektov, {\em {Optimising sensitivity to
  $\gamma$ with $B^0 \to DK^+\pi^-$, $D \to K_{\rm S}^0\pi^+\pi^-$ double
  Dalitz plot analysis}\/},
  \href{http://dx.doi.org/10.1103/PhysRevD.97.056002}{Phys. Rev. {\bf D97}
  (2018)  056002},
\href{http://arxiv.org/abs/1712.07853}{{\tt arXiv:1712.07853 [hep-ph]}}.

\bibitem{LHCb-PAPER-2015-020}
{LHCb collaboration}, R.~Aaij et al., {\em {Study of $\Bm\to \D\Km\pip\pim$ and
  $\Bm\to \D\pim\pip\pim$ decays and determination of the CKM angle
  $\gamma$}\/},  Phys. Rev. {\bf D92} (2015)  112005,
  \href{http://arxiv.org/abs/1505.07044}{{\tt arXiv:1505.07044 [hep-ex]}}.

\bibitem{LHCb-PAPER-2017-040}
{LHCb collaboration}, R.~Aaij et al., {\em {Studies of the resonance structure
  in $\Dz\to K^{\mp}\pi^{\pm}\pi^+\pi^-$ decays}\/},  Eur. Phys. J. {\bf C78}
  (2018)  443, \href{http://arxiv.org/abs/1712.08609}{{\tt arXiv:1712.08609
  [hep-ex]}}.

\bibitem{Harnew:2017tlp}
S.~Harnew, P.~Naik, C.~Prouve, J.~Rademacker, and D.~Asner, {\em
  {Model-independent determination of the strong phase difference between $D^0$
  and $\bar{D}^0 \to\pi^+\pi^-\pi^+\pi^-$ amplitudes}\/},
  \href{http://dx.doi.org/10.1007/JHEP01(2018)144}{JHEP {\bf 01} (2018)  144},
\href{http://arxiv.org/abs/1709.03467}{{\tt arXiv:1709.03467 [hep-ex]}}.

\bibitem{LHCb-PAPER-2013-068}
{LHCb collaboration}, R.~Aaij et al., {\em {A study of \CP violation in
  $\Bpm\to \D\Kpm$ and $\Bpm\to \D\pipm$ decays with $\D\to \KS\Kpm\pimp$ final
  states}\/},  Phys. Lett. {\bf B733} (2014)  36,
  \href{http://arxiv.org/abs/1402.2982}{{\tt arXiv:1402.2982 [hep-ex]}}.

\bibitem{Ciuchini:2006kv}
M.~Ciuchini, M.~Pierini, and L.~Silvestrini, {\em {New bounds on the CKM matrix
  from $B \to K \pi \pi$ Dalitz plot analyses}\/},
  \href{http://dx.doi.org/10.1103/PhysRevD.74.051301}{Phys. Rev. {\bf D74}
  (2006)  051301},
\href{http://arxiv.org/abs/hep-ph/0601233}{{\tt arXiv:hep-ph/0601233
  [hep-ph]}}.

\bibitem{Ciuchini:2006st}
M.~Ciuchini, M.~Pierini, and L.~Silvestrini, {\em {Hunting the CKM weak phase
  with time-integrated Dalitz analyses of $B_s \to K \pi \pi$ decays}\/},
  \href{http://dx.doi.org/10.1016/j.physletb.2006.12.043}{Phys. Lett. {\bf
  B645} (2007)  201--203},
\href{http://arxiv.org/abs/hep-ph/0602207}{{\tt arXiv:hep-ph/0602207
  [hep-ph]}}.

\bibitem{Gronau:2006qn}
M.~Gronau, D.~Pirjol, A.~Soni, and J.~Zupan, {\em {Improved method for CKM
  constraints in charmless three-body \B and \Bs decays}\/},
  \href{http://dx.doi.org/10.1103/PhysRevD.75.014002}{Phys.Rev. {\bf D75}
  (2007)  014002},
\href{http://arxiv.org/abs/hep-ph/0608243}{{\tt arXiv:hep-ph/0608243
  [hep-ph]}}.

\bibitem{Gronau:2007vr}
M.~Gronau, D.~Pirjol, A.~Soni, and J.~Zupan, {\em {Constraint on $\bar{\rho},
  \bar{\eta}$ from $B \to K^* \pi$}\/},
  \href{http://dx.doi.org/10.1103/PhysRevD.77.057504}{Phys.Rev. {\bf D77}
  (2008)  057504},
\href{http://arxiv.org/abs/0712.3751}{{\tt arXiv:0712.3751 [hep-ph]}}.

\bibitem{Bediaga:2006jk}
I.~Bediaga, G.~Guerrer, and J.~M. de~Miranda, {\em {Extracting the quark mixing
  phase $\gamma$ from $B^\pm \to K^\pm \pi^+ \pi^-$, $B^0 \to K_S \pi^+ \pi^-$,
  and $\bar{B}^0 \to K_S \pi^+ \pi^-$}\/},
  \href{http://dx.doi.org/10.1103/PhysRevD.76.073011}{Phys.Rev. {\bf D76}
  (2007)  073011},
\href{http://arxiv.org/abs/hep-ph/0608268}{{\tt arXiv:hep-ph/0608268
  [hep-ph]}}.

\bibitem{Charles:2017ptc}
J.~Charles, S.~Descotes-Genon, J.~Ocariz, and A.~Perez~Perez, {\em
  {Disentangling weak and strong interactions in $B\to K^*(\to K\pi)\pi$
  Dalitz-plot analyses}\/},
  \href{http://dx.doi.org/10.1140/epjc/s10052-017-5133-x}{Eur. Phys. J. {\bf
  C77} (2017)  561},
\href{http://arxiv.org/abs/1704.01596}{{\tt arXiv:1704.01596 [hep-ph]}}.

\bibitem{ReyLeLorier:2011ww}
N.~Rey-Le~Lorier and D.~London, {\em {Measuring $\gamma$ with $B \to K \pi \pi$
  and $B \to K K \Kbar$ Decays}\/},
  \href{http://dx.doi.org/10.1103/PhysRevD.85.016010}{Phys. Rev. {\bf D85}
  (2012)  016010},
\href{http://arxiv.org/abs/1109.0881}{{\tt arXiv:1109.0881 [hep-ph]}}.

\bibitem{Bhattacharya:2013cla}
B.~Bhattacharya, M.~Imbeault, and D.~London, {\em {Extraction of the
  \CP-violating phase $\gamma$ using $B \to K \pi \pi$ and $B \to K K \Kbar$
  decays}\/},  \href{http://dx.doi.org/10.1016/j.physletb.2013.11.038}{Phys.
  Lett. {\bf B728} (2014)  206--209},
\href{http://arxiv.org/abs/1303.0846}{{\tt arXiv:1303.0846 [hep-ph]}}.

\bibitem{Bhattacharya:2015uua}
B.~Bhattacharya and D.~London, {\em {Using U spin to extract $\gamma$ from
  charmless $B \to PPP$ decays}\/},
  \href{http://dx.doi.org/10.1007/JHEP04(2015)154}{JHEP {\bf 04} (2015)  154},
\href{http://arxiv.org/abs/1503.00737}{{\tt arXiv:1503.00737 [hep-ph]}}.

\bibitem{Bediaga:1998ma}
I.~Bediaga, R.~E. Blanco, C.~Gobel, and R.~Mendez-Galain, {\em {A Direct
  measurement of the CKM angle gamma}\/},
  \href{http://dx.doi.org/10.1103/PhysRevLett.81.4067}{Phys. Rev. Lett. {\bf
  81} (1998)  4067--4070},
\href{http://arxiv.org/abs/hep-ph/9804222}{{\tt arXiv:hep-ph/9804222
  [hep-ph]}}.

\bibitem{Blanco:2000gw}
R.~E. Blanco, C.~Gobel, and R.~Mendez-Galain, {\em {Measuring the CP violating
  phase gamma using \decay{\Bp}{\pip\pip\pim} and \decay{\Bp}{\Kp\pip\pim}
  decays}\/},  \href{http://dx.doi.org/10.1103/PhysRevLett.86.2720}{Phys. Rev.
  Lett. {\bf 86} (2001)  2720--2723},
\href{http://arxiv.org/abs/hep-ph/0007105}{{\tt arXiv:hep-ph/0007105
  [hep-ph]}}.

\bibitem{LHCb-PAPER-2014-044}
{LHCb collaboration}, R.~Aaij et al., {\em {Measurement of \CP violation in the
  three-body phase space of charmless $\Bpm$ decays}\/},  Phys. Rev. {\bf D90}
  (2014)  112004, \href{http://arxiv.org/abs/1408.5373}{{\tt arXiv:1408.5373
  [hep-ex]}}.

\bibitem{Aubert:2008bj}
{\babar\ collaboration}, B.~Aubert et al., {\em {Evidence for direct \CP
  violation from Dalitz-plot analysis of $B^\pm \to K^\pm \pi^\mp \pi^\pm$}\/},
   \href{http://dx.doi.org/10.1103/PhysRevD.78.012004}{Phys. Rev. {\bf D78}
  (2008)  012004},
\href{http://arxiv.org/abs/0803.4451}{{\tt arXiv:0803.4451 [hep-ex]}}.

\bibitem{Hsu:2017kir}
{Belle collaboration}, C.~L. Hsu et al., {\em {Measurement of branching
  fraction and direct \CP asymmetry in charmless $B^+ \to K^+K^- \pi^+$ decays
  at Belle}\/},  \href{http://dx.doi.org/10.1103/PhysRevD.96.031101}{Phys. Rev.
  {\bf D96} (2017)  031101},
\href{http://arxiv.org/abs/1705.02640}{{\tt arXiv:1705.02640 [hep-ex]}}.

\bibitem{LHCb-PAPER-2018-009}
{LHCb collaboration}, R.~Aaij et al., {\em {Measurement of \CP violation in
  \decay{\Bz}{D^\pm \pi^\mp} decays}\/},  JHEP {\bf 06} (2018)  084,
  \href{http://arxiv.org/abs/1805.03448}{{\tt arXiv:1805.03448 [hep-ex]}}.

\bibitem{Aubert:2006tw}
{BaBar collaboration}, B.~Aubert et al., {\em {Measurement of time-dependent
  \CP asymmetries in $B^0 \to D^{(*)\pm}\pi^\mp$ and $B^0 \to D^\pm \rho^\mp$
  decays}\/},  \href{http://dx.doi.org/10.1103/PhysRevD.73.111101}{Phys. Rev.
  {\bf D73} (2006)  111101},
\href{http://arxiv.org/abs/hep-ex/0602049}{{\tt arXiv:hep-ex/0602049
  [hep-ex]}}.

\bibitem{Ronga:2006hv}
{Belle collaboration}, F.~Ronga et al., {\em {Measurements of \CP violation in
  $\Bz \to D^{*-} \pi^+$ and $\Bz \to D^- \pi^+$ decays}\/},
  \href{http://dx.doi.org/10.1103/PhysRevD.73.092003}{Phys. Rev. {\bf D73}
  (2006)  092003},
\href{http://arxiv.org/abs/hep-ex/0604013}{{\tt arXiv:hep-ex/0604013
  [hep-ex]}}.

\bibitem{DeBruyn:2012jp}
K.~De~Bruyn, R.~Fleischer, R.~Knegjens, M.~Merk, M.~Schiller, and N.~Tuning,
  {\em {Exploring $B_s \to D_s^{(*)\pm} K^\mp$ decays in the presence of a
  sizable width difference $\Delta\Gamma_s$}\/},
  \href{http://dx.doi.org/10.1016/j.nuclphysb.2012.11.012}{Nucl. Phys. {\bf
  B868} (2013)  351--367},
\href{http://arxiv.org/abs/1208.6463}{{\tt arXiv:1208.6463 [hep-ph]}}.

\bibitem{LHCb-PAPER-2018-006}
{LHCb collaboration}, R.~Aaij et al., {\em {Measurement of \CP asymmetries in
  two-body \BdorBs-meson decays to charged pions and kaons}\/},  Phys. Rev.
  {\bf D98} (2018)  032004, \href{http://arxiv.org/abs/1805.06759}{{\tt
  arXiv:1805.06759 [hep-ex]}}.

\bibitem{LHCb-PAPER-2018-017}
{LHCb collaboration}, R.~Aaij et al., {\em {Measurement of the CKM angle
  $\gamma$ using \decay{\Bpm}{DK^\pm} with \decay{D}{K_S^0 \pi^+ \pi^-,\ \KS
  K^+K^-} decays}\/},  JHEP {\bf 08} (2018)  176,
  \href{http://arxiv.org/abs/1806.01202}{{\tt arXiv:1806.01202 [hep-ex]}}.

\bibitem{LHCb-PAPER-2015-031}
{LHCb collaboration}, R.~Aaij et al., {\em {A precise measurement of the $\Bz$
  meson oscillation frequency}\/},  Eur. Phys. J. {\bf C76} (2016)  412,
  \href{http://arxiv.org/abs/1604.03475}{{\tt arXiv:1604.03475 [hep-ex]}}.

\bibitem{LHCb-PAPER-2013-006}
{LHCb collaboration}, R.~Aaij et al., {\em {Precision measurement of the
  $\Bs$--$\Bsb$ oscillation frequency in the decay $\Bs\to \Dsm\pip$}\/},  New
  J. Phys. {\bf 15} (2013)  053021, \href{http://arxiv.org/abs/1304.4741}{{\tt
  arXiv:1304.4741 [hep-ex]}}.

\bibitem{Lenz:2006hd}
A.~Lenz and U.~Nierste, {\em {Theoretical update of $\Bs-\Bsb$ mixing}\/},
  \href{http://dx.doi.org/10.1088/1126-6708/2007/06/072}{JHEP {\bf 06} (2007)
  072},
\href{http://arxiv.org/abs/hep-ph/0612167}{{\tt arXiv:hep-ph/0612167
  [hep-ph]}}.

\bibitem{LHCb-PAPER-2014-003}
{LHCb collaboration}, R.~Aaij et al., {\em {Precision measurement of the ratio
  of the $\Lb$ to $\Bzb$ lifetimes}\/},  Phys. Lett. {\bf B734} (2014)  122,
  \href{http://arxiv.org/abs/1402.6242}{{\tt arXiv:1402.6242 [hep-ex]}}.

\bibitem{LHCb-PAPER-2014-053}
{LHCb collaboration}, R.~Aaij et al., {\em {Measurement of the semileptonic \CP
  asymmetry in $\Bz$--$\Bzb$ mixing}\/},  Phys. Rev. Lett. {\bf 114} (2015)
  041601, \href{http://arxiv.org/abs/1409.8586}{{\tt arXiv:1409.8586
  [hep-ex]}}.

\bibitem{Keum:1998fd}
Y.-Y. Keum and U.~Nierste, {\em {Probing penguin coefficients with the lifetime
  ratio tau (B-s) / tau (B-d)}\/},
  \href{http://dx.doi.org/10.1103/PhysRevD.57.4282}{Phys. Rev. {\bf D57} (1998)
   4282--4289},
\href{http://arxiv.org/abs/hep-ph/9710512}{{\tt arXiv:hep-ph/9710512
  [hep-ph]}}.

\bibitem{Davis:2310213}
A.~Davis, L.~Dufour, F.~Ferrari, S.~Stahl, M.~A. Vesterinen, and
  J.~Van~Tilburg, {\em {Measurement of the instrumental asymmetry for
  $K^{-}\pi^{+}$-pairs at LHCb in Run 2}\/},  Mar, 2018.
\newblock \url{https://cds.cern.ch/record/2310213}.

\bibitem{Davis:2284097}
A.~Davis, L.~Dufour, F.~Ferrari, S.~Stahl, M.~A. Vesterinen, and
  J.~Van~Tilburg, {\em {Measurement of the $K^{-}\pi^{+}$ two-track detection
  asymmetry in Run 2 using the Turbo stream}\/},  Sep, 2017.
\newblock \url{https://cds.cern.ch/record/2284097}.

\bibitem{Vesterinen:1642153}
{M.~Vesterinen, on behalf of the LHCb collaboration}, {\em {Considerations on
  the LHCb dipole magnet polarity reversal}\/},  Apr, 2014.
\newblock \url{https://cds.cern.ch/record/1642153}.

\bibitem{Dufour:2304546}
L.~Dufour and J.~Van~Tilburg, {\em {Decomposition of simulated detection
  asymmetries in LHCb}\/},  Feb, 2018.
\newblock \url{https://cds.cern.ch/record/2304546}.

\bibitem{Artuso:2015swg}
M.~Artuso, G.~Borissov, and A.~Lenz, {\em {CP violation in the $B_s^0$
  system}\/},  \href{http://dx.doi.org/10.1103/RevModPhys.88.045002}{Rev. Mod.
  Phys. {\bf 88} (2016)  045002},
\href{http://arxiv.org/abs/1511.09466}{{\tt arXiv:1511.09466 [hep-ph]}}.

\bibitem{Gershon:2010wx}
T.~Gershon, {\em {$\Delta \Gamma_d$: A forgotten null test of the Standard
  Model}\/},  \href{http://dx.doi.org/10.1088/0954-3899/38/1/015007}{J. Phys.
  {\bf G38} (2011)  015007},
\href{http://arxiv.org/abs/1007.5135}{{\tt arXiv:1007.5135 [hep-ph]}}.

\bibitem{LHCb-PAPER-2013-065}
{LHCb collaboration}, R.~Aaij et al., {\em {Measurements of the $\Bp$, $\Bz$,
  $\Bs$ meson and $\Lb$ baryon lifetimes}\/},  JHEP {\bf 04} (2014)  114,
  \href{http://arxiv.org/abs/1402.2554}{{\tt arXiv:1402.2554 [hep-ex]}}.

\bibitem{Aaboud:2016bro}
{ATLAS collaboration}, M.~Aaboud et al., {\em {Measurement of the relative
  width difference of the $\Bd$--$\Bdb$ system with the ATLAS detector}\/},
  \href{http://dx.doi.org/10.1007/JHEP06(2016)081}{JHEP {\bf 06} (2016)  081},
\href{http://arxiv.org/abs/1605.07485}{{\tt arXiv:1605.07485 [hep-ex]}}.

\bibitem{Sirunyan:2017nbv}
{CMS collaboration}, A.~M. Sirunyan et al., {\em {Measurement of \bquark\
  hadron lifetimes in $pp$ collisions at $\sqrt{s} =$ 8 TeV}\/},
  \href{http://dx.doi.org/10.1140/epjc/s10052-018-5929-3}{Eur. Phys. J. {\bf
  C78} (2018) no.~6, 457},
\href{http://arxiv.org/abs/1710.08949}{{\tt arXiv:1710.08949 [hep-ex]}}.

\bibitem{Faller:2008gt}
S.~Faller, R.~Fleischer, and T.~Mannel, {\em {Precision physics with $B^0_s \to
  J/\psi \phi$ at the LHC: The quest for new physics}\/},
  \href{http://dx.doi.org/10.1103/PhysRevD.79.014005}{Phys.Rev. {\bf D79}
  (2009)  014005},
\href{http://arxiv.org/abs/0810.4248}{{\tt arXiv:0810.4248 [hep-ph]}}.

\bibitem{Bhattacharya:2012ph}
B.~Bhattacharya, A.~Datta, and D.~London, {\em {Reducing penguin pollution}\/},
   \href{http://dx.doi.org/10.1142/S0217751X13500632}{Int. J. Mod. Phys. {\bf
  A28} (2013)  1350063},
\href{http://arxiv.org/abs/1209.1413}{{\tt arXiv:1209.1413 [hep-ph]}}.

\bibitem{CMS-PAS-FTR-18-041}
{CMS Collaboration}, {\em CP-Violation studies at the HL-LHC with CMS using
  $B^0_s$ decays to $J/\psi\phi(1020)$\/},  CMS Physics Analysis Summary
  CMS-PAS-FTR-18-041, 2018.
\newblock \url{http://cdsweb.cern.ch/record/2650772}.

\bibitem{CMSPhis}
{CMS Collaboration}, V.~Khachatryanand et al., {\em {Measurement of the
  CP-violating weak phase $\phi_s$ and the decay width difference
  $\Delta\Gamma_s$ using the $B^0_s \to J/\psi \phi(1020)$ decay channel in pp
  collisions at $\sqrt s = 8$ TeV}\/},
  \href{http://dx.doi.org/10.1016/j.physletb.2016.03.046}{Phys. Lett. {\bf
  B757} (2016)  97},
\href{http://arxiv.org/abs/1507.07527}{{\tt arXiv:1507.07527 [hep-ex]}}.

\bibitem{ATL-PHYS-PUB-2018-041}
{ATLAS Collaboration}, {\em {CP-violation measurement prospects in the $B^0_s
  \to J/\psi\phi$ channel with the upgraded ATLAS detector at the HL-LHC}\/},
  ATL-PHYS-PUB-2018-041, CERN, Geneva, Dec, 2018.
\newblock \url{http://cds.cern.ch/record/2649881}.

\bibitem{ATL-PHYS-PUB-2013-010}
{\em {ATLAS B-physics studies at increased LHC luminosity, potential for
  CP-violation measurement in the B0s → J/ψφ decay}\/},
  ATL-PHYS-PUB-2013-010, CERN, Geneva, Sep, 2013.
\newblock \url{http://cds.cern.ch/record/1604429}.

\bibitem{Aad:2016tdj}
{ATLAS Collaboration}, G.~Aad et al., {\em {Measurement of the CP-violating
  phase $\phi_s$ and the $B^0_s$ meson decay width difference with $B^0_s \to
  J/\psi\phi$ decays in ATLAS}\/},
  \href{http://dx.doi.org/10.1007/JHEP08(2016)147}{JHEP {\bf 08} (2016)  147},
\href{http://arxiv.org/abs/1601.03297}{{\tt arXiv:1601.03297 [hep-ex]}}.

\bibitem{HFLAVPDG18}
{Heavy Flavour Averaging Group Collaboration}, Y.~Amhisand et al., {\em
  {Averages of b-hadron, c-hadron, and $\tau$-lepton properties as of summer
  2016}\/},  \href{http://dx.doi.org/10.1016/j.physletb.2016.03.046}{Eur. Phys.
  {\bf J77} (2017)  895},
\href{http://arxiv.org/abs/arXiv:1612.07233}{{\tt arXiv:arXiv:1612.07233
  [hep-ex]}}.
and online update http://www.slac.stanford.edu/xorg/hfag.

\bibitem{LHCb-PAPER-2015-004}
{LHCb collaboration}, R.~Aaij et al., {\em {Measurement of \CP violation in
  $\Bz\to \jpsi\KS$ decays}\/},  Phys. Rev. Lett. {\bf 115} (2015)  031601,
  \href{http://arxiv.org/abs/1503.07089}{{\tt arXiv:1503.07089 [hep-ex]}}.

\bibitem{LHCb-PAPER-2017-008}
{LHCb collaboration}, R.~Aaij et al., {\em {Resonances and \CP-violation in
  $\Bsb$ and $\Bs\to \jpsi\Kp\Km$ decays in the mass region above the
  $\phiz(1020)$}\/},  JHEP {\bf 08} (2017)  037,
  \href{http://arxiv.org/abs/1704.08217}{{\tt arXiv:1704.08217 [hep-ex]}}.

\bibitem{LHCb-PAPER-2014-019}
{LHCb collaboration}, R.~Aaij et al., {\em {Measurement of the \CP-violating
  phase $\phis$ in $\Bsb\to \jpsi\pim\pim$ decays}\/},  Phys. Lett. {\bf B736}
  (2014)  186, \href{http://arxiv.org/abs/1405.4140}{{\tt arXiv:1405.4140
  [hep-ex]}}.

\bibitem{LHCb-PAPER-2016-027}
{LHCb collaboration}, R.~Aaij et al., {\em {Measurement of the \CP violating
  phase and decay-width difference in $\Bs\to \psitwos\phi$ decays}\/},  Phys.
  Lett. {\bf B762} (2016)  253, \href{http://arxiv.org/abs/1608.04855}{{\tt
  arXiv:1608.04855 [hep-ex]}}.

\bibitem{LHCb-PAPER-2014-056}
{LHCb collaboration}, R.~Aaij et al., {\em {Study of $\etaz$--$\etapr$ mixing
  from measurement of $\BdorBs \to \jpsi\eta^{(\prime)}$ decay rates}\/},  JHEP
  {\bf 01} (2015)  024, \href{http://arxiv.org/abs/1411.0943}{{\tt
  arXiv:1411.0943 [hep-ex]}}.

\bibitem{LHCb-PAPER-2016-017}
{LHCb collaboration}, R.~Aaij et al., {\em {Measurement of the $\Bs\to
  \jpsi\eta$ lifetime}\/},  Phys. Lett. {\bf B762} (2016)  484,
  \href{http://arxiv.org/abs/1607.06314}{{\tt arXiv:1607.06314 [hep-ex]}}.

\bibitem{Adachi:2012et}
I.~Adachi et al., {\em {Precise measurement of the CP violation parameter
  $\sin2\phi_1$ in $\Bz \to (c\bar{c})K^0$ decays}\/},
  \href{http://dx.doi.org/10.1103/PhysRevLett.108.171802}{Phys. Rev. Lett. {\bf
  108} (2012)  171802},
\href{http://arxiv.org/abs/1201.4643}{{\tt arXiv:1201.4643 [hep-ex]}}.

\bibitem{Gaz:2017fgl}
A.~Gaz, {\em {Physics prospects at SuperKEKB/Belle II}\/},
PoS {\bf KMI2017} (2017)  005.

\bibitem{Abdesselam:2015gha}
{Belle and BaBar collaborations}, A.~Abdesselam et al., {\em {First observation
  of \CP violation in $\overline{B}^0 \to D^{(*)}_{\CP} h^0$ decays by a
  combined time-dependent analysis of BaBar and Belle data}\/},
  \href{http://dx.doi.org/10.1103/PhysRevLett.115.121604}{Phys. Rev. Lett. {\bf
  115} (2015) no.~12, 121604},
\href{http://arxiv.org/abs/1505.04147}{{\tt arXiv:1505.04147 [hep-ex]}}.

\bibitem{Adachi:2018itz}
{Belle and BaBar collaborations}, I.~Adachi et al., {\em {First evidence for
  $\cos 2\beta>0$ and resolution of the CKM Unitarity Triangle ambiguity by a
  time-dependent Dalitz plot analysis of $B^{0} \to D^{(*)} h^{0}$ with $D \to
  K_{S}^{0} \pi^{+} \pi^{-}$ decays}\/},
\href{http://arxiv.org/abs/1804.06152}{{\tt arXiv:1804.06152 [hep-ex]}}.

\bibitem{Adachi:2018jqe}
{Belle and BaBar collaboration}, I.~Adachi et al., {\em {Measurement of
  $\cos{2\beta}$ in $B^{0} \to D^{(*)} h^{0}$ with $D \to K_{S}^{0} \pi^{+}
  \pi^{-}$ decays by a combined time-dependent Dalitz plot analysis of BaBar
  and Belle data}\/},
\href{http://arxiv.org/abs/1804.06153}{{\tt arXiv:1804.06153 [hep-ex]}}.

\bibitem{Kuzmin:2006mw}
{Belle collaboration}, A.~Kuzmin et al., {\em {Study of $\Bzb \to D^0 \pi^+
  \pi^-$ decays}\/},  \href{http://dx.doi.org/10.1103/PhysRevD.76.012006}{Phys.
  Rev. {\bf D76} (2007)  012006},
\href{http://arxiv.org/abs/hep-ex/0611054}{{\tt arXiv:hep-ex/0611054
  [hep-ex]}}.

\bibitem{delAmoSanchez:2010ad}
{BaBar collaboration}, P.~del Amo~Sanchez et al., {\em {Dalitz plot analysis of
  $B^0 \to \Dzb \pi^+ \pi^-$}\/},  PoS {\bf ICHEP2010} (2010)  250,
\href{http://arxiv.org/abs/1007.4464}{{\tt arXiv:1007.4464 [hep-ex]}}.

\bibitem{LHCb-PAPER-2014-070}
{LHCb collaboration}, R.~Aaij et al., {\em {Dalitz plot analysis of $\Bz\to
  \Dzb\pip\pim$ decays}\/},  Phys. Rev. {\bf D92} (2015)  032002,
  \href{http://arxiv.org/abs/1505.01710}{{\tt arXiv:1505.01710 [hep-ex]}}.

\bibitem{Charles:1998vf}
J.~Charles, A.~Le~Yaouanc, L.~Oliver, O.~Pene, and J.~Raynal, {\em
  {$B^{0}_{d}(t) \to DPP$ time dependent Dalitz plots, \CP-violating angles
  $2\beta$, $2\beta + \gamma$, and discrete ambiguities}\/},
  \href{http://dx.doi.org/10.1016/S0370-2693(98)00250-0}{Phys.Lett. {\bf B425}
  (1998)  375},
\href{http://arxiv.org/abs/hep-ph/9801363}{{\tt arXiv:hep-ph/9801363
  [hep-ph]}}.

\bibitem{Latham:2008zs}
T.~Latham and T.~Gershon, {\em {A method to measure $\cos(2\beta)$ using
  time-dependent Dalitz plot analysis of $B^0 \to D_{\CP} \pi^+ \pi^-$}\/},
  \href{http://dx.doi.org/10.1088/0954-3899/36/2/025006}{J. Phys. {\bf G36}
  (2009)  025006},
\href{http://arxiv.org/abs/0809.0872}{{\tt arXiv:0809.0872 [hep-ph]}}.

\bibitem{Bondar:2018gpb}
A.~Bondar, A.~Kuzmin, and V.~Vorobyev, {\em {A method for model-independent
  measurement of the CKM angle $\beta$ via time-dependent analysis of the
  $B^0\to D\pi^+\pi^-$, $D\to K_S^0\pi^+\pi^-$ decays}\/},
  \href{http://dx.doi.org/10.1007/JHEP03(2018)195}{JHEP {\bf 03} (2018)  195},
\href{http://arxiv.org/abs/1802.00200}{{\tt arXiv:1802.00200 [hep-ph]}}.

\bibitem{LHCb-PAPER-2014-058}
{LHCb collaboration}, R.~Aaij et al., {\em {Measurement of the \CP-violating
  phase $\beta$ in $\Bzb\to \jpsi\pip\pim$ decays and limits on penguin
  effects}\/},  Phys. Lett. {\bf B742} (2015)  38,
  \href{http://arxiv.org/abs/1411.1634}{{\tt arXiv:1411.1634 [hep-ex]}}.

\bibitem{LHCb-PAPER-2015-034}
{LHCb collaboration}, R.~Aaij et al., {\em {Measurement of \CP violation
  parameters and polarisation fractions in $\Bs\to \jpsi\Kstarzb$ decays}\/},
  JHEP {\bf 11} (2015)  082, \href{http://arxiv.org/abs/1509.00400}{{\tt
  arXiv:1509.00400 [hep-ex]}}.

\bibitem{Fleischer:2007zn}
R.~Fleischer, {\em {Exploring CP violation and penguin effects through $B^0_{d}
  \to D^{+} D^{-}$ and $B^0_{s} \to D^+_{s} D^-_{s}$}\/},
  \href{http://dx.doi.org/10.1140/epjc/s10052-007-0341-4}{Eur. Phys. J. {\bf
  C51} (2007)  849--858},
\href{http://arxiv.org/abs/0705.4421}{{\tt arXiv:0705.4421 [hep-ph]}}.

\bibitem{Jung:2014jfa}
M.~Jung and S.~Schacht, {\em {Standard model predictions and new physics
  sensitivity in $B \to DD$ decays}\/},
  \href{http://dx.doi.org/10.1103/PhysRevD.91.034027}{Phys. Rev. {\bf D91}
  (2015) no.~3, 034027},
\href{http://arxiv.org/abs/1410.8396}{{\tt arXiv:1410.8396 [hep-ph]}}.

\bibitem{Bel:2015wha}
L.~Bel, K.~De~Bruyn, R.~Fleischer, M.~Mulder, and N.~Tuning, {\em {Anatomy of $
  B\to D\overline{D} $ decays}\/},
  \href{http://dx.doi.org/10.1007/JHEP07(2015)108}{JHEP {\bf 07} (2015)  108},
\href{http://arxiv.org/abs/1505.01361}{{\tt arXiv:1505.01361 [hep-ph]}}.

\bibitem{LHCb-PAPER-2016-037}
{LHCb collaboration}, R.~Aaij et al., {\em {Measurement of \CP violation in
  $\B\to \Dp\Dm$ decays}\/},  Phys. Rev. Lett. {\bf 117} (2016)  261801,
  \href{http://arxiv.org/abs/1608.06620}{{\tt arXiv:1608.06620 [hep-ex]}}.

\bibitem{LHCb-PAPER-2015-005}
{LHCb collaboration}, R.~Aaij et al., {\em {Measurement of the time-dependent
  \CP asymmetries in $\Bs\to \jpsi\KS$}\/},  JHEP {\bf 06} (2015)  131,
  \href{http://arxiv.org/abs/1503.07055}{{\tt arXiv:1503.07055 [hep-ex]}}.

\bibitem{Aubert:2008bs}
{BaBar collaboration}, B.~Aubert et al., {\em {Evidence for \CP violation in
  $B^0 \to J/\psi \pi^0$ decays}\/},
  \href{http://dx.doi.org/10.1103/PhysRevLett.101.021801}{Phys. Rev. Lett. {\bf
  101} (2008)  021801},
\href{http://arxiv.org/abs/0804.0896}{{\tt arXiv:0804.0896 [hep-ex]}}.

\bibitem{Lee:2007wd}
{Belle collaboration}, S.~E. Lee et al., {\em {Improved measurement of
  time-dependent \CP violation in $\Bz \to \jpsi \piz$ decays}\/},
  \href{http://dx.doi.org/10.1103/PhysRevD.77.071101}{Phys. Rev. {\bf D77}
  (2008)  071101},
\href{http://arxiv.org/abs/0708.0304}{{\tt arXiv:0708.0304 [hep-ex]}}.

\bibitem{Bartsch:2008ps}
M.~Bartsch, G.~Buchalla, and C.~Kraus, {\em {$B \to V_{\textrm{L}}
  V_{\textrm{L}}$ decays at next-to-leading order in QCD}\/},
\href{http://arxiv.org/abs/0810.0249}{{\tt arXiv:0810.0249 [hep-ph]}}.

\bibitem{Beneke:2006hg}
M.~Beneke, J.~Rohrer, and D.~Yang, {\em {Branching fractions, polarisation and
  asymmetries of $B\to VV$ decays}\/},
  \href{http://dx.doi.org/10.1016/j.nuclphysb.2007.03.020}{Nucl. Phys. {\bf
  B774} (2007)  64--101},
\href{http://arxiv.org/abs/hep-ph/0612290}{{\tt arXiv:hep-ph/0612290
  [hep-ph]}}.

\bibitem{PhysRevD.80.114026}
H.-Y. Cheng and C.-K. Chua, {\em {QCD factorization for charmless hadronic
  $B_{s}$ decays revisited}\/},
  \href{http://dx.doi.org/10.1103/PhysRevD.80.114026}{Phys. Rev. {\bf D80}
  (2009)  114026},
\href{http://arxiv.org/abs/0910.5237}{{\tt arXiv:0910.5237 [hep-ph]}}.

\bibitem{CMSTDR:trk}
{CMS Collaboration}, {\em {The Phase-2 Upgrade of the CMS Tracker}\/},
  CERN-LHCC-2017-009 ; CMS-TDR-014, CERN, Geneva, 2017.
\newblock \url{https://cds.cern.ch/record/2272264}.

\bibitem{LHCb-PAPER-2017-048}
{LHCb collaboration}, R.~Aaij et al., {\em {First measurement of the
  \CP-violating phase $\phi_s^{d\dquarkbar}$ in $\Bs\to(K^+\pi^-)(K^-\pi^+)$
  decays}\/},  JHEP {\bf 03} (2018)  140,
  \href{http://arxiv.org/abs/1712.08683}{{\tt arXiv:1712.08683 [hep-ex]}}.

\bibitem{Gershon:2014yma}
T.~Gershon, T.~Latham, and R.~Silva~Coutinho, {\em {Probing CP violation in
  $B^0_s \rightarrow K^{0}_{\rm S} \pi^{+}\pi^{-}$ decays}\/},
  \href{http://dx.doi.org/10.1016/j.nuclphysbps.2015.09.229}{Nucl. Part. Phys.
  Proc. {\bf 273-275} (2016)  1417--1422},
\href{http://arxiv.org/abs/1411.2018}{{\tt arXiv:1411.2018 [hep-ph]}}.

\bibitem{SilvaCoutinho:2045786}
R.~Silva~Coutinho, {\em {Studies of charmless three-body $b$-hadron decays at
  LHCb}\/},  Apr, 2015.
\newblock \url{http://cds.cern.ch/record/2045786}.

\bibitem{LHCb-PAPER-2017-010}
{LHCb collaboration}, R.~Aaij et al., {\em {Updated branching fraction
  measurements of $\BdorBs\to \KS h^+ h^{\prime-}$ decays}\/},  JHEP {\bf 11}
  (2017)  027, \href{http://arxiv.org/abs/1707.01665}{{\tt arXiv:1707.01665
  [hep-ex]}}.

\bibitem{LHCb-PAPER-2017-033}
{LHCb collaboration}, R.~Aaij et al., {\em {Amplitude analysis of the decay
  $\Bzb\to \KS \pi^+\pi^-$ and first observation of \CP asymmetry in $\Bzb\to
  K^{*}(892)^{-} \pi^+$}\/},  Phys. Rev. Lett. {\bf 120} (2018)  261801,
  \href{http://arxiv.org/abs/1712.09320}{{\tt arXiv:1712.09320 [hep-ex]}}.

\bibitem{LHCb-CONF-2015-001}
{{LHCb collaboration}}, {\em {Study of the decay $B^+\to K^+\pi^0$ at LHCb}\/},
   {Mar}, {2015}.

\bibitem{Dunietz:1990cj}
I.~Dunietz, H.~R. Quinn, A.~Snyder, W.~Toki, and H.~J. Lipkin, {\em {How to
  extract CP violating asymmetries from angular correlations}\/},
\href{http://dx.doi.org/10.1103/PhysRevD.43.2193}{Phys. Rev. {\bf D43} (1991)
  2193--2208}.

\bibitem{Falk:2003uq}
A.~F. Falk, Z.~Ligeti, Y.~Nir, and H.~Quinn, {\em {Comment on extracting
  $\alpha$ from $\B \rightarrow \rho\rho$}\/},
  \href{http://dx.doi.org/10.1103/PhysRevD.69.011502}{Phys. Rev. {\bf D69}
  (2004)  011502},
\href{http://arxiv.org/abs/hep-ph/0310242}{{\tt arXiv:hep-ph/0310242
  [hep-ph]}}.

\bibitem{Beneke:2006rb}
M.~Beneke, M.~Gronau, J.~Rohrer, and M.~Spranger, {\em {A Precise determination
  of $\alpha$ using $\Bz \rightarrow \rho^+\rho^-$ and $\Bz \rightarrow
  \Kstarp\rho^-$}\/},
  \href{http://dx.doi.org/10.1016/j.physletb.2006.05.012}{Phys. Lett. {\bf
  B638} (2006)  68--73},
\href{http://arxiv.org/abs/hep-ph/0604005}{{\tt arXiv:hep-ph/0604005
  [hep-ph]}}.

\bibitem{LHCb-PAPER-2015-006}
{LHCb collaboration}, R.~Aaij et al., {\em {Observation of the $\Bz\to
  \rhoz\rhoz$ decay from an amplitude analysis of $\Bz\to (\pip\pim)(\pip\pim)$
  decays}\/},  Phys. Lett. {\bf B747} (2015)  468,
  \href{http://arxiv.org/abs/1503.07770}{{\tt arXiv:1503.07770 [hep-ex]}}.

\bibitem{Lees:2013nwa}
{BaBar collaboration}, J.~P. Lees et al., {\em {Measurement of CP-violating
  asymmetries in $B^0 \to (\rho \pi)^0$ decays using a time-dependent Dalitz
  plot analysis}\/},  \href{http://dx.doi.org/10.1103/PhysRevD.88.012003}{Phys.
  Rev. {\bf D88} (2013) no.~1, 012003},
\href{http://arxiv.org/abs/1304.3503}{{\tt arXiv:1304.3503 [hep-ex]}}.

\bibitem{Kusaka:2007dv}
{Belle collaboration}, A.~Kusaka et al., {\em {Measurement of \CP asymmetry in
  a time-dependent Dalitz analysis of $B^0 \to (\rho \pi)^0$ and a constraint
  on the CKM angle $\phi_2$}\/},
  \href{http://dx.doi.org/10.1103/PhysRevLett.98.221602}{Phys. Rev. Lett. {\bf
  98} (2007)  221602},
\href{http://arxiv.org/abs/hep-ex/0701015}{{\tt arXiv:hep-ex/0701015
  [hep-ex]}}.

\bibitem{Kusaka:2007mj}
{Belle collaboration}, A.~Kusaka et al., {\em {Measurement of \CP asymmetries
  and branching fractions in a time-dependent Dalitz analysis of $B^0 \to (\rho
  \pi)^0$ and a constraint on the quark mixing angle $\phi_{2}$}\/},
  \href{http://dx.doi.org/10.1103/PhysRevD.77.072001}{Phys. Rev. {\bf D77}
  (2008)  072001},
\href{http://arxiv.org/abs/0710.4974}{{\tt arXiv:0710.4974 [hep-ex]}}.

\bibitem{Aaltonen:2014vra}
{CDF collaboration}, T.~A. Aaltonen et al., {\em {Measurements of direct
  \CP-violating asymmetries in charmless decays of bottom baryons}\/},
  \href{http://dx.doi.org/10.1103/PhysRevLett.113.242001}{Phys. Rev. Lett. {\bf
  113} (2014)  242001},
\href{http://arxiv.org/abs/1403.5586}{{\tt arXiv:1403.5586 [hep-ex]}}.

\bibitem{LHCb-PAPER-2013-061}
{LHCb collaboration}, R.~Aaij et al., {\em {Searches for $\Lb$ and $\Xibz$
  decays to $\KS\proton\pim$ and $\KS\proton\Km$ final states with first
  observation of the $\Lb \to \KS\proton\pim$ decay}\/},
  \href{http://dx.doi.org/10.1007/JHEP04(2014)087}{JHEP {\bf 04} (2014)  087}
  LHCb-PAPER-2013-061, CERN-PH-EP-2014-012,
\href{http://arxiv.org/abs/1402.0770}{{\tt arXiv:1402.0770 [hep-ex]}}.

\bibitem{LHCb-PAPER-2016-004}
{LHCb collaboration}, R.~Aaij et al., {\em {Observations of $\Lb\to \Lz\Kp\pim$
  and $\Lb\to \Lz\Kp\Km$ decays and searches for other $\Lb$ and $\Xibz$ decays
  to $\Lz h^+h^-$ final states}\/},  JHEP {\bf 05} (2016)  081,
  \href{http://arxiv.org/abs/1603.00413}{{\tt arXiv:1603.00413 [hep-ex]}}.

\bibitem{LHCb-PAPER-2017-034}
{LHCb collaboration}, R.~Aaij et al., {\em {Measurement of branching fractions
  of charmless four-body $\Lb$ and $\Xires_b^0$ decays}\/},  JHEP {\bf 02}
  (2018)  098, \href{http://arxiv.org/abs/1711.05490}{{\tt arXiv:1711.05490
  [hep-ex]}}.

\bibitem{LHCb-PAPER-2016-059}
{LHCb collaboration}, R.~Aaij et al., {\em {Observation of the decay $\Lb\to
  \proton\Km\mup\mun$ and search for \CP violation}\/},  JHEP {\bf 06} (2017)
  108, \href{http://arxiv.org/abs/1703.00256}{{\tt arXiv:1703.00256 [hep-ex]}}.

\bibitem{Dunietz:1992ti}
I.~Dunietz, {\em {\CP violation with beautiful baryons}\/},
\href{http://dx.doi.org/10.1007/BF01589716}{Z. Phys. {\bf C56} (1992)
  129--144}.

\bibitem{Fayyazuddin:1998aa}
Fayyazuddin, {\em {\decay{\Lb}{\Lz+\Dz(\Dzb)} decays and \CP violation}\/},
  \href{http://dx.doi.org/10.1142/S0217732399000092}{Mod. Phys. Lett. {\bf A14}
  (1999)  63--70},
\href{http://arxiv.org/abs/hep-ph/9806393}{{\tt arXiv:hep-ph/9806393
  [hep-ph]}}.

\bibitem{Giri:2001ju}
A.~K. Giri, R.~Mohanta, and M.~P. Khanna, {\em {Possibility of extracting the
  weak phase $\gamma$ from \decay{\Lb}{\Lambda\Dz} decays}\/},
  \href{http://dx.doi.org/10.1103/PhysRevD.65.073029}{Phys. Rev. {\bf D65}
  (2002)  073029},
\href{http://arxiv.org/abs/hep-ph/0112220}{{\tt arXiv:hep-ph/0112220
  [hep-ph]}}.

\bibitem{LHCb-PAPER-2013-056}
{LHCb collaboration}, R.~Aaij et al., {\em {Study of beauty baryon decays to
  $\Dz\proton h^-$ and $\Lc h^-$ final states}\/},
  \href{http://dx.doi.org/10.1103/PhysRevD.89.032001}{Phys. Rev. {\bf D89}
  (2014)  032001} CERN-PH-EP-2013-207, LHCb-PAPER-2013-056,
\href{http://arxiv.org/abs/1311.4823}{{\tt arXiv:1311.4823 [hep-ex]}}.

\bibitem{LHCb-PAPER-2016-030}
{LHCb collaboration}, R.~Aaij et al., {\em {Measurement of matter-antimatter
  differences in beauty baryon decays}\/},  Nature Physics {\bf 13} (2017)
  391, \href{http://arxiv.org/abs/1609.05216}{{\tt arXiv:1609.05216 [hep-ex]}}.

\bibitem{LHCb-PAPER-2018-001}
{LHCb collaboration}, R.~Aaij et al., {\em {Search for \CP violation using
  triple product asymmetries in \decay{\Lb}{pK^-\pi^+\pi^-},
  \decay{\Lb}{pK^-K^+K^-}, and \decay{\Xires^0_b}{pK^-K^-\pi^+} decays}\/},
  JHEP {\bf 08} (2018)  039, \href{http://arxiv.org/abs/1805.03941}{{\tt
  arXiv:1805.03941 [hep-ex]}}.

\bibitem{LHCb-PAPER-2016-062}
{LHCb collaboration}, R.~Aaij et al., {\em {Measurement of \Bd, \Bs, \Bp and
  \Lb production asymmetries in $7$ and $8$\tev\ \proton\proton collisions}\/},
   Phys. Lett. {\bf B774} (2017)  139,
  \href{http://arxiv.org/abs/1703.08464}{{\tt arXiv:1703.08464 [hep-ex]}}.

\bibitem{LHCb-PAPER-2017-044}
{LHCb collaboration}, R.~Aaij et al., {\em {Search for \CP violation in $\Lc\to
  p K^- K^+$ and $\Lc p\pi^-\pi^+$ decays}\/},  JHEP {\bf 03} (2018)  182,
  \href{http://arxiv.org/abs/1712.07051}{{\tt arXiv:1712.07051 [hep-ex]}}.

\bibitem{LHCb-PAPER-2014-054}
{LHCb collaboration}, R.~Aaij et al., {\em {Search for \CP violation in $\Dz\to
  \pim\pip\piz$ decays with the energy test}\/},  Phys. Lett. {\bf B740} (2015)
   158, \href{http://arxiv.org/abs/1410.4170}{{\tt arXiv:1410.4170 [hep-ex]}}.

\bibitem{Kou:2018nap}
{Belle II Collaboration}, E.~Kou et al., {\em {The Belle II Physics Book}\/},
\href{http://arxiv.org/abs/1808.10567}{{\tt arXiv:1808.10567 [hep-ex]}}.

\bibitem{Amhis:2016xyh}
{HFLAV Collaboration}, Y.~Amhis et al., {\em {Averages of $b$-hadron,
  $c$-hadron, and $\tau$-lepton properties as of summer 2016}\/},
  \href{http://dx.doi.org/10.1140/epjc/s10052-017-5058-4}{Eur. Phys. J. {\bf
  C77} (2017) no.~12, 895},
\href{http://arxiv.org/abs/1612.07233}{{\tt arXiv:1612.07233 [hep-ex]}}.

\bibitem{Bona:2007vi}
{UTfit Collaboration}, M.~Bona et al., {\em {Model-independent constraints on
  $\Delta F=2$ operators and the scale of new physics}\/},
  \href{http://dx.doi.org/10.1088/1126-6708/2008/03/049}{JHEP {\bf 03} (2008)
  049},
\href{http://arxiv.org/abs/0707.0636}{{\tt arXiv:0707.0636 [hep-ph]}}.

\bibitem{Laplace:2002ik}
S.~Laplace, Z.~Ligeti, Y.~Nir, and G.~Perez, {\em {Implications of the CP
  asymmetry in semileptonic B decay}\/},
  \href{http://dx.doi.org/10.1103/PhysRevD.65.094040}{Phys. Rev. {\bf D65}
  (2002)  094040},
\href{http://arxiv.org/abs/hep-ph/0202010}{{\tt arXiv:hep-ph/0202010
  [hep-ph]}}.

\bibitem{Gabbiani:1996hi}
F.~Gabbiani, E.~Gabrielli, A.~Masiero, and L.~Silvestrini, {\em {A Complete
  analysis of FCNC and CP constraints in general SUSY extensions of the
  standard model}\/},
  \href{http://dx.doi.org/10.1016/0550-3213(96)00390-2}{Nucl. Phys. {\bf B477}
  (1996)  321--352},
\href{http://arxiv.org/abs/hep-ph/9604387}{{\tt arXiv:hep-ph/9604387
  [hep-ph]}}.

\bibitem{LHCbimplications}
L.~Silvestrini.
\newblock {talk at Implications of LHCb measurements and future prospects,
  CERN, Nov 3-5, 2015}.

\bibitem{Nierste:2009wg}
U.~Nierste, {\em {Three Lectures on Meson Mixing and CKM phenomenology}\/},  in
  {\em {Heavy quark physics. Proceedings, Helmholtz International School,
  HQP08, Dubna, Russia, August 11-21, 2008}}, pp.~1--38.
\newblock 2009.
\newblock \href{http://arxiv.org/abs/0904.1869}{{\tt arXiv:0904.1869
  [hep-ph]}}.
\newblock
\url{http://inspirehep.net/record/817820/files/arXiv:0904.1869.pdf}.
\newblock

\bibitem{Jubb:2016mvq}
T.~Jubb, M.~Kirk, A.~Lenz, and G.~Tetlalmatzi-Xolocotzi, {\em {On the ultimate
  precision of meson mixing observables}\/},
  \href{http://dx.doi.org/10.1016/j.nuclphysb.2016.12.020}{Nucl. Phys. {\bf
  B915} (2017)  431--453},
\href{http://arxiv.org/abs/1603.07770}{{\tt arXiv:1603.07770 [hep-ph]}}.

\bibitem{Falk:2001hx}
A.~F. Falk, Y.~Grossman, Z.~Ligeti, and A.~A. Petrov, {\em {SU(3) breaking and
  D0 - anti-D0 mixing}\/},
  \href{http://dx.doi.org/10.1103/PhysRevD.65.054034}{Phys. Rev. {\bf D65}
  (2002)  054034},
\href{http://arxiv.org/abs/hep-ph/0110317}{{\tt arXiv:hep-ph/0110317
  [hep-ph]}}.

\bibitem{Gronau:2012kq}
M.~Gronau and J.~L. Rosner, {\em {Revisiting D0-D0bar mixing using U-spin}\/},
  \href{http://dx.doi.org/10.1103/PhysRevD.86.114029}{Phys. Rev. {\bf D86}
  (2012)  114029},
\href{http://arxiv.org/abs/1209.1348}{{\tt arXiv:1209.1348 [hep-ph]}}.

\bibitem{Georgi:1992as}
H.~Georgi, {\em {D - anti-D mixing in heavy quark effective field theory}\/},
  \href{http://dx.doi.org/10.1016/0370-2693(92)91274-D}{Phys. Lett. {\bf B297}
  (1992)  353--357},
\href{http://arxiv.org/abs/hep-ph/9209291}{{\tt arXiv:hep-ph/9209291
  [hep-ph]}}.

\bibitem{Ohl:1992sr}
T.~Ohl, G.~Ricciardi, and E.~H. Simmons, {\em {D - anti-D mixing in heavy quark
  effective field theory: The Sequel}\/},
  \href{http://dx.doi.org/10.1016/0550-3213(93)90364-U}{Nucl. Phys. {\bf B403}
  (1993)  605--632},
\href{http://arxiv.org/abs/hep-ph/9301212}{{\tt arXiv:hep-ph/9301212
  [hep-ph]}}.

\bibitem{Bigi:2000wn}
I.~I.~Y. Bigi and N.~G. Uraltsev, {\em {D0 - anti-D0 oscillations as a probe of
  quark hadron duality}\/},
  \href{http://dx.doi.org/10.1016/S0550-3213(00)00604-0}{Nucl. Phys. {\bf B592}
  (2001)  92--106},
\href{http://arxiv.org/abs/hep-ph/0005089}{{\tt arXiv:hep-ph/0005089
  [hep-ph]}}.

\bibitem{Bobrowski:2010xg}
M.~Bobrowski, A.~Lenz, J.~Riedl, and J.~Rohrwild, {\em {How Large Can the SM
  Contribution to CP Violation in $D^0-\bar D^0$ Mixing Be?}\/},
  \href{http://dx.doi.org/10.1007/JHEP03(2010)009}{JHEP {\bf 03} (2010)  009},
\href{http://arxiv.org/abs/1002.4794}{{\tt arXiv:1002.4794 [hep-ph]}}.

\bibitem{Carrasco:2014uya}
N.~Carrasco et al., {\em {$D^0$$-\bar{D}^0$ mixing in the standard model and
  beyond from $N_f$ =2 twisted mass QCD}\/},
  \href{http://dx.doi.org/10.1103/PhysRevD.90.014502}{Phys. Rev. {\bf D90}
  (2014) no.~1, 014502},
\href{http://arxiv.org/abs/1403.7302}{{\tt arXiv:1403.7302 [hep-lat]}}.

\bibitem{Carrasco:2015pra}
{ETM Collaboration}, N.~Carrasco, P.~Dimopoulos, R.~Frezzotti, V.~Lubicz, G.~C.
  Rossi, S.~Simula, and C.~Tarantino, {\em {$\Delta S=2$ and $\Delta C=2$ bag
  parameters in the standard model and beyond from N$_f$=2+1+1 twisted-mass
  lattice QCD}\/},  \href{http://dx.doi.org/10.1103/PhysRevD.92.034516}{Phys.
  Rev. {\bf D92} (2015) no.~3, 034516},
\href{http://arxiv.org/abs/1505.06639}{{\tt arXiv:1505.06639 [hep-lat]}}.

\bibitem{Bazavov:2017weg}
A.~Bazavov et al., {\em {Short-distance matrix elements for $D^0$-meson mixing
  for $N_f=2+1$ lattice QCD}\/},
  \href{http://dx.doi.org/10.1103/PhysRevD.97.034513}{Phys. Rev. {\bf D97}
  (2018) no.~3, 034513},
\href{http://arxiv.org/abs/1706.04622}{{\tt arXiv:1706.04622 [hep-lat]}}.

\bibitem{Kirk:2017juj}
M.~Kirk, A.~Lenz, and T.~Rauh, {\em {Dimension-six matrix elements for meson
  mixing and lifetimes from sum rules}\/},
  \href{http://dx.doi.org/10.1007/JHEP12(2017)068}{JHEP {\bf 12} (2017)  068},
\href{http://arxiv.org/abs/1711.02100}{{\tt arXiv:1711.02100 [hep-ph]}}.

\bibitem{Lenz:2016fcv}
A.~Lenz, {\em {Theory Overview}\/},  PoS {\bf CHARM2016} (2017)  003,
\href{http://arxiv.org/abs/1610.07943}{{\tt arXiv:1610.07943 [hep-ph]}}.

\bibitem{Lenz:2013aua}
A.~Lenz and T.~Rauh, {\em {D-meson lifetimes within the heavy quark
  expansion}\/},  \href{http://dx.doi.org/10.1103/PhysRevD.88.034004}{Phys.
  Rev. {\bf D88} (2013)  034004},
\href{http://arxiv.org/abs/1305.3588}{{\tt arXiv:1305.3588 [hep-ph]}}.

\bibitem{Bobrowski:2012jf}
M.~Bobrowski, A.~Lenz, and T.~Rauh, {\em {Short distance D-Dbar mixing}\/},  in
  {\em {Proceedings, 5th International Workshop on Charm Physics (Charm 2012):
  Honolulu, Hawaii, USA, May 14-17, 2012}}.
\newblock 2012.
\newblock \href{http://arxiv.org/abs/1208.6438}{{\tt arXiv:1208.6438
  [hep-ph]}}.
\newblock
\url{http://inspirehep.net/record/1184026/files/arXiv:1208.6438.pdf}.
\newblock

\bibitem{Falk:2004wg}
A.~F. Falk, Y.~Grossman, Z.~Ligeti, Y.~Nir, and A.~A. Petrov, {\em {The D0 -
  anti-D0 mass difference from a dispersion relation}\/},
  \href{http://dx.doi.org/10.1103/PhysRevD.69.114021}{Phys. Rev. {\bf D69}
  (2004)  114021},
\href{http://arxiv.org/abs/hep-ph/0402204}{{\tt arXiv:hep-ph/0402204
  [hep-ph]}}.

\bibitem{Cheng:2010rv}
H.-Y. Cheng and C.-W. Chiang, {\em {Long-Distance Contributions to
  $D^0-\bar{D}^0$ Mixing Parameters}\/},
  \href{http://dx.doi.org/10.1103/PhysRevD.81.114020}{Phys. Rev. {\bf D81}
  (2010)  114020},
\href{http://arxiv.org/abs/1005.1106}{{\tt arXiv:1005.1106 [hep-ph]}}.

\bibitem{Jiang:2017zwr}
H.-Y. Jiang, F.-S. Yu, Q.~Qin, H.-n. Li, and C.-D. Lu, {\em
  {$D^0$-$\overline{D}^0$ mixing parameter $y$ in the factorization-assisted
  topological-amplitude approach}\/},
  \href{http://arxiv.org/abs/1705.07335}{{\tt arXiv:1705.07335 [hep-ph]}}.
[Chin. Phys.C42,063101(2018)].

\bibitem{Nir:1992uv}
Y.~Nir, {\em {CP violation}\/},  Conf. Proc. {\bf C9207131} (1992)  81--136.
[,81(1992)].

\bibitem{Hansen:2012tf}
M.~T. Hansen and S.~R. Sharpe, {\em {Multiple-channel generalization of
  Lellouch-Luscher formula}\/},
  \href{http://dx.doi.org/10.1103/PhysRevD.86.016007}{Phys. Rev. {\bf D86}
  (2012)  016007},
\href{http://arxiv.org/abs/1204.0826}{{\tt arXiv:1204.0826 [hep-lat]}}.

\bibitem{GKLPPS}
Y.~Grossman, A.~Kagan, Z.~Ligeti, G.~Perez, A.~Petrov, and L.~Silvestrini.
\newblock in preparation.

\bibitem{Branco:1999fs}
G.~C. Branco, L.~Lavoura, and J.~P. Silva, {\em {CP Violation}\/},
Int. Ser. Monogr. Phys. {\bf 103} (1999)  1--536.

\bibitem{Bigi:2000yz}
I.~I. Bigi and A.~I. Sanda, {\em {CP violation}\/}, .
[Camb. Monogr. Part. Phys. Nucl. Phys. Cosmol.9,1(2009)].

\bibitem{Brod:2012ud}
J.~Brod, Y.~Grossman, A.~L. Kagan, and J.~Zupan, {\em {A consistent picture for
  large penguins in $D to\pip\pim, \Kp\Km$}\/},
  \href{http://dx.doi.org/10.1007/JHEP10(2012)161}{JHEP {\bf 10} (2012)  161},
\href{http://arxiv.org/abs/1203.6659}{{\tt arXiv:1203.6659 [hep-ph]}}.

\bibitem{Nierste:2015zra}
U.~Nierste and S.~Schacht, {\em {CP Violation in $D^0\rightarrow K_SK_S$}\/},
  \href{http://dx.doi.org/10.1103/PhysRevD.92.054036}{Phys. Rev. {\bf D92}
  (2015) no.~5, 054036},
\href{http://arxiv.org/abs/1508.00074}{{\tt arXiv:1508.00074 [hep-ph]}}.

\bibitem{Nierste:2017cua}
U.~Nierste and S.~Schacht, {\em {Neutral $D\rightarrow K K^*$ decays as
  discovery channels for charm CP violation}\/},
  \href{http://dx.doi.org/10.1103/PhysRevLett.119.251801}{Phys. Rev. Lett. {\bf
  119} (2017) no.~25, 251801},
\href{http://arxiv.org/abs/1708.03572}{{\tt arXiv:1708.03572 [hep-ph]}}.

\bibitem{Bergmann:1999pm}
S.~Bergmann and Y.~Nir, {\em {New physics effects in doubly Cabibbo suppressed
  D decays}\/},  \href{http://dx.doi.org/10.1088/1126-6708/1999/09/031}{JHEP
  {\bf 09} (1999)  031},
\href{http://arxiv.org/abs/hep-ph/9909391}{{\tt arXiv:hep-ph/9909391
  [hep-ph]}}.

\bibitem{Peng:2014oda}
{Belle collaboration}, T.~Peng et al., {\em {Measurement of $\Dz-\Dzb$ mixing
  and search for indirect CP violation using $D^0\to K_S^0\pi^+\pi^-$
  decays}\/},  \href{http://dx.doi.org/10.1103/PhysRevD.89.091103}{Phys. Rev.
  {\bf D89} (2014)  091103},
\href{http://arxiv.org/abs/1404.2412}{{\tt arXiv:1404.2412 [hep-ex]}}.

\bibitem{Blum:2009sk}
K.~Blum, Y.~Grossman, Y.~Nir, and G.~Perez, {\em {Combining $K^0 - \bar K^0$
  mixing and $D^0 - \bar D^0$ mixing to constrain the flavor structure of new
  physics}\/},  \href{http://dx.doi.org/10.1103/PhysRevLett.102.211802}{Phys.
  Rev. Lett. {\bf 102} (2009)  211802},
\href{http://arxiv.org/abs/0903.2118}{{\tt arXiv:0903.2118 [hep-ph]}}.

\bibitem{Kagan:2009gb}
A.~L. Kagan and M.~D. Sokoloff, {\em {On Indirect CP Violation and Implications
  for $D^0 - \bar D^0$ and $B_{(s)} - \bar B_{(s)}$ mixing}\/},
  \href{http://dx.doi.org/10.1103/PhysRevD.80.076008}{Phys. Rev. {\bf D80}
  (2009)  076008},
\href{http://arxiv.org/abs/0907.3917}{{\tt arXiv:0907.3917 [hep-ph]}}.

\bibitem{Savage:1991wu}
M.~J. Savage, {\em {SU(3) violations in the nonleptonic decay of charmed
  hadrons}\/},
\href{http://dx.doi.org/10.1016/0370-2693(91)91917-K}{Phys. Lett. {\bf B257}
  (1991)  414--418}.

\bibitem{Pirtskhalava:2011va}
D.~Pirtskhalava and P.~Uttayarat, {\em {\CP violation and flavor $SU(3)$
  breaking in $D$-meson decays}\/},
  \href{http://dx.doi.org/10.1016/j.physletb.2012.04.039}{Phys. Lett. {\bf
  B712} (2012)  81--86},
\href{http://arxiv.org/abs/1112.5451}{{\tt arXiv:1112.5451 [hep-ph]}}.

\bibitem{Feldmann:2012js}
T.~Feldmann, S.~Nandi, and A.~Soni, {\em {Repercussions of flavour symmetry
  breaking on \CP violation in $D$-meson decays}\/},
  \href{http://dx.doi.org/10.1007/JHEP06(2012)007}{JHEP {\bf 06} (2012)  007},
\href{http://arxiv.org/abs/1202.3795}{{\tt arXiv:1202.3795 [hep-ph]}}.

\bibitem{Franco:2012ck}
E.~Franco, S.~Mishima, and L.~Silvestrini, {\em {The Standard Model confronts
  CP violation in $D^0 \to \pi^+\pi^-$ and $D^0 \to K^+K^-$}\/},
  \href{http://dx.doi.org/10.1007/JHEP05(2012)140}{JHEP {\bf 05} (2012)  140},
\href{http://arxiv.org/abs/1203.3131}{{\tt arXiv:1203.3131 [hep-ph]}}.

\bibitem{Hiller:2012xm}
G.~Hiller, M.~Jung, and S.~Schacht, {\em {SU(3)-flavor anatomy of nonleptonic
  charm decays}\/},  \href{http://dx.doi.org/10.1103/PhysRevD.87.014024}{Phys.
  Rev. {\bf D87} (2013) no.~1, 014024},
\href{http://arxiv.org/abs/1211.3734}{{\tt arXiv:1211.3734 [hep-ph]}}.

\bibitem{Atwood:2012ac}
D.~Atwood and A.~Soni, {\em {Searching for the Origin of CP violation in
  Cabibbo Suppressed D-meson Decays}\/},
  \href{http://dx.doi.org/10.1093/ptep/ptt065}{PTEP {\bf 2013} (2013) no.~9,
  093B05},
\href{http://arxiv.org/abs/1211.1026}{{\tt arXiv:1211.1026 [hep-ph]}}.

\bibitem{Brod:2011re}
J.~Brod, A.~L. Kagan, and J.~Zupan, {\em {Size of direct CP violation in singly
  Cabibbo-suppressed D decays}\/},
  \href{http://dx.doi.org/10.1103/PhysRevD.86.014023}{Phys. Rev. {\bf D86}
  (2012)  014023},
\href{http://arxiv.org/abs/1111.5000}{{\tt arXiv:1111.5000 [hep-ph]}}.

\bibitem{Khodjamirian:2017zdu}
A.~Khodjamirian and A.~A. Petrov, {\em {Direct CP asymmetry in $D\to
  \pi^-\pi^+$ and $D\to K^-K^+$ in QCD-based approach}\/},
  \href{http://dx.doi.org/10.1016/j.physletb.2017.09.070}{Phys. Lett. {\bf
  B774} (2017)  235--242},
\href{http://arxiv.org/abs/1706.07780}{{\tt arXiv:1706.07780 [hep-ph]}}.

\bibitem{Grossman:2006jg}
Y.~Grossman, A.~L. Kagan, and Y.~Nir, {\em {New physics and CP violation in
  singly Cabibbo suppressed D decays}\/},
  \href{http://dx.doi.org/10.1103/PhysRevD.75.036008}{Phys. Rev. {\bf D75}
  (2007)  036008},
\href{http://arxiv.org/abs/hep-ph/0609178}{{\tt arXiv:hep-ph/0609178
  [hep-ph]}}.

\bibitem{Ryd:2009uf}
A.~Ryd and A.~A. Petrov, {\em {Hadronic D and D(s) Meson Decays}\/},
  \href{http://dx.doi.org/10.1103/RevModPhys.84.65}{Rev. Mod. Phys. {\bf 84}
  (2012)  65--117},
\href{http://arxiv.org/abs/0910.1265}{{\tt arXiv:0910.1265 [hep-ph]}}.

\bibitem{deBoer:2015boa}
S.~de~Boer and G.~Hiller, {\em {Flavor and new physics opportunities with rare
  charm decays into leptons}\/},
  \href{http://dx.doi.org/10.1103/PhysRevD.93.074001}{Phys. Rev. {\bf D93}
  (2016) no.~7, 074001},
\href{http://arxiv.org/abs/1510.00311}{{\tt arXiv:1510.00311 [hep-ph]}}.

\bibitem{deBoer:2017que}
S.~de~Boer and G.~Hiller, {\em {Rare radiative charm decays within the standard
  model and beyond}\/},  \href{http://dx.doi.org/10.1007/JHEP08(2017)091}{JHEP
  {\bf 08} (2017)  091},
\href{http://arxiv.org/abs/1701.06392}{{\tt arXiv:1701.06392 [hep-ph]}}.

\bibitem{deBoer:2018buv}
S.~de~Boer and G.~Hiller, {\em {Null tests from angular distributions in $D \to
  P_1 P_2 l^+l^-$, $l=e,\mu$ decays on and off peak}\/},
\href{http://arxiv.org/abs/1805.08516}{{\tt arXiv:1805.08516 [hep-ph]}}.

\bibitem{LHCb-PAPER-2018-020}
{LHCb collaboration}, R.~Aaij et al., {\em {Measurement of angular and \CP
  asymmetries in \decay{\Dz}{\pi^+\pi^-\mumu} and \decay{\Dz}{K^+K^-\mumu}
  decays}\/},  Phys. Rev. Lett. {\bf 121} (2018)  091801,
  \href{http://arxiv.org/abs/1806.10793}{{\tt arXiv:1806.10793 [hep-ex]}}.

\bibitem{Datta:2003mj}
A.~Datta and D.~London, {\em {Triple-product correlations in $B \rightarrow
  V_{1} V_{2}$ decays and new physics}\/},
  \href{http://dx.doi.org/10.1142/S0217751X04018300}{Int. J. Mod. Phys. {\bf
  A19} (2004)  2505--2544},
\href{http://arxiv.org/abs/hep-ph/0303159}{{\tt arXiv:hep-ph/0303159
  [hep-ph]}}.

\bibitem{Grossman:2012eb}
Y.~Grossman, A.~L. Kagan, and J.~Zupan, {\em {Testing for new physics in singly
  Cabibbo suppressed D decays}\/},
  \href{http://dx.doi.org/10.1103/PhysRevD.85.114036}{Phys. Rev. {\bf D85}
  (2012)  114036},
\href{http://arxiv.org/abs/1204.3557}{{\tt arXiv:1204.3557 [hep-ph]}}.

\bibitem{Grossman:2012ry}
Y.~Grossman and D.~J. Robinson, {\em {SU(3) Sum Rules for Charm Decay}\/},
  \href{http://dx.doi.org/10.1007/JHEP04(2013)067}{JHEP {\bf 04} (2013)  067},
\href{http://arxiv.org/abs/1211.3361}{{\tt arXiv:1211.3361 [hep-ph]}}.

\bibitem{Grossman:2013lya}
Y.~Grossman, Z.~Ligeti, and D.~J. Robinson, {\em {More Flavor SU(3) Tests for
  New Physics in CP Violating B Decays}\/},
  \href{http://dx.doi.org/10.1007/JHEP01(2014)066}{JHEP {\bf 01} (2014)  066},
\href{http://arxiv.org/abs/1308.4143}{{\tt arXiv:1308.4143 [hep-ph]}}.

\bibitem{Muller:2015rna}
S.~Muller, U.~Nierste, and S.~Schacht, {\em {Sum Rules of Charm CP Asymmetries
  beyond the SU(3)$_F$ Limit}\/},
  \href{http://dx.doi.org/10.1103/PhysRevLett.115.251802}{Phys. Rev. Lett. {\bf
  115} (2015) no.~25, 251802},
\href{http://arxiv.org/abs/1506.04121}{{\tt arXiv:1506.04121 [hep-ph]}}.

\bibitem{deBoer:2016dcg}
S.~de~Boer, B.~Muller, and D.~Seidel, {\em {Higher-order Wilson coefficients
  for $c \to u$ transitions in the standard model}\/},
  \href{http://dx.doi.org/10.1007/JHEP08(2016)091}{JHEP {\bf 08} (2016)  091},
\href{http://arxiv.org/abs/1606.05521}{{\tt arXiv:1606.05521 [hep-ph]}}.

\bibitem{Greub:1996wn}
C.~Greub, T.~Hurth, M.~Misiak, and D.~Wyler, {\em {The $c\to u$ gamma
  contribution to weak radiative charm decay}\/},
  \href{http://dx.doi.org/10.1016/0370-2693(96)00694-6}{Phys. Lett. {\bf B382}
  (1996)  415--420},
\href{http://arxiv.org/abs/hep-ph/9603417}{{\tt arXiv:hep-ph/9603417
  [hep-ph]}}.

\bibitem{Fajfer:2002gp}
S.~Fajfer, P.~Singer, and J.~Zupan, {\em {The Radiative leptonic decays $D^0
  \to e^+ e^- \gamma, \mu^+ \mu^- \gamma$ in the standard model and beyond}\/},
   \href{http://dx.doi.org/10.1140/epjc/s2002-01090-5}{Eur. Phys. J. {\bf C27}
  (2003)  201--218},
\href{http://arxiv.org/abs/hep-ph/0209250}{{\tt arXiv:hep-ph/0209250
  [hep-ph]}}.

\bibitem{Fajfer:1997bh}
S.~Fajfer and P.~Singer, {\em {Long distance $c\to u$ gamma effects in weak
  radiative decays of D mesons}\/},
  \href{http://dx.doi.org/10.1103/PhysRevD.56.4302}{Phys. Rev. {\bf D56} (1997)
   4302--4310},
\href{http://arxiv.org/abs/hep-ph/9705327}{{\tt arXiv:hep-ph/9705327
  [hep-ph]}}.

\bibitem{Fajfer:1998dv}
S.~Fajfer, S.~Prelovsek, and P.~Singer, {\em {Long distance contributions in
  $D\to V \gamma$ decays}\/},  \href{http://dx.doi.org/10.1007/s100520050356,
  10.1007/s100529800914}{Eur. Phys. J. {\bf C6} (1999)  471--476},
\href{http://arxiv.org/abs/hep-ph/9801279}{{\tt arXiv:hep-ph/9801279
  [hep-ph]}}.

\bibitem{Tanabashi:2018oca}
{Particle Data Group Collaboration}, M.~Tanabashi et al., {\em {Review of
  Particle Physics}\/},
\href{http://dx.doi.org/10.1103/PhysRevD.98.030001}{Phys. Rev. {\bf D98} (2018)
  no.~3, 030001}.

\bibitem{deBoer:2017rzd}
S.~de~Boer, {\em {Rare radiative charm decays in the standard model and
  beyond}\/},  \href{http://dx.doi.org/10.22323/1.314.0209}{PoS {\bf
  EPS-HEP2017} (2017)  209},
\href{http://arxiv.org/abs/1710.06670}{{\tt arXiv:1710.06670 [hep-ph]}}.

\bibitem{deBoer:2018zhz}
S.~de~Boer and G.~Hiller, {\em {The photon polarization in radiative $D$
  decays, phenomenologically}\/},
  \href{http://dx.doi.org/10.1140/epjc/s10052-018-5682-7}{Eur. Phys. J. {\bf
  C78} (2018) no.~3, 188},
\href{http://arxiv.org/abs/1802.02769}{{\tt arXiv:1802.02769 [hep-ph]}}.

\bibitem{Fajfer:2015mia}
S.~Fajfer and N.~Ko{\v s}nik, {\em {Prospects of discovering new physics in
  rare charm decays}\/},
  \href{http://dx.doi.org/10.1140/epjc/s10052-015-3801-2}{Eur. Phys. J. {\bf
  C75} (2015) no.~12, 567},
\href{http://arxiv.org/abs/1510.00965}{{\tt arXiv:1510.00965 [hep-ph]}}.

\bibitem{LHCb-PAPER-2013-013}
{LHCb collaboration}, R.~Aaij et al., {\em {Search for the rare decay $\Dz\to
  \mup\mun$}\/},  Phys. Lett. {\bf B725} (2013)  15,
  \href{http://arxiv.org/abs/1305.5059}{{\tt arXiv:1305.5059 [hep-ex]}}.

\bibitem{LHCb-PAPER-2012-051}
{LHCb collaboration}, R.~Aaij et al., {\em {Search for $D^+_{(s)}\to
  \pip\mup\mun$ and $D^+_{(s)}\to \pim\mup\mup$ decays}\/},  Phys. Lett. {\bf
  B724} (2013)  203, \href{http://arxiv.org/abs/1304.6365}{{\tt arXiv:1304.6365
  [hep-ex]}}.

\bibitem{Meinel:2017ggx}
S.~Meinel, {\em {$\Lambda_c \to N$ form factors from lattice QCD and
  phenomenology of $\Lambda_c \to n \ell^+ \nu_\ell$ and $\Lambda_c \to p \mu^+
  \mu^-$ decays}\/},  \href{http://dx.doi.org/10.1103/PhysRevD.97.034511}{Phys.
  Rev. {\bf D97} (2018) no.~3, 034511},
\href{http://arxiv.org/abs/1712.05783}{{\tt arXiv:1712.05783 [hep-lat]}}.

\bibitem{LHCb-PAPER-2017-039}
{LHCb collaboration}, R.~Aaij et al., {\em {Search for the rare decay $\Lc\to
  p\mumu$}\/},  Phys. Rev. {\bf D97} (2018)  091101,
  \href{http://arxiv.org/abs/1712.07938}{{\tt arXiv:1712.07938 [hep-ex]}}.

\bibitem{Dorsner:2016wpm}
I.~Dor\v{s}ner, S.~Fajfer, A.~Greljo, J.~F. Kamenik, and N.~Ko\v{s}nik, {\em
  {Physics of leptoquarks in precision experiments and at particle
  colliders}\/},  \href{http://dx.doi.org/10.1016/j.physrep.2016.06.001}{Phys.
  Rept. {\bf 641} (2016)  1--68},
\href{http://arxiv.org/abs/1603.04993}{{\tt arXiv:1603.04993 [hep-ph]}}.

\bibitem{Buttazzo:2017ixm}
D.~Buttazzo, A.~Greljo, G.~Isidori, and D.~Marzocca, {\em {$B$-physics
  anomalies: a guide to combined explanations}\/},
  \href{http://dx.doi.org/10.1007/JHEP11(2017)044}{JHEP {\bf 11} (2017)  044},
\href{http://arxiv.org/abs/1706.07808}{{\tt arXiv:1706.07808 [hep-ph]}}.

\bibitem{Fajfer:2012nr}
S.~Fajfer and N.~Ko{\v s}nik, {\em {Resonance catalyzed CP asymmetries in $D\to
  P\ell^+\ell^-$}\/},
  \href{http://dx.doi.org/10.1103/PhysRevD.87.054026}{Phys. Rev. {\bf D87}
  (2013) no.~5, 054026},
\href{http://arxiv.org/abs/1208.0759}{{\tt arXiv:1208.0759 [hep-ph]}}.

\bibitem{Golowich:2009ii}
E.~Golowich, J.~Hewett, S.~Pakvasa, and A.~A. Petrov, {\em {Relating D0-anti-D0
  Mixing and D0 ---> l+ l- with New Physics}\/},
  \href{http://dx.doi.org/10.1103/PhysRevD.79.114030}{Phys. Rev. {\bf D79}
  (2009)  114030},
\href{http://arxiv.org/abs/0903.2830}{{\tt arXiv:0903.2830 [hep-ph]}}.

\bibitem{LHCb-PAPER-2017-019}
{LHCb collaboration}, R.~Aaij et al., {\em {Observation of $\Dz$ meson decays
  to $\pi^+\pi^-\mumu$ and $K^+K^-\mumu$ final states}\/},  Phys. Rev. Lett.
  {\bf 119} (2017)  181805, \href{http://arxiv.org/abs/1707.08377}{{\tt
  arXiv:1707.08377 [hep-ex]}}.

\bibitem{Ablikim:2018gro}
{BESIII Collaboration}, M.~Ablikim et al., {\em {Search for the rare decays
  $D\to h(h')e^+e^-$}\/},
  \href{http://dx.doi.org/10.1103/PhysRevD.97.072015}{Phys. Rev. {\bf D97}
  (2018) no.~7, 072015},
\href{http://arxiv.org/abs/1802.09752}{{\tt arXiv:1802.09752 [hep-ex]}}.

\bibitem{Martone:2012nj}
M.~Martone and J.~Zupan, {\em {$B^\pm \to D K^\pm$ with direct CP violation in
  charm}\/},  \href{http://dx.doi.org/10.1103/PhysRevD.87.034005}{Phys. Rev.
  {\bf D87} (2013) no.~3, 034005},
\href{http://arxiv.org/abs/1212.0165}{{\tt arXiv:1212.0165 [hep-ph]}}.

\bibitem{Wang:2012ie}
W.~Wang, {\em {CP Violation Effects on the Measurement of the
  Cabibbo-Kobayashi-Maskawa Angle $\gamma$ from B $\to$ D K}\/},
  \href{http://dx.doi.org/10.1103/PhysRevLett.110.061802}{Phys. Rev. Lett. {\bf
  110} (2013) no.~6, 061802},
\href{http://arxiv.org/abs/1211.4539}{{\tt arXiv:1211.4539 [hep-ph]}}.

\bibitem{Bhattacharya:2013vc}
B.~Bhattacharya, D.~London, M.~Gronau, and J.~L. Rosner, {\em {Shift in weak
  phase $\gamma$ due to CP asymmetries in D decays to two pseudoscalar
  mesons}\/},  \href{http://dx.doi.org/10.1103/PhysRevD.87.074002}{Phys. Rev.
  {\bf D87} (2013) no.~7, 074002},
\href{http://arxiv.org/abs/1301.5631}{{\tt arXiv:1301.5631 [hep-ph]}}.

\bibitem{Asner:2005sz}
{CLEO collaboration}, D.~M. Asner et al., {\em {Search for $\Dz-\Dzb$ mixing in
  the Dalitz plot analysis of $D^0\to\KS\pip\pim$}\/},
  \href{http://dx.doi.org/10.1103/PhysRevD.72.012001}{Phys. Rev. {\bf D72}
  (2005)  012001},
\href{http://arxiv.org/abs/hep-ex/0503045}{{\tt arXiv:hep-ex/0503045
  [hep-ex]}}.

\bibitem{delAmoSanchez:2010xz}
{BaBar collaboration}, P.~del Amo~Sanchez et al., {\em {Measurement of
  $\Dz-\Dzb$ mixing parameters using $D^0\to\KS\pip\pim$ and $D^0\to\KS\Kp\Km$
  decays}\/},  \href{http://dx.doi.org/10.1103/PhysRevLett.105.081803}{Phys.
  Rev. Lett. {\bf 105} (2010)  081803},
\href{http://arxiv.org/abs/1004.5053}{{\tt arXiv:1004.5053 [hep-ex]}}.

\bibitem{LHCb-PAPER-2015-042}
{LHCb collaboration}, R.~Aaij et al., {\em {Model-independent measurement of
  mixing parameters in $\Dz\to \KS\pip\pim$ decays}\/},  JHEP {\bf 04} (2016)
  033, \href{http://arxiv.org/abs/1510.01664}{{\tt arXiv:1510.01664 [hep-ex]}}.

\bibitem{Libby:2010nu}
{CLEO collaboration}, J.~Libby et al., {\em {Model-independent determination of
  the strong-phase difference between $D^0$ and $\bar{D}^0 \to K^0_{S,L} h^+
  h^-$ ($h=\pi,K$) and its impact on the measurement of the CKM angle
  $\gamma/\phi_3$}\/},
  \href{http://dx.doi.org/10.1103/PhysRevD.82.112006}{Phys. Rev. {\bf D82}
  (2010)  112006},
\href{http://arxiv.org/abs/1010.2817}{{\tt arXiv:1010.2817 [hep-ex]}}.

\bibitem{DiCanto:2018tsd}
A.~Di~Canto, J.~G. Tic{\'o}, T.~Gershon, N.~Jurik, M.~Martinelli,
  T.~Pila{\v{r}}, S.~Stahl, and D.~Tonelli, {\em {A novel method for measuring
  charm-mixing parameters using multibody decays}\/},
\href{http://arxiv.org/abs/1811.01032}{{\tt arXiv:1811.01032 [hep-ex]}}.

\bibitem{LHCb-PAPER-2015-057}
{LHCb collaboration}, R.~Aaij et al., {\em {First observation of $\Dz-\Dzb$
  oscillations in $\Dz\to \Kp\pip\pim\pim$ decays and a measurement of the
  associated coherence parameters}\/},  Phys. Rev. Lett. {\bf 116} (2016)
  241801, \href{http://arxiv.org/abs/1602.07224}{{\tt arXiv:1602.07224
  [hep-ex]}}.

\bibitem{Atwood:2003mj}
D.~Atwood and A.~Soni, {\em {Role of charm factory in extracting CKM phase
  information via $B \to DK$}\/},
  \href{http://dx.doi.org/10.1103/PhysRevD.68.033003}{Phys. Rev. {\bf D68}
  (2003)  033003},
\href{http://arxiv.org/abs/hep-ph/0304085}{{\tt arXiv:hep-ph/0304085
  [hep-ph]}}.

\bibitem{Evans:2016tlp}
T.~Evans, S.~Harnew, J.~Libby, S.~Malde, J.~Rademacker, and G.~Wilkinson, {\em
  {Improved determination of the $D \to K^-\pi^+\pi^+\pi^-$ coherence factor
  and associated hadronic parameters from a combination of $e^+e^-\to
  \psi(3770)\to c\bar{c}$ and $pp \to c \bar{c} X$ data}\/},
  \href{http://dx.doi.org/10.1016/j.physletb.2016.04.037,
  10.1016/j.physletb.2016.11.021}{Phys. Lett. {\bf B757} (2016)  520--527},
  \href{http://arxiv.org/abs/1602.07430}{{\tt arXiv:1602.07430 [hep-ex]}}.
[Erratum: Phys. Lett.B765,402(2017)].

\bibitem{Malde:2223391}
S.~S. Malde, {\em {Synergy of BESIII and LHCb physics programmes}\/},  Oct,
  2016.
\newblock \url{http://cds.cern.ch/record/2223391}.

\bibitem{Muller:2297069}
D.~M{\"u}ller, M.~Gersabeck, and C.~Parkes, {\em {Measurements of production
  cross-sections and mixing of charm mesons at LHCb}\/},  Nov, 2017.
\newblock \url{http://cds.cern.ch/record/2297069}.

\bibitem{LHCb-PAPER-2014-069}
{LHCb collaboration}, R.~Aaij et al., {\em {Measurement of indirect \CP
  asymmetries in $\Dz\to \Km\Kp$ and $\Dz\to \pim\pip$ decays}\/},  JHEP {\bf
  04} (2015)  043, \href{http://arxiv.org/abs/1501.06777}{{\tt arXiv:1501.06777
  [hep-ex]}}.

\bibitem{LHCb-PAPER-2016-032}
{LHCb collaboration}, R.~Aaij et al., {\em {Measurement of the CKM angle
  $\gamma$ from a combination of LHCb results}\/},  JHEP {\bf 12} (2016)  087,
  \href{http://arxiv.org/abs/1611.03076}{{\tt arXiv:1611.03076 [hep-ex]}}.

\bibitem{LHCb-PAPER-2016-035}
{LHCb collaboration}, R.~Aaij et al., {\em {Measurement of \CP asymmetry in
  $\Dz\to \Kp\Km$ decays}\/},  Phys. Lett. {\bf B767} (2017)  177,
  \href{http://arxiv.org/abs/1610.09476}{{\tt arXiv:1610.09476 [hep-ex]}}.

\bibitem{HFLAV16}
{Heavy Flavor Averaging Group Collaboration}, Y.~Amhis et al., {\em {Averages
  of $b$-hadron, $c$-hadron, and $\tau$-lepton properties as of summer
  2016}\/},  \href{http://dx.doi.org/10.1140/epjc/s10052-017-5058-4}{Eur. Phys.
  J. {\bf C77} (2017)  895}, \href{http://arxiv.org/abs/1612.07233}{{\tt
  arXiv:1612.07233 [hep-ex]}}.
{updated results and plots available at
  \href{https://hflav.web.cern.ch}{{\texttt{https://hflav.web.cern.ch}}}}.

\bibitem{Durieux:2015zwa}
G.~Durieux and Y.~Grossman, {\em {Probing CP violation systematically in
  differential distributions}\/},
  \href{http://dx.doi.org/10.1103/PhysRevD.92.076013}{Phys. Rev. {\bf D92}
  (2015)  076013},
\href{http://arxiv.org/abs/1508.03054}{{\tt arXiv:1508.03054 [hep-ph]}}.

\bibitem{LHCB-PAPER-2014-046}
{LHCb collaboration}, R.~Aaij et al., {\em {Search for \CP violation using
  $T$-odd correlations in $\Dz\to \Kp\Km\pip\pim$ decays}\/},  JHEP {\bf 10}
  (2014)  005, \href{http://arxiv.org/abs/1408.1299}{{\tt arXiv:1408.1299
  [hep-ex]}}.

\bibitem{LHCb-PAPER-2016-044}
{LHCb collaboration}, R.~Aaij et al., {\em {Search for \CP violation in the
  phase space of $\Dz\to \pip\pim\pip\pim$ decays}\/},  Phys. Lett. {\bf B769}
  (2017)  345, \href{http://arxiv.org/abs/1612.03207}{{\tt arXiv:1612.03207
  [hep-ex]}}.

\bibitem{Bhattacharya:2012ah}
B.~Bhattacharya, M.~Gronau, and J.~L. Rosner, {\em {\CP asymmetries in
  singly-Cabibbo-suppressed $D$ decays to two pseudoscalar mesons}\/},
  \href{http://dx.doi.org/10.1103/PhysRevD.85.079901}{Phys. Rev. {\bf D85}
  (2012)  079901},
\href{http://arxiv.org/abs/1201.2351}{{\tt arXiv:1201.2351 [hep-ph]}}.

\bibitem{TheBABAR:2016gom}
{BaBar collaboration}, J.~P. Lees et al., {\em {Measurement of the neutral $D$
  meson mixing parameters in a time-dependent amplitude analysis of the
  $D^0\to\pi^+\pi^-\pi^0$ decay}\/},
  \href{http://dx.doi.org/10.1103/PhysRevD.93.112014}{Phys. Rev. {\bf D93}
  (2016)  112014},
\href{http://arxiv.org/abs/1604.00857}{{\tt arXiv:1604.00857 [hep-ex]}}.

\bibitem{LHCb-PAPER-2016-041}
{LHCb collaboration}, R.~Aaij et al., {\em {Measurement of \CP asymmetries in
  $\Dpm\to \etapr\pipm$ and $\Dspm \to \etapr\pipm$ decays}\/},  Phys. Lett.
  {\bf B771} (2017)  21, \href{http://arxiv.org/abs/1701.01871}{{\tt
  arXiv:1701.01871 [hep-ex]}}.

\bibitem{LHCb-PAPER-2015-043}
{LHCb collaboration}, R.~Aaij et al., {\em {First observation of the decay
  $\Dz\to \Km\pip\mup\mun$ in the $\rhoz-\omegaz$ region of the dimuon mass
  spectrum}\/},  Phys. Lett. {\bf B757} (2016)  558,
  \href{http://arxiv.org/abs/1510.08367}{{\tt arXiv:1510.08367 [hep-ex]}}.

\bibitem{Bigi:2012ev}
I.~I. Bigi, {\em {Probing CP asymmetries in charm baryons decays}\/},
\href{http://arxiv.org/abs/1206.4554}{{\tt arXiv:1206.4554 [hep-ph]}}.

\bibitem{LHCb-PAPER-2018-026}
{LHCb collaboration}, R.~Aaij et al., {\em {First observation of the doubly
  charmed baryon decay \decay{\Xires_{cc}^{++}}{\Xires_c^+\pi^+} decay}\/},
  Phys. Rev. Lett. {\bf 121} (2018)  162002,
  \href{http://arxiv.org/abs/1807.01919}{{\tt arXiv:1807.01919 [hep-ex]}}.

\bibitem{LHCb-PAPER-2012-023}
{LHCb collaboration}, R.~Aaij et al., {\em {Search for the rare decay $\KS\to
  \mup\mun$}\/},  JHEP {\bf 01} (2013)  090,
  \href{http://arxiv.org/abs/1209.4029}{{\tt arXiv:1209.4029 [hep-ex]}}.

\bibitem{Dettori:2297352}
F.~Dettori, D.~Martinez~Santos, and J.~Prisciandaro, {\em {Low-$p_T$ dimuon
  triggers at LHCb in Run 2}\/},   LHCb-PUB-2017-023. CERN-LHCb-PUB-2017-023,
  CERN, Geneva, Dec, 2017.
\newblock \url{http://cds.cern.ch/record/2297352}.

\bibitem{kaon_paper}
A.~A. Alves~Junior et al., {\em {Prospects for Measurements with Strange
  Hadrons at LHCb}\/},
\href{http://arxiv.org/abs/1808.03477}{{\tt arXiv:1808.03477 [hep-ex]}}.

\bibitem{dmsFPCP}
D.~Martinez~Santos.
\newblock
  \url{https://cds.cern.ch/record/2270191/files/fpcp2017-MartinezSantos.pdf}.
  LHCb-TALK-2017-164, at FPCP 2017.

\bibitem{Chobanova:2195218}
V.~G. Chobanova, X.~Cid~Vidal, J.~P. Dalseno, M.~Lucio~Martinez,
  D.~Martinez~Santos, and V.~Renaudin, {\em {Sensitivity of LHCb and its
  upgrade in the measurement of
  $\mathcal{B}(K_S^0\rightarrow\pi^0\mu^+\mu^-)$}\/},   LHCb-PUB-2016-017,
  CERN, Geneva, Oct, 2016.
\newblock \url{http://cds.cern.ch/record/2195218}.

\bibitem{Borsato:2018tcz}
M.~Borsato, V.~V. Gligorov, D.~Guadagnoli, D.~Martinez~Santos, and
  O.~Sumensari, {\em {The strange side of LHCb}\/},
\href{http://arxiv.org/abs/1808.02006}{{\tt arXiv:1808.02006 [hep-ph]}}.

\bibitem{Ecker:1991ru}
G.~Ecker and A.~Pich, {\em {The Longitudinal muon polarization in $K_L \to
  \mu^+ \mu^-$}\/},
\href{http://dx.doi.org/10.1016/0550-3213(91)90056-4}{Nucl. Phys. {\bf B366}
  (1991)  189--205}.

\bibitem{Isidori:2003ts}
G.~Isidori and R.~Unterdorfer, {\em {On the short distance constraints from
  $K_{L,S}\to \mu^+ \mu^-$}\/},
  \href{http://dx.doi.org/10.1088/1126-6708/2004/01/009}{JHEP {\bf 01} (2004)
  009},
\href{http://arxiv.org/abs/hep-ph/0311084}{{\tt arXiv:hep-ph/0311084
  [hep-ph]}}.

\bibitem{DAmbrosio:2017klp}
G.~D'Ambrosio and T.~Kitahara, {\em {Direct $CP$ Violation in $K \to \mu^+
  \mu^-$}\/},  \href{http://dx.doi.org/10.1103/PhysRevLett.119.201802}{Phys.
  Rev. Lett. {\bf 119} (2017) no.~20, 201802},
\href{http://arxiv.org/abs/1707.06999}{{\tt arXiv:1707.06999 [hep-ph]}}.

\bibitem{Colangelo:2016ruc}
G.~Colangelo, R.~Stucki, and L.~C. Tunstall, {\em {Dispersive treatment of
  $K_S\rightarrow \gamma \gamma $ and $K_S\rightarrow \gamma \ell ^+\ell
  ^-$}\/},  \href{http://dx.doi.org/10.1140/epjc/s10052-016-4449-2}{Eur. Phys.
  J. {\bf C76} (2016) no.~11, 604},
\href{http://arxiv.org/abs/1609.03574}{{\tt arXiv:1609.03574 [hep-ph]}}.

\bibitem{AmelinoCamelia:2010me}
G.~Amelino-Camelia et al., {\em {Physics with the KLOE-2 experiment at the
  upgraded DA$\phi$NE}\/},
  \href{http://dx.doi.org/10.1140/epjc/s10052-010-1351-1}{Eur. Phys. J. {\bf
  C68} (2010)  619--681},
\href{http://arxiv.org/abs/1003.3868}{{\tt arXiv:1003.3868 [hep-ex]}}.

\bibitem{Bobeth:2017ecx}
C.~Bobeth and A.~J. Buras, {\em {Leptoquarks meet $\varepsilon'/\varepsilon$
  and rare Kaon processes}\/},
  \href{http://dx.doi.org/10.1007/JHEP02(2018)101}{JHEP {\bf 02} (2018)  101},
\href{http://arxiv.org/abs/1712.01295}{{\tt arXiv:1712.01295 [hep-ph]}}.

\bibitem{Chobanova:2017rkj}
V.~Chobanova, G.~D'Ambrosio, T.~Kitahara, M.~Lucio~Martinez,
  D.~Martinez~Santos, I.~S. Fernandez, and K.~Yamamoto, {\em {Probing SUSY
  effects in $K_S^0\rightarrow\mu^+\mu^-$}\/},
  \href{http://dx.doi.org/10.1007/JHEP05(2018)024}{JHEP {\bf 05} (2018)  024},
\href{http://arxiv.org/abs/1711.11030}{{\tt arXiv:1711.11030 [hep-ph]}}.

\bibitem{LHCb-PAPER-2017-009}
{LHCb collaboration}, R.~Aaij et al., {\em {Improved limit on the branching
  fraction of the rare decay $\KS\to \muon\muon$}\/},  Eur. Phys. J. {\bf C77}
  (2017)  678, \href{http://arxiv.org/abs/1706.00758}{{\tt arXiv:1706.00758
  [hep-ex]}}.

\bibitem{PDG2018}
{Particle Data Group Collaboration}, M.~Tanabashi et al., {\em {Review of
  Particle Physics}\/},  Phys. Rev. {\bf D98} (2018)  030001.

\bibitem{Gorbahn:2006bm}
M.~Gorbahn and U.~Haisch, {\em {Charm Quark Contribution to $K_L \to \mu^+
  \mu^-$ at Next-to-Next-to-Leading}\/},
  \href{http://dx.doi.org/10.1103/PhysRevLett.97.122002}{Phys. Rev. Lett. {\bf
  97} (2006)  122002},
\href{http://arxiv.org/abs/hep-ph/0605203}{{\tt arXiv:hep-ph/0605203
  [hep-ph]}}.

\bibitem{Mescia:2006jd}
F.~Mescia, C.~Smith, and S.~Trine, {\em {$K_L \to \pi^0 e^+ e^-$ and $K_L \to
  \pi^0 \mu^+ \mu^-$: A Binary star on the stage of flavor physics}\/},
  \href{http://dx.doi.org/10.1088/1126-6708/2006/08/088}{JHEP {\bf 08} (2006)
  088},
\href{http://arxiv.org/abs/hep-ph/0606081}{{\tt arXiv:hep-ph/0606081
  [hep-ph]}}.

\bibitem{Buras:2015yba}
A.~J. Buras, M.~Gorbahn, S.~J{\"a}ger, and M.~Jamin, {\em {Improved anatomy of
  $\varepsilon'/\varepsilon$ in the Standard Model}\/},
  \href{http://dx.doi.org/10.1007/JHEP11(2015)202}{JHEP {\bf 11} (2015)  202},
\href{http://arxiv.org/abs/1507.06345}{{\tt arXiv:1507.06345 [hep-ph]}}.

\bibitem{Kitahara:2016nld}
T.~Kitahara, U.~Nierste, and P.~Tremper, {\em {Singularity-free next-to-leading
  order $\Delta$S = 1 renormalization group evolution and
  $\epsilon_K'/\epsilon_K$ in the Standard Model and beyond}\/},
  \href{http://dx.doi.org/10.1007/JHEP12(2016)078}{JHEP {\bf 12} (2016)  078},
\href{http://arxiv.org/abs/1607.06727}{{\tt arXiv:1607.06727 [hep-ph]}}.

\bibitem{Bai:2015nea}
{RBC, UKQCD Collaboration}, Z.~Bai et al., {\em {Standard Model Prediction for
  Direct $CP$ Violation in $K \to \pi \pi$ Decay}\/},
  \href{http://dx.doi.org/10.1103/PhysRevLett.115.212001}{Phys. Rev. Lett. {\bf
  115} (2015) no.~21, 212001},
\href{http://arxiv.org/abs/1505.07863}{{\tt arXiv:1505.07863 [hep-lat]}}.

\bibitem{Buras:2015xba}
A.~J. Buras and J.-M. G\'erard, {\em {Upper bounds on
  $\varepsilon^\prime/\varepsilon$ parameters B$_{6}^{(1/2)}$ and
  B$_{8}^{(3/2)}$ from large N QCD and other news}\/},
  \href{http://dx.doi.org/10.1007/JHEP12(2015)008}{JHEP {\bf 12} (2015)  008},
\href{http://arxiv.org/abs/1507.06326}{{\tt arXiv:1507.06326 [hep-ph]}}.

\bibitem{Pallante:1999qf}
E.~Pallante and A.~Pich, {\em {Strong enhancement of epsilon-prime / epsilon
  through final state interactions}\/},
  \href{http://dx.doi.org/10.1103/PhysRevLett.84.2568}{Phys. Rev. Lett. {\bf
  84} (2000)  2568--2571},
\href{http://arxiv.org/abs/hep-ph/9911233}{{\tt arXiv:hep-ph/9911233
  [hep-ph]}}.

\bibitem{Gisbert:2017vvj}
H.~Gisbert and A.~Pich, {\em {Direct CP violation in $K^0\to\pi\pi$: Standard
  Model Status}\/},  \href{http://dx.doi.org/10.1088/1361-6633/aac18e}{Rept.
  Prog. Phys. {\bf 81} (2018) no.~7, 076201},
\href{http://arxiv.org/abs/1712.06147}{{\tt arXiv:1712.06147 [hep-ph]}}.

\bibitem{Endo:2017ums}
M.~Endo, T.~Goto, T.~Kitahara, S.~Mishima, D.~Ueda, and K.~Yamamoto, {\em
  {Gluino-mediated electroweak penguin with flavor-violating trilinear
  couplings}\/},  \href{http://dx.doi.org/10.1007/JHEP04(2018)019}{JHEP {\bf
  04} (2018)  019},
\href{http://arxiv.org/abs/1712.04959}{{\tt arXiv:1712.04959 [hep-ph]}}.

\bibitem{DeBruyn:2012wk}
K.~De~Bruyn, R.~Fleischer, R.~Knegjens, P.~Koppenburg, M.~Merk, A.~Pellegrino,
  and N.~Tuning, {\em {Probing New Physics via the $B^0_s\to \mu^+\mu^-$
  Effective Lifetime}\/},
  \href{http://dx.doi.org/10.1103/PhysRevLett.109.041801}{Phys. Rev. Lett. {\bf
  109} (2012)  041801},
\href{http://arxiv.org/abs/1204.1737}{{\tt arXiv:1204.1737 [hep-ph]}}.

\bibitem{Buras:2013uqa}
A.~J. Buras, R.~Fleischer, J.~Girrbach, and R.~Knegjens, {\em {Probing New
  Physics with the $B_s \to \mu^+ \mu^-$ Time-Dependent Rate}\/},
  \href{http://dx.doi.org/10.1007/JHEP07(2013)077}{JHEP {\bf 07} (2013)  77},
\href{http://arxiv.org/abs/1303.3820}{{\tt arXiv:1303.3820 [hep-ph]}}.

\bibitem{GomezDumm:1998gw}
D.~Gomez~Dumm and A.~Pich, {\em {Long distance contributions to the $K_L \to
  \mu^+ \mu^-$ decay width}\/},
  \href{http://dx.doi.org/10.1103/PhysRevLett.80.4633}{Phys. Rev. Lett. {\bf
  80} (1998)  4633--4636},
\href{http://arxiv.org/abs/hep-ph/9801298}{{\tt arXiv:hep-ph/9801298
  [hep-ph]}}.

\bibitem{Cirigliano:2011ny}
V.~Cirigliano, G.~Ecker, H.~Neufeld, A.~Pich, and J.~Portoles, {\em {Kaon
  Decays in the Standard Model}\/},
  \href{http://dx.doi.org/10.1103/RevModPhys.84.399}{Rev. Mod. Phys. {\bf 84}
  (2012)  399},
\href{http://arxiv.org/abs/1107.6001}{{\tt arXiv:1107.6001 [hep-ph]}}.

\bibitem{DAmbrosio:2013qmd}
G.~D'Ambrosio, D.~Greynat, and G.~Vulvert, {\em {Standard Model and New Physics
  contributions to $K_L$ and $K_S$ into four leptons}\/},
  \href{http://dx.doi.org/10.1140/epjc/s10052-013-2678-1}{Eur. Phys. J. {\bf
  C73} (2013) no.~12, 2678},
\href{http://arxiv.org/abs/1309.5736}{{\tt arXiv:1309.5736 [hep-ph]}}.

\bibitem{DAmbrosio:1986zin}
G.~D'Ambrosio and D.~Espriu, {\em {Rare Decay Modes of the K Mesons in the
  Chiral Lagrangian}\/},
\href{http://dx.doi.org/10.1016/0370-2693(86)90724-0}{Phys. Lett. {\bf B175}
  (1986)  237--242}.

\bibitem{DAmbrosio:1998gur}
G.~D'Ambrosio, G.~Ecker, G.~Isidori, and J.~Portoles, {\em {The Decays $K \to
  \pi \ell^+ \ell^-$ beyond leading order in the chiral expansion}\/},
  \href{http://dx.doi.org/10.1088/1126-6708/1998/08/004}{JHEP {\bf 08} (1998)
  004},
\href{http://arxiv.org/abs/hep-ph/9808289}{{\tt arXiv:hep-ph/9808289
  [hep-ph]}}.

\bibitem{DAmbrosio:2018ytt}
G.~D'Ambrosio, D.~Greynat, and M.~Knecht, {\em {On the amplitudes for the
  CP-conserving $K^\pm(K_S)\to\pi^\pm(\pi^0)\ell^+\ell^-$ rare decay modes}\/},
\href{http://arxiv.org/abs/1812.00735}{{\tt arXiv:1812.00735 [hep-ph]}}.

\bibitem{Appel:1999yq}
{E865 Collaboration}, R.~Appel et al., {\em {A New measurement of the
  properties of the rare decay $K^+ \to \pi^+ e^+ e^-$}\/},
  \href{http://dx.doi.org/10.1103/PhysRevLett.83.4482}{Phys. Rev. Lett. {\bf
  83} (1999)  4482--4485},
\href{http://arxiv.org/abs/hep-ex/9907045}{{\tt arXiv:hep-ex/9907045
  [hep-ex]}}.

\bibitem{Batley:2009aa}
{NA48/2 Collaboration}, J.~R. Batley et al., {\em {Precise measurement of the
  $K^{\pm} \to \pi^{\pm} e^+e^-$ decay}\/},
  \href{http://dx.doi.org/10.1016/j.physletb.2009.05.040}{Phys. Lett. {\bf
  B677} (2009)  246--254},
\href{http://arxiv.org/abs/0903.3130}{{\tt arXiv:0903.3130 [hep-ex]}}.

\bibitem{Batley:2011zz}
{NA48/2 Collaboration}, J.~R. Batley et al., {\em {New measurement of the
  $K^{\pm} \to \pi^{\pm} \mu^+ \mu^-$ decay}\/},
  \href{http://dx.doi.org/10.1016/j.physletb.2011.01.042}{Phys. Lett. {\bf
  B697} (2011)  107--115},
\href{http://arxiv.org/abs/1011.4817}{{\tt arXiv:1011.4817 [hep-ex]}}.

\bibitem{Batley:2003mu}
{NA48/1 Collaboration}, J.~R. Batley et al., {\em {Observation of the rare
  decay $K_S \to \pi^0 e^+ e^-$}\/},
  \href{http://dx.doi.org/10.1016/j.physletb.2003.10.001}{Phys. Lett. {\bf
  B576} (2003)  43--54},
\href{http://arxiv.org/abs/hep-ex/0309075}{{\tt arXiv:hep-ex/0309075
  [hep-ex]}}.

\bibitem{Batley:2004wg}
{NA48/1 Collaboration}, J.~R. Batley et al., {\em {Observation of the rare
  decay $K_S \to \pi^0 \mu^+ \mu^-$}\/},
  \href{http://dx.doi.org/10.1016/j.physletb.2004.08.058}{Phys. Lett. {\bf
  B599} (2004)  197--211},
\href{http://arxiv.org/abs/hep-ex/0409011}{{\tt arXiv:hep-ex/0409011
  [hep-ex]}}.

\bibitem{Crivellin:2016vjc}
A.~Crivellin, G.~D'Ambrosio, M.~Hoferichter, and L.~C. Tunstall, {\em
  {Violation of lepton flavor and lepton flavor universality in rare kaon
  decays}\/},  \href{http://dx.doi.org/10.1103/PhysRevD.93.074038}{Phys. Rev.
  {\bf D93} (2016) no.~7, 074038},
\href{http://arxiv.org/abs/1601.00970}{{\tt arXiv:1601.00970 [hep-ph]}}.

\bibitem{Descotes-Genon:2015uva}
S.~Descotes-Genon, L.~Hofer, J.~Matias, and J.~Virto, {\em {Global analysis of
  $b\to s\ell\ell$ anomalies}\/},
  \href{http://dx.doi.org/10.1007/JHEP06(2016)092}{JHEP {\bf 06} (2016)  092},
\href{http://arxiv.org/abs/1510.04239}{{\tt arXiv:1510.04239 [hep-ph]}}.

\bibitem{Cappiello:2017ilv}
L.~Cappiello, O.~Cat{\`a}, and G.~D'Ambrosio, {\em {Closing in on the radiative
  weak chiral couplings}\/},
  \href{http://dx.doi.org/10.1140/epjc/s10052-018-5748-6}{Eur. Phys. J. {\bf
  C78} (2018) no.~3, 265},
\href{http://arxiv.org/abs/1712.10270}{{\tt arXiv:1712.10270 [hep-ph]}}.

\bibitem{MarinBenito:2193358}
C.~Marin~Benito, L.~Garrido~Beltran, and X.~Cid~Vidal, {\em {Feasibility study
  of $K^0_{\rm S} \to \pi^+\pi^-e^+e^-$ at LHCb }\/},   LHCb-PUB-2016-016.
  CERN-LHCb-PUB-2016-016, CERN, Geneva, Oct, 2016.
\newblock \url{https://cds.cern.ch/record/2193358}.

\bibitem{Glashow:2014iga}
S.~L. Glashow, D.~Guadagnoli, and K.~Lane, {\em {Lepton Flavor Violation in $B$
  Decays?}\/},  \href{http://dx.doi.org/10.1103/PhysRevLett.114.091801}{Phys.
  Rev. Lett. {\bf 114} (2015)  091801},
\href{http://arxiv.org/abs/1411.0565}{{\tt arXiv:1411.0565 [hep-ph]}}.

\bibitem{Guadagnoli:2015nra}
D.~Guadagnoli and K.~Lane, {\em {Charged-Lepton Mixing and Lepton Flavor
  Violation}\/},  \href{http://dx.doi.org/10.1016/j.physletb.2015.10.010}{Phys.
  Lett. {\bf B751} (2015)  54--58},
\href{http://arxiv.org/abs/1507.01412}{{\tt arXiv:1507.01412 [hep-ph]}}.

\bibitem{Boucenna:2015raa}
S.~M. Boucenna, J.~W.~F. Valle, and A.~Vicente, {\em {Are the B decay anomalies
  related to neutrino oscillations?}\/},
  \href{http://dx.doi.org/10.1016/j.physletb.2015.09.040}{Phys. Lett. {\bf
  B750} (2015)  367--371},
\href{http://arxiv.org/abs/1503.07099}{{\tt arXiv:1503.07099 [hep-ph]}}.

\bibitem{Celis:2015ara}
A.~Celis, J.~Fuentes-Martin, M.~Jung, and H.~Serodio, {\em {Family nonuniversal
  $Z^{\prime}$ models with protected flavor-changing interactions}\/},
  \href{http://dx.doi.org/10.1103/PhysRevD.92.015007}{Phys. Rev. {\bf D92}
  (2015) no.~1, 015007},
\href{http://arxiv.org/abs/1505.03079}{{\tt arXiv:1505.03079 [hep-ph]}}.

\bibitem{Alonso:2015sja}
R.~Alonso, B.~Grinstein, and J.~Martin~Camalich, {\em {Lepton universality
  violation and lepton flavor conservation in $B$-meson decays}\/},  JHEP {\bf
  10} (2015)  184, \href{http://arxiv.org/abs/1505.05164}{{\tt arXiv:1505.05164
  [hep-ph]}}.

\bibitem{Gripaios:2015gra}
B.~Gripaios, M.~Nardecchia, and S.~A. Renner, {\em {Linear flavour violation
  and anomalies in B physics}\/},
  \href{http://dx.doi.org/10.1007/JHEP06(2016)083}{JHEP {\bf 06} (2016)  083},
\href{http://arxiv.org/abs/1509.05020}{{\tt arXiv:1509.05020 [hep-ph]}}.

\bibitem{Becirevic:2016zri}
D.~Be{\v c}irevi{\'c}, O.~Sumensari, and R.~Zukanovich~Funchal, {\em {Lepton
  flavor violation in exclusive $b\rightarrow s$ decays}\/},
  \href{http://dx.doi.org/10.1140/epjc/s10052-016-3985-0}{Eur. Phys. J. {\bf
  C76} (2016) no.~3, 134},
\href{http://arxiv.org/abs/1602.00881}{{\tt arXiv:1602.00881 [hep-ph]}}.

\bibitem{Becirevic:2016oho}
D.~Be{\v c}irevi{\' c}, N.~Ko{\v s}nik, O.~Sumensari, and
  R.~Zukanovich~Funchal, {\em {Palatable Leptoquark Scenarios for Lepton Flavor
  Violation in Exclusive $b\to s\ell_1\ell_2$ modes}\/},  JHEP {\bf 11} (2016)
  035, \href{http://arxiv.org/abs/1608.07583}{{\tt arXiv:1608.07583 [hep-ph]}}.

\bibitem{Hiller:2016kry}
G.~Hiller, D.~Loose, and K.~Sch{\"o}nwald, {\em {Leptoquark Flavor Patterns \&
  B Decay Anomalies}\/},  \href{http://dx.doi.org/10.1007/JHEP12(2016)027}{JHEP
  {\bf 12} (2016)  027},
\href{http://arxiv.org/abs/1609.08895}{{\tt arXiv:1609.08895 [hep-ph]}}.

\bibitem{Becirevic:2017jtw}
D.~Be{\v c}irevi{\'c} and O.~Sumensari, {\em {A leptoquark model to accommodate
  $R_K^\mathrm{exp} < R_K^\mathrm{SM}$ and $R_{K^\ast}^\mathrm{exp} <
  R_{K^\ast}^\mathrm{SM}$}\/},
  \href{http://dx.doi.org/10.1007/JHEP08(2017)104}{JHEP {\bf 08} (2017)  104},
\href{http://arxiv.org/abs/1704.05835}{{\tt arXiv:1704.05835 [hep-ph]}}.

\bibitem{King:2017anf}
S.~F. King, {\em {Flavourful Z$^{′}$ models for $ {R}_{K^{\left(\ast
  \right)}} $}\/},  \href{http://dx.doi.org/10.1007/JHEP08(2017)019}{JHEP {\bf
  08} (2017)  019},
\href{http://arxiv.org/abs/1706.06100}{{\tt arXiv:1706.06100 [hep-ph]}}.

\bibitem{Bordone:2018nbg}
M.~Bordone, C.~Cornella, J.~Fuentes-Mart{\'i}n, and G.~Isidori, {\em
  {Low-energy signatures of the $\mathrm{PS}^3$ model: from $B$-physics
  anomalies to LFV}\/},  \href{http://dx.doi.org/10.1007/JHEP10(2018)148}{JHEP
  {\bf 10} (2018)  148},
\href{http://arxiv.org/abs/1805.09328}{{\tt arXiv:1805.09328 [hep-ph]}}.

\bibitem{Ambrose:1998us}
{BNL Collaboration}, D.~Ambrose et al., {\em {New limit on muon and electron
  lepton number violation from $K^0_L \to \mu^{\pm} e^{\mp}$ decay}\/},
  \href{http://dx.doi.org/10.1103/PhysRevLett.81.5734}{Phys. Rev. Lett. {\bf
  81} (1998)  5734--5737},
\href{http://arxiv.org/abs/hep-ex/9811038}{{\tt arXiv:hep-ex/9811038
  [hep-ex]}}.

\bibitem{Abouzaid:2007aa}
{KTeV Collaboration}, E.~Abouzaid et al., {\em {Search for lepton flavor
  violating decays of the neutral kaon}\/},
  \href{http://dx.doi.org/10.1103/PhysRevLett.100.131803}{Phys. Rev. Lett. {\bf
  100} (2008)  131803},
\href{http://arxiv.org/abs/0711.3472}{{\tt arXiv:0711.3472 [hep-ex]}}.

\bibitem{Sher:2005sp}
A.~Sher et al., {\em {An Improved upper limit on the decay $K^+ \to \pi^+ \mu^+
  e^-$}\/},  \href{http://dx.doi.org/10.1103/PhysRevD.72.012005}{Phys. Rev.
  {\bf D72} (2005)  012005},
\href{http://arxiv.org/abs/hep-ex/0502020}{{\tt arXiv:hep-ex/0502020
  [hep-ex]}}.

\bibitem{Appel:2000tc}
R.~Appel et al., {\em {Search for lepton flavor violation in K+ decays}\/},
  \href{http://dx.doi.org/10.1103/PhysRevLett.85.2877}{Phys. Rev. Lett. {\bf
  85} (2000)  2877--2880},
\href{http://arxiv.org/abs/hep-ex/0006003}{{\tt arXiv:hep-ex/0006003
  [hep-ex]}}.

\bibitem{NA62:2312430}
{NA62 Collaboration}, C.~NA62, {\em {2018 NA62 Status Report to the CERN
  SPSC}\/},   CERN-SPSC-2018-010. SPSC-SR-229, CERN, Geneva, Apr, 2018.
\newblock \url{http://cds.cern.ch/record/2312430}.

\bibitem{PBCTalk}
G.~Wilkinson.
\newblock
  \url{https://indico.cern.ch/event/706741/contributions/3017537/attachments/1667814/2703428/TauFV_PBC.pdf}.
  Talk at Physics Beyond Colliders, June 2018.

\bibitem{Junior:2018odx}
A.~A. Alves~Junior et al., {\em {Prospects for Measurements with Strange
  Hadrons at LHCb}\/},
\href{http://arxiv.org/abs/1808.03477}{{\tt arXiv:1808.03477 [hep-ex]}}.

\bibitem{Chang:2014iba}
H.-M. Chang, M.~Gonz{\'a}lez-Alonso, and J.~Martin~Camalich, {\em {Nonstandard
  Semileptonic Hyperon Decays}\/},
  \href{http://dx.doi.org/10.1103/PhysRevLett.114.161802}{Phys. Rev. Lett. {\bf
  114} (2015) no.~16, 161802},
\href{http://arxiv.org/abs/1412.8484}{{\tt arXiv:1412.8484 [hep-ph]}}.

\bibitem{Cabibbo:2003cu}
N.~Cabibbo, E.~C. Swallow, and R.~Winston, {\em {Semileptonic hyperon
  decays}\/},
  \href{http://dx.doi.org/10.1146/annurev.nucl.53.013103.155258}{Ann. Rev.
  Nucl. Part. Sci. {\bf 53} (2003)  39--75},
\href{http://arxiv.org/abs/hep-ph/0307298}{{\tt arXiv:hep-ph/0307298
  [hep-ph]}}.

\bibitem{Cabibbo:2003ea}
N.~Cabibbo, E.~C. Swallow, and R.~Winston, {\em {Semileptonic hyperon decays
  and CKM unitarity}\/},
  \href{http://dx.doi.org/10.1103/PhysRevLett.92.251803}{Phys. Rev. Lett. {\bf
  92} (2004)  251803},
\href{http://arxiv.org/abs/hep-ph/0307214}{{\tt arXiv:hep-ph/0307214
  [hep-ph]}}.

\bibitem{Mateu:2005wi}
V.~Mateu and A.~Pich, {\em {$V_{us}$ determination from hyperon semileptonic
  decays}\/},  \href{http://dx.doi.org/10.1088/1126-6708/2005/10/041}{JHEP {\bf
  10} (2005)  041},
\href{http://arxiv.org/abs/hep-ph/0509045}{{\tt arXiv:hep-ph/0509045
  [hep-ph]}}.

\bibitem{Sasaki:2017jue}
S.~Sasaki, {\em {Continuum limit of hyperon vector coupling $f_1(0)$ from 2+1
  flavor domain wall QCD}\/},
  \href{http://dx.doi.org/10.1103/PhysRevD.96.074509}{Phys. Rev. {\bf D96}
  (2017) no.~7, 074509},
\href{http://arxiv.org/abs/1708.04008}{{\tt arXiv:1708.04008 [hep-lat]}}.

\bibitem{LHCb-PAPER-2017-049}
{LHCb collaboration}, R.~Aaij et al., {\em {Evidence for the rare decay
  \decay{\Sigma^+}{p \mup\mun}}\/},  Phys. Rev. Lett. {\bf 120} (2018)  221803,
  \href{http://arxiv.org/abs/1712.08606}{{\tt arXiv:1712.08606 [hep-ex]}}.

\bibitem{Park:2005eka}
{HyperCP Collaboration}, H.~Park et al., {\em {Evidence for the decay $\Sigma^+
  \to p \mu^+ \mu^-$}\/},
  \href{http://dx.doi.org/10.1103/PhysRevLett.94.021801}{Phys. Rev. Lett. {\bf
  94} (2005)  021801},
\href{http://arxiv.org/abs/hep-ex/0501014}{{\tt arXiv:hep-ex/0501014
  [hep-ex]}}.

\bibitem{He:2018yzu}
X.-G. He, J.~Tandean, and G.~Valencia, {\em {Decay rate and asymmetries of
  $\Sigma^+\to p\mu^+\mu^-$}\/},
  \href{http://dx.doi.org/10.1007/JHEP10(2018)040}{JHEP {\bf 10} (2018)  040},
\href{http://arxiv.org/abs/1806.08350}{{\tt arXiv:1806.08350 [hep-ph]}}.

\bibitem{Pich:2013lsa}
A.~Pich, {\em {Precision Tau Physics}\/},
  \href{http://dx.doi.org/10.1016/j.ppnp.2013.11.002}{Prog. Part. Nucl. Phys.
  {\bf 75} (2014)  41--85},
\href{http://arxiv.org/abs/1310.7922}{{\tt arXiv:1310.7922 [hep-ph]}}.

\bibitem{Schael:2005am}
{ALEPH Collaboration}, S.~Schael et al., {\em {Branching ratios and spectral
  functions of tau decays: Final ALEPH measurements and physics
  implications}\/},
  \href{http://dx.doi.org/10.1016/j.physrep.2005.06.007}{Phys. Rept. {\bf 421}
  (2005)  191--284},
\href{http://arxiv.org/abs/hep-ex/0506072}{{\tt arXiv:hep-ex/0506072}}.

\bibitem{Buchmuller:1985jz}
W.~Buchmuller and D.~Wyler, {\em {Effective Lagrangian Analysis of New
  Interactions and Flavor Conservation}\/},
\href{http://dx.doi.org/10.1016/0550-3213(86)90262-2}{Nucl.Phys. {\bf B268}
  (1986)  621}.

\bibitem{Cirigliano:2012ab}
V.~Cirigliano, M.~Gonzalez-Alonso, and M.~L. Graesser, {\em {Non-standard
  Charged Current Interactions: beta decays versus the LHC}\/},
  \href{http://dx.doi.org/10.1007/JHEP02(2013)046}{JHEP {\bf 02} (2013)  046},
\href{http://arxiv.org/abs/1210.4553}{{\tt arXiv:1210.4553 [hep-ph]}}.

\bibitem{deBlas:2013qqa}
J.~de~Blas, M.~Chala, and J.~Santiago, {\em {Global Constraints on Lepton-Quark
  Contact Interactions}\/},
  \href{http://dx.doi.org/10.1103/PhysRevD.88.095011}{Phys. Rev. {\bf D88}
  (2013)  095011},
\href{http://arxiv.org/abs/1307.5068}{{\tt arXiv:1307.5068 [hep-ph]}}.

\bibitem{Faroughy:2016osc}
D.~A. Faroughy, A.~Greljo, and J.~F. Kamenik, {\em {Confronting lepton flavor
  universality violation in B decays with high-$p_T$ tau lepton searches at
  LHC}\/},  \href{http://dx.doi.org/10.1016/j.physletb.2016.11.011}{Phys. Lett.
  {\bf B764} (2017)  126--134},
\href{http://arxiv.org/abs/1609.07138}{{\tt arXiv:1609.07138 [hep-ph]}}.

\bibitem{Cirigliano:2018dyk}
V.~Cirigliano, A.~Falkowski, M.~Gonz\'alez-Alonso, and
  A.~Rodr\'iguez-S\'anchez, {\em {Hadronic tau decays as New Physics probes in
  the LHC era}\/},
\href{http://arxiv.org/abs/1809.01161}{{\tt arXiv:1809.01161 [hep-ph]}}.

\bibitem{Aaboud:2018vgh}
{ATLAS Collaboration}, M.~Aaboud et al., {\em {A search for high-mass
  resonances decaying to $\tau\nu$ in $pp$ collisions at $\sqrt{s}$ = 13 TeV
  with the ATLAS detector}\/},
\href{http://arxiv.org/abs/1801.06992}{{\tt arXiv:1801.06992 [hep-ex]}}.

\bibitem{Alwall:2014hca}
J.~Alwall, R.~Frederix, S.~Frixione, V.~Hirschi, F.~Maltoni, O.~Mattelaer,
  H.~S. Shao, T.~Stelzer, P.~Torrielli, and M.~Zaro, {\em {The automated
  computation of tree-level and next-to-leading order differential cross
  sections, and their matching to parton shower simulations}\/},
  \href{http://dx.doi.org/10.1007/JHEP07(2014)079}{JHEP {\bf 07} (2014)  079},
\href{http://arxiv.org/abs/1405.0301}{{\tt arXiv:1405.0301 [hep-ph]}}.

\bibitem{Sjostrand:2014zea}
T.~Sj{\"o}strand, S.~Ask, J.~R. Christiansen, R.~Corke, N.~Desai, P.~Ilten,
  S.~Mrenna, S.~Prestel, C.~O. Rasmussen, and P.~Z. Skands, {\em {An
  Introduction to PYTHIA 8.2}\/},
  \href{http://dx.doi.org/10.1016/j.cpc.2015.01.024}{Comput. Phys. Commun. {\bf
  191} (2015)  159--177},
\href{http://arxiv.org/abs/1410.3012}{{\tt arXiv:1410.3012 [hep-ph]}}.

\bibitem{deFavereau:2013fsa}
J.~de~Favereau, C.~Delaere, P.~Demin, A.~Giammanco, V.~Lema{\^ i}tre, et al.,
  {\em {DELPHES 3, A modular framework for fast simulation of a generic
  collider experiment}\/},
\href{http://arxiv.org/abs/1307.6346}{{\tt arXiv:1307.6346 [hep-ex]}}.

\bibitem{Grzadkowski:2010es}
B.~Grzadkowski, M.~Iskrzynski, M.~Misiak, and J.~Rosiek, {\em {Dimension-Six
  Terms in the Standard Model Lagrangian}\/},  JHEP {\bf 10} (2010)  085,
  \href{http://arxiv.org/abs/1008.4884}{{\tt arXiv:1008.4884 [hep-ph]}}.

\bibitem{Contino:2016jqw}
R.~Contino, A.~Falkowski, F.~Goertz, C.~Grojean, and F.~Riva, {\em {On the
  Validity of the Effective Field Theory Approach to SM Precision Tests}\/},
  \href{http://dx.doi.org/10.1007/JHEP07(2016)144}{JHEP {\bf 07} (2016)  144},
\href{http://arxiv.org/abs/1604.06444}{{\tt arXiv:1604.06444 [hep-ph]}}.

\bibitem{Altmannshofer:2017poe}
W.~Altmannshofer, P.~Bhupal~Dev, and A.~Soni, {\em {$R_{D^{(*)}}$ anomaly: A
  possible hint for natural supersymmetry with $R$-parity violation}\/},
  \href{http://dx.doi.org/10.1103/PhysRevD.96.095010}{Phys. Rev. {\bf D96}
  (2017) no.~9, 095010},
\href{http://arxiv.org/abs/1704.06659}{{\tt arXiv:1704.06659 [hep-ph]}}.

\bibitem{Greljo:2018tzh}
A.~Greljo, J.~Martin~Camalich, and J.~D. Ruiz-{\'A}lvarez, {\em {The Mono-Tau
  Menace: From $B$ Decays to High-$p_T$ Tails}\/},
\href{http://arxiv.org/abs/1811.07920}{{\tt arXiv:1811.07920 [hep-ph]}}.

\bibitem{Lees:2013uzd}
{BaBar Collaboration}, J.~P. Lees et al., {\em {Measurement of an Excess of
  $\bar{B} \to D^{(*)}\tau^- \bar{\nu}_\tau$ Decays and Implications for
  Charged Higgs Bosons}\/},  Phys. Rev. {\bf D88} (2013) no.~7, 072012,
  \href{http://arxiv.org/abs/1303.0571}{{\tt arXiv:1303.0571 [hep-ex]}}.

\bibitem{Huschle:2015rga}
{Belle Collaboration}, M.~Huschle et al., {\em {Measurement of the branching
  ratio of $\bar{B} \to D^{(\ast)} \tau^- \bar{\nu}_\tau$ relative to $\bar{B}
  \to D^{(\ast)} \ell^- \bar{\nu}_\ell$ decays with hadronic tagging at
  Belle}\/},  Phys. Rev. {\bf D92} (2015) no.~7, 072014,
  \href{http://arxiv.org/abs/1507.03233}{{\tt arXiv:1507.03233 [hep-ex]}}.

\bibitem{Sato:2016svk}
{Belle Collaboration}, Y.~Sato et al., {\em {Measurement of the branching ratio
  of $\bar{B}^0 \rightarrow D^{*+} \tau^- \bar{\nu}_{\tau}$ relative to
  $\bar{B}^0 \rightarrow D^{*+} \ell^- \bar{\nu}_{\ell}$ decays with a
  semileptonic tagging method}\/},  Phys. Rev. {\bf D94} (2016) no.~7, 072007,
  \href{http://arxiv.org/abs/1607.07923}{{\tt arXiv:1607.07923 [hep-ex]}}.

\bibitem{Patrignani:2016xqp}
{Particle Data Group Collaboration}, C.~Patrignani et al., {\em {Review of
  Particle Physics}\/},
\href{http://dx.doi.org/10.1088/1674-1137/40/10/100001}{Chin. Phys. {\bf C40}
  (2016) no.~10, 100001}.

\bibitem{Filipuzzi:2012mg}
A.~Filipuzzi, J.~Portoles, and M.~Gonzalez-Alonso, {\em {U(2)$^5$ flavor
  symmetry and lepton universality violation in $W \to \tau \nu_\tau$}\/},
  \href{http://dx.doi.org/10.1103/PhysRevD.85.116010}{Phys. Rev. {\bf D85}
  (2012)  116010},
\href{http://arxiv.org/abs/1203.2092}{{\tt arXiv:1203.2092 [hep-ph]}}.

\bibitem{Abbott:1999pk}
{D0 Collaboration}, B.~Abbott et al., {\em {A measurement of the $W \to \tau
  \nu$ production cross section in $p\bar{p}$ collisions at $\sqrt{s} = 1.8$
  TeV}\/},  \href{http://dx.doi.org/10.1103/PhysRevLett.84.5710}{Phys. Rev.
  Lett. {\bf 84} (2000)  5710--5715},
\href{http://arxiv.org/abs/hep-ex/9912065}{{\tt arXiv:hep-ex/9912065
  [hep-ex]}}.

\bibitem{Cheng:1977nv}
T.-P. Cheng and L.-F. Li, {\em {Muon Number Nonconservation Effects in a Gauge
  Theory with V A Currents and Heavy Neutral Leptons}\/},
\href{http://dx.doi.org/10.1103/PhysRevD.16.1425}{Phys. Rev. {\bf D16} (1977)
  1425}.

\bibitem{Lee:1977tib}
B.~W. Lee and R.~E. Shrock, {\em {Natural Suppression of Symmetry Violation in
  Gauge Theories: Muon - Lepton and Electron Lepton Number Nonconservation}\/},
\href{http://dx.doi.org/10.1103/PhysRevD.16.1444}{Phys. Rev. {\bf D16} (1977)
  1444}.

\bibitem{Petcov:1976ff}
S.~T. Petcov, {\em {The Processes $mu \to e \gamma$, $\mu \to e e \bar e$,
  $\nu' \to \nu\gamma$ in the Weinberg-Salam Model with Neutrino Mixing}\/},
  Sov. J. Nucl. Phys. {\bf 25} (1977)  340.
[Erratum: Yad. Fiz.25,1336(1977)].

\bibitem{Hernandez-Tome:2018fbq}
G.~Hern{\'a}ndez-Tom{\'e}, G.~L{\'o}pez~Castro, and P.~Roig, {\em {Flavor
  violating leptonic decays of $\tau$ and $\mu$ leptons in the Standard Model
  with massive neutrinos}\/},
\href{http://arxiv.org/abs/1807.06050}{{\tt arXiv:1807.06050 [hep-ph]}}.

\bibitem{Calibbi:2017uvl}
L.~Calibbi and G.~Signorelli, {\em {Charged Lepton Flavour Violation: An
  Experimental and Theoretical Introduction}\/},
  \href{http://dx.doi.org/10.1393/ncr/i2018-10144-0}{Riv. Nuovo Cim. {\bf 41}
  (2018) no.~2, 1},
\href{http://arxiv.org/abs/1709.00294}{{\tt arXiv:1709.00294 [hep-ph]}}.

\bibitem{Bernstein:2013hba}
R.~H. Bernstein and P.~S. Cooper, {\em {Charged Lepton Flavor Violation: An
  Experimenter's Guide}\/},
  \href{http://dx.doi.org/10.1016/j.physrep.2013.07.002}{Phys. Rept. {\bf 532}
  (2013)  27--64},
\href{http://arxiv.org/abs/1307.5787}{{\tt arXiv:1307.5787 [hep-ex]}}.

\bibitem{TheMEG:2016wtm}
{MEG Collaboration}, A.~M. Baldini et al., {\em {Search for the lepton flavour
  violating decay $\mu ^+ \rightarrow \mathrm {e}^+ \gamma $ with the full
  dataset of the MEG experiment}\/},
  \href{http://dx.doi.org/10.1140/epjc/s10052-016-4271-x}{Eur. Phys. J. {\bf
  C76} (2016) no.~8, 434},
\href{http://arxiv.org/abs/1605.05081}{{\tt arXiv:1605.05081 [hep-ex]}}.

\bibitem{Baldini:2013ke}
A.~M. Baldini et al., {\em {MEG Upgrade Proposal}\/},
\href{http://arxiv.org/abs/1301.7225}{{\tt arXiv:1301.7225 [physics.ins-det]}}.

\bibitem{Baldini:2018nnn}
{MEG II Collaboration}, A.~M. Baldini et al., {\em {The design of the MEG II
  experiment}\/},  \href{http://dx.doi.org/10.1140/epjc/s10052-018-5845-6}{Eur.
  Phys. J. {\bf C78} (2018) no.~5, 380},
\href{http://arxiv.org/abs/1801.04688}{{\tt arXiv:1801.04688
  [physics.ins-det]}}.

\bibitem{Blondel:2013ia}
A.~Blondel et al., {\em {Research Proposal for an Experiment to Search for the
  Decay $\mu \to eee$}\/},
\href{http://arxiv.org/abs/1301.6113}{{\tt arXiv:1301.6113 [physics.ins-det]}}.

\bibitem{Berger:2014vba}
{Mu3e Collaboration}, N.~Berger, {\em {The Mu3e Experiment}\/},
\href{http://dx.doi.org/10.1016/j.nuclphysbps.2014.02.007}{Nucl. Phys. Proc.
  Suppl. {\bf 248-250} (2014)  35--40}.

\bibitem{Bartoszek:2014mya}
{Mu2e Collaboration}, L.~Bartoszek et al., {\em {Mu2e Technical Design
  Report}\/},
\href{http://arxiv.org/abs/1501.05241}{{\tt arXiv:1501.05241
  [physics.ins-det]}}.

\bibitem{COMET}
{COMET Collaboration}, {\em {COMET Phase-I technical design report}\/},
  \href{http://dx.doi.org/http://comet.kek.jp/Documents_files/IPNS-Review-2014.pdf}{
  (2014)  }.

\bibitem{DeeMe}
{DeeMe Collaboration}, {\em {DeeMe KEK J-PARC Proposal}\/},
  \href{http://dx.doi.org/http://deeme.hep.sci.osaka-u.ac.jp/documents/deeme-proposal-r28.pdf}{
  (2010)  }.

\bibitem{Nguyen:2015vkk}
{DeeMe Collaboration}, T.~M. Nguyen, {\em {Search for $\mu \to e$ conversion
  with DeeMe experiment at J-PARC MLF}\/},
\href{http://dx.doi.org/10.22323/1.248.0060}{PoS {\bf FPCP2015} (2015)  060}.

\bibitem{Hazard:2017udp}
D.~E. Hazard and A.~A. Petrov, {\em {Radiative lepton flavor violating B, D,
  and K decays}\/},  \href{http://dx.doi.org/10.1103/PhysRevD.98.015027}{Phys.
  Rev. {\bf D98} (2018) no.~1, 015027},
\href{http://arxiv.org/abs/1711.05314}{{\tt arXiv:1711.05314 [hep-ph]}}.

\bibitem{Hazard:2016fnc}
D.~E. Hazard and A.~A. Petrov, {\em {Lepton flavor violating quarkonium
  decays}\/},  \href{http://dx.doi.org/10.1103/PhysRevD.94.074023}{Phys. Rev.
  {\bf D94} (2016) no.~7, 074023},
\href{http://arxiv.org/abs/1607.00815}{{\tt arXiv:1607.00815 [hep-ph]}}.

\bibitem{Davidson:2012wn}
S.~Davidson, S.~Lacroix, and P.~Verdier, {\em {LHC sensitivity to lepton
  flavour violating Z boson decays}\/},
  \href{http://dx.doi.org/10.1007/JHEP09(2012)092}{JHEP {\bf 09} (2012)  092},
\href{http://arxiv.org/abs/1207.4894}{{\tt arXiv:1207.4894 [hep-ph]}}.

\bibitem{Decamp:1991uy}
{ALEPH Collaboration}, D.~Decamp et al., {\em {Searches for new particles in
  $Z$ decays using the ALEPH detector}\/},
\href{http://dx.doi.org/10.1016/0370-1573(92)90177-2}{Phys. Rept. {\bf 216}
  (1992)  253--340}.

\bibitem{Adriani:1993sy}
{L3 Collaboration}, O.~Adriani et al., {\em {Search for lepton flavor violation
  in Z decays}\/},
\href{http://dx.doi.org/10.1016/0370-2693(93)90348-L}{Phys. Lett. {\bf B316}
  (1993)  427--434}.

\bibitem{Akers:1995gz}
{OPAL Collaboration}, R.~Akers et al., {\em {A Search for lepton flavor
  violating $Z^0$ decays}\/},
\href{http://dx.doi.org/10.1007/BF01553981}{Z. Phys. {\bf C67} (1995)
  555--564}.

\bibitem{Abreu:1996mj}
{DELPHI Collaboration}, P.~Abreu et al., {\em {Search for lepton flavor number
  violating $Z^0$ decays}\/},
\href{http://dx.doi.org/10.1007/s002880050313}{Z. Phys. {\bf C73} (1997)
  243--251}.

\bibitem{Aaboud:2018cxn}
{ATLAS Collaboration}, M.~Aaboud et al., {\em {A search for
  lepton-flavor-violating decays of the $Z$ boson into a $\tau$-lepton and a
  light lepton with the ATLAS detector}\/},  Submitted to: Phys. Rev. (2018)  ,
\href{http://arxiv.org/abs/1804.09568}{{\tt arXiv:1804.09568 [hep-ex]}}.

\bibitem{Abada:2014nwa}
A.~Abada, V.~De~Romeri, and A.~M. Teixeira, {\em {Effect of steriles states on
  lepton magnetic moments and neutrinoless double beta decay}\/},
  \href{http://dx.doi.org/10.1007/JHEP09(2014)074}{JHEP {\bf 09} (2014)  074},
\href{http://arxiv.org/abs/1406.6978}{{\tt arXiv:1406.6978 [hep-ph]}}.

\bibitem{DeRomeri:2016gum}
V.~De~Romeri, M.~J. Herrero, X.~Marcano, and F.~Scarcella, {\em {Lepton flavor
  violating Z decays: A promising window to low scale seesaw neutrinos}\/},
  \href{http://dx.doi.org/10.1103/PhysRevD.95.075028}{Phys. Rev. {\bf D95}
  (2017) no.~7, 075028},
\href{http://arxiv.org/abs/1607.05257}{{\tt arXiv:1607.05257 [hep-ph]}}.

\bibitem{Bhattacharya:2018ryy}
B.~Bhattacharya, R.~Morgan, J.~Osborne, and A.~A. Petrov, {\em {Studies of
  Lepton Flavor Violation at the LHC}\/},
  \href{http://dx.doi.org/10.1016/j.physletb.2018.08.037}{Phys. Lett. {\bf
  B785} (2018)  165--170},
\href{http://arxiv.org/abs/1802.06082}{{\tt arXiv:1802.06082 [hep-ph]}}.

\bibitem{Cai:2018cog}
Y.~Cai, M.~A. Schmidt, and G.~Valencia, {\em {Lepton-flavour-violating gluonic
  operators: constraints from the LHC and low energy experiments}\/},
  \href{http://dx.doi.org/10.1007/JHEP05(2018)143}{JHEP {\bf 05} (2018)  143},
\href{http://arxiv.org/abs/1802.09822}{{\tt arXiv:1802.09822 [hep-ph]}}.

\bibitem{Aubert:2006cz}
{BaBar Collaboration}, B.~Aubert et al., {\em {Search for Lepton Flavor
  Violating Decays $\tau^\pm \to \ell^\pm \pi^0$, $\ell^\pm \eta$, $\ell^\pm
  \eta^\prime$}\/},
  \href{http://dx.doi.org/10.1103/PhysRevLett.98.061803}{Phys. Rev. Lett. {\bf
  98} (2007)  061803},
\href{http://arxiv.org/abs/hep-ex/0610067}{{\tt arXiv:hep-ex/0610067
  [hep-ex]}}.

\bibitem{Aubert:2007kx}
{BaBar Collaboration}, B.~Aubert et al., {\em {Search for lepton flavor
  violating decays $\tau^\pm \to \ell^\pm \omega$ ($\ell = e,~\mu$)}\/},
  \href{http://dx.doi.org/10.1103/PhysRevLett.100.071802}{Phys. Rev. Lett. {\bf
  100} (2008)  071802},
\href{http://arxiv.org/abs/0711.0980}{{\tt arXiv:0711.0980 [hep-ex]}}.

\bibitem{Aubert:2009ag}
{BaBar Collaboration}, B.~Aubert et al., {\em {Searches for Lepton Flavor
  Violation in the Decays $\tau^{\pm} \to e^\pm \gamma$ and $\tau^\pm \to
  \mu^\pm \gamma$}\/},
  \href{http://dx.doi.org/10.1103/PhysRevLett.104.021802}{Phys. Rev. Lett. {\bf
  104} (2010)  021802},
\href{http://arxiv.org/abs/0908.2381}{{\tt arXiv:0908.2381 [hep-ex]}}.

\bibitem{Aubert:2009ys}
{BaBar Collaboration}, B.~Aubert et al., {\em {Search for Lepton Flavor
  Violating Decays $\tau \to \ell^- K_S^0$ with the BABAR Experiment}\/},
  \href{http://dx.doi.org/10.1103/PhysRevD.79.012004}{Phys. Rev. {\bf D79}
  (2009)  012004},
\href{http://arxiv.org/abs/0812.3804}{{\tt arXiv:0812.3804 [hep-ex]}}.

\bibitem{Aubert:2009ap}
{BaBar Collaboration}, B.~Aubert et al., {\em {Improved limits on lepton flavor
  violating tau decays to $\ell \phi$, $\ell \rho$, $\ell K^*$ and $\ell {\bar
  K^*}$}\/},  \href{http://dx.doi.org/10.1103/PhysRevLett.103.021801}{Phys.
  Rev. Lett. {\bf 103} (2009)  021801},
\href{http://arxiv.org/abs/0904.0339}{{\tt arXiv:0904.0339 [hep-ex]}}.

\bibitem{Lees:2010ez}
{BaBar Collaboration}, J.~P. Lees et al., {\em {Limits on tau Lepton-Flavor
  Violating Decays in three charged leptons}\/},
  \href{http://dx.doi.org/10.1103/PhysRevD.81.111101}{Phys. Rev. {\bf D81}
  (2010)  111101},
\href{http://arxiv.org/abs/1002.4550}{{\tt arXiv:1002.4550 [hep-ex]}}.

\bibitem{Miyazaki:2005ng}
{Belle Collaboration}, Y.~Miyazaki et al., {\em {Search for lepton and baryon
  number violating tau- decays into anti-Lambda pi- and Lambda pi-}\/},
  \href{http://dx.doi.org/10.1016/j.physletb.2005.10.024}{Phys. Lett. {\bf
  B632} (2006)  51--57},
\href{http://arxiv.org/abs/hep-ex/0508044}{{\tt arXiv:hep-ex/0508044
  [hep-ex]}}.

\bibitem{Hayasaka:2007vc}
{Belle Collaboration}, K.~Hayasaka et al., {\em {New Search for $\tau \to \mu
  \gamma$ and $\tau \to e \gamma$ Decays at Belle}\/},
  \href{http://dx.doi.org/10.1016/j.physletb.2008.06.056}{Phys. Lett. {\bf
  B666} (2008)  16--22},
\href{http://arxiv.org/abs/0705.0650}{{\tt arXiv:0705.0650 [hep-ex]}}.

\bibitem{Miyazaki:2007jp}
{Belle Collaboration}, Y.~Miyazaki et al., {\em {Search for lepton flavor
  violating $\tau^-$ decays into $\ell^- \eta$, $\ell^- \eta^\prime$ and
  $\ell^- \pi^0$}\/},
  \href{http://dx.doi.org/10.1016/j.physletb.2007.03.027}{Phys. Lett. {\bf
  B648} (2007)  341--350},
\href{http://arxiv.org/abs/hep-ex/0703009}{{\tt arXiv:hep-ex/0703009
  [HEP-EX]}}.

\bibitem{Miyazaki:2008mw}
{Belle Collaboration}, Y.~Miyazaki et al., {\em {Search for
  Lepton-Flavor-Violating tau Decays into Lepton and $f_0(980)$ Meson}\/},
  \href{http://dx.doi.org/10.1016/j.physletb.2009.01.058}{Phys. Lett. {\bf
  B672} (2009)  317--322},
\href{http://arxiv.org/abs/0810.3519}{{\tt arXiv:0810.3519 [hep-ex]}}.

\bibitem{Miyazaki:2010qb}
{Belle Collaboration}, Y.~Miyazaki et al., {\em {Search for Lepton Flavor
  Violating $tau^-$ Decays into $\ell^- K^0_S$ and $\ell^- K^0_S K^0_S$}\/},
  \href{http://dx.doi.org/10.1016/j.physletb.2010.07.012}{Phys. Lett. {\bf
  B692} (2010)  4--9},
\href{http://arxiv.org/abs/1003.1183}{{\tt arXiv:1003.1183 [hep-ex]}}.

\bibitem{Hayasaka:2010np}
K.~Hayasaka et al., {\em {Search for Lepton Flavor Violating Tau Decays into
  Three Leptons with 719 Million Produced $\tau^+\tau^-$ Pairs}\/},
  \href{http://dx.doi.org/10.1016/j.physletb.2010.03.037}{Phys. Lett. {\bf
  B687} (2010)  139--143},
\href{http://arxiv.org/abs/1001.3221}{{\tt arXiv:1001.3221 [hep-ex]}}.

\bibitem{Miyazaki:2011xe}
{Belle Collaboration}, Y.~Miyazaki et al., {\em {Search for
  Lepton-Flavor-Violating tau Decays into a Lepton and a Vector Meson}\/},
  \href{http://dx.doi.org/10.1016/j.physletb.2011.04.011}{Phys. Lett. {\bf
  B699} (2011)  251--257},
\href{http://arxiv.org/abs/1101.0755}{{\tt arXiv:1101.0755 [hep-ex]}}.

\bibitem{Bowcock:1989mq}
{CLEO Collaboration}, T.~J.~V. Bowcock et al., {\em {Search for Neutrinoless
  Decays of the $\tau$ Lepton}\/},
\href{http://dx.doi.org/10.1103/PhysRevD.41.805}{Phys. Rev. {\bf D41} (1990)
  805}.

\bibitem{Bonvicini:1997bw}
{CLEO Collaboration}, G.~Bonvicini et al., {\em {Search for neutrinoless tau
  decays involving pi0 or eta mesons}\/},
  \href{http://dx.doi.org/10.1103/PhysRevLett.79.1221}{Phys. Rev. Lett. {\bf
  79} (1997)  1221--1224},
\href{http://arxiv.org/abs/hep-ex/9704010}{{\tt arXiv:hep-ex/9704010
  [hep-ex]}}.

\bibitem{Chen:2002ug}
{CLEO Collaboration}, S.~Chen et al., {\em {Search for neutrinoless tau decays
  involving the $K^0_S$ meson}\/},
  \href{http://dx.doi.org/10.1103/PhysRevD.66.071101}{Phys. Rev. {\bf D66}
  (2002)  071101},
\href{http://arxiv.org/abs/hep-ex/0208019}{{\tt arXiv:hep-ex/0208019
  [hep-ex]}}.

\bibitem{LHCb-PAPER-2014-052}
{LHCb collaboration}, R.~Aaij et al., {\em {Search for the lepton flavour
  violating decay $\taum\to \mun\mup\mun$}\/},  JHEP {\bf 02} (2015)  121,
  \href{http://arxiv.org/abs/1409.8548}{{\tt arXiv:1409.8548 [hep-ex]}}.

\bibitem{Aad:2016wce}
{ATLAS Collaboration}, G.~Aad et al., {\em {Probing lepton flavour violation
  via neutrinoless $\tau \to 3\mu$ decays with the ATLAS detector}\/},
  \href{http://dx.doi.org/10.1140/epjc/s10052-016-4041-9}{Eur. Phys. J. {\bf
  C76} (2016) no.~5, 232},
\href{http://arxiv.org/abs/1601.03567}{{\tt arXiv:1601.03567 [hep-ex]}}.

\bibitem{Celis:2014asa}
A.~Celis, V.~Cirigliano, and E.~Passemar, {\em {Model-discriminating power of
  lepton flavor violating $\tau$ decays}\/},
  \href{http://dx.doi.org/10.1103/PhysRevD.89.095014}{Phys. Rev. {\bf D89}
  (2014) no.~9, 095014},
\href{http://arxiv.org/abs/1403.5781}{{\tt arXiv:1403.5781 [hep-ph]}}.

\bibitem{Dassinger:2007ru}
B.~M. Dassinger, T.~Feldmann, T.~Mannel, and S.~Turczyk, {\em
  {Model-independent analysis of lepton flavour violating tau decays}\/},
  \href{http://dx.doi.org/10.1088/1126-6708/2007/10/039}{JHEP {\bf 10} (2007)
  039},
\href{http://arxiv.org/abs/0707.0988}{{\tt arXiv:0707.0988 [hep-ph]}}.

\bibitem{Matsuzaki:2007hh}
A.~Matsuzaki and A.~I. Sanda, {\em {Analysis of lepton flavor violating
  $\tau^\pm \to \mu^\pm \mu^\pm \mu^\mp$ decays}\/},
  \href{http://dx.doi.org/10.1103/PhysRevD.77.073003}{Phys. Rev. {\bf D77}
  (2008)  073003},
\href{http://arxiv.org/abs/0711.0792}{{\tt arXiv:0711.0792 [hep-ph]}}.

\bibitem{Paradisi:2005tk}
P.~Paradisi, {\em {Higgs-mediated $\tau \to \mu$ and $\tau \to e$ transitions
  in II Higgs doublet model and supersymmetry}\/},
  \href{http://dx.doi.org/10.1088/1126-6708/2006/02/050}{JHEP {\bf 02} (2006)
  050},
\href{http://arxiv.org/abs/hep-ph/0508054}{{\tt arXiv:hep-ph/0508054
  [hep-ph]}}.

\bibitem{Celis:2013xja}
A.~Celis, V.~Cirigliano, and E.~Passemar, {\em {Lepton flavor violation in the
  Higgs sector and the role of hadronic $\tau$-lepton decays}\/},
  \href{http://dx.doi.org/10.1103/PhysRevD.89.013008}{Phys. Rev. {\bf D89}
  (2014)  013008},
\href{http://arxiv.org/abs/1309.3564}{{\tt arXiv:1309.3564 [hep-ph]}}.

\bibitem{Petrov:2013vka}
A.~A. Petrov and D.~V. Zhuridov, {\em {Lepton flavor-violating transitions in
  effective field theory and gluonic operators}\/},
  \href{http://dx.doi.org/10.1103/PhysRevD.89.033005}{Phys. Rev. {\bf D89}
  (2014) no.~3, 033005},
\href{http://arxiv.org/abs/1308.6561}{{\tt arXiv:1308.6561 [hep-ph]}}.

\bibitem{CMSTDR:muon}
{CMS Collaboration}, {\em {The Phase-2 Upgrade of the CMS Muon Detectors }\/},
   CERN-LHCC-2017-012 ; CMS-TDR-016, CERN, Geneva, 2017.
\newblock \url{https://cds.cern.ch/record/2283189/}.

\bibitem{ATL-PHYS-PUB-2018-032}
{ATLAS Collaboration}, {\em {Prospects for lepton flavour violation
  measurements in $\tau\rightarrow 3\mu$ decays with the ATLAS detector at the
  HL-LHC}\/},   ATL-PHYS-PUB-2018-032, CERN, Geneva, Nov, 2018.
\newblock \url{http://cds.cern.ch/record/2647956}.

\bibitem{LHCb-PAPER-2017-023}
{LHCb collaboration}, R.~Aaij et al., {\em {Search for baryon-number-violating
  $\Xires_b^0$ oscillations}\/},  Phys. Rev. Lett. {\bf 119} (2017)  181807,
  \href{http://arxiv.org/abs/1708.05808}{{\tt arXiv:1708.05808 [hep-ex]}}.

\bibitem{LHCb-PAPER-2016-039}
{LHCb collaboration}, R.~Aaij et al., {\em {New algorithms for identifying the
  flavour of $\Bz$ mesons using pions and protons}\/},  Eur. Phys. J. {\bf C77}
  (2017)  238, \href{http://arxiv.org/abs/1610.06019}{{\tt arXiv:1610.06019
  [hep-ex]}}.

\bibitem{Chen:2016qju}
H.-X. Chen, W.~Chen, X.~Liu, and S.-L. Zhu, {\em {The hidden-charm pentaquark
  and tetraquark states}\/},
  \href{http://dx.doi.org/10.1016/j.physrep.2016.05.004}{Phys. Rept. {\bf 639}
  (2016)  1--121},
\href{http://arxiv.org/abs/1601.02092}{{\tt arXiv:1601.02092 [hep-ph]}}.

\bibitem{Hosaka:2016pey}
A.~Hosaka, T.~Iijima, K.~Miyabayashi, Y.~Sakai, and S.~Yasui, {\em {Exotic
  hadrons with heavy flavors: X, Y, Z, and related states}\/},
  \href{http://dx.doi.org/10.1093/ptep/ptw045}{PTEP {\bf 2016} (2016) no.~6,
  062C01},
\href{http://arxiv.org/abs/1603.09229}{{\tt arXiv:1603.09229 [hep-ph]}}.

\bibitem{Lebed:2016hpi}
R.~F. Lebed, R.~E. Mitchell, and E.~S. Swanson, {\em {Heavy-quark QCD
  exotica}\/},  \href{http://dx.doi.org/10.1016/j.ppnp.2016.11.003}{Prog. Part.
  Nucl. Phys. {\bf 93} (2017)  143--194},
\href{http://arxiv.org/abs/1610.04528}{{\tt arXiv:1610.04528 [hep-ph]}}.

\bibitem{Esposito:2016noz}
A.~Esposito, A.~Pilloni, and A.~D. Polosa, {\em {Multiquark Resonances}\/},
  \href{http://dx.doi.org/10.1016/j.physrep.2016.11.002}{Phys. Rept. {\bf 668}
  (2016)  1--97},
\href{http://arxiv.org/abs/1611.07920}{{\tt arXiv:1611.07920 [hep-ph]}}.

\bibitem{Guo:2017jvc}
F.-K. Guo, C.~Hanhart, U.-G. Mei{\ss}ner, Q.~Wang, Q.~Zhao, and B.-S. Zou, {\em
  {Hadronic molecules}\/},
  \href{http://dx.doi.org/10.1103/RevModPhys.90.015004}{Rev. Mod. Phys. {\bf
  90} (2018) no.~1, 015004},
\href{http://arxiv.org/abs/1705.00141}{{\tt arXiv:1705.00141 [hep-ph]}}.

\bibitem{Ali:2017jda}
A.~Ali, J.~S. Lange, and S.~Stone, {\em {Exotics: Heavy pentaquarks and
  tetraquarks}\/},  \href{http://dx.doi.org/10.1016/j.ppnp.2017.08.003}{Prog.
  Part. Nucl. Phys. {\bf 97} (2017)  123--198},
\href{http://arxiv.org/abs/1706.00610}{{\tt arXiv:1706.00610 [hep-ph]}}.

\bibitem{Olsen:2017bmm}
S.~L. Olsen, T.~Skwarnicki, and D.~Zieminska, {\em {Nonstandard heavy mesons
  and baryons: Experimental evidence}\/},
  \href{http://dx.doi.org/10.1103/RevModPhys.90.015003}{Rev. Mod. Phys. {\bf
  90} (2018) no.~1, 015003},
\href{http://arxiv.org/abs/1708.04012}{{\tt arXiv:1708.04012 [hep-ph]}}.

\bibitem{Karliner:2017qhf}
M.~Karliner, J.~L. Rosner, and T.~Skwarnicki, {\em {Multiquark states}\/},
\href{http://arxiv.org/abs/1711.10626}{{\tt arXiv:1711.10626 [hep-ph]}}.

\bibitem{Yuan:2018inv}
C.-Z. Yuan, {\em {The XYZ states revisited}\/},
  \href{http://dx.doi.org/10.1142/S0217751X18300181}{Int. J. Mod. Phys. {\bf
  A33} (2018) no.~21, 1830018},
\href{http://arxiv.org/abs/1808.01570}{{\tt arXiv:1808.01570 [hep-ex]}}.

\bibitem{Eichten:1978tg}
E.~Eichten, K.~Gottfried, T.~Kinoshita, K.~D. Lane, and T.-M. Yan, {\em
  {Charmonium: The model}\/},
\href{http://dx.doi.org/10.1103/PhysRevD.17.3090}{Phys. Rev. {\bf D17} (1978)
  3090}.

\bibitem{Eichten:1979ms}
E.~Eichten, K.~Gottfried, T.~Kinoshita, K.~D. Lane, and T.-M. Yan, {\em
  {Charmonium: Comparison with Experiment}\/},
\href{http://dx.doi.org/10.1103/PhysRevD.21.203}{Phys. Rev. {\bf D21} (1980)
  203}.

\bibitem{Godfrey:1985xj}
S.~Godfrey and N.~Isgur, {\em {Mesons in a Relativized Quark Model with
  Chromodynamics}\/},
\href{http://dx.doi.org/10.1103/PhysRevD.32.189}{Phys. Rev. {\bf D32} (1985)
  189--231}.

\bibitem{Caswell:1985ui}
W.~E. Caswell and G.~P. Lepage, {\em {Effective lagrangians for bound state
  problems in QED, QCD, and other field theories}\/},
\href{http://dx.doi.org/10.1016/0370-2693(86)91297-9}{Phys. Lett. {\bf 167B}
  (1986)  437--442}.

\bibitem{Bodwin:1994jh}
G.~T. Bodwin, E.~Braaten, and G.~P. Lepage, {\em {Rigorous QCD analysis of
  inclusive annihilation and production of heavy quarkonium}\/},
  \href{http://dx.doi.org/10.1103/PhysRevD.51.1125}{Phys. Rev. {\bf D51} (1995)
   1125--1171},
\href{http://arxiv.org/abs/hep-ph/9407339}{{\tt arXiv:hep-ph/9407339
  [hep-ph]}}.

\bibitem{Brambilla:2004jw}
N.~Brambilla, A.~Pineda, J.~Soto, and A.~Vairo, {\em {Effective field theories
  for heavy quarkonium}\/},
  \href{http://dx.doi.org/10.1103/RevModPhys.77.1423}{Rev. Mod. Phys. {\bf 77}
  (2005)  1423},
\href{http://arxiv.org/abs/hep-ph/0410047}{{\tt arXiv:hep-ph/0410047
  [hep-ph]}}.

\bibitem{Pineda:2011dg}
A.~Pineda, {\em {Review of Heavy Quarkonium at weak coupling}\/},
  \href{http://dx.doi.org/10.1016/j.ppnp.2012.01.038}{Prog. Part. Nucl. Phys.
  {\bf 67} (2012)  735--785},
\href{http://arxiv.org/abs/1111.0165}{{\tt arXiv:1111.0165 [hep-ph]}}.

\bibitem{GellMann:1964nj}
M.~Gell-Mann, {\em {A Schematic Model of Baryons and Mesons}\/},
\href{http://dx.doi.org/10.1016/S0031-9163(64)92001-3}{Phys. Lett. {\bf 8}
  (1964)  214--215}.

\bibitem{Zweig:1964jf}
G.~Zweig, {\em {An SU(3) model for strong interaction symmetry and its
  breaking. Version 2}\/},  in {\em DEVELOPMENTS IN THE QUARK THEORY OF
  HADRONS. VOL. 1. 1964 - 1978}, D.~Lichtenberg and S.~P. Rosen, eds.,
  pp.~22--101.
\newblock
1964.
\newblock

\bibitem{Lipkin:1987xs}
H.~J. Lipkin, {\em {NEW POSSIBILITIES FOR EXOTIC HADRONS}\/},  in {\em
  {HADRONS, QUARKS AND GLUONS. PROCEEDINGS, HADRONIC SESSION OF THE 22ND
  RENCONTRES DE MORIOND, LES ARCS, FRANCE, MARCH 15-21, 1987}}, pp.~691--696.
\newblock
1987.
\newblock

\bibitem{Aaij:2015tga}
{LHCb Collaboration}, R.~Aaij et al., {\em {Observation of $J/\psi p$
  Resonances Consistent with Pentaquark States in $\Lambda_b^0 \to J/\psi K^-
  p$ Decays}\/},  \href{http://dx.doi.org/10.1103/PhysRevLett.115.072001}{Phys.
  Rev. Lett. {\bf 115} (2015)  072001},
\href{http://arxiv.org/abs/1507.03414}{{\tt arXiv:1507.03414 [hep-ex]}}.

\bibitem{Maiani:2004vq}
L.~Maiani, F.~Piccinini, A.~D. Polosa, and V.~Riquer, {\em
  {Diquark-antidiquarks with hidden or open charm and the nature of
  X(3872)}\/},  \href{http://dx.doi.org/10.1103/PhysRevD.71.014028}{Phys. Rev.
  {\bf D71} (2005)  014028},
\href{http://arxiv.org/abs/hep-ph/0412098}{{\tt arXiv:hep-ph/0412098
  [hep-ph]}}.

\bibitem{Maiani:2014aja}
L.~Maiani, F.~Piccinini, A.~D. Polosa, and V.~Riquer, {\em {The Z(4430) and a
  New Paradigm for Spin Interactions in Tetraquarks}\/},
  \href{http://dx.doi.org/10.1103/PhysRevD.89.114010}{Phys. Rev. {\bf D89}
  (2014)  114010},
\href{http://arxiv.org/abs/1405.1551}{{\tt arXiv:1405.1551 [hep-ph]}}.

\bibitem{Maiani:2017kyi}
L.~Maiani, A.~D. Polosa, and V.~Riquer, {\em {A Theory of X and Z Multiquark
  Resonances}\/},
  \href{http://dx.doi.org/10.1016/j.physletb.2018.01.039}{Phys. Lett. {\bf
  B778} (2018)  247--251},
\href{http://arxiv.org/abs/1712.05296}{{\tt arXiv:1712.05296 [hep-ph]}}.

\bibitem{Rossi:2004yr}
G.~C. Rossi and G.~Veneziano, {\em {Isospin mixing of narrow pentaquark
  states}\/},  \href{http://dx.doi.org/10.1016/j.physletb.2004.07.042}{Phys.
  Lett. {\bf B597} (2004)  338--345},
\href{http://arxiv.org/abs/hep-ph/0404262}{{\tt arXiv:hep-ph/0404262
  [hep-ph]}}.

\bibitem{Maiani:2008zz}
L.~Maiani, A.~D. Polosa, and V.~Riquer, {\em {The charged Z(4430) in the
  diquark-antidiquark picture}\/},
\href{http://dx.doi.org/10.1088/1367-2630/10/7/073004}{New J. Phys. {\bf 10}
  (2008)  073004}.

\bibitem{Maiani:2007wz}
L.~Maiani, A.~D. Polosa, and V.~Riquer, {\em {The Charged Z(4433): Towards a
  new spectroscopy}\/},
\href{http://arxiv.org/abs/0708.3997}{{\tt arXiv:0708.3997 [hep-ph]}}.

\bibitem{Cotugno:2009ys}
G.~Cotugno, R.~Faccini, A.~D. Polosa, and C.~Sabelli, {\em {Charmed
  Baryonium}\/},  \href{http://dx.doi.org/10.1103/PhysRevLett.104.132005}{Phys.
  Rev. Lett. {\bf 104} (2010)  132005},
\href{http://arxiv.org/abs/0911.2178}{{\tt arXiv:0911.2178 [hep-ph]}}.

\bibitem{Esposito:2014rxa}
A.~Esposito, A.~L. Guerrieri, F.~Piccinini, A.~Pilloni, and A.~D. Polosa, {\em
  {Four-Quark Hadrons: an Updated Review}\/},
  \href{http://dx.doi.org/10.1142/S0217751X15300021}{Int. J. Mod. Phys. {\bf
  A30} (2015)  1530002},
\href{http://arxiv.org/abs/1411.5997}{{\tt arXiv:1411.5997 [hep-ph]}}.

\bibitem{Esposito:2016itg}
A.~Esposito, A.~Pilloni, and A.~D. Polosa, {\em {Hybridized Tetraquarks}\/},
  \href{http://dx.doi.org/10.1016/j.physletb.2016.05.028}{Phys. Lett. {\bf
  B758} (2016)  292--295},
\href{http://arxiv.org/abs/1603.07667}{{\tt arXiv:1603.07667 [hep-ph]}}.

\bibitem{Jaffe:2003sg}
R.~L. Jaffe and F.~Wilczek, {\em {Diquarks and exotic spectroscopy}\/},
  \href{http://dx.doi.org/10.1103/PhysRevLett.91.232003}{Phys. Rev. Lett. {\bf
  91} (2003)  232003},
\href{http://arxiv.org/abs/hep-ph/0307341}{{\tt arXiv:hep-ph/0307341
  [hep-ph]}}.

\bibitem{Selem:2006nd}
A.~Selem and F.~Wilczek,
  \href{http://dx.doi.org/10.1142/9789812773524_0030}{{\em {Hadron systematics
  and emergent diquarks}\/}, } in {\em {Proceedings, Ringberg Workshop on New
  Trends in HERA Physics 2005: Ringberg Castle, Tegernsee, Germany, October
  2-7, 2005}}, pp.~337--356.
\newblock 2006.
\newblock
\href{http://arxiv.org/abs/hep-ph/0602128}{{\tt arXiv:hep-ph/0602128
  [hep-ph]}}.
\newblock

\bibitem{Esposito:2018cwh}
A.~Esposito and A.~D. Polosa, {\em {A $bb\bar b\bar b$di-bottomonium at the
  LHC?}\/},
\href{http://arxiv.org/abs/1807.06040}{{\tt arXiv:1807.06040 [hep-ph]}}.

\bibitem{Brodsky:2014xia}
S.~J. Brodsky, D.~S. Hwang, and R.~F. Lebed, {\em {Dynamical Picture for the
  Formation and Decay of the Exotic XYZ Mesons}\/},
  \href{http://dx.doi.org/10.1103/PhysRevLett.113.112001}{Phys. Rev. Lett. {\bf
  113} (2014) no.~11, 112001},
\href{http://arxiv.org/abs/1406.7281}{{\tt arXiv:1406.7281 [hep-ph]}}.

\bibitem{Lebed:2017min}
R.~F. Lebed, {\em {Spectroscopy of Exotic Hadrons Formed from Dynamical
  Diquarks}\/},  \href{http://dx.doi.org/10.1103/PhysRevD.96.116003}{Phys. Rev.
  {\bf D96} (2017) no.~11, 116003},
\href{http://arxiv.org/abs/1709.06097}{{\tt arXiv:1709.06097 [hep-ph]}}.

\bibitem{Lebed:2015tna}
R.~F. Lebed, {\em {The Pentaquark Candidates in the Dynamical Diquark
  Picture}\/},  \href{http://dx.doi.org/10.1016/j.physletb.2015.08.032}{Phys.
  Lett. {\bf B749} (2015)  454--457},
\href{http://arxiv.org/abs/1507.05867}{{\tt arXiv:1507.05867 [hep-ph]}}.

\bibitem{Braaten:2013boa}
E.~Braaten, {\em {How the $Z_c(3900)$ Reveals the Spectra of Quarkonium Hybrid
  and Tetraquark Mesons}\/},
  \href{http://dx.doi.org/10.1103/PhysRevLett.111.162003}{Phys. Rev. Lett. {\bf
  111} (2013)  162003},
\href{http://arxiv.org/abs/1305.6905}{{\tt arXiv:1305.6905 [hep-ph]}}.

\bibitem{Brodsky:2015wza}
S.~J. Brodsky and R.~F. Lebed, {\em {QCD dynamics of tetraquark production}\/},
   \href{http://dx.doi.org/10.1103/PhysRevD.91.114025}{Phys. Rev. {\bf D91}
  (2015)  114025},
\href{http://arxiv.org/abs/1505.00803}{{\tt arXiv:1505.00803 [hep-ph]}}.

\bibitem{Lebed:2018jcr}
R.~F. Lebed, {\em {Constituent Counting Rules and Exotic Hadrons}\/},
  \href{http://dx.doi.org/10.1007/s00601-018-1427-2}{Few Body Syst. {\bf 59}
  (2018) no.~5, 106},
\href{http://arxiv.org/abs/1807.01650}{{\tt arXiv:1807.01650 [hep-ph]}}.

\bibitem{Kalashnikova:2018vkv}
Y.~S. Kalashnikova and A.~V. Nefediev, {\em {$X(3872)$ in the molecular
  model}\/},
\href{http://arxiv.org/abs/1811.01324}{{\tt arXiv:1811.01324 [hep-ph]}}.

\bibitem{Weinberg:1965zz}
S.~Weinberg, {\em {Evidence That the Deuteron Is Not an Elementary
  Particle}\/},
\href{http://dx.doi.org/10.1103/PhysRev.137.B672}{Phys. Rev. {\bf 137} (1965)
  B672--B678}.

\bibitem{Sekihara:2014kya}
T.~Sekihara, T.~Hyodo, and D.~Jido, {\em {Comprehensive analysis of the wave
  function of a hadronic resonance and its compositeness}\/},
  \href{http://dx.doi.org/10.1093/ptep/ptv081}{PTEP {\bf 2015} (2015)  063D04},
\href{http://arxiv.org/abs/1411.2308}{{\tt arXiv:1411.2308 [hep-ph]}}.

\bibitem{Molina:2016pbg}
R.~Molina, M.~D{\"o}ring, and E.~Oset, {\em {Determination of the compositeness
  of resonances from decays: the case of the $B^0_s\to J/\psi f_1(1285)$}\/},
  \href{http://dx.doi.org/10.1103/PhysRevD.93.114004}{Phys. Rev. {\bf D93}
  (2016) no.~11, 114004},
\href{http://arxiv.org/abs/1604.02574}{{\tt arXiv:1604.02574 [hep-ph]}}.

\bibitem{Oset:2016lyh}
E.~Oset et al., {\em {Weak decays of heavy hadrons into dynamically generated
  resonances}\/},  \href{http://dx.doi.org/10.1142/S0218301316300010}{Int. J.
  Mod. Phys. {\bf E25} (2016)  1630001},
\href{http://arxiv.org/abs/1601.03972}{{\tt arXiv:1601.03972 [hep-ph]}}.

\bibitem{Eichten:2005ga}
E.~J. Eichten, K.~Lane, and C.~Quigg, {\em {New states above charm
  threshold}\/},  \href{http://dx.doi.org/10.1103/PhysRevD.73.014014,
  10.1103/PhysRevD.73.079903}{Phys. Rev. {\bf D73} (2006)  014014},
  \href{http://arxiv.org/abs/hep-ph/0511179}{{\tt arXiv:hep-ph/0511179
  [hep-ph]}}.
[Erratum: Phys. Rev.D73,079903(2006)].

\bibitem{Kalashnikova:2005ui}
{\relax Yu}.~S. Kalashnikova, {\em {Coupled-channel model for charmonium levels
  and an option for X(3872)}\/},
  \href{http://dx.doi.org/10.1103/PhysRevD.72.034010}{Phys. Rev. {\bf D72}
  (2005)  034010},
\href{http://arxiv.org/abs/hep-ph/0506270}{{\tt arXiv:hep-ph/0506270
  [hep-ph]}}.

\bibitem{Baru:2010ww}
V.~Baru, C.~Hanhart, {\relax Yu}.~S. Kalashnikova, A.~E. Kudryavtsev, and A.~V.
  Nefediev, {\em {Interplay of quark and meson degrees of freedom in a
  near-threshold resonance}\/},
  \href{http://dx.doi.org/10.1140/epja/i2010-10929-7}{Eur. Phys. J. {\bf A44}
  (2010)  93--103},
\href{http://arxiv.org/abs/1001.0369}{{\tt arXiv:1001.0369 [hep-ph]}}.

\bibitem{Tornqvist:1979hx}
N.~A. T{\"o}rnqvist, {\em {The Meson Mass Spectrum and Unitarity}\/},
\href{http://dx.doi.org/10.1016/0003-4916(79)90262-8}{Annals Phys. {\bf 123}
  (1979)  1}.

\bibitem{Ortega:2012rs}
P.~G. Ortega, D.~R. Entem, and F.~Fernandez, {\em {Molecular Structures in
  Charmonium Spectrum: The $XYZ$ Puzzle}\/},
  \href{http://dx.doi.org/10.1088/0954-3899/40/6/065107}{J. Phys. {\bf G40}
  (2013)  065107},
\href{http://arxiv.org/abs/1205.1699}{{\tt arXiv:1205.1699 [hep-ph]}}.

\bibitem{Segovia:2016xqb}
J.~Segovia, P.~G. Ortega, D.~R. Entem, and F.~Fern{\'a}ndez, {\em {Bottomonium
  spectrum revisited}\/},
  \href{http://dx.doi.org/10.1103/PhysRevD.93.074027}{Phys. Rev. {\bf D93}
  (2016) no.~7, 074027},
\href{http://arxiv.org/abs/1601.05093}{{\tt arXiv:1601.05093 [hep-ph]}}.

\bibitem{Vijande:2004he}
J.~Vijande, F.~Fernandez, and A.~Valcarce, {\em {Constituent quark model study
  of the meson spectra}\/},
  \href{http://dx.doi.org/10.1088/0954-3899/31/5/017}{J. Phys. {\bf G31} (2005)
   481},
\href{http://arxiv.org/abs/hep-ph/0411299}{{\tt arXiv:hep-ph/0411299
  [hep-ph]}}.

\bibitem{Segovia:2013wma}
J.~Segovia, D.~R. Entem, F.~Fernandez, and E.~Hernandez, {\em {Constituent
  quark model description of charmonium phenomenology}\/},
  \href{http://dx.doi.org/10.1142/S0218301313300269}{Int. J. Mod. Phys. {\bf
  E22} (2013)  1330026},
\href{http://arxiv.org/abs/1309.6926}{{\tt arXiv:1309.6926 [hep-ph]}}.

\bibitem{Swanson:2005rc}
E.~S. Swanson, {\em {Unquenching the quark model and screened potentials}\/},
  \href{http://dx.doi.org/10.1088/0954-3899/31/7/025}{J. Phys. {\bf G31} (2005)
   845--854},
\href{http://arxiv.org/abs/hep-ph/0504097}{{\tt arXiv:hep-ph/0504097
  [hep-ph]}}.

\bibitem{Barnes:2007xu}
T.~Barnes and E.~S. Swanson, {\em {Hadron loops: General theorems and
  application to charmonium}\/},
  \href{http://dx.doi.org/10.1103/PhysRevC.77.055206}{Phys. Rev. {\bf C77}
  (2008)  055206},
\href{http://arxiv.org/abs/0711.2080}{{\tt arXiv:0711.2080 [hep-ph]}}.

\bibitem{Tang:1978zz}
Y.~C. Tang, M.~Lemere, and D.~R. Thompson, {\em {Resonating-group method for
  nuclear many-body problems}\/},
\href{http://dx.doi.org/10.1016/0370-1573(78)90175-8}{Phys. Rept. {\bf 47}
  (1978)  167--223}.

\bibitem{Entem:2000mq}
D.~R. Entem, F.~Fernandez, and A.~Valcarce, {\em {Chiral quark model of the N N
  system within a Lippmann-Schwinger resonating group method}\/},
\href{http://dx.doi.org/10.1103/PhysRevC.62.034002}{Phys. Rev. {\bf C62} (2000)
   034002}.

\bibitem{Micu:1968mk}
L.~Micu, {\em {Decay rates of meson resonances in a quark model}\/},
\href{http://dx.doi.org/10.1016/0550-3213(69)90039-X}{Nucl. Phys. {\bf B10}
  (1969)  521--526}.

\bibitem{LeYaouanc:1972vsx}
A.~Le~Yaouanc, L.~Oliver, O.~Pene, and J.~C. Raynal, {\em {Naive quark pair
  creation model of strong interaction vertices}\/},
\href{http://dx.doi.org/10.1103/PhysRevD.8.2223}{Phys. Rev. {\bf D8} (1973)
  2223--2234}.

\bibitem{LeYaouanc:1973ldf}
A.~Le~Yaouanc, L.~Oliver, O.~Pene, and J.~C. Raynal, {\em {Naive quark pair
  creation model and baryon decays}\/},
\href{http://dx.doi.org/10.1103/PhysRevD.9.1415}{Phys. Rev. {\bf D9} (1974)
  1415--1419}.

\bibitem{Ortega:2016hde}
P.~G. Ortega, J.~Segovia, D.~R. Entem, and F.~Fernandez, {\em {Canonical
  description of the new LHCb resonances}\/},
  \href{http://dx.doi.org/10.1103/PhysRevD.94.114018}{Phys. Rev. {\bf D94}
  (2016) no.~11, 114018},
\href{http://arxiv.org/abs/1608.01325}{{\tt arXiv:1608.01325 [hep-ph]}}.

\bibitem{Godfrey:2003kg}
S.~Godfrey, {\em {Testing the nature of the D(sJ)*(2317)+ and D(sJ)(2463)+
  states using radiative transitions}\/},
  \href{http://dx.doi.org/10.1016/j.physletb.2003.06.049}{Phys. Lett. {\bf
  B568} (2003)  254--260},
\href{http://arxiv.org/abs/hep-ph/0305122}{{\tt arXiv:hep-ph/0305122
  [hep-ph]}}.

\bibitem{Colangelo:2003vg}
P.~Colangelo and F.~De~Fazio, {\em {Understanding D(sJ)(2317)}\/},
  \href{http://dx.doi.org/10.1016/j.physletb.2003.08.003}{Phys. Lett. {\bf
  B570} (2003)  180--184},
\href{http://arxiv.org/abs/hep-ph/0305140}{{\tt arXiv:hep-ph/0305140
  [hep-ph]}}.

\bibitem{Mehen:2005hc}
T.~Mehen and R.~P. Springer, {\em {Even- and odd-parity charmed meson masses in
  heavy hadron chiral perturbation theory}\/},
  \href{http://dx.doi.org/10.1103/PhysRevD.72.034006}{Phys. Rev. {\bf D72}
  (2005)  034006},
\href{http://arxiv.org/abs/hep-ph/0503134}{{\tt arXiv:hep-ph/0503134
  [hep-ph]}}.

\bibitem{Lakhina:2006fy}
O.~Lakhina and E.~S. Swanson, {\em {A Canonical Ds(2317)?}\/},
  \href{http://dx.doi.org/10.1016/j.physletb.2007.01.075}{Phys. Lett. {\bf
  B650} (2007)  159--165},
\href{http://arxiv.org/abs/hep-ph/0608011}{{\tt arXiv:hep-ph/0608011
  [hep-ph]}}.

\bibitem{Bardeen:2003kt}
W.~A. Bardeen, E.~J. Eichten, and C.~T. Hill, {\em {Chiral multiplets of heavy
  - light mesons}\/},
  \href{http://dx.doi.org/10.1103/PhysRevD.68.054024}{Phys. Rev. {\bf D68}
  (2003)  054024},
\href{http://arxiv.org/abs/hep-ph/0305049}{{\tt arXiv:hep-ph/0305049
  [hep-ph]}}.

\bibitem{Nowak:2003ra}
M.~A. Nowak, M.~Rho, and I.~Zahed, {\em {Chiral doubling of heavy light
  hadrons: BABAR 2317 MeV$/c^2$ and CLEO 2463 MeV$/c^2$ discoveries}\/},  Acta
  Phys. Polon. {\bf B35} (2004)  2377--2392,
\href{http://arxiv.org/abs/hep-ph/0307102}{{\tt arXiv:hep-ph/0307102
  [hep-ph]}}.

\bibitem{Godfrey:2015dva}
S.~Godfrey and K.~Moats, {\em {Properties of Excited Charm and Charm-Strange
  Mesons}\/},  \href{http://dx.doi.org/10.1103/PhysRevD.93.034035}{Phys. Rev.
  {\bf D93} (2016) no.~3, 034035},
\href{http://arxiv.org/abs/1510.08305}{{\tt arXiv:1510.08305 [hep-ph]}}.

\bibitem{Browder:2003fk}
T.~E. Browder, S.~Pakvasa, and A.~A. Petrov, {\em {Comment on the new $D_s^{*+}
  \pi^0$ resonances}\/},
  \href{http://dx.doi.org/10.1016/j.physletb.2003.10.067}{Phys. Lett. {\bf
  B578} (2004)  365--368},
\href{http://arxiv.org/abs/hep-ph/0307054}{{\tt arXiv:hep-ph/0307054
  [hep-ph]}}.

\bibitem{Barnes:2003dj}
T.~Barnes, F.~E. Close, and H.~J. Lipkin, {\em {Implications of a DK molecule
  at 2.32-GeV}\/},  \href{http://dx.doi.org/10.1103/PhysRevD.68.054006}{Phys.
  Rev. {\bf D68} (2003)  054006},
\href{http://arxiv.org/abs/hep-ph/0305025}{{\tt arXiv:hep-ph/0305025
  [hep-ph]}}.

\bibitem{vanBeveren:2003kd}
E.~van Beveren and G.~Rupp, {\em {Observed D(s)(2317) and tentative D(2030) as
  the charmed cousins of the light scalar nonet}\/},
  \href{http://dx.doi.org/10.1103/PhysRevLett.91.012003}{Phys. Rev. Lett. {\bf
  91} (2003)  012003},
\href{http://arxiv.org/abs/hep-ph/0305035}{{\tt arXiv:hep-ph/0305035
  [hep-ph]}}.

\bibitem{Kolomeitsev:2003ac}
E.~E. Kolomeitsev and M.~F.~M. Lutz, {\em {On Heavy light meson resonances and
  chiral symmetry}\/},
  \href{http://dx.doi.org/10.1016/j.physletb.2003.10.118}{Phys. Lett. {\bf
  B582} (2004)  39--48},
\href{http://arxiv.org/abs/hep-ph/0307133}{{\tt arXiv:hep-ph/0307133
  [hep-ph]}}.

\bibitem{Guo:2006fu}
F.-K. Guo, P.-N. Shen, H.-C. Chiang, R.-G. Ping, and B.-S. Zou, {\em
  {Dynamically generated $0^+$ heavy mesons in a heavy chiral unitary
  approach}\/},  \href{http://dx.doi.org/10.1016/j.physletb.2006.08.064}{Phys.
  Lett. {\bf B641} (2006)  278--285},
\href{http://arxiv.org/abs/hep-ph/0603072}{{\tt arXiv:hep-ph/0603072
  [hep-ph]}}.

\bibitem{Guo:2006rp}
F.-K. Guo, P.-N. Shen, and H.-C. Chiang, {\em {Dynamically generated $1^+$
  heavy mesons}\/},
  \href{http://dx.doi.org/10.1016/j.physletb.2007.01.050}{Phys. Lett. {\bf
  B647} (2007)  133--139},
\href{http://arxiv.org/abs/hep-ph/0610008}{{\tt arXiv:hep-ph/0610008
  [hep-ph]}}.

\bibitem{Gamermann:2006nm}
D.~Gamermann, E.~Oset, D.~Strottman, and M.~J. Vicente~Vacas, {\em {Dynamically
  generated open and hidden charm meson systems}\/},
  \href{http://dx.doi.org/10.1103/PhysRevD.76.074016}{Phys. Rev. {\bf D76}
  (2007)  074016},
\href{http://arxiv.org/abs/hep-ph/0612179}{{\tt arXiv:hep-ph/0612179
  [hep-ph]}}.

\bibitem{Liu:2012zya}
L.~Liu, K.~Orginos, F.-K. Guo, C.~Hanhart, and U.-G. Mei{\ss}ner, {\em
  {Interactions of charmed mesons with light pseudoscalar mesons from lattice
  QCD and implications on the nature of the $D_{s0}^*(2317)$}\/},
  \href{http://dx.doi.org/10.1103/PhysRevD.87.014508}{Phys. Rev. {\bf D87}
  (2013) no.~1, 014508},
\href{http://arxiv.org/abs/1208.4535}{{\tt arXiv:1208.4535 [hep-lat]}}.

\bibitem{Bali:2017pdv}
G.~S. Bali, S.~Collins, A.~Cox, and A.~Sch{\"a}fer, {\em {Masses and decay
  constants of the $D_{s0}^*(2317)$ and $D_{s1}(2460)$ from $N_f=2$ lattice QCD
  close to the physical point}\/},
  \href{http://dx.doi.org/10.1103/PhysRevD.96.074501}{Phys. Rev. {\bf D96}
  (2017) no.~7, 074501},
\href{http://arxiv.org/abs/1706.01247}{{\tt arXiv:1706.01247 [hep-lat]}}.

\bibitem{Lang:2014yfa}
C.~B. Lang, L.~Leskovec, D.~Mohler, S.~Prelovsek, and R.~M. Woloshyn, {\em {Ds
  mesons with DK and D*K scattering near threshold}\/},
  \href{http://dx.doi.org/10.1103/PhysRevD.90.034510}{Phys. Rev. {\bf D90}
  (2014) no.~3, 034510},
\href{http://arxiv.org/abs/1403.8103}{{\tt arXiv:1403.8103 [hep-lat]}}.

\bibitem{Torres:2014vna}
A.~Mart{\'i}nez~Torres, E.~Oset, S.~Prelovsek, and A.~Ramos, {\em {Reanalysis
  of lattice QCD spectra leading to the $D_{s0}^*(2317)$ and
  $D_{s1}^*(2460)$}\/},  \href{http://dx.doi.org/10.1007/JHEP05(2015)153}{JHEP
  {\bf 05} (2015)  153},
\href{http://arxiv.org/abs/1412.1706}{{\tt arXiv:1412.1706 [hep-lat]}}.

\bibitem{Albaladejo:2018mhb}
M.~Albaladejo, P.~Fernandez-Soler, J.~Nieves, and P.~G. Ortega, {\em
  {Contribution of constituent quark model $c\bar{s}$ states to the dynamics of
  the $D_{s0}^*(2317)$ and $D_{s1}(2460)$ resonances}\/},
  \href{http://dx.doi.org/10.1140/epjc/s10052-018-6176-3}{Eur. Phys. J. {\bf
  C78} (2018) no.~9, 722},
\href{http://arxiv.org/abs/1805.07104}{{\tt arXiv:1805.07104 [hep-ph]}}.

\bibitem{Aaij:2017vck}
{LHCb Collaboration}, R.~Aaij et al., {\em {$\chi_{c1}$ and $\chi_{c2}$
  Resonance Parameters with the Decays $\chi_{c1,c2}\to J/\psi\mu^+\mu^-$}\/},
  \href{http://dx.doi.org/10.1103/PhysRevLett.119.221801}{Phys. Rev. Lett. {\bf
  119} (2017) no.~22, 221801},
\href{http://arxiv.org/abs/1709.04247}{{\tt arXiv:1709.04247 [hep-ex]}}.

\bibitem{Albaladejo:2016lbb}
M.~Albaladejo, P.~Fernandez-Soler, F.-K. Guo, and J.~Nieves, {\em {Two-pole
  structure of the $D^\ast_0(2400)$}\/},
  \href{http://dx.doi.org/10.1016/j.physletb.2017.02.036}{Phys. Lett. {\bf
  B767} (2017)  465--469},
\href{http://arxiv.org/abs/1610.06727}{{\tt arXiv:1610.06727 [hep-ph]}}.

\bibitem{Du:2017zvv}
M.-L. Du, M.~Albaladejo, P.~Fernandez-Soler, F.-K. Guo, C.~Hanhart, U.-G.
  Mei{\ss}ner, J.~Nieves, and D.-L. Yao, {\em {A new paradigm for heavy-light
  meson spectroscopy}\/},
\href{http://arxiv.org/abs/1712.07957}{{\tt arXiv:1712.07957 [hep-ph]}}.

\bibitem{Aaij:2016fma}
{LHCb Collaboration}, R.~Aaij et al., {\em {Amplitude analysis of $B^{-} \to
  D^{+} \pi^{-} \pi^{-}$ decays}\/},
  \href{http://dx.doi.org/10.1103/PhysRevD.94.072001}{Phys. Rev. {\bf D94}
  (2016) no.~7, 072001},
\href{http://arxiv.org/abs/1608.01289}{{\tt arXiv:1608.01289 [hep-ex]}}.

\bibitem{Aubert:2007xma}
{BaBar Collaboration}, B.~Aubert et al., {\em {Observation of tree-level $B$
  decays with s anti-s production from gluon radiation.}\/},
  \href{http://dx.doi.org/10.1103/PhysRevLett.100.171803}{Phys. Rev. Lett. {\bf
  100} (2008)  171803},
\href{http://arxiv.org/abs/0707.1043}{{\tt arXiv:0707.1043 [hep-ex]}}.

\bibitem{Wiechczynski:2009rg}
{Belle Collaboration}, J.~Wiechczynski et al., {\em {Measurement of $B \to
  D^{(*)}_s K \pi$ branching fractions}\/},
  \href{http://dx.doi.org/10.1103/PhysRevD.80.052005}{Phys. Rev. {\bf D80}
  (2009)  052005},
\href{http://arxiv.org/abs/0903.4956}{{\tt arXiv:0903.4956 [hep-ex]}}.

\bibitem{Wiechczynski:2014kxh}
{Belle Collaboration}, J.~Wiechczynski et al., {\em {Measurement of $B^0 \to
  D_s^- K^0_S\pi^+$ and $B^+ \to D_s^- K^+K^+$ branching fractions}\/},
  \href{http://dx.doi.org/10.1103/PhysRevD.91.032008}{Phys. Rev. {\bf D91}
  (2015) no.~3, 032008},
\href{http://arxiv.org/abs/1411.2035}{{\tt arXiv:1411.2035 [hep-ex]}}.

\bibitem{Lang:2015hza}
C.~B. Lang, D.~Mohler, S.~Prelovsek, and R.~M. Woloshyn, {\em {Predicting
  positive parity B$_s$ mesons from lattice QCD}\/},
  \href{http://dx.doi.org/10.1016/j.physletb.2015.08.038}{Phys. Lett. {\bf
  B750} (2015)  17--21},
\href{http://arxiv.org/abs/1501.01646}{{\tt arXiv:1501.01646 [hep-lat]}}.

\bibitem{Cleven:2014oka}
M.~Cleven, H.~W. Grie{\ss}hammer, F.-K. Guo, C.~Hanhart, and U.-G. Mei{\ss}ner,
  {\em {Strong and radiative decays of the $D^*_{s0}(2317)$ and
  $D_{s1}(2460)$}\/},  \href{http://dx.doi.org/10.1140/epja/i2014-14149-y}{Eur.
  Phys. J. {\bf A50} (2014)  149},
\href{http://arxiv.org/abs/1405.2242}{{\tt arXiv:1405.2242 [hep-ph]}}.

\bibitem{Aaij:2012uva}
{LHCb Collaboration}, R.~Aaij et al., {\em {First observation of the decay
  $B_{s2}^*(5840)^0 \to B^{*+} K^-$ and studies of excited $B^0_s$ mesons}\/},
  \href{http://dx.doi.org/10.1103/PhysRevLett.110.151803}{Phys. Rev. Lett. {\bf
  110} (2013) no.~15, 151803},
\href{http://arxiv.org/abs/1211.5994}{{\tt arXiv:1211.5994 [hep-ex]}}.

\bibitem{Godfrey:2016nwn}
S.~Godfrey, K.~Moats, and E.~S. Swanson, {\em {$B$ and $B_s$ Meson
  Spectroscopy}\/},  \href{http://dx.doi.org/10.1103/PhysRevD.94.054025}{Phys.
  Rev. {\bf D94} (2016) no.~5, 054025},
\href{http://arxiv.org/abs/1607.02169}{{\tt arXiv:1607.02169 [hep-ph]}}.

\bibitem{Aubert:2006mh}
{BaBar Collaboration}, B.~Aubert et al., {\em {Observation of a New $D_s$ Meson
  Decaying to $DK$ at a Mass of 2.86 GeV/$c^2$}\/},
  \href{http://dx.doi.org/10.1103/PhysRevLett.97.222001}{Phys. Rev. Lett. {\bf
  97} (2006)  222001},
\href{http://arxiv.org/abs/hep-ex/0607082}{{\tt arXiv:hep-ex/0607082
  [hep-ex]}}.

\bibitem{Aaij:2014xza}
{LHCb Collaboration}, R.~Aaij et al., {\em {Observation of overlapping spin-1
  and spin-3 $\bar{D}^0 K^-$ resonances at mass $2.86 {\rm GeV}/c^2$}\/},
  \href{http://dx.doi.org/10.1103/PhysRevLett.113.162001}{Phys. Rev. Lett. {\bf
  113} (2014)  162001},
\href{http://arxiv.org/abs/1407.7574}{{\tt arXiv:1407.7574 [hep-ex]}}.

\bibitem{Aubert:2009ah}
{BaBar Collaboration}, B.~Aubert et al., {\em {Study of $D_{sJ}$ decays to
  $D^*K$ in inclusive $e^+ e^-$ interactions}\/},
  \href{http://dx.doi.org/10.1103/PhysRevD.80.092003}{Phys. Rev. {\bf D80}
  (2009)  092003},
\href{http://arxiv.org/abs/0908.0806}{{\tt arXiv:0908.0806 [hep-ex]}}.

\bibitem{Colangelo:2006rq}
P.~Colangelo, F.~De~Fazio, and S.~Nicotri, {\em {$D_{sJ}(2860)$ resonance and
  the $s^P_l = 5/2^-$ $c\bar s (c\bar q)$ doublet}\/},
  \href{http://dx.doi.org/10.1016/j.physletb.2006.09.018}{Phys. Lett. {\bf
  B642} (2006)  48--52},
\href{http://arxiv.org/abs/hep-ph/0607245}{{\tt arXiv:hep-ph/0607245
  [hep-ph]}}.

\bibitem{Segovia:2015dia}
J.~Segovia, D.~R. Entem, and F.~Fernandez, {\em {Charmed-strange Meson
  Spectrum: Old and New Problems}\/},
  \href{http://dx.doi.org/10.1103/PhysRevD.91.094020}{Phys. Rev. {\bf D91}
  (2015) no.~9, 094020},
\href{http://arxiv.org/abs/1502.03827}{{\tt arXiv:1502.03827 [hep-ph]}}.

\bibitem{Guo:2011dd}
F.-K. Guo and U.-G. Mei{\ss}ner, {\em {More kaonic bound states and a
  comprehensive interpretation of the $D_{sJ}$ states}\/},
  \href{http://dx.doi.org/10.1103/PhysRevD.84.014013}{Phys. Rev. {\bf D84}
  (2011)  014013},
\href{http://arxiv.org/abs/1102.3536}{{\tt arXiv:1102.3536 [hep-ph]}}.

\bibitem{Cheung:2016bym}
{Hadron Spectrum Collaboration}, G.~K.~C. Cheung, C.~O'Hara, G.~Moir,
  M.~Peardon, S.~M. Ryan, C.~E. Thomas, and D.~Tims, {\em {Excited and exotic
  charmonium, $D_s$ and $D$ meson spectra for two light quark masses from
  lattice QCD}\/},  \href{http://dx.doi.org/10.1007/JHEP12(2016)089}{JHEP {\bf
  12} (2016)  089},
\href{http://arxiv.org/abs/1610.01073}{{\tt arXiv:1610.01073 [hep-lat]}}.

\bibitem{Luscher:1990ux}
M.~Luscher, {\em {Two particle states on a torus and their relation to the
  scattering matrix}\/},
\href{http://dx.doi.org/10.1016/0550-3213(91)90366-6}{Nucl. Phys. {\bf B354}
  (1991)  531--578}.

\bibitem{Mohler:2012na}
D.~Mohler, S.~Prelovsek, and R.~M. Woloshyn, {\em {$D \pi$ scattering and $D$
  meson resonances from lattice QCD}\/},
  \href{http://dx.doi.org/10.1103/PhysRevD.87.034501}{Phys. Rev. {\bf D87}
  (2013) no.~3, 034501},
\href{http://arxiv.org/abs/1208.4059}{{\tt arXiv:1208.4059 [hep-lat]}}.

\bibitem{Moir:2016srx}
G.~Moir, M.~Peardon, S.~M. Ryan, C.~E. Thomas, and D.~J. Wilson, {\em
  {Coupled-Channel $D\pi$, $D\eta$ and $D_{s}\bar{K}$ Scattering from Lattice
  QCD}\/},  \href{http://dx.doi.org/10.1007/JHEP10(2016)011}{JHEP {\bf 10}
  (2016)  011},
\href{http://arxiv.org/abs/1607.07093}{{\tt arXiv:1607.07093 [hep-lat]}}.

\bibitem{Padmanath:2013bla}
M.~Padmanath, R.~G. Edwards, N.~Mathur, and M.~Peardon, {\em {Excited-state
  spectroscopy of singly, doubly and triply-charmed baryons from lattice
  QCD}\/},  in {\em {Proceedings, 6th International Workshop on Charm Physics
  (Charm 2013): Manchester, UK, August 31-September 4, 2013}}.
\newblock 2013.
\newblock \href{http://arxiv.org/abs/1311.4806}{{\tt arXiv:1311.4806
  [hep-lat]}}.
\newblock
\url{http://inspirehep.net/record/1265083/files/arXiv:1311.4806.pdf}.
\newblock

\bibitem{Aaij:2017nav}
{LHCb Collaboration}, R.~Aaij et al., {\em {Observation of five new narrow
  $\Omega_c^0$ states decaying to $\Xi_c^+ K^-$}\/},
  \href{http://dx.doi.org/10.1103/PhysRevLett.118.182001}{Phys. Rev. Lett. {\bf
  118} (2017) no.~18, 182001},
\href{http://arxiv.org/abs/1703.04639}{{\tt arXiv:1703.04639 [hep-ex]}}.

\bibitem{Padmanath:2017lng}
M.~Padmanath and N.~Mathur, {\em {Quantum Numbers of Recently Discovered
  $\Omega^{0}_{c}$ Baryons from Lattice QCD}\/},
  \href{http://dx.doi.org/10.1103/PhysRevLett.119.042001}{Phys. Rev. Lett. {\bf
  119} (2017) no.~4, 042001},
\href{http://arxiv.org/abs/1704.00259}{{\tt arXiv:1704.00259 [hep-ph]}}.

\bibitem{Karliner:2017kfm}
M.~Karliner and J.~L. Rosner, {\em {Very narrow excited $\Omega_c$ baryons}\/},
   \href{http://dx.doi.org/10.1103/PhysRevD.95.114012}{Phys. Rev. {\bf D95}
  (2017) no.~11, 114012},
\href{http://arxiv.org/abs/1703.07774}{{\tt arXiv:1703.07774 [hep-ph]}}.

\bibitem{Wang:2017hej}
K.-L. Wang, L.-Y. Xiao, X.-H. Zhong, and Q.~Zhao, {\em {Understanding the newly
  observed $\Omega_c$ states through their decays}\/},
  \href{http://dx.doi.org/10.1103/PhysRevD.95.116010}{Phys. Rev. {\bf D95}
  (2017) no.~11, 116010},
\href{http://arxiv.org/abs/1703.09130}{{\tt arXiv:1703.09130 [hep-ph]}}.

\bibitem{Chen:2017gnu}
B.~Chen and X.~Liu, {\em {New $\Omega_c^0$ baryons discovered by LHCb as the
  members of $1P$ and $2S$ states}\/},
  \href{http://dx.doi.org/10.1103/PhysRevD.96.094015}{Phys. Rev. {\bf D96}
  (2017) no.~9, 094015},
\href{http://arxiv.org/abs/1704.02583}{{\tt arXiv:1704.02583 [hep-ph]}}.

\bibitem{Cheng:2017ove}
H.-Y. Cheng and C.-W. Chiang, {\em {Quantum numbers of $\Omega_c$ states and
  other charmed baryons}\/},
  \href{http://dx.doi.org/10.1103/PhysRevD.95.094018}{Phys. Rev. {\bf D95}
  (2017) no.~9, 094018},
\href{http://arxiv.org/abs/1704.00396}{{\tt arXiv:1704.00396 [hep-ph]}}.

\bibitem{Wang:2017vnc}
W.~Wang and R.-L. Zhu, {\em {Interpretation of the newly observed $\Omega_c^0$
  resonances}\/},  \href{http://dx.doi.org/10.1103/PhysRevD.96.014024}{Phys.
  Rev. {\bf D96} (2017) no.~1, 014024},
\href{http://arxiv.org/abs/1704.00179}{{\tt arXiv:1704.00179 [hep-ph]}}.

\bibitem{Yang:2017rpg}
G.~Yang and J.~Ping, {\em {Dynamical study of $\Omega_c^0$ in the chiral quark
  model}\/},  \href{http://dx.doi.org/10.1103/PhysRevD.97.034023}{Phys. Rev.
  {\bf D97} (2018) no.~3, 034023},
\href{http://arxiv.org/abs/1703.08845}{{\tt arXiv:1703.08845 [hep-ph]}}.

\bibitem{Ali:2017wsf}
A.~Ali, L.~Maiani, A.~V. Borisov, I.~Ahmed, M.~Jamil~Aslam, A.~{\relax Ya}.
  Parkhomenko, A.~D. Polosa, and A.~Rehman, {\em {A new look at the Y
  tetraquarks and $\Omega _c$ baryons in the diquark model}\/},
  \href{http://dx.doi.org/10.1140/epjc/s10052-017-5501-6}{Eur. Phys. J. {\bf
  C78} (2018) no.~1, 29},
\href{http://arxiv.org/abs/1708.04650}{{\tt arXiv:1708.04650 [hep-ph]}}.

\bibitem{Kim:2017jpx}
H.-C. Kim, M.~V. Polyakov, and M.~Prasza{\l}owicz, {\em {Possibility of the
  existence of charmed exotica}\/},
  \href{http://dx.doi.org/10.1103/PhysRevD.96.039902,
  10.1103/PhysRevD.96.014009}{Phys. Rev. {\bf D96} (2017) no.~1, 014009},
  \href{http://arxiv.org/abs/1704.04082}{{\tt arXiv:1704.04082 [hep-ph]}}.
[Addendum: Phys. Rev.D96,no.3,039902(2017)].

\bibitem{Nieves:2017jjx}
J.~Nieves, R.~Pavao, and L.~Tolos, {\em {$\Omega _c$ excited states within a
  $\mathrm{SU(6)}_{\mathrm{lsf}}\times $ HQSS model}\/},
  \href{http://dx.doi.org/10.1140/epjc/s10052-018-5597-3}{Eur. Phys. J. {\bf
  C78} (2018) no.~2, 114},
\href{http://arxiv.org/abs/1712.00327}{{\tt arXiv:1712.00327 [hep-ph]}}.

\bibitem{Montana:2017kjw}
G.~Monta\~na, A.~Feijoo, and A.~Ramos, {\em {A meson-baryon molecular
  interpretation for some $\Omega_{c}$ excited states}\/},
  \href{http://dx.doi.org/10.1140/epja/i2018-12498-1}{Eur. Phys. J. {\bf A54}
  (2018) no.~4, 64},
\href{http://arxiv.org/abs/1709.08737}{{\tt arXiv:1709.08737 [hep-ph]}}.

\bibitem{Debastiani:2017ewu}
V.~R. Debastiani, J.~M. Dias, W.~H. Liang, and E.~Oset, {\em {Molecular
  $\Omega_c$ states generated from coupled meson-baryon channels}\/},
  \href{http://dx.doi.org/10.1103/PhysRevD.97.094035}{Phys. Rev. {\bf D97}
  (2018) no.~9, 094035},
\href{http://arxiv.org/abs/1710.04231}{{\tt arXiv:1710.04231 [hep-ph]}}.

\bibitem{GarciaRecio:2008dp}
C.~Garcia-Recio, V.~K. Magas, T.~Mizutani, J.~Nieves, A.~Ramos, L.~L. Salcedo,
  and L.~Tolos, {\em {The s-wave charmed baryon resonances from a
  coupled-channel approach with heavy quark symmetry}\/},
  \href{http://dx.doi.org/10.1103/PhysRevD.79.054004}{Phys. Rev. {\bf D79}
  (2009)  054004},
\href{http://arxiv.org/abs/0807.2969}{{\tt arXiv:0807.2969 [hep-ph]}}.

\bibitem{Romanets:2012hm}
O.~Romanets, L.~Tolos, C.~Garcia-Recio, J.~Nieves, L.~L. Salcedo, and R.~G.~E.
  Timmermans, {\em {Charmed and strange baryon resonances with heavy-quark spin
  symmetry}\/},  \href{http://dx.doi.org/10.1103/PhysRevD.85.114032}{Phys. Rev.
  {\bf D85} (2012)  114032},
\href{http://arxiv.org/abs/1202.2239}{{\tt arXiv:1202.2239 [hep-ph]}}.

\bibitem{Liang:2014kra}
W.~H. Liang, T.~Uchino, C.~W. Xiao, and E.~Oset, {\em {Baryon states with open
  charm in the extended local hidden gauge approach}\/},
  \href{http://dx.doi.org/10.1140/epja/i2015-15016-1}{Eur. Phys. J. {\bf A51}
  (2015) no.~2, 16},
\href{http://arxiv.org/abs/1402.5293}{{\tt arXiv:1402.5293 [hep-ph]}}.

\bibitem{Liang:2014eba}
W.~H. Liang, C.~W. Xiao, and E.~Oset, {\em {Baryon states with open beauty in
  the extended local hidden gauge approach}\/},
  \href{http://dx.doi.org/10.1103/PhysRevD.89.054023}{Phys. Rev. {\bf D89}
  (2014) no.~5, 054023},
\href{http://arxiv.org/abs/1401.1441}{{\tt arXiv:1401.1441 [hep-ph]}}.

\bibitem{GarciaRecio:2012db}
C.~Garcia-Recio, J.~Nieves, O.~Romanets, L.~L. Salcedo, and L.~Tolos, {\em {Odd
  parity bottom-flavored baryon resonances}\/},
  \href{http://dx.doi.org/10.1103/PhysRevD.87.034032}{Phys. Rev. {\bf D87}
  (2013) no.~3, 034032},
\href{http://arxiv.org/abs/1210.4755}{{\tt arXiv:1210.4755 [hep-ph]}}.

\bibitem{Yoshida:2015tia}
T.~Yoshida, E.~Hiyama, A.~Hosaka, M.~Oka, and K.~Sadato, {\em {Spectrum of
  heavy baryons in the quark model}\/},
  \href{http://dx.doi.org/10.1103/PhysRevD.92.114029}{Phys. Rev. {\bf D92}
  (2015) no.~11, 114029},
\href{http://arxiv.org/abs/1510.01067}{{\tt arXiv:1510.01067 [hep-ph]}}.

\bibitem{Nagahiro:2016nsx}
H.~Nagahiro, S.~Yasui, A.~Hosaka, M.~Oka, and H.~Noumi, {\em {Structure of
  charmed baryons studied by pionic decays}\/},
  \href{http://dx.doi.org/10.1103/PhysRevD.95.014023}{Phys. Rev. {\bf D95}
  (2017) no.~1, 014023},
\href{http://arxiv.org/abs/1609.01085}{{\tt arXiv:1609.01085 [hep-ph]}}.

\bibitem{Liang:2016exm}
W.-H. Liang, E.~Oset, and Z.-S. Xie, {\em {Semileptonic $\Lambda_b \to \bar
  \nu_l l \Lambda_c(2595)$ and $\Lambda_b \to \bar \nu_l l \Lambda_c(2625)$
  decays in the molecular picture of $\Lambda_c(2595)$ and
  $\Lambda_c(2625)$}\/},
  \href{http://dx.doi.org/10.1103/PhysRevD.95.014015}{Phys. Rev. {\bf D95}
  (2017) no.~1, 014015},
\href{http://arxiv.org/abs/1611.07334}{{\tt arXiv:1611.07334 [hep-ph]}}.

\bibitem{Liang:2017ejq}
W.-H. Liang, J.~M. Dias, V.~R. Debastiani, and E.~Oset, {\em {Molecular
  $\Omega_b$ states}\/},
  \href{http://dx.doi.org/10.1016/j.nuclphysb.2018.03.008}{Nucl. Phys. {\bf
  B930} (2018)  524--532},
\href{http://arxiv.org/abs/1711.10623}{{\tt arXiv:1711.10623 [hep-ph]}}.

\bibitem{Aaij:2018yqz}
{LHCb Collaboration}, R.~Aaij et al., {\em {Observation of a new $\Xi_b^-$
  resonance}\/},  \href{http://dx.doi.org/10.1103/PhysRevLett.121.072002}{Phys.
  Rev. Lett. {\bf 121} (2018) no.~7, 072002},
\href{http://arxiv.org/abs/1805.09418}{{\tt arXiv:1805.09418 [hep-ex]}}.

\bibitem{Lu:2014ina}
J.-X. Lu, Y.~Zhou, H.-X. Chen, J.-J. Xie, and L.-S. Geng, {\em {Dynamically
  generated $J^P=1/2^-(3/2^-)$ singly charmed and bottom heavy baryons}\/},
  \href{http://dx.doi.org/10.1103/PhysRevD.92.014036}{Phys. Rev. {\bf D92}
  (2015) no.~1, 014036},
\href{http://arxiv.org/abs/1409.3133}{{\tt arXiv:1409.3133 [hep-ph]}}.

\bibitem{Huang:2018bed}
Y.~Huang, C.-j. Xiao, L.-S. Geng, and J.~He, {\em {Strong decays of the
  $\Xi_b(6227)$ as a $\Sigma_b\bar{K}$ molecule}\/},
\href{http://arxiv.org/abs/1811.10769}{{\tt arXiv:1811.10769 [hep-ph]}}.

\bibitem{Yu:2018yxl}
Q.~X. Yu, R.~Pavao, V.~R. Debastiani, and E.~Oset, {\em {Description of the
  $\Xi_c$ and $\Xi_b$ states as molecular states}\/},
\href{http://arxiv.org/abs/1811.11738}{{\tt arXiv:1811.11738 [hep-ph]}}.

\bibitem{Chen:2016spr}
H.-X. Chen, W.~Chen, X.~Liu, Y.-R. Liu, and S.-L. Zhu, {\em {A review of the
  open charm and open bottom systems}\/},
  \href{http://dx.doi.org/10.1088/1361-6633/aa6420}{Rept. Prog. Phys. {\bf 80}
  (2017) no.~7, 076201},
\href{http://arxiv.org/abs/1609.08928}{{\tt arXiv:1609.08928 [hep-ph]}}.

\bibitem{Olsen:2015zcy}
S.~L. Olsen, {\em {XYZ Meson Spectroscopy}\/},  in {\em {Proceedings, 53rd
  International Winter Meeting on Nuclear Physics (Bormio 2015): Bormio, Italy,
  January 26-30, 2015}}.
\newblock 2015.
\newblock \href{http://arxiv.org/abs/1511.01589}{{\tt arXiv:1511.01589
  [hep-ex]}}.
\newblock
\url{https://inspirehep.net/record/1402960/files/arXiv:1511.01589.pdf}.
\newblock

\bibitem{Aaij:2018bla}
{LHCb Collaboration}, R.~Aaij et al., {\em {Evidence for an $\eta_c(1S) \pi^-$
  resonance in $B^0 \to \eta_c(1S) K^+\pi^-$ decays}\/},  Submitted to: Eur.
  Phys. J. (2018)  ,
\href{http://arxiv.org/abs/1809.07416}{{\tt arXiv:1809.07416 [hep-ex]}}.

\bibitem{Tornqvist:2004qy}
N.~A. T{\"o}rnqvist, {\em {Isospin breaking of the narrow charmonium state of
  Belle at 3872-MeV as a deuson}\/},
  \href{http://dx.doi.org/10.1016/j.physletb.2004.03.077}{Phys. Lett. {\bf
  B590} (2004)  209--215},
\href{http://arxiv.org/abs/hep-ph/0402237}{{\tt arXiv:hep-ph/0402237
  [hep-ph]}}.

\bibitem{Swanson:2003tb}
E.~S. Swanson, {\em {Short range structure in the X(3872)}\/},
  \href{http://dx.doi.org/10.1016/j.physletb.2004.03.033}{Phys. Lett. {\bf
  B588} (2004)  189--195},
\href{http://arxiv.org/abs/hep-ph/0311229}{{\tt arXiv:hep-ph/0311229
  [hep-ph]}}.

\bibitem{AlFiky:2005jd}
M.~T. AlFiky, F.~Gabbiani, and A.~A. Petrov, {\em {X(3872): Hadronic molecules
  in effective field theory}\/},
  \href{http://dx.doi.org/10.1016/j.physletb.2006.07.069}{Phys. Lett. {\bf
  B640} (2006)  238--245},
\href{http://arxiv.org/abs/hep-ph/0506141}{{\tt arXiv:hep-ph/0506141
  [hep-ph]}}.

\bibitem{Gamermann:2009fv}
D.~Gamermann and E.~Oset, {\em {Isospin breaking effects in the X(3872)
  resonance}\/},  \href{http://dx.doi.org/10.1103/PhysRevD.80.014003}{Phys.
  Rev. {\bf D80} (2009)  014003},
\href{http://arxiv.org/abs/0905.0402}{{\tt arXiv:0905.0402 [hep-ph]}}.

\bibitem{Hanhart:2011tn}
C.~Hanhart, {\relax Yu}.~S. Kalashnikova, A.~E. Kudryavtsev, and A.~V.
  Nefediev, {\em {Remarks on the quantum numbers of X(3872) from the invariant
  mass distributions of the rho J/psi and omega J/psi final states}\/},
  \href{http://dx.doi.org/10.1103/PhysRevD.85.011501}{Phys. Rev. {\bf D85}
  (2012)  011501},
\href{http://arxiv.org/abs/1111.6241}{{\tt arXiv:1111.6241 [hep-ph]}}.

\bibitem{Li:2012cs}
N.~Li and S.-L. Zhu, {\em {Isospin breaking, Coupled-channel effects and
  Diagnosis of X(3872)}\/},
  \href{http://dx.doi.org/10.1103/PhysRevD.86.074022}{Phys. Rev. {\bf D86}
  (2012)  074022},
\href{http://arxiv.org/abs/1207.3954}{{\tt arXiv:1207.3954 [hep-ph]}}.

\bibitem{Hanhart:2007yq}
C.~Hanhart, {\relax Yu}.~S. Kalashnikova, A.~E. Kudryavtsev, and A.~V.
  Nefediev, {\em {Reconciling the X(3872) with the near-threshold enhancement
  in the D0 anti-D*0 final state}\/},
  \href{http://dx.doi.org/10.1103/PhysRevD.76.034007}{Phys. Rev. {\bf D76}
  (2007)  034007},
\href{http://arxiv.org/abs/0704.0605}{{\tt arXiv:0704.0605 [hep-ph]}}.

\bibitem{Kalashnikova:2009gt}
{\relax Yu}.~S. Kalashnikova and A.~V. Nefediev, {\em {Nature of X(3872) from
  data}\/},  \href{http://dx.doi.org/10.1103/PhysRevD.80.074004}{Phys. Rev.
  {\bf D80} (2009)  074004},
\href{http://arxiv.org/abs/0907.4901}{{\tt arXiv:0907.4901 [hep-ph]}}.

\bibitem{Artoisenet:2010va}
P.~Artoisenet, E.~Braaten, and D.~Kang, {\em {Using Line Shapes to Discriminate
  between Binding Mechanisms for the X(3872)}\/},
  \href{http://dx.doi.org/10.1103/PhysRevD.82.014013}{Phys. Rev. {\bf D82}
  (2010)  014013},
\href{http://arxiv.org/abs/1005.2167}{{\tt arXiv:1005.2167 [hep-ph]}}.

\bibitem{Baru:2011rs}
V.~Baru, A.~A. Filin, C.~Hanhart, Y.~S. Kalashnikova, A.~E. Kudryavtsev, and
  A.~V. Nefediev, {\em {Three-body $D\bar{D}\pi$ dynamics for the X(3872)}\/},
  \href{http://dx.doi.org/10.1103/PhysRevD.84.074029}{Phys. Rev. {\bf D84}
  (2011)  074029},
\href{http://arxiv.org/abs/1108.5644}{{\tt arXiv:1108.5644 [hep-ph]}}.

\bibitem{Kang:2016jxw}
X.-W. Kang and J.~A. Oller, {\em {Different pole structures in line shapes of
  the $X(3872)$}\/},
  \href{http://dx.doi.org/10.1140/epjc/s10052-017-4961-z}{Eur. Phys. J. {\bf
  C77} (2017) no.~6, 399},
\href{http://arxiv.org/abs/1612.08420}{{\tt arXiv:1612.08420 [hep-ph]}}.

\bibitem{Albaladejo:2015dsa}
M.~Albaladejo, F.~K. Guo, C.~Hidalgo-Duque, J.~Nieves, and M.~P. Valderrama,
  {\em {Decay widths of the spin-2 partners of the X(3872)}\/},
  \href{http://dx.doi.org/10.1140/epjc/s10052-015-3753-6}{Eur. Phys. J. {\bf
  C75} (2015) no.~11, 547},
\href{http://arxiv.org/abs/1504.00861}{{\tt arXiv:1504.00861 [hep-ph]}}.

\bibitem{Baru:2016iwj}
V.~Baru, E.~Epelbaum, A.~A. Filin, C.~Hanhart, U.-G. Mei{\ss}ner, and A.~V.
  Nefediev, {\em {Heavy-quark spin symmetry partners of the X (3872)
  revisited}\/},  \href{http://dx.doi.org/10.1016/j.physletb.2016.10.008}{Phys.
  Lett. {\bf B763} (2016)  20--28},
\href{http://arxiv.org/abs/1605.09649}{{\tt arXiv:1605.09649 [hep-ph]}}.

\bibitem{Cincioglu:2016fkm}
E.~Cincioglu, J.~Nieves, A.~Ozpineci, and A.~U. Yilmazer, {\em {Quarkonium
  Contribution to Meson Molecules}\/},
  \href{http://dx.doi.org/10.1140/epjc/s10052-016-4413-1}{Eur. Phys. J. {\bf
  C76} (2016) no.~10, 576},
\href{http://arxiv.org/abs/1606.03239}{{\tt arXiv:1606.03239 [hep-ph]}}.

\bibitem{Ortega:2017qmg}
P.~G. Ortega, J.~Segovia, D.~R. Entem, and F.~Fern\'andez, {\em {Charmonium
  resonances in the 3.9 GeV/$c^2$ energy region and the $X(3915)/X(3930)$
  puzzle}\/},  \href{http://dx.doi.org/10.1016/j.physletb.2018.01.005}{Phys.
  Lett. {\bf B778} (2018)  1--5},
\href{http://arxiv.org/abs/1706.02639}{{\tt arXiv:1706.02639 [hep-ph]}}.

\bibitem{Guo:2012tv}
F.-K. Guo and U.-G. Mei{\ss}ner, {\em {Where is the $\chi_{c0}(2P)$?}\/},
  \href{http://dx.doi.org/10.1103/PhysRevD.86.091501}{Phys. Rev. {\bf D86}
  (2012)  091501},
\href{http://arxiv.org/abs/1208.1134}{{\tt arXiv:1208.1134 [hep-ph]}}.

\bibitem{Chilikin:2017evr}
{Belle Collaboration}, K.~Chilikin et al., {\em {Observation of an alternative
  $\chi_{c0}(2P)$ candidate in $e^+ e^- \rightarrow J/\psi D \bar{D}$}\/},
  \href{http://dx.doi.org/10.1103/PhysRevD.95.112003}{Phys. Rev. {\bf D95}
  (2017)  112003},
\href{http://arxiv.org/abs/1704.01872}{{\tt arXiv:1704.01872 [hep-ex]}}.

\bibitem{Zhou:2017dwj}
Z.-Y. Zhou and Z.~Xiao, {\em {Understanding $X(3862)$, $X(3872)$, and $X(3930)$
  in a Friedrichs-model-like scheme}\/},
  \href{http://dx.doi.org/10.1103/PhysRevD.96.099905,
  10.1103/PhysRevD.96.054031}{Phys. Rev. {\bf D96} (2017) no.~5, 054031},
  \href{http://arxiv.org/abs/1704.04438}{{\tt arXiv:1704.04438 [hep-ph]}}.
[Erratum: Phys. Rev.D96,no.9,099905(2017)].

\bibitem{Lebed:2017yme}
R.~F. Lebed and E.~S. Swanson, {\em {Quarkonium $h$ States As Arbiters of
  Exoticity}\/},  \href{http://dx.doi.org/10.1103/PhysRevD.96.056015}{Phys.
  Rev. {\bf D96} (2017) no.~5, 056015},
\href{http://arxiv.org/abs/1705.03140}{{\tt arXiv:1705.03140 [hep-ph]}}.

\bibitem{Aghasyan:2017utv}
{COMPASS Collaboration}, M.~Aghasyan et al., {\em {Search for muoproduction of
  $X (3872)$ at COMPASS and indication of a new state
  $\widetilde{X}(3872)$}\/},
  \href{http://dx.doi.org/10.1016/j.physletb.2018.07.008}{Phys. Lett. {\bf
  B783} (2018)  334--340},
\href{http://arxiv.org/abs/1707.01796}{{\tt arXiv:1707.01796 [hep-ex]}}.

\bibitem{Ortega:2009hj}
P.~G. Ortega, J.~Segovia, D.~R. Entem, and F.~Fernandez, {\em {Coupled channel
  approach to the structure of the X(3872)}\/},
  \href{http://dx.doi.org/10.1103/PhysRevD.81.054023}{Phys. Rev. {\bf D81}
  (2010)  054023},
\href{http://arxiv.org/abs/0907.3997}{{\tt arXiv:0907.3997 [hep-ph]}}.

\bibitem{Danilkin:2009hr}
I.~V. Danilkin and {\relax Yu}.~A. Simonov, {\em {Channel coupling in heavy
  quarkonia: Energy levels, mixing, widths and new states}\/},
  \href{http://dx.doi.org/10.1103/PhysRevD.81.074027}{Phys. Rev. {\bf D81}
  (2010)  074027},
\href{http://arxiv.org/abs/0907.1088}{{\tt arXiv:0907.1088 [hep-ph]}}.

\bibitem{Takizawa:2012hy}
M.~Takizawa and S.~Takeuchi, {\em {X(3872) as a hybrid state of charmonium and
  the hadronic molecule}\/},  \href{http://dx.doi.org/10.1093/ptep/ptt063}{PTEP
  {\bf 2013} (2013)  093D01},
\href{http://arxiv.org/abs/1206.4877}{{\tt arXiv:1206.4877 [hep-ph]}}.

\bibitem{Meng:2013gga}
C.~Meng, H.~Han, and K.-T. Chao, {\em {$X(3872)$ and its production at hadron
  colliders}\/},  \href{http://dx.doi.org/10.1103/PhysRevD.96.074014}{Phys.
  Rev. {\bf D96} (2013) no.~7, 074014},
\href{http://arxiv.org/abs/1304.6710}{{\tt arXiv:1304.6710 [hep-ph]}}.

\bibitem{Takeuchi:2014rsa}
S.~Takeuchi, K.~Shimizu, and M.~Takizawa, {\em {On the origin of the narrow
  peak and the isospin symmetry breaking of the $X$(3872)}\/},
  \href{http://dx.doi.org/10.1093/ptep/ptu160, 10.1093/ptep/ptv104}{PTEP {\bf
  2014} (2014) no.~12, 123D01}, \href{http://arxiv.org/abs/1408.0973}{{\tt
  arXiv:1408.0973 [hep-ph]}}.
[Erratum: PTEP2015,no.7,079203(2015)].

\bibitem{Zhou:2017txt}
Z.-Y. Zhou and Z.~Xiao, {\em {Comprehending Isospin breaking effects of
  $X(3872)$ in a Friedrichs-model-like scheme}\/},
  \href{http://dx.doi.org/10.1103/PhysRevD.97.034011}{Phys. Rev. {\bf D97}
  (2018) no.~3, 034011},
\href{http://arxiv.org/abs/1711.01930}{{\tt arXiv:1711.01930 [hep-ph]}}.

\bibitem{Zhou:2015uva}
Z.-Y. Zhou, Z.~Xiao, and H.-Q. Zhou, {\em {Could the $X(3915)$ and the
  $X(3930)$ Be the Same Tensor State?}\/},
  \href{http://dx.doi.org/10.1103/PhysRevLett.115.022001}{Phys. Rev. Lett. {\bf
  115} (2015) no.~2, 022001},
\href{http://arxiv.org/abs/1501.00879}{{\tt arXiv:1501.00879 [hep-ph]}}.

\bibitem{Guo:2013sya}
F.-K. Guo, C.~Hidalgo-Duque, J.~Nieves, and M.~P. Valderrama, {\em
  {Consequences of Heavy Quark Symmetries for Hadronic Molecules}\/},
  \href{http://dx.doi.org/10.1103/PhysRevD.88.054007}{Phys. Rev. {\bf D88}
  (2013)  054007},
\href{http://arxiv.org/abs/1303.6608}{{\tt arXiv:1303.6608 [hep-ph]}}.

\bibitem{Guo:2014sca}
F.-K. Guo, U.-G. Mei{\ss}ner, W.~Wang, and Z.~Yang, {\em {Production of the
  bottom analogs and the spin partner of the $X$(3872) at hadron colliders}\/},
   \href{http://dx.doi.org/10.1140/epjc/s10052-014-3063-4}{Eur. Phys. J. {\bf
  C74} (2014) no.~9, 3063},
\href{http://arxiv.org/abs/1402.6236}{{\tt arXiv:1402.6236 [hep-ph]}}.

\bibitem{Karliner:2014lta}
M.~Karliner and J.~L. Rosner, {\em {$X(3872)$, $X_b$, and the $\chi_{b1}(3P)$
  state}\/},  \href{http://dx.doi.org/10.1103/PhysRevD.91.014014}{Phys. Rev.
  {\bf D91} (2015) no.~1, 014014},
\href{http://arxiv.org/abs/1410.7729}{{\tt arXiv:1410.7729 [hep-ph]}}.

\bibitem{Chatrchyan:2013mea}
{CMS Collaboration}, S.~Chatrchyan et al., {\em {Search for a new bottomonium
  state decaying to $\Upsilon(1S)\pi^+\pi^-$ in pp collisions at $\sqrt{s}$ = 8
  TeV}\/},  \href{http://dx.doi.org/10.1016/j.physletb.2013.10.016}{Phys. Lett.
  {\bf B727} (2013)  57--76},
\href{http://arxiv.org/abs/1309.0250}{{\tt arXiv:1309.0250 [hep-ex]}}.

\bibitem{Aad:2014ama}
{ATLAS Collaboration}, G.~Aad et al., {\em {Search for the $X_b$ and other
  hidden-beauty states in the $\pi^+ \pi^- \Upsilon(1 \rm S)$ channel at
  ATLAS}\/},  \href{http://dx.doi.org/10.1016/j.physletb.2014.11.055}{Phys.
  Lett. {\bf B740} (2015)  199--217},
\href{http://arxiv.org/abs/1410.4409}{{\tt arXiv:1410.4409 [hep-ex]}}.

\bibitem{Guo:2008ns}
F.-K. Guo, C.~Hanhart, and U.-G. Mei{\ss}ner, {\em {Mass splittings within
  heavy baryon isospin multiplets in chiral perturbation theory}\/},
  \href{http://dx.doi.org/10.1088/1126-6708/2008/09/136}{JHEP {\bf 09} (2008)
  136},
\href{http://arxiv.org/abs/0809.2359}{{\tt arXiv:0809.2359 [hep-ph]}}.

\bibitem{He:2014sqj}
{Belle Collaboration}, X.~H. He et al., {\em {Observation of $e^+e^- \to \pi^+
  \pi^- \pi^0 \chi_{bJ}$ and Search for $X_b \to \omega \Upsilon(1S)$ at
  $\sqrt{s}=10.867$ GeV}\/},
  \href{http://dx.doi.org/10.1103/PhysRevLett.113.142001}{Phys. Rev. Lett. {\bf
  113} (2014) no.~14, 142001},
\href{http://arxiv.org/abs/1408.0504}{{\tt arXiv:1408.0504 [hep-ex]}}.

\bibitem{Aubert:2005rm}
{BaBar Collaboration}, B.~Aubert et al., {\em {Observation of a broad structure
  in the $\pi^+ \pi^- J/\psi$ mass spectrum around 4.26-GeV/c$^2$}\/},
  \href{http://dx.doi.org/10.1103/PhysRevLett.95.142001}{Phys. Rev. Lett. {\bf
  95} (2005)  142001},
\href{http://arxiv.org/abs/hep-ex/0506081}{{\tt arXiv:hep-ex/0506081
  [hep-ex]}}.

\bibitem{Liu:2013dau}
{Belle Collaboration}, Z.~Q. Liu et al., {\em {Study of $e^+e^- \to \pi^+ \pi^-
  J/\psi$ and Observation of a Charged Charmoniumlike State at Belle}\/},
  \href{http://dx.doi.org/10.1103/PhysRevLett.110.252002}{Phys. Rev. Lett. {\bf
  110} (2013)  252002},
\href{http://arxiv.org/abs/1304.0121}{{\tt arXiv:1304.0121 [hep-ex]}}.

\bibitem{Ablikim:2013dyn}
{BESIII Collaboration}, M.~Ablikim et al., {\em {Observation of $e^+e^- \to
  \gamma X$(3872) at BESIII}\/},
  \href{http://dx.doi.org/10.1103/PhysRevLett.112.092001}{Phys. Rev. Lett. {\bf
  112} (2014) no.~9, 092001},
\href{http://arxiv.org/abs/1310.4101}{{\tt arXiv:1310.4101 [hep-ex]}}.

\bibitem{Wang:2014hta}
{Belle Collaboration}, X.~L. Wang et al., {\em {Measurement of $e^+e^- \to
  \pi^+\pi^-\psi(2S)$ via Initial State Radiation at Belle}\/},
  \href{http://dx.doi.org/10.1103/PhysRevD.91.112007}{Phys. Rev. {\bf D91}
  (2015)  112007},
\href{http://arxiv.org/abs/1410.7641}{{\tt arXiv:1410.7641 [hep-ex]}}.

\bibitem{Lees:2012pv}
{BaBar Collaboration}, J.~P. Lees et al., {\em {Study of the reaction
  $e^{+}e^{-}\to \psi(2S)\pi^{-}\pi^{-}$ via initial-state radiation at
  BaBar}\/},  \href{http://dx.doi.org/10.1103/PhysRevD.89.111103}{Phys. Rev.
  {\bf D89} (2014) no.~11, 111103},
\href{http://arxiv.org/abs/1211.6271}{{\tt arXiv:1211.6271 [hep-ex]}}.

\bibitem{Wang:2018ntv}
Z.-G. Wang, {\em {Lowest vector tetraquark states: $Y(4260/4220)$ or
  $Z_c(4100)$}\/},
  \href{http://dx.doi.org/10.1140/epjc/s10052-018-6417-5}{Eur. Phys. J. {\bf
  C78} (2018) no.~11, 933},
\href{http://arxiv.org/abs/1809.10299}{{\tt arXiv:1809.10299 [hep-ph]}}.

\bibitem{Cleven:2015era}
M.~Cleven, F.-K. Guo, C.~Hanhart, Q.~Wang, and Q.~Zhao, {\em {Employing spin
  symmetry to disentangle different models for the XYZ states}\/},
  \href{http://dx.doi.org/10.1103/PhysRevD.92.014005}{Phys. Rev. {\bf D92}
  (2015) no.~1, 014005},
\href{http://arxiv.org/abs/1505.01771}{{\tt arXiv:1505.01771 [hep-ph]}}.

\bibitem{Wang:2013cya}
Q.~Wang, C.~Hanhart, and Q.~Zhao, {\em {Decoding the riddle of $Y(4260)$ and
  $Z_c(3900)$}\/},
  \href{http://dx.doi.org/10.1103/PhysRevLett.111.132003}{Phys. Rev. Lett. {\bf
  111} (2013) no.~13, 132003},
\href{http://arxiv.org/abs/1303.6355}{{\tt arXiv:1303.6355 [hep-ph]}}.

\bibitem{Guo:2008zg}
F.-K. Guo, C.~Hanhart, and U.-G. Mei{\ss}ner, {\em {Evidence that the $Y(4660)$
  is a $f_0(980)\psi'$ bound state}\/},
  \href{http://dx.doi.org/10.1016/j.physletb.2008.05.057}{Phys. Lett. {\bf
  B665} (2008)  26--29},
\href{http://arxiv.org/abs/0803.1392}{{\tt arXiv:0803.1392 [hep-ph]}}.

\bibitem{Lu:2017yhl}
Y.~Lu, M.~N. Anwar, and B.-S. Zou, {\em {$X(4260)$ Revisited: A Coupled Channel
  Perspective}\/},  \href{http://dx.doi.org/10.1103/PhysRevD.96.114022}{Phys.
  Rev. {\bf D96} (2017) no.~11, 114022},
\href{http://arxiv.org/abs/1705.00449}{{\tt arXiv:1705.00449 [hep-ph]}}.

\bibitem{Zhu:2005hp}
S.-L. Zhu, {\em {The Possible interpretations of Y(4260)}\/},
  \href{http://dx.doi.org/10.1016/j.physletb.2005.08.068}{Phys. Lett. {\bf
  B625} (2005)  212},
\href{http://arxiv.org/abs/hep-ph/0507025}{{\tt arXiv:hep-ph/0507025
  [hep-ph]}}.

\bibitem{Kalashnikova:2008qr}
Y.~S. Kalashnikova and A.~V. Nefediev, {\em {Spectra and decays of hybrid
  charmonia}\/},  \href{http://dx.doi.org/10.1103/PhysRevD.77.054025}{Phys.
  Rev. {\bf D77} (2008)  054025},
\href{http://arxiv.org/abs/0801.2036}{{\tt arXiv:0801.2036 [hep-ph]}}.

\bibitem{Berwein:2015vca}
M.~Berwein, N.~Brambilla, J.~Tarrus~Castella, and A.~Vairo, {\em {Quarkonium
  Hybrids with Nonrelativistic Effective Field Theories}\/},
  \href{http://dx.doi.org/10.1103/PhysRevD.92.114019}{Phys. Rev. {\bf D92}
  (2015) no.~11, 114019},
\href{http://arxiv.org/abs/1510.04299}{{\tt arXiv:1510.04299 [hep-ph]}}.

\bibitem{Chen:2016ejo}
Y.~Chen, W.-F. Chiu, M.~Gong, L.-C. Gui, and Z.~Liu, {\em {Exotic vector
  charmonium and its leptonic decay width}\/},
  \href{http://dx.doi.org/10.1088/1674-1137/40/8/081002}{Chin. Phys. {\bf C40}
  (2016) no.~8, 081002},
\href{http://arxiv.org/abs/1604.03401}{{\tt arXiv:1604.03401 [hep-lat]}}.

\bibitem{Oncala:2017hop}
R.~Oncala and J.~Soto, {\em {Heavy Quarkonium Hybrids: Spectrum, Decay and
  Mixing}\/},  \href{http://dx.doi.org/10.1103/PhysRevD.96.014004}{Phys. Rev.
  {\bf D96} (2017) no.~1, 014004},
\href{http://arxiv.org/abs/1702.03900}{{\tt arXiv:1702.03900 [hep-ph]}}.

\bibitem{Li:2013ssa}
X.~Li and M.~B. Voloshin, {\em {$Y$(4260) and $Y$(4360) as mixed
  hadrocharmonium}\/},  \href{http://dx.doi.org/10.1142/S0217732314500606}{Mod.
  Phys. Lett. {\bf A29} (2014) no.~12, 1450060},
\href{http://arxiv.org/abs/1309.1681}{{\tt arXiv:1309.1681 [hep-ph]}}.

\bibitem{Gao:2017sqa}
X.~Y. Gao, C.~P. Shen, and C.~Z. Yuan, {\em {Resonant parameters of the
  $Y(4220)$}\/},  \href{http://dx.doi.org/10.1103/PhysRevD.95.092007}{Phys.
  Rev. {\bf D95} (2017) no.~9, 092007},
\href{http://arxiv.org/abs/1703.10351}{{\tt arXiv:1703.10351 [hep-ex]}}.

\bibitem{Wurtz:2015mqa}
M.~Wurtz, R.~Lewis, and R.~M. Woloshyn, {\em {Free-form smearing for
  bottomonium and B meson spectroscopy}\/},
  \href{http://dx.doi.org/10.1103/PhysRevD.92.054504}{Phys. Rev. {\bf D92}
  (2015) no.~5, 054504},
\href{http://arxiv.org/abs/1505.04410}{{\tt arXiv:1505.04410 [hep-lat]}}.

\bibitem{Lang:2015sba}
C.~B. Lang, L.~Leskovec, D.~Mohler, and S.~Prelovsek, {\em {Vector and scalar
  charmonium resonances with lattice QCD}\/},
  \href{http://dx.doi.org/10.1007/JHEP09(2015)089}{JHEP {\bf 09} (2015)  089},
\href{http://arxiv.org/abs/1503.05363}{{\tt arXiv:1503.05363 [hep-lat]}}.

\bibitem{Prelovsek:2013cra}
S.~Prelovsek and L.~Leskovec, {\em {Evidence for X(3872) from $DD^*$ scattering
  on the lattice}\/},
  \href{http://dx.doi.org/10.1103/PhysRevLett.111.192001}{Phys.Rev.Lett. {\bf
  111} (2013)  192001},
\href{http://arxiv.org/abs/1307.5172}{{\tt arXiv:1307.5172 [hep-lat]}}.

\bibitem{Padmanath:2015era}
M.~Padmanath, C.~B. Lang, and S.~Prelovsek, {\em {X(3872) and Y(4140) using
  diquark-antidiquark operators with lattice QCD}\/},
  \href{http://dx.doi.org/10.1103/PhysRevD.92.034501}{Phys. Rev. {\bf D92}
  (2015) no.~3, 034501},
\href{http://arxiv.org/abs/1503.03257}{{\tt arXiv:1503.03257 [hep-lat]}}.

\bibitem{Ali:2011ug}
A.~Ali, C.~Hambrock, and W.~Wang, {\em {Tetraquark Interpretation of the
  Charged Bottomonium-like states $Z_b^{+-}(10610)$ and $Z_b^{+-}(10650)$ and
  Implications}\/},  \href{http://dx.doi.org/10.1103/PhysRevD.85.054011}{Phys.
  Rev. {\bf D85} (2012)  054011},
\href{http://arxiv.org/abs/1110.1333}{{\tt arXiv:1110.1333 [hep-ph]}}.

\bibitem{Esposito:2014hsa}
A.~Esposito, A.~L. Guerrieri, and A.~Pilloni, {\em {Probing the nature of
  $Z_c^{(′)}$ states via the $η_cρ$ decay}\/},
  \href{http://dx.doi.org/10.1016/j.physletb.2015.04.057}{Phys. Lett. {\bf
  B746} (2015)  194--201},
\href{http://arxiv.org/abs/1409.3551}{{\tt arXiv:1409.3551 [hep-ph]}}.

\bibitem{Agaev:2016dev}
S.~S. Agaev, K.~Azizi, and H.~Sundu, {\em {Strong $Z_c^{+}(3900)\rightarrow
  J/\psi \pi^{+}; \eta_{c} \rho^{+}$ decays in QCD}\/},
  \href{http://dx.doi.org/10.1103/PhysRevD.93.074002}{Phys. Rev. {\bf D93}
  (2016) no.~7, 074002},
\href{http://arxiv.org/abs/1601.03847}{{\tt arXiv:1601.03847 [hep-ph]}}.

\bibitem{Agaev:2017tzv}
S.~S. Agaev, K.~Azizi, and H.~Sundu, {\em {Treating $Z_c(3900)$ and $Z(4430)$
  as the ground-state and first radially excited tetraquarks}\/},
  \href{http://dx.doi.org/10.1103/PhysRevD.96.034026}{Phys. Rev. {\bf D96}
  (2017) no.~3, 034026},
\href{http://arxiv.org/abs/1706.01216}{{\tt arXiv:1706.01216 [hep-ph]}}.

\bibitem{Xiao:2018kfx}
C.-J. Xiao, D.-Y. Chen, Y.-B. Dong, W.~Zuo, and T.~Matsuki, {\em {Understanding
  the $\eta_c\rho$ decay mode of $Z_c^{(\prime)}$ via final state
  interactions}\/},
\href{http://arxiv.org/abs/1811.04688}{{\tt arXiv:1811.04688 [hep-ph]}}.

\bibitem{Pilloni:2016obd}
{JPAC Collaboration}, A.~Pilloni, C.~Fernandez-Ramirez, A.~Jackura, V.~Mathieu,
  M.~Mikhasenko, J.~Nys, and A.~P. Szczepaniak, {\em {Amplitude analysis and
  the nature of the Z$_c$(3900)}\/},
  \href{http://dx.doi.org/10.1016/j.physletb.2017.06.030}{Phys. Lett. {\bf
  B772} (2017)  200--209},
\href{http://arxiv.org/abs/1612.06490}{{\tt arXiv:1612.06490 [hep-ph]}}.

\bibitem{Albaladejo:2015lob}
M.~Albaladejo, F.-K. Guo, C.~Hidalgo-Duque, and J.~Nieves, {\em {$Z_c(3900)$:
  What has been really seen?}\/},
  \href{http://dx.doi.org/10.1016/j.physletb.2016.02.025}{Phys. Lett. {\bf
  B755} (2016)  337--342},
\href{http://arxiv.org/abs/1512.03638}{{\tt arXiv:1512.03638 [hep-ph]}}.

\bibitem{Albaladejo:2016jsg}
M.~Albaladejo, P.~Fernandez-Soler, and J.~Nieves, {\em {$Z_c(3900)$:
  Confronting theory and lattice simulations}\/},
  \href{http://dx.doi.org/10.1140/epjc/s10052-016-4427-8}{Eur. Phys. J. {\bf
  C76} (2016) no.~10, 573},
\href{http://arxiv.org/abs/1606.03008}{{\tt arXiv:1606.03008 [hep-ph]}}.

\bibitem{Prelovsek:2014swa}
S.~Prelovsek, C.~B. Lang, L.~Leskovec, and D.~Mohler, {\em {Study of the
  $Z\_c^+$ channel using lattice QCD}\/},
  \href{http://dx.doi.org/10.1103/PhysRevD.91.014504}{Phys. Rev. {\bf D91}
  (2015) no.~1, 014504},
\href{http://arxiv.org/abs/1405.7623}{{\tt arXiv:1405.7623 [hep-lat]}}.

\bibitem{Guo:2016bjq}
F.~K. Guo, C.~Hanhart, {\relax Yu}.~S. Kalashnikova, P.~Matuschek, R.~V. Mizuk,
  A.~V. Nefediev, Q.~Wang, and J.~L. Wynen, {\em {Interplay of quark and meson
  degrees of freedom in near-threshold states: A practical parametrization for
  line shapes}\/},  \href{http://dx.doi.org/10.1103/PhysRevD.93.074031}{Phys.
  Rev. {\bf D93} (2016) no.~7, 074031},
\href{http://arxiv.org/abs/1602.00940}{{\tt arXiv:1602.00940 [hep-ph]}}.

\bibitem{Wang:2018jlv}
Q.~Wang, V.~Baru, A.~A. Filin, C.~Hanhart, A.~V. Nefediev, and J.~L. Wynen,
  {\em {The line shapes of the $Z_b(10610)$ and $Z_b(10650)$ in the elastic and
  inelastic channels revisited}\/},
\href{http://arxiv.org/abs/1805.07453}{{\tt arXiv:1805.07453 [hep-ph]}}.

\bibitem{Guo:2014iya}
F.-K. Guo, C.~Hanhart, Q.~Wang, and Q.~Zhao, {\em {Could the near-threshold
  $XYZ$ states be simply kinematic effects?}\/},
  \href{http://dx.doi.org/10.1103/PhysRevD.91.051504}{Phys. Rev. {\bf D91}
  (2015) no.~5, 051504},
\href{http://arxiv.org/abs/1411.5584}{{\tt arXiv:1411.5584 [hep-ph]}}.

\bibitem{Wang:2013hga}
Q.~Wang, C.~Hanhart, and Q.~Zhao, {\em {Systematic study of the singularity
  mechanism in heavy quarkonium decays}\/},
  \href{http://dx.doi.org/10.1016/j.physletb.2013.06.049}{Phys. Lett. {\bf
  B725} (2013) no.~1-3, 106--110},
\href{http://arxiv.org/abs/1305.1997}{{\tt arXiv:1305.1997 [hep-ph]}}.

\bibitem{Szczepaniak:2015eza}
A.~P. Szczepaniak, {\em {Triangle Singularities and XYZ Quarkonium Peaks}\/},
  \href{http://dx.doi.org/10.1016/j.physletb.2015.06.029}{Phys. Lett. {\bf
  B747} (2015)  410--416},
\href{http://arxiv.org/abs/1501.01691}{{\tt arXiv:1501.01691 [hep-ph]}}.

\bibitem{Gong:2016jzb}
Q.-R. Gong, J.-L. Pang, Y.-F. Wang, and H.-Q. Zheng, {\em {The $Z_c(3900)$ peak
  does not come from the triangle singularity}\/},
  \href{http://dx.doi.org/10.1140/epjc/s10052-018-5690-7}{Eur. Phys. J. {\bf
  C78} (2018) no.~4, 276},
\href{http://arxiv.org/abs/1612.08159}{{\tt arXiv:1612.08159 [hep-ph]}}.

\bibitem{Bondar:2016pox}
A.~E. Bondar and M.~B. Voloshin, {\em {$\Upsilon(6S)$ and triangle singularity
  in $e^+e^- \to B_1(5721) \bar B \to Z_b(10610) \, \pi$}\/},
  \href{http://dx.doi.org/10.1103/PhysRevD.93.094008}{Phys. Rev. {\bf D93}
  (2016) no.~9, 094008},
\href{http://arxiv.org/abs/1603.08436}{{\tt arXiv:1603.08436 [hep-ph]}}.

\bibitem{Abazov:2018cyu}
{D0 Collaboration}, V.~M. Abazov et al., {\em {Evidence for $Z_c^{\pm}(3900)$
  in semi-inclusive decays of $b$-flavored hadrons}\/},
\href{http://arxiv.org/abs/1807.00183}{{\tt arXiv:1807.00183 [hep-ex]}}.

\bibitem{Voloshin:2011qa}
M.~B. Voloshin, {\em {Radiative transitions from Upsilon(5S) to molecular
  bottomonium}\/},  \href{http://dx.doi.org/10.1103/PhysRevD.84.031502}{Phys.
  Rev. {\bf D84} (2011)  031502},
\href{http://arxiv.org/abs/1105.5829}{{\tt arXiv:1105.5829 [hep-ph]}}.

\bibitem{Ikeda:2016zwx}
{HAL QCD Collaboration}, Y.~Ikeda, S.~Aoki, T.~Doi, S.~Gongyo, T.~Hatsuda,
  T.~Inoue, T.~Iritani, N.~Ishii, K.~Murano, and K.~Sasaki, {\em {Fate of the
  Tetraquark Candidate $Z_c$(3900) from Lattice QCD}\/},
  \href{http://dx.doi.org/10.1103/PhysRevLett.117.242001}{Phys. Rev. Lett. {\bf
  117} (2016) no.~24, 242001},
\href{http://arxiv.org/abs/1602.03465}{{\tt arXiv:1602.03465 [hep-lat]}}.

\bibitem{Cheung:2017tnt}
{Hadron Spectrum Collaboration}, G.~K.~C. Cheung, C.~E. Thomas, J.~J. Dudek,
  and R.~G. Edwards, {\em {Tetraquark operators in lattice QCD and exotic
  flavour states in the charm sector}\/},
  \href{http://dx.doi.org/10.1007/JHEP11(2017)033}{JHEP {\bf 11} (2017)  033},
\href{http://arxiv.org/abs/1709.01417}{{\tt arXiv:1709.01417 [hep-lat]}}.

\bibitem{Peters:2017hon}
A.~Peters, P.~Bicudo, and M.~Wagner, {\em {$b\bar b u\bar d$ four-quark systems
  in the Born-Oppenheimer approximation: prospects and challenges}\/},
  \href{http://dx.doi.org/10.1051/epjconf/201817514018}{EPJ Web Conf. {\bf 175}
  (2018)  14018},
\href{http://arxiv.org/abs/1709.03306}{{\tt arXiv:1709.03306 [hep-lat]}}.

\bibitem{Maiani:2015vwa}
L.~Maiani, A.~D. Polosa, and V.~Riquer, {\em {The New Pentaquarks in the
  Diquark Model}\/},
  \href{http://dx.doi.org/10.1016/j.physletb.2015.08.008}{Phys. Lett. {\bf
  B749} (2015)  289--291},
\href{http://arxiv.org/abs/1507.04980}{{\tt arXiv:1507.04980 [hep-ph]}}.

\bibitem{Maiani:2015iaa}
L.~Maiani, A.~D. Polosa, and V.~Riquer, {\em {From pentaquarks to dibaryons in
  $\Lambda_b$(5620) decays}\/},
  \href{http://dx.doi.org/10.1016/j.physletb.2015.08.049}{Phys. Lett. {\bf
  B750} (2015)  37--38},
\href{http://arxiv.org/abs/1508.04459}{{\tt arXiv:1508.04459 [hep-ph]}}.

\bibitem{Ali:2016dkf}
A.~Ali, I.~Ahmed, M.~J. Aslam, and A.~Rehman, {\em {Heavy quark symmetry and
  weak decays of the $b$-baryons in pentaquarks with a $c\bar{c}$
  component}\/},  \href{http://dx.doi.org/10.1103/PhysRevD.94.054001}{Phys.
  Rev. {\bf D94} (2016) no.~5, 054001},
\href{http://arxiv.org/abs/1607.00987}{{\tt arXiv:1607.00987 [hep-ph]}}.

\bibitem{Ali:2017ebb}
A.~Ali, I.~Ahmed, M.~J. Aslam, and A.~Rehman, {\em {Mass spectrum of spin-1/2
  pentaquarks with a $c\bar{c}$ component and their anticipated discovery modes
  in $b$-baryon decays}\/},
\href{http://arxiv.org/abs/1704.05419}{{\tt arXiv:1704.05419 [hep-ph]}}.

\bibitem{Wu:2010jy}
J.-J. Wu, R.~Molina, E.~Oset, and B.~S. Zou, {\em {Prediction of narrow $N^*$
  and $\Lambda^*$ resonances with hidden charm above 4 GeV}\/},
  \href{http://dx.doi.org/10.1103/PhysRevLett.105.232001}{Phys. Rev. Lett. {\bf
  105} (2010)  232001},
\href{http://arxiv.org/abs/1007.0573}{{\tt arXiv:1007.0573 [nucl-th]}}.

\bibitem{Wang:2011rga}
W.~L. Wang, F.~Huang, Z.~Y. Zhang, and B.~S. Zou, {\em {$\Sigma_c \bar{D}$ and
  $\Lambda_c \bar{D}$ states in a chiral quark model}\/},
  \href{http://dx.doi.org/10.1103/PhysRevC.84.015203}{Phys. Rev. {\bf C84}
  (2011)  015203},
\href{http://arxiv.org/abs/1101.0453}{{\tt arXiv:1101.0453 [nucl-th]}}.

\bibitem{Yang:2011wz}
Z.-C. Yang, Z.-F. Sun, J.~He, X.~Liu, and S.-L. Zhu, {\em {The possible
  hidden-charm molecular baryons composed of anti-charmed meson and charmed
  baryon}\/},  \href{http://dx.doi.org/10.1088/1674-1137/36/1/002,
  10.1088/1674-1137/36/3/006}{Chin. Phys. {\bf C36} (2012)  6--13},
\href{http://arxiv.org/abs/1105.2901}{{\tt arXiv:1105.2901 [hep-ph]}}.

\bibitem{Wu:2012md}
J.-J. Wu, T.~S.~H. Lee, and B.~S. Zou, {\em {Nucleon Resonances with Hidden
  Charm in Coupled-Channel Models}\/},
  \href{http://dx.doi.org/10.1103/PhysRevC.85.044002}{Phys. Rev. {\bf C85}
  (2012)  044002},
\href{http://arxiv.org/abs/1202.1036}{{\tt arXiv:1202.1036 [nucl-th]}}.

\bibitem{Karliner:2015ina}
M.~Karliner and J.~L. Rosner, {\em {New Exotic Meson and Baryon Resonances from
  Doubly-Heavy Hadronic Molecules}\/},
  \href{http://dx.doi.org/10.1103/PhysRevLett.115.122001}{Phys. Rev. Lett. {\bf
  115} (2015) no.~12, 122001},
\href{http://arxiv.org/abs/1506.06386}{{\tt arXiv:1506.06386 [hep-ph]}}.

\bibitem{Roca:2015dva}
L.~Roca, J.~Nieves, and E.~Oset, {\em {LHCb pentaquark as a
  $\bar{D}^*\Sigma_c-\bar{D}^*\Sigma_c^*$ molecular state}\/},
  \href{http://dx.doi.org/10.1103/PhysRevD.92.094003}{Phys. Rev. {\bf D92}
  (2015) no.~9, 094003},
\href{http://arxiv.org/abs/1507.04249}{{\tt arXiv:1507.04249 [hep-ph]}}.

\bibitem{Roca:2016tdh}
L.~Roca and E.~Oset, {\em {On the hidden charm pentaquarks in $\Lambda _b
  \rightarrow J/\psi K^- p$ decay}\/},
  \href{http://dx.doi.org/10.1140/epjc/s10052-016-4407-z}{Eur. Phys. J. {\bf
  C76} (2016) no.~11, 591},
\href{http://arxiv.org/abs/1602.06791}{{\tt arXiv:1602.06791 [hep-ph]}}.

\bibitem{Xiao:2013yca}
C.~W. Xiao, J.~Nieves, and E.~Oset, {\em {Combining heavy quark spin and local
  hidden gauge symmetries in the dynamical generation of hidden charm
  baryons}\/},  \href{http://dx.doi.org/10.1103/PhysRevD.88.056012}{Phys. Rev.
  {\bf D88} (2013)  056012},
\href{http://arxiv.org/abs/1304.5368}{{\tt arXiv:1304.5368 [hep-ph]}}.

\bibitem{Guo:2015umn}
F.-K. Guo, U.-G. Mei{\ss}ner, W.~Wang, and Z.~Yang, {\em {How to reveal the
  exotic nature of the P$_c$(4450)}\/},
  \href{http://dx.doi.org/10.1103/PhysRevD.92.071502}{Phys. Rev. {\bf D92}
  (2015) no.~7, 071502},
\href{http://arxiv.org/abs/1507.04950}{{\tt arXiv:1507.04950 [hep-ph]}}.

\bibitem{Liu:2015fea}
X.-H. Liu, Q.~Wang, and Q.~Zhao, {\em {Understanding the newly observed heavy
  pentaquark candidates}\/},
  \href{http://dx.doi.org/10.1016/j.physletb.2016.03.089}{Phys. Lett. {\bf
  B757} (2016)  231--236},
\href{http://arxiv.org/abs/1507.05359}{{\tt arXiv:1507.05359 [hep-ph]}}.

\bibitem{Bayar:2016ftu}
M.~Bayar, F.~Aceti, F.-K. Guo, and E.~Oset, {\em {A Discussion on Triangle
  Singularities in the $\Lambda_b \to J/\psi K^{-} p$ Reaction}\/},
  \href{http://dx.doi.org/10.1103/PhysRevD.94.074039}{Phys. Rev. {\bf D94}
  (2016) no.~7, 074039},
\href{http://arxiv.org/abs/1609.04133}{{\tt arXiv:1609.04133 [hep-ph]}}.

\bibitem{Jurik:2016bdm}
N.~P. Jurik, {\em {Observation of $J/\psi$ p resonances consistent with
  pentaquark states in$\Lambda_ b^0\to J/\psi K^-p$ decays}}.
\newblock PhD thesis, Syracuse U.,
2016-08-08.
\newblock

\bibitem{Kubarovsky:2015aaa}
V.~Kubarovsky and M.~B. Voloshin, {\em {Formation of hidden-charm pentaquarks
  in photon-nucleon collisions}\/},
  \href{http://dx.doi.org/10.1103/PhysRevD.92.031502}{Phys. Rev. {\bf D92}
  (2015) no.~3, 031502},
\href{http://arxiv.org/abs/1508.00888}{{\tt arXiv:1508.00888 [hep-ph]}}.

\bibitem{Blin:2016dlf}
A.~N. Hiller~Blin, C.~Fern\'andez-Ram\'irez, A.~Jackura, V.~Mathieu, V.~I.
  Mokeev, A.~Pilloni, and A.~P. Szczepaniak, {\em {Studying the P$_c$(4450)
  resonance in J/$\psi$ photoproduction off protons}\/},
  \href{http://dx.doi.org/10.1103/PhysRevD.94.034002}{Phys. Rev. {\bf D94}
  (2016) no.~3, 034002},
\href{http://arxiv.org/abs/1606.08912}{{\tt arXiv:1606.08912 [hep-ph]}}.

\bibitem{Karliner:2015voa}
M.~Karliner and J.~L. Rosner, {\em {Photoproduction of Exotic Baryon
  Resonances}\/},
  \href{http://dx.doi.org/10.1016/j.physletb.2015.11.068}{Phys. Lett. {\bf
  B752} (2016)  329--332},
\href{http://arxiv.org/abs/1508.01496}{{\tt arXiv:1508.01496 [hep-ph]}}.

\bibitem{Shen:2016tzq}
C.-W. Shen, F.-K. Guo, J.-J. Xie, and B.-S. Zou, {\em {Disentangling the
  hadronic molecule nature of the $P_c(4380)$ pentaquark-like structure}\/},
  \href{http://dx.doi.org/10.1016/j.nuclphysa.2016.04.034}{Nucl. Phys. {\bf
  A954} (2016)  393--405},
\href{http://arxiv.org/abs/1603.04672}{{\tt arXiv:1603.04672 [hep-ph]}}.

\bibitem{Lin:2017mtz}
Y.-H. Lin, C.-W. Shen, F.-K. Guo, and B.-S. Zou, {\em {Decay behaviors of the
  $P_c$ hadronic molecules}\/},
  \href{http://dx.doi.org/10.1103/PhysRevD.95.114017}{Phys. Rev. {\bf D95}
  (2017) no.~11, 114017},
\href{http://arxiv.org/abs/1703.01045}{{\tt arXiv:1703.01045 [hep-ph]}}.

\bibitem{Eides:2018lqg}
M.~I. Eides and V.~{\relax Yu}. Petrov, {\em {Decays of Pentaquarks in
  Hadrocharmonium and Molecular Pictures}\/},
\href{http://arxiv.org/abs/1811.01691}{{\tt arXiv:1811.01691 [hep-ph]}}.

\bibitem{Wu:2010rv}
J.-J. Wu and B.~S. Zou, {\em {Prediction of super-heavy $N^*$ and $\Lambda^*$
  resonances with hidden beauty}\/},
  \href{http://dx.doi.org/10.1016/j.physletb.2012.01.068}{Phys. Lett. {\bf
  B709} (2012)  70--76},
\href{http://arxiv.org/abs/1011.5743}{{\tt arXiv:1011.5743 [hep-ph]}}.

\bibitem{Wu:2017weo}
J.~Wu, Y.-R. Liu, K.~Chen, X.~Liu, and S.-L. Zhu, {\em {Hidden-charm
  pentaquarks and their hidden-bottom and $B_c$-like partner states}\/},
  \href{http://dx.doi.org/10.1103/PhysRevD.95.034002}{Phys. Rev. {\bf D95}
  (2017) no.~3, 034002},
\href{http://arxiv.org/abs/1701.03873}{{\tt arXiv:1701.03873 [hep-ph]}}.

\bibitem{Yamaguchi:2017zmn}
Y.~Yamaguchi, A.~Giachino, A.~Hosaka, E.~Santopinto, S.~Takeuchi, and
  M.~Takizawa, {\em {Hidden-charm and bottom meson-baryon molecules coupled
  with five-quark states}\/},
  \href{http://dx.doi.org/10.1103/PhysRevD.96.114031}{Phys. Rev. {\bf D96}
  (2017) no.~11, 114031},
\href{http://arxiv.org/abs/1709.00819}{{\tt arXiv:1709.00819 [hep-ph]}}.

\bibitem{Shen:2017ayv}
C.-W. Shen, D.~R{\"o}nchen, U.-G. Mei{\ss}ner, and B.-S. Zou, {\em {Exploratory
  study of possible resonances in heavy meson - heavy baryon coupled-channel
  interactions}\/},
  \href{http://dx.doi.org/10.1088/1674-1137/42/2/023106}{Chin. Phys. {\bf C42}
  (2018) no.~2, 023106},
\href{http://arxiv.org/abs/1710.03885}{{\tt arXiv:1710.03885 [hep-ph]}}.

\bibitem{Lin:2018kcc}
Y.-H. Lin, C.-W. Shen, and B.-S. Zou, {\em {Decay behavior of the strange and
  beauty partners of $P_c$ hadronic molecules}\/},
  \href{http://dx.doi.org/10.1016/j.nuclphysa.2018.10.001}{Nucl. Phys. {\bf
  A980} (2018)  21--31},
\href{http://arxiv.org/abs/1805.06843}{{\tt arXiv:1805.06843 [hep-ph]}}.

\bibitem{Yang:2018oqd}
G.~Yang, J.~Ping, and J.~Segovia, {\em {Hidden-Bottom Pentaquarks}\/},
\href{http://arxiv.org/abs/1809.06193}{{\tt arXiv:1809.06193 [hep-ph]}}.

\bibitem{Gershtein:2000nx}
S.~S. Gershtein, V.~V. Kiselev, A.~K. Likhoded, and A.~I. Onishchenko, {\em
  {Spectroscopy of doubly heavy baryons}\/},
\href{http://dx.doi.org/10.1103/PhysRevD.62.054021}{Phys. Rev. {\bf D62} (2000)
   054021}.

\bibitem{Kiselev:2001fw}
V.~V. Kiselev and A.~K. Likhoded, {\em {Baryons with two heavy quarks}\/},
  \href{http://dx.doi.org/10.1070/PU2002v045n05ABEH000958}{Phys. Usp. {\bf 45}
  (2002)  455--506}, \href{http://arxiv.org/abs/hep-ph/0103169}{{\tt
  arXiv:hep-ph/0103169 [hep-ph]}}.
[Usp. Fiz. Nauk172,497(2002)].

\bibitem{Ebert:2002ig}
D.~Ebert, R.~N. Faustov, V.~O. Galkin, and A.~P. Martynenko, {\em {Mass spectra
  of doubly heavy baryons in the relativistic quark model}\/},
  \href{http://dx.doi.org/10.1103/PhysRevD.66.014008}{Phys. Rev. {\bf D66}
  (2002)  014008},
\href{http://arxiv.org/abs/hep-ph/0201217}{{\tt arXiv:hep-ph/0201217
  [hep-ph]}}.

\bibitem{Kiselev:2002iy}
V.~V. Kiselev, A.~K. Likhoded, O.~N. Pakhomova, and V.~A. Saleev, {\em {Mass
  spectra of doubly heavy Omega $Q Q^\prime$ baryons}\/},
  \href{http://dx.doi.org/10.1103/PhysRevD.66.034030}{Phys. Rev. {\bf D66}
  (2002)  034030},
\href{http://arxiv.org/abs/hep-ph/0206140}{{\tt arXiv:hep-ph/0206140
  [hep-ph]}}.

\bibitem{Albertus:2006ya}
C.~Albertus, E.~Hernandez, J.~Nieves, and J.~M. Verde-Velasco, {\em {Static
  properties and semileptonic decays of doubly heavy baryons in a
  nonrelativistic quark model}\/},
  \href{http://dx.doi.org/10.1140/epja/i2007-10364-y,
  10.1140/epja/i2008-10547-0}{Eur. Phys. J. {\bf A32} (2007)  183--199},
  \href{http://arxiv.org/abs/hep-ph/0610030}{{\tt arXiv:hep-ph/0610030
  [hep-ph]}}.
[Erratum: Eur. Phys. J.A36,119(2008)].

\bibitem{Padmanath:2015jea}
M.~Padmanath, R.~G. Edwards, N.~Mathur, and M.~Peardon, {\em {Spectroscopy of
  doubly-charmed baryons from lattice QCD}\/},
  \href{http://dx.doi.org/10.1103/PhysRevD.91.094502}{Phys. Rev. {\bf D91}
  (2015) no.~9, 094502},
\href{http://arxiv.org/abs/1502.01845}{{\tt arXiv:1502.01845 [hep-lat]}}.

\bibitem{Karliner:2014gca}
M.~Karliner and J.~L. Rosner, {\em {Baryons with two heavy quarks: Masses,
  production, decays, and detection}\/},
  \href{http://dx.doi.org/10.1103/PhysRevD.90.094007}{Phys. Rev. {\bf D90}
  (2014) no.~9, 094007},
\href{http://arxiv.org/abs/1408.5877}{{\tt arXiv:1408.5877 [hep-ph]}}.

\bibitem{Yu:2017zst}
F.-S. Yu, H.-Y. Jiang, R.-H. Li, C.-D. Lü, W.~Wang, and Z.-X. Zhao, {\em
  {Discovery Potentials of Doubly Charmed Baryons}\/},
  \href{http://dx.doi.org/10.1088/1674-1137/42/5/051001}{Chin. Phys. {\bf C42}
  (2018) no.~5, 051001},
\href{http://arxiv.org/abs/1703.09086}{{\tt arXiv:1703.09086 [hep-ph]}}.

\bibitem{Wang:2017mqp}
W.~Wang, F.-S. Yu, and Z.-X. Zhao, {\em {Weak decays of doubly heavy baryons:
  the $1/2\rightarrow 1/2$ case}\/},
  \href{http://dx.doi.org/10.1140/epjc/s10052-017-5360-1}{Eur. Phys. J. {\bf
  C77} (2017) no.~11, 781},
\href{http://arxiv.org/abs/1707.02834}{{\tt arXiv:1707.02834 [hep-ph]}}.

\bibitem{Xiao:2017udy}
L.-Y. Xiao, K.-L. Wang, Q.-f. Lu, X.-H. Zhong, and S.-L. Zhu, {\em {Strong and
  radiative decays of the doubly charmed baryons}\/},
  \href{http://dx.doi.org/10.1103/PhysRevD.96.094005}{Phys. Rev. {\bf D96}
  (2017) no.~9, 094005},
\href{http://arxiv.org/abs/1708.04384}{{\tt arXiv:1708.04384 [hep-ph]}}.

\bibitem{Cui:2017udv}
E.-L. Cui, H.-X. Chen, W.~Chen, X.~Liu, and S.-L. Zhu, {\em {Suggested search
  for doubly charmed baryons of $J^P=3/2^+$ via their electromagnetic
  transitions}\/},  \href{http://dx.doi.org/10.1103/PhysRevD.97.034018}{Phys.
  Rev. {\bf D97} (2018) no.~3, 034018},
\href{http://arxiv.org/abs/1712.03615}{{\tt arXiv:1712.03615 [hep-ph]}}.

\bibitem{Xiao:2017dly}
L.-Y. Xiao, Q.-F. L{\"u}, and S.-L. Zhu, {\em {Strong decays of the 1P and 2D
  doubly charmed states}\/},
  \href{http://dx.doi.org/10.1103/PhysRevD.97.074005}{Phys. Rev. {\bf D97}
  (2018) no.~7, 074005},
\href{http://arxiv.org/abs/1712.07295}{{\tt arXiv:1712.07295 [hep-ph]}}.

\bibitem{Li:2017pxa}
H.-S. Li, L.~Meng, Z.-W. Liu, and S.-L. Zhu, {\em {Radiative decays of the
  doubly charmed baryons in chiral perturbation theory}\/},
  \href{http://dx.doi.org/10.1016/j.physletb.2017.12.031}{Phys. Lett. {\bf
  B777} (2018)  169--176},
\href{http://arxiv.org/abs/1708.03620}{{\tt arXiv:1708.03620 [hep-ph]}}.

\bibitem{Aaij:2017ueg}
{LHCb Collaboration}, R.~Aaij et al., {\em {Observation of the doubly charmed
  baryon $\Xi_{cc}^{++}$}\/},
  \href{http://dx.doi.org/10.1103/PhysRevLett.119.112001}{Phys. Rev. Lett. {\bf
  119} (2017) no.~11, 112001},
\href{http://arxiv.org/abs/1707.01621}{{\tt arXiv:1707.01621 [hep-ex]}}.

\bibitem{Mattson:2002vu}
{SELEX Collaboration}, M.~Mattson et al., {\em {First Observation of the Doubly
  Charmed Baryon $\Xi^+_{cc}$}\/},
  \href{http://dx.doi.org/10.1103/PhysRevLett.89.112001}{Phys. Rev. Lett. {\bf
  89} (2002)  112001},
\href{http://arxiv.org/abs/hep-ex/0208014}{{\tt arXiv:hep-ex/0208014
  [hep-ex]}}.

\bibitem{Brodsky:2011zs}
S.~J. Brodsky, F.-K. Guo, C.~Hanhart, and U.-G. Mei{\ss}ner, {\em {Isospin
  splittings of doubly heavy baryons}\/},
  \href{http://dx.doi.org/10.1016/j.physletb.2011.03.014}{Phys. Lett. {\bf
  B698} (2011)  251--255},
\href{http://arxiv.org/abs/1101.1983}{{\tt arXiv:1101.1983 [hep-ph]}}.

\bibitem{Karliner:2017gml}
M.~Karliner and J.~L. Rosner, {\em {Isospin splittings in baryons with two
  heavy quarks}\/},  \href{http://dx.doi.org/10.1103/PhysRevD.96.033004}{Phys.
  Rev. {\bf D96} (2017) no.~3, 033004},
\href{http://arxiv.org/abs/1706.06961}{{\tt arXiv:1706.06961 [hep-ph]}}.

\bibitem{Yan:2018zdt}
M.-J. Yan, X.-H. Liu, S.~Gonz{\`a}lez-Sol{\'i}s, F.-K. Guo, C.~Hanhart, U.-G.
  Mei{\ss}ner, and B.-S. Zou, {\em {New spectrum of negative parity doubly
  charmed baryons: Possibility of two quasi-stable states}\/},
\href{http://arxiv.org/abs/1805.10972}{{\tt arXiv:1805.10972 [hep-ph]}}.

\bibitem{Guo:2017vcf}
Z.-H. Guo, {\em {Prediction of exotic doubly charmed baryons within chiral
  effective field theory}\/},
  \href{http://dx.doi.org/10.1103/PhysRevD.96.074004}{Phys. Rev. {\bf D96}
  (2017) no.~7, 074004},
\href{http://arxiv.org/abs/1708.04145}{{\tt arXiv:1708.04145 [hep-ph]}}.

\bibitem{Dias:2018qhp}
J.~M. Dias, V.~R. Debastiani, J.~J. Xie, and E.~Oset, {\em {Doubly charmed
  $\Xi_{cc}$ molecular states from meson-baryon interaction}\/},
  \href{http://dx.doi.org/10.1103/PhysRevD.98.094017}{Phys. Rev. {\bf D98}
  (2018) no.~9, 094017},
\href{http://arxiv.org/abs/1805.03286}{{\tt arXiv:1805.03286 [hep-ph]}}.

\bibitem{Karliner:2018hos}
M.~Karliner and J.~L. Rosner, {\em {Strange baryons with two heavy quarks}\/},
  \href{http://dx.doi.org/10.1103/PhysRevD.97.094006}{Phys. Rev. {\bf D97}
  (2018) no.~9, 094006},
\href{http://arxiv.org/abs/1803.01657}{{\tt arXiv:1803.01657 [hep-ph]}}.

\bibitem{Mathur:2018rwu}
N.~Mathur and M.~Padmanath, {\em {On the discovery of next doubly charmed
  baryon}\/},
\href{http://arxiv.org/abs/1807.00174}{{\tt arXiv:1807.00174 [hep-lat]}}.

\bibitem{Mathur:2018epb}
N.~Mathur, M.~Padmanath, and S.~Mondal, {\em {Precise predictions of
  charmed-bottom hadrons from lattice QCD}\/},
\href{http://arxiv.org/abs/1806.04151}{{\tt arXiv:1806.04151 [hep-lat]}}.

\bibitem{Eichten:2017ffp}
E.~J. Eichten and C.~Quigg, {\em {Heavy-quark symmetry implies stable heavy
  tetraquark mesons $Q_iQ_j \bar q_k \bar q_l$}\/},
  \href{http://dx.doi.org/10.1103/PhysRevLett.119.202002}{Phys. Rev. Lett. {\bf
  119} (2017) no.~20, 202002},
\href{http://arxiv.org/abs/1707.09575}{{\tt arXiv:1707.09575 [hep-ph]}}.

\bibitem{Czarnecki:2017vco}
A.~Czarnecki, B.~Leng, and M.~B. Voloshin, {\em {Stability of tetrons}\/},
  \href{http://dx.doi.org/10.1016/j.physletb.2018.01.034}{Phys. Lett. {\bf
  B778} (2018)  233--238},
\href{http://arxiv.org/abs/1708.04594}{{\tt arXiv:1708.04594 [hep-ph]}}.

\bibitem{Karliner:2017qjm}
M.~Karliner and J.~L. Rosner, {\em {Discovery of doubly-charmed $\Xi_{cc}$
  baryon implies a stable ($b b \bar{u} \bar{d}$) tetraquark}\/},
  \href{http://dx.doi.org/10.1103/PhysRevLett.119.202001}{Phys. Rev. Lett. {\bf
  119} (2017) no.~20, 202001},
\href{http://arxiv.org/abs/1707.07666}{{\tt arXiv:1707.07666 [hep-ph]}}.

\bibitem{Ader:1981db}
J.~P. Ader, J.~M. Richard, and P.~Taxil, {\em {Do narrow heavy multi-quark
  states exist?}\/},
\href{http://dx.doi.org/10.1103/PhysRevD.25.2370}{Phys. Rev. {\bf D25} (1982)
  2370}.

\bibitem{Manohar:1992nd}
A.~V. Manohar and M.~B. Wise, {\em {Exotic $QQ\bar q\bar q$ states in QCD}\/},
  \href{http://dx.doi.org/10.1016/0550-3213(93)90614-U}{Nucl. Phys. {\bf B399}
  (1993)  17--33},
\href{http://arxiv.org/abs/hep-ph/9212236}{{\tt arXiv:hep-ph/9212236
  [hep-ph]}}.

\bibitem{Mehen:2017nrh}
T.~Mehen, {\em {Implications of Heavy Quark-Diquark Symmetry for Excited Doubly
  Heavy Baryons and Tetraquarks}\/},
  \href{http://dx.doi.org/10.1103/PhysRevD.96.094028}{Phys. Rev. {\bf D96}
  (2017) no.~9, 094028},
\href{http://arxiv.org/abs/1708.05020}{{\tt arXiv:1708.05020 [hep-ph]}}.

\bibitem{Esposito:2013fma}
A.~Esposito, M.~Papinutto, A.~Pilloni, A.~D. Polosa, and N.~Tantalo, {\em
  {Doubly charmed tetraquarks in $B_c$ and $\Xi_{bc}$ decays}\/},
  \href{http://dx.doi.org/10.1103/PhysRevD.88.054029}{Phys. Rev. {\bf D88}
  (2013) no.~5, 054029},
\href{http://arxiv.org/abs/1307.2873}{{\tt arXiv:1307.2873 [hep-ph]}}.

\bibitem{Luo:2017eub}
S.-Q. Luo, K.~Chen, X.~Liu, Y.-R. Liu, and S.-L. Zhu, {\em {Exotic tetraquark
  states with the $qq\bar{Q}\bar{Q}$ configuration}\/},
  \href{http://dx.doi.org/10.1140/epjc/s10052-017-5297-4}{Eur. Phys. J. {\bf
  C77} (2017) no.~10, 709},
\href{http://arxiv.org/abs/1707.01180}{{\tt arXiv:1707.01180 [hep-ph]}}.

\bibitem{Yuqi:2011gm}
Y.-q. Chen and S.-z. Wu, {\em {Production of four-quark states with double
  heavy quarks at LHC}\/},
  \href{http://dx.doi.org/10.1016/j.physletb.2011.09.096}{Phys. Lett. {\bf
  B705} (2011)  93--97},
\href{http://arxiv.org/abs/1101.4568}{{\tt arXiv:1101.4568 [hep-ph]}}.

\bibitem{Ali:2018xfq}
A.~Ali, Q.~Qin, and W.~Wang, {\em {Discovery potential of stable and
  near-threshold doubly heavy tetraquarks at the LHC}\/},
\href{http://arxiv.org/abs/1806.09288}{{\tt arXiv:1806.09288 [hep-ph]}}.

\bibitem{Gershon:2018gda}
T.~Gershon and A.~Poluektov, {\em {Displaced $B_c^-$ mesons as an inclusive
  signature of weakly decaying double beauty hadrons}\/},
\href{http://arxiv.org/abs/1810.06657}{{\tt arXiv:1810.06657 [hep-ph]}}.

\bibitem{Bicudo:2012qt}
{European Twisted Mass Collaboration}, P.~Bicudo and M.~Wagner, {\em {Lattice
  QCD signal for a bottom-bottom tetraquark}\/},
  \href{http://dx.doi.org/10.1103/PhysRevD.87.114511}{Phys. Rev. {\bf D87}
  (2013) no.~11, 114511},
\href{http://arxiv.org/abs/1209.6274}{{\tt arXiv:1209.6274 [hep-ph]}}.

\bibitem{Francis:2016hui}
A.~Francis, R.~J. Hudspith, R.~Lewis, and K.~Maltman, {\em {Lattice Prediction
  for Deeply Bound Doubly Heavy Tetraquarks}\/},
  \href{http://dx.doi.org/10.1103/PhysRevLett.118.142001}{Phys. Rev. Lett. {\bf
  118} (2017) no.~14, 142001},
\href{http://arxiv.org/abs/1607.05214}{{\tt arXiv:1607.05214 [hep-lat]}}.

\bibitem{Bicudo:2015vta}
P.~Bicudo, K.~Cichy, A.~Peters, B.~Wagenbach, and M.~Wagner, {\em {Evidence for
  the existence of $u d \bar{b} \bar{b}$ and the non-existence of $s s \bar{b}
  \bar{b}$ and $c c \bar{b} \bar{b}$ tetraquarks from lattice QCD}\/},
  \href{http://dx.doi.org/10.1103/PhysRevD.92.014507}{Phys. Rev. {\bf D92}
  (2015) no.~1, 014507},
\href{http://arxiv.org/abs/1505.00613}{{\tt arXiv:1505.00613 [hep-lat]}}.

\bibitem{Hughes:2017xie}
C.~Hughes, E.~Eichten, and C.~T.~H. Davies, {\em {Searching for beauty-fully
  bound tetraquarks using lattice nonrelativistic QCD}\/},
  \href{http://dx.doi.org/10.1103/PhysRevD.97.054505}{Phys. Rev. {\bf D97}
  (2018) no.~5, 054505},
\href{http://arxiv.org/abs/1710.03236}{{\tt arXiv:1710.03236 [hep-lat]}}.

\bibitem{Heller:1985cb}
L.~Heller and J.~A. Tjon, {\em {On Bound States of Heavy $Q^2 \bar{Q}^2$
  Systems}\/},
\href{http://dx.doi.org/10.1103/PhysRevD.32.755}{Phys. Rev. {\bf D32} (1985)
  755}.

\bibitem{Berezhnoy:2011xn}
A.~V. Berezhnoy, A.~V. Luchinsky, and A.~A. Novoselov, {\em {Heavy tetraquarks
  production at the LHC}\/},
  \href{http://dx.doi.org/10.1103/PhysRevD.86.034004}{Phys. Rev. {\bf D86}
  (2012)  034004},
\href{http://arxiv.org/abs/1111.1867}{{\tt arXiv:1111.1867 [hep-ph]}}.

\bibitem{Wu:2016vtq}
J.~Wu, Y.-R. Liu, K.~Chen, X.~Liu, and S.-L. Zhu, {\em {Heavy-flavored
  tetraquark states with the $QQ\bar{Q}\bar{Q}$ configuration}\/},
  \href{http://dx.doi.org/10.1103/PhysRevD.97.094015}{Phys. Rev. {\bf D97}
  (2018)  094015},
\href{http://arxiv.org/abs/1605.01134}{{\tt arXiv:1605.01134 [hep-ph]}}.

\bibitem{Chen:2016jxd}
W.~Chen, H.-X. Chen, X.~Liu, T.~G. Steele, and S.-L. Zhu, {\em {Hunting for
  exotic doubly hidden-charm/bottom tetraquark states}\/},
  \href{http://dx.doi.org/10.1016/j.physletb.2017.08.034}{Phys. Lett. {\bf
  B773} (2017)  247--251},
\href{http://arxiv.org/abs/1605.01647}{{\tt arXiv:1605.01647 [hep-ph]}}.

\bibitem{Karliner:2016zzc}
M.~Karliner, S.~Nussinov, and J.~L. Rosner, {\em {$Q Q \bar Q \bar Q$ states:
  masses, production, and decays}\/},
  \href{http://dx.doi.org/10.1103/PhysRevD.95.034011}{Phys. Rev. {\bf D95}
  (2017) no.~3, 034011},
\href{http://arxiv.org/abs/1611.00348}{{\tt arXiv:1611.00348 [hep-ph]}}.

\bibitem{Bai:2016int}
Y.~Bai, S.~Lu, and J.~Osborne, {\em {Beauty-full Tetraquarks}\/},
\href{http://arxiv.org/abs/1612.00012}{{\tt arXiv:1612.00012 [hep-ph]}}.

\bibitem{Wang:2017jtz}
Z.-G. Wang, {\em {Analysis of the $QQ\bar{Q}\bar{Q}$ tetraquark states with QCD
  sum rules}\/},  \href{http://dx.doi.org/10.1140/epjc/s10052-017-4997-0}{Eur.
  Phys. J. {\bf C77} (2017)  432},
\href{http://arxiv.org/abs/1701.04285}{{\tt arXiv:1701.04285 [hep-ph]}}.

\bibitem{Richard:2017vry}
J.-M. Richard, A.~Valcarce, and J.~Vijande, {\em {String dynamics and
  metastability of all-heavy tetraquarks}\/},
  \href{http://dx.doi.org/10.1103/PhysRevD.95.054019}{Phys. Rev. {\bf D95}
  (2017)  054019},
\href{http://arxiv.org/abs/1703.00783}{{\tt arXiv:1703.00783 [hep-ph]}}.

\bibitem{Anwar:2017toa}
M.~N. Anwar, J.~Ferretti, F.-K. Guo, E.~Santopinto, and B.-S. Zou, {\em
  {Spectroscopy and decays of the fully-heavy tetraquarks}\/},
\href{http://arxiv.org/abs/1710.02540}{{\tt arXiv:1710.02540 [hep-ph]}}.

\bibitem{Vega-Morales:2017pmm}
Y.~Chen and R.~Vega-Morales, {\em {Golden probe of the di-$\PUpsilon$
  threshold}\/},
\href{http://arxiv.org/abs/1710.02738}{{\tt arXiv:1710.02738 [hep-ph]}}.

\bibitem{Eichten:2017ual}
E.~Eichten and Z.~Liu, {\em {Would a Deeply Bound $b\bar b b\bar b$ Tetraquark
  Meson be Observed at the LHC?}\/},
\href{http://arxiv.org/abs/1709.09605}{{\tt arXiv:1709.09605 [hep-ph]}}.

\bibitem{Chao:1980dv}
K.-T. Chao, {\em {The $cc\bar c\bar c$ (diquark-antidiquark) states in $e^+
  e^-$ annihilation}\/},
\href{http://dx.doi.org/10.1007/BF01431564}{Z. Phys. {\bf C7} (1981)  317}.

\bibitem{Padmanath:2013zfa}
M.~Padmanath, R.~G. Edwards, N.~Mathur, and M.~Peardon, {\em {Spectroscopy of
  triply-charmed baryons from lattice QCD}\/},
  \href{http://dx.doi.org/10.1103/PhysRevD.90.074504}{Phys. Rev. {\bf D90}
  (2014) no.~7, 074504},
\href{http://arxiv.org/abs/1307.7022}{{\tt arXiv:1307.7022 [hep-lat]}}.

\bibitem{Bignamini:2009sk}
C.~Bignamini, B.~Grinstein, F.~Piccinini, A.~D. Polosa, and C.~Sabelli, {\em
  {Is the X(3872) Production Cross Section at Tevatron Compatible with a Hadron
  Molecule Interpretation?}\/},
  \href{http://dx.doi.org/10.1103/PhysRevLett.103.162001}{Phys. Rev. Lett. {\bf
  103} (2009)  162001},
\href{http://arxiv.org/abs/0906.0882}{{\tt arXiv:0906.0882 [hep-ph]}}.

\bibitem{Artoisenet:2009wk}
P.~Artoisenet and E.~Braaten, {\em {Production of the X(3872) at the Tevatron
  and the LHC}\/},  \href{http://dx.doi.org/10.1103/PhysRevD.81.114018}{Phys.
  Rev. {\bf D81} (2010)  114018},
\href{http://arxiv.org/abs/0911.2016}{{\tt arXiv:0911.2016 [hep-ph]}}.

\bibitem{Bignamini:2009fn}
C.~Bignamini, B.~Grinstein, F.~Piccinini, A.~D. Polosa, V.~Riquer, and
  C.~Sabelli, {\em {More loosely bound hadron molecules at CDF?}\/},
  \href{http://dx.doi.org/10.1016/j.physletb.2010.01.037}{Phys. Lett. {\bf
  B684} (2010)  228--230},
\href{http://arxiv.org/abs/0912.5064}{{\tt arXiv:0912.5064 [hep-ph]}}.

\bibitem{Artoisenet:2010uu}
P.~Artoisenet and E.~Braaten, {\em {Estimating the Production Rate of
  Loosely-bound Hadronic Molecules using Event Generators}\/},
  \href{http://dx.doi.org/10.1103/PhysRevD.83.014019}{Phys. Rev. {\bf D83}
  (2011)  014019},
\href{http://arxiv.org/abs/1007.2868}{{\tt arXiv:1007.2868 [hep-ph]}}.

\bibitem{Esposito:2013ada}
A.~Esposito, F.~Piccinini, A.~Pilloni, and A.~D. Polosa, {\em {A Mechanism for
  Hadron Molecule Production in p pbar(p) Collisions}\/},
  \href{http://dx.doi.org/10.4236/jmp.2013.412193}{J. Mod. Phys. {\bf 4} (2013)
   1569--1573},
\href{http://arxiv.org/abs/1305.0527}{{\tt arXiv:1305.0527 [hep-ph]}}.

\bibitem{Guerrieri:2014gfa}
A.~Guerrieri, F.~Piccinini, A.~Pilloni, and A.~Polosa, {\em {Production of
  Tetraquarks at the LHC}\/},
  \href{http://dx.doi.org/10.1103/PhysRevD.90.034003}{Phys. Rev. {\bf D90}
  (2014)  034003},
\href{http://arxiv.org/abs/1405.7929}{{\tt arXiv:1405.7929 [hep-ph]}}.

\bibitem{Albaladejo:2017blx}
M.~Albaladejo, F.-K. Guo, C.~Hanhart, U.-G. Mei{\ss}ner, J.~Nieves, A.~Nogga,
  and Z.~Yang, {\em {Note on $X(3872)$ production at hadron colliders and its
  molecular structure}\/},
  \href{http://dx.doi.org/10.1088/1674-1137/41/12/121001}{Chin. Phys. {\bf C41}
  (2017)  121001},
\href{http://arxiv.org/abs/1709.09101}{{\tt arXiv:1709.09101 [hep-ph]}}.

\bibitem{Esposito:2017qef}
A.~Esposito, B.~Grinstein, L.~Maiani, F.~Piccinini, A.~Pilloni, A.~D. Polosa,
  and V.~Riquer, {\em {Comment on `Note on $X(3872)$ production at hadron
  colliders and its molecular structure'}\/},
\href{http://arxiv.org/abs/1709.09631}{{\tt arXiv:1709.09631 [hep-ph]}}.

\bibitem{Braaten:2018eov}
E.~Braaten, L.-P. He, and K.~Ingles, {\em {Predictive Solution to the $X(3872)$
  Collider Production Puzzle}\/},
\href{http://arxiv.org/abs/1811.08876}{{\tt arXiv:1811.08876 [hep-ph]}}.

\bibitem{Esposito:2015fsa}
A.~Esposito, A.~L. Guerrieri, L.~Maiani, F.~Piccinini, A.~Pilloni, A.~D.
  Polosa, and V.~Riquer, {\em {Observation of light nuclei at ALICE and the
  $X(3872)$ conundrum}\/},
  \href{http://dx.doi.org/10.1103/PhysRevD.92.034028}{Phys. Rev. {\bf D92}
  (2015) no.~3, 034028},
\href{http://arxiv.org/abs/1508.00295}{{\tt arXiv:1508.00295 [hep-ph]}}.

\bibitem{Wang:2017gay}
W.~Wang, {\em {On the production of hidden-flavored hadronic states at high
  energy}\/},  \href{http://dx.doi.org/10.1088/1674-1137/42/4/043103}{Chin.
  Phys. {\bf 42} (2018)  043103},
\href{http://arxiv.org/abs/1709.10382}{{\tt arXiv:1709.10382 [hep-ph]}}.

\bibitem{Guo:2013ufa}
F.-K. Guo, U.-G. Mei{\ss}ner, and W.~Wang, {\em {Production of charged heavy
  quarkonium-like states at the LHC and the Tevatron}\/},
  \href{http://dx.doi.org/10.1088/0253-6102/61/3/14}{Commun. Theor. Phys. {\bf
  61} (2014)  354--358},
\href{http://arxiv.org/abs/1308.0193}{{\tt arXiv:1308.0193 [hep-ph]}}.

\bibitem{Guo:2014ppa}
F.-K. Guo, U.-G. Mei{\ss}ner, W.~Wang, and Z.~Yang, {\em {Production of
  charm-strange hadronic molecules at the LHC}\/},
  \href{http://dx.doi.org/10.1007/JHEP05(2014)138}{JHEP {\bf 05} (2014)  138},
\href{http://arxiv.org/abs/1403.4032}{{\tt arXiv:1403.4032 [hep-ph]}}.

\bibitem{LHCb-PAPER-2014-014}
{LHCb collaboration}, R.~Aaij et al., {\em {Observation of the resonant
  character of the $Z(4430)^-$ state}\/},  Phys. Rev. Lett. {\bf 112} (2014)
  222002, \href{http://arxiv.org/abs/1404.1903}{{\tt arXiv:1404.1903
  [hep-ex]}}.

\bibitem{LHCb-PAPER-2016-018}
{LHCb collaboration}, R.~Aaij et al., {\em {Observation of exotic $\jpsi\phi$
  structures from amplitude analysis of $\Bp\to \jpsi\phi\Kp$ decays}\/},
  Phys. Rev. Lett. {\bf 118} (2016)  022003,
  \href{http://arxiv.org/abs/1606.07895}{{\tt arXiv:1606.07895 [hep-ex]}}.

\bibitem{Chatrchyan:2013dma}
{CMS collaboration}, S.~Chatrchyan et al., {\em {Observation of a peaking
  structure in the $J/\psi \phi$ mass spectrum from $B^{\pm} \to J/\psi \phi
  K^{\pm}$ decays}\/},
  \href{http://dx.doi.org/10.1016/j.physletb.2014.05.055}{Phys. Lett. {\bf
  B734} (2014)  261--281},
\href{http://arxiv.org/abs/1309.6920}{{\tt arXiv:1309.6920 [hep-ex]}}.

\bibitem{Aaltonen:2009tz}
{CDF collaboration}, T.~Aaltonen et al., {\em {Evidence for a narrow
  near-threshold structure in the $J/\psi\phi$ mass spectrum in $B^+\to
  J/\psi\phi K^+$ Decays}\/},
  \href{http://dx.doi.org/10.1103/PhysRevLett.102.242002}{Phys. Rev. Lett. {\bf
  102} (2009)  242002},
\href{http://arxiv.org/abs/0903.2229}{{\tt arXiv:0903.2229 [hep-ex]}}.

\bibitem{LHCb-PAPER-2017-011}
{LHCb collaboration}, R.~Aaij et al., {\em {Observation of the decays $\Lb\to
  \chicone\proton\Km$ and $\Lb\to \chictwo\proton\Km$}\/},  Phys. Rev. Lett.
  {\bf 119} (2017)  062001, \href{http://arxiv.org/abs/1704.07900}{{\tt
  arXiv:1704.07900 [hep-ex]}}.

\bibitem{Nieves:2012tt}
J.~Nieves and M.~P. Valderrama, {\em {The heavy quark spin symmetry partners of
  the X(3872)}\/},  \href{http://dx.doi.org/10.1103/PhysRevD.86.056004}{Phys.
  Rev. {\bf D86} (2012)  056004},
\href{http://arxiv.org/abs/1204.2790}{{\tt arXiv:1204.2790 [hep-ph]}}.

\bibitem{Wu:2010vk}
J.-J. Wu, R.~Molina, E.~Oset, and B.~S. Zou, {\em {Dynamically generated
  $N^{*}$ and $\Lambda^*$ resonances in the hidden charm sector around 4.3
  GeV}\/},  \href{http://dx.doi.org/10.1103/PhysRevC.84.015202}{Phys. Rev. {\bf
  C84} (2011)  015202},
\href{http://arxiv.org/abs/1011.2399}{{\tt arXiv:1011.2399 [nucl-th]}}.

\bibitem{LHCB-PAPER-2016-053}
{LHCb collaboration}, R.~Aaij et al., {\em {Observation of the $\Xibm\to
  \jpsi\Lz\Km$ decay}\/},  Phys. Lett. {\bf B772} (2017)  265,
  \href{http://arxiv.org/abs/1701.05274}{{\tt arXiv:1701.05274 [hep-ex]}}.

\bibitem{Moinester:1995fk}
M.~A. Moinester, {\em {How to search for doubly charmed baryons and
  tetraquarks}\/},  \href{http://dx.doi.org/10.1007/s002180050123}{Z. Phys.
  {\bf A355} (1996)  349--362},
\href{http://arxiv.org/abs/hep-ph/9506405}{{\tt arXiv:hep-ph/9506405
  [hep-ph]}}.

\bibitem{DelFabbro:2004ta}
A.~Del~Fabbro, D.~Janc, M.~Rosina, and D.~Treleani, {\em {Production and
  detection of doubly charmed tetraquarks}\/},
  \href{http://dx.doi.org/10.1103/PhysRevD.71.014008}{Phys. Rev. {\bf D71}
  (2005)  014008},
\href{http://arxiv.org/abs/hep-ph/0408258}{{\tt arXiv:hep-ph/0408258
  [hep-ph]}}.

\bibitem{Carames:2011zz}
T.~F. Caram{\'e}s, A.~Valcarce, and J.~Vijande, {\em {Doubly charmed exotic
  mesons: A gift of nature?}\/},
\href{http://dx.doi.org/10.1016/j.physletb.2011.04.023}{Phys. Lett. {\bf B699}
  (2011)  291--295}.

\bibitem{Hyodo:2012pm}
T.~Hyodo, Y.-R. Liu, M.~Oka, K.~Sudoh, and S.~Yasui, {\em {Production of doubly
  charmed tetraquarks with exotic color configurations in electron-positron
  collisions}\/},
  \href{http://dx.doi.org/10.1016/j.physletb.2013.02.045}{Phys. Lett. {\bf
  B721} (2013)  56--60},
\href{http://arxiv.org/abs/1209.6207}{{\tt arXiv:1209.6207 [hep-ph]}}.

\bibitem{Ikeda:2013vwa}
Y.~Ikeda, B.~Charron, S.~Aoki, T.~Doi, T.~Hatsuda, T.~Inoue, N.~Ishii,
  K.~Murano, H.~Nemura, and K.~Sasaki, {\em {Charmed tetraquarks $T_{cc}$ and
  $T_{cs}$ from dynamical lattice QCD simulations}\/},
  \href{http://dx.doi.org/10.1016/j.physletb.2014.01.002}{Phys. Lett. {\bf
  B729} (2014)  85--90},
\href{http://arxiv.org/abs/1311.6214}{{\tt arXiv:1311.6214 [hep-lat]}}.

\bibitem{Guerrieri:2014nxa}
A.~L. Guerrieri, M.~Papinutto, A.~Pilloni, A.~D. Polosa, and N.~Tantalo, {\em
  {Flavored tetraquark spectroscopy}\/},  PoS {\bf LATTICE2014} (2015)  106,
\href{http://arxiv.org/abs/1411.2247}{{\tt arXiv:1411.2247 [hep-lat]}}.

\bibitem{Maciula:2016wci}
R.~Maciu{\l}a, V.~A. Saleev, A.~V. Shipilova, and A.~Szczurek, {\em {New
  mechanisms for double charmed meson production at the LHCb}\/},
  \href{http://dx.doi.org/10.1016/j.physletb.2016.05.052}{Phys. Lett. {\bf
  B758} (2016)  458--464},
\href{http://arxiv.org/abs/1601.06981}{{\tt arXiv:1601.06981 [hep-ph]}}.

\bibitem{Richard:2016eis}
J.-M. Richard, {\em {Exotic hadrons: review and perspectives}\/},
  \href{http://dx.doi.org/10.1007/s00601-016-1159-0}{Few Body Syst. {\bf 57}
  (2016) no.~12, 1185--1212},
\href{http://arxiv.org/abs/1606.08593}{{\tt arXiv:1606.08593 [hep-ph]}}.

\bibitem{Hyodo:2017hue}
T.~Hyodo, Y.-R. Liu, M.~Oka, and S.~Yasui, {\em {Spectroscopy and production of
  doubly charmed tetraquarks}\/},
\href{http://arxiv.org/abs/1708.05169}{{\tt arXiv:1708.05169 [hep-ph]}}.

\bibitem{Wang:2017dtg}
Z.-G. Wang and Z.-H. Yan, {\em {Analysis of the scalar, axialvector, vector,
  tensor doubly charmed tetraquark states with QCD sum rules}\/},
  \href{http://dx.doi.org/10.1140/epjc/s10052-017-5507-0}{Eur. Phys. J. {\bf
  C78} (2018) no.~1, 19},
\href{http://arxiv.org/abs/1710.02810}{{\tt arXiv:1710.02810 [hep-ph]}}.

\bibitem{Yan:2018gik}
X.~Yan, B.~Zhong, and R.~Zhu, {\em {Doubly charmed tetraquarks in a
  diquark-antidiquark model}\/},
\href{http://arxiv.org/abs/1804.06761}{{\tt arXiv:1804.06761 [hep-ph]}}.

\bibitem{LHCb-PAPER-2017-018}
{LHCb collaboration}, R.~Aaij et al., {\em {Observation of the doubly charmed
  baryon $\Xires_{cc}^{++}$}\/},  Phys. Rev. Lett. {\bf 119} (2017)  112001,
  \href{http://arxiv.org/abs/1707.01621}{{\tt arXiv:1707.01621 [hep-ex]}}.

\bibitem{LHCb-PAPER-2015-029}
{LHCb collaboration}, R.~Aaij et al., {\em {Observation of $\jpsi\proton$
  resonances consistent with pentaquark states in $\Lb\to \jpsi\proton\Km$
  decays}\/},  Phys. Rev. Lett. {\bf 115} (2015)  072001,
  \href{http://arxiv.org/abs/1507.03414}{{\tt arXiv:1507.03414 [hep-ex]}}.

\bibitem{LHCb-PAPER-2013-010}
{LHCb collaboration}, R.~Aaij et al., {\em {Observation of $\Bcp\to \jpsi\Dsp$
  and $\Bcp\to \jpsi\Dssp$ decays}\/},  Phys. Rev. {\bf D87} (2013)  112012,
  \href{http://arxiv.org/abs/1304.4530}{{\tt arXiv:1304.4530 [hep-ex]}}.

\bibitem{LHCb-PAPER-2012-003}
{LHCb collaboration}, R.~Aaij et al., {\em {Observation of double charm
  production involving open charm in $\proton\proton$ collisions at
  $\sqrt{s}=7$\tev}\/},  JHEP {\bf 06} (2012)  141,
  \href{http://arxiv.org/abs/1205.0975}{{\tt arXiv:1205.0975 [hep-ex]}}.

\bibitem{LHCb-PAPER-2018-027}
{LHCb collaboration}, R.~Aaij et al., {\em {Search for beautiful tetraquarks in
  the $\PUpsilon\mu^+\mu^-$ invariant-mass spectrum}\/},  JHEP {\bf 10} (2018)
  086, \href{http://arxiv.org/abs/1806.09707}{{\tt arXiv:1806.09707 [hep-ex]}}.

\bibitem{Stewart:2004pd}
I.~W. Stewart, M.~E. Wessling, and M.~B. Wise, {\em {Stable heavy pentaquark
  states}\/},  \href{http://dx.doi.org/10.1016/j.physletb.2004.03.087}{Phys.
  Lett. {\bf B590} (2004)  185--189},
\href{http://arxiv.org/abs/hep-ph/0402076}{{\tt arXiv:hep-ph/0402076
  [hep-ph]}}.

\bibitem{Oh:1994np}
Y.-s. Oh, B.-Y. Park, and D.-P. Min, {\em {Pentaquark exotic baryons in the
  Skyrme model}\/},
  \href{http://dx.doi.org/10.1016/0370-2693(94)91065-0}{Phys. Lett. {\bf B331}
  (1994)  362--370},
\href{http://arxiv.org/abs/hep-ph/9405297}{{\tt arXiv:hep-ph/9405297
  [hep-ph]}}.

\bibitem{LHCb-PAPER-2017-043}
{LHCb collaboration}, R.~Aaij et al., {\em {Search for weakly decaying
  $b$-flavored pentaquarks}\/},  Phys. Rev. {\bf D97} (2018)  032010,
  \href{http://arxiv.org/abs/1712.08086}{{\tt arXiv:1712.08086 [hep-ex]}}.

\bibitem{LHCb-PAPER-2017-002}
{LHCb collaboration}, R.~Aaij et al., {\em {Observation of five new narrow
  $\Omegac$ states decaying to $\Xicp\Km$}\/},  Phys. Rev. Lett. {\bf 118}
  (2017)  182001, \href{http://arxiv.org/abs/1703.04639}{{\tt arXiv:1703.04639
  [hep-ex]}}.

\bibitem{LHCb-TDR-012}
{LHCb collaboration}, {\em {Framework TDR for the LHCb Upgrade: Technical
  Design Report}\/},  2012.
\newblock LHCb-TDR-012.

\bibitem{LHCb-PAPER-2012-031}
{LHCb collaboration}, {R. Aaij \emph{et al.}, and A. Bharucha} et al., {\em
  {Implications of LHCb measurements and future prospects}\/},  Eur. Phys. J.
  {\bf C73} (2013)  2373, \href{http://arxiv.org/abs/1208.3355}{{\tt
  arXiv:1208.3355 [hep-ex]}}.

\bibitem{Gutsche:2017hux}
T.~Gutsche, M.~A. Ivanov, J.~G. K{\"o}rner, and V.~E. Lyubovitskij, {\em {Decay
  chain information on the newly discovered double charm baryon state
  $\PXi_{cc}^{++}$}\/},
  \href{http://dx.doi.org/10.1103/PhysRevD.96.054013}{Phys. Rev. {\bf D96}
  (2017)  054013},
\href{http://arxiv.org/abs/1708.00703}{{\tt arXiv:1708.00703 [hep-ph]}}.

\bibitem{Sharma:2017txj}
N.~Sharma and R.~Dhir, {\em {Estimates of W-exchange contributions to
  $\PXi_{cc}$ decays}\/},
  \href{http://dx.doi.org/10.1103/PhysRevD.96.113006}{Phys. Rev. {\bf D96}
  (2017)  113006},
\href{http://arxiv.org/abs/1709.08217}{{\tt arXiv:1709.08217 [hep-ph]}}.

\bibitem{LHCb-PAPER-2013-049}
{LHCb collaboration}, R.~Aaij et al., {\em {Search for the doubly charmed
  baryon $\Xi_{cc}^+$}\/},  JHEP {\bf 12} (2013)  090,
  \href{http://arxiv.org/abs/1310.2538}{{\tt arXiv:1310.2538 [hep-ex]}}.

\bibitem{Fleck:1989mb}
S.~Fleck and J.-M. Richard, {\em {Baryons with double charm}\/},
\href{http://dx.doi.org/10.1143/PTP.82.760}{Prog. Theor. Phys. {\bf 82} (1989)
  760--774}.

\bibitem{Guberina:1999mx}
B.~Guberina, B.~Meli\'{c}, and H.~\v{S}tefan\v{c}i\'c, {\em {Inclusive decays
  and lifetimes of doubly charmed baryons}\/},
  \href{http://dx.doi.org/10.1007/s100529900039}{Eur.Phys.J. {\bf C9} (1999)
  213--219},
\href{http://arxiv.org/abs/hep-ph/9901323}{{\tt arXiv:hep-ph/9901323
  [hep-ph]}}.

\bibitem{Kiselev:1998sy}
V.~Kiselev, A.~Likhoded, and A.~Onishchenko, {\em {Lifetimes of doubly charmed
  baryons: \Xiccp and \Xiccpp}\/},
  \href{http://dx.doi.org/10.1103/PhysRevD.60.014007}{Phys.Rev. {\bf D60}
  (1999)  014007},
\href{http://arxiv.org/abs/hep-ph/9807354}{{\tt arXiv:hep-ph/9807354
  [hep-ph]}}.

\bibitem{Chang:2007xa}
C.-H. Chang, T.~Li, X.-Q. Li, and Y.-M. Wang, {\em {Lifetime of doubly charmed
  baryons}\/},
  \href{http://dx.doi.org/10.1088/0253-6102/49/4/38}{Commun.Theor.Phys. {\bf
  49} (2008)  993--1000},
\href{http://arxiv.org/abs/0704.0016}{{\tt arXiv:0704.0016 [hep-ph]}}.

\bibitem{Berezhnoy:2016wix}
A.~V. Berezhnoy and A.~K. Likhoded, {\em {Doubly heavy baryons}\/},
\href{http://dx.doi.org/10.1134/S1063778816010087}{Phys. Atom. Nucl. {\bf 79}
  (2016)  260--265}.

\bibitem{LHCb-PAPER-2018-019}
{LHCb collaboration}, R.~Aaij et al., {\em {Measurement of the lifetime of the
  doubly charmed baryon $\Xires_{cc}^{++}$}\/},  Phys. Rev. Lett. {\bf 121}
  (2018)  052002, \href{http://arxiv.org/abs/1806.02744}{{\tt arXiv:1806.02744
  [hep-ex]}}.

\bibitem{Zhang:2011hi}
J.-W. Zhang, X.-G. Wu, T.~Zhong, Y.~Yu, and Z.-Y. Fang, {\em {Production of the
  doubly heavy baryon $\PXi_{bc}$ at LHC}\/},
  \href{http://dx.doi.org/10.1103/PhysRevD.83.034026}{Phys. Rev. {\bf D83}
  (2011)  034026},
\href{http://arxiv.org/abs/1101.1130}{{\tt arXiv:1101.1130 [hep-ph]}}.

\bibitem{Chang:2003cr}
C.-H. Chang and X.-G. Wu, {\em {Uncertainties in estimating hadronic production
  of the meson $B_c$ and comparisons between TEVATRON and LHC}\/},
  \href{http://dx.doi.org/10.1140/epjc/s2004-02015-0}{Eur. Phys. J. {\bf C38}
  (2004)  267--276},
\href{http://arxiv.org/abs/hep-ph/0309121}{{\tt arXiv:hep-ph/0309121
  [hep-ph]}}.

\bibitem{Gao:2010zzc}
Y.-N. Gao, J.~He, P.~Robbe, M.-H. Schune, and Z.-W. Yang, {\em {Experimental
  prospects of the $B_c$ studies of the LHCb experiment}\/},
\href{http://dx.doi.org/10.1088/0256-307X/27/6/061302}{Chin. Phys. Lett. {\bf
  27} (2010)  061302}.

\bibitem{LHCb-PAPER-2011-018}
{LHCb collaboration}, R.~Aaij et al., {\em {Measurement of $\bquark$ hadron
  production fractions in 7\,TeV $\proton\proton$ collisions}\/},  Phys. Rev.
  {\bf D85} (2012)  032008, \href{http://arxiv.org/abs/1111.2357}{{\tt
  arXiv:1111.2357 [hep-ex]}}.

\bibitem{LHCb-PAPER-2014-004}
{LHCb collaboration}, R.~Aaij et al., {\em {Study of the kinematic dependences
  of $\Lb$ production in $\proton\proton$ collisions and a measurement of the
  $\Lb\to \Lc\pim$ branching fraction}\/},  JHEP {\bf 08} (2014)  143,
  \href{http://arxiv.org/abs/1405.6842}{{\tt arXiv:1405.6842 [hep-ex]}}.

\bibitem{LHCb-PAPER-2018-032}
{LHCb collaboration}, R.~Aaij et al., {\em {Observation of two resonances in
  the $\Lb\pipm$ systems and precise measurement of $\Sigmares_b^\pm$ and
  $\Sigmares_b^{\ast\pm}$ properties}\/},
  \href{http://arxiv.org/abs/1809.07752}{{\tt arXiv:1809.07752 [hep-ex]}}.

\bibitem{LHCb-PAPER-2011-030}
{LHCb collaboration}, R.~Aaij et al., {\em {Measurement of the ratio of prompt
  $\chi_{c}$ to $\jpsi$ production in $\proton\proton$ collisions at
  $\sqrt{s}=7$\tev}\/},  Phys. Lett. {\bf B718} (2012)  431,
  \href{http://arxiv.org/abs/1204.1462}{{\tt arXiv:1204.1462 [hep-ex]}}.

\bibitem{LHCb-PAPER-2014-031}
{LHCb collaboration}, R.~Aaij et al., {\em {Study of $\chi_b$ meson production
  in $\proton\proton$ collisions at $\sqrt{s}=7$ and $8$\tev and observation of
  the decay $\chi_b\to \ThreeS\gamma$}\/},  Eur. Phys. J. {\bf C74} (2014)
  3092, \href{http://arxiv.org/abs/1407.7734}{{\tt arXiv:1407.7734 [hep-ex]}}.

\bibitem{LHCb-PAPER-2017-036}
{LHCb collaboration}, R.~Aaij et al., {\em {$\chi_{c1}$ and $\chi_{c2}$
  resonance parameters with the decays $\chi_{c1,c2}\to \jpsi\mumu$}\/},  Phys.
  Rev. Lett. {\bf 119} (2017)  221801,
  \href{http://arxiv.org/abs/1709.04247}{{\tt arXiv:1709.04247 [hep-ex]}}.

\bibitem{LHCb-PAPER-2011-019}
{LHCb collaboration}, R.~Aaij et al., {\em {Measurement of the cross-section
  ratio $\sigma(\chictwo)/\sigma(\chicone)$ for prompt $\chi_c$ production at
  $\sqrt{s}=7$\tev}\/},  Phys. Lett. {\bf B714} (2012)  215--223,
  \href{http://arxiv.org/abs/1202.1080}{{\tt arXiv:1202.1080 [hep-ex]}}.

\bibitem{LHCb-PAPER-2012-015}
{LHCb collaboration}, R.~Aaij et al., {\em {Measurement of the fraction of
  $\OneS$ originating from $\chi_b(1P)$ decays in $\proton\proton$ collisions
  at $\sqrt{s}=7$\tev}\/},  JHEP {\bf 11} (2012)  031,
  \href{http://arxiv.org/abs/1209.0282}{{\tt arXiv:1209.0282 [hep-ex]}}.

\bibitem{LHCb-PAPER-2013-028}
{LHCb collaboration}, R.~Aaij et al., {\em {Measurement of the relative rate of
  prompt $\chiczero$, $\chicone$ and $\chictwo$ production at
  $\sqrt{s}=7$\tev}\/},  JHEP {\bf 10} (2013)  115,
  \href{http://arxiv.org/abs/1307.4285}{{\tt arXiv:1307.4285 [hep-ex]}}.

\bibitem{LHCb-PAPER-2014-040}
{LHCb collaboration}, R.~Aaij et al., {\em {Measurement of the $\chi_b(3P)$
  mass and of the relative rate of $\chi_{b1}(1P)$ and $\chi_{b2}(1P)$
  production}\/},  JHEP {\bf 10} (2014)  088,
  \href{http://arxiv.org/abs/1409.1408}{{\tt arXiv:1409.1408 [hep-ex]}}.

\bibitem{Faessler:1999de}
A.~Faessler, C.~Fuchs, and M.~I. Krivoruchenko, {\em {Dilepton spectra from
  decays of light unflavored mesons}\/},
  \href{http://dx.doi.org/10.1103/PhysRevC.61.035206}{Phys. Rev. {\bf C61}
  (2000)  035206},
\href{http://arxiv.org/abs/nucl-th/9904024}{{\tt arXiv:nucl-th/9904024
  [nucl-th]}}.

\bibitem{Luchinsky:2017pby}
A.~V. Luchinsky, {\em {Muon pair production in radiative decays of heavy
  quarkonia}\/},  \href{http://dx.doi.org/10.1142/S0217732318500013}{Mod. Phys.
  Lett. {\bf A33} (2017)  1850001},
\href{http://arxiv.org/abs/1709.02444}{{\tt arXiv:1709.02444 [hep-ph]}}.

\bibitem{LHCb-PAPER-2011-013}
{LHCb collaboration}, R.~Aaij et al., {\em {Observation of $\jpsi$-pair
  production in $\proton\proton$ collisions at $\sqrt{s}=7$\tev}\/},  Phys.
  Lett. {\bf B707} (2012)  52, \href{http://arxiv.org/abs/1109.0963}{{\tt
  arXiv:1109.0963 [hep-ex]}}.

\bibitem{LHCb-PAPER-2016-057}
{LHCb collaboration}, R.~Aaij et al., {\em {Measurement of the \jpsi pair
  production cross-section in $\proton\proton$ collisions at
  $\sqrt{s}=13$\,TeV}\/},  JHEP {\bf 06} (2017)  047,
  \href{http://arxiv.org/abs/1612.07451}{{\tt arXiv:1612.07451 [hep-ex]}}.

\bibitem{Sun:2014gca}
L.-P. Sun, H.~Han, and K.-T. Chao, {\em {Impact of $\jpsi$ pair production at
  the LHC and predictions in nonrelativistic QCD}\/},
  \href{http://dx.doi.org/10.1103/PhysRevD.94.074033}{Phys. Rev. {\bf D94}
  (2016)  074033},
\href{http://arxiv.org/abs/1404.4042}{{\tt arXiv:1404.4042 [hep-ph]}}.

\bibitem{Likhoded:2016zmk}
A.~K. Likhoded, A.~V. Luchinsky, and S.~V. Poslavsky, {\em {Production of
  $\jpsi + \chi_c$ and $\jpsi + \jpsi$ with real gluon emission at LHC}\/},
  \href{http://dx.doi.org/10.1103/PhysRevD.94.054017}{Phys. Rev. {\bf D94}
  (2016)  054017},
\href{http://arxiv.org/abs/1606.06767}{{\tt arXiv:1606.06767 [hep-ph]}}.

\bibitem{Lansberg:2013qka}
J.-P. Lansberg and H.-S. Shao, {\em {Production of $\jpsi +
  \Peta_{\cquark}$~versus $\jpsi + \jpsi$~at the LHC: Importance of real
  $\upalpha^{5}_{\mathrm{s}}$~corrections}\/},
  \href{http://dx.doi.org/10.1103/PhysRevLett.111.122001}{Phys.\ Rev.\ Lett.
  {\bf 111} (2013)  122001},
\href{http://arxiv.org/abs/1308.0474}{{\tt arXiv:1308.0474 [hep-ph]}}.

\bibitem{Lansberg:2014swa}
J.-P. Lansberg and H.-S. Shao, {\em {\jpsi-pair production at large momenta:
  indications for double parton scatterings and large
  $\upalpha_{\mathrm{s}}^5$~contributions}\/},
  \href{http://dx.doi.org/10.1016/j.physletb.2015.10.083}{Phys.\ Lett. {\bf
  B751} (2015)  479},
\href{http://arxiv.org/abs/1410.8822}{{\tt arXiv:1410.8822 [hep-ph]}}.

\bibitem{Lansberg:2015lva}
J.-P. Lansberg and H.-S. Shao, {\em {Double-quarkonium production at a
  fixed-target experiment at the LHC (AFTER@LHC)}\/},
  \href{http://dx.doi.org/10.1016/j.nuclphysb.2015.09.005}{Nucl. Phys. {\bf
  B900} (2015)  273--294},
\href{http://arxiv.org/abs/1504.06531}{{\tt arXiv:1504.06531 [hep-ph]}}.

\bibitem{Shao:2012iz}
H.-S. Shao, {\em {{\sc{HELAC-Onia}}: An automatic matrix element generator for
  heavy quarkonium physics}\/},
  \href{http://dx.doi.org/10.1016/j.cpc.2013.05.023}{Comput. Phys. Commun. {\bf
  184} (2013)  2562},
\href{http://arxiv.org/abs/1212.5293}{{\tt arXiv:1212.5293 [hep-ph]}}.

\bibitem{Shao:2015vga}
H.-S. Shao, {\em {{\sc{HELAC-Onia~2.0}}: An~upgraded matrix-element and event
  generator for heavy quarkonium physics}\/},
  \href{http://dx.doi.org/10.1016/j.cpc.2015.09.011}{Comput.\ Phys.\ Commun.
  {\bf 198} (2016)  238},
\href{http://arxiv.org/abs/1507.03435}{{\tt arXiv:1507.03435 [hep-ph]}}.

\bibitem{Baranov:2011zz}
S.~P. Baranov, {\em {Pair production of \jpsi~mesons in the
  $k_{\mathrm{T}}$-factorization approach}\/},
\href{http://dx.doi.org/10.1103/PhysRevD.84.054012}{Phys.\ Rev. {\bf D84}
  (2011)  054012}.

\bibitem{Baranov:1993qv}
S.~P. Baranov and H.~Jung, {\em {Double \jpsi~production: A~probe of gluon
  polarization?}\/},
\href{http://dx.doi.org/10.1007/BF01579639}{Z.\ Phys. {\bf C66} (1995)  647}.

\bibitem{Bansal:2014paa}
S.~Bansal et al., {\em {Progress in double parton scattering studies}\/},  in
  {\em {Workshop on Multi-Parton Interactions at the LHC (MPI @ LHC 2013)
  Antwerp, Belgium, December 2-6, 2013}}.
\newblock \href{http://arxiv.org/abs/1410.6664}{{\tt arXiv:1410.6664
  [hep-ph]}}.
\newblock
\url{https://inspirehep.net/record/1323623/files/arXiv:1410.6664.pdf}.
\newblock

\bibitem{Belyaev:2017sws}
I.~Belyaev and D.~Savrina, {\em {Study of double parton scattering processes
  with heavy quarks}\/}, .
\newblock \href{http://arxiv.org/abs/1711.10877}{{\tt arXiv:1711.10877
  [hep-ex]}}.
\newblock
\url{http://inspirehep.net/record/1639442/files/arXiv:1711.10877.pdf}.
\newblock

\bibitem{Hurth:2017sqw}
T.~Hurth, C.~Langenbruch, and F.~Mahmoudi, {\em {Direct determination of Wilson
  coefficients using $B^0\to K^{*0}\mu^+\mu^-$ decays}\/},
  \href{http://dx.doi.org/10.1007/JHEP11(2017)176}{JHEP {\bf 11} (2017)  176},
  \href{http://arxiv.org/abs/1708.04474}{{\tt arXiv:1708.04474 [hep-ph]}}.

\bibitem{Blake:2017fyh}
T.~Blake, U.~Egede, P.~Owen, K.~A. Petridis, and G.~Pomery, {\em {An empirical
  model to determine the hadronic resonance contributions to $\overline{B}{} ^0
  \!\rightarrow \overline{K}{} ^{*0} \mu ^+ \mu ^- $ transitions}\/},
  \href{http://dx.doi.org/10.1140/epjc/s10052-018-5937-3}{Eur. Phys. J. {\bf
  C78} (2018) no.~6, 453}, \href{http://arxiv.org/abs/1709.03921}{{\tt
  arXiv:1709.03921 [hep-ph]}}.

\bibitem{Chrzaszcz:2018yza}
M.~Chrzaszcz, A.~Mauri, N.~Serra, R.~Silva~Coutinho, and D.~van Dyk, {\em
  {Prospects for disentangling long- and short-distance effects in the decays
  $B\to K^* \mu^+\mu^-$}\/},  \href{http://arxiv.org/abs/1805.06378}{{\tt
  arXiv:1805.06378 [hep-ph]}}.

\bibitem{Aaij:2244311}
{LHCb Collaboration}, R.~Aaij et al., {\em {Expression of Interest for a
  Phase-II LHCb Upgrade: Opportunities in flavour physics, and beyond, in the
  HL-LHC era}\/},   CERN-LHCC-2017-003, CERN, Geneva, Feb, 2017.
\newblock \url{http://cds.cern.ch/record/2244311}.

\bibitem{Buchalla:1995vs}
G.~Buchalla, A.~J. Buras, and M.~E. Lautenbacher, {\em {Weak decays beyond
  leading logarithms}\/},  Rev. Mod. Phys. {\bf 68} (1996)  1125--1144,
  \href{http://arxiv.org/abs/hep-ph/9512380}{{\tt arXiv:hep-ph/9512380
  [hep-ph]}}.

\bibitem{Bobeth:2007dw}
C.~Bobeth, G.~Hiller, and G.~Piranishvili, {\em {Angular distributions of
  $\bar{B} \to \bar{K} \ell^+\ell^-$ decays}\/},  JHEP {\bf 12} (2007)  040,
  \href{http://arxiv.org/abs/0709.4174}{{\tt arXiv:0709.4174 [hep-ph]}}.

\bibitem{1304.6325}
{LHCb Collaboration}, R.~Aaij et al., {\em {Differential branching fraction and
  angular analysis of the decay $B^{0} \to K^{*0} \mu^{+}\mu^{-}$}\/},  JHEP
  {\bf 08} (2013)  131, \href{http://arxiv.org/abs/1304.6325}{{\tt
  arXiv:1304.6325 [hep-ex]}}.

\bibitem{1307.5683}
S.~Descotes-Genon, J.~Matias, and J.~Virto, {\em {Understanding the $B\to
  K^*\mu^+\mu^-$ Anomaly}\/},  Phys. Rev. {\bf D88} (2013)  074002,
  \href{http://arxiv.org/abs/1307.5683}{{\tt arXiv:1307.5683 [hep-ph]}}.

\bibitem{1308.1501}
W.~Altmannshofer and D.~M. Straub, {\em {New Physics in $B \to K^*\mu\mu$?}\/},
   Eur. Phys. J. {\bf C73} (2013)  2646,
  \href{http://arxiv.org/abs/1308.1501}{{\tt arXiv:1308.1501 [hep-ph]}}.

\bibitem{Beaujean:2013soa}
F.~Beaujean, C.~Bobeth, and D.~van Dyk, {\em {Comprehensive Bayesian analysis
  of rare (semi)leptonic and radiative $B$ decays}\/},  Eur. Phys. J. {\bf C74}
  (2014)  2897, \href{http://arxiv.org/abs/1310.2478}{{\tt arXiv:1310.2478
  [hep-ph]}}.

\bibitem{Hurth:2013ssa}
T.~Hurth and F.~Mahmoudi, {\em {On the LHCb anomaly in $B\to
  K^*\ell^+\ell^-$}\/},  JHEP {\bf 04} (2014)  097,
  \href{http://arxiv.org/abs/1312.5267}{{\tt arXiv:1312.5267 [hep-ph]}}.

\bibitem{Altmannshofer:2017fio}
W.~Altmannshofer, C.~Niehoff, P.~Stangl, and D.~M. Straub, {\em {Status of the
  $B\rightarrow K^*\mu ^+\mu ^-$ anomaly after Moriond 2017}\/},  Eur. Phys. J.
  {\bf C77} (2017) no.~6, 377, \href{http://arxiv.org/abs/1703.09189}{{\tt
  arXiv:1703.09189 [hep-ph]}}.

\bibitem{Capdevila:2017bsm}
B.~Capdevila, A.~Crivellin, S.~Descotes-Genon, J.~Matias, and J.~Virto, {\em
  {Patterns of New Physics in $b\to s\ell^+\ell^-$ transitions in the light of
  recent data}\/},  JHEP {\bf 01} (2018)  093,
  \href{http://arxiv.org/abs/1704.05340}{{\tt arXiv:1704.05340 [hep-ph]}}.

\bibitem{Altmannshofer:2017yso}
W.~Altmannshofer, P.~Stangl, and D.~M. Straub, {\em {Interpreting Hints for
  Lepton Flavor Universality Violation}\/},  Phys. Rev. {\bf D96} (2017) no.~5,
  055008, \href{http://arxiv.org/abs/1704.05435}{{\tt arXiv:1704.05435
  [hep-ph]}}.

\bibitem{Hurth:2017hxg}
T.~Hurth, F.~Mahmoudi, D.~Martinez~Santos, and S.~Neshatpour, {\em {Lepton
  nonuniversality in exclusive $b{\rightarrow}s{\ell}{\ell}$ decays}\/},  Phys.
  Rev. {\bf D96} (2017) no.~9, 095034,
  \href{http://arxiv.org/abs/1705.06274}{{\tt arXiv:1705.06274 [hep-ph]}}.

\bibitem{Ciuchini:2017mik}
M.~Ciuchini, A.~M. Coutinho, M.~Fedele, E.~Franco, A.~Paul, L.~Silvestrini, and
  M.~Valli, {\em {On Flavourful Easter eggs for New Physics hunger and Lepton
  Flavour Universality violation}\/},  Eur. Phys. J. {\bf C77} (2017) no.~10,
  688, \href{http://arxiv.org/abs/1704.05447}{{\tt arXiv:1704.05447 [hep-ph]}}.

\bibitem{Geng:2017svp}
L.-S. Geng, B.~Grinstein, S.~J{\"a}ger, J.~Martin~Camalich, X.-L. Ren, and
  R.-X. Shi, {\em {Towards the discovery of new physics with
  lepton-universality ratios of $b\to s\ell\ell$ decays}\/},
  \href{http://arxiv.org/abs/1704.05446}{{\tt arXiv:1704.05446 [hep-ph]}}.

\bibitem{1708.09152}
M.~Beneke, C.~Bobeth, and R.~Szafron, {\em {Enhanced electromagnetic correction
  to the rare $B$-meson decay $B_{s,d} \to \mu^+ \mu^-$}\/},  Phys. Rev. Lett.
  {\bf 120} (2018) no.~1, 011801, \href{http://arxiv.org/abs/1708.09152}{{\tt
  arXiv:1708.09152 [hep-ph]}}.

\bibitem{Bazavov:2017lyh}
A.~Bazavov et al., {\em {$B$- and $D$-meson leptonic decay constants from
  four-flavor lattice QCD}\/},
\href{http://arxiv.org/abs/1712.09262}{{\tt arXiv:1712.09262 [hep-lat]}}.

\bibitem{Bussone:2016iua}
{ETM Collaboration}, A.~Bussone et al., {\em {Mass of the b quark and B -meson
  decay constants from N$_f$=2+1+1 twisted-mass lattice QCD}\/},
  \href{http://dx.doi.org/10.1103/PhysRevD.93.114505}{Phys. Rev. {\bf D93}
  (2016) no.~11, 114505},
\href{http://arxiv.org/abs/1603.04306}{{\tt arXiv:1603.04306 [hep-lat]}}.

\bibitem{Hughes:2017spc}
C.~Hughes, C.~T.~H. Davies, and C.~J. Monahan, {\em {New methods for B meson
  decay constants and form factors from lattice NRQCD}\/},
  \href{http://dx.doi.org/10.1103/PhysRevD.97.054509}{Phys. Rev. {\bf D97}
  (2018) no.~5, 054509},
\href{http://arxiv.org/abs/1711.09981}{{\tt arXiv:1711.09981 [hep-lat]}}.

\bibitem{Aaboud:2018mst}
{ATLAS Collaboration}, M.~Aaboud et al., {\em {Study of the rare decays of
  $B^0_s$ and $B^0$ mesons into muon pairs using data collected during 2015 and
  2016 with the ATLAS detector}\/},
\href{http://arxiv.org/abs/1812.03017}{{\tt arXiv:1812.03017 [hep-ex]}}.

\bibitem{Alonso:2014csa}
R.~Alonso, B.~Grinstein, and J.~Martin~Camalich, {\em {$SU(2)\times U(1)$ gauge
  invariance and the shape of new physics in rare $B$ decays}\/},  Phys. Rev.
  Lett. {\bf 113} (2014)  241802, \href{http://arxiv.org/abs/1407.7044}{{\tt
  arXiv:1407.7044 [hep-ph]}}.

\bibitem{Arbey:2018ics}
A.~Arbey, T.~Hurth, F.~Mahmoudi, and S.~Neshatpour, {\em {Hadronic and New
  Physics Contributions to $b \to s$ Transitions}\/},
  \href{http://dx.doi.org/10.1103/PhysRevD.98.095027}{Phys. Rev. {\bf D98}
  (2018) no.~9, 095027},
\href{http://arxiv.org/abs/1806.02791}{{\tt arXiv:1806.02791 [hep-ph]}}.

\bibitem{Calibbi:2015kma}
L.~Calibbi, A.~Crivellin, and T.~Ota, {\em {Effective Field Theory Approach to
  $b\to \ell\ell^{(\prime)}, B\to K^{(*)}\nu\overline{\nu}$ and $B\to
  D^{(*)}\tau\nu$ with Third Generation Couplings}\/},
  \href{http://dx.doi.org/10.1103/PhysRevLett.115.181801}{Phys. Rev. Lett. {\bf
  115} (2015)  181801}, \href{http://arxiv.org/abs/1506.02661}{{\tt
  arXiv:1506.02661 [hep-ph]}}.

\bibitem{Crivellin:2015era}
A.~Crivellin, L.~Hofer, J.~Matias, U.~Nierste, S.~Pokorski, and J.~Rosiek, {\em
  {Lepton-flavour violating $B$ decays in generic $Z'$ models}\/},  Phys. Rev.
  {\bf D92} (2015)  054013, \href{http://arxiv.org/abs/1504.07928}{{\tt
  arXiv:1504.07928 [hep-ph]}}.

\bibitem{DeBruyn:2012wj}
K.~De~Bruyn, R.~Fleischer, R.~Knegjens, P.~Koppenburg, M.~Merk, and N.~Tuning,
  {\em {Branching Ratio Measurements of $B_s$ Decays}\/},
  \href{http://dx.doi.org/10.1103/PhysRevD.86.014027}{Phys. Rev. {\bf D86}
  (2012)  014027},
\href{http://arxiv.org/abs/1204.1735}{{\tt arXiv:1204.1735 [hep-ph]}}.

\bibitem{Altmannshofer:2017wqy}
W.~Altmannshofer, C.~Niehoff, and D.~M. Straub, {\em {$B_s\to\mu^+\mu^-$ as
  current and future probe of new physics}\/},
  \href{http://dx.doi.org/10.1007/JHEP05(2017)076}{JHEP {\bf 05} (2017)  076},
  \href{http://arxiv.org/abs/1702.05498}{{\tt arXiv:1702.05498 [hep-ph]}}.

\bibitem{Bobeth:2013uxa}
C.~Bobeth, M.~Gorbahn, T.~Hermann, M.~Misiak, E.~Stamou, et al., {\em
  {$B_{s,d}\to\ell^+\ell^-$ in the Standard Model with reduced theoretical
  uncertainty}\/},  Phys. Rev. Lett. {\bf 112} (2014)  101801,
  \href{http://arxiv.org/abs/1311.0903}{{\tt arXiv:1311.0903 [hep-ph]}}.

\bibitem{Buras:2003td}
A.~J. Buras, {\em {Relations between $\Delta M_{s,d}$ and $B_{s, d} \to \mu
  \bar{\mu}$ in models with minimal flavor violation}\/},
  \href{http://dx.doi.org/10.1016/S0370-2693(03)00561-6}{Phys. Lett. {\bf B566}
  (2003)  115--119},
\href{http://arxiv.org/abs/hep-ph/0303060}{{\tt arXiv:hep-ph/0303060
  [hep-ph]}}.

\bibitem{Dettori:2016zff}
F.~Dettori, D.~Guadagnoli, and M.~Reboud, {\em {$B^{0}_{s} \to
  \mu^{+}\mu^{-}\gamma$ from $B^{0}_{s} \to \mu^{+}\mu^{-}$}\/},  Phys. Lett.
  {\bf B768} (2017)  163--167, \href{http://arxiv.org/abs/1610.00629}{{\tt
  arXiv:1610.00629 [hep-ph]}}.

\bibitem{Guadagnoli:2017quo}
D.~Guadagnoli, M.~Reboud, and R.~Zwicky, {\em {$B_{s}^{0}\to\ell^{+} \ell^{−}
  \gamma$ as a test of lepton flavor universality}\/},
  \href{http://dx.doi.org/10.1007/JHEP11(2017)184}{JHEP {\bf 11} (2017)  184},
  \href{http://arxiv.org/abs/1708.02649}{{\tt arXiv:1708.02649 [hep-ph]}}.

\bibitem{Melikhov:1998cd}
D.~Melikhov, N.~Nikitin, and S.~Simula, {\em {Probing right-handed currents in
  $B \to K^* \ell^+\ell^-$ transitions}\/},
  \href{http://dx.doi.org/10.1016/S0370-2693(98)01271-4}{Phys. Lett. {\bf B442}
  (1998)  381--389},
\href{http://arxiv.org/abs/hep-ph/9807464}{{\tt arXiv:hep-ph/9807464
  [hep-ph]}}.

\bibitem{0811.1214}
W.~Altmannshofer, P.~Ball, A.~Bharucha, A.~J. Buras, D.~M. Straub, and M.~Wick,
  {\em {Symmetries and Asymmetries of $B \to K^{*} \mu^{+} \mu^{-}$ Decays in
  the Standard Model and Beyond}\/},  JHEP {\bf 01} (2009)  019,
  \href{http://arxiv.org/abs/0811.1214}{{\tt arXiv:0811.1214 [hep-ph]}}.

\bibitem{1202.4266}
J.~Matias, F.~Mescia, M.~Ramon, and J.~Virto, {\em {Complete Anatomy of
  $\bar{B}_d \to \bar{K}^{* 0} (\to K \pi)l^+l^-$ and its angular
  distribution}\/},  JHEP {\bf 04} (2012)  104,
  \href{http://arxiv.org/abs/1202.4266}{{\tt arXiv:1202.4266 [hep-ph]}}.

\bibitem{Bobeth:2008ij}
C.~Bobeth, G.~Hiller, and G.~Piranishvili, {\em {CP Asymmetries in bar $B \to
  \bar{K}^* (\to \bar{K} \pi) \bar{\ell} \ell$ and Untagged $\bar{B}_s$, $B_s
  \to \phi (\to K^{+} K^-) \bar{\ell} \ell$ Decays at NLO}\/},
  \href{http://dx.doi.org/10.1088/1126-6708/2008/07/106}{JHEP {\bf 07} (2008)
  106},
\href{http://arxiv.org/abs/0805.2525}{{\tt arXiv:0805.2525 [hep-ph]}}.

\bibitem{1207.2753}
S.~Descotes-Genon, J.~Matias, M.~Ramon, and J.~Virto, {\em {Implications from
  clean observables for the binned analysis of $B \to K^*\mu^+\mu^-$ at large
  recoil}\/},  JHEP {\bf 01} (2013)  048,
  \href{http://arxiv.org/abs/1207.2753}{{\tt arXiv:1207.2753 [hep-ph]}}.

\bibitem{1303.5794}
S.~Descotes-Genon, T.~Hurth, J.~Matias, and J.~Virto, {\em {Optimizing the
  basis of $B\to K^*ll$ observables in the full kinematic range}\/},  JHEP {\bf
  05} (2013)  137, \href{http://arxiv.org/abs/1303.5794}{{\tt arXiv:1303.5794
  [hep-ph]}}.

\bibitem{1006.5013}
C.~Bobeth, G.~Hiller, and D.~van Dyk, {\em {The Benefits of $\bar{B} \to
  \bar{K}^* l^+ l^-$ Decays at Low Recoil}\/},  JHEP {\bf 07} (2010)  098,
  \href{http://arxiv.org/abs/1006.5013}{{\tt arXiv:1006.5013 [hep-ph]}}.

\bibitem{Jager:2012uw}
S.~J{\"a}ger and J.~Martin~Camalich, {\em {On $B \to V \ell \ell$ at small
  dilepton invariant mass, power corrections, and new physics}\/},
  \href{http://dx.doi.org/10.1007/JHEP05(2013)043}{JHEP {\bf 05} (2013)  043},
  \href{http://arxiv.org/abs/1212.2263}{{\tt arXiv:1212.2263 [hep-ph]}}.

\bibitem{Jager:2014rwa}
{J{\"a}ger, Sebastian and Martin Camalich, Jorge}, {\em {Reassessing the
  discovery potential of the $B \to K^{*} \ell^+\ell^-$ decays in the
  large-recoil region: SM challenges and BSM opportunities}\/},  Phys. Rev.
  {\bf D93} (2016) no.~1, 014028, \href{http://arxiv.org/abs/1412.3183}{{\tt
  arXiv:1412.3183 [hep-ph]}}.

\bibitem{1707.07305}
C.~Bobeth, M.~Chrzaszcz, D.~van Dyk, and J.~Virto, {\em {Long-distance effects
  in $B\rightarrow K^*\ell \ell $ from analyticity}\/},  Eur. Phys. J. {\bf
  C78} (2018) no.~6, 451, \href{http://arxiv.org/abs/1707.07305}{{\tt
  arXiv:1707.07305 [hep-ph]}}.

\bibitem{hep-ph/0611193}
A.~Khodjamirian, T.~Mannel, and N.~Offen, {\em {Form-factors from light-cone
  sum rules with B-meson distribution amplitudes}\/},  Phys. Rev. {\bf D75}
  (2007)  054013, \href{http://arxiv.org/abs/hep-ph/0611193}{{\tt
  arXiv:hep-ph/0611193 [hep-ph]}}.

\bibitem{1503.05534}
A.~Bharucha, D.~M. Straub, and R.~Zwicky, {\em {$B\to V\ell^+\ell^-$ in the
  Standard Model from light-cone sum rules}\/},  JHEP {\bf 08} (2016)  098,
  \href{http://arxiv.org/abs/1503.05534}{{\tt arXiv:1503.05534 [hep-ph]}}.

\bibitem{Khodjamirian:2017fxg}
A.~Khodjamirian and A.~V. Rusov, {\em {$B_{s}\to K \ell \nu_\ell$ and $B_{(s)}
  \to \pi (K) \ell^+\ell^-$ decays at large recoil and CKM matrix elements}\/},
   \href{http://dx.doi.org/10.1007/JHEP08(2017)112}{JHEP {\bf 08} (2017)  112},
\href{http://arxiv.org/abs/1703.04765}{{\tt arXiv:1703.04765 [hep-ph]}}.

\bibitem{Gubernari:2018wyi}
N.~Gubernari, A.~Kokulu, and D.~van Dyk, {\em {$B\to P$ and $B\to V$ Form
  Factors from $B$-Meson Light-Cone Sum Rules beyond Leading Twist}\/},
\href{http://arxiv.org/abs/1811.00983}{{\tt arXiv:1811.00983 [hep-ph]}}.

\bibitem{1509.06235}
J.~A. Bailey et al., {\em {$B\to Kl^+l^-$ decay form factors from three-flavor
  lattice QCD}\/},  Phys. Rev. {\bf D93} (2016) no.~2, 025026,
  \href{http://arxiv.org/abs/1509.06235}{{\tt arXiv:1509.06235 [hep-lat]}}.

\bibitem{1310.3722}
R.~R. Horgan, Z.~Liu, S.~Meinel, and M.~Wingate, {\em {Lattice QCD calculation
  of form factors describing the rare decays $B \to K^* \ell^+ \ell^-$ and $B_s
  \to \phi \ell^+ \ell^-$}\/},  Phys. Rev. {\bf D89} (2014) no.~9, 094501,
  \href{http://arxiv.org/abs/1310.3722}{{\tt arXiv:1310.3722 [hep-lat]}}.

\bibitem{Meiman:1963}
N.~Meiman, {\em {\em Analytic Expressions for Upper Limits of Coupling
  Constants in Quantum Field Theory}\/},  Zh.\ Eksp.\ Teor.\ Fiz. {\bf {\bf
  44}} (1963)  1228. [Sov.\ Phys.\ JETP {\bf 17}, 830 (1963)].

\bibitem{Boyd:1994tt}
C.~G. Boyd, B.~Grinstein, and R.~F. Lebed, {\em {Constraints on form-factors
  for exclusive semileptonic heavy to light meson decays}\/},
  \href{http://dx.doi.org/10.1103/PhysRevLett.74.4603}{Phys. Rev. Lett. {\bf
  74} (1995)  4603--4606},
\href{http://arxiv.org/abs/hep-ph/9412324}{{\tt arXiv:hep-ph/9412324
  [hep-ph]}}.

\bibitem{0807.2722}
C.~Bourrely, I.~Caprini, and L.~Lellouch, {\em {Model-independent description
  of $B \to\pi l \nu$ decays and a determination of $|V_{ub}|$}\/},
  \href{http://dx.doi.org/10.1103/PhysRevD.82.099902,
  10.1103/PhysRevD.79.013008}{Phys. Rev. {\bf D79} (2009)  013008},
  \href{http://arxiv.org/abs/0807.2722}{{\tt arXiv:0807.2722 [hep-ph]}}.
  [Erratum: Phys. Rev.D82,099902(2010)].

\bibitem{1701.01633}
S.~Cheng, A.~Khodjamirian, and J.~Virto, {\em {$B\to\pi\pi$ Form Factors from
  Light-Cone Sum Rules with $B$-meson Distribution Amplitudes}\/},  JHEP {\bf
  05} (2017)  157, \href{http://arxiv.org/abs/1701.01633}{{\tt arXiv:1701.01633
  [hep-ph]}}.

\bibitem{1709.00173}
S.~Cheng, A.~Khodjamirian, and J.~Virto, {\em {Timelike-helicity $B\to \pi\pi$
  form factor from light-cone sum rules with dipion distribution
  amplitudes}\/},  Phys. Rev. {\bf D96} (2017) no.~5, 051901,
  \href{http://arxiv.org/abs/1709.00173}{{\tt arXiv:1709.00173 [hep-ph]}}.

\bibitem{1704.05439}
C.~Alexandrou, L.~Leskovec, S.~Meinel, J.~Negele, S.~Paul, M.~Petschlies,
  A.~Pochinsky, G.~Rendon, and S.~Syritsyn, {\em {$P$-wave $\pi\pi$ scattering
  and the $\rho$ resonance from lattice QCD}\/},  Phys. Rev. {\bf D96} (2017)
  no.~3, 034525, \href{http://arxiv.org/abs/1704.05439}{{\tt arXiv:1704.05439
  [hep-lat]}}.

\bibitem{hep-ph/0106067}
M.~Beneke, T.~Feldmann, and D.~Seidel, {\em {Systematic approach to exclusive
  $B \to V l^+ l^-$, $V \gamma$ decays}\/},  Nucl. Phys. {\bf B612} (2001)
  25--58, \href{http://arxiv.org/abs/hep-ph/0106067}{{\tt arXiv:hep-ph/0106067
  [hep-ph]}}.

\bibitem{1006.4945}
A.~Khodjamirian, T.~Mannel, A.~A. Pivovarov, and Y.~M. Wang, {\em {Charm-loop
  effect in $B \to K^{(*)} \ell^{+} \ell^{-}$ and $B\to K^*\gamma$}\/},  JHEP
  {\bf 09} (2010)  089, \href{http://arxiv.org/abs/1006.4945}{{\tt
  arXiv:1006.4945 [hep-ph]}}.

\bibitem{1101.5118}
M.~Beylich, G.~Buchalla, and T.~Feldmann, {\em {Theory of $B \to K^{(*)}\ell^+
  \ell^-$ decays at high $q^2$: OPE and quark-hadron duality}\/},  Eur. Phys.
  J. {\bf C71} (2011)  1635, \href{http://arxiv.org/abs/1101.5118}{{\tt
  arXiv:1101.5118 [hep-ph]}}.

\bibitem{1805.06378}
M.~Chrzaszcz, A.~Mauri, N.~Serra, R.~Silva~Coutinho, and D.~van Dyk, {\em
  {Prospects for disentangling long- and short-distance effects in the decays
  $B\to K^* \mu^+\mu^-$}\/},  \href{http://arxiv.org/abs/1805.06378}{{\tt
  arXiv:1805.06378 [hep-ph]}}.

\bibitem{hep-ph/0404250}
B.~Grinstein and D.~Pirjol, {\em {Exclusive rare $B \to K^*\ell^+\ell^-$ decays
  at low recoil: Controlling the long-distance effects}\/},  Phys. Rev. {\bf
  D70} (2004)  114005, \href{http://arxiv.org/abs/hep-ph/0404250}{{\tt
  arXiv:hep-ph/0404250 [hep-ph]}}.

\bibitem{hep-ph/9603237}
F.~Kruger and L.~M. Sehgal, {\em {Lepton polarization in the decays $b \to X_s
  \mu^+ \mu^-$ and $B \to X_s \tau^+ \tau^-$}\/},  Phys. Lett. {\bf B380}
  (1996)  199--204, \href{http://arxiv.org/abs/hep-ph/9603237}{{\tt
  arXiv:hep-ph/9603237 [hep-ph]}}.

\bibitem{1406.0566}
J.~Lyon and R.~Zwicky, {\em {Resonances gone topsy turvy - the charm of QCD or
  new physics in $b \to s \ell^+ \ell^-$?}\/},
  \href{http://arxiv.org/abs/1406.0566}{{\tt arXiv:1406.0566 [hep-ph]}}.

\bibitem{1606.00775}
S.~Bra{\ss}, G.~Hiller, and I.~Nisandzic, {\em {Zooming in on $B\rightarrow
  K^*\ell \ell $ decays at low recoil}\/},  Eur. Phys. J. {\bf C77} (2017)
  no.~1, 16, \href{http://arxiv.org/abs/1606.00775}{{\tt arXiv:1606.00775
  [hep-ph]}}.

\bibitem{hep-ph/0310219}
G.~Hiller and F.~Kruger, {\em {More model-independent analysis of $b \to s$
  processes}\/},  Phys. Rev. {\bf D69} (2004)  074020,
  \href{http://arxiv.org/abs/hep-ph/0310219}{{\tt arXiv:hep-ph/0310219
  [hep-ph]}}.

\bibitem{Bordone:2016gaq}
M.~Bordone, G.~Isidori, and A.~Pattori, {\em {On the Standard Model predictions
  for $R_K$ and $R_{K^*}$}\/},  Eur. Phys. J. {\bf C76} (2016)  440,
  \href{http://arxiv.org/abs/1605.07633}{{\tt arXiv:1605.07633 [hep-ph]}}.

\bibitem{1406.6482}
{LHCb Collaboration}, R.~Aaij et al., {\em {Test of lepton universality using
  $B^{+}\rightarrow K^{+}\ell^{+}\ell^{-}$ decays}\/},  Phys. Rev. Lett. {\bf
  113} (2014)  151601, \href{http://arxiv.org/abs/1406.6482}{{\tt
  arXiv:1406.6482 [hep-ex]}}.

\bibitem{1705.05802}
{LHCb Collaboration}, R.~Aaij et al., {\em {Test of lepton universality with
  $B^{0} \rightarrow K^{*0}\ell^{+}\ell^{-}$ decays}\/},  JHEP {\bf 08} (2017)
  055, \href{http://arxiv.org/abs/1705.05802}{{\tt arXiv:1705.05802 [hep-ex]}}.

\bibitem{1605.03156}
B.~Capdevila, S.~Descotes-Genon, J.~Matias, and J.~Virto, {\em {Assessing
  lepton-flavour non-universality from $B\to K^*\ell\ell$ angular analyses}\/},
   JHEP {\bf 10} (2016)  075, \href{http://arxiv.org/abs/1605.03156}{{\tt
  arXiv:1605.03156 [hep-ph]}}.

\bibitem{1612.05014}
{Belle Collaboration}, S.~Wehle et al., {\em {Lepton-Flavor-Dependent Angular
  Analysis of $B\to K^\ast \ell^+\ell^-$}\/},  Phys. Rev. Lett. {\bf 118}
  (2017) no.~11, 111801, \href{http://arxiv.org/abs/1612.05014}{{\tt
  arXiv:1612.05014 [hep-ex]}}.

\bibitem{Capdevila:2017iqn}
B.~Capdevila, A.~Crivellin, S.~Descotes-Genon, L.~Hofer, and J.~Matias, {\em
  {Searching for New Physics with $b\to s\tau^+\tau^-$ processes}\/},
  \href{http://dx.doi.org/10.1103/PhysRevLett.120.181802}{Phys. Rev. Lett. {\bf
  120} (2018) no.~18, 181802},
\href{http://arxiv.org/abs/1712.01919}{{\tt arXiv:1712.01919 [hep-ph]}}.

\bibitem{Kamenik:2017ghi}
J.~F. Kamenik, S.~Monteil, A.~Semkiv, and L.~V. Silva, {\em {Lepton
  polarization asymmetries in rare semi-tauonic $ b \rightarrow s $ exclusive
  decays at FCC-$ee$}\/},
  \href{http://dx.doi.org/10.1140/epjc/s10052-017-5272-0}{Eur. Phys. J. {\bf
  C77} (2017) no.~10, 701},
\href{http://arxiv.org/abs/1705.11106}{{\tt arXiv:1705.11106 [hep-ph]}}.

\bibitem{Das:2018iap}
D.~Das, {\em {On the angular distribution of $\Lambda_b\to\Lambda(\to
  N\pi)\tau^+\tau^-$ decay}\/},
  \href{http://dx.doi.org/10.1007/JHEP07(2018)063}{JHEP {\bf 07} (2018)  063},
\href{http://arxiv.org/abs/1804.08527}{{\tt arXiv:1804.08527 [hep-ph]}}.

\bibitem{1805.06401}
A.~Mauri, N.~Serra, and R.~Silva~Coutinho, {\em {Towards establishing Lepton
  Flavour Universality violation in $\bar{B}\to \bar{K}^*\ell^+\ell^-$
  decays}\/},  \href{http://arxiv.org/abs/1805.06401}{{\tt arXiv:1805.06401
  [hep-ph]}}.

\bibitem{Boer:2014kda}
P.~Böer, T.~Feldmann, and D.~van Dyk, {\em {Angular Analysis of the Decay
  $\Lambda_b \to \Lambda (\to N \pi) \ell^+\ell^-$}\/},
  \href{http://dx.doi.org/10.1007/JHEP01(2015)155}{JHEP {\bf 01} (2015)  155},
\href{http://arxiv.org/abs/1410.2115}{{\tt arXiv:1410.2115 [hep-ph]}}.

\bibitem{Das:2018sms}
D.~Das, {\em {Model independent New Physics analysis in
  $\Lambda_b\to\Lambda\mu^+\mu^-$ decay}\/},
  \href{http://dx.doi.org/10.1140/epjc/s10052-018-5731-2}{Eur. Phys. J. {\bf
  C78} (2018) no.~3, 230},
\href{http://arxiv.org/abs/1802.09404}{{\tt arXiv:1802.09404 [hep-ph]}}.

\bibitem{Blake:2017une}
T.~Blake and M.~Kreps, {\em {Angular distribution of polarised $\Lambda_b$
  baryons decaying to $\Lambda \ell^+\ell^-$}\/},
  \href{http://dx.doi.org/10.1007/JHEP11(2017)138}{JHEP {\bf 11} (2017)  138},
\href{http://arxiv.org/abs/1710.00746}{{\tt arXiv:1710.00746 [hep-ph]}}.

\bibitem{1502.05509}
S.~Descotes-Genon and J.~Virto, {\em {Time dependence in $B \to V\ell\ell$
  decays}\/},  \href{http://dx.doi.org/10.1007/JHEP04(2015)045,
  10.1007/JHEP07(2015)049}{JHEP {\bf 04} (2015)  045},
  \href{http://arxiv.org/abs/1502.05509}{{\tt arXiv:1502.05509 [hep-ph]}}.
  [Erratum: JHEP07,049(2015)].

\bibitem{Hambrock:2015wka}
C.~Hambrock, A.~Khodjamirian, and A.~Rusov, {\em {Hadronic effects and
  observables in $B\to \pi\ell^{+}\ell^{-}$ decay at large recoil}\/},  Phys.
  Rev. {\bf D92} (2015)  074020, \href{http://arxiv.org/abs/1506.07760}{{\tt
  arXiv:1506.07760 [hep-ph]}}.

\bibitem{Atwood:1997zr}
D.~Atwood, M.~Gronau, and A.~Soni, {\em {Mixing induced CP asymmetries in
  radiative B decays in and beyond the standard model}\/},  Phys. Rev. Lett.
  {\bf 79} (1997)  185--188, \href{http://arxiv.org/abs/hep-ph/9704272}{{\tt
  arXiv:hep-ph/9704272 [hep-ph]}}.

\bibitem{Gronau:2001ng}
M.~Gronau, Y.~Grossman, D.~Pirjol, and A.~Ryd, {\em {Measuring the photon
  polarization in $B \to K \pi \pi \gamma$}\/},
  \href{http://dx.doi.org/10.1103/PhysRevLett.88.051802}{Phys. Rev. Lett. {\bf
  88} (2002)  051802}, \href{http://arxiv.org/abs/hep-ph/0107254}{{\tt
  arXiv:hep-ph/0107254 [hep-ph]}}.

\bibitem{Gronau:2002rz}
M.~Gronau and D.~Pirjol, {\em {Photon polarization in radiative B decays}\/},
  Phys. Rev. {\bf D66} (2002)  054008,
  \href{http://arxiv.org/abs/hep-ph/0205065}{{\tt arXiv:hep-ph/0205065
  [hep-ph]}}.

\bibitem{Ball:2006eu}
P.~Ball, G.~W. Jones, and R.~Zwicky, {\em {$B \to V \gamma$ beyond QCD
  factorisation}\/},  \href{http://dx.doi.org/10.1103/PhysRevD.75.054004}{Phys.
  Rev. {\bf D75} (2007)  054004},
  \href{http://arxiv.org/abs/hep-ph/0612081}{{\tt arXiv:hep-ph/0612081
  [hep-ph]}}.

\bibitem{Kou:2010kn}
E.~Kou, A.~Le~Yaouanc, and A.~Tayduganov, {\em {Determining the photon
  polarization of the $b \to s \gamma$ using the $B \to K_1(1270) \gamma \to (K
  \pi \pi) \gamma$ decay}\/},
  \href{http://dx.doi.org/10.1103/PhysRevD.83.094007}{Phys. Rev. {\bf D83}
  (2011)  094007}, \href{http://arxiv.org/abs/1011.6593}{{\tt arXiv:1011.6593
  [hep-ph]}}.

\bibitem{Becirevic:2012dx}
D.~Becirevic, E.~Kou, A.~Le~Yaouanc, and A.~Tayduganov, {\em {Future prospects
  for the determination of the Wilson coefficient $C_{7\gamma}^\prime$}\/},
  \href{http://dx.doi.org/10.1007/JHEP08(2012)090}{JHEP {\bf 08} (2012)  090},
  \href{http://arxiv.org/abs/1206.1502}{{\tt arXiv:1206.1502 [hep-ph]}}.

\bibitem{DescotesGenon:2011yn}
S.~Descotes-Genon, D.~Ghosh, J.~Matias, and M.~Ramon, {\em {Exploring New
  Physics in the C7-C7' plane}\/},
  \href{http://dx.doi.org/10.1007/JHEP06(2011)099}{JHEP {\bf 06} (2011)  099},
\href{http://arxiv.org/abs/1104.3342}{{\tt arXiv:1104.3342 [hep-ph]}}.

\bibitem{Muheim:2008vu}
F.~Muheim, Y.~Xie, and R.~Zwicky, {\em {Exploiting the width difference in $B_s
  \to \phi \gamma$}\/},  Phys. Lett. {\bf B664} (2008)  174--179,
  \href{http://arxiv.org/abs/0802.0876}{{\tt arXiv:0802.0876 [hep-ph]}}.

\bibitem{Aubert:2005bu}
{BaBar Collaboration}, B.~Aubert et al., {\em {Measurement of the
  time-dependent CP-violating asymmetry in $B^0 \to K^0_S \pi^0 \gamma$
  decays}\/},  \href{http://dx.doi.org/10.1103/PhysRevD.72.051103}{Phys. Rev.
  {\bf D72} (2005)  051103}, \href{http://arxiv.org/abs/hep-ex/0507038}{{\tt
  arXiv:hep-ex/0507038 [hep-ex]}}.

\bibitem{Ushiroda:2006fi}
{Belle Collaboration}, Y.~Ushiroda et al., {\em {Time-Dependent CP Asymmetries
  in $B^0 \to K^0_S \pi^0 \gamma$ transitions}\/},
  \href{http://dx.doi.org/10.1103/PhysRevD.74.111104}{Phys. Rev. {\bf D74}
  (2006)  111104}, \href{http://arxiv.org/abs/hep-ex/0608017}{{\tt
  arXiv:hep-ex/0608017 [hep-ex]}}.

\bibitem{LHCb-PAPER-2016-034}
{LHCb collaboration}, R.~Aaij et al., {\em {First experimental study of photon
  polarization in radiative $\Bs$ decays}\/},  Phys. Rev. Lett. {\bf 118}
  (2017)  021801, \href{http://arxiv.org/abs/1609.02032}{{\tt arXiv:1609.02032
  [hep-ex]}}.

\bibitem{1501.03038}
{LHCb Collaboration}, R.~Aaij et al., {\em {Angular analysis of the $B^{0} \to
  K^{*0} e^{+} e^{-}$ decay in the low-q$^{2}$ region}\/},  JHEP {\bf 04}
  (2015)  064, \href{http://arxiv.org/abs/1501.03038}{{\tt arXiv:1501.03038
  [hep-ex]}}.

\bibitem{1003.5012}
M.~Benzke, S.~J. Lee, M.~Neubert, and G.~Paz, {\em {Factorization at Subleading
  Power and Irreducible Uncertainties in $\bar B\to X_s\gamma$ Decay}\/},  JHEP
  {\bf 08} (2010)  099, \href{http://arxiv.org/abs/1003.5012}{{\tt
  arXiv:1003.5012 [hep-ph]}}.

\bibitem{1705.10366}
M.~Benzke, T.~Hurth, and S.~Turczyk, {\em {Subleading power factorization in $
  \bar{B}\to {X}_s{\ell}^{+}{\ell}^{-} $}\/},  JHEP {\bf 10} (2017)  031,
  \href{http://arxiv.org/abs/1705.10366}{{\tt arXiv:1705.10366 [hep-ph]}}.

\bibitem{1503.01789}
M.~Misiak et al., {\em {Updated NNLO QCD predictions for the weak radiative
  B-meson decays}\/},  Phys. Rev. Lett. {\bf 114} (2015) no.~22, 221801,
  \href{http://arxiv.org/abs/1503.01789}{{\tt arXiv:1503.01789 [hep-ph]}}.

\bibitem{1503.04849}
T.~Huber, T.~Hurth, and E.~Lunghi, {\em {Inclusive $ \overline{B}\to
  {X}_s{\ell}^{+}{\ell}^{-} $ : complete angular analysis and a thorough study
  of collinear photons}\/},  JHEP {\bf 06} (2015)  176,
  \href{http://arxiv.org/abs/1503.04849}{{\tt arXiv:1503.04849 [hep-ph]}}.

\bibitem{1409.4557}
A.~J. Buras, J.~Girrbach-Noe, C.~Niehoff, and D.~M. Straub, {\em {$ B\to
  {K}^{\left(\ast \right)}\nu \overline{\nu} $ decays in the Standard Model and
  beyond}\/},  JHEP {\bf 02} (2015)  184,
  \href{http://arxiv.org/abs/1409.4557}{{\tt arXiv:1409.4557 [hep-ph]}}.

\bibitem{Buras:2014fpa}
A.~J. Buras, J.~Girrbach-Noe, C.~Niehoff, and D.~M. Straub, {\em {$B\to
  K^{(*)}\nu\bar\nu$ decays in the Standard Model and beyond}\/},
\href{http://arxiv.org/abs/1409.4557}{{\tt arXiv:1409.4557 [hep-ph]}}.

\bibitem{LHCb-PAPER-2015-051}
{LHCb collaboration}, R.~Aaij et al., {\em {Angular analysis of the $\Bz\to
  \Kstarz\mup\mun$ decay using $3\invfb$ of integrated luminosity}\/},  JHEP
  {\bf 02} (2016)  104, \href{http://arxiv.org/abs/1512.04442}{{\tt
  arXiv:1512.04442 [hep-ex]}}.

\bibitem{LHCb-PAPER-2015-023}
{LHCb collaboration}, R.~Aaij et al., {\em {Angular analysis and differential
  branching fraction of the decay $\Bs\to \phi\mup\mun$}\/},  JHEP {\bf 09}
  (2015)  179, \href{http://arxiv.org/abs/1506.08777}{{\tt arXiv:1506.08777
  [hep-ex]}}.

\bibitem{Jager:2017gal}
S.~Jager, K.~Leslie, M.~Kirk, and A.~Lenz, {\em {Charming new physics in rare
  B-decays and mixing?}\/},  Phys. Rev. {\bf D97} (2018) no.~1, 015021,
  \href{http://arxiv.org/abs/1701.09183}{{\tt arXiv:1701.09183 [hep-ph]}}.

\bibitem{Alguero:2018nvb}
M.~Alguer{\' o}, B.~Capdevila, S.~Descotes-Genon, P.~Masjuan, and J.~Matias,
  {\em {Are we overlooking Lepton Flavour Universal New Physics in $b\to
  s\ell\ell$ ?}\/},
\href{http://arxiv.org/abs/1809.08447}{{\tt arXiv:1809.08447 [hep-ph]}}.

\bibitem{Straub:2018kue}
D.~M. Straub, {\em {flavio: a Python package for flavour and precision
  phenomenology in the Standard Model and beyond}\/},
\href{http://arxiv.org/abs/1810.08132}{{\tt arXiv:1810.08132 [hep-ph]}}.

\bibitem{Aebischer:2015fzz}
J.~Aebischer, A.~Crivellin, M.~Fael, and C.~Greub, {\em {Matching of gauge
  invariant dimension-six operators for $b\to s$ and $b\to c$ transitions}\/},
  \href{http://dx.doi.org/10.1007/JHEP05(2016)037}{JHEP {\bf 05} (2016)  037},
\href{http://arxiv.org/abs/1512.02830}{{\tt arXiv:1512.02830 [hep-ph]}}.

\bibitem{Belanger:2015nma}
G.~Belanger, C.~Delaunay, and S.~Westhoff, {\em {A Dark Matter Relic From Muon
  Anomalies}\/},  Phys. Rev. {\bf D92} (2015)  055021,
  \href{http://arxiv.org/abs/1507.06660}{{\tt arXiv:1507.06660 [hep-ph]}}.

\bibitem{Arnan:2016cpy}
P.~Arnan, L.~Hofer, F.~Mescia, and A.~Crivellin, {\em {Loop effects of heavy
  new scalars and fermions in $b\to s\mu^+\mu^-$}\/},  JHEP {\bf 04} (2017)
  043, \href{http://arxiv.org/abs/1608.07832}{{\tt arXiv:1608.07832 [hep-ph]}}.

\bibitem{Kamenik:2017tnu}
J.~F. Kamenik, Y.~Soreq, and J.~Zupan, {\em {Lepton flavor universality
  violation without new sources of quark flavor violation}\/},
  \href{http://dx.doi.org/10.1103/PhysRevD.97.035002}{Phys. Rev. {\bf D97}
  (2018) no.~3, 035002},
\href{http://arxiv.org/abs/1704.06005}{{\tt arXiv:1704.06005 [hep-ph]}}.

\bibitem{Crivellin:2018yvo}
A.~Crivellin, C.~Greub, F.~Saturnino, and D.~M{\" u}ller, {\em {Importance of
  Loop Effects in Explaining the Accumulated Evidence for New Physics in B
  Decays with a Vector Leptoquark}\/},
\href{http://arxiv.org/abs/1807.02068}{{\tt arXiv:1807.02068 [hep-ph]}}.

\bibitem{Altmannshofer:2014rta}
W.~Altmannshofer and D.~M. Straub, {\em {New physics in $b\rightarrow s$
  transitions after LHC run 1}\/},  Eur. Phys. J. {\bf C75} (2015) no.~8, 382,
  \href{http://arxiv.org/abs/1411.3161}{{\tt arXiv:1411.3161 [hep-ph]}}.

\bibitem{DiLuzio:2017fdq}
L.~Di~Luzio, M.~Kirk, and A.~Lenz, {\em {One constraint to kill them all?}\/},
  \href{http://arxiv.org/abs/1712.06572}{{\tt arXiv:1712.06572 [hep-ph]}}.

\bibitem{DiLuzio:2018wch}
L.~Di~Luzio, M.~Kirk, and A.~Lenz, {\em {$B_s$-$\bar B_s$ mixing interplay with
  $B$ anomalies}\/},  in {\em {10th International Workshop on the CKM Unitarity
  Triangle (CKM 2018) Heidelberg, Germany, September 17-21, 2018}}.
\newblock 2018.
\newblock
\href{http://arxiv.org/abs/1811.12884}{{\tt arXiv:1811.12884 [hep-ph]}}.
\newblock

\bibitem{Datta:2017pfz}
A.~Datta, J.~Liao, and D.~Marfatia, {\em {A light $Z^\prime$ for the $R_K$
  puzzle and nonstandard neutrino interactions}\/},  Phys. Lett. {\bf B768}
  (2017)  265--269, \href{http://arxiv.org/abs/1702.01099}{{\tt
  arXiv:1702.01099 [hep-ph]}}.

\bibitem{Sala:2017ihs}
F.~Sala and D.~M. Straub, {\em {A New Light Particle in B Decays?}\/},  Phys.
  Lett. {\bf B774} (2017)  205--209,
  \href{http://arxiv.org/abs/1704.06188}{{\tt arXiv:1704.06188 [hep-ph]}}.

\bibitem{Altmannshofer:2017bsz}
W.~Altmannshofer, M.~J. Baker, S.~Gori, R.~Harnik, M.~Pospelov, E.~Stamou, and
  A.~Thamm, {\em {Light resonances and the low-q$^{2}$ bin of $ {R}_{K^{*}}
  $}\/},  JHEP {\bf 03} (2018)  188,
  \href{http://arxiv.org/abs/1711.07494}{{\tt arXiv:1711.07494 [hep-ph]}}.

\bibitem{Altmannshofer:2014cfa}
W.~Altmannshofer, S.~Gori, M.~Pospelov, and I.~Yavin, {\em {Quark flavor
  transitions in $L_\mu-L_\tau$ models}\/},  Phys. Rev. {\bf D89} (2014)
  095033, \href{http://arxiv.org/abs/1403.1269}{{\tt arXiv:1403.1269
  [hep-ph]}}.

\bibitem{Crivellin:2015mga}
A.~Crivellin, G.~D'Ambrosio, and J.~Heeck, {\em {Explaining
  $h\to\mu^\pm\tau^\mp$, $B\to K^* \mu^+\mu^-$ and $B\to K \mu^+\mu^-/B\to K
  e^+e^-$ in a two-Higgs-doublet model with gauged $L_\mu-L_\tau$}\/},  Phys.
  Rev. Lett. {\bf 114} (2015)  151801,
  \href{http://arxiv.org/abs/1501.00993}{{\tt arXiv:1501.00993 [hep-ph]}}.

\bibitem{Altmannshofer:2015mqa}
W.~Altmannshofer and I.~Yavin, {\em {Predictions for lepton flavor universality
  violation in rare B decays in models with gauged $L_\mu - L_\tau$}\/},  Phys.
  Rev. {\bf D92} (2015) no.~7, 075022,
  \href{http://arxiv.org/abs/1508.07009}{{\tt arXiv:1508.07009 [hep-ph]}}.

\bibitem{Fuyuto:2015gmk}
K.~Fuyuto, W.-S. Hou, and M.~Kohda, {\em {Z' -induced FCNC decays of top,
  beauty, and strange quarks}\/},  Phys. Rev. {\bf D93} (2016) no.~5, 054021,
  \href{http://arxiv.org/abs/1512.09026}{{\tt arXiv:1512.09026 [hep-ph]}}.

\bibitem{Altmannshofer:2016jzy}
W.~Altmannshofer, S.~Gori, S.~Profumo, and F.~S. Queiroz, {\em {Explaining dark
  matter and B decay anomalies with an $L_\mu - L_\tau$ model}\/},  JHEP {\bf
  12} (2016)  106, \href{http://arxiv.org/abs/1609.04026}{{\tt arXiv:1609.04026
  [hep-ph]}}.

\bibitem{Baek:2017sew}
S.~Baek, {\em {Dark matter contribution to $b\to s \mu^+ \mu^-$ anomaly in
  local $U(1)_{L_\mu-L_\tau}$ model}\/},
  \href{http://arxiv.org/abs/1707.04573}{{\tt arXiv:1707.04573 [hep-ph]}}.

\bibitem{Chen:2017usq}
C.-H. Chen and T.~Nomura, {\em {Penguin $b \to s\ell'^+ \ell'^-$ and $B$-meson
  anomalies in a gauged ${L_\mu -L_\tau}$}\/},  Phys. Lett. {\bf B777} (2018)
  420--427, \href{http://arxiv.org/abs/1707.03249}{{\tt arXiv:1707.03249
  [hep-ph]}}.

\bibitem{Altmannshofer:2016oaq}
W.~Altmannshofer, M.~Carena, and A.~Crivellin, {\em {$L_\mu - L_\tau$ theory of
  Higgs flavor violation and $(g-2)_\mu$}\/},
  \href{http://dx.doi.org/10.1103/PhysRevD.94.095026}{Phys. Rev. {\bf D94}
  (2016) no.~9, 095026},
\href{http://arxiv.org/abs/1604.08221}{{\tt arXiv:1604.08221 [hep-ph]}}.

\bibitem{Crivellin:2015lwa}
A.~Crivellin, G.~D'Ambrosio, and J.~Heeck, {\em {Addressing the LHC flavor
  anomalies with horizontal gauge symmetries}\/},  Phys. Rev. {\bf D91} (2015)
  no.~7, 075006, \href{http://arxiv.org/abs/1503.03477}{{\tt arXiv:1503.03477
  [hep-ph]}}.

\bibitem{Falkowski:2015zwa}
A.~Falkowski, M.~Nardecchia, and R.~Ziegler, {\em {Lepton Flavor
  Non-Universality in B-meson Decays from a U(2) Flavor Model}\/},  JHEP {\bf
  11} (2015)  173, \href{http://arxiv.org/abs/1509.01249}{{\tt arXiv:1509.01249
  [hep-ph]}}.

\bibitem{Boucenna:2016wpr}
S.~M. Boucenna, A.~Celis, J.~Fuentes-Martin, A.~Vicente, and J.~Virto, {\em
  {Non-abelian gauge extensions for B-decay anomalies}\/},  Phys. Lett. {\bf
  B760} (2016)  214--219, \href{http://arxiv.org/abs/1604.03088}{{\tt
  arXiv:1604.03088 [hep-ph]}}.

\bibitem{Boucenna:2016qad}
S.~M. Boucenna, A.~Celis, J.~Fuentes-Martin, A.~Vicente, and J.~Virto, {\em
  {Phenomenology of an $SU(2) \times SU(2) \times U(1)$ model with
  lepton-flavour non-universality}\/},  JHEP {\bf 12} (2016)  059,
  \href{http://arxiv.org/abs/1608.01349}{{\tt arXiv:1608.01349 [hep-ph]}}.

\bibitem{Celis:2016ayl}
A.~Celis, W.-Z. Feng, and M.~Vollmann, {\em {Dirac dark matter and $b \to s
  \ell^+ \ell^-$ with $\mathrm{U(1)}$ gauge symmetry}\/},  Phys. Rev. {\bf D95}
  (2017) no.~3, 035018, \href{http://arxiv.org/abs/1608.03894}{{\tt
  arXiv:1608.03894 [hep-ph]}}.

\bibitem{Crivellin:2016ejn}
A.~Crivellin, J.~Fuentes-Martin, A.~Greljo, and G.~Isidori, {\em {Lepton Flavor
  Non-Universality in B decays from Dynamical Yukawas}\/},  Phys. Lett. {\bf
  B766} (2017)  77--85, \href{http://arxiv.org/abs/1611.02703}{{\tt
  arXiv:1611.02703 [hep-ph]}}.

\bibitem{Alonso:2017bff}
R.~Alonso, P.~Cox, C.~Han, and T.~T. Yanagida, {\em {Anomaly-free local
  horizontal symmetry and anomaly-full rare B-decays}\/},  Phys. Rev. {\bf D96}
  (2017) no.~7, 071701, \href{http://arxiv.org/abs/1704.08158}{{\tt
  arXiv:1704.08158 [hep-ph]}}.

\bibitem{Ellis:2017nrp}
J.~Ellis, M.~Fairbairn, and P.~Tunney, {\em {Anomaly-Free Models for Flavour
  Anomalies}\/},  \href{http://arxiv.org/abs/1705.03447}{{\tt arXiv:1705.03447
  [hep-ph]}}.

\bibitem{Alonso:2017uky}
R.~Alonso, P.~Cox, C.~Han, and T.~T. Yanagida, {\em {Flavoured $B-L$ local
  symmetry and anomalous rare $B$ decays}\/},  Phys. Lett. {\bf B774} (2017)
  643--648, \href{http://arxiv.org/abs/1705.03858}{{\tt arXiv:1705.03858
  [hep-ph]}}.

\bibitem{Bonilla:2017lsq}
C.~Bonilla, T.~Modak, R.~Srivastava, and J.~W.~F. Valle, {\em
  {$U(1)_{B_3-3L_\mu}$ gauge symmetry as the simplest description of $b\to s$
  anomalies}\/},  \href{http://arxiv.org/abs/1705.00915}{{\tt arXiv:1705.00915
  [hep-ph]}}.

\bibitem{Babu:2017olk}
K.~S. Babu, A.~Friedland, P.~A.~N. Machado, and I.~Mocioiu, {\em {Flavor Gauge
  Models Below the Fermi Scale}\/},  JHEP {\bf 12} (2017)  096,
  \href{http://arxiv.org/abs/1705.01822}{{\tt arXiv:1705.01822 [hep-ph]}}.

\bibitem{Bian:2017rpg}
L.~Bian, S.-M. Choi, Y.-J. Kang, and H.~M. Lee, {\em {A minimal flavored
  $U(1)'$ for $B$-meson anomalies}\/},  Phys. Rev. {\bf D96} (2017) no.~7,
  075038, \href{http://arxiv.org/abs/1707.04811}{{\tt arXiv:1707.04811
  [hep-ph]}}.

\bibitem{Tang:2017gkz}
Y.~Tang and Y.-L. Wu, {\em {Flavor non-universal gauge interactions and
  anomalies in B-meson decays}\/},  Chin. Phys. {\bf C42} (2018) no.~3, 033104,
  \href{http://arxiv.org/abs/1705.05643}{{\tt arXiv:1705.05643 [hep-ph]}}.

\bibitem{Cline:2017ihf}
J.~M. Cline and J.~Martin~Camalich, {\em {$B$ decay anomalies from nonabelian
  local horizontal symmetry}\/},  Phys. Rev. {\bf D96} (2017) no.~5, 055036,
  \href{http://arxiv.org/abs/1706.08510}{{\tt arXiv:1706.08510 [hep-ph]}}.

\bibitem{Sierra:2015fma}
D.~Aristizabal~Sierra, F.~Staub, and A.~Vicente, {\em {Shedding light on the
  $b\to s$ anomalies with a dark sector}\/},  Phys. Rev. {\bf D92} (2015)
  no.~1, 015001, \href{http://arxiv.org/abs/1503.06077}{{\tt arXiv:1503.06077
  [hep-ph]}}.

\bibitem{Fuyuto:2017sys}
K.~Fuyuto, H.-L. Li, and J.-H. Yu, {\em {Implications of hidden gauged $U(1)$
  model for $B$ anomalies}\/},  \href{http://arxiv.org/abs/1712.06736}{{\tt
  arXiv:1712.06736 [hep-ph]}}.

\bibitem{Niehoff:2015bfa}
C.~Niehoff, P.~Stangl, and D.~M. Straub, {\em {Violation of lepton flavour
  universality in composite Higgs models}\/},  Phys. Lett. {\bf B747} (2015)
  182--186, \href{http://arxiv.org/abs/1503.03865}{{\tt arXiv:1503.03865
  [hep-ph]}}.

\bibitem{Carmona:2015ena}
A.~Carmona and F.~Goertz, {\em {Lepton Flavor and Nonuniversality from Minimal
  Composite Higgs Setups}\/},  Phys. Rev. Lett. {\bf 116} (2016) no.~25,
  251801, \href{http://arxiv.org/abs/1510.07658}{{\tt arXiv:1510.07658
  [hep-ph]}}.

\bibitem{Megias:2016bde}
E.~Megias, G.~Panico, O.~Pujolas, and M.~Quiros, {\em {A Natural origin for the
  LHCb anomalies}\/},  JHEP {\bf 09} (2016)  118,
  \href{http://arxiv.org/abs/1608.02362}{{\tt arXiv:1608.02362 [hep-ph]}}.

\bibitem{Carmona:2017fsn}
A.~Carmona and F.~Goertz, {\em {Recent $\boldsymbol{B}$ Physics Anomalies - a
  First Hint for Compositeness?}\/},
  \href{http://arxiv.org/abs/1712.02536}{{\tt arXiv:1712.02536 [hep-ph]}}.

\bibitem{Megias:2017ove}
E.~Megias, M.~Quiros, and L.~Salas, {\em {Lepton-flavor universality violation
  in R$_{K}$ and $ {R}_{D^{{\left(\ast \right)}}} $ from warped space}\/},
  JHEP {\bf 07} (2017)  102, \href{http://arxiv.org/abs/1703.06019}{{\tt
  arXiv:1703.06019 [hep-ph]}}.

\bibitem{Sannino:2017utc}
F.~Sannino, P.~Stangl, D.~M. Straub, and A.~E. Thomsen, {\em {Flavor Physics
  and Flavor Anomalies in Minimal Fundamental Partial Compositeness}\/},
  \href{http://arxiv.org/abs/1712.07646}{{\tt arXiv:1712.07646 [hep-ph]}}.

\bibitem{Bhattacharya:2014wla}
B.~Bhattacharya, A.~Datta, D.~London, and S.~Shivashankara, {\em {Simultaneous
  Explanation of the $R_K$ and $R(D^{(*)})$ Puzzles}\/},  Phys. Lett. {\bf
  B742} (2015)  370--374, \href{http://arxiv.org/abs/1412.7164}{{\tt
  arXiv:1412.7164 [hep-ph]}}.

\bibitem{Greljo:2015mma}
A.~Greljo, G.~Isidori, and D.~Marzocca, {\em {On the breaking of Lepton Flavor
  Universality in B decays}\/},  JHEP {\bf 07} (2015)  142,
  \href{http://arxiv.org/abs/1506.01705}{{\tt arXiv:1506.01705 [hep-ph]}}.

\bibitem{Bhattacharya:2016mcc}
B.~Bhattacharya, A.~Datta, J.-P. Guevin, D.~London, and R.~Watanabe, {\em
  {Simultaneous Explanation of the $R_K$ and $R_{D^{(*)}}$ Puzzles: a Model
  Analysis}\/},  JHEP {\bf 01} (2017)  015,
  \href{http://arxiv.org/abs/1609.09078}{{\tt arXiv:1609.09078 [hep-ph]}}.

\bibitem{Kumar:2018kmr}
J.~Kumar, D.~London, and R.~Watanabe, {\em {Combined Explanations of the $b \to
  s \mu^+ \mu^-$ and $b \to c \tau^- {\bar\nu}$ Anomalies: a General Model
  Analysis}\/},  \href{http://arxiv.org/abs/1806.07403}{{\tt arXiv:1806.07403
  [hep-ph]}}.

\bibitem{Diaz:2017lit}
B.~Diaz, M.~Schmaltz, and Y.-M. Zhong, {\em {The leptoquark Hunter's guide:
  Pair production}\/},  \href{http://dx.doi.org/10.1007/JHEP10(2017)097}{JHEP
  {\bf 10} (2017)  097},
\href{http://arxiv.org/abs/1706.05033}{{\tt arXiv:1706.05033 [hep-ph]}}.

\bibitem{Bauer:2015knc}
M.~Bauer and M.~Neubert, {\em {Minimal Leptoquark Explanation for the
  R$_{D^{(*)}}$ , R$_K$ , and $(g-2)_g$ Anomalies}\/},  Phys. Rev. Lett. {\bf
  116} (2016) no.~14, 141802, \href{http://arxiv.org/abs/1511.01900}{{\tt
  arXiv:1511.01900 [hep-ph]}}.

\bibitem{Hiller:2014yaa}
G.~Hiller and M.~Schmaltz, {\em {$R_K$ and future $b \to s \ell \ell$ physics
  beyond the standard model opportunities}\/},  Phys. Rev. {\bf D90} (2014)
  054014, \href{http://arxiv.org/abs/1408.1627}{{\tt arXiv:1408.1627
  [hep-ph]}}.

\bibitem{Fajfer:2015ycq}
S.~Fajfer and N.~Kosnik, {\em {Vector leptoquark resolution of $R_K$ and
  $R_{D^{(*)}}$ puzzles}\/},  Phys. Lett. {\bf B755} (2016)  270--274,
  \href{http://arxiv.org/abs/1511.06024}{{\tt arXiv:1511.06024 [hep-ph]}}.

\bibitem{Hiller:2017bzc}
G.~Hiller and I.~Nisandzic, {\em {$R_K$ and $R_{K^{\ast}}$ beyond the standard
  model}\/},  Phys. Rev. {\bf D96} (2017) no.~3, 035003,
  \href{http://arxiv.org/abs/1704.05444}{{\tt arXiv:1704.05444 [hep-ph]}}.

\bibitem{Chen:2017hir}
C.-H. Chen, T.~Nomura, and H.~Okada, {\em {Excesses of muon $g-2$,
  $R_{D^{(\ast)}}$, and $R_K$ in a leptoquark model}\/},  Phys. Lett. {\bf
  B774} (2017)  456--464, \href{http://arxiv.org/abs/1703.03251}{{\tt
  arXiv:1703.03251 [hep-ph]}}.

\bibitem{Crivellin:2017zlb}
A.~Crivellin, D.~Mueller, and T.~Ota, {\em {Simultaneous explanation of
  $R_{D^{({\^a})}}$ and $b \to s \mu^{+} \mu^{-}$: the last scalar
  leptoquarks standing}\/},  JHEP {\bf 09} (2017)  040,
  \href{http://arxiv.org/abs/1703.09226}{{\tt arXiv:1703.09226 [hep-ph]}}.

\bibitem{Aloni:2017ixa}
D.~Aloni, A.~Dery, C.~Frugiuele, and Y.~Nir, {\em {Testing minimal flavor
  violation in leptoquark models of the $ {R_K}_{{}^{\left(\ast \right)}} $
  anomaly}\/},  JHEP {\bf 11} (2017)  109,
  \href{http://arxiv.org/abs/1708.06161}{{\tt arXiv:1708.06161 [hep-ph]}}.

\bibitem{Gripaios:2014tna}
B.~Gripaios, M.~Nardecchia, and S.~A. Renner, {\em {Composite leptoquarks and
  anomalies in $B$-meson decays}\/},  JHEP {\bf 05} (2015)  006,
  \href{http://arxiv.org/abs/1412.1791}{{\tt arXiv:1412.1791 [hep-ph]}}.

\bibitem{Barbieri:2016las}
R.~Barbieri, C.~W. Murphy, and F.~Senia, {\em {B-decay Anomalies in a Composite
  Leptoquark Model}\/},  Eur. Phys. J. {\bf C77} (2017) no.~1, 8,
  \href{http://arxiv.org/abs/1611.04930}{{\tt arXiv:1611.04930 [hep-ph]}}.

\bibitem{Das:2016vkr}
D.~Das, C.~Hati, G.~Kumar, and N.~Mahajan, {\em {Towards a unified explanation
  of $R_{D^{(\ast)}}$, $R_{K}$ and $(g-2)_{\mu}$ anomalies in a left-right
  model with leptoquarks}\/},  Phys. Rev. {\bf D94} (2016)  055034,
  \href{http://arxiv.org/abs/1605.06313}{{\tt arXiv:1605.06313 [hep-ph]}}.

\bibitem{Fornal:2018dqn}
B.~Fornal, S.~A. Gadam, and B.~Grinstein, {\em {Left-Right SU(4) Vector
  Leptoquark Model for Flavor Anomalies}\/},
\href{http://arxiv.org/abs/1812.01603}{{\tt arXiv:1812.01603 [hep-ph]}}.

\bibitem{Assad:2017iib}
N.~Assad, B.~Fornal, and B.~Grinstein, {\em {Baryon Number and Lepton
  Universality Violation in Leptoquark and Diquark Models}\/},
  \href{http://dx.doi.org/10.1016/j.physletb.2017.12.042}{Phys. Lett. {\bf
  B777} (2018)  324--331},
\href{http://arxiv.org/abs/1708.06350}{{\tt arXiv:1708.06350 [hep-ph]}}.

\bibitem{Bordone:2017bld}
M.~Bordone, C.~Cornella, J.~Fuentes-Martin, and G.~Isidori, {\em {A three-site
  gauge model for flavor hierarchies and flavor anomalies}\/},  Phys. Lett.
  {\bf B779} (2018)  317--323, \href{http://arxiv.org/abs/1712.01368}{{\tt
  arXiv:1712.01368 [hep-ph]}}.

\bibitem{DiLuzio:2017vat}
L.~Di~Luzio, A.~Greljo, and M.~Nardecchia, {\em {Gauge leptoquark as the origin
  of B-physics anomalies}\/},  Phys. Rev. {\bf D96} (2017) no.~11, 115011,
  \href{http://arxiv.org/abs/1708.08450}{{\tt arXiv:1708.08450 [hep-ph]}}.

\bibitem{Blanke:2018sro}
M.~Blanke and A.~Crivellin, {\em {$B$ Meson Anomalies in a Pati-Salam Model
  within the Randall-Sundrum Background}\/},  Phys. Rev. Lett. {\bf 121} (2018)
  no.~1, 011801, \href{http://arxiv.org/abs/1801.07256}{{\tt arXiv:1801.07256
  [hep-ph]}}.

\bibitem{Calibbi:2017qbu}
L.~Calibbi, A.~Crivellin, and T.~Li, {\em {A model of vector leptoquarks in
  view of the $B$-physics anomalies}\/},
\href{http://arxiv.org/abs/1709.00692}{{\tt arXiv:1709.00692 [hep-ph]}}.

\bibitem{Das:2017kfo}
D.~Das, C.~Hati, G.~Kumar, and N.~Mahajan, {\em {Scrutinizing $R$-parity
  violating interactions in light of $R_{K^{(\ast)}}$ data}\/},  Phys. Rev.
  {\bf D96} (2017) no.~9, 095033, \href{http://arxiv.org/abs/1705.09188}{{\tt
  arXiv:1705.09188 [hep-ph]}}.

\bibitem{Earl:2018snx}
K.~Earl and T.~Gregoire, {\em {Contributions to ${b \rightarrow s \ell \ell}$
  Anomalies from ${R}$-Parity Violating Interactions}\/},
  \href{http://arxiv.org/abs/1806.01343}{{\tt arXiv:1806.01343 [hep-ph]}}.

\bibitem{Becirevic:2018afm}
D.~Be{\v c}irevic, I.~Dor{\v s}ner, S.~Fajfer, N.~Ko{\v s}nik, D.~A. Faroughy,
  and O.~Sumensari, {\em {Scalar leptoquarks from grand unified theories to
  accommodate the $B$-physics anomalies}\/},
  \href{http://dx.doi.org/10.1103/PhysRevD.98.055003}{Phys. Rev. {\bf D98}
  (2018) no.~5, 055003}, \href{http://arxiv.org/abs/1806.05689}{{\tt
  arXiv:1806.05689 [hep-ph]}}.

\bibitem{Feruglio:2016gvd}
F.~Feruglio, P.~Paradisi, and A.~Pattori, {\em {Revisiting Lepton Flavor
  Universality in B Decays}\/},  Phys. Rev. Lett. {\bf 118} (2017) no.~1,
  011801, \href{http://arxiv.org/abs/1606.00524}{{\tt arXiv:1606.00524
  [hep-ph]}}.

\bibitem{Feruglio:2017rjo}
F.~Feruglio, P.~Paradisi, and A.~Pattori, {\em {On the Importance of
  Electroweak Corrections for B Anomalies}\/},  JHEP {\bf 09} (2017)  061,
  \href{http://arxiv.org/abs/1705.00929}{{\tt arXiv:1705.00929 [hep-ph]}}.

\bibitem{Cornella:2018tfd}
C.~Cornella, F.~Feruglio, and P.~Paradisi, {\em {Low-energy Effects of Lepton
  Flavour Universality Violation}\/},
  \href{http://arxiv.org/abs/1803.00945}{{\tt arXiv:1803.00945 [hep-ph]}}.

\bibitem{Cirigliano:2009wk}
V.~Cirigliano, J.~Jenkins, and M.~Gonzalez-Alonso, {\em {Semileptonic decays of
  light quarks beyond the Standard Model}\/},
  \href{http://dx.doi.org/10.1016/j.nuclphysb.2009.12.020}{Nucl. Phys. {\bf
  B830} (2010)  95--115}, \href{http://arxiv.org/abs/0908.1754}{{\tt
  arXiv:0908.1754 [hep-ph]}}.

\bibitem{Cata:2015lta}
O.~Cat{\'a} and M.~Jung, {\em {Signatures of a nonstandard Higgs boson from
  flavor physics}\/},  Phys. Rev. {\bf D92} (2015) no.~5, 055018,
  \href{http://arxiv.org/abs/1505.05804}{{\tt arXiv:1505.05804 [hep-ph]}}.

\bibitem{Jung:2018lfu}
M.~Jung and D.~M. Straub, {\em {Constraining new physics in $b\to c\ell\nu$
  transitions}\/},  \href{http://arxiv.org/abs/1801.01112}{{\tt
  arXiv:1801.01112 [hep-ph]}}.

\bibitem{Caprini:1997mu}
I.~Caprini, L.~Lellouch, and M.~Neubert, {\em {Dispersive bounds on the shape
  of anti-B \to D(*) lepton anti-neutrino form-factors}\/},  Nucl. Phys. {\bf
  B530} (1998)  153--181, \href{http://arxiv.org/abs/hep-ph/9712417}{{\tt
  arXiv:hep-ph/9712417 [hep-ph]}}.

\bibitem{Boyd:1997kz}
C.~G. Boyd, B.~Grinstein, and R.~F. Lebed, {\em {Precision corrections to
  dispersive bounds on form-factors}\/},  Phys. Rev. {\bf D56} (1997)
  6895--6911, \href{http://arxiv.org/abs/hep-ph/9705252}{{\tt
  arXiv:hep-ph/9705252 [hep-ph]}}.

\bibitem{Lattice:2015rga}
{MILC Collaboration}, J.~A. Bailey et al., {\em {$B\rightarrow D l \nu$ form
  factors at nonzero recoil and |V$_{cb}$| from 2+1-flavor lattice QCD}\/},
  Phys. Rev. {\bf D92} (2015) no.~3, 034506,
  \href{http://arxiv.org/abs/1503.07237}{{\tt arXiv:1503.07237 [hep-lat]}}.

\bibitem{Na:2015kha}
{HPQCD Collaboration}, H.~Na, C.~M. Bouchard, G.~P. Lepage, C.~Monahan, and
  J.~Shigemitsu, {\em {$B \rightarrow D l \nu$ form factors at nonzero recoil
  and extraction of $|V_{cb}|$}\/},
  \href{http://dx.doi.org/10.1103/PhysRevD.93.119906,
  10.1103/PhysRevD.92.054510}{Phys. Rev. {\bf D92} (2015) no.~5, 054510},
  \href{http://arxiv.org/abs/1505.03925}{{\tt arXiv:1505.03925 [hep-lat]}}.
  [Erratum: Phys. Rev.D93,no.11,119906(2016)].

\bibitem{deBoer:2018ipi}
S.~de~Boer, T.~Kitahara, and I.~Nisandzic, {\em {Soft-Photon Corrections to
  $\bar{B} \to D \tau^{-} \bar{\nu}_{\tau}$ Relative to $\bar{B} \to D \mu^{-}
  \bar{\nu}_{\mu}$}\/},
  \href{http://dx.doi.org/10.1103/PhysRevLett.120.261804}{Phys. Rev. Lett. {\bf
  120} (2018) no.~26, 261804},
\href{http://arxiv.org/abs/1803.05881}{{\tt arXiv:1803.05881 [hep-ph]}}.

\bibitem{Bailey:2014tva}
{Fermilab Lattice, MILC Collaboration}, J.~A. Bailey et al., {\em {Update of
  $|V_{cb}|$ from the $\bar{B}\to D^*\ell\bar{\nu}$ form factor at zero recoil
  with three-flavor lattice QCD}\/},  Phys. Rev. {\bf D89} (2014) no.~11,
  114504, \href{http://arxiv.org/abs/1403.0635}{{\tt arXiv:1403.0635
  [hep-lat]}}.

\bibitem{Harrison:2017fmw}
{HPQCD Collaboration}, J.~Harrison, C.~Davies, and M.~Wingate, {\em {Lattice
  QCD calculation of the ${{B}_{(s)}\to D_{(s)}^{*}\ell{\nu}}$ form factors at
  zero recoil and implications for ${|V_{cb}|}$}\/},  Phys. Rev. {\bf D97}
  (2018) no.~5, 054502, \href{http://arxiv.org/abs/1711.11013}{{\tt
  arXiv:1711.11013 [hep-lat]}}.

\bibitem{Bernlochner:2017jka}
F.~U. Bernlochner, Z.~Ligeti, M.~Papucci, and D.~J. Robinson, {\em {Combined
  analysis of semileptonic $B$ decays to $D$ and $D^*$: $R(D^{(*)})$,
  $|V_{cb}|$, and new physics}\/},
  \href{http://dx.doi.org/10.1103/PhysRevD.95.115008,
  10.1103/PhysRevD.97.059902}{Phys. Rev. {\bf D95} (2017) no.~11, 115008},
  \href{http://arxiv.org/abs/1703.05330}{{\tt arXiv:1703.05330 [hep-ph]}}.
  [Erratum: Phys. Rev.D97,no.5,059902(2018)].

\bibitem{Luke:1990eg}
M.~E. Luke, {\em {Effects of subleading operators in the heavy quark effective
  theory}\/},  Phys. Lett. {\bf B252} (1990)  447--455.

\bibitem{Neubert:1991xw}
M.~Neubert and V.~Rieckert, {\em {New approach to the universal form-factors in
  decays of heavy mesons}\/},  Nucl. Phys. {\bf B382} (1992)  97--119.

\bibitem{Neubert:1993mb}
M.~Neubert, {\em {Heavy quark symmetry}\/},  Phys. Rept. {\bf 245} (1994)
  259--396, \href{http://arxiv.org/abs/hep-ph/9306320}{{\tt
  arXiv:hep-ph/9306320 [hep-ph]}}.

\bibitem{Neubert:1992wq}
M.~Neubert, Z.~Ligeti, and Y.~Nir, {\em {QCD sum rule analysis of the
  subleading Isgur-Wise form-factor $\chi_2 (v-v')$}\/},  Phys. Lett. {\bf
  B301} (1993)  101--107, \href{http://arxiv.org/abs/hep-ph/9209271}{{\tt
  arXiv:hep-ph/9209271 [hep-ph]}}.

\bibitem{Neubert:1992pn}
M.~Neubert, Z.~Ligeti, and Y.~Nir, {\em {The Subleading Isgur-Wise form-factor
  $\chi_3 (v,v')$ to order $\alpha_s$ in QCD sum rules}\/},  Phys. Rev. {\bf
  D47} (1993)  5060--5066, \href{http://arxiv.org/abs/hep-ph/9212266}{{\tt
  arXiv:hep-ph/9212266 [hep-ph]}}.

\bibitem{Ligeti:1993hw}
Z.~Ligeti, Y.~Nir, and M.~Neubert, {\em {The Subleading Isgur-Wise form-factor
  $xi_3 (v - v')$ and its implications for the decays $\bar B \to D* l\bar \nu$
  }\/},  Phys. Rev. {\bf D49} (1994)  1302--1309,
  \href{http://arxiv.org/abs/hep-ph/9305304}{{\tt arXiv:hep-ph/9305304
  [hep-ph]}}.

\bibitem{Jaiswal:2017rve}
S.~Jaiswal, S.~Nandi, and S.~K. Patra, {\em {Extraction of $|V_{cb}|$ from
  $B\to D^{(*)}\ell\nu_\ell$ and the Standard Model predictions of
  $R(D^{(*)})$}\/},  JHEP {\bf 12} (2017)  060,
  \href{http://arxiv.org/abs/1707.09977}{{\tt arXiv:1707.09977 [hep-ph]}}.

\bibitem{Faller:2008tr}
S.~Faller, A.~Khodjamirian, C.~Klein, and T.~Mannel, {\em {$B\to D^{(*)}$ Form
  Factors from QCD Light-Cone Sum Rules}\/},  Eur. Phys. J. {\bf C60} (2009)
  603--615, \href{http://arxiv.org/abs/0809.0222}{{\tt arXiv:0809.0222
  [hep-ph]}}.

\bibitem{Abdesselam:2017kjf}
{Belle Collaboration}, A.~Abdesselam et al., {\em {Precise determination of the
  CKM matrix element $\left| V_{cb}\right|$ with $\bar B^0 \to D^{*\,+} \,
  \ell^- \, \bar \nu_\ell$ decays with hadronic tagging at Belle}\/},
  \href{http://arxiv.org/abs/1702.01521}{{\tt arXiv:1702.01521 [hep-ex]}}.

\bibitem{Schacht:2017vfd}
S.~Schacht, {\em {The role of theory input for exclusive $V_{cb}$
  determinations}\/},  PoS {\bf EPS-HEP2017} (2017)  241,
  \href{http://arxiv.org/abs/1710.07948}{{\tt arXiv:1710.07948 [hep-ph]}}.

\bibitem{Aviles-Casco:2017nge}
A.~Vaquero Avil{\'e}s-Casco, C.~DeTar, D.~Du, A.~El-Khadra, A.~S. Kronfeld,
  J.~Laiho, and R.~S. Van~de Water, {\em {$\overline{B}\rightarrow
  D^\ast\ell\overline{\nu}$ at Non-Zero Recoil}\/},  EPJ Web Conf. {\bf 175}
  (2018)  13003, \href{http://arxiv.org/abs/1710.09817}{{\tt arXiv:1710.09817
  [hep-lat]}}.

\bibitem{Lees:2012xj}
{BaBar Collaboration}, J.~P. Lees et al., {\em {Evidence for an excess of
  $\bar{B} \to D^{(*)} \tau^-\bar{\nu}_\tau$ decays}\/},  Phys. Rev. Lett. {\bf
  109} (2012)  101802, \href{http://arxiv.org/abs/1205.5442}{{\tt
  arXiv:1205.5442 [hep-ex]}}.

\bibitem{Hirose:2016wfn}
{Belle Collaboration}, S.~Hirose et al., {\em {Measurement of the $\tau$ lepton
  polarization and $R(D^*)$ in the decay $\bar{B} \to D^* \tau^-
  \bar{\nu}_\tau$}\/},  Phys. Rev. Lett. {\bf 118} (2017) no.~21, 211801,
  \href{http://arxiv.org/abs/1612.00529}{{\tt arXiv:1612.00529 [hep-ex]}}.

\bibitem{Hirose:2017dxl}
{Belle Collaboration}, S.~Hirose et al., {\em {Measurement of the $\tau$ lepton
  polarization and $R(D^*)$ in the decay $\bar{B} \rightarrow D^* \tau^-
  \bar{\nu}_\tau$ with one-prong hadronic $\tau$ decays at Belle}\/},  Phys.
  Rev. {\bf D97} (2018) no.~1, 012004,
  \href{http://arxiv.org/abs/1709.00129}{{\tt arXiv:1709.00129 [hep-ex]}}.

\bibitem{LHCb-PAPER-2017-017}
{LHCb collaboration}, R.~Aaij et al., {\em {Measurement of the ratio of the
  $\mathcal{B}(\Bz\to D^{\ast-} \tau^+ \nu_{\tau})$ and
  \hbox{$\mathcal{B}(\Bz\to D^{\ast-}\mu^+\nu_{\mu})$} branching fractions
  using three-prong $\tau$-lepton decays}\/},  Phys. Rev. Lett. {\bf 120}
  (2018)  171802, \href{http://arxiv.org/abs/1708.08856}{{\tt arXiv:1708.08856
  [hep-ex]}}.

\bibitem{Fajfer:2012vx}
S.~Fajfer, J.~F. Kamenik, and I.~Nisandzic, {\em {On the $B \to D^* \tau \bar
  \nu_{\tau}$ Sensitivity to New Physics}\/},  Phys. Rev. {\bf D85} (2012)
  094025, \href{http://arxiv.org/abs/1203.2654}{{\tt arXiv:1203.2654
  [hep-ph]}}.

\bibitem{Celis:2012dk}
A.~Celis, M.~Jung, X.-Q. Li, and A.~Pich, {\em {Sensitivity to charged scalars
  in $\boldsymbol{B\to D^{(*)}\tau\nu_\tau}$ and
  $\boldsymbol{B\to\tau\nu_\tau}$ decays}\/},  JHEP {\bf 01} (2013)  054,
  \href{http://arxiv.org/abs/1210.8443}{{\tt arXiv:1210.8443 [hep-ph]}}.

\bibitem{Tanaka:2012nw}
M.~Tanaka and R.~Watanabe, {\em {New physics in the weak interaction of $\bar
  B\to D^{(*)}\tau\bar\nu$}\/},  Phys. Rev. {\bf D87} (2013) no.~3, 034028,
  \href{http://arxiv.org/abs/1212.1878}{{\tt arXiv:1212.1878 [hep-ph]}}.

\bibitem{Bigi:2016mdz}
D.~Bigi and P.~Gambino, {\em {Revisiting $B\to D \ell \nu$}\/},  Phys. Rev.
  {\bf D94} (2016) no.~9, 094008, \href{http://arxiv.org/abs/1606.08030}{{\tt
  arXiv:1606.08030 [hep-ph]}}.

\bibitem{Leibovich:1997tu}
A.~K. Leibovich, Z.~Ligeti, I.~W. Stewart, and M.~B. Wise, {\em {Model
  independent results for $B \to D_1(2420) \ell \bar \nu$ and $B \to D_2^*
  (2460) \ell \bar\nu$ at order $\Lambda_{\rm QCD} / m{c,b}$}\/},  Phys. Rev.
  Lett. {\bf 78} (1997)  3995--3998,
  \href{http://arxiv.org/abs/hep-ph/9703213}{{\tt arXiv:hep-ph/9703213
  [hep-ph]}}.

\bibitem{Leibovich:1997em}
A.~K. Leibovich, Z.~Ligeti, I.~W. Stewart, and M.~B. Wise, {\em {Semileptonic B
  decays to excited charmed mesons}\/},  Phys. Rev. {\bf D57} (1998)  308--330,
  \href{http://arxiv.org/abs/hep-ph/9705467}{{\tt arXiv:hep-ph/9705467
  [hep-ph]}}.

\bibitem{Bernlochner:2012bc}
F.~U. Bernlochner, Z.~Ligeti, and S.~Turczyk, {\em {A Proposal to solve some
  puzzles in semileptonic B decays}\/},  Phys. Rev. {\bf D85} (2012)  094033,
  \href{http://arxiv.org/abs/1202.1834}{{\tt arXiv:1202.1834 [hep-ph]}}.

\bibitem{Bernlochner:2016bci}
F.~U. Bernlochner and Z.~Ligeti, {\em {Semileptonic $B_{(s)}$ decays to excited
  charmed mesons with $e,\mu,\tau$ and searching for new physics with
  $R(D^{**})$}\/},  Phys. Rev. {\bf D95} (2017) no.~1, 014022,
  \href{http://arxiv.org/abs/1606.09300}{{\tt arXiv:1606.09300 [hep-ph]}}.

\bibitem{Bernlochner:2017jxt}
F.~U. Bernlochner, Z.~Ligeti, and D.~J. Robinson, {\em {Model independent
  analysis of semileptonic $B$ decays to $D^{**}$ for arbitrary new
  physics}\/},  Phys. Rev. {\bf D97} (2018) no.~7, 075011,
  \href{http://arxiv.org/abs/1711.03110}{{\tt arXiv:1711.03110 [hep-ph]}}.

\bibitem{Aloni:2018ipm}
D.~Aloni, Y.~Grossman, and A.~Soffer, {\em {Measuring CP violation in $b\to
  c\tau^-\bar{\nu}_\tau$ using excited charm mesons}\/},
  \href{http://dx.doi.org/10.1103/PhysRevD.98.035022}{Phys. Rev. {\bf D98}
  (2018) no.~3, 035022},
\href{http://arxiv.org/abs/1806.04146}{{\tt arXiv:1806.04146 [hep-ph]}}.

\bibitem{LHCb-PAPER-2013-004}
{LHCb collaboration}, R.~Aaij et al., {\em {Measurement of $\B$ meson
  production cross-sections in proton-proton collisions at $\sqrt{s} =
  7$\tev}\/},  \href{http://dx.doi.org/10.1007/JHEP08(2013)117}{JHEP {\bf 08}
  (2013)  117} CERN-PH-EP-2013-095, LHCb-PAPER-2013-004,
  \href{http://arxiv.org/abs/1306.3663}{{\tt arXiv:1306.3663 [hep-ex]}}.

\bibitem{LHCb-PAPER-2013-044}
{LHCb collaboration}, R.~Aaij et al., {\em {Observation of the decay $\Bcp\to
  \Bs\pip$}\/},  \href{http://dx.doi.org/10.1103/PhysRevLett.111.181801}{Phys.
  Rev. Lett. {\bf 111} (2013)  181801} CERN-PH-EP-2013-136,
  LHCb-PAPER-2013-044, \href{http://arxiv.org/abs/1308.4544}{{\tt
  arXiv:1308.4544 [hep-ex]}}.

\bibitem{Anisimov:1998uk}
A.~{\relax Yu}. Anisimov, I.~M. Narodetsky, C.~Semay, and B.~Silvestre-Brac,
  {\em {The $B_c$ meson lifetime in the light front constituent quark
  model}\/},  Phys. Lett. {\bf B452} (1999)  129--136,
  \href{http://arxiv.org/abs/hep-ph/9812514}{{\tt arXiv:hep-ph/9812514
  [hep-ph]}}.

\bibitem{Kiselev:1999sc}
V.~V. Kiselev, A.~K. Likhoded, and A.~I. Onishchenko, {\em {Semileptonic $B_c$
  meson decays in sum rules of QCD and NRQCD}\/},  Nucl. Phys. {\bf B569}
  (2000)  473--504, \href{http://arxiv.org/abs/hep-ph/9905359}{{\tt
  arXiv:hep-ph/9905359 [hep-ph]}}.

\bibitem{Ivanov:2000aj}
M.~A. Ivanov, J.~G. Korner, and P.~Santorelli, {\em {The Semileptonic decays of
  the $B_c$ meson}\/},  Phys. Rev. {\bf D63} (2001)  074010,
  \href{http://arxiv.org/abs/hep-ph/0007169}{{\tt arXiv:hep-ph/0007169
  [hep-ph]}}.

\bibitem{Kiselev:2002vz}
V.~V. Kiselev, {\em {Exclusive decays and lifetime of $B_c$ meson in QCD sum
  rules}\/},  \href{http://arxiv.org/abs/hep-ph/0211021}{{\tt
  arXiv:hep-ph/0211021 [hep-ph]}}.

\bibitem{Hernandez:2006gt}
E.~Hernandez, J.~Nieves, and J.~M. Verde-Velasco, {\em {Study of exclusive
  semileptonic and non-leptonic decays of $B_c$ - in a nonrelativistic quark
  model}\/},  Phys. Rev. {\bf D74} (2006)  074008,
  \href{http://arxiv.org/abs/hep-ph/0607150}{{\tt arXiv:hep-ph/0607150
  [hep-ph]}}.

\bibitem{Ivanov:2006ni}
M.~A. Ivanov, J.~G. Korner, and P.~Santorelli, {\em {Exclusive semileptonic and
  nonleptonic decays of the $B_c$ meson}\/},  Phys. Rev. {\bf D73} (2006)
  054024, \href{http://arxiv.org/abs/hep-ph/0602050}{{\tt arXiv:hep-ph/0602050
  [hep-ph]}}.

\bibitem{Wen-Fei:2013uea}
W.-F. Wang, Y.-Y. Fan, and Z.-J. Xiao, {\em {Semileptonic decays
  $B_c\to(\eta_c,J/\Psi)l\nu$ in the perturbative QCD approach}\/},  Chin.
  Phys. {\bf C37} (2013)  093102, \href{http://arxiv.org/abs/1212.5903}{{\tt
  arXiv:1212.5903 [hep-ph]}}.

\bibitem{Qiao:2012vt}
C.-F. Qiao and R.-L. Zhu, {\em {Estimation of semileptonic decays of $B_c$
  meson to S-wave charmonia with nonrelativistic QCD}\/},  Phys. Rev. {\bf D87}
  (2013) no.~1, 014009, \href{http://arxiv.org/abs/1208.5916}{{\tt
  arXiv:1208.5916 [hep-ph]}}.

\bibitem{Rui:2016opu}
Z.~Rui, H.~Li, G.-x. Wang, and Y.~Xiao, {\em {Semileptonic decays of $B_c$
  meson to S-wave charmonium states in the perturbative QCD approach}\/},  Eur.
  Phys. J. {\bf C76} (2016) no.~10, 564,
  \href{http://arxiv.org/abs/1602.08918}{{\tt arXiv:1602.08918 [hep-ph]}}.

\bibitem{Dutta:2017xmj}
R.~Dutta and A.~Bhol, {\em {$B_c \to (J/\psi,\,\eta_c)\tau\nu$ semileptonic
  decays within the standard model and beyond}\/},  Phys. Rev. {\bf D96} (2017)
  no.~7, 076001, \href{http://arxiv.org/abs/1701.08598}{{\tt arXiv:1701.08598
  [hep-ph]}}.

\bibitem{Tran:2018kuv}
C.-T. Tran, M.~A. Ivanov, J.~G. K{\"o}rner, and P.~Santorelli, {\em
  {Implications of new physics in the decays $B_c \to
  (J/\psi,\eta_c)\tau\nu$}\/},  Phys. Rev. {\bf D97} (2018) no.~5, 054014,
  \href{http://arxiv.org/abs/1801.06927}{{\tt arXiv:1801.06927 [hep-ph]}}.

\bibitem{Issadykov:2018myx}
A.~Issadykov and M.~A. Ivanov, {\em {The decays $B_{c}\to
  J/\psi+\bar\ell\nu_\ell$ and $B_{c}\to J/\psi + \pi(K)$ in covariant confined
  quark model}\/},  Phys. Lett. {\bf B783} (2018)  178--182,
  \href{http://arxiv.org/abs/1804.00472}{{\tt arXiv:1804.00472 [hep-ph]}}.

\bibitem{Watanabe:2017mip}
R.~Watanabe, {\em {New Physics effect on $B_c \to J/\psi \tau\bar\nu$ in
  relation to the $R_{D^{(*)}}$ anomaly}\/},  Phys. Lett. {\bf B776} (2018)
  5--9, \href{http://arxiv.org/abs/1709.08644}{{\tt arXiv:1709.08644
  [hep-ph]}}.

\bibitem{Cohen:2018dgz}
T.~D. Cohen, H.~Lamm, and R.~F. Lebed, {\em {Model-Independent Bounds on
  $R(J/\psi)$}\/},  \href{http://arxiv.org/abs/1807.02730}{{\tt
  arXiv:1807.02730 [hep-ph]}}.

\bibitem{Bernlochner:2018kxh}
F.~U. Bernlochner, Z.~Ligeti, D.~J. Robinson, and W.~L. Sutcliffe, {\em {New
  predictions for $\Lambda_b\to\Lambda_c$ semileptonic decays and tests of
  heavy quark symmetry}\/},  \href{http://arxiv.org/abs/1808.09464}{{\tt
  arXiv:1808.09464 [hep-ph]}}.

\bibitem{Leibovich:1997az}
A.~K. Leibovich and I.~W. Stewart, {\em {Semileptonic Lambda(b) decay to
  excited Lambda(c) baryons at order Lambda(QCD) / m(Q)}\/},
  \href{http://dx.doi.org/10.1103/PhysRevD.57.5620}{Phys. Rev. {\bf D57} (1998)
   5620--5631},
\href{http://arxiv.org/abs/hep-ph/9711257}{{\tt arXiv:hep-ph/9711257
  [hep-ph]}}.

\bibitem{Boer:2018vpx}
P.~B{\"o}er, M.~Bordone, E.~Graverini, P.~Owen, M.~Rotondo, and D.~Van~Dyk,
  {\em {Testing lepton flavour universality in semileptonic $\Lambda_b \to
  \Lambda_c^*$ decays}\/},
  \href{http://dx.doi.org/10.1007/JHEP06(2018)155}{JHEP {\bf 06} (2018)  155},
\href{http://arxiv.org/abs/1801.08367}{{\tt arXiv:1801.08367 [hep-ph]}}.

\bibitem{Ligeti:2014kia}
Z.~Ligeti and F.~J. Tackmann, {\em {Precise predictions for $B \to X_c \tau
  \bar \nu$ decay distributions}\/},
  \href{http://dx.doi.org/10.1103/PhysRevD.90.034021}{Phys. Rev. {\bf D90}
  (2014) no.~3, 034021}, \href{http://arxiv.org/abs/1406.7013}{{\tt
  arXiv:1406.7013 [hep-ph]}}.

\bibitem{Freytsis:2015qca}
M.~Freytsis, Z.~Ligeti, and J.~T. Ruderman, {\em {Flavor models for $\bar{B}
  \to D^{(*)} \tau \bar{\nu}$}\/},
  \href{http://dx.doi.org/10.1103/PhysRevD.92.054018}{Phys. Rev. {\bf D92}
  (2015) no.~5, 054018}, \href{http://arxiv.org/abs/1506.08896}{{\tt
  arXiv:1506.08896 [hep-ph]}}.

\bibitem{Mannel:2017jfk}
T.~Mannel, A.~V. Rusov, and F.~Shahriaran, {\em {Inclusive semitauonic $B$
  decays to order ${\cal O} (\Lambda_{QCD}^3/m_b^3)$}\/},
  \href{http://dx.doi.org/10.1016/j.nuclphysb.2017.05.016}{Nucl. Phys. {\bf
  B921} (2017)  211--224}, \href{http://arxiv.org/abs/1702.01089}{{\tt
  arXiv:1702.01089 [hep-ph]}}.

\bibitem{Bhattacharya:2018kig}
S.~Bhattacharya, S.~Nandi, and S.~Kumar~Patra, {\em {$b \to c \tau \nu_{\tau}$
  Decays: A Catalogue to Compare, Constrain, and Correlate New Physics
  Effects}\/},
\href{http://arxiv.org/abs/1805.08222}{{\tt arXiv:1805.08222 [hep-ph]}}.

\bibitem{Celis:2016azn}
A.~Celis, M.~Jung, X.-Q. Li, and A.~Pich, {\em {Scalar contributions to $b\to c
  (u) \tau \nu$ transitions}\/},
  \href{http://dx.doi.org/10.1016/j.physletb.2017.05.037}{Phys. Lett. {\bf
  B771} (2017)  168--179}, \href{http://arxiv.org/abs/1612.07757}{{\tt
  arXiv:1612.07757 [hep-ph]}}.

\bibitem{Crivellin:2012ye}
A.~Crivellin, C.~Greub, and A.~Kokulu, {\em {Explaining $B\to D\tau\nu$, $B\to
  D^*\tau\nu$ and $B\to \tau\nu$ in a 2HDM of type III}\/},
  \href{http://dx.doi.org/10.1103/PhysRevD.86.054014}{Phys. Rev. {\bf D86}
  (2012)  054014}, \href{http://arxiv.org/abs/1206.2634}{{\tt arXiv:1206.2634
  [hep-ph]}}.

\bibitem{Crivellin:2015hha}
A.~Crivellin, J.~Heeck, and P.~Stoffer, {\em {A perturbed lepton-specific
  two-Higgs-doublet model facing experimental hints for physics beyond the
  Standard Model}\/},
  \href{http://dx.doi.org/10.1103/PhysRevLett.116.081801}{Phys. Rev. Lett. {\bf
  116} (2016) no.~8, 081801}, \href{http://arxiv.org/abs/1507.07567}{{\tt
  arXiv:1507.07567 [hep-ph]}}.

\bibitem{Chen:2017eby}
C.-H. Chen and T.~Nomura, {\em {Charged-Higgs on $R_{D^{(*)}}$, $\tau$
  polarization, and FBA}\/},
  \href{http://dx.doi.org/10.1140/epjc/s10052-017-5198-6}{Eur. Phys. J. {\bf
  C77} (2017) no.~9, 631}, \href{http://arxiv.org/abs/1703.03646}{{\tt
  arXiv:1703.03646 [hep-ph]}}.

\bibitem{Iguro:2017ysu}
S.~Iguro and K.~Tobe, {\em {$R(D^{(*)})$ in a general two Higgs doublet
  model}\/},  \href{http://dx.doi.org/10.1016/j.nuclphysb.2017.10.014}{Nucl.
  Phys. {\bf B925} (2017)  560--606},
  \href{http://arxiv.org/abs/1708.06176}{{\tt arXiv:1708.06176 [hep-ph]}}.

\bibitem{Chen:2018hqy}
C.-H. Chen and T.~Nomura, {\em {Charged-Higgs on $B^-_{q} \to \ell \bar \nu$
  and $\bar B\to (P, V) \ell \bar\nu$ in a generic two-Higgs doublet model}\/},
   \href{http://arxiv.org/abs/1803.00171}{{\tt arXiv:1803.00171 [hep-ph]}}.

\bibitem{Li:2018rax}
S.-P. Li, X.-Q. Li, Y.-D. Yang, and X.~Zhang, {\em {$R_{D^{(\ast)}}$,
  $R_{K^{(\ast)}}$ and neutrino mass in the 2HDM-III with right-handed
  neutrinos}\/},  \href{http://dx.doi.org/10.1007/JHEP09(2018)149}{JHEP {\bf
  09} (2018)  149}, \href{http://arxiv.org/abs/1807.08530}{{\tt
  arXiv:1807.08530 [hep-ph]}}.

\bibitem{Fajfer:2012jt}
S.~Fajfer, J.~F. Kamenik, I.~Nisandzic, and J.~Zupan, {\em {Implications of
  Lepton Flavor Universality Violations in B Decays}\/},
  \href{http://dx.doi.org/10.1103/PhysRevLett.109.161801}{Phys. Rev. Lett. {\bf
  109} (2012)  161801}, \href{http://arxiv.org/abs/1206.1872}{{\tt
  arXiv:1206.1872 [hep-ph]}}.

\bibitem{Deshpande:2012rr}
N.~G. Deshpande and A.~Menon, {\em {Hints of R-parity violation in B decays
  into $\tau \nu$}\/},  \href{http://dx.doi.org/10.1007/JHEP01(2013)025}{JHEP
  {\bf 01} (2013)  025}, \href{http://arxiv.org/abs/1208.4134}{{\tt
  arXiv:1208.4134 [hep-ph]}}.

\bibitem{Sakaki:2013bfa}
Y.~Sakaki, M.~Tanaka, A.~Tayduganov, and R.~Watanabe, {\em {Testing leptoquark
  models in $\bar B \to D^{(*)} \tau \bar\nu$}\/},
  \href{http://dx.doi.org/10.1103/PhysRevD.88.094012}{Phys. Rev. {\bf D88}
  (2013) no.~9, 094012}, \href{http://arxiv.org/abs/1309.0301}{{\tt
  arXiv:1309.0301 [hep-ph]}}.

\bibitem{Duraisamy:2014sna}
M.~Duraisamy, P.~Sharma, and A.~Datta, {\em {Azimuthal $B \to D^{*} \tau^{-}
  \bar{\nu_\tau}$ angular distribution with tensor operators}\/},
  \href{http://dx.doi.org/10.1103/PhysRevD.90.074013}{Phys. Rev. {\bf D90}
  (2014) no.~7, 074013}, \href{http://arxiv.org/abs/1405.3719}{{\tt
  arXiv:1405.3719 [hep-ph]}}.

\bibitem{Barbieri:2015yvd}
R.~Barbieri, G.~Isidori, A.~Pattori, and F.~Senia, {\em {Anomalies in
  $B$-decays and $U(2)$ flavour symmetry}\/},
  \href{http://dx.doi.org/10.1140/epjc/s10052-016-3905-3}{Eur. Phys. J. {\bf
  C76} (2016) no.~2, 67}, \href{http://arxiv.org/abs/1512.01560}{{\tt
  arXiv:1512.01560 [hep-ph]}}.

\bibitem{Deshpand:2016cpw}
N.~G. Deshpande and X.-G. He, {\em {Consequences of R-parity violating
  interactions for anomalies in $\bar B\to D^{(*)} \tau \bar \nu$ and $b\to s
  \mu^+\mu^-$}\/},
  \href{http://dx.doi.org/10.1140/epjc/s10052-017-4707-y}{Eur. Phys. J. {\bf
  C77} (2017) no.~2, 134}, \href{http://arxiv.org/abs/1608.04817}{{\tt
  arXiv:1608.04817 [hep-ph]}}.

\bibitem{Sahoo:2016pet}
S.~Sahoo, R.~Mohanta, and A.~K. Giri, {\em {Explaining the $R_{K}$ and
  $R_{D^{(*)}}$ anomalies with vector leptoquarks}\/},
  \href{http://dx.doi.org/10.1103/PhysRevD.95.035027}{Phys. Rev. {\bf D95}
  (2017) no.~3, 035027}, \href{http://arxiv.org/abs/1609.04367}{{\tt
  arXiv:1609.04367 [hep-ph]}}.

\bibitem{Dumont:2016xpj}
B.~Dumont, K.~Nishiwaki, and R.~Watanabe, {\em {LHC constraints and prospects
  for $S_1$ scalar leptoquark explaining the $\bar B \to D^{(*)} \tau \bar\nu$
  anomaly}\/},  \href{http://dx.doi.org/10.1103/PhysRevD.94.034001}{Phys. Rev.
  {\bf D94} (2016) no.~3, 034001}, \href{http://arxiv.org/abs/1603.05248}{{\tt
  arXiv:1603.05248 [hep-ph]}}.

\bibitem{Li:2016vvp}
X.-Q. Li, Y.-D. Yang, and X.~Zhang, {\em {Revisiting the one leptoquark
  solution to the R(D$^{(∗)}$) anomalies and its phenomenological
  implications}\/},  \href{http://dx.doi.org/10.1007/JHEP08(2016)054}{JHEP {\bf
  08} (2016)  054}, \href{http://arxiv.org/abs/1605.09308}{{\tt
  arXiv:1605.09308 [hep-ph]}}.

\bibitem{Becirevic:2016yqi}
D.~Be\v{c}irevi\'c, S.~Fajfer, N.~Ko\v{s}nik, and O.~Sumensari, {\em
  {Leptoquark model to explain the $B$-physics anomalies, $R_K$ and $R_D$}\/},
  Phys. Rev. {\bf D94} (2016)  115021,
  \href{http://arxiv.org/abs/1608.08501}{{\tt arXiv:1608.08501 [hep-ph]}}.

\bibitem{Iguro:2018vqb}
S.~Iguro, T.~Kitahara, Y.~Omura, R.~Watanabe, and K.~Yamamoto, {\em {$D^{\ast}$
  polarization vs. $R_{D^{(\ast)}}$ anomalies in the leptoquark models}\/},
\href{http://arxiv.org/abs/1811.08899}{{\tt arXiv:1811.08899 [hep-ph]}}.

\bibitem{Bhattacharya:2016zcw}
S.~Bhattacharya, S.~Nandi, and S.~K. Patra, {\em {Looking for possible new
  physics in $B\to D^{(\ast)}\tau\nu_{\tau}$ in light of recent data}\/},
  \href{http://dx.doi.org/10.1103/PhysRevD.95.075012}{Phys. Rev. {\bf D95}
  (2017) no.~7, 075012}, \href{http://arxiv.org/abs/1611.04605}{{\tt
  arXiv:1611.04605 [hep-ph]}}.

\bibitem{Ivanov:2017mrj}
M.~A. Ivanov, J.~G. K{\" o}rner, and C.-T. Tran, {\em {Probing new physics in
  $\bar{B}^0 \to D^{(\ast)} \tau^- \bar\nu_{\tau}$ using the longitudinal,
  transverse, and normal polarization components of the tau lepton}\/},
  \href{http://dx.doi.org/10.1103/PhysRevD.95.036021}{Phys. Rev. {\bf D95}
  (2017) no.~3, 036021}, \href{http://arxiv.org/abs/1701.02937}{{\tt
  arXiv:1701.02937 [hep-ph]}}.

\bibitem{Alok:2017qsi}
A.~K. Alok, D.~Kumar, J.~Kumar, S.~Kumbhakar, and S.~U. Sankar, {\em {New
  physics solutions for $R_D$ and $R_{D^*}$}\/},
  \href{http://dx.doi.org/10.1007/JHEP09(2018)152}{JHEP {\bf 09} (2018)  152},
  \href{http://arxiv.org/abs/1710.04127}{{\tt arXiv:1710.04127 [hep-ph]}}.

\bibitem{Bifani:2018zmi}
S.~Bifani, S.~Descotes-Genon, A.~Romero~Vidal, and M.-H. Schune, {\em {Review
  of Lepton Universality tests in $B$ decays}\/},
  \href{http://arxiv.org/abs/1809.06229}{{\tt arXiv:1809.06229 [hep-ex]}}.

\bibitem{Blanke:2018yud}
M.~Blanke, A.~Crivellin, S.~de~Boer, T.~Kitahara, M.~Moscati, U.~Nierste, and
  I.~Ni{\v s}and{\v z}i{\' c}, {\em {Impact of polarization observables and $
  B_c\to \tau \nu$ on new physics explanations of the $b\to c \tau \nu$
  anomaly}\/},
\href{http://arxiv.org/abs/1811.09603}{{\tt arXiv:1811.09603 [hep-ph]}}.

\bibitem{He:2012zp}
X.-G. He and G.~Valencia, {\em {$B$ decays with $\tau$ leptons in nonuniversal
  left-right models}\/},
  \href{http://dx.doi.org/10.1103/PhysRevD.87.014014}{Phys. Rev. {\bf D87}
  (2013) no.~1, 014014}, \href{http://arxiv.org/abs/1211.0348}{{\tt
  arXiv:1211.0348 [hep-ph]}}.

\bibitem{He:2017bft}
X.-G. He and G.~Valencia, {\em {Lepton universality violation and right-handed
  currents in $b \to c \tau \nu$}\/},
  \href{http://dx.doi.org/10.1016/j.physletb.2018.01.073}{Phys. Lett. {\bf
  B779} (2018)  52--57}, \href{http://arxiv.org/abs/1711.09525}{{\tt
  arXiv:1711.09525 [hep-ph]}}.

\bibitem{Greljo:2018ogz}
A.~Greljo, D.~J. Robinson, B.~Shakya, and J.~Zupan, {\em {$R(D^{(*)})$ from
  $W'$ and right-handed neutrinos}\/},
  \href{http://dx.doi.org/10.1007/JHEP09(2018)169}{JHEP {\bf 09} (2018)  169},
  \href{http://arxiv.org/abs/1804.04642}{{\tt arXiv:1804.04642 [hep-ph]}}.

\bibitem{Asadi:2018wea}
P.~Asadi, M.~R. Buckley, and D.~Shih, {\em {It's all right(-handed neutrinos):
  a new $W'$ model for the $ {R}_{D^{{\left(\ast \right)}}} $ anomaly}\/},
  \href{http://dx.doi.org/10.1007/JHEP09(2018)010}{JHEP {\bf 09} (2018)  010},
  \href{http://arxiv.org/abs/1804.04135}{{\tt arXiv:1804.04135 [hep-ph]}}.

\bibitem{Robinson:2018gza}
D.~J. Robinson, B.~Shakya, and J.~Zupan, {\em {Right-handed Neutrinos and
  $R(D^{(*)})$}\/},  \href{http://arxiv.org/abs/1807.04753}{{\tt
  arXiv:1807.04753 [hep-ph]}}.

\bibitem{Azatov:2018kzb}
A.~Azatov, D.~Barducci, D.~Ghosh, D.~Marzocca, and L.~Ubaldi, {\em {Combined
  explanations of B-physics anomalies: the sterile neutrino solution}\/},
  \href{http://arxiv.org/abs/1807.10745}{{\tt arXiv:1807.10745 [hep-ph]}}.

\bibitem{Heeck:2018ntp}
J.~Heeck and D.~Teresi, {\em {Pati-Salam explanations of the $B$-meson
  anomalies}\/},  \href{http://arxiv.org/abs/1808.07492}{{\tt arXiv:1808.07492
  [hep-ph]}}.

\bibitem{Bhattacharya:2015ida}
S.~Bhattacharya, S.~Nandi, and S.~K. Patra, {\em {Optimal-observable analysis
  of possible new physics in $B\to D^{(\ast)}\tau\nu_{\tau}$}\/},
  \href{http://dx.doi.org/10.1103/PhysRevD.93.034011}{Phys. Rev. {\bf D93}
  (2016) no.~3, 034011},
\href{http://arxiv.org/abs/1509.07259}{{\tt arXiv:1509.07259 [hep-ph]}}.

\bibitem{Abbaneo:2001bv}
{CDF, DELPHI, ALEPH, SLD, OPAL, L3 Collaboration}, D.~Abbaneo et al., {\em
  {Combined results on $b$ hadron production rates and decay properties}\/},
  \href{http://arxiv.org/abs/hep-ex/0112028}{{\tt arXiv:hep-ex/0112028
  [hep-ex]}}.

\bibitem{Alonso:2016oyd}
R.~Alonso, B.-n. Grinstein, and J.~Martin~Camalich, {\em {Lifetime of $B_c^-$
  Constrains Explanations for Anomalies in $B\to D^{(*)}\tau\nu$}\/},  Phys.
  Rev. Lett. {\bf 118} (2017) no.~8, 081802,
  \href{http://arxiv.org/abs/1611.06676}{{\tt arXiv:1611.06676 [hep-ph]}}.

\bibitem{Akeroyd:2017mhr}
A.~G. Akeroyd and C.-H. Chen, {\em {Constraint on the branching ratio of $B_c
  \to \tau \bar{\nu}$ from LEP1 and consequences for $R(D^{(*)})$ anomaly}\/},
  \href{http://dx.doi.org/10.1103/PhysRevD.96.075011}{Phys. Rev. {\bf D96}
  (2017) no.~7, 075011},
\href{http://arxiv.org/abs/1708.04072}{{\tt arXiv:1708.04072 [hep-ph]}}.

\bibitem{Feruglio:2018fxo}
F.~Feruglio, P.~Paradisi, and O.~Sumensari, {\em {Implications of scalar and
  tensor explanations of $R_{D^{(\ast)}}$}\/},
  \href{http://arxiv.org/abs/1806.10155}{{\tt arXiv:1806.10155 [hep-ph]}}.

\bibitem{Dekens:2018bci}
W.~Dekens, J.~de~Vries, M.~Jung, and K.~K. Vos, {\em {The phenomenology of
  electric dipole moments in models of scalar leptoquarks}\/},
  \href{http://arxiv.org/abs/1809.09114}{{\tt arXiv:1809.09114 [hep-ph]}}.

\bibitem{Aloni:2017eny}
D.~Aloni, A.~Efrati, Y.~Grossman, and Y.~Nir, {\em {$\Upsilon$ and $\psi$
  leptonic decays as probes of solutions to the $R_D^{(*)}$ puzzle}\/},
  \href{http://dx.doi.org/10.1007/JHEP06(2017)019}{JHEP {\bf 06} (2017)  019},
  \href{http://arxiv.org/abs/1702.07356}{{\tt arXiv:1702.07356 [hep-ph]}}.

\bibitem{Lees:2013kla}
{BaBar Collaboration}, J.~P. Lees et al., {\em {Search for $B \to K^{(*)} \nu
  \overline \nu$ and invisible quarkonium decays}\/},
  \href{http://dx.doi.org/10.1103/PhysRevD.87.112005}{Phys. Rev. {\bf D87}
  (2013) no.~11, 112005}, \href{http://arxiv.org/abs/1303.7465}{{\tt
  arXiv:1303.7465 [hep-ex]}}.

\bibitem{Grygier:2017tzo}
{Belle Collaboration}, J.~Grygier et al., {\em {Search for $\boldsymbol{B\to
  h\nu\bar{\nu}}$ decays with semileptonic tagging at Belle}\/},
  \href{http://dx.doi.org/10.1103/PhysRevD.97.099902,
  10.1103/PhysRevD.96.091101}{Phys. Rev. {\bf D96} (2017) no.~9, 091101},
  \href{http://arxiv.org/abs/1702.03224}{{\tt arXiv:1702.03224 [hep-ex]}}.
  [Addendum: Phys. Rev.D97,no.9,099902(2018)].

\bibitem{Blake:2016olu}
T.~Blake, G.~Lanfranchi, and D.~M. Straub, {\em {Rare $B$ Decays as Tests of
  the Standard Model}\/},
  \href{http://dx.doi.org/10.1016/j.ppnp.2016.10.001}{Prog. Part. Nucl. Phys.
  {\bf 92} (2017)  50--91}, \href{http://arxiv.org/abs/1606.00916}{{\tt
  arXiv:1606.00916 [hep-ph]}}.

\bibitem{Cai:2017wry}
Y.~Cai, J.~Gargalionis, M.~A. Schmidt, and R.~R. Volkas, {\em {Reconsidering
  the One Leptoquark solution: flavor anomalies and neutrino mass}\/},
  \href{http://dx.doi.org/10.1007/JHEP10(2017)047}{JHEP {\bf 10} (2017)  047},
\href{http://arxiv.org/abs/1704.05849}{{\tt arXiv:1704.05849 [hep-ph]}}.

\bibitem{Korner:1987kd}
J.~G. K{\"o}rner and G.~A. Schuler, {\em {Exclusive Semileptonic Decays of
  Bottom Mesons in the Spectator Quark Model}\/},
  \href{http://dx.doi.org/10.1007/BF01584403}{Z. Phys. {\bf C38} (1988)  511}.
  [Erratum: Z. Phys.C41,690(1989)].

\bibitem{Hagiwara:1989gza}
K.~Hagiwara, A.~D. Martin, and M.~F. Wade, {\em {Helicity Amplitude Analysis of
  $B \to D^* \ell$ Neutrino Decays}\/},
  \href{http://dx.doi.org/10.1016/0370-2693(89)90541-8}{Phys. Lett. {\bf B228}
  (1989)  144--148}.

\bibitem{Tanaka:1994ay}
M.~Tanaka, {\em {Charged Higgs effects on exclusive semitauonic $B$ decays}\/},
   \href{http://dx.doi.org/10.1007/BF01571294}{Z. Phys. {\bf C67} (1995)
  321--326}, \href{http://arxiv.org/abs/hep-ph/9411405}{{\tt
  arXiv:hep-ph/9411405 [hep-ph]}}.

\bibitem{Chen:2005gr}
C.-H. Chen and C.-Q. Geng, {\em {Lepton angular asymmetries in semileptonic
  charmful B decays}\/},
  \href{http://dx.doi.org/10.1103/PhysRevD.71.077501}{Phys. Rev. {\bf D71}
  (2005)  077501}, \href{http://arxiv.org/abs/hep-ph/0503123}{{\tt
  arXiv:hep-ph/0503123 [hep-ph]}}.

\bibitem{Chen:2006nua}
C.-H. Chen and C.-Q. Geng, {\em {Charged Higgs on $B^- \to \tau \bar\nu_\tau$
  and $\bar B \to P(V) \ell \bar \nu_\ell$}\/},
  \href{http://dx.doi.org/10.1088/1126-6708/2006/10/053}{JHEP {\bf 10} (2006)
  053}, \href{http://arxiv.org/abs/hep-ph/0608166}{{\tt arXiv:hep-ph/0608166
  [hep-ph]}}.

\bibitem{Nierste:2008qe}
U.~Nierste, S.~Trine, and S.~Westhoff, {\em {Charged-Higgs effects in a new $B
  \to D \tau \nu$ differential decay distribution}\/},
  \href{http://dx.doi.org/10.1103/PhysRevD.78.015006}{Phys. Rev. {\bf D78}
  (2008)  015006}, \href{http://arxiv.org/abs/0801.4938}{{\tt arXiv:0801.4938
  [hep-ph]}}.

\bibitem{Tanaka:2010se}
M.~Tanaka and R.~Watanabe, {\em {Tau longitudinal polarization in B -> D tau nu
  and its role in the search for charged Higgs boson}\/},
  \href{http://dx.doi.org/10.1103/PhysRevD.82.034027}{Phys. Rev. {\bf D82}
  (2010)  034027}, \href{http://arxiv.org/abs/1005.4306}{{\tt arXiv:1005.4306
  [hep-ph]}}.

\bibitem{Datta:2012qk}
A.~Datta, M.~Duraisamy, and D.~Ghosh, {\em {Diagnosing New Physics in $b \to c
  \, \tau \, \nu_\tau$ decays in the light of the recent BaBar result}\/},
  \href{http://dx.doi.org/10.1103/PhysRevD.86.034027}{Phys. Rev. {\bf D86}
  (2012)  034027}, \href{http://arxiv.org/abs/1206.3760}{{\tt arXiv:1206.3760
  [hep-ph]}}.

\bibitem{Sakaki:2012ft}
Y.~Sakaki and H.~Tanaka, {\em {Constraints on the charged scalar effects using
  the forward-backward asymmetry on $B^-\to D^{(*)}\tau^-\nu_\tau$}\/},
  \href{http://dx.doi.org/10.1103/PhysRevD.87.054002}{Phys. Rev. {\bf D87}
  (2013) no.~5, 054002}, \href{http://arxiv.org/abs/1205.4908}{{\tt
  arXiv:1205.4908 [hep-ph]}}.

\bibitem{Duraisamy:2013kcw}
M.~Duraisamy and A.~Datta, {\em {The Full $B \to D^{*} \tau^{-} \bar{\nu_\tau}$
  Angular Distribution and CP violating Triple Products}\/},
  \href{http://dx.doi.org/10.1007/JHEP09(2013)059}{JHEP {\bf 09} (2013)  059},
  \href{http://arxiv.org/abs/1302.7031}{{\tt arXiv:1302.7031 [hep-ph]}}.

\bibitem{Alonso:2016gym}
R.~Alonso, A.~Kobach, and J.~Martin~Camalich, {\em {New physics in the
  kinematic distributions of $\bar B\to
  D^{(*)}\tau^-(\to\ell^-\bar\nu_\ell\nu_\tau)\bar\nu_\tau$}\/},
  \href{http://dx.doi.org/10.1103/PhysRevD.94.094021}{Phys. Rev. {\bf D94}
  (2016) no.~9, 094021}, \href{http://arxiv.org/abs/1602.07671}{{\tt
  arXiv:1602.07671 [hep-ph]}}.

\bibitem{Ligeti:2016npd}
Z.~Ligeti, M.~Papucci, and D.~J. Robinson, {\em {New Physics in the Visible
  Final States of $B\to D^{(*)}\tau\nu$}\/},
  \href{http://dx.doi.org/10.1007/JHEP01(2017)083}{JHEP {\bf 01} (2017)  083},
  \href{http://arxiv.org/abs/1610.02045}{{\tt arXiv:1610.02045 [hep-ph]}}.

\bibitem{Alonso:2017ktd}
R.~Alonso, J.~Martin~Camalich, and S.~Westhoff, {\em {Tau properties in $B\to
  D\tau\nu$ from visible final-state kinematics}\/},
  \href{http://dx.doi.org/10.1103/PhysRevD.95.093006}{Phys. Rev. {\bf D95}
  (2017) no.~9, 093006}, \href{http://arxiv.org/abs/1702.02773}{{\tt
  arXiv:1702.02773 [hep-ph]}}.

\bibitem{Duell:2016maj}
S.~Duell, F.~Bernlochner, Z.~Ligeti, M.~Papucci, and D.~Robinson, {\em {HAMMER:
  Reweighting tool for simulated data samples}\/},  PoS {\bf ICHEP2016} (2017)
  1074.

\bibitem{CMS:2014xfa}
{LHCb, CMS Collaboration}, V.~Khachatryan et al., {\em {Observation of the rare
  $B^0_s\to\mu^+\mu^-$ decay from the combined analysis of CMS and LHCb
  data}\/},  Nature {\bf 522} (2015)  68--72,
  \href{http://arxiv.org/abs/1411.4413}{{\tt arXiv:1411.4413 [hep-ex]}}.

\bibitem{ATLAS-CONF-2018-046}
{ATLAS Collaboration}, {\em {Study of the rare decays of B0s and B0 into muon
  pairs from data collected during 2015 and 2016 with the ATLAS detector}\/},
  ATLAS-CONF-2018-046, CERN, Geneva, Sep, 2018.
\newblock \url{https://cds.cern.ch/record/2639673}.

\bibitem{CMS-PAS-FTR-18-013}
{CMS Collaboration}, {\em Measurement of rare $B \to \mu^+ \mu^-$ decays with
  the Phase-2 upgraded CMS detector at the HL-LHC\/},  CMS Physics Analysis
  Summary CMS-PAS-FTR-18-013, 2018.
\newblock \url{http://cdsweb.cern.ch/record/2650545}.

\bibitem{fsfd}
{LHCb collaboration}, R.~Aaij et al., {\em {Measurement of the fragmentation
  fraction ratio $f_s/f_d$ and its dependence on $B$ meson kinematics}\/},
  \href{http://dx.doi.org/10.1007/JHEP04(2013)001}{JHEP {\bf 04} (2013)  001},
  \href{http://arxiv.org/abs/1301.5286}{{\tt arXiv:1301.5286 [hep-ex]}}.
  $f_s/f_d$ value updated in
  \href{https://cds.cern.ch/record/1559262}{LHCb-CONF-2013-011}.

\bibitem{ATL-PHYS-PUB-2018-005}
{ATLAS Collaboration}, {\em {Prospects for the $\cal{B}$$(B^0_{(s)} \to \mu^+
  \mu^-)$ measurements with the ATLAS detector in the Run 2 and HL-LHC data
  campaigns}\/},   ATL-PHYS-PUB-2018-005, CERN, Geneva, May, 2018.
\newblock \url{https://cds.cern.ch/record/2317211}.

\bibitem{CMS-PAS-FTR-14-015}
{CMS Collaboration}, {\em {B Physics analyses for the Phase-II Upgrade
  Technical Proposal}\/},  CMS Physics Analysis Summary CMS-PAS-FTR-14-015,
  2014.
\newblock \url{https://cds.cern.ch/record/2036007}.

\bibitem{Aaboud:2016ire}
{ATLAS Collaboration}, M.~Aaboud et al., {\em { Study of the rare decays of
  $B^0_s$ and $B^0$ into muon pairs from data collected during the LHC Run 1
  with the ATLAS detector}\/},  Eur. Phys. J {\bf C76} (2016)  513,
  \href{http://arxiv.org/abs/1604.04263}{{\tt arXiv:1604.04263 [hep-ex]}}.

\bibitem{Bobeth}
C.~Bobeth, M.~Gorbahn, T.~Hermann, M.~Misiak, E.~Stamou, and M.~Steinhauser,
  {\em {$B_{s,d} \to l^+ l^-$ in the Standard Model with Reduced Theoretical
  Uncertainty}\/},
  \href{http://dx.doi.org/10.1103/PhysRevLett.112.101801}{Phys. Rev. Lett. {\bf
  112} (2014)  101801}, \href{http://arxiv.org/abs/1311.0903}{{\tt
  arXiv:1311.0903 [hep-ph]}}.

\bibitem{Dutta:2015dla}
B.~Dutta and Y.~Mimura, {\em {Enhancement of $Br(B_d \to \mu^+\mu^-)/Br(B_s \to
  \mu^+\mu^-)$ in supersymmetric unified models}\/},
  \href{http://dx.doi.org/10.1103/PhysRevD.91.095011}{Phys. Rev. {\bf D91}
  (2015) no.~9, 095011},
\href{http://arxiv.org/abs/1501.02044}{{\tt arXiv:1501.02044 [hep-ph]}}.

\bibitem{splot}
M.~Pivk and F.~R. Le~Diberder, {\em { sPlot: a statistical tool to unfold data
  distributions}\/},  Nucl. Instrum. Method. {\bf A555} (2005)  356,
  \href{http://arxiv.org/abs/0402083}{{\tt arXiv:0402083 [physics]}}.

\bibitem{Kruger:2002gf}
F.~Kruger and D.~Melikhov, {\em {Gauge invariance and form-factors for the
  decay $B\to\gamma\ellp\ellm$}\/},  Phys. Rev. {\bf D67} (2003)  034002,
  \href{http://arxiv.org/abs/hep-ph/0208256}{{\tt arXiv:hep-ph/0208256
  [hep-ph]}}.

\bibitem{Melikhov:2004mk}
D.~Melikhov and N.~Nikitin, {\em {Rare radiative leptonic decays
  $B_{d,s}\to\ellp\ellm\gamma$}\/},  Phys. Rev. {\bf D70} (2004)  114028,
  \href{http://arxiv.org/abs/hep-ph/0410146}{{\tt arXiv:hep-ph/0410146
  [hep-ph]}}.

\bibitem{Aubert:2007up}
{BaBar collaboration}, B.~Aubert et al., {\em {Search for the decays $B^0 \to
  e^{+} e^{-} \gamma$ and $B^0 \to \mu^{+} \mu^{-} \gamma$}\/},  Phys. Rev.
  {\bf D77} (2008)  011104, \href{http://arxiv.org/abs/0706.2870}{{\tt
  arXiv:0706.2870 [hep-ex]}}.

\bibitem{CMS-PAS-FTR-18-033}
{CMS Collaboration}, {\em Study of the expected sensitivity to the
  $P_5^{\prime}$ parameter in the $B^0 \to K^{*0} \mu^+\mu^-$ decay at the
  HL-LHC\/},  CMS Physics Analysis Summary CMS-PAS-FTR-18-033, 2018.
\newblock \url{http://cds.cern.ch/record/2651298}.

\bibitem{Sirunyan:2017dhj}
{CMS collaboration}, A.~M. Sirunyan et al., {\em {Measurement of angular
  parameters from the decay $\mathrm{B}^0 \to \mathrm{K}^{*0} \mu^+ \mu^-$ in
  proton-proton collisions at $\sqrt{s} = $ 8 TeV}\/},  Phys. Lett. {\bf B781}
  (2018)  517--541, \href{http://arxiv.org/abs/1710.02846}{{\tt
  arXiv:1710.02846 [hep-ex]}}.

\bibitem{Aaboud:2018krd}
{ATLAS collaboration}, M.~Aaboud et al., {\em {Angular analysis of $B^0_d
  \rightarrow K^{*}\mu^+\mu^-$ decays in $pp$ collisions at $\sqrt{s}= 8$ TeV
  with the ATLAS detector}\/},  \href{http://arxiv.org/abs/1805.04000}{{\tt
  arXiv:1805.04000 [hep-ex]}}.

\bibitem{ATLAS:2285585}
A.~Collaboration, {\em {Technical Design Report for the ATLAS Inner Tracker
  Pixel Detector}\/},   CERN-LHCC-2017-021. ATLAS-TDR-030, CERN, Geneva, Sep,
  2017.
\newblock \url{https://cds.cern.ch/record/2285585}.

\bibitem{ATL-PHYS-PUB-2018-003}
{ATLAS Collaboration}, {\em {Studies of radial distortions of the ATLAS Inner
  Detector}\/},   ATL-PHYS-PUB-2018-003, CERN, Geneva, Mar, 2018.
\newblock \url{https://cds.cern.ch/record/2309785}.

\bibitem{LHCb-PAPER-2013-037}
{LHCb collaboration}, R.~Aaij et al., {\em {Measurement of
  form-factor-independent observables in the decay $\Bz\to
  \Kstarz\mup\mun$}\/},  Phys. Rev. Lett. {\bf 111} (2013)  191801,
  \href{http://arxiv.org/abs/1308.1707}{{\tt arXiv:1308.1707 [hep-ex]}}.

\bibitem{LHCb-PAPER-2014-006}
{LHCb collaboration}, R.~Aaij et al., {\em {Differential branching fractions
  and isospin asymmetries of $\B\to \Kstar\mup\mun$ decays}\/},  JHEP {\bf 06}
  (2014)  133, \href{http://arxiv.org/abs/1403.8044}{{\tt arXiv:1403.8044
  [hep-ex]}}.

\bibitem{LHCb-PAPER-2013-019}
{LHCb collaboration}, R.~Aaij et al., {\em {Differential branching fraction and
  angular analysis of the decay $\Bz\to \Kstarz\mup\mun$}\/},  JHEP {\bf 08}
  (2013)  131, \href{http://arxiv.org/abs/1304.6325}{{\tt arXiv:1304.6325
  [hep-ex]}}.

\bibitem{Aubert:2006vb}
{BaBar collaboration}, B.~Aubert et al., {\em {Measurements of branching
  fractions, rate asymmetries, and angular distributions in the rare decays $B
  \to K \ell^{+} \ell^{-}$ and $B \to K^{*} \ell^{+} \ell^{-}$}\/},  Phys. Rev.
  {\bf D73} (2006)  092001, \href{http://arxiv.org/abs/hep-ex/0604007}{{\tt
  arXiv:hep-ex/0604007 [hep-ex]}}.

\bibitem{Lees:2015ymt}
{BaBar collaboration}, J.~P. Lees et al., {\em {Measurement of angular
  asymmetries in the decays $B \to K^\ast \ell^+\ell^-$}\/},  Phys. Rev. {\bf
  D93} (2016)  052015, \href{http://arxiv.org/abs/1508.07960}{{\tt
  arXiv:1508.07960 [hep-ex]}}.

\bibitem{Wei:2009zv}
{Belle Collaboration}, J.~T. Wei et al., {\em {Measurement of the Differential
  Branching Fraction and Forward-Backword Asymmetry for $B \to
  K^{(*)}\ell^+\ell^-$}\/},  Phys. Rev. Lett. {\bf 103} (2009)  171801,
  \href{http://arxiv.org/abs/0904.0770}{{\tt arXiv:0904.0770 [hep-ex]}}.

\bibitem{Aaltonen:2011ja}
{CDF Collaboration}, T.~Aaltonen et al., {\em {Measurements of the Angular
  Distributions in the Decays $B \to K^{(*)} \mu^+ \mu^-$ at CDF}\/},  Phys.
  Rev. Lett. {\bf 108} (2012)  081807,
  \href{http://arxiv.org/abs/1108.0695}{{\tt arXiv:1108.0695 [hep-ex]}}.

\bibitem{Chatrchyan:2013cda}
{CMS Collaboration}, S.~Chatrchyan et al., {\em {Angular analysis and branching
  fraction measurement of the decay $B^0 \to K^{*0} \mu^+\mu^-$}\/},  Phys.
  Lett. {\bf B727} (2013)  77--100, \href{http://arxiv.org/abs/1308.3409}{{\tt
  arXiv:1308.3409 [hep-ex]}}.

\bibitem{Khachatryan:2015isa}
{CMS collaboration}, V.~Khachatryan et al., {\em {Angular analysis of the decay
  $B^0 \to K^{*0} \mu^+ \mu^-$ from pp collisions at $\sqrt s = 8$ TeV}\/},
  Phys. Lett. {\bf B753} (2016)  424--448,
  \href{http://arxiv.org/abs/1507.08126}{{\tt arXiv:1507.08126 [hep-ex]}}.

\bibitem{Beneke:2004dp}
M.~Beneke, T.~Feldmann, and D.~Seidel, {\em {Exclusive radiative and
  electroweak $b \to d$ and $b \to s$ penguin decays at NLO}\/},  Eur. Phys. J.
  {\bf C41} (2005)  173--188, \href{http://arxiv.org/abs/hep-ph/0412400}{{\tt
  arXiv:hep-ph/0412400 [hep-ph]}}.

\bibitem{Egede:2015kha}
U.~Egede, M.~Patel, and K.~A. Petridis, {\em {Method for an unbinned
  measurement of the q$^{2}$ dependent decay amplitudes of $\Bzb\to {K}^{\ast
  0}{\mu}^{+}{\mu}^{-}$ decays}\/},  JHEP {\bf 06} (2015)  084,
  \href{http://arxiv.org/abs/1504.00574}{{\tt arXiv:1504.00574 [hep-ph]}}.

\bibitem{Kruger:2005ep}
F.~Kruger and J.~Matias, {\em {Probing new physics via the transverse
  amplitudes of $B^0\to K^{*0} (\to K^- \pi^+) l^+l^-$ at large recoil}\/},
  Phys. Rev. {\bf D71} (2005)  094009,
  \href{http://arxiv.org/abs/hep-ph/0502060}{{\tt arXiv:hep-ph/0502060
  [hep-ph]}}.

\bibitem{Lyon:2014hpa}
J.~Lyon and R.~Zwicky, {\em {Resonances gone topsy turvy - the charm of QCD or
  new physics in $b \to s \ell^+ \ell^-$?}\/},
  \href{http://arxiv.org/abs/1406.0566}{{\tt arXiv:1406.0566 [hep-ph]}}.

\bibitem{Altmannshofer:2013foa}
W.~Altmannshofer and D.~M. Straub, {\em {New Physics in $B \to K^*\mu\mu$?}\/},
   Eur. Phys. J. {\bf C73} (2013)  2646,
  \href{http://arxiv.org/abs/1308.1501}{{\tt arXiv:1308.1501 [hep-ph]}}.

\bibitem{Descotes-Genon:2013wba}
S.~Descotes-Genon, J.~Matias, and J.~Virto, {\em {Understanding the $B\to
  K^*\mu^+\mu^-$ anomaly}\/},  Phys. Rev. {\bf D88} (2013)  074002,
  \href{http://arxiv.org/abs/1307.5683}{{\tt arXiv:1307.5683 [hep-ph]}}.

\bibitem{ATL-PHYS-PUB-2019-003}
{ATLAS Collaboration}, {\em $B^0_d \to K^{*0}\mu\mu$ angular analysis prospects
  with the upgraded ATLAS detector at the HL-LHC\/},   ATL-PHYS-PUB-2019-003,
  CERN, Geneva, Jan, 2019.
\newblock \url{https://cds.cern.ch/record/2654519}.

\bibitem{Ciuchini:2015qxb}
M.~Ciuchini, M.~Fedele, E.~Franco, S.~Mishima, A.~Paul, L.~Silvestrini, and
  M.~Valli, {\em {$B\to K^* \ell^+ \ell^-$ decays at large recoil in the
  Standard Model: a theoretical reappraisal}\/},  JHEP {\bf 06} (2016)  116,
  \href{http://arxiv.org/abs/1512.07157}{{\tt arXiv:1512.07157 [hep-ph]}}.

\bibitem{Descotes-Genon:2014uoa}
S.~Descotes-Genon, L.~Hofer, J.~Matias, and J.~Virto, {\em {On the impact of
  power corrections in the prediction of $B \to K^*\mu^+\mu^-$ observables}\/},
   \href{http://dx.doi.org/10.1007/JHEP12(2014)125}{JHEP {\bf 12} (2014)  125},
\href{http://arxiv.org/abs/1407.8526}{{\tt arXiv:1407.8526 [hep-ph]}}.

\bibitem{LHCb-PAPER-2016-045}
{LHCb collaboration}, R.~Aaij et al., {\em {Measurement of the phase difference
  between the short- and long-distance amplitudes in the $\Bp\to \Kp\mup\mun$
  decay}\/},  Eur. Phys. J. {\bf C77} (2017)  161,
  \href{http://arxiv.org/abs/1612.06764}{{\tt arXiv:1612.06764 [hep-ex]}}.

\bibitem{Bobeth:2017vxj}
C.~Bobeth, M.~Chrzaszcz, D.~van Dyk, and J.~Virto, {\em {Long-distance effects
  in $B\to K^*\ell\ell$ from Analyticity}\/},
  \href{http://arxiv.org/abs/1707.07305}{{\tt arXiv:1707.07305 [hep-ph]}}.

\bibitem{LHCb-PAPER-2012-020}
{LHCb collaboration}, R.~Aaij et al., {\em {First observation of the decay
  $\Bp\to \pip\mup\mun$}\/},  JHEP {\bf 12} (2012)  125,
  \href{http://arxiv.org/abs/1210.2645}{{\tt arXiv:1210.2645 [hep-ex]}}.

\bibitem{LHCb-PAPER-2015-035}
{LHCb collaboration}, R.~Aaij et al., {\em {First measurement of the
  differential branching fraction and \CP asymmetry of the $\Bp\to
  \pip\mup\mun$ decay}\/},  JHEP {\bf 10} (2015)  034,
  \href{http://arxiv.org/abs/1509.00414}{{\tt arXiv:1509.00414 [hep-ex]}}.

\bibitem{LHCb-PAPER-2016-049}
{LHCb collaboration}, R.~Aaij et al., {\em {Observation of the suppressed decay
  $\Lb\to \proton\pim\mup\mun$}\/},  JHEP {\bf 04} (2017)  029,
  \href{http://arxiv.org/abs/1701.08705}{{\tt arXiv:1701.08705 [hep-ex]}}.

\bibitem{LHCb-PAPER-2018-004}
{LHCb collaboration}, R.~Aaij et al., {\em {Evidence for the decay
  \decay{\Bs}{\Kstarzb \mumu}}\/},  JHEP {\bf 07} (2018)  020,
  \href{http://arxiv.org/abs/1804.07167}{{\tt arXiv:1804.07167 [hep-ex]}}.

\bibitem{Du:2015tda}
D.~Du, A.~X. El-Khadra, S.~Gottlieb, A.~S. Kronfeld, J.~Laiho, E.~Lunghi, R.~S.
  Van~de Water, and R.~Zhou, {\em {Phenomenology of semileptonic B-meson decays
  with form factors from lattice QCD}\/},  Phys. Rev. {\bf D93} (2016)  034005,
  \href{http://arxiv.org/abs/1510.02349}{{\tt arXiv:1510.02349 [hep-ph]}}.

\bibitem{Capdevila:2016ivx}
B.~Capdevila, S.~Descotes-Genon, J.~Matias, and J.~Virto, {\em {Assessing
  lepton-flavour non-universality from $B\to K^*\ell\ell$ angular analyses}\/},
   JHEP {\bf 10} (2016)  075, \href{http://arxiv.org/abs/1605.03156}{{\tt
  arXiv:1605.03156 [hep-ph]}}.

\bibitem{Serra:2016ivr}
N.~Serra, R.~Silva~Coutinho, and D.~van Dyk, {\em {Measuring the breaking of
  lepton flavor universality in $B\to K^*\ell^+\ell^-$}\/},
  \href{http://dx.doi.org/10.1103/PhysRevD.95.035029}{Phys. Rev. {\bf D95}
  (2017) no.~3, 035029},
\href{http://arxiv.org/abs/1610.08761}{{\tt arXiv:1610.08761 [hep-ph]}}.

\bibitem{Mauri:2018vbg}
A.~Mauri, N.~Serra, and R.~Silva~Coutinho, {\em {Towards establishing Lepton
  Flavour Universality violation in $\bar{B}\to \bar{K}^*\ell^+\ell^-$
  decays}\/},
\href{http://arxiv.org/abs/1805.06401}{{\tt arXiv:1805.06401 [hep-ph]}}.

\bibitem{Descotes-Genon:2015hea}
S.~Descotes-Genon and J.~Virto, {\em {Time dependence in $B \to V\ell^+\ell^-$
  decays}\/},  JHEP {\bf 04} (2015)  045,
  \href{http://arxiv.org/abs/1502.05509}{{\tt arXiv:1502.05509 [hep-ph]}}.

\bibitem{Ball:2006cva}
P.~Ball and R.~Zwicky, {\em {Time-dependent \CP\ asymmetry in $B\to K^*\gamma$
  as a (quasi) null test of the Standard Model}\/},  Phys. Lett. {\bf B642}
  (2006)  478--486, \href{http://arxiv.org/abs/hep-ph/0609037}{{\tt
  arXiv:hep-ph/0609037 [hep-ph]}}.

\bibitem{Matsumori:2005ax}
M.~Matsumori and A.~I. Sanda, {\em {The mixing-induced $CP$ asymmetry in $B \to
  K^* \gamma$ decays with perturbative QCD approach}\/},  Phys. Rev. {\bf D73}
  (2006)  114022, \href{http://arxiv.org/abs/hep-ph/0512175}{{\tt
  arXiv:hep-ph/0512175 [hep-ph]}}.

\bibitem{LHCb-PAPER-2014-001}
{LHCb collaboration}, R.~Aaij et al., {\em {Observation of photon polarization
  in the $\bquark\to \squark\g$ transition}\/},  Phys. Rev. Lett. {\bf 112}
  (2014)  161801, \href{http://arxiv.org/abs/1402.6852}{{\tt arXiv:1402.6852
  [hep-ex]}}.

\bibitem{Bellee:2018}
V.~Bell{\'e}e, F.~Blanc, P.~Pais, A.~Puig, O.~Schneider, and K.~Trabelsi, {\em
  {Measuring photon polarisation in $B\to K\pi\pi\gamma$ decays}\/},  2018.
\newblock in preparation.

\bibitem{LHCb-PAPER-2014-066}
{LHCb collaboration}, R.~Aaij et al., {\em {Angular analysis of the $\Bz\to
  \Kstar\epem$ decay in the low-$q^2$ region}\/},  JHEP {\bf 04} (2015)  064,
  \href{http://arxiv.org/abs/1501.03038}{{\tt arXiv:1501.03038 [hep-ex]}}.

\bibitem{Becirevic:2011bp}
D.~Becirevic and E.~Schneider, {\em {On transverse asymmetries in $B \to K^*
  \ell^+\ell^-$}\/},
  \href{http://dx.doi.org/10.1016/j.nuclphysb.2011.09.004}{Nucl. Phys. {\bf
  B854} (2012)  321--339},
\href{http://arxiv.org/abs/1106.3283}{{\tt arXiv:1106.3283 [hep-ph]}}.

\bibitem{Acosta:2002fh}
{CDF collaboration}, D.~Acosta et al., {\em {Search for radiative b-hadron
  decays in $p\bar{p}$ collisions at $\sqrt{s} = 1.8$ TeV}\/},  Phys. Rev. {\bf
  D66} (2002)  112002, \href{http://arxiv.org/abs/hep-ex/0208035}{{\tt
  arXiv:hep-ex/0208035 [hep-ex]}}.

\bibitem{LHCb-PAPER-2012-057}
{LHCb collaboration}, R.~Aaij et al., {\em {Measurements of the $\Lb\to
  \jpsi\Lz$ decay amplitudes and the $\Lb$ polarisation in $\proton\proton$
  collisions at $\sqrt{s} = 7$\tev}\/},  Phys. Lett. {\bf B724} (2013)  27,
  \href{http://arxiv.org/abs/1302.5578}{{\tt arXiv:1302.5578 [hep-ex]}}.

\bibitem{Becirevic:2016hea}
D.~Becirevic, S.~Fajfer, I.~Nisandzic, and A.~Tayduganov, {\em {Angular
  distributions of $\bar B \to D^{(\ast)}\ell\bar \nu_\ell$ decays and search
  of New Physics}\/},
\href{http://arxiv.org/abs/1602.03030}{{\tt arXiv:1602.03030 [hep-ph]}}.

\bibitem{Azatov:2018knx}
A.~Azatov, D.~Bardhan, D.~Ghosh, F.~Sgarlata, and E.~Venturini, {\em {Anatomy
  of $b \to c \tau \nu$ anomalies}\/},
\href{http://arxiv.org/abs/1805.03209}{{\tt arXiv:1805.03209 [hep-ph]}}.

\bibitem{LHCb-PAPER-2017-031}
{LHCb collaboration}, R.~Aaij et al., {\em {Search for the lepton-flavour
  violating decays $\BdorBs\to e^\pm\mu^\mp$}\/},  JHEP {\bf 03} (2018)  078,
  \href{http://arxiv.org/abs/1710.04111}{{\tt arXiv:1710.04111}}.

\bibitem{Aubert:2008cu}
{BaBar collaboration}, B.~Aubert et al., {\em {Searches for the decays $B^0 \to
  \ell^\pm \tau^\mp$ and $B^{+} \to \ell^{+} \nu$ (l=e, $\mu$) using hadronic
  tag reconstruction}\/},  Phys. Rev. {\bf D77} (2008)  091104,
  \href{http://arxiv.org/abs/0801.0697}{{\tt arXiv:0801.0697 [hep-ex]}}.

\bibitem{Lees:2012zz}
{BaBar collaboration}, J.~P. Lees et al., {\em {A search for the decay modes
  $B^{+-} \to h^{+-} \tau^{+-}l$}\/},  Phys. Rev. {\bf D86} (2012)  012004,
  \href{http://arxiv.org/abs/1204.2852}{{\tt arXiv:1204.2852 [hep-ex]}}.

\bibitem{Atre:2009rg}
A.~Atre, T.~Han, S.~Pascoli, and B.~Zhang, {\em {The search for heavy Majorana
  neutrinos}\/},  JHEP {\bf 05} (2009)  030,
  \href{http://arxiv.org/abs/0901.3589}{{\tt arXiv:0901.3589 [hep-ph]}}.

\bibitem{LHCb-PAPER-2011-009}
{LHCb collaboration}, R.~Aaij et al., {\em {Search for lepton number violating
  decays $\Bp\to \pim\mup\mup$ and $\Bp\to \Km\mup\mup$}\/},  Phys. Rev. Lett.
  {\bf 108} (2012)  101601, \href{http://arxiv.org/abs/1110.0730}{{\tt
  arXiv:1110.0730 [hep-ex]}}.

\bibitem{LHCb-PAPER-2011-038}
{LHCb collaboration}, R.~Aaij et al., {\em {Searches for Majorana neutrinos in
  $\Bm$ decays}\/},  Phys. Rev. {\bf D85} (2012)  112004,
  \href{http://arxiv.org/abs/1201.5600}{{\tt arXiv:1201.5600 [hep-ex]}}.

\bibitem{LHCb-PAPER-2013-064}
{LHCb collaboration}, R.~Aaij et al., {\em {Search for Majorana neutrinos in
  $\Bm\to \pip\mun\mun$ decays}\/},  Phys. Rev. Lett. {\bf 112} (2014)  131802,
  \href{http://arxiv.org/abs/1401.5361}{{\tt arXiv:1401.5361 [hep-ex]}}.

\bibitem{Smith:2011rp}
C.~Smith, {\em {Proton stability from a fourth family}\/},  Phys.Rev. {\bf D85}
  (2012)  036005, \href{http://arxiv.org/abs/1105.1723}{{\tt arXiv:1105.1723
  [hep-ph]}}.

\bibitem{Durieux:2012gj}
G.~Durieux, J.-M. G\'erard, F.~Maltoni, and C.~Smith, {\em {Three-generation
  baryon and lepton number violation at the LHC}\/},  Phys. Lett. {\bf B721}
  (2013)  82--85, \href{http://arxiv.org/abs/1210.6598}{{\tt arXiv:1210.6598
  [hep-ph]}}.

\bibitem{Lees:2011hb}
{BaBar collaboration}, J.~P. Lees et al., {\em {Searches for rare or forbidden
  semileptonic charm decays}\/},  Phys. Rev. {\bf D84} (2011)  072006,
  \href{http://arxiv.org/abs/1107.4465}{{\tt arXiv:1107.4465 [hep-ex]}}.

\bibitem{Gedalia:2012sx}
O.~Gedalia, G.~Isidori, F.~Maltoni, G.~Perez, M.~Selvaggi, and Y.~Soreq, {\em
  {Top B Physics at the LHC}\/},
  \href{http://dx.doi.org/10.1103/PhysRevLett.110.232002}{Phys. Rev. Lett. {\bf
  110} (2013) no.~23, 232002},
\href{http://arxiv.org/abs/1212.4611}{{\tt arXiv:1212.4611 [hep-ph]}}.

\bibitem{Aaboud:2016bmk}
{ATLAS Collaboration}, M.~Aaboud et al., {\em {Measurements of charge and CP
  asymmetries in $b$-hadron decays using top-quark events collected by the
  ATLAS detector in $pp$ collisions at $\sqrt{s}=8$ TeV}\/},
  \href{http://dx.doi.org/10.1007/JHEP02(2017)071}{JHEP {\bf 02} (2017)  071},
\href{http://arxiv.org/abs/1610.07869}{{\tt arXiv:1610.07869 [hep-ex]}}.

\bibitem{Degrande:2012wf}
C.~Degrande, N.~Greiner, W.~Kilian, O.~Mattelaer, H.~Mebane, T.~Stelzer,
  S.~Willenbrock, and C.~Zhang, {\em {Effective Field Theory: A Modern Approach
  to Anomalous Couplings}\/},
  \href{http://dx.doi.org/10.1016/j.aop.2013.04.016}{Annals Phys. {\bf 335}
  (2013)  21--32},
\href{http://arxiv.org/abs/1205.4231}{{\tt arXiv:1205.4231 [hep-ph]}}.

\bibitem{AguilarSaavedra:2018nen}
D.~Barducci et al., {\em {Interpreting top-quark LHC measurements in the
  standard-model effective field theory}\/},
\href{http://arxiv.org/abs/1802.07237}{{\tt arXiv:1802.07237 [hep-ph]}}.

\bibitem{Durieux:2014xla}
G.~Durieux, F.~Maltoni, and C.~Zhang, {\em {Global approach to top-quark
  flavor-changing interactions}\/},
  \href{http://dx.doi.org/10.1103/PhysRevD.91.074017}{Phys. Rev. {\bf D91}
  (2015) no.~7, 074017},
\href{http://arxiv.org/abs/1412.7166}{{\tt arXiv:1412.7166 [hep-ph]}}.

\bibitem{Achard:2002vv}
{L3 Collaboration}, P.~Achard et al., {\em {Search for single top production at
  LEP}\/},  \href{http://dx.doi.org/10.1016/S0370-2693(02)02933-7}{Phys.Lett.
  {\bf B549} (2002)  290--300},
\href{http://arxiv.org/abs/hep-ex/0210041}{{\tt arXiv:hep-ex/0210041
  [hep-ex]}}.

\bibitem{Kidonakis:2003sc}
N.~Kidonakis and A.~Belyaev, {\em {FCNC top quark production via anomalous tqV
  couplings beyond leading order}\/},
  \href{http://dx.doi.org/10.1088/1126-6708/2003/12/004}{JHEP {\bf 0312} (2003)
   004},
\href{http://arxiv.org/abs/hep-ph/0310299}{{\tt arXiv:hep-ph/0310299
  [hep-ph]}}.

\bibitem{Zhang:2008yn}
J.~J. Zhang, C.~S. Li, J.~Gao, H.~Zhang, Z.~Li, et al., {\em {Next-to-leading
  order QCD corrections to the top quark decay via model-independent FCNC
  couplings}\/},
  \href{http://dx.doi.org/10.1103/PhysRevLett.102.072001}{Phys.Rev.Lett. {\bf
  102} (2009)  072001},
\href{http://arxiv.org/abs/0810.3889}{{\tt arXiv:0810.3889 [hep-ph]}}.

\bibitem{Drobnak:2010wh}
J.~Drobnak, S.~Fajfer, and J.~F. Kamenik, {\em {Flavor Changing Neutral
  Coupling Mediated Radiative Top Quark Decays at Next-to-Leading Order in
  QCD}\/},
  \href{http://dx.doi.org/10.1103/PhysRevLett.104.252001}{Phys.Rev.Lett. {\bf
  104} (2010)  252001},
\href{http://arxiv.org/abs/1004.0620}{{\tt arXiv:1004.0620 [hep-ph]}}.

\bibitem{Drobnak:2010by}
J.~Drobnak, S.~Fajfer, and J.~F. Kamenik, {\em {QCD Corrections to Flavor
  Changing Neutral Coupling Mediated Rare Top Quark Decays}\/},
  \href{http://dx.doi.org/10.1103/PhysRevD.82.073016}{Phys.Rev. {\bf D82}
  (2010)  073016},
\href{http://arxiv.org/abs/1007.2551}{{\tt arXiv:1007.2551 [hep-ph]}}.

\bibitem{Zhang:2010bm}
J.~J. Zhang, C.~S. Li, J.~Gao, H.~X. Zhu, C.-P. Yuan, et al., {\em
  {Next-to-leading order QCD corrections to the top quark decay via the
  Flavor-Changing Neutral-Current operators with mixing effects}\/},
  \href{http://dx.doi.org/10.1103/PhysRevD.82.073005}{Phys.Rev. {\bf D82}
  (2010)  073005},
\href{http://arxiv.org/abs/1004.0898}{{\tt arXiv:1004.0898 [hep-ph]}}.

\bibitem{Zhang:2014rja}
C.~Zhang, {\em {Effective field theory approach to top-quark decay at
  next-to-leading order in QCD}\/},
  \href{http://dx.doi.org/10.1103/PhysRevD.90.014008}{Phys. Rev. {\bf D90}
  (2014) no.~1, 014008},
\href{http://arxiv.org/abs/1404.1264}{{\tt arXiv:1404.1264 [hep-ph]}}.

\bibitem{Liu:2005dp}
J.~J. Liu, C.~S. Li, L.~L. Yang, and L.~G. Jin, {\em {Next-to-leading order QCD
  corrections to the direct top quark production via model-independent FCNC
  couplings at hadron colliders}\/},
  \href{http://dx.doi.org/10.1103/PhysRevD.72.074018}{Phys.Rev. {\bf D72}
  (2005)  074018},
\href{http://arxiv.org/abs/hep-ph/0508016}{{\tt arXiv:hep-ph/0508016
  [hep-ph]}}.

\bibitem{Gao:2009rf}
J.~Gao, C.~S. Li, J.~J. Zhang, and H.~X. Zhu, {\em {Next-to-leading order QCD
  corrections to the single top quark production via model-independent t-q-g
  flavor-changing neutral-current couplings at hadron colliders}\/},
  \href{http://dx.doi.org/10.1103/PhysRevD.80.114017}{Phys.Rev. {\bf D80}
  (2009)  114017},
\href{http://arxiv.org/abs/0910.4349}{{\tt arXiv:0910.4349 [hep-ph]}}.

\bibitem{Zhang:2011gh}
Y.~Zhang, B.~H. Li, C.~S. Li, J.~Gao, and H.~X. Zhu, {\em {Next-to-leading
  order QCD corrections to the top quark associated with $\gamma$ production
  via model-independent flavor-changing neutral-current couplings at hadron
  colliders}\/},  \href{http://dx.doi.org/10.1103/PhysRevD.83.094003}{Phys.Rev.
  {\bf D83} (2011)  094003},
\href{http://arxiv.org/abs/1101.5346}{{\tt arXiv:1101.5346 [hep-ph]}}.

\bibitem{Li:2011ek}
B.~H. Li, Y.~Zhang, C.~S. Li, J.~Gao, and H.~X. Zhu, {\em {Next-to-leading
  order QCD corrections to $tZ$ associated production via the flavor-changing
  neutral-current couplings at hadron colliders}\/},
  \href{http://dx.doi.org/10.1103/PhysRevD.83.114049}{Phys.Rev. {\bf D83}
  (2011)  114049},
\href{http://arxiv.org/abs/1103.5122}{{\tt arXiv:1103.5122 [hep-ph]}}.

\bibitem{Wang:2012gp}
Y.~Wang, F.~P. Huang, C.~S. Li, B.~H. Li, D.~Y. Shao, et al., {\em {Constraints
  on flavor-changing neutral-current $Htq$ couplings from the signal of $tH$
  associated production with QCD next-to-leading order accuracy at the LHC}\/},
   \href{http://dx.doi.org/10.1103/PhysRevD.86.094014}{Phys.Rev. {\bf D86}
  (2012)  094014},
\href{http://arxiv.org/abs/1208.2902}{{\tt arXiv:1208.2902 [hep-ph]}}.

\bibitem{Alloul:2013bka}
A.~Alloul, N.~D. Christensen, C.~Degrande, C.~Duhr, and B.~Fuks, {\em
  {FeynRules 2.0 - A complete toolbox for tree-level phenomenology}\/},
  \href{http://dx.doi.org/10.1016/j.cpc.2014.04.012}{Comput.Phys.Commun. {\bf
  185} (2014)  2250--2300},
\href{http://arxiv.org/abs/1310.1921}{{\tt arXiv:1310.1921 [hep-ph]}}.

\bibitem{Degrande:2011ua}
C.~Degrande, C.~Duhr, B.~Fuks, D.~Grellscheid, O.~Mattelaer, et al., {\em {UFO
  - The Universal FeynRules Output}\/},
  \href{http://dx.doi.org/10.1016/j.cpc.2012.01.022}{Comput.Phys.Commun. {\bf
  183} (2012)  1201--1214},
\href{http://arxiv.org/abs/1108.2040}{{\tt arXiv:1108.2040 [hep-ph]}}.

\bibitem{Degrande:2014tta}
C.~Degrande, F.~Maltoni, J.~Wang, and C.~Zhang, {\em {Automatic computations at
  next-to-leading order in QCD for top-quark flavor-changing neutral
  processes}\/},  \href{http://dx.doi.org/10.1103/PhysRevD.91.034024}{Phys.
  Rev. {\bf D91} (2015)  034024},
\href{http://arxiv.org/abs/1412.5594}{{\tt arXiv:1412.5594 [hep-ph]}}.

\bibitem{Durieux:2018tev}
G.~Durieux, M.~Perell{\'o}, M.~Vos, and C.~Zhang, {\em {Global and optimal
  probes for the top-quark effective field theory at future lepton
  colliders}\/},  \href{http://dx.doi.org/10.1007/JHEP10(2018)168}{JHEP {\bf
  10} (2018)  168},
\href{http://arxiv.org/abs/1807.02121}{{\tt arXiv:1807.02121 [hep-ph]}}.

\bibitem{directtop}
C.~Degrande, A.~S. Papanastasiou, and C.~Zhang , work in progress.

\bibitem{Khachatryan:2015att}
{CMS Collaboration}, V.~Khachatryan et al., {\em {Search for anomalous single
  top quark production in association with a photon in pp collisions at
  $\sqrt{s}=8\,$TeV}\/},  \href{http://dx.doi.org/10.1007/JHEP04(2016)035}{JHEP
  {\bf 04} (2016)  035},
\href{http://arxiv.org/abs/1511.03951}{{\tt arXiv:1511.03951 [hep-ex]}}.

\bibitem{Aaboud:2018nyl}
{ATLAS Collaboration}, M.~Aaboud et al., {\em {Search for flavour-changing
  neutral current top-quark decays $t\to qZ$ in proton-proton collisions at
  $\sqrt{s}=13$ TeV with the ATLAS detector}\/},
  \href{http://dx.doi.org/10.1007/JHEP07(2018)176}{JHEP {\bf 07} (2018)  176},
\href{http://arxiv.org/abs/1803.09923}{{\tt arXiv:1803.09923 [hep-ex]}}.

\bibitem{CMS:2017twu}
{CMS Collaboration}, {\em {Search for flavour changing neutral currents in top
  quark production and decays with three-lepton final state using the data
  collected at $\sqrt{s} = 13\,$TeV}\/},  CMS-PAS-TOP-17-017  .
\url{http://cds.cern.ch/record/2292045}.

\bibitem{Sirunyan:2017kkr}
{CMS Collaboration}, A.~M. Sirunyan et al., {\em {Search for associated
  production of a $Z$ boson with a single top quark and for $tZ$
  flavour-changing interactions in pp collisions at $\sqrt{s}=8\,$TeV}\/},
  \href{http://dx.doi.org/10.1007/JHEP07(2017)003}{JHEP {\bf 07} (2017)  003},
\href{http://arxiv.org/abs/1702.01404}{{\tt arXiv:1702.01404 [hep-ex]}}.

\bibitem{Aad:2015gea}
{ATLAS Collaboration}, G.~Aad et al., {\em {Search for single top-quark
  production via flavour-changing neutral currents at 8 TeV with the ATLAS
  detector}\/},  \href{http://dx.doi.org/10.1140/epjc/s10052-016-3876-4}{Eur.
  Phys. J. {\bf C76} (2016) no.~2, 55},
\href{http://arxiv.org/abs/1509.00294}{{\tt arXiv:1509.00294 [hep-ex]}}.

\bibitem{Khachatryan:2016sib}
{CMS Collaboration}, V.~Khachatryan et al., {\em {Search for anomalous Wtb
  couplings and flavour-changing neutral currents in t-channel single top quark
  production in pp collisions at $\sqrt{s} =$ 7 and 8 TeV}\/},
  \href{http://dx.doi.org/10.1007/JHEP02(2017)028}{JHEP {\bf 02} (2017)  028},
\href{http://arxiv.org/abs/1610.03545}{{\tt arXiv:1610.03545 [hep-ex]}}.

\bibitem{Aaboud:2018pob}
{ATLAS Collaboration}, M.~Aaboud et al., {\em {Search for flavor-changing
  neutral currents in top quark decays $t\to Hc$ and $t \to Hu$ in multilepton
  final states in proton-proton collisions at $\sqrt{s}= 13$ TeV with the ATLAS
  detector}\/},  \href{http://dx.doi.org/10.1103/PhysRevD.98.032002}{Phys. Rev.
  {\bf D98} (2018) no.~3, 032002},
\href{http://arxiv.org/abs/1805.03483}{{\tt arXiv:1805.03483 [hep-ex]}}.

\bibitem{Khachatryan:2016atv}
{CMS Collaboration}, V.~Khachatryan et al., {\em {Search for top quark decays
  via Higgs-boson-mediated flavor-changing neutral currents in pp collisions at
  $ \sqrt{s}=8\,$TeV}\/},
  \href{http://dx.doi.org/10.1007/JHEP02(2017)079}{JHEP {\bf 02} (2017)  079},
\href{http://arxiv.org/abs/1610.04857}{{\tt arXiv:1610.04857 [hep-ex]}}.

\bibitem{Sirunyan:2017uae}
{{CMS} Collaboration}, A.~M. Sirunyan et al., {\em {Search for the
  flavor-changing neutral current interactions of the top quark and the Higgs
  boson which decays into a pair of b quarks at $\sqrt{s}=$ 13 TeV}\/},
  \href{http://dx.doi.org/10.1007/JHEP06(2018)102}{JHEP {\bf 06} (2018)  102},
\href{http://arxiv.org/abs/1712.02399}{{\tt arXiv:1712.02399 [hep-ex]}}.

\bibitem{Aleph:2001dzz}
{Aleph, Delphi, L3, Opal Collaborations, and the LEP Exotica Working Group},
  {\em {Search for single top production via flavour changing neutral currents:
  preliminary combined results of the LEP experiments}\/},  DELPHI-2001-119
  CONF 542 (2001)  .
\url{https://cds.cern.ch/record/1006392}.

\bibitem{CMS:2017gvo}
{CMS Collaboration}, {\em {ECFA 2016: Prospects for selected standard model
  measurements with the CMS experiment at the High-Luminosity LHC}\/},
  CMS-PAS-FTR-16-006 (2017)  .
\url{http://cds.cern.ch/record/2262606}.

\bibitem{Fox:2007in}
P.~J. Fox, Z.~Ligeti, M.~Papucci, G.~Perez, and M.~D. Schwartz, {\em
  {Deciphering top flavor violation at the LHC with $B$ factories}\/},
  \href{http://dx.doi.org/10.1103/PhysRevD.78.054008}{Phys. Rev. {\bf D78}
  (2008)  054008},
\href{http://arxiv.org/abs/0704.1482}{{\tt arXiv:0704.1482 [hep-ph]}}.

\bibitem{Drobnak:2011aa}
J.~Drobnak, S.~Fajfer, and J.~F. Kamenik, {\em {Probing anomalous tWb
  interactions with rare B decays}\/},
  \href{http://dx.doi.org/10.1016/j.nuclphysb.2011.10.004}{Nucl. Phys. {\bf
  B855} (2012)  82--99},
\href{http://arxiv.org/abs/1109.2357}{{\tt arXiv:1109.2357 [hep-ph]}}.

\bibitem{Brod:2014hsa}
J.~Brod, A.~Greljo, E.~Stamou, and P.~Uttayarat, {\em {Probing anomalous $
  t\overline{t}Z $ interactions with rare meson decays}\/},
  \href{http://dx.doi.org/10.1007/JHEP02(2015)141}{JHEP {\bf 02} (2015)  141},
\href{http://arxiv.org/abs/1408.0792}{{\tt arXiv:1408.0792 [hep-ph]}}.

\bibitem{Endo:2018gdn}
M.~Endo, T.~Kitahara, and D.~Ueda, {\em {SMEFT top-quark effects on $\Delta
  F=2$ observables}\/},
\href{http://arxiv.org/abs/1811.04961}{{\tt arXiv:1811.04961 [hep-ph]}}.

\bibitem{Mangano:2006rw}
M.~L. Mangano, M.~Moretti, F.~Piccinini, and M.~Treccani, {\em {Matching matrix
  elements and shower evolution for top-quark production in hadronic
  collisions}\/},  \href{http://dx.doi.org/10.1088/1126-6708/2007/01/013}{JHEP
  {\bf 01} (2007)  013},
\href{http://arxiv.org/abs/hep-ph/0611129}{{\tt arXiv:hep-ph/0611129
  [hep-ph]}}.

\bibitem{Coimbra:2012ys}
R.~Coimbra, A.~Onofre, R.~Santos, and M.~Won, {\em {MEtop - a generator for
  single top production via FCNC interactions}\/},
  \href{http://dx.doi.org/10.1140/epjc/s10052-012-2222-8}{Eur. Phys. J. {\bf
  C72} (2012)  2222},
\href{http://arxiv.org/abs/1207.7026}{{\tt arXiv:1207.7026 [hep-ph]}}.

\bibitem{Boos:2004kh}
{CompHEP Collaboration}, E.~Boos, V.~Bunichev, M.~Dubinin, L.~Dudko, V.~Ilyin,
  A.~Kryukov, V.~Edneral, V.~Savrin, A.~Semenov, and A.~Sherstnev, {\em
  {CompHEP 4.4: Automatic computations from Lagrangians to events}\/},
  \href{http://dx.doi.org/10.1016/j.nima.2004.07.096}{Nucl. Instrum. Meth. {\bf
  A534} (2004)  250--259},
\href{http://arxiv.org/abs/hep-ph/0403113}{{\tt arXiv:hep-ph/0403113
  [hep-ph]}}.

\bibitem{AguilarSaavedra:2010rx}
J.~Aguilar-Saavedra, {\em {Zt, gamma t and t production at hadron colliders via
  strong flavour-changing neutral couplings}\/},
  \href{http://dx.doi.org/10.1016/j.nuclphysb.2010.05.005}{Nucl.Phys. {\bf
  B837} (2010)  122--136},
\href{http://arxiv.org/abs/1003.3173}{{\tt arXiv:1003.3173 [hep-ph]}}.

\bibitem{CMS:2018kwi}
{CMS Collaboration}, {\em {Prospects for the search for gluon-mediated FCNC in
  top quark production with the CMS Phase-2 detector at the HL-LHC}\/},  CMS
  Physics Analysis Summary CMS-PAS-FTR-18-004, 2018.
\newblock \url{https://cds.cern.ch/record/2638815}.

\bibitem{ATL-PHYS-PUB-2016-019}
{ATLAS Collaboration}, {\em {Expected sensitivity of ATLAS to FCNC top quark
  decays $t \rightarrow Zu$ and $t \rightarrow Hq$ at the High Luminosity
  LHC}\/},   ATL-PHYS-PUB-2016-019, CERN, Geneva, Aug, 2016.
\newblock \url{http://cds.cern.ch/record/2209126}.

\bibitem{Collaboration:2293646}
{CMS Collaboration}, {\em {The Phase-2 Upgrade of the CMS Endcap
  Calorimeter}\/},   CERN-LHCC-2017-023. CMS-TDR-019, CERN, Geneva, Nov, 2017.
\newblock \url{https://cds.cern.ch/record/2293646}.
\newblock Technical Design Report of the endcap calorimeter for the Phase-2
  upgrade of the CMS experiment, in view of the HL-LHC run.

\bibitem{ATL-PHYS-PUB-2019-001}
{ATLAS Collaboration}, {\em {Sensitivity of searches for the flavour-changing
  neutral current decay $t\rightarrow qZ$ using the upgraded ATLAS experiment
  at the High Luminosity LHC}\/},  {ATL-PHYS-PUB-2019-001}, 2019.
\newblock \url{https://cds.cern.ch/record/2653389}.

\bibitem{Aaboud:2017mfd}
{ATLAS Collaboration}, M.~Aaboud et al., {\em {Search for top quark decays
  $t\rightarrow qH$, with $H\to\gamma\gamma$, in $\sqrt{s}=13$ TeV $pp$
  collisions using the ATLAS detector}\/},
  \href{http://dx.doi.org/10.1007/JHEP10(2017)129}{JHEP {\bf 10} (2017)  129},
\href{http://arxiv.org/abs/1707.01404}{{\tt arXiv:1707.01404 [hep-ex]}}.

\bibitem{ATL-PHYS-PUB-2013-012}
{\em {Sensitivity of ATLAS at HL-LHC to flavour changing neutral currents in
  top quark decays $t\to cH$, with $H\to \gamma\gamma$ }\/},
  ATL-PHYS-PUB-2013-012, CERN, Geneva, Sep, 2013.
\newblock \url{https://cds.cern.ch/record/1604506}.

\bibitem{Grzadkowski:2008mf}
B.~Grzadkowski and M.~Misiak, {\em {Anomalous Wtb coupling effects in the weak
  radiative B-meson decay}\/},
  \href{http://dx.doi.org/10.1103/PhysRevD.84.059903,
  10.1103/PhysRevD.78.077501}{Phys. Rev. D {\bf 78} (2008)  077501},
\href{http://arxiv.org/abs/0802.1413}{{\tt arXiv:0802.1413 [hep-ph]}}.

\bibitem{Drobnak:2010ej}
J.~Drobnak, S.~Fajfer, and J.~F. Kamenik, {\em {New physics in $t\rightarrow b
  W$ decay at next-to-leading order in QCD}\/},
  \href{http://dx.doi.org/10.1103/PhysRevD.82.114008}{Phys. Rev. {\bf D82}
  (2010)  114008},
\href{http://arxiv.org/abs/1010.2402}{{\tt arXiv:1010.2402 [hep-ph]}}.

\bibitem{AguilarSaavedra:2008zc}
J.~A. Aguilar-Saavedra, {\em {A Minimal set of top anomalous couplings}\/},
  \href{http://dx.doi.org/10.1016/j.nuclphysb.2008.12.012}{Nucl. Phys. {\bf
  B812} (2009)  181--204},
\href{http://arxiv.org/abs/0811.3842}{{\tt arXiv:0811.3842 [hep-ph]}}.

\bibitem{GonzalezSprinberg:2011kx}
G.~A. Gonzalez-Sprinberg, R.~Martinez, and J.~Vidal, {\em {Top quark tensor
  couplings}\/},  \href{http://dx.doi.org/10.1007/JHEP07(2011)094,
  10.1007/JHEP05(2013)117}{JHEP {\bf 07} (2011)  094},
  \href{http://arxiv.org/abs/1105.5601}{{\tt arXiv:1105.5601 [hep-ph]}}.
[Erratum: JHEP05,117(2013)].

\bibitem{Cao:2015doa}
Q.-H. Cao, B.~Yan, J.-H. Yu, and C.~Zhang, {\em {A General Analysis of $Wtb$
  anomalous Couplings}\/},
\href{http://arxiv.org/abs/1504.03785}{{\tt arXiv:1504.03785 [hep-ph]}}.

\bibitem{Hioki:2015env}
Z.~Hioki and K.~Ohkuma, {\em {Full analysis of general non-standard tbW
  couplings}\/},  \href{http://dx.doi.org/10.1016/j.physletb.2015.11.029}{Phys.
  Lett. {\bf B752} (2016)  128--130},
\href{http://arxiv.org/abs/1511.03437}{{\tt arXiv:1511.03437 [hep-ph]}}.

\bibitem{Schulze:2016qas}
M.~Schulze and Y.~Soreq, {\em {Pinning down electroweak dipole operators of the
  top quark}\/},
\href{http://arxiv.org/abs/1603.08911}{{\tt arXiv:1603.08911 [hep-ph]}}.

\bibitem{Kamenik:2011dk}
J.~F. Kamenik, M.~Papucci, and A.~Weiler, {\em {Constraining the dipole moments
  of the top quark}\/},  \href{http://dx.doi.org/10.1103/PhysRevD.88.039903,
  10.1103/PhysRevD.85.071501}{Phys. Rev. D {\bf 85} (2012)  071501},
  \href{http://arxiv.org/abs/1107.3143}{{\tt arXiv:1107.3143 [hep-ph]}}.
[Erratum: Phys. Rev.D88,no.3,039903(2013)].

\bibitem{Zhang:2012cd}
C.~Zhang, N.~Greiner, and S.~Willenbrock, {\em {Constraints on Non-standard Top
  Quark Couplings}\/},
  \href{http://dx.doi.org/10.1103/PhysRevD.86.014024}{Phys. Rev. {\bf D86}
  (2012)  014024},
\href{http://arxiv.org/abs/1201.6670}{{\tt arXiv:1201.6670 [hep-ph]}}.

\bibitem{deBlas:2015aea}
J.~de~Blas, M.~Chala, and J.~Santiago, {\em {Renormalization Group Constraints
  on New Top Interactions from Electroweak Precision Data}\/},
  \href{http://dx.doi.org/10.1007/JHEP09(2015)189}{JHEP {\bf 09} (2015)  189},
\href{http://arxiv.org/abs/1507.00757}{{\tt arXiv:1507.00757 [hep-ph]}}.

\bibitem{Buckley:2015nca}
A.~Buckley, C.~Englert, J.~Ferrando, D.~J. Miller, L.~Moore, M.~Russell, and
  C.~D. White, {\em {Global fit of top quark effective theory to data}\/},
  \href{http://dx.doi.org/10.1103/PhysRevD.92.091501}{Phys. Rev. {\bf D92}
  (2015) no.~9, 091501},
\href{http://arxiv.org/abs/1506.08845}{{\tt arXiv:1506.08845 [hep-ph]}}.

\bibitem{Buckley:2015lku}
A.~Buckley, C.~Englert, J.~Ferrando, D.~J. Miller, L.~Moore, M.~Russell, and
  C.~D. White, {\em {Constraining top quark effective theory in the LHC Run II
  era}\/},  \href{http://dx.doi.org/10.1007/JHEP04(2016)015}{JHEP {\bf 04}
  (2016)  015},
\href{http://arxiv.org/abs/1512.03360}{{\tt arXiv:1512.03360 [hep-ph]}}.

\bibitem{Bylund:2016phk}
O.~B. Bylund, F.~Maltoni, I.~Tsinikos, E.~Vryonidou, and C.~Zhang, {\em
  {Probing top quark neutral couplings in the Standard Model Effective Field
  Theory at NLO QCD}\/},
\href{http://arxiv.org/abs/1601.08193}{{\tt arXiv:1601.08193 [hep-ph]}}.

\bibitem{Castro:2016jjv}
N.~Castro, J.~Erdmann, C.~Grunwald, K.~Kr\"oninger, and N.-A. Rosien, {\em
  {EFTfitter---A tool for interpreting measurements in the context of effective
  field theories}\/},
  \href{http://dx.doi.org/10.1140/epjc/s10052-016-4280-9}{Eur. Phys. J. {\bf
  C76} (2016) no.~8, 432},
\href{http://arxiv.org/abs/1605.05585}{{\tt arXiv:1605.05585 [hep-ex]}}.

\bibitem{Deliot:2017byp}
F.~D\'eliot, R.~Faria, M.~C.~N. Fiolhais, P.~Lagarelhos, A.~Onofre, C.~M.
  Pease, and A.~Vasconcelos, {\em {Global Constraints on Top Quark Anomalous
  Couplings}\/},  \href{http://dx.doi.org/10.1103/PhysRevD.97.013007}{Phys.
  Rev. {\bf D97} (2018) no.~1, 013007},
\href{http://arxiv.org/abs/1711.04847}{{\tt arXiv:1711.04847 [hep-ph]}}.

\bibitem{AguilarSaavedra:2006fy}
J.~A. Aguilar-Saavedra, J.~Carvalho, N.~F. Castro, F.~Veloso, and A.~Onofre,
  {\em {Probing anomalous Wtb couplings in top pair decays}\/},
  \href{http://dx.doi.org/10.1140/epjc/s10052-007-0289-4}{Eur. Phys. J. {\bf
  C50} (2007)  519--533},
\href{http://arxiv.org/abs/hep-ph/0605190}{{\tt arXiv:hep-ph/0605190
  [hep-ph]}}.

\bibitem{Aad:2014fwa}
{ATLAS Collaboration}, G.~Aad et al., {\em {Comprehensive measurements of
  $t$-channel single top-quark production cross sections at $\sqrt{s} = 7$ TeV
  with the ATLAS detector}\/},
  \href{http://dx.doi.org/10.1103/PhysRevD.90.112006}{Phys. Rev. {\bf D90}
  (2014) no.~11, 112006},
\href{http://arxiv.org/abs/1406.7844}{{\tt arXiv:1406.7844 [hep-ex]}}.

\bibitem{Aaboud:2016ymp}
{ATLAS Collaboration}, M.~Aaboud et al., {\em {Measurement of the inclusive
  cross-sections of single top-quark and top-antiquark $t$-channel production
  in $pp$ collisions at $\sqrt{s}$ = 13 TeV with the ATLAS detector}\/},
\href{http://arxiv.org/abs/1609.03920}{{\tt arXiv:1609.03920 [hep-ex]}}.

\bibitem{Aaboud:2017pdi}
{ATLAS Collaboration}, M.~Aaboud et al., {\em {Fiducial, total and differential
  cross-section measurements of $t$-channel single top-quark production in $pp$
  collisions at 8 TeV using data collected by the ATLAS detector}\/},
\href{http://arxiv.org/abs/1702.02859}{{\tt arXiv:1702.02859 [hep-ex]}}.

\bibitem{Chatrchyan:2012ep}
{CMS Collaboration}, S.~Chatrchyan et al., {\em {Measurement of the
  single-top-quark $t$-channel cross section in $pp$ collisions at $\sqrt{s}=7$
  TeV}\/},  \href{http://dx.doi.org/10.1007/JHEP12(2012)035}{JHEP {\bf 12}
  (2012)  035},
\href{http://arxiv.org/abs/1209.4533}{{\tt arXiv:1209.4533 [hep-ex]}}.

\bibitem{Khachatryan:2014iya}
{CMS Collaboration}, V.~Khachatryan et al., {\em {Measurement of the t-channel
  single-top-quark production cross section and of the $\mid V_{tb} \mid$ CKM
  matrix element in pp collisions at $\sqrt{s}$= 8 TeV}\/},
  \href{http://dx.doi.org/10.1007/JHEP06(2014)090}{JHEP {\bf 06} (2014)  090},
\href{http://arxiv.org/abs/1403.7366}{{\tt arXiv:1403.7366 [hep-ex]}}.

\bibitem{Sirunyan:2016cdg}
{CMS Collaboration}, A.~M. Sirunyan et al., {\em {Cross section measurement of
  t-channel single top quark production in pp collisions at $\sqrt{s} = $ 13
  TeV}\/},
\href{http://arxiv.org/abs/1610.00678}{{\tt arXiv:1610.00678 [hep-ex]}}.

\bibitem{Brucherseifer:2014ama}
M.~Brucherseifer, F.~Caola, and K.~Melnikov, {\em {On the NNLO QCD corrections
  to single-top production at the LHC}\/},
  \href{http://dx.doi.org/10.1016/j.physletb.2014.06.075}{Phys. Lett. {\bf
  B736} (2014)  58--63},
\href{http://arxiv.org/abs/1404.7116}{{\tt arXiv:1404.7116 [hep-ph]}}.

\bibitem{Zhang:2016omx}
C.~Zhang, {\em {Single Top Production at Next-to-Leading Order in the Standard
  Model Effective Field Theory}\/},
  \href{http://dx.doi.org/10.1103/PhysRevLett.116.162002}{Phys. Rev. Lett. {\bf
  116} (2016) no.~16, 162002},
\href{http://arxiv.org/abs/1601.06163}{{\tt arXiv:1601.06163 [hep-ph]}}.

\bibitem{Cirigliano:2016nyn}
V.~Cirigliano, W.~Dekens, J.~de~Vries, and E.~Mereghetti, {\em {Constraining
  the top-Higgs sector of the Standard Model Effective Field Theory}\/},
  \href{http://dx.doi.org/10.1103/PhysRevD.94.034031}{Phys. Rev. {\bf D94}
  (2016) no.~3, 034031},
\href{http://arxiv.org/abs/1605.04311}{{\tt arXiv:1605.04311 [hep-ph]}}.

\bibitem{Alioli:2017ce}
S.~Alioli, V.~Cirigliano, W.~Dekens, J.~de~Vries, and E.~Mereghetti, {\em
  {Right-handed charged currents in the era of the Large Hadron Collider}\/},
  \href{http://dx.doi.org/10.1007/JHEP05(2017)086}{JHEP {\bf 05} (2017)  086},
\href{http://arxiv.org/abs/1703.04751}{{\tt arXiv:1703.04751 [hep-ph]}}.

\bibitem{deBeurs:2018pvs}
M.~de~Beurs, E.~Laenen, M.~Vreeswijk, and E.~Vryonidou, {\em {Effective
  operators in $t$-channel single top production and decay}\/},
  \href{http://dx.doi.org/10.1140/epjc/s10052-018-6399-3}{Eur. Phys. J. {\bf
  C78} (2018) no.~11, 919},
\href{http://arxiv.org/abs/1807.03576}{{\tt arXiv:1807.03576 [hep-ph]}}.

\bibitem{Aaltonen:2012rz}
{CDF, D0 Collaboration}, T.~Aaltonen et al., {\em {Combination of CDF and D0
  measurements of the $W$ boson helicity in top quark decays}\/},
  \href{http://dx.doi.org/10.1103/PhysRevD.85.071106}{Phys. Rev. {\bf D85}
  (2012)  071106},
\href{http://arxiv.org/abs/1202.5272}{{\tt arXiv:1202.5272 [hep-ex]}}.

\bibitem{Aad:2012ky}
{ATLAS Collaboration}, G.~Aad et al., {\em {Measurement of the W boson
  polarization in top quark decays with the ATLAS detector}\/},
  \href{http://dx.doi.org/10.1007/JHEP06(2012)088}{JHEP {\bf 06} (2012)  088},
\href{http://arxiv.org/abs/1205.2484}{{\tt arXiv:1205.2484 [hep-ex]}}.

\bibitem{Chatrchyan:2013jna}
{CMS Collaboration}, S.~Chatrchyan et al., {\em {Measurement of the W-boson
  helicity in top-quark decays from $t\bar{t}$ production in lepton+jets events
  in pp collisions at $\sqrt{s} =$ 7 TeV}\/},
  \href{http://dx.doi.org/10.1007/JHEP10(2013)167}{JHEP {\bf 10} (2013)  167},
\href{http://arxiv.org/abs/1308.3879}{{\tt arXiv:1308.3879 [hep-ex]}}.

\bibitem{Aad:2015yem}
{ATLAS Collaboration}, G.~Aad et al., {\em {Search for anomalous couplings in
  the $Wtb$ vertex from the measurement of double differential angular decay
  rates of single top quarks produced in the $t$-channel with the ATLAS
  detector}\/},  \href{http://dx.doi.org/10.1007/JHEP04(2016)023}{JHEP {\bf 04}
  (2016)  023},
\href{http://arxiv.org/abs/1510.03764}{{\tt arXiv:1510.03764 [hep-ex]}}.

\bibitem{Khachatryan:2014vma}
{CMS Collaboration}, V.~Khachatryan et al., {\em {Measurement of the W boson
  helicity in events with a single reconstructed top quark in pp collisions at
  $ \sqrt{s}=8 $ TeV}\/},
  \href{http://dx.doi.org/10.1007/JHEP01(2015)053}{JHEP {\bf 01} (2015)  053},
\href{http://arxiv.org/abs/1410.1154}{{\tt arXiv:1410.1154 [hep-ex]}}.

\bibitem{Aaboud:2016hsq}
{ATLAS Collaboration}, M.~Aaboud et al., {\em {Measurement of the $W$ boson
  polarisation in $t\bar{t}$ events from $pp$ collisions at $\sqrt{s}$ = 8 TeV
  in the lepton+jets channel with ATLAS}\/},
\href{http://arxiv.org/abs/1612.02577}{{\tt arXiv:1612.02577 [hep-ex]}}.

\bibitem{Boudreau:2013yna}
J.~Boudreau, C.~Escobar, J.~Mueller, K.~Sapp, and J.~Su, {\em {Single top quark
  differential decay rate formulae including detector effects}\/},
\href{http://arxiv.org/abs/1304.5639}{{\tt arXiv:1304.5639 [hep-ex]}}.

\bibitem{Czarnecki:2010gb}
A.~Czarnecki, J.~G. Korner, and J.~H. Piclum, {\em {Helicity fractions of W
  bosons from top quark decays at NNLO in QCD}\/},
  \href{http://dx.doi.org/10.1103/PhysRevD.81.111503}{Phys. Rev. {\bf D81}
  (2010)  111503},
\href{http://arxiv.org/abs/1005.2625}{{\tt arXiv:1005.2625 [hep-ph]}}.

\bibitem{Aguilar-Saavedra:2015yza}
J.~A. Aguilar-Saavedra and J.~Bernabeu, {\em {Breaking down the entire W boson
  spin observables from its decay}\/},
  \href{http://dx.doi.org/10.1103/PhysRevD.93.011301}{Phys. Rev. {\bf D93}
  (2016) no.~1, 011301},
\href{http://arxiv.org/abs/1508.04592}{{\tt arXiv:1508.04592 [hep-ph]}}.

\bibitem{AguilarSaavedra:2010nx}
J.~A. Aguilar-Saavedra and J.~Bernabeu, {\em {W polarisation beyond helicity
  fractions in top quark decays}\/},
  \href{http://dx.doi.org/10.1016/j.nuclphysb.2010.07.012}{Nucl. Phys. {\bf
  B840} (2010)  349--378},
\href{http://arxiv.org/abs/1005.5382}{{\tt arXiv:1005.5382 [hep-ph]}}.

\bibitem{Aaboud:2017yqf}
{ATLAS Collaboration}, M.~Aaboud et al., {\em {Analysis of the $Wtb$ vertex
  from the measurement of triple-differential angular decay rates of single top
  quarks produced in the $t$-channel at $\sqrt{s}$ = 8 TeV with the ATLAS
  detector}\/},
\href{http://arxiv.org/abs/1707.05393}{{\tt arXiv:1707.05393 [hep-ex]}}.

\bibitem{Deliot:2018jts}
F.~D{\' e}liot, M.~C.~N. Fiolhais, and A.~Onofre, {\em {Top Quark Anomalous
  Couplings at the High-Luminosity Phase of the LHC}\/},
\href{http://arxiv.org/abs/1811.02492}{{\tt arXiv:1811.02492 [hep-ph]}}.

\bibitem{CMS-PAS-TOP-17-017}
{CMS Collaboration}, {\em {Search for flavour changing neutral currents in top
  quark production and decays with three-lepton final state using the data
  collected at sqrt(s) = 13 TeV}\/},   CMS-PAS-TOP-17-017, CERN, Geneva, 2017.
\newblock \url{https://cds.cern.ch/record/2292045}.

\bibitem{Harrison:2018bqi}
P.~F. Harrison and V.~E. Vladimirov, {\em {A Method to Determine $|V_{cb}|$ at
  the Weak Scale in Top Decays at the LHC}\/},
\href{http://arxiv.org/abs/1810.09424}{{\tt arXiv:1810.09424 [hep-ph]}}.

\bibitem{Dekens:2013zca}
W.~Dekens and J.~de~Vries, {\em {Renormalization Group Running of Dimension-Six
  Sources of Parity and Time-Reversal Violation}\/},
  \href{http://dx.doi.org/10.1007/JHEP05(2013)149}{JHEP {\bf 1305} (2013)
  149},
\href{http://arxiv.org/abs/1303.3156}{{\tt arXiv:1303.3156 [hep-ph]}}.

\bibitem{Alonso:2013hga}
R.~Alonso, E.~E. Jenkins, A.~V. Manohar, and M.~Trott, {\em {Renormalization
  Group Evolution of the Standard Model Dimension Six Operators III: Gauge
  Coupling Dependence and Phenomenology}\/},
  \href{http://dx.doi.org/10.1007/JHEP04(2014)159}{JHEP {\bf 04} (2014)  159},
\href{http://arxiv.org/abs/1312.2014}{{\tt arXiv:1312.2014 [hep-ph]}}.

\bibitem{Dicus:1989va}
D.~A. Dicus, {\em {Neutron Electric Dipole Moment From Charged Higgs
  Exchange}\/},
\href{http://dx.doi.org/10.1103/PhysRevD.41.999}{Phys.Rev. {\bf D41} (1990)
  999}.

\bibitem{Weinberg:1989dx}
S.~Weinberg, {\em {Larger Higgs Exchange Terms in the Neutron Electric Dipole
  Moment}\/},
\href{http://dx.doi.org/10.1103/PhysRevLett.63.2333}{Phys. Rev. Lett. {\bf 63}
  (1989)  2333}.

\bibitem{BraatenPRL}
E.~Braaten, C.-S. Li, and T.-C. Yuan, {\em {The Evolution of Weinberg's Gluonic
  {CP} Violation Operator}\/},
\href{http://dx.doi.org/10.1103/PhysRevLett.64.1709}{Phys. Rev. Lett. {\bf 64}
  (1990)  1709}.

\bibitem{Boyd:1990bx}
G.~Boyd, A.~K. Gupta, S.~P. Trivedi, and M.~B. Wise, {\em {Effective
  Hamiltonian for the Electric Dipole Moment of the Neutron}\/},
\href{http://dx.doi.org/10.1016/0370-2693(90)91874-B}{Phys.Lett. {\bf B241}
  (1990)  584}.

\bibitem{Pospelov_Weinberg}
D.~A. Demir, M.~Pospelov, and A.~Ritz, {\em {Hadronic EDMs, the Weinberg
  operator, and light gluinos}\/},
  \href{http://dx.doi.org/10.1103/PhysRevD.67.015007}{Phys. Rev. D {\bf 67}
  (2003)  015007},
\href{http://arxiv.org/abs/hep-ph/0208257}{{\tt arXiv:hep-ph/0208257
  [hep-ph]}}.

\bibitem{deVries:2010ah}
J.~de~Vries, R.~G.~E. Timmermans, E.~Mereghetti, and U.~van Kolck, {\em {The
  Nucleon Electric Dipole Form Factor From Dimension-Six Time-Reversal
  Violation}\/},  \href{http://dx.doi.org/10.1016/j.physletb.2010.11.042}{Phys.
  Lett. {\bf B695} (2011)  268--274},
\href{http://arxiv.org/abs/1006.2304}{{\tt arXiv:1006.2304 [hep-ph]}}.

\bibitem{Cirigliano:2016njn}
V.~Cirigliano, W.~Dekens, J.~de~Vries, and E.~Mereghetti, {\em {Is there room
  for CP violation in the top-Higgs sector?}\/},
  \href{http://dx.doi.org/10.1103/PhysRevD.94.016002}{Phys. Rev. {\bf D94}
  (2016) no.~1, 016002},
\href{http://arxiv.org/abs/1603.03049}{{\tt arXiv:1603.03049 [hep-ph]}}.

\bibitem{Fuyuto:2017xup}
K.~Fuyuto and M.~Ramsey-Musolf, {\em {Top Down Electroweak Dipole
  Operators}\/},
\href{http://arxiv.org/abs/1706.08548}{{\tt arXiv:1706.08548 [hep-ph]}}.

\bibitem{Afach:2015sja}
J.~Pendlebury et al., {\em {Revised experimental upper limit on the electric
  dipole moment of the neutron}\/},
  \href{http://dx.doi.org/10.1103/PhysRevD.92.092003}{Phys. Rev. {\bf D92}
  (2015) no.~9, 092003},
\href{http://arxiv.org/abs/1509.04411}{{\tt arXiv:1509.04411 [hep-ex]}}.

\bibitem{Baker:2006ts}
C.~A. Baker, D.~D. Doyle, P.~Geltenbort, K.~Green, M.~G.~D. van~der Grinten, et
  al., {\em {An Improved experimental limit on the electric dipole moment of
  the neutron}\/},
  \href{http://dx.doi.org/10.1103/PhysRevLett.97.131801}{Phys. Rev. Lett. {\bf
  97} (2006)  131801},
\href{http://arxiv.org/abs/hep-ex/0602020}{{\tt arXiv:hep-ex/0602020
  [hep-ex]}}.

\bibitem{Andreev:2018ayy}
{ACME Collaboration}, V.~Andreev et al., {\em {Improved limit on the electric
  dipole moment of the electron}\/},
\href{http://dx.doi.org/10.1038/s41586-018-0599-8}{Nature {\bf 562} (2018)
  no.~7727, 355--360}.

\bibitem{Chupp:2017rkp}
T.~Chupp, P.~Fierlinger, M.~Ramsey-Musolf, and J.~Singh, {\em {Electric Dipole
  Moments of the Atoms, Molecules, Nuclei and Particles}\/},
\href{http://arxiv.org/abs/1710.02504}{{\tt arXiv:1710.02504
  [physics.atom-ph]}}.

\bibitem{Kozyryev:2017cwq}
I.~Kozyryev and N.~R. Hutzler, {\em {Precision Measurement of Time-Reversal
  Symmetry Violation with Laser-Cooled Polyatomic Molecules}\/},
  \href{http://dx.doi.org/10.1103/PhysRevLett.119.133002}{Phys. Rev. Lett. {\bf
  119} (2017) no.~13, 133002},
\href{http://arxiv.org/abs/1705.11020}{{\tt arXiv:1705.11020
  [physics.atom-ph]}}.

\bibitem{Vutha:2017pej}
A.~C. Vutha, M.~Horbatsch, and E.~A. Hessels, {\em {Oriented polar molecules in
  a solid inert-gas matrix: a proposed method for measuring the electric dipole
  moment of the electron}\/},
\href{http://arxiv.org/abs/1710.08785}{{\tt arXiv:1710.08785
  [physics.atom-ph]}}.

\bibitem{PhysRevLett.120.123201}
J.~Lim, J.~R. Almond, M.~A. Trigatzis, J.~A. Devlin, N.~J. Fitch, B.~E. Sauer,
  M.~R. Tarbutt, and E.~A. Hinds,
  \href{http://dx.doi.org/10.1103/PhysRevLett.120.123201}{{\em Laser Cooled YbF
  Molecules for Measuring the Electron's Electric Dipole Moment\/}, Phys. Rev.
  Lett. {\bf 120} (Mar, 2018)  123201}.
  \url{https://link.aps.org/doi/10.1103/PhysRevLett.120.123201}.

\bibitem{Hurth:2003dk}
T.~Hurth, E.~Lunghi, and W.~Porod, {\em {Untagged $\bar B \to X_{s+d} \gamma$
  CP asymmetry as a probe for new physics}\/},
  \href{http://dx.doi.org/10.1016/j.nuclphysb.2004.10.024}{Nucl. Phys. B {\bf
  704} (2005)  56--74},
\href{http://arxiv.org/abs/hep-ph/0312260}{{\tt arXiv:hep-ph/0312260
  [hep-ph]}}.

\bibitem{Benzke:2010tq}
M.~Benzke, S.~J. Lee, M.~Neubert, and G.~Paz, {\em {Long-Distance Dominance of
  the CP Asymmetry in $B\to X_{s,d}+\gamma$ Decays}\/},
  \href{http://dx.doi.org/10.1103/PhysRevLett.106.141801}{Phys. Rev. Lett. {\bf
  106} (2011)  141801},
\href{http://arxiv.org/abs/1012.3167}{{\tt arXiv:1012.3167 [hep-ph]}}.

\bibitem{Peskin:1990zt}
M.~E. Peskin and T.~Takeuchi, {\em {A New constraint on a strongly interacting
  Higgs sector}\/},
\href{http://dx.doi.org/10.1103/PhysRevLett.65.964}{Phys. Rev. Lett. {\bf 65}
  (1990)  964--967}.

\bibitem{Peskin:1991sw}
M.~E. Peskin and T.~Takeuchi, {\em {Estimation of oblique electroweak
  corrections}\/},
\href{http://dx.doi.org/10.1103/PhysRevD.46.381}{Phys. Rev. {\bf D46} (1992)
  381--409}.

\bibitem{Barbieri:2004qk}
R.~Barbieri, A.~Pomarol, R.~Rattazzi, and A.~Strumia, {\em {Electroweak
  symmetry breaking after LEP-1 and LEP-2}\/},
  \href{http://dx.doi.org/10.1016/j.nuclphysb.2004.10.014}{Nucl. Phys. {\bf
  B703} (2004)  127--146},
\href{http://arxiv.org/abs/hep-ph/0405040}{{\tt arXiv:hep-ph/0405040
  [hep-ph]}}.

\bibitem{Greiner:2011tt}
N.~Greiner, S.~Willenbrock, and C.~Zhang, {\em {Effective Field Theory for
  Nonstandard Top Quark Couplings}\/},
  \href{http://dx.doi.org/10.1016/j.physletb.2011.09.026}{Phys. Lett. {\bf
  B704} (2011)  218--222},
\href{http://arxiv.org/abs/1104.3122}{{\tt arXiv:1104.3122 [hep-ph]}}.

\bibitem{Khachatryan:2016vau}
{ATLAS, CMS Collaboration}, G.~Aad et al., {\em {Measurements of the Higgs
  boson production and decay rates and constraints on its couplings from a
  combined ATLAS and CMS analysis of the LHC pp collision data at $ \sqrt{s}=7
  $ and 8 TeV}\/},  \href{http://dx.doi.org/10.1007/JHEP08(2016)045}{JHEP {\bf
  08} (2016)  045},
\href{http://arxiv.org/abs/1606.02266}{{\tt arXiv:1606.02266 [hep-ex]}}.

\bibitem{Khachatryan:2014nda}
{CMS Collaboration}, V.~Khachatryan et al., {\em {Measurement of the ratio
  $\mathcal B(t \to Wb)/\mathcal B(t \to Wq)$ in pp collisions at $\sqrt{s}$ =
  8 TeV}\/},  \href{http://dx.doi.org/10.1016/j.physletb.2014.06.076}{Phys.
  Lett. {\bf B736} (2014)  33--57},
\href{http://arxiv.org/abs/1404.2292}{{\tt arXiv:1404.2292 [hep-ex]}}.

\bibitem{Alvarez:2017ybk}
E.~Alvarez, L.~Da~Rold, M.~Estevez, and J.~F. Kamenik, {\em {Measuring
  $|V_{td}|$ at the LHC}\/},
  \href{http://dx.doi.org/10.1103/PhysRevD.97.033002}{Phys. Rev. {\bf D97}
  (2018) no.~3, 033002},
\href{http://arxiv.org/abs/1709.07887}{{\tt arXiv:1709.07887 [hep-ph]}}.

\bibitem{Khachatryan:2016ysn}
{CMS Collaboration}, V.~Khachatryan et al., {\em {Measurements of $t \bar t$
  charge asymmetry using dilepton final states in pp collisions at $\sqrt s=8$
  TeV}\/},  \href{http://dx.doi.org/10.1016/j.physletb.2016.07.006}{Phys. Lett.
  {\bf B760} (2016)  365--386},
\href{http://arxiv.org/abs/1603.06221}{{\tt arXiv:1603.06221 [hep-ex]}}.

\bibitem{Chatrchyan:2014tua}
{CMS Collaboration}, S.~Chatrchyan et al., {\em {Observation of the associated
  production of a single top quark and a $W$ boson in $pp$ collisions at $\sqrt
  s = $8 TeV}\/},
  \href{http://dx.doi.org/10.1103/PhysRevLett.112.231802}{Phys. Rev. Lett. {\bf
  112} (2014) no.~23, 231802},
\href{http://arxiv.org/abs/1401.2942}{{\tt arXiv:1401.2942 [hep-ex]}}.

\bibitem{Aad:2015eto}
{ATLAS Collaboration}, G.~Aad et al., {\em {Measurement of the production
  cross-section of a single top quark in association with a $W$ boson at 8 TeV
  with the ATLAS experiment}\/},
  \href{http://dx.doi.org/10.1007/JHEP01(2016)064}{JHEP {\bf 01} (2016)  064},
\href{http://arxiv.org/abs/1510.03752}{{\tt arXiv:1510.03752 [hep-ex]}}.

\bibitem{Aaboud:2016lpj}
{ATLAS Collaboration}, M.~Aaboud et al., {\em {Measurement of the cross-section
  for producing a W boson in association with a single top quark in pp
  collisions at $ \sqrt{s}=13 $ TeV with ATLAS}\/},
  \href{http://dx.doi.org/10.1007/JHEP01(2018)063}{JHEP {\bf 01} (2018)  063},
\href{http://arxiv.org/abs/1612.07231}{{\tt arXiv:1612.07231 [hep-ex]}}.

\bibitem{Gallicchio:2011xq}
J.~Gallicchio and M.~D. Schwartz, {\em {Quark and Gluon Tagging at the LHC}\/},
   \href{http://dx.doi.org/10.1103/PhysRevLett.107.172001}{Phys. Rev. Lett.
  {\bf 107} (2011)  172001},
\href{http://arxiv.org/abs/1106.3076}{{\tt arXiv:1106.3076 [hep-ph]}}.

\bibitem{Larkoski:2013eya}
A.~J. Larkoski, G.~P. Salam, and J.~Thaler, {\em {Energy Correlation Functions
  for Jet Substructure}\/},
  \href{http://dx.doi.org/10.1007/JHEP06(2013)108}{JHEP {\bf 06} (2013)  108},
\href{http://arxiv.org/abs/1305.0007}{{\tt arXiv:1305.0007 [hep-ph]}}.

\bibitem{Moult:2016cvt}
I.~Moult, L.~Necib, and J.~Thaler, {\em {New Angles on Energy Correlation
  Functions}\/},  \href{http://dx.doi.org/10.1007/JHEP12(2016)153}{JHEP {\bf
  12} (2016)  153},
\href{http://arxiv.org/abs/1609.07483}{{\tt arXiv:1609.07483 [hep-ph]}}.

\bibitem{thePaper}
D.~Faroughy, J.~F.~Kamenik, M.~Patra, and J.~Zupan, {\em {}\/},  {to appear}
  (2018)  .

\bibitem{Silva:2010qt}
P.~Silva and M.~Gallinaro, {\em {Probing the flavor of the top quark decay}\/},
   \href{http://dx.doi.org/10.1393/ncb/i2010-10896-0}{Nuovo Cim. {\bf B125}
  (2010)  983--998},
\href{http://arxiv.org/abs/1010.2994}{{\tt arXiv:1010.2994 [hep-ph]}}.

\bibitem{Aaboud:2017jvq}
{ATLAS Collaboration}, M.~Aaboud et al., {\em {Evidence for the associated
  production of the Higgs boson and a top quark pair with the ATLAS
  detector}\/},  \href{http://dx.doi.org/10.1103/PhysRevD.97.072003}{Phys. Rev.
  {\bf D97} (2018) no.~7, 072003},
\href{http://arxiv.org/abs/1712.08891}{{\tt arXiv:1712.08891 [hep-ex]}}.

\bibitem{Aaboud:2018urx}
{ATLAS Collaboration}, M.~Aaboud et al., {\em {Observation of Higgs boson
  production in association with a top quark pair at the LHC with the ATLAS
  detector}\/},  \href{http://dx.doi.org/10.1016/j.physletb.2018.07.035}{Phys.
  Lett. {\bf B784} (2018)  173--191},
\href{http://arxiv.org/abs/1806.00425}{{\tt arXiv:1806.00425 [hep-ex]}}.

\bibitem{Sirunyan:2018shy}
{CMS Collaboration}, A.~M. Sirunyan et al., {\em {Evidence for associated
  production of a Higgs boson with a top quark pair in final states with
  electrons, muons, and hadronically decaying $\tau$ leptons at $\sqrt{s} =$ 13
  TeV}\/},  \href{http://dx.doi.org/10.1007/JHEP08(2018)066}{JHEP {\bf 08}
  (2018)  066},
\href{http://arxiv.org/abs/1803.05485}{{\tt arXiv:1803.05485 [hep-ex]}}.

\bibitem{Aad:2015vsa}
{ATLAS Collaboration}, G.~Aad et al., {\em {Evidence for the Higgs-boson Yukawa
  coupling to tau leptons with the ATLAS detector}\/},
  \href{http://dx.doi.org/10.1007/JHEP04(2015)117}{JHEP {\bf 04} (2015)  117},
\href{http://arxiv.org/abs/1501.04943}{{\tt arXiv:1501.04943 [hep-ex]}}.

\bibitem{Aaboud:2018pen}
{ATLAS Collaboration}, M.~Aaboud et al., {\em {Cross-section measurements of
  the Higgs boson decaying into a pair of tau-leptons in proton-proton
  collisions at $\sqrt{s}=13$ TeV with the ATLAS detector}\/},
\href{http://arxiv.org/abs/1811.08856}{{\tt arXiv:1811.08856 [hep-ex]}}.

\bibitem{Sirunyan:2018kst}
{CMS Collaboration}, A.~M. Sirunyan et al., {\em {Observation of Higgs boson
  decay to bottom quarks}\/},
  \href{http://dx.doi.org/10.1103/PhysRevLett.121.121801}{Phys. Rev. Lett. {\bf
  121} (2018) no.~12, 121801},
\href{http://arxiv.org/abs/1808.08242}{{\tt arXiv:1808.08242 [hep-ex]}}.

\bibitem{Aaboud:2018zhk}
{ATLAS Collaboration}, M.~Aaboud et al., {\em {Observation of $H \rightarrow
  b\bar{b}$ decays and $VH$ production with the ATLAS detector}\/},
  \href{http://dx.doi.org/10.1016/j.physletb.2018.09.013}{Phys. Lett. {\bf
  B786} (2018)  59--86},
\href{http://arxiv.org/abs/1808.08238}{{\tt arXiv:1808.08238 [hep-ex]}}.

\bibitem{Aaboud:2018fhh}
{ATLAS Collaboration}, M.~Aaboud et al., {\em {Search for the Decay of the
  Higgs Boson to Charm Quarks with the ATLAS Experiment}\/},
  \href{http://dx.doi.org/10.1103/PhysRevLett.120.211802}{Phys. Rev. Lett. {\bf
  120} (2018) no.~21, 211802},
\href{http://arxiv.org/abs/1802.04329}{{\tt arXiv:1802.04329 [hep-ex]}}.

\bibitem{Aaboud:2017ojs}
{ATLAS Collaboration}, M.~Aaboud et al., {\em {Search for the dimuon decay of
  the Higgs boson in $pp$ collisions at $\sqrt{s}$ = 13 TeV with the ATLAS
  detector}\/},  \href{http://dx.doi.org/10.1103/PhysRevLett.119.051802}{Phys.
  Rev. Lett. {\bf 119} (2017) no.~5, 051802},
\href{http://arxiv.org/abs/1705.04582}{{\tt arXiv:1705.04582 [hep-ex]}}.

\bibitem{Perez:2015aoa}
G.~Perez, Y.~Soreq, E.~Stamou, and K.~Tobioka, {\em {Constraining the charm
  Yukawa and Higgs-quark coupling universality}\/},
  \href{http://dx.doi.org/10.1103/PhysRevD.92.033016}{Phys. Rev. {\bf D92}
  (2015) no.~3, 033016},
\href{http://arxiv.org/abs/1503.00290}{{\tt arXiv:1503.00290 [hep-ph]}}.

\bibitem{Altmannshofer:2015qra}
W.~Altmannshofer, J.~Brod, and M.~Schmaltz, {\em {Experimental constraints on
  the coupling of the Higgs boson to electrons}\/},
  \href{http://dx.doi.org/10.1007/JHEP05(2015)125}{JHEP {\bf 05} (2015)  125},
\href{http://arxiv.org/abs/1503.04830}{{\tt arXiv:1503.04830 [hep-ph]}}.

\bibitem{Kagan:2014ila}
A.~L. Kagan, G.~Perez, F.~Petriello, Y.~Soreq, S.~Stoynev, and J.~Zupan, {\em
  {Exclusive Window onto Higgs Yukawa Couplings}\/},
  \href{http://dx.doi.org/10.1103/PhysRevLett.114.101802}{Phys. Rev. Lett. {\bf
  114} (2015) no.~10, 101802},
\href{http://arxiv.org/abs/1406.1722}{{\tt arXiv:1406.1722 [hep-ph]}}.

\bibitem{Nir:2016zkd}
Y.~Nir, \href{http://dx.doi.org/10.5170/CERN-2015-001.123}{{\em {Flavour
  Physics and CP Violation}\/}, } in {\em {Proceedings, 7th CERN Latin-American
  School of High-Energy Physics (CLASHEP2013): Arequipa, Peru, March 6-19,
  2013}}, pp.~123--156.
\newblock 2015.
\newblock \href{http://arxiv.org/abs/1605.00433}{{\tt arXiv:1605.00433
  [hep-ph]}}.
\newblock
\url{http://inspirehep.net/record/1454240/files/arXiv:1605.00433.pdf}.
\newblock

\bibitem{Aad:2015sda}
{ATLAS Collaboration}, G.~Aad et al., {\em {Search for Higgs and Z Boson Decays
  to $J/\psi\gamma$ and $\Upsilon(nS)\gamma$ with the ATLAS Detector}\/},
  \href{http://dx.doi.org/10.1103/PhysRevLett.114.121801}{Phys. Rev. Lett. {\bf
  114} (2015) no.~12, 121801},
\href{http://arxiv.org/abs/1501.03276}{{\tt arXiv:1501.03276 [hep-ex]}}.

\bibitem{Khachatryan:2015lga}
{CMS Collaboration}, V.~Khachatryan et al., {\em {Search for a Higgs boson
  decaying into $\gamma^* \gamma \to \ell \ell \gamma$ with low dilepton mass
  in pp collisions at $\sqrt s = $ 8 TeV}\/},
  \href{http://dx.doi.org/10.1016/j.physletb.2015.12.039}{Phys. Lett. {\bf
  B753} (2016)  341--362},
\href{http://arxiv.org/abs/1507.03031}{{\tt arXiv:1507.03031 [hep-ex]}}.

\bibitem{Aaboud:2018txb}
{ATLAS Collaboration}, M.~Aaboud et al., {\em {Searches for exclusive Higgs and
  $Z$ boson decays into $J/\psi\gamma$, $\psi(2S)\gamma$, and
  $\Upsilon(nS)\gamma$ at $\sqrt{s}=13$ TeV with the ATLAS detector}\/},
  \href{http://dx.doi.org/10.1016/j.physletb.2018.09.024}{Phys. Lett. {\bf
  B786} (2018)  134--155},
\href{http://arxiv.org/abs/1807.00802}{{\tt arXiv:1807.00802 [hep-ex]}}.

\bibitem{Aaboud:2017xnb}
{ATLAS Collaboration}, M.~Aaboud et al., {\em {Search for exclusive Higgs and
  $Z$ boson decays to $\phi\gamma$ and $\rho\gamma$ with the ATLAS
  detector}\/},
\href{http://arxiv.org/abs/1712.02758}{{\tt arXiv:1712.02758 [hep-ex]}}.

\bibitem{Soreq:2016rae}
Y.~Soreq, H.~X. Zhu, and J.~Zupan, {\em {Light quark Yukawa couplings from
  Higgs kinematics}\/},  \href{http://dx.doi.org/10.1007/JHEP12(2016)045}{JHEP
  {\bf 12} (2016)  045},
\href{http://arxiv.org/abs/1606.09621}{{\tt arXiv:1606.09621 [hep-ph]}}.

\bibitem{Bishara:2016jga}
F.~Bishara, U.~Haisch, P.~F. Monni, and E.~Re, {\em {Constraining Light-Quark
  Yukawa Couplings from Higgs Distributions}\/},
  \href{http://dx.doi.org/10.1103/PhysRevLett.118.121801}{Phys. Rev. Lett. {\bf
  118} (2017) no.~12, 121801},
\href{http://arxiv.org/abs/1606.09253}{{\tt arXiv:1606.09253 [hep-ph]}}.

\bibitem{Sirunyan:2018koj}
{{CMS} Collaboration}, A.~M. Sirunyan et al., {\em {Combined measurements of
  Higgs boson couplings in proton-proton collisions at $\sqrt{s}=$ 13 TeV}\/},
  Submitted to: Eur. Phys. J. (2018)  ,
\href{http://arxiv.org/abs/1809.10733}{{\tt arXiv:1809.10733 [hep-ex]}}.

\bibitem{Khachatryan:2014aep}
{CMS Collaboration}, V.~Khachatryan et al., {\em {Search for a standard
  model-like Higgs boson in the $\mu^+\mu^-��$ and $e^+e^-��$ decay
  channels at the LHC}\/},
  \href{http://dx.doi.org/10.1016/j.physletb.2015.03.048}{Phys. Lett. {\bf
  B744} (2015)  184--207},
\href{http://arxiv.org/abs/1410.6679}{{\tt arXiv:1410.6679 [hep-ex]}}.

\bibitem{Perez:2015lra}
G.~Perez, Y.~Soreq, E.~Stamou, and K.~Tobioka, {\em {Prospects for measuring
  the Higgs boson coupling to light quarks}\/},
  \href{http://dx.doi.org/10.1103/PhysRevD.93.013001}{Phys. Rev. {\bf D93}
  (2016) no.~1, 013001},
\href{http://arxiv.org/abs/1505.06689}{{\tt arXiv:1505.06689 [hep-ph]}}.

\bibitem{Brivio:2015fxa}
I.~Brivio, F.~Goertz, and G.~Isidori, {\em {Probing the Charm Quark Yukawa
  Coupling in Higgs+Charm Production}\/},
  \href{http://dx.doi.org/10.1103/PhysRevLett.115.211801}{Phys. Rev. Lett. {\bf
  115} (2015) no.~21, 211801},
\href{http://arxiv.org/abs/1507.02916}{{\tt arXiv:1507.02916 [hep-ph]}}.

\bibitem{Koenig:2015pha}
M.~Koenig and M.~Neubert, {\em {Exclusive Radiative Higgs Decays as Probes of
  Light-Quark Yukawa Couplings}\/},
  \href{http://dx.doi.org/10.1007/JHEP08(2015)012}{JHEP {\bf 08} (2015)  012},
\href{http://arxiv.org/abs/1505.03870}{{\tt arXiv:1505.03870 [hep-ph]}}.

\bibitem{Bodwin:2014bpa}
G.~T. Bodwin, H.~S. Chung, J.-H. Ee, J.~Lee, and F.~Petriello, {\em
  {Relativistic corrections to Higgs boson decays to quarkonia}\/},
  \href{http://dx.doi.org/10.1103/PhysRevD.90.113010}{Phys. Rev. {\bf D90}
  (2014) no.~11, 113010},
\href{http://arxiv.org/abs/1407.6695}{{\tt arXiv:1407.6695 [hep-ph]}}.

\bibitem{Bodwin:2013gca}
G.~T. Bodwin, F.~Petriello, S.~Stoynev, and M.~Velasco, {\em {Higgs boson
  decays to quarkonia and the $H\bar{c}c$ coupling}\/},
  \href{http://dx.doi.org/10.1103/PhysRevD.88.053003}{Phys. Rev. {\bf D88}
  (2013) no.~5, 053003},
\href{http://arxiv.org/abs/1306.5770}{{\tt arXiv:1306.5770 [hep-ph]}}.

\bibitem{Bishara:2015cha}
F.~Bishara, J.~Brod, P.~Uttayarat, and J.~Zupan, {\em {Nonstandard Yukawa
  Couplings and Higgs Portal Dark Matter}\/},
  \href{http://dx.doi.org/10.1007/JHEP01(2016)010}{JHEP {\bf 01} (2016)  010},
\href{http://arxiv.org/abs/1504.04022}{{\tt arXiv:1504.04022 [hep-ph]}}.

\bibitem{Dery:2014kxa}
A.~Dery, A.~Efrati, Y.~Nir, Y.~Soreq, and V.~Susic, {\em {Model building for
  flavor changing Higgs couplings}\/},
  \href{http://dx.doi.org/10.1103/PhysRevD.90.115022}{Phys. Rev. {\bf D90}
  (2014)  115022},
\href{http://arxiv.org/abs/1408.1371}{{\tt arXiv:1408.1371 [hep-ph]}}.

\bibitem{Dery:2013aba}
A.~Dery, A.~Efrati, G.~Hiller, Y.~Hochberg, and Y.~Nir, {\em {Higgs couplings
  to fermions: 2HDM with MFV}\/},
  \href{http://dx.doi.org/10.1007/JHEP08(2013)006}{JHEP {\bf 08} (2013)  006},
\href{http://arxiv.org/abs/1304.6727}{{\tt arXiv:1304.6727 [hep-ph]}}.

\bibitem{Dery:2013rta}
A.~Dery, A.~Efrati, Y.~Hochberg, and Y.~Nir, {\em {What if
  $BR(h\to\mu\mu)/BR(h\to\tau\tau)$ does not equal $m_\mu^2/m_\tau^2$?}\/},
  \href{http://dx.doi.org/10.1007/JHEP05(2013)039}{JHEP {\bf 05} (2013)  039},
\href{http://arxiv.org/abs/1302.3229}{{\tt arXiv:1302.3229 [hep-ph]}}.

\bibitem{Bauer:2015kzy}
M.~Bauer, M.~Carena, and K.~Gemmler, {\em {Creating the fermion mass
  hierarchies with multiple Higgs bosons}\/},
  \href{http://dx.doi.org/10.1103/PhysRevD.94.115030}{Phys. Rev. {\bf D94}
  (2016) no.~11, 115030},
\href{http://arxiv.org/abs/1512.03458}{{\tt arXiv:1512.03458 [hep-ph]}}.

\bibitem{Glashow:1976nt}
S.~L. Glashow and S.~Weinberg, {\em {Natural Conservation Laws for Neutral
  Currents}\/},
\href{http://dx.doi.org/10.1103/PhysRevD.15.1958}{Phys. Rev. {\bf D15} (1977)
  1958}.

\bibitem{Paschos:1976ay}
E.~A. Paschos, {\em {Diagonal Neutral Currents}\/},
\href{http://dx.doi.org/10.1103/PhysRevD.15.1966}{Phys. Rev. {\bf D15} (1977)
  1966}.

\bibitem{Altmannshofer:2015esa}
W.~Altmannshofer, S.~Gori, A.~L. Kagan, L.~Silvestrini, and J.~Zupan, {\em
  {Uncovering Mass Generation Through Higgs Flavor Violation}\/},
  \href{http://dx.doi.org/10.1103/PhysRevD.93.031301}{Phys. Rev. {\bf D93}
  (2016) no.~3, 031301},
\href{http://arxiv.org/abs/1507.07927}{{\tt arXiv:1507.07927 [hep-ph]}}.

\bibitem{Altmannshofer:2016zrn}
W.~Altmannshofer, J.~Eby, S.~Gori, M.~Lotito, M.~Martone, and D.~Tuckler, {\em
  {Collider Signatures of Flavorful Higgs Bosons}\/},
  \href{http://dx.doi.org/10.1103/PhysRevD.94.115032}{Phys. Rev. {\bf D94}
  (2016) no.~11, 115032},
\href{http://arxiv.org/abs/1610.02398}{{\tt arXiv:1610.02398 [hep-ph]}}.

\bibitem{Altmannshofer:2017uvs}
W.~Altmannshofer, S.~Gori, D.~J. Robinson, and D.~Tuckler, {\em {The
  Flavor-locked Flavorful Two Higgs Doublet Model}\/},
  \href{http://dx.doi.org/10.1007/JHEP03(2018)129}{JHEP {\bf 03} (2018)  129},
\href{http://arxiv.org/abs/1712.01847}{{\tt arXiv:1712.01847 [hep-ph]}}.

\bibitem{Altmannshofer:2018bch}
W.~Altmannshofer and B.~Maddock, {\em {Flavorful Two Higgs Doublet Models with
  a Twist}\/},  \href{http://dx.doi.org/10.1103/PhysRevD.98.075005}{Phys. Rev.
  {\bf D98} (2018) no.~7, 075005},
\href{http://arxiv.org/abs/1805.08659}{{\tt arXiv:1805.08659 [hep-ph]}}.

\bibitem{Froggatt:1978nt}
C.~D. Froggatt and H.~B. Nielsen, {\em {Hierarchy of Quark Masses, Cabibbo
  Angles and CP Violation}\/},
\href{http://dx.doi.org/10.1016/0550-3213(79)90316-X}{Nucl. Phys. {\bf B147}
  (1979)  277--298}.

\bibitem{Giudice:2008uua}
G.~F. Giudice and O.~Lebedev, {\em {Higgs-dependent Yukawa couplings}\/},
  \href{http://dx.doi.org/10.1016/j.physletb.2008.05.062}{Phys. Lett. {\bf
  B665} (2008)  79--85},
\href{http://arxiv.org/abs/0804.1753}{{\tt arXiv:0804.1753 [hep-ph]}}.

\bibitem{D'Ambrosio:2002ex}
G.~D'Ambrosio, G.~F. Giudice, G.~Isidori, and A.~Strumia, {\em {Minimal flavor
  violation: An Effective field theory approach}\/},
  \href{http://dx.doi.org/10.1016/S0550-3213(02)00836-2}{Nucl. Phys. {\bf B645}
  (2002)  155--187},
\href{http://arxiv.org/abs/hep-ph/0207036}{{\tt arXiv:hep-ph/0207036
  [hep-ph]}}.

\bibitem{Randall:1999ee}
L.~Randall and R.~Sundrum, {\em {A Large mass hierarchy from a small extra
  dimension}\/},  \href{http://dx.doi.org/10.1103/PhysRevLett.83.3370}{Phys.
  Rev. Lett. {\bf 83} (1999)  3370--3373},
\href{http://arxiv.org/abs/hep-ph/9905221}{{\tt arXiv:hep-ph/9905221
  [hep-ph]}}.

\bibitem{Dugan:1984hq}
M.~J. Dugan, H.~Georgi, and D.~B. Kaplan, {\em {Anatomy of a Composite Higgs
  Model}\/},
\href{http://dx.doi.org/10.1016/0550-3213(85)90221-4}{Nucl. Phys. {\bf B254}
  (1985)  299--326}.

\bibitem{Georgi:1984ef}
H.~Georgi, D.~B. Kaplan, and P.~Galison, {\em {Calculation of the Composite
  Higgs Mass}\/},
\href{http://dx.doi.org/10.1016/0370-2693(84)90823-2}{Phys. Lett. {\bf 143B}
  (1984)  152--154}.

\bibitem{Kaplan:1983sm}
D.~B. Kaplan, H.~Georgi, and S.~Dimopoulos, {\em {Composite Higgs Scalars}\/},
\href{http://dx.doi.org/10.1016/0370-2693(84)91178-X}{Phys. Lett. {\bf 136B}
  (1984)  187--190}.

\bibitem{Kaplan:1983fs}
D.~B. Kaplan and H.~Georgi, {\em {SU(2) x U(1) Breaking by Vacuum
  Misalignment}\/},
\href{http://dx.doi.org/10.1016/0370-2693(84)91177-8}{Phys. Lett. {\bf 136B}
  (1984)  183--186}.

\bibitem{Botella:2016krk}
F.~J. Botella, G.~C. Branco, M.~N. Rebelo, and J.~I. Silva-Marcos, {\em {What
  if the masses of the first two quark families are not generated by the
  standard model Higgs boson?}\/},
  \href{http://dx.doi.org/10.1103/PhysRevD.94.115031}{Phys. Rev. {\bf D94}
  (2016) no.~11, 115031},
\href{http://arxiv.org/abs/1602.08011}{{\tt arXiv:1602.08011 [hep-ph]}}.

\bibitem{Ghosh:2015gpa}
D.~Ghosh, R.~S. Gupta, and G.~Perez, {\em {Is the Higgs Mechanism of Fermion
  Mass Generation a Fact? A Yukawa-less First-Two-Generation Model}\/},
  \href{http://dx.doi.org/10.1016/j.physletb.2016.02.059}{Phys. Lett. {\bf
  B755} (2016)  504--508},
\href{http://arxiv.org/abs/1508.01501}{{\tt arXiv:1508.01501 [hep-ph]}}.

\bibitem{Das:1995df}
A.~K. Das and C.~Kao, {\em {A Two Higgs doublet model for the top quark}\/},
  \href{http://dx.doi.org/10.1016/0370-2693(96)00031-7}{Phys. Lett. {\bf B372}
  (1996)  106--112},
\href{http://arxiv.org/abs/hep-ph/9511329}{{\tt arXiv:hep-ph/9511329
  [hep-ph]}}.

\bibitem{Blechman:2010cs}
A.~E. Blechman, A.~A. Petrov, and G.~Yeghiyan, {\em {The Flavor puzzle in
  multi-Higgs models}\/},
  \href{http://dx.doi.org/10.1007/JHEP11(2010)075}{JHEP {\bf 11} (2010)  075},
\href{http://arxiv.org/abs/1009.1612}{{\tt arXiv:1009.1612 [hep-ph]}}.

\bibitem{Gherghetta:2000qt}
T.~Gherghetta and A.~Pomarol, {\em {Bulk fields and supersymmetry in a slice of
  AdS}\/},  \href{http://dx.doi.org/10.1016/S0550-3213(00)00392-8}{Nucl. Phys.
  {\bf B586} (2000)  141--162},
\href{http://arxiv.org/abs/hep-ph/0003129}{{\tt arXiv:hep-ph/0003129
  [hep-ph]}}.

\bibitem{Grossman:1999ra}
Y.~Grossman and M.~Neubert, {\em {Neutrino masses and mixings in
  nonfactorizable geometry}\/},
  \href{http://dx.doi.org/10.1016/S0370-2693(00)00054-X}{Phys. Lett. {\bf B474}
  (2000)  361--371},
\href{http://arxiv.org/abs/hep-ph/9912408}{{\tt arXiv:hep-ph/9912408
  [hep-ph]}}.

\bibitem{Huber:2000ie}
S.~J. Huber and Q.~Shafi, {\em {Fermion masses, mixings and proton decay in a
  Randall-Sundrum model}\/},
  \href{http://dx.doi.org/10.1016/S0370-2693(00)01399-X}{Phys. Lett. {\bf B498}
  (2001)  256--262},
\href{http://arxiv.org/abs/hep-ph/0010195}{{\tt arXiv:hep-ph/0010195
  [hep-ph]}}.

\bibitem{Huber:2003tu}
S.~J. Huber, {\em {Flavor violation and warped geometry}\/},
  \href{http://dx.doi.org/10.1016/S0550-3213(03)00502-9}{Nucl. Phys. {\bf B666}
  (2003)  269--288},
\href{http://arxiv.org/abs/hep-ph/0303183}{{\tt arXiv:hep-ph/0303183
  [hep-ph]}}.

\bibitem{Panico:2015jxa}
G.~Panico and A.~Wulzer, {\em {The Composite Nambu-Goldstone Higgs}\/},
  \href{http://dx.doi.org/10.1007/978-3-319-22617-0}{Lect. Notes Phys. {\bf
  913} (2016)  pp.1--316},
\href{http://arxiv.org/abs/1506.01961}{{\tt arXiv:1506.01961 [hep-ph]}}.

\bibitem{Aad:2014nra}
{ATLAS Collaboration}, G.~Aad et al., {\em {Search for pair-produced
  third-generation squarks decaying via charm quarks or in compressed
  supersymmetric scenarios in $pp$ collisions at $\sqrt{s}=8~$TeV with the
  ATLAS detector}\/},
  \href{http://dx.doi.org/10.1103/PhysRevD.90.052008}{Phys. Rev. {\bf D90}
  (2014) no.~5, 052008},
\href{http://arxiv.org/abs/1407.0608}{{\tt arXiv:1407.0608 [hep-ex]}}.

\bibitem{Aad:2015gna}
{ATLAS Collaboration}, G.~Aad et al., {\em {Search for Scalar Charm Quark Pair
  Production in $pp$ Collisions at $\sqrt{s}=~$8 TeV with the ATLAS
  Detector}\/},  \href{http://dx.doi.org/10.1103/PhysRevLett.114.161801}{Phys.
  Rev. Lett. {\bf 114} (2015) no.~16, 161801},
\href{http://arxiv.org/abs/1501.01325}{{\tt arXiv:1501.01325 [hep-ex]}}.

\bibitem{Delaunay:2013pja}
C.~Delaunay, T.~Golling, G.~Perez, and Y.~Soreq, {\em {Enhanced Higgs boson
  coupling to charm pairs}\/},
  \href{http://dx.doi.org/10.1103/PhysRevD.89.033014}{Phys. Rev. {\bf D89}
  (2014) no.~3, 033014},
\href{http://arxiv.org/abs/1310.7029}{{\tt arXiv:1310.7029 [hep-ph]}}.

\bibitem{ATL-PHYS-PUB-2018-016}
{ATLAS Collaboration}, {\em {Prospects for $H\rightarrow c\bar c$ using Charm
  Tagging with the ATLAS Experiment at the HL-LHC}\/},   ATL-PHYS-PUB-2018-016,
  CERN, Geneva, Aug, 2018.
\newblock \url{https://cds.cern.ch/record/2633635}.

\bibitem{ATLAS-collaboration:2012iza}
{ATLAS Collaboration}, {\em {Physics at a High-Luminosity LHC with ATLAS}\/},
  2012.
\newblock \href{http://arxiv.org/abs/ATL-PHYS-PUB-2012-004,
  ATL-COM-PHYS-2012-1455}{{\tt ATL-PHYS-PUB-2012-004, ATL-COM-PHYS-2012-1455}}.

\bibitem{LHCb-PAPER-2015-016}
{LHCb collaboration}, R.~Aaij et al., {\em {Identification of beauty and charm
  quark jets at LHCb}\/},  JINST {\bf 10} (2015)  P06013,
  \href{http://arxiv.org/abs/1504.07670}{{\tt arXiv:1504.07670 [hep-ex]}}.

\bibitem{LHCb:2016yxg}
{LHCb Collaboration}, T.~L. Collaboration,
{\em {Search for $H^0 \rightarrow b \bar{b}$ or $c \bar{c}$ in association with
  a $W$ or $Z$ boson in the forward region of $pp$ collisions}\/}, .

\bibitem{Duarte-Campderros:2018ouv}
J.~Duarte-Campderros, G.~Perez, M.~Schlaffer, and A.~Soffer, {\em {Probing the
  strange Higgs coupling at lepton colliders using light-jet flavor
  tagging}\/},
\href{http://arxiv.org/abs/1811.09636}{{\tt arXiv:1811.09636 [hep-ph]}}.

\bibitem{Boudinov:1998fao}
E.~Boudinov, P.~Kluit, F.~Cossutti, K.~Huet, M.~G{u}nther, and O.~Botner,
{\em {Measurement of the strange quark forward- backward asymmetry around the Z
  peak}\/}, .

\bibitem{Kalelkar:2000ig}
{SLD Collaboration}, M.~Kalelkar et al., {\em {Light quark fragmentation in
  polarized Z0 decays at SLD}\/},
  \href{http://dx.doi.org/10.1016/S0920-5632(01)01103-3}{Nucl. Phys. Proc.
  Suppl. {\bf 96} (2001)  31--35},
  \href{http://arxiv.org/abs/hep-ex/0008032}{{\tt arXiv:hep-ex/0008032
  [hep-ex]}}.
[,31(2000)].

\bibitem{Sjostrand:2006za}
T.~Sj{\"o}strand, S.~Mrenna, and P.~Z. Skands, {\em {PYTHIA 6.4 Physics and
  Manual}\/},  \href{http://dx.doi.org/10.1088/1126-6708/2006/05/026}{JHEP {\bf
  05} (2006)  026},
\href{http://arxiv.org/abs/hep-ph/0603175}{{\tt arXiv:hep-ph/0603175
  [hep-ph]}}.

\bibitem{Aaboud:2016rug}
{ATLAS Collaboration}, M.~Aaboud et al., {\em {Search for Higgs and $Z$ Boson
  Decays to $\phi\,\gamma$ with the ATLAS Detector}\/},
  \href{http://dx.doi.org/10.1103/PhysRevLett.117.111802}{Phys. Rev. Lett. {\bf
  117} (2016) no.~11, 111802},
\href{http://arxiv.org/abs/1607.03400}{{\tt arXiv:1607.03400 [hep-ex]}}.

\bibitem{Keung:1983ac}
W.-Y. Keung, {\em {THE DECAY OF THE HIGGS BOSON INTO HEAVY QUARKONIUM
  STATES}\/},
\href{http://dx.doi.org/10.1103/PhysRevD.27.2762}{Phys. Rev. {\bf D27} (1983)
  2762}.

\bibitem{ATL-PHYS-PUB-2015-043}
{\em {Search for the Standard Model Higgs and Z Boson decays to
  $J/\psi\,\gamma$: HL-LHC projections}\/},   ATL-PHYS-PUB-2015-043, CERN,
  Geneva, Sep, 2015.
\newblock \url{http://cds.cern.ch/record/2054550}.

\bibitem{Blankenburg:2012ex}
G.~Blankenburg, J.~Ellis, and G.~Isidori, {\em {Flavour-Changing Decays of a
  125 GeV Higgs-like Particle}\/},
  \href{http://dx.doi.org/10.1016/j.physletb.2012.05.007}{Phys. Lett. {\bf
  B712} (2012)  386--390},
\href{http://arxiv.org/abs/1202.5704}{{\tt arXiv:1202.5704 [hep-ph]}}.

\bibitem{Harnik:2012pb}
R.~Harnik, J.~Kopp, and J.~Zupan, {\em {Flavor-violating Higgs decays}\/},
  JHEP {\bf 03} (2013)  026, \href{http://arxiv.org/abs/1209.1397}{{\tt
  arXiv:1209.1397 [hep-ph]}}.

\bibitem{Alte:2016yuw}
S.~Alte, M.~K\"{o}nig, and M.~Neubert, {\em {Exclusive Weak Radiative Higgs
  Decays in the Standard Model and Beyond}\/},
  \href{http://dx.doi.org/10.1007/JHEP12(2016)037}{JHEP {\bf 12} (2016)  037},
\href{http://arxiv.org/abs/1609.06310}{{\tt arXiv:1609.06310 [hep-ph]}}.

\bibitem{Arnesen:2008fb}
C.~Arnesen, I.~Z. Rothstein, and J.~Zupan, {\em {Smoking Guns for On-Shell New
  Physics at the LHC}\/},
  \href{http://dx.doi.org/10.1103/PhysRevLett.103.151801}{Phys. Rev. Lett. {\bf
  103} (2009)  151801},
\href{http://arxiv.org/abs/0809.1429}{{\tt arXiv:0809.1429 [hep-ph]}}.

\bibitem{Biekotter:2016ecg}
A.~Biek{\" o}tter, J.~Brehmer, and T.~Plehn, {\em {Extending the limits of
  Higgs effective theory}\/},
  \href{http://dx.doi.org/10.1103/PhysRevD.94.055032}{Phys. Rev. {\bf D94}
  (2016) no.~5, 055032},
\href{http://arxiv.org/abs/1602.05202}{{\tt arXiv:1602.05202 [hep-ph]}}.

\bibitem{Brehmer:2015rna}
J.~Brehmer, A.~Freitas, D.~Lopez-Val, and T.~Plehn, {\em {Pushing Higgs
  Effective Theory to its Limits}\/},
  \href{http://dx.doi.org/10.1103/PhysRevD.93.075014}{Phys. Rev. {\bf D93}
  (2016) no.~7, 075014},
\href{http://arxiv.org/abs/1510.03443}{{\tt arXiv:1510.03443 [hep-ph]}}.

\bibitem{Dawson:2015gka}
S.~Dawson, I.~M. Lewis, and M.~Zeng, {\em {Usefulness of effective field theory
  for boosted Higgs production}\/},
  \href{http://dx.doi.org/10.1103/PhysRevD.91.074012}{Phys. Rev. {\bf D91}
  (2015)  074012},
\href{http://arxiv.org/abs/1501.04103}{{\tt arXiv:1501.04103 [hep-ph]}}.

\bibitem{Schlaffer:2014osa}
M.~Schlaffer, M.~Spannowsky, M.~Takeuchi, A.~Weiler, and C.~Wymant, {\em
  {Boosted Higgs Shapes}\/},
  \href{http://dx.doi.org/10.1140/epjc/s10052-014-3120-z}{Eur. Phys. J. {\bf
  C74} (2014) no.~10, 3120},
\href{http://arxiv.org/abs/1405.4295}{{\tt arXiv:1405.4295 [hep-ph]}}.

\bibitem{Grojean:2013nya}
C.~Grojean, E.~Salvioni, M.~Schlaffer, and A.~Weiler, {\em {Very boosted Higgs
  in gluon fusion}\/},  \href{http://dx.doi.org/10.1007/JHEP05(2014)022}{JHEP
  {\bf 05} (2014)  022},
\href{http://arxiv.org/abs/1312.3317}{{\tt arXiv:1312.3317 [hep-ph]}}.

\bibitem{Langenegger:2015lra}
U.~Langenegger, M.~Spira, and I.~Strebel, {\em {Testing the Higgs Boson
  Coupling to Gluons}\/},
\href{http://arxiv.org/abs/1507.01373}{{\tt arXiv:1507.01373 [hep-ph]}}.

\bibitem{Bramante:2014hua}
J.~Bramante, A.~Delgado, L.~Lehman, and A.~Martin, {\em {Boosted Higgses from
  chromomagnetic $b$'s: $b\bar{b}h$ at high luminosity}\/},
  \href{http://dx.doi.org/10.1103/PhysRevD.93.053001}{Phys. Rev. {\bf D93}
  (2016) no.~5, 053001},
\href{http://arxiv.org/abs/1410.3484}{{\tt arXiv:1410.3484 [hep-ph]}}.

\bibitem{Buschmann:2014twa}
M.~Buschmann, C.~Englert, D.~Goncalves, T.~Plehn, and M.~Spannowsky, {\em
  {Resolving the Higgs-Gluon Coupling with Jets}\/},
  \href{http://dx.doi.org/10.1103/PhysRevD.90.013010}{Phys. Rev. {\bf D90}
  (2014) no.~1, 013010},
\href{http://arxiv.org/abs/1405.7651}{{\tt arXiv:1405.7651 [hep-ph]}}.

\bibitem{Azatov:2013xha}
A.~Azatov and A.~Paul, {\em {Probing Higgs couplings with high $p_T$ Higgs
  production}\/},  \href{http://dx.doi.org/10.1007/JHEP01(2014)014}{JHEP {\bf
  01} (2014)  014},
\href{http://arxiv.org/abs/1309.5273}{{\tt arXiv:1309.5273 [hep-ph]}}.

\bibitem{Banfi:2013yoa}
A.~Banfi, A.~Martin, and V.~Sanz, {\em {Probing top-partners in Higgs+jets}\/},
   \href{http://dx.doi.org/10.1007/JHEP08(2014)053}{JHEP {\bf 08} (2014)  053},
\href{http://arxiv.org/abs/1308.4771}{{\tt arXiv:1308.4771 [hep-ph]}}.

\bibitem{Buschmann:2014sia}
M.~Buschmann, D.~Goncalves, S.~Kuttimalai, M.~Schonherr, F.~Krauss, and
  T.~Plehn, {\em {Mass Effects in the Higgs-Gluon Coupling: Boosted vs
  Off-Shell Production}\/},
  \href{http://dx.doi.org/10.1007/JHEP02(2015)038}{JHEP {\bf 02} (2015)  038},
\href{http://arxiv.org/abs/1410.5806}{{\tt arXiv:1410.5806 [hep-ph]}}.

\bibitem{Collins:1984kg}
J.~C. Collins, D.~E. Soper, and G.~F. Sterman, {\em {Transverse Momentum
  Distribution in Drell-Yan Pair and W and Z Boson Production}\/},
\href{http://dx.doi.org/10.1016/0550-3213(85)90479-1}{Nucl. Phys. {\bf B250}
  (1985)  199--224}.

\bibitem{Baur:1989cm}
U.~Baur and E.~W.~N. Glover, {\em {Higgs Boson Production at Large Transverse
  Momentum in Hadronic Collisions}\/},
\href{http://dx.doi.org/10.1016/0550-3213(90)90532-I}{Nucl. Phys. {\bf B339}
  (1990)  38--66}.

\bibitem{Aad:2015lha}
{ATLAS Collaboration}, G.~Aad et al., {\em {Measurements of the Total and
  Differential Higgs Boson Production Cross Sections Combining the
  $H\to\gamma\gamma$ and $H\to ZZ^*\to 4\ell$ Decay Channels at $\sqrt{s}=8$TeV
  with the ATLAS Detector}\/},
  \href{http://dx.doi.org/10.1103/PhysRevLett.115.091801}{Phys. Rev. Lett. {\bf
  115} (2015) no.~9, 091801},
\href{http://arxiv.org/abs/1504.05833}{{\tt arXiv:1504.05833 [hep-ex]}}.

\bibitem{Sirunyan:2018sgc}
{CMS Collaboration}, A.~M. Sirunyan et al., {\em {Measurement and
  interpretation of differential cross sections for Higgs boson production at
  $\sqrt{s}=$ 13 TeV}\/},  Submitted to: Phys. Lett. (2018)  ,
\href{http://arxiv.org/abs/1812.06504}{{\tt arXiv:1812.06504 [hep-ex]}}.

\bibitem{CMS:2018hhg}
{CMS Collaboration}, {\em {Combined measurement and interpretation of
  differential Higgs boson production cross sections at $\sqrt{s}$=13 TeV}\/},
  .
\url{http://cdsweb.cern.ch/record/2628757}.

\bibitem{Sirunyan:2018kta}
{CMS Collaboration}, A.~M. Sirunyan et al., {\em {Measurement of inclusive and
  differential Higgs boson production cross sections in the diphoton decay
  channel in proton-proton collisions at $\sqrt{s}=$ 13 TeV}\/},
\href{http://arxiv.org/abs/1807.03825}{{\tt arXiv:1807.03825 [hep-ex]}}.

\bibitem{Sirunyan:2017exp}
{CMS Collaboration}, A.~M. Sirunyan et al., {\em {M}easurements of properties
  of the {H}iggs boson decaying into the four-lepton final state in pp
  collisions at {$ \sqrt{s}$} = 13 {TeV}\/},
  \href{http://dx.doi.org/10.1007/JHEP11(2017)047}{JHEP {\bf 11} (2017)  047},
\href{http://arxiv.org/abs/1706.09936}{{\tt arXiv:1706.09936 [hep-ex]}}.

\bibitem{Sirunyan:2017dgc}
{CMS Collaboration}, A.~M. Sirunyan et al., {\em {I}nclusive search for a
  highly boosted {H}iggs boson decaying to a bottom quark-antiquark pair\/},
  \href{http://dx.doi.org/10.1103/PhysRevLett.120.071802}{Phys. Rev. Lett. {\bf
  120} (2018)  071802},
\href{http://arxiv.org/abs/1709.05543}{{\tt arXiv:1709.05543 [hep-ex]}}.

\bibitem{Dasgupta:2013ihk}
M.~Dasgupta, A.~Fregoso, S.~Marzani, and G.~P. Salam, {\em {Towards an
  understanding of jet substructure}\/},
  \href{http://dx.doi.org/10.1007/JHEP09(2013)029}{JHEP {\bf 09} (2013)  029},
\href{http://arxiv.org/abs/1307.0007}{{\tt arXiv:1307.0007 [hep-ph]}}.

\bibitem{Larkoski:2014wba}
A.~J. Larkoski, S.~Marzani, G.~Soyez, and J.~Thaler, {\em {Soft Drop}\/},
  \href{http://dx.doi.org/10.1007/JHEP05(2014)146}{JHEP {\bf 05} (2014)  146},
\href{http://arxiv.org/abs/1402.2657}{{\tt arXiv:1402.2657 [hep-ph]}}.

\bibitem{CMS-PAS-FTR-18-011}
{CMS Collaboration}, {\em Sensitivity projections for Higgs boson properties
  measurements at the HL-LHC\/},   CMS-PAS-FTR-18-011, {CERN}, Geneva, 2018.
\newblock \url{http://cds.cern.ch/record/2647699?ln=en}.

\bibitem{Yu:2016rvv}
F.~Yu, {\em {Phenomenology of Enhanced Light Quark Yukawa Couplings and the
  $W^\pm h$ Charge Asymmetry}\/},
  \href{http://dx.doi.org/10.1007/JHEP02(2017)083}{JHEP {\bf 02} (2017)  083},
\href{http://arxiv.org/abs/1609.06592}{{\tt arXiv:1609.06592 [hep-ph]}}.

\bibitem{Gorbahn:2014sha}
M.~Gorbahn and U.~Haisch, {\em {Searching for $t \to c(u)h$ with dipole
  moments}\/},  \href{http://dx.doi.org/10.1007/JHEP06(2014)033}{JHEP {\bf 06}
  (2014)  033},
\href{http://arxiv.org/abs/1404.4873}{{\tt arXiv:1404.4873 [hep-ph]}}.

\bibitem{Khachatryan:2016rke}
{CMS Collaboration}, V.~Khachatryan et al., {\em {Search for lepton flavour
  violating decays of the Higgs boson to $e \tau$ and $e \mu$ in proton-proton
  collisions at $\sqrt s=$ 8 TeV}\/},
  \href{http://dx.doi.org/10.1016/j.physletb.2016.09.062}{Phys. Lett. {\bf
  B763} (2016)  472--500},
\href{http://arxiv.org/abs/1607.03561}{{\tt arXiv:1607.03561 [hep-ex]}}.

\bibitem{Sirunyan:2017xzt}
{CMS Collaboration}, A.~M. Sirunyan et al., {\em {Search for lepton flavour
  violating decays of the Higgs boson to $\mu\tau$ and e$\tau$ in proton-proton
  collisions at $\sqrt{s}=$ 13 TeV}\/},
  \href{http://dx.doi.org/10.1007/JHEP06(2018)001}{JHEP {\bf 06} (2018)  001},
\href{http://arxiv.org/abs/1712.07173}{{\tt arXiv:1712.07173 [hep-ex]}}.

\bibitem{Aad:2016blu}
{ATLAS Collaboration}, G.~Aad et al., {\em {Search for lepton-flavour-violating
  decays of the Higgs and $Z$ bosons with the ATLAS detector}\/},
  \href{http://dx.doi.org/10.1140/epjc/s10052-017-4624-0}{Eur. Phys. J. {\bf
  C77} (2017) no.~2, 70},
\href{http://arxiv.org/abs/1604.07730}{{\tt arXiv:1604.07730 [hep-ex]}}.

\bibitem{Bressler:2014jta}
S.~Bressler, A.~Dery, and A.~Efrati, {\em {Asymmetric lepton-flavor violating
  Higgs boson decays}\/},
  \href{http://dx.doi.org/10.1103/PhysRevD.90.015025}{Phys. Rev. {\bf D90}
  (2014) no.~1, 015025},
\href{http://arxiv.org/abs/1405.4545}{{\tt arXiv:1405.4545 [hep-ph]}}.

\bibitem{Aad:2015pja}
{ATLAS Collaboration}, G.~Aad et al., {\em {Search for flavour-changing neutral
  current top quark decays $t\to Hq$ in $pp$ collisions at $\sqrt{s}=8$ TeV
  with the ATLAS detector}\/},
  \href{http://dx.doi.org/10.1007/JHEP12(2015)061}{JHEP {\bf 12} (2015)  061},
\href{http://arxiv.org/abs/1509.06047}{{\tt arXiv:1509.06047 [hep-ex]}}.

\bibitem{Aad:2014dya}
{ATLAS Collaboration}, G.~Aad et al., {\em {Search for top quark decays $t \to
  qH$ with $H \to \gamma\gamma$ using the ATLAS detector}\/},
  \href{http://dx.doi.org/10.1007/JHEP06(2014)008}{JHEP {\bf 1406} (2014)
  008},
\href{http://arxiv.org/abs/1403.6293}{{\tt arXiv:1403.6293 [hep-ex]}}.

\bibitem{Brod:2013cka}
J.~Brod, U.~Haisch, and J.~Zupan, {\em {Constraints on CP-violating Higgs
  couplings to the third generation}\/},
  \href{http://dx.doi.org/10.1007/JHEP11(2013)180}{JHEP {\bf 11} (2013)  180},
\href{http://arxiv.org/abs/1310.1385}{{\tt arXiv:1310.1385 [hep-ph]}}.

\bibitem{Chien:2015xha}
Y.~T. Chien, V.~Cirigliano, W.~Dekens, J.~de~Vries, and E.~Mereghetti, {\em
  {Direct and indirect constraints on CP-violating Higgs-quark and Higgs-gluon
  interactions}\/},  \href{http://dx.doi.org/10.1007/JHEP02(2016)011}{JHEP {\bf
  02} (2016)  011}, \href{http://arxiv.org/abs/1510.00725}{{\tt
  arXiv:1510.00725 [hep-ph]}}.
[JHEP02,011(2016)].

\bibitem{Egana-Ugrinovic:2018fpy}
D.~Egana-Ugrinovic and S.~Thomas, {\em {Higgs Boson Contributions to the
  Electron Electric Dipole Moment}\/},
\href{http://arxiv.org/abs/1810.08631}{{\tt arXiv:1810.08631 [hep-ph]}}.

\bibitem{Brod:2018pli}
J.~Brod and E.~Stamou, {\em {Electric dipole moment constraints on CP-violating
  heavy-quark Yukawas at next-to-leading order}\/},
\href{http://arxiv.org/abs/1810.12303}{{\tt arXiv:1810.12303 [hep-ph]}}.

\bibitem{Baron:2013eja}
{ACME Collaboration}, J.~Baron et al., {\em {Order of Magnitude Smaller Limit
  on the Electric Dipole Moment of the Electron}\/},
  \href{http://dx.doi.org/10.1126/science.1248213}{Science {\bf 343} (2014)
  no.~6168, 269--272},
\href{http://arxiv.org/abs/1310.7534}{{\tt arXiv:1310.7534 [physics.atom-ph]}}.

\bibitem{Brod:2018xyz}
J.~Brod and D.~Skodras, {\em {Electric dipole moment constraints on
  CP-violating light-quark Yukawas}\/},
\href{http://arxiv.org/abs/1811.05480}{{\tt arXiv:1811.05480 [hep-ph]}}.

\bibitem{Buckley:2015vsa}
M.~R. Buckley and D.~Goncalves, {\em {Boosting the Direct CP Measurement of the
  Higgs-Top Coupling}\/},
  \href{http://dx.doi.org/10.1103/PhysRevLett.116.091801}{Phys. Rev. Lett. {\bf
  116} (2016) no.~9, 091801},
\href{http://arxiv.org/abs/1507.07926}{{\tt arXiv:1507.07926 [hep-ph]}}.

\bibitem{Boudjema:2015nda}
F.~Boudjema, R.~M. Godbole, D.~Guadagnoli, and K.~A. Mohan, {\em {Lab-frame
  observables for probing the top-Higgs interaction}\/},
  \href{http://dx.doi.org/10.1103/PhysRevD.92.015019}{Phys. Rev. {\bf D92}
  (2015) no.~1, 015019},
\href{http://arxiv.org/abs/1501.03157}{{\tt arXiv:1501.03157 [hep-ph]}}.

\bibitem{Mahlon:1995zn}
G.~Mahlon and S.~J. Parke, {\em {Angular correlations in top quark pair
  production and decay at hadron colliders}\/},
  \href{http://dx.doi.org/10.1103/PhysRevD.53.4886}{Phys. Rev. {\bf D53} (1996)
   4886--4896},
\href{http://arxiv.org/abs/hep-ph/9512264}{{\tt arXiv:hep-ph/9512264
  [hep-ph]}}.

\bibitem{Artoisenet:2013puc}
P.~Artoisenet et al., {\em {A framework for Higgs characterisation}\/},
  \href{http://dx.doi.org/10.1007/JHEP11(2013)043}{JHEP {\bf 11} (2013)  043},
\href{http://arxiv.org/abs/1306.6464}{{\tt arXiv:1306.6464 [hep-ph]}}.

\bibitem{AmorDosSantos:2017ayi}
S.~Amor Dos~Santos et al., {\em {Probing the CP nature of the Higgs coupling in
  $t{\bar t}h$ events at the LHC}\/},
  \href{http://dx.doi.org/10.1103/PhysRevD.96.013004}{Phys. Rev. {\bf D96}
  (2017) no.~1, 013004},
\href{http://arxiv.org/abs/1704.03565}{{\tt arXiv:1704.03565 [hep-ph]}}.

\bibitem{Demartin:2014fia}
F.~Demartin, F.~Maltoni, K.~Mawatari, B.~Page, and M.~Zaro, {\em {Higgs
  characterisation at NLO in QCD: CP properties of the top-quark Yukawa
  interaction}\/},
  \href{http://dx.doi.org/10.1140/epjc/s10052-014-3065-2}{Eur. Phys. J. {\bf
  C74} (2014) no.~9, 3065},
\href{http://arxiv.org/abs/1407.5089}{{\tt arXiv:1407.5089 [hep-ph]}}.

\bibitem{Bower:2002zx}
G.~R. Bower, T.~Pierzchala, Z.~Was, and M.~Worek, {\em {Measuring the Higgs
  boson's parity using $\tau \to \rho \nu$}\/},
  \href{http://dx.doi.org/10.1016/S0370-2693(02)02445-0}{Phys. Lett. {\bf B543}
  (2002)  227--234},
\href{http://arxiv.org/abs/hep-ph/0204292}{{\tt arXiv:hep-ph/0204292
  [hep-ph]}}.

\bibitem{Desch:2003mw}
K.~Desch, Z.~Was, and M.~Worek, {\em {Measuring the Higgs boson parity at a
  linear collider using the tau impact parameter and $\tau \to \rho \nu$
  decay}\/},  \href{http://dx.doi.org/10.1140/epjc/s2003-01231-4}{Eur. Phys. J.
  {\bf C29} (2003)  491--496},
\href{http://arxiv.org/abs/hep-ph/0302046}{{\tt arXiv:hep-ph/0302046
  [hep-ph]}}.

\bibitem{Desch:2003rw}
K.~Desch, A.~Imhof, Z.~Was, and M.~Worek, {\em {Probing the CP nature of the
  Higgs boson at linear colliders with tau spin correlations: The Case of mixed
  scalar - pseudoscalar couplings}\/},
  \href{http://dx.doi.org/10.1016/j.physletb.2003.10.074}{Phys. Lett. {\bf
  B579} (2004)  157--164},
\href{http://arxiv.org/abs/hep-ph/0307331}{{\tt arXiv:hep-ph/0307331
  [hep-ph]}}.

\bibitem{Harnik:2013aja}
R.~Harnik, A.~Martin, T.~Okui, R.~Primulando, and F.~Yu, {\em {Measuring CP
  violation in $h \to \tau^+ \tau^-$ at colliders}\/},
  \href{http://dx.doi.org/10.1103/PhysRevD.88.076009}{Phys. Rev. {\bf D88}
  (2013) no.~7, 076009},
\href{http://arxiv.org/abs/1308.1094}{{\tt arXiv:1308.1094 [hep-ph]}}.

\bibitem{Askew:2015mda}
A.~Askew, P.~Jaiswal, T.~Okui, H.~B. Prosper, and N.~Sato, {\em {Prospect for
  measuring the CP phase in the $h\tau\tau$ coupling at the LHC}\/},
  \href{http://dx.doi.org/10.1103/PhysRevD.91.075014}{Phys. Rev. {\bf D91}
  (2015) no.~7, 075014},
\href{http://arxiv.org/abs/1501.03156}{{\tt arXiv:1501.03156 [hep-ph]}}.

\bibitem{Jozefowicz:2016kvz}
R.~Józefowicz, E.~Richter-Was, and Z.~Was, {\em {Potential for optimizing the
  Higgs boson CP measurement in H $\to \tau \tau$ decays at the LHC including
  machine learning techniques}\/},
  \href{http://dx.doi.org/10.1103/PhysRevD.94.093001}{Phys. Rev. {\bf D94}
  (2016) no.~9, 093001},
\href{http://arxiv.org/abs/1608.02609}{{\tt arXiv:1608.02609 [hep-ph]}}.

\bibitem{Berge:2008wi}
S.~Berge, W.~Bernreuther, and J.~Ziethe, {\em {Determining the CP parity of
  Higgs bosons at the LHC in their tau decay channels}\/},
  \href{http://dx.doi.org/10.1103/PhysRevLett.100.171605}{Phys. Rev. Lett. {\bf
  100} (2008)  171605},
\href{http://arxiv.org/abs/0801.2297}{{\tt arXiv:0801.2297 [hep-ph]}}.

\bibitem{Berge:2008dr}
S.~Berge and W.~Bernreuther, {\em {Determining the CP parity of Higgs bosons at
  the LHC in the tau to 1-prong decay channels}\/},
  \href{http://dx.doi.org/10.1016/j.physletb.2008.12.065}{Phys. Lett. {\bf
  B671} (2009)  470--476},
\href{http://arxiv.org/abs/0812.1910}{{\tt arXiv:0812.1910 [hep-ph]}}.

\bibitem{Berge:2011ij}
S.~Berge, W.~Bernreuther, B.~Niepelt, and H.~Spiesberger, {\em {How to pin down
  the CP quantum numbers of a Higgs boson in its tau decays at the LHC}\/},
  \href{http://dx.doi.org/10.1103/PhysRevD.84.116003}{Phys. Rev. {\bf D84}
  (2011)  116003},
\href{http://arxiv.org/abs/1108.0670}{{\tt arXiv:1108.0670 [hep-ph]}}.

\bibitem{Agashe:2004cp}
K.~Agashe, G.~Perez, and A.~Soni, {\em {Flavor structure of warped extra
  dimension models}\/},
  \href{http://dx.doi.org/10.1103/PhysRevD.71.016002}{Phys. Rev. {\bf D71}
  (2005)  016002},
\href{http://arxiv.org/abs/hep-ph/0408134}{{\tt arXiv:hep-ph/0408134
  [hep-ph]}}.

\bibitem{Agashe:2004ay}
K.~Agashe, G.~Perez, and A.~Soni, {\em {B-factory signals for a warped extra
  dimension}\/},  \href{http://dx.doi.org/10.1103/PhysRevLett.93.201804}{Phys.
  Rev. Lett. {\bf 93} (2004)  201804},
\href{http://arxiv.org/abs/hep-ph/0406101}{{\tt arXiv:hep-ph/0406101
  [hep-ph]}}.

\bibitem{Agashe:2003zs}
K.~Agashe, A.~Delgado, M.~J. May, and R.~Sundrum, {\em {RS1, custodial isospin
  and precision tests}\/},
  \href{http://dx.doi.org/10.1088/1126-6708/2003/08/050}{JHEP {\bf 08} (2003)
  050},
\href{http://arxiv.org/abs/hep-ph/0308036}{{\tt arXiv:hep-ph/0308036
  [hep-ph]}}.

\bibitem{Agashe:2013kxa}
K.~Agashe, M.~Bauer, F.~Goertz, S.~J. Lee, L.~Vecchi, L.-T. Wang, and F.~Yu,
  {\em {Constraining RS Models by Future Flavor and Collider Measurements: A
  Snowmass Whitepaper}\/},
\href{http://arxiv.org/abs/1310.1070}{{\tt arXiv:1310.1070 [hep-ph]}}.

\bibitem{Blanke:2008yr}
M.~Blanke, A.~J. Buras, B.~Duling, K.~Gemmler, and S.~Gori, {\em {Rare K and B
  Decays in a Warped Extra Dimension with Custodial Protection}\/},
  \href{http://dx.doi.org/10.1088/1126-6708/2009/03/108}{JHEP {\bf 03} (2009)
  108},
\href{http://arxiv.org/abs/0812.3803}{{\tt arXiv:0812.3803 [hep-ph]}}.

\bibitem{Albrecht:2009xr}
M.~E. Albrecht, M.~Blanke, A.~J. Buras, B.~Duling, and K.~Gemmler, {\em
  {Electroweak and Flavour Structure of a Warped Extra Dimension with Custodial
  Protection}\/},  \href{http://dx.doi.org/10.1088/1126-6708/2009/09/064}{JHEP
  {\bf 09} (2009)  064},
\href{http://arxiv.org/abs/0903.2415}{{\tt arXiv:0903.2415 [hep-ph]}}.

\bibitem{Casagrande:2010si}
S.~Casagrande, F.~Goertz, U.~Haisch, M.~Neubert, and T.~Pfoh, {\em {The
  Custodial Randall-Sundrum Model: From Precision Tests to Higgs Physics}\/},
  \href{http://dx.doi.org/10.1007/JHEP09(2010)014}{JHEP {\bf 09} (2010)  014},
\href{http://arxiv.org/abs/1005.4315}{{\tt arXiv:1005.4315 [hep-ph]}}.

\bibitem{Casagrande:2008hr}
S.~Casagrande, F.~Goertz, U.~Haisch, M.~Neubert, and T.~Pfoh, {\em {Flavor
  Physics in the Randall-Sundrum Model: I. Theoretical Setup and Electroweak
  Precision Tests}\/},
  \href{http://dx.doi.org/10.1088/1126-6708/2008/10/094}{JHEP {\bf 10} (2008)
  094},
\href{http://arxiv.org/abs/0807.4937}{{\tt arXiv:0807.4937 [hep-ph]}}.

\bibitem{Csaki:2009wc}
C.~Csaki, G.~Perez, Z.~Surujon, and A.~Weiler, {\em {Flavor Alignment via
  Shining in RS}\/},  \href{http://dx.doi.org/10.1103/PhysRevD.81.075025}{Phys.
  Rev. {\bf D81} (2010)  075025},
\href{http://arxiv.org/abs/0907.0474}{{\tt arXiv:0907.0474 [hep-ph]}}.

\bibitem{Allanach:2009vz}
B.~C. Allanach, F.~Mahmoudi, J.~P. Skittrall, and K.~Sridhar, {\em
  {Gluon-initiated production of a Kaluza-Klein gluon in a Bulk Randall-Sundrum
  model}\/},  \href{http://dx.doi.org/10.1007/JHEP03(2010)014}{JHEP {\bf 03}
  (2010)  014},
\href{http://arxiv.org/abs/0910.1350}{{\tt arXiv:0910.1350 [hep-ph]}}.

\bibitem{Agashe:2006hk}
K.~Agashe, A.~Belyaev, T.~Krupovnickas, G.~Perez, and J.~Virzi, {\em {LHC
  Signals from Warped Extra Dimensions}\/},
  \href{http://dx.doi.org/10.1103/PhysRevD.77.015003}{Phys. Rev. {\bf D77}
  (2008)  015003},
\href{http://arxiv.org/abs/hep-ph/0612015}{{\tt arXiv:hep-ph/0612015
  [hep-ph]}}.

\bibitem{Lillie:2007yh}
B.~Lillie, L.~Randall, and L.-T. Wang, {\em {The Bulk RS KK-gluon at the
  LHC}\/},  \href{http://dx.doi.org/10.1088/1126-6708/2007/09/074}{JHEP {\bf
  09} (2007)  074},
\href{http://arxiv.org/abs/hep-ph/0701166}{{\tt arXiv:hep-ph/0701166
  [hep-ph]}}.

\bibitem{Sirunyan:2018ryr}
{CMS Collaboration}, A.~M. Sirunyan et al., {\em {Search for resonant
  $\mathrm{t \bar{t}}$ production in proton-proton collisions at $\sqrt{s}=$ 13
  TeV}\/},  Submitted to: JHEP (2018)  ,
\href{http://arxiv.org/abs/1810.05905}{{\tt arXiv:1810.05905 [hep-ex]}}.

\bibitem{Kaplan:1991dc}
D.~B. Kaplan, {\em {Flavor at SSC energies: A New mechanism for dynamically
  generated fermion masses}\/},
\href{http://dx.doi.org/10.1016/S0550-3213(05)80021-5}{Nucl. Phys. {\bf B365}
  (1991)  259--278}.

\bibitem{Grinstein:2010ve}
B.~Grinstein, M.~Redi, and G.~Villadoro, {\em {Low Scale Flavor Gauge
  Symmetries}\/},  \href{http://dx.doi.org/10.1007/JHEP11(2010)067}{JHEP {\bf
  11} (2010)  067},
\href{http://arxiv.org/abs/1009.2049}{{\tt arXiv:1009.2049 [hep-ph]}}.

\bibitem{Buras:2011wi}
A.~J. Buras, M.~V. Carlucci, L.~Merlo, and E.~Stamou, {\em {Phenomenology of a
  Gauged $SU(3)^3$ Flavour Model}\/},
  \href{http://dx.doi.org/10.1007/JHEP03(2012)088}{JHEP {\bf 03} (2012)  088},
\href{http://arxiv.org/abs/1112.4477}{{\tt arXiv:1112.4477 [hep-ph]}}.

\bibitem{Bishara:2015mha}
F.~Bishara, A.~Greljo, J.~F. Kamenik, E.~Stamou, and J.~Zupan, {\em {Dark
  Matter and Gauged Flavor Symmetries}\/},
  \href{http://dx.doi.org/10.1007/JHEP12(2015)130}{JHEP {\bf 12} (2015)  130},
\href{http://arxiv.org/abs/1505.03862}{{\tt arXiv:1505.03862 [hep-ph]}}.

\bibitem{Bauer:2015fxa}
M.~Bauer, M.~Carena, and K.~Gemmler, {\em {Flavor from the Electroweak
  Scale}\/},  \href{http://dx.doi.org/10.1007/JHEP11(2015)016}{JHEP {\bf 11}
  (2015)  016},
\href{http://arxiv.org/abs/1506.01719}{{\tt arXiv:1506.01719 [hep-ph]}}.

\bibitem{Dery:2016fyj}
A.~Dery and Y.~Nir, {\em {FN-2HDM: Two Higgs Doublet Models with
  Froggatt-Nielsen Symmetry}\/},
  \href{http://dx.doi.org/10.1007/JHEP04(2017)003}{JHEP {\bf 04} (2017)  003},
\href{http://arxiv.org/abs/1612.05219}{{\tt arXiv:1612.05219 [hep-ph]}}.

\bibitem{Bauer:2017cov}
M.~Bauer, M.~Carena, and A.~Carmona, {\em {Higgs Pair Production as a Signal of
  Enhanced Yukawa Couplings}\/},
  \href{http://dx.doi.org/10.1103/PhysRevLett.121.021801}{Phys. Rev. Lett. {\bf
  121} (2018) no.~2, 021801},
\href{http://arxiv.org/abs/1801.00363}{{\tt arXiv:1801.00363 [hep-ph]}}.

\bibitem{Choi:2015fiu}
K.~Choi and S.~H. Im, {\em {Realizing the relaxion from multiple axions and its
  UV completion with high scale supersymmetry}\/},
  \href{http://dx.doi.org/10.1007/JHEP01(2016)149}{JHEP {\bf 01} (2016)  149},
\href{http://arxiv.org/abs/1511.00132}{{\tt arXiv:1511.00132 [hep-ph]}}.

\bibitem{Kaplan:2015fuy}
D.~E. Kaplan and R.~Rattazzi, {\em {Large field excursions and approximate
  discrete symmetries from a clockwork axion}\/},
  \href{http://dx.doi.org/10.1103/PhysRevD.93.085007}{Phys. Rev. {\bf D93}
  (2016) no.~8, 085007},
\href{http://arxiv.org/abs/1511.01827}{{\tt arXiv:1511.01827 [hep-ph]}}.

\bibitem{Giudice:2016yja}
G.~F. Giudice and M.~McCullough, {\em {A Clockwork Theory}\/},
  \href{http://dx.doi.org/10.1007/JHEP02(2017)036}{JHEP {\bf 02} (2017)  036},
\href{http://arxiv.org/abs/1610.07962}{{\tt arXiv:1610.07962 [hep-ph]}}.

\bibitem{Alonso:2018bcg}
R.~Alonso, A.~Carmona, B.~M. Dillon, J.~F. Kamenik, J.~Martin~Camalich, and
  J.~Zupan, {\em {A clockwork solution to the flavor puzzle}\/},
\href{http://arxiv.org/abs/1807.09792}{{\tt arXiv:1807.09792 [hep-ph]}}.

\bibitem{Aaboud:2018uek}
{ATLAS Collaboration}, M.~Aaboud et al., {\em {Search for pair production of
  heavy vector-like quarks decaying into high-$p_T$ $W$ bosons and top quarks
  in the lepton-plus-jets final state in $pp$ collisions at $\sqrt{s}=13$ TeV
  with the ATLAS detector}\/},
\href{http://arxiv.org/abs/1806.01762}{{\tt arXiv:1806.01762 [hep-ex]}}.

\bibitem{Aaboud:2018xuw}
{ATLAS Collaboration}, M.~Aaboud et al., {\em {Search for pair production of
  up-type vector-like quarks and for four-top-quark events in final states with
  multiple $b$-jets with the ATLAS detector}\/},
\href{http://arxiv.org/abs/1803.09678}{{\tt arXiv:1803.09678 [hep-ex]}}.

\bibitem{Agashe:2009di}
K.~Agashe and R.~Contino, {\em {Composite Higgs-Mediated FCNC}\/},
  \href{http://dx.doi.org/10.1103/PhysRevD.80.075016}{Phys. Rev. {\bf D80}
  (2009)  075016},
\href{http://arxiv.org/abs/0906.1542}{{\tt arXiv:0906.1542 [hep-ph]}}.

\bibitem{Mele:1998ag}
B.~Mele, S.~Petrarca, and A.~Soddu, {\em {A New evaluation of the $t\to cH$
  decay width in the standard model}\/},
  \href{http://dx.doi.org/10.1016/S0370-2693(98)00822-3}{Phys.Lett. {\bf B435}
  (1998)  401--406},
\href{http://arxiv.org/abs/hep-ph/9805498}{{\tt arXiv:hep-ph/9805498
  [hep-ph]}}.

\bibitem{Greljo:2014dka}
A.~Greljo, J.~F. Kamenik, and J.~Kopp, {\em {Disentangling Flavor Violation in
  the Top-Higgs Sector at the LHC}\/},
  \href{http://dx.doi.org/10.1007/JHEP07(2014)046}{JHEP {\bf 1407} (2014)
  046},
\href{http://arxiv.org/abs/1404.1278}{{\tt arXiv:1404.1278 [hep-ph]}}.

\bibitem{Azatov:2014lha}
A.~Azatov, G.~Panico, G.~Perez, and Y.~Soreq, {\em {On the Flavor Structure of
  Natural Composite Higgs Models \& Top Flavor Violation}\/},
  \href{http://dx.doi.org/10.1007/JHEP12(2014)082}{JHEP {\bf 12} (2014)  082},
\href{http://arxiv.org/abs/1408.4525}{{\tt arXiv:1408.4525 [hep-ph]}}.

\bibitem{Botella:2015hoa}
F.~J. Botella, G.~C. Branco, M.~Nebot, and M.~N. Rebelo, {\em {Flavour Changing
  Higgs Couplings in a Class of Two Higgs Doublet Models}\/},
  \href{http://dx.doi.org/10.1140/epjc/s10052-016-3993-0}{Eur. Phys. J. {\bf
  C76} (2016) no.~3, 161},
\href{http://arxiv.org/abs/1508.05101}{{\tt arXiv:1508.05101 [hep-ph]}}.

\bibitem{Bardhan:2016txk}
D.~Bardhan, G.~Bhattacharyya, D.~Ghosh, M.~Patra, and S.~Raychaudhuri, {\em
  {Detailed analysis of flavor-changing decays of top quarks as a probe of new
  physics at the LHC}\/},
  \href{http://dx.doi.org/10.1103/PhysRevD.94.015026}{Phys. Rev. {\bf D94}
  (2016) no.~1, 015026},
\href{http://arxiv.org/abs/1601.04165}{{\tt arXiv:1601.04165 [hep-ph]}}.

\bibitem{Badziak:2017wxn}
M.~Badziak and K.~Harigaya, {\em {Asymptotically Free Natural SUSY Twin
  Higgs}\/},  \href{http://dx.doi.org/10.1103/PhysRevLett.120.211803}{Phys.
  Rev. Lett. {\bf 120} (2018)  211803},
\href{http://arxiv.org/abs/1711.11040}{{\tt arXiv:1711.11040 [hep-ph]}}.

\bibitem{Gabrielli:2016cut}
E.~Gabrielli, B.~Mele, M.~Raidal, and E.~Venturini, {\em {FCNC decays of
  standard model fermions into a dark photon}\/},
  \href{http://dx.doi.org/10.1103/PhysRevD.94.115013}{Phys. Rev. {\bf D94}
  (2016) no.~11, 115013},
\href{http://arxiv.org/abs/1607.05928}{{\tt arXiv:1607.05928 [hep-ph]}}.

\bibitem{Papaefstathiou:2017xuv}
A.~Papaefstathiou and G.~Tetlalmatzi-Xolocotzi, {\em {Rare top quark decays at
  a 100 TeV proton–proton collider: $t \rightarrow bWZ$ and $t\rightarrow
  hc$}\/},  \href{http://dx.doi.org/10.1140/epjc/s10052-018-5701-8}{Eur. Phys.
  J. {\bf C78} (2018) no.~3, 214},
\href{http://arxiv.org/abs/1712.06332}{{\tt arXiv:1712.06332 [hep-ph]}}.

\bibitem{Abbas:2015cua}
G.~Abbas, A.~Celis, X.-Q. Li, J.~Lu, and A.~Pich, {\em {Flavour-changing top
  decays in the aligned two-Higgs-doublet model}\/},
  \href{http://dx.doi.org/10.1007/JHEP06(2015)005}{JHEP {\bf 06} (2015)  005},
\href{http://arxiv.org/abs/1503.06423}{{\tt arXiv:1503.06423 [hep-ph]}}.

\bibitem{Ellwanger:2009dp}
U.~Ellwanger, C.~Hugonie, and A.~M. Teixeira, {\em {The Next-to-Minimal
  Supersymmetric Standard Model}\/},
  \href{http://dx.doi.org/10.1016/j.physrep.2010.07.001}{Phys. Rept. {\bf 496}
  (2010)  1--77},
\href{http://arxiv.org/abs/0910.1785}{{\tt arXiv:0910.1785 [hep-ph]}}.

\bibitem{Dimopoulos:1981xc}
S.~Dimopoulos and J.~Preskill, {\em {Massless Composites With Massive
  Constituents}\/},
\href{http://dx.doi.org/10.1016/0550-3213(82)90345-5}{Nucl. Phys. {\bf B199}
  (1982)  206--222}.

\bibitem{Zhang:2013xya}
C.~Zhang and F.~Maltoni, {\em {Top-quark decay into Higgs boson and a light
  quark at next-to-leading order in QCD}\/},
  \href{http://dx.doi.org/10.1103/PhysRevD.88.054005}{Phys.Rev. {\bf D88}
  (2013)  054005},
\href{http://arxiv.org/abs/1305.7386}{{\tt arXiv:1305.7386 [hep-ph]}}.

\bibitem{Khachatryan:2016ctc}
{CMS Collaboration}, V.~Khachatryan et al., {\em {Search for Higgs boson
  off-shell production in proton-proton collisions at 7 and 8 TeV and
  derivation of constraints on its total decay width}\/},
  \href{http://dx.doi.org/10.1007/JHEP09(2016)051}{JHEP {\bf 09} (2016)  051},
\href{http://arxiv.org/abs/1605.02329}{{\tt arXiv:1605.02329 [hep-ex]}}.

\bibitem{Agashe:2013hma}
{Top Quark Working Group Collaboration}, K.~Agashe et al., {\em {Working Group
  Report: Top Quark}\/},  in {\em {Proceedings, 2013 Community Summer Study on
  the Future of U.S. Particle Physics: Snowmass on the Mississippi (CSS2013):
  Minneapolis, MN, USA, July 29-August 6, 2013}}.
\newblock 2013.
\newblock \href{http://arxiv.org/abs/1311.2028}{{\tt arXiv:1311.2028
  [hep-ph]}}.
\newblock
\url{https://inspirehep.net/record/1263763/files/arXiv:1311.2028.pdf}.
\newblock

\bibitem{Banerjee:2018fsx}
S.~Banerjee, M.~Chala, and M.~Spannowsky, {\em {Top quark FCNCs in extended
  Higgs sectors}\/},
\href{http://arxiv.org/abs/1806.02836}{{\tt arXiv:1806.02836 [hep-ph]}}.

\bibitem{Sirunyan:2018ikr}
{CMS Collaboration}, A.~M. Sirunyan et al., {\em {Search for low-mass
  resonances decaying into bottom quark-antiquark pairs in proton-proton
  collisions at $\sqrt{s} =$ 13 TeV}\/},  Submitted to: Phys. Rev. (2018)  ,
\href{http://arxiv.org/abs/1810.11822}{{\tt arXiv:1810.11822 [hep-ex]}}.

\bibitem{Boudjema:2001ii}
F.~Boudjema and A.~Semenov, {\em {Measurements of the SUSY Higgs selfcouplings
  and the reconstruction of the Higgs potential}\/},
  \href{http://dx.doi.org/10.1103/PhysRevD.66.095007}{Phys. Rev. {\bf D66}
  (2002)  095007},
\href{http://arxiv.org/abs/hep-ph/0201219}{{\tt arXiv:hep-ph/0201219
  [hep-ph]}}.

\bibitem{Gunion:2002zf}
J.~F. Gunion and H.~E. Haber, {\em {The CP conserving two Higgs doublet model:
  The Approach to the decoupling limit}\/},
  \href{http://dx.doi.org/10.1103/PhysRevD.67.075019}{Phys. Rev. {\bf D67}
  (2003)  075019},
\href{http://arxiv.org/abs/hep-ph/0207010}{{\tt arXiv:hep-ph/0207010
  [hep-ph]}}.

\bibitem{ATLAS:2018kbw}
{ATLAS Collaboration}, T.~A. collaboration,
{\em {A search for the rare decay of the Standard Model Higgs boson to dimuons
  in $pp$ collisions at $\sqrt{s} = 13$ TeV with the ATLAS Detector}\/}, .

\bibitem{Cai:2017mow}
Y.~Cai, T.~Han, T.~Li, and R.~Ruiz, {\em {Lepton Number Violation: Seesaw
  Models and Their Collider Tests}\/},
  \href{http://dx.doi.org/10.3389/fphy.2018.00040}{Front.in Phys. {\bf 6}
  (2018)  40},
\href{http://arxiv.org/abs/1711.02180}{{\tt arXiv:1711.02180 [hep-ph]}}.

\bibitem{Weiland:2013wha}
C.~Weiland, {\em {Effects of fermionic singlet neutrinos on high- and
  low-energy observables}}.
\newblock PhD thesis, Orsay, LPT, 2013.
\newblock
\href{http://arxiv.org/abs/1311.5860}{{\tt arXiv:1311.5860 [hep-ph]}}.
\newblock

\bibitem{Ruiz:2015gsa}
R.~E. Ruiz, {\em {Hadron Collider Tests of Neutrino Mass-Generating
  Mechanisms}}.
\newblock PhD thesis, Pittsburgh U., 2015.
\newblock
\href{http://arxiv.org/abs/1509.06375}{{\tt arXiv:1509.06375 [hep-ph]}}.
\newblock

\bibitem{Marcano:2017ucg}
X.~Marcano~Imaz, {\em {Lepton flavor violation from low scale seesaw neutrinos
  with masses reachable at the LHC}}.
\newblock PhD thesis, U. Autonoma, Madrid (main), Cham, 2017-06.
\newblock \href{http://arxiv.org/abs/1710.08032}{{\tt arXiv:1710.08032
  [hep-ph]}}.
\newblock
\url{https://repositorio.uam.es/handle/10486/681399}.
\newblock

\bibitem{Kersten:2007vk}
J.~Kersten and A.~{\relax Yu}. Smirnov, {\em {Right-Handed Neutrinos at CERN
  LHC and the Mechanism of Neutrino Mass Generation}\/},
  \href{http://dx.doi.org/10.1103/PhysRevD.76.073005}{Phys. Rev. {\bf D76}
  (2007)  073005},
\href{http://arxiv.org/abs/0705.3221}{{\tt arXiv:0705.3221 [hep-ph]}}.

\bibitem{Moffat:2017feq}
K.~Moffat, S.~Pascoli, and C.~Weiland, {\em {Equivalence between massless
  neutrinos and lepton number conservation in fermionic singlet extensions of
  the Standard Model}\/},
\href{http://arxiv.org/abs/1712.07611}{{\tt arXiv:1712.07611 [hep-ph]}}.

\bibitem{Pascoli:2018rsg}
S.~Pascoli, R.~Ruiz, and C.~Weiland, {\em {Safe Jet Vetoes}\/},
  \href{http://dx.doi.org/10.1016/j.physletb.2018.08.060}{Phys. Lett. {\bf
  B786} (2018)  106},
\href{http://arxiv.org/abs/1805.09335}{{\tt arXiv:1805.09335 [hep-ph]}}.

\bibitem{Arganda:2015ija}
E.~Arganda, M.~J. Herrero, X.~Marcano, and C.~Weiland, {\em {Exotic $\mu\tau
  jj$ events from heavy ISS neutrinos at the LHC}\/},
  \href{http://dx.doi.org/10.1016/j.physletb.2015.11.013}{Phys. Lett. {\bf
  B752} (2016)  46--50},
\href{http://arxiv.org/abs/1508.05074}{{\tt arXiv:1508.05074 [hep-ph]}}.

\bibitem{Adam:2013mnn}
{MEG Collaboration}, J.~Adam et al., {\em {New constraint on the existence of
  the $\mu^+ \to e^+\gamma$ decay}\/},
  \href{http://dx.doi.org/10.1103/PhysRevLett.110.201801}{Phys. Rev. Lett. {\bf
  110} (2013)  201801},
\href{http://arxiv.org/abs/1303.0754}{{\tt arXiv:1303.0754 [hep-ex]}}.

\bibitem{Alva:2014gxa}
D.~Alva, T.~Han, and R.~Ruiz, {\em {Heavy Majorana neutrinos from $W\gamma$
  fusion at hadron colliders}\/},
  \href{http://dx.doi.org/10.1007/JHEP02(2015)072}{JHEP {\bf 02} (2015)  072},
\href{http://arxiv.org/abs/1411.7305}{{\tt arXiv:1411.7305 [hep-ph]}}.

\bibitem{Ruiz:2017nip}
R.~Ruiz, {\em {Lepton Number Violation at Colliders from Kinematically
  Inaccessible Gauge Bosons}\/},
  \href{http://dx.doi.org/10.1140/epjc/s10052-017-4950-2}{Eur. Phys. J. {\bf
  C77} (2017) no.~6, 375},
\href{http://arxiv.org/abs/1703.04669}{{\tt arXiv:1703.04669 [hep-ph]}}.

\bibitem{Keung:1983uu}
W.-Y. Keung and G.~Senjanovic, {\em {Majorana Neutrinos and the Production of
  the Right-handed Charged Gauge Boson}\/},
\href{http://dx.doi.org/10.1103/PhysRevLett.50.1427}{Phys. Rev. Lett. {\bf 50}
  (1983)  1427}.

\bibitem{Mitra:2016kov}
M.~Mitra, R.~Ruiz, D.~J. Scott, and M.~Spannowsky, {\em {Neutrino Jets from
  High-Mass $W_R$ Gauge Bosons in TeV-Scale Left-Right Symmetric Models}\/},
  \href{http://dx.doi.org/10.1103/PhysRevD.94.095016}{Phys. Rev. {\bf D94}
  (2016) no.~9, 095016},
\href{http://arxiv.org/abs/1607.03504}{{\tt arXiv:1607.03504 [hep-ph]}}.

\bibitem{Han:2012vk}
T.~Han, I.~Lewis, R.~Ruiz, and Z.-g. Si, {\em {Lepton Number Violation and
  $W^\prime$ Chiral Couplings at the LHC}\/},
  \href{http://dx.doi.org/10.1103/PhysRevD.87.035011,
  10.1103/PhysRevD.87.039906}{Phys. Rev. {\bf D87} (2013) no.~3, 035011},
  \href{http://arxiv.org/abs/1211.6447}{{\tt arXiv:1211.6447 [hep-ph]}}.
[Erratum: Phys. Rev.D87,no.3,039906(2013)].

\bibitem{Ferrari:2000sp}
A.~Ferrari, J.~Collot, M.-L. Andrieux, B.~Belhorma, P.~de~Saintignon, J.-Y.
  Hostachy, P.~Martin, and M.~Wielers, {\em {Sensitivity study for new gauge
  bosons and right-handed Majorana neutrinos in $p p$ collisions at $s$ =
  14-TeV}\/},
\href{http://dx.doi.org/10.1103/PhysRevD.62.013001}{Phys. Rev. {\bf D62} (2000)
   013001}.

\bibitem{Mattelaer:2016ynf}
O.~Mattelaer, M.~Mitra, and R.~Ruiz, {\em {Automated Neutrino Jet and Top Jet
  Predictions at Next-to-Leading-Order with Parton Shower Matching in Effective
  Left-Right Symmetric Models}\/},
\href{http://arxiv.org/abs/1610.08985}{{\tt arXiv:1610.08985 [hep-ph]}}.

\bibitem{Perez:2008ha}
P.~Fileviez~Perez, T.~Han, G.-y. Huang, T.~Li, and K.~Wang, {\em {Neutrino
  Masses and the CERN LHC: Testing Type II Seesaw}\/},
  \href{http://dx.doi.org/10.1103/PhysRevD.78.015018}{Phys. Rev. {\bf D78}
  (2008)  015018},
\href{http://arxiv.org/abs/0805.3536}{{\tt arXiv:0805.3536 [hep-ph]}}.

\bibitem{Li:2018jns}
T.~Li, {\em {Type II Seesaw and tau lepton at the HL-LHC, HE-LHC and
  FCC-hh}\/},  \href{http://dx.doi.org/10.1007/JHEP09(2018)079}{JHEP {\bf 09}
  (2018)  079},
\href{http://arxiv.org/abs/1802.00945}{{\tt arXiv:1802.00945 [hep-ph]}}.

\bibitem{Sugiyama:2012yw}
H.~Sugiyama, K.~Tsumura, and H.~Yokoya, {\em {Discrimination of models
  including doubly charged scalar bosons by using tau lepton decay
  distributions}\/},
  \href{http://dx.doi.org/10.1016/j.physletb.2012.09.044}{Phys. Lett. {\bf
  B717} (2012)  229--234},
\href{http://arxiv.org/abs/1207.0179}{{\tt arXiv:1207.0179 [hep-ph]}}.

\bibitem{Ruiz:2015zca}
R.~Ruiz, {\em {QCD Corrections to Pair Production of Type III Seesaw Leptons at
  Hadron Colliders}\/},  \href{http://dx.doi.org/10.1007/JHEP12(2015)165}{JHEP
  {\bf 12} (2015)  165},
\href{http://arxiv.org/abs/1509.05416}{{\tt arXiv:1509.05416 [hep-ph]}}.

\bibitem{Arhrib:2009mz}
A.~Arhrib, B.~Bajc, D.~K. Ghosh, T.~Han, G.-Y. Huang, I.~Puljak, and
  G.~Senjanovic, {\em {Collider Signatures for Heavy Lepton Triplet in Type
  I+III Seesaw}\/},  \href{http://dx.doi.org/10.1103/PhysRevD.82.053004}{Phys.
  Rev. {\bf D82} (2010)  053004},
\href{http://arxiv.org/abs/0904.2390}{{\tt arXiv:0904.2390 [hep-ph]}}.

\bibitem{Li:2009mw}
T.~Li and X.-G. He, {\em {Neutrino Masses and Heavy Triplet Leptons at the LHC:
  Testability of Type III Seesaw}\/},
  \href{http://dx.doi.org/10.1103/PhysRevD.80.093003}{Phys. Rev. {\bf D80}
  (2009)  093003},
\href{http://arxiv.org/abs/0907.4193}{{\tt arXiv:0907.4193 [hep-ph]}}.

\bibitem{Baldes:2016rqn}
I.~Baldes, T.~Konstandin, and G.~Servant, {\em {A First-Order Electroweak Phase
  Transition in the Standard Model from Varying Yukawas}\/},
\href{http://arxiv.org/abs/1604.04526}{{\tt arXiv:1604.04526 [hep-ph]}}.

\bibitem{Baldes:2016gaf}
I.~Baldes, T.~Konstandin, and G.~Servant, {\em {Flavor Cosmology: Dynamical
  Yukawas in the Froggatt-Nielsen Mechanism}\/},
  \href{http://dx.doi.org/10.1007/JHEP12(2016)073}{JHEP {\bf 12} (2016)  073},
\href{http://arxiv.org/abs/1608.03254}{{\tt arXiv:1608.03254 [hep-ph]}}.

\bibitem{vonHarling:2016vhf}
B.~von Harling and G.~Servant, {\em {Cosmological evolution of Yukawa
  couplings: the 5D perspective}\/},
  \href{http://dx.doi.org/10.1007/JHEP05(2017)077}{JHEP {\bf 05} (2017)  077},
\href{http://arxiv.org/abs/1612.02447}{{\tt arXiv:1612.02447 [hep-ph]}}.

\bibitem{Bruggisser:2017lhc}
S.~Bruggisser, T.~Konstandin, and G.~Servant, {\em {CP-violation for
  Electroweak Baryogenesis from Dynamical CKM Matrix}\/},
  \href{http://dx.doi.org/10.1088/1475-7516/2017/11/034}{JCAP {\bf 1711} (2017)
  no.~11, 034},
\href{http://arxiv.org/abs/1706.08534}{{\tt arXiv:1706.08534 [hep-ph]}}.

\bibitem{Bruggisser:2018mus}
S.~Bruggisser, B.~Von~Harling, O.~Matsedonskyi, and G.~Servant, {\em {The
  Baryon Asymmetry from a Composite Higgs}\/},
  \href{http://dx.doi.org/10.1103/PhysRevLett.121.131801}{Phys. Rev. Lett. {\bf
  121} (2018) no.~13, 131801},
\href{http://arxiv.org/abs/1803.08546}{{\tt arXiv:1803.08546 [hep-ph]}}.

\bibitem{Bruggisser:2018mrt}
S.~Bruggisser, B.~Von~Harling, O.~Matsedonskyi, and G.~Servant, {\em
  {Electroweak Phase Transition and Baryogenesis in Composite Higgs Models}\/},
\href{http://arxiv.org/abs/1804.07314}{{\tt arXiv:1804.07314 [hep-ph]}}.

\bibitem{Baldes:2018nel}
I.~Baldes and G.~Servant, {\em {High Scale Electroweak Phase Transition:
  Baryogenesis \&amp; Symmetry Non-Restoration}\/},
\href{http://arxiv.org/abs/1807.08770}{{\tt arXiv:1807.08770 [hep-ph]}}.

\bibitem{Servant:2018xcs}
G.~Servant, {\em {The serendipity of electroweak baryogenesis}\/},
  \href{http://dx.doi.org/10.1098/rsta.2017.0124}{Phil. Trans. Roy. Soc. Lond.
  {\bf A376} (2018) no.~2114, 20170124},
\href{http://arxiv.org/abs/1807.11507}{{\tt arXiv:1807.11507 [hep-ph]}}.

\bibitem{Espinosa:2011eu}
J.~R. Espinosa, B.~Gripaios, T.~Konstandin, and F.~Riva, {\em {Electroweak
  Baryogenesis in Non-minimal Composite Higgs Models}\/},
  \href{http://dx.doi.org/10.1088/1475-7516/2012/01/012}{JCAP {\bf 1201} (2012)
   012},
\href{http://arxiv.org/abs/1110.2876}{{\tt arXiv:1110.2876 [hep-ph]}}.

\bibitem{Chala:2016ykx}
M.~Chala, G.~Nardini, and I.~Sobolev, {\em {Unified explanation for dark matter
  and electroweak baryogenesis with direct detection and gravitational wave
  signatures}\/},  \href{http://dx.doi.org/10.1103/PhysRevD.94.055006}{Phys.
  Rev. {\bf D94} (2016) no.~5, 055006},
\href{http://arxiv.org/abs/1605.08663}{{\tt arXiv:1605.08663 [hep-ph]}}.

\bibitem{Megias:2018sxv}
E.~Megias, G.~Nardini, and M.~Quiros, {\em {Cosmological Phase Transitions in
  Warped Space: Gravitational Waves and Collider Signatures}\/},
\href{http://arxiv.org/abs/1806.04877}{{\tt arXiv:1806.04877 [hep-ph]}}.

\bibitem{Coradeschi:2013gda}
F.~Coradeschi, P.~Lodone, D.~Pappadopulo, R.~Rattazzi, and L.~Vitale, {\em {A
  naturally light dilaton}\/},
  \href{http://dx.doi.org/10.1007/JHEP11(2013)057}{JHEP {\bf 11} (2013)  057},
\href{http://arxiv.org/abs/1306.4601}{{\tt arXiv:1306.4601 [hep-th]}}.

\bibitem{Kumar:2013qya}
K.~Kumar, Z.-T. Lu, and M.~J. Ramsey-Musolf, {\em {Working Group Report:
  Nucleons, Nuclei, and Atoms}\/},  in {\em {Fundamental Physics at the
  Intensity Frontier}}, pp.~159--214.
\newblock 2013.
\newblock \href{http://arxiv.org/abs/1312.5416}{{\tt arXiv:1312.5416
  [hep-ph]}}.
\newblock
\url{https://inspirehep.net/record/1272872/files/arXiv:1312.5416.pdf}.
\newblock

\bibitem{Matsedonskyi:2012ym}
O.~Matsedonskyi, G.~Panico, and A.~Wulzer, {\em {Light Top Partners for a Light
  Composite Higgs}\/},  \href{http://dx.doi.org/10.1007/JHEP01(2013)164}{JHEP
  {\bf 01} (2013)  164},
\href{http://arxiv.org/abs/1204.6333}{{\tt arXiv:1204.6333 [hep-ph]}}.

\bibitem{Panico:2012uw}
G.~Panico, M.~Redi, A.~Tesi, and A.~Wulzer, {\em {On the Tuning and the Mass of
  the Composite Higgs}\/},
  \href{http://dx.doi.org/10.1007/JHEP03(2013)051}{JHEP {\bf 03} (2013)  051},
\href{http://arxiv.org/abs/1210.7114}{{\tt arXiv:1210.7114 [hep-ph]}}.

\bibitem{Caprini:2015zlo}
C.~Caprini et al., {\em {Science with the space-based interferometer eLISA. II:
  Gravitational waves from cosmological phase transitions}\/},
  \href{http://dx.doi.org/10.1088/1475-7516/2016/04/001}{JCAP {\bf 1604} (2016)
  no.~04, 001},
\href{http://arxiv.org/abs/1512.06239}{{\tt arXiv:1512.06239 [astro-ph.CO]}}.

\bibitem{DiLuzio:2017chi}
L.~Di~Luzio and M.~Nardecchia, {\em {What is the scale of new physics behind
  the $B$-flavour anomalies?}\/},
  \href{http://dx.doi.org/10.1140/epjc/s10052-017-5118-9}{Eur. Phys. J. {\bf
  C77} (2017) no.~8, 536},
\href{http://arxiv.org/abs/1706.01868}{{\tt arXiv:1706.01868 [hep-ph]}}.

\bibitem{Aaboud:2016cre}
{ATLAS Collaboration}, M.~Aaboud et al., {\em {Search for Minimal
  Supersymmetric Standard Model Higgs bosons $H/A$ and for a $Z^{\prime}$ boson
  in the $\tau \tau$ final state produced in $pp$ collisions at $\sqrt{s}=13$
  TeV with the ATLAS Detector}\/},
  \href{http://dx.doi.org/10.1140/epjc/s10052-016-4400-6}{Eur. Phys. J. {\bf
  C76} (2016) no.~11, 585},
\href{http://arxiv.org/abs/1608.00890}{{\tt arXiv:1608.00890 [hep-ex]}}.

\bibitem{Greljo:2017vvb}
A.~Greljo and D.~Marzocca, {\em {High-$p_T$ dilepton tails and flavor
  physics}\/},  \href{http://dx.doi.org/10.1140/epjc/s10052-017-5119-8}{Eur.
  Phys. J. {\bf C77} (2017) no.~8, 548},
\href{http://arxiv.org/abs/1704.09015}{{\tt arXiv:1704.09015 [hep-ph]}}.

\bibitem{ATLAS:2017wce}
{ATLAS Collaboration}, T.~A. collaboration,
{\em {Search for new high-mass phenomena in the dilepton final state using 36.1
  fb$^{-1}$ of proton-proton collision data at $\sqrt{s} =$ 13 TeV with the
  ATLAS detector}\/}, .

\bibitem{Aaboud:2017buh}
{ATLAS Collaboration}, M.~Aaboud et al., {\em {Search for new high-mass
  phenomena in the dilepton final state using 36 fb$^{-1}$ of proton-proton
  collision data at $ \sqrt{s}=13 $ TeV with the ATLAS detector}\/},
  \href{http://dx.doi.org/10.1007/JHEP10(2017)182}{JHEP {\bf 10} (2017)  182},
\href{http://arxiv.org/abs/1707.02424}{{\tt arXiv:1707.02424 [hep-ex]}}.

\bibitem{Afik:2018nlr}
Y.~Afik, J.~Cohen, E.~Gozani, E.~Kajomovitz, and Y.~Rozen, {\em {Establishing a
  Search for $b \rightarrow s \ell^{+} \ell^{-}$ Anomalies at the LHC}\/},
\href{http://arxiv.org/abs/1805.11402}{{\tt arXiv:1805.11402 [hep-ph]}}.

\bibitem{Abdullah:2018ets}
M.~Abdullah, J.~Calle, B.~Dutta, A.~Florez, and D.~Restrepo, {\em {Probing a
  simplified, $W^{\prime}$ model of $R(D^{(\ast)})$ anomalies using $b$-tags,
  $\tau$ leptons and missing energy}\/},
\href{http://arxiv.org/abs/1805.01869}{{\tt arXiv:1805.01869 [hep-ph]}}.

\bibitem{Dorsner:2018ynv}
I.~Dorsner and A.~Greljo, {\em {Leptoquark toolbox for precision collider
  studies}\/},  \href{http://dx.doi.org/10.1007/JHEP05(2018)126}{JHEP {\bf 05}
  (2018)  126},
\href{http://arxiv.org/abs/1801.07641}{{\tt arXiv:1801.07641 [hep-ph]}}.

\bibitem{DiLuzio:2018zxy}
L.~Di~Luzio, J.~Fuentes-Martin, A.~Greljo, M.~Nardecchia, and S.~Renner, {\em
  {Maximal Flavour Violation: a Cabibbo mechanism for leptoquarks}\/},
  \href{http://dx.doi.org/10.1007/JHEP11(2018)081}{JHEP {\bf 11} (2018)  081},
\href{http://arxiv.org/abs/1808.00942}{{\tt arXiv:1808.00942 [hep-ph]}}.

\bibitem{Kramer:2004df}
M.~Kramer, T.~Plehn, M.~Spira, and P.~M. Zerwas, {\em {Pair production of
  scalar leptoquarks at the CERN LHC}\/},
  \href{http://dx.doi.org/10.1103/PhysRevD.71.057503}{Phys. Rev. {\bf D71}
  (2005)  057503},
\href{http://arxiv.org/abs/hep-ph/0411038}{{\tt arXiv:hep-ph/0411038
  [hep-ph]}}.

\bibitem{Allanach:2017bta}
B.~C. Allanach, B.~Gripaios, and T.~You, {\em {The case for future hadron
  colliders from $B \to K^{(*)} \mu^+ \mu^-$ decays}\/},
  \href{http://dx.doi.org/10.1007/JHEP03(2018)021}{JHEP {\bf 03} (2018)  021},
\href{http://arxiv.org/abs/1710.06363}{{\tt arXiv:1710.06363 [hep-ph]}}.

\bibitem{Vignaroli:2018lpq}
N.~Vignaroli, {\em {Seeking LQs in the $\bf t\bar{t}$ plus missing energy
  channel at the high-luminosity LHC}\/},
\href{http://arxiv.org/abs/1808.10309}{{\tt arXiv:1808.10309 [hep-ph]}}.

\bibitem{Bazavov:2016nty}
{Fermilab Lattice, MILC Collaboration}, A.~Bazavov et al., {\em
  {$B^0_{(s)}$-mixing matrix elements from lattice QCD for the Standard Model
  and beyond}\/},  Phys. Rev. {\bf D93} (2016)  113016,
  \href{http://arxiv.org/abs/1602.03560}{{\tt arXiv:1602.03560 [hep-lat]}}.

\bibitem{Allanach:2018odd}
B.~C. Allanach, T.~Corbett, M.~J. Dolan, and T.~You, {\em {Hadron Collider
  Sensitivity to Fat Flavourful $Z^\prime$s for $R_{K^{(\ast)}}$}\/},
\href{http://arxiv.org/abs/1810.02166}{{\tt arXiv:1810.02166 [hep-ph]}}.

\bibitem{Falkowski:2017pss}
A.~Falkowski, M.~Gonzalez-Alonso, and K.~Mimouni, {\em {Compilation of
  low-energy constraints on 4-fermion operators in the SMEFT}\/},
  \href{http://dx.doi.org/10.1007/JHEP08(2017)123}{JHEP {\bf 08} (2017)  123},
\href{http://arxiv.org/abs/1706.03783}{{\tt arXiv:1706.03783 [hep-ph]}}.

\bibitem{Falkowski:2018dsl}
A.~Falkowski, S.~F. King, E.~Perdomo, and M.~Pierre, {\em {Flavourful $Z'$
  portal for vector-like neutrino Dark Matter and $R_{K^{(*)}}$}\/},
  \href{http://dx.doi.org/10.1007/JHEP08(2018)061}{JHEP {\bf 08} (2018)  061},
\href{http://arxiv.org/abs/1803.04430}{{\tt arXiv:1803.04430 [hep-ph]}}.

\bibitem{Celis:2017doq}
A.~Celis, J.~Fuentes-Martin, A.~Vicente, and J.~Virto, {\em {Gauge-invariant
  implications of the LHCb measurements on lepton-flavor nonuniversality}\/},
  \href{http://dx.doi.org/10.1103/PhysRevD.96.035026}{Phys. Rev. {\bf D96}
  (2017) no.~3, 035026},
\href{http://arxiv.org/abs/1704.05672}{{\tt arXiv:1704.05672 [hep-ph]}}.

\bibitem{Fox:2018ldq}
P.~J. Fox, I.~Low, and Y.~Zhang, {\em {Top-philic $Z'$ Forces at the LHC}\/},
  \href{http://dx.doi.org/10.1007/JHEP03(2018)074}{JHEP {\bf 03} (2018)  074},
\href{http://arxiv.org/abs/1801.03505}{{\tt arXiv:1801.03505 [hep-ph]}}.

\bibitem{Camargo-Molina:2018cwu}
J.~E. Camargo-Molina, A.~Celis, and D.~A. Faroughy, {\em {Anomalies in Bottom
  from new physics in Top}\/},
\href{http://arxiv.org/abs/1805.04917}{{\tt arXiv:1805.04917 [hep-ph]}}.

\bibitem{CMS-PAS-FTR-18-008}
{CMS Collaboration}, {\em Projection of searches for pair production of scalar
  leptoquarks decaying to a top quark and a charged lepton at the {HL-LHC}\/},
   CMS-PAS-FTR-18-008, {CERN}, Geneva, 2018.
\newblock \url{https://cds.cern.ch/record/2645611}.

\bibitem{Sirunyan:2018ruf}
{CMS Collaboration}, A.~M. Sirunyan et al., {\em Search for leptoquarks coupled
  to third-generation quarks in proton-proton collisions at $\sqrt{s}=$ 13
  {TeV}\/},  Phys. Rev. Lett. {\bf 121} (2018)  241802,
\href{http://arxiv.org/abs/1809.05558}{{\tt arXiv:1809.05558 [hep-ex]}}.

\bibitem{Sirunyan:2018nkj}
{{CMS} Collaboration}, A.~M. Sirunyan et al., {\em Search for third-generation
  scalar leptoquarks decaying to a top quark and a $\tau$ lepton at
  $\sqrt{s}=13$ {TeV}\/},
  \href{http://dx.doi.org/10.1140/epjc/s10052-018-6143-z}{Eur. Phys. J. C {\bf
  78} (2018)  707},
\href{http://arxiv.org/abs/1803.02864}{{\tt arXiv:1803.02864 [hep-ex]}}.

\bibitem{CMS-PAS-FTR-18-028}
{CMS Collaboration}, {\em {Prospects for exclusion or discovery of a third
  generation leptoquark decaying into a $\tau$ lepton and a $b$ quark at
  CMS}\/},   CMS-PAS-FTR-18-028, CERN, Geneva, 2018.
\newblock \url{http://cds.cern.ch/record/2652363}.

\bibitem{Plehn:1997az}
T.~Plehn, H.~Spiesberger, M.~Spira, and P.~M. Zerwas, {\em {Formation and decay
  of scalar leptoquarks / squarks in ep collisions}\/},
  \href{http://dx.doi.org/10.1007/s002880050426}{Z. Phys. C {\bf 74} (1997)
  611}, \href{http://arxiv.org/abs/hep-ph/9703433}{{\tt arXiv:hep-ph/9703433
  [hep-ph]}}.

\bibitem{Cacciari:2011ma}
M.~Cacciari, G.~P. Salam, and G.~Soyez, {\em {FastJet User Manual}\/},
  \href{http://dx.doi.org/10.1140/epjc/s10052-012-1896-2}{Eur. Phys. J. {\bf
  C72} (2012)  1896},
\href{http://arxiv.org/abs/1111.6097}{{\tt arXiv:1111.6097 [hep-ph]}}.

\bibitem{Cacciari:2008gp}
M.~Cacciari, G.~P. Salam, and G.~Soyez, {\em {The Anti-k(t) jet clustering
  algorithm}\/},  \href{http://dx.doi.org/10.1088/1126-6708/2008/04/063}{JHEP
  {\bf 04} (2008)  063},
\href{http://arxiv.org/abs/0802.1189}{{\tt arXiv:0802.1189 [hep-ph]}}.

\bibitem{Cowan:2010js}
G.~Cowan, K.~Cranmer, E.~Gross, and O.~Vitells, {\em {Asymptotic formulae for
  likelihood-based tests of new physics}\/},
  \href{http://dx.doi.org/10.1140/epjc/s10052-011-1554-0,
  10.1140/epjc/s10052-013-2501-z}{Eur. Phys. J. {\bf C71} (2011)  1554},
  \href{http://arxiv.org/abs/1007.1727}{{\tt arXiv:1007.1727
  [physics.data-an]}}.
[Erratum: Eur. Phys. J.C73,2501(2013)].

\bibitem{Junk:1999kv}
T.~Junk, {\em {Confidence level computation for combining searches with small
  statistics}\/},  \href{http://dx.doi.org/10.1016/S0168-9002(99)00498-2}{Nucl.
  Instrum. Meth. {\bf A434} (1999)  435--443},
\href{http://arxiv.org/abs/hep-ex/9902006}{{\tt arXiv:hep-ex/9902006
  [hep-ex]}}.

\bibitem{Read:2002hq}
A.~L. Read, {\em {Presentation of search results: The CL(s) technique}\/},
  \href{http://dx.doi.org/10.1088/0954-3899/28/10/313}{J. Phys. {\bf G28}
  (2002)  2693--2704}.
[,11(2002)].

\bibitem{ATLAS:2011tau}
{ATLAS, CMS, LHC Higgs Combination Group Collaboration},
{\em {Procedure for the LHC Higgs boson search combination in summer 2011}\/},
  .

\bibitem{CMS-PAS-FTR-18-030}
{CMS Collaboration}, {\em Sensitivity study for $W’ \to \tau \nu$ at the
  HL-LHC\/},  CMS Physics Analysis Summary CMS-PAS-FTR-18-030, 2018.
\newblock \url{http://cdsweb.cern.ch/record/2655312}.

\bibitem{Sirunyan:2018lbg}
{CMS Collaboration}, A.~M. Sirunyan et al., {\em {Search for a W' boson
  decaying to a $\tau$ lepton and a neutrino in proton-proton collisions at
  $\sqrt{s} =$ 13 TeV}\/},  Submitted to: Phys. Lett. (2018)  ,
\href{http://arxiv.org/abs/1807.11421}{{\tt arXiv:1807.11421 [hep-ex]}}.

\bibitem{Altarelli:1989ff}
G.~Altarelli, B.~Mele, and M.~Ruiz-Altaba, {\em {Searching for New Heavy Vector
  Bosons in $p \bar{p}$ Colliders}\/},
  \href{http://dx.doi.org/10.1007/BF01552335, 10.1007/BF01556677}{Z. Phys. {\bf
  C45} (1989)  109}. [Erratum: Z. Phys.C47,676(1990)].

\bibitem{Sirunyan:2017yrk}
{CMS Collaboration}, A.~M. Sirunyan et al., {\em {Search for third-generation
  scalar leptoquarks and heavy right-handed neutrinos in final states with two
  tau leptons and two jets in proton-proton collisions at $ \sqrt{s}=13 $
  TeV}\/},  \href{http://dx.doi.org/10.1007/JHEP07(2017)121}{JHEP {\bf 07}
  (2017)  121},
\href{http://arxiv.org/abs/1703.03995}{{\tt arXiv:1703.03995 [hep-ex]}}.

\bibitem{Carrasco:2015xwa}
N.~Carrasco, V.~Lubicz, G.~Martinelli, C.~T. Sachrajda, N.~Tantalo,
  C.~Tarantino, and M.~Testa, {\em {QED Corrections to Hadronic Processes in
  Lattice QCD}\/},  \href{http://dx.doi.org/10.1103/PhysRevD.91.074506}{Phys.
  Rev. {\bf D91} (2015) no.~7, 074506},
\href{http://arxiv.org/abs/1502.00257}{{\tt arXiv:1502.00257 [hep-lat]}}.

\bibitem{Patella:2017fgk}
A.~Patella, {\em {QED Corrections to Hadronic Observables}\/},  PoS {\bf
  LATTICE2016} (2017)  020,
\href{http://arxiv.org/abs/1702.03857}{{\tt arXiv:1702.03857 [hep-lat]}}.

\bibitem{DellaMorte:2017dyu}
M.~Della~Morte, A.~Francis, V.~Guelpers, G.~Herdoi�za, G.~von Hippel,
  H.~Horch, B.~Jaeger, H.~B. Meyer, A.~Nyffeler, and H.~Wittig, {\em {The
  hadronic vacuum polarization contribution to the muon $g-2$ from lattice
  QCD}\/},  \href{http://dx.doi.org/10.1007/JHEP10(2017)020}{JHEP {\bf 10}
  (2017)  020},
\href{http://arxiv.org/abs/1705.01775}{{\tt arXiv:1705.01775 [hep-lat]}}.

\bibitem{Giusti:2018guw}
D.~Giusti, V.~Lubicz, G.~Martinelli, C.~Sachrajda, F.~Sanfilippo, S.~Simula,
  and N.~Tantalo, {\em {Radiative corrections to decay amplitudes in lattice
  QCD}\/},  in {\em {36th International Symposium on Lattice Field Theory
  (Lattice 2018) East Lansing, MI, United States, July 22-28, 2018}}.
\newblock 2018.
\newblock
\href{http://arxiv.org/abs/1811.06364}{{\tt arXiv:1811.06364 [hep-lat]}}.
\newblock

\bibitem{Dimopoulos:2007ht}
{ALPHA Collaboration}, P.~Dimopoulos, G.~Herdoiza, F.~Palombi, M.~Papinutto,
  C.~Pena, A.~Vladikas, and H.~Wittig, {\em {Non-perturbative renormalisation
  of Delta F=2 four-fermion operators in two-flavour QCD}\/},
  \href{http://dx.doi.org/10.1088/1126-6708/2008/05/065}{JHEP {\bf 05} (2008)
  065},
\href{http://arxiv.org/abs/0712.2429}{{\tt arXiv:0712.2429 [hep-lat]}}.

\bibitem{Durr:2011ap}
S.~Durr et al., {\em {Precision computation of the kaon bag parameter}\/},
  \href{http://dx.doi.org/10.1016/j.physletb.2011.10.043}{Phys. Lett. {\bf
  B705} (2011)  477--481},
\href{http://arxiv.org/abs/1106.3230}{{\tt arXiv:1106.3230 [hep-lat]}}.

\bibitem{Christ:2012se}
{RBC, UKQCD Collaboration}, N.~H. Christ, T.~Izubuchi, C.~T. Sachrajda,
  A.~Soni, and J.~Yu, {\em {Long distance contribution to the KL-KS mass
  difference}\/},  \href{http://dx.doi.org/10.1103/PhysRevD.88.014508}{Phys.
  Rev. {\bf D88} (2013)  014508},
\href{http://arxiv.org/abs/1212.5931}{{\tt arXiv:1212.5931 [hep-lat]}}.

\bibitem{Bai:2016gzv}
Z.~Bai, {\em {Long distance part of $\epsilon_K$ from lattice QCD}\/},  PoS
  {\bf LATTICE2016} (2017)  309,
\href{http://arxiv.org/abs/1611.06601}{{\tt arXiv:1611.06601 [hep-lat]}}.

\bibitem{Bailey:2018feb}
J.~A. Bailey, S.~Lee, W.~Lee, J.~Leem, and S.~Park, {\em {Updated evaluation of
  $\epsilon_K$ in the standard model with lattice QCD inputs}\/},
  \href{http://dx.doi.org/10.1103/PhysRevD.98.094505}{Phys. Rev. {\bf D98}
  (2018) no.~9, 094505},
\href{http://arxiv.org/abs/1808.09657}{{\tt arXiv:1808.09657 [hep-lat]}}.

\bibitem{Davies:2017jbi}
C.~Davies, J.~Harrison, G.~P. Lepage, C.~Monahan, J.~Shigemitsu, and
  M.~Wingate, {\em {Improving the theoretical prediction for the
  $B_s-\bar{B}_s$ width difference: matrix elements of next-to-leading order
  $\Delta B=2$ operators}\/},
  \href{http://dx.doi.org/10.1051/epjconf/201817513023}{EPJ Web Conf. {\bf 175}
  (2018)  13023},
\href{http://arxiv.org/abs/1712.09934}{{\tt arXiv:1712.09934 [hep-lat]}}.

\bibitem{Bazavov:2018kjg}
A.~Bazavov et al., {\em {$|V_{us}|$ from $K_{\ell 3}$ decay and four-flavor
  lattice QCD}\/},
\href{http://arxiv.org/abs/1809.02827}{{\tt arXiv:1809.02827 [hep-lat]}}.

\bibitem{Moulson:2017ive}
M.~Moulson, {\em {Experimental determination of $V_{us}$ from kaon decays}\/},
  PoS {\bf CKM2016} (2017)  033,
\href{http://arxiv.org/abs/1704.04104}{{\tt arXiv:1704.04104 [hep-ex]}}.

\bibitem{Carrasco:2016kpy}
N.~Carrasco, P.~Lami, V.~Lubicz, L.~Riggio, S.~Simula, and C.~Tarantino, {\em
  {$K \to \pi$ semileptonic form factors with $N_f=2+1+1$ twisted mass
  fermions}\/},  \href{http://dx.doi.org/10.1103/PhysRevD.93.114512}{Phys. Rev.
  {\bf D93} (2016) no.~11, 114512},
\href{http://arxiv.org/abs/1602.04113}{{\tt arXiv:1602.04113 [hep-lat]}}.

\bibitem{Briceno:2014uqa}
R.~A. Briceño, M.~T. Hansen, and A.~Walker-Loud, {\em {Multichannel 1
  $\rightarrow$ 2 transition amplitudes in a finite volume}\/},
  \href{http://dx.doi.org/10.1103/PhysRevD.91.034501}{Phys. Rev. {\bf D91}
  (2015) no.~3, 034501},
\href{http://arxiv.org/abs/1406.5965}{{\tt arXiv:1406.5965 [hep-lat]}}.

\bibitem{Lattice:2015tia}
{Fermilab Lattice, MILC Collaboration}, J.~A. Bailey et al., {\em {$|V_{ub}|$
  from $B\to\pi\ell\nu$ decays and (2+1)-flavor lattice QCD}\/},
  \href{http://dx.doi.org/10.1103/PhysRevD.92.014024}{Phys. Rev. {\bf D92}
  (2015) no.~1, 014024},
\href{http://arxiv.org/abs/1503.07839}{{\tt arXiv:1503.07839 [hep-lat]}}.

\bibitem{Bouchard:2013pna}
{HPQCD Collaboration}, C.~Bouchard, G.~P. Lepage, C.~Monahan, H.~Na, and
  J.~Shigemitsu, {\em {Rare decay $B \to K \ell^+ \ell^-$ form factors from
  lattice QCD}\/},  \href{http://dx.doi.org/10.1103/PhysRevD.88.079901,
  10.1103/PhysRevD.88.054509}{Phys. Rev. {\bf D88} (2013) no.~5, 054509},
  \href{http://arxiv.org/abs/1306.2384}{{\tt arXiv:1306.2384 [hep-lat]}}.
[Erratum: Phys. Rev.D88,no.7,079901(2013)].

\bibitem{Bailey:2015dka}
J.~A. Bailey et al., {\em {$B\to Kl^+l^-$ decay form factors from three-flavor
  lattice QCD}\/},  \href{http://dx.doi.org/10.1103/PhysRevD.93.025026}{Phys.
  Rev. {\bf D93} (2016) no.~2, 025026},
\href{http://arxiv.org/abs/1509.06235}{{\tt arXiv:1509.06235 [hep-lat]}}.

\bibitem{Bouchard:2014ypa}
C.~M. Bouchard, G.~P. Lepage, C.~Monahan, H.~Na, and J.~Shigemitsu, {\em {$B_s
  \to K \ell \nu$ form factors from lattice QCD}\/},
  \href{http://dx.doi.org/10.1103/PhysRevD.90.054506}{Phys. Rev. {\bf D90}
  (2014)  054506},
\href{http://arxiv.org/abs/1406.2279}{{\tt arXiv:1406.2279 [hep-lat]}}.

\bibitem{Monahan:2018lzv}
C.~J. Monahan, C.~M. Bouchard, G.~P. Lepage, H.~Na, and J.~Shigemitsu, {\em
  {Form factor ratios for $B_s \rightarrow K \, \ell \, \nu$ and $B_s
  \rightarrow D_s \, \ell \, \nu$ semileptonic decays and
  $|V_{ub}/V_{cb}|$}\/},
\href{http://arxiv.org/abs/1808.09285}{{\tt arXiv:1808.09285 [hep-lat]}}.

\bibitem{Horgan:2013hoa}
R.~R. Horgan, Z.~Liu, S.~Meinel, and M.~Wingate, {\em {Lattice QCD calculation
  of form factors describing the rare decays $B \to K^* \ell^+ \ell^-$ and $B_s
  \to \phi \ell^+ \ell^-$}\/},
  \href{http://dx.doi.org/10.1103/PhysRevD.89.094501}{Phys. Rev. {\bf D89}
  (2014) no.~9, 094501},
\href{http://arxiv.org/abs/1310.3722}{{\tt arXiv:1310.3722 [hep-lat]}}.

\bibitem{Horgan:2013pva}
R.~R. Horgan, Z.~Liu, S.~Meinel, and M.~Wingate, {\em {Calculation of $B^0 \to
  K^{*0} \mu^+ \mu^-$ and $B_s^0 \to \phi \mu^+ \mu^-$ observables using form
  factors from lattice QCD}\/},
  \href{http://dx.doi.org/10.1103/PhysRevLett.112.212003}{Phys. Rev. Lett. {\bf
  112} (2014)  212003},
\href{http://arxiv.org/abs/1310.3887}{{\tt arXiv:1310.3887 [hep-ph]}}.

\bibitem{Luscher:1986pf}
M.~Luscher, {\em {Volume Dependence of the Energy Spectrum in Massive Quantum
  Field Theories. 2. Scattering States}\/},
\href{http://dx.doi.org/10.1007/BF01211097}{Commun. Math. Phys. {\bf 105}
  (1986)  153--188}.

\bibitem{Luscher:1991cf}
M.~Luscher, {\em {Signatures of unstable particles in finite volume}\/},
\href{http://dx.doi.org/10.1016/0550-3213(91)90584-K}{Nucl. Phys. {\bf B364}
  (1991)  237--251}.

\bibitem{Lage:2009zv}
M.~Lage, U.-G. Meissner, and A.~Rusetsky, {\em {A Method to measure the
  antikaon-nucleon scattering length in lattice QCD}\/},
  \href{http://dx.doi.org/10.1016/j.physletb.2009.10.055}{Phys. Lett. {\bf
  B681} (2009)  439--443},
\href{http://arxiv.org/abs/0905.0069}{{\tt arXiv:0905.0069 [hep-lat]}}.

\bibitem{Bernard:2010fp}
V.~Bernard, M.~Lage, U.~G. Meissner, and A.~Rusetsky, {\em {Scalar mesons in a
  finite volume}\/},  \href{http://dx.doi.org/10.1007/JHEP01(2011)019}{JHEP
  {\bf 01} (2011)  019},
\href{http://arxiv.org/abs/1010.6018}{{\tt arXiv:1010.6018 [hep-lat]}}.

\bibitem{Doring:2011vk}
M.~Doring, U.-G. Meissner, E.~Oset, and A.~Rusetsky, {\em {Unitarized Chiral
  Perturbation Theory in a finite volume: Scalar meson sector}\/},
  \href{http://dx.doi.org/10.1140/epja/i2011-11139-7}{Eur. Phys. J. {\bf A47}
  (2011)  139},
\href{http://arxiv.org/abs/1107.3988}{{\tt arXiv:1107.3988 [hep-lat]}}.

\bibitem{Briceno:2012yi}
R.~A. Briceno and Z.~Davoudi, {\em {Moving multichannel systems in a finite
  volume with application to proton-proton fusion}\/},
  \href{http://dx.doi.org/10.1103/PhysRevD.88.094507}{Phys. Rev. {\bf D88}
  (2013) no.~9, 094507},
\href{http://arxiv.org/abs/1204.1110}{{\tt arXiv:1204.1110 [hep-lat]}}.

\bibitem{Dudek:2014qha}
{Hadron Spectrum Collaboration}, J.~J. Dudek, R.~G. Edwards, C.~E. Thomas, and
  D.~J. Wilson, {\em {Resonances in coupled $\pi K -\eta K$ scattering from
  quantum chromodynamics}\/},
  \href{http://dx.doi.org/10.1103/PhysRevLett.113.182001}{Phys. Rev. Lett. {\bf
  113} (2014) no.~18, 182001},
\href{http://arxiv.org/abs/1406.4158}{{\tt arXiv:1406.4158 [hep-ph]}}.

\bibitem{Meinel:2018MITP}
S.~Meinel, {\em $\Lambda_b \to \Lambda_c^{(*)}$ form factors from lattice
  QCD\/},
\newblock
  \url{https://indico.mitp.uni-mainz.de/event/129/session/0/contribution/2/material/slides/0.pdf}.
  Talk at the workshop "Challenges in Semileptonic B Decays" held 9-13 April
  2018 at the Mainz Institute for Theoretical Physics, Johannes Gutenberg
  University.

\bibitem{Christ:2015aha}
{RBC, UKQCD Collaboration}, N.~H. Christ, X.~Feng, A.~Portelli, and C.~T.
  Sachrajda, {\em {Prospects for a lattice computation of rare kaon decay
  amplitudes: $K\to\pi\ell^+\ell^-$ decays}\/},
  \href{http://dx.doi.org/10.1103/PhysRevD.92.094512}{Phys. Rev. {\bf D92}
  (2015) no.~9, 094512},
\href{http://arxiv.org/abs/1507.03094}{{\tt arXiv:1507.03094 [hep-lat]}}.

\bibitem{Christ:2016mmq}
N.~H. Christ, X.~Feng, A.~Juttner, A.~Lawson, A.~Portelli, and C.~T. Sachrajda,
  {\em {First exploratory calculation of the long-distance contributions to the
  rare kaon decays $K\to\pi\ell^+\ell^-$}\/},
  \href{http://dx.doi.org/10.1103/PhysRevD.94.114516}{Phys. Rev. {\bf D94}
  (2016) no.~11, 114516},
\href{http://arxiv.org/abs/1608.07585}{{\tt arXiv:1608.07585 [hep-lat]}}.

\bibitem{fastjet}
M.~Cacciari, G.~P. Salam, and G.~Soyez, {\em {FastJet User Manual}\/},
  \href{http://dx.doi.org/10.1140/epjc/s10052-012-1896-2}{Eur. Phys. J. {\bf
  C72} (2012)  1896},
\href{http://arxiv.org/abs/1111.6097}{{\tt arXiv:1111.6097 [hep-ph]}}.

\bibitem{ATLAS_PERF_Note}
{ATLAS Collaboration}, {ATLAS Collaboration}, {\em {Expected performance of the
  ATLAS detector at HL-LHC}\/},   in progress, CERN, Geneva, Dec, 2018.

\bibitem{ATLAS_TDAQ_TDR}
A.~Collaboration, {\em {Technical Design Report for the Phase-II Upgrade of the
  ATLAS TDAQ System}\/},   CERN-LHCC-2017-020. ATLAS-TDR-029, CERN, Geneva,
  Sep, 2017.
\newblock \url{http://cds.cern.ch/record/2285584}.

\bibitem{CMSL1interim}
{CMS Collaboration}, {\em {The Phase-2 Upgrade of the CMS L1 Trigger Interim
  Technical Design Report}\/},   CERN-LHCC-2017-013. CMS-TDR-017, CERN, Geneva,
  Sep, 2017.
\newblock \url{https://cds.cern.ch/record/2283192}.
\newblock This is the CMS Interim TDR devoted to the upgrade of the CMS L1
  trigger in view of the HL-LHC running, as approved by the LHCC.

\bibitem{ATLAS_Pixel_TDR}
A.~Collaboration, {\em {Technical Design Report for the ATLAS Inner Tracker
  Pixel Detector}\/},   CERN-LHCC-2017-021. ATLAS-TDR-030, CERN, Geneva, Sep,
  2017.
\newblock \url{http://cds.cern.ch/record/2285585}.

\bibitem{ATLAS_Strip_TDR}
A.~Collaboration, {\em {Technical Design Report for the ATLAS Inner Tracker
  Strip Detector}\/},   CERN-LHCC-2017-005. ATLAS-TDR-025, CERN, Geneva, Apr,
  2017.
\newblock \url{http://cds.cern.ch/record/2257755}.

\bibitem{ATLAS_LAr_TDR}
A.~Collaboration, {\em {Technical Design Report for the Phase-II Upgrade of the
  ATLAS LAr Calorimeter}\/},   CERN-LHCC-2017-018. ATLAS-TDR-027, CERN, Geneva,
  Sep, 2017.
\newblock \url{http://cds.cern.ch/record/2285582}.

\bibitem{CMS_Barrel_TDR}
{CMS Collaboration}, {\em {The Phase-2 Upgrade of the CMS Barrel Calorimeters
  Technical Design Report}\/},   CERN-LHCC-2017-011. CMS-TDR-015, CERN, 2017.
\newblock \url{https://cds.cern.ch/record/2283187}.

\bibitem{CMS_HGCAL_TDR}
{CMS Collaboration}, {\em {The Phase-2 Upgrade of the CMS Endcap
  Calorimeter}\/},   CERN-LHCC-2017-023. CMS-TDR-019, 2017.
\newblock \url{https://cds.cern.ch/record/2293646}.

\bibitem{ATLAS_Muon_TDR}
A.~Collaboration, {\em {Technical Design Report for the Phase-II Upgrade of the
  ATLAS Muon Spectrometer}\/},   CERN-LHCC-2017-017. ATLAS-TDR-026, CERN,
  Geneva, Sep, 2017.
\newblock \url{http://cds.cern.ch/record/2285580}.

\bibitem{ATLAS_Tile_TDR}
A.~Collaboration, {\em {Technical Design Report for the Phase-II Upgrade of the
  ATLAS Tile Calorimeter}\/},   CERN-LHCC-2017-019. ATLAS-TDR-028, CERN,
  Geneva, Sep, 2017.
\newblock \url{http://cds.cern.ch/record/2285583}.

\bibitem{Contardo:2020886}
{CMS Collaboration}, D.~Contardo, M.~Klute, J.~Mans, L.~Silvestris, and
  J.~Butler, {\em {Technical Proposal for the Phase-II Upgrade of the CMS
  Detector}\/},   CERN-LHCC-2015-010. LHCC-P-008. CMS-TDR-15-02, 2015.
\newblock \url{https://cds.cern.ch/record/2020886}.

\bibitem{ATLAS_TP_HGTD}
A.~Collaboration, {\em {Technical Proposal: A High-Granularity Timing Detector
  for the ATLAS Phase-II Upgrade}\/},   CERN-LHCC-2018-023. LHCC-P-012, CERN,
  Geneva, Jun, 2018.
\newblock \url{http://cds.cern.ch/record/2623663}.

\bibitem{Sjostrand:2007gs}
T.~Sjostrand, S.~Mrenna, and P.~Z. Skands, {\em {A Brief Introduction to PYTHIA
  8.1}\/},  \href{http://dx.doi.org/10.1016/j.cpc.2008.01.036}{Comput. Phys.
  Commun. {\bf 178} (2008)  852--867},
\href{http://arxiv.org/abs/0710.3820}{{\tt arXiv:0710.3820 [hep-ph]}}.

\bibitem{LHCb-PROC-2010-056}
I.~Belyaev et al., {\em {Handling of the generation of primary events in Gauss,
  the LHCb simulation framework}\/},
  \href{http://dx.doi.org/10.1088/1742-6596/331/3/032047}{J. Phys. Conf. Ser.
  {\bf 331} (2011)  032047}.

\bibitem{cteq6}
J.~Pumplin, D.~Stump, J.~Huston, H.~Lai, P.~M. Nadolsky, et al., {\em {New
  generation of parton distributions with uncertainties from global QCD
  analysis}\/},  \href{http://dx.doi.org/10.1088/1126-6708/2002/07/012}{JHEP
  {\bf 07} (2002)  012},
\href{http://arxiv.org/abs/hep-ph/0201195}{{\tt arXiv:hep-ph/0201195
  [hep-ph]}}.

\bibitem{Allison:2006ve}
J.~Allison et al., {\em {Geant4 developments and applications}\/},
\href{http://dx.doi.org/10.1109/TNS.2006.869826}{IEEE Trans. Nucl. Sci. {\bf
  53} (2006)  270}.

\bibitem{Agostinelli:2002hh}
{GEANT4 Collaboration}, S.~Agostinelli et al., {\em {GEANT4: A Simulation
  toolkit}\/},
\href{http://dx.doi.org/10.1016/S0168-9002(03)01368-8}{Nucl. Instrum. Meth.
  {\bf A506} (2003)  250--303}.

\bibitem{LHCb-PROC-2011-006}
M.~Clemencic et al., {\em {The LHCb simulation application, Gauss: Design,
  evolution and experience}\/},
  \href{http://dx.doi.org/10.1088/1742-6596/331/3/032023}{J. Phys. Conf. Ser.
  {\bf 331} (2011)  032023}.

\bibitem{Khalek:2018mdn}
R.~A. Khalek, S.~Bailey, J.~Gao, L.~Harland-Lang, and J.~Rojo, {\em {Towards
  Ultimate Parton Distributions at the High-Luminosity LHC}\/},
  \href{http://dx.doi.org/10.1140/epjc/s10052-018-6448-y}{Eur. Phys. J. {\bf
  C78} (2018) no.~11, 962},
\href{http://arxiv.org/abs/1810.03639}{{\tt arXiv:1810.03639 [hep-ph]}}.

\end{thebibliography}\endgroup
